%% file: main.tex
\newif \ifdraft \draftfalse
\newif \ifreview \reviewfalse

\newif \iffull \fullfalse

\PassOptionsToPackage{linktocpage}{hyperref}
\usepackage[hyperindex,breaklinks]{hyperref}

\ifreview
\documentclass[acmsmall,screen,review,anonymous]{acmart}
\else
\documentclass[acmsmall,screen]{acmart}
\fi

\AtBeginDocument{%
  \providecommand\BibTeX{{%
    \normalfont B\kern-0.5em{\scshape i\kern-0.25em b}\kern-0.8em\TeX}}}

\setcopyright{acmcopyright}
\copyrightyear{2018}
\acmYear{2018}
\acmDOI{XXXXXXX.XXXXXXX}

\acmJournal{PACMPL}
\acmVolume{37}
\acmNumber{ICFP}
\acmArticle{111}
\acmMonth{8}

\usepackage{bussproofs}
\usepackage{multirow}
\usepackage{stackengine}
\usepackage{listings}
\usepackage{xcolor}

\input{macro}

\input{tce}

\renewcommand{\ottdrule}[4][]{{\displaystyle\frac{\begin{array}{c}#2\end{array}}{#3}\ \ottdrulename{#4}}}

\begin{document}

\title{Temporal Verification with Answer-Effect Modification}
\subtitle{ Dependent Temporal Type-and-Effect System with Delimited Continuations}

\author{Taro Sekiyama}
\email{tsekiyama@acm.org}
\orcid{0000-0001-9286-230X}
\affiliation{%
  \institution{National Institute of Informatics}
  \country{Japan}
}

\author{Hiroshi Unno}
\email{uhiro@cs.tsukuba.ac.jp}
\orcid{0000-0002-4225-8195}
\affiliation{%
  \institution{University of Tsukuba \& RIKEN AIP}
  \country{Japan}
}


\input{sections/abstract}




\maketitle

\input{sections/intro}
\input{sections/overview}

\input{sections/lang}
\input{sections/eff}
\input{sections/lr}
\input{sections/relwork}
\input{sections/conclusion}


\bibliographystyle{ACM-Reference-Format}
\bibliography{main}


\end{document}


\maketitle

\tableofcontents

\listoftheorems

\section{Outline}

This is the supplementary material of the paper titled ``Temporal Verification
with Answer-Effect Modification.''

The presentation of the paper has the following changes to reduce the space of the
paper, clarify the presentation, and avoid the confusion.
First, the language presented in the supplementary material additionally
supports predicate polymorphism.  This is not presented in the paper.
Second, the semantics in the paper uses evaluation contexts to reduce subterms,
but the semantics defined in this supplementary material defines reduction of
subterms using inferences rules.
This is just a superficial difference.
Third, the fixpoint logic in the supplementary material supports second-order quantifiers
to formalize the extension of the validity checking to open formulas rigorously.
The formalization only needs the universal quantifiers.
Therefore, we can easily enforce the convention that the second-order
existential quantifiers never appear and the second-order universal quantifiers
only appear at the top-level of formulas.
Fourth, the language in the supplementary material supports a special typing
rule for discarding continuations.  We thought that this typing rule is useful
for typing exception raising, but it is actually unnecessary.  This rule can be
eliminated safely from the language.

Type safety is proven in \reflem{ts-type-sound-terminating} and
soundness for infinite traces is proven in \reflem{lr-sound}.

\input{defn}

\clearpage
\input{proof}

%% file: macro.tex
\usepackage{amsmath}
\usepackage{enumitem}
\usepackage{color}
\usepackage{proof}

\def\case#1:{\item[\textmd{Case {#1}:}]}
\def\otherwise:{\item[\textmd{Otherwise:}]}

\newcommand{\reffig}[1]{Figure~\ref{fig:#1}}
\newcommand{\refsec}[1]{Section~\ref{sec:#1}}

\newcommand{\refthm}[1]{Theorem~\ref{thm:#1}}

\newtheorem{defn}{Definition}

\newtheorem{corollary}{Corollary}

\newtheorem{lemm}{Lemma}

\newtheorem{thm}{Theorem}

\newcommand{\Rule}[2]{\ensuremath{\text{({\sc{{#1}\_{#2}}})}}}
\newcommand{\RuleWoP}[2]{\ensuremath{\text{{\sc{{#1}\_{#2}}}}}}

\newcommand{\WF}[1]{\Rule{WF}{#1}}

\newcommand{\WFTwop}[1]{\RuleWoP{WFT}{#1}}
\newcommand{\Srule}[1]{\Rule{S}{#1}}
\newcommand{\Srulewop}[1]{\RuleWoP{S}{#1}}
\newcommand{\T}[1]{\Rule{T}{#1}}
\newcommand{\Twop}[1]{\RuleWoP{T}{#1}}

\newcommand{\TS}[1]{\ifdraft\textcolor{red}{TS:#1}\fi}

\makeatletter
\let\MakeLowercaseUnsupportedInPdfStrings\@firstofone
\makeatother
\newcommand{\mylang}{\ensuremath{\lambda^{\mathcal{S}_0}_{\mathsf{\MakeLowercase{ev}}}}}

\newcommand{\defeq}{\stackrel{\rm \tiny def}{=}}

\newcommand{\shift}{\textsf{shift}}
\newcommand{\reset}{\textsf{reset}}
\newcommand{\shiftzero}{\textsf{shift0}}
\newcommand{\resetzero}{\textsf{reset0}}
\newcommand{\control}{\textsf{control}}
\newcommand{\prompt}{\textsf{prompt}}
\newcommand{\callcc}{\textsf{call/cc}}
\newcommand{\abort}{\textsf{abort}}

%% file: tce.tex
\newcommand{\ottdrule}[4][]{{\displaystyle\frac{\begin{array}{l}#2\end{array}}{#3}\quad\ottdrulename{#4}}}

\newcommand{\ottpremise}[1]{ #1 \\}
\newenvironment{ottdefnblock}[3][]{ \framebox{\mbox{#2}} \quad #3 \\[0pt]}{}

\newcommand{\ottnt}[1]{\mathit{#1}}
\newcommand{\ottmv}[1]{\mathit{#1}}

\newcommand{\ottsym}[1]{#1}

\newcommand{\ottdrulename}[1]{\textsc{#1}}

\usepackage{bm}

\DeclareSymbolFont{stmry}{U}{stmry}{m}{n}
\DeclareMathSymbol\Lbag\mathopen{stmry}{"48} 
\DeclareMathSymbol\Rbag\mathclose{stmry}{"49} 

\makeatletter
\@ifundefined{bigsqcap}{%
\DeclareMathSymbol\bigsqcap\mathop{stmry}{"64}
}{}%
\makeatother

\newcommand{\tceappendop}{\mathord{\cdot}%
}

\newcommand{\tceseq}[1]{\ensuremath{\overline{#1} }%
}
\newcommand{\LRRel}[2]{\ensuremath{#1 \, [\![ {#2} ]\!]}%
}

\newcommand{\resetcnstr}[1]{\ensuremath{\langle #1 \rangle}%
}

\newcommand{\cpstransty}[1]{\ensuremath{[\![ #1 ]\!]}%
}

\newcommand{\ottdruleWFTXXRefine}[1]{\ottdrule[#1]{%
\ottpremise{\Gamma  \ottsym{,}  \mathit{x} \,  {:}  \, \iota  \vdash  \phi}%
}{
\Gamma  \vdash  \ottsym{\{}  \mathit{x} \,  {:}  \, \iota \,  |  \, \phi  \ottsym{\}}}{%
{\ottdrulename{WFT\_Refine}}{}%
}}

\newcommand{\ottdruleWFTXXFun}[1]{\ottdrule[#1]{%
\ottpremise{\Gamma  \ottsym{,}  \mathit{x} \,  {:}  \, \ottnt{T}  \vdash  \ottnt{C}}%
}{
\Gamma  \vdash  \ottsym{(}  \mathit{x} \,  {:}  \, \ottnt{T}  \ottsym{)}  \rightarrow  \ottnt{C}}{%
{\ottdrulename{WFT\_Fun}}{}%
}}

\newcommand{\ottdruleWFTXXForall}[1]{\ottdrule[#1]{%
\ottpremise{\Gamma  \ottsym{,}  \mathit{X} \,  {:}  \,  \tceseq{ \ottnt{s} }   \vdash  \ottnt{C}}%
}{
\Gamma  \vdash   \forall   \ottsym{(}  \mathit{X} \,  {:}  \,  \tceseq{ \ottnt{s} }   \ottsym{)}  \ottsym{.}  \ottnt{C}}{%
{\ottdrulename{WFT\_Forall}}{}%
}}

\newcommand{\ottdruleWFTXXComp}[1]{\ottdrule[#1]{%
\ottpremise{ \Gamma  \vdash  \ottnt{T}  \quad   \Gamma  \vdash  \Phi  \quad  \Gamma \,  |  \, \ottnt{T}  \vdash  \ottnt{S}  }%
}{
\Gamma  \vdash   \ottnt{T}  \,  \,\&\,  \,  \Phi  \, / \,  \ottnt{S} }{%
{\ottdrulename{WFT\_Comp}}{}%
}}

\newcommand{\ottdruleWFTXXEmpty}[1]{\ottdrule[#1]{%
\ottpremise{\vdash  \Gamma}%
}{
\Gamma \,  |  \, \ottnt{T}  \vdash   \square }{%
{\ottdrulename{WFT\_Empty}}{}%
}}

\newcommand{\ottdruleWFTXXAns}[1]{\ottdrule[#1]{%
\ottpremise{ \Gamma  \ottsym{,}  \mathit{x} \,  {:}  \, \ottnt{T}  \vdash  \ottnt{C_{{\mathrm{1}}}}  \quad  \Gamma  \vdash  \ottnt{C_{{\mathrm{2}}}} }%
}{
\Gamma \,  |  \, \ottnt{T}  \vdash   ( \forall  \mathit{x}  .  \ottnt{C_{{\mathrm{1}}}}  )  \Rightarrow   \ottnt{C_{{\mathrm{2}}}} }{%
{\ottdrulename{WFT\_Ans}}{}%
}}

\newcommand{\ottdruleWFTXXEff}[1]{\ottdrule[#1]{%
\ottpremise{ \Gamma  \ottsym{,}  \mathit{x} \,  {:}  \,  \Sigma ^\ast   \vdash  \phi_{{\mathrm{1}}}  \quad  \Gamma  \ottsym{,}  \mathit{y} \,  {:}  \,  \Sigma ^\omega   \vdash  \phi_{{\mathrm{2}}} }%
}{
\Gamma  \vdash   (   \lambda_\mu\!   \,  \mathit{x}  .\,  \phi_{{\mathrm{1}}}  ,   \lambda_\nu\!   \,  \mathit{y}  .\,  \phi_{{\mathrm{2}}}  ) }{%
{\ottdrulename{WFT\_Eff}}{}%
}}

\newcommand{\ottdruleTXXCVar}[1]{\ottdrule[#1]{%
\ottpremise{ \vdash  \Gamma  \quad  \Gamma  \ottsym{(}  \mathit{x}  \ottsym{)} \,  =  \, \ottsym{\{}  \mathit{y} \,  {:}  \, \iota \,  |  \, \phi  \ottsym{\}} }%
}{
\Gamma  \vdash  \mathit{x}  \ottsym{:}  \ottsym{\{}  \mathit{y} \,  {:}  \, \iota \,  |  \,  \mathit{x}    =    \mathit{y}   \ottsym{\}}}{%
{\ottdrulename{T\_CVar}}{}%
}}

\newcommand{\ottdruleTXXVar}[1]{\ottdrule[#1]{%
\ottpremise{ \vdash  \Gamma  \quad    \forall  \, { \mathit{y}  \ottsym{,}  \iota  \ottsym{,}  \phi } . \  \Gamma  \ottsym{(}  \mathit{x}  \ottsym{)} \,  \not=  \, \ottsym{\{}  \mathit{y} \,  {:}  \, \iota \,  |  \, \phi  \ottsym{\}}  }%
}{
\Gamma  \vdash  \mathit{x}  \ottsym{:}  \Gamma  \ottsym{(}  \mathit{x}  \ottsym{)}}{%
{\ottdrulename{T\_Var}}{}%
}}

\newcommand{\ottdruleTXXConst}[1]{\ottdrule[#1]{%
\ottpremise{ \vdash  \Gamma  \quad   \mathit{ty}  (  \ottnt{c}  )  \,  =  \, \iota }%
}{
\Gamma  \vdash  \ottnt{c}  \ottsym{:}  \ottsym{\{}  \mathit{x} \,  {:}  \, \iota \,  |  \,  \ottnt{c}    =    \mathit{x}   \ottsym{\}}}{%
{\ottdrulename{T\_Const}}{}%
}}

\newcommand{\ottdruleTXXAbs}[1]{\ottdrule[#1]{%
\ottpremise{\Gamma  \ottsym{,}   \tceseq{ \mathit{x}  \,  {:}  \,  \ottnt{T} }   \vdash  \ottnt{e}  \ottsym{:}  \ottnt{C}}%
}{
\Gamma  \vdash   \lambda\!  \,  \tceseq{ \mathit{x} }   \ottsym{.}  \ottnt{e}  \ottsym{:}  \ottsym{(}   \tceseq{ \mathit{x}  \,  {:}  \,  \ottnt{T} }   \ottsym{)}  \rightarrow  \ottnt{C}}{%
{\ottdrulename{T\_Abs}}{}%
}}

\newcommand{\ottdruleTXXFun}[1]{\ottdrule[#1]{%
\ottpremise{  \tceseq{ \mathit{z_{{\mathrm{0}}}}  \,  {:}  \,  \iota_{{\mathrm{0}}} }  \,  =  \,  \gamma  (   \tceseq{ \mathit{z}  \,  {:}  \,  \ottnt{T_{{\mathrm{1}}}} }   )   \quad    \{   \tceseq{ \mathit{X} }   \ottsym{,}   \tceseq{ \mathit{Y} }   \}  \,  \mathbin{\cap}  \,  \mathit{fpv}  (   \tceseq{ \ottnt{T_{{\mathrm{1}}}} }   )  \,  =  \,  \emptyset   \quad    \tceseq{ \ottsym{(}  \mathit{X}  \ottsym{,}  \mathit{Y}  \ottsym{)} }   \ottsym{;}   \tceseq{ \mathit{z_{{\mathrm{0}}}}  \,  {:}  \,  \iota_{{\mathrm{0}}} }   \vdash  \ottnt{C_{{\mathrm{0}}}}  \succ  \ottnt{C}  \quad   \emptyset   \ottsym{;}   \tceseq{ \ottsym{(}  \mathit{X}  \ottsym{,}  \mathit{Y}  \ottsym{)} }  \,  \vdash ^{\circ}  \, \ottnt{C}   }%
\ottpremise{  \tceseq{ \ottsym{(}  \mathit{X}  \ottsym{,}  \mathit{Y}  \ottsym{)} }   \ottsym{;}   \tceseq{ \mathit{z_{{\mathrm{0}}}}  \,  {:}  \,  \iota_{{\mathrm{0}}} }  \,  |  \, \ottnt{C}  \vdash  \sigma  \quad  \Gamma  \ottsym{,}   \tceseq{ \mathit{X} }  \,  {:}  \,  (   \Sigma ^\ast   ,   \tceseq{ \iota_{{\mathrm{0}}} }   )   \ottsym{,}   \tceseq{ \mathit{Y} }  \,  {:}  \,  (   \Sigma ^\omega   ,   \tceseq{ \iota_{{\mathrm{0}}} }   )   \ottsym{,}  \mathit{f} \,  {:}  \, \ottsym{(}   \tceseq{ \mathit{z}  \,  {:}  \,  \ottnt{T_{{\mathrm{1}}}} }   \ottsym{)}  \rightarrow  \ottnt{C_{{\mathrm{0}}}}  \ottsym{,}   \tceseq{ \mathit{z}  \,  {:}  \,  \ottnt{T_{{\mathrm{1}}}} }   \vdash  \ottnt{e}  \ottsym{:}  \ottnt{C} }%
}{
\Gamma  \vdash   \mathsf{rec}  \, (  \tceseq{  \mathit{X} ^{ \sigma  \ottsym{(}  \mathit{X}  \ottsym{)} },  \mathit{Y} ^{ \sigma  \ottsym{(}  \mathit{Y}  \ottsym{)} }  }  ,  \mathit{f} ,   \tceseq{ \mathit{z} }  ) . \,  \ottnt{e}   \ottsym{:}  \ottsym{(}   \tceseq{ \mathit{z}  \,  {:}  \,  \ottnt{T_{{\mathrm{1}}}} }   \ottsym{)}  \rightarrow  \sigma  \ottsym{(}  \ottnt{C_{{\mathrm{0}}}}  \ottsym{)}}{%
{\ottdrulename{T\_Fun}}{}%
}}

\newcommand{\ottdruleTXXPAbs}[1]{\ottdrule[#1]{%
\ottpremise{\Gamma  \ottsym{,}  \mathit{X} \,  {:}  \,  \tceseq{ \ottnt{s} }   \vdash  \ottnt{e}  \ottsym{:}  \ottnt{C}}%
}{
\Gamma  \vdash   \Lambda   \ottsym{(}  \mathit{X} \,  {:}  \,  \tceseq{ \ottnt{s} }   \ottsym{)}  \ottsym{.}  \ottnt{e}  \ottsym{:}   \forall   \ottsym{(}  \mathit{X} \,  {:}  \,  \tceseq{ \ottnt{s} }   \ottsym{)}  \ottsym{.}  \ottnt{C}}{%
{\ottdrulename{T\_PAbs}}{}%
}}

\newcommand{\ottdruleTXXOp}[1]{\ottdrule[#1]{%
\ottpremise{ \vdash  \Gamma  \quad    \tceseq{ \Gamma  \vdash  \ottnt{v}  \ottsym{:}  \ottnt{T} }   \quad  \mathit{ty} \, \ottsym{(}  \ottnt{o}  \ottsym{)} \,  =  \,  \tceseq{(  \mathit{x}  \,  {:}  \,  \ottnt{T}  )}  \rightarrow   \ottnt{T_{{\mathrm{0}}}}   }%
}{
\Gamma  \vdash  \ottnt{o}  \ottsym{(}   \tceseq{ \ottnt{v} }   \ottsym{)}  \ottsym{:}   \ottnt{T_{{\mathrm{0}}}}    [ \tceseq{ \ottnt{v}  /  \mathit{x} } ]  }{%
{\ottdrulename{T\_Op}}{}%
}}

\newcommand{\ottdruleTXXSub}[1]{\ottdrule[#1]{%
\ottpremise{\Gamma  \vdash  \ottnt{C}  \ottsym{<:}  \ottnt{C'}}%
\ottpremise{ \Gamma  \vdash  \ottnt{e}  \ottsym{:}  \ottnt{C}  \quad  \Gamma  \vdash  \ottnt{C'} }%
}{
\Gamma  \vdash  \ottnt{e}  \ottsym{:}  \ottnt{C'}}{%
{\ottdrulename{T\_Sub}}{}%
}}

\newcommand{\ottdruleTXXApp}[1]{\ottdrule[#1]{%
\ottpremise{ \Gamma  \vdash  \ottnt{v_{{\mathrm{1}}}}  \ottsym{:}  \ottsym{(}  \mathit{x} \,  {:}  \, \ottnt{T}  \ottsym{)}  \rightarrow  \ottnt{C}  \quad  \Gamma  \vdash  \ottnt{v_{{\mathrm{2}}}}  \ottsym{:}  \ottnt{T} }%
}{
\Gamma  \vdash  \ottnt{v_{{\mathrm{1}}}} \, \ottnt{v_{{\mathrm{2}}}}  \ottsym{:}   \ottnt{C}    [  \ottnt{v_{{\mathrm{2}}}}  /  \mathit{x}  ]  }{%
{\ottdrulename{T\_App}}{}%
}}

\newcommand{\ottdruleTXXPApp}[1]{\ottdrule[#1]{%
\ottpremise{ \Gamma  \vdash  \ottnt{v}  \ottsym{:}   \forall   \ottsym{(}  \mathit{X} \,  {:}  \,  \tceseq{ \ottnt{s} }   \ottsym{)}  \ottsym{.}  \ottnt{C}  \quad  \Gamma  \vdash  \ottnt{A}  \ottsym{:}   \tceseq{ \ottnt{s} }  }%
}{
\Gamma  \vdash  \ottnt{v} \, \ottnt{A}  \ottsym{:}   \ottnt{C}    [  \ottnt{A}  /  \mathit{X}  ]  }{%
{\ottdrulename{T\_PApp}}{}%
}}

\newcommand{\ottdruleTXXIf}[1]{\ottdrule[#1]{%
\ottpremise{\Gamma  \vdash  \ottnt{v}  \ottsym{:}  \ottsym{\{}  \mathit{x} \,  {:}  \,  \mathsf{bool}  \,  |  \, \phi  \ottsym{\}}}%
\ottpremise{ \Gamma  \ottsym{,}    \ottnt{v}     =     \mathsf{true}    \vdash  \ottnt{e_{{\mathrm{1}}}}  \ottsym{:}  \ottnt{C}  \quad  \Gamma  \ottsym{,}    \ottnt{v}     =     \mathsf{false}    \vdash  \ottnt{e_{{\mathrm{2}}}}  \ottsym{:}  \ottnt{C} }%
}{
\Gamma  \vdash  \mathsf{if} \, \ottnt{v} \, \mathsf{then} \, \ottnt{e_{{\mathrm{1}}}} \, \mathsf{else} \, \ottnt{e_{{\mathrm{2}}}}  \ottsym{:}  \ottnt{C}}{%
{\ottdrulename{T\_If}}{}%
}}

\newcommand{\ottdruleTXXEvent}[1]{\ottdrule[#1]{%
\ottpremise{\vdash  \Gamma}%
}{
\Gamma  \vdash   \mathsf{ev}  [  \mathbf{a}  ]   \ottsym{:}   \ottsym{\{}  \mathit{x} \,  {:}  \,  \mathsf{unit}  \,  |  \,  \top   \ottsym{\}}  \,  \,\&\,  \,   (  \mathbf{a}  , \bot )   \, / \,   \square  }{%
{\ottdrulename{T\_Event}}{}%
}}

\newcommand{\ottdruleTXXLet}[1]{\ottdrule[#1]{%
\ottpremise{ \Gamma  \vdash  \ottnt{e_{{\mathrm{1}}}}  \ottsym{:}   \ottnt{T_{{\mathrm{1}}}}  \,  \,\&\,  \,  \Phi_{{\mathrm{1}}}  \, / \,  \ottnt{S_{{\mathrm{1}}}}   \quad   \Gamma  \ottsym{,}  \mathit{x_{{\mathrm{1}}}} \,  {:}  \, \ottnt{T_{{\mathrm{1}}}}  \vdash  \ottnt{e_{{\mathrm{2}}}}  \ottsym{:}   \ottnt{T_{{\mathrm{2}}}}  \,  \,\&\,  \,  \Phi_{{\mathrm{2}}}  \, / \,  \ottnt{S_{{\mathrm{2}}}}   \quad  \mathit{x_{{\mathrm{1}}}} \,  \not\in  \,  \mathit{fv}  (  \ottnt{T_{{\mathrm{2}}}}  )  \,  \mathbin{\cup}  \,  \mathit{fv}  (  \Phi_{{\mathrm{2}}}  )   }%
}{
\Gamma  \vdash  \mathsf{let} \, \mathit{x_{{\mathrm{1}}}}  \ottsym{=}  \ottnt{e_{{\mathrm{1}}}} \,  \mathsf{in}  \, \ottnt{e_{{\mathrm{2}}}}  \ottsym{:}   \ottnt{T_{{\mathrm{2}}}}  \,  \,\&\,  \,  \Phi_{{\mathrm{1}}} \,  \tceappendop  \, \Phi_{{\mathrm{2}}}  \, / \,  \ottsym{(}  \ottnt{S_{{\mathrm{1}}}}  \gg\!\!=  \ottsym{(}   \lambda\!  \, \mathit{x_{{\mathrm{1}}}}  \ottsym{.}  \ottnt{S_{{\mathrm{2}}}}  \ottsym{)}  \ottsym{)} }{%
{\ottdrulename{T\_Let}}{}%
}}

\newcommand{\ottdruleTXXLetV}[1]{\ottdrule[#1]{%
\ottpremise{ \Gamma  \vdash  \ottnt{v_{{\mathrm{1}}}}  \ottsym{:}  \ottnt{T_{{\mathrm{1}}}}  \quad  \Gamma  \ottsym{,}  \mathit{x_{{\mathrm{1}}}} \,  {:}  \, \ottnt{T_{{\mathrm{1}}}}  \vdash  \ottnt{e_{{\mathrm{2}}}}  \ottsym{:}  \ottnt{C_{{\mathrm{2}}}} }%
}{
\Gamma  \vdash  \mathsf{let} \, \mathit{x_{{\mathrm{1}}}}  \ottsym{=}  \ottnt{v_{{\mathrm{1}}}} \,  \mathsf{in}  \, \ottnt{e_{{\mathrm{2}}}}  \ottsym{:}   \ottnt{C_{{\mathrm{2}}}}    [  \ottnt{v_{{\mathrm{1}}}}  /  \mathit{x_{{\mathrm{1}}}}  ]  }{%
{\ottdrulename{T\_LetV}}{}%
}}

\newcommand{\ottdruleTXXShift}[1]{\ottdrule[#1]{%
\ottpremise{\Gamma  \ottsym{,}  \mathit{x} \,  {:}  \, \ottsym{(}  \mathit{y} \,  {:}  \, \ottnt{T}  \ottsym{)}  \rightarrow  \ottnt{C_{{\mathrm{1}}}}  \vdash  \ottnt{e}  \ottsym{:}  \ottnt{C_{{\mathrm{2}}}}}%
}{
\Gamma  \vdash  \mathcal{S}_0 \, \mathit{x}  \ottsym{.}  \ottnt{e}  \ottsym{:}   \ottnt{T}  \,  \,\&\,  \,   \Phi_\mathsf{val}   \, / \,   ( \forall  \mathit{y}  .  \ottnt{C_{{\mathrm{1}}}}  )  \Rightarrow   \ottnt{C_{{\mathrm{2}}}}  }{%
{\ottdrulename{T\_Shift}}{}%
}}

\newcommand{\ottdruleTXXReset}[1]{\ottdrule[#1]{%
\ottpremise{\mathit{x} \,  \not\in  \,  \mathit{fv}  (  \ottnt{T}  ) }%
\ottpremise{\Gamma  \vdash  \ottnt{e}  \ottsym{:}   \ottnt{T}  \,  \,\&\,  \,  \Phi  \, / \,   ( \forall  \mathit{x}  .   \ottnt{T}  \,  \,\&\,  \,   \Phi_\mathsf{val}   \, / \,   \square    )  \Rightarrow   \ottnt{C}  }%
}{
\Gamma  \vdash   \resetcnstr{ \ottnt{e} }   \ottsym{:}  \ottnt{C}}{%
{\ottdrulename{T\_Reset}}{}%
}}

\newcommand{\ottdruleTXXDiv}[1]{\ottdrule[#1]{%
\ottpremise{ \Gamma  \vdash  \ottnt{C}  \quad  \Gamma  \models   { \ottnt{C} }^\nu (  \pi  )  }%
}{
\Gamma  \vdash   \bullet ^{ \pi }   \ottsym{:}  \ottnt{C}}{%
{\ottdrulename{T\_Div}}{}%
}}

\newcommand{\ottdruleSXXRefine}[1]{\ottdrule[#1]{%
\ottpremise{\Gamma  \ottsym{,}  \mathit{x} \,  {:}  \, \iota  \models   \phi_{{\mathrm{1}}}    \Rightarrow    \phi_{{\mathrm{2}}} }%
}{
\Gamma  \vdash  \ottsym{\{}  \mathit{x} \,  {:}  \, \iota \,  |  \, \phi_{{\mathrm{1}}}  \ottsym{\}}  \ottsym{<:}  \ottsym{\{}  \mathit{x} \,  {:}  \, \iota \,  |  \, \phi_{{\mathrm{2}}}  \ottsym{\}}}{%
{\ottdrulename{S\_Refine}}{}%
}}

\newcommand{\ottdruleSXXFun}[1]{\ottdrule[#1]{%
\ottpremise{ \Gamma  \vdash  \ottnt{T_{{\mathrm{2}}}}  \ottsym{<:}  \ottnt{T_{{\mathrm{1}}}}  \quad  \Gamma  \ottsym{,}  \mathit{x} \,  {:}  \, \ottnt{T_{{\mathrm{2}}}}  \vdash  \ottnt{C_{{\mathrm{1}}}}  \ottsym{<:}  \ottnt{C_{{\mathrm{2}}}} }%
}{
\Gamma  \vdash  \ottsym{(}  \mathit{x} \,  {:}  \, \ottnt{T_{{\mathrm{1}}}}  \ottsym{)}  \rightarrow  \ottnt{C_{{\mathrm{1}}}}  \ottsym{<:}  \ottsym{(}  \mathit{x} \,  {:}  \, \ottnt{T_{{\mathrm{2}}}}  \ottsym{)}  \rightarrow  \ottnt{C_{{\mathrm{2}}}}}{%
{\ottdrulename{S\_Fun}}{}%
}}

\newcommand{\ottdruleSXXForall}[1]{\ottdrule[#1]{%
\ottpremise{\Gamma  \ottsym{,}  \mathit{X} \,  {:}  \,  \tceseq{ \ottnt{s} }   \vdash  \ottnt{C_{{\mathrm{1}}}}  \ottsym{<:}  \ottnt{C_{{\mathrm{2}}}}}%
}{
\Gamma  \vdash   \forall   \ottsym{(}  \mathit{X} \,  {:}  \,  \tceseq{ \ottnt{s} }   \ottsym{)}  \ottsym{.}  \ottnt{C_{{\mathrm{1}}}}  \ottsym{<:}   \forall   \ottsym{(}  \mathit{X} \,  {:}  \,  \tceseq{ \ottnt{s} }   \ottsym{)}  \ottsym{.}  \ottnt{C_{{\mathrm{2}}}}}{%
{\ottdrulename{S\_Forall}}{}%
}}

\newcommand{\ottdruleSXXComp}[1]{\ottdrule[#1]{%
\ottpremise{ \Gamma  \vdash  \ottnt{T_{{\mathrm{1}}}}  \ottsym{<:}  \ottnt{T_{{\mathrm{2}}}}  \quad  \Gamma  \vdash  \Phi_{{\mathrm{1}}}  \ottsym{<:}  \Phi_{{\mathrm{2}}} }%
\ottpremise{\Gamma \,  |  \, \ottnt{T_{{\mathrm{1}}}} \,  \,\&\,  \, \Phi_{{\mathrm{1}}}  \vdash  \ottnt{S_{{\mathrm{1}}}}  \ottsym{<:}  \ottnt{S_{{\mathrm{2}}}}}%
}{
\Gamma  \vdash   \ottnt{T_{{\mathrm{1}}}}  \,  \,\&\,  \,  \Phi_{{\mathrm{1}}}  \, / \,  \ottnt{S_{{\mathrm{1}}}}   \ottsym{<:}   \ottnt{T_{{\mathrm{2}}}}  \,  \,\&\,  \,  \Phi_{{\mathrm{2}}}  \, / \,  \ottnt{S_{{\mathrm{2}}}} }{%
{\ottdrulename{S\_Comp}}{}%
}}

\newcommand{\ottdruleSXXEmpty}[1]{\ottdrule[#1]{%
}{
\Gamma \,  |  \, \ottnt{T} \,  \,\&\,  \, \Phi  \vdash   \square   \ottsym{<:}   \square }{%
{\ottdrulename{S\_Empty}}{}%
}}

\newcommand{\ottdruleSXXAns}[1]{\ottdrule[#1]{%
\ottpremise{ \Gamma  \ottsym{,}  \mathit{x} \,  {:}  \, \ottnt{T}  \vdash  \ottnt{C_{{\mathrm{21}}}}  \ottsym{<:}  \ottnt{C_{{\mathrm{11}}}}  \quad  \Gamma  \vdash  \ottnt{C_{{\mathrm{12}}}}  \ottsym{<:}  \ottnt{C_{{\mathrm{22}}}} }%
}{
\Gamma \,  |  \, \ottnt{T} \,  \,\&\,  \, \Phi  \vdash   ( \forall  \mathit{x}  .  \ottnt{C_{{\mathrm{11}}}}  )  \Rightarrow   \ottnt{C_{{\mathrm{12}}}}   \ottsym{<:}   ( \forall  \mathit{x}  .  \ottnt{C_{{\mathrm{21}}}}  )  \Rightarrow   \ottnt{C_{{\mathrm{22}}}} }{%
{\ottdrulename{S\_Ans}}{}%
}}

\newcommand{\ottdruleSXXEmbed}[1]{\ottdrule[#1]{%
\ottpremise{ \Gamma  \ottsym{,}  \mathit{x} \,  {:}  \, \ottnt{T}  \vdash  \Phi \,  \tceappendop  \, \ottnt{C_{{\mathrm{1}}}}  \ottsym{<:}  \ottnt{C_{{\mathrm{2}}}}  \quad   \Gamma  \ottsym{,}  \mathit{y} \,  {:}  \,  \Sigma ^\omega   \vdash    { \Phi }^\nu   \ottsym{(}  \mathit{y}  \ottsym{)}    \Rightarrow     { \ottnt{C_{{\mathrm{2}}}} }^\nu (  \mathit{y}  )    \quad  \mathit{x}  \ottsym{,}  \mathit{y} \,  \not\in  \,  \mathit{fv}  (  \Phi  )  \,  \mathbin{\cup}  \,  \mathit{fv}  (  \ottnt{C_{{\mathrm{2}}}}  )   }%
}{
\Gamma \,  |  \, \ottnt{T} \,  \,\&\,  \, \Phi  \vdash   \square   \ottsym{<:}   ( \forall  \mathit{x}  .  \ottnt{C_{{\mathrm{1}}}}  )  \Rightarrow   \ottnt{C_{{\mathrm{2}}}} }{%
{\ottdrulename{S\_Embed}}{}%
}}

\newcommand{\ottdruleSXXEff}[1]{\ottdrule[#1]{%
\ottpremise{\mathit{x} \,  \not\in  \,  \mathit{fv}  (  \Phi_{{\mathrm{1}}}  )  \,  \mathbin{\cup}  \,  \mathit{fv}  (  \Phi_{{\mathrm{2}}}  ) }%
\ottpremise{\Gamma  \ottsym{,}  \mathit{x} \,  {:}  \,  \Sigma ^\ast   \models    { \Phi_{{\mathrm{1}}} }^\mu   \ottsym{(}  \mathit{x}  \ottsym{)}    \Rightarrow     { \Phi_{{\mathrm{2}}} }^\mu   \ottsym{(}  \mathit{x}  \ottsym{)} }%
\ottpremise{\Gamma  \ottsym{,}  \mathit{x} \,  {:}  \,  \Sigma ^\omega   \models    { \Phi_{{\mathrm{1}}} }^\nu   \ottsym{(}  \mathit{x}  \ottsym{)}    \Rightarrow     { \Phi_{{\mathrm{2}}} }^\nu   \ottsym{(}  \mathit{x}  \ottsym{)} }%
}{
\Gamma  \vdash  \Phi_{{\mathrm{1}}}  \ottsym{<:}  \Phi_{{\mathrm{2}}}}{%
{\ottdrulename{S\_Eff}}{}%
}}



%% file: sections/abstract.tex
\begin{abstract}
 Type-and-effect systems are a widely-used approach to program verification,
 verifying the result of a computation using types, and the behavior using
 effects.
 This paper extends an effect system for verifying temporal, value-dependent
 properties on event sequences yielded by programs to the delimited control
 operators {\shiftzero}/{\resetzero}.
 While these delimited control operators enable useful and powerful programming
 techniques, they hinder reasoning about the behavior of programs because of
 their ability to suspend, resume, discard, and duplicate delimited
 continuations.
 This problem is more serious in effect systems for temporal properties because
 these systems must be capable of identifying what event sequences are yielded by captured continuations.
 Our key observation for achieving effective reasoning in the presence of the
 delimited control operators is that their use modifies answer effects, which
 are temporal effects of the continuations.
 Based on this observation, we extend an effect system for temporal
 verification to accommodate answer-effect modification.
 Allowing answer-effect modification enables easily reasoning about traces that
 captured continuations yield.
 Another novel feature of our effect system is the support for dependently-typed
 continuations, which allows us to reason about programs more precisely.
 We prove soundness of the effect system for finite event sequences via type
 safety and that for infinite event sequences using a logical relation.
\end{abstract}

%% file: sections/intro.tex
\section{Introduction}

\subsection{Background: Type-and-Effect System for Temporal Verification}
\label{sec:intro:background}

Type-and-effect (or, simply effect) systems are a widely-used approach to
program verification, verifying the result of a computation using types, and the
behavior using effects.
Their usefulness and applicability have been proven in broad areas, such as
memory management~\cite{Tofte/Talpin_1997_IC},
deadlock-freedom~\cite{Padovani/Novara_2015_FORTE}, and safe use of
exceptions~\cite{Marino/Millstein_2009_TLDI} and
continuations~\cite{Danvy/Filinski_1990_LFP}.
A benefit shared among these existing, various effect systems is
\emph{compositionality}: the verification of an expression rests only on the types
and effects of its subexpressions.

\emph{Safety} and \emph{liveness} are major classes of verification properties
addressed by many effect systems developed thus far.
Safety means that nothing ``bad'' happens during the program execution.
For example, the effect systems cited above focus on safety.
A safety property can be ensured by formalizing ``bad'' things as ``stuck''
program states and then by proving that a well-typed program does not get
stuck~\cite{Milner_1978_JCSS}.
Liveness properties state that something good will happen eventually.
For instance, termination is a liveness property that can be ensured by effect
systems~\cite{Boudol_2010_IC}.

Safety and liveness generalize to \emph{temporal properties},\footnote{In this
paper, we mean \emph{linear-time} temporal properties by temporal properties.}
which specify sets of possibly infinite sequences (called \emph{traces}) of
events that programs yield.
For instance, resource usage safety and starvation freedom are formulated as
temporal properties.

Several works have proposed type and effect systems that expand the benefit of
compositionality to the verification of temporal properties.
%
%
\citet{Igarashi/Kobayashi_2002_POPL} addressed the problem of resource usage
analysis, which is a temporal safety property.
Typestate-oriented programming~\cite{Aldrich/Sunshine/Saini/Sparks_2009_OOPSLA}
is an approach to resource usage analysis and has a success in object-oriented
programming.
\citet{Skalka/Smith_2004_APLAS} provided an effect system for reasoning about
finite traces, and \citet{Gordon_2017_ECOOP,Gordon_2021_TOPLAS} defined a generic framework that
can model a variety of effect systems, including that proposed by
Skalka and Smith, for verifying temporal safety properties.
\citet{Kobayashi/Ong_2009_LICS} proposed an expressive type system for general
temporal properties.\footnote{More precisely, Kobayashi and Ong
addressed \emph{branching-time} temporal properties, which subsume linear-time
ones.}  Their system supports only finite data domains, and infinite
data domains such as integers cannot be handled directly.
The effect system of \citet{Hofmann/Chen_2014_CSL-LICS} can verify temporal
properties expressed in an $\omega$-regular language.
\citet{Koskinen/Terauchi_2014_CSL-LICS} proposed the notion of \emph{temporal
effects}, which manage the sets of finite and infinite event traces
separately as effects, and presented a temporal effect system for higher-order
functional programs with recursion.  Their system is restricted by neither data
domains nor the class of temporal properties.
\citet{Nanjo/Unno/Koskinen/Terauchi_2018_LICS} refined and generalized Koskinen and Terauchi's
temporal effect system to \emph{dependent} temporal effects,
which can specify finite and infinite event traces using value-dependent
predicates.
%

While these systems address higher-order recursive programs,
they are not yet expressive enough to verify temporal properties of programs in
practical, real languages.
Especially crucial among a number of missing features are \emph{control
operators}, which are capable of suspending, resuming, discarding, and
duplicating a part of running computation, called a \emph{continuation}.
This ability of control operators enables useful and powerful programming
techniques such as exceptions, generators, backtracking, modeling of
nondeterminism and state, and any other monadic effects~\cite{Filinski_1994_POPL}.
Therefore, support for control operators leads to a general verification
framework that accommodates these features.
On the other hand, the use of the control operators hinders reasoning about
the behavior of programs because reasoning methods must be aware of the
manipulation of continuations.
A few studies have focused on verification of temporal safety properties in the
presence of control effects.
\citet{Iwama/Igarashi/Kobayashi_2006_PEPM} extended
\citet{Igarashi/Kobayashi_2002_POPL}'s type system to exceptions, and
\citet{Gordon_2020_ECOOP} extended the previous work~\cite{Gordon_2017_ECOOP} to
certain tagged delimited control operators.
However, to the best of our knowledge, there has been no work on verification of
general temporal (especially, liveness) properties in the presence of control
operators.

\subsection{This Work}

The aim of this work is to develop an effect system that verifies temporal
properties of higher-order functional programs with control operators.
Specifically, we focus on \emph{delimited} control operators.  As the name
suggests, continuations captured by this kind of operators are delimited and
behave as functions that execute the computations up to the closest delimiters.
Therefore, the captured delimited continuations are composable.  This is in
contrast to the behavior of \emph{undelimited} continuations captured by
undelimited control operators such as
{\callcc}~\cite{Clinger/Friedman/Wand_1985} because undelimited continuations
never return the control to call sites.
The composability is a key factor for delimited control operators to facilitate
functional programming and become mainstream in control operators.

In general, temporal effect systems should be flow sensitive because they need
to track event traces that arise according to the order of computation.
However, it is challenging to build a flow sensitive effect system for control
operators as their use makes control flow much complicated.
For example, consider a program $\ottnt{e}  \ottsym{;}   \mathsf{ev}  [  \mathbf{a}  ] $, which executes an expression
$\ottnt{e}$ and raises an event $\mathbf{a}$ using the operation $ \mathsf{ev}  [  \mathbf{a}  ] $.
The behavior of this program depends on how the expression $\ottnt{e}$ operates its
continuation. If $\ottnt{e}$ captures and discards the continuation, the event
$\mathbf{a}$ will never happen because the operation $ \mathsf{ev}  [  \mathbf{a}  ] $ is involved in the
continuation.  By contrast, if $\ottnt{e}$ invokes the captured continuation twice,
the event $\mathbf{a}$ will also be raised twice.  This indicates that precise
temporal verification requires a methodology to reason about not only how expressions
manipulate continuations, but also what traces the continuations generate.

Our key idea to track the information about continuations precisely is to extend
the notion of answer types to \emph{answer effects}.
Answer types are the types of values returned by contexts up to closest
delimiters~\cite{Danvy/Filinski_1990_LFP}.  Because answer types tell what captured continuations
return, the tracking of answer types is crucial to ensure safe use of delimited
continuations.
Because this work is interested in temporal verification, we introduce a new
notion of answer (temporal) effects, which represent traces yielded by the
delimited contexts.
The use of answer effects makes it possible to reason about the traces that
captured continuations generate.

Correct and precise verification with answer effects, however, poses two
challenges.
First, we need a means to verify that an answer effect assumed by a call site of
the continuation-capture operation specifies correctly what event will happen in
what order in the remaining context.
Second, for precise approximation, we need to address \emph{Answer-Effect
Modification} (AEM), which is a variant of Answer-Type Modification
(ATM)~\cite{Danvy/Filinski_1990_LFP} for temporal effects.
Briefly speaking, AEM, as well as ATM, is caused by the ability of delimited
control operators to manipulate captured continuations in a flexible manner.
Therefore, on one hand, this ability is a source of expressivity of delimited
control operators.  However, on the other hand, it allows modifying answer
effects and hinders reasoning.
We need a means to track how answer effects are modified for precise
verification of programs utilizing delimited control operators.


In this paper, we present a dependent temporal effect system that accommodates
AEM for temporal effects.
Our system is an extension of \citet{Materzok/Biernacki_2011_ICFP}'s effect
system (MB), which allows ATM in the presence of the delimited control operators
{\shiftzero}/{\resetzero}~\cite{Danvy/Filinski_1989_TR,Shan_2004_SFP}, to temporal verification.
Inspired by MB (and \citet{Danvy/Filinski_1990_LFP}, which is the first work that
proposes an effect system accommodating ATM), our effect system enriches typing
judgments with both answer (types and) effects before and after modification,
and propagates how answer effects are modified to outer contexts.
Intuitively, answer effects before modification represent requirements for
contexts, and answer effects after modification represent guarantees for
\emph{meta}-contexts (that is, the contexts of the closest delimited
contexts).
This require-and-guarantee view of AEM is also useful to address the first
challenge mentioned above.
In fact, we represent the assumption on continuations imposed by a call site of the
continuation-capture operator {\shiftzero} as an answer effect before
modification.
Then, our effect system checks that the delimited context of the call site
satisfies the assumption.

In our system, answer effects are temporal effects in the style of
\citet{Nanjo/Unno/Koskinen/Terauchi_2018_LICS} and represent finite and infinite
traces yielded by programs.
%
%
Nanjo et al.'s system can reason about what traces recursive functions yield.
In addition to this ability, our effect system can address subtle interaction
between recursive functions and contexts.
Except for this point, we can adapt MB smoothly to reason
about finite traces.  This is unsurprising because predicting finite traces is a
safety property (that is, predicting what happens when programs terminate) and
MB is designed for type safety.
Indeed, we prove soundness of the effect system with respect to finite traces as
type safety.
By contrast, reasoning about infinite traces involves a subtlety.  In designing
the effect system, we need to take into account the fact that an infinite trace
is observable at the top-level of a program wherever it happens (e.g., even when it happens under
one or more delimiters).

Our effect system is dependent in that it can address value-dependent predicates
over traces (and first-order values) as in
\citet{Nanjo/Unno/Koskinen/Terauchi_2018_LICS}.
In general, one must be careful to address effectful features in dependent type
systems because their integration without restriction leads to
inconsistency~\cite{Herbelin_2005_TLCA}.
Fortunately, it has been shown in literature that \emph{value}-dependent
(or, more generally, pure-computation-dependent) type systems can safely address
various effectful features, including control
operators~\cite{Herbelin_2012_LICS,Swamy-et.al_2016_POPL,Lepigre_2016_ESOP,Miquey_2017_ESOP,Ahman_2018_POPL,Cong/Asai_2018_ICFP}.

A novelty of our system as a dependent type system is in the typing of
captured continuations: it enables the types of captured continuations to depend on
arguments.
This new dependency empowers the effect system to express the traces yielded by a
continuation using the arguments passed to it.
The full use of this ability leads to more precise analysis of programs.

The contributions of this paper are summarized as follows.

\begin{itemize}
 \item We provide a dependent temporal effect system for the control operators
       {\shiftzero}/{\resetzero}.  By accommodating AEM,
       our effect system can effectively track finite and infinite traces of
       higher-order programs that use delimited continuations fully.
       We also demonstrate the usefulness of allowing AEM via examples.

 \item Our system assigns dependent function types to captured continuations.
       This type assignment enables expressing a precise relationship between
       the input and output of the continuations.

 \item We prove type safety of our language via progress and subject
       reduction~\cite{Felleisen/Hieb_1992_TCS}.
       It implies not only that well-typed expressions never get stuck, but
       also that the effect system is sound with respect to reasoning about
       finite traces.

 \item We prove soundness of the effect system with respect to infinite traces
       by defining a logical relation that relates only expressions yielding
       infinite traces specified by temporal effects, and then showing that it
       contains all well-typed expressions.
       {}

 \item We have implemented the proposed effect system as a tool that can
       generate the constraints for temporal effects and can generate and solve
       the constraints for refinement types.

\end{itemize}

\paragraph{Organization of the remainder of this paper.}
The rest of this paper is organized as follows.
\refsec{overview} provides an overview of our work with motivating examples.
\refsec{lang} defines the syntax and semantics of our language.
\refsec{eff} formalizes our effect system, presents typing examples, and states type safety.
\refsec{lr} presents our logical relation%
{\iffull
\else %
{ and }%
\fi}%
shows the soundness property for
infinite traces%
{}.
\refsec{relwork} discusses related work, and \refsec{conclusion} concludes.
\TS{Add the section for prototype implementation.}
This paper only states key properties of the metatheory and omits the
formal definitions of some well-known notions, auxiliary lemmas, and detailed
proofs.
The full definitions, lemmas, and proofs are found in the supplementary
material.
The supplementary material also includes an extension of the calculus to
predicate polymorphism.
%
%
The implementation of the effect system will be available as an artifact.

%% file: sections/overview.tex
\section{Overview}
\label{sec:overview}

This section reviews dependent temporal effects and the delimited control
operators {\shiftzero}/{\resetzero} and then presents challenges involved in
integrating them via a few examples.

\subsection{Temporal Effects}
\label{sec:overview:temporal-eff}
Temporal effects specify finite and infinite event traces yielded by
expressions.
In \citet{Nanjo/Unno/Koskinen/Terauchi_2018_LICS}, a (dependent) temporal effect
$\Phi$ is a pair $ (   \lambda_\mu\!   \,  \mathit{x}  .\,  \phi_{{\mathrm{1}}}  ,   \lambda_\nu\!   \,  \mathit{y}  .\,  \phi_{{\mathrm{2}}}  ) $ of a predicate $\phi_{{\mathrm{1}}}$ on finite traces
denoted by $\mathit{x}$ and predicate $\phi_{{\mathrm{2}}}$ on infinite traces denoted by $\mathit{y}$.
An expression is assigned a type $\ottnt{T} \,  \,\&\,  \, \Phi$ if: its terminating run produces a
value of type $\ottnt{T}$ and generates a finite trace $\varpi$ such that
$ \phi_{{\mathrm{1}}}    [  \varpi  /  \mathit{x}  ]  $ is true; and its diverging run generates an infinite trace
$\pi$ such that $ \phi_{{\mathrm{2}}}    [  \pi  /  \mathit{x}  ]  $ is true.

For example, consider the following function \lstinline{f} in an ML-like
language.
\begin{lstlisting}
 let rec f n = if n = 0 then () else (ev[$\mathbf{a}$]; f (n-1))
\end{lstlisting}
This function uses the construct \lstinline{ev[$\mathbf{a}$]}, which raises the event
$\mathbf{a}$ and then returns the unit value \lstinline{()}.
Given a nonnegative number $n$, the function \lstinline{f} raises the event
$\mathbf{a}$ $n$-times and then terminates.  Therefore, the run of \lstinline{f n}
terminates with the sequence of event $\mathbf{a}$ with length $\ottmv{n}$, $ \mathbf{a} ^{ \ottmv{n} } $ for
short.
Otherwise, if $n$ is negative, the application \lstinline{f n} diverges and
yields the sequence of infinite repetition of event $\mathbf{a}$; we write
$ { \mathbf{a} }^\omega $ for it.
Therefore, the behavior of the function \lstinline{f} can be specified by a dependent function type
$\ottsym{(}   \mathit{n}  \,  {:}  \,  \mathsf{int}   \ottsym{)}  \rightarrow   \mathsf{unit}  \,  \,\&\,  \,  (   \lambda_\mu\!   \,  \mathit{x}  .\,     \mathit{n}     \geq     0      \Rightarrow    \ottsym{(}   \mathit{x}    =     \mathbf{a} ^{  \mathit{n}  }    \ottsym{)}   ,   \lambda_\nu\!   \,  \mathit{y}  .\,     \mathit{n}     <     0      \Rightarrow    \ottsym{(}   \mathit{y}    =     { \mathbf{a} }^\omega    \ottsym{)}   ) $.\footnote{The traces $ \mathbf{a} ^{  \mathit{n}  } $ and $ { \mathbf{a} }^\omega $ can be
expressed using the least and great fixpoint operators in the formalization.  See \refsec{lang:syntax}. \TS{?}}

Temporal effects enable temporal verification.
Consider the following program.
\begin{lstlisting}
 let rec wait ready =
   if ready () then (ev[$ \mathbf{Ready} $]; ()) else (ev[$ \mathbf{Wait} $]; wait ready)

 let rec send x ready receiver =
   wait ready; ev[$ \mathbf{Send} $]; receiver x; send (x+1) ready receiver

 let ready () = if * then true else false in
 let receiver x = print_int x in
 send 0 ready receiver
\end{lstlisting}
The function \lstinline{wait} calls a given function
\lstinline{ready} repeatedly and finishes only when it returns
\lstinline{true}.
Because \lstinline{wait} raises the event $ \mathbf{Ready} $ if \lstinline{ready}
returns \lstinline{true}, and the event $ \mathbf{Wait} $ otherwise, \lstinline{wait}
can be assigned a temporal effect
$ (   \lambda_\mu\!   \,  \mathit{x}  .\,  \mathit{x} \,  \in  \,   \mathbf{Wait}  ^\ast  \,  \tceappendop  \,  \mathbf{Ready}   ,   \lambda_\nu\!   \,  \mathit{y}  .\,   \mathit{y}    =     {  \mathbf{Wait}  }^\omega    ) $,
where the notation $ \varpi ^\ast $ denotes the set of finite repetitions of $\varpi$,
and, for a set $\mathit{L}$ of finite traces and a finite or infinite trace $t$,
$\mathit{L} \,\tceappendop\, t$ denotes the set
$\{ \varpi \,\tceappendop\, t \mid \varpi \,  \in  \, \mathit{L} \}$.
The function \lstinline{send} repeats the process of waiting for a receiver to
be ready, raising the event $ \mathbf{Send} $, and sending a value to the receiver.
The last three lines implement the functions \lstinline{ready} and
\lstinline{receiver}, and then call \lstinline{send} with them.
The implementation of \lstinline{ready} uses the nondeterministic Boolean
choice \lstinline{*}.
The program \lstinline{send 0 ready receiver} diverges with two possibilities.
First, after finitely repeating waiting and sending, a call to the function
\lstinline{wait} diverges.
Second, the waiting and sending actions are repeated infinitely.
Therefore, a temporal effect of the program is
\begin{equation}
  (   \lambda_\mu\!   \,  \mathit{x}  .\,   \bot   ,   \lambda_\nu\!   \,  \mathit{y}  .\,  \mathit{y} \,  \in  \, \ottsym{(}   \ottsym{(}    \mathbf{Wait}  ^\ast  \,  \tceappendop  \,  \mathbf{Ready}  \,  \tceappendop  \,  \mathbf{Send}   \ottsym{)} ^\ast  \,  \tceappendop  \,  {  \mathbf{Wait}  }^\omega   \ottsym{)} \,  \mathbin{\cup}  \,  \ottsym{(}    \mathbf{Wait}  ^\ast  \,  \tceappendop  \,  \mathbf{Ready}  \,  \tceappendop  \,  \mathbf{Send}   \ottsym{)} ^\omega   )  ~,
  \label{eqn:overview:send}
\end{equation}
which indicates that the program diverges because no finite trace satisfies the
false $ \bot $ and that the event $ \mathbf{Send} $ is always raised (i.e., some value
is sent to the receiver) when $ \mathbf{Ready} $ has been raised (i.e., the receiver has
become ready).

\subsection{Delimited Control Operators {\shiftzero}/{\resetzero} and Answer-Type Modification}
\label{sec:overview:delim-control-op}
This section explains the behavior of the delimited control operators
{\shiftzero}/{\resetzero}.
We suppose that {\resetzero} is implemented by constructs of the form
\lstinline{$\langle$e$\rangle$}, which evaluates expression
\lstinline{e} under a delimited context, and that {\shiftzero} is by
\lstinline{$\mathcal{S}_0$k.e}, which binds variable \lstinline{k} to the
delimited continuation up to the closest {\resetzero} construct and then
``shifts'' the control from the {\resetzero} construct to expression
\lstinline{e}.

For example, consider program \lstinline{f 0} with the function \lstinline{f} defined
as follows.
\begin{lstlisting}
 let raise x = $\mathcal{S}_0$k.x
 let div x y = if y = 0 then raise "div_by_0" else x / y
 let f y = $\langle$let z = div 42 y in if z mod 2 = 0 then "even" else "odd"$\rangle$
\end{lstlisting}
The function \lstinline{raise} implements exception raising, and the function
\lstinline{div} raises an exception if divisor \lstinline{y} is zero.
The evaluation of the program \lstinline{f 0} starts with reducing the body of
the {\resetzero} construct:
\[\begin{array}{lll}
 \text{\lstinline{f 0}}
 & \longrightarrow &
 \langle\text{\lstinline{let z = div 42 0 in if z mod 2 = 0 then "even" else "odd"}}\rangle \\
 & \mathrel{  \longrightarrow  ^*} &
 \langle\text{\lstinline{let z =}\ \,}\mathcal{S}_0\text{\lstinline{k."div_by_0" in if z mod 2 = 0 then "even" else "odd"}}\rangle
  \end{array}
\]
The {\shiftzero} expression \lstinline{$\mathcal{S}_0$k."div_by_0"} is at the redex
position.
Then, the {\resetzero} expression is replaced with the body \lstinline{"div_by_0"}.
Therefore, the program \lstinline{f 0} finally evaluates to the value
\lstinline{"div_by_0"}.

As an example using continuations, consider an implementation of the
nondeterministic choice.
\begin{lstlisting}
 let choice () = $\mathcal{S}_0$k.(k true) @ (k false) in
 $\langle$let x = choice () in let y = choice () in [x && y]$\rangle$
\end{lstlisting}
The operator \lstinline{@} concatenates given two lists.
The first call to the function \lstinline{choice} in the second line captures
the delimited continuation
$\ottnt{E} \defeq$ \lstinline{$\langle$let x = $\blacklozenge$ in let y = choice () in [x && y]$\rangle$}
(where $\blacklozenge$ denotes the hole) and binds the variable \lstinline{k} in the
body \lstinline{(k true)  @ (k false)} of the {\shiftzero} expression to the
continuation.
Then, \lstinline{k true} represents the expression
\lstinline{$\langle$let x = true in let y = choice () in [x && y]$\rangle$}
obtained by filling the hole in $\ottnt{E}$ with the argument \lstinline{true}.
The second call to \lstinline{choice} in \lstinline{k true} captures the continuation
\lstinline{$\langle$let y = $\blacklozenge$ in [true && y]$\rangle$} and concatenates
the results of applying the continuation to \lstinline{true} and
\lstinline{false}.
Therefore, \lstinline{k true} evaluates to
\lstinline{[true && true; true && false]}, i.e., \lstinline{[true; false]}.
Similarly, \lstinline{k false} evaluates to
\lstinline{[false && true; false && false]}, i.e., \lstinline{[false; false]},
because \lstinline{x} is replaced by \lstinline{false}.
Because the entire {\resetzero} expression evaluates to
\lstinline{(k true) @ (k false)} with the substitution of $\ottnt{E}$ for
\lstinline{k}, its result is \lstinline{[true; false] @ [false; false]}.
%
This is the list of all the possible outcomes of the expression \lstinline{x && y}
for any $\text{\lstinline{x}}, \text{\lstinline{y}} \in \{
\text{\lstinline{true}}, \text{\lstinline{false}} \}$.

In the examples we have shown thus far, the type of a {\reset0} construct
matches the type of its body.
However, there exist programs that do not conform to this convention, as follows:
\begin{lstlisting}
 let get_int () = $\mathcal{S}_0$k.fun (x:int) -> k (string_of_int x) in
 $\langle$"Input number is " ^ (get_int ())$\rangle$ 42
\end{lstlisting}
where the function \lstinline{string_of_int} converts integers to strings, and
the operator \lstinline{^} concatenates given strings.
The program invokes {\shiftzero} via \lstinline{get_int}.  Therefore, the entire
{\resetzero} expression evaluates to the body of {\shiftzero}, which is a
function that invokes the captured continuation \lstinline{$\langle$"Input number is " ^ $\blacklozenge$$\rangle$} with the string representation
of a given integer \lstinline{x}.
Because the {\resetzero} expression is applied to number \lstinline{42},
it evaluates to \lstinline{"Input number is 42"}.
The notable point of this example is that the body of the
{\resetzero} expression is of the type \lstinline{string}, but it returns a
function.
This type mismatch between the return values of {\resetzero} constructs and their bodies is known as
\emph{Answer-Type Modification} (ATM).\footnote{ATM was first discussed for
{\shift}/{\reset}~\cite{Danvy/Filinski_1990_LFP}.  Because, unlike {\shiftzero},
{\shift} delimit its body when invoked, ATM for {\shift}/{\reset} means that
the types of {\reset} bodies are modified at run time.
}

\subsection{Temporal Effects Extended to Delimited Control Operators}
\label{sec:overview:this-work}
\subsubsection{Answer Effects and Answer-Effect Modification}
\label{sec:overview:this-work:answer}
The use of the delimited control operators makes reasoning about traces more
complicated.
For example, consider a program $\langle$\lstinline{send 0 ready f}$\rangle$
with some function \lstinline{f}, and \lstinline{send} and \lstinline{ready} defined in \refsec{overview:temporal-eff}.
As aforementioned, the program should diverge if the function \lstinline{f} does not use {\shiftzero}.
However, otherwise, the program may terminate.
For example, the program $\langle$\lstinline{send 0 ready raise}$\rangle$ with
the function \lstinline{raise} in \refsec{overview:delim-control-op}
diverges with the infinite trace $ {  \mathbf{Wait}  }^\omega $ \emph{or} terminates with a
finite trace in $  \mathbf{Wait}  ^\ast  \,  \tceappendop  \,  \mathbf{Ready}  \,  \tceappendop  \,  \mathbf{Send} $ because: once \lstinline{ready}
returns \lstinline{true}, the program calls \lstinline{raise} after raising the
events $ \mathbf{Ready} $ and $ \mathbf{Send} $; the call to \lstinline{raise} discards the
delimited continuation; as a result, the run of the
{\resetzero} expression terminates.

To reason about the temporal behavior of such a program, we introduce the
notion of \emph{answer effects}, which are temporal effects of the delimited
continuations.
For example, consider a simple program
$\langle$\lstinline{$\mathcal{S}_0$k.$\ottnt{e}$; ev[$\mathbf{a}$]}$\rangle$.
For the {\shiftzero} construct in the program, the answer effect is
$ (   \lambda_\mu\!   \,  \mathit{x}  .\,   \mathit{x}    =    \mathbf{a}   ,   \lambda_\nu\!   \,  \mathit{y}  .\,   \bot   ) $ because its delimited continuation
\lstinline{$\langle$$\blacklozenge$; ev[$\mathbf{a}$]$\rangle$}
termiantes after raising the event $\mathbf{a}$.
The answer effect is used as the latent temporal effect of the continuation
variable $\ottmv{k}$ (i.e., the tempral effect caused by calling $\ottmv{k}$).
Hence, if the body $\ottnt{e}$ only calls $\ottmv{k}$ $n$-times, the answer effect tells that the
program yields trace $ \mathbf{a} ^{ \ottmv{n} } $.

It is noteworhy that the temporal effect of the entire {\resetzero} expression
is modified from the temporal effect of its body.  The latter looks as if only
$\mathbf{a}$ is raised once, while the former may raise $\mathbf{a}$ zero or more times
and even may generate any other finite or infinite traces (depending on
$\ottnt{e}$).
Inspired by ATM, we call this phenomenon \emph{Answer-Effect Modification}
(AEM).

Our effect system introduced in \refsec{eff} accommodates AEM.
This ability enables reasoning about our motivating example
$\langle$\lstinline{send 0 ready raise}$\rangle$.
The body (\lstinline{send 0 ready raise}) looks as if it yields a trace
conforming to the temporal effect (\ref{eqn:overview:send}).
However, the function \texttt{raise} discards the delimited continuation
and modifies the temporal effect of the entire {\resetzero} to
$ (   \lambda_\mu\!   \,  \mathit{x}  .\,  \mathit{x} \,  \in  \,   \mathbf{Wait}  ^\ast  \,  \tceappendop  \,  \mathbf{Ready}  \,  \tceappendop  \,  \mathbf{Send}   ,   \lambda_\nu\!   \,  \mathit{y}  .\,   \mathit{y}    =     {  \mathbf{Wait}  }^\omega    ) $,
which exactly specifies the traces that may be yielded before calling
\lstinline{raise}.
We will detail this verification process in \refsec{eff:example} after
defining the effect system.

\TS{Is the following paragraph necessary?}
\paragraph{Remark.}
\TS{Perhaps there is a better wording than ``ubiquitous.''}
AEM is ubiquitous in temporal verification for delimited control operators.
It is caused by two features of them.
One is the capability of capturing continuations.
It leads to the necessity of knowing what happens in invoking captured
continuations.
The other is the ability to replace the continuations with other computation.
AEM arises by using together these features indispensable to delimited control
operators.
Because temporal verification aims to reason about the intermediate states of
program execution, addressing AEM seems unavoidable for precise verification.
Note that it is not mandatory to support ATM for our aim.  Indeed, ATM does not
happen in all the above examples except for the one presented to explain ATM.
In \refsec{eff:example}, we will introduce notation for verifying programs
without ATM, which reduces complexity and improves readability.

\subsubsection{Dependently Typed Continuations.}
Our effect system also allows the types of captured continuations to
depend on arguments.
To see its usefulness, consider the
program $\langle$\lstinline{[wait choice]}$\rangle$ with the functions
\lstinline{wait} and \lstinline{choice} defined in
Section~\ref{sec:overview:temporal-eff} and \ref{sec:overview:delim-control-op},
respectively.
This program calls \lstinline{choice} to ask if a receiver is
ready.
The function \lstinline{choice} captures the delimited continuation, binds it to
variable \lstinline{k}, and passes the value \lstinline{true} and then
\lstinline{false} to it.
The invocation of the continuation \lstinline{k} with \lstinline{true}
immediately finishes after raising the event $ \mathbf{Ready} $.
The invocation with \lstinline{false} calls \lstinline{wait} recursively after
raising $ \mathbf{Wait} $, and then the same process will be repeated.
Thus, the program diverges with the infinite trace $ { \ottsym{(}   \mathbf{Ready}  \,  \tceappendop  \,  \mathbf{Wait}   \ottsym{)} }^\omega $.
%
%
To verify this behavior, the effect system needs to ensure that the application
\lstinline{k true} terminates with finite trace \lstinline{$ \mathbf{Ready} $} and
\lstinline{k false} diverges with infinite trace \lstinline{$ {  \mathbf{Wait}  \,  \tceappendop  \, \ottsym{(}   \mathbf{Ready}  \,  \tceappendop  \,  \mathbf{Wait}   \ottsym{)} }^\omega $}.
The ability of our effect system to allow captured continuations to be dependently typed
enables this reasoning by specifying the behavior of \lstinline{k} depending on
which value, \lstinline{true} or \lstinline{false}, is passed.

{}

%% file: sections/lang.tex
\section{{\mylang}: An event-raising $\lambda$-calculus with {\shiftzero}/{\reset0}}
\label{sec:lang}

This section presents the syntax and semantics of our language {\mylang}, which
is a call-by-value $\lambda$-calculus equipped with the control operators
{\shiftzero}/{\resetzero}, the event-raising operator,
{\iffull
\else
and recursion.
\fi}
In \citet{Nanjo/Unno/Koskinen/Terauchi_2018_LICS}, all computations are
sequentialized by the $ \mathsf{let} $ construct, and their language has a big-step semantics.
We follow the former convention because it easily enforces predicates to be
value-dependent.
However, the semantics of our language is small-step, which enables us to prove
type safety via progress and subject reduction.

\subsection{Syntax}
\label{sec:lang:syntax}
\begin{figure}[t]
 \[
 \begin{array}{rc@{\ \ }l@{\ \ }l}
  \multicolumn{4}{l}{
   \textbf{Variables} \quad \mathit{x}, \mathit{y}, \mathit{z}, \mathit{f} \qquad
   \textbf{Predicate variables} \quad \mathit{X}, \mathit{Y} \qquad
   \textbf{Term functions} \quad \mathit{F}
  } \\
  \multicolumn{4}{l}{
   \textbf{Events} \quad \mathbf{a} \,  \in  \, \Sigma \qquad
   \textbf{Finite traces} \quad \varpi \,  \in  \,  \Sigma ^\ast  \qquad
   \textbf{Infinite traces} \quad \pi \,  \in  \,  \Sigma ^\omega  \qquad
  } \\[1ex]
  \multicolumn{4}{l}{
  \textbf{Base types} \ \  %
   \iota \ ::= \  \mathsf{unit}  \mid  \mathsf{bool}  \mid  \mathsf{int}  \mid \cdots
   \qquad
  \textbf{Sorts} \ \  %
   \ottnt{s} \ ::= \ \iota \mid  \Sigma ^\ast  \mid  \Sigma ^\omega 
  \qquad
  \textbf{Terms} \ \  %
   \ottnt{t} \ ::= \ \mathit{x} \mid \mathit{F}  \ottsym{(}   \tceseq{ \ottnt{t} }   \ottsym{)}
 }
 \\
 \multicolumn{4}{l}{
  \textbf{Primitive predicates} \ \ %
   \ottnt{P} \ ::= \ { (=) } \mid \cdots
  \qquad\ \ \,%
  \textbf{Constants} \ \ %
   \ottnt{c} \ ::= \  ()  \mid  \mathsf{true}  \mid  \mathsf{false}  \mid \ottsym{0} \mid \cdots
 }
  \\[1ex]
  \textbf{Predicates} &
   \ottnt{A} & ::= & \mathit{X} \mid \ottsym{(}   \mu\!  \, \mathit{X}  \ottsym{(}   \tceseq{ \mathit{x}  \,  {:}  \,  \ottnt{s} }   \ottsym{)}  \ottsym{.}  \phi  \ottsym{)} \mid \ottsym{(}   \nu\!  \, \mathit{X}  \ottsym{(}   \tceseq{ \mathit{x}  \,  {:}  \,  \ottnt{s} }   \ottsym{)}  \ottsym{.}  \phi  \ottsym{)} \mid \ottnt{P}
  \\
  \textbf{Formulas} &
   \phi & ::= &  \top  \mid  \bot  \mid \ottnt{A}  \ottsym{(}   \tceseq{ \ottnt{t} }   \ottsym{)} \mid
                  \neg  \phi  \mid  \phi_{{\mathrm{1}}}    \vee    \phi_{{\mathrm{2}}}  \mid  \phi_{{\mathrm{1}}}    \wedge    \phi_{{\mathrm{2}}}  \mid  \phi_{{\mathrm{1}}}    \Rightarrow    \phi_{{\mathrm{2}}}  \mid
                   \forall  \,  \mathit{x}  \mathrel{  \in  }  \ottnt{s}  . \  \phi  \mid   \exists  \,  \mathit{x}  \mathrel{  \in  }  \ottnt{s}  . \  \phi 
  \\[1ex]
  \textbf{Values} & \ottnt{v} & ::= &
   \mathit{x} \mid \ottnt{c} \mid
   \ottsym{(}   \lambda\!  \,  \tceseq{ \mathit{x} }   \ottsym{.}  \ottnt{e}  \ottsym{)} \,  \tceseq{ \ottnt{v} }  \mid \ottsym{(}   \mathsf{rec}  \, (  \tceseq{  \mathit{X} ^{  \ottnt{A} _{  \mu  }  },  \mathit{Y} ^{  \ottnt{A} _{  \nu  }  }  }  ,  \mathit{f} ,   \tceseq{ \mathit{x} }  ) . \,  \ottnt{e}   \ottsym{)} \,  \tceseq{ \ottnt{v} } 
   \quad \text{(where $\ottsym{\mbox{$\mid$}}   \tceseq{ \ottnt{v} }   \ottsym{\mbox{$\mid$}} \,  <  \, \ottsym{\mbox{$\mid$}}   \tceseq{ \mathit{x} }   \ottsym{\mbox{$\mid$}}$)}
  \\
  \textbf{Expressions} & \ottnt{e} & ::= &
   \ottnt{v} \mid \ottnt{o}  \ottsym{(}   \tceseq{ \ottnt{v} }   \ottsym{)} \mid \ottnt{v_{{\mathrm{1}}}} \, \ottnt{v_{{\mathrm{2}}}} \mid
   \mathsf{if} \, \ottnt{v} \, \mathsf{then} \, \ottnt{e_{{\mathrm{1}}}} \, \mathsf{else} \, \ottnt{e_{{\mathrm{2}}}} \mid
   \mathsf{let} \, \mathit{x}  \ottsym{=}  \ottnt{e_{{\mathrm{1}}}} \,  \mathsf{in}  \, \ottnt{e_{{\mathrm{2}}}} \mid \\ &&&
    \mathsf{ev}  [  \mathbf{a}  ]  \mid
   \mathcal{S}_0 \, \mathit{x}  \ottsym{.}  \ottnt{e} \mid  \resetcnstr{ \ottnt{e} } 
   \mid  \bullet ^{ \pi } 
 \end{array}
 \]
 \caption{Syntax.}
 \label{fig:syntax}
\end{figure}

The syntax of {\mylang} is shown in \reffig{syntax}.
We use the metavariables $\mathit{x}, \mathit{y}, \mathit{z}, \mathit{f}, \ottmv{k}$ for variables
ranging over values and terms, $\mathit{X}, \mathit{Y}$ for predicate variables, $\mathit{F}$
for functions of terms, $\mathbf{a}$ for events, $\varpi$ for finite traces, and
$\pi$ for infinite traces.
Let $ \Sigma $ be the finite set of all events.
Then, $ \Sigma ^\ast $ and $ \Sigma ^\omega $ denote the sets of all the finite and infinite
traces, respectively, of events in $ \Sigma $.
We write $ \epsilon $ for the empty trace, and $\varpi \,  \tceappendop  \, \varpi'$ and
$\varpi \,  \tceappendop  \, \pi'$ for the concatenation of $\varpi$ and $\varpi'$ and that of
$\varpi$ and $\pi'$, respectively.
Throughout this paper, we employ the overline notation for representing possibly
empty finite sequences, and write $|-|$ for the length of a finite sequence.
For example, $ \tceseq{ \mathit{x} } $ is a finite sequence of variables, and $\ottsym{\mbox{$\mid$}}   \tceseq{ \mathit{x} }   \ottsym{\mbox{$\mid$}}$ is its
length.

We first introduce a logic to represent temporal effects.
It is a 
fixpoint logic over finite and infinite traces as
well as first-order values such as integers.
Terms are either variables or applications of term functions.
We suppose that constants, finite and infinite traces, and their operations can
be expressed as terms.
Sorts, ranged over by $\ottnt{s}$, represent the types of terms and include base
types $\iota$.
Predicates, ranged over by $\ottnt{A}$, consist of predicate variables, the least
fixpoint operator $\ottsym{(}   \mu\!  \, \mathit{X}  \ottsym{(}   \tceseq{ \mathit{x}  \,  {:}  \,  \ottnt{s} }   \ottsym{)}  \ottsym{.}  \phi  \ottsym{)}$, the greatest fixpoint operator
$\ottsym{(}   \nu\!  \, \mathit{X}  \ottsym{(}   \tceseq{ \mathit{x}  \,  {:}  \,  \ottnt{s} }   \ottsym{)}  \ottsym{.}  \phi  \ottsym{)}$, and primitive predicates $\ottnt{P}$ such as the equality of
constants and traces.
We assume that $\mathit{X}$ in $\ottsym{(}   \mu\!  \, \mathit{X}  \ottsym{(}   \tceseq{ \mathit{x}  \,  {:}  \,  \ottnt{s} }   \ottsym{)}  \ottsym{.}  \phi  \ottsym{)}$ and $\ottsym{(}   \nu\!  \, \mathit{X}  \ottsym{(}   \tceseq{ \mathit{x}  \,  {:}  \,  \ottnt{s} }   \ottsym{)}  \ottsym{.}  \phi  \ottsym{)}$ occurs
only positively in $\phi$.
Formulas, ranged over by $\phi$, are standard.
%
%
For example, the formula $\mathit{x} \,  \in  \,   \mathbf{Wait}  ^\ast  \,  \tceappendop  \,  \mathbf{Ready} $ presented in \refsec{overview:temporal-eff}
is expressed as
$
 \ottsym{(}   \mu\!  \, \mathit{X}  \ottsym{(}  \mathit{x_{{\mathrm{0}}}} \,  {:}  \,  \Sigma ^\ast   \ottsym{)}  \ottsym{.}    \mathit{x_{{\mathrm{0}}}}    =     \mathbf{Ready}      \vee    \ottsym{(}    \exists  \,  \mathit{x'_{{\mathrm{0}}}}  \mathrel{  \in  }   \Sigma ^\ast   . \    \mathit{x_{{\mathrm{0}}}}    =     \mathbf{Wait}  \,  \tceappendop  \, \mathit{x'_{{\mathrm{0}}}}     \wedge    \mathit{X}  \ottsym{(}  \mathit{x'_{{\mathrm{0}}}}  \ottsym{)}    \ottsym{)}   \ottsym{)}  \ottsym{(}  \mathit{x}  \ottsym{)}
$
and the formula $ \mathit{y}    =     {  \mathbf{Wait}  }^\omega  $ is expressed as
$
 \ottsym{(}   \nu\!  \, \mathit{Y}  \ottsym{(}  \mathit{y_{{\mathrm{0}}}} \,  {:}  \,  \Sigma ^\omega   \ottsym{)}  \ottsym{.}    \exists  \,  \mathit{y'_{{\mathrm{0}}}}  \mathrel{  \in  }   \Sigma ^\omega   . \    \mathit{y_{{\mathrm{0}}}}    =     \mathbf{Wait}  \,  \tceappendop  \, \mathit{y'_{{\mathrm{0}}}}     \wedge    \mathit{Y}  \ottsym{(}  \mathit{y'_{{\mathrm{0}}}}  \ottsym{)}    \ottsym{)}  \ottsym{(}  \mathit{y}  \ottsym{)}
$.

Programs are represented by expressions, ranged over by $\ottnt{e}$, and values,
ranged over by $\ottnt{v}$.
Expressions consist of:
values;
primitive operations $\ottnt{o}  \ottsym{(}   \tceseq{ \ottnt{v} }   \ottsym{)}$ with arguments $ \tceseq{ \ottnt{v} } $;
function applications $\ottnt{v_{{\mathrm{1}}}} \, \ottnt{v_{{\mathrm{2}}}}$;
$ \mathsf{if} $-expressions $\mathsf{if} \, \ottnt{v} \, \mathsf{then} \, \ottnt{e_{{\mathrm{1}}}} \, \mathsf{else} \, \ottnt{e_{{\mathrm{2}}}}$;
$ \mathsf{let} $-expressions $\mathsf{let} \, \mathit{x}  \ottsym{=}  \ottnt{e_{{\mathrm{1}}}} \,  \mathsf{in}  \, \ottnt{e_{{\mathrm{2}}}}$, which bind $\mathit{x}$ in $\ottnt{e_{{\mathrm{2}}}}$;
event-raising expressions $ \mathsf{ev}  [  \mathbf{a}  ] $, which yield the event $\mathbf{a}$;
{\shiftzero} expressions $\mathcal{S}_0 \, \mathit{x}  \ottsym{.}  \ottnt{e}$, which bind $\mathit{x}$ in $\ottnt{e}$ to
continuations captured at run time;
{\resetzero} expressions $ \resetcnstr{ \ottnt{e} } $, which delimit the context of expression
$\ottnt{e}$; and
divergence $ \bullet ^{ \pi } $ at infinite trace $\pi$, which is introduced as a
technical device to prove soundness for infinite traces (see
\refsec{lr} for detail).
Values include
variables,
constants (ranged over by $\ottnt{c}$),
and
possibly partially applied $\lambda$-abstractions and recursive functions.
%
%
%
We suppose that constants include Boolean values $ \mathsf{true} $ and $ \mathsf{false} $.
A $\lambda$-abstraction $ \lambda\!  \,  \tceseq{ \mathit{x} }   \ottsym{.}  \ottnt{e}$ may take multiple arguments, binding
argument variables $ \tceseq{ \mathit{x} } $ in $\ottnt{e}$.
We write $\ottnt{v_{{\mathrm{1}}}} \,  \tceseq{ \ottnt{v_{{\mathrm{2}}}} } $ for the function application $ \ottnt{v_{{\mathrm{1}}}} \, \ottnt{v_{{\mathrm{21}}}}  \cdots  \ottnt{v} _{  \ottsym{2}   \ottmv{n}  } $
when $ \tceseq{ \ottnt{v_{{\mathrm{2}}}} }  = \ottnt{v_{{\mathrm{21}}}}, \cdots,  \ottnt{v} _{  \ottsym{2}   \ottmv{n}  } $.
Then, a function application $\ottsym{(}   \lambda\!  \,  \tceseq{ \mathit{x} }   \ottsym{.}  \ottnt{e}  \ottsym{)} \,  \tceseq{ \ottnt{v} } $ is a value if and only if it is
partial, that is, $\ottsym{\mbox{$\mid$}}   \tceseq{ \ottnt{v} }   \ottsym{\mbox{$\mid$}} \,  <  \, \ottsym{\mbox{$\mid$}}   \tceseq{ \mathit{x} }   \ottsym{\mbox{$\mid$}}$; the same convention is also applied to recursive functions.
For a recursive function $ \mathsf{rec}  \, (  \tceseq{  \mathit{X} ^{  \ottnt{A} _{  \mu  }  },  \mathit{Y} ^{  \ottnt{A} _{  \nu  }  }  }  ,  \mathit{f} ,   \tceseq{ \mathit{x} }  ) . \,  \ottnt{e} $, variables
$ \tceseq{ \mathit{x} } $ and $\mathit{f}$ denote arguments and the recursive function itself,
respectively.
The predicates $ \tceseq{  \ottnt{A} _{  \mu  }  } $ and $ \tceseq{  \ottnt{A} _{  \nu  }  } $ represent the finite and infinite
traces yielded by the body $\ottnt{e}$, respectively.
If $ \tceseq{  \ottnt{A} _{  \mu  }  }  = \ottnt{A_{{\mathrm{0}}}}, \cdots, \ottnt{A_{\ottmv{n}}}$, the predicate $\ottnt{A_{\ottmv{i}}}$ expresses the
finite traces observable by the $i$-th closest meta-context of a call site of the
function (traces in $\ottnt{A_{{\mathrm{0}}}}$ are observed by the caller).
Similarly, each predicate in $ \tceseq{  \ottnt{A} _{  \nu  }  } $ stands for infinite traces observable for a certain
(meta-)context.
For example, the function \lstinline{wait} in \refsec{overview:temporal-eff} is
given the predicates corresponding to $\mathit{x} \,  \in  \,   \mathbf{Wait}  ^\ast  \,  \tceappendop  \,  \mathbf{Ready} $ and $ \mathit{y}    =     {  \mathbf{Wait}  }^\omega  $
as the ones for the caller.
If \lstinline{wait} interacts with meta-contexts via argument functions,
$ \tceseq{  \ottnt{A} _{  \mu  }  } $ and $ \tceseq{  \ottnt{A} _{  \nu  }  } $ can contain more predicates.
We will present such examples in \refsec{eff}.
We use predicate variables $ \tceseq{ \mathit{X} } $ and $ \tceseq{ \mathit{Y} } $, together with the least and greatest fixpoint operators, to identify $ \tceseq{  \ottnt{A} _{  \mu  }  } $ and
$ \tceseq{  \ottnt{A} _{  \nu  }  } $, respectively.
These predicate variables are replaced by the corresponding predicates at run time.
Although our calculus is equipped with those kinds of the information about predicates (i.e.,
$ \tceseq{ \mathit{X} } $, $ \tceseq{ \mathit{Y} } $, $ \tceseq{  \ottnt{A} _{  \mu  }  } $, and $ \tceseq{  \ottnt{A} _{  \nu  }  } $) as annotations for the
metatheory, it can be inferred automatically by constraint generation
implemented in our tool.
\TS{Really?}
%
%
%

This work distinguishes between $\lambda$-abstractions and
recursive functions because it is convenient to prove soundness of the effect
system for infinite traces.
As shown in \refsec{lr}, our proof for the soundness property uses
$\lambda$-abstractions to approximate a recursive function.
In the proof, we need to distinguish between other recursive functions and the
approximations.
Introducing $\lambda$-abstractions as a different constructor enables it easily.

We define the notions of free variables, free predicate variables, and
substitution of values, terms, and predicates as usual.
We suppose that the metafunctions $ \mathit{fv} $ and $ \mathit{fpv} $ return the free
variables and free predicate variables, respectively, in a given syntactic
entity (expressions, formulas, terms, etc.\ as well as types and effects
introduced in \refsec{eff}).
The entity is closed if $ \mathit{fv} $ and $ \mathit{fpv} $ both return the empty set.
Otherwise, it is open.
We write $ \ottnt{e}    [ \tceseq{ \ottnt{v}  /  \mathit{x} } ]  $ (resp.\ $ \ottnt{e}    [  \ottnt{A}  /  \mathit{X}  ]  $ and $ \ottnt{e}    [  \ottnt{t}  /  \mathit{x}  ]  $) for the
expression obtained by substituting values $ \tceseq{ \ottnt{v} } $ (resp.\ predicate $\ottnt{A}$
and term $\ottnt{t}$) for the corresponding variables $ \tceseq{ \mathit{x} } $ (resp.\ $\mathit{X}$ and
$\mathit{x}$) in $\ottnt{e}$ in a capture-avoiding manner.
%
%
We use the similar notation for other syntax categories such as formulas,
types, and effects.

\subsection{Semantics}

\begin{figure}[t]
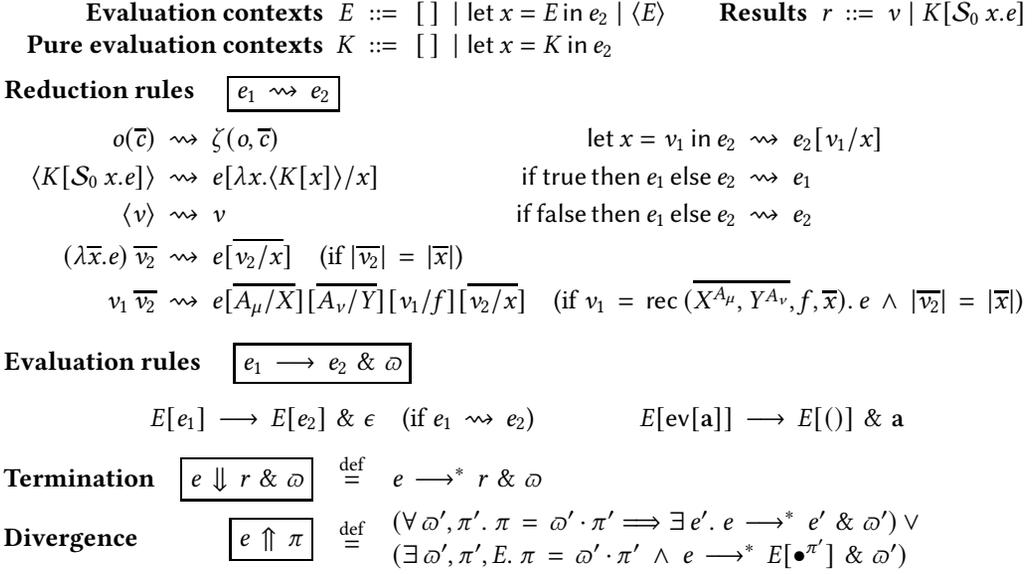

 \[
 \begin{array}{r@{\ \ }c@{\ \ }c@{\ \ }l@{\qquad}r@{\ \ }c@{\ \ }c@{\ \ }l}
  \textbf{Evaluation contexts} & \ottnt{E} & ::= &
    \,[\,]\,  \mid \mathsf{let} \, \mathit{x}  \ottsym{=}  \ottnt{E} \,  \mathsf{in}  \, \ottnt{e_{{\mathrm{2}}}} \mid  \resetcnstr{ \ottnt{E} } 
  &
  \textbf{Results} & \ottnt{r} & ::= & \ottnt{v} \mid  \ottnt{K}  [  \mathcal{S}_0 \, \mathit{x}  \ottsym{.}  \ottnt{e}  ] 
  \\
  \textbf{Pure evaluation contexts} & \ottnt{K} & ::= &
    \,[\,]\,  \mid \mathsf{let} \, \mathit{x}  \ottsym{=}  \ottnt{K} \,  \mathsf{in}  \, \ottnt{e_{{\mathrm{2}}}}
 \end{array}
 \]
 \begin{flushleft}
  \textbf{Reduction rules} \quad \framebox{$\ottnt{e_{{\mathrm{1}}}} \,  \rightsquigarrow  \, \ottnt{e_{{\mathrm{2}}}}$}
 \end{flushleft}
 \mbox{} \\[-5ex]
 \[\begin{array}{r@{\ \ }c@{\ \ }l@{\qquad}r@{\ \ }c@{\ \ }l}
  \ottnt{o}  \ottsym{(}   \tceseq{ \ottnt{c} }   \ottsym{)} & \rightsquigarrow &  \zeta  (  \ottnt{o}  ,   \tceseq{ \ottnt{c} }   ) 
  &
  \mathsf{let} \, \mathit{x}  \ottsym{=}  \ottnt{v_{{\mathrm{1}}}} \,  \mathsf{in}  \, \ottnt{e_{{\mathrm{2}}}} & \rightsquigarrow &  \ottnt{e_{{\mathrm{2}}}}    [  \ottnt{v_{{\mathrm{1}}}}  /  \mathit{x}  ]  
  \\[.5ex]
   \resetcnstr{  \ottnt{K}  [  \mathcal{S}_0 \, \mathit{x}  \ottsym{.}  \ottnt{e}  ]  }       &  \rightsquigarrow  &  \ottnt{e}    [   \lambda\!  \, \mathit{x}  \ottsym{.}   \resetcnstr{  \ottnt{K}  [  \mathit{x}  ]  }   /  \mathit{x}  ]  
  &
  \mathsf{if} \,  \mathsf{true}  \, \mathsf{then} \, \ottnt{e_{{\mathrm{1}}}} \, \mathsf{else} \, \ottnt{e_{{\mathrm{2}}}}  &  \rightsquigarrow  & \ottnt{e_{{\mathrm{1}}}}
  \\[.5ex]
   \resetcnstr{ \ottnt{v} }                  &  \rightsquigarrow  & \ottnt{v}
  &
  \mathsf{if} \,  \mathsf{false}  \, \mathsf{then} \, \ottnt{e_{{\mathrm{1}}}} \, \mathsf{else} \, \ottnt{e_{{\mathrm{2}}}} &  \rightsquigarrow  & \ottnt{e_{{\mathrm{2}}}}
  \\[.5ex]
  \iffull\else
  \ottsym{(}   \lambda\!  \,  \tceseq{ \mathit{x} }   \ottsym{.}  \ottnt{e}  \ottsym{)} \,  \tceseq{ \ottnt{v_{{\mathrm{2}}}} }  & \rightsquigarrow &  \ottnt{e}    [ \tceseq{ \ottnt{v_{{\mathrm{2}}}}  /  \mathit{x} } ]  
    \quad \text{(if $\ottsym{\mbox{$\mid$}}   \tceseq{ \ottnt{v_{{\mathrm{2}}}} }   \ottsym{\mbox{$\mid$}} \,  =  \, \ottsym{\mbox{$\mid$}}   \tceseq{ \mathit{x} }   \ottsym{\mbox{$\mid$}}$)} \\[.5ex]
  \fi
  \ottnt{v_{{\mathrm{1}}}} \,  \tceseq{ \ottnt{v_{{\mathrm{2}}}} }  & \rightsquigarrow &
   \multicolumn{4}{@{}l}{
       \ottnt{e}    [ \tceseq{  \ottnt{A} _{  \mu  }   /  \mathit{X} } ]      [ \tceseq{  \ottnt{A} _{  \nu  }   /  \mathit{Y} } ]      [  \ottnt{v_{{\mathrm{1}}}}  /  \mathit{f}  ]      [ \tceseq{ \ottnt{v_{{\mathrm{2}}}}  /  \mathit{x} } ]  
   \quad
   \text{(if $ { \ottnt{v_{{\mathrm{1}}}} \,  =  \,  \mathsf{rec}  \, (  \tceseq{  \mathit{X} ^{  \ottnt{A} _{  \mu  }  },  \mathit{Y} ^{  \ottnt{A} _{  \nu  }  }  }  ,  \mathit{f} ,   \tceseq{ \mathit{x} }  ) . \,  \ottnt{e}  } \,\mathrel{\wedge}\, { \ottsym{\mbox{$\mid$}}   \tceseq{ \ottnt{v_{{\mathrm{2}}}} }   \ottsym{\mbox{$\mid$}} \,  =  \, \ottsym{\mbox{$\mid$}}   \tceseq{ \mathit{x} }   \ottsym{\mbox{$\mid$}} } $)}
  }
   \end{array}\]
 \mbox{}
 \begin{flushleft}
  \textbf{Evaluation rules} \quad \framebox{$\ottnt{e_{{\mathrm{1}}}} \,  \mathrel{  \longrightarrow  }  \, \ottnt{e_{{\mathrm{2}}}} \,  \,\&\,  \, \varpi$}
 \end{flushleft}
 \mbox{} \\[-5ex]
 \[
   \ottnt{E}  [  \ottnt{e_{{\mathrm{1}}}}  ]  \,  \mathrel{  \longrightarrow  }  \,  \ottnt{E}  [  \ottnt{e_{{\mathrm{2}}}}  ]  \,  \,\&\,  \,  \epsilon 
   \quad \text{(if $\ottnt{e_{{\mathrm{1}}}} \,  \rightsquigarrow  \, \ottnt{e_{{\mathrm{2}}}}$)}
  \qquad\qquad
  { \ottnt{E}  [   \mathsf{ev}  [  \mathbf{a}  ]   ]  \,  \mathrel{  \longrightarrow  }  \,  \ottnt{E}  [   ()   ]  \,  \,\&\,  \, \mathbf{a}}
 \]
 \begin{flushleft}
 \hspace{-1.8ex}
 $
 \begin{array}{lrcl}
   \\[-2ex]
   \textbf{Termination} &
   \framebox{$\ottnt{e} \,  \Downarrow  \, \ottnt{r} \,  \,\&\,  \, \varpi$} &\defeq& \ottnt{e} \,  \mathrel{  \longrightarrow  ^*}  \, \ottnt{r} \,  \,\&\,  \, \varpi
   \\[1ex]
   \multirow{2}{*}{\textbf{Divergence}} & \multirow{2}{*}{\framebox{$\ottnt{e} \,  \Uparrow  \, \pi$}} &\multirow{2}{*}{$\defeq$}&
    \ottsym{(}     \forall  \, { \varpi'  \ottsym{,}  \pi' } . \  \pi \,  =  \, \varpi' \,  \tceappendop  \, \pi'   \mathrel{\Longrightarrow}    \exists  \, { \ottnt{e'} } . \  \ottnt{e} \,  \mathrel{  \longrightarrow  ^*}  \, \ottnt{e'} \,  \,\&\,  \, \varpi'    \ottsym{)} \, \vee
    \\ &&&
    \ottsym{(}   {   \exists  \, { \varpi'  \ottsym{,}  \pi'  \ottsym{,}  \ottnt{E} } . \  \pi \,  =  \, \varpi' \,  \tceappendop  \, \pi'  } \,\mathrel{\wedge}\, { \ottnt{e} \,  \mathrel{  \longrightarrow  ^*}  \,  \ottnt{E}  [   \bullet ^{ \pi' }   ]  \,  \,\&\,  \, \varpi' }   \ottsym{)}
 \end{array}
  $
 \end{flushleft}
 \caption{Semantics.}
 \label{fig:semantics}
\end{figure}

\reffig{semantics} presents the call-by-value, small-step semantics of
{\mylang}.  It is a straightforward variant of the semantics of the languages in
the previous
work~\cite{Materzok/Biernacki_2011_ICFP,Nanjo/Unno/Koskinen/Terauchi_2018_LICS}.
Our semantics is defined by two relations:
the reduction relation $\ottnt{e_{{\mathrm{1}}}} \,  \rightsquigarrow  \, \ottnt{e_{{\mathrm{2}}}}$, which means that expression $\ottnt{e_{{\mathrm{1}}}}$
reduces to expression $\ottnt{e_{{\mathrm{2}}}}$ without raising events, and
the evaluation relation $\ottnt{e_{{\mathrm{1}}}} \,  \mathrel{  \longrightarrow  }  \, \ottnt{e_{{\mathrm{2}}}} \,  \,\&\,  \, \varpi$, which means that $\ottnt{e_{{\mathrm{1}}}}$
evaluates to $\ottnt{e_{{\mathrm{2}}}}$ with finite trace $\varpi$.

Reduction is defined in a straightforward manner, following the
previous work.
The reduction of primitive operation $\ottnt{o}  \ottsym{(}   \tceseq{ \ottnt{c} }   \ottsym{)}$ depends on the metafunction
$ \zeta $, which maps tuples of operation $\ottnt{o}$ and arguments $ \tceseq{ \ottnt{c} } $
to constants. 
Application of a $\lambda$-abstraction and recursive function reduces only
when they are fully applied. 
In the application of a recursive function $ \mathsf{rec}  \, (  \tceseq{  \mathit{X} ^{  \ottnt{A} _{  \mu  }  },  \mathit{Y} ^{  \ottnt{A} _{  \nu  }  }  }  ,  \mathit{f} ,   \tceseq{ \mathit{x} }  ) . \,  \ottnt{e} $,
the predicates $ \tceseq{  \ottnt{A} _{  \mu  }  } $ are substituted for $ \tceseq{ \mathit{X} } $ in the body $\ottnt{e}$ because
$ \tceseq{ \mathit{X} } $ represent the finite parts $ \tceseq{  \ottnt{A} _{  \mu  }  } $
of the temporal effects of $\ottnt{e}$, as mentioned in \refsec{lang:syntax}.
Similarly, the predicates $ \tceseq{  \ottnt{A} _{  \nu  }  } $ on infinite traces are substituted for
$ \tceseq{ \mathit{Y} } $.
%
{}
%

The behavior of a {\resetzero} expression $ \resetcnstr{ \ottnt{e} } $ depends on the
evaluation result of the body $\ottnt{e}$.
If $\ottnt{e}$ evaluates to a value $\ottnt{v}$, then $\ottnt{v}$ is the result of the
entire {\resetzero} expression. 
If $\ottnt{e}$ invokes the {\shiftzero} operator, its context up to the closest {\resetzero}
construct is captured as the continuation. 
Such a context is formalized as pure evaluation contexts, ranged over by
$\ottnt{K}$, which are evaluation contexts that contain no {\reset0} construct
enclosing the hole $ \,[\,]\, $.
Evaluation contexts, ranged over by $\ottnt{E}$, and pure evaluation contexts
are both defined at the top of \reffig{semantics}.
Notice that only $ \mathsf{let} $-expressions and {\resetzero} expressions allow
non-value expressions to be placed at redex positions.
We write $ \ottnt{E}  [  \ottnt{e}  ] $ and $ \ottnt{K}  [  \ottnt{e}  ] $ for the expressions obtained by filling the
holes in $\ottnt{E}$ and $\ottnt{K}$ with expression $\ottnt{e}$, respectively.

We can now formalize the interaction between {\shiftzero} and {\resetzero}.
The reduction of an expression $ \resetcnstr{  \ottnt{K}  [  \mathcal{S}_0 \, \mathit{x}  \ottsym{.}  \ottnt{e}  ]  } $ proceeds as follows.
First, the pure evaluation context $\ottnt{K}$ from the {\shiftzero} construct up to
the closest {\reset0} construct is captured.
Subsequently, the body $\ottnt{e}$ is evaluated with the binding of $\mathit{x}$ to
$ \lambda\!  \, \mathit{x}  \ottsym{.}   \resetcnstr{  \ottnt{K}  [  \mathit{x}  ]  } $, which is a functional representation of the remaining
context $\ottnt{K}$.
%

Evaluation is defined as the relation satisfying the rules shown at
the middle in \reffig{semantics}.  These rules imply that an
expression evaluates by reducing its subterm or raising an event at a
redex position.
We write $\ottnt{e_{{\mathrm{1}}}} \,  \mathrel{  \longrightarrow  ^*}  \, \ottnt{e_{{\mathrm{2}}}} \,  \,\&\,  \, \varpi$ when expression $\ottnt{e_{{\mathrm{1}}}}$ evaluates to
$\ottnt{e_{{\mathrm{2}}}}$ with finite trace $\varpi$ in a finite number of steps.
Formally, $\ottnt{e_{{\mathrm{1}}}} \,  \mathrel{  \longrightarrow  ^*}  \, \ottnt{e_{{\mathrm{2}}}} \,  \,\&\,  \, \varpi$ if and only if there exist some $\ottnt{e'_{{\mathrm{0}}}},
\cdots, \ottnt{e'_{\ottmv{n}}}, \varpi'_{{\mathrm{1}}}, \cdots, \varpi'_{\ottmv{n}}$ such that
$ {  {  { \ottnt{e_{{\mathrm{1}}}} \,  =  \, \ottnt{e'_{{\mathrm{0}}}} } \,\mathrel{\wedge}\, {   \forall  \, { \ottmv{i} \,  <  \, \ottmv{n} } . \  \ottnt{e'_{\ottmv{i}}} \,  \mathrel{  \longrightarrow  }  \,  \ottnt{e} '_{ \ottmv{i}  \ottsym{+}  \ottsym{1} }  \,  \,\&\,  \,  \varpi' _{ \ottmv{i}  \ottsym{+}  \ottsym{1} }   }  } \,\mathrel{\wedge}\, { \ottnt{e_{{\mathrm{2}}}} \,  =  \, \ottnt{e'_{\ottmv{n}}} }  } \,\mathrel{\wedge}\, { \varpi \,  =  \,  \varpi'_{{\mathrm{1}}}  \,   \tceappendop   \ \, \cdots \ \,   \tceappendop   \,  \varpi'_{\ottmv{n}}  } $
(if $\ottmv{n} \,  =  \, \ottsym{0}$, $ \varpi    =     \epsilon  $).

Finally, we define (valid) termination and divergence of an expression.
We define valid evaluation results, ranged over by $\ottnt{r}$, as either values or
the call to {\shiftzero} not enclosed by {\resetzero}.
We write $\ottnt{e} \,  \Downarrow  \, \ottnt{r} \,  \,\&\,  \, \varpi$ if and only if the evaluation of expression $\ottnt{e}$
terminates at result $\ottnt{r}$ with finite trace $\varpi$.
We also define the divergence of expressions.
In this work,
as \citet{Koskinen/Terauchi_2014_CSL-LICS} and
\citet{Nanjo/Unno/Koskinen/Terauchi_2018_LICS}, we assume that
non-terminating evaluation yields infinite traces.  This assumption can be
enforced easily by, e.g., inserting an event-raising operation for every redex
in a program.
Under this assumption, an expression $\ottnt{e}$ diverges at infinite trace
$\pi$, defined at the bottom of \reffig{semantics} as $\ottnt{e} \,  \Uparrow  \, \pi$, if and
only if either of the following holds:
for any finite prefix $\varpi'$ of $\pi$, the evaluation of $\ottnt{e}$ yields
$\varpi'$;
or, the evaluation of $\ottnt{e}$ terminates at expression $ \ottnt{E}  [   \bullet ^{ \pi' }   ] $ with
finite trace $\varpi'$ such that $ \pi    =    \varpi' \,  \tceappendop  \, \pi' $ (In this manner, the
expression $ \bullet ^{ \pi' } $ behaves as the ``divergence'' with $\pi'$).

%% file: sections/eff.tex
\section{Dependent Temporal Effect System for {\mylang}}
\label{sec:eff}

This section introduces a dependent temporal effect system for {\mylang}.
It is based on the effect systems in the two previous works:
\citet{Materzok/Biernacki_2011_ICFP}, which enabled ATM and subtyping for
{\shiftzero}/{\resetzero}, and
\citet{Nanjo/Unno/Koskinen/Terauchi_2018_LICS}, which introduced dependent
temporal effects.
Our effect system extends these systems to handle AEM and allow dependent typing of continuations%
{}.

\subsection{Types and Effects}
\label{sec:eff:ty-eff}

\begin{figure}[t]
 \[
 \begin{array}{@{}r@{\ \ }c@{\ \ }c@{\ \ }lr@{\ \ }c@{\ \ }c@{\ \ }l}
  \textbf{Value types} &
   \ottnt{T} & ::= & \ottsym{\{}  \mathit{x} \,  {:}  \, \iota \,  |  \, \phi  \ottsym{\}} \mid \ottsym{(}  \mathit{x} \,  {:}  \, \ottnt{T}  \ottsym{)}  \rightarrow  \ottnt{C}
   {}
  &
  \textbf{Control effects} &
   \ottnt{S} & ::= &  \square  \mid  ( \forall  \mathit{x}  .  \ottnt{C_{{\mathrm{1}}}}  )  \Rightarrow   \ottnt{C_{{\mathrm{2}}}} 
  \\
  \textbf{Computation types} &
   \ottnt{C} & ::= &  \ottnt{T}  \,  \,\&\,  \,  \Phi  \, / \,  \ottnt{S} 
  &
  \multicolumn{1}{@{\,}l@{\ \ }}{\textbf{Temporal effects}} &
   \Phi & ::= &  (   \lambda_\mu\!   \,  \mathit{x}  .\,  \phi_{{\mathrm{1}}}  ,   \lambda_\nu\!   \,  \mathit{y}  .\,  \phi_{{\mathrm{2}}}  ) 
  \\
  \textbf{Typing contexts} & \Gamma & ::= &
   \multicolumn{5}{l}{
    \emptyset  \mid \Gamma  \ottsym{,}  \mathit{x} \,  {:}  \, \ottnt{T} \mid \Gamma  \ottsym{,}  \mathit{X} \,  {:}  \,  \tceseq{ \ottnt{s} }  \mid \Gamma  \ottsym{,}  \mathit{x} \,  {:}  \, \ottnt{s}
   }
 \end{array}
 \]
 \caption{Type syntax.}
 \label{fig:type-eff-syntax}
\end{figure}

\reffig{type-eff-syntax} presents the syntax of types and effects in this work.

Value types, ranged over by $\ottnt{T}$, specify values.
A refinement type $\ottsym{\{}  \mathit{x} \,  {:}  \, \iota \,  |  \, \phi  \ottsym{\}}$, which binds $\mathit{x}$ in $\phi$, specifies
constants $\ottnt{c}$ satisfying the predicate $\phi$ (i.e., $ \phi    [  \ottnt{c}  /  \mathit{x}  ]  $ is
true).  We write $\iota$ simply if $\phi$ is not important.
A dependent function type $\ottsym{(}  \mathit{x} \,  {:}  \, \ottnt{T}  \ottsym{)}  \rightarrow  \ottnt{C}$, which binds $\mathit{x}$ in $\ottnt{C}$,
specifies (partially applied) $\lambda$-abstractions and recursive functions
that, given a value $\ottnt{v}$ of $\ottnt{T}$, perform the computation specified by the
type $ \ottnt{C}    [  \ottnt{v}  /  \mathit{x}  ]  $ depending on the argument $\ottnt{v}$.
We write $\ottnt{T}  \rightarrow  \ottnt{C}$ if $\mathit{x}$ does not occur free in $\ottnt{C}$.
{}

A computation type $\ottnt{C}$ specifies the behavior of expressions, consisting of
three components: a value type, which specifies the value of an expression (if
its evaluation terminates); a temporal effect, which specifies finite and infinite
traces yielded by the expression; and a \emph{control
effect},
which specifies the usage of the control operators in the
expression.

Temporal effects, ranged over by $\Phi$, take the form of $ (   \lambda_\mu\!   \,  \mathit{x}  .\,  \phi_{{\mathrm{1}}}  ,   \lambda_\nu\!   \,  \mathit{y}  .\,  \phi_{{\mathrm{2}}}  ) $,
where $\phi_{{\mathrm{1}}}$ is a predicate on finite traces $\mathit{x}$ and $\phi_{{\mathrm{2}}}$ is on
infinite traces $\mathit{y}$.
Our effect system ensures that, if an expression $\ottnt{e}$ is assigned temporal
effect $ (   \lambda_\mu\!   \,  \mathit{x}  .\,  \phi_{{\mathrm{1}}}  ,   \lambda_\nu\!   \,  \mathit{y}  .\,  \phi_{{\mathrm{2}}}  ) $, the finite trace $\varpi$ yielded by the
terminating evaluation of $\ottnt{e}$ satisfies $\phi_{{\mathrm{1}}}$ (i.e., $ \phi_{{\mathrm{1}}}    [  \varpi  /  \mathit{x}  ]  $ is true)
and the infinite trace $\pi$ yielded by its diverging evaluation satisfies
$\phi_{{\mathrm{2}}}$ (i.e., $ \phi_{{\mathrm{2}}}    [  \pi  /  \mathit{y}  ]  $ is true).
Our formalism with the small-step semantics also implies that
finite traces yielded on the course of evaluating the expression $\ottnt{e}$ are
prefixes of some trace contained in $\phi_{{\mathrm{1}}}$ or $\phi_{{\mathrm{2}}}$.
%
We write $ (  \varpi  , \bot ) $ simply to mean the temporal effect
$ (   \lambda_\mu\!   \,  \mathit{x}  .\,   \mathit{x}    =    \varpi   ,   \lambda_\nu\!   \,  \mathit{y}  .\,   \bot   ) $.

Control effects, ranged over by $\ottnt{S}$, characterize the use of the control
operators in expressions.
They take either the form $ \square $ or $ ( \forall  \mathit{x}  .  \ottnt{C_{{\mathrm{1}}}}  )  \Rightarrow   \ottnt{C_{{\mathrm{2}}}} $.
The control effect $ \square $, which we call pure, means that an expression
never invokes the {\shiftzero} operator.
An expression that operates {\shiftzero} is assigned an \emph{impure} control
effect $ ( \forall  \mathit{x}  .  \ottnt{C_{{\mathrm{1}}}}  )  \Rightarrow   \ottnt{C_{{\mathrm{2}}}} $, which binds the variable $\mathit{x}$ in the type
$\ottnt{C_{{\mathrm{1}}}}$; we write $\ottnt{C_{{\mathrm{1}}}}  \Rightarrow  \ottnt{C_{{\mathrm{2}}}}$ if $\mathit{x}$ does not occur free in $\ottnt{C_{{\mathrm{1}}}}$.
Control effects are reminiscent of type representations for ATM in the previous
work.  Control effects without temporal effects correspond to effect annotations
in \citet{Materzok/Biernacki_2011_ICFP}, and dropping both temporal and control
effects from $\ottnt{C_{{\mathrm{1}}}}  \Rightarrow  \ottnt{C_{{\mathrm{2}}}}$ results in pairs of initial and final answer
types considered in \citet{Danvy/Filinski_1990_LFP}.

First, we explain what control effects of the simple form $\ottnt{C_{{\mathrm{1}}}}  \Rightarrow  \ottnt{C_{{\mathrm{2}}}}$ mean;
we call control effects of this form \emph{nondependent}.
Roughly speaking, the type $\ottnt{C_{{\mathrm{1}}}}$ represents what an invocation of
{\shiftzero} requires for the context up to the closest {\resetzero} construct,
and $\ottnt{C_{{\mathrm{2}}}}$ represents what the invocation guarantees for the meta-context, i.e., the context of
the {\resetzero} construct.
%
%
%
For detail, consider an expression $ \resetcnstr{  \ottnt{K}  [  \mathcal{S}_0 \, \mathit{x}  \ottsym{.}  \ottnt{e}  ]  } $ where
the expression $\mathcal{S}_0 \, \mathit{x}  \ottsym{.}  \ottnt{e}$ is assigned a control effect $\ottnt{C_{{\mathrm{1}}}}  \Rightarrow  \ottnt{C_{{\mathrm{2}}}}$.
Then, the type $\ottnt{C_{{\mathrm{1}}}}$ expresses the requirement for the context
$ \resetcnstr{ \ottnt{K} } $.  Namely, given an appropriate value $\ottnt{v}$, the resulting expression $ \resetcnstr{  \ottnt{K}  [  \ottnt{v}  ]  } $ of filling the hole with $\ottnt{v}$ must follow the type $\ottnt{C_{{\mathrm{1}}}}$.
Thus, the body $\ottnt{e}$ supposes the
variable $\mathit{x}$, which is bound to $ \lambda\!  \, \mathit{x}  \ottsym{.}   \resetcnstr{  \ottnt{K}  [  \mathit{x}  ]  } $, to be of a function
type $\ottnt{T}  \rightarrow  \ottnt{C_{{\mathrm{1}}}}$, where the type $\ottnt{T}$ is of the appropriate values $\ottnt{v}$ (it is
decided by the context $\ottnt{K}$).
The type $\ottnt{C_{{\mathrm{2}}}}$ guarantees that the result of operating {\shiftzero} behaves
as expected by the context of the {\resetzero} construct.  Because
$ \resetcnstr{  \ottnt{K}  [  \mathcal{S}_0 \, \mathit{x}  \ottsym{.}  \ottnt{e}  ]  } $ reduces to $ \ottnt{e}    [   \lambda\!  \, \mathit{x}  \ottsym{.}   \resetcnstr{  \ottnt{K}  [  \mathit{x}  ]  }   /  \mathit{x}  ]  $, the
expression $ \ottnt{e}    [   \lambda\!  \, \mathit{x}  \ottsym{.}   \resetcnstr{  \ottnt{K}  [  \mathit{x}  ]  }   /  \mathit{x}  ]  $ must follow the expectation of the outer
context, that is, it must follow the type $\ottnt{C_{{\mathrm{2}}}}$.
%
%
For example, recall the program $ \resetcnstr{ \ottsym{(}  \mathcal{S}_0 \, \mathit{f}  \ottsym{.}  \ottnt{e}  \ottsym{)}  \ottsym{;}   \mathsf{ev}  [  \mathbf{a}  ]  } $ given in
\refsec{overview:this-work} (note that the sequential composition $\ottnt{e_{{\mathrm{1}}}}  \ottsym{;}  \ottnt{e_{{\mathrm{2}}}}$ is
encoded using the $ \mathsf{let} $-constructor).
Assume that the expression $\ottnt{e}$ of type $\ottnt{T}$ only calls $\mathit{f}$ $n$-times.
Then, the {\shiftzero} expression has a control effect
$\ottsym{(}    \mathsf{unit}   \,  \,\&\,  \,   (  \mathbf{a}  , \bot )   \, / \,   \square    \ottsym{)}  \Rightarrow  \ottsym{(}   \ottnt{T}  \,  \,\&\,  \,   (   \mathbf{a} ^{ \ottmv{n} }   , \bot )   \, / \,   \square    \ottsym{)}$.
The temporal effects $ (  \mathbf{a}  , \bot ) $ and $ (   \mathbf{a} ^{ \ottmv{n} }   , \bot ) $ represent the fact that the delimited continuation
$ \resetcnstr{ \,[\,]\,  \ottsym{;}   \mathsf{ev}  [  \mathbf{a}  ]  } $ raises the event $\mathbf{a}$ only once and the entire {\resetzero}
construct yields the trace $ \mathbf{a} ^{ \ottmv{n} } $, respectively.


While nondependent control effects are expressive enough to accommodate AEM,
they cannot utilize the expressivity of dependent typing fully.
Specifically, it is critical that the requirement $\ottnt{C_{{\mathrm{1}}}}$ for contexts cannot
depend on values passed to the contexts.
To see the problem,
consider an expression
$ \resetcnstr{ \mathsf{let} \, \mathit{x}  \ottsym{=}  \ottsym{(}  \mathcal{S}_0 \, \mathit{f}  \ottsym{.}  \mathit{f} \,  3   \ottsym{;}  \mathit{f} \,  5   \ottsym{)} \,  \mathsf{in}  \,  \mathsf{ev}  [  \mathbf{a}  ]^{ \mathit{x} }  } $,
where $ \mathsf{ev}  [  \mathbf{a}  ]^{ \mathit{x} } $ repeats $ \mathsf{ev}  [  \mathbf{a}  ] $ $\mathit{x}$-times.
This expression reduces to $\ottsym{(}  \ottnt{v} \,  3   \ottsym{;}  \ottnt{v} \,  5   \ottsym{)}$ with $\ottnt{v} \,  =  \,  \lambda\!  \, \mathit{y}  \ottsym{.}   \resetcnstr{ \mathsf{let} \, \mathit{x}  \ottsym{=}  \mathit{y} \,  \mathsf{in}  \,  \mathsf{ev}  [  \mathbf{a}  ]^{ \mathit{x} }  } $
and finally generates the trace $ \mathbf{a} ^{  8  } $.
%
Therefore, we expect that the effect system assigns the temporal effect
$ (   \mathbf{a} ^{  8  }   , \bot ) $ to the expression.
This assignment is possible if the continuation variable $\mathit{f}$ is of the type
$\ottsym{(}  \mathit{x} \,  {:}  \,  \mathsf{int}   \ottsym{)}  \rightarrow    \mathsf{unit}   \,  \,\&\,  \,   (   \mathbf{a} ^{ \mathit{x} }   , \bot )   \, / \,  \ottnt{S} $, which states that the
finite trace yielded by the continuation depends on the argument
$\mathit{x}$.
However, nondependent control effects do not allow $\mathit{f}$ to have such a type;
a control effect $\ottnt{C_{{\mathrm{1}}}}  \Rightarrow  \ottnt{C_{{\mathrm{2}}}}$ given to {\shiftzero} only allows continuation
variables to nondependent function types of the form $\ottnt{T}  \rightarrow  \ottnt{C_{{\mathrm{1}}}}$.

We solve this problem using \emph{dependent} control effects of the form $ ( \forall  \mathit{x}  .  \ottnt{C_{{\mathrm{1}}}}  )  \Rightarrow   \ottnt{C_{{\mathrm{2}}}} $.
The variable $\mathit{x}$ in this form stands for values passed to contexts.
Because an expression with a value type $\ottnt{T}$ passes a value of $\ottnt{T}$ to its
context when terminating, the type of $\mathit{x}$ is determined by the computation
type involving the control effect.
In general, a computation type $ \ottnt{T}  \,  \,\&\,  \,  \Phi  \, / \,   ( \forall  \mathit{x}  .  \ottnt{C_{{\mathrm{1}}}}  )  \Rightarrow   \ottnt{C_{{\mathrm{2}}}}  $ represents that: the
value of an expression follows type $\ottnt{T}$; the traces yielded by the
expression follow $\Phi$; the context of the expression up to the closest
{\resetzero} construct must follow type $ \ottnt{C_{{\mathrm{1}}}}    [  \ottnt{v}  /  \mathit{x}  ]  $ when a value $\ottnt{v}$
of $\ottnt{T}$ is passed; and the closest {\resetzero} construct behaves as
specified by type $\ottnt{C_{{\mathrm{2}}}}$.
Dependent control effects allow the types of continuations to depend on
arguments.
For a {\shiftzero} expression with a control effect $ ( \forall  \mathit{x}  .  \ottnt{C_{{\mathrm{1}}}}  )  \Rightarrow   \ottnt{C_{{\mathrm{2}}}} $, if its
context expects values of type $\ottnt{T}$ to be passed, the continuation variable
is given the \emph{dependent} function type $\ottsym{(}  \mathit{x} \,  {:}  \, \ottnt{T}  \ottsym{)}  \rightarrow  \ottnt{C_{{\mathrm{1}}}}$.


Let us now revisit the above example.
Let $\ottnt{K}$ be the delimited context under which the {\shiftzero} expression
$\mathcal{S}_0 \, \mathit{f}  \ottsym{.}  \mathit{f} \,  3   \ottsym{;}  \mathit{f} \,  5 $ is placed, that is,
$\ottnt{K} \,  =  \,  \resetcnstr{ \mathsf{let} \, \mathit{x}  \ottsym{=}  \,[\,]\, \,  \mathsf{in}  \,  \mathsf{ev}  [  \mathbf{a}  ]^{ \mathit{x} }  } $.
Given an integer value $\ottnt{v}$, the context $ \resetcnstr{ \ottnt{K} } $ raises event
$\mathbf{a}$ $\ottnt{v}$-times.
Dependent control effects can describe this behavior as the requirement for the
context.
For example, the {\shiftzero} expression can be assigned the
control effect $ ( \forall  \mathit{x}  .    \mathsf{unit}   \,  \,\&\,  \,   (   \mathbf{a} ^{ \mathit{x} }   , \bot )   \, / \,  \ottnt{S}   )  \Rightarrow   \ottnt{C_{{\mathrm{2}}}} $ for some
$\ottnt{S}$ and $\ottnt{C_{{\mathrm{2}}}}$.
Then, the continuation variable $\mathit{f}$ is assigned the dependent function type
$\ottsym{(}  \mathit{x} \,  {:}  \,  \mathsf{int}   \ottsym{)}  \rightarrow    \mathsf{unit}   \,  \,\&\,  \,   (   \mathbf{a} ^{ \mathit{x} }   , \bot )   \, / \,  \ottnt{S} $.
This dependent type implies that, when terminating, the expressions
$\mathit{f} \,  3 $ and $\mathit{f} \,  5 $ yield the finite traces $ \mathbf{a} ^{  3  } $ and $ \mathbf{a} ^{  5  } $,
respectively.
Therefore, the body $\ottsym{(}  \mathit{f} \,  3   \ottsym{;}  \mathit{f} \,  5   \ottsym{)}$ is assigned the temporal effect
$ (   \mathbf{a} ^{  8  }   , \bot ) $.
The type of the body corresponds to the type $\ottnt{C_{{\mathrm{2}}}}$ in the control effect of
the {\shiftzero} expression, and it specifies how the closest {\resetzero}
construct behaves.
Therefore, the entire expression $ \resetcnstr{  \ottnt{K}  [  \mathcal{S}_0 \, \mathit{f}  \ottsym{.}  \mathit{f} \,  3   \ottsym{;}  \mathit{f} \,  5   ]  } $ has the temporal
effect $ (   \mathbf{a} ^{  8  }   , \bot ) $, from which we can deduce that its evaluation
terminates with the finite trace $ \mathbf{a} ^{  8  } $.

\begin{sloppypar}
It is notable that the dependency of the control effects is added
systematically, not ad-hoc, from Continuation Passing Style (CPS) transformation.
To see it, we consider computation types $\ottnt{T}  \ottsym{/}  \ottnt{S}$ which omit temporal effects.
By adapting the CPS transformation $\cpstransty{-}$ for
{\shiftzero}/{\resetzero}~\cite{Materzok/Biernacki_2011_ICFP}, a
computation type $\ottnt{T}  \ottsym{/}  \ottnt{C_{{\mathrm{1}}}}  \Rightarrow  \ottnt{C_{{\mathrm{2}}}}$ is
transformed to $\ottsym{(}   \cpstransty{ \ottnt{T} }   \rightarrow   \cpstransty{ \ottnt{C_{{\mathrm{1}}}} }   \ottsym{)}  \rightarrow   \cpstransty{ \ottnt{C_{{\mathrm{2}}}} } $.
A computation type with a dependent control effect $ ( \forall  \mathit{x}  .  \ottnt{C_{{\mathrm{1}}}}  )  \Rightarrow   \ottnt{C_{{\mathrm{2}}}} $ can be
obtained by making the argument function type of $ \cpstransty{ \ottnt{T}  \ottsym{/}  \ottnt{C_{{\mathrm{1}}}}  \Rightarrow  \ottnt{C_{{\mathrm{2}}}} } $
dependent: $ \cpstransty{ \ottnt{T}  \ottsym{/}   ( \forall  \mathit{x}  .  \ottnt{C_{{\mathrm{1}}}}  )  \Rightarrow   \ottnt{C_{{\mathrm{2}}}}  }  = \ottsym{(}  \ottsym{(}  \mathit{x} \,  {:}  \,  \cpstransty{ \ottnt{T} }   \ottsym{)}  \rightarrow   \cpstransty{ \ottnt{C_{{\mathrm{1}}}} }   \ottsym{)}  \rightarrow   \cpstransty{ \ottnt{C_{{\mathrm{2}}}} } $.
Furthermore, ones might consider that the entire function type can similarly be made dependent,
as $\ottsym{(}  \mathit{y} \,  {:}  \, \ottsym{(}  \ottsym{(}  \mathit{x} \,  {:}  \,  \cpstransty{ \ottnt{T} }   \ottsym{)}  \rightarrow   \cpstransty{ \ottnt{C_{{\mathrm{1}}}} }   \ottsym{)}  \ottsym{)}  \rightarrow   \cpstransty{ \ottnt{C_{{\mathrm{2}}}} } $.
This idea might be worth considering in general dependent type systems.  However, it is
meaningless in our system because our types and effects can depend only on
first-order values, following the convention in many refinement type
systems~\cite{Rondon/Kawaguchi/Jhala_2008_PLDI,Bengtson/Bhargavan/Fournet/Gordon/Maffeis_2011_TOPLAS,Unno/Kobayashi_2009_PPDP}.
The formal investigation of the relationship between dependent control effects
and their CPS transformation is beyond the scope of this work, and is left for
future work.
\end{sloppypar}

Typing contexts, ranged over by $\Gamma$, are finite sequences of bindings of
the form $\mathit{x} \,  {:}  \, \ottnt{T}$ (variable $\mathit{x}$ is of value type $\ottnt{T}$),
$\mathit{X} \,  {:}  \,  \tceseq{ \ottnt{s} } $ ($\mathit{X}$ denotes predicates on terms of sorts $ \tceseq{ \ottnt{s} } $), or
$\mathit{x} \,  {:}  \, \ottnt{s}$ ($\mathit{x}$ is of sort $\ottnt{s}$).

We introduce certain notation in what follows.
We write $ { \Phi }^\mu   \ottsym{(}  \ottnt{t}  \ottsym{)}$ and $ { \Phi }^\nu   \ottsym{(}  \ottnt{t}  \ottsym{)}$ for the formulas
$ \phi_{{\mathrm{1}}}    [  \ottnt{t}  /  \mathit{x}  ]  $ and $ \phi_{{\mathrm{2}}}    [  \ottnt{t}  /  \mathit{y}  ]  $ when $\Phi \,  =  \,  (   \lambda_\mu\!   \,  \mathit{x}  .\,  \phi_{{\mathrm{1}}}  ,   \lambda_\nu\!   \,  \mathit{y}  .\,  \phi_{{\mathrm{2}}}  ) $.
%
%
We also write $ \Phi_\mathsf{val} $ for $ (   \epsilon   , \bot ) $.
When $\ottnt{C} \,  =  \,  \ottnt{T}  \,  \,\&\,  \,  \Phi  \, / \,  \ottnt{S} $, $ \ottnt{C}  . \ottnt{T} $, $ \ottnt{C}  . \Phi $, and $ \ottnt{C}  . \ottnt{S} $ denote
$\ottnt{T}$, $\Phi$, and $\ottnt{S}$, respectively.
We write $\ottsym{(}   \tceseq{ \mathit{x}  \,  {:}  \,  \ottnt{T} }   \ottsym{)}  \rightarrow  \ottnt{C}$ for
$ \ottsym{(}  \mathit{x_{{\mathrm{1}}}} \,  {:}  \, \ottnt{T_{{\mathrm{1}}}}  \ottsym{)}  \rightarrow  \ottsym{(}  \cdots  \rightarrow  \ottsym{(}  \ottsym{(}   \mathit{x} _{ \ottmv{n}  \ottsym{-}  \ottsym{1} }  \,  {:}  \,  \ottnt{T} _{ \ottmv{n}  \ottsym{-}  \ottsym{1} }   \ottsym{)}  \rightarrow   \ottsym{(}  \ottsym{(}  \mathit{x_{\ottmv{n}}} \,  {:}  \, \ottnt{T_{\ottmv{n}}}  \ottsym{)}  \rightarrow  \ottnt{C}  \ottsym{)}  \,  \,\&\,  \,   \Phi_\mathsf{val}   \, / \,   \square    \ottsym{)}  \cdots  \ottsym{)}  \,  \,\&\,  \,   \Phi_\mathsf{val}   \, / \,   \square  $
where $ \tceseq{ \mathit{x}  \,  {:}  \,  \ottnt{T} }  = \mathit{x_{{\mathrm{1}}}} \,  {:}  \, \ottnt{T_{{\mathrm{1}}}}, \cdots, \mathit{x_{\ottmv{n}}} \,  {:}  \, \ottnt{T_{\ottmv{n}}}$.
%
%
A typing context $\Gamma  \ottsym{,}  \mathit{x} \,  {:}  \, \ottsym{\{}  \mathit{y} \,  {:}  \,  \mathsf{unit}  \,  |  \, \phi  \ottsym{\}}$, where $\mathit{x}$ and $\mathit{y}$ are fresh, is expressed as $\Gamma  \ottsym{,}  \phi$.
Furthermore, $ \mathit{dom}  (  \Gamma  ) $ denotes the set of variables and predicate variables
bound by $\Gamma$.
We write $ { \models  \,  \phi } $ when a closed formula $\phi$ is valid.
It is extended to the validity $\Gamma  \vdash  \phi$ of a formula $\phi$ with free
variables bound in $\Gamma$ by quantifying the free variables with their sorts
and assuming the refinement predicates.
See the supplementary material for the formal semantics of the fixpoint logic
and the extension of the validity to open formulas.

\subsection{Type-and-Effect System}

Our effect system consists of three kinds of judgments:
well-formedness judgments for type-level constructs;
subtyping judgments for types and effects;
and typing judgments for expressions.

\subsubsection{Well-formedness}

{}

Well-formedness is defined for every type-level construct.
{\iffull\else
This paper only presents the judgment forms for well-formedness and omits the rules for
them because most of the rules are standard or easy to understand;
the full definitions are found in the supplementary material.
The only exception is the rule for dependent control effects, which we explain in what follows.
\fi}
A judgment $\vdash  \Gamma$ states that typing context $\Gamma$ is well formed.
Judgments $\Gamma  \vdash  \ottnt{T}$ and $\Gamma  \vdash  \phi$ state that type $\ottnt{T}$ and formula
$\phi$, respectively, are well formed under $\Gamma$.
A judgment $\Gamma \,  |  \, \ottnt{T}  \vdash  \ottnt{S}$ states that control effect $\ottnt{S}$ with value type
$\ottnt{T}$ is well formed under $\Gamma$.
{\iffull\else
For a dependent control effect $ ( \forall  \mathit{x}  .  \ottnt{C_{{\mathrm{1}}}}  )  \Rightarrow   \ottnt{C_{{\mathrm{2}}}} $,
a judgment $\Gamma \,  |  \, \ottnt{T}  \vdash   ( \forall  \mathit{x}  .  \ottnt{C_{{\mathrm{1}}}}  )  \Rightarrow   \ottnt{C_{{\mathrm{2}}}} $ can be derived
if $\Gamma  \ottsym{,}  \mathit{x} \,  {:}  \, \ottnt{T}  \vdash  \ottnt{C_{{\mathrm{1}}}}$ and $\Gamma  \vdash  \ottnt{C_{{\mathrm{2}}}}$.
This rule indicates that continuations can depend on arguments of the type $\ottnt{T}$.
\fi}
A judgment $\Gamma  \vdash  \ottnt{A}  \ottsym{:}   \tceseq{ \ottnt{s} } $ states that predicate $\ottnt{A}$ is a well-formed
predicate on terms of sorts $ \tceseq{ \ottnt{s} } $ under $\Gamma$.
A judgment $\Gamma  \vdash  \ottnt{t}  \ottsym{:}  \ottnt{s}$ states that term $\ottnt{t}$ is a well-formed
term of sort $\ottnt{s}$ under $\Gamma$.
{}


{}

\subsubsection{Typing}
\label{sec:eff:typing}

\begin{figure}[t]
 \begin{flushleft}
  \textbf{Typing rules} \quad
  \framebox{$\Gamma  \vdash  \ottnt{e}  \ottsym{:}  \ottnt{C}$} \qquad
  \framebox{$\Gamma  \vdash  \ottnt{e}  \ottsym{:}  \ottnt{T}$} $ \ \defeq\ \Gamma  \vdash  \ottnt{e}  \ottsym{:}   \ottnt{T}  \,  \,\&\,  \,   \Phi_\mathsf{val}   \, / \,   \square  $
 \end{flushleft}
 \begin{center}
  $\ottdruleTXXCVar{}$ \hfil
  $\ottdruleTXXVar{}$ \\[1.5ex]
  $\ottdruleTXXConst{}$ \hfil
  $\ottdruleTXXOp{}$ \\[1.5ex]
  {}
  $\ottdruleTXXAbs{}$ \hfil
  $\ottdruleTXXApp{}$ \\[1.5ex]
  $\ottdruleTXXSub{}$ \hfil
  $\ottdruleTXXIf{}$ \\[1.5ex]
  $\ottdruleTXXShift{}$ \\[1.5ex]
  $\ottdruleTXXReset{}$ \hfil
  $\ottdruleTXXLetV{}$ \\[1.5ex]
  $\ottdruleTXXLet{}$ \\[1.5ex]
  $\ottdruleTXXEvent{}$ \hfil
  $\ottdruleTXXDiv{}$ \\[1.5ex]
  $\ottdruleTXXFun{}$ \\[1.5ex]
 \end{center}
 \caption{Type-and-effect system.}
 \label{fig:typing-exp}
\end{figure}

Typing judgments take the form $\Gamma  \vdash  \ottnt{e}  \ottsym{:}  \ottnt{C}$.
Because the temporal effect $ \Phi_\mathsf{val} $ implies that an expression terminates
with the empty trace, and the control effect $ \square $ implies that an
expression does not invoke the {\shiftzero} operator, the purity of an
expression $\ottnt{e}$ is represented by a judgment $\Gamma  \vdash  \ottnt{e}  \ottsym{:}   \ottnt{T}  \,  \,\&\,  \,   \Phi_\mathsf{val}   \, / \,   \square  $.
We simply express this judgment as $\Gamma  \vdash  \ottnt{e}  \ottsym{:}  \ottnt{T}$.
The typing rules are shown in \reffig{typing-exp}.
Subtyping in the rule \T{Sub} will be explained in the next section.
%
%
We will show some examples of typing derivations in \refsec{eff:example}.

The rules for variables, constants,
primitive operations, {}
$\lambda$-abstractions, {}
function applications, and $ \mathsf{if} $-expressions are
standard or easy to understand.
%
The rule \T{CVar} gives a variable $\mathit{x}$ of a refinement type the most
precise type.
A variable of a non-refinement type is given the type itself (\Twop{Var}).
The rules \T{Const} and \T{Op} use
the metafunction $ \mathit{ty} $, which assigns a base type to every constant and a
first-order closed type of the form $\ottsym{(}  \mathit{x_{{\mathrm{1}}}} \,  {:}  \, \ottnt{T_{{\mathrm{1}}}}  \ottsym{)}  \rightarrow  \cdots  \rightarrow  \ottsym{(}  \mathit{x_{\ottmv{n}}} \,  {:}  \, \ottnt{T_{\ottmv{n}}}  \ottsym{)}  \rightarrow  \ottnt{T_{{\mathrm{0}}}}$, for which we
simply write $ \tceseq{(  \mathit{x}  \,  {:}  \,  \ottnt{T}  )}  \rightarrow   \ottnt{T_{{\mathrm{0}}}} $, to every primitive operation.
For an $ \mathsf{if} $-expression $\mathsf{if} \, \ottnt{v} \, \mathsf{then} \, \ottnt{e_{{\mathrm{1}}}} \, \mathsf{else} \, \ottnt{e_{{\mathrm{2}}}}$, the rule \T{If} allows
the $ \mathsf{then} $-expression $\ottnt{e_{{\mathrm{1}}}}$ and $ \mathsf{else} $-expression $\ottnt{e_{{\mathrm{2}}}}$ to assume
the conditional value $\ottnt{v}$ to be $ \mathsf{true} $ and $ \mathsf{false} $, respectively.
These rules clarify the purity of values and primitive operations.

The typing rules for the control operators are \T{Shift} and \T{Reset}.
The rule \T{Shift} follows the intuition described in \refsec{eff:ty-eff}.
One might be concerned that the continuation variable $\mathit{x}$ may occur in the type
$\ottnt{C_{{\mathrm{2}}}}$ in a control effect.  However, it does not happen because variables of
function types never occur in well-formed types and effects.
%
%
The rule \T{Reset} is for {\resetzero} expressions.
To understand this rule, consider a {\resetzero} expression $ \resetcnstr{ \ottnt{e} } $ and
suppose that $\ottnt{e}$ is assigned a control effect $ ( \forall  \mathit{x}  .  \ottnt{C'}  )  \Rightarrow   \ottnt{C} $.
Because the delimited context of $\ottnt{e}$ is the hole $ \,[\,]\, $, the type $\ottnt{C'}$
can only require that the trace yielded by the context is the empty
sequence $ \epsilon $, the {\shiftzero} operator is not invoked, and
the context simply returns a passed value.
This requirement is expressed as $ \ottnt{T}  \,  \,\&\,  \,   \Phi_\mathsf{val}   \, / \,   \square  $
using the value type $\ottnt{T}$ of the body $\ottnt{e}$.
The side condition $\mathit{x} \,  \not\in  \,  \mathit{fv}  (  \ottnt{T}  ) $ is necessary to prevent capturing free
variables in $\ottnt{T}$ accidentally.
The entire expression is assigned the computation type $\ottnt{C}$ because the
control effect $ ( \forall  \mathit{x}  .  \ottnt{C'}  )  \Rightarrow   \ottnt{C} $ of the body $\ottnt{e}$ states that the behavior
of the closest {\resetzero} construct is specified by the type $\ottnt{C}$.
Although the rule \T{Reset} requires the body $\ottnt{e}$ to have a dependent control effect,
we can typecheck it even if its control effect is $ \square $ because
$ \square $ can be converted to a dependent control effect $ ( \forall  \mathit{x}  .   \ottnt{T}  \,  \,\&\,  \,   \Phi_\mathsf{val}   \, / \,   \square    )  \Rightarrow    \ottnt{T}  \,  \,\&\,  \,  \Phi  \, / \,   \square   $ (where $\Phi$ is the temporal effect of $\ottnt{e}$)
via subtyping, as discussed in the next section.

The rules \T{LetV} and \T{Let} are for $ \mathsf{let} $-expressions.
To \T{LetV} expresses that $\mathsf{let} \, \mathit{x_{{\mathrm{1}}}}  \ottsym{=}  \ottnt{v_{{\mathrm{1}}}} \,  \mathsf{in}  \, \ottnt{e_{{\mathrm{2}}}}$ is the same as $\ottsym{(}   \lambda\!  \, \mathit{x_{{\mathrm{1}}}}  \ottsym{.}  \ottnt{e_{{\mathrm{2}}}}  \ottsym{)} \, \ottnt{v_{{\mathrm{1}}}}$.
To understand \T{Let}, consider expression $\mathsf{let} \, \mathit{x_{{\mathrm{1}}}}  \ottsym{=}  \ottnt{e_{{\mathrm{1}}}} \,  \mathsf{in}  \, \ottnt{e_{{\mathrm{2}}}}$.
First, as usual in refinement type
systems~\cite{Rondon/Kawaguchi/Jhala_2008_PLDI,Bengtson/Bhargavan/Fournet/Gordon/Maffeis_2011_TOPLAS,Unno/Kobayashi_2009_PPDP},
the third condition in the premise of \T{Let} is given to prevent the bound variable
$\mathit{x_{{\mathrm{1}}}}$ from escaping its valid scope.
%
Because the $ \mathsf{let} $-expression sequentializes the two computations $\ottnt{e_{{\mathrm{1}}}}$
and $\ottnt{e_{{\mathrm{2}}}}$, we need operations to ``sequentialize'' their temporal and control
effects, respectively.
Suppose that $\ottnt{e_{{\mathrm{1}}}}$ is of $ \ottnt{T_{{\mathrm{1}}}}  \,  \,\&\,  \,  \Phi_{{\mathrm{1}}}  \, / \,  \ottnt{S_{{\mathrm{1}}}} $ and $\ottnt{e_{{\mathrm{2}}}}$ is of $ \ottnt{T_{{\mathrm{2}}}}  \,  \,\&\,  \,  \Phi_{{\mathrm{2}}}  \, / \,  \ottnt{S_{{\mathrm{2}}}} $.
%
%
%
%
The temporal effect $\Phi_{{\mathrm{1}}} \,  \tceappendop  \, \Phi_{{\mathrm{2}}}$ of the $ \mathsf{let} $-expression is defined to
accept all the traces that may be yielded when $\ottnt{e_{{\mathrm{1}}}}$ and then
$\ottnt{e_{{\mathrm{2}}}}$ are run sequentially.
\begin{defn}[Temporal Effect Concatenation]
 For temporal effects $\Phi_{{\mathrm{1}}}$ and $\Phi_{{\mathrm{2}}}$, we define the temporal effect
 $\Phi_{{\mathrm{1}}} \,  \tceappendop  \, \Phi_{{\mathrm{2}}}$ as $ (   \lambda_\mu\!   \,  \mathit{x}  .\,  \phi_{{\mathrm{1}}}  ,   \lambda_\nu\!   \,  \mathit{y}  .\,  \phi_{{\mathrm{2}}}  ) $ where:
 \[\begin{array}{llll}
  \phi_{{\mathrm{1}}} &\defeq&   \exists  \,  \mathit{x_{{\mathrm{1}}}}  \mathrel{  \in  }   \Sigma ^\ast   . \    \exists  \,  \mathit{x_{{\mathrm{2}}}}  \mathrel{  \in  }   \Sigma ^\ast   . \     \mathit{x}    =    \mathit{x_{{\mathrm{1}}}} \,  \tceappendop  \, \mathit{x_{{\mathrm{2}}}}     \wedge     { \Phi_{{\mathrm{1}}} }^\mu   \ottsym{(}  \mathit{x_{{\mathrm{1}}}}  \ottsym{)}     \wedge     { \Phi_{{\mathrm{2}}} }^\mu   \ottsym{(}  \mathit{x_{{\mathrm{2}}}}  \ottsym{)}    & \text{and} \\
  \phi_{{\mathrm{2}}} &\defeq&   { \Phi_{{\mathrm{1}}} }^\nu   \ottsym{(}  \mathit{y}  \ottsym{)}    \vee    \ottsym{(}    \exists  \,  \mathit{x'}  \mathrel{  \in  }   \Sigma ^\ast   . \    \exists  \,  \mathit{y'}  \mathrel{  \in  }   \Sigma ^\omega   . \     \mathit{y}    =    \mathit{x'} \,  \tceappendop  \, \mathit{y'}     \wedge     { \Phi_{{\mathrm{1}}} }^\mu   \ottsym{(}  \mathit{x'}  \ottsym{)}     \wedge     { \Phi_{{\mathrm{2}}} }^\nu   \ottsym{(}  \mathit{y'}  \ottsym{)}     \ottsym{)} 
   \end{array}
 \]
\end{defn}
It is easy to check that this definition satisfies the following desired
properties.
If the runs of $\ottnt{e_{{\mathrm{1}}}}$ and $\ottnt{e_{{\mathrm{2}}}}$ terminate with traces $\varpi_{{\mathrm{1}}}$ in
$ { \Phi_{{\mathrm{1}}} }^\mu $ and $\varpi_{{\mathrm{2}}}$ in $ { \Phi_{{\mathrm{2}}} }^\mu $, respectively, then the
$ \mathsf{let} $-expression generates the trace $\varpi_{{\mathrm{1}}} \,  \tceappendop  \, \varpi_{{\mathrm{2}}}$, which can be accepted by
$ { \ottsym{(}  \Phi_{{\mathrm{1}}} \,  \tceappendop  \, \Phi_{{\mathrm{2}}}  \ottsym{)} }^\mu $.
The $ \mathsf{let} $-expression diverges with trace $\pi$ if either $\ottnt{e_{{\mathrm{1}}}}$ diverges with
$\pi$ in $ { \Phi_{{\mathrm{1}}} }^\nu $, or $\ottnt{e_{{\mathrm{1}}}}$ terminates with a trace $\varpi_{{\mathrm{1}}}$ in
$ { \Phi_{{\mathrm{1}}} }^\mu $ and then $\ottnt{e_{{\mathrm{2}}}}$ diverges with a trace $\pi_{{\mathrm{2}}}$ in $ { \Phi_{{\mathrm{2}}} }^\nu $
such that $ \pi    =    \varpi_{{\mathrm{1}}} \,  \tceappendop  \, \pi_{{\mathrm{2}}} $.
In either case, the trace $\pi$ can be accepted by $ { \ottsym{(}  \Phi_{{\mathrm{1}}} \,  \tceappendop  \, \Phi_{{\mathrm{2}}}  \ottsym{)} }^\nu $.
For control effects, the effect $\ottnt{S_{{\mathrm{1}}}}  \gg\!\!=  \ottsym{(}   \lambda\!  \, \mathit{x}  \ottsym{.}  \ottnt{S_{{\mathrm{2}}}}  \ottsym{)}$ of the $ \mathsf{let} $-expression
is defined as follows.
%
\begin{defn}[Control Effects Composition]
 For control effects $\ottnt{S_{{\mathrm{1}}}}$ and $\ottnt{S_{{\mathrm{2}}}}$, we define the control effect
 $\ottnt{S_{{\mathrm{1}}}}  \gg\!\!=  \ottsym{(}   \lambda\!  \, \mathit{x_{{\mathrm{1}}}}  \ottsym{.}  \ottnt{S_{{\mathrm{2}}}}  \ottsym{)}$ as follows:
 \[\begin{array}{rcl}
   \square   \gg\!\!=  \ottsym{(}   \lambda\!  \, \mathit{x_{{\mathrm{1}}}}  \ottsym{.}   \square   \ottsym{)} &\defeq&  \square  \\
   ( \forall  \mathit{x_{{\mathrm{1}}}}  .  \ottnt{C}  )  \Rightarrow   \ottnt{C_{{\mathrm{1}}}}   \gg\!\!=  \ottsym{(}   \lambda\!  \, \mathit{x_{{\mathrm{1}}}}  \ottsym{.}   ( \forall  \mathit{x_{{\mathrm{2}}}}  .  \ottnt{C_{{\mathrm{2}}}}  )  \Rightarrow   \ottnt{C}   \ottsym{)} &\defeq&  ( \forall  \mathit{x_{{\mathrm{2}}}}  .  \ottnt{C_{{\mathrm{2}}}}  )  \Rightarrow   \ottnt{C_{{\mathrm{1}}}}  \quad \text{(if $\mathit{x_{{\mathrm{1}}}} \,  \not\in  \, \ottsym{(}   \mathit{fv}  (  \ottnt{C_{{\mathrm{2}}}}  )  \,  \mathbin{\backslash}  \,  \{  \mathit{x_{{\mathrm{2}}}}  \}   \ottsym{)}$)} ~.
   \end{array}
 \]
\end{defn}
This definition considers only two cases: $\ottnt{S_{{\mathrm{1}}}} \,  =  \,  \square $ and
$\ottnt{S_{{\mathrm{2}}}} \,  =  \,  \square $, or $\ottnt{S_{{\mathrm{1}}}} \,  =  \,  ( \forall  \mathit{x_{{\mathrm{1}}}}  .  \ottnt{C'_{{\mathrm{1}}}}  )  \Rightarrow   \ottnt{C_{{\mathrm{1}}}} $ and $\ottnt{S_{{\mathrm{2}}}} \,  =  \,  ( \forall  \mathit{x_{{\mathrm{2}}}}  .  \ottnt{C'_{{\mathrm{2}}}}  )  \Rightarrow   \ottnt{C_{{\mathrm{2}}}} $.
If only one of $\ottnt{S_{{\mathrm{1}}}}$ and $\ottnt{S_{{\mathrm{2}}}}$ is $ \square $, then it can be converted to
a dependent temporal effect via subtyping.
The definition is justified as follows.
If $\ottnt{S_{{\mathrm{1}}}} \,  =  \,  \square $ and $\ottnt{S_{{\mathrm{2}}}} \,  =  \,  \square $, i.e., neither $\ottnt{e_{{\mathrm{1}}}}$ nor $\ottnt{e_{{\mathrm{2}}}}$
invokes the {\shiftzero} operator, then the entire expression does not invoke it either.
Therefore, the assigned control effect $\ottnt{S_{{\mathrm{1}}}}  \gg\!\!=  \ottsym{(}   \lambda\!  \, \mathit{x_{{\mathrm{1}}}}  \ottsym{.}  \ottnt{S_{{\mathrm{2}}}}  \ottsym{)}$ is $ \square $.
If $\ottnt{S_{{\mathrm{1}}}} \,  =  \,  ( \forall  \mathit{x_{{\mathrm{1}}}}  .  \ottnt{C'_{{\mathrm{1}}}}  )  \Rightarrow   \ottnt{C_{{\mathrm{1}}}} $ and $\ottnt{S_{{\mathrm{2}}}} \,  =  \,  ( \forall  \mathit{x_{{\mathrm{2}}}}  .  \ottnt{C_{{\mathrm{2}}}}  )  \Rightarrow   \ottnt{C'_{{\mathrm{2}}}} $, i.e., both
$\ottnt{e_{{\mathrm{1}}}}$ and $\ottnt{e_{{\mathrm{2}}}}$ may invoke the {\shiftzero} operator, then the entire
expression may also do.
Recall that the bound variable $\mathit{x_{{\mathrm{1}}}}$ in $\ottnt{S_{{\mathrm{1}}}}$ denotes values passed to
contexts.
Because the value of $\ottnt{e_{{\mathrm{1}}}}$ is passed, and the $ \mathsf{let} $-expression names it
$\mathit{x_{{\mathrm{1}}}}$, we can suppose that the bound variables in $\ottnt{S_{{\mathrm{1}}}}$ and the
$ \mathsf{let} $-expression have the same name $\mathit{x_{{\mathrm{1}}}}$.
Because the type $\ottnt{C'_{{\mathrm{1}}}}$ in $\ottnt{S_{{\mathrm{1}}}}$ is the requirement for the context up to
the closest {\resetzero} construct, and the type $\ottnt{C'_{{\mathrm{2}}}}$ in $\ottnt{S_{{\mathrm{2}}}}$ is the
guarantee for the behavior of the {\resetzero} construct, $\ottnt{C'_{{\mathrm{2}}}}$ must imply
$\ottnt{C'_{{\mathrm{1}}}}$, that is, $\ottnt{C'_{{\mathrm{2}}}}$ must be a subtype of $\ottnt{C'_{{\mathrm{1}}}}$.
This is ensured by requiring that $\ottnt{C'_{{\mathrm{1}}}}$ and $\ottnt{C'_{{\mathrm{2}}}}$ be the same: if
$\ottnt{C'_{{\mathrm{2}}}}$ is a subtype of $\ottnt{C'_{{\mathrm{1}}}}$, the expression $\ottnt{e_{{\mathrm{2}}}}$ with control effect
$ ( \forall  \mathit{x_{{\mathrm{2}}}}  .  \ottnt{C_{{\mathrm{2}}}}  )  \Rightarrow   \ottnt{C'_{{\mathrm{2}}}} $ can be assigned the effect $ ( \forall  \mathit{x_{{\mathrm{2}}}}  .  \ottnt{C_{{\mathrm{2}}}}  )  \Rightarrow   \ottnt{C'_{{\mathrm{1}}}} $.
Therefore, the composition operation $ \gg\!\!= $ supposes the types $\ottnt{C'_{{\mathrm{1}}}}$ and
$\ottnt{C'_{{\mathrm{2}}}}$ to be the same.
Furthermore, because the type $\ottnt{C_{{\mathrm{1}}}}$ in $\ottnt{S_{{\mathrm{1}}}}$ represents the guarantee for the
closest {\resetzero} construct and the type $\ottnt{C_{{\mathrm{2}}}}$ in $\ottnt{S_{{\mathrm{2}}}}$ represents the
requirement for the context up to it, the composition operation $ \gg\!\!= $ assigns
the control effect $ ( \forall  \mathit{x_{{\mathrm{2}}}}  .  \ottnt{C_{{\mathrm{2}}}}  )  \Rightarrow   \ottnt{C_{{\mathrm{1}}}} $ to the $ \mathsf{let} $-expression.
Finally, the side condition $\mathit{x_{{\mathrm{1}}}} \,  \not\in  \,  \mathit{fv}  (  \ottnt{C_{{\mathrm{2}}}}  )  \,  \mathbin{\backslash}  \,  \{  \mathit{x_{{\mathrm{2}}}}  \} $ is imposed for avoiding
the problem of the scope escaping.

The event-raising operator is typechecked by \T{Event}.  An expression
$ \mathsf{ev}  [  \mathbf{a}  ] $ is given the computation type $ \ottsym{\{}  \mathit{x} \,  {:}  \,  \mathsf{unit}  \,  |  \,  \top   \ottsym{\}}  \,  \,\&\,  \,   (  \mathbf{a}  , \bot )   \, / \,   \square  $
because
it evaluates to the unit value $ () $, 
yields the finite trace $\mathbf{a}$, and
does not invoke the {\shiftzero} operator.
If the {\shiftzero} operator is invoked before, we need to check that raising
the event $\mathbf{a}$ is consistent with the assumption on the continuations
captured by the invocation.
This check is done via subtyping presented in \refsec{eff:type-and-eff:subtyping}.

A diverging expression $ \bullet ^{ \pi } $ is typechecked by \T{Div}.
Because it diverges with the infinite trace $\pi$, the temporal effect
$ \ottnt{C}  . \Phi $ in its computation type $\ottnt{C}$ has to include $\pi$.
Furthermore, if $ \ottnt{C}  . \ottnt{S}  \,  =  \,  ( \forall  \mathit{x_{{\mathrm{1}}}}  .  \ottnt{C_{{\mathrm{1}}}}  )  \Rightarrow   \ottnt{C_{{\mathrm{2}}}} $ for some $\mathit{x}$, $\ottnt{C_{{\mathrm{1}}}}$, and $\ottnt{C_{{\mathrm{2}}}}$,
the temporal effect $ \ottnt{C_{{\mathrm{2}}}}  . \Phi $ in the type $\ottnt{C_{{\mathrm{2}}}}$ also has to include
$\pi$ because the type $\ottnt{C_{{\mathrm{2}}}}$ must capture the behavior of the
closest {\resetzero} construct enclosing $ \bullet ^{ \pi } $, that is, the divergence
with $\pi$.
In general, all the temporal effects in $\ottnt{C}$ that specify the behavior of the
enclosing {\resetzero} constructs have to include $\pi$.
The rule \T{Div} formalizes this requirement by checking the validity of the
formula $ { \ottnt{C} }^\nu (  \pi  ) $, which is defined as follows.
\begin{defn}
 For a type $\ottnt{C}$ and control effect $\ottnt{S}$,
 we define the formulas $ { \ottnt{C} }^\nu (  \ottnt{t}  ) $ and $ { \ottnt{S} }^\nu (  \ottnt{t}  ) $ as follows.
 \[\begin{array}{l@{\ \ }l@{\ \ }l@{\qquad}l@{\ \ }l@{\ \ }l@{\qquad}l@{\ \ }l@{\ \ }l}
   { \ottnt{C} }^\nu (  \ottnt{t}  )     &\defeq&   { \ottsym{(}   \ottnt{C}  . \Phi   \ottsym{)} }^\nu   \ottsym{(}  \ottnt{t}  \ottsym{)}    \wedge     { \ottsym{(}   \ottnt{C}  . \ottnt{S}   \ottsym{)} }^\nu (  \ottnt{t}  )   &
   {  \square  }^\nu (  \ottnt{t}  )            &\defeq&  \top  &
   { \ottsym{(}   ( \forall  \mathit{x}  .  \ottnt{C_{{\mathrm{1}}}}  )  \Rightarrow   \ottnt{C_{{\mathrm{2}}}}   \ottsym{)} }^\nu (  \ottnt{t}  )  &\defeq&  { \ottnt{C_{{\mathrm{2}}}} }^\nu (  \ottnt{t}  ) 
   \end{array}
 \]
\end{defn}

The rule \T{Fun} for recursive functions is obtained by adapting a typing rule
for recursive functions in the previous
work~\cite{Nanjo/Unno/Koskinen/Terauchi_2018_LICS} to the control operators and
clarifying implicit assumptions in it.
In this rule, argument variables that can occur in formulas, and their sorts are
supposed to be $ \tceseq{ \mathit{z_{{\mathrm{0}}}} } $ and $ \tceseq{ \iota_{{\mathrm{0}}} } $, respectively.  The metafunction
$ \gamma $ filters out a pair $(\mathit{z},\ottnt{T_{{\mathrm{1}}}})$ of an argument variable $\mathit{z}$
and its type $\ottnt{T_{{\mathrm{1}}}}$ if $\ottnt{T_{{\mathrm{1}}}}$ is not a refinement type---as variables of
non-refinement types cannot occur in well-formed types and effects---and maps
the other pairs to $(\mathit{z},\iota)$ where $\iota$ is the underlying base type of the
refinement type $\ottnt{T_{{\mathrm{1}}}}$.
The body $\ottnt{e}$ of a recursive function is typechecked under the typing context
augmented with the bindings for $ \tceseq{ \mathit{X} } $, $ \tceseq{ \mathit{Y} } $, variable $\mathit{f}$, which
denotes the recursive function itself, and argument variables $ \tceseq{ \mathit{z} } $.
This premise in \T{Fun} is expressed using the notation
$ \tceseq{ \mathit{X} }  \,  {:}  \,  \tceseq{ \ottnt{s} } $ that denotes a typing context
$\mathit{X_{{\mathrm{1}}}} \,  {:}  \,  \tceseq{ \ottnt{s} }   \ottsym{,}   \cdots   \ottsym{,}  \mathit{X_{\ottmv{n}}} \,  {:}  \,  \tceseq{ \ottnt{s} } $ when $ \tceseq{ \mathit{X} }  = \mathit{X_{{\mathrm{1}}}}, \cdots, \mathit{X_{\ottmv{n}}}$.
The sorts of $ \tceseq{ \mathit{X} } $ and $ \tceseq{ \mathit{Y} } $ state that they are predicates on finite
and infinite traces, respectively, and given arguments.

A key idea in Nanjo et al.'s system is that, during typechecking the body $\ottnt{e}$, it assumes that
traces yielded by the recursive function (denoted by $\mathit{f}$) are
specified by predicate variables $\mathit{X}$ and $\mathit{Y}$.
%
%
Then, the typechecking process collects the information about what trace $\ottnt{e}$
yields using the ``assumptions'' $\mathit{X}$ and $\mathit{Y}$.
For example, consider expression $\mathsf{if} \, \ottnt{v} \, \mathsf{then} \,  \mathsf{ev}  [  \mathbf{a_{{\mathrm{1}}}}  ]  \, \mathsf{else} \, \ottsym{(}   \mathsf{ev}  [  \mathbf{a_{{\mathrm{2}}}}  ]   \ottsym{;}  \mathit{f} \,  ()   \ottsym{)}$ as
the body, and assume that a temporal effect $ (   \lambda_\mu\!   \,  \mathit{x}  .\,  \mathit{X}  \ottsym{(}  \mathit{x}  \ottsym{)}  ,   \lambda_\nu\!   \,  \mathit{y}  .\,  \mathit{Y}  \ottsym{(}  \mathit{y}  \ottsym{)}  ) $ is
assigned to $\mathit{f}$ during the typechecking.
%
%
The typechecking can assign to the body a temporal effect
$\Phi \defeq  (   \lambda_\mu\!   \,  \mathit{x}  .\,    \mathit{x}    =    \mathbf{a_{{\mathrm{1}}}}     \vee    \ottsym{(}    \exists  \,  \mathit{x'}  . \    \mathit{x}    =    \mathbf{a_{{\mathrm{2}}}} \,  \tceappendop  \, \mathit{x'}     \wedge    \mathit{X}  \ottsym{(}  \mathit{x'}  \ottsym{)}    \ottsym{)}   ,   \lambda_\nu\!   \,  \mathit{y}  .\,    \exists  \,  \mathit{y'}  . \    \mathit{y}    =    \mathbf{a_{{\mathrm{2}}}} \,  \tceappendop  \, \mathit{y'}     \wedge    \mathit{Y}  \ottsym{(}  \mathit{y'}  \ottsym{)}    ) $.
The finite part means that a yielded finite trace is $\mathbf{a_{{\mathrm{1}}}}$, or
$\mathbf{a_{{\mathrm{2}}}}$ followed by one in $\mathit{X}$.
The infinite part means that a yielded infinite trace is $\mathbf{a_{{\mathrm{2}}}}$ followed by
one in $\mathit{Y}$.
Because $\mathit{X}$ and $\mathit{Y}$ represent traces yielded by the recursive function
itself, the entire finite and infinite traces can be expressed by
$\ottsym{(}   \mu\!  \, \mathit{X}  \ottsym{(}  \mathit{x} \,  {:}  \,  \Sigma ^\ast   \ottsym{)}  \ottsym{.}   { \Phi }^\mu   \ottsym{(}  \mathit{x}  \ottsym{)}  \ottsym{)}$ and $\ottsym{(}   \nu\!  \, \mathit{Y}  \ottsym{(}  \mathit{y} \,  {:}  \,  \Sigma ^\ast   \ottsym{)}  \ottsym{.}   { \Phi }^\nu   \ottsym{(}  \mathit{y}  \ottsym{)}  \ottsym{)}$, respectively.

We extend this idea to delimited control operators.
Our novelty is to prepare different ``assumptions'' for different
kinds of contexts.
Consider expression
$\ottsym{(}  \mathcal{S}_0 \, \mathit{x}  \ottsym{.}   \mathsf{ev}  [  \mathbf{a_{{\mathrm{1}}}}  ]   \ottsym{;}  \mathit{x} \,  ()   \ottsym{)}  \ottsym{;}   \mathsf{ev}  [  \mathbf{a_{{\mathrm{2}}}}  ]   \ottsym{;}  \mathsf{if} \, \ottnt{v} \, \mathsf{then} \,  ()  \, \mathsf{else} \, \mathit{f} \,  () $
as the body of a recursive function.
For a caller of the function, the body looks as if it yields a finite or
infinite trace only containing the event $\mathbf{a_{{\mathrm{2}}}}$.  Note that the caller cannot
know what happens from operating {\shiftzero} to, if any, the invocation of the captured
continuation.
However, for the meta-context---i.e., the context enclosing the closest
{\resetzero} construct---the trace yielded by the call to this function contains
$\mathbf{a_{{\mathrm{1}}}}$ followed by $\mathbf{a_{{\mathrm{2}}}}$.
The use of different, multiple predicate variables enables representing the
different views of the behavior for the different contexts.

To implement this idea, \T{Fun} uses three judgments.
First, $ \tceseq{ \ottsym{(}  \mathit{X}  \ottsym{,}  \mathit{Y}  \ottsym{)} }   \ottsym{;}   \tceseq{ \mathit{z_{{\mathrm{0}}}}  \,  {:}  \,  \iota_{{\mathrm{0}}} }   \vdash  \ottnt{C_{{\mathrm{0}}}}  \succ  \ottnt{C}$ checks that the return
type $\ottnt{C_{{\mathrm{0}}}}$ of $\mathit{f}$ appropriately represents the assumptions on traces
for different contexts using $ \tceseq{ \mathit{X} } $ and $ \tceseq{ \mathit{Y} } $, and the other
parts are the same as the type $\ottnt{C}$ inferred from the body $\ottnt{e}$.
\begin{defn}[Consistency]
 The judgments
 $ \tceseq{ \ottsym{(}  \mathit{X}  \ottsym{,}  \mathit{Y}  \ottsym{)} }   \ottsym{;}   \tceseq{ \mathit{z_{{\mathrm{0}}}}  \,  {:}  \,  \iota_{{\mathrm{0}}} }   \vdash  \ottnt{C_{{\mathrm{0}}}}  \succ  \ottnt{C}$ and
 $ \tceseq{ \ottsym{(}  \mathit{X}  \ottsym{,}  \mathit{Y}  \ottsym{)} }   \ottsym{;}   \tceseq{ \mathit{z_{{\mathrm{0}}}}  \,  {:}  \,  \iota_{{\mathrm{0}}} }   \vdash  \ottnt{S_{{\mathrm{0}}}}  \succ  \ottnt{S}$
 are defined as the smallest relations satisfying the following rules.
 \\[-1ex]
 \begin{center}
  $\ottdrule[]{%
    { \Phi_{{\mathrm{0}}} \,  =  \,  (   \lambda_\mu\!   \,  \mathit{x}  .\,  \mathit{X_{{\mathrm{0}}}}  \ottsym{(}  \mathit{x}  \ottsym{,}   \tceseq{ \mathit{z_{{\mathrm{0}}}} }   \ottsym{)}  ,   \lambda_\nu\!   \,  \mathit{y}  .\,  \mathit{Y_{{\mathrm{0}}}}  \ottsym{(}  \mathit{y}  \ottsym{,}   \tceseq{ \mathit{z_{{\mathrm{0}}}} }   \ottsym{)}  )  } \,\mathrel{\wedge}\, {  \tceseq{ \ottsym{(}  \mathit{X}  \ottsym{,}  \mathit{Y}  \ottsym{)} }   \ottsym{;}   \tceseq{ \mathit{z_{{\mathrm{0}}}}  \,  {:}  \,  \iota_{{\mathrm{0}}} }   \vdash  \ottnt{S_{{\mathrm{0}}}}  \succ  \ottnt{S} } 
  }{
   \ottsym{(}  \mathit{X_{{\mathrm{0}}}}  \ottsym{,}  \mathit{Y_{{\mathrm{0}}}}  \ottsym{)}  \ottsym{,}   \tceseq{ \ottsym{(}  \mathit{X}  \ottsym{,}  \mathit{Y}  \ottsym{)} }   \ottsym{;}   \tceseq{ \mathit{z_{{\mathrm{0}}}}  \,  {:}  \,  \iota_{{\mathrm{0}}} }   \vdash   \ottnt{T}  \,  \,\&\,  \,  \Phi_{{\mathrm{0}}}  \, / \,  \ottnt{S_{{\mathrm{0}}}}   \succ   \ottnt{T}  \,  \,\&\,  \,  \Phi  \, / \,  \ottnt{S} 
  }{%
  }$
  \\[1.5ex]
  $\ottdrule[]{%
  }{
    \emptyset   \ottsym{;}   \tceseq{ \mathit{z_{{\mathrm{0}}}}  \,  {:}  \,  \iota_{{\mathrm{0}}} }   \vdash   \square   \succ   \square 
  }{%
  }$
  \hfil
  $\ottdrule[]{%
    \tceseq{ \ottsym{(}  \mathit{X}  \ottsym{,}  \mathit{Y}  \ottsym{)} }   \ottsym{;}   \tceseq{ \mathit{z_{{\mathrm{0}}}}  \,  {:}  \,  \iota_{{\mathrm{0}}} }   \vdash  \ottnt{C_{{\mathrm{02}}}}  \succ  \ottnt{C_{{\mathrm{2}}}}
  }{
    \tceseq{ \ottsym{(}  \mathit{X}  \ottsym{,}  \mathit{Y}  \ottsym{)} }   \ottsym{;}   \tceseq{ \mathit{z_{{\mathrm{0}}}}  \,  {:}  \,  \iota_{{\mathrm{0}}} }   \vdash   ( \forall  \mathit{x}  .  \ottnt{C}  )  \Rightarrow   \ottnt{C_{{\mathrm{02}}}}   \succ   ( \forall  \mathit{x}  .  \ottnt{C}  )  \Rightarrow   \ottnt{C_{{\mathrm{2}}}} 
  }{%
  }$
 \end{center}
\end{defn}
The second judgment $ \emptyset   \ottsym{;}   \tceseq{ \ottsym{(}  \mathit{X}  \ottsym{,}  \mathit{Y}  \ottsym{)} }  \,  \vdash ^{\circ}  \, \ottnt{C}$ checks that there is
no occurrence of $ \tceseq{ \mathit{X} } $ and $ \tceseq{ \mathit{Y} } $ at inappropriate positions in type
$\ottnt{C}$.
For example, it ensures that $ \tceseq{ \mathit{X} } $ and $ \tceseq{ \mathit{Y} } $ do not occur negatively in
the formulas of the temporal effects.
This condition is imposed for our proof of soundness.
See the supplementary material for the detail of the occurrence checking.
The third judgment $ \tceseq{ \ottsym{(}  \mathit{X}  \ottsym{,}  \mathit{Y}  \ottsym{)} }   \ottsym{;}   \tceseq{ \mathit{z_{{\mathrm{0}}}}  \,  {:}  \,  \iota_{{\mathrm{0}}} }  \,  |  \, \ottnt{C}  \vdash  \sigma$ infers the predicates for $ \tceseq{ \mathit{X} } $ and $ \tceseq{ \mathit{Y} } $ from the type $\ottnt{C}$.
The metavariable $\sigma$ ranges over predicate substitutions.
For substitutions $\sigma$ and $\sigma'$ with disjoint domains,
substitution $\sigma  \mathbin{\uplus}  \sigma'$ maps predicate variables $\mathit{X}$ to $\sigma  \ottsym{(}  \mathit{X}  \ottsym{)}$ or $\sigma'  \ottsym{(}  \mathit{X}  \ottsym{)}$
depending on whether $\mathit{X} \,  \in  \, \mathit{dom} \, \ottsym{(}  \sigma  \ottsym{)}$ or $\mathit{X} \,  \in  \, \mathit{dom} \, \ottsym{(}  \sigma'  \ottsym{)}$.
We write $\sigma  \ottsym{(}  \sigma'  \ottsym{)}$ for the substitution obtained by applying $\sigma$ to every predicate in the codomain of $\sigma'$.
\begin{defn}[Predicate Fixpoints]
 We write
 $  \mu\!   \, (  \mathit{X} ,  \mathit{Y} ,   \tceseq{ \mathit{z_{{\mathrm{0}}}}  \,  {:}  \,  \iota_{{\mathrm{0}}} }  ,  \Phi  )  \defeq \ottsym{(}   \mu\!  \, \mathit{X}  \ottsym{(}  \mathit{x} \,  {:}  \,  \Sigma ^\ast   \ottsym{,}   \tceseq{ \mathit{z_{{\mathrm{0}}}}  \,  {:}  \,  \iota_{{\mathrm{0}}} }   \ottsym{)}  \ottsym{.}   { \Phi }^\mu   \ottsym{(}  \mathit{x}  \ottsym{)}  \ottsym{)}$ and
 $  \nu\!   \, (  \mathit{X} ,  \mathit{Y} ,   \tceseq{ \mathit{z_{{\mathrm{0}}}}  \,  {:}  \,  \iota_{{\mathrm{0}}} }  ,  \Phi  )  \defeq  \ottsym{(}   \nu\!  \, \mathit{Y}  \ottsym{(}  \mathit{y} \,  {:}  \,  \Sigma ^\omega   \ottsym{,}   \tceseq{ \mathit{z_{{\mathrm{0}}}}  \,  {:}  \,  \iota_{{\mathrm{0}}} }   \ottsym{)}  \ottsym{.}   { \Phi }^\nu   \ottsym{(}  \mathit{y}  \ottsym{)}  \ottsym{)}    [    \mu\!   \, (  \mathit{X} ,  \mathit{Y} ,   \tceseq{ \mathit{z_{{\mathrm{0}}}}  \,  {:}  \,  \iota_{{\mathrm{0}}} }  ,  \Phi  )   /  \mathit{X}  ]  $
 for fresh $\mathit{x}, \mathit{y}$.
 The judgments $ \tceseq{ \ottsym{(}  \mathit{X}  \ottsym{,}  \mathit{Y}  \ottsym{)} }   \ottsym{;}   \tceseq{ \mathit{z_{{\mathrm{0}}}}  \,  {:}  \,  \iota_{{\mathrm{0}}} }  \,  |  \, \ottnt{C}  \vdash  \sigma$ and $ \tceseq{ \ottsym{(}  \mathit{X}  \ottsym{,}  \mathit{Y}  \ottsym{)} }   \ottsym{;}   \tceseq{ \mathit{z_{{\mathrm{0}}}}  \,  {:}  \,  \iota_{{\mathrm{0}}} }  \,  |  \, \ottnt{S}  \vdash  \sigma$ are the smallest relations satisfying the following rules.
 \\[-1ex]
 \begin{center}
  $\ottdrule[]{
     \tceseq{ \ottsym{(}  \mathit{X}  \ottsym{,}  \mathit{Y}  \ottsym{)} }   \ottsym{;}   \tceseq{ \mathit{z_{{\mathrm{0}}}}  \,  {:}  \,  \iota_{{\mathrm{0}}} }  \,  |  \, \ottnt{S}  \vdash  \sigma  \quad  \sigma' \,  =  \,  [    \mu\!   \, (  \mathit{X_{{\mathrm{0}}}} ,  \mathit{Y_{{\mathrm{0}}}} ,   \tceseq{ \mathit{z_{{\mathrm{0}}}}  \,  {:}  \,  \iota_{{\mathrm{0}}} }  ,  \Phi  )   /  \mathit{X_{{\mathrm{0}}}}  ]   \mathbin{\uplus}   [    \nu\!   \, (  \mathit{X_{{\mathrm{0}}}} ,  \mathit{Y_{{\mathrm{0}}}} ,   \tceseq{ \mathit{z_{{\mathrm{0}}}}  \,  {:}  \,  \iota_{{\mathrm{0}}} }  ,  \Phi  )   /  \mathit{Y_{{\mathrm{0}}}}  ]  
  }{
   \ottsym{(}  \mathit{X_{{\mathrm{0}}}}  \ottsym{,}  \mathit{Y_{{\mathrm{0}}}}  \ottsym{)}  \ottsym{,}   \tceseq{ \ottsym{(}  \mathit{X}  \ottsym{,}  \mathit{Y}  \ottsym{)} }   \ottsym{;}   \tceseq{ \mathit{z_{{\mathrm{0}}}}  \,  {:}  \,  \iota_{{\mathrm{0}}} }  \,  |  \,  \ottnt{T}  \,  \,\&\,  \,  \Phi  \, / \,  \ottnt{S}   \vdash  \sigma'  \ottsym{(}  \sigma  \ottsym{)}  \mathbin{\uplus}  \sigma'
  }{
  }$
  \\[1.5ex]
  $\ottdrule[]{
  }{
    \emptyset   \ottsym{;}   \tceseq{ \mathit{z_{{\mathrm{0}}}}  \,  {:}  \,  \iota_{{\mathrm{0}}} }  \,  |  \,  \square   \vdash   \emptyset 
  }{
  }$
  \hfil
  $\ottdrule[]{
    \tceseq{ \ottsym{(}  \mathit{X}  \ottsym{,}  \mathit{Y}  \ottsym{)} }   \ottsym{;}   \tceseq{ \mathit{z_{{\mathrm{0}}}}  \,  {:}  \,  \iota_{{\mathrm{0}}} }  \,  |  \, \ottnt{C_{{\mathrm{2}}}}  \vdash  \sigma
  }{
    \tceseq{ \ottsym{(}  \mathit{X}  \ottsym{,}  \mathit{Y}  \ottsym{)} }   \ottsym{;}   \tceseq{ \mathit{z_{{\mathrm{0}}}}  \,  {:}  \,  \iota_{{\mathrm{0}}} }  \,  |  \,  ( \forall  \mathit{x}  .  \ottnt{C_{{\mathrm{1}}}}  )  \Rightarrow   \ottnt{C_{{\mathrm{2}}}}   \vdash  \sigma
  }{
  }$
 \end{center}
\end{defn}
If predicate variables in $ \tceseq{ \mathit{X} } $ occur in the infinite parts of some temporal effects in $\ottnt{C}$,
they are replaced by the least fixpoints.
The occurrence checking ensures that $ \tceseq{ \mathit{Y} } $ never occur in the finite parts.
The return type of the recursive function is determined by substituting the
inferred predicates for predicate variables in $\ottnt{C_{{\mathrm{0}}}}$.

\subsubsection{Subtyping}
\label{sec:eff:type-and-eff:subtyping}
\begin{figure}[t]
 \begin{flushleft}
  \textbf{Subtyping rules}
  \quad \framebox{$\Gamma  \vdash  \ottnt{T_{{\mathrm{1}}}}  \ottsym{<:}  \ottnt{T_{{\mathrm{2}}}}$}
  \     \framebox{$\Gamma  \vdash  \ottnt{C_{{\mathrm{1}}}}  \ottsym{<:}  \ottnt{C_{{\mathrm{2}}}}$}
  \     \framebox{$\Gamma \,  |  \, \ottnt{T} \,  \,\&\,  \, \Phi  \vdash  \ottnt{S_{{\mathrm{1}}}}  \ottsym{<:}  \ottnt{S_{{\mathrm{2}}}}$}
  \     \framebox{$\Gamma  \vdash  \Phi_{{\mathrm{1}}}  \ottsym{<:}  \Phi_{{\mathrm{2}}}$}
 \end{flushleft}
 \begin{center}
  $\ottdruleSXXRefine{}$ \hfil
  $\ottdruleSXXFun{}$ \\[2ex]
  {}
  $\ottdruleSXXComp{}$ \hfil
  $\ottdruleSXXEff{}$ \\[2ex]
  $\ottdruleSXXEmpty{}$ \hfil
  $\ottdruleSXXAns{}$ \\[2ex]
  $\ottdruleSXXEmbed{}$ \hfil
 \end{center}
 \caption{Subtyping.}
 \label{fig:subtyping}
\end{figure}
Subtyping is defined for every type-level construct.
Judgment $\Gamma  \vdash  \ottnt{T_{{\mathrm{1}}}}  \ottsym{<:}  \ottnt{T_{{\mathrm{2}}}}$ for value types states that a value of
type $\ottnt{T_{{\mathrm{1}}}}$ can be used as type $\ottnt{T_{{\mathrm{2}}}}$.
Judgment $\Gamma  \vdash  \ottnt{C_{{\mathrm{1}}}}  \ottsym{<:}  \ottnt{C_{{\mathrm{2}}}}$ for computation types states that an
expression of type $\ottnt{C_{{\mathrm{1}}}}$ behaves as type $\ottnt{C_{{\mathrm{2}}}}$.
Judgment $\Gamma \,  |  \, \ottnt{T} \,  \,\&\,  \, \Phi  \vdash  \ottnt{S_{{\mathrm{1}}}}  \ottsym{<:}  \ottnt{S_{{\mathrm{2}}}}$ for control effects of expressions with
value type $\ottnt{T}$ and temporal effect $\Phi$ states that the requirement of
$\ottnt{S_{{\mathrm{1}}}}$ for contexts is strengthened to that of $\ottnt{S_{{\mathrm{2}}}}$, and that the
guarantee of $\ottnt{S_{{\mathrm{1}}}}$ for closest {\resetzero} constructs is weakened to that of
$\ottnt{S_{{\mathrm{2}}}}$.
Judgment $\Gamma  \vdash  \Phi_{{\mathrm{1}}}  \ottsym{<:}  \Phi_{{\mathrm{2}}}$ for temporal effects states that
finite and infinite traces in $\Phi_{{\mathrm{1}}}$ are contained in $\Phi_{{\mathrm{2}}}$.
The subtyping rules are presented in \reffig{subtyping}.

The rules \Srule{Refine} and \Srule{Eff} weaken the predicates of refinement types
and temporal effects, respectively.
These rules reduce subtype checking to validity checking of formulas.
{\iffull
\else
The subtyping rule for function types (\Srulewop{Fun}) is standard.
%
\fi}
Subtyping between computation types is determined by subtyping between their
components (\Srulewop{Comp}).
The rule \Srule{Empty} only states reflexivity of $ \square $.
%
%
The rule \Srule{Ans} states that a dependent control effect $ ( \forall  \mathit{x}  .  \ottnt{C_{{\mathrm{11}}}}  )  \Rightarrow   \ottnt{C_{{\mathrm{12}}}} $ is
a subeffect of another one $ ( \forall  \mathit{x}  .  \ottnt{C_{{\mathrm{21}}}}  )  \Rightarrow   \ottnt{C_{{\mathrm{22}}}} $ if the supereffect side
requires more for contexts and guarantees less for closest {\resetzero}
constructs than the subeffect side.

The rule \Srule{Embed}, which is the most interesting part of our subtyping,
allows the control effect $ \square $ to be a subeffect of a dependent control
effect $ ( \forall  \mathit{x}  .  \ottnt{C_{{\mathrm{1}}}}  )  \Rightarrow   \ottnt{C_{{\mathrm{2}}}} $.
This rule is based on the idea in \citet{Materzok/Biernacki_2011_ICFP} that
$ \square $ can be a subeffect of an impure effect $\ottnt{C_{{\mathrm{1}}}}  \Rightarrow  \ottnt{C_{{\mathrm{2}}}}$ if $\ottnt{C_{{\mathrm{1}}}}$ is
a subtype of $\ottnt{C_{{\mathrm{2}}}}$ (note that the prior work addresses neither temporal
effects nor dependent typing).
This idea is justified as follows.
Because expressions with the pure effect $ \square $ never invokes the
{\shiftzero} operator, what they can guarantee for contexts is only what they require for the contexts.
Thus, if the requirement $\ottnt{C_{{\mathrm{1}}}}$ implies the guarantee $\ottnt{C_{{\mathrm{2}}}}$,
expressions with $ \square $ can be assigned the effect $\ottnt{C_{{\mathrm{1}}}}  \Rightarrow  \ottnt{C_{{\mathrm{2}}}}$.
We extend this idea to temporal effects and dependent typing.
Consider an expression of a computation type $ \ottnt{T}  \,  \,\&\,  \,  \Phi  \, / \,   \square  $.
In our language, this expression never invokes the {\shiftzero} operator, but
it may yield traces specified by $\Phi$.
As indicated by the semantics, events raised under a {\resetzero}
construct are propagated to its outer context.
Therefore, for assigning a control effect $ ( \forall  \mathit{x}  .  \ottnt{C_{{\mathrm{1}}}}  )  \Rightarrow   \ottnt{C_{{\mathrm{2}}}} $ to the expression, the
type $\ottnt{C_{{\mathrm{2}}}}$ must involve traces specified by $\Phi$.
After yielding a trace in $\Phi$, if the evaluation of the expression terminates,
the computation specified by $\ottnt{C_{{\mathrm{1}}}}$ is executed.
This behavior is captured by the operation $\Phi \,  \tceappendop  \, \ottnt{C_{{\mathrm{1}}}}$, which guarantees for
every enclosing {\resetzero} construct that traces specified by $\Phi$
are yielded first, and then traces specified by $\ottnt{C_{{\mathrm{1}}}}$ are yielded.
\begin{defn}
 We define the composition $\Phi \,  \tceappendop  \, \ottnt{C}$ of a temporal effect $\Phi$ and computation
 type $\ottnt{C}$, and the composition $\Phi \,  \tceappendop  \, \ottnt{S}$ of a temporal effect $\Phi$ and
 control effect $\ottnt{S}$ as follows.
 \[\begin{array}{l@{\ \,}l@{\ \,}l@{\quad\ \ }l@{\ \,}l@{\ \,}l@{\quad\ \ }l@{\ \,}l@{\ \,}l}
  \Phi \,  \tceappendop  \, \ottnt{C} &\defeq&   \ottnt{C}  . \ottnt{T}   \,  \,\&\,  \,  \Phi \,  \tceappendop  \, \ottsym{(}   \ottnt{C}  . \Phi   \ottsym{)}  \, / \,  \Phi \,  \tceappendop  \, \ottsym{(}   \ottnt{C}  . \ottnt{S}   \ottsym{)}  &
  \Phi \,  \tceappendop  \,  \square            &\defeq&  \square  &
  \Phi \,  \tceappendop  \,  ( \forall  \mathit{x}  .  \ottnt{C_{{\mathrm{1}}}}  )  \Rightarrow   \ottnt{C_{{\mathrm{2}}}}    &\defeq&  ( \forall  \mathit{x}  .  \ottnt{C_{{\mathrm{1}}}}  )  \Rightarrow   \Phi \,  \tceappendop  \, \ottnt{C_{{\mathrm{2}}}} 
   \end{array}
 \]
\end{defn}
\noindent
The rule checks $\Phi \,  \tceappendop  \, \ottnt{C_{{\mathrm{1}}}}  \ottsym{<:}  \ottnt{C_{{\mathrm{2}}}}$ under the typing context
augmented with $\mathit{x} \,  {:}  \, \ottnt{T}$ because $\ottnt{C_{{\mathrm{1}}}}$ may refer to $\mathit{x}$.
The side condition $\mathit{x} \,  \not\in  \,  \mathit{fv}  (  \Phi  )  \,  \mathbin{\cup}  \,  \mathit{fv}  (  \ottnt{C_{{\mathrm{2}}}}  ) $ is imposed to clarify that
$\mathit{x}$ is local in $\ottnt{C_{{\mathrm{1}}}}$.
\TS{The rest of this paragraph needs a check by Hiroshi.}
We also need to ensure that the infinite traces in $ { \Phi }^\nu $ are
appropriately propagated to temporal effects in $\ottnt{C_{{\mathrm{2}}}}$.
Formally, the predicate $ { \ottnt{C_{{\mathrm{2}}}} }^\nu $, which specifies the infinite behavior of
enclosing {\resetzero} constructs, must hold on any infinite traces in
$ { \Phi }^\nu $ because, when an expression diverges with an infinite trace
$\pi$ in $ { \Phi }^\nu $, the {\resetzero} constructs enclosing it also
diverges with $\pi$.
This requirement is not ensured by $\Gamma  \ottsym{,}  \mathit{x} \,  {:}  \, \ottnt{T}  \vdash  \Phi \,  \tceappendop  \, \ottnt{C_{{\mathrm{1}}}}  \ottsym{<:}  \ottnt{C_{{\mathrm{2}}}}$ because
$\Gamma  \ottsym{,}  \mathit{x} \,  {:}  \, \ottnt{T}$ may be inconsistent, i.e., there may not exist value assignments for
$\Gamma  \ottsym{,}  \mathit{x} \,  {:}  \, \ottnt{T}$.  For instance, $\Gamma  \ottsym{,}  \mathit{x} \,  {:}  \, \ottnt{T}$ is inconsistent when $\ottnt{T} \,  =  \, \ottsym{\{}  \mathit{z} \,  {:}  \, \iota \,  |  \,  \bot   \ottsym{\}}$.
If $\Gamma  \ottsym{,}  \mathit{x} \,  {:}  \, \ottnt{T}$ is inconsistent, the truth of the requirement is not ensured
by the check of $\Gamma  \ottsym{,}  \mathit{x} \,  {:}  \, \ottnt{T}  \vdash  \Phi \,  \tceappendop  \, \ottnt{C_{{\mathrm{1}}}}  \ottsym{<:}  \ottnt{C_{{\mathrm{2}}}}$.
Instead, \Srule{Embed} explicitly enforces the desired property
$\Gamma  \ottsym{,}  \mathit{y} \,  {:}  \,  \Sigma ^\omega   \models    { \Phi }^\nu   \ottsym{(}  \mathit{y}  \ottsym{)}    \Rightarrow     { \ottnt{C_{{\mathrm{2}}}} }^\nu (  \mathit{y}  )  $.
\TS{CHECK!}

Readers may wonder if \Srule{Embed} should also require that the
finite parts of the temporal effects in $\ottnt{C_{{\mathrm{2}}}}$ for {\resetzero} constructs
include the finite traces in $ { \Phi }^\mu $, as the infinite parts include the
infinite traces in $ { \Phi }^\nu $.
However, this is unnecessary.
The above problem happens when considering a computation having a type
$ \ottnt{T}  \,  \,\&\,  \,  \Phi  \, / \,   \square  $ under $\Gamma$.
If $\Gamma  \ottsym{,}  \mathit{x} \,  {:}  \, \ottnt{T}$ is inconsistent, it is guaranteed that the computation is not
executed or diverges (the pure effect $ \square $
indicates that the computation does not invoke the {\shiftzero} operator).
That is, the problem happens only when the computation diverges.  Therefore,
only the infinite traces must be propagated.


\subsection{Typing Example}
\label{sec:eff:example}
This section illustrates typechecking in our effect system via a few examples.
For readability, we use the following notation.
First, we write $\ottnt{T} \,  \,\&\,  \, \Phi$ for the computation type $ \ottnt{T}  \,  \,\&\,  \,  \Phi  \, / \,   \square  $.
Second, we write $ \ottnt{T}  \,  \,\&\,  \, ( \forall  \mathit{x}  .  \Phi_{{\mathrm{1}}}  )  \Rightarrow   \Phi_{{\mathrm{2}}} $ to express the control effect
$ ( \forall  \mathit{x}  .   \ottnt{T}  \,  \,\&\,  \,  \Phi_{{\mathrm{1}}}  \, / \,   \square    )  \Rightarrow    \ottnt{T}  \,  \,\&\,  \,  \Phi_{{\mathrm{2}}}  \, / \,   \square   $
for value type $\ottnt{T}$ in which $\mathit{x}$ does not occur.
This notation means that the value type is not modified and the underlying
control effect is pure.
We write $ \ottnt{T}  \,  \,\&\,  \,  \Phi_{{\mathrm{1}}}   \Rightarrow   \Phi_{{\mathrm{2}}} $ simply if $\mathit{x}$ does not occur in $\Phi_{{\mathrm{1}}}$.
The constraints found in this section can also be generated by our
tool implementing the effect system.

\paragraph{Example 1.}
The first example is the expression
$ \resetcnstr{ \mathsf{let} \, \mathit{x}  \ottsym{=}  \ottsym{(}  \mathcal{S}_0 \, \mathit{f}  \ottsym{.}  \mathit{f} \,  3   \ottsym{;}  \mathit{f} \,  5   \ottsym{)} \,  \mathsf{in}  \,  \mathsf{ev}  [  \mathbf{a}  ]^{ \mathit{x} }  } $ presented in \refsec{eff:ty-eff}.
The most precise type of this expression would be
$ \mathsf{unit}  \,  \,\&\,  \,  (   \mathbf{a} ^{  8  }   , \bot ) $.
To prove this type assignment possible, \T{Reset} indicates that it suffices to derive
the following judgment:
\begin{equation}
  \emptyset   \vdash  \mathsf{let} \, \mathit{x}  \ottsym{=}  \ottsym{(}  \mathcal{S}_0 \, \mathit{f}  \ottsym{.}  \mathit{f} \,  3   \ottsym{;}  \mathit{f} \,  5   \ottsym{)} \,  \mathsf{in}  \,  \mathsf{ev}  [  \mathbf{a}  ]^{ \mathit{x} }   \ottsym{:}    \mathsf{unit}   \,  \,\&\,  \,  \Phi  \, / \,    \mathsf{unit}   \,  \,\&\,  \,   \Phi_\mathsf{val}    \Rightarrow    (   \mathbf{a} ^{  8  }   , \bot )   
  \label{sec:eff:ex1:let}
\end{equation}
where $\Phi \defeq  (   \lambda_\mu\!   \,  \mathit{x_{{\mathrm{0}}}}  .\,   \top   ,   \lambda_\nu\!   \,  \mathit{y_{{\mathrm{0}}}}  .\,   \bot   ) $.
From \T{Let} and \T{Shift}, we can find that the body $\mathit{f} \,  3   \ottsym{;}  \mathit{f} \,  5 $ of the
{\shiftzero} construct should be of the type $ \mathsf{unit}  \,  \,\&\,  \,  (   \mathbf{a} ^{  8  }   , \bot ) $.
It is possible if the continuation variable $\mathit{f}$ is assigned a dependent
function type $\ottsym{(}  \mathit{x} \,  {:}  \,  \mathsf{int}   \ottsym{)}  \rightarrow   \mathsf{unit}  \,  \,\&\,  \,  (   \mathbf{a} ^{ \mathit{x} }   , \bot ) $.
To achieve this type assignment, \T{Shift} states that the following judgment should hold:
\[
  \emptyset   \vdash  \mathcal{S}_0 \, \mathit{f}  \ottsym{.}  \mathit{f} \,  3  \, \mathit{f} \,  5   \ottsym{:}    \mathsf{int}   \,  \,\&\,  \,   \Phi_\mathsf{val}   \, / \,    \mathsf{unit}   \,  \,\&\,  \, ( \forall  \mathit{x}  .   (   \mathbf{a} ^{ \mathit{x} }   , \bot )   )  \Rightarrow    (   \mathbf{a} ^{  8  }   , \bot )    ~.
\]
Then, to derive the judgment (\ref{sec:eff:ex1:let}),
\T{Let} indicates that it suffices to derive the following judgment:
\begin{equation}
 \mathit{x} \,  {:}  \,  \mathsf{int}   \vdash   \mathsf{ev}  [  \mathbf{a}  ]^{ \mathit{x} }   \ottsym{:}    \mathsf{unit}   \,  \,\&\,  \,  \Phi  \, / \,    \mathsf{unit}   \,  \,\&\,  \,   \Phi_\mathsf{val}    \Rightarrow    (   \mathbf{a} ^{ \mathit{x} }   , \bot )    ~.
  \label{sec:eff:ex1:ev}
\end{equation}
Note that $ \Phi_\mathsf{val}  \,  \tceappendop  \, \Phi$ is a subeffect of $\Phi$, and
that the composition of the control effects given above to the {\shiftzero} construct and $ \mathsf{ev}  [  \mathbf{a}  ]^{ \mathit{x} } $
returns the type $  \mathsf{unit}   \,  \,\&\,  \,   \Phi_\mathsf{val}    \Rightarrow    (   \mathbf{a} ^{  8  }   , \bot )  $.
%
Because
\[
 \mathit{x} \,  {:}  \,  \mathsf{int}   \vdash   \mathsf{ev}  [  \mathbf{a}  ]^{ \mathit{x} }   \ottsym{:}    \mathsf{unit}  \,  \,\&\,  \,  (   \mathbf{a} ^{ \mathit{x} }   , \bot )   ~,
\]
the judgment (\ref{sec:eff:ex1:ev}) can be derived via \T{Sub}, by proving the subtyping judgment
\[
 \mathit{x} \,  {:}  \,  \mathsf{int}   \vdash    \mathsf{unit}   \,  \,\&\,  \,   (   \mathbf{a} ^{ \mathit{x} }   , \bot )   \, / \,   \square    \ottsym{<:}    \mathsf{unit}   \,  \,\&\,  \,  \Phi  \, / \,    \mathsf{unit}   \,  \,\&\,  \,   \Phi_\mathsf{val}    \Rightarrow    (   \mathbf{a} ^{ \mathit{x} }   , \bot )    ~.
\]
As $\Phi \,  =  \,  (   \lambda_\mu\!   \,  \mathit{x_{{\mathrm{0}}}}  .\,   \top   ,   \lambda_\nu\!   \,  \mathit{y_{{\mathrm{0}}}}  .\,   \bot   ) $,
$ (   \mathbf{a} ^{ \mathit{x} }   , \bot ) $ is a subeffect of $\Phi$.
Thus, by \Srule{Comp}, the remaining proof obligation is to derive
the subtyping judgment for the control effects:
\[
 \mathit{x} \,  {:}  \,  \mathsf{int}  \,  |  \,  \mathsf{unit}  \,  \,\&\,  \,  (   \mathbf{a} ^{ \mathit{x} }   , \bot )   \vdash   \square   \ottsym{<:}    \mathsf{unit}   \,  \,\&\,  \,   \Phi_\mathsf{val}    \Rightarrow    (   \mathbf{a} ^{ \mathit{x} }   , \bot )   ~.
\]
This is achieved using \Srule{Embed} with the fact that
$ (   \mathbf{a} ^{ \mathit{x} }   , \bot )  \,  \tceappendop  \, \ottsym{(}    \mathsf{unit}  \,  \,\&\,  \,  \Phi_\mathsf{val}    \ottsym{)}  \ottsym{<:}    \mathsf{unit}  \,  \,\&\,  \,  (   \mathbf{a} ^{ \mathit{x} }   , \bot )  $.

\paragraph{Example 2.}
The second example is the program
$\langle$\lstinline{send 0 ready raise}$\rangle$
presented in \refsec{overview:this-work}.
First, the function \lstinline{raise} is expressed as $ \lambda\!  \, \mathit{x}  \ottsym{.}  \mathcal{S}_0 \, \mathit{f}  \ottsym{.}  \mathit{x}$ in our
language.
By \T{Shift},
this function can be given a type
$ \mathsf{int}   \rightarrow  \ottsym{(}   \ottnt{T}  \,  \,\&\,  \,   \Phi_\mathsf{val}   \, / \,  \ottnt{C}  \Rightarrow  \ottsym{(}    \mathsf{int}  \,  \,\&\,  \,  \Phi_\mathsf{val}    \ottsym{)}   \ottsym{)}$
(here, we do not consider dependency for simplicity).
Note that $\ottnt{T}$ and $\ottnt{C}$ can be arbitrary.

The function \lstinline{send} is implemented by
\[
  \mathsf{rec}  \, (  \tceseq{  \mathit{X} ^{  \ottnt{A} _{  \mu  }  },  \mathit{Y} ^{  \ottnt{A} _{  \nu  }  }  }  ,  \mathit{f} ,  \mathit{x}  \ottsym{,}   \textit{rdy}   \ottsym{,}   \textit{rec}  ) . \,   \textit{wait}  \,  \textit{rdy}   \ottsym{;}   \mathsf{ev}  [   \mathbf{Send}   ]   \ottsym{;}   \textit{rec}  \, \mathit{x}  \ottsym{;}  \mathit{f} \, \mathit{x} \,  \textit{rdy}  \,  \textit{rec}  
\]
for some $ \tceseq{ \mathit{X} } $, $ \tceseq{  \ottnt{A} _{  \mu  }  } $, $ \tceseq{ \mathit{Y} } $, and $ \tceseq{  \ottnt{A} _{  \nu  }  } $, which
are not important in this example because the function is not called recursively at run time.
%
%
We assume that $ \textit{rec} $ has the same type as \lstinline{raise}.
Recall that the type of $ \textit{wait}  \,  \textit{rdy} $ is $ \mathsf{unit}  \,  \,\&\,  \,  \Phi _{  \textit{wait}  } $
where $ \Phi _{  \textit{wait}  }  \,  =  \,  (   \lambda_\mu\!   \,  \mathit{x_{{\mathrm{0}}}}  .\,  \mathit{x_{{\mathrm{0}}}} \,  \in  \,   \mathbf{Wait}  ^\ast  \,  \tceappendop  \,  \mathbf{Ready}   ,   \lambda_\nu\!   \,  \mathit{y_{{\mathrm{0}}}}  .\,   \mathit{y_{{\mathrm{0}}}}    =     {  \mathbf{Wait}  }^\omega    ) $.

Consider the typechecking of the body.
%
It infers the type $\ottnt{C}$ of the body as follows.
First, the type assigned to the first two subexpressions
$ \textit{wait}  \,  \textit{rdy}   \ottsym{;}   \mathsf{ev}  [   \mathbf{Send}   ] $ is $ \mathsf{unit}  \,  \,\&\,  \,  \Phi _{  \textit{wait}  }  \,  \tceappendop  \,  (   \mathbf{Send}   , \bot ) $.
The third subexpression $ \textit{rec}  \, \mathit{x}$ involves the control effect
$\ottnt{C}  \Rightarrow  \ottsym{(}    \mathsf{int}  \,  \,\&\,  \,  \Phi_\mathsf{val}    \ottsym{)}$.
To combine this with the control effect of the preceding expression $ \textit{wait}  \,  \textit{rdy}   \ottsym{;}   \mathsf{ev}  [   \mathbf{Send}   ] $, we
need to convert the control effect of $ \textit{wait}  \,  \textit{rdy}   \ottsym{;}   \mathsf{ev}  [   \mathbf{Send}   ] $ to an
impure one of the form $\ottsym{(}    \mathsf{int}  \,  \,\&\,  \,  \Phi_\mathsf{val}    \ottsym{)}  \Rightarrow  \ottnt{C_{{\mathrm{1}}}}$ for some $\ottnt{C_{{\mathrm{1}}}}$.
This is possible by the subtyping rule \Srule{Embed}, which allows converting the
control effect of $ \textit{wait}  \,  \textit{rdy}   \ottsym{;}   \mathsf{ev}  [   \mathbf{Send}   ] $ to
$\ottsym{(}    \mathsf{int}  \,  \,\&\,  \,  \Phi_\mathsf{val}    \ottsym{)}  \Rightarrow  \ottnt{C_{{\mathrm{1}}}}$ with the constraint
$ \Phi _{  \textit{wait}  }  \,  \tceappendop  \,  (   \mathbf{Send}   , \bot )  \,  \tceappendop  \, \ottsym{(}    \mathsf{int}  \,  \,\&\,  \,  \Phi_\mathsf{val}    \ottsym{)}  \ottsym{<:}  \ottnt{C_{{\mathrm{1}}}}$.
Furthermore, the type $\ottnt{C_{{\mathrm{1}}}}$ guarantees what traces the body yields for the
meta-context.  Namely, the control effect of the inferred type $\ottnt{C}$ takes the
form $\ottnt{C'}  \Rightarrow  \ottnt{C_{{\mathrm{1}}}}$ for some $\ottnt{C'}$.
(By further analysis, we can infer the type $\ottnt{C'}$ from the fact that the last
subexpression is the call to the recursive function, but it is not
important for reasoning about the behavior of this example.)

The constraint on $\ottnt{C_{{\mathrm{1}}}}$ suggests $\ottnt{C_{{\mathrm{1}}}} \,  =  \,   \mathsf{int}  \,  \,\&\,  \,  \Phi _{  \textit{wait}  }  \,  \tceappendop  \,  (   \mathbf{Send}   , \bot )  $
as a solution.
Because the temporal effect of the program
$\langle$\lstinline{send 0 ready raise}$\rangle$
corresponds to that of $\ottnt{C_{{\mathrm{1}}}}$, the effect system indicates that
the evaluation of the program terminates
with a finite trace $\varpi \,  \in  \,   \mathbf{Wait}  ^\ast  \,  \tceappendop  \,  \mathbf{Ready}  \,  \tceappendop  \,  \mathbf{Send} $ or it diverges with
infinite trace $ {  \mathbf{Wait}  }^\omega $.
This result exactly matches with the reasoning conducted in \refsec{overview:this-work:answer}.

\paragraph{Example 3.}
The third example is the program $\langle$\lstinline{[wait choice]}$\rangle$ given
in \refsec{overview:this-work}.  This example demonstrates the expressivity of
\T{Fun} and the usefulness of dependently typed continuations.
The function \lstinline{choice} is expressed as
$ \lambda\!  \, \mathit{z}  \ottsym{.}   \mathcal{S}_0 \, \mathit{f}  \ottsym{.}  \mathsf{let} \, \mathit{y_{{\mathrm{1}}}}  \ottsym{=}  \ottsym{(}  \mathit{f} \,  \mathsf{true}   \ottsym{)} \,  \mathsf{in}  \, \mathsf{let} \, \mathit{y_{{\mathrm{2}}}}  \ottsym{=}  \ottsym{(}  \mathit{f} \,  \mathsf{false}   \ottsym{)} \,  \mathsf{in}  \, \mathit{y_{{\mathrm{1}}}}  @  \mathit{y_{{\mathrm{2}}}} $ in our language, and it can have the type
\[
   \mathsf{unit}   \rightarrow    \mathsf{bool}   \,  \,\&\,  \,   \Phi_\mathsf{val}   \, / \,     \mathsf{unit}   \ \mathsf{list}   \,  \,\&\,  \, ( \forall  \mathit{x}  .   \Phi _{ \mathit{x} }   )  \Rightarrow   \Phi   _{  \text{choice}  }  ~.
\]
We suppose that
\[\begin{array}{lll}
  {  \Phi _{ \mathit{x} }  }^\mu   \ottsym{(}  \mathit{x_{{\mathrm{0}}}}  \ottsym{)} &\defeq&     \mathit{x}    =     \mathsf{true}      \Rightarrow     {  \Phi _{  \mathsf{true}  }  }^\mu   \ottsym{(}  \mathit{x_{{\mathrm{0}}}}  \ottsym{)}     \wedge     \mathit{x}    =     \mathsf{false}       \Rightarrow     {  \Phi _{  \mathsf{false}  }  }^\mu   \ottsym{(}  \mathit{x_{{\mathrm{0}}}}  \ottsym{)}  \\
  {  \Phi _{ \mathit{x} }  }^\nu   \ottsym{(}  \mathit{y_{{\mathrm{0}}}}  \ottsym{)} &\defeq&     \mathit{x}    =     \mathsf{true}      \Rightarrow     {  \Phi _{  \mathsf{true}  }  }^\nu   \ottsym{(}  \mathit{y_{{\mathrm{0}}}}  \ottsym{)}     \wedge     \mathit{x}    =     \mathsf{false}       \Rightarrow     {  \Phi _{  \mathsf{false}  }  }^\nu   \ottsym{(}  \mathit{y_{{\mathrm{0}}}}  \ottsym{)}  \\
 \end{array}
\]
for some $ \Phi _{  \mathsf{true}  } $ and $ \Phi _{  \mathsf{false}  } $, which specify the behavior of the
continuations when $ \mathsf{true} $ and $ \mathsf{false} $, respectively, are passed.
Then, $ \Phi _{  \text{choice}  } $ is defined as $  \Phi _{  \mathsf{true}  }  \,  \tceappendop  \, \Phi _{  \mathsf{false}  } $.

The body of the expression $ \textit{wait} $ can be expressed by
\[
  \mathsf{rec}  \, (  \tceseq{  \mathit{X} ^{  \ottnt{A} _{  \mu  }  },  \mathit{Y} ^{  \ottnt{A} _{  \nu  }  }  }  ,  \mathit{f} ,   \textit{rdy}  ) . \,  \mathsf{let} \, \mathit{x}  \ottsym{=}   \textit{rdy}  \,  ()  \,  \mathsf{in}  \, \mathsf{if} \, \mathit{x} \, \mathsf{then} \,  \mathsf{ev}  [   \mathbf{Ready}   ]  \, \mathsf{else} \, \ottsym{(}   \mathsf{ev}  [   \mathbf{Wait}   ]   \ottsym{;}  \mathit{f} \,  \textit{rdy}   \ottsym{)} 
\]
where $ \tceseq{  \mathit{X} ^{  \ottnt{A} _{  \mu  }  },  \mathit{Y} ^{  \ottnt{A} _{  \nu  }  }  } $ is a list of the form $\ottsym{(}   \mathit{X_{{\mathrm{0}}}} ^{  \ottnt{A} _{  \mu\!  , \ottsym{0} }  },  \mathit{Y_{{\mathrm{0}}}} ^{  \ottnt{A} _{  \nu\!  , \ottsym{0} }  }   \ottsym{)}  \ottsym{,}  \ottsym{(}   \mathit{X_{{\mathrm{1}}}} ^{  \ottnt{A} _{  \mu\!  , \ottsym{1} }  },  \mathit{Y_{{\mathrm{1}}}} ^{  \ottnt{A} _{  \nu\!  , \ottsym{1} }  }   \ottsym{)}$.
The first and second pair specify traces guaranteed to yield for the caller and the meta-context, respectively.
We also assume that $ \textit{rdy} $ has the same type as \lstinline{choice}.

Let the assumption type $\ottnt{C_{{\mathrm{0}}}}$ for typechecking the body be $  \mathsf{unit}   \,  \,\&\,  \,  \Phi_{{\mathrm{0}}}  \, / \,     \mathsf{unit}   \ \mathsf{list}   \,  \,\&\,  \,  \Phi   \Rightarrow   \Phi_{{\mathrm{1}}}  $
for some $\Phi$, $\Phi_{{\mathrm{0}}}$, and $\Phi_{{\mathrm{1}}}$.
The temporal effects $\Phi_{{\mathrm{0}}}$ and $\Phi_{{\mathrm{1}}}$ represent the assumptions on traces
observable for the caller and the meta-context and take the form
$ (   \lambda_\mu\!   \,  \mathit{x}  .\,  \mathit{X_{{\mathrm{0}}}}  \ottsym{(}  \mathit{x}  \ottsym{)}  ,   \lambda_\nu\!   \,  \mathit{y}  .\,  \mathit{Y_{{\mathrm{0}}}}  \ottsym{(}  \mathit{y}  \ottsym{)}  ) $ and $ (   \lambda_\mu\!   \,  \mathit{x}  .\,  \mathit{X_{{\mathrm{1}}}}  \ottsym{(}  \mathit{x}  \ottsym{)}  ,   \lambda_\nu\!   \,  \mathit{y}  .\,  \mathit{Y_{{\mathrm{1}}}}  \ottsym{(}  \mathit{y}  \ottsym{)}  ) $,
respectively.
The temporal effect $\Phi$ is the requirement for the contexts and inferred
below.

Consider the typechecking of the body.
Assume that its inferred type is a type $\ottnt{C}$ with control effect
$   \mathsf{unit}   \ \mathsf{list}  \,  \,\&\,  \, \Phi   \Rightarrow  \ottnt{C_{{\mathrm{2}}}}$ for some $\ottnt{C_{{\mathrm{2}}}}$.
Because the conditional branching is placed after the call to $ \textit{rdy} $ and
it is at the end of the body, the branching must have the control effect $    \mathsf{unit}   \ \mathsf{list}  \,  \,\&\,  \, \Phi   \Rightarrow  \ottnt{C} _{ \mathit{x} } $,
where $ \ottnt{C} _{ \mathit{x} }  \defeq    \mathsf{unit}   \ \mathsf{list}  \,  \,\&\,  \,  \Phi _{ \mathit{x} }  $.
Then, by \Srule{Embed},
the $ \mathsf{then} $-branch generates the constraints
$  \mathit{x}     =     \mathsf{true}    \vdash   (   \mathbf{Ready}   , \bot )  \,  \tceappendop  \,    \mathsf{unit}   \ \mathsf{list}  \,  \,\&\,  \, \Phi   \ottsym{<:}   \ottnt{C} _{ \mathit{x} } $, which is equivalent to
(I) $ (   \mathbf{Ready}   , \bot )  \,  \tceappendop  \, \Phi  \ottsym{<:}   \Phi _{  \mathsf{true}  } $.
In the $ \mathsf{else} $-branch, the fact that the last subexpression is the call to
the recursive function $\mathit{f}$ generates the constraint that the control effect
of $ \mathsf{ev}  [   \mathbf{Wait}   ] $ should be $    \mathsf{unit}   \ \mathsf{list}  \,  \,\&\,  \, \Phi_{{\mathrm{1}}}   \Rightarrow  \ottnt{C} _{ \mathit{x} } $.
By \Srule{Embed}, this constraint is converted to
$  \mathit{x}     =     \mathsf{false}    \vdash   (   \mathbf{Wait}   , \bot )  \,  \tceappendop  \, \Phi_{{\mathrm{1}}}  \ottsym{<:}   \Phi _{ \mathit{x} } $, that is,
(II) $ (   \mathbf{Wait}   , \bot )  \,  \tceappendop  \, \Phi_{{\mathrm{1}}}  \ottsym{<:}   \Phi _{  \mathsf{false}  } $.
Next, because the call to $ \textit{rdy} $ is the first subexpression of the body,
the constraint (III) $   \mathsf{unit}   \ \mathsf{list}  \,  \,\&\,  \,  \Phi _{  \text{choice}  }    \ottsym{<:}  \ottnt{C_{{\mathrm{2}}}}$ should be
satisfied.
Furthermore, for simplicity, assume that the typechecking of $ \textit{wait} $
supposes that its application is surrounded by {\resetzero} constructs.\footnote{We can address more general contexts, but it makes the type of $ \textit{wait} $ more complicated.}
Then, \T{Reset} imposes the constraint (IV) $ \Phi_\mathsf{val}   \ottsym{<:}  \Phi$.

Now, we identify the type $\ottnt{C_{{\mathrm{2}}}}$ by solving the constraints (I)--(IV).
By the constraints (IV) and (I), we can take
$ \Phi_\mathsf{val} $ as $\Phi$ and $ (   \mathbf{Ready}   , \bot ) $ as $ \Phi _{  \mathsf{true}  } $.
The constraint (II) suggests $ \Phi _{  \mathsf{false}  }  \,  =  \,  (   \mathbf{Wait}   , \bot )  \,  \tceappendop  \, \Phi_{{\mathrm{1}}}$.
Then, by taking $ \Phi _{  \text{choice}  }  \,  =  \,   \Phi _{  \mathsf{true}  }  \,  \tceappendop  \, \Phi _{  \mathsf{false}  } $ into account,
we can find that the constraint (III) suggests $\ottnt{C_{{\mathrm{2}}}} \,  =  \,    \mathsf{unit}   \ \mathsf{list}  \,  \,\&\,  \,  (   \mathbf{Ready}   , \bot )  \,  \tceappendop  \,  (   \mathbf{Wait}   , \bot )  \,  \tceappendop  \, \Phi_{{\mathrm{1}}} $.

Finally, the rule \T{Fun} gives temporal effects to the function by binding
$\mathit{X_{{\mathrm{1}}}}$ and $\mathit{Y_{{\mathrm{1}}}}$ in $\ottnt{C_{{\mathrm{2}}}}$ using the fixpoint operators.
Note that the constraints on $ \ottnt{C}  . \Phi $ can be generated and solved in a way
similar to the previous work~\cite{Nanjo/Unno/Koskinen/Terauchi_2018_LICS}.
As a result, the function is given the type
\[
   \mathsf{unit}   \,  \,\&\,  \,   \Phi _{  \textit{wait}  }   \, / \,     \mathsf{unit}   \ \mathsf{list}   \,  \,\&\,  \,   \Phi_\mathsf{val}    \Rightarrow    (   \lambda_\mu\!   \,  \mathit{x}  .\,   \ottnt{A} _{  \mu  }   \ottsym{(}  \mathit{x}  \ottsym{)}  ,   \lambda_\nu\!   \,  \mathit{y}  .\,   \ottnt{A} _{  \nu  }   \ottsym{(}  \mathit{y}  \ottsym{)}  )   
\]
where
$ \ottnt{A} _{  \mu  }  \defeq \ottsym{(}   \mu\!  \, \mathit{X_{{\mathrm{1}}}}  \ottsym{(}  \mathit{x} \,  {:}  \,  \Sigma ^\ast   \ottsym{)}  \ottsym{.}    \exists  \,  \mathit{x'}  . \    \mathit{x}    =     \mathbf{Ready}  \,  \tceappendop  \,  \mathbf{Wait}  \,  \tceappendop  \, \mathit{x'}     \wedge    \mathit{X_{{\mathrm{1}}}}  \ottsym{(}  \mathit{x'}  \ottsym{)}    \ottsym{)}$ and
$ \ottnt{A} _{  \nu  }  \defeq \ottsym{(}   \nu\!  \, \mathit{Y_{{\mathrm{1}}}}  \ottsym{(}  \mathit{y} \,  {:}  \,  \Sigma ^\omega   \ottsym{)}  \ottsym{.}    \exists  \,  \mathit{y'}  . \    \mathit{y}    =     \mathbf{Ready}  \,  \tceappendop  \,  \mathbf{Wait}  \,  \tceappendop  \, \mathit{y'}     \wedge    \mathit{Y_{{\mathrm{1}}}}  \ottsym{(}  \mathit{y'}  \ottsym{)}    \ottsym{)}$.
It can be simplified into
\[
   \mathsf{unit}   \,  \,\&\,  \,   \Phi _{  \textit{wait}  }   \, / \,     \mathsf{unit}   \ \mathsf{list}   \,  \,\&\,  \,   \Phi_\mathsf{val}    \Rightarrow    (   \lambda_\mu\!   \,  \mathit{x}  .\,  \mathit{x} \,  \in  \,  \ottsym{(}   \mathbf{Ready}  \,  \tceappendop  \,  \mathbf{Wait}   \ottsym{)} ^\ast   ,   \lambda_\nu\!   \,  \mathit{y}  .\,   \mathit{y}    =     { \ottsym{(}   \mathbf{Ready}  \,  \tceappendop  \,  \mathbf{Wait}   \ottsym{)} }^\omega    )    ~.
\]
This type implies that the entire program has the temporal effect
$ (   \lambda_\mu\!   \,  \mathit{x}  .\,  \mathit{x} \,  \in  \,  \ottsym{(}   \mathbf{Ready}  \,  \tceappendop  \,  \mathbf{Wait}   \ottsym{)} ^\ast   ,   \lambda_\nu\!   \,  \mathit{y}  .\,   \mathit{y}    =     { \ottsym{(}   \mathbf{Ready}  \,  \tceappendop  \,  \mathbf{Wait}   \ottsym{)} }^\omega    ) $.
Therefore, we can find that, when diverging, the program generates the infinite trace $ { \ottsym{(}   \mathbf{Ready}  \,  \tceappendop  \,  \mathbf{Wait}   \ottsym{)} }^\omega $.

\subsection{Properties}
\label{sec:eff:prop}
In this section, we show type safety, which implies soundness of the effect
system for finite traces.
Type safety is proven via progress and subject reduction.
\begin{lemm}[Progress]
 If $ \emptyset   \vdash  \ottnt{e}  \ottsym{:}  \ottnt{C}$,
 then one of the following holds:
 $\ottnt{e} \,  =  \, \ottnt{v}$ for some $\ottnt{v}$;
 $\ottnt{e} \,  =  \,  \ottnt{K}  [  \mathcal{S}_0 \, \mathit{x}  \ottsym{.}  \ottnt{e'}  ] $ for some $\ottnt{K}$, $\mathit{x}$, and $\ottnt{e'}$;
 $\ottnt{e} \,  =  \,  \ottnt{E}  [   \bullet ^{ \pi }   ] $ for some $\ottnt{E}$ and $\pi$; or
 $\ottnt{e} \,  \mathrel{  \longrightarrow  }  \, \ottnt{e'} \,  \,\&\,  \, \varpi$ for some $\ottnt{e'}$ and $\varpi$.
\end{lemm}
For subject reduction, we define an operation to remove yielded finite traces
from types.
\begin{defn}[Effect Quotient]
 We define
 the quotient $ { \varpi }^{-1}  \,  \tceappendop  \, \Phi$ of a temporal effect $\Phi$ with respect to a finite trace $\varpi$
 to be the temporal effect $ (   \lambda_\mu\!   \,  \mathit{x}  .\,   { \Phi }^\mu   \ottsym{(}  \varpi \,  \tceappendop  \, \mathit{x}  \ottsym{)}  ,   \lambda_\nu\!   \,  \mathit{y}  .\,   { \Phi }^\nu   \ottsym{(}  \varpi \,  \tceappendop  \, \mathit{y}  \ottsym{)}  ) $
 using variables $\mathit{x}, \mathit{y} \,  \not\in  \,  \mathit{fv}  (  \Phi  ) $.
 We define
 the quotient $ { \varpi }^{-1}  \,  \tceappendop  \, \ottnt{C}$ of computation type $\ottnt{C}$ and
 quotient $ { \varpi }^{-1}  \,  \tceappendop  \, \ottnt{S}$ of control effect $\ottnt{S}$
 with respect to finite trace $\varpi$, as follows.
 \[\begin{array}{rcl}
   { \varpi }^{-1}  \,  \tceappendop  \, \ottnt{C}             &\defeq&   \ottnt{C}  . \ottnt{T}   \,  \,\&\,  \,   { \varpi }^{-1}  \,  \tceappendop  \, \ottsym{(}   \ottnt{C}  . \Phi   \ottsym{)}  \, / \,   { \varpi }^{-1}  \,  \tceappendop  \, \ottsym{(}   \ottnt{C}  . \ottnt{S}   \ottsym{)}  \\
   { \varpi }^{-1}  \,  \tceappendop  \,  \square          &\defeq&  \square  \\
   { \varpi }^{-1}  \,  \tceappendop  \,  ( \forall  \mathit{x}  .  \ottnt{C_{{\mathrm{1}}}}  )  \Rightarrow   \ottnt{C_{{\mathrm{2}}}}  &\defeq&  ( \forall  \mathit{x}  .  \ottnt{C_{{\mathrm{1}}}}  )  \Rightarrow    { \varpi }^{-1}  \,  \tceappendop  \, \ottnt{C_{{\mathrm{2}}}} 
   \end{array}
 \]
\end{defn}
\begin{lemm}[Subject Reduction]
 If $ \emptyset   \vdash  \ottnt{e}  \ottsym{:}  \ottnt{C}$ and $\ottnt{e} \,  \mathrel{  \longrightarrow  }  \, \ottnt{e'} \,  \,\&\,  \, \varpi$,
 then $ \emptyset   \vdash  \ottnt{e'}  \ottsym{:}   { \varpi }^{-1}  \,  \tceappendop  \, \ottnt{C}$.
\end{lemm}
\begin{corollary}[Type Safety]
 \label{thm:type-safety}
 If $ \emptyset   \vdash  \ottnt{e}  \ottsym{:}  \ottnt{C}$ and $\ottnt{e} \,  \mathrel{  \longrightarrow  ^*}  \, \ottnt{e'} \,  \,\&\,  \, \varpi$ and $\ottnt{e'}$ cannot be evaluated further,
 then $ \emptyset   \vdash  \ottnt{e'}  \ottsym{:}   { \varpi }^{-1}  \,  \tceappendop  \, \ottnt{C}$, and either of the following holds:
 \begin{itemize}
  \item $\ottnt{e'} \,  =  \, \ottnt{v}$ for some $\ottnt{v}$ and $\ottnt{T}$ such that
        $ \emptyset   \vdash  \ottnt{v}  \ottsym{:}  \ottnt{T}$ and $ \emptyset   \vdash   \ottnt{T}  \,  \,\&\,  \,   (  \varpi  , \bot )   \, / \,   \square    \ottsym{<:}  \ottnt{C}$;
  \item $\ottnt{e'} \,  =  \,  \ottnt{K}  [  \mathcal{S}_0 \, \mathit{x}  \ottsym{.}  \ottnt{e_{{\mathrm{0}}}}  ] $ for some $\ottnt{K}$, $\mathit{x}$, and $\ottnt{e_{{\mathrm{0}}}}$, and
        $ \ottnt{C}  . \ottnt{S}  \,  \not=  \,  \square $; or
  \item $\ottnt{e'} \,  =  \,  \ottnt{E}  [   \bullet ^{ \pi }   ] $ for some $\ottnt{E}$ and $\pi$.
 \end{itemize}
\end{corollary}
This type safety property implies soundness of the effect system for finite
traces.
First, it ensures that, if a closed expression of type $\ottnt{C}$
terminates at a value with a finite trace $\varpi$, a subtyping judgment
$ \emptyset   \vdash   \ottnt{T}  \,  \,\&\,  \,   (  \varpi  , \bot )   \, / \,   \square    \ottsym{<:}  \ottnt{C}$ can be derived.
Then, the subtyping judgment implies that $\varpi$ is in $ { \ottsym{(}   \ottnt{C}  . \Phi   \ottsym{)} }^\mu $.

%% file: sections/lr.tex
\section{Soundness for Infinite Traces}
\label{sec:lr}

\begin{figure}[t]
 \[\begin{array}{lll}
   \LRRel{ \mathcal{V} }{ \ottsym{\{}  \mathit{x} \,  {:}  \, \iota \,  |  \, \phi  \ottsym{\}} }   &\defeq& \{ \ottnt{v} \mid  \emptyset   \vdash  \ottnt{v}  \ottsym{:}  \ottsym{\{}  \mathit{x} \,  {:}  \, \iota \,  |  \, \phi  \ottsym{\}} \}
  \\
   \LRRel{ \mathcal{V} }{ \ottsym{(}  \mathit{x} \,  {:}  \, \ottnt{T}  \ottsym{)}  \rightarrow  \ottnt{C} }    &\defeq&
    \{ \ottnt{v} \mid   \forall  \, { \ottnt{v'} \,  \in  \,  \LRRel{ \mathcal{V} }{ \ottnt{T} }  } . \  \ottnt{v} \, \ottnt{v'} \,  \in  \,  \LRRel{ \mathcal{E} }{  \ottnt{C}    [  \ottnt{v'}  /  \mathit{x}  ]   }   \}
  \\
  {}
   \LRRel{ \mathcal{E} }{ \ottnt{C} }  &\defeq& \{ \ottnt{e} \mid
   (   \forall  \, { \ottnt{v}  \ottsym{,}  \varpi } . \  \ottnt{e} \,  \Downarrow  \, \ottnt{v} \,  \,\&\,  \, \varpi   \mathrel{\Longrightarrow}  \ottnt{v} \,  \in  \,  \LRRel{ \mathcal{V} }{  \ottnt{C}  . \ottnt{T}  }  )
   \, \wedge
   \\ && \qquad
   (   \forall  \, { \ottnt{K_{{\mathrm{0}}}}  \ottsym{,}  \mathit{x_{{\mathrm{0}}}}  \ottsym{,}  \ottnt{e_{{\mathrm{0}}}}  \ottsym{,}  \varpi } . \  \ottnt{e} \,  \Downarrow  \,  \ottnt{K_{{\mathrm{0}}}}  [  \mathcal{S}_0 \, \mathit{x_{{\mathrm{0}}}}  \ottsym{.}  \ottnt{e_{{\mathrm{0}}}}  ]  \,  \,\&\,  \, \varpi   \mathrel{\Longrightarrow}    \exists  \, { \mathit{x}  \ottsym{,}  \ottnt{C_{{\mathrm{1}}}}  \ottsym{,}  \ottnt{C_{{\mathrm{2}}}} } . \       \\ && \qquad\quad
     {  \ottnt{C}  . \ottnt{S}  \,  =  \,  ( \forall  \mathit{x}  .  \ottnt{C_{{\mathrm{1}}}}  )  \Rightarrow   \ottnt{C_{{\mathrm{2}}}}  } \,\mathrel{\wedge}\, {    }   \\ && \qquad\quad
      \forall  \, { \ottnt{K} \,  \in  \,  \LRRel{ \mathcal{K} }{   \ottnt{C}  . \ottnt{T}   \, / \, ( \forall  \mathit{x}  .  \ottnt{C_{{\mathrm{1}}}}  )  }  } . \   \resetcnstr{  \ottnt{K}  [  \ottnt{e}  ]  }  \,  \in  \,  \LRRel{ \mathcal{E} }{ \ottnt{C_{{\mathrm{2}}}} }  )
   \, \wedge
   \\ && \qquad
   (   \forall  \, { \pi } . \  \ottnt{e} \,  \Uparrow  \, \pi   \mathrel{\Longrightarrow}   { \models  \,   { \ottnt{C} }^\nu (  \pi  )  }  )
   \, \}
   \\
  %
   \LRRel{ \mathcal{K} }{  \ottnt{T}  \, / \, ( \forall  \mathit{x}  .  \ottnt{C}  )  }  &\defeq& \{ \ottnt{K} \mid
     { \ottnt{K}  \ottsym{:}   \ottnt{T}  \, / \, ( \forall  \mathit{x}  .  \ottnt{C}  )  } \,\mathrel{\wedge}\, {  \lambda\!  \, \mathit{x}  \ottsym{.}   \resetcnstr{  \ottnt{K}  [  \mathit{x}  ]  }  \,  \in  \,  \LRRel{ \mathcal{V} }{ \ottsym{(}  \mathit{x} \,  {:}  \, \ottnt{T}  \ottsym{)}  \rightarrow  \ottnt{C} }  } 
   \}
   \\
  %
  %
   \LRRel{ \mathcal{O} }{ \Gamma  \, \vdash \,  \ottnt{C} }      &\defeq& \{ \ottnt{e} \mid  { \Gamma  \vdash  \ottnt{e}  \ottsym{:}  \ottnt{C} } \,\mathrel{\wedge}\, {   \forall  \, { \sigma \,  \in  \,  \LRRel{ \mathcal{G} }{ \Gamma }  } . \  \sigma  \ottsym{(}  \ottnt{e}  \ottsym{)} \,  \in  \,  \LRRel{ \mathcal{E} }{ \sigma  \ottsym{(}  \ottnt{C}  \ottsym{)} }   }  \}
   \end{array}
 \]
 \caption{Logical relation.}
 \label{fig:LR}
\end{figure}

This section proves soundness of the effect system for infinite traces.
To this end, we define a logical relation by following the previous
work~\cite{Nanjo/Unno/Koskinen/Terauchi_2018_LICS}.
Our work is different from the previous work in that our logical relation takes
control effects into account.

Our logical relation, defined in \reffig{LR}, consists of value relations,
expression relations, and pure evaluation context relations.
A value relation $ \LRRel{ \mathcal{V} }{ \ottnt{T} } $ is over closed values of the type $\ottnt{T}$.
An expression relation $ \LRRel{ \mathcal{E} }{ \ottnt{C} } $ is over closed expression of the type
$\ottnt{C}$.
A pure evaluation context relation $ \LRRel{ \mathcal{K} }{  \ottnt{T}  \, / \, ( \forall  \mathit{x}  .  \ottnt{C}  )  } $ is over closed pure
evaluation contexts that can be reified into functions of the type
$\ottsym{(}  \mathit{x} \,  {:}  \, \ottnt{T}  \ottsym{)}  \rightarrow  \ottnt{C}$.
We implicitly assume the typability of inhabitants in these relations.
\TS{''Closed'' has been introduced?}

The definition of value relations is standard.
All the values of a refinement type $\ottsym{\{}  \mathit{x} \,  {:}  \, \iota \,  |  \, \phi  \ottsym{\}}$ are contained in
$ \LRRel{ \mathcal{V} }{ \ottsym{\{}  \mathit{x} \,  {:}  \, \iota \,  |  \, \phi  \ottsym{\}} } $.
The value relation for dependent function type $\ottsym{(}  \mathit{x} \,  {:}  \, \ottnt{T}  \ottsym{)}  \rightarrow  \ottnt{C}$
contains values that map arguments $\ottnt{v}$ in $ \LRRel{ \mathcal{V} }{ \ottnt{T} } $ to expressions in
$ \LRRel{ \mathcal{E} }{  \ottnt{C}    [  \ottnt{v}  /  \mathit{x}  ]   } $.
{}

The expression relation for computation type $\ottnt{C}$ contains expressions that
behave as specified by the type $\ottnt{C}$.
If an expression $\ottnt{e}$ in $ \LRRel{ \mathcal{E} }{ \ottnt{C} } $ terminates at a value,
then that value must be contained in the value relation $ \LRRel{ \mathcal{V} }{  \ottnt{C}  . \ottnt{T}  } $.
If the expression $\ottnt{e}$ diverges with infinite trace $\pi$, then the formula
$ { \ottnt{C} }^\nu (  \pi  ) $ must be true.
%
Note that only requiring $ { \models  \,   { \ottsym{(}   \ottnt{C}  . \Phi   \ottsym{)} }^\nu   \ottsym{(}  \pi  \ottsym{)} } $ is insufficient because the control
effects in type $\ottnt{C}$ have to propagate infinite traces yielded by inner
expressions to meta-contexts. 
If the expression $\ottnt{e}$ invokes the {\shiftzero} operator, its behavior
depends on a delimited context enclosing the invocation.
First, the evaluation relation can suppose $ \ottnt{C}  . \ottnt{S} $ to be a dependent control
effect $ ( \forall  \mathit{x}  .  \ottnt{C_{{\mathrm{1}}}}  )  \Rightarrow   \ottnt{C_{{\mathrm{2}}}} $ by type safety (\refthm{type-safety}).
Then, delimited contexts up to the {\shiftzero} construct are
restricted to be those in the relation $ \LRRel{ \mathcal{K} }{  \ottnt{T}  \, / \, ( \forall  \mathit{x}  .  \ottnt{C_{{\mathrm{1}}}}  )  } $.
For any $\ottnt{K} \,  \in  \,  \LRRel{ \mathcal{K} }{  \ottnt{T}  \, / \, ( \forall  \mathit{x}  .  \ottnt{C_{{\mathrm{1}}}}  )  } $, the functional representation of
$ \resetcnstr{ \ottnt{K} } $ behaves as specified by the {\shiftzero} construct.
Given such a context $\ottnt{K}$, the expression $ \resetcnstr{  \ottnt{K}  [  \ottnt{e}  ]  } $ must behave as
specified by the type $\ottnt{C_{{\mathrm{2}}}}$.


A pure evaluation context $\ottnt{K}$ is contained in $ \LRRel{ \mathcal{K} }{  \ottnt{T}  \, / \, ( \forall  \mathit{x}  .  \ottnt{C}  )  } $ if and only if
$\ottnt{K}$ is well typed and the $\lambda$-abstraction
$ \lambda\!  \, \mathit{x}  \ottsym{.}   \resetcnstr{  \ottnt{K}  [  \mathit{x}  ]  } $
behaves as specified by the type $\ottsym{(}  \mathit{x} \,  {:}  \, \ottnt{T}  \ottsym{)}  \rightarrow  \ottnt{C}$.
A typing judgment $\ottnt{K}  \ottsym{:}   \ottnt{T}  \, / \, ( \forall  \mathit{x}  .  \ottnt{C_{{\mathrm{1}}}}  ) $ for context $\ottnt{K}$ ensures
that, for any expression $\ottnt{e}$ of a type $ \ottnt{T}  \,  \,\&\,  \,  \Phi  \, / \,   ( \forall  \mathit{x}  .  \ottnt{C_{{\mathrm{1}}}}  )  \Rightarrow   \ottnt{C_{{\mathrm{2}}}}  $, the
expression $ \resetcnstr{  \ottnt{K}  [  \ottnt{e}  ]  } $ can be of type $\ottnt{C_{{\mathrm{2}}}}$.
See the supplementary material for the inference rules.

The expression relations for closed expressions are extended to open expressions.
%
For this purpose, we define a relation $ \LRRel{ \mathcal{G} }{ \Gamma } $, which contains substitutions $\sigma$
that map a variable of type $\ottnt{T}$ in $\Gamma$ to a value in
$ \LRRel{ \mathcal{V} }{ \ottnt{T} } $, a variable of sort $\ottnt{s}$ in $\Gamma$ to a closed term of
$\ottnt{s}$, and a predicate variable on sorts $ \tceseq{ \ottnt{s} } $ in $\Gamma$ to a closed
predicate on $ \tceseq{ \ottnt{s} } $.
%
%
We then define an open expression relation $ \LRRel{ \mathcal{O} }{ \Gamma  \, \vdash \,  \ottnt{C} } $, which
contains possibly open expressions $\ottnt{e}$ such that $\Gamma  \vdash  \ottnt{e}  \ottsym{:}  \ottnt{C}$ and
$\sigma  \ottsym{(}  \ottnt{e}  \ottsym{)} \,  \in  \,  \LRRel{ \mathcal{E} }{ \sigma  \ottsym{(}  \ottnt{C}  \ottsym{)} } $ for any $\sigma \,  \in  \,  \LRRel{ \mathcal{G} }{ \Gamma } $.

Next, we show that a well-typed expression is contained in the logical relation, which
is called the Fundamental Property and implies soundness of the effect system
for infinite traces.
\begin{thm}[Fundamental Property]
 If $\Gamma  \vdash  \ottnt{e}  \ottsym{:}  \ottnt{C}$, then $\ottnt{e} \,  \in  \,  \LRRel{ \mathcal{O} }{ \Gamma  \, \vdash \,  \ottnt{C} } $.
\end{thm}
\begin{corollary}[Soundness for Infinite Traces]
 If $ \emptyset   \vdash  \ottnt{e}  \ottsym{:}  \ottnt{C}$ and $\ottnt{e} \,  \Uparrow  \, \pi$,
 then $ { \models  \,   { \ottnt{C} }^\nu (  \pi  )  } $.
\end{corollary}

{}

We remark on the use of the divergence constructs $ \bullet ^{ \pi } $ in
our development.
They are used to define the finite unrollings of a recursive function.
The finite unrollings are indexed by natural numbers $\ottmv{i}$ which denote the
limits of the approximations: if an approximation function calls itself
recursively over its limit $\ottmv{i}$, then the recursive call diverges.
Because we are interested in infinite traces yielded by divergence, the finite
unrollings need to specify infinite traces yielded when they diverge.
However, it is not clear how to \emph{construct} an expression that
diverges with a given infinite trace; \citet{Nanjo/Unno/Koskinen/Terauchi_2018_TR}
only supposed that the construction is possible.
Furthermore, constructing desired diverging expressions seems much harder in our
setting because the diverging expressions have to be of the return type of the
recursive function, and the return type may be equipped with complicated control
effects in our language.
The use of the divergence constructs enables us to avoid constructing
complicated diverging expressions.
For example, to approximate the diverging behavior of a recursive function
$\ottnt{v} \,  =  \,  \mathsf{rec}  \, (  \tceseq{  \mathit{X} ^{  \ottnt{A} _{  \mu  }  },  \mathit{Y} ^{  \ottnt{A} _{  \nu  }  }  }  ,  \mathit{f} ,   \tceseq{ \mathit{z} }  ) . \,  \ottnt{e} $, we define the finite unrollings
$ \ottnt{v} _{  \nu   \ottsym{;}  \ottmv{i}  \ottsym{;}  \pi } $ of $\ottnt{v}$ for every limit $\ottmv{i}$ as follows:
%
\[\begin{array}{l@{\ \ }l@{\ \ }l@{\qquad}l@{\ \ }l@{\ \ }l}
 %
  \ottnt{v} _{  \nu   \ottsym{;}  \ottsym{0}  \ottsym{;}  \pi }    &\defeq&  \lambda\!  \,  \tceseq{ \mathit{z} }   \ottsym{.}   \bullet ^{ \pi }  &
  \ottnt{v} _{  \nu   \ottsym{;}  \ottmv{i}  \ottsym{+}  \ottsym{1}  \ottsym{;}  \pi }  &\defeq&  \lambda\!  \,  \tceseq{ \mathit{z} }   \ottsym{.}     \ottnt{e}    [ \tceseq{  \ottnt{A} _{  \mu  }   /  \mathit{X} } ]      [ \tceseq{  \ottnt{A} _{  \nu\!  , \ottmv{i} }   /  \mathit{Y} } ]      [   \ottnt{v} _{  \nu   \ottsym{;}  \ottmv{i}  \ottsym{;}  \pi }   /  \mathit{f}  ]   \\
  \end{array}
\]
where $ \tceseq{  \ottnt{A} _{  \nu\!  , \ottmv{i} }  } $ is the finite unrolling of $ \tceseq{  \ottnt{A} _{  \nu  }  } $ (see the supplementary material for detail).
The body of the function $ \ottnt{v} _{  \nu   \ottsym{;}  \ottsym{0}  \ottsym{;}  \pi } $ clarifies that the call diverges with
the infinite trace $\pi$.

\TS{Hiroshi, please check this paragraph.}
To prove soundness, we assume cocontinuity on the fixpoint logic (see the
supplementary material for its formal definition).
\citet{Nanjo/Unno/Koskinen/Terauchi_2018_TR} proved soundness for finite and
infinite traces using a logical relation, which results in requiring continuity
as well as cocontinuity for their logic.
By contrast, we use the logical relation to prove only soundness for infinite
traces because soundness for finite traces is implied by type safety.
Thus, our proof requires cocontinuity but not continuity.

%% file: sections/relwork.tex
\section{Related Work}
\label{sec:relwork}

\subsection{Type-and-Effect Systems for Temporal Verification of Traces}
\begin{sloppypar}
As described in \refsec{intro:background}, several researchers proposed
effect systems for verification of temporal safety
properties~\cite{Igarashi/Kobayashi_2002_POPL,Aldrich/Sunshine/Saini/Sparks_2009_OOPSLA,Skalka/Smith_2004_APLAS,Gordon_2017_ECOOP,Iwama/Igarashi/Kobayashi_2006_PEPM,Gordon_2020_ECOOP}
or general temporal
properties~\cite{Kobayashi/Ong_2009_LICS,Hofmann/Chen_2014_CSL-LICS,Koskinen/Terauchi_2014_CSL-LICS,Nanjo/Unno/Koskinen/Terauchi_2018_LICS}.
In this section, we discuss the previous work that are closely related to the present work.
A brief comparison with other works is found in \refsec{intro:background}.
\end{sloppypar}

\citet{Skalka/Smith_2004_APLAS} developed an effect system to reason about event
traces.
They proposed the notion of history effects, which express sets of traces, and
introduced an operation to concatenate history effects for sequentialized
computation, and the least fixpoint operator to describe traces recursively.
Their effect system can statically verify assertions on past events.
Such assertions can be viewed as temporal safety properties, but they did not
address general temporal properties.

\citet{Hofmann/Chen_2014_CSL-LICS} and \citet{Koskinen/Terauchi_2014_CSL-LICS}
proposed effect systems for general temporal properties.
Their effect systems manage \TS{manage?} finite and infinite traces in a similar manner, which
is followed by our work.  They both introduced effects that intuitively denote
pairs comprising a set of finite traces and a set of infinite traces, and
defined similar operations to concatenate the effects.
The effects in \citet{Hofmann/Chen_2014_CSL-LICS} are constructed from a
B{\"{u}}chi automaton, which enables specifying the trace sets using an
$\omega$-regular language.
\citet{Koskinen/Terauchi_2014_CSL-LICS} did not provide a concrete means to
reason about infinite traces.
Instead, they assumed the existence of an oracle that tells the effect system
temporal properties of programs.  Without an oracle, the effect system can only
verify that recursive functions may yield any infinite trace.
Given an appropriate oracle, their effect system can reason about nontrivial
infinite traces.

Recently, \citet{Nanjo/Unno/Koskinen/Terauchi_2018_LICS} introduced the
first-order fixpoint logic to the effect system of
\citet{Koskinen/Terauchi_2014_CSL-LICS}.
The use of the fixpoint logic enables specifying trace sets in a value-dependent
manner and provides a means to construct predicates on infinite traces without
an oracle.
Our work is an extension of their effect system to the control operators
{\shiftzero}/{\resetzero}.

\citet{Gordon_2017_ECOOP} proposed a generic framework for sequential effect
systems, where the order of effects is important.
This framework is parameterized over effect quantales, the class of effects
satisfying some algebraic properties desired to define sequential effect
systems.
%
%
\citet{Gordon_2020_ECOOP} extended this framework to the tagged versions of
the control operators {\abort} and {\callcc}.
%
Gordon's effect system can also track
the use of the control operators (abortion, capture of continuations, and their
invocation) in a flow-sensitive manner.
As we express requirements for delimited contexts using dependent control
effects, Gordon introduced the notion of \emph{prophecies} to predict
effects yielded by captured continuations.
However, the representation of guarantees for enclosing delimiter constructs in
the previous work seems more complicated.
The complexity might arise from the employed control operators.
In our work, continuations are captured only by {\shiftzero}.
By contrast, Gordon adopted the two operations {\abort} and {\callcc} to manipulate
continuations, which necessitates different effects for the effect system to
capture their different behavior.
Other differences between Gordon's work and ours include the following.
First, Gordon addressed only temporal safety properties, whereas we address
general temporal properties.
Second, Gordon addressed effect quantales, whereas we focus only on traces.
Third, Gordon did not support dependent
typing,\footnote{\citet{Gordon_2021_TOPLAS} defined effect quantales extended to
value dependency, but it is not clear how they can be applied to languages with
control operators.} whereas we utilize the fixpoint logic to define the sound
effect system for verifying the infinite behavior of programs, and allow
captured continuations to be dependently typed.
%

\subsection{Other Approaches to Temporal Verification}
\begin{sloppypar}
There is a long line of work on automated temporal liveness verification of finite-state programs~\cite{ClarkeES83,KupfermanVW00}, and recent developments of software model checking~\cite{DietschHLP15,MuraseT0SU16,CookKP17} and abstract interpretation~\cite{UrbanU018} enabled temporal liveness verification of infinite-state programs.  Compared to these approaches, our type-and-effect-based approach to temporal verification enables to handle complex control structures caused by {\shiftzero}/{\resetzero} as well as higher-order functions, in a compositional manner.  Though the current level of automation is lower than that of the previous approaches, we plan to automate type-and-effect reconstruction by leveraging recent advances in fixpoint logic validity checking~\cite{KNIU19,journals/corr/abs-2007-03656}.
\end{sloppypar}
\subsection{Type Systems for Control Operators}
Our effect system utilizes the ability to accommodate ATM for allowing AEM.
The first type system allowing ATM was proposed by
\citet{Danvy/Filinski_1990_LFP} for the delimited control operators
{\shift}/{\reset}.
Based on this seminal system, \citet{Materzok/Biernacki_2011_ICFP} proposed a type
system that allows ATM in the presence of {\shiftzero}/{\resetzero} and
introduced subtyping to embed pure expressions (i.e., without control effects)
into impure contexts (i.e., possibly with control effects).
We believe that it is not difficult to adapt our effect system to other control
operators equipped with a type system for ATM, such as
{\control}/{\prompt}~\cite{Kameyama/Yonezawa_2008_FLOPS}.
%
%
However, it is left for future work how to address control operators for which
type systems that allow ATM have not been provided.
For example, such operators include tagged control
operators~\cite{Gunter/Remy/Riecke_1995_FPCA} and algebraic effect
handlers~\cite{Plotkin/Pretnar_2013_LMCS}.
We plan to start from building effect systems that support ATM for these operators.

\begin{sloppypar}
Several researchers proposed dependent type systems for control
operators~\cite{Herbelin_2012_LICS,Lepigre_2016_ESOP,Miquey_2017_ESOP,Ahman_2018_POPL,Cong/Asai_2018_ICFP}.
These type systems allow types to depend on pure expressions in programs.
Similarly, we restrict program expressions that can appear in types to be only
first-order values because we suppose that the logic to describe predicates can
handle only them.
Unlike our work, these systems do not allow the types of continuations to
depend on arguments.
\citet{Miquey_2017_ESOP} showed that a CPS-transformed expressions can be
assigned a CPS-transformed type in which the return type of continuations can be
dependent on arguments, but it is not reflected to the type system for typing
programs in direct style.
\end{sloppypar}

%% file: sections/conclusion.tex
\section{Conclusion}
\label{sec:conclusion}

This work extended a dependent temporal effect system to the delimited control
operators {\shiftzero}/{\resetzero}.
A key observation for this extension is that the
{\shiftzero} operator modifies answer effects.
Based on this observation, we define a temporal effect system that accommodates
AEM.
Our effect system addresses subtle interaction between recursive functions and
delimited control operators by providing different predicate variables for
different kinds of contexts.
Our system also allows types of captured continuations to be dependent on
arguments, which is crucial for reasoning about the behavior of the
continuations precisely.
Soundness of our effect system is implied by two properties: type safety, which
implies soundness for finite traces, and the fundamental property of the logical
relation, which implies soundness for infinite traces.
We also implemented a tool that can generate constraints on temporal effects.
We believe that this work will serve on compositional
verification of temporal properties of programs in practical languages.

There are several directions for future work.
A key step for practice is to automate verification of the reasoning with our
effect system.  We have implemented constraint generation for temporal effects,
but we also plan to implement a constraint solver using techniques of
automata-based approximation.
%
%
We are also interested in extending the present work to other control operators,
such as tagged control operators.
%
%
%
Other interesting features not supported yet are recursive types and
higher-order state, which are expressive enough to implement recursive
functions.
%


%% file: defn.tex
\section{Definition}
\label{sec:defn}

\input{defn/lang}

%% file: defn/lang.tex
\subsection{Language}
\label{sec:defn:lang}

\subsubsection{Syntax}

\[
 \begin{array}{rcll}
  \multicolumn{4}{l}{
   \textbf{Variables} \quad \mathit{x}, \mathit{y}, \mathit{z}, \mathit{f} \qquad
   \textbf{Predicate variables} \quad \mathit{X}, \mathit{Y} \qquad
   \textbf{Term functions} \quad \mathit{F}
  } \\
  \multicolumn{4}{l}{
   \textbf{Events} \quad \mathbf{a} \,  \in  \, \Sigma \qquad
   \textbf{Finite traces} \quad \varpi \,  \in  \,  \Sigma ^\ast  \qquad
   \textbf{Infinite traces} \quad \pi \,  \in  \,  \Sigma ^\omega  \qquad
  } \\[2ex]
  \textbf{Base types} &
   \iota & ::= &  \mathsf{unit}  \mid  \mathsf{bool}  \mid  \mathsf{int}  \mid \cdots
  \\
  \textbf{Value types} &
   \ottnt{T} & ::= & \ottsym{\{}  \mathit{x} \,  {:}  \, \iota \,  |  \, \phi  \ottsym{\}} \mid \ottsym{(}  \mathit{x} \,  {:}  \, \ottnt{T}  \ottsym{)}  \rightarrow  \ottnt{C} \mid  \forall   \ottsym{(}  \mathit{X} \,  {:}  \,  \tceseq{ \ottnt{s} }   \ottsym{)}  \ottsym{.}  \ottnt{C}
  \\
  \textbf{Computation types} &
   \ottnt{C} & ::= &  \ottnt{T}  \,  \,\&\,  \,  \Phi  \, / \,  \ottnt{S} 
  \\
  \textbf{Control effects} &
   \ottnt{S} & ::= &  \square  \mid  ( \forall  \mathit{x}  .  \ottnt{C_{{\mathrm{1}}}}  )  \Rightarrow   \ottnt{C_{{\mathrm{2}}}} 
  \\
  \textbf{Temporal effects} &
   \Phi & ::= &  (   \lambda_\mu\!   \,  \mathit{x}  .\,  \phi_{{\mathrm{1}}}  ,   \lambda_\nu\!   \,  \mathit{y}  .\,  \phi_{{\mathrm{2}}}  ) 
  \\[2ex]
  \textbf{Sorts} &
   \ottnt{s} & ::= & \iota \mid  \Sigma ^\ast  \mid  \Sigma ^\omega 
  \\
  \textbf{Terms} &
   \ottnt{t} & ::= & \mathit{x} \mid \mathit{F}  \ottsym{(}   \tceseq{ \ottnt{t} }   \ottsym{)}
  \\
  \textbf{Primitive predicates} &
   \ottnt{P} & ::= & { (=) } \mid \cdots
  \\
  \textbf{Predicates} &
   \ottnt{A} & ::= & \mathit{X} \mid \ottsym{(}   \mu\!  \, \mathit{X}  \ottsym{(}   \tceseq{ \mathit{x}  \,  {:}  \,  \ottnt{s} }   \ottsym{)}  \ottsym{.}  \phi  \ottsym{)} \mid \ottsym{(}   \nu\!  \, \mathit{X}  \ottsym{(}   \tceseq{ \mathit{x}  \,  {:}  \,  \ottnt{s} }   \ottsym{)}  \ottsym{.}  \phi  \ottsym{)} \mid \ottnt{P}
  \\
  \textbf{Formulas} &
   \phi & ::= &  \top  \mid  \bot  \mid \ottnt{A}  \ottsym{(}   \tceseq{ \ottnt{t} }   \ottsym{)} \mid
                  \neg  \phi  \mid  \phi_{{\mathrm{1}}}    \vee    \phi_{{\mathrm{2}}}  \mid  \phi_{{\mathrm{1}}}    \wedge    \phi_{{\mathrm{2}}}  \mid  \phi_{{\mathrm{1}}}    \Rightarrow    \phi_{{\mathrm{2}}}  \mid \\ &&&
                   \forall  \,  \mathit{x}  \mathrel{  \in  }  \ottnt{s}  . \  \phi  \mid   \exists  \,  \mathit{x}  \mathrel{  \in  }  \ottnt{s}  . \  \phi  \mid
                   \forall  \,  \mathit{X} \,  {:}  \,  \tceseq{ \ottnt{s} }   . \  \phi  \mid   \exists  \,  \mathit{X} \,  {:}  \,  \tceseq{ \ottnt{s} }   . \  \phi 
  \\[2ex]
  \textbf{Constants} & \ottnt{c} & ::= &
    ()  \mid  \mathsf{true}  \mid  \mathsf{false}  \mid \ottsym{0} \mid \cdots \\
  \textbf{Values} & \ottnt{v} & ::= &
   \mathit{x} \mid \ottnt{c} \mid  \Lambda   \ottsym{(}  \mathit{X} \,  {:}  \,  \tceseq{ \ottnt{s} }   \ottsym{)}  \ottsym{.}  \ottnt{e} \mid
   \ottsym{(}   \lambda\!  \,  \tceseq{ \mathit{x} }   \ottsym{.}  \ottnt{e}  \ottsym{)} \,  \tceseq{ \ottnt{v} }  \mid \ottsym{(}   \mathsf{rec}  \, (  \tceseq{  \mathit{X} ^{  \ottnt{A} _{  \mu  }  },  \mathit{Y} ^{  \ottnt{A} _{  \nu  }  }  }  ,  \mathit{f} ,   \tceseq{ \mathit{x} }  ) . \,  \ottnt{e}   \ottsym{)} \,  \tceseq{ \ottnt{v} }  \\ &&&
   \quad \text{(where $\ottsym{\mbox{$\mid$}}   \tceseq{ \ottnt{v} }   \ottsym{\mbox{$\mid$}} \,  <  \, \ottsym{\mbox{$\mid$}}   \tceseq{ \mathit{x} }   \ottsym{\mbox{$\mid$}}$)}
  \\
  \textbf{Expressions} & \ottnt{e} & ::= &
   \ottnt{v} \mid \ottnt{o}  \ottsym{(}   \tceseq{ \ottnt{v} }   \ottsym{)} \mid \ottnt{v_{{\mathrm{1}}}} \, \ottnt{v_{{\mathrm{2}}}} \mid \ottnt{v} \, \ottnt{A} \mid
   \mathsf{if} \, \ottnt{v} \, \mathsf{then} \, \ottnt{e_{{\mathrm{1}}}} \, \mathsf{else} \, \ottnt{e_{{\mathrm{2}}}} \mid
   \mathsf{let} \, \mathit{x}  \ottsym{=}  \ottnt{e_{{\mathrm{1}}}} \,  \mathsf{in}  \, \ottnt{e_{{\mathrm{2}}}} \mid \\ &&&
    \mathsf{ev}  [  \mathbf{a}  ]  \mid
   \mathcal{S}_0 \, \mathit{x}  \ottsym{.}  \ottnt{e} \mid  \resetcnstr{ \ottnt{e} }  \mid
    \bullet ^{ \pi } 
  \\
  \textbf{Evaluation contexts} & \ottnt{E} & ::= &
    \,[\,]\,  \mid \mathsf{let} \, \mathit{x}  \ottsym{=}  \ottnt{E} \,  \mathsf{in}  \, \ottnt{e_{{\mathrm{2}}}} \mid  \resetcnstr{ \ottnt{E} } 
  \\
  \textbf{Pure evaluation contexts} & \ottnt{K} & ::= &
    \,[\,]\,  \mid \mathsf{let} \, \mathit{x}  \ottsym{=}  \ottnt{K} \,  \mathsf{in}  \, \ottnt{e_{{\mathrm{2}}}}
  \\
  \textbf{Results} & \ottnt{r} & ::= & \ottnt{v} \mid  \ottnt{K}  [  \mathcal{S}_0 \, \mathit{x}  \ottsym{.}  \ottnt{e}  ] 
  \\
  \textbf{Typing contexts} & \Gamma & ::= &
    \emptyset  \mid \Gamma  \ottsym{,}  \mathit{x} \,  {:}  \, \ottnt{T} \mid \Gamma  \ottsym{,}  \mathit{X} \,  {:}  \,  \tceseq{ \ottnt{s} }  \mid \Gamma  \ottsym{,}  \mathit{x} \,  {:}  \, \ottnt{s}
  \\
 \end{array}
\]

\begin{defn}[Free variables and substitution]
 The notions of free variables and free predicate variables are defined as usual.
 The function $ \mathit{fv} (-)$ returns the set of free (term) variables in a given entity (for
 example, $ \mathit{fv}  (  \ottnt{e}  ) $ is the set of free variables in an expression $\ottnt{e}$).
 The function $ \mathit{fpv} (-)$ returns the set of free predicate variables in a given
 entity.
 The function $ \mathit{afv} (-)$ returns the union of the sets $ \mathit{fv} (-)$ and $ \mathit{fpv} (-)$
 for a given entity.

 Substitution $ \ottnt{e}    [  \ottnt{v}  /  \mathit{x}  ]  $ of $\ottnt{v}$ for $\mathit{x}$ in $\ottnt{e}$ is defined in a
 capture-avoiding manner as usual.
 We apply the same notation to other syntax categories including types and temporal effects.
 We also write
 $ \phi    [  \ottnt{t}  /  \mathit{x}  ]  $ for substitution of $\ottnt{t}$ for $\mathit{x}$ in $\phi$ and
 $ \ottnt{e}    [  \ottnt{A}  /  \mathit{X}  ]  $ (resp.\ $ \ottnt{C}    [  \ottnt{A}  /  \mathit{X}  ]  $ and $ \phi    [  \ottnt{A}  /  \mathit{X}  ]  $) for substitution of $\ottnt{A}$
 for $\mathit{X}$ in $\ottnt{e}$ (resp.\ $\ottnt{C}$ and $\phi$).
 For substitutions $\sigma$ and $\sigma'$,
 we write $\sigma'  \ottsym{(}  \sigma  \ottsym{)}$ for the substitution that maps
 $\mathit{X} \,  \in  \, \mathit{dom} \, \ottsym{(}  \sigma  \ottsym{)}$ to $\sigma'  \ottsym{(}  \sigma  \ottsym{(}  \mathit{X}  \ottsym{)}  \ottsym{)}$ and
 $\mathit{x} \,  \in  \, \mathit{dom} \, \ottsym{(}  \sigma  \ottsym{)}$ to $\sigma'  \ottsym{(}  \sigma  \ottsym{(}  \mathit{x}  \ottsym{)}  \ottsym{)}$.
\end{defn}

\begin{defn}[Arity]
 The function $ \mathit{arity}  (  \ottnt{v}  ) $ returns the arities of
 $\lambda$-abstractions and recursive functions $\ottnt{v}$.
 It is defined as follows.
 \[\begin{array}{lll}
   \mathit{arity}  (   \lambda\!  \,  \tceseq{ \mathit{x} }   \ottsym{.}  \ottnt{e}  )  &\defeq& \ottsym{\mbox{$\mid$}}   \tceseq{ \mathit{x} }   \ottsym{\mbox{$\mid$}} \\
   \mathit{arity}  (   \mathsf{rec}  \, (  \tceseq{  \mathit{X} ^{  \ottnt{A} _{  \mu  }  },  \mathit{Y} ^{  \ottnt{A} _{  \nu  }  }  }  ,  \mathit{f} ,   \tceseq{ \mathit{x} }  ) . \,  \ottnt{e}   )  &\defeq& \ottsym{\mbox{$\mid$}}   \tceseq{ \mathit{x} }   \ottsym{\mbox{$\mid$}}
   \end{array}
 \]
\end{defn}

\begin{conv}[Notation]
 \noindent
 \begin{itemize}
  \item We use the overline notation for (possibly empty) sequences;
        for example, $ \tceseq{ \ottnt{v} } $ denotes a sequence of values.
        We write $ \langle \rangle $ for the empty sequence, and
        write $\ottsym{\mbox{$\mid$}}   \tceseq{ \ottnt{v} }   \ottsym{\mbox{$\mid$}}$ for the length of the sequence $ \tceseq{ \ottnt{v} } $.

  \item For any sets $S_1, S_2, \cdots, S_n$, we write $S_1 \,\#\, S_2 \,\#\, \cdots \,\#\, S_n$
        to denote that, for every pair $(S_i, S_j)$ ($1 \leq i < j \leq n$), $S_i$ and $S_j$ are disjoint.

  \item We write $ \epsilon $ for the empty trace.  We write $\varpi_{{\mathrm{1}}} \,  \tceappendop  \, \varpi_{{\mathrm{2}}}$
        and $\varpi \,  \tceappendop  \, \pi$ for the concatenation of $\varpi_{{\mathrm{1}}}$
        and $\varpi_{{\mathrm{2}}}$ and that of $\varpi$ and $\pi$, respectively.

  \item For $\Phi \,  =  \,  (   \lambda_\mu\!   \,  \mathit{x}  .\,   \phi _{  \mu  }   ,   \lambda_\nu\!   \,  \mathit{y}  .\,   \phi _{  \nu  }   ) $,
        we write $ { \Phi }^\mu   \ottsym{(}  \ottnt{t}  \ottsym{)}$ and $ { \Phi }^\nu   \ottsym{(}  \ottnt{t}  \ottsym{)}$ for
        the formulas $  \phi _{  \mu  }     [  \ottnt{t}  /  \mathit{x}  ]  $ and $  \phi _{  \nu  }     [  \ottnt{t}  /  \mathit{y}  ]  $, respectively.

  \item We write $ (  \varpi  , \bot ) $ for the effect $ (   \lambda_\mu\!   \,  \mathit{x}  .\,   \mathit{x}    =    \varpi   ,   \lambda_\nu\!   \,  \mathit{y}  .\,   \bot   ) $.
        In particular, we write $ \Phi_\mathsf{val} $ for $ (   \epsilon   , \bot ) $.


  \item We write $\Gamma_{{\mathrm{1}}}  \ottsym{,}  \Gamma_{{\mathrm{2}}}$ for the concatenation of $\Gamma_{{\mathrm{1}}}$ and $\Gamma_{{\mathrm{2}}}$.

  \item We write $\Gamma  \ottsym{,}  \phi$ for the typing context $\Gamma  \ottsym{,}  \mathit{x} \,  {:}  \, \ottsym{\{}  \mathit{y} \,  {:}  \,  \mathsf{unit}  \,  |  \, \phi  \ottsym{\}}$
        for fresh variables $\mathit{x}$ and $\mathit{y}$.

  \item We write $ \tceseq{ \mathit{X} }  \,  {:}  \,  \tceseq{ \ottnt{s} } $ for the typing context $\mathit{X_{{\mathrm{1}}}} \,  {:}  \,  \tceseq{ \ottnt{s} }   \ottsym{,}   \cdots   \ottsym{,}  \mathit{X_{\ottmv{n}}} \,  {:}  \,  \tceseq{ \ottnt{s} } $
        for $ \tceseq{ \mathit{X} }  = \mathit{X_{{\mathrm{1}}}}, \cdots, \mathit{X_{\ottmv{n}}}$.

  \item For $\ottnt{C} \,  =  \,  \ottnt{T}  \,  \,\&\,  \,  \Phi  \, / \,  \ottnt{S} $,
        we write $ \ottnt{C}  . \ottnt{T} $, $ \ottnt{C}  . \Phi $, and $ \ottnt{C}  . \ottnt{S} $ for
        $\ottnt{T}$, $\Phi$, and $\ottnt{S}$, respectively.

  \item We write $\ottnt{T}  \rightarrow  \ottnt{C}$ for a function type
        $\ottsym{(}  \mathit{x} \,  {:}  \, \ottnt{T}  \ottsym{)}  \rightarrow  \ottnt{C}$ where $\mathit{x} \,  \not\in  \,  \mathit{fv}  (  \ottnt{C}  ) $.

  \item We write $\ottsym{(}  \mathit{x} \,  {:}   \ottsym{\{}  \iota \,  |  \, \phi  \ottsym{\}}  \ottsym{)}  \rightarrow  \ottnt{C}$ for the function type
        $\ottsym{(}  \mathit{x} \,  {:}  \, \ottsym{\{}  \mathit{x} \,  {:}  \, \iota \,  |  \, \phi  \ottsym{\}}  \ottsym{)}  \rightarrow  \ottnt{C}$.

  \item We write $\ottnt{C_{{\mathrm{1}}}}  \Rightarrow  \ottnt{C_{{\mathrm{2}}}}$ for a control effect
        $ ( \forall  \mathit{x}  .  \ottnt{C_{{\mathrm{1}}}}  )  \Rightarrow   \ottnt{C_{{\mathrm{2}}}} $ where $\mathit{x} \,  \not\in  \,  \mathit{fv}  (  \ottnt{C_{{\mathrm{1}}}}  ) $.

  \item We write $  \lambda\!   \, (   \tceseq{ \mathit{x}  \,  {:}  \,  \ottnt{s} }   ) .  \phi $ for $\ottsym{(}   \mu\!  \, \mathit{X}  \ottsym{(}   \tceseq{ \mathit{x}  \,  {:}  \,  \ottnt{s} }   \ottsym{)}  \ottsym{.}  \phi  \ottsym{)}$ or $\ottsym{(}   \nu\!  \, \mathit{X}  \ottsym{(}   \tceseq{ \mathit{x}  \,  {:}  \,  \ottnt{s} }   \ottsym{)}  \ottsym{.}  \phi  \ottsym{)}$
        where $\mathit{X}$ does not occur free in $\phi$.

  \item We write $ \ottnt{E}  [  \ottnt{e}  ] $ for the expression obtained by filling the hole in
        evaluation context $\ottnt{E}$ with expression $\ottnt{e}$.
        We also write $ \ottnt{E_{{\mathrm{1}}}}  [  \ottnt{E_{{\mathrm{2}}}}  ] $ for the evaluation context obtained by
        filling the hole in evaluation context $\ottnt{E_{{\mathrm{1}}}}$ with evaluation context
        $\ottnt{E_{{\mathrm{2}}}}$.
        We use the same notation for pure evaluation contexts because
        they are a subclass of evaluation contexts.


  \item We write $\ottsym{(}   \tceseq{ \mathit{x}  \,  {:}  \,  \ottnt{T} }   \ottsym{)}  \rightarrow  \ottnt{C}$ for
        $ \ottsym{(}  \mathit{x_{{\mathrm{1}}}} \,  {:}  \, \ottnt{T_{{\mathrm{1}}}}  \ottsym{)}  \rightarrow  \ottsym{(}  \cdots  \rightarrow  \ottsym{(}  \ottsym{(}   \mathit{x} _{ \ottmv{n}  \ottsym{-}  \ottsym{1} }  \,  {:}  \,  \ottnt{T} _{ \ottmv{n}  \ottsym{-}  \ottsym{1} }   \ottsym{)}  \rightarrow   \ottsym{(}  \ottsym{(}  \mathit{x_{\ottmv{n}}} \,  {:}  \, \ottnt{T_{\ottmv{n}}}  \ottsym{)}  \rightarrow  \ottnt{C}  \ottsym{)}  \,  \,\&\,  \,   \Phi_\mathsf{val}   \, / \,   \square    \ottsym{)}  \cdots  \ottsym{)}  \,  \,\&\,  \,   \Phi_\mathsf{val}   \, / \,   \square  $
        where $ \tceseq{ \mathit{x}  \,  {:}  \,  \ottnt{T} }  = \mathit{x_{{\mathrm{1}}}} \,  {:}  \, \ottnt{T_{{\mathrm{1}}}}, \cdots, \mathit{x_{\ottmv{n}}} \,  {:}  \, \ottnt{T_{\ottmv{n}}}$.

 \end{itemize}
\end{conv}

\subsubsection{Semantics of The Fixpoint Logic}

\begin{assum}[Argument Sorts of Predicates]
 We assume that the function $ \mathit{argsorts}  (  \ottnt{P}  ) $ returns the sequence of the
 argument sorts of a primitive predicate $\ottnt{P}$.
\end{assum}

\begin{defn}[Semantics of The Logic]
 We define predicate sorts $\mathcal{T}$ as follows.
 \[\begin{array}{lll}
  \mathcal{T} & ::= &  \mathbb{B}  \mid \ottnt{s}  \rightarrow  \mathcal{T} \mid  \mathcal{T}_{{\mathrm{1}}}  \times  \mathcal{T}_{{\mathrm{2}}}  \\
   \end{array}
 \]
 We define a partially ordered set $( \mathcal{D}_{ \mathcal{T} } ,  \sqsubseteq_{ \mathcal{T} } )$ by induction on
 $\mathcal{T}$ as follows.
 \[\begin{array}{l@{\ }c@{\ }l@{\qquad}l@{\ }c@{\ }l}
   \mathcal{D}_{  \mathbb{B}  }  & \defeq & \{  \top ,  \bot  \} &
   \sqsubseteq_{  \mathbb{B}  }  & \defeq & \{ ( \top , \top ), ( \bot , \bot ), ( \bot , \top ) \}
  \\
   \mathcal{D}_{ \iota }  & \defeq & \{ \ottnt{c} \mid  \mathit{ty}  (  \ottnt{c}  )  \,  =  \, \iota \} &
   \sqsubseteq_{ \iota }  & \defeq & \{ (\ottnt{c},\ottnt{c}) \mid \ottnt{c} \,  \in  \,  \mathcal{D}_{ \iota }  \}
  \\
   \mathcal{D}_{  \Sigma ^\ast  }  & \defeq & \{ \varpi \mid \varpi \,  \in  \,  \Sigma ^\ast  \} &
   \sqsubseteq_{  \Sigma ^\ast  }  & \defeq & \{ (\varpi,\varpi) \mid \varpi \,  \in  \,  \Sigma ^\ast  \}
  \\
   \mathcal{D}_{  \Sigma ^\omega  }  & \defeq & \{ \pi \mid \pi \,  \in  \,  \Sigma ^\omega  \} &
   \sqsubseteq_{  \Sigma ^\omega  }  & \defeq & \{ (\pi,\pi) \mid \pi \,  \in  \,  \Sigma ^\omega  \}
  \\
   \mathcal{D}_{  \mathcal{T}_{{\mathrm{1}}}  \times  \mathcal{T}_{{\mathrm{2}}}  }  & \defeq & \{ (\ottnt{d_{{\mathrm{1}}}},\ottnt{d_{{\mathrm{2}}}}) \mid  { \ottnt{d_{{\mathrm{1}}}} \,  \in  \,  \mathcal{D}_{ \mathcal{T}_{{\mathrm{1}}} }  } \,\mathrel{\wedge}\, { \ottnt{d_{{\mathrm{2}}}} \,  \in  \,  \mathcal{D}_{ \mathcal{T}_{{\mathrm{2}}} }  }  \} \\
   \sqsubseteq_{  \mathcal{T}_{{\mathrm{1}}}  \times  \mathcal{T}_{{\mathrm{2}}}  }  & \defeq &
   \multicolumn{4}{l}{
    \{ ((\ottnt{d_{{\mathrm{11}}}},\ottnt{d_{{\mathrm{12}}}}), (\ottnt{d_{{\mathrm{21}}}},\ottnt{d_{{\mathrm{22}}}})) \mid  { \ottnt{d_{{\mathrm{11}}}} \,  \sqsubseteq_{ \mathcal{T}_{{\mathrm{1}}} }  \, \ottnt{d_{{\mathrm{21}}}} } \,\mathrel{\wedge}\, { \ottnt{d_{{\mathrm{12}}}} \,  \sqsubseteq_{ \mathcal{T}_{{\mathrm{2}}} }  \, \ottnt{d_{{\mathrm{22}}}} }  \}
   }
  \\
   \mathcal{D}_{ \ottnt{s}  \rightarrow  \mathcal{T} }  & \defeq &
    \multicolumn{4}{l}{
     \{ \mathbf{\mathit{f} } \in  \mathcal{D}_{ \ottnt{s} }  \rightarrow  \mathcal{D}_{ \mathcal{T} }  \mid    \forall  \, { \ottnt{d_{{\mathrm{1}}}} }, { \ottnt{d_{{\mathrm{2}}}} \,  \in  \,  \mathcal{D}_{ \ottnt{s} }  } . \  \ottnt{d_{{\mathrm{1}}}} \,  \sqsubseteq_{ \ottnt{s} }  \, \ottnt{d_{{\mathrm{2}}}}   \mathrel{\Longrightarrow}  \mathbf{\mathit{f} }  \ottsym{(}  \ottnt{d_{{\mathrm{1}}}}  \ottsym{)} \,  \sqsubseteq_{ \mathcal{T} }  \, \mathbf{\mathit{f} }  \ottsym{(}  \ottnt{d_{{\mathrm{2}}}}  \ottsym{)}  \}
    }
    \\
   \sqsubseteq_{ \ottnt{s}  \rightarrow  \mathcal{T} }  & \defeq &
    \multicolumn{4}{l}{
     \{ (\mathbf{\mathit{f} }_{{\mathrm{1}}},\mathbf{\mathit{f} }_{{\mathrm{2}}}) \mid   \forall  \, { \ottnt{d} \,  \in  \,  \mathcal{D}_{ \ottnt{s} }  } . \  \mathbf{\mathit{f} }_{{\mathrm{1}}}  \ottsym{(}  \ottnt{d}  \ottsym{)} \,  \sqsubseteq_{ \mathcal{T} }  \, \mathbf{\mathit{f} }_{{\mathrm{2}}}  \ottsym{(}  \ottnt{d}  \ottsym{)}  \}%
    }
   \end{array}
 \]
 We define conjunction $ \ottnt{d_{{\mathrm{1}}}}  \mathbin{\wedge_{ \mathcal{T} } }  \ottnt{d_{{\mathrm{2}}}} $ and disjunction of $ \ottnt{d_{{\mathrm{1}}}}  \mathbin{\vee_{ \mathcal{T} } }  \ottnt{d_{{\mathrm{2}}}} $
 by induction on $\mathcal{T}$ as follows.
 \[\begin{array}{lcl}
   \ottnt{d_{{\mathrm{1}}}}  \mathbin{\wedge_{  \mathbb{B}  } }  \ottnt{d_{{\mathrm{2}}}}  &\defeq& \ottnt{d_{{\mathrm{1}}}} \,  \wedge  \, \ottnt{d_{{\mathrm{2}}}}
   \quad \text{(if $\ottnt{d_{{\mathrm{1}}}},\ottnt{d_{{\mathrm{2}}}} \in \{  \top ,  \bot  \}$)}
  \\
   \ottsym{(}  \ottnt{d_{{\mathrm{11}}}}  \ottsym{,}  \ottnt{d_{{\mathrm{12}}}}  \ottsym{)}  \mathbin{\wedge_{  \mathcal{T}_{{\mathrm{1}}}  \times  \mathcal{T}_{{\mathrm{2}}}  } }  \ottsym{(}  \ottnt{d_{{\mathrm{21}}}}  \ottsym{,}  \ottnt{d_{{\mathrm{22}}}}  \ottsym{)}  &\defeq& \ottsym{(}   \ottnt{d_{{\mathrm{11}}}}  \mathbin{\wedge_{ \mathcal{T}_{{\mathrm{1}}} } }  \ottnt{d_{{\mathrm{21}}}}   \ottsym{,}   \ottnt{d_{{\mathrm{12}}}}  \mathbin{\wedge_{ \mathcal{T}_{{\mathrm{2}}} } }  \ottnt{d_{{\mathrm{22}}}}   \ottsym{)}
  \\
   \mathbf{\mathit{f} }_{{\mathrm{1}}}  \mathbin{\wedge_{ \ottnt{s}  \rightarrow  \mathcal{T} } }  \mathbf{\mathit{f} }_{{\mathrm{2}}}  &\defeq&  \lambda\!  \, \mathit{x} \,  {:}  \, \ottnt{s}  \ottsym{.}   \mathbf{\mathit{f} }_{{\mathrm{1}}}  \ottsym{(}  \mathit{x}  \ottsym{)}  \mathbin{\wedge_{ \mathcal{T} } }  \mathbf{\mathit{f} }_{{\mathrm{2}}}  \ottsym{(}  \mathit{x}  \ottsym{)} 
  \\[2ex]
   \ottnt{d_{{\mathrm{1}}}}  \mathbin{\vee_{  \mathbb{B}  } }  \ottnt{d_{{\mathrm{2}}}}  &\defeq& \ottnt{d_{{\mathrm{1}}}} \,  \vee  \, \ottnt{d_{{\mathrm{2}}}}
   \quad \text{(if $\ottnt{d_{{\mathrm{1}}}},\ottnt{d_{{\mathrm{2}}}} \in \{  \top ,  \bot  \}$)}
  \\
   \ottsym{(}  \ottnt{d_{{\mathrm{11}}}}  \ottsym{,}  \ottnt{d_{{\mathrm{12}}}}  \ottsym{)}  \mathbin{\vee_{  \mathcal{T}_{{\mathrm{1}}}  \times  \mathcal{T}_{{\mathrm{2}}}  } }  \ottsym{(}  \ottnt{d_{{\mathrm{21}}}}  \ottsym{,}  \ottnt{d_{{\mathrm{22}}}}  \ottsym{)}  &\defeq& \ottsym{(}   \ottnt{d_{{\mathrm{11}}}}  \mathbin{\vee_{ \mathcal{T}_{{\mathrm{1}}} } }  \ottnt{d_{{\mathrm{21}}}}   \ottsym{,}   \ottnt{d_{{\mathrm{12}}}}  \mathbin{\vee_{ \mathcal{T}_{{\mathrm{2}}} } }  \ottnt{d_{{\mathrm{22}}}}   \ottsym{)}
  \\
   \mathbf{\mathit{f} }_{{\mathrm{1}}}  \mathbin{\vee_{ \ottnt{s}  \rightarrow  \mathcal{T} } }  \mathbf{\mathit{f} }_{{\mathrm{2}}}  &\defeq&  \lambda\!  \, \mathit{x} \,  {:}  \, \ottnt{s}  \ottsym{.}   \mathbf{\mathit{f} }_{{\mathrm{1}}}  \ottsym{(}  \mathit{x}  \ottsym{)}  \mathbin{\vee_{ \mathcal{T} } }  \mathbf{\mathit{f} }_{{\mathrm{2}}}  \ottsym{(}  \mathit{x}  \ottsym{)} 
  \\[2ex]
   \end{array}
 \]
 We write $ \otimes {  \tceseq{ \mathcal{T} }  } $ for $ \mathcal{T}_{{\mathrm{1}}}  \times   \cdots \times  \mathcal{T}_{\ottmv{n}}  $
 when $ \tceseq{ \mathcal{T} }  = \mathcal{T}_{{\mathrm{1}}}, \cdots, \mathcal{T}_{\ottmv{n}}$ and $\ottmv{n} \,  \geq  \, \ottsym{1}$.
 We also write $ \otimes {  \tceseq{ \ottnt{d} }  } $ for $(\ottnt{d_{{\mathrm{1}}}}, (\cdots, \ottnt{d_{\ottmv{n}}}) \cdots )$
 when $ \tceseq{ \ottnt{d} }  = \ottnt{d_{{\mathrm{1}}}}, \cdots, \ottnt{d_{\ottmv{n}}}$ and $\ottmv{n} \,  \geq  \, \ottsym{1}$.

 Let $\theta$ be the metavariable ranging over functions that map variables
 to $\ottnt{d} \,  \in  \,  \mathcal{D}_{ \ottnt{s} } $ for some $\ottnt{s}$, and predicate variables to
 $\mathbf{\mathit{f} } \,  \in  \,  \mathcal{D}_{  \tceseq{ \ottnt{s} }   \rightarrow   \mathbb{B}  } $ for some $ \tceseq{ \ottnt{s} } $.
 We write $\theta_{{\mathrm{1}}}  \mathbin{\uplus}  \theta_{{\mathrm{2}}}$ for the union of the graphs of $\theta_{{\mathrm{1}}}$ and
 $\theta_{{\mathrm{2}}}$ when their domains are disjoint.
 We write $\mathit{dom} \, \ottsym{(}  \theta  \ottsym{)}$ for the set of variables and predicate variables
 mapped by $\theta$.

 We suppose that, for each term function $\mathit{F}$ such that $\mathit{ty} \, \ottsym{(}  \mathit{F}  \ottsym{)} \,  =  \,  \tceseq{ \ottnt{s} }   \rightarrow  \ottnt{s_{{\mathrm{0}}}}$,
 there exists a function $ { [\![  \mathit{F}  ]\!] } $ that maps $ \tceseq{ \ottnt{d} \,  \in  \,  \mathcal{D}_{ \ottnt{s} }  } $ to some $\ottnt{d_{{\mathrm{0}}}} \,  \in  \,  \mathcal{D}_{ \ottnt{s_{{\mathrm{0}}}} } $.
 We define the valuation $ { [\![  \ottnt{t}  ]\!]_{ \theta } } $ of term $\ottnt{t}$ with mapping
 $\theta$ by induction on $\ottnt{t}$ as follows.
 \[\begin{array}{lcl}
   { [\![  \mathit{x}  ]\!]_{ \theta } }  &\defeq& \theta  \ottsym{(}  \mathit{x}  \ottsym{)}
  \\
   { [\![  \mathit{F}  \ottsym{(}   \tceseq{ \ottnt{t} }   \ottsym{)}  ]\!]_{ \theta } }  &\defeq&  { [\![  \mathit{F}  ]\!] }   \ottsym{(}   \tceseq{  { [\![  \ottnt{t}  ]\!]_{ \theta } }  }   \ottsym{)}
  \\
   \end{array}
 \]

 For a subset $\mathit{D}$ of $ \mathcal{D}_{  \otimes {  \tceseq{  \tceseq{ \ottnt{s} }   \rightarrow   \mathbb{B}  }  }  } $, we use the notation
 $  \bigsqcap\nolimits_{  \otimes {  \tceseq{  \tceseq{ \ottnt{s} }   \rightarrow   \mathbb{B}  }  }  }    \mathit{D} $ and
 $  \bigsqcup\nolimits_{  \otimes {  \tceseq{  \tceseq{ \ottnt{s} }   \rightarrow   \mathbb{B}  }  }  }    \mathit{D} $
 for the greatest lower bound and least upper bound of $\mathit{D}$
 with respect to $ \sqsubseteq_{  \otimes {  \tceseq{  \tceseq{ \ottnt{s} }   \rightarrow   \mathbb{B}  }  }  } $, respectively.
 See \reflem{ts-logic-domain-lattice} for the proof that every partially ordered
 set $( \mathcal{D}_{ \mathcal{T} } ,  \sqsubseteq_{ \mathcal{T} } )$ is a complete lattice.
 We define the least and greatest fixpoint operators $ \mathsf{lfp}^{  \otimes {  \tceseq{  \tceseq{ \ottnt{s} }   \rightarrow   \mathbb{B}  }  }  } $ and
 $ \mathsf{gfp}^{  \otimes {  \tceseq{  \tceseq{ \ottnt{s} }   \rightarrow   \mathbb{B}  }  }  } $ as follows.
 \[\begin{array}{lcl}
   \mathsf{lfp}^{  \otimes {  \tceseq{  \tceseq{ \ottnt{s} }   \rightarrow   \mathbb{B}  }  }  }   \ottsym{(}  \mathbf{\mathit{F} }  \ottsym{)} &\defeq&   \bigsqcap\nolimits_{  \otimes {  \tceseq{  \tceseq{ \ottnt{s} }   \rightarrow   \mathbb{B}  }  }  }     \{ \,   \otimes {  \tceseq{ \mathbf{\mathit{f} } }  }   \mathrel{  \in  }   \mathcal{D}_{  \otimes {  \tceseq{  \tceseq{ \ottnt{s} }   \rightarrow   \mathbb{B}  }  }  }   \mid  \mathbf{\mathit{F} }  \ottsym{(}   \tceseq{ \mathbf{\mathit{f} } }   \ottsym{)}  \mathrel{  \sqsubseteq_{  \otimes {  \tceseq{  \tceseq{ \ottnt{s} }   \rightarrow   \mathbb{B}  }  }  }  }   \otimes {  \tceseq{ \mathbf{\mathit{f} } }  }   \, \}   \\
   \mathsf{gfp}^{  \otimes {  \tceseq{  \tceseq{ \ottnt{s} }   \rightarrow   \mathbb{B}  }  }  }   \ottsym{(}  \mathbf{\mathit{F} }  \ottsym{)} &\defeq&   \bigsqcup\nolimits_{  \otimes {  \tceseq{  \tceseq{ \ottnt{s} }   \rightarrow   \mathbb{B}  }  }  }     \{ \,   \otimes {  \tceseq{ \mathbf{\mathit{f} } }  }   \mathrel{  \in  }   \mathcal{D}_{  \otimes {  \tceseq{  \tceseq{ \ottnt{s} }   \rightarrow   \mathbb{B}  }  }  }   \mid   \otimes {  \tceseq{ \mathbf{\mathit{f} } }  }   \mathrel{  \sqsubseteq_{  \otimes {  \tceseq{  \tceseq{ \ottnt{s} }   \rightarrow   \mathbb{B}  }  }  }  }  \mathbf{\mathit{F} }  \ottsym{(}   \tceseq{ \mathbf{\mathit{f} } }   \ottsym{)}  \, \}   \\
   \end{array}
 \]

 We define the function $ \mathit{argsorts}  (  \ottnt{A}  )_{ \theta } $ that returns the sequence of the
 argument sorts of a predicate $\ottnt{A}$, as follows; we write $ \mathit{argsorts}  (  \mathbf{\mathit{f} }  ) $
 for the sequence of argument sorts of a function $\mathbf{\mathit{f} }$.
 \[\begin{array}{lcl}
   \mathit{argsorts}  (  \ottnt{P}  )_{ \theta }  &\defeq&  \mathit{argsorts}  (  \ottnt{P}  )  \\
   \mathit{argsorts}  (  \mathit{X}  )_{ \theta }  &\defeq&  \mathit{argsorts}  (  \theta  \ottsym{(}  \mathit{X}  \ottsym{)}  )  \\
   \mathit{argsorts}  (  \ottsym{(}   \mu\!  \, \mathit{X}  \ottsym{(}   \tceseq{ \mathit{x}  \,  {:}  \,  \ottnt{s} }   \ottsym{)}  \ottsym{.}  \phi  \ottsym{)}  )_{ \theta }  &\defeq&  \tceseq{ \ottnt{s} }  \\
   \mathit{argsorts}  (  \ottsym{(}   \nu\!  \, \mathit{X}  \ottsym{(}   \tceseq{ \mathit{x}  \,  {:}  \,  \ottnt{s} }   \ottsym{)}  \ottsym{.}  \phi  \ottsym{)}  )_{ \theta }  &\defeq&  \tceseq{ \ottnt{s} }  \\
   \end{array}
 \]

 We suppose that, for each $\ottnt{P}$ such that $ \mathit{argsorts}  (  \ottnt{P}  )  \,  =  \,  \tceseq{ \ottnt{s} } $, there exists
 a function $ { [\![  \ottnt{P}  ]\!]_{ \theta } } $ that maps $ \tceseq{ \ottnt{d} \,  \in  \,  \mathcal{D}_{ \ottnt{s} }  } $ to some
 $\ottnt{d_{{\mathrm{0}}}} \,  \in  \,  \mathcal{D}_{  \mathbb{B}  } $.
 In particular, if $\ottnt{P}$ is the equality, then
 $ { [\![  \ottnt{P}  ]\!]_{ \theta } } $ maps pairs of $(\ottnt{d},\ottnt{d})$ to $ \top $.
 We further define the valuation $ { [\![  \ottnt{A}  ]\!]_{ \theta } } $ of predicate $\ottnt{A}$ with
 mapping $\theta$ as follows.
 \[\begin{array}{lcl}
   { [\![  \mathit{X}  ]\!]_{ \theta } }  &\defeq& \theta  \ottsym{(}  \mathit{X}  \ottsym{)}
  \\
   { [\![  \ottsym{(}   \mu\!  \, \mathit{X}  \ottsym{(}   \tceseq{ \mathit{x}  \,  {:}  \,  \ottnt{s} }   \ottsym{)}  \ottsym{.}  \phi  \ottsym{)}  ]\!]_{ \theta } }  &\defeq&  \mathsf{lfp}^{  \tceseq{ \ottnt{s} }   \rightarrow   \mathbb{B}  }   \ottsym{(}   \lambda\!  \, \mathit{X}  \ottsym{.}   \lambda\!  \,  \tceseq{ \mathit{x}  \,  {:}  \,  \ottnt{s} }   \ottsym{.}   { [\![  \phi  ]\!]_{ \theta  \mathbin{\uplus}   \{  \mathit{X}  \mapsto  \mathit{X}  \}   \mathbin{\uplus}   \{ \tceseq{ \mathit{x}  \mapsto  \mathit{x} } \}  } }   \ottsym{)}
  \\
   { [\![  \ottsym{(}   \nu\!  \, \mathit{X}  \ottsym{(}   \tceseq{ \mathit{x}  \,  {:}  \,  \ottnt{s} }   \ottsym{)}  \ottsym{.}  \phi  \ottsym{)}  ]\!]_{ \theta } }  &\defeq&  \mathsf{gfp}^{  \tceseq{ \ottnt{s} }   \rightarrow   \mathbb{B}  }   \ottsym{(}   \lambda\!  \, \mathit{X}  \ottsym{.}   \lambda\!  \,  \tceseq{ \mathit{x}  \,  {:}  \,  \ottnt{s} }   \ottsym{.}   { [\![  \phi  ]\!]_{ \theta  \mathbin{\uplus}   \{  \mathit{X}  \mapsto  \mathit{X}  \}   \mathbin{\uplus}   \{ \tceseq{ \mathit{x}  \mapsto  \mathit{x} } \}  } }   \ottsym{)}
   \end{array}
 \]

 We define the valuation $ { [\![  \phi  ]\!]_{ \theta } } $ of formula $\phi$ with
 mapping $\theta$ by induction on $\phi$ as follows.
 \[\begin{array}{lcl}
   { [\![   \top   ]\!]_{ \theta } }  &\defeq&  \top 
  \\
   { [\![   \bot   ]\!]_{ \theta } }  &\defeq&  \bot 
  \\
   { [\![   \neg  \phi   ]\!]_{ \theta } }  &\defeq&  \neg  \ottsym{(}   { [\![  \phi  ]\!]_{ \theta } }   \ottsym{)} 
  \\
   { [\![   \phi_{{\mathrm{1}}}   \oplus   \phi_{{\mathrm{2}}}   ]\!]_{ \theta } }  &\defeq& \ottsym{(}   { [\![  \phi_{{\mathrm{1}}}  ]\!]_{ \theta } }   \ottsym{)} \, \oplus \, \ottsym{(}   { [\![  \phi_{{\mathrm{2}}}  ]\!]_{ \theta } }   \ottsym{)}
   \quad \text{(where $\oplus \,  \in  \, \ottsym{\{}   \wedge   \ottsym{,}   \vee   \ottsym{,}   \Rightarrow   \ottsym{\}}$)}
  \\
   { [\![    \forall  \,  \mathit{x}  \mathrel{  \in  }  \ottnt{s}  . \  \phi   ]\!]_{ \theta } }  &\defeq&  \top 
   \quad \text{(if $  \forall  \, { \ottnt{d} \,  \in  \,  \mathcal{D}_{ \ottnt{s} }  } . \   { [\![  \phi  ]\!]_{ \theta  \mathbin{\uplus}   \{  \mathit{x}  \mapsto  \ottnt{d}  \}  } }  \,  =  \,  \top  $)} \\
   { [\![    \forall  \,  \mathit{x}  \mathrel{  \in  }  \ottnt{s}  . \  \phi   ]\!]_{ \theta } }  &\defeq&  \bot 
   \quad \text{(otherwise)}
  \\
   { [\![    \exists  \,  \mathit{x}  \mathrel{  \in  }  \ottnt{s}  . \  \phi   ]\!]_{ \theta } }  &\defeq&  \top 
   \quad \text{(if $  \exists  \, { \ottnt{d} \,  \in  \,  \mathcal{D}_{ \ottnt{s} }  } . \   { [\![  \phi  ]\!]_{ \theta  \mathbin{\uplus}   \{  \mathit{x}  \mapsto  \ottnt{d}  \}  } }  \,  =  \,  \top  $)} \\
   { [\![    \exists  \,  \mathit{x}  \mathrel{  \in  }  \ottnt{s}  . \  \phi   ]\!]_{ \theta } }  &\defeq&  \bot 
   \quad \text{(otherwise)}
  \\
   { [\![    \forall  \,  \mathit{X} \,  {:}  \,  \tceseq{ \ottnt{s} }   . \  \phi   ]\!]_{ \theta } }  &\defeq&  \top 
   \quad \text{(if $  \forall  \, { \mathbf{\mathit{f} } \,  \in  \,  \mathcal{D}_{  \tceseq{ \ottnt{s} }   \rightarrow   \mathbb{B}  }  } . \   { [\![  \phi  ]\!]_{ \theta  \mathbin{\uplus}   \{  \mathit{X}  \mapsto  \mathbf{\mathit{f} }  \}  } }  \,  =  \,  \top  $)} \\
   { [\![    \forall  \,  \mathit{X} \,  {:}  \,  \tceseq{ \ottnt{s} }   . \  \phi   ]\!]_{ \theta } }  &\defeq&  \bot 
   \quad \text{(otherwise)}
  \\
   { [\![    \exists  \,  \mathit{X} \,  {:}  \,  \tceseq{ \ottnt{s} }   . \  \phi   ]\!]_{ \theta } }  &\defeq&  \top 
   \quad \text{(if $  \exists  \, { \mathbf{\mathit{f} } \,  \in  \,  \mathcal{D}_{  \tceseq{ \ottnt{s} }   \rightarrow   \mathbb{B}  }  } . \   { [\![  \phi  ]\!]_{ \theta  \mathbin{\uplus}   \{  \mathit{X}  \mapsto  \mathbf{\mathit{f} }  \}  } }  \,  =  \,  \top  $)} \\
   { [\![    \exists  \,  \mathit{X} \,  {:}  \,  \tceseq{ \ottnt{s} }   . \  \phi   ]\!]_{ \theta } }  &\defeq&  \bot 
   \quad \text{(otherwise)}
  \\
   { [\![  \ottnt{A}  \ottsym{(}   \tceseq{ \ottnt{t} }   \ottsym{)}  ]\!]_{ \theta } }  &\defeq&  \top 
   \quad \text{(if $ {  {  \mathit{argsorts}  (  \ottnt{A}  )_{ \theta }  \,  =  \,  \tceseq{ \ottnt{s} }  } \,\mathrel{\wedge}\, {  \tceseq{  { [\![  \ottnt{t}  ]\!]_{ \theta } }  \,  \in  \,  \mathcal{D}_{ \ottnt{s} }  }  }  } \,\mathrel{\wedge}\, {  { [\![  \ottnt{A}  ]\!]_{ \theta } }   \ottsym{(}   \tceseq{  { [\![  \ottnt{t}  ]\!]_{ \theta } }  }   \ottsym{)} \,  =  \,  \top  } $)} \\
   { [\![  \ottnt{A}  \ottsym{(}   \tceseq{ \ottnt{t} }   \ottsym{)}  ]\!]_{ \theta } }  &\defeq&  \bot 
   \quad \text{(otherwise)}
  \\
   \end{array}
 \]

 A formula $\phi$ is valid if and only if $ { [\![  \phi  ]\!]_{  \emptyset  } }  \,  =  \,  \top $.
\end{defn}

\begin{defn}[Cocontinuous Functions]
 An endofunction $\mathbf{\mathit{F} }$ on $ \mathcal{D}_{  \otimes {  \tceseq{ \mathcal{T} }  }  } $ is cocontinuous if and only if
 \[
    \forall  \, { \mathit{D} \,  \subseteq  \,  \mathcal{D}_{  \otimes {  \tceseq{ \mathcal{T} }  }  }  } . \  \mathbf{\mathit{F} }  \ottsym{(}    \bigsqcap\nolimits_{  \otimes {  \tceseq{ \mathcal{T} }  }  }    \mathit{D}   \ottsym{)} \,  =  \,   \bigsqcap\nolimits_{  \otimes {  \tceseq{ \mathcal{T} }  }  }     \{ \,  \mathbf{\mathit{F} }  \ottsym{(}   \otimes {  \tceseq{ \mathbf{\mathit{f} } }  }   \ottsym{)}  \mid   \otimes {  \tceseq{ \mathbf{\mathit{f} } }  }   \mathrel{  \in  }  \mathit{D}  \, \}    ~.
 \]
 \TS{TODO: Check by Hiroshi}
\end{defn}

\begin{assum}[Cocontinuity on Infinite Traces] \label{assum:cocontinuity-inf}
 For any $ \tceseq{ \iota_{{\mathrm{0}}} } $, we assume that endofunctions $  \lambda\!   \, \tceseq{ \mathit{X} } .\,   \otimes {  \tceseq{  \lambda\!  \, \mathit{y} \,  {:}  \,  \Sigma ^\omega   \ottsym{.}   \lambda\!  \,  \tceseq{ \mathit{z_{{\mathrm{0}}}}  \,  {:}  \,  \iota_{{\mathrm{0}}} }   \ottsym{.}   { [\![  \phi  ]\!]_{ \theta } }  }  }  $
 on $ \mathcal{D}_{  \otimes {  \tceseq{  \Sigma ^\omega   \rightarrow   \tceseq{ \iota_{{\mathrm{0}}} }   \rightarrow   \mathbb{B}  }  }  } $ are cocontinuous.
\end{assum}

\begin{remark}[Background Theory]
  Assumption~\ref{assum:cocontinuity-inf} is a technical condition needed to prove the soundness for infinite traces. 
  Though we plan to remove the condition by developing a new proof technique for infinite traces (see the brief remark in the conclusion of the main body),
  the condition can be satisfied with a reasonable cost.  We here introduce one of the concrete ways.  Note that the cause of violating the condition is a predicate vairable application within an existential quantifier.
  We therefore eliminate existential quantifiers as follows.  Our fixpoint logic is powerful enough to eliminate quantifiers over a recursively enumerable domain.
  For example, $  \exists  \,  \mathit{x}  \mathrel{  \in  }   \mathsf{int}   . \  \phi $ has an equivalent formula $\ottsym{(}   \mu\!  \, \mathit{X}  \ottsym{(}  \mathit{x} \,  {:}  \,  \mathsf{int}   \ottsym{)}  \ottsym{.}   \phi    \vee      \phi    [   -  \mathit{x}   /  \mathit{x}  ]      \vee    \mathit{X}  \ottsym{(}   \mathit{x}  +   1    \ottsym{)}    \ottsym{)}  \ottsym{(}   0   \ottsym{)}$.  We thus can eliminate existential quantifiers over $ \Sigma ^\ast $.
  Because $ \Sigma ^\omega $ is not recursively enumerable, we assume that any formula of the form $  \exists  \,  \mathit{x}  \mathrel{  \in  }   \Sigma ^\omega   . \  \phi $ has an equivalent form $  \exists  \,  \mathit{x}  \mathrel{  \in  }   \Sigma ^\omega   . \    \mathit{x}    =    \ottnt{t}     \wedge    \phi'  $ where $\mathit{x}$ does not occur in $t$.
  Then, the quantifier in $  \exists  \,  \mathit{x}  \mathrel{  \in  }   \Sigma ^\omega   . \    \mathit{x}    =    \ottnt{t}     \wedge    \phi'  $ can be eliminated as $ \phi'    [  \ottnt{t}  /  \mathit{x}  ]  $.  Recall that the composition of temporal effects (see Definition~\ref{def:eff_comp}) uses an existential quantifier over $ \Sigma ^\omega $.
  This can be removed by introducing and using a left-quotient operator on traces.
\end{remark}

\begin{defn}[Validity of Formulas]
 We write $ { \models  \,  \phi } $ to denote a closed formula $\phi$ is valid.
 We define $\Gamma  \models  \phi$ as $ { \models  \,   \lfloor  \Gamma  \, \vdash \,  \phi  \rfloor  } $, which is defined
 as follows by induction on $\Gamma$:
 \[\begin{array}{rcl}
   \lfloor   \emptyset   \, \vdash \,  \phi  \rfloor  &\defeq& \phi \\
   \lfloor  \mathit{x} \,  {:}  \, \ottsym{\{}  \mathit{x} \,  {:}  \, \iota \,  |  \, \phi_{{\mathrm{0}}}  \ottsym{\}}  \ottsym{,}  \Gamma  \, \vdash \,  \phi  \rfloor 
   &\defeq&
     \forall  \,  \mathit{x}  \mathrel{  \in  }  \iota  . \   \phi_{{\mathrm{0}}}    \Rightarrow     \lfloor  \Gamma  \, \vdash \,  \phi  \rfloor   
   \\
   \lfloor  \mathit{x} \,  {:}  \, \ottsym{(}  \mathit{y} \,  {:}  \, \ottnt{T}  \ottsym{)}  \rightarrow  \ottnt{C}  \ottsym{,}  \Gamma  \, \vdash \,  \phi  \rfloor 
   &\defeq&
    \lfloor  \Gamma  \, \vdash \,  \phi  \rfloor 
   \\
   \lfloor  \mathit{x} \,  {:}  \,  \forall   \ottsym{(}  \mathit{X} \,  {:}  \,  \tceseq{ \ottnt{s} }   \ottsym{)}  \ottsym{.}  \ottnt{C}  \ottsym{,}  \Gamma  \, \vdash \,  \phi  \rfloor 
   &\defeq&
    \lfloor  \Gamma  \, \vdash \,  \phi  \rfloor 
   \\
   \lfloor  \mathit{x} \,  {:}  \, \ottnt{s}  \ottsym{,}  \Gamma  \, \vdash \,  \phi  \rfloor 
   &\defeq&
     \forall  \,  \mathit{x}  \mathrel{  \in  }  \ottnt{s}  . \   \lfloor  \Gamma  \, \vdash \,  \phi  \rfloor  
   \\
   \lfloor  \mathit{X} \,  {:}  \,  \tceseq{ \ottnt{s} }   \ottsym{,}  \Gamma  \, \vdash \,  \phi  \rfloor 
   &\defeq&
     \forall  \,  \mathit{X} \,  {:}  \,  \tceseq{ \ottnt{s} }   . \   \lfloor  \Gamma  \, \vdash \,  \phi  \rfloor  
   ~. \\
   \end{array}
 \]
\end{defn}

\begin{assum}[Term Functions] \label{assum:termfun}
 We suppose that a metafunction $ \mathit{ty} $ assigns to each term function $\mathit{F}$
 a functional sort $ \tceseq{ \ottnt{s} }   \rightarrow  \ottnt{s_{{\mathrm{0}}}}$.
 We also suppose that, for any constant $\ottnt{c}$, there exists a term function
 $\mathit{F}$ that represents $\ottnt{c}$; therefore, $\mathit{ty} \, \ottsym{(}  \mathit{F}  \ottsym{)} \,  =  \, \langle \rangle  \rightarrow   \mathit{ty}  (  \ottnt{c}  ) $.
\end{assum}

\subsubsection{Semantics of The Language}

\begin{assum}[Constants] \label{assum:const}
 We suppose that a metafunction $ \mathit{ty} $ assigns to each constant $\ottnt{c}$
 a base type $\iota$, and
 to each operation $\ottnt{o}$
 a closed, well-formed first-order type $ \tceseq{(  \mathit{x}  \,  {:}  \,  \ottsym{\{}  \mathit{x} \,  {:}  \, \iota \,  |  \, \phi  \ottsym{\}}  )}  \rightarrow   \ottsym{\{}  \mathit{x_{{\mathrm{0}}}} \,  {:}  \, \iota_{{\mathrm{0}}} \,  |  \, \phi_{{\mathrm{0}}}  \ottsym{\}} $.
 We assume the following on the metafunction $ \mathit{ty} $ for operations:
 For any operation $\ottnt{o}$,
 $\mathit{ty} \, \ottsym{(}  \ottnt{o}  \ottsym{)} \,  =  \,  \tceseq{(  \mathit{x}  \,  {:}  \,  \ottsym{\{}  \mathit{x} \,  {:}  \, \iota \,  |  \, \phi  \ottsym{\}}  )}  \rightarrow   \ottsym{\{}  \mathit{x_{{\mathrm{0}}}} \,  {:}  \, \iota_{{\mathrm{0}}} \,  |  \, \phi_{{\mathrm{0}}}  \ottsym{\}} $ if and only if,
 for any $ \tceseq{ \ottnt{c} } $,
 $ \tceseq{  \mathit{ty}  (  \ottnt{c}  )  \,  =  \, \iota } $ and $ \tceseq{  { \models  \,   \phi    [  \ottnt{c}  /  \mathit{x}  ]   }  } $
 imply
 $ \mathit{ty}  (   \zeta  (  \ottnt{o}  ,   \tceseq{ \ottnt{c} }   )   )  \,  =  \, \iota_{{\mathrm{0}}}$ and
 $ { \models  \,    \phi_{{\mathrm{0}}}    [ \tceseq{ \ottnt{c}  /  \mathit{x} } ]      [   \zeta  (  \ottnt{o}  ,   \tceseq{ \ottnt{c} }   )   /  \mathit{x_{{\mathrm{0}}}}  ]   } $.

 For Boolean, we suppose that:
 $ \mathit{ty}  (  \ottnt{c}  )  \,  =  \,  \mathsf{bool} $ if and only if
 $ \ottnt{c}    =     \mathsf{true}  $ or $ \ottnt{c}    =     \mathsf{false}  $.
\end{assum}

\begin{figure}[t!]
 \begin{flushleft}
  \textbf{Reduction rules} \quad \framebox{$\ottnt{e_{{\mathrm{1}}}} \,  \rightsquigarrow  \, \ottnt{e_{{\mathrm{2}}}}$}
 \end{flushleft}
 \[\begin{array}{rcll}
  \ottnt{o}  \ottsym{(}   \tceseq{ \ottnt{c} }   \ottsym{)}                   &  \rightsquigarrow  &  \zeta  (  \ottnt{o}  ,   \tceseq{ \ottnt{c} }   )          & \R{Const} \\
  \ottsym{(}   \lambda\!  \,  \tceseq{ \mathit{x} }   \ottsym{.}  \ottnt{e}  \ottsym{)} \,  \tceseq{ \ottnt{v_{{\mathrm{2}}}} }             &  \rightsquigarrow  &  \ottnt{e}    [ \tceseq{ \ottnt{v_{{\mathrm{2}}}}  /  \mathit{x} } ]   \quad \hfill
                                            \text{(where $\ottsym{\mbox{$\mid$}}   \tceseq{ \ottnt{v_{{\mathrm{2}}}} }   \ottsym{\mbox{$\mid$}} \,  =  \, \ottsym{\mbox{$\mid$}}   \tceseq{ \mathit{x} }   \ottsym{\mbox{$\mid$}}$)}
                                                                  & \R{Beta} \\
  \ottnt{v_{{\mathrm{1}}}} \,  \tceseq{ \ottnt{v_{{\mathrm{2}}}} }                   &  \rightsquigarrow  &     \ottnt{e}    [ \tceseq{  \ottnt{A} _{  \mu  }   /  \mathit{X} } ]      [ \tceseq{  \ottnt{A} _{  \nu  }   /  \mathit{Y} } ]      [  \ottnt{v_{{\mathrm{1}}}}  /  \mathit{f}  ]      [ \tceseq{ \ottnt{v_{{\mathrm{2}}}}  /  \mathit{x} } ]  
                                                                  & \R{RBeta} \\
  \multicolumn{4}{r}{
   \text{(where $\ottnt{v_{{\mathrm{1}}}} \,  =  \,  \mathsf{rec}  \, (  \tceseq{  \mathit{X} ^{  \ottnt{A} _{  \mu  }  },  \mathit{Y} ^{  \ottnt{A} _{  \nu  }  }  }  ,  \mathit{f} ,   \tceseq{ \mathit{x} }  ) . \,  \ottnt{e} $ and $\ottsym{\mbox{$\mid$}}   \tceseq{ \ottnt{v_{{\mathrm{2}}}} }   \ottsym{\mbox{$\mid$}} \,  =  \, \ottsym{\mbox{$\mid$}}   \tceseq{ \mathit{x} }   \ottsym{\mbox{$\mid$}}$)}
  } \\[1ex]
  \ottsym{(}   \Lambda   \ottsym{(}  \mathit{X} \,  {:}  \,  \tceseq{ \ottnt{s} }   \ottsym{)}  \ottsym{.}  \ottnt{e}  \ottsym{)} \, \ottnt{A}           &  \rightsquigarrow  &  \ottnt{e}    [  \ottnt{A}  /  \mathit{X}  ]                & \R{PBeta} \\
  \mathsf{if} \,  \mathsf{true}  \, \mathsf{then} \, \ottnt{e_{{\mathrm{1}}}} \, \mathsf{else} \, \ottnt{e_{{\mathrm{2}}}}  &  \rightsquigarrow  & \ottnt{e_{{\mathrm{1}}}}                  & \R{IfT} \\
  \mathsf{if} \,  \mathsf{false}  \, \mathsf{then} \, \ottnt{e_{{\mathrm{1}}}} \, \mathsf{else} \, \ottnt{e_{{\mathrm{2}}}} &  \rightsquigarrow  & \ottnt{e_{{\mathrm{2}}}}                  & \R{IfF} \\
  \mathsf{let} \, \mathit{x}  \ottsym{=}  \ottnt{v_{{\mathrm{1}}}} \,  \mathsf{in}  \, \ottnt{e_{{\mathrm{2}}}}         &  \rightsquigarrow  &  \ottnt{e_{{\mathrm{2}}}}    [  \ottnt{v_{{\mathrm{1}}}}  /  \mathit{x}  ]              & \R{Let} \\
   \resetcnstr{  \ottnt{K}  [  \mathcal{S}_0 \, \mathit{x}  \ottsym{.}  \ottnt{e}  ]  }       &  \rightsquigarrow  &  \ottnt{e}    [   \lambda\!  \, \mathit{x}  \ottsym{.}   \resetcnstr{  \ottnt{K}  [  \mathit{x}  ]  }   /  \mathit{x}  ]   & \R{Shift} \\
   \resetcnstr{ \ottnt{v} }                  &  \rightsquigarrow  & \ottnt{v}                   & \R{Reset} \\
   \end{array}\]
 \begin{flushleft}
  \textbf{Evaluation rules} \quad \framebox{$\ottnt{e_{{\mathrm{1}}}} \,  \mathrel{  \longrightarrow  }  \, \ottnt{e_{{\mathrm{2}}}} \,  \,\&\,  \, \varpi$}
 \end{flushleft}
 \begin{center}
  $\ottdruleEXXRed{}$ \hfil
  $\ottdruleEXXEv{}$ \\[2ex]
  $\ottdruleEXXLet{}$ \hfil
  $\ottdruleEXXReset{}$ \hfil
 \end{center}
 %
 %

 \caption{Semantics.}
 \label{fig:app-semantics}
\end{figure}

\begin{defn}[Reduction and Evaluation]
 The reduction relation $\ottnt{e_{{\mathrm{1}}}} \,  \rightsquigarrow  \, \ottnt{e_{{\mathrm{2}}}}$ and
 evaluation relation $\ottnt{e_{{\mathrm{1}}}} \,  \mathrel{  \longrightarrow  }  \, \ottnt{e_{{\mathrm{2}}}} \,  \,\&\,  \, \varpi$ are
 the smallest relations satisfying the rules in
 \reffig{app-semantics}.
 We write $\ottnt{e_{{\mathrm{1}}}} \,  \mathrel{  \longrightarrow  }^{ \ottmv{n} }  \, \ottnt{e_{{\mathrm{2}}}} \,  \,\&\,  \, \varpi$ if and only if
 there exist some $\ottnt{e'_{{\mathrm{0}}}}, \cdots, \ottnt{e'_{\ottmv{n}}}$ and $\varpi'_{{\mathrm{1}}}, \cdots, \varpi'_{\ottmv{n}}$
 such that
 $\ottnt{e_{{\mathrm{1}}}} \,  =  \, \ottnt{e'_{{\mathrm{0}}}}$ and $  \forall  \, { \ottmv{i} \,  <  \, \ottmv{n} } . \  \ottnt{e'_{\ottmv{i}}} \,  \mathrel{  \longrightarrow  }  \,  \ottnt{e} '_{ \ottmv{i}  \ottsym{+}  \ottsym{1} }  \,  \,\&\,  \,  \varpi' _{ \ottmv{i}  \ottsym{+}  \ottsym{1} }  $ and $\ottnt{e_{{\mathrm{2}}}} \,  =  \, \ottnt{e'_{\ottmv{n}}}$
 and $ \varpi    =     \varpi'_{{\mathrm{1}}}  \,   \tceappendop   \ \, \cdots \ \,   \tceappendop   \,  \varpi'_{\ottmv{n}}  $.
 Note that $\ottnt{e_{{\mathrm{1}}}} \,  \mathrel{  \longrightarrow  }^{ \ottsym{0} }  \, \ottnt{e_{{\mathrm{2}}}} \,  \,\&\,  \, \varpi$ if and only if $\ottnt{e_{{\mathrm{1}}}} \,  =  \, \ottnt{e_{{\mathrm{2}}}}$ and $ \varpi    =     \epsilon  $.
 We write $\ottnt{e_{{\mathrm{1}}}} \,  \mathrel{  \longrightarrow  ^*}  \, \ottnt{e_{{\mathrm{2}}}} \,  \,\&\,  \, \varpi$ if and only if $\ottnt{e_{{\mathrm{1}}}} \,  \mathrel{  \longrightarrow  }^{ \ottmv{n} }  \, \ottnt{e_{{\mathrm{2}}}} \,  \,\&\,  \, \varpi$ for some $\ottmv{n}$.
 The terminating relation $\ottnt{e} \,  \Downarrow  \, \ottnt{r} \,  \,\&\,  \, \varpi$ means that $\ottnt{e} \,  \mathrel{  \longrightarrow  ^*}  \, \ottnt{r} \,  \,\&\,  \, \varpi$.
 The diverging relation $\ottnt{e} \,  \Uparrow  \, \pi$ means either of the following:
 $   \forall  \, { \varpi'  \ottsym{,}  \pi' } . \  \pi \,  =  \, \varpi' \,  \tceappendop  \, \pi'   \mathrel{\Longrightarrow}    \exists  \, { \ottnt{e'} } . \  \ottnt{e} \,  \mathrel{  \longrightarrow  ^*}  \, \ottnt{e'} \,  \,\&\,  \, \varpi'  $; or
 $ {   \exists  \, { \varpi'  \ottsym{,}  \pi'  \ottsym{,}  \ottnt{E} } . \  \pi \,  =  \, \varpi' \,  \tceappendop  \, \pi'  } \,\mathrel{\wedge}\, { \ottnt{e} \,  \mathrel{  \longrightarrow  ^*}  \,  \ottnt{E}  [   \bullet ^{ \pi' }   ]  \,  \,\&\,  \, \varpi' } $.
\end{defn}

\subsubsection{Type System}

\begin{defn}[Typing Contexts as Functions]
 We view $\Gamma$ as a function that maps term variables to types or sorts and
 maps predicate variables to sequences of sorts, as follows:
 \begin{itemize}
  \item $\Gamma  \ottsym{(}  \mathit{x}  \ottsym{)} \,  =  \, \ottnt{T}$ if and only if $\mathit{x} \,  {:}  \, \ottnt{T} \,  \in  \, \Gamma$;
  \item $\Gamma  \ottsym{(}  \mathit{x}  \ottsym{)} \,  =  \, \ottnt{s}$ if and only if $\mathit{x} \,  {:}  \, \ottnt{s} \,  \in  \, \Gamma$; and
  \item $\Gamma  \ottsym{(}  \mathit{X}  \ottsym{)} \,  =  \,  \tceseq{ \ottnt{s} } $ if and only if $\mathit{X} \,  {:}  \,  \tceseq{ \ottnt{s} }  \,  \in  \, \Gamma$.
 \end{itemize}
 Given a typing context $\Gamma$, its domain $ \mathit{dom}  (  \Gamma  ) $ is defined by induction
 on $\Gamma$ as follows.
 \[\begin{array}{lll}
   \mathit{dom}  (   \emptyset   )        & \defeq &  \emptyset  \\
   \mathit{dom}  (  \Gamma  \ottsym{,}  \mathit{x} \,  {:}  \, \ottnt{T}  )    & \defeq &  \{  \mathit{x}  \}  \,  \mathbin{\cup}  \,  \mathit{dom}  (  \Gamma  )  \\
   \mathit{dom}  (  \Gamma  \ottsym{,}  \mathit{x} \,  {:}  \, \ottnt{s}  )    & \defeq &  \{  \mathit{x}  \}  \,  \mathbin{\cup}  \,  \mathit{dom}  (  \Gamma  )  \\
   \mathit{dom}  (  \Gamma  \ottsym{,}  \mathit{X} \,  {:}  \,  \tceseq{ \ottnt{s} }   )  & \defeq &  \{  \mathit{X}  \}  \,  \mathbin{\cup}  \,  \mathit{dom}  (  \Gamma  )  \\
   \end{array}\]
 We also define a set $ \mathit{fv}  (  \Gamma  ) $ of free variables in bindings of $\Gamma$, as follows:
 \[\begin{array}{lll}
   \mathit{fv}  (   \emptyset   )         & \defeq &  \emptyset  \\
   \mathit{fv}  (  \Gamma  \ottsym{,}  \mathit{x} \,  {:}  \, \ottnt{T}  )     & \defeq &  \mathit{fv}  (  \ottnt{T}  )  \,  \mathbin{\cup}  \,  \mathit{fv}  (  \Gamma  )  \\
   \mathit{fv}  (  \Gamma  \ottsym{,}  \mathit{x} \,  {:}  \, \ottnt{s}  )     & \defeq &  \mathit{fv}  (  \Gamma  )  \\
   \mathit{fv}  (  \Gamma  \ottsym{,}  \mathit{X} \,  {:}  \,  \tceseq{ \ottnt{s} }   )   & \defeq &  \mathit{fv}  (  \Gamma  )  ~. \\
   \end{array}\]
 Similarly, a set $ \mathit{fpv}  (  \Gamma  ) $ of free predicate variables in bindings of $\Gamma$ is defined as follows:
 \[\begin{array}{lll}
   \mathit{fpv}  (   \emptyset   )        & \defeq &  \emptyset  \\
   \mathit{fpv}  (  \Gamma  \ottsym{,}  \mathit{x} \,  {:}  \, \ottnt{T}  )    & \defeq &  \mathit{fpv}  (  \ottnt{T}  )  \,  \mathbin{\cup}  \,  \mathit{fpv}  (  \Gamma  )  \\
   \mathit{fpv}  (  \Gamma  \ottsym{,}  \mathit{x} \,  {:}  \, \ottnt{s}  )    & \defeq &  \mathit{fpv}  (  \Gamma  )  \\
   \mathit{fpv}  (  \Gamma  \ottsym{,}  \mathit{X} \,  {:}  \,  \tceseq{ \ottnt{s} }   )  & \defeq &  \mathit{fpv}  (  \Gamma  )  ~. \\
   \end{array}\]
 We write $ \mathit{afv}  (  \Gamma  ) $ for the set $ \mathit{fv}  (  \Gamma  )  \,  \mathbin{\cup}  \,  \mathit{fpv}  (  \Gamma  ) $.
\end{defn}

\begin{defn}[Temporal Effect Composition]
\label{def:eff_comp}
 For temporal effects $\Phi_{{\mathrm{1}}}$ and $\Phi_{{\mathrm{2}}}$, we define the temporal effect
 $\Phi_{{\mathrm{1}}} \,  \tceappendop  \, \Phi_{{\mathrm{2}}}$ as $ (   \lambda_\mu\!   \,  \mathit{x}  .\,  \phi_{{\mathrm{1}}}  ,   \lambda_\nu\!   \,  \mathit{y}  .\,  \phi_{{\mathrm{2}}}  ) $ where:
 \[\begin{array}{llll}
  \phi_{{\mathrm{1}}} &\defeq&   \exists  \,  \mathit{x_{{\mathrm{1}}}}  \mathrel{  \in  }   \Sigma ^\ast   . \    \exists  \,  \mathit{x_{{\mathrm{2}}}}  \mathrel{  \in  }   \Sigma ^\ast   . \     \mathit{x}    =    \mathit{x_{{\mathrm{1}}}} \,  \tceappendop  \, \mathit{x_{{\mathrm{2}}}}     \wedge     { \Phi_{{\mathrm{1}}} }^\mu   \ottsym{(}  \mathit{x_{{\mathrm{1}}}}  \ottsym{)}     \wedge     { \Phi_{{\mathrm{2}}} }^\mu   \ottsym{(}  \mathit{x_{{\mathrm{2}}}}  \ottsym{)}    & \text{and} \\
  \phi_{{\mathrm{2}}} &\defeq&   { \Phi_{{\mathrm{1}}} }^\nu   \ottsym{(}  \mathit{y}  \ottsym{)}    \vee    \ottsym{(}    \exists  \,  \mathit{x'}  \mathrel{  \in  }   \Sigma ^\ast   . \    \exists  \,  \mathit{y'}  \mathrel{  \in  }   \Sigma ^\omega   . \     \mathit{y}    =    \mathit{x'} \,  \tceappendop  \, \mathit{y'}     \wedge     { \Phi_{{\mathrm{1}}} }^\mu   \ottsym{(}  \mathit{x'}  \ottsym{)}     \wedge     { \Phi_{{\mathrm{2}}} }^\nu   \ottsym{(}  \mathit{y'}  \ottsym{)}     \ottsym{)} 
   \end{array}
 \]
 for variables $\mathit{x}, \mathit{y}, \mathit{x_{{\mathrm{1}}}}, \mathit{x_{{\mathrm{2}}}}, \mathit{x'}, \mathit{y'} \,  \not\in  \,  \mathit{fv}  (  \Phi_{{\mathrm{1}}}  )  \,  \mathbin{\cup}  \,  \mathit{fv}  (  \Phi_{{\mathrm{2}}}  ) $.
 We define composition $\Phi \,  \tceappendop  \, \ottnt{C}$ of temporal effect $\Phi$ and computation type
 $\ottnt{C}$, and composition $\Phi \,  \tceappendop  \, \ottnt{S}$ of temporal effect $\Phi$ and control effect $\ottnt{S}$
 by simultaneous induction on $\ottnt{C}$ and $\ottnt{S}$, as follows.
 \[\begin{array}{rcl}
  \Phi \,  \tceappendop  \, \ottsym{(}   \ottnt{T_{{\mathrm{0}}}}  \,  \,\&\,  \,  \Phi_{{\mathrm{0}}}  \, / \,  \ottnt{S_{{\mathrm{0}}}}   \ottsym{)} &\defeq&  \ottnt{T_{{\mathrm{0}}}}  \,  \,\&\,  \,  \Phi \,  \tceappendop  \, \Phi_{{\mathrm{0}}}  \, / \,  \Phi \,  \tceappendop  \, \ottnt{S_{{\mathrm{0}}}}  \\
  \Phi \,  \tceappendop  \,  \square            &\defeq&  \square  \\
  \Phi \,  \tceappendop  \,  ( \forall  \mathit{x}  .  \ottnt{C_{{\mathrm{1}}}}  )  \Rightarrow   \ottnt{C_{{\mathrm{2}}}}    &\defeq&  ( \forall  \mathit{x}  .  \ottnt{C_{{\mathrm{1}}}}  )  \Rightarrow   \Phi \,  \tceappendop  \, \ottnt{C_{{\mathrm{2}}}}  \\
   \end{array}
 \]
\end{defn}

\begin{defn}[Effect Quotient]
 We define
 the quotient $ { \varpi }^{-1}  \,  \tceappendop  \, \Phi$ of temporal effect $\Phi$ with respect to finite trace $\varpi$
 to be the temporal effect $ (   \lambda_\mu\!   \,  \mathit{x}  .\,   { \Phi }^\mu   \ottsym{(}  \varpi \,  \tceappendop  \, \mathit{x}  \ottsym{)}  ,   \lambda_\nu\!   \,  \mathit{y}  .\,   { \Phi }^\nu   \ottsym{(}  \varpi \,  \tceappendop  \, \mathit{y}  \ottsym{)}  ) $
 using variables $\mathit{x}, \mathit{y} \,  \not\in  \,  \mathit{fv}  (  \Phi  ) $.
 We define
 the quotient $ { \varpi }^{-1}  \,  \tceappendop  \, \ottnt{C}$ of computation type $\ottnt{C}$ and
 the quotient $ { \varpi }^{-1}  \,  \tceappendop  \, \ottnt{S}$ of control effect $\ottnt{S}$
 with respect to finite trace $\varpi$, as follows.
 \[\begin{array}{rcl}
   { \varpi }^{-1}  \,  \tceappendop  \, \ottsym{(}   \ottnt{T}  \,  \,\&\,  \,  \Phi  \, / \,  \ottnt{S}   \ottsym{)}  &\defeq&  \ottnt{T}  \,  \,\&\,  \,   { \varpi }^{-1}  \,  \tceappendop  \, \Phi  \, / \,   { \varpi }^{-1}  \,  \tceappendop  \, \ottnt{S}  \\
   { \varpi }^{-1}  \,  \tceappendop  \,  \square          &\defeq&  \square  \\
   { \varpi }^{-1}  \,  \tceappendop  \,  ( \forall  \mathit{x}  .  \ottnt{C_{{\mathrm{1}}}}  )  \Rightarrow   \ottnt{C_{{\mathrm{2}}}}  &\defeq&  ( \forall  \mathit{x}  .  \ottnt{C_{{\mathrm{1}}}}  )  \Rightarrow    { \varpi }^{-1}  \,  \tceappendop  \, \ottnt{C_{{\mathrm{2}}}}  \\
   \end{array}
 \]
\end{defn}

\begin{defn}[Binding of Control Effects]
 For control effects $\ottnt{S_{{\mathrm{1}}}}$ and $\ottnt{S_{{\mathrm{2}}}}$, we define the control effect
 $\ottnt{S_{{\mathrm{1}}}}  \gg\!\!=  \ottsym{(}   \lambda\!  \, \mathit{x_{{\mathrm{1}}}}  \ottsym{.}  \ottnt{S_{{\mathrm{2}}}}  \ottsym{)}$ as follows:
 \[\begin{array}{rcl}
   \square   \gg\!\!=  \ottsym{(}   \lambda\!  \, \mathit{x_{{\mathrm{1}}}}  \ottsym{.}   \square   \ottsym{)} &\defeq&  \square  \\
   ( \forall  \mathit{x_{{\mathrm{1}}}}  .  \ottnt{C_{{\mathrm{1}}}}  )  \Rightarrow   \ottnt{C}   \gg\!\!=  \ottsym{(}   \lambda\!  \, \mathit{x_{{\mathrm{1}}}}  \ottsym{.}   ( \forall  \mathit{x_{{\mathrm{2}}}}  .  \ottnt{C_{{\mathrm{2}}}}  )  \Rightarrow   \ottnt{C_{{\mathrm{1}}}}   \ottsym{)} &\defeq&  ( \forall  \mathit{x_{{\mathrm{2}}}}  .  \ottnt{C_{{\mathrm{2}}}}  )  \Rightarrow   \ottnt{C}  \quad \text{(if $\mathit{x_{{\mathrm{1}}}} \,  \not\in  \, \ottsym{(}   \mathit{fv}  (  \ottnt{C_{{\mathrm{2}}}}  )  \,  \mathbin{\backslash}  \,  \{  \mathit{x_{{\mathrm{2}}}}  \}   \ottsym{)}$)} ~.
   \end{array}
 \]
\end{defn}

\begin{figure}[t!]
 \begin{flushleft}
  \textbf{Subtyping rules}
  \quad \framebox{$\Gamma  \vdash  \ottnt{T_{{\mathrm{1}}}}  \ottsym{<:}  \ottnt{T_{{\mathrm{2}}}}$}
  \     \framebox{$\Gamma  \vdash  \ottnt{C_{{\mathrm{1}}}}  \ottsym{<:}  \ottnt{C_{{\mathrm{2}}}}$}
  \     \framebox{$\Gamma \,  |  \, \ottnt{T} \,  \,\&\,  \, \Phi  \vdash  \ottnt{S_{{\mathrm{1}}}}  \ottsym{<:}  \ottnt{S_{{\mathrm{2}}}}$}
  \     \framebox{$\Gamma  \vdash  \Phi_{{\mathrm{1}}}  \ottsym{<:}  \Phi_{{\mathrm{2}}}$}
 \end{flushleft}
 \begin{center}
  $\ottdruleSXXRefine{}$ \hfil
  $\ottdruleSXXFun{}$ \\[2ex]
  $\ottdruleSXXForall{}$ \hfil
  $\ottdruleSXXComp{}$ \\[2ex]
  $\ottdruleSXXEmpty{}$ \hfil
  $\ottdruleSXXAns{}$ \\[2ex]
  $\ottdruleSXXEmbed{}$ \hfil
  $\ottdruleSXXEff{}$ \hfil
 \end{center}
 \caption{Subtyping.}
 \label{fig:app-subtyping}
\end{figure}

\begin{figure}[t!]
 \begin{flushleft}
  \textbf{Well-formedness rules for typing contexts} \quad \framebox{$\vdash  \Gamma$}
 \end{flushleft}
 \begin{center}
  $\ottdruleWFCXXEmpty{}$ \hfil
  $\ottdruleWFCXXVar{}$ \\[2ex]
  $\ottdruleWFCXXSVar{}$ \hfil
  $\ottdruleWFCXXPVar{}$ \hfil
 \end{center}
 \begin{flushleft}
  \textbf{Well-formedness rules for types and effects}
   \quad \framebox{$\Gamma  \vdash  \ottnt{T}$}
   \quad \framebox{$\Gamma  \vdash  \ottnt{C}$}
   \quad \framebox{$\Gamma \,  |  \, \ottnt{T}  \vdash  \ottnt{S}$}
   \quad \framebox{$\Gamma  \vdash  \Phi$}
 \end{flushleft}
 \begin{center}
  $\ottdruleWFTXXRefine{}$ \hfil
  $\ottdruleWFTXXFun{}$ \\[2ex]
  $\ottdruleWFTXXForall{}$ \hfil
  $\ottdruleWFTXXComp{}$ \\[2ex]
  $\ottdruleWFTXXEff{}$ \hfil
  $\ottdruleWFTXXEmpty{}$ \\[2ex]
  $\ottdruleWFTXXAns{}$ \\[2ex]
 \end{center}
 \caption{Well-formedness.}
 \label{fig:app-wf}
\end{figure}

\begin{figure}[t!]
 \begin{flushleft}
  \textbf{Typing rules} \quad \framebox{$\Gamma  \vdash  \ottnt{e}  \ottsym{:}  \ottnt{C}$}
 \end{flushleft}
 \begin{center}
  $\ottdruleTXXCVar{}$ \hfil
  $\ottdruleTXXVar{}$ \\[1.8ex]
  $\ottdruleTXXConst{}$ \hfil
  $\ottdruleTXXPAbs{}$ \hfil
  $\ottdruleTXXAbs{}$ \\[1.8ex]
  $\ottdruleTXXFun{}$ \\[1.8ex]
  $\ottdruleTXXSub{}$ \hfil
  $\ottdruleTXXOp{}$ \\[1.8ex]
  $\ottdruleTXXApp{}$ \hfil
  $\ottdruleTXXPApp{}$ \\[1.8ex]
  $\ottdruleTXXIf{}$ \\[1.8ex]
  $\ottdruleTXXLet{}$ \\[1.8ex]
  $\ottdruleTXXLetV{}$ \\[1.8ex]
  $\ottdruleTXXShift{}$ \\[1.8ex]
  $\ottdruleTXXShiftD{}$ \\[1.8ex]
  $\ottdruleTXXReset{}$ \\[1.8ex]
  $\ottdruleTXXEvent{}$ \hfil
  $\ottdruleTXXDiv{}$
 \end{center}
 \caption{Type system for expressions.}
 \label{fig:app-typing-exp}
\end{figure}

\begin{figure}[t!]
 \begin{flushleft}
  \textbf{Typing rules for terms} \quad \framebox{$\Gamma  \vdash  \ottnt{t}  \ottsym{:}  \ottnt{s}$}
 \end{flushleft}
 \begin{center}
  $\ottdruleTfXXVarT{}$ \hfil
  $\ottdruleTfXXVarS{}$ \\[2ex]
  $\ottdruleTfXXFun{}$ \hfil
 \end{center}
 \begin{flushleft}
  \textbf{Typing rules for predicates} \quad \framebox{$\Gamma  \vdash  \ottnt{A}  \ottsym{:}   \tceseq{ \ottnt{s} } $}
 \end{flushleft}
 \begin{center}
  $\ottdruleTfXXPrim{}$ \hfil
  $\ottdruleTfXXPVar{}$ \\[2ex]
  $\ottdruleTfXXMu{}$ \hfil
  $\ottdruleTfXXNu{}$ \hfil
 \end{center}
 \begin{flushleft}
  \textbf{Typing rules for formulas} \quad \framebox{$\Gamma  \vdash  \phi$}
 \end{flushleft}
 \begin{center}
  $\ottdruleTfXXBot{}$ \hfil
  $\ottdruleTfXXTop{}$ \hfil
  $\ottdruleTfXXPred{}$ \hfil
  $\ottdruleTfXXNeg{}$ \\[2ex]
  $\ottdruleTfXXBinOp{}$ \hfil
  $\ottdruleTfXXForallS{}$ \hfil
  $\ottdruleTfXXExistS{}$ \\[2ex]
  $\ottdruleTfXXForallP{}$ \hfil
  $\ottdruleTfXXExistP{}$ \\[2ex]
 \end{center}
 \caption{Type system for terms and formulas.}
 \label{fig:app-typing-formula}
\end{figure}

\begin{defn}[Choice]
 We define a choice function $ \gamma  (   \tceseq{ \mathit{x}  \,  {:}  \,  \ottnt{T} }   ) $, which returns a sequence
 $ \tceseq{ \mathit{x_{{\mathrm{0}}}}  \,  {:}  \,  \iota } $ of pairs of a variable and a base type, as follows.
 \[\begin{array}{lll}
   \gamma  (  \langle \rangle  )                  &\defeq&  \langle \rangle  \\
   \gamma  (  \mathit{x} \,  {:}  \, \ottsym{\{}  \mathit{y} \,  {:}  \, \iota \,  |  \, \phi  \ottsym{\}}  \ottsym{,}   \tceseq{ \mathit{z}  \,  {:}  \,  \ottnt{T'} }   )  &\defeq& \mathit{x} \,  {:}  \, \iota  \ottsym{,}   \gamma  (   \tceseq{ \mathit{z}  \,  {:}  \,  \ottnt{T'} }   )  \\
   \gamma  (  \mathit{x} \,  {:}  \, \ottnt{T}  \ottsym{,}   \tceseq{ \mathit{z}  \,  {:}  \,  \ottnt{T'} }   )          &\defeq&  \gamma  (   \tceseq{ \mathit{z}  \,  {:}  \,  \ottnt{T'} }   ) 
   \quad \text{(if $  \forall  \, { \mathit{y}  \ottsym{,}  \iota  \ottsym{,}  \phi } . \  \ottnt{T} \,  \not=  \, \ottsym{\{}  \mathit{y} \,  {:}  \, \iota \,  |  \, \phi  \ottsym{\}} $)}
   \\
   \end{array}
 \]
\end{defn}

\begin{defn}[Strictly Positive $\nu$-Effects]
 We define $ { \ottnt{C} }^\nu (  \ottnt{t}  ) $ and $ { \ottnt{S} }^\nu (  \ottnt{t}  ) $, which are formulas to state
 that $\nu$-effects occurring in strictly positive positions in $\ottnt{C}$ and
 $\ottnt{S}$ satisfy infinite trace $\ottnt{t}$, respectively, as follows.
 \[\begin{array}{lll}
   { \ottsym{(}   \ottnt{T}  \,  \,\&\,  \,  \Phi  \, / \,  \ottnt{S}   \ottsym{)} }^\nu (  \ottnt{t}  )     &\defeq&   { \Phi }^\nu   \ottsym{(}  \ottnt{t}  \ottsym{)}    \wedge     { \ottnt{S} }^\nu (  \ottnt{t}  )   \\
   {  \square  }^\nu (  \ottnt{t}  )            &\defeq&  \top  \\
   { \ottsym{(}   ( \forall  \mathit{x}  .  \ottnt{C_{{\mathrm{1}}}}  )  \Rightarrow   \ottnt{C_{{\mathrm{2}}}}   \ottsym{)} }^\nu (  \ottnt{t}  )  &\defeq&  { \ottnt{C_{{\mathrm{2}}}} }^\nu (  \ottnt{t}  ) 
   \end{array}
 \]
\end{defn}


\begin{defn}[Consistency of Computation Types]
 The predicates $ \tceseq{ \ottsym{(}  \mathit{X}  \ottsym{,}  \mathit{Y}  \ottsym{)} }   \ottsym{;}   \tceseq{ \mathit{z_{{\mathrm{0}}}}  \,  {:}  \,  \iota_{{\mathrm{0}}} }   \vdash  \ottnt{C_{{\mathrm{0}}}}  \succ  \ottnt{C}$ and $ \tceseq{ \ottsym{(}  \mathit{X}  \ottsym{,}  \mathit{Y}  \ottsym{)} }   \ottsym{;}   \tceseq{ \mathit{z_{{\mathrm{0}}}}  \,  {:}  \,  \iota_{{\mathrm{0}}} }   \vdash  \ottnt{S_{{\mathrm{0}}}}  \succ  \ottnt{S}$ are the smallest relations satisfying the following rules.
 \[\begin{array}{rcl}
   { \Phi_{{\mathrm{0}}} \,  =  \,  (   \lambda_\mu\!   \,  \mathit{x}  .\,  \mathit{X_{{\mathrm{0}}}}  \ottsym{(}  \mathit{x}  \ottsym{,}   \tceseq{ \mathit{z_{{\mathrm{0}}}} }   \ottsym{)}  ,   \lambda_\nu\!   \,  \mathit{y}  .\,  \mathit{Y_{{\mathrm{0}}}}  \ottsym{(}  \mathit{y}  \ottsym{,}   \tceseq{ \mathit{z_{{\mathrm{0}}}} }   \ottsym{)}  )  } \,\mathrel{\wedge}\, {  \tceseq{ \ottsym{(}  \mathit{X}  \ottsym{,}  \mathit{Y}  \ottsym{)} }   \ottsym{;}   \tceseq{ \mathit{z_{{\mathrm{0}}}}  \,  {:}  \,  \iota_{{\mathrm{0}}} }   \vdash  \ottnt{S_{{\mathrm{0}}}}  \succ  \ottnt{S} } 
  & \Rightarrow &
  \ottsym{(}  \mathit{X_{{\mathrm{0}}}}  \ottsym{,}  \mathit{Y_{{\mathrm{0}}}}  \ottsym{)}  \ottsym{,}   \tceseq{ \ottsym{(}  \mathit{X}  \ottsym{,}  \mathit{Y}  \ottsym{)} }   \ottsym{;}   \tceseq{ \mathit{z_{{\mathrm{0}}}}  \,  {:}  \,  \iota_{{\mathrm{0}}} }   \vdash   \ottnt{T}  \,  \,\&\,  \,  \Phi_{{\mathrm{0}}}  \, / \,  \ottnt{S_{{\mathrm{0}}}}   \succ   \ottnt{T}  \,  \,\&\,  \,  \Phi  \, / \,  \ottnt{S}  \\[2ex]
   \emptyset   \ottsym{;}   \tceseq{ \mathit{z_{{\mathrm{0}}}}  \,  {:}  \,  \iota_{{\mathrm{0}}} }   \vdash   \square   \succ   \square  \\[2ex]
   \tceseq{ \ottsym{(}  \mathit{X}  \ottsym{,}  \mathit{Y}  \ottsym{)} }   \ottsym{;}   \tceseq{ \mathit{z_{{\mathrm{0}}}}  \,  {:}  \,  \iota_{{\mathrm{0}}} }   \vdash  \ottnt{C_{{\mathrm{02}}}}  \succ  \ottnt{C_{{\mathrm{2}}}}
  & \Rightarrow &
   \tceseq{ \ottsym{(}  \mathit{X}  \ottsym{,}  \mathit{Y}  \ottsym{)} }   \ottsym{;}   \tceseq{ \mathit{z_{{\mathrm{0}}}}  \,  {:}  \,  \iota_{{\mathrm{0}}} }   \vdash   ( \forall  \mathit{x}  .  \ottnt{C}  )  \Rightarrow   \ottnt{C_{{\mathrm{02}}}}   \succ   ( \forall  \mathit{x}  .  \ottnt{C}  )  \Rightarrow   \ottnt{C_{{\mathrm{2}}}}  ~.
   \end{array}
 \]
\end{defn}

\begin{defn}[Occurrence Checking]
 The predicates $ \tceseq{ \ottsym{(}  \mathit{X'}  \ottsym{,}  \mathit{Y'}  \ottsym{)} }   \ottsym{;}   \tceseq{ \ottsym{(}  \mathit{X}  \ottsym{,}  \mathit{Y}  \ottsym{)} }  \,  \vdash ^{\circ}  \, \ottnt{C}$ and $ \tceseq{ \ottsym{(}  \mathit{X'}  \ottsym{,}  \mathit{Y'}  \ottsym{)} }   \ottsym{;}   \tceseq{ \ottsym{(}  \mathit{X}  \ottsym{,}  \mathit{Y}  \ottsym{)} }  \,  \vdash ^{\circ}  \, \ottnt{S}$ are the smallest relations satisfying the following rules.
 \[\begin{array}{rcl}
   {   \{   \tceseq{ \mathit{X'} }   \ottsym{,}   \tceseq{ \mathit{Y'} }   \ottsym{,}  \mathit{X_{{\mathrm{0}}}}  \ottsym{,}  \mathit{Y_{{\mathrm{0}}}}  \ottsym{,}   \tceseq{ \mathit{X} }   \ottsym{,}   \tceseq{ \mathit{Y} }   \}   \, \ottsym{\#} \,   \mathit{fpv}  (   \ottnt{C}  . \ottnt{T}   )   } \,\mathrel{\wedge}\, {    }  \\
   {  {   \{   \tceseq{ \mathit{Y'} }   \ottsym{,}  \mathit{Y_{{\mathrm{0}}}}  \}   \, \ottsym{\#} \,  \ottsym{(}   \mathit{fpv}^\mathbf{n}  (   { \ottsym{(}   \ottnt{C}  . \Phi   \ottsym{)} }^\nu   )  \,  \mathbin{\cup}  \,  \mathit{fpv}  (   { \ottsym{(}   \ottnt{C}  . \Phi   \ottsym{)} }^\mu   )   \ottsym{)}  } \,\mathrel{\wedge}\, { \mathit{X_{{\mathrm{0}}}} \,  \not\in  \,  \mathit{fpv}^\mathbf{n}  (   { \ottsym{(}   \ottnt{C}  . \Phi   \ottsym{)} }^\mu   )  }  } \,\mathrel{\wedge}\, {    } 
  & \Rightarrow &
   \tceseq{ \ottsym{(}  \mathit{X'}  \ottsym{,}  \mathit{Y'}  \ottsym{)} }   \ottsym{;}  \ottsym{(}  \mathit{X_{{\mathrm{0}}}}  \ottsym{,}  \mathit{Y_{{\mathrm{0}}}}  \ottsym{)}  \ottsym{,}   \tceseq{ \ottsym{(}  \mathit{X}  \ottsym{,}  \mathit{Y}  \ottsym{)} }  \,  \vdash ^{\circ}  \, \ottnt{C} \\
   {   \{   \tceseq{ \mathit{X} }   \ottsym{,}   \tceseq{ \mathit{Y} }   \}   \, \ottsym{\#} \,   \mathit{fpv}  (   \ottnt{C}  . \Phi   )   } \,\mathrel{\wedge}\, {  \tceseq{ \ottsym{(}  \mathit{X'}  \ottsym{,}  \mathit{Y'}  \ottsym{)} }   \ottsym{,}  \ottsym{(}  \mathit{X_{{\mathrm{0}}}}  \ottsym{,}  \mathit{Y_{{\mathrm{0}}}}  \ottsym{)}  \ottsym{;}   \tceseq{ \ottsym{(}  \mathit{X}  \ottsym{,}  \mathit{Y}  \ottsym{)} }  \,  \vdash ^{\circ}  \,  \ottnt{C}  . \ottnt{S}  }  \quad\ \ %
  \\[2ex]
   \tceseq{ \ottsym{(}  \mathit{X'}  \ottsym{,}  \mathit{Y'}  \ottsym{)} }   \ottsym{;}   \emptyset  \,  \vdash ^{\circ}  \,  \square 
  \\[2ex]
   {   \{   \tceseq{ \mathit{X'} }   \ottsym{,}   \tceseq{ \mathit{Y'} }   \ottsym{,}   \tceseq{ \mathit{X} }   \ottsym{,}   \tceseq{ \mathit{Y} }   \}   \, \ottsym{\#} \,   \mathit{fpv}  (  \ottnt{C_{{\mathrm{1}}}}  )   } \,\mathrel{\wedge}\, {  \tceseq{ \ottsym{(}  \mathit{X'}  \ottsym{,}  \mathit{Y'}  \ottsym{)} }   \ottsym{;}   \tceseq{ \ottsym{(}  \mathit{X}  \ottsym{,}  \mathit{Y}  \ottsym{)} }  \,  \vdash ^{\circ}  \, \ottnt{C_{{\mathrm{2}}}} } 
  & \Rightarrow &
   \tceseq{ \ottsym{(}  \mathit{X'}  \ottsym{,}  \mathit{Y'}  \ottsym{)} }   \ottsym{;}   \tceseq{ \ottsym{(}  \mathit{X}  \ottsym{,}  \mathit{Y}  \ottsym{)} }  \,  \vdash ^{\circ}  \,  ( \forall  \mathit{x}  .  \ottnt{C_{{\mathrm{1}}}}  )  \Rightarrow   \ottnt{C_{{\mathrm{2}}}} 
   \end{array}
 \]
\end{defn}

\begin{defn}[Consistency and Occurrence Checking]
 Te defined the following predicates:
 \[\begin{array}{rcl}
   \tceseq{ \ottsym{(}  \mathit{X}  \ottsym{,}  \mathit{Y}  \ottsym{)} }   \ottsym{;}   \tceseq{ \mathit{z_{{\mathrm{0}}}}  \,  {:}  \,  \iota_{{\mathrm{0}}} }   \vdash  \ottnt{C_{{\mathrm{0}}}}  \stackrel{\succ}{\circ}  \ottnt{C} &\defeq&
    {  \tceseq{ \ottsym{(}  \mathit{X}  \ottsym{,}  \mathit{Y}  \ottsym{)} }   \ottsym{;}   \tceseq{ \mathit{z_{{\mathrm{0}}}}  \,  {:}  \,  \iota_{{\mathrm{0}}} }   \vdash  \ottnt{C_{{\mathrm{0}}}}  \succ  \ottnt{C} } \,\mathrel{\wedge}\, {  \emptyset   \ottsym{;}   \tceseq{ \ottsym{(}  \mathit{X}  \ottsym{,}  \mathit{Y}  \ottsym{)} }  \,  \vdash ^{\circ}  \, \ottnt{C} } 
  \\
   \tceseq{ \ottsym{(}  \mathit{X}  \ottsym{,}  \mathit{Y}  \ottsym{)} }   \ottsym{;}   \tceseq{ \mathit{z_{{\mathrm{0}}}}  \,  {:}  \,  \iota_{{\mathrm{0}}} }   \vdash  \ottnt{S_{{\mathrm{0}}}}  \stackrel{\succ}{\circ}  \ottnt{S} &\defeq&
   {  \tceseq{ \ottsym{(}  \mathit{X}  \ottsym{,}  \mathit{Y}  \ottsym{)} }   \ottsym{;}   \tceseq{ \mathit{z_{{\mathrm{0}}}}  \,  {:}  \,  \iota_{{\mathrm{0}}} }   \vdash  \ottnt{S_{{\mathrm{0}}}}  \succ  \ottnt{S} } \,\mathrel{\wedge}\, {  \emptyset   \ottsym{;}   \tceseq{ \ottsym{(}  \mathit{X}  \ottsym{,}  \mathit{Y}  \ottsym{)} }  \,  \vdash ^{\circ}  \, \ottnt{S} } 
   \end{array}\]
\end{defn}

\begin{defn}[Substitution of Predicate Fixpoint]
 The atomic predicates $  \mu\!   \, (  \mathit{X} ,  \mathit{Y} ,   \tceseq{ \mathit{z_{{\mathrm{0}}}}  \,  {:}  \,  \iota_{{\mathrm{0}}} }  ,  \Phi  ) $ and $  \nu\!   \, (  \mathit{X} ,  \mathit{Y} ,   \tceseq{ \mathit{z_{{\mathrm{0}}}}  \,  {:}  \,  \iota_{{\mathrm{0}}} }  ,  \Phi  ) $ are defined as follows.
 \[\begin{array}{lll}
    \mu\!   \, (  \mathit{X} ,  \mathit{Y} ,   \tceseq{ \mathit{z_{{\mathrm{0}}}}  \,  {:}  \,  \iota_{{\mathrm{0}}} }  ,  \Phi  )  &\defeq& \ottsym{(}   \mu\!  \, \mathit{X}  \ottsym{(}  \mathit{x} \,  {:}  \,  \Sigma ^\ast   \ottsym{,}   \tceseq{ \mathit{z_{{\mathrm{0}}}}  \,  {:}  \,  \iota_{{\mathrm{0}}} }   \ottsym{)}  \ottsym{.}   { \Phi }^\mu   \ottsym{(}  \mathit{x}  \ottsym{)}  \ottsym{)}
    \quad \text{(if $\mathit{x} \,  \not\in  \,  \mathit{fv}  (  \Phi  ) $)} \\
    \nu\!   \, (  \mathit{X} ,  \mathit{Y} ,   \tceseq{ \mathit{z_{{\mathrm{0}}}}  \,  {:}  \,  \iota_{{\mathrm{0}}} }  ,  \Phi  )  &\defeq&  \ottsym{(}   \nu\!  \, \mathit{Y}  \ottsym{(}  \mathit{y} \,  {:}  \,  \Sigma ^\omega   \ottsym{,}   \tceseq{ \mathit{z_{{\mathrm{0}}}}  \,  {:}  \,  \iota_{{\mathrm{0}}} }   \ottsym{)}  \ottsym{.}   { \Phi }^\nu   \ottsym{(}  \mathit{y}  \ottsym{)}  \ottsym{)}    [    \mu\!   \, (  \mathit{X} ,  \mathit{Y} ,   \tceseq{ \mathit{z_{{\mathrm{0}}}}  \,  {:}  \,  \iota_{{\mathrm{0}}} }  ,  \Phi  )   /  \mathit{X}  ]  
    \quad \text{(if $\mathit{y} \,  \not\in  \,  \mathit{fv}  (  \Phi  ) $)}
   \end{array}
 \]
 The predicates $ \tceseq{ \ottsym{(}  \mathit{X}  \ottsym{,}  \mathit{Y}  \ottsym{)} }   \ottsym{;}   \tceseq{ \mathit{z_{{\mathrm{0}}}}  \,  {:}  \,  \iota_{{\mathrm{0}}} }  \,  |  \, \ottnt{C}  \vdash  \sigma$ and $ \tceseq{ \ottsym{(}  \mathit{X}  \ottsym{,}  \mathit{Y}  \ottsym{)} }   \ottsym{;}   \tceseq{ \mathit{z_{{\mathrm{0}}}}  \,  {:}  \,  \iota_{{\mathrm{0}}} }  \,  |  \, \ottnt{S}  \vdash  \sigma$ are the smallest relations satisfying the following rules.
 %
 \[\begin{array}{rcl}
   \tceseq{ \ottsym{(}  \mathit{X}  \ottsym{,}  \mathit{Y}  \ottsym{)} }   \ottsym{;}   \tceseq{ \mathit{z_{{\mathrm{0}}}}  \,  {:}  \,  \iota_{{\mathrm{0}}} }  \,  |  \, \ottnt{S}  \vdash  \sigma
  & \Rightarrow &
  \ottsym{(}  \mathit{X_{{\mathrm{0}}}}  \ottsym{,}  \mathit{Y_{{\mathrm{0}}}}  \ottsym{)}  \ottsym{,}   \tceseq{ \ottsym{(}  \mathit{X}  \ottsym{,}  \mathit{Y}  \ottsym{)} }   \ottsym{;}   \tceseq{ \mathit{z_{{\mathrm{0}}}}  \,  {:}  \,  \iota_{{\mathrm{0}}} }  \,  |  \,  \ottnt{T}  \,  \,\&\,  \,  \Phi  \, / \,  \ottnt{S}   \vdash  \sigma'  \ottsym{(}  \sigma  \ottsym{)}  \mathbin{\uplus}  \sigma' \\ &&
  \quad \text{where $\sigma' \,  =  \,  [    \mu\!   \, (  \mathit{X_{{\mathrm{0}}}} ,  \mathit{Y_{{\mathrm{0}}}} ,   \tceseq{ \mathit{z_{{\mathrm{0}}}}  \,  {:}  \,  \iota_{{\mathrm{0}}} }  ,  \Phi  )   /  \mathit{X_{{\mathrm{0}}}}  ]   \mathbin{\uplus}   [    \nu\!   \, (  \mathit{X_{{\mathrm{0}}}} ,  \mathit{Y_{{\mathrm{0}}}} ,   \tceseq{ \mathit{z_{{\mathrm{0}}}}  \,  {:}  \,  \iota_{{\mathrm{0}}} }  ,  \Phi  )   /  \mathit{Y_{{\mathrm{0}}}}  ] $}
  \\[2ex]
   \emptyset   \ottsym{;}   \tceseq{ \mathit{z_{{\mathrm{0}}}}  \,  {:}  \,  \iota_{{\mathrm{0}}} }  \,  |  \,  \square   \vdash   \emptyset 
  \\[2ex]
   \tceseq{ \ottsym{(}  \mathit{X}  \ottsym{,}  \mathit{Y}  \ottsym{)} }   \ottsym{;}   \tceseq{ \mathit{z_{{\mathrm{0}}}}  \,  {:}  \,  \iota_{{\mathrm{0}}} }  \,  |  \, \ottnt{C_{{\mathrm{2}}}}  \vdash  \sigma
  & \Rightarrow &
   \tceseq{ \ottsym{(}  \mathit{X}  \ottsym{,}  \mathit{Y}  \ottsym{)} }   \ottsym{;}   \tceseq{ \mathit{z_{{\mathrm{0}}}}  \,  {:}  \,  \iota_{{\mathrm{0}}} }  \,  |  \,  ( \forall  \mathit{x}  .  \ottnt{C_{{\mathrm{1}}}}  )  \Rightarrow   \ottnt{C_{{\mathrm{2}}}}   \vdash  \sigma
   \end{array}
 \]
 %
 %
\end{defn}

\begin{defn}[Type System]
 \noindent
 \begin{itemize}
  \item Subtyping judgments $\Gamma  \vdash  \ottnt{T_{{\mathrm{1}}}}  \ottsym{<:}  \ottnt{T_{{\mathrm{2}}}}$, $\Gamma  \vdash  \ottnt{C_{{\mathrm{1}}}}  \ottsym{<:}  \ottnt{C_{{\mathrm{2}}}}$,
        $\Gamma \,  |  \, \ottnt{T} \,  \,\&\,  \, \Phi  \vdash  \ottnt{S_{{\mathrm{1}}}}  \ottsym{<:}  \ottnt{S_{{\mathrm{2}}}}$, and $\Gamma  \vdash  \Phi_{{\mathrm{1}}}  \ottsym{<:}  \Phi_{{\mathrm{2}}}$ are the
        smallest relations satisfying the
        rules in \reffig{app-subtyping}.

  \item Well-formedness judgments for
        typing contexts $\vdash  \Gamma$,
        types $\Gamma  \vdash  \ottnt{T}$, $\Gamma  \vdash  \ottnt{C}$, and $\Gamma \,  |  \, \ottnt{T}  \vdash  \ottnt{S}$, and
        effects $\Gamma  \vdash  \Phi$
        are the smallest relations satisfying the rules in \reffig{app-wf}.

  \item Typing judgment $\Gamma  \vdash  \ottnt{e}  \ottsym{:}  \ottnt{C}$ is the smallest relation satisfying
        the rules in \reffig{app-typing-exp}.
        We write $\Gamma  \vdash  \ottnt{e}  \ottsym{:}  \ottnt{T}$ for $\Gamma  \vdash  \ottnt{e}  \ottsym{:}   \ottnt{T}  \,  \,\&\,  \,   \Phi_\mathsf{val}   \, / \,   \square  $.

  \item Typing judgments for terms $\Gamma  \vdash  \ottnt{t}  \ottsym{:}  \ottnt{s}$,
        predicates $\Gamma  \vdash  \ottnt{A}  \ottsym{:}   \tceseq{ \ottnt{s} } $, and formulas $\Gamma  \vdash  \phi$
        are the smallest relations satisfying the rules in
        \reffig{app-typing-formula}.
 \end{itemize}
\end{defn}

\subsubsection{Logical Relation}



\begin{figure}[t]
 \begin{flushleft}
  \textbf{Typing rules for pure evaluation contexts} \quad \framebox{$\ottnt{K}  \ottsym{:}    \ottnt{T}  \, / \, ( \forall  \mathit{x}  .  \ottnt{C}  )   \mathrel{ \Rrightarrow }  \ottnt{T_{{\mathrm{0}}}}  \,  \,\&\,  \,  \Phi_{{\mathrm{0}}}  \, / \, ( \forall  \mathit{x_{{\mathrm{0}}}}  .  \ottnt{C_{{\mathrm{0}}}}  ) $}
 \end{flushleft}
 \begin{center}
  $\ottdruleTKXXEmpty{}$ \\[2ex]
  $\ottdruleTKXXSub{}$ \\[2ex]
  $\ottdruleTKXXLet{}$
 \end{center}
 \caption{Type system for pure evaluation contexts.}
 \label{fig:app-typing-ectx}
\end{figure}

\begin{defn}[Typing of Pure Evaluation Contexts]
 Typing judgment $\ottnt{K}  \ottsym{:}    \ottnt{T}  \, / \, ( \forall  \mathit{x}  .  \ottnt{C}  )   \mathrel{ \Rrightarrow }  \ottnt{T_{{\mathrm{0}}}}  \,  \,\&\,  \,  \Phi_{{\mathrm{0}}}  \, / \, ( \forall  \mathit{x_{{\mathrm{0}}}}  .  \ottnt{C_{{\mathrm{0}}}}  ) $ for pure evaluation
 contexts is the smallest relation satisfying the rules in
 \reffig{app-typing-ectx}.
 We write $\ottnt{K}  \ottsym{:}   \ottnt{T}  \, / \, ( \forall  \mathit{x}  .  \ottnt{C}  ) $ if and only if
 $\ottnt{K}  \ottsym{:}    \ottnt{T}  \, / \, ( \forall  \mathit{x}  .  \ottnt{C}  )   \mathrel{ \Rrightarrow }  \ottnt{T_{{\mathrm{0}}}}  \,  \,\&\,  \,  \Phi_{{\mathrm{0}}}  \, / \, ( \forall  \mathit{x_{{\mathrm{0}}}}  .   \ottnt{T_{{\mathrm{0}}}}  \,  \,\&\,  \,   \Phi_\mathsf{val}   \, / \,   \square    ) $
 for some $\ottnt{T_{{\mathrm{0}}}}$, $\Phi_{{\mathrm{0}}}$, and $\mathit{x_{{\mathrm{0}}}}$ such that $\mathit{x_{{\mathrm{0}}}} \,  \not\in  \,  \mathit{fv}  (  \ottnt{T_{{\mathrm{0}}}}  ) $.
 \TS{There is no typing rule for evaluation contexts that corresponds to \T{LetV}
 because shift expressions are not values.}
\end{defn}

\begin{figure}[t!]
 \[\begin{array}{lll}
   \LRRel{ \mathrm{Atom} }{ \ottnt{T} }              &\defeq& \{ \ottnt{v} \mid  \emptyset   \vdash  \ottnt{v}  \ottsym{:}  \ottnt{T} \} \\
   \LRRel{ \mathrm{Atom} }{ \ottnt{C} }              &\defeq& \{ \ottnt{e} \mid  \emptyset   \vdash  \ottnt{e}  \ottsym{:}  \ottnt{C} \} \\[2ex]
   \LRRel{ \mathcal{V} }{ \ottsym{\{}  \mathit{x} \,  {:}  \, \iota \,  |  \, \phi  \ottsym{\}} }   &\defeq&  \LRRel{ \mathrm{Atom} }{ \ottsym{\{}  \mathit{x} \,  {:}  \, \iota \,  |  \, \phi  \ottsym{\}} } 
  \\
   \LRRel{ \mathcal{V} }{ \ottsym{(}  \mathit{x} \,  {:}  \, \ottnt{T}  \ottsym{)}  \rightarrow  \ottnt{C} }    &\defeq&
    \{ \ottnt{v} \,  \in  \,  \LRRel{ \mathrm{Atom} }{ \ottsym{(}  \mathit{x} \,  {:}  \, \ottnt{T}  \ottsym{)}  \rightarrow  \ottnt{C} }  \mid \\ && \quad
         \forall  \, { \ottnt{v'} \,  \in  \,  \LRRel{ \mathcal{V} }{ \ottnt{T} }  } . \  \ottnt{v} \, \ottnt{v'} \,  \in  \,  \LRRel{ \mathcal{E} }{  \ottnt{C}    [  \ottnt{v'}  /  \mathit{x}  ]   }   \}
  \\
   \LRRel{ \mathcal{V} }{  \forall   \ottsym{(}  \mathit{X} \,  {:}  \,  \tceseq{ \ottnt{s} }   \ottsym{)}  \ottsym{.}  \ottnt{C} }  &\defeq&
    \{ \ottnt{v} \,  \in  \,  \LRRel{ \mathrm{Atom} }{  \forall   \ottsym{(}  \mathit{X} \,  {:}  \,  \tceseq{ \ottnt{s} }   \ottsym{)}  \ottsym{.}  \ottnt{C} }  \mid    \forall  \, { \ottnt{A} } . \   \emptyset   \vdash  \ottnt{A}  \ottsym{:}   \tceseq{ \ottnt{s} }    \mathrel{\Longrightarrow}  \ottnt{v} \, \ottnt{A} \,  \in  \,  \LRRel{ \mathcal{E} }{  \ottnt{C}    [  \ottnt{A}  /  \mathit{X}  ]   }   \}
  \\[2ex]
   \LRRel{ \mathcal{E} }{ \ottnt{C} }  &\defeq& \{ \ottnt{e} \,  \in  \,  \LRRel{ \mathrm{Atom} }{ \ottnt{C} }  \mid \\ && \quad
   (   \forall  \, { \ottnt{v}  \ottsym{,}  \varpi } . \  \ottnt{e} \,  \Downarrow  \, \ottnt{v} \,  \,\&\,  \, \varpi   \mathrel{\Longrightarrow}  \ottnt{v} \,  \in  \,  \LRRel{ \mathcal{V} }{  \ottnt{C}  . \ottnt{T}  }  )
   \, \wedge
   \\ && \quad
   (   \forall  \, { \ottnt{K_{{\mathrm{0}}}}  \ottsym{,}  \mathit{x_{{\mathrm{0}}}}  \ottsym{,}  \ottnt{e_{{\mathrm{0}}}}  \ottsym{,}  \varpi } . \  \ottnt{e} \,  \Downarrow  \,  \ottnt{K_{{\mathrm{0}}}}  [  \mathcal{S}_0 \, \mathit{x_{{\mathrm{0}}}}  \ottsym{.}  \ottnt{e_{{\mathrm{0}}}}  ]  \,  \,\&\,  \, \varpi   \mathrel{\Longrightarrow}    \exists  \, { \mathit{x}  \ottsym{,}  \ottnt{C_{{\mathrm{1}}}}  \ottsym{,}  \ottnt{C_{{\mathrm{2}}}} } . \       \\ && \quad\quad
     {  \ottnt{C}  . \ottnt{S}  \,  =  \,  ( \forall  \mathit{x}  .  \ottnt{C_{{\mathrm{1}}}}  )  \Rightarrow   \ottnt{C_{{\mathrm{2}}}}  } \,\mathrel{\wedge}\, {    }   \\ && \quad\quad
      \forall  \, { \ottnt{K} \,  \in  \,  \LRRel{ \mathcal{K} }{   \ottnt{C}  . \ottnt{T}   \, / \, ( \forall  \mathit{x}  .  \ottnt{C_{{\mathrm{1}}}}  )  }  } . \   \resetcnstr{  \ottnt{K}  [  \ottnt{e}  ]  }  \,  \in  \,  \LRRel{ \mathcal{E} }{ \ottnt{C_{{\mathrm{2}}}} }  )
   \, \wedge
   \\ && \quad
   (   \forall  \, { \pi } . \  \ottnt{e} \,  \Uparrow  \, \pi   \mathrel{\Longrightarrow}   { \models  \,   { \ottnt{C} }^\nu (  \pi  )  }  )
   \, \}
   \\[2ex]
  %
   \LRRel{ \mathcal{K} }{  \ottnt{T}  \, / \, ( \forall  \mathit{x}  .  \ottnt{C}  )  }  &\defeq& \{ \ottnt{K} \mid
     { \ottnt{K}  \ottsym{:}   \ottnt{T}  \, / \, ( \forall  \mathit{x}  .  \ottnt{C}  )  } \,\mathrel{\wedge}\, {  \lambda\!  \, \mathit{x}  \ottsym{.}   \resetcnstr{  \ottnt{K}  [  \mathit{x}  ]  }  \,  \in  \,  \LRRel{ \mathcal{V} }{ \ottsym{(}  \mathit{x} \,  {:}  \, \ottnt{T}  \ottsym{)}  \rightarrow  \ottnt{C} }  } 
   \}
   \\[2ex]
   \LRRel{ \mathcal{G} }{  \emptyset  }        &\defeq& \{  \emptyset  \}
  \\
   \LRRel{ \mathcal{G} }{ \Gamma  \ottsym{,}  \mathit{x} \,  {:}  \, \ottnt{T} }       &\defeq& \{ \sigma  \mathbin{\uplus}   [  \ottnt{v}  /  \mathit{x}  ]  \mid  { \sigma \,  \in  \,  \LRRel{ \mathcal{G} }{ \Gamma }  } \,\mathrel{\wedge}\, { \ottnt{v} \,  \in  \,  \LRRel{ \mathcal{V} }{ \sigma  \ottsym{(}  \ottnt{T}  \ottsym{)} }  }  \}
  \\
   \LRRel{ \mathcal{G} }{ \Gamma  \ottsym{,}  \mathit{x} \,  {:}  \, \ottnt{s} }       &\defeq& \{ \sigma  \mathbin{\uplus}   [  \ottnt{t}  /  \mathit{x}  ]  \mid  { \sigma \,  \in  \,  \LRRel{ \mathcal{G} }{ \Gamma }  } \,\mathrel{\wedge}\, {  \emptyset   \vdash  \ottnt{t}  \ottsym{:}  \ottnt{s} }  \}
  \\
   \LRRel{ \mathcal{G} }{ \Gamma  \ottsym{,}  \mathit{X} \,  {:}  \,  \tceseq{ \ottnt{s} }  }     &\defeq& \{ \sigma  \mathbin{\uplus}   [  \ottnt{A}  /  \mathit{X}  ]  \mid  { \sigma \,  \in  \,  \LRRel{ \mathcal{G} }{ \Gamma }  } \,\mathrel{\wedge}\, {  \emptyset   \vdash  \ottnt{A}  \ottsym{:}   \tceseq{ \ottnt{s} }  }  \}
  \\[2ex]
   \LRRel{ \mathcal{O} }{ \Gamma  \, \vdash \,  \ottnt{C} }      &\defeq& \{ \ottnt{e} \mid  { \Gamma  \vdash  \ottnt{e}  \ottsym{:}  \ottnt{C} } \,\mathrel{\wedge}\, {   \forall  \, { \sigma \,  \in  \,  \LRRel{ \mathcal{G} }{ \Gamma }  } . \  \sigma  \ottsym{(}  \ottnt{e}  \ottsym{)} \,  \in  \,  \LRRel{ \mathcal{E} }{ \sigma  \ottsym{(}  \ottnt{C}  \ottsym{)} }   }  \}
   \end{array}
 \]
 \caption{Logical relation.}
 \label{fig:app-LR}
\end{figure}

\begin{defn}[Substitution]
 We use the metavariable $\sigma$ to denote substitutions that map variables
 $\mathit{x}$ to values, variables $\mathit{x}$ to terms, and predicate variables $\mathit{X}$
 to predicates.
 We write $\mathit{dom} \, \ottsym{(}  \sigma  \ottsym{)}$ for the set of variables $\mathit{x}$ and $\mathit{X}$
 mapped by $\sigma$.
 We write $\sigma(-)$ to stand for the result of applying substitution
 $\sigma$ to a given entity in a capture-avoiding manner.
 We write $\sigma  \mathbin{\uplus}   [  \ottnt{v}  /  \mathit{x}  ] $ (resp.\ $\sigma  \mathbin{\uplus}   [  \ottnt{t}  /  \mathit{x}  ] $ and $\sigma  \mathbin{\uplus}   [  \ottnt{A}  /  \mathit{X}  ] $)
 for the same substitution as $\sigma$ except that it maps $\mathit{x}$ to $\ottnt{v}$
 (resp.\ $\mathit{x}$ to $\ottnt{t}$ and $\mathit{X}$ to $\ottnt{A}$).
\end{defn}

\begin{defn}[Logical Relation]
 \reffig{app-LR} defines relations over
 closed values $ \LRRel{ \mathcal{V} }{ \ottnt{T} } $,
 closed expressions $ \LRRel{ \mathcal{E} }{ \ottnt{C} } $,
 closed pure evaluation contexts $ \LRRel{ \mathcal{K} }{  \ottnt{T}  \, / \, ( \forall  \mathit{x}  .  \ottnt{C}  )  } $
 by simultaneous induction on
 $\ottnt{T}$, $\ottnt{C}$, $\ottsym{(}  \mathit{x} \,  {:}  \, \ottnt{T}  \ottsym{)}  \rightarrow  \ottnt{C}$, respectively.
 It also defines
 the relation $ \LRRel{ \mathcal{G} }{ \Gamma } $ over substitutions by induction on $\Gamma$, and
 the relation $ \LRRel{ \mathcal{O} }{ \Gamma  \, \vdash \,  \ottnt{C} } $ over open expressions.
 We use the notation $ \LRRel{ \mathcal{O} }{ \Gamma  \, \vdash \,  \ottnt{T} } $ for denoting $ \LRRel{ \mathcal{O} }{ \Gamma  \, \vdash \,   \ottnt{T}  \,  \,\&\,  \,   \Phi_\mathsf{val}   \, / \,   \square   } $.
\end{defn}

\begin{defn}[Finite Approximation of Predicate Fixpoints]
 \label{def:approx-fixpoint-pred}

 The predicates $\ottnt{C'}  \triangleright   \tceseq{ \ottsym{(}  \mathit{X'}  \ottsym{,}  \mathit{Y'}  \ottsym{)} }   \ottsym{;}   \tceseq{ \ottsym{(}  \mathit{X}  \ottsym{,}  \mathit{Y}  \ottsym{)} }   \ottsym{;}   \tceseq{ \mathit{z_{{\mathrm{0}}}}  \,  {:}  \,  \iota_{{\mathrm{0}}} }  \,  |  \, \ottnt{C} \,  \vdash _{ \ottmv{i} }  \, \sigma$ and $\ottnt{C'}  \triangleright   \tceseq{ \ottsym{(}  \mathit{X'}  \ottsym{,}  \mathit{Y'}  \ottsym{)} }   \ottsym{;}   \tceseq{ \ottsym{(}  \mathit{X}  \ottsym{,}  \mathit{Y}  \ottsym{)} }   \ottsym{;}   \tceseq{ \mathit{z_{{\mathrm{0}}}}  \,  {:}  \,  \iota_{{\mathrm{0}}} }  \,  |  \, \ottnt{S} \,  \vdash _{ \ottmv{i} }  \, \sigma$ are defined by simultaneous induction as follows.
 \[\begin{array}{cl} &
  \ottnt{C'}  \triangleright   \tceseq{ \ottsym{(}  \mathit{X'}  \ottsym{,}  \mathit{Y'}  \ottsym{)} }   \ottsym{;}   \emptyset   \ottsym{;}   \tceseq{ \mathit{z_{{\mathrm{0}}}}  \,  {:}  \,  \iota_{{\mathrm{0}}} }  \,  |  \, \ottnt{C} \,  \vdash _{ \ottmv{i} }  \,  \emptyset 
  \\[2ex] &
  \ottnt{C'}  \triangleright   \tceseq{ \ottsym{(}  \mathit{X'}  \ottsym{,}  \mathit{Y'}  \ottsym{)} }   \ottsym{,}  \ottsym{(}  \mathit{X_{{\mathrm{0}}}}  \ottsym{,}  \mathit{Y_{{\mathrm{0}}}}  \ottsym{)}  \ottsym{;}   \tceseq{ \ottsym{(}  \mathit{X}  \ottsym{,}  \mathit{Y}  \ottsym{)} }   \ottsym{;}   \tceseq{ \mathit{z_{{\mathrm{0}}}}  \,  {:}  \,  \iota_{{\mathrm{0}}} }  \,  |  \, \ottnt{S} \,  \vdash _{ \ottsym{0} }  \, \sigma
  \\  \Rightarrow &
  \ottnt{C'}  \triangleright   \tceseq{ \ottsym{(}  \mathit{X'}  \ottsym{,}  \mathit{Y'}  \ottsym{)} }   \ottsym{;}  \ottsym{(}  \mathit{X_{{\mathrm{0}}}}  \ottsym{,}  \mathit{Y_{{\mathrm{0}}}}  \ottsym{)}  \ottsym{,}   \tceseq{ \ottsym{(}  \mathit{X}  \ottsym{,}  \mathit{Y}  \ottsym{)} }   \ottsym{;}   \tceseq{ \mathit{z_{{\mathrm{0}}}}  \,  {:}  \,  \iota_{{\mathrm{0}}} }  \,  |  \,  \ottnt{T}  \,  \,\&\,  \,  \Phi  \, / \,  \ottnt{S}  \,  \vdash _{ \ottsym{0} }  \, \sigma  \mathbin{\uplus}   [  \ottsym{(}   \nu\!  \, \mathit{X_{{\mathrm{0}}}}  \ottsym{(}  \mathit{x} \,  {:}  \,  \Sigma ^\ast   \ottsym{,}   \tceseq{ \mathit{z_{{\mathrm{0}}}}  \,  {:}  \,  \iota_{{\mathrm{0}}} }   \ottsym{)}  \ottsym{.}   \bot   \ottsym{)}  /  \mathit{X_{{\mathrm{0}}}}  ]   \mathbin{\uplus}   [  \ottsym{(}   \mu\!  \, \mathit{Y_{{\mathrm{0}}}}  \ottsym{(}  \mathit{y} \,  {:}  \,  \Sigma ^\omega   \ottsym{,}   \tceseq{ \mathit{z_{{\mathrm{0}}}}  \,  {:}  \,  \iota_{{\mathrm{0}}} }   \ottsym{)}  \ottsym{.}   \top   \ottsym{)}  /  \mathit{Y_{{\mathrm{0}}}}  ] 
  \\[2ex] &
  \ottnt{C'}  \triangleright   \tceseq{ \ottsym{(}  \mathit{X'}  \ottsym{,}  \mathit{Y'}  \ottsym{)} }   \ottsym{,}  \ottsym{(}  \mathit{X_{{\mathrm{0}}}}  \ottsym{,}  \mathit{Y_{{\mathrm{0}}}}  \ottsym{)}  \ottsym{;}   \tceseq{ \ottsym{(}  \mathit{X}  \ottsym{,}  \mathit{Y}  \ottsym{)} }   \ottsym{;}   \tceseq{ \mathit{z_{{\mathrm{0}}}}  \,  {:}  \,  \iota_{{\mathrm{0}}} }  \,  |  \, \ottnt{S} \,  \vdash _{ \ottmv{i}  \ottsym{+}  \ottsym{1} }  \, \sigma
  \\  \Rightarrow &
  \ottnt{C'}  \triangleright   \tceseq{ \ottsym{(}  \mathit{X'}  \ottsym{,}  \mathit{Y'}  \ottsym{)} }   \ottsym{;}  \ottsym{(}  \mathit{X_{{\mathrm{0}}}}  \ottsym{,}  \mathit{Y_{{\mathrm{0}}}}  \ottsym{)}  \ottsym{,}   \tceseq{ \ottsym{(}  \mathit{X}  \ottsym{,}  \mathit{Y}  \ottsym{)} }   \ottsym{;}   \tceseq{ \mathit{z_{{\mathrm{0}}}}  \,  {:}  \,  \iota_{{\mathrm{0}}} }  \,  |  \,  \ottnt{T}  \,  \,\&\,  \,  \Phi  \, / \,  \ottnt{S}  \,  \vdash _{ \ottmv{i}  \ottsym{+}  \ottsym{1} }  \, \sigma  \mathbin{\uplus}   [   \ottnt{A} _{  \mu  }   /  \mathit{X_{{\mathrm{0}}}}  ]   \mathbin{\uplus}   [   \ottnt{A} _{  \nu  }   /  \mathit{Y_{{\mathrm{0}}}}  ] 
  \\ & \quad
  \text{where $\ottnt{C'}  \triangleright   \emptyset   \ottsym{;}   \tceseq{ \ottsym{(}  \mathit{X'}  \ottsym{,}  \mathit{Y'}  \ottsym{)} }   \ottsym{;}   \tceseq{ \mathit{z_{{\mathrm{0}}}}  \,  {:}  \,  \iota_{{\mathrm{0}}} }  \,  |  \, \ottnt{C'} \,  \vdash _{ \ottmv{i} }  \, \sigma'_{\ottmv{i}}$ and}
  \\ & \qquad\qquad
  \text{$\ottnt{C'}  \triangleright   \tceseq{ \ottsym{(}  \mathit{X'}  \ottsym{,}  \mathit{Y'}  \ottsym{)} }   \ottsym{;}  \ottsym{(}  \mathit{X_{{\mathrm{0}}}}  \ottsym{,}  \mathit{Y_{{\mathrm{0}}}}  \ottsym{)}  \ottsym{;}   \tceseq{ \mathit{z_{{\mathrm{0}}}}  \,  {:}  \,  \iota_{{\mathrm{0}}} }  \,  |  \,  \ottnt{T}  \,  \,\&\,  \,  \Phi  \, / \,  \ottnt{S}  \,  \vdash _{ \ottmv{i} }  \,  { \sigma }_{ \ottmv{i}  \ottsym{;}  \ottsym{0} } $ and}
  \\ & \qquad\qquad
  \text{$ \tceseq{ \ottsym{(}  \mathit{X'}  \ottsym{,}  \mathit{Y'}  \ottsym{)} }   \ottsym{,}  \ottsym{(}  \mathit{X_{{\mathrm{0}}}}  \ottsym{,}  \mathit{Y_{{\mathrm{0}}}}  \ottsym{)}  \ottsym{,}   \tceseq{ \ottsym{(}  \mathit{X}  \ottsym{,}  \mathit{Y}  \ottsym{)} }   \ottsym{;}   \tceseq{ \mathit{z_{{\mathrm{0}}}}  \,  {:}  \,  \iota_{{\mathrm{0}}} }  \,  |  \, \ottnt{C'}  \vdash  \sigma'$ and}
  \\ & \qquad\qquad
  \text{$ \ottnt{A} _{  \mu  }  \,  =  \, \ottsym{(}   \nu\!  \, \mathit{X_{{\mathrm{0}}}}  \ottsym{(}  \mathit{x} \,  {:}  \,  \Sigma ^\ast   \ottsym{,}   \tceseq{ \mathit{z_{{\mathrm{0}}}}  \,  {:}  \,  \iota_{{\mathrm{0}}} }   \ottsym{)}  \ottsym{.}     { \Phi }^\mu   \ottsym{(}  \mathit{x}  \ottsym{)}    [ \tceseq{ \sigma'_{\ottmv{i}}  \ottsym{(}  \mathit{X'}  \ottsym{)}  /  \mathit{X'} } ]      [   { \sigma }_{ \ottmv{i}  \ottsym{;}  \ottsym{0} }   \ottsym{(}  \mathit{X_{{\mathrm{0}}}}  \ottsym{)}  /  \mathit{X_{{\mathrm{0}}}}  ]    \ottsym{)}$, and}
  \\ & \qquad\qquad
  \text{$ \ottnt{A} _{  \nu  }  \,  =  \, \ottsym{(}   \mu\!  \, \mathit{Y_{{\mathrm{0}}}}  \ottsym{(}  \mathit{y} \,  {:}  \,  \Sigma ^\omega   \ottsym{,}   \tceseq{ \mathit{z_{{\mathrm{0}}}}  \,  {:}  \,  \iota_{{\mathrm{0}}} }   \ottsym{)}  \ottsym{.}       { \Phi }^\nu   \ottsym{(}  \mathit{y}  \ottsym{)}    [ \tceseq{ \sigma'  \ottsym{(}  \mathit{X'}  \ottsym{)}  /  \mathit{X'} } ]      [  \sigma'  \ottsym{(}  \mathit{X_{{\mathrm{0}}}}  \ottsym{)}  /  \mathit{X_{{\mathrm{0}}}}  ]      [ \tceseq{ \sigma'_{\ottmv{i}}  \ottsym{(}  \mathit{Y'}  \ottsym{)}  /  \mathit{Y'} } ]      [   { \sigma }_{ \ottmv{i}  \ottsym{;}  \ottsym{0} }   \ottsym{(}  \mathit{Y_{{\mathrm{0}}}}  \ottsym{)}  /  \mathit{Y_{{\mathrm{0}}}}  ]    \ottsym{)}$}
  \\[2ex] &
  \ottnt{C'}  \triangleright   \tceseq{ \ottsym{(}  \mathit{X'}  \ottsym{,}  \mathit{Y'}  \ottsym{)} }   \ottsym{;}   \tceseq{ \ottsym{(}  \mathit{X}  \ottsym{,}  \mathit{Y}  \ottsym{)} }   \ottsym{;}   \tceseq{ \mathit{z_{{\mathrm{0}}}}  \,  {:}  \,  \iota_{{\mathrm{0}}} }  \,  |  \,  \square  \,  \vdash _{ \ottmv{i} }  \,  \emptyset 
  \\[2ex] &
  \ottnt{C'}  \triangleright   \tceseq{ \ottsym{(}  \mathit{X'}  \ottsym{,}  \mathit{Y'}  \ottsym{)} }   \ottsym{;}   \tceseq{ \ottsym{(}  \mathit{X}  \ottsym{,}  \mathit{Y}  \ottsym{)} }   \ottsym{;}   \tceseq{ \mathit{z_{{\mathrm{0}}}}  \,  {:}  \,  \iota_{{\mathrm{0}}} }  \,  |  \, \ottnt{C_{{\mathrm{2}}}} \,  \vdash _{ \ottmv{i} }  \, \sigma
  \\  \Rightarrow &
  \ottnt{C'}  \triangleright   \tceseq{ \ottsym{(}  \mathit{X'}  \ottsym{,}  \mathit{Y'}  \ottsym{)} }   \ottsym{;}   \tceseq{ \ottsym{(}  \mathit{X}  \ottsym{,}  \mathit{Y}  \ottsym{)} }   \ottsym{;}   \tceseq{ \mathit{z_{{\mathrm{0}}}}  \,  {:}  \,  \iota_{{\mathrm{0}}} }  \,  |  \,  ( \forall  \mathit{x}  .  \ottnt{C_{{\mathrm{1}}}}  )  \Rightarrow   \ottnt{C_{{\mathrm{2}}}}  \,  \vdash _{ \ottmv{i} }  \, \sigma
   \end{array}
 \]
\end{defn}

\begin{defn}[Finite Approximation of Recursive Functions]
 \label{def:munu-sim}

 Let
 \begin{itemize}
  \item $\ottnt{v} \,  =  \,  \mathsf{rec}  \, (  \tceseq{  \mathit{X} ^{ \sigma  \ottsym{(}  \mathit{X}  \ottsym{)} },  \mathit{Y} ^{ \sigma  \ottsym{(}  \mathit{Y}  \ottsym{)} }  }  ,  \mathit{f} ,   \tceseq{ \mathit{z} }  ) . \,  \ottnt{e} $,
  \item $ \tceseq{ \mathit{z_{{\mathrm{0}}}}  \,  {:}  \,  \iota_{{\mathrm{0}}} }  \,  =  \,  \gamma  (   \tceseq{ \mathit{z}  \,  {:}  \,  \ottnt{T_{{\mathrm{1}}}} }   ) $,
  \item $\sigma$ be a substitution such that $ \tceseq{ \ottsym{(}  \mathit{X}  \ottsym{,}  \mathit{Y}  \ottsym{)} }   \ottsym{;}   \tceseq{ \mathit{z_{{\mathrm{0}}}}  \,  {:}  \,  \iota_{{\mathrm{0}}} }  \,  |  \, \ottnt{C}  \vdash  \sigma$,
  \item for every natural number $\ottmv{i}$, $\sigma_{\ottmv{i}}$ be a substitution such that $\ottnt{C}  \triangleright   \emptyset   \ottsym{;}   \tceseq{ \ottsym{(}  \mathit{X}  \ottsym{,}  \mathit{Y}  \ottsym{)} }   \ottsym{;}   \tceseq{ \mathit{z_{{\mathrm{0}}}}  \,  {:}  \,  \iota_{{\mathrm{0}}} }  \,  |  \, \ottnt{C} \,  \vdash _{ \ottmv{i} }  \, \sigma_{\ottmv{i}}$, and
  \item $ \top^P  = \ottsym{(}   \mu\!  \, \mathit{Y}  \ottsym{(}  \mathit{y} \,  {:}  \,  \Sigma ^\omega   \ottsym{,}   \tceseq{ \mathit{z_{{\mathrm{0}}}}  \,  {:}  \,  \iota_{{\mathrm{0}}} }   \ottsym{)}  \ottsym{.}   \top   \ottsym{)}$.
 \end{itemize}
 %
 %
 %

 We define a series of values $ \ottnt{v} _{  \nu   \ottsym{;}  \ottmv{i}  \ottsym{;}  \pi } $ by induction on natural number $\ottmv{i}$.
 \[\begin{array}{lll}
   \ottnt{v} _{  \nu   \ottsym{;}  \ottsym{0}  \ottsym{;}  \pi }    &\defeq&  \lambda\!  \,  \tceseq{ \mathit{z} }   \ottsym{.}   \bullet ^{ \pi }  \\
   \ottnt{v} _{  \nu   \ottsym{;}  \ottmv{i}  \ottsym{+}  \ottsym{1}  \ottsym{;}  \pi }  &\defeq&  \lambda\!  \,  \tceseq{ \mathit{z} }   \ottsym{.}     \ottnt{e}    [ \tceseq{ \sigma  \ottsym{(}  \mathit{X}  \ottsym{)}  /  \mathit{X} } ]      [ \tceseq{ \sigma_{\ottmv{i}}  \ottsym{(}  \mathit{Y}  \ottsym{)}  /  \mathit{Y} } ]      [   \ottnt{v} _{  \nu   \ottsym{;}  \ottmv{i}  \ottsym{;}  \pi }   /  \mathit{f}  ]   \\
   \end{array}
 \]
 We also define a series of values $ \ottnt{v} _{  \mu   \ottsym{;}  \ottmv{i}  \ottsym{;}  \pi } $
 by induction on natural number $\ottmv{i}$.
 \[\begin{array}{lll}
   \ottnt{v} _{  \mu   \ottsym{;}  \ottsym{0}  \ottsym{;}  \pi }    &\defeq&  \lambda\!  \,  \tceseq{ \mathit{z} }   \ottsym{.}   \bullet ^{ \pi }  \\
   \ottnt{v} _{  \mu   \ottsym{;}  \ottmv{i}  \ottsym{+}  \ottsym{1}  \ottsym{;}  \pi }  &\defeq&  \lambda\!  \,  \tceseq{ \mathit{z} }   \ottsym{.}     \ottnt{e}    [ \tceseq{ \sigma_{\ottmv{i}}  \ottsym{(}  \mathit{X}  \ottsym{)}  /  \mathit{X} } ]      [ \tceseq{  \top^P   /  \mathit{Y} } ]      [   \ottnt{v} _{  \mu   \ottsym{;}  \ottmv{i}  \ottsym{;}  \pi }   /  \mathit{f}  ]   \\
   \end{array}
 \]
\end{defn}

\begin{defn}[Composition with Finite Approximation]
 Let $\ottnt{v} \,  =  \,  \mathsf{rec}  \, (  \tceseq{  \mathit{X} ^{  \ottnt{A} _{  \mu  }  },  \mathit{Y} ^{  \ottnt{A} _{  \nu  }  }  }  ,  \mathit{f_{{\mathrm{0}}}} ,   \tceseq{ \mathit{x_{{\mathrm{0}}}} }  ) . \,  \ottnt{e_{{\mathrm{0}}}} $ be a closed, well-typed value.
 For every $\ottmv{i}$,
 let $\sigma_{\ottmv{i}}$ be the substitution determined as in \refdef{munu-sim} (we can
 determine it uniquely).
 We define
 $\ottnt{e_{{\mathrm{1}}}} \,  \prec_{ \ottnt{v} }  \, \ottnt{e_{{\mathrm{2}}}} \,  \uparrow \, (  \ottmv{i}  ,  \pi  ) $,
 $\ottnt{A_{{\mathrm{1}}}} \,  \prec_{ \ottnt{v} }  \, \ottnt{A_{{\mathrm{2}}}} \,  \uparrow \, (  \ottmv{i}  ,  \pi  ) $, and
 $\phi_{{\mathrm{1}}} \,  \prec_{ \ottnt{v} }  \, \phi_{{\mathrm{2}}} \,  \uparrow \, (  \ottmv{i}  ,  \pi  ) $
 as follows.
 \[\begin{array}{lll}
  \ottnt{v} \,  \prec_{ \ottnt{v} }  \,  \ottnt{v} _{  \mu   \ottsym{;}  \ottmv{j}  \ottsym{;}  \pi_{{\mathrm{0}}} }  \,  \uparrow \, (  \ottmv{i}  ,  \pi  )  &\defeq&  \ottnt{v} _{  \mu   \ottsym{;}  \ottmv{j}  \ottsym{+}  \ottmv{i}  \ottsym{;}  \pi }  \\
  \mathit{x} \,  \prec_{ \ottnt{v} }  \, \mathit{x} \,  \uparrow \, (  \ottmv{i}  ,  \pi  )  &\defeq& \mathit{x} \\
  \ottnt{c} \,  \prec_{ \ottnt{v} }  \, \ottnt{c} \,  \uparrow \, (  \ottmv{i}  ,  \pi  )  &\defeq& \ottnt{c} \\
  \ottsym{(}   \lambda\!  \,  \tceseq{ \mathit{x} }   \ottsym{.}  \ottnt{e}  \ottsym{)} \,  \prec_{ \ottnt{v} }  \, \ottsym{(}   \lambda\!  \,  \tceseq{ \mathit{x} }   \ottsym{.}  \ottnt{e'}  \ottsym{)} \,  \uparrow \, (  \ottmv{i}  ,  \pi  )  &\defeq&  \lambda\!  \,  \tceseq{ \mathit{x} }   \ottsym{.}  \ottsym{(}  \ottnt{e} \,  \prec_{ \ottnt{v} }  \, \ottnt{e'} \,  \uparrow \, (  \ottmv{i}  ,  \pi  )   \ottsym{)} \\
  \ottsym{(}   \Lambda   \ottsym{(}  \mathit{X} \,  {:}  \,  \tceseq{ \ottnt{s} }   \ottsym{)}  \ottsym{.}  \ottnt{e}  \ottsym{)} \,  \prec_{ \ottnt{v} }  \, \ottsym{(}   \Lambda   \ottsym{(}  \mathit{X} \,  {:}  \,  \tceseq{ \ottnt{s} }   \ottsym{)}  \ottsym{.}  \ottnt{e'}  \ottsym{)} \,  \uparrow \, (  \ottmv{i}  ,  \pi  )  &\defeq&  \Lambda   \ottsym{(}  \mathit{X} \,  {:}  \,  \tceseq{ \ottnt{s} }   \ottsym{)}  \ottsym{.}  \ottsym{(}  \ottnt{e} \,  \prec_{ \ottnt{v} }  \, \ottnt{e'} \,  \uparrow \, (  \ottmv{i}  ,  \pi  )   \ottsym{)} \\
  \multicolumn{3}{l}{
   \ottsym{(}   \mathsf{rec}  \, (  \tceseq{  \mathit{X} ^{ \ottnt{A_{{\mathrm{1}}}} },  \mathit{Y} ^{ \ottnt{A_{{\mathrm{2}}}} }  }  ,  \mathit{f} ,   \tceseq{ \mathit{x} }  ) . \,  \ottnt{e}   \ottsym{)} \,  \prec_{ \ottnt{v} }  \, \ottsym{(}   \mathsf{rec}  \, (  \tceseq{  \mathit{X} ^{ \ottnt{A'_{{\mathrm{1}}}} },  \mathit{Y} ^{ \ottnt{A'_{{\mathrm{2}}}} }  }  ,  \mathit{f} ,   \tceseq{ \mathit{x} }  ) . \,  \ottnt{e'}   \ottsym{)} \,  \uparrow \, (  \ottmv{i}  ,  \pi  ) 
   \ \defeq
  } \\
  \multicolumn{3}{r}{
    \mathsf{rec}  \, (  \tceseq{  \mathit{X} ^{ \ottsym{(}  \ottnt{A_{{\mathrm{1}}}} \,  \prec_{ \ottnt{v} }  \, \ottnt{A'_{{\mathrm{1}}}} \,  \uparrow \, (  \ottmv{i}  ,  \pi  )   \ottsym{)} },  \mathit{Y} ^{ \ottsym{(}  \ottnt{A_{{\mathrm{2}}}} \,  \prec_{ \ottnt{v} }  \, \ottnt{A'_{{\mathrm{2}}}} \,  \uparrow \, (  \ottmv{i}  ,  \pi  )   \ottsym{)} }  }  ,  \mathit{f} ,   \tceseq{ \mathit{x} }  ) . \,  \ottsym{(}  \ottnt{e} \,  \prec_{ \ottnt{v} }  \, \ottnt{e'} \,  \uparrow \, (  \ottmv{i}  ,  \pi  )   \ottsym{)} 
  } \\
  \ottnt{o}  \ottsym{(}   \tceseq{ \ottnt{v_{{\mathrm{0}}}} }   \ottsym{)} \,  \prec_{ \ottnt{v} }  \, \ottnt{o}  \ottsym{(}   \tceseq{ \ottnt{v'_{{\mathrm{0}}}} }   \ottsym{)} \,  \uparrow \, (  \ottmv{i}  ,  \pi  )  &\defeq& \ottnt{o}  \ottsym{(}   \tceseq{ \ottnt{v_{{\mathrm{0}}}} \,  \prec_{ \ottnt{v} }  \, \ottnt{v'_{{\mathrm{0}}}} \,  \uparrow \, (  \ottmv{i}  ,  \pi  )  }   \ottsym{)} \\
  \ottsym{(}  \ottnt{v_{{\mathrm{1}}}} \, \ottnt{v_{{\mathrm{2}}}}  \ottsym{)} \,  \prec_{ \ottnt{v} }  \, \ottsym{(}  \ottnt{v'_{{\mathrm{1}}}} \, \ottnt{v'_{{\mathrm{2}}}}  \ottsym{)} \,  \uparrow \, (  \ottmv{i}  ,  \pi  )  &\defeq& \ottsym{(}  \ottnt{v_{{\mathrm{1}}}} \,  \prec_{ \ottnt{v} }  \, \ottnt{v'_{{\mathrm{1}}}} \,  \uparrow \, (  \ottmv{i}  ,  \pi  )   \ottsym{)} \, \ottsym{(}  \ottnt{v_{{\mathrm{2}}}} \,  \prec_{ \ottnt{v} }  \, \ottnt{v'_{{\mathrm{2}}}} \,  \uparrow \, (  \ottmv{i}  ,  \pi  )   \ottsym{)} \\
  \ottsym{(}  \ottnt{v_{{\mathrm{1}}}} \, \ottnt{A}  \ottsym{)} \,  \prec_{ \ottnt{v} }  \, \ottsym{(}  \ottnt{v'_{{\mathrm{1}}}} \, \ottnt{A'}  \ottsym{)} \,  \uparrow \, (  \ottmv{i}  ,  \pi  )  &\defeq& \ottsym{(}  \ottnt{v_{{\mathrm{1}}}} \,  \prec_{ \ottnt{v} }  \, \ottnt{v'_{{\mathrm{1}}}} \,  \uparrow \, (  \ottmv{i}  ,  \pi  )   \ottsym{)} \, \ottsym{(}  \ottnt{A} \,  \prec_{ \ottnt{v} }  \, \ottnt{A'} \,  \uparrow \, (  \ottmv{i}  ,  \pi  )   \ottsym{)} \\
  \multicolumn{3}{l}{
   \ottsym{(}  \mathsf{if} \, \ottnt{v_{{\mathrm{0}}}} \, \mathsf{then} \, \ottnt{e_{{\mathrm{1}}}} \, \mathsf{else} \, \ottnt{e_{{\mathrm{2}}}}  \ottsym{)} \,  \prec_{ \ottnt{v} }  \, \ottsym{(}  \mathsf{if} \, \ottnt{v'_{{\mathrm{0}}}} \, \mathsf{then} \, \ottnt{e'_{{\mathrm{1}}}} \, \mathsf{else} \, \ottnt{e'_{{\mathrm{2}}}}  \ottsym{)} \,  \uparrow \, (  \ottmv{i}  ,  \pi  ) 
   \ \defeq
  } \\
  \multicolumn{3}{r}{
   \mathsf{if} \, \ottsym{(}  \ottnt{v_{{\mathrm{0}}}} \,  \prec_{ \ottnt{v} }  \, \ottnt{v'_{{\mathrm{0}}}} \,  \uparrow \, (  \ottmv{i}  ,  \pi  )   \ottsym{)} \, \mathsf{then} \, \ottsym{(}  \ottnt{e_{{\mathrm{1}}}} \,  \prec_{ \ottnt{v} }  \, \ottnt{e'_{{\mathrm{1}}}} \,  \uparrow \, (  \ottmv{i}  ,  \pi  )   \ottsym{)} \, \mathsf{else} \, \ottsym{(}  \ottnt{e_{{\mathrm{2}}}} \,  \prec_{ \ottnt{v} }  \, \ottnt{e'_{{\mathrm{2}}}} \,  \uparrow \, (  \ottmv{i}  ,  \pi  )   \ottsym{)}
  }
  \\
  \ottsym{(}  \mathsf{let} \, \mathit{x}  \ottsym{=}  \ottnt{e_{{\mathrm{1}}}} \,  \mathsf{in}  \, \ottnt{e_{{\mathrm{2}}}}  \ottsym{)} \,  \prec_{ \ottnt{v} }  \, \ottsym{(}  \mathsf{let} \, \mathit{x}  \ottsym{=}  \ottnt{e'_{{\mathrm{1}}}} \,  \mathsf{in}  \, \ottnt{e'_{{\mathrm{2}}}}  \ottsym{)} \,  \uparrow \, (  \ottmv{i}  ,  \pi  )  &\defeq& \mathsf{let} \, \mathit{x}  \ottsym{=}  \ottsym{(}  \ottnt{e_{{\mathrm{1}}}} \,  \prec_{ \ottnt{v} }  \, \ottnt{e'_{{\mathrm{1}}}} \,  \uparrow \, (  \ottmv{i}  ,  \pi  )   \ottsym{)} \,  \mathsf{in}  \, \ottsym{(}  \ottnt{e_{{\mathrm{2}}}} \,  \prec_{ \ottnt{v} }  \, \ottnt{e'_{{\mathrm{2}}}} \,  \uparrow \, (  \ottmv{i}  ,  \pi  )   \ottsym{)} \\
   \mathsf{ev}  [  \mathbf{a}  ]  \,  \prec_{ \ottnt{v} }  \,  \mathsf{ev}  [  \mathbf{a}  ]  \,  \uparrow \, (  \ottmv{i}  ,  \pi  )  &\defeq&  \mathsf{ev}  [  \mathbf{a}  ]  \\
  \ottsym{(}  \mathcal{S}_0 \, \mathit{x}  \ottsym{.}  \ottnt{e}  \ottsym{)} \,  \prec_{ \ottnt{v} }  \, \ottsym{(}  \mathcal{S}_0 \, \mathit{x}  \ottsym{.}  \ottnt{e'}  \ottsym{)} \,  \uparrow \, (  \ottmv{i}  ,  \pi  )  &\defeq& \mathcal{S}_0 \, \mathit{x}  \ottsym{.}  \ottsym{(}  \ottnt{e} \,  \prec_{ \ottnt{v} }  \, \ottnt{e'} \,  \uparrow \, (  \ottmv{i}  ,  \pi  )   \ottsym{)} \\
   \resetcnstr{ \ottnt{e} }  \,  \prec_{ \ottnt{v} }  \,  \resetcnstr{ \ottnt{e'} }  \,  \uparrow \, (  \ottmv{i}  ,  \pi  )  &\defeq&  \resetcnstr{ \ottnt{e} \,  \prec_{ \ottnt{v} }  \, \ottnt{e'} \,  \uparrow \, (  \ottmv{i}  ,  \pi  )  }  \\
   \bullet ^{ \pi_{{\mathrm{0}}} }  \,  \prec_{ \ottnt{v} }  \,  \bullet ^{ \pi_{{\mathrm{0}}} }  \,  \uparrow \, (  \ottmv{i}  ,  \pi  )  &\defeq&  \bullet ^{ \pi_{{\mathrm{0}}} } 
  \\[2ex]
   \ottnt{A} _{  \mu\!  , \ottsym{0} }  \,  \prec_{ \ottnt{v} }  \, \sigma_{\ottmv{j}}  \ottsym{(}  \mathit{X_{{\mathrm{0}}}}  \ottsym{)} \,  \uparrow \, (  \ottmv{i}  ,  \pi  )  &\defeq&  { \sigma }_{ \ottmv{i}  \ottsym{+}  \ottmv{j} }   \ottsym{(}  \mathit{X_{{\mathrm{0}}}}  \ottsym{)}
   \\ \multicolumn{3}{l}{
       \qquad
       \text{(where $\mathit{X_{{\mathrm{0}}}}$ is in $ \tceseq{ \mathit{X} } $, and $ \ottnt{A} _{  \mu\!  , \ottsym{0} } $ is in $ \tceseq{  \ottnt{A} _{  \mu  }  } $ and corresponds to $\mathit{X_{{\mathrm{0}}}}$)}
       }
  \\[2ex]
   \ottnt{A} _{  \nu\!  , \ottsym{0} }  \,  \prec_{ \ottnt{v} }  \,  \top^P  \,  \uparrow \, (  \ottmv{i}  ,  \pi  )  &\defeq&  \top^P  \\
   \\ \multicolumn{3}{l}{
       \qquad
       \text{(where $ \ottnt{A} _{  \nu\!  , \ottsym{0} } $ is in $ \tceseq{  \ottnt{A} _{  \nu  }  } $)}
       }
  \\[2ex]
  \mathit{X} \,  \prec_{ \ottnt{v} }  \, \mathit{X} \,  \uparrow \, (  \ottmv{i}  ,  \pi  )  &\defeq& \mathit{X} \\
  \ottsym{(}   \mu\!  \, \mathit{X}  \ottsym{(}   \tceseq{ \mathit{x}  \,  {:}  \,  \ottnt{s} }   \ottsym{)}  \ottsym{.}  \phi  \ottsym{)} \,  \prec_{ \ottnt{v} }  \, \ottsym{(}   \mu\!  \, \mathit{X}  \ottsym{(}   \tceseq{ \mathit{x}  \,  {:}  \,  \ottnt{s} }   \ottsym{)}  \ottsym{.}  \phi'  \ottsym{)} \,  \uparrow \, (  \ottmv{i}  ,  \pi  )  &\defeq& \ottsym{(}   \mu\!  \, \mathit{X}  \ottsym{(}   \tceseq{ \mathit{x}  \,  {:}  \,  \ottnt{s} }   \ottsym{)}  \ottsym{.}  \ottsym{(}  \phi \,  \prec_{ \ottnt{v} }  \, \phi' \,  \uparrow \, (  \ottmv{i}  ,  \pi  )   \ottsym{)}  \ottsym{)} \\
  \ottsym{(}   \nu\!  \, \mathit{Y}  \ottsym{(}   \tceseq{ \mathit{x}  \,  {:}  \,  \ottnt{s} }   \ottsym{)}  \ottsym{.}  \phi  \ottsym{)} \,  \prec_{ \ottnt{v} }  \, \ottsym{(}   \nu\!  \, \mathit{Y}  \ottsym{(}   \tceseq{ \mathit{x}  \,  {:}  \,  \ottnt{s} }   \ottsym{)}  \ottsym{.}  \phi'  \ottsym{)} \,  \uparrow \, (  \ottmv{i}  ,  \pi  )  &\defeq& \ottsym{(}   \nu\!  \, \mathit{Y}  \ottsym{(}   \tceseq{ \mathit{x}  \,  {:}  \,  \ottnt{s} }   \ottsym{)}  \ottsym{.}  \ottsym{(}  \phi \,  \prec_{ \ottnt{v} }  \, \phi' \,  \uparrow \, (  \ottmv{i}  ,  \pi  )   \ottsym{)}  \ottsym{)} \\
  \ottnt{P} \,  \prec_{ \ottnt{v} }  \, \ottnt{P} \,  \uparrow \, (  \ottmv{i}  ,  \pi  )  &\defeq& \ottnt{P}
  \\[2ex]
   \top  \,  \prec_{ \ottnt{v} }  \,  \top  \,  \uparrow \, (  \ottmv{i}  ,  \pi  )  &\defeq&  \top  \\
   \bot  \,  \prec_{ \ottnt{v} }  \,  \bot  \,  \uparrow \, (  \ottmv{i}  ,  \pi  )  &\defeq&  \bot  \\
  \ottnt{A}  \ottsym{(}   \tceseq{ \ottnt{t} }   \ottsym{)} \,  \prec_{ \ottnt{v} }  \, \ottnt{A'}  \ottsym{(}   \tceseq{ \ottnt{t} }   \ottsym{)} \,  \uparrow \, (  \ottmv{i}  ,  \pi  )  &\defeq& \ottsym{(}  \ottnt{A} \,  \prec_{ \ottnt{v} }  \, \ottnt{A'} \,  \uparrow \, (  \ottmv{i}  ,  \pi  )   \ottsym{)}  \ottsym{(}   \tceseq{ \ottnt{t} }   \ottsym{)} \\
  \ottsym{(}   \neg  \phi   \ottsym{)} \,  \prec_{ \ottnt{v} }  \, \ottsym{(}   \neg  \phi'   \ottsym{)} \,  \uparrow \, (  \ottmv{i}  ,  \pi  )  &\defeq&  \neg  \ottsym{(}  \phi \,  \prec_{ \ottnt{v} }  \, \phi' \,  \uparrow \, (  \ottmv{i}  ,  \pi  )   \ottsym{)}  \\
  \ottsym{(}   \phi_{{\mathrm{1}}}   \oplus   \phi_{{\mathrm{2}}}   \ottsym{)} \,  \prec_{ \ottnt{v} }  \, \ottsym{(}   \phi'_{{\mathrm{1}}}   \oplus   \phi'_{{\mathrm{2}}}   \ottsym{)} \,  \uparrow \, (  \ottmv{i}  ,  \pi  )  &\defeq&  \ottsym{(}  \phi_{{\mathrm{1}}} \,  \prec_{ \ottnt{v} }  \, \phi'_{{\mathrm{1}}} \,  \uparrow \, (  \ottmv{i}  ,  \pi  )   \ottsym{)}   \oplus   \ottsym{(}  \phi_{{\mathrm{2}}} \,  \prec_{ \ottnt{v} }  \, \phi'_{{\mathrm{2}}} \,  \uparrow \, (  \ottmv{i}  ,  \pi  )   \ottsym{)}  \\
    && \quad \text{(where $\oplus \,  \in  \, \ottsym{\{}   \wedge   \ottsym{,}   \vee   \ottsym{,}   \Rightarrow   \ottsym{\}}$)}
  \\
  \ottsym{(}    \forall  \,  \mathit{x}  \mathrel{  \in  }  \ottnt{s}  . \  \phi   \ottsym{)} \,  \prec_{ \ottnt{v} }  \, \ottsym{(}    \forall  \,  \mathit{x}  \mathrel{  \in  }  \ottnt{s}  . \  \phi'   \ottsym{)} \,  \uparrow \, (  \ottmv{i}  ,  \pi  )  &\defeq&   \forall  \,  \mathit{x}  \mathrel{  \in  }  \ottnt{s}  . \  \ottsym{(}  \phi \,  \prec_{ \ottnt{v} }  \, \phi' \,  \uparrow \, (  \ottmv{i}  ,  \pi  )   \ottsym{)}  \\
  \ottsym{(}    \exists  \,  \mathit{x}  \mathrel{  \in  }  \ottnt{s}  . \  \phi   \ottsym{)} \,  \prec_{ \ottnt{v} }  \, \ottsym{(}    \exists  \,  \mathit{x}  \mathrel{  \in  }  \ottnt{s}  . \  \phi'   \ottsym{)} \,  \uparrow \, (  \ottmv{i}  ,  \pi  )  &\defeq&   \exists  \,  \mathit{x}  \mathrel{  \in  }  \ottnt{s}  . \  \ottsym{(}  \phi \,  \prec_{ \ottnt{v} }  \, \phi' \,  \uparrow \, (  \ottmv{i}  ,  \pi  )   \ottsym{)}  \\
  \ottsym{(}    \forall  \,  \mathit{X} \,  {:}  \,  \tceseq{ \ottnt{s} }   . \  \phi   \ottsym{)} \,  \prec_{ \ottnt{v} }  \, \ottsym{(}    \forall  \,  \mathit{X} \,  {:}  \,  \tceseq{ \ottnt{s} }   . \  \phi'   \ottsym{)} \,  \uparrow \, (  \ottmv{i}  ,  \pi  )  &\defeq&   \forall  \,  \mathit{X} \,  {:}  \,  \tceseq{ \ottnt{s} }   . \  \ottsym{(}  \phi \,  \prec_{ \ottnt{v} }  \, \phi' \,  \uparrow \, (  \ottmv{i}  ,  \pi  )   \ottsym{)}  \\
  \ottsym{(}    \exists  \,  \mathit{X} \,  {:}  \,  \tceseq{ \ottnt{s} }   . \  \phi   \ottsym{)} \,  \prec_{ \ottnt{v} }  \, \ottsym{(}    \exists  \,  \mathit{X} \,  {:}  \,  \tceseq{ \ottnt{s} }   . \  \phi'   \ottsym{)} \,  \uparrow \, (  \ottmv{i}  ,  \pi  )  &\defeq&   \exists  \,  \mathit{X} \,  {:}  \,  \tceseq{ \ottnt{s} }   . \  \ottsym{(}  \phi \,  \prec_{ \ottnt{v} }  \, \phi' \,  \uparrow \, (  \ottmv{i}  ,  \pi  )   \ottsym{)}  \\
   \end{array}
 \]

 We also define $\ottnt{E_{{\mathrm{1}}}} \,  \prec_{ \ottnt{v} }  \, \ottnt{E_{{\mathrm{2}}}} \,  \uparrow \, (  \ottmv{i}  ,  \pi  ) $ as follows.
 \[\begin{array}{lll}
  \,[\,]\, \,  \prec_{ \ottnt{v} }  \, \,[\,]\, \,  \uparrow \, (  \ottmv{i}  ,  \pi  )  &\defeq&  \,[\,]\,  \\
  \ottsym{(}  \mathsf{let} \, \mathit{x_{{\mathrm{1}}}}  \ottsym{=}  \ottnt{E_{{\mathrm{1}}}} \,  \mathsf{in}  \, \ottnt{e_{{\mathrm{2}}}}  \ottsym{)} \,  \prec_{ \ottnt{v} }  \, \ottsym{(}  \mathsf{let} \, \mathit{x_{{\mathrm{1}}}}  \ottsym{=}  \ottnt{E'_{{\mathrm{1}}}} \,  \mathsf{in}  \, \ottnt{e'_{{\mathrm{2}}}}  \ottsym{)} \,  \uparrow \, (  \ottmv{i}  ,  \pi  )  &\defeq& \mathsf{let} \, \mathit{x_{{\mathrm{1}}}}  \ottsym{=}  \ottsym{(}  \ottnt{E_{{\mathrm{1}}}} \,  \prec_{ \ottnt{v} }  \, \ottnt{E'_{{\mathrm{1}}}} \,  \uparrow \, (  \ottmv{i}  ,  \pi  )   \ottsym{)} \,  \mathsf{in}  \, \ottsym{(}  \ottnt{e_{{\mathrm{2}}}} \,  \prec_{ \ottnt{v} }  \, \ottnt{e'_{{\mathrm{2}}}} \,  \uparrow \, (  \ottmv{i}  ,  \pi  )   \ottsym{)} \\
   \resetcnstr{ \ottnt{E_{{\mathrm{1}}}} }  \,  \prec_{ \ottnt{v} }  \,  \resetcnstr{ \ottnt{E_{{\mathrm{2}}}} }  \,  \uparrow \, (  \ottmv{i}  ,  \pi  )  &\defeq&  \resetcnstr{ \ottnt{E_{{\mathrm{1}}}} \,  \prec_{ \ottnt{v} }  \, \ottnt{E_{{\mathrm{2}}}} \,  \uparrow \, (  \ottmv{i}  ,  \pi  )  }  ~.
   \end{array}\]

 We write $\ottnt{e} \,  \prec_{ \ottnt{v} }  \, \ottnt{e'}$, $\ottnt{A} \,  \prec_{ \ottnt{v} }  \, \ottnt{A'}$, $\phi \,  \prec_{ \ottnt{v} }  \, \phi'$, and $\ottnt{E} \,  \prec_{ \ottnt{v} }  \, \ottnt{E'}$
 if and only if, for any $\pi$,
 $\ottnt{e} \,  \prec_{ \ottnt{v} }  \, \ottnt{e'} \,  \uparrow \, (  \ottsym{0}  ,  \pi  ) $, $\ottnt{A} \,  \prec_{ \ottnt{v} }  \, \ottnt{A'} \,  \uparrow \, (  \ottsym{0}  ,  \pi  ) $, $\phi \,  \prec_{ \ottnt{v} }  \, \phi' \,  \uparrow \, (  \ottsym{0}  ,  \pi  ) $, and $\ottnt{E} \,  \prec_{ \ottnt{v} }  \, \ottnt{E'} \,  \uparrow \, (  \ottsym{0}  ,  \pi  ) $
 are well defined, respectively.
 %
\end{defn}

\subsubsection{Auxiliary Definitions for Proofs}

\begin{defn}[Closing Domain Substitution]
 We define $\theta_{{\mathrm{0}}}  \ottsym{;}  \Gamma  \vdash  \theta$ as the conjunction of the following:
 \begin{itemize}
  \item $ {    \forall  \, { \mathit{x} } . \  \mathit{x} \,  \in  \, \mathit{dom} \, \ottsym{(}  \theta  \ottsym{)}   \mathrel{\Longrightarrow}  \ottsym{(}    \exists  \, { \iota  \ottsym{,}  \phi } . \  \mathit{x} \,  {:}  \, \ottsym{\{}  \mathit{x} \,  {:}  \, \iota \,  |  \, \phi  \ottsym{\}} \,  \in  \, \Gamma   \ottsym{)}  } \,\mathrel{\vee}\, { \ottsym{(}    \exists  \, { \ottnt{s} } . \  \mathit{x} \,  {:}  \, \ottnt{s} \,  \in  \, \Gamma   \ottsym{)} } $;
  \item $ {    \forall  \, { \mathit{x}  \ottsym{,}  \iota  \ottsym{,}  \phi } . \  \mathit{x} \,  {:}  \, \ottsym{\{}  \mathit{x} \,  {:}  \, \iota \,  |  \, \phi  \ottsym{\}} \,  \in  \, \Gamma   \mathrel{\Longrightarrow}  \theta  \ottsym{(}  \mathit{x}  \ottsym{)} \,  \in  \,  \mathcal{D}_{ \iota }   } \,\mathrel{\wedge}\, {  { [\![  \phi  ]\!]_{ \theta_{{\mathrm{0}}}  \mathbin{\uplus}  \theta } }  \,  =  \,  \top  } $; and
  \item $   \forall  \, { \mathit{x}  \ottsym{,}  \ottnt{s} } . \  \mathit{x} \,  {:}  \, \ottnt{s} \,  \in  \, \Gamma   \mathrel{\Longrightarrow}  \theta  \ottsym{(}  \mathit{x}  \ottsym{)} \,  \in  \,  \mathcal{D}_{ \ottnt{s} }  $;
  \item $   \forall  \, { \mathit{X} } . \  \mathit{X} \,  \in  \, \mathit{dom} \, \ottsym{(}  \theta  \ottsym{)}   \mathrel{\Longrightarrow}    \exists  \, {  \tceseq{ \ottnt{s} }  } . \  \mathit{X} \,  {:}  \,  \tceseq{ \ottnt{s} }  \,  \in  \, \Gamma  $; and
  \item $   \forall  \, { \mathit{X}  \ottsym{,}   \tceseq{ \ottnt{s} }  } . \  \mathit{X} \,  {:}  \,  \tceseq{ \ottnt{s} }  \,  \in  \, \Gamma   \mathrel{\Longrightarrow}  \theta  \ottsym{(}  \mathit{X}  \ottsym{)} \,  \in  \,  \mathcal{D}_{  \tceseq{ \ottnt{s} }   \rightarrow   \mathbb{B}  }  $.
 \end{itemize}
 We write $\Gamma  \vdash  \theta$ for $ \emptyset   \ottsym{;}  \Gamma  \vdash  \theta$.
\end{defn}

\begin{defn}[Equivalent Predicates]
 \noindent
 \begin{itemize}

  \item We write $ \ottnt{A_{{\mathrm{1}}}}  \, \Rightarrow \,  \ottnt{A_{{\mathrm{2}}}}  :   \tceseq{ \ottnt{s} }  $ if and only if:
        $ \emptyset   \vdash  \ottnt{A_{{\mathrm{1}}}}  \ottsym{:}   \tceseq{ \ottnt{s} } $;
        $ \emptyset   \vdash  \ottnt{A_{{\mathrm{2}}}}  \ottsym{:}   \tceseq{ \ottnt{s} } $; and
        $   \forall  \, { \Gamma  \ottsym{,}   \tceseq{ \ottnt{t} }  } . \   \tceseq{ \Gamma  \vdash  \ottnt{t}  \ottsym{:}  \ottnt{s} }    \mathrel{\Longrightarrow}  \Gamma  \models   \ottnt{A_{{\mathrm{1}}}}  \ottsym{(}   \tceseq{ \ottnt{t} }   \ottsym{)}    \Rightarrow    \ottnt{A_{{\mathrm{2}}}}  \ottsym{(}   \tceseq{ \ottnt{t} }   \ottsym{)}  $.

  \item We write $\ottnt{A_{{\mathrm{1}}}} \, \Leftrightarrow \, \ottnt{A_{{\mathrm{2}}}}  \ottsym{:}   \tceseq{ \ottnt{s} } $ if and only if
        $ \ottnt{A_{{\mathrm{1}}}}  \, \Rightarrow \,  \ottnt{A_{{\mathrm{2}}}}  :   \tceseq{ \ottnt{s} }  $ and $ \ottnt{A_{{\mathrm{2}}}}  \, \Rightarrow \,  \ottnt{A_{{\mathrm{1}}}}  :   \tceseq{ \ottnt{s} }  $.

  \item We write $ \ottnt{T_{{\mathrm{1}}}}  \mathrel{ \Leftrightarrow _{( \ottnt{A_{{\mathrm{1}}}} , \ottnt{A_{{\mathrm{2}}}} ) /  \mathit{X} } }  \ottnt{T_{{\mathrm{2}}}} $ if and only if
        there exist some $\ottnt{T}$ and $\ottmv{i} \in \{1,2\}$ such that
        $\ottnt{T_{{\mathrm{1}}}} \,  =  \,  \ottnt{T}    [  \ottnt{A_{\ottmv{i}}}  /  \mathit{X}  ]  $ and $\ottnt{T_{{\mathrm{2}}}} \,  =  \,  \ottnt{T}    [   \ottnt{A} _{ \ottsym{3}  \ottsym{-}  \ottmv{i} }   /  \mathit{X}  ]  $.

  \item We write $ \ottnt{C_{{\mathrm{1}}}}  \mathrel{ \Leftrightarrow _{( \ottnt{A_{{\mathrm{1}}}} , \ottnt{A_{{\mathrm{2}}}} ) /  \mathit{X} } }  \ottnt{C_{{\mathrm{2}}}} $ if and only if
        there exist some $\ottnt{C}$ and $\ottmv{i} \in \{1,2\}$ such that
        $\ottnt{C_{{\mathrm{1}}}} \,  =  \,  \ottnt{C}    [  \ottnt{A_{\ottmv{i}}}  /  \mathit{X}  ]  $ and $\ottnt{C_{{\mathrm{2}}}} \,  =  \,  \ottnt{C}    [   \ottnt{A} _{ \ottsym{3}  \ottsym{-}  \ottmv{i} }   /  \mathit{X}  ]  $.

  \item We write $ \ottnt{S_{{\mathrm{1}}}}  \mathrel{ \Leftrightarrow _{( \ottnt{A_{{\mathrm{1}}}} , \ottnt{A_{{\mathrm{2}}}} ) /  \mathit{X} } }  \ottnt{S_{{\mathrm{2}}}} $ if and only if
        there exist some $\ottnt{S}$ and $\ottmv{i} \in \{1,2\}$ such that
        $\ottnt{S_{{\mathrm{1}}}} \,  =  \,  \ottnt{S}    [  \ottnt{A_{\ottmv{i}}}  /  \mathit{X}  ]  $ and $\ottnt{S_{{\mathrm{2}}}} \,  =  \,  \ottnt{S}    [   \ottnt{A} _{ \ottsym{3}  \ottsym{-}  \ottmv{i} }   /  \mathit{X}  ]  $.

  \item We write $ \Phi_{{\mathrm{1}}}  \mathrel{ \Leftrightarrow _{( \ottnt{A_{{\mathrm{1}}}} , \ottnt{A_{{\mathrm{2}}}} ) /  \mathit{X} } }  \Phi_{{\mathrm{2}}} $ if and only if
        there exist some $\Phi$ and $\ottmv{i} \in \{1,2\}$ such that
        $\Phi_{{\mathrm{1}}} \,  =  \,  \Phi    [  \ottnt{A_{\ottmv{i}}}  /  \mathit{X}  ]  $ and $\Phi_{{\mathrm{2}}} \,  =  \,  \Phi    [   \ottnt{A} _{ \ottsym{3}  \ottsym{-}  \ottmv{i} }   /  \mathit{X}  ]  $.

  \item We write $ \phi_{{\mathrm{1}}}  \mathrel{ \Leftrightarrow _{( \ottnt{A_{{\mathrm{1}}}} , \ottnt{A_{{\mathrm{2}}}} ) /  \mathit{X} } }  \phi_{{\mathrm{2}}} $ if and only if
        there exist some $\phi$ and $\ottmv{i}$ such that
        $\phi_{{\mathrm{1}}} \,  =  \,  \phi    [  \ottnt{A_{\ottmv{i}}}  /  \mathit{X}  ]  $ and $\phi_{{\mathrm{2}}} \,  =  \,  \phi    [   \ottnt{A} _{ \ottsym{3}  \ottsym{-}  \ottmv{i} }   /  \mathit{X}  ]  $.

  \item We write $ \Gamma_{{\mathrm{1}}}  \mathrel{ \Leftrightarrow _{( \ottnt{A_{{\mathrm{1}}}} , \ottnt{A_{{\mathrm{2}}}} ) /  \mathit{X} } }  \Gamma_{{\mathrm{2}}} $ for the smallest relation satisfying
        the following rules.
        \begin{center}
         $\ottdruleIffXXEmpty{}$ \hfil
         $\ottdruleIffXXVar{}$ \\[2ex]
         $\ottdruleIffXXSVar{}$ \hfil
         $\ottdruleIffXXPVar{}$ \hfil
        \end{center}
 \end{itemize}
\end{defn}

\begin{defn}[Well-Typed Evaluation Contexts]
 We define $\ottsym{(}  \ottnt{E}  \ottsym{,}  \sigma  \ottsym{)}  \ottsym{:}  \ottnt{C_{{\mathrm{1}}}}  \Rrightarrow  \ottnt{C_{{\mathrm{2}}}}$ as the smallest relation satisfying the following rules.
 \begin{center}
  $\ottdruleTCXXEmpty{}$ \hfil
  $\ottdruleTCXXCons{}$
 \end{center}
\end{defn}

\begin{defn}[Path on Strictly Positive Temporal Effects]
 The predicates $ \tceseq{ \ottsym{(}  \mathit{X}  \ottsym{,}  \mathit{Y}  \ottsym{)} }   \vdash  \ottnt{C}  \Rrightarrow  \ottnt{C'}$ and $ \tceseq{ \ottsym{(}  \mathit{X}  \ottsym{,}  \mathit{Y}  \ottsym{)} }   \vdash  \ottnt{S}  \Rrightarrow  \ottnt{C'}$ are the smallest relations satisfying the following rules.
 \[\begin{array}{rcl}
   \emptyset   \vdash  \ottnt{C}  \Rrightarrow  \ottnt{C}
  \\[2ex]
   \tceseq{ \ottsym{(}  \mathit{X}  \ottsym{,}  \mathit{Y}  \ottsym{)} }   \vdash  \ottnt{S}  \Rrightarrow  \ottnt{C'}
  & \Rightarrow &
  \ottsym{(}  \mathit{X_{{\mathrm{0}}}}  \ottsym{,}  \mathit{Y_{{\mathrm{0}}}}  \ottsym{)}  \ottsym{,}   \tceseq{ \ottsym{(}  \mathit{X}  \ottsym{,}  \mathit{Y}  \ottsym{)} }   \vdash   \ottnt{T}  \,  \,\&\,  \,  \Phi  \, / \,  \ottnt{S}   \Rrightarrow  \ottnt{C'}
  \\[2ex]
   \tceseq{ \ottsym{(}  \mathit{X}  \ottsym{,}  \mathit{Y}  \ottsym{)} }   \vdash  \ottnt{C_{{\mathrm{2}}}}  \Rrightarrow  \ottnt{C'}
  & \Rightarrow &
   \tceseq{ \ottsym{(}  \mathit{X}  \ottsym{,}  \mathit{Y}  \ottsym{)} }   \vdash   ( \forall  \mathit{x}  .  \ottnt{C_{{\mathrm{1}}}}  )  \Rightarrow   \ottnt{C_{{\mathrm{2}}}}   \Rrightarrow  \ottnt{C'}
   \end{array}
 \]
\end{defn}

%% file: proof.tex
\section{Proofs}
\label{sec:proof}

\input{proof/type_soundness}

\clearpage
\input{proof/lr}

%% file: proof/type_soundness.tex
\subsection{Type Safety}

\begin{lemmap}{Weakening Well-Formedness and Formula Typing}{ts-weak-wf-ftyping}
 Suppose that $\vdash  \Gamma_{{\mathrm{1}}}  \ottsym{,}  \Gamma_{{\mathrm{2}}}$ and $  \mathit{dom}  (  \Gamma_{{\mathrm{2}}}  )   \, \ottsym{\#} \,   \mathit{dom}  (  \Gamma_{{\mathrm{3}}}  )  $.
 \begin{enumerate}
  \item If $\vdash  \Gamma_{{\mathrm{1}}}  \ottsym{,}  \Gamma_{{\mathrm{3}}}$, then $\vdash  \Gamma_{{\mathrm{1}}}  \ottsym{,}  \Gamma_{{\mathrm{2}}}  \ottsym{,}  \Gamma_{{\mathrm{3}}}$.
  \item If $\Gamma_{{\mathrm{1}}}  \ottsym{,}  \Gamma_{{\mathrm{3}}}  \vdash  \ottnt{T}$, then $\Gamma_{{\mathrm{1}}}  \ottsym{,}  \Gamma_{{\mathrm{2}}}  \ottsym{,}  \Gamma_{{\mathrm{3}}}  \vdash  \ottnt{T}$.
  \item If $\Gamma_{{\mathrm{1}}}  \ottsym{,}  \Gamma_{{\mathrm{3}}}  \vdash  \ottnt{C}$, then $\Gamma_{{\mathrm{1}}}  \ottsym{,}  \Gamma_{{\mathrm{2}}}  \ottsym{,}  \Gamma_{{\mathrm{3}}}  \vdash  \ottnt{C}$.
  \item If $\Gamma_{{\mathrm{1}}}  \ottsym{,}  \Gamma_{{\mathrm{3}}} \,  |  \, \ottnt{T}  \vdash  \ottnt{S}$, then $\Gamma_{{\mathrm{1}}}  \ottsym{,}  \Gamma_{{\mathrm{2}}}  \ottsym{,}  \Gamma_{{\mathrm{3}}} \,  |  \, \ottnt{T}  \vdash  \ottnt{S}$.
  \item If $\Gamma_{{\mathrm{1}}}  \ottsym{,}  \Gamma_{{\mathrm{3}}}  \vdash  \Phi$, then $\Gamma_{{\mathrm{1}}}  \ottsym{,}  \Gamma_{{\mathrm{2}}}  \ottsym{,}  \Gamma_{{\mathrm{3}}}  \vdash  \Phi$.
  \item If $\Gamma_{{\mathrm{1}}}  \ottsym{,}  \Gamma_{{\mathrm{3}}}  \vdash  \ottnt{t}  \ottsym{:}  \ottnt{s}$, then $\Gamma_{{\mathrm{1}}}  \ottsym{,}  \Gamma_{{\mathrm{2}}}  \ottsym{,}  \Gamma_{{\mathrm{3}}}  \vdash  \ottnt{t}  \ottsym{:}  \ottnt{s}$.
  \item If $\Gamma_{{\mathrm{1}}}  \ottsym{,}  \Gamma_{{\mathrm{3}}}  \vdash  \ottnt{A}  \ottsym{:}   \tceseq{ \ottnt{s} } $, then $\Gamma_{{\mathrm{1}}}  \ottsym{,}  \Gamma_{{\mathrm{2}}}  \ottsym{,}  \Gamma_{{\mathrm{3}}}  \vdash  \ottnt{A}  \ottsym{:}   \tceseq{ \ottnt{s} } $.
  \item If $\Gamma_{{\mathrm{1}}}  \ottsym{,}  \Gamma_{{\mathrm{3}}}  \vdash  \phi$, then $\Gamma_{{\mathrm{1}}}  \ottsym{,}  \Gamma_{{\mathrm{2}}}  \ottsym{,}  \Gamma_{{\mathrm{3}}}  \vdash  \phi$.
 \end{enumerate}
\end{lemmap}
\begin{proof}
 Straightforward by simultaneous induction on the derivations.
\end{proof}

\begin{lemmap}{Weakening Validity}{ts-weak-valid}
 Suppose that $  \mathit{dom}  (  \Gamma_{{\mathrm{2}}}  )   \, \ottsym{\#} \,   \mathit{dom}  (  \Gamma_{{\mathrm{1}}}  \ottsym{,}  \Gamma_{{\mathrm{3}}}  )  $.
 If $\Gamma_{{\mathrm{1}}}  \ottsym{,}  \Gamma_{{\mathrm{3}}}  \models  \phi$, then $\Gamma_{{\mathrm{1}}}  \ottsym{,}  \Gamma_{{\mathrm{2}}}  \ottsym{,}  \Gamma_{{\mathrm{3}}}  \models  \phi$.
\end{lemmap}
\begin{proof}
 \TS{Validity}
 It suffices to show that
 \[
     \forall  \, { \theta } . \   { [\![   \lfloor  \Gamma_{{\mathrm{1}}}  \ottsym{,}  \Gamma_{{\mathrm{3}}}  \, \vdash \,  \phi  \rfloor   ]\!]_{ \theta } }  \,  =  \,  \top    \mathrel{\Longrightarrow}   { [\![   \lfloor  \Gamma_{{\mathrm{1}}}  \ottsym{,}  \Gamma_{{\mathrm{2}}}  \ottsym{,}  \Gamma_{{\mathrm{3}}}  \, \vdash \,  \phi  \rfloor   ]\!]_{ \theta } }  \,  =  \,  \top   ~.
 \]
 It is easily proven by induction on $\Gamma_{{\mathrm{1}}}$.
\end{proof}

\begin{lemmap}{Weakening Subtyping}{ts-weak-subtyping}
 Suppose that $  \mathit{dom}  (  \Gamma_{{\mathrm{2}}}  )   \, \ottsym{\#} \,   \mathit{dom}  (  \Gamma_{{\mathrm{1}}}  \ottsym{,}  \Gamma_{{\mathrm{3}}}  )  $.
 \begin{enumerate}
  \item If $\Gamma_{{\mathrm{1}}}  \ottsym{,}  \Gamma_{{\mathrm{3}}}  \vdash  \ottnt{T_{{\mathrm{1}}}}  \ottsym{<:}  \ottnt{T_{{\mathrm{2}}}}$, then $\Gamma_{{\mathrm{1}}}  \ottsym{,}  \Gamma_{{\mathrm{2}}}  \ottsym{,}  \Gamma_{{\mathrm{3}}}  \vdash  \ottnt{T_{{\mathrm{1}}}}  \ottsym{<:}  \ottnt{T_{{\mathrm{2}}}}$.
  \item If $\Gamma_{{\mathrm{1}}}  \ottsym{,}  \Gamma_{{\mathrm{3}}}  \vdash  \ottnt{C_{{\mathrm{1}}}}  \ottsym{<:}  \ottnt{C_{{\mathrm{2}}}}$, then $\Gamma_{{\mathrm{1}}}  \ottsym{,}  \Gamma_{{\mathrm{2}}}  \ottsym{,}  \Gamma_{{\mathrm{3}}}  \vdash  \ottnt{C_{{\mathrm{1}}}}  \ottsym{<:}  \ottnt{C_{{\mathrm{2}}}}$.
  \item If $\Gamma_{{\mathrm{1}}}  \ottsym{,}  \Gamma_{{\mathrm{3}}} \,  |  \, \ottnt{T} \,  \,\&\,  \, \Phi  \vdash  \ottnt{S_{{\mathrm{1}}}}  \ottsym{<:}  \ottnt{S_{{\mathrm{2}}}}$, then $\Gamma_{{\mathrm{1}}}  \ottsym{,}  \Gamma_{{\mathrm{2}}}  \ottsym{,}  \Gamma_{{\mathrm{3}}} \,  |  \, \ottnt{T} \,  \,\&\,  \, \Phi  \vdash  \ottnt{S_{{\mathrm{1}}}}  \ottsym{<:}  \ottnt{S_{{\mathrm{2}}}}$.
  \item If $\Gamma_{{\mathrm{1}}}  \ottsym{,}  \Gamma_{{\mathrm{3}}}  \vdash  \Phi_{{\mathrm{1}}}  \ottsym{<:}  \Phi_{{\mathrm{2}}}$, then $\Gamma_{{\mathrm{1}}}  \ottsym{,}  \Gamma_{{\mathrm{2}}}  \ottsym{,}  \Gamma_{{\mathrm{3}}}  \vdash  \Phi_{{\mathrm{1}}}  \ottsym{<:}  \Phi_{{\mathrm{2}}}$.
 \end{enumerate}
\end{lemmap}
\begin{proof}
 Straightforward by simultaneous induction on the derivations.
 The cases for \Srule{Refine} and \Srule{Eff} use \reflem{ts-weak-valid}.
\end{proof}

\begin{lemmap}{Weakening Typing}{ts-weak-typing}
 Suppose that $\vdash  \Gamma_{{\mathrm{1}}}  \ottsym{,}  \Gamma_{{\mathrm{2}}}$ and $  \mathit{dom}  (  \Gamma_{{\mathrm{2}}}  )   \, \ottsym{\#} \,   \mathit{dom}  (  \Gamma_{{\mathrm{3}}}  )  $.
 If $\Gamma_{{\mathrm{1}}}  \ottsym{,}  \Gamma_{{\mathrm{3}}}  \vdash  \ottnt{e}  \ottsym{:}  \ottnt{C}$, then $\Gamma_{{\mathrm{1}}}  \ottsym{,}  \Gamma_{{\mathrm{2}}}  \ottsym{,}  \Gamma_{{\mathrm{3}}}  \vdash  \ottnt{e}  \ottsym{:}  \ottnt{C}$.
 Especially, if $\Gamma_{{\mathrm{1}}}  \ottsym{,}  \Gamma_{{\mathrm{3}}}  \vdash  \ottnt{e}  \ottsym{:}  \ottnt{T}$, then $\Gamma_{{\mathrm{1}}}  \ottsym{,}  \Gamma_{{\mathrm{2}}}  \ottsym{,}  \Gamma_{{\mathrm{3}}}  \vdash  \ottnt{e}  \ottsym{:}  \ottnt{T}$.
\end{lemmap}
\begin{proof}
 Straightforward by induction on the typing derivation
 with Lemmas~\ref{lem:ts-weak-wf-ftyping}, \ref{lem:ts-weak-valid}, and \ref{lem:ts-weak-subtyping}.
 Note that $  \mathit{dom}  (  \Gamma_{{\mathrm{2}}}  )   \, \ottsym{\#} \,   \mathit{dom}  (  \Gamma_{{\mathrm{1}}}  \ottsym{,}  \Gamma_{{\mathrm{3}}}  )  $ because
 $\vdash  \Gamma_{{\mathrm{1}}}  \ottsym{,}  \Gamma_{{\mathrm{2}}}$ implies $  \mathit{dom}  (  \Gamma_{{\mathrm{1}}}  )   \, \ottsym{\#} \,   \mathit{dom}  (  \Gamma_{{\mathrm{2}}}  )  $ and $ \mathit{dom}  (  \Gamma_{{\mathrm{1}}}  \ottsym{,}  \Gamma_{{\mathrm{3}}}  )  \,  =  \,  \mathit{dom}  (  \Gamma_{{\mathrm{1}}}  )  \,  \mathbin{\cup}  \,  \mathit{dom}  (  \Gamma_{{\mathrm{3}}}  ) $.
\end{proof}

\begin{lemmap}{Types of Partially Applied Functions}{ts-val-applied-fun}
 Let $\ottnt{v_{{\mathrm{1}}}}$ be a $\lambda$-abstraction $ \lambda\!  \,  \tceseq{ \mathit{z} }   \ottsym{.}  \ottnt{e}$ or recursive function $ \mathsf{rec}  \, (  \tceseq{  \mathit{X} ^{  \ottnt{A} _{  \mu  }  },  \mathit{Y} ^{  \ottnt{A} _{  \nu  }  }  }  ,  \mathit{f} ,   \tceseq{ \mathit{z} }  ) . \,  \ottnt{e} $.
 If $\Gamma  \vdash  \ottnt{v_{{\mathrm{1}}}} \,  \tceseq{ \ottnt{v_{{\mathrm{2}}}} }   \ottsym{:}  \ottnt{C}$ and $\ottsym{\mbox{$\mid$}}   \tceseq{ \ottnt{v_{{\mathrm{2}}}} }   \ottsym{\mbox{$\mid$}} \,  <  \, \ottsym{\mbox{$\mid$}}   \tceseq{ \mathit{z} }   \ottsym{\mbox{$\mid$}}$,
 then $ \ottnt{C}  . \ottnt{T} $ is a function type with $(\ottsym{\mbox{$\mid$}}   \tceseq{ \mathit{z} }   \ottsym{\mbox{$\mid$}}  \ottsym{-}  \ottsym{\mbox{$\mid$}}   \tceseq{ \ottnt{v_{{\mathrm{2}}}} }   \ottsym{\mbox{$\mid$}})$ argument types,
 that is,
 \[
   \ottnt{C}  . \ottnt{T}  \,  =  \, \ottsym{(}  \mathit{x_{{\mathrm{1}}}} \,  {:}  \, \ottnt{T_{{\mathrm{1}}}}  \ottsym{)}  \rightarrow   \ottsym{(}  \cdots  \rightarrow  \ottsym{(}  \ottsym{(}   \mathit{x} _{ \ottmv{n}  \ottsym{-}  \ottsym{1} }  \,  {:}  \,  \ottnt{T} _{ \ottmv{n}  \ottsym{-}  \ottsym{1} }   \ottsym{)}  \rightarrow   \ottsym{(}  \ottsym{(}  \mathit{x_{\ottmv{n}}} \,  {:}  \, \ottnt{T_{\ottmv{n}}}  \ottsym{)}  \rightarrow  \ottnt{C_{{\mathrm{0}}}}  \ottsym{)}  \,  \,\&\,  \,   \Phi _{ \ottmv{n}  \ottsym{-}  \ottsym{1} }   \, / \,   \ottnt{S} _{ \ottmv{n}  \ottsym{-}  \ottsym{1} }    \ottsym{)}  \cdots  \ottsym{)}  \,  \,\&\,  \,  \Phi_{{\mathrm{1}}}  \, / \,  \ottnt{S_{{\mathrm{1}}}} 
 \]
 for some $\mathit{x_{{\mathrm{1}}}}, \cdots, \mathit{x_{\ottmv{n}}}$, $\ottnt{T_{{\mathrm{1}}}}, \cdots, \ottnt{T_{\ottmv{n}}}$, $\Phi_{{\mathrm{1}}}, \cdots,  \Phi _{ \ottmv{n}  \ottsym{-}  \ottsym{1} } $, $\ottnt{S_{{\mathrm{1}}}}, \cdots,  \ottnt{S} _{ \ottmv{n}  \ottsym{-}  \ottsym{1} } $, and $\ottnt{C_{{\mathrm{0}}}}$, where
 $\ottmv{n} \,  =  \, \ottsym{\mbox{$\mid$}}   \tceseq{ \mathit{z} }   \ottsym{\mbox{$\mid$}}  \ottsym{-}  \ottsym{\mbox{$\mid$}}   \tceseq{ \ottnt{v_{{\mathrm{2}}}} }   \ottsym{\mbox{$\mid$}}$.
\end{lemmap}
\begin{proof}
 By induction on the derivation of $\Gamma  \vdash  \ottnt{v_{{\mathrm{1}}}} \,  \tceseq{ \ottnt{v_{{\mathrm{2}}}} }   \ottsym{:}  \ottnt{C}$.
 \begin{caseanalysis}
  \case \T{Abs} and \T{Fun}: Obvious.

  \case \T{Sub}: We are given
   \begin{prooftree}
    \AxiomC{$\Gamma  \vdash  \ottnt{v_{{\mathrm{1}}}} \,  \tceseq{ \ottnt{v_{{\mathrm{2}}}} }   \ottsym{:}  \ottnt{C'}$}
    \AxiomC{$\Gamma  \vdash  \ottnt{C'}  \ottsym{<:}  \ottnt{C}$}
    \AxiomC{$\Gamma  \vdash  \ottnt{C}$}
    \TrinaryInfC{$\Gamma  \vdash  \ottnt{v_{{\mathrm{1}}}} \,  \tceseq{ \ottnt{v_{{\mathrm{2}}}} }   \ottsym{:}  \ottnt{C}$}
   \end{prooftree}
   for some $\ottnt{C'}$.
   Let $\ottmv{n} \,  =  \, \ottsym{\mbox{$\mid$}}   \tceseq{ \mathit{z} }   \ottsym{\mbox{$\mid$}}  \ottsym{-}  \ottsym{\mbox{$\mid$}}   \tceseq{ \ottnt{v_{{\mathrm{2}}}} }   \ottsym{\mbox{$\mid$}}$.
   By the IH, $ \ottnt{C'}  . \ottnt{T} $ is a function type with $\ottmv{n}$ argument types.
   Because $\Gamma  \vdash  \ottnt{C'}  \ottsym{<:}  \ottnt{C}$ implies $\Gamma  \vdash   \ottnt{C'}  . \ottnt{T}   \ottsym{<:}   \ottnt{C}  . \ottnt{T} $, the inversion
   implies that $ \ottnt{C}  . \ottnt{T} $ is a function type with $\ottmv{n}$ argument
   types.
   Therefore, we have the conclusion.

  \case \T{App}:
   We are given
   \begin{prooftree}
    \AxiomC{$\Gamma  \vdash  \ottnt{v_{{\mathrm{1}}}} \,  \tceseq{ \ottnt{v_{{\mathrm{21}}}} }   \ottsym{:}  \ottsym{(}  \mathit{x_{{\mathrm{22}}}} \,  {:}  \, \ottnt{T_{{\mathrm{22}}}}  \ottsym{)}  \rightarrow  \ottnt{C'}$}
    \AxiomC{$\Gamma  \vdash  \ottnt{v_{{\mathrm{22}}}}  \ottsym{:}  \ottnt{T_{{\mathrm{22}}}}$}
    \BinaryInfC{$\Gamma  \vdash  \ottnt{v_{{\mathrm{1}}}} \,  \tceseq{ \ottnt{v_{{\mathrm{21}}}} }  \, \ottnt{v_{{\mathrm{22}}}}  \ottsym{:}   \ottnt{C'}    [  \ottnt{v_{{\mathrm{22}}}}  /  \mathit{x_{{\mathrm{22}}}}  ]  $}
   \end{prooftree}
   for some $ \tceseq{ \ottnt{v_{{\mathrm{21}}}} } $, $\ottnt{v_{{\mathrm{22}}}}$, $\mathit{x_{{\mathrm{22}}}}$, $\ottnt{T_{{\mathrm{22}}}}$, and $\ottnt{C'}$ such that
   $ \tceseq{ \ottnt{v_{{\mathrm{2}}}} }  \,  =  \,  \tceseq{ \ottnt{v_{{\mathrm{21}}}} }   \ottsym{,}  \ottnt{v_{{\mathrm{22}}}}$ and $\ottnt{C} \,  =  \,  \ottnt{C'}    [  \ottnt{v_{{\mathrm{22}}}}  /  \mathit{x_{{\mathrm{22}}}}  ]  $.
   Because $\ottsym{\mbox{$\mid$}}   \tceseq{ \ottnt{v_{{\mathrm{21}}}} }   \ottsym{\mbox{$\mid$}}  \ottsym{+}  \ottsym{1} \,  =  \, \ottsym{\mbox{$\mid$}}   \tceseq{ \ottnt{v_{{\mathrm{2}}}} }   \ottsym{\mbox{$\mid$}}$ and $\ottsym{\mbox{$\mid$}}   \tceseq{ \ottnt{v_{{\mathrm{2}}}} }   \ottsym{\mbox{$\mid$}} \,  <  \, \ottsym{\mbox{$\mid$}}   \tceseq{ \mathit{z} }   \ottsym{\mbox{$\mid$}}$,
   we have $\ottsym{\mbox{$\mid$}}   \tceseq{ \ottnt{v_{{\mathrm{21}}}} }   \ottsym{\mbox{$\mid$}} \,  <  \, \ottsym{\mbox{$\mid$}}   \tceseq{ \mathit{z} }   \ottsym{\mbox{$\mid$}}$.
   Therefore, the IH implies that
   $\ottsym{(}  \mathit{x_{{\mathrm{22}}}} \,  {:}  \, \ottnt{T_{{\mathrm{22}}}}  \ottsym{)}  \rightarrow  \ottnt{C'}$ is a function type with $(\ottsym{\mbox{$\mid$}}   \tceseq{ \mathit{z} }   \ottsym{\mbox{$\mid$}}  \ottsym{-}  \ottsym{\mbox{$\mid$}}   \tceseq{ \ottnt{v_{{\mathrm{21}}}} }   \ottsym{\mbox{$\mid$}})$ argument types.
   Because
   $\ottnt{C} \,  =  \,  \ottnt{C'}    [  \ottnt{v_{{\mathrm{22}}}}  /  \mathit{x_{{\mathrm{22}}}}  ]  $,
   we can find that $ \ottnt{C}  . \ottnt{T} $ is a function type with $(\ottsym{\mbox{$\mid$}}   \tceseq{ \mathit{z} }   \ottsym{\mbox{$\mid$}}  \ottsym{-}  \ottsym{\mbox{$\mid$}}   \tceseq{ \ottnt{v_{{\mathrm{21}}}} }   \ottsym{\mbox{$\mid$}}  \ottsym{-}  \ottsym{1})$ argument types.
   Because $\ottsym{\mbox{$\mid$}}   \tceseq{ \mathit{z} }   \ottsym{\mbox{$\mid$}}  \ottsym{-}  \ottsym{\mbox{$\mid$}}   \tceseq{ \ottnt{v_{{\mathrm{21}}}} }   \ottsym{\mbox{$\mid$}}  \ottsym{-}  \ottsym{1} = \ottsym{\mbox{$\mid$}}   \tceseq{ \mathit{z} }   \ottsym{\mbox{$\mid$}}  \ottsym{-}  \ottsym{\mbox{$\mid$}}   \tceseq{ \ottnt{v_{{\mathrm{2}}}} }   \ottsym{\mbox{$\mid$}}$, we have the conclusion.

  \otherwise: Contradictory.
 \end{caseanalysis}
\end{proof}

\begin{lemmap}{Reflexivity of Constant Equality in Validity}{ts-refl-const-eq}
 For any term $\ottnt{t}$ such that $ \emptyset   \vdash  \ottnt{t}  \ottsym{:}  \ottnt{s}$, $ { \models  \,   \ottnt{t}    =    \ottnt{t}  } $.
 Especially, $ { \models  \,   \ottnt{c}    =    \ottnt{c}  } $ for any $\ottnt{c}$.
\end{lemmap}
\begin{proof}
 \TS{Validity}
 Obvious.
\end{proof}

\begin{lemmap}{Open and Closed Validity}{ts-logic-closing-dsubs}
 \begin{enumerate}
  \item \label{lem:ts-logic-closing-dsubs:case1}
        $ { [\![   \lfloor  \Gamma  \, \vdash \,  \phi  \rfloor   ]\!]_{ \theta_{{\mathrm{0}}} } }  \,  =  \,  \top $
        if and only if
        $   \forall  \, { \theta } . \  \theta_{{\mathrm{0}}}  \ottsym{;}  \Gamma  \vdash  \theta   \mathrel{\Longrightarrow}   { [\![  \phi  ]\!]_{ \theta_{{\mathrm{0}}}  \mathbin{\uplus}  \theta } }  \,  =  \,  \top  $.
  \item $\Gamma  \models  \phi$
        if and only if
        $   \forall  \, { \theta } . \  \Gamma  \vdash  \theta   \mathrel{\Longrightarrow}   { [\![  \phi  ]\!]_{ \theta } }  \,  =  \,  \top  $.
 \end{enumerate}
\end{lemmap}
\begin{proof}
 \noindent
 \begin{enumerate}
  \item By induction on $\Gamma$.
        We proceed by case analysis on $\Gamma$.
        \begin{caseanalysis}
         \case $\Gamma \,  =  \,  \emptyset $:
          For any $\theta$,
          $\theta_{{\mathrm{0}}}  \ottsym{;}   \emptyset   \vdash  \theta$ if and only if $\theta \,  =  \,  \emptyset $.
          Therefore, it suffices to show that
          $ { [\![   \lfloor   \emptyset   \, \vdash \,  \phi  \rfloor   ]\!]_{ \theta_{{\mathrm{0}}} } }  \,  =  \,  \top $ if and only if
          $ { [\![  \phi  ]\!]_{ \theta_{{\mathrm{0}}} } }  \,  =  \,  \top $, which are equivalent by definition.

         \case $  \exists  \, { \mathit{x}  \ottsym{,}  \ottnt{T}  \ottsym{,}  \Gamma' } . \  \Gamma \,  =  \, \mathit{x} \,  {:}  \, \ottnt{T}  \ottsym{,}  \Gamma' $:
          By case analysis on $\ottnt{T}$.
          \begin{caseanalysis}
           \case $  \exists  \, { \iota_{{\mathrm{0}}}  \ottsym{,}  \phi_{{\mathrm{0}}} } . \  \ottnt{T} \,  =  \, \ottsym{\{}  \mathit{x} \,  {:}  \, \iota_{{\mathrm{0}}} \,  |  \, \phi_{{\mathrm{0}}}  \ottsym{\}} $:
            Consider the left-to-right direction.
            Suppose $ { [\![   \lfloor  \mathit{x} \,  {:}  \, \ottsym{\{}  \mathit{x} \,  {:}  \, \iota_{{\mathrm{0}}} \,  |  \, \phi_{{\mathrm{0}}}  \ottsym{\}}  \ottsym{,}  \Gamma'  \, \vdash \,  \phi  \rfloor   ]\!]_{ \theta_{{\mathrm{0}}} } }  \,  =  \,  \top $, and
            let $\theta$ be any substitution such that
            $\theta_{{\mathrm{0}}}  \ottsym{;}  \mathit{x} \,  {:}  \, \ottsym{\{}  \mathit{x} \,  {:}  \, \iota_{{\mathrm{0}}} \,  |  \, \phi_{{\mathrm{0}}}  \ottsym{\}}  \ottsym{,}  \Gamma'  \vdash  \theta$.
            Then, it suffices to show that
            \[
              { [\![  \phi  ]\!]_{ \theta_{{\mathrm{0}}}  \mathbin{\uplus}  \theta } }  \,  =  \,  \top  ~.
            \]
            Because $\theta_{{\mathrm{0}}}  \ottsym{;}  \mathit{x} \,  {:}  \, \ottsym{\{}  \mathit{x} \,  {:}  \, \iota_{{\mathrm{0}}} \,  |  \, \phi_{{\mathrm{0}}}  \ottsym{\}}  \ottsym{,}  \Gamma'  \vdash  \theta$,
            we have
            $\theta \,  =  \,  \{  \mathit{x}  \mapsto  \ottnt{c}  \}   \mathbin{\uplus}  \theta'$ for some $\ottnt{c}$ and $\theta'$ such that
            \begin{itemize}
             \item $ \mathit{ty}  (  \ottnt{c}  )  \,  =  \, \iota_{{\mathrm{0}}}$,
             \item $ { [\![  \phi_{{\mathrm{0}}}  ]\!]_{  \{  \mathit{x}  \mapsto  \ottnt{c}  \}   \mathbin{\uplus}  \theta_{{\mathrm{0}}} } }  \,  =  \,  \top $, and
             \item $ \{  \mathit{x}  \mapsto  \ottnt{c}  \}   \mathbin{\uplus}  \theta_{{\mathrm{0}}}  \ottsym{;}  \Gamma'  \vdash  \theta'$.
            \end{itemize}
            Therefore, the IH implies that it suffices to show that
            \[
              { [\![   \lfloor  \Gamma'  \, \vdash \,  \phi  \rfloor   ]\!]_{  \{  \mathit{x}  \mapsto  \ottnt{c}  \}   \mathbin{\uplus}  \theta_{{\mathrm{0}}} } }  \,  =  \,  \top  ~.
            \]
            Because
            \begin{itemize}
             \item $ { [\![   \lfloor  \mathit{x} \,  {:}  \, \ottsym{\{}  \mathit{x} \,  {:}  \, \iota_{{\mathrm{0}}} \,  |  \, \phi_{{\mathrm{0}}}  \ottsym{\}}  \ottsym{,}  \Gamma'  \, \vdash \,  \phi  \rfloor   ]\!]_{ \theta_{{\mathrm{0}}} } }  \,  =  \,  \top $,
             \item $ \mathit{ty}  (  \ottnt{c}  )  \,  =  \, \iota_{{\mathrm{0}}}$, and
             \item $ { [\![  \phi_{{\mathrm{0}}}  ]\!]_{  \{  \mathit{x}  \mapsto  \ottnt{c}  \}   \mathbin{\uplus}  \theta_{{\mathrm{0}}} } }  \,  =  \,  \top $,
            \end{itemize}
            we have the conclusion.

            Next, consider the right-to-left direction.
            Suppose $   \forall  \, { \theta } . \  \theta_{{\mathrm{0}}}  \ottsym{;}  \mathit{x} \,  {:}  \, \ottsym{\{}  \mathit{x} \,  {:}  \, \iota_{{\mathrm{0}}} \,  |  \, \phi_{{\mathrm{0}}}  \ottsym{\}}  \ottsym{,}  \Gamma'  \vdash  \theta   \mathrel{\Longrightarrow}   { [\![  \phi  ]\!]_{ \theta_{{\mathrm{0}}}  \mathbin{\uplus}  \theta } }  \,  =  \,  \top  $.
            We show that
            \[
              { [\![   \lfloor  \mathit{x} \,  {:}  \, \ottsym{\{}  \mathit{x} \,  {:}  \, \iota_{{\mathrm{0}}} \,  |  \, \phi_{{\mathrm{0}}}  \ottsym{\}}  \ottsym{,}  \Gamma'  \, \vdash \,  \phi  \rfloor   ]\!]_{ \theta_{{\mathrm{0}}} } }  \,  =  \,  \top  ~.
            \]
            Let $\ottnt{c}$ be any constant such that
            $ \mathit{ty}  (  \ottnt{c}  )  \,  =  \, \iota_{{\mathrm{0}}}$ and $ { [\![  \phi  ]\!]_{  \{  \mathit{x}  \mapsto  \ottnt{c}  \}   \mathbin{\uplus}  \theta_{{\mathrm{0}}} } }  \,  =  \,  \top $.
            By definition, it suffices to show that
            $ { [\![   \lfloor  \Gamma'  \, \vdash \,  \phi  \rfloor   ]\!]_{  \{  \mathit{x}  \mapsto  \ottnt{c}  \}   \mathbin{\uplus}  \theta_{{\mathrm{0}}} } }  \,  =  \,  \bot $.
            By the IH, it suffices to show that
            \[
                \forall  \, { \theta' } . \   \{  \mathit{x}  \mapsto  \ottnt{c}  \}   \mathbin{\uplus}  \theta_{{\mathrm{0}}}  \ottsym{;}  \Gamma'  \vdash  \theta'   \mathrel{\Longrightarrow}   { [\![  \phi  ]\!]_{  \{  \mathit{x}  \mapsto  \ottnt{c}  \}   \mathbin{\uplus}  \theta_{{\mathrm{0}}}  \mathbin{\uplus}  \theta' } }  \,  =  \,  \top   ~.
            \]
            Let $\theta'$ be any substitution such that
            $ \{  \mathit{x}  \mapsto  \ottnt{c}  \}   \mathbin{\uplus}  \theta_{{\mathrm{0}}}  \ottsym{;}  \Gamma'  \vdash  \theta'$.
            By definition,
            \[
             \theta_{{\mathrm{0}}}  \ottsym{;}  \mathit{x} \,  {:}  \, \ottsym{\{}  \mathit{x} \,  {:}  \, \iota_{{\mathrm{0}}} \,  |  \, \phi_{{\mathrm{0}}}  \ottsym{\}}  \ottsym{,}  \Gamma'  \vdash   \{  \mathit{x}  \mapsto  \ottnt{c}  \}   \mathbin{\uplus}  \theta'
            \]
            because $ \mathit{ty}  (  \ottnt{c}  )  \,  =  \, \iota_{{\mathrm{0}}}$ and $ { [\![  \phi_{{\mathrm{0}}}  ]\!]_{  \{  \mathit{x}  \mapsto  \ottnt{c}  \}   \mathbin{\uplus}  \theta_{{\mathrm{0}}} } }  \,  =  \,  \top $.
            Therefore, by the assumption, we have the conclusion
            $ { [\![  \phi  ]\!]_{  \{  \mathit{x}  \mapsto  \ottnt{c}  \}   \mathbin{\uplus}  \theta_{{\mathrm{0}}}  \mathbin{\uplus}  \theta' } }  \,  =  \,  \top $.

           \case $\ottnt{T}$ is not a refinement type:
            Obvious by the IH because:
            \begin{itemize}
             \item $ { [\![   \lfloor  \Gamma  \, \vdash \,  \phi  \rfloor   ]\!]_{ \theta_{{\mathrm{0}}} } }  \,  =  \,  \top $ if and only if
                   $ { [\![   \lfloor  \Gamma'  \, \vdash \,  \phi  \rfloor   ]\!]_{ \theta_{{\mathrm{0}}} } }  \,  =  \,  \top $; and
             \item for any $\theta$,
                   $\theta_{{\mathrm{0}}}  \ottsym{;}  \Gamma  \vdash  \theta$ if and only if
                   $\theta_{{\mathrm{0}}}  \ottsym{;}  \Gamma'  \vdash  \theta$.
            \end{itemize}
          \end{caseanalysis}

         \case $  \exists  \, { \mathit{x}  \ottsym{,}  \ottnt{s}  \ottsym{,}  \Gamma' } . \  \Gamma \,  =  \, \mathit{x} \,  {:}  \, \ottnt{s}  \ottsym{,}  \Gamma' $:
          Consider the left-to-right direction.
          Suppose $ { [\![   \lfloor  \mathit{x} \,  {:}  \, \ottnt{s}  \ottsym{,}  \Gamma'  \, \vdash \,  \phi  \rfloor   ]\!]_{ \theta_{{\mathrm{0}}} } }  \,  =  \,  \top $, and
          let $\theta$ be any substitution such that
          $\theta_{{\mathrm{0}}}  \ottsym{;}  \mathit{x} \,  {:}  \, \ottnt{s}  \ottsym{,}  \Gamma'  \vdash  \theta$.
          Then, it suffices to show that
          \[
            { [\![  \phi  ]\!]_{ \theta_{{\mathrm{0}}}  \mathbin{\uplus}  \theta } }  \,  =  \,  \top  ~.
          \]
          Because $\theta_{{\mathrm{0}}}  \ottsym{;}  \mathit{x} \,  {:}  \, \ottnt{s}  \ottsym{,}  \Gamma'  \vdash  \theta$, we have
          $\theta \,  =  \,  \{  \mathit{x}  \mapsto  \ottnt{d}  \}   \mathbin{\uplus}  \theta'$ for some $\ottnt{d} \,  \in  \,  \mathcal{D}_{ \ottnt{s} } $ such that
          $ \{  \mathit{x}  \mapsto  \ottnt{d}  \}   \mathbin{\uplus}  \theta_{{\mathrm{0}}}  \ottsym{;}  \Gamma'  \vdash  \theta'$.
          Therefore, the IH implies that it suffices to show that
          \[
            { [\![   \lfloor  \Gamma'  \, \vdash \,  \phi  \rfloor   ]\!]_{  \{  \mathit{x}  \mapsto  \ottnt{d}  \}   \mathbin{\uplus}  \theta_{{\mathrm{0}}} } }  \,  =  \,  \top  ~.
          \]
          Because
          $ { [\![   \lfloor  \mathit{x} \,  {:}  \, \ottnt{s}  \ottsym{,}  \Gamma'  \, \vdash \,  \phi  \rfloor   ]\!]_{ \theta_{{\mathrm{0}}} } }  \,  =  \,  \top $ and
          $\ottnt{d} \,  \in  \,  \mathcal{D}_{ \ottnt{s} } $,
          we have the conclusion.

          Next, consider the right-to-left direction.
          Suppose $   \forall  \, { \theta } . \  \theta_{{\mathrm{0}}}  \ottsym{;}  \mathit{x} \,  {:}  \, \ottnt{s}  \ottsym{,}  \Gamma'  \vdash  \theta   \mathrel{\Longrightarrow}   { [\![  \phi  ]\!]_{ \theta_{{\mathrm{0}}}  \mathbin{\uplus}  \theta } }  \,  =  \,  \top  $.
          We show that
          \[
            { [\![   \lfloor  \mathit{x} \,  {:}  \, \ottnt{s}  \ottsym{,}  \Gamma'  \, \vdash \,  \phi  \rfloor   ]\!]_{ \theta_{{\mathrm{0}}} } }  \,  =  \,  \top  ~.
          \]
          Let $\ottnt{d} \,  \in  \,  \mathcal{D}_{ \ottnt{s} } $.
          By definition, it suffices to show that
          $ { [\![   \lfloor  \Gamma'  \, \vdash \,  \phi  \rfloor   ]\!]_{  \{  \mathit{x}  \mapsto  \ottnt{d}  \}   \mathbin{\uplus}  \theta_{{\mathrm{0}}} } }  \,  =  \,  \bot $.
          By the IH, it suffices to show that
          \[
              \forall  \, { \theta' } . \   \{  \mathit{x}  \mapsto  \ottnt{d}  \}   \mathbin{\uplus}  \theta_{{\mathrm{0}}}  \ottsym{;}  \Gamma'  \vdash  \theta'   \mathrel{\Longrightarrow}   { [\![  \phi  ]\!]_{  \{  \mathit{x}  \mapsto  \ottnt{d}  \}   \mathbin{\uplus}  \theta_{{\mathrm{0}}}  \mathbin{\uplus}  \theta' } }  \,  =  \,  \top   ~.
          \]
          Let $\theta'$ be any substitution such that
          $ \{  \mathit{x}  \mapsto  \ottnt{d}  \}   \mathbin{\uplus}  \theta_{{\mathrm{0}}}  \ottsym{;}  \Gamma'  \vdash  \theta'$.
          By definition, $\theta_{{\mathrm{0}}}  \ottsym{;}  \mathit{x} \,  {:}  \, \ottnt{s}  \ottsym{,}  \Gamma'  \vdash   \{  \mathit{x}  \mapsto  \ottnt{d}  \}   \mathbin{\uplus}  \theta'$
          because $\ottnt{d} \,  \in  \,  \mathcal{D}_{ \ottnt{s} } $.
          Therefore, by the assumption, we have the conclusion
          $ { [\![  \phi  ]\!]_{  \{  \mathit{x}  \mapsto  \ottnt{d}  \}   \mathbin{\uplus}  \theta_{{\mathrm{0}}}  \mathbin{\uplus}  \theta' } }  \,  =  \,  \top $.

         \case $  \exists  \, { \mathit{X}  \ottsym{,}   \tceseq{ \ottnt{s} }   \ottsym{,}  \Gamma' } . \  \Gamma \,  =  \, \mathit{X} \,  {:}  \,  \tceseq{ \ottnt{s} }   \ottsym{,}  \Gamma' $:
          Consider the left-to-right direction.
          Suppose $ { [\![   \lfloor  \mathit{X} \,  {:}  \,  \tceseq{ \ottnt{s} }   \ottsym{,}  \Gamma'  \, \vdash \,  \phi  \rfloor   ]\!]_{ \theta_{{\mathrm{0}}} } }  \,  =  \,  \top $, and
          let $\theta$ be any substitution such that
          $\theta_{{\mathrm{0}}}  \ottsym{;}  \mathit{X} \,  {:}  \,  \tceseq{ \ottnt{s} }   \ottsym{,}  \Gamma'  \vdash  \theta$.
          Then, it suffices to show that
          \[
            { [\![  \phi  ]\!]_{ \theta_{{\mathrm{0}}}  \mathbin{\uplus}  \theta } }  \,  =  \,  \top  ~.
          \]
          Because $\theta_{{\mathrm{0}}}  \ottsym{;}  \mathit{X} \,  {:}  \,  \tceseq{ \ottnt{s} }   \ottsym{,}  \Gamma'  \vdash  \theta$, we have
          $\theta \,  =  \,  \{  \mathit{X}  \mapsto  \mathbf{\mathit{f} }  \}   \mathbin{\uplus}  \theta'$ for some $\mathbf{\mathit{f} } \,  \in  \,  \mathcal{D}_{  \tceseq{ \ottnt{s} }   \rightarrow   \mathbb{B}  } $
          such that
          $ \{  \mathit{X}  \mapsto  \mathbf{\mathit{f} }  \}   \mathbin{\uplus}  \theta_{{\mathrm{0}}}  \ottsym{;}  \Gamma'  \vdash  \theta'$.
          Therefore, the IH implies that it suffices to show that
          \[
            { [\![   \lfloor  \Gamma'  \, \vdash \,  \phi  \rfloor   ]\!]_{  \{  \mathit{X}  \mapsto  \mathbf{\mathit{f} }  \}   \mathbin{\uplus}  \theta_{{\mathrm{0}}} } }  \,  =  \,  \top  ~.
          \]
          Because
          $ { [\![   \lfloor  \mathit{X} \,  {:}  \,  \tceseq{ \ottnt{s} }   \ottsym{,}  \Gamma'  \, \vdash \,  \phi  \rfloor   ]\!]_{ \theta_{{\mathrm{0}}} } }  \,  =  \,  \top $ and
          $\mathbf{\mathit{f} } \,  \in  \,  \mathcal{D}_{  \tceseq{ \ottnt{s} }   \rightarrow   \mathbb{B}  } $,
          we have the conclusion.

          Next, consider the right-to-left direction.
          Suppose $   \forall  \, { \theta } . \  \theta_{{\mathrm{0}}}  \ottsym{;}  \mathit{X} \,  {:}  \,  \tceseq{ \ottnt{s} }   \ottsym{,}  \Gamma'  \vdash  \theta   \mathrel{\Longrightarrow}   { [\![  \phi  ]\!]_{ \theta_{{\mathrm{0}}}  \mathbin{\uplus}  \theta } }  \,  =  \,  \top  $.
          We show that
          \[
            { [\![   \lfloor  \mathit{X} \,  {:}  \,  \tceseq{ \ottnt{s} }   \ottsym{,}  \Gamma'  \, \vdash \,  \phi  \rfloor   ]\!]_{ \theta_{{\mathrm{0}}} } }  \,  =  \,  \top  ~.
          \]
          Let $\mathbf{\mathit{f} } \,  \in  \,  \mathcal{D}_{  \tceseq{ \ottnt{s} }   \rightarrow   \mathbb{B}  } $.
          By definition, it suffices to show that
          $ { [\![   \lfloor  \Gamma'  \, \vdash \,  \phi  \rfloor   ]\!]_{  \{  \mathit{X}  \mapsto  \mathbf{\mathit{f} }  \}   \mathbin{\uplus}  \theta_{{\mathrm{0}}} } }  \,  =  \,  \bot $.
          By the IH, it suffices to show that
          \[
              \forall  \, { \theta' } . \   \{  \mathit{X}  \mapsto  \mathbf{\mathit{f} }  \}   \mathbin{\uplus}  \theta_{{\mathrm{0}}}  \ottsym{;}  \Gamma'  \vdash  \theta'   \mathrel{\Longrightarrow}   { [\![  \phi  ]\!]_{  \{  \mathit{X}  \mapsto  \mathbf{\mathit{f} }  \}   \mathbin{\uplus}  \theta_{{\mathrm{0}}}  \mathbin{\uplus}  \theta' } }  \,  =  \,  \top   ~.
          \]
          Let $\theta'$ be any substitution such that
          $ \{  \mathit{X}  \mapsto  \mathbf{\mathit{f} }  \}   \mathbin{\uplus}  \theta_{{\mathrm{0}}}  \ottsym{;}  \Gamma'  \vdash  \theta'$.
          By definition, $\theta_{{\mathrm{0}}}  \ottsym{;}  \mathit{X} \,  {:}  \,  \tceseq{ \ottnt{s} }   \ottsym{,}  \Gamma'  \vdash   \{  \mathit{X}  \mapsto  \mathbf{\mathit{f} }  \}   \mathbin{\uplus}  \theta'$
          because $\mathbf{\mathit{f} } \,  \in  \,  \mathcal{D}_{  \tceseq{ \ottnt{s} }   \rightarrow   \mathbb{B}  } $.
          Therefore, by the assumption, we have the conclusion
          $ { [\![  \phi  ]\!]_{  \{  \mathit{X}  \mapsto  \mathbf{\mathit{f} }  \}   \mathbin{\uplus}  \theta_{{\mathrm{0}}}  \mathbin{\uplus}  \theta' } }  \,  =  \,  \top $.
        \end{caseanalysis}

  \item Obvious by the case (\ref{lem:ts-logic-closing-dsubs:case1})
        with $\theta_{{\mathrm{0}}} \,  =  \,  \emptyset $.
 \end{enumerate}
\end{proof}

\begin{lemmap}{Term Substitution in Valuation}{ts-logic-val-subst}
 $ { [\![   \phi    [  \ottnt{t}  /  \mathit{x}  ]    ]\!]_{ \theta } }  \,  =  \,  { [\![  \phi  ]\!]_{  \{  \mathit{x}  \mapsto   { [\![  \ottnt{t}  ]\!]_{ \theta } }   \}   \mathbin{\uplus}  \theta } } $.
\end{lemmap}
\begin{proof}
 \TS{Validity}
 Straightforward by induction on $\phi$.
\end{proof}

\begin{lemmap}{Predicate Substitution in Valuation}{ts-logic-pred-subst}
 $ { [\![   \phi    [  \ottnt{A}  /  \mathit{X}  ]    ]\!]_{ \theta } }  \,  =  \,  { [\![  \phi  ]\!]_{  \{  \mathit{X}  \mapsto   { [\![  \ottnt{A}  ]\!]_{ \theta } }   \}   \mathbin{\uplus}  \theta } } $.
\end{lemmap}
\begin{proof}
 \TS{Validity}
 Straightforward by induction on $\phi$.
\end{proof}

\begin{lemmap}{Term Substitution in Validity}{ts-term-subst-valid}
 If $\Gamma_{{\mathrm{1}}}  \vdash  \ottnt{t}  \ottsym{:}  \ottnt{s}$ and $\Gamma_{{\mathrm{1}}}  \ottsym{,}  \mathit{x} \,  {:}  \, \ottnt{s}  \ottsym{,}  \Gamma_{{\mathrm{2}}}  \models  \phi$, then $ \Gamma_{{\mathrm{1}}}  \ottsym{,}  \Gamma_{{\mathrm{2}}}    [  \ottnt{t}  /  \mathit{x}  ]    \models   \phi    [  \ottnt{t}  /  \mathit{x}  ]  $.
\end{lemmap}
\begin{proof}
 Let $\theta$ be any substitution such that
 $ \Gamma_{{\mathrm{1}}}  \ottsym{,}  \Gamma_{{\mathrm{2}}}    [  \ottnt{t}  /  \mathit{x}  ]    \vdash  \theta$.
 By \reflem{ts-logic-closing-dsubs},
 it suffices to show that $ { [\![   \phi    [  \ottnt{t}  /  \mathit{x}  ]    ]\!]_{ \theta } }  \,  =  \,  \top $.
 Then, by \reflem{ts-logic-val-subst}, it suffices to show that
 $ { [\![  \phi  ]\!]_{  \{  \mathit{x}  \mapsto   { [\![  \ottnt{t}  ]\!]_{ \theta } }   \}   \mathbin{\uplus}  \theta } }  \,  =  \,  \top $.
 By \reflem{ts-logic-closing-dsubs} with $\Gamma_{{\mathrm{1}}}  \ottsym{,}  \mathit{x} \,  {:}  \, \ottnt{s}  \ottsym{,}  \Gamma_{{\mathrm{2}}}  \models  \phi$,
 it suffices to show that
 $\Gamma_{{\mathrm{1}}}  \ottsym{,}  \mathit{x} \,  {:}  \, \ottnt{s}  \ottsym{,}  \Gamma_{{\mathrm{2}}}  \vdash   \{  \mathit{x}  \mapsto   { [\![  \ottnt{t}  ]\!]_{ \theta } }   \}   \mathbin{\uplus}  \theta$.
 $ \Gamma_{{\mathrm{1}}}  \ottsym{,}  \Gamma_{{\mathrm{2}}}    [  \ottnt{t}  /  \mathit{x}  ]    \vdash  \theta$ and $\Gamma_{{\mathrm{1}}}  \vdash  \ottnt{t}  \ottsym{:}  \ottnt{s}$
 implies $ { [\![  \ottnt{t}  ]\!]_{ \theta } }  \,  \in  \,  \mathcal{D}_{ \ottnt{s} } $.
 Therefore, as $ \Gamma_{{\mathrm{1}}}  \ottsym{,}  \Gamma_{{\mathrm{2}}}    [  \ottnt{t}  /  \mathit{x}  ]    \vdash  \theta$,
 we have the conclusion
 $\Gamma_{{\mathrm{1}}}  \ottsym{,}  \mathit{x} \,  {:}  \, \ottnt{s}  \ottsym{,}  \Gamma_{{\mathrm{2}}}  \vdash   \{  \mathit{x}  \mapsto   { [\![  \ottnt{t}  ]\!]_{ \theta } }   \}   \mathbin{\uplus}  \theta$
 by definition and \reflem{ts-logic-val-subst}.
\end{proof}

\begin{lemmap}{Discharging Assumptions in Validity}{ts-discharge-assumption-valid}
 If $\Gamma  \models   \phi_{{\mathrm{1}}}    \Rightarrow    \phi_{{\mathrm{2}}} $ and $\Gamma  \models  \phi_{{\mathrm{1}}}$, then $\Gamma  \models  \phi_{{\mathrm{2}}}$.
\end{lemmap}
\begin{proof}
 \TS{Validity}
 Obvious by \reflem{ts-logic-closing-dsubs}.
\end{proof}

\begin{lemmap}{Validity is Reflexive}{ts-refl-valid}
 If $\Gamma  \vdash  \phi$, then $\Gamma  \models   \phi    \Rightarrow    \phi $.
\end{lemmap}
\begin{proof}
 \TS{Validity}
 Obvious by \reflem{ts-logic-closing-dsubs}.
\end{proof}

\begin{lemmap}{Transitivity of Implication}{ts-transitivity-valid}
 If $\Gamma  \models   \phi_{{\mathrm{1}}}    \Rightarrow    \phi_{{\mathrm{2}}} $ and $\Gamma  \models   \phi_{{\mathrm{2}}}    \Rightarrow    \phi_{{\mathrm{3}}} $, then $\Gamma  \models   \phi_{{\mathrm{1}}}    \Rightarrow    \phi_{{\mathrm{3}}} $.
\end{lemmap}
\begin{proof}
 \TS{Validity}
 Obvious by \reflem{ts-logic-closing-dsubs}.
\end{proof}

\begin{lemmap}{Open Values of Refinement Types}{ts-open-val-inv-refine-ty}
 If $\Gamma  \vdash  \ottnt{v}  \ottsym{:}   \ottsym{\{}  \mathit{x} \,  {:}  \, \iota \,  |  \, \phi  \ottsym{\}}  \,  \,\&\,  \,  \Phi  \, / \,  \ottnt{S} $,
 then either of the following holds:
 \begin{itemize}
  \item $ {  {   \exists  \, { \mathit{y}  \ottsym{,}  \phi' } . \  \ottnt{v} \,  =  \, \mathit{y}  } \,\mathrel{\wedge}\, { \Gamma  \ottsym{(}  \mathit{y}  \ottsym{)} \,  =  \, \ottsym{\{}  \mathit{x} \,  {:}  \, \iota \,  |  \, \phi'  \ottsym{\}} }  } \,\mathrel{\wedge}\, { \Gamma  \ottsym{,}  \mathit{x} \,  {:}  \, \iota  \models   \ottsym{(}   \mathit{y}    =    \mathit{x}   \ottsym{)}    \Rightarrow    \phi  } $; or
  \item $ {  {   \exists  \, { \ottnt{c} } . \  \ottnt{v} \,  =  \, \ottnt{c}  } \,\mathrel{\wedge}\, {  \mathit{ty}  (  \ottnt{c}  )  \,  =  \, \iota }  } \,\mathrel{\wedge}\, { \Gamma  \models   \phi    [  \ottnt{c}  /  \mathit{x}  ]   } $.
 \end{itemize}
\end{lemmap}
\begin{proof}
 By induction on the typing derivation.
 \begin{caseanalysis}
  \case \T{CVar}:
   We are given
   \begin{itemize}
    \item $\ottnt{v} \,  =  \, \mathit{y}$,
    \item $\Gamma  \ottsym{(}  \mathit{y}  \ottsym{)} \,  =  \, \ottsym{\{}  \mathit{x} \,  {:}  \, \iota \,  |  \, \phi'  \ottsym{\}}$, and
    \item $\phi \,  =  \, \ottsym{(}   \mathit{y}    =    \mathit{x}   \ottsym{)}$
   \end{itemize}
   for some $\mathit{y}$ and $\phi'$.
   Because $\Gamma  \ottsym{(}  \mathit{y}  \ottsym{)} \,  =  \, \ottsym{\{}  \mathit{x} \,  {:}  \, \iota \,  |  \, \phi'  \ottsym{\}}$,
   we have $\Gamma \,  =  \, \Gamma_{{\mathrm{1}}}  \ottsym{,}  \mathit{y} \,  {:}  \, \ottsym{\{}  \mathit{x} \,  {:}  \, \iota \,  |  \, \phi'  \ottsym{\}}  \ottsym{,}  \Gamma_{{\mathrm{2}}}$
   for some $\Gamma_{{\mathrm{1}}}$ and $\Gamma_{{\mathrm{2}}}$.
   Therefore, it suffices to show that
   \[
     \Gamma_{{\mathrm{1}}}  \ottsym{,}  \mathit{y} \,  {:}  \, \ottsym{\{}  \mathit{x} \,  {:}  \, \iota \,  |  \, \phi'  \ottsym{\}}  \ottsym{,}  \Gamma_{{\mathrm{2}}}  \ottsym{,}  \mathit{x} \,  {:}  \, \iota  \models  \ottsym{(}   \mathit{y}    =    \mathit{x}   \ottsym{)}  \mathrel{\Longrightarrow}  \ottsym{(}  \mathit{y} \,  =  \, \mathit{x}  \ottsym{)}  ~,
   \]
   which is prove by \reflem{ts-refl-valid}.
   Note that $ \Gamma_{{\mathrm{1}}}  \ottsym{,}  \mathit{y} \,  {:}  \, \ottsym{\{}  \mathit{x} \,  {:}  \, \iota \,  |  \, \phi'  \ottsym{\}}  \ottsym{,}  \Gamma_{{\mathrm{2}}}  \ottsym{,}  \mathit{x} \,  {:}  \, \iota  \vdash  \ottsym{(}   \mathit{y}    =    \mathit{x}   \ottsym{)}  \mathrel{\Longrightarrow}  \ottsym{(}  \mathit{y} \,  =  \, \mathit{x}  \ottsym{)} $
   is derived with $\vdash  \Gamma$, which is obtained by inversion of \T{CVar}.

  \case \T{Const}:
   We are given $\ottnt{v} \,  =  \, \ottnt{c}$ and $\phi \,  =  \, \ottsym{(}   \ottnt{c}    =    \mathit{x}   \ottsym{)}$ for some $\ottnt{c}$
   such that $ \mathit{ty}  (  \ottnt{c}  )  \,  =  \, \iota$.
   It suffices to show that $\Gamma  \models   \ottnt{c}    =    \ottnt{c} $.
   By \reflem{ts-refl-const-eq}, $ { \models  \,   \ottnt{c}    =    \ottnt{c}  } $.
   Therefore, \reflem{ts-weak-valid} implies the conclusion $\Gamma  \models   \ottnt{c}    =    \ottnt{c} $.

  \case \T{Sub}:
   We are given
   \begin{prooftree}
    \AxiomC{$\Gamma  \vdash  \ottnt{v}  \ottsym{:}  \ottnt{C}$}
    \AxiomC{$\Gamma  \vdash  \ottnt{C}  \ottsym{<:}   \ottsym{\{}  \mathit{x} \,  {:}  \, \iota \,  |  \, \phi  \ottsym{\}}  \,  \,\&\,  \,  \Phi  \, / \,  \ottnt{S} $}
    \AxiomC{$\Gamma  \vdash   \ottsym{\{}  \mathit{x} \,  {:}  \, \iota \,  |  \, \phi  \ottsym{\}}  \,  \,\&\,  \,  \Phi  \, / \,  \ottnt{S} $}
    \TrinaryInfC{$\Gamma  \vdash  \ottnt{v}  \ottsym{:}   \ottsym{\{}  \mathit{x} \,  {:}  \, \iota \,  |  \, \phi  \ottsym{\}}  \,  \,\&\,  \,  \Phi  \, / \,  \ottnt{S} $}
   \end{prooftree}
   for some $\ottnt{C}$.
   The judgment $\Gamma  \vdash  \ottnt{C}  \ottsym{<:}   \ottsym{\{}  \mathit{x} \,  {:}  \, \iota \,  |  \, \phi  \ottsym{\}}  \,  \,\&\,  \,  \Phi  \, / \,  \ottnt{S} $ can be derived only by \Srule{Comp} followed by \Srule{Refine},
   so we can find $ \ottnt{C}  . \ottnt{T}  \,  =  \, \ottsym{\{}  \mathit{x} \,  {:}  \, \iota \,  |  \, \phi''  \ottsym{\}}$ for some $\phi''$ such that
   $\Gamma  \ottsym{,}  \mathit{x} \,  {:}  \, \iota  \models   \phi''    \Rightarrow    \phi $.
   Thus, we have $\Gamma  \vdash  \ottnt{v}  \ottsym{:}   \ottsym{\{}  \mathit{x} \,  {:}  \, \iota \,  |  \, \phi''  \ottsym{\}}  \,  \,\&\,  \,   \ottnt{C}  . \Phi   \, / \,   \ottnt{C}  . \ottnt{S}  $.
   By case analysis on the result of the IH.
   \begin{caseanalysis}
    \case $ {  {   \exists  \, { \mathit{y}  \ottsym{,}  \phi' } . \  \ottnt{v} \,  =  \, \mathit{y}  } \,\mathrel{\wedge}\, { \Gamma  \ottsym{(}  \mathit{y}  \ottsym{)} \,  =  \, \ottsym{\{}  \mathit{x} \,  {:}  \, \iota \,  |  \, \phi'  \ottsym{\}} }  } \,\mathrel{\wedge}\, { \Gamma  \ottsym{,}  \mathit{x} \,  {:}  \, \iota  \models   \ottsym{(}   \mathit{y}    =    \mathit{x}   \ottsym{)}    \Rightarrow    \phi''  } $:
     We have the conclusion by \reflem{ts-transitivity-valid}.

    \case $ {  {   \exists  \, { \ottnt{c} } . \  \ottnt{v} \,  =  \, \ottnt{c}  } \,\mathrel{\wedge}\, {  \mathit{ty}  (  \ottnt{c}  )  \,  =  \, \iota }  } \,\mathrel{\wedge}\, { \Gamma  \models   \phi''    [  \ottnt{c}  /  \mathit{x}  ]   } $:
     It suffices to show that $\Gamma  \models   \phi    [  \ottnt{c}  /  \mathit{x}  ]  $.
     Because $\Gamma  \ottsym{,}  \mathit{x} \,  {:}  \, \iota  \models   \phi''    \Rightarrow    \phi $ and $\Gamma  \vdash  \ottnt{c}  \ottsym{:}  \ottnt{s}$ by \refasm{const},
     we have $\Gamma  \models     \phi''    [  \ottnt{c}  /  \mathit{x}  ]      \Rightarrow    \phi     [  \ottnt{c}  /  \mathit{x}  ]  $ by \reflem{ts-term-subst-valid}.
     Because $\Gamma  \models   \phi''    [  \ottnt{c}  /  \mathit{x}  ]  $,
     \reflem{ts-discharge-assumption-valid} implies the conclusion $\Gamma  \models   \phi    [  \ottnt{c}  /  \mathit{x}  ]  $.
   \end{caseanalysis}

  \case \T{App}:
   We are given $\ottnt{v} \,  =  \, \ottnt{v_{{\mathrm{1}}}} \,  \tceseq{ \ottnt{v_{{\mathrm{2}}}} } $ for some $\ottnt{v_{{\mathrm{1}}}}$ and $ \tceseq{ \ottnt{v_{{\mathrm{2}}}} } $ such that
   \begin{itemize}
    \item $\ottnt{v_{{\mathrm{1}}}}$ is a $\lambda$-abstraction or recursive function of $n$ arguments, and
    \item $\ottsym{\mbox{$\mid$}}   \tceseq{ \ottnt{v_{{\mathrm{2}}}} }   \ottsym{\mbox{$\mid$}} \,  <  \, \ottmv{n}$.
   \end{itemize}
   Then, \reflem{ts-val-applied-fun} implies that
   the refinement type $\ottsym{\{}  \mathit{x} \,  {:}  \, \iota \,  |  \, \phi  \ottsym{\}}$ must be a function type; it is contradictory.

  \otherwise: Contradictory.
 \end{caseanalysis}
\end{proof}

\begin{lemmap}{Value Substitution in Well-Formedness and Formula Typing}{ts-val-subst-wf-ftyping}
 Suppose that $\Gamma_{{\mathrm{1}}}  \vdash  \ottnt{v}  \ottsym{:}  \ottnt{T}$.
 \begin{enumerate}
  \item If $\vdash  \Gamma_{{\mathrm{1}}}  \ottsym{,}  \mathit{x} \,  {:}  \, \ottnt{T}  \ottsym{,}  \Gamma_{{\mathrm{2}}}$, then $\vdash   \Gamma_{{\mathrm{1}}}  \ottsym{,}  \Gamma_{{\mathrm{2}}}    [  \ottnt{v}  /  \mathit{x}  ]  $.
  \item If $\Gamma_{{\mathrm{1}}}  \ottsym{,}  \mathit{x} \,  {:}  \, \ottnt{T}  \ottsym{,}  \Gamma_{{\mathrm{2}}}  \vdash  \ottnt{T_{{\mathrm{0}}}}$, then $ \Gamma_{{\mathrm{1}}}  \ottsym{,}  \Gamma_{{\mathrm{2}}}    [  \ottnt{v}  /  \mathit{x}  ]    \vdash   \ottnt{T_{{\mathrm{0}}}}    [  \ottnt{v}  /  \mathit{x}  ]  $.
  \item If $\Gamma_{{\mathrm{1}}}  \ottsym{,}  \mathit{x} \,  {:}  \, \ottnt{T}  \ottsym{,}  \Gamma_{{\mathrm{2}}}  \vdash  \ottnt{C}$, then $ \Gamma_{{\mathrm{1}}}  \ottsym{,}  \Gamma_{{\mathrm{2}}}    [  \ottnt{v}  /  \mathit{x}  ]    \vdash   \ottnt{C}    [  \ottnt{v}  /  \mathit{x}  ]  $.
  \item If $\Gamma_{{\mathrm{1}}}  \ottsym{,}  \mathit{x} \,  {:}  \, \ottnt{T}  \ottsym{,}  \Gamma_{{\mathrm{2}}} \,  |  \, \ottnt{T_{{\mathrm{0}}}}  \vdash  \ottnt{S}$, then $ \Gamma_{{\mathrm{1}}}  \ottsym{,}  \Gamma_{{\mathrm{2}}}    [  \ottnt{v}  /  \mathit{x}  ]   \,  |  \,  \ottnt{T_{{\mathrm{0}}}}    [  \ottnt{v}  /  \mathit{x}  ]    \vdash   \ottnt{S}    [  \ottnt{v}  /  \mathit{x}  ]  $.
  \item If $\Gamma_{{\mathrm{1}}}  \ottsym{,}  \mathit{x} \,  {:}  \, \ottnt{T}  \ottsym{,}  \Gamma_{{\mathrm{2}}}  \vdash  \Phi$, then $ \Gamma_{{\mathrm{1}}}  \ottsym{,}  \Gamma_{{\mathrm{2}}}    [  \ottnt{v}  /  \mathit{x}  ]    \vdash   \Phi    [  \ottnt{v}  /  \mathit{x}  ]  $.
  \item If $\Gamma_{{\mathrm{1}}}  \ottsym{,}  \mathit{x} \,  {:}  \, \ottnt{T}  \ottsym{,}  \Gamma_{{\mathrm{2}}}  \vdash  \ottnt{t}  \ottsym{:}  \ottnt{s}$, then $ \Gamma_{{\mathrm{1}}}  \ottsym{,}  \Gamma_{{\mathrm{2}}}    [  \ottnt{v}  /  \mathit{x}  ]    \vdash   \ottnt{t}    [  \ottnt{v}  /  \mathit{x}  ]    \ottsym{:}  \ottnt{s}$.
  \item If $\Gamma_{{\mathrm{1}}}  \ottsym{,}  \mathit{x} \,  {:}  \, \ottnt{T}  \ottsym{,}  \Gamma_{{\mathrm{2}}}  \vdash  \ottnt{A}  \ottsym{:}   \tceseq{ \ottnt{s} } $, then $ \Gamma_{{\mathrm{1}}}  \ottsym{,}  \Gamma_{{\mathrm{2}}}    [  \ottnt{v}  /  \mathit{x}  ]    \vdash   \ottnt{A}    [  \ottnt{v}  /  \mathit{x}  ]    \ottsym{:}   \tceseq{ \ottnt{s} } $.
  \item If $\Gamma_{{\mathrm{1}}}  \ottsym{,}  \mathit{x} \,  {:}  \, \ottnt{T}  \ottsym{,}  \Gamma_{{\mathrm{2}}}  \vdash  \phi$, then $ \Gamma_{{\mathrm{1}}}  \ottsym{,}  \Gamma_{{\mathrm{2}}}    [  \ottnt{v}  /  \mathit{x}  ]    \vdash   \phi    [  \ottnt{v}  /  \mathit{x}  ]  $.
 \end{enumerate}
\end{lemmap}
\begin{proof}
 By simultaneous induction on the derivations.
 The interesting cases are only of \Tf{VarT} and \Tf{VarS}.
 \begin{caseanalysis}
  \case \Tf{VarT}:
   We are given
   \begin{prooftree}
    \AxiomC{$\vdash  \Gamma_{{\mathrm{1}}}  \ottsym{,}  \mathit{x} \,  {:}  \, \ottnt{T}  \ottsym{,}  \Gamma_{{\mathrm{2}}}$}
    \AxiomC{$\ottsym{(}  \Gamma_{{\mathrm{1}}}  \ottsym{,}  \mathit{x} \,  {:}  \, \ottnt{T}  \ottsym{,}  \Gamma_{{\mathrm{2}}}  \ottsym{)}  \ottsym{(}  \mathit{y}  \ottsym{)} \,  =  \, \ottsym{\{}  \mathit{z} \,  {:}  \, \iota \,  |  \, \phi  \ottsym{\}}$}
    \BinaryInfC{$\Gamma_{{\mathrm{1}}}  \ottsym{,}  \mathit{x} \,  {:}  \, \ottnt{T}  \ottsym{,}  \Gamma_{{\mathrm{2}}}  \vdash  \mathit{y}  \ottsym{:}  \iota$}
   \end{prooftree}
   for some $\mathit{y}$, $\iota$, $\phi$ such that
   $ \ottnt{t}    =    \mathit{y} $ and $\ottnt{s} \,  =  \, \iota$.
   By the IH, $\vdash   \Gamma_{{\mathrm{1}}}  \ottsym{,}  \Gamma_{{\mathrm{2}}}    [  \ottnt{v}  /  \mathit{x}  ]  $.

   By case analysis on whether $ \mathit{x}    =    \mathit{y} $ or not.
   \begin{caseanalysis}
    \case $ \mathit{x}    =    \mathit{y} $:
     We must show that
     \[
       \Gamma_{{\mathrm{1}}}  \ottsym{,}  \Gamma_{{\mathrm{2}}}    [  \ottnt{v}  /  \mathit{x}  ]    \vdash   \ottnt{v}   \ottsym{:}  \iota ~.
     \]
     Because $\ottsym{(}  \Gamma_{{\mathrm{1}}}  \ottsym{,}  \mathit{x} \,  {:}  \, \ottnt{T}  \ottsym{,}  \Gamma_{{\mathrm{2}}}  \ottsym{)}  \ottsym{(}  \mathit{y}  \ottsym{)} \,  =  \, \ottsym{\{}  \mathit{z} \,  {:}  \, \iota \,  |  \, \phi  \ottsym{\}}$, we find $\ottnt{T} \,  =  \, \ottsym{\{}  \mathit{z} \,  {:}  \, \iota \,  |  \, \phi  \ottsym{\}}$, so
     the assumption $\Gamma_{{\mathrm{1}}}  \vdash  \ottnt{v}  \ottsym{:}  \ottnt{T}$ implies $\Gamma_{{\mathrm{1}}}  \vdash  \ottnt{v}  \ottsym{:}  \ottsym{\{}  \mathit{z} \,  {:}  \, \iota \,  |  \, \phi  \ottsym{\}}$.
     By case analysis on $\ottnt{v}$ by \reflem{ts-open-val-inv-refine-ty}.
     \begin{caseanalysis}
      \case $ {   \exists  \, { \mathit{y'}  \ottsym{,}  \phi' } . \  \ottnt{v} \,  =  \, \mathit{y'}  } \,\mathrel{\wedge}\, { \Gamma_{{\mathrm{1}}}  \ottsym{(}  \mathit{y'}  \ottsym{)} \,  =  \, \ottsym{\{}  \mathit{z} \,  {:}  \, \iota \,  |  \, \phi'  \ottsym{\}} } $:
       We must show that
       \[
         \Gamma_{{\mathrm{1}}}  \ottsym{,}  \Gamma_{{\mathrm{2}}}    [  \ottnt{v}  /  \mathit{x}  ]    \vdash  \mathit{y'}  \ottsym{:}  \iota ~.
       \]
       Because $\vdash   \Gamma_{{\mathrm{1}}}  \ottsym{,}  \Gamma_{{\mathrm{2}}}    [  \ottnt{v}  /  \mathit{x}  ]  $ and $\ottsym{(}   \Gamma_{{\mathrm{1}}}  \ottsym{,}  \Gamma_{{\mathrm{2}}}    [  \ottnt{v}  /  \mathit{x}  ]    \ottsym{)}  \ottsym{(}  \mathit{y'}  \ottsym{)} \,  =  \, \ottsym{\{}  \mathit{z} \,  {:}  \, \iota \,  |  \, \phi'  \ottsym{\}}$,
       \Tf{VarT} implies the conclusion.

      \case $ {   \exists  \, { \ottnt{c} } . \  \ottnt{v} \,  =  \, \ottnt{c}  } \,\mathrel{\wedge}\, {  \mathit{ty}  (  \ottnt{c}  )  \,  =  \, \iota } $:
       We must show that
       \[
         \Gamma_{{\mathrm{1}}}  \ottsym{,}  \Gamma_{{\mathrm{2}}}    [  \ottnt{v}  /  \mathit{x}  ]    \vdash  \ottnt{c}  \ottsym{:}  \iota ~.
       \]
       By \refasm{termfun}, there exists some $\mathit{F}$ that corresponds to $\ottnt{c}$.
       Therefore, \Tf{Fun} with $\vdash   \Gamma_{{\mathrm{1}}}  \ottsym{,}  \Gamma_{{\mathrm{2}}}    [  \ottnt{v}  /  \mathit{x}  ]  $ implies the conclusion.
      \end{caseanalysis}

    \case $ \mathit{x}    \not=    \mathit{y} $:
     We must show that
     \[
       \Gamma_{{\mathrm{1}}}  \ottsym{,}  \Gamma_{{\mathrm{2}}}    [  \ottnt{v}  /  \mathit{x}  ]    \vdash  \mathit{y}  \ottsym{:}  \iota ~.
     \]
     Because $\ottsym{(}  \Gamma_{{\mathrm{1}}}  \ottsym{,}  \mathit{x} \,  {:}  \, \ottnt{T}  \ottsym{,}  \Gamma_{{\mathrm{2}}}  \ottsym{)}  \ottsym{(}  \mathit{y}  \ottsym{)} \,  =  \, \ottsym{\{}  \mathit{z} \,  {:}  \, \iota \,  |  \, \phi  \ottsym{\}}$ and $ \mathit{x}    \not=    \mathit{y} $,
     we have $\ottsym{(}   \Gamma_{{\mathrm{1}}}  \ottsym{,}  \Gamma_{{\mathrm{2}}}    [  \ottnt{v}  /  \mathit{x}  ]    \ottsym{)}  \ottsym{(}  \mathit{y}  \ottsym{)} \,  =  \,  \ottsym{\{}  \mathit{z} \,  {:}  \, \iota \,  |  \, \phi  \ottsym{\}}    [  \ottnt{v}  /  \mathit{x}  ]  $.
     Therefore, \Tf{VarT} implies the conclusion.
   \end{caseanalysis}

  \case \Tf{VarS}:
   We are given
   \begin{prooftree}
    \AxiomC{$\vdash  \Gamma_{{\mathrm{1}}}  \ottsym{,}  \mathit{x} \,  {:}  \, \ottnt{T}  \ottsym{,}  \Gamma_{{\mathrm{2}}}$}
    \AxiomC{$\ottsym{(}  \Gamma_{{\mathrm{1}}}  \ottsym{,}  \mathit{x} \,  {:}  \, \ottnt{T}  \ottsym{,}  \Gamma_{{\mathrm{2}}}  \ottsym{)}  \ottsym{(}  \mathit{y}  \ottsym{)} \,  =  \, \ottnt{s}$}
    \BinaryInfC{$\Gamma_{{\mathrm{1}}}  \ottsym{,}  \mathit{x} \,  {:}  \, \ottnt{T}  \ottsym{,}  \Gamma_{{\mathrm{2}}}  \vdash  \mathit{y}  \ottsym{:}  \ottnt{s}$}
   \end{prooftree}
   for some $\mathit{y}$ such that $ \ottnt{t}    =    \mathit{y} $.
   Because $\ottsym{(}  \Gamma_{{\mathrm{1}}}  \ottsym{,}  \mathit{x} \,  {:}  \, \ottnt{T}  \ottsym{,}  \Gamma_{{\mathrm{2}}}  \ottsym{)}  \ottsym{(}  \mathit{y}  \ottsym{)} \,  =  \, \ottnt{s}$, we have $ \mathit{x}    \not=    \mathit{y} $, so
   $\ottsym{(}   \Gamma_{{\mathrm{1}}}  \ottsym{,}  \Gamma_{{\mathrm{2}}}    [  \ottnt{v}  /  \mathit{x}  ]    \ottsym{)}  \ottsym{(}  \mathit{y}  \ottsym{)} \,  =  \, \ottnt{s}$.
   By the IH, $\vdash   \Gamma_{{\mathrm{1}}}  \ottsym{,}  \Gamma_{{\mathrm{2}}}    [  \ottnt{v}  /  \mathit{x}  ]  $.
   Therefore, \Tf{VarS} implies $ \Gamma_{{\mathrm{1}}}  \ottsym{,}  \Gamma_{{\mathrm{2}}}    [  \ottnt{v}  /  \mathit{x}  ]    \vdash  \mathit{y}  \ottsym{:}  \ottnt{s}$, which we must show.
 \end{caseanalysis}
\end{proof}

\begin{lemmap}{Substitution Preserves Occurrence Checking}{lr-subst-occur}
 Assume that $ \tceseq{ \mathit{X'} }   \ottsym{,}   \tceseq{ \mathit{X} } $ and $ \tceseq{ \mathit{Y'} }   \ottsym{,}   \tceseq{ \mathit{Y} } $ do not occur in $\sigma$.
 Then, $ \tceseq{ \ottsym{(}  \mathit{X'}  \ottsym{,}  \mathit{Y'}  \ottsym{)} }   \ottsym{;}   \tceseq{ \ottsym{(}  \mathit{X}  \ottsym{,}  \mathit{Y}  \ottsym{)} }  \,  \vdash ^{\circ}  \, \ottnt{C}$ implies $ \tceseq{ \ottsym{(}  \mathit{X'}  \ottsym{,}  \mathit{Y'}  \ottsym{)} }   \ottsym{;}   \tceseq{ \ottsym{(}  \mathit{X}  \ottsym{,}  \mathit{Y}  \ottsym{)} }  \,  \vdash ^{\circ}  \, \sigma  \ottsym{(}  \ottnt{C}  \ottsym{)}$.
\end{lemmap}
\begin{proof}
 \TS{CHECKED: 2022/07/07}
 Straightforward by induction on the derivation of $ \tceseq{ \ottsym{(}  \mathit{X'}  \ottsym{,}  \mathit{Y'}  \ottsym{)} }   \ottsym{;}   \tceseq{ \ottsym{(}  \mathit{X}  \ottsym{,}  \mathit{Y}  \ottsym{)} }  \,  \vdash ^{\circ}  \, \ottnt{C}$.
\end{proof}

\begin{lemmap}{Substitution Preserves Consistency}{lr-subst-consist}
 Assume that $ \tceseq{ \mathit{X} } $ and $ \tceseq{ \mathit{Y} } $ do not occur in $\sigma$.
 Then, $ \tceseq{ \ottsym{(}  \mathit{X}  \ottsym{,}  \mathit{Y}  \ottsym{)} }   \ottsym{;}   \tceseq{ \mathit{z_{{\mathrm{0}}}}  \,  {:}  \,  \iota_{{\mathrm{0}}} }   \vdash  \ottnt{C_{{\mathrm{0}}}}  \succ  \ottnt{C}$ implies $ \tceseq{ \ottsym{(}  \mathit{X}  \ottsym{,}  \mathit{Y}  \ottsym{)} }   \ottsym{;}   \tceseq{ \mathit{z_{{\mathrm{0}}}}  \,  {:}  \,  \iota_{{\mathrm{0}}} }   \vdash  \sigma  \ottsym{(}  \ottnt{C_{{\mathrm{0}}}}  \ottsym{)}  \succ  \sigma  \ottsym{(}  \ottnt{C}  \ottsym{)}$.
\end{lemmap}
\begin{proof}
 \TS{CHECKED: 2022/07/07}
 Straightforward by induction on the derivation of $ \tceseq{ \ottsym{(}  \mathit{X}  \ottsym{,}  \mathit{Y}  \ottsym{)} }   \ottsym{;}   \tceseq{ \mathit{z_{{\mathrm{0}}}}  \,  {:}  \,  \iota_{{\mathrm{0}}} }   \vdash  \ottnt{C_{{\mathrm{0}}}}  \succ  \ottnt{C}$.
\end{proof}

\begin{lemmap}{Substitution of Effect Substitution}{lr-subst-eff-subst}
 Assume that $ \tceseq{ \mathit{X} } $ and $ \tceseq{ \mathit{Y} } $ do not occur in $\sigma$.
 Then, $ \tceseq{ \ottsym{(}  \mathit{X}  \ottsym{,}  \mathit{Y}  \ottsym{)} }   \ottsym{;}   \tceseq{ \mathit{z_{{\mathrm{0}}}}  \,  {:}  \,  \iota_{{\mathrm{0}}} }  \,  |  \, \ottnt{C}  \vdash  \sigma'$ implies $ \tceseq{ \ottsym{(}  \mathit{X}  \ottsym{,}  \mathit{Y}  \ottsym{)} }   \ottsym{;}   \tceseq{ \mathit{z_{{\mathrm{0}}}}  \,  {:}  \,  \iota_{{\mathrm{0}}} }  \,  |  \, \sigma  \ottsym{(}  \ottnt{C}  \ottsym{)}  \vdash  \sigma  \ottsym{(}  \sigma'  \ottsym{)}$.
\end{lemmap}
\begin{proof}
 \TS{CHECKED: 2022/07/07}
 Straightforward by induction on the derivation of $ \tceseq{ \ottsym{(}  \mathit{X}  \ottsym{,}  \mathit{Y}  \ottsym{)} }   \ottsym{;}   \tceseq{ \mathit{z_{{\mathrm{0}}}}  \,  {:}  \,  \iota_{{\mathrm{0}}} }  \,  |  \, \ottnt{C}  \vdash  \sigma'$.
\end{proof}

\begin{lemmap}{Substitution is Commutative with $ \nu $-Effect Extraction}{ts-subst-nu-cty}
 $  \forall  \, { \ottnt{C} } . \   { \sigma  \ottsym{(}  \ottnt{C}  \ottsym{)} }^\nu (  \sigma  \ottsym{(}  \ottnt{t}  \ottsym{)}  )  \,  =  \, \sigma  \ottsym{(}   { \ottnt{C} }^\nu (  \ottnt{t}  )   \ottsym{)} $.
\end{lemmap}
\begin{proof}
 By induction on $\ottnt{C}$.
 Suppose that $\ottnt{C} \,  =  \,  \ottnt{T}  \,  \,\&\,  \,  \Phi  \, / \,  \ottnt{S} $ for some $\ottnt{T}$, $\Phi$, and $\ottnt{S}$.
 By definition,
 $ { \sigma  \ottsym{(}  \ottnt{C}  \ottsym{)} }^\nu (  \sigma  \ottsym{(}  \ottnt{t}  \ottsym{)}  )  \,  =  \,   { \sigma  \ottsym{(}  \Phi  \ottsym{)} }^\nu   \ottsym{(}  \sigma  \ottsym{(}  \ottnt{t}  \ottsym{)}  \ottsym{)}    \wedge     { \sigma  \ottsym{(}  \ottnt{S}  \ottsym{)} }^\nu (  \sigma  \ottsym{(}  \ottnt{t}  \ottsym{)}  )  $.
 Because $\sigma  \ottsym{(}   { \ottnt{C} }^\nu (  \ottnt{t}  )   \ottsym{)} = \sigma  \ottsym{(}    { \Phi }^\nu   \ottsym{(}  \ottnt{t}  \ottsym{)}    \wedge     { \ottnt{S} }^\nu (  \ottnt{t}  )    \ottsym{)} =   { \sigma  \ottsym{(}  \Phi  \ottsym{)} }^\nu   \ottsym{(}  \sigma  \ottsym{(}  \ottnt{t}  \ottsym{)}  \ottsym{)}    \wedge    \sigma  \ottsym{(}   { \ottnt{S} }^\nu (  \ottnt{t}  )   \ottsym{)} $,
 it suffices to show that $ { \sigma  \ottsym{(}  \ottnt{S}  \ottsym{)} }^\nu (  \sigma  \ottsym{(}  \ottnt{t}  \ottsym{)}  )  \,  =  \, \sigma  \ottsym{(}   { \ottnt{S} }^\nu (  \ottnt{t}  )   \ottsym{)}$.
 We proceed by case analysis on $\ottnt{S}$.
 \begin{caseanalysis}
  \case $\ottnt{S} \,  =  \,  \square $: By definition,
   $ { \sigma  \ottsym{(}  \ottnt{S}  \ottsym{)} }^\nu (  \sigma  \ottsym{(}  \ottnt{t}  \ottsym{)}  )  =  \top  = \sigma  \ottsym{(}   { \ottnt{S} }^\nu (  \ottnt{t}  )   \ottsym{)}$.

  \case $  \exists  \, { \mathit{x}  \ottsym{,}  \ottnt{C_{{\mathrm{1}}}}  \ottsym{,}  \ottnt{C_{{\mathrm{2}}}} } . \  \ottnt{S} \,  =  \,  ( \forall  \mathit{x}  .  \ottnt{C_{{\mathrm{1}}}}  )  \Rightarrow   \ottnt{C_{{\mathrm{2}}}}  $:
   We have the conclusion by:
   \[\begin{array}{llll}
     { \sigma  \ottsym{(}  \ottnt{S}  \ottsym{)} }^\nu (  \sigma  \ottsym{(}  \ottnt{t}  \ottsym{)}  )  &=&
      { \sigma  \ottsym{(}  \ottnt{C_{{\mathrm{2}}}}  \ottsym{)} }^\nu (  \sigma  \ottsym{(}  \ottnt{t}  \ottsym{)}  )  & \text{(by definition)} \\ &&
     \sigma  \ottsym{(}   { \ottnt{C_{{\mathrm{2}}}} }^\nu (  \ottnt{t}  )   \ottsym{)} & \text{(by the IH)} \\ &&
     \sigma  \ottsym{(}   { \ottnt{S} }^\nu (  \ottnt{t}  )   \ottsym{)} & \text{(by definition)} ~.
     \end{array}
   \]
 \end{caseanalysis}
\end{proof}

\begin{lemmap}{Value Substitution in Validity}{ts-val-subst-valid}
 Suppose that $\Gamma_{{\mathrm{1}}}  \vdash  \ottnt{v}  \ottsym{:}  \ottnt{T}$.
 If $\Gamma_{{\mathrm{1}}}  \ottsym{,}  \mathit{x} \,  {:}  \, \ottnt{T}  \ottsym{,}  \Gamma_{{\mathrm{2}}}  \models  \phi$, then $ \Gamma_{{\mathrm{1}}}  \ottsym{,}  \Gamma_{{\mathrm{2}}}    [  \ottnt{v}  /  \mathit{x}  ]    \models   \phi    [  \ottnt{v}  /  \mathit{x}  ]  $.
\end{lemmap}
\begin{proof}
 \TS{Validity}
 If $\ottnt{T}$ is not a refinement type,
 then $\Gamma_{{\mathrm{1}}}  \ottsym{,}  \mathit{x} \,  {:}  \, \ottnt{T}  \ottsym{,}  \Gamma_{{\mathrm{2}}}  \models  \phi$ is equivalent to $ \Gamma_{{\mathrm{1}}}  \ottsym{,}  \Gamma_{{\mathrm{2}}}    [  \ottnt{v}  /  \mathit{x}  ]    \models   \phi    [  \ottnt{v}  /  \mathit{x}  ]  $
 because $\mathit{x}$ is not quantified in the formula $ \lfloor  \Gamma_{{\mathrm{1}}}  \ottsym{,}  \mathit{x} \,  {:}  \, \ottnt{T}  \ottsym{,}  \Gamma_{{\mathrm{2}}}  \, \vdash \,  \phi  \rfloor $.

 In what follows, we suppose that
 $\ottnt{T} \,  =  \, \ottsym{\{}  \mathit{x} \,  {:}  \, \iota \,  |  \, \phi  \ottsym{\}}$ for some $\iota$, and $\phi$.
 By \reflem{ts-logic-closing-dsubs},
 it suffices to show that
 \[
     \forall  \, { \theta } . \   \Gamma_{{\mathrm{1}}}  \ottsym{,}  \Gamma_{{\mathrm{2}}}    [  \ottnt{v}  /  \mathit{x}  ]    \vdash  \theta   \mathrel{\Longrightarrow}   { [\![   \phi    [  \ottnt{v}  /  \mathit{x}  ]    ]\!]_{ \theta } }  \,  =  \,  \top   ~.
 \]
 Let $\theta$ be any substitution such that $ \Gamma_{{\mathrm{1}}}  \ottsym{,}  \Gamma_{{\mathrm{2}}}    [  \ottnt{v}  /  \mathit{x}  ]    \vdash  \theta$.
 We show that
 \[
   { [\![   \phi    [  \ottnt{v}  /  \mathit{x}  ]    ]\!]_{ \theta } }  \,  =  \,  \top  ~.
 \]
 By case analysis on the result of applying \reflem{ts-open-val-inv-refine-ty}
 with $\Gamma_{{\mathrm{1}}}  \vdash  \ottnt{v}  \ottsym{:}  \ottsym{\{}  \mathit{x} \,  {:}  \, \iota \,  |  \, \phi  \ottsym{\}}$.
 \begin{caseanalysis}
  \case $ {  {   \exists  \, { \mathit{y}  \ottsym{,}  \phi' } . \  \ottnt{v} \,  =  \, \mathit{y}  } \,\mathrel{\wedge}\, { \Gamma_{{\mathrm{1}}}  \ottsym{(}  \mathit{y}  \ottsym{)} \,  =  \, \ottsym{\{}  \mathit{x} \,  {:}  \, \iota \,  |  \, \phi'  \ottsym{\}} }  } \,\mathrel{\wedge}\, { \Gamma_{{\mathrm{1}}}  \ottsym{,}  \mathit{x} \,  {:}  \, \iota  \models   \ottsym{(}   \mathit{y}    =    \mathit{x}   \ottsym{)}    \Rightarrow    \phi  } $:
   Because $\Gamma_{{\mathrm{1}}}  \ottsym{(}  \mathit{y}  \ottsym{)} \,  =  \, \ottsym{\{}  \mathit{x} \,  {:}  \, \iota \,  |  \, \phi'  \ottsym{\}}$ and $ \Gamma_{{\mathrm{1}}}  \ottsym{,}  \Gamma_{{\mathrm{2}}}    [  \ottnt{v}  /  \mathit{x}  ]    \vdash  \theta$,
   we have $\theta  \ottsym{(}  \mathit{y}  \ottsym{)} \,  =  \, \ottnt{c}$ for some $\ottnt{c}$ such that
   $ \mathit{ty}  (  \ottnt{c}  )  \,  =  \, \iota$ and $ { [\![  \phi'  ]\!]_{ \theta } }  \,  =  \,  \top $.
   Because $ \Gamma_{{\mathrm{1}}}  \ottsym{,}  \Gamma_{{\mathrm{2}}}    [  \ottnt{v}  /  \mathit{x}  ]    \vdash  \theta$,
   we have $\theta \,  =  \, \theta_{{\mathrm{1}}}  \mathbin{\uplus}  \theta_{{\mathrm{2}}}$ for some $\theta_{{\mathrm{1}}}$ and $\theta_{{\mathrm{2}}}$
   such that $\Gamma_{{\mathrm{1}}}  \vdash  \theta_{{\mathrm{1}}}$.
   By \reflem{ts-logic-closing-dsubs} with
   $\Gamma_{{\mathrm{1}}}  \ottsym{,}  \mathit{x} \,  {:}  \, \iota  \models   \ottsym{(}   \mathit{y}    =    \mathit{x}   \ottsym{)}    \Rightarrow    \phi $ and
   $\Gamma_{{\mathrm{1}}}  \ottsym{,}  \mathit{x} \,  {:}  \, \iota  \vdash  \theta_{{\mathrm{1}}}  \mathbin{\uplus}   \{  \mathit{x}  \mapsto  \ottnt{c}  \} $,
   we have $ { [\![   \ottsym{(}   \mathit{y}    =    \mathit{x}   \ottsym{)}    \Rightarrow    \phi   ]\!]_{ \theta_{{\mathrm{1}}}  \mathbin{\uplus}   \{  \mathit{x}  \mapsto  \ottnt{c}  \}  } }  \,  =  \,  \top $.
   By \reflem{ts-logic-val-subst},
   $ { [\![     \ottsym{(}   \ottnt{c}    =    \ottnt{c}   \ottsym{)}    \Rightarrow    \phi     [  \ottnt{c}  /  \mathit{y}  ]      [  \ottnt{c}  /  \mathit{x}  ]    ]\!]_{ \theta'_{{\mathrm{1}}} } }  \,  =  \,  \top $
   where $\theta'_{{\mathrm{1}}}$ is obtained by removing $ \{  \mathit{y}  \mapsto  \ottnt{c}  \} $ from $\theta_{{\mathrm{1}}}$.
   It implies
   $ { [\![    \phi    [  \ottnt{c}  /  \mathit{y}  ]      [  \ottnt{c}  /  \mathit{x}  ]    ]\!]_{ \theta'_{{\mathrm{1}}} } }  \,  =  \,  \top $.
   Because $  \phi    [  \ottnt{c}  /  \mathit{y}  ]      [  \ottnt{c}  /  \mathit{x}  ]   \,  =  \,   \phi    [  \mathit{y}  /  \mathit{x}  ]      [  \ottnt{c}  /  \mathit{y}  ]  $,
   we have $ { [\![   \phi    [  \mathit{y}  /  \mathit{x}  ]    ]\!]_{ \theta_{{\mathrm{1}}} } }  \,  =  \,  \top $
   by \reflem{ts-logic-val-subst}.
   Because $\Gamma_{{\mathrm{1}}}  \vdash  \ottnt{v}  \ottsym{:}  \ottsym{\{}  \mathit{x} \,  {:}  \, \iota \,  |  \, \phi  \ottsym{\}}$,
   variables and predicate variables occurring free in $\phi$
   are only $ \mathit{dom}  (  \Gamma_{{\mathrm{1}}}  )  = \mathit{dom} \, \ottsym{(}  \theta_{{\mathrm{1}}}  \ottsym{)}$ and $\mathit{x}$.
   Therefore, as $\mathit{y} \,  =  \, \ottnt{v}$ and $\theta \,  =  \, \theta_{{\mathrm{1}}}  \mathbin{\uplus}  \theta_{{\mathrm{2}}}$,
   we have the conclusion.

  \case $ {  {   \exists  \, { \ottnt{c} } . \  \ottnt{v} \,  =  \, \ottnt{c}  } \,\mathrel{\wedge}\, {  \mathit{ty}  (  \ottnt{c}  )  \,  =  \, \iota }  } \,\mathrel{\wedge}\, { \Gamma_{{\mathrm{1}}}  \models   \phi    [  \ottnt{c}  /  \mathit{x}  ]   } $:
   Because $ \Gamma_{{\mathrm{1}}}  \ottsym{,}  \Gamma_{{\mathrm{2}}}    [  \ottnt{v}  /  \mathit{x}  ]    \vdash  \theta$,
   we have $\theta \,  =  \, \theta_{{\mathrm{1}}}  \mathbin{\uplus}  \theta_{{\mathrm{2}}}$ for some $\theta_{{\mathrm{1}}}$ and $\theta_{{\mathrm{2}}}$
   such that $\Gamma_{{\mathrm{1}}}  \vdash  \theta_{{\mathrm{1}}}$.
   By \reflem{ts-logic-closing-dsubs} with $\Gamma_{{\mathrm{1}}}  \models   \phi    [  \ottnt{c}  /  \mathit{x}  ]  $ and $\Gamma_{{\mathrm{1}}}  \vdash  \theta_{{\mathrm{1}}}$,
   we have $ { [\![   \phi    [  \ottnt{c}  /  \mathit{x}  ]    ]\!]_{ \theta_{{\mathrm{1}}} } }  \,  =  \,  \top $.
   Because $\Gamma_{{\mathrm{1}}}  \vdash  \ottnt{v}  \ottsym{:}  \ottsym{\{}  \mathit{x} \,  {:}  \, \iota \,  |  \, \phi  \ottsym{\}}$ implies that
   variables and predicate variables occurring free in $\phi$
   are only $ \mathit{dom}  (  \Gamma_{{\mathrm{1}}}  )  = \mathit{dom} \, \ottsym{(}  \theta_{{\mathrm{1}}}  \ottsym{)}$ and $\mathit{x}$,
   we have the conclusion $ { [\![   \phi    [  \ottnt{c}  /  \mathit{x}  ]    ]\!]_{ \theta } }  \,  =  \,  \top $.
 \end{caseanalysis}
\end{proof}

\begin{lemmap}{Value Substitution in Type Composition}{ts-val-subst-ty-compose}
 \begin{enumerate}
  \item $   \forall  \, { \ottnt{S_{{\mathrm{1}}}}  \ottsym{,}  \ottnt{S_{{\mathrm{2}}}}  \ottsym{,}  \ottnt{v}  \ottsym{,}  \mathit{x}  \ottsym{,}  \mathit{y} } . \  \mathit{y} \,  \not\in  \,  \mathit{fv}  (  \ottnt{v}  )  \,  \mathbin{\cup}  \,  \{  \mathit{x}  \}    \mathrel{\Longrightarrow}  \ottsym{(}   \ottnt{S_{{\mathrm{1}}}}    [  \ottnt{v}  /  \mathit{x}  ]    \gg\!\!=  \ottsym{(}   \lambda\!  \, \mathit{y}  \ottsym{.}   \ottnt{S_{{\mathrm{2}}}}    [  \ottnt{v}  /  \mathit{x}  ]    \ottsym{)}  \ottsym{)} \,  =  \,  \ottsym{(}  \ottnt{S_{{\mathrm{1}}}}  \gg\!\!=  \ottsym{(}   \lambda\!  \, \mathit{y}  \ottsym{.}  \ottnt{S_{{\mathrm{2}}}}  \ottsym{)}  \ottsym{)}    [  \ottnt{v}  /  \mathit{x}  ]   $.
  \item $  \forall  \, { \Phi_{{\mathrm{1}}}  \ottsym{,}  \Phi_{{\mathrm{2}}}  \ottsym{,}  \ottnt{v}  \ottsym{,}  \mathit{x} } . \    \Phi_{{\mathrm{1}}}    [  \ottnt{v}  /  \mathit{x}  ]   \,  \tceappendop  \, \Phi_{{\mathrm{2}}}    [  \ottnt{v}  /  \mathit{x}  ]   \,  =  \,  \ottsym{(}  \Phi_{{\mathrm{1}}} \,  \tceappendop  \, \Phi_{{\mathrm{2}}}  \ottsym{)}    [  \ottnt{v}  /  \mathit{x}  ]   $.
  \item $  \forall  \, { \Phi  \ottsym{,}  \ottnt{C}  \ottsym{,}  \ottnt{v}  \ottsym{,}  \mathit{x} } . \    \Phi    [  \ottnt{v}  /  \mathit{x}  ]   \,  \tceappendop  \, \ottnt{C}    [  \ottnt{v}  /  \mathit{x}  ]   \,  =  \,  \ottsym{(}  \Phi \,  \tceappendop  \, \ottnt{C}  \ottsym{)}    [  \ottnt{v}  /  \mathit{x}  ]   $.
  \item $  \forall  \, { \Phi  \ottsym{,}  \ottnt{S}  \ottsym{,}  \ottnt{v}  \ottsym{,}  \mathit{x} } . \    \Phi    [  \ottnt{v}  /  \mathit{x}  ]   \,  \tceappendop  \, \ottnt{S}    [  \ottnt{v}  /  \mathit{x}  ]   \,  =  \,  \ottsym{(}  \Phi \,  \tceappendop  \, \ottnt{S}  \ottsym{)}    [  \ottnt{v}  /  \mathit{x}  ]   $.
 \end{enumerate}
\end{lemmap}
\begin{proof}
 The first and second cases are by definition.
 The third and fourth cases are straightforward by simultaneous induction on $\ottnt{C}$ and $\ottnt{S}$ with the second case.
\end{proof}

\begin{lemmap}{Value Substitution in Subtyping}{ts-val-subst-subtyping}
 Suppose that $\Gamma_{{\mathrm{1}}}  \vdash  \ottnt{v}  \ottsym{:}  \ottnt{T}$.
 \begin{enumerate}
  \item If $\Gamma_{{\mathrm{1}}}  \ottsym{,}  \mathit{x} \,  {:}  \, \ottnt{T}  \ottsym{,}  \Gamma_{{\mathrm{2}}}  \vdash  \ottnt{T_{{\mathrm{1}}}}  \ottsym{<:}  \ottnt{T_{{\mathrm{2}}}}$, then $ \Gamma_{{\mathrm{1}}}  \ottsym{,}  \Gamma_{{\mathrm{2}}}    [  \ottnt{v}  /  \mathit{x}  ]    \vdash   \ottnt{T_{{\mathrm{1}}}}    [  \ottnt{v}  /  \mathit{x}  ]    \ottsym{<:}   \ottnt{T_{{\mathrm{2}}}}    [  \ottnt{v}  /  \mathit{x}  ]  $.
  \item If $\Gamma_{{\mathrm{1}}}  \ottsym{,}  \mathit{x} \,  {:}  \, \ottnt{T}  \ottsym{,}  \Gamma_{{\mathrm{2}}}  \vdash  \ottnt{C_{{\mathrm{1}}}}  \ottsym{<:}  \ottnt{C_{{\mathrm{2}}}}$, then $ \Gamma_{{\mathrm{1}}}  \ottsym{,}  \Gamma_{{\mathrm{2}}}    [  \ottnt{v}  /  \mathit{x}  ]    \vdash   \ottnt{C_{{\mathrm{1}}}}    [  \ottnt{v}  /  \mathit{x}  ]    \ottsym{<:}   \ottnt{C_{{\mathrm{2}}}}    [  \ottnt{v}  /  \mathit{x}  ]  $.
  \item If $\Gamma_{{\mathrm{1}}}  \ottsym{,}  \mathit{x} \,  {:}  \, \ottnt{T}  \ottsym{,}  \Gamma_{{\mathrm{2}}} \,  |  \, \ottnt{T_{{\mathrm{0}}}} \,  \,\&\,  \, \Phi_{{\mathrm{0}}}  \vdash  \ottnt{S_{{\mathrm{1}}}}  \ottsym{<:}  \ottnt{S_{{\mathrm{2}}}}$, then $ \Gamma_{{\mathrm{1}}}  \ottsym{,}  \Gamma_{{\mathrm{2}}}    [  \ottnt{v}  /  \mathit{x}  ]   \,  |  \,  \ottnt{T_{{\mathrm{0}}}}    [  \ottnt{v}  /  \mathit{x}  ]   \,  \,\&\,  \,  \Phi_{{\mathrm{0}}}    [  \ottnt{v}  /  \mathit{x}  ]    \vdash   \ottnt{S_{{\mathrm{1}}}}    [  \ottnt{v}  /  \mathit{x}  ]    \ottsym{<:}   \ottnt{S_{{\mathrm{2}}}}    [  \ottnt{v}  /  \mathit{x}  ]  $.
  \item If $\Gamma_{{\mathrm{1}}}  \ottsym{,}  \mathit{x} \,  {:}  \, \ottnt{T}  \ottsym{,}  \Gamma_{{\mathrm{2}}}  \vdash  \Phi_{{\mathrm{1}}}  \ottsym{<:}  \Phi_{{\mathrm{2}}}$, then $ \Gamma_{{\mathrm{1}}}  \ottsym{,}  \Gamma_{{\mathrm{2}}}    [  \ottnt{v}  /  \mathit{x}  ]    \vdash   \Phi_{{\mathrm{1}}}    [  \ottnt{v}  /  \mathit{x}  ]    \ottsym{<:}   \Phi_{{\mathrm{2}}}    [  \ottnt{v}  /  \mathit{x}  ]  $.
 \end{enumerate}
\end{lemmap}
\begin{proof}
 Straightforward by simultaneous induction on the derivations.
 The cases for \Srule{Refine}, \Srule{Eff}, and \Srule{Embed} use \reflem{ts-val-subst-valid}, and
 the case for \Srule{Embed} also uses Lemmas~\ref{lem:ts-subst-nu-cty} and \ref{lem:ts-val-subst-ty-compose}.
\end{proof}

\begin{lemmap}{Value Substitution in Typing}{ts-val-subst-typing}
 Suppose that $\Gamma_{{\mathrm{1}}}  \vdash  \ottnt{v}  \ottsym{:}  \ottnt{T}$.
 If $\Gamma_{{\mathrm{1}}}  \ottsym{,}  \mathit{x} \,  {:}  \, \ottnt{T}  \ottsym{,}  \Gamma_{{\mathrm{2}}}  \vdash  \ottnt{e}  \ottsym{:}  \ottnt{C}$, then $ \Gamma_{{\mathrm{1}}}  \ottsym{,}  \Gamma_{{\mathrm{2}}}    [  \ottnt{v}  /  \mathit{x}  ]    \vdash   \ottnt{e}    [  \ottnt{v}  /  \mathit{x}  ]    \ottsym{:}   \ottnt{C}    [  \ottnt{v}  /  \mathit{x}  ]  $.
 Especially, if $\Gamma_{{\mathrm{1}}}  \ottsym{,}  \mathit{x} \,  {:}  \, \ottnt{T}  \ottsym{,}  \Gamma_{{\mathrm{2}}}  \vdash  \ottnt{e}  \ottsym{:}  \ottnt{T_{{\mathrm{0}}}}$, then $ \Gamma_{{\mathrm{1}}}  \ottsym{,}  \Gamma_{{\mathrm{2}}}    [  \ottnt{v}  /  \mathit{x}  ]    \vdash   \ottnt{e}    [  \ottnt{v}  /  \mathit{x}  ]    \ottsym{:}   \ottnt{T_{{\mathrm{0}}}}    [  \ottnt{v}  /  \mathit{x}  ]  $.
\end{lemmap}
\begin{proof}
 By induction on the typing derivation with \reflem{ts-val-subst-wf-ftyping}.
 We mention only the interesting cases.
 \begin{caseanalysis}
  \case \T{CVar}:
   We are given $\ottnt{e} \,  =  \, \mathit{y}$ and $\ottnt{C} \,  =  \,  \ottsym{\{}  \mathit{z} \,  {:}  \, \iota \,  |  \,  \mathit{y}    =    \mathit{z}   \ottsym{\}}  \,  \,\&\,  \,   \Phi_\mathsf{val}   \, / \,   \square  $
   for some $\mathit{y}$, $\mathit{z}$, and $\iota$
   (i.e., $\Gamma_{{\mathrm{1}}}  \ottsym{,}  \mathit{x} \,  {:}  \, \ottnt{T}  \ottsym{,}  \Gamma_{{\mathrm{2}}}  \vdash  \mathit{y}  \ottsym{:}  \ottsym{\{}  \mathit{z} \,  {:}  \, \iota \,  |  \,  \mathit{y}    =    \mathit{z}   \ottsym{\}}$).
   Without loss of generality, we can suppose that $\mathit{z} \,  \not\in  \,  \mathit{fv}  (  \ottnt{v}  ) $.
   By inversion,
   $\vdash  \Gamma_{{\mathrm{1}}}  \ottsym{,}  \mathit{x} \,  {:}  \, \ottnt{T}  \ottsym{,}  \Gamma_{{\mathrm{2}}}$ and $\ottsym{(}  \Gamma_{{\mathrm{1}}}  \ottsym{,}  \mathit{x} \,  {:}  \, \ottnt{T}  \ottsym{,}  \Gamma_{{\mathrm{2}}}  \ottsym{)}  \ottsym{(}  \mathit{y}  \ottsym{)} \,  =  \, \ottsym{\{}  \mathit{z} \,  {:}  \, \iota \,  |  \, \phi  \ottsym{\}}$ for some $\phi$.
   Because $\vdash  \Gamma_{{\mathrm{1}}}  \ottsym{,}  \mathit{x} \,  {:}  \, \ottnt{T}  \ottsym{,}  \Gamma_{{\mathrm{2}}}$ and $\Gamma_{{\mathrm{1}}}  \vdash  \ottnt{v}  \ottsym{:}  \ottnt{T}$,
   \reflem{ts-val-subst-wf-ftyping} implies $ \Gamma_{{\mathrm{1}}}  \ottsym{,}  \Gamma_{{\mathrm{2}}}    [  \ottnt{v}  /  \mathit{x}  ]  $.
   By case analysis on whether $ \mathit{x}    =    \mathit{y} $ or not.
   \begin{caseanalysis}
    \case $ \mathit{x}    =    \mathit{y} $:
     We must show that
     \[
       \Gamma_{{\mathrm{1}}}  \ottsym{,}  \Gamma_{{\mathrm{2}}}    [  \ottnt{v}  /  \mathit{x}  ]    \vdash  \ottnt{v}  \ottsym{:}  \ottsym{\{}  \mathit{z} \,  {:}  \, \iota \,  |  \,   \ottnt{v}     =    \mathit{z}   \ottsym{\}} ~.
     \]
     Because $\ottsym{(}  \Gamma_{{\mathrm{1}}}  \ottsym{,}  \mathit{x} \,  {:}  \, \ottnt{T}  \ottsym{,}  \Gamma_{{\mathrm{2}}}  \ottsym{)}  \ottsym{(}  \mathit{y}  \ottsym{)} \,  =  \, \ottsym{\{}  \mathit{z} \,  {:}  \, \iota \,  |  \, \phi  \ottsym{\}}$ and $ \mathit{x}    =    \mathit{y} $,
     we find $\ottnt{T} \,  =  \, \ottsym{\{}  \mathit{z} \,  {:}  \, \iota \,  |  \, \phi  \ottsym{\}}$, so
     the assumption $\Gamma_{{\mathrm{1}}}  \vdash  \ottnt{v}  \ottsym{:}  \ottnt{T}$ implies $\Gamma_{{\mathrm{1}}}  \vdash  \ottnt{v}  \ottsym{:}  \ottsym{\{}  \mathit{z} \,  {:}  \, \iota \,  |  \, \phi  \ottsym{\}}$.
     By case analysis on $\ottnt{v}$ by \reflem{ts-open-val-inv-refine-ty}.
     \begin{caseanalysis}
      \case $ {   \exists  \, { \mathit{y'}  \ottsym{,}  \phi' } . \  \ottnt{v} \,  =  \, \mathit{y'}  } \,\mathrel{\wedge}\, { \Gamma_{{\mathrm{1}}}  \ottsym{(}  \mathit{y'}  \ottsym{)} \,  =  \, \ottsym{\{}  \mathit{z} \,  {:}  \, \iota \,  |  \, \phi'  \ottsym{\}} } $:
       We must show that
       \[
         \Gamma_{{\mathrm{1}}}  \ottsym{,}  \Gamma_{{\mathrm{2}}}    [  \ottnt{v}  /  \mathit{x}  ]    \vdash  \mathit{y'}  \ottsym{:}  \ottsym{\{}  \mathit{z} \,  {:}  \, \iota \,  |  \,  \mathit{y'}    =    \mathit{z}   \ottsym{\}} ~.
       \]
       By \T{CVar} with $\vdash   \Gamma_{{\mathrm{1}}}  \ottsym{,}  \Gamma_{{\mathrm{2}}}    [  \ottnt{v}  /  \mathit{x}  ]  $ and $\ottsym{(}   \Gamma_{{\mathrm{1}}}  \ottsym{,}  \Gamma_{{\mathrm{2}}}    [  \ottnt{v}  /  \mathit{x}  ]    \ottsym{)}  \ottsym{(}  \mathit{y'}  \ottsym{)} \,  =  \, \ottsym{\{}  \mathit{z} \,  {:}  \, \iota \,  |  \, \phi'  \ottsym{\}}$,
       we have the conclusion.

      \case $ {   \exists  \, { \ottnt{c} } . \  \ottnt{v} \,  =  \, \ottnt{c}  } \,\mathrel{\wedge}\, {  \mathit{ty}  (  \ottnt{c}  )  \,  =  \, \iota } $:
       We must show that
       \[
         \Gamma_{{\mathrm{1}}}  \ottsym{,}  \Gamma_{{\mathrm{2}}}    [  \ottnt{v}  /  \mathit{x}  ]    \vdash  \ottnt{c}  \ottsym{:}  \ottsym{\{}  \mathit{z} \,  {:}  \, \iota \,  |  \,  \ottnt{c}    =    \mathit{z}   \ottsym{\}} ~.
       \]
       By \T{Const} with $\vdash   \Gamma_{{\mathrm{1}}}  \ottsym{,}  \Gamma_{{\mathrm{2}}}    [  \ottnt{v}  /  \mathit{x}  ]  $,
       we have the conclusion.
     \end{caseanalysis}

    \case $ \mathit{x}    \not=    \mathit{y} $:
     We must show that
     \[
       \Gamma_{{\mathrm{1}}}  \ottsym{,}  \Gamma_{{\mathrm{2}}}    [  \ottnt{v}  /  \mathit{x}  ]    \vdash  \mathit{y}  \ottsym{:}  \ottsym{\{}  \mathit{z} \,  {:}  \, \iota \,  |  \,  \mathit{y}    =    \mathit{z}   \ottsym{\}} ~.
     \]
     By \T{CVar} with $\vdash   \Gamma_{{\mathrm{1}}}  \ottsym{,}  \Gamma_{{\mathrm{2}}}    [  \ottnt{v}  /  \mathit{x}  ]  $ and $\ottsym{(}   \Gamma_{{\mathrm{1}}}  \ottsym{,}  \Gamma_{{\mathrm{2}}}    [  \ottnt{v}  /  \mathit{x}  ]    \ottsym{)}  \ottsym{(}  \mathit{y}  \ottsym{)} \,  =  \, \ottsym{\{}  \mathit{z} \,  {:}  \, \iota \,  |  \,  \phi    [  \ottnt{v}  /  \mathit{x}  ]    \ottsym{\}}$,
     we have the conclusion.
   \end{caseanalysis}

  \case \T{Var}:
   We are given $\ottnt{e} \,  =  \, \mathit{y}$ and $\ottnt{C} \,  =  \,  \ottsym{(}  \Gamma_{{\mathrm{1}}}  \ottsym{,}  \mathit{x} \,  {:}  \, \ottnt{T}  \ottsym{,}  \Gamma_{{\mathrm{2}}}  \ottsym{)}  \ottsym{(}  \mathit{y}  \ottsym{)}  \,  \,\&\,  \,   \Phi_\mathsf{val}   \, / \,   \square  $ for some $\mathit{y}$
   (i.e., $\Gamma_{{\mathrm{1}}}  \ottsym{,}  \mathit{x} \,  {:}  \, \ottnt{T}  \ottsym{,}  \Gamma_{{\mathrm{2}}}  \vdash  \mathit{y}  \ottsym{:}  \ottsym{(}  \Gamma_{{\mathrm{1}}}  \ottsym{,}  \mathit{x} \,  {:}  \, \ottnt{T}  \ottsym{,}  \Gamma_{{\mathrm{2}}}  \ottsym{)}  \ottsym{(}  \mathit{y}  \ottsym{)}$).
   By inversion,
   $\vdash  \Gamma_{{\mathrm{1}}}  \ottsym{,}  \mathit{x} \,  {:}  \, \ottnt{T}  \ottsym{,}  \Gamma_{{\mathrm{2}}}$ and $\ottsym{(}  \Gamma_{{\mathrm{1}}}  \ottsym{,}  \mathit{x} \,  {:}  \, \ottnt{T}  \ottsym{,}  \Gamma_{{\mathrm{2}}}  \ottsym{)}  \ottsym{(}  \mathit{y}  \ottsym{)}$ is not a refinement type.
   Because $\vdash  \Gamma_{{\mathrm{1}}}  \ottsym{,}  \mathit{x} \,  {:}  \, \ottnt{T}  \ottsym{,}  \Gamma_{{\mathrm{2}}}$ and $\Gamma_{{\mathrm{1}}}  \vdash  \ottnt{v}  \ottsym{:}  \ottnt{T}$,
   \reflem{ts-val-subst-wf-ftyping} implies $\vdash   \Gamma_{{\mathrm{1}}}  \ottsym{,}  \Gamma_{{\mathrm{2}}}    [  \ottnt{v}  /  \mathit{x}  ]  $.
   By case analysis on $ \mathit{x}    =    \mathit{y} $ or not.
   \begin{caseanalysis}
    \case $ \mathit{x}    =    \mathit{y} $:
     We must show that
     \[
       \Gamma_{{\mathrm{1}}}  \ottsym{,}  \Gamma_{{\mathrm{2}}}    [  \ottnt{v}  /  \mathit{x}  ]    \vdash  \ottnt{v}  \ottsym{:}   \ottsym{(}  \Gamma_{{\mathrm{1}}}  \ottsym{,}  \mathit{x} \,  {:}  \, \ottnt{T}  \ottsym{,}  \Gamma_{{\mathrm{2}}}  \ottsym{)}  \ottsym{(}  \mathit{y}  \ottsym{)}    [  \ottnt{v}  /  \mathit{x}  ]   ~.
     \]
     Because $ \mathit{x}    =    \mathit{y} $, we find $\ottsym{(}  \Gamma_{{\mathrm{1}}}  \ottsym{,}  \mathit{x} \,  {:}  \, \ottnt{T}  \ottsym{,}  \Gamma_{{\mathrm{2}}}  \ottsym{)}  \ottsym{(}  \mathit{y}  \ottsym{)} \,  =  \, \ottnt{T}$.
     $\vdash  \Gamma_{{\mathrm{1}}}  \ottsym{,}  \mathit{x} \,  {:}  \, \ottnt{T}  \ottsym{,}  \Gamma_{{\mathrm{2}}}$ implies that $\mathit{x}$ does not occur in $\ottnt{T}$.
     Therefore, it suffices to show that
     \[
       \Gamma_{{\mathrm{1}}}  \ottsym{,}  \Gamma_{{\mathrm{2}}}    [  \ottnt{v}  /  \mathit{x}  ]    \vdash  \ottnt{v}  \ottsym{:}  \ottnt{T} ~.
     \]
     By \reflem{ts-weak-typing} with $\Gamma_{{\mathrm{1}}}  \vdash  \ottnt{v}  \ottsym{:}  \ottnt{T}$ and $\vdash   \Gamma_{{\mathrm{1}}}  \ottsym{,}  \Gamma_{{\mathrm{2}}}    [  \ottnt{v}  /  \mathit{x}  ]  $,
     we have the conclusion.

    \case $ \mathit{x}    \not=    \mathit{y} $:
     We must show that
     \[
       \Gamma_{{\mathrm{1}}}  \ottsym{,}  \Gamma_{{\mathrm{2}}}    [  \ottnt{v}  /  \mathit{x}  ]    \vdash  \mathit{y}  \ottsym{:}   \ottsym{(}  \Gamma_{{\mathrm{1}}}  \ottsym{,}  \mathit{x} \,  {:}  \, \ottnt{T}  \ottsym{,}  \Gamma_{{\mathrm{2}}}  \ottsym{)}  \ottsym{(}  \mathit{y}  \ottsym{)}    [  \ottnt{v}  /  \mathit{x}  ]   ~.
     \]
     By \T{Var} with $\vdash   \Gamma_{{\mathrm{1}}}  \ottsym{,}  \Gamma_{{\mathrm{2}}}    [  \ottnt{v}  /  \mathit{x}  ]  $ and $\ottsym{(}   \Gamma_{{\mathrm{1}}}  \ottsym{,}  \Gamma_{{\mathrm{2}}}    [  \ottnt{v}  /  \mathit{x}  ]    \ottsym{)}  \ottsym{(}  \mathit{y}  \ottsym{)} \,  =  \,  \ottsym{(}  \Gamma_{{\mathrm{1}}}  \ottsym{,}  \mathit{x} \,  {:}  \, \ottnt{T}  \ottsym{,}  \Gamma_{{\mathrm{2}}}  \ottsym{)}  \ottsym{(}  \mathit{y}  \ottsym{)}    [  \ottnt{v}  /  \mathit{x}  ]  $
     (note that we can suppose that $\mathit{x}$ does not occur in $\Gamma_{{\mathrm{1}}}$),
     we have the conclusion.
   \end{caseanalysis}

  \case \T{Fun}: By the IH and Lemmas~\ref{lem:lr-subst-occur}, \ref{lem:lr-subst-consist}, and \ref{lem:lr-subst-eff-subst}.

  \case \T{Sub}: By the IH and Lemmas~\ref{lem:ts-val-subst-wf-ftyping} and \ref{lem:ts-val-subst-subtyping}.

  \case \T{Op}:  By the IHs and \reflem{ts-val-subst-wf-ftyping}.
   Note that $\mathit{ty} \, \ottsym{(}  \ottnt{o}  \ottsym{)}$ is closed (see \refasm{const}).

  \case \T{Let}: By the IHs and \reflem{ts-val-subst-ty-compose}.

  \case \T{Div}: By Lemmas~\ref{lem:ts-val-subst-wf-ftyping}, \ref{lem:ts-val-subst-valid}, and \ref{lem:ts-subst-nu-cty}.
 \end{caseanalysis}
\end{proof}

\begin{lemmap}{Predicate Substitution in Well-Formedness and Formula Typing}{ts-pred-subst-wf-ftyping}
 Suppose that $\Gamma_{{\mathrm{1}}}  \vdash  \ottnt{A}  \ottsym{:}   \tceseq{ \ottnt{s} } $.
 \begin{enumerate}
  \item If $\vdash  \Gamma_{{\mathrm{1}}}  \ottsym{,}  \mathit{X} \,  {:}  \,  \tceseq{ \ottnt{s} }   \ottsym{,}  \Gamma_{{\mathrm{2}}}$, then $\vdash   \Gamma_{{\mathrm{1}}}  \ottsym{,}  \Gamma_{{\mathrm{2}}}    [  \ottnt{A}  /  \mathit{X}  ]  $.
  \item If $\Gamma_{{\mathrm{1}}}  \ottsym{,}  \mathit{X} \,  {:}  \,  \tceseq{ \ottnt{s} }   \ottsym{,}  \Gamma_{{\mathrm{2}}}  \vdash  \ottnt{T}$, then $ \Gamma_{{\mathrm{1}}}  \ottsym{,}  \Gamma_{{\mathrm{2}}}    [  \ottnt{A}  /  \mathit{X}  ]    \vdash   \ottnt{T}    [  \ottnt{A}  /  \mathit{X}  ]  $.
  \item If $\Gamma_{{\mathrm{1}}}  \ottsym{,}  \mathit{X} \,  {:}  \,  \tceseq{ \ottnt{s} }   \ottsym{,}  \Gamma_{{\mathrm{2}}}  \vdash  \ottnt{C}$, then $ \Gamma_{{\mathrm{1}}}  \ottsym{,}  \Gamma_{{\mathrm{2}}}    [  \ottnt{A}  /  \mathit{X}  ]    \vdash   \ottnt{C}    [  \ottnt{A}  /  \mathit{X}  ]  $.
  \item If $\Gamma_{{\mathrm{1}}}  \ottsym{,}  \mathit{X} \,  {:}  \,  \tceseq{ \ottnt{s} }   \ottsym{,}  \Gamma_{{\mathrm{2}}} \,  |  \, \ottnt{T}  \vdash  \ottnt{S}$, then $ \Gamma_{{\mathrm{1}}}  \ottsym{,}  \Gamma_{{\mathrm{2}}}    [  \ottnt{A}  /  \mathit{X}  ]   \,  |  \,  \ottnt{T}    [  \ottnt{A}  /  \mathit{X}  ]    \vdash   \ottnt{S}    [  \ottnt{A}  /  \mathit{X}  ]  $.
  \item If $\Gamma_{{\mathrm{1}}}  \ottsym{,}  \mathit{X} \,  {:}  \,  \tceseq{ \ottnt{s} }   \ottsym{,}  \Gamma_{{\mathrm{2}}}  \vdash  \Phi$, then $ \Gamma_{{\mathrm{1}}}  \ottsym{,}  \Gamma_{{\mathrm{2}}}    [  \ottnt{A}  /  \mathit{X}  ]    \vdash   \Phi    [  \ottnt{A}  /  \mathit{X}  ]  $.
  \item If $\Gamma_{{\mathrm{1}}}  \ottsym{,}  \mathit{X} \,  {:}  \,  \tceseq{ \ottnt{s} }   \ottsym{,}  \Gamma_{{\mathrm{2}}}  \vdash  \ottnt{t}  \ottsym{:}  \ottnt{s_{{\mathrm{0}}}}$, then $ \Gamma_{{\mathrm{1}}}  \ottsym{,}  \Gamma_{{\mathrm{2}}}    [  \ottnt{A}  /  \mathit{X}  ]    \vdash   \ottnt{t}    [  \ottnt{A}  /  \mathit{X}  ]    \ottsym{:}  \ottnt{s_{{\mathrm{0}}}}$.
  \item If $\Gamma_{{\mathrm{1}}}  \ottsym{,}  \mathit{X} \,  {:}  \,  \tceseq{ \ottnt{s} }   \ottsym{,}  \Gamma_{{\mathrm{2}}}  \vdash  \ottnt{A_{{\mathrm{0}}}}  \ottsym{:}   \tceseq{ \ottnt{s_{{\mathrm{0}}}} } $, then $ \Gamma_{{\mathrm{1}}}  \ottsym{,}  \Gamma_{{\mathrm{2}}}    [  \ottnt{A}  /  \mathit{X}  ]    \vdash   \ottnt{A_{{\mathrm{0}}}}    [  \ottnt{A}  /  \mathit{X}  ]    \ottsym{:}   \tceseq{ \ottnt{s_{{\mathrm{0}}}} } $.
  \item If $\Gamma_{{\mathrm{1}}}  \ottsym{,}  \mathit{X} \,  {:}  \,  \tceseq{ \ottnt{s} }   \ottsym{,}  \Gamma_{{\mathrm{2}}}  \vdash  \phi$, then $ \Gamma_{{\mathrm{1}}}  \ottsym{,}  \Gamma_{{\mathrm{2}}}    [  \ottnt{A}  /  \mathit{X}  ]    \vdash   \phi    [  \ottnt{A}  /  \mathit{X}  ]  $.
 \end{enumerate}
\end{lemmap}
\begin{proof}
 By simultaneous induction on the derivations.
 The only interesting case is of \Tf{PVar}.

 In the case of \Tf{PVar}, we are given
 $\ottnt{A_{{\mathrm{0}}}} \,  =  \, \mathit{Y}$ and $ \tceseq{ \ottnt{s_{{\mathrm{0}}}} }  \,  =  \, \ottsym{(}  \Gamma_{{\mathrm{1}}}  \ottsym{,}  \mathit{X} \,  {:}  \,  \tceseq{ \ottnt{s} }   \ottsym{,}  \Gamma_{{\mathrm{2}}}  \ottsym{)}  \ottsym{(}  \mathit{Y}  \ottsym{)}$
 for some $\mathit{Y}$ (i.e., $\Gamma_{{\mathrm{1}}}  \ottsym{,}  \mathit{X} \,  {:}  \,  \tceseq{ \ottnt{s} }   \ottsym{,}  \Gamma_{{\mathrm{2}}}  \vdash  \mathit{Y}  \ottsym{:}  \ottsym{(}  \Gamma_{{\mathrm{1}}}  \ottsym{,}  \mathit{X} \,  {:}  \,  \tceseq{ \ottnt{s} }   \ottsym{,}  \Gamma_{{\mathrm{2}}}  \ottsym{)}  \ottsym{(}  \mathit{Y}  \ottsym{)}$).
 By inversion, $\vdash  \Gamma_{{\mathrm{1}}}  \ottsym{,}  \mathit{X} \,  {:}  \,  \tceseq{ \ottnt{s} }   \ottsym{,}  \Gamma_{{\mathrm{2}}}$.
 By the IH, $\vdash   \Gamma_{{\mathrm{1}}}  \ottsym{,}  \Gamma_{{\mathrm{2}}}    [  \ottnt{A}  /  \mathit{X}  ]  $.
 By case analysis on whether $\mathit{X} \,  =  \, \mathit{Y}$ or not.
 \begin{caseanalysis}
  \case $\mathit{X} \,  =  \, \mathit{Y}$:
   Because $ \tceseq{ \ottnt{s_{{\mathrm{0}}}} }  \,  =  \, \ottsym{(}  \Gamma_{{\mathrm{1}}}  \ottsym{,}  \mathit{X} \,  {:}  \,  \tceseq{ \ottnt{s} }   \ottsym{,}  \Gamma_{{\mathrm{2}}}  \ottsym{)}  \ottsym{(}  \mathit{Y}  \ottsym{)} =  \tceseq{ \ottnt{s} } $,
   we must show that
   \[
     \Gamma_{{\mathrm{1}}}  \ottsym{,}  \Gamma_{{\mathrm{2}}}    [  \ottnt{A}  /  \mathit{X}  ]    \vdash  \ottnt{A}  \ottsym{:}   \tceseq{ \ottnt{s} }  ~,
   \]
   which is derived by \reflem{ts-weak-wf-ftyping} with $\vdash   \Gamma_{{\mathrm{1}}}  \ottsym{,}  \Gamma_{{\mathrm{2}}}    [  \ottnt{A}  /  \mathit{X}  ]  $ and
   the assumption $\Gamma_{{\mathrm{1}}}  \vdash  \ottnt{A}  \ottsym{:}   \tceseq{ \ottnt{s} } $.

  \case $\mathit{X} \,  \not=  \, \mathit{Y}$:
   We must show that
   \[
     \Gamma_{{\mathrm{1}}}  \ottsym{,}  \Gamma_{{\mathrm{2}}}    [  \ottnt{A}  /  \mathit{X}  ]    \vdash  \mathit{Y}  \ottsym{:}   \tceseq{ \ottnt{s_{{\mathrm{0}}}} }  ~,
   \]
   which is derived by
  \Tf{PVar} with $\vdash   \Gamma_{{\mathrm{1}}}  \ottsym{,}  \Gamma_{{\mathrm{2}}}    [  \ottnt{A}  /  \mathit{X}  ]  $ and the fact that
   $ \tceseq{ \ottnt{s_{{\mathrm{0}}}} }  \,  =  \, \ottsym{(}  \Gamma_{{\mathrm{1}}}  \ottsym{,}  \mathit{X} \,  {:}  \,  \tceseq{ \ottnt{s} }   \ottsym{,}  \Gamma_{{\mathrm{2}}}  \ottsym{)}  \ottsym{(}  \mathit{Y}  \ottsym{)} = \ottsym{(}   \Gamma_{{\mathrm{1}}}  \ottsym{,}  \Gamma_{{\mathrm{2}}}    [  \ottnt{A}  /  \mathit{X}  ]    \ottsym{)}  \ottsym{(}  \mathit{Y}  \ottsym{)}$.
 \end{caseanalysis}
\end{proof}

\begin{lemmap}{Predicate Substitution in Validity}{ts-pred-subst-valid}
 Suppose that $\Gamma_{{\mathrm{1}}}  \vdash  \ottnt{A}  \ottsym{:}   \tceseq{ \ottnt{s} } $.
 If $\Gamma_{{\mathrm{1}}}  \ottsym{,}  \mathit{X} \,  {:}  \,  \tceseq{ \ottnt{s} }   \ottsym{,}  \Gamma_{{\mathrm{2}}}  \models  \phi$, then $ \Gamma_{{\mathrm{1}}}  \ottsym{,}  \Gamma_{{\mathrm{2}}}    [  \ottnt{A}  /  \mathit{X}  ]    \models   \phi    [  \ottnt{A}  /  \mathit{X}  ]  $.
\end{lemmap}
\begin{proof}
 \TS{Validity}
 Let $\theta$ be any substitution such that $ \Gamma_{{\mathrm{1}}}  \ottsym{,}  \Gamma_{{\mathrm{2}}}    [  \ottnt{A}  /  \mathit{X}  ]    \vdash  \theta$.
 By \reflem{ts-logic-closing-dsubs}, it suffices to show that
 $ { [\![   \phi    [  \ottnt{A}  /  \mathit{X}  ]    ]\!]_{ \theta } }  \,  =  \,  \top $.
 By \reflem{ts-logic-pred-subst}, it suffices to show that
 $ { [\![  \phi  ]\!]_{ \theta  \mathbin{\uplus}   \{  \mathit{X}  \mapsto   { [\![  \ottnt{A}  ]\!]_{ \theta } }   \}  } }  \,  =  \,  \top $.
 By \reflem{ts-logic-closing-dsubs} with $\Gamma_{{\mathrm{1}}}  \ottsym{,}  \mathit{X} \,  {:}  \,  \tceseq{ \ottnt{s} }   \ottsym{,}  \Gamma_{{\mathrm{2}}}  \models  \phi$,
 it suffices to show that
 $\Gamma_{{\mathrm{1}}}  \ottsym{,}  \mathit{X} \,  {:}  \,  \tceseq{ \ottnt{s} }   \ottsym{,}  \Gamma_{{\mathrm{2}}}  \vdash  \theta  \mathbin{\uplus}   \{  \mathit{X}  \mapsto   { [\![  \ottnt{A}  ]\!]_{ \theta } }   \} $.
 $ \Gamma_{{\mathrm{1}}}  \ottsym{,}  \Gamma_{{\mathrm{2}}}    [  \ottnt{A}  /  \mathit{X}  ]    \vdash  \theta$ and $\Gamma_{{\mathrm{1}}}  \vdash  \ottnt{A}  \ottsym{:}   \tceseq{ \ottnt{s} } $
 implies $ { [\![  \ottnt{A}  ]\!]_{ \theta } }  \,  \in  \,  \mathcal{D}_{  \tceseq{ \ottnt{s} }   \rightarrow   \mathbb{B}  } $.
 Therefore, as $ \Gamma_{{\mathrm{1}}}  \ottsym{,}  \Gamma_{{\mathrm{2}}}    [  \ottnt{A}  /  \mathit{X}  ]    \vdash  \theta$,
 we have the conclusion
 $\Gamma_{{\mathrm{1}}}  \ottsym{,}  \mathit{X} \,  {:}  \,  \tceseq{ \ottnt{s} }   \ottsym{,}  \Gamma_{{\mathrm{2}}}  \vdash   \{  \mathit{X}  \mapsto   { [\![  \ottnt{A}  ]\!]_{ \theta } }   \}   \mathbin{\uplus}  \theta$
 by definition and \reflem{ts-logic-pred-subst}.
\end{proof}

\begin{lemmap}{Predicate Substitution in Type Composition}{ts-pred-subst-ty-compose}
 \begin{enumerate}
  \item $   \forall  \, { \ottnt{S_{{\mathrm{1}}}}  \ottsym{,}  \ottnt{S_{{\mathrm{2}}}}  \ottsym{,}  \ottnt{A}  \ottsym{,}  \mathit{X}  \ottsym{,}  \mathit{x} } . \  \mathit{x} \,  \not\in  \,  \mathit{fv}  (  \ottnt{A}  )    \mathrel{\Longrightarrow}  \ottsym{(}   \ottnt{S_{{\mathrm{1}}}}    [  \ottnt{A}  /  \mathit{X}  ]    \gg\!\!=  \ottsym{(}   \lambda\!  \, \mathit{x}  \ottsym{.}   \ottnt{S_{{\mathrm{2}}}}    [  \ottnt{A}  /  \mathit{X}  ]    \ottsym{)}  \ottsym{)} \,  =  \,  \ottsym{(}  \ottnt{S_{{\mathrm{1}}}}  \gg\!\!=  \ottsym{(}   \lambda\!  \, \mathit{x}  \ottsym{.}  \ottnt{S_{{\mathrm{2}}}}  \ottsym{)}  \ottsym{)}    [  \ottnt{A}  /  \mathit{X}  ]   $.
  \item $  \forall  \, { \Phi_{{\mathrm{1}}}  \ottsym{,}  \Phi_{{\mathrm{2}}}  \ottsym{,}  \ottnt{A}  \ottsym{,}  \mathit{X} } . \    \Phi_{{\mathrm{1}}}    [  \ottnt{A}  /  \mathit{X}  ]   \,  \tceappendop  \, \Phi_{{\mathrm{2}}}    [  \ottnt{A}  /  \mathit{X}  ]   \,  =  \,  \ottsym{(}  \Phi_{{\mathrm{1}}} \,  \tceappendop  \, \Phi_{{\mathrm{2}}}  \ottsym{)}    [  \ottnt{A}  /  \mathit{X}  ]   $.
  \item $  \forall  \, { \Phi  \ottsym{,}  \ottnt{C}  \ottsym{,}  \ottnt{A}  \ottsym{,}  \mathit{X} } . \    \Phi    [  \ottnt{A}  /  \mathit{X}  ]   \,  \tceappendop  \, \ottnt{C}    [  \ottnt{A}  /  \mathit{X}  ]   \,  =  \,  \ottsym{(}  \Phi \,  \tceappendop  \, \ottnt{C}  \ottsym{)}    [  \ottnt{A}  /  \mathit{X}  ]   $.
  \item $  \forall  \, { \Phi  \ottsym{,}  \ottnt{S}  \ottsym{,}  \ottnt{A}  \ottsym{,}  \mathit{X} } . \    \Phi    [  \ottnt{A}  /  \mathit{X}  ]   \,  \tceappendop  \, \ottnt{S}    [  \ottnt{A}  /  \mathit{X}  ]   \,  =  \,  \ottsym{(}  \Phi \,  \tceappendop  \, \ottnt{S}  \ottsym{)}    [  \ottnt{A}  /  \mathit{X}  ]   $.
 \end{enumerate}
\end{lemmap}
\begin{proof}
 The first and second cases are by definition.
 The third and fourth cases are straightforward by simultaneous induction on $\ottnt{C}$ and $\ottnt{S}$ with the second case.
\end{proof}

\begin{lemmap}{Predicate Substitution in Subtyping}{ts-pred-subst-subtyping}
 Suppose that $\Gamma_{{\mathrm{1}}}  \vdash  \ottnt{A}  \ottsym{:}   \tceseq{ \ottnt{s} } $.
 \begin{enumerate}
  \item If $\Gamma_{{\mathrm{1}}}  \ottsym{,}  \mathit{X} \,  {:}  \,  \tceseq{ \ottnt{s} }   \ottsym{,}  \Gamma_{{\mathrm{2}}}  \vdash  \ottnt{T_{{\mathrm{1}}}}  \ottsym{<:}  \ottnt{T_{{\mathrm{2}}}}$, then $ \Gamma_{{\mathrm{1}}}  \ottsym{,}  \Gamma_{{\mathrm{2}}}    [  \ottnt{A}  /  \mathit{X}  ]    \vdash   \ottnt{T_{{\mathrm{1}}}}    [  \ottnt{A}  /  \mathit{X}  ]    \ottsym{<:}   \ottnt{T_{{\mathrm{2}}}}    [  \ottnt{A}  /  \mathit{X}  ]  $.
  \item If $\Gamma_{{\mathrm{1}}}  \ottsym{,}  \mathit{X} \,  {:}  \,  \tceseq{ \ottnt{s} }   \ottsym{,}  \Gamma_{{\mathrm{2}}}  \vdash  \ottnt{C_{{\mathrm{1}}}}  \ottsym{<:}  \ottnt{C_{{\mathrm{2}}}}$, then $ \Gamma_{{\mathrm{1}}}  \ottsym{,}  \Gamma_{{\mathrm{2}}}    [  \ottnt{A}  /  \mathit{X}  ]    \vdash   \ottnt{C_{{\mathrm{1}}}}    [  \ottnt{A}  /  \mathit{X}  ]    \ottsym{<:}   \ottnt{C_{{\mathrm{2}}}}    [  \ottnt{A}  /  \mathit{X}  ]  $.
  \item If $\Gamma_{{\mathrm{1}}}  \ottsym{,}  \mathit{X} \,  {:}  \,  \tceseq{ \ottnt{s} }   \ottsym{,}  \Gamma_{{\mathrm{2}}} \,  |  \, \ottnt{T} \,  \,\&\,  \, \Phi  \vdash  \ottnt{S_{{\mathrm{1}}}}  \ottsym{<:}  \ottnt{S_{{\mathrm{2}}}}$, then $ \Gamma_{{\mathrm{1}}}  \ottsym{,}  \Gamma_{{\mathrm{2}}}    [  \ottnt{A}  /  \mathit{X}  ]   \,  |  \,  \ottnt{T}    [  \ottnt{A}  /  \mathit{X}  ]   \,  \,\&\,  \,  \Phi    [  \ottnt{A}  /  \mathit{X}  ]    \vdash   \ottnt{S_{{\mathrm{1}}}}    [  \ottnt{A}  /  \mathit{X}  ]    \ottsym{<:}   \ottnt{S_{{\mathrm{2}}}}    [  \ottnt{A}  /  \mathit{X}  ]  $.
  \item If $\Gamma_{{\mathrm{1}}}  \ottsym{,}  \mathit{X} \,  {:}  \,  \tceseq{ \ottnt{s} }   \ottsym{,}  \Gamma_{{\mathrm{2}}}  \vdash  \Phi_{{\mathrm{1}}}  \ottsym{<:}  \Phi_{{\mathrm{2}}}$, then $ \Gamma_{{\mathrm{1}}}  \ottsym{,}  \Gamma_{{\mathrm{2}}}    [  \ottnt{A}  /  \mathit{X}  ]    \vdash   \Phi_{{\mathrm{1}}}    [  \ottnt{A}  /  \mathit{X}  ]    \ottsym{<:}   \Phi_{{\mathrm{2}}}    [  \ottnt{A}  /  \mathit{X}  ]  $.
 \end{enumerate}
\end{lemmap}
\begin{proof}
 Straightforward by simultaneous induction on the derivations.
 The cases for \Srule{Refine}, \Srule{Eff}, and \Srule{Embed} use \reflem{ts-pred-subst-valid}, and
 the case for \Srule{Embed} also uses Lemmas~\ref{lem:ts-subst-nu-cty} and \ref{lem:ts-pred-subst-ty-compose}.
\end{proof}

\begin{lemmap}{Predicate Substitution in Typing}{ts-pred-subst-typing}
 Suppose that $\Gamma_{{\mathrm{1}}}  \vdash  \ottnt{A}  \ottsym{:}   \tceseq{ \ottnt{s} } $.
 If $\Gamma_{{\mathrm{1}}}  \ottsym{,}  \mathit{X} \,  {:}  \,  \tceseq{ \ottnt{s} }   \ottsym{,}  \Gamma_{{\mathrm{2}}}  \vdash  \ottnt{e}  \ottsym{:}  \ottnt{C}$, then $ \Gamma_{{\mathrm{1}}}  \ottsym{,}  \Gamma_{{\mathrm{2}}}    [  \ottnt{A}  /  \mathit{X}  ]    \vdash   \ottnt{e}    [  \ottnt{A}  /  \mathit{X}  ]    \ottsym{:}   \ottnt{C}    [  \ottnt{A}  /  \mathit{X}  ]  $.
 Especially, if $\Gamma_{{\mathrm{1}}}  \ottsym{,}  \mathit{X} \,  {:}  \,  \tceseq{ \ottnt{s} }   \ottsym{,}  \Gamma_{{\mathrm{2}}}  \vdash  \ottnt{e}  \ottsym{:}  \ottnt{T}$, then $ \Gamma_{{\mathrm{1}}}  \ottsym{,}  \Gamma_{{\mathrm{2}}}    [  \ottnt{A}  /  \mathit{X}  ]    \vdash   \ottnt{e}    [  \ottnt{A}  /  \mathit{X}  ]    \ottsym{:}   \ottnt{T}    [  \ottnt{A}  /  \mathit{X}  ]  $.
\end{lemmap}
\begin{proof}
 By induction on the typing derivation with \reflem{ts-pred-subst-wf-ftyping}.
 We mention only the interesting cases.
 \begin{caseanalysis}
  \case \T{Fun}: By the IH and Lemmas~\ref{lem:lr-subst-occur}, \ref{lem:lr-subst-consist}, and \ref{lem:lr-subst-eff-subst}.
  \case \T{Sub}: By the IH and Lemmas~\ref{lem:ts-pred-subst-wf-ftyping} and \ref{lem:ts-pred-subst-subtyping}.
  \case \T{Op}: By the IHs and \reflem{ts-pred-subst-wf-ftyping}.
   Note that $\mathit{ty} \, \ottsym{(}  \ottnt{o}  \ottsym{)}$ is closed (see \refasm{const}).
  \case \T{Let}: By the IHs and \reflem{ts-pred-subst-ty-compose}.
  \case \T{Div}: By Lemmas~\ref{lem:ts-pred-subst-wf-ftyping}, \ref{lem:ts-pred-subst-valid}, and \ref{lem:ts-subst-nu-cty}.
 \end{caseanalysis}
\end{proof}

\begin{lemmap}{Strengthening Type Hypothesis in Validity}{ts-strength-ty-hypo-valid}
 Suppose that $\Gamma_{{\mathrm{1}}}  \vdash  \ottnt{T_{{\mathrm{01}}}}  \ottsym{<:}  \ottnt{T_{{\mathrm{02}}}}$.
 If $\Gamma_{{\mathrm{1}}}  \ottsym{,}  \mathit{x} \,  {:}  \, \ottnt{T_{{\mathrm{02}}}}  \ottsym{,}  \Gamma_{{\mathrm{2}}}  \models  \phi$, then $\Gamma_{{\mathrm{1}}}  \ottsym{,}  \mathit{x} \,  {:}  \, \ottnt{T_{{\mathrm{01}}}}  \ottsym{,}  \Gamma_{{\mathrm{2}}}  \models  \phi$.
\end{lemmap}
\begin{proof}
 \TS{Validity}
 If $\ottnt{T_{{\mathrm{01}}}}$ is not a refinement type, then we have the conclusion
 because the formulas $\Gamma_{{\mathrm{1}}}  \ottsym{,}  \mathit{x} \,  {:}  \, \ottnt{T_{{\mathrm{02}}}}  \ottsym{,}  \Gamma_{{\mathrm{2}}}  \models  \phi$ and $\Gamma_{{\mathrm{1}}}  \ottsym{,}  \mathit{x} \,  {:}  \, \ottnt{T_{{\mathrm{01}}}}  \ottsym{,}  \Gamma_{{\mathrm{2}}}  \models  \phi$ are equivalent;
 note that $\ottnt{T_{{\mathrm{02}}}}$ is not a refinement type either.

 Otherwise, suppose that $\ottnt{T_{{\mathrm{01}}}} \,  =  \, \ottsym{\{}  \mathit{x} \,  {:}  \, \iota \,  |  \, \phi_{{\mathrm{01}}}  \ottsym{\}}$ for some $\mathit{x}$, $\iota$, and $\phi_{{\mathrm{01}}}$.
 Then, $\Gamma_{{\mathrm{1}}}  \vdash  \ottnt{T_{{\mathrm{01}}}}  \ottsym{<:}  \ottnt{T_{{\mathrm{02}}}}$ implies that
 $\ottnt{T_{{\mathrm{02}}}} \,  =  \, \ottsym{\{}  \mathit{x} \,  {:}  \, \iota \,  |  \, \phi_{{\mathrm{02}}}  \ottsym{\}}$ and $\Gamma_{{\mathrm{1}}}  \ottsym{,}  \mathit{x} \,  {:}  \, \iota  \models   \phi_{{\mathrm{01}}}    \Rightarrow    \phi_{{\mathrm{02}}} $ for some $\phi_{{\mathrm{02}}}$.
 Let $\theta$ be any substitution such that $\Gamma_{{\mathrm{1}}}  \ottsym{,}  \mathit{x} \,  {:}  \, \ottnt{T_{{\mathrm{01}}}}  \ottsym{,}  \Gamma_{{\mathrm{2}}}  \vdash  \theta$.
 By \reflem{ts-logic-closing-dsubs},
 it suffices to show that $ { [\![  \phi  ]\!]_{ \theta } }  \,  =  \,  \top $.
 Again by \reflem{ts-logic-closing-dsubs} with $\Gamma_{{\mathrm{1}}}  \ottsym{,}  \mathit{x} \,  {:}  \, \ottnt{T_{{\mathrm{02}}}}  \ottsym{,}  \Gamma_{{\mathrm{2}}}  \models  \phi$,
 it suffices to show that $\Gamma_{{\mathrm{1}}}  \ottsym{,}  \mathit{x} \,  {:}  \, \ottnt{T_{{\mathrm{02}}}}  \ottsym{,}  \Gamma_{{\mathrm{2}}}  \vdash  \theta$.
 Because $\Gamma_{{\mathrm{1}}}  \ottsym{,}  \mathit{x} \,  {:}  \, \ottnt{T_{{\mathrm{01}}}}  \ottsym{,}  \Gamma_{{\mathrm{2}}}  \vdash  \theta$,
 it suffices to show that $ { [\![  \phi_{{\mathrm{02}}}  ]\!]_{ \theta } }  \,  =  \,  \top $.
 Further, $\Gamma_{{\mathrm{1}}}  \ottsym{,}  \mathit{x} \,  {:}  \, \ottnt{T_{{\mathrm{01}}}}  \ottsym{,}  \Gamma_{{\mathrm{2}}}  \vdash  \theta$ implies
 $ { [\![  \phi_{{\mathrm{01}}}  ]\!]_{ \theta } }  \,  =  \,  \top $ and $\Gamma_{{\mathrm{1}}}  \ottsym{,}  \mathit{x} \,  {:}  \, \iota  \ottsym{,}  \Gamma_{{\mathrm{2}}}  \vdash  \theta$.
 Because $\Gamma_{{\mathrm{1}}}  \ottsym{,}  \mathit{x} \,  {:}  \, \iota  \models   \phi_{{\mathrm{01}}}    \Rightarrow    \phi_{{\mathrm{02}}} $,
 Lemmas~\ref{lem:ts-weak-valid} and \ref{lem:ts-logic-closing-dsubs}
 imply the conclusion.
\end{proof}

\begin{lemmap}{Strengthening Type Hypothesis in Subtyping}{ts-strength-ty-hypo-subtyping}
 Suppose that $\Gamma_{{\mathrm{1}}}  \vdash  \ottnt{T_{{\mathrm{01}}}}  \ottsym{<:}  \ottnt{T_{{\mathrm{02}}}}$.
 \begin{enumerate}
  \item If $\Gamma_{{\mathrm{1}}}  \ottsym{,}  \mathit{x} \,  {:}  \, \ottnt{T_{{\mathrm{02}}}}  \ottsym{,}  \Gamma_{{\mathrm{2}}}  \vdash  \ottnt{T_{{\mathrm{1}}}}  \ottsym{<:}  \ottnt{T_{{\mathrm{2}}}}$, then $\Gamma_{{\mathrm{1}}}  \ottsym{,}  \mathit{x} \,  {:}  \, \ottnt{T_{{\mathrm{01}}}}  \ottsym{,}  \Gamma_{{\mathrm{2}}}  \vdash  \ottnt{T_{{\mathrm{1}}}}  \ottsym{<:}  \ottnt{T_{{\mathrm{2}}}}$.
  \item If $\Gamma_{{\mathrm{1}}}  \ottsym{,}  \mathit{x} \,  {:}  \, \ottnt{T_{{\mathrm{02}}}}  \ottsym{,}  \Gamma_{{\mathrm{2}}}  \vdash  \ottnt{C_{{\mathrm{1}}}}  \ottsym{<:}  \ottnt{C_{{\mathrm{2}}}}$, then $\Gamma_{{\mathrm{1}}}  \ottsym{,}  \mathit{x} \,  {:}  \, \ottnt{T_{{\mathrm{01}}}}  \ottsym{,}  \Gamma_{{\mathrm{2}}}  \vdash  \ottnt{C_{{\mathrm{1}}}}  \ottsym{<:}  \ottnt{C_{{\mathrm{2}}}}$.
  \item If $\Gamma_{{\mathrm{1}}}  \ottsym{,}  \mathit{x} \,  {:}  \, \ottnt{T_{{\mathrm{02}}}}  \ottsym{,}  \Gamma_{{\mathrm{2}}} \,  |  \, \ottnt{T} \,  \,\&\,  \, \Phi  \vdash  \ottnt{S_{{\mathrm{1}}}}  \ottsym{<:}  \ottnt{S_{{\mathrm{2}}}}$, then $\Gamma_{{\mathrm{1}}}  \ottsym{,}  \mathit{x} \,  {:}  \, \ottnt{T_{{\mathrm{01}}}}  \ottsym{,}  \Gamma_{{\mathrm{2}}} \,  |  \, \ottnt{T} \,  \,\&\,  \, \Phi  \vdash  \ottnt{S_{{\mathrm{1}}}}  \ottsym{<:}  \ottnt{S_{{\mathrm{2}}}}$.
  \item If $\Gamma_{{\mathrm{1}}}  \ottsym{,}  \mathit{x} \,  {:}  \, \ottnt{T_{{\mathrm{02}}}}  \ottsym{,}  \Gamma_{{\mathrm{2}}}  \vdash  \Phi_{{\mathrm{1}}}  \ottsym{<:}  \Phi_{{\mathrm{2}}}$, then $\Gamma_{{\mathrm{1}}}  \ottsym{,}  \mathit{x} \,  {:}  \, \ottnt{T_{{\mathrm{01}}}}  \ottsym{,}  \Gamma_{{\mathrm{2}}}  \vdash  \Phi_{{\mathrm{1}}}  \ottsym{<:}  \Phi_{{\mathrm{2}}}$.
 \end{enumerate}
\end{lemmap}
\begin{proof}
 Straightforward by simultaneous induction on the derivations.
 The cases for \Srule{Refine}, \Srule{Eff}, and \Srule{Embed} use \reflem{ts-strength-ty-hypo-valid}.
\end{proof}

\begin{lemmap}{Strengthening Type Hypothesis in Stack Subtyping}{ts-strength-ty-hypo-stack-subtyping}
 If $\Gamma  \vdash  \ottnt{T_{{\mathrm{1}}}}  \ottsym{<:}  \ottnt{T_{{\mathrm{2}}}}$ and $\Gamma \,  |  \, \ottnt{T_{{\mathrm{2}}}} \,  \,\&\,  \, \Phi  \vdash  \ottnt{S_{{\mathrm{1}}}}  \ottsym{<:}  \ottnt{S_{{\mathrm{2}}}}$, then
 $\Gamma \,  |  \, \ottnt{T_{{\mathrm{1}}}} \,  \,\&\,  \, \Phi  \vdash  \ottnt{S_{{\mathrm{1}}}}  \ottsym{<:}  \ottnt{S_{{\mathrm{2}}}}$.
\end{lemmap}
\begin{proof}
 By case analysis on the inference rule applied last to derive $\Gamma \,  |  \, \ottnt{T_{{\mathrm{2}}}} \,  \,\&\,  \, \Phi  \vdash  \ottnt{S_{{\mathrm{1}}}}  \ottsym{<:}  \ottnt{S_{{\mathrm{2}}}}$.
 \begin{caseanalysis}
  \case \Srule{Empty}: Trivial by \Srule{Empty}.
  \case \Srule{Ans}: By \reflem{ts-strength-ty-hypo-subtyping} and \Srule{Ans}.
  \case \Srule{Embed}: By \reflem{ts-strength-ty-hypo-subtyping} and \Srule{Embed}.
 \end{caseanalysis}
\end{proof}

\begin{lemma}{ts-imp-strength-eff-hypo-formula-aux}
 If
 $\Gamma  \ottsym{,}  \mathit{x} \,  {:}  \, \ottnt{s}  \models   \phi_{{\mathrm{1}}}    \Rightarrow    \phi_{{\mathrm{2}}} $,
 then
 \[
  \Gamma  \models   \ottsym{(}    \exists  \,  \mathit{x}  \mathrel{  \in  }  \ottnt{s}  . \    \exists  \,  \mathit{x'}  \mathrel{  \in  }  \ottnt{s'}  . \    \phi    \wedge    \phi_{{\mathrm{1}}}     \wedge    \phi'     \ottsym{)}    \Rightarrow    \ottsym{(}    \exists  \,  \mathit{x}  \mathrel{  \in  }  \ottnt{s}  . \    \exists  \,  \mathit{x'}  \mathrel{  \in  }  \ottnt{s'}  . \    \phi    \wedge    \phi_{{\mathrm{2}}}     \wedge    \phi'     \ottsym{)}  ~.
 \]
\end{lemma}
\begin{proof}
 \TS{Validity}
 Obvious by \reflem{ts-logic-closing-dsubs}.
\end{proof}

\begin{lemma}{ts-imp-or}
 If
 $\Gamma  \vdash   \phi_{{\mathrm{11}}}    \Rightarrow    \phi_{{\mathrm{21}}} $ and $\Gamma  \vdash   \phi_{{\mathrm{12}}}    \Rightarrow    \phi_{{\mathrm{22}}} $,
 then
 $\Gamma  \vdash     \phi_{{\mathrm{11}}}    \vee    \phi_{{\mathrm{12}}}     \Rightarrow    \phi_{{\mathrm{21}}}     \vee    \phi_{{\mathrm{22}}} $.
\end{lemma}
\begin{proof}
 \TS{Validity}
 Obvious by \reflem{ts-logic-closing-dsubs}.
\end{proof}

\begin{lemmap}{Monotonicity of Effect Composition w.r.t.\ Pre-Effects}{ts-mono-eff-compose-preeff}
 If $\Gamma  \vdash  \Phi_{{\mathrm{1}}}  \ottsym{<:}  \Phi_{{\mathrm{2}}}$,
 then $\Gamma  \vdash  \Phi_{{\mathrm{1}}} \,  \tceappendop  \, \Phi  \ottsym{<:}  \Phi_{{\mathrm{2}}} \,  \tceappendop  \, \Phi$ for any $\Phi$.
\end{lemmap}
\begin{proof}
 Let $\mathit{x}, \mathit{x_{{\mathrm{1}}}}, \mathit{x_{{\mathrm{2}}}}, \mathit{y}, \mathit{x'}, \mathit{y'}$ be fresh variables, and
 $\Phi_{{\mathrm{0}}}$ be either of $\Phi_{{\mathrm{1}}}$ and $\Phi_{{\mathrm{2}}}$.
 Then,
 \[
  \Phi_{{\mathrm{0}}} \,  \tceappendop  \, \Phi \,  =  \,  (   \lambda_\mu\!   \,  \mathit{x}  .\,   \mathit{H} _{  \mu  } (  \Phi_{{\mathrm{0}}}  )   ,   \lambda_\nu\!   \,  \mathit{y}  .\,   \mathit{H} _{  \nu  } (  \Phi_{{\mathrm{0}}}  )   ) 
 \]
 where
 \begin{itemize}
  \item $ \mathit{H} _{  \mu  } (  \Phi_{{\mathrm{0}}}  )  \,  =  \,   \exists  \,  \mathit{x_{{\mathrm{1}}}}  \mathrel{  \in  }   \Sigma ^\ast   . \    \exists  \,  \mathit{x_{{\mathrm{2}}}}  \mathrel{  \in  }   \Sigma ^\ast   . \     \mathit{x}    =    \mathit{x_{{\mathrm{1}}}} \,  \tceappendop  \, \mathit{x_{{\mathrm{2}}}}     \wedge     { \Phi_{{\mathrm{0}}} }^\mu   \ottsym{(}  \mathit{x_{{\mathrm{1}}}}  \ottsym{)}     \wedge     { \Phi }^\mu   \ottsym{(}  \mathit{x_{{\mathrm{2}}}}  \ottsym{)}   $ and
  \item $ \mathit{H} _{  \nu  } (  \Phi_{{\mathrm{0}}}  )  \,  =  \,   { \Phi_{{\mathrm{0}}} }^\nu   \ottsym{(}  \mathit{y}  \ottsym{)}    \vee    \ottsym{(}    \exists  \,  \mathit{x'}  \mathrel{  \in  }   \Sigma ^\ast   . \    \exists  \,  \mathit{y'}  \mathrel{  \in  }   \Sigma ^\omega   . \     \mathit{y}    =    \mathit{x'} \,  \tceappendop  \, \mathit{y'}     \wedge     { \Phi_{{\mathrm{0}}} }^\mu   \ottsym{(}  \mathit{x'}  \ottsym{)}     \wedge     { \Phi }^\nu   \ottsym{(}  \mathit{y'}  \ottsym{)}     \ottsym{)} $.
 \end{itemize}
 We have the conclusion by the following.
 \begin{itemize}
  \item We show that
        \[
         \Gamma  \ottsym{,}  \mathit{x} \,  {:}  \,  \Sigma ^\ast   \models    \mathit{H} _{  \mu  } (  \Phi_{{\mathrm{1}}}  )     \Rightarrow     \mathit{H} _{  \mu  } (  \Phi_{{\mathrm{2}}}  )   ~.
        \]
        The assumption $\Gamma  \vdash  \Phi_{{\mathrm{1}}}  \ottsym{<:}  \Phi_{{\mathrm{2}}}$ with \reflem{ts-weak-valid} implies
        \[
         \Gamma  \ottsym{,}  \mathit{x} \,  {:}  \,  \Sigma ^\ast   \ottsym{,}  \mathit{x_{{\mathrm{1}}}} \,  {:}  \,  \Sigma ^\ast   \models    { \Phi_{{\mathrm{1}}} }^\mu   \ottsym{(}  \mathit{x_{{\mathrm{1}}}}  \ottsym{)}    \Rightarrow     { \Phi_{{\mathrm{2}}} }^\mu   \ottsym{(}  \mathit{x_{{\mathrm{1}}}}  \ottsym{)}  ~.
        \]
        Then, \reflem{ts-imp-strength-eff-hypo-formula-aux}
        implies the conclusion
        \[
         \Gamma  \ottsym{,}  \mathit{x} \,  {:}  \,  \Sigma ^\ast   \models    \mathit{H} _{  \mu  } (  \Phi_{{\mathrm{1}}}  )     \Rightarrow     \mathit{H} _{  \mu  } (  \Phi_{{\mathrm{2}}}  )   ~.
        \]

  \item We show that
        \[
         \Gamma  \ottsym{,}  \mathit{y} \,  {:}  \,  \Sigma ^\omega   \models    \mathit{H} _{  \nu  } (  \Phi_{{\mathrm{1}}}  )     \Rightarrow     \mathit{H} _{  \nu  } (  \Phi_{{\mathrm{2}}}  )   ~.
        \]
        The assumption $\Gamma  \vdash  \Phi_{{\mathrm{1}}}  \ottsym{<:}  \Phi_{{\mathrm{2}}}$ with \reflem{ts-weak-valid} implies
        \[
         \Gamma  \ottsym{,}  \mathit{y} \,  {:}  \,  \Sigma ^\omega   \ottsym{,}  \mathit{x'} \,  {:}  \,  \Sigma ^\ast   \models    { \Phi_{{\mathrm{1}}} }^\mu   \ottsym{(}  \mathit{x'}  \ottsym{)}    \Rightarrow     { \Phi_{{\mathrm{2}}} }^\mu   \ottsym{(}  \mathit{x'}  \ottsym{)} 
        \]
        and
        \[
         \Gamma  \ottsym{,}  \mathit{y} \,  {:}  \,  \Sigma ^\omega   \models    { \Phi_{{\mathrm{1}}} }^\nu   \ottsym{(}  \mathit{y}  \ottsym{)}    \Rightarrow     { \Phi_{{\mathrm{2}}} }^\nu   \ottsym{(}  \mathit{y}  \ottsym{)}  ~.
        \]
        Then, Lemmas~\ref{lem:ts-imp-strength-eff-hypo-formula-aux} and \ref{lem:ts-imp-or} imply the conclusion
        \[
         \Gamma  \ottsym{,}  \mathit{y} \,  {:}  \,  \Sigma ^\omega   \models    \mathit{H} _{  \nu  } (  \Phi_{{\mathrm{1}}}  )     \Rightarrow     \mathit{H} _{  \nu  } (  \Phi_{{\mathrm{2}}}  )   ~.
        \]
 \end{itemize}
\end{proof}

\begin{lemmap}{Transitivity of Effect Subtyping}{ts-transitivity-eff-subtyping}
 If $\Gamma  \vdash  \Phi_{{\mathrm{1}}}  \ottsym{<:}  \Phi_{{\mathrm{2}}}$ and $\Gamma  \vdash  \Phi_{{\mathrm{2}}}  \ottsym{<:}  \Phi_{{\mathrm{3}}}$, then $\Gamma  \vdash  \Phi_{{\mathrm{1}}}  \ottsym{<:}  \Phi_{{\mathrm{3}}}$.
\end{lemmap}
\begin{proof}
 We must show that
 \begin{itemize}
  \item $\Gamma  \ottsym{,}  \mathit{x} \,  {:}  \,  \Sigma ^\ast   \models    { \Phi_{{\mathrm{1}}} }^\mu   \ottsym{(}  \mathit{x}  \ottsym{)}    \Rightarrow     { \Phi_{{\mathrm{3}}} }^\mu   \ottsym{(}  \mathit{x}  \ottsym{)} $ and
  \item $\Gamma  \ottsym{,}  \mathit{x} \,  {:}  \,  \Sigma ^\omega   \models    { \Phi_{{\mathrm{1}}} }^\nu   \ottsym{(}  \mathit{x}  \ottsym{)}    \Rightarrow     { \Phi_{{\mathrm{3}}} }^\nu   \ottsym{(}  \mathit{x}  \ottsym{)} $
 \end{itemize}
 for $\mathit{x} \,  \not\in  \,  \mathit{fv}  (  \Phi_{{\mathrm{1}}}  )  \,  \mathbin{\cup}  \,  \mathit{fv}  (  \Phi_{{\mathrm{3}}}  ) $.
 Without loss of generality, we can suppose that $\mathit{x} \,  \not\in  \,  \mathit{fv}  (  \Phi_{{\mathrm{2}}}  ) $.

 The assumption $\Gamma  \vdash  \Phi_{{\mathrm{1}}}  \ottsym{<:}  \Phi_{{\mathrm{2}}}$ implies
 \begin{itemize}
  \item $\Gamma  \ottsym{,}  \mathit{x} \,  {:}  \,  \Sigma ^\ast   \models    { \Phi_{{\mathrm{1}}} }^\mu   \ottsym{(}  \mathit{x}  \ottsym{)}    \Rightarrow     { \Phi_{{\mathrm{2}}} }^\mu   \ottsym{(}  \mathit{x}  \ottsym{)} $ and
  \item $\Gamma  \ottsym{,}  \mathit{x} \,  {:}  \,  \Sigma ^\omega   \models    { \Phi_{{\mathrm{1}}} }^\nu   \ottsym{(}  \mathit{x}  \ottsym{)}    \Rightarrow     { \Phi_{{\mathrm{2}}} }^\nu   \ottsym{(}  \mathit{x}  \ottsym{)} $.
 \end{itemize}
 The assumption $\Gamma  \vdash  \Phi_{{\mathrm{2}}}  \ottsym{<:}  \Phi_{{\mathrm{3}}}$ implies
 \begin{itemize}
  \item $\Gamma  \ottsym{,}  \mathit{x} \,  {:}  \,  \Sigma ^\ast   \models    { \Phi_{{\mathrm{2}}} }^\mu   \ottsym{(}  \mathit{x}  \ottsym{)}    \Rightarrow     { \Phi_{{\mathrm{3}}} }^\mu   \ottsym{(}  \mathit{x}  \ottsym{)} $ and
  \item $\Gamma  \ottsym{,}  \mathit{x} \,  {:}  \,  \Sigma ^\omega   \models    { \Phi_{{\mathrm{2}}} }^\nu   \ottsym{(}  \mathit{x}  \ottsym{)}    \Rightarrow     { \Phi_{{\mathrm{3}}} }^\nu   \ottsym{(}  \mathit{x}  \ottsym{)} $.
 \end{itemize}
 Therefore, \reflem{ts-transitivity-valid} implies the conclusion.
\end{proof}

\begin{lemmap}{Strengthening Pre-Effects in Subtyping}{ts-strength-preeff-subtyping}
 Suppose that $\Gamma  \vdash  \Phi_{{\mathrm{1}}}  \ottsym{<:}  \Phi_{{\mathrm{2}}}$.
 \begin{enumerate}
  \item \label{lem:ts-strength-preeff-subtyping:cty}
        If $\Gamma  \vdash  \Phi_{{\mathrm{2}}} \,  \tceappendop  \, \ottnt{C_{{\mathrm{1}}}}  \ottsym{<:}  \ottnt{C_{{\mathrm{2}}}}$, then $\Gamma  \vdash  \Phi_{{\mathrm{1}}} \,  \tceappendop  \, \ottnt{C_{{\mathrm{1}}}}  \ottsym{<:}  \ottnt{C_{{\mathrm{2}}}}$.

  \item \label{lem:ts-strength-preeff-subtyping:sty}
        If $\Gamma \,  |  \, \ottnt{T} \,  \,\&\,  \, \Phi_{{\mathrm{2}}} \,  \tceappendop  \, \Phi  \vdash  \Phi_{{\mathrm{2}}} \,  \tceappendop  \, \ottnt{S_{{\mathrm{1}}}}  \ottsym{<:}  \ottnt{S_{{\mathrm{2}}}}$,
        then $\Gamma \,  |  \, \ottnt{T} \,  \,\&\,  \, \Phi_{{\mathrm{1}}} \,  \tceappendop  \, \Phi  \vdash  \Phi_{{\mathrm{1}}} \,  \tceappendop  \, \ottnt{S_{{\mathrm{1}}}}  \ottsym{<:}  \ottnt{S_{{\mathrm{2}}}}$.
 \end{enumerate}
\end{lemmap}
\begin{proof}
 By simultaneous induction on the derivations.
 \begin{enumerate}
  \item We are given $\Gamma  \vdash  \Phi_{{\mathrm{2}}} \,  \tceappendop  \, \ottnt{C_{{\mathrm{1}}}}  \ottsym{<:}  \ottnt{C_{{\mathrm{2}}}}$.
   We have $\Phi_{{\mathrm{2}}} \,  \tceappendop  \, \ottnt{C_{{\mathrm{1}}}} \,  =  \,   \ottnt{C_{{\mathrm{1}}}}  . \ottnt{T}   \,  \,\&\,  \,  \Phi_{{\mathrm{2}}} \,  \tceappendop  \, \ottsym{(}   \ottnt{C_{{\mathrm{1}}}}  . \Phi   \ottsym{)}  \, / \,  \Phi_{{\mathrm{2}}} \,  \tceappendop  \, \ottsym{(}   \ottnt{C_{{\mathrm{1}}}}  . \ottnt{S}   \ottsym{)} $.
   Because $\Gamma  \vdash  \Phi_{{\mathrm{2}}} \,  \tceappendop  \, \ottnt{C_{{\mathrm{1}}}}  \ottsym{<:}  \ottnt{C_{{\mathrm{2}}}}$ can be derived only by \Srule{Comp},
   its inversion implies
   \begin{itemize}
    \item $\Gamma  \vdash   \ottnt{C_{{\mathrm{1}}}}  . \ottnt{T}   \ottsym{<:}   \ottnt{C_{{\mathrm{2}}}}  . \ottnt{T} $,
    \item $\Gamma  \vdash  \Phi_{{\mathrm{2}}} \,  \tceappendop  \, \ottsym{(}   \ottnt{C_{{\mathrm{1}}}}  . \Phi   \ottsym{)}  \ottsym{<:}   \ottnt{C_{{\mathrm{2}}}}  . \Phi $, and
    \item $\Gamma \,  |  \,  \ottnt{C_{{\mathrm{1}}}}  . \ottnt{T}  \,  \,\&\,  \, \Phi_{{\mathrm{2}}} \,  \tceappendop  \, \ottsym{(}   \ottnt{C_{{\mathrm{1}}}}  . \Phi   \ottsym{)}  \vdash  \Phi_{{\mathrm{2}}} \,  \tceappendop  \, \ottsym{(}   \ottnt{C_{{\mathrm{1}}}}  . \ottnt{S}   \ottsym{)}  \ottsym{<:}   \ottnt{C_{{\mathrm{2}}}}  . \ottnt{S} $.
   \end{itemize}
   \reflem{ts-mono-eff-compose-preeff} with $\Gamma  \vdash  \Phi_{{\mathrm{1}}}  \ottsym{<:}  \Phi_{{\mathrm{2}}}$ implies
   $\Gamma  \vdash  \Phi_{{\mathrm{1}}} \,  \tceappendop  \, \ottsym{(}   \ottnt{C_{{\mathrm{1}}}}  . \Phi   \ottsym{)}  \ottsym{<:}  \Phi_{{\mathrm{2}}} \,  \tceappendop  \, \ottsym{(}   \ottnt{C_{{\mathrm{1}}}}  . \Phi   \ottsym{)}$.
   Thus, \reflem{ts-transitivity-eff-subtyping} with $\Gamma  \vdash  \Phi_{{\mathrm{2}}} \,  \tceappendop  \, \ottsym{(}   \ottnt{C_{{\mathrm{1}}}}  . \Phi   \ottsym{)}  \ottsym{<:}   \ottnt{C_{{\mathrm{2}}}}  . \Phi $
   implies
   \[
    \Gamma  \vdash  \Phi_{{\mathrm{1}}} \,  \tceappendop  \, \ottsym{(}   \ottnt{C_{{\mathrm{1}}}}  . \Phi   \ottsym{)}  \ottsym{<:}   \ottnt{C_{{\mathrm{2}}}}  . \Phi  ~.
   \]
   Further, the IH implies
   \[
    \Gamma \,  |  \,  \ottnt{C_{{\mathrm{1}}}}  . \ottnt{T}  \,  \,\&\,  \, \Phi_{{\mathrm{1}}} \,  \tceappendop  \, \ottsym{(}   \ottnt{C_{{\mathrm{1}}}}  . \Phi   \ottsym{)}  \vdash  \Phi_{{\mathrm{1}}} \,  \tceappendop  \, \ottsym{(}   \ottnt{C_{{\mathrm{1}}}}  . \ottnt{S}   \ottsym{)}  \ottsym{<:}   \ottnt{C_{{\mathrm{2}}}}  . \ottnt{S}  ~.
   \]
   Then, \Srule{Comp} implies the conclusion
   \[
    \Gamma  \vdash  \Phi_{{\mathrm{1}}} \,  \tceappendop  \, \ottnt{C_{{\mathrm{1}}}}  \ottsym{<:}  \ottnt{C_{{\mathrm{2}}}} ~.
   \]

  \item By case analysis on the inference rule applied last to derive
        $\Gamma \,  |  \, \ottnt{T} \,  \,\&\,  \, \Phi_{{\mathrm{2}}} \,  \tceappendop  \, \Phi  \vdash  \Phi_{{\mathrm{2}}} \,  \tceappendop  \, \ottnt{S_{{\mathrm{1}}}}  \ottsym{<:}  \ottnt{S_{{\mathrm{2}}}}$.
        \begin{caseanalysis}
         \case \Srule{Empty}: Trivial by \Srule{Empty}
          because $\ottnt{S_{{\mathrm{1}}}} = \ottnt{S_{{\mathrm{2}}}} =  \square $.

         \case \Srule{Ans}:
          We are given
          \begin{prooftree}
           \AxiomC{$\Gamma  \ottsym{,}  \mathit{x} \,  {:}  \, \ottnt{T}  \vdash  \ottnt{C_{{\mathrm{21}}}}  \ottsym{<:}  \ottnt{C_{{\mathrm{11}}}}$}
           \AxiomC{$\Gamma  \vdash  \Phi_{{\mathrm{2}}} \,  \tceappendop  \, \ottnt{C_{{\mathrm{12}}}}  \ottsym{<:}  \ottnt{C_{{\mathrm{22}}}}$}
           \BinaryInfC{$\Gamma \,  |  \, \ottnt{T} \,  \,\&\,  \, \Phi_{{\mathrm{2}}} \,  \tceappendop  \, \Phi  \vdash   ( \forall  \mathit{x}  .  \ottnt{C_{{\mathrm{11}}}}  )  \Rightarrow   \Phi_{{\mathrm{2}}} \,  \tceappendop  \, \ottnt{C_{{\mathrm{12}}}}   \ottsym{<:}   ( \forall  \mathit{x}  .  \ottnt{C_{{\mathrm{21}}}}  )  \Rightarrow   \ottnt{C_{{\mathrm{22}}}} $}
          \end{prooftree}
          for some $\mathit{x}$, $\ottnt{C_{{\mathrm{11}}}}$, $\ottnt{C_{{\mathrm{12}}}}$, $\ottnt{C_{{\mathrm{21}}}}$, and $\ottnt{C_{{\mathrm{22}}}}$ such that
          $\ottnt{S_{{\mathrm{1}}}} \,  =  \,  ( \forall  \mathit{x}  .  \ottnt{C_{{\mathrm{11}}}}  )  \Rightarrow   \ottnt{C_{{\mathrm{12}}}} $ and $\ottnt{S_{{\mathrm{2}}}} \,  =  \,  ( \forall  \mathit{x}  .  \ottnt{C_{{\mathrm{21}}}}  )  \Rightarrow   \ottnt{C_{{\mathrm{22}}}} $.
          The IH implies $\Gamma  \vdash  \Phi_{{\mathrm{1}}} \,  \tceappendop  \, \ottnt{C_{{\mathrm{12}}}}  \ottsym{<:}  \ottnt{C_{{\mathrm{22}}}}$.
          Thus, \Srule{Ans} implies
          \[
           \Gamma \,  |  \, \ottnt{T} \,  \,\&\,  \, \Phi_{{\mathrm{1}}} \,  \tceappendop  \, \Phi  \vdash   ( \forall  \mathit{x}  .  \ottnt{C_{{\mathrm{11}}}}  )  \Rightarrow   \Phi_{{\mathrm{1}}} \,  \tceappendop  \, \ottnt{C_{{\mathrm{12}}}}   \ottsym{<:}   ( \forall  \mathit{x}  .  \ottnt{C_{{\mathrm{21}}}}  )  \Rightarrow   \ottnt{C_{{\mathrm{22}}}}  ~,
          \]
          which is equivalent to the conclusion $\Gamma \,  |  \, \ottnt{T} \,  \,\&\,  \, \Phi_{{\mathrm{1}}} \,  \tceappendop  \, \Phi  \vdash  \Phi_{{\mathrm{1}}} \,  \tceappendop  \, \ottnt{S_{{\mathrm{1}}}}  \ottsym{<:}  \ottnt{S_{{\mathrm{2}}}}$.

         \case \Srule{Embed}:
          We are given
          \begin{prooftree}
           \AxiomC{$\Gamma  \ottsym{,}  \mathit{x} \,  {:}  \, \ottnt{T}  \vdash  \ottsym{(}  \Phi_{{\mathrm{2}}} \,  \tceappendop  \, \Phi  \ottsym{)} \,  \tceappendop  \, \ottnt{C_{{\mathrm{21}}}}  \ottsym{<:}  \ottnt{C_{{\mathrm{22}}}}$}
           \AxiomC{$\Gamma  \ottsym{,}  \mathit{y} \,  {:}  \,  \Sigma ^\omega   \models    { \ottsym{(}  \Phi_{{\mathrm{2}}} \,  \tceappendop  \, \Phi  \ottsym{)} }^\nu   \ottsym{(}  \mathit{y}  \ottsym{)}    \Rightarrow     { \ottnt{C_{{\mathrm{22}}}} }^\nu (  \mathit{y}  )  $}
           \AxiomC{$\mathit{x}  \ottsym{,}  \mathit{y} \,  \not\in  \,  \mathit{fv}  (  \ottnt{C_{{\mathrm{22}}}}  )  \,  \mathbin{\cup}  \,  \mathit{fv}  (  \Phi_{{\mathrm{2}}} \,  \tceappendop  \, \Phi  ) $}
           \TrinaryInfC{$\Gamma \,  |  \, \ottnt{T} \,  \,\&\,  \, \Phi_{{\mathrm{2}}} \,  \tceappendop  \, \Phi  \vdash   \square   \ottsym{<:}   ( \forall  \mathit{x}  .  \ottnt{C_{{\mathrm{21}}}}  )  \Rightarrow   \ottnt{C_{{\mathrm{22}}}} $}
          \end{prooftree}
          for some $\mathit{x}$, $\mathit{y}$, $\ottnt{C_{{\mathrm{21}}}}$, and $\ottnt{C_{{\mathrm{22}}}}$ such that
          $\ottnt{S_{{\mathrm{2}}}} \,  =  \,  ( \forall  \mathit{x}  .  \ottnt{C_{{\mathrm{21}}}}  )  \Rightarrow   \ottnt{C_{{\mathrm{22}}}} $.
          We also have $\ottnt{S_{{\mathrm{1}}}} \,  =  \,  \square $.
          Without loss of generality, we can suppose that $\mathit{x}  \ottsym{,}  \mathit{y} \,  \not\in  \,  \mathit{fv}  (  \Phi_{{\mathrm{1}}} \,  \tceappendop  \, \Phi  ) $.
          \reflem{ts-mono-eff-compose-preeff} with the assumption $\Gamma  \vdash  \Phi_{{\mathrm{1}}}  \ottsym{<:}  \Phi_{{\mathrm{2}}}$
          implies $\Gamma  \vdash  \Phi_{{\mathrm{1}}} \,  \tceappendop  \, \Phi  \ottsym{<:}  \Phi_{{\mathrm{2}}} \,  \tceappendop  \, \Phi$.
          Thus, the IH implies
          \[
           \Gamma  \ottsym{,}  \mathit{x} \,  {:}  \, \ottnt{T}  \vdash  \ottsym{(}  \Phi_{{\mathrm{1}}} \,  \tceappendop  \, \Phi  \ottsym{)} \,  \tceappendop  \, \ottnt{C_{{\mathrm{21}}}}  \ottsym{<:}  \ottnt{C_{{\mathrm{22}}}} ~.
          \]
          Further, $\Gamma  \vdash  \Phi_{{\mathrm{1}}} \,  \tceappendop  \, \Phi  \ottsym{<:}  \Phi_{{\mathrm{2}}} \,  \tceappendop  \, \Phi$ implies that
          $\Gamma  \ottsym{,}  \mathit{y} \,  {:}  \,  \Sigma ^\omega   \models    { \ottsym{(}  \Phi_{{\mathrm{1}}} \,  \tceappendop  \, \Phi  \ottsym{)} }^\nu   \ottsym{(}  \mathit{y}  \ottsym{)}    \Rightarrow     { \ottsym{(}  \Phi_{{\mathrm{2}}} \,  \tceappendop  \, \Phi  \ottsym{)} }^\nu   \ottsym{(}  \mathit{y}  \ottsym{)} $.
          Therefore, by \reflem{ts-transitivity-valid} with $\Gamma  \ottsym{,}  \mathit{y} \,  {:}  \,  \Sigma ^\omega   \models    { \ottsym{(}  \Phi_{{\mathrm{2}}} \,  \tceappendop  \, \Phi  \ottsym{)} }^\nu   \ottsym{(}  \mathit{y}  \ottsym{)}    \Rightarrow     { \ottnt{C_{{\mathrm{22}}}} }^\nu (  \mathit{y}  )  $,
          we have
          \[
           \Gamma  \ottsym{,}  \mathit{y} \,  {:}  \,  \Sigma ^\omega   \models    { \ottsym{(}  \Phi_{{\mathrm{1}}} \,  \tceappendop  \, \Phi  \ottsym{)} }^\nu   \ottsym{(}  \mathit{y}  \ottsym{)}    \Rightarrow     { \ottnt{C_{{\mathrm{22}}}} }^\nu (  \mathit{y}  )   ~.
          \]
          Then, \Srule{Embed} implies
          \[
           \Gamma \,  |  \, \ottnt{T} \,  \,\&\,  \, \Phi_{{\mathrm{1}}} \,  \tceappendop  \, \Phi  \vdash   \square   \ottsym{<:}   ( \forall  \mathit{x}  .  \ottnt{C_{{\mathrm{21}}}}  )  \Rightarrow   \ottnt{C_{{\mathrm{22}}}}  ~.
          \]
          Because $ \square  \,  =  \, \Phi_{{\mathrm{1}}} \,  \tceappendop  \, \ottnt{S_{{\mathrm{1}}}}$, we have the conclusion.
        \end{caseanalysis}
 \end{enumerate}
\end{proof}

\begin{lemmap}{Strengthening Pre-Effects in Stack Subtyping}{ts-strength-preeff-stack-subtyping}
 If
 $\Gamma  \vdash  \Phi_{{\mathrm{1}}}  \ottsym{<:}  \Phi_{{\mathrm{2}}}$ and
 $\Gamma \,  |  \, \ottnt{T} \,  \,\&\,  \, \Phi_{{\mathrm{2}}}  \vdash  \ottnt{S_{{\mathrm{1}}}}  \ottsym{<:}  \ottnt{S_{{\mathrm{2}}}}$,
 then
 $\Gamma \,  |  \, \ottnt{T} \,  \,\&\,  \, \Phi_{{\mathrm{1}}}  \vdash  \ottnt{S_{{\mathrm{1}}}}  \ottsym{<:}  \ottnt{S_{{\mathrm{2}}}}$.
\end{lemmap}
\begin{proof}
 By case analysis on the inference rule applied last to derive
 $\Gamma \,  |  \, \ottnt{T} \,  \,\&\,  \, \Phi_{{\mathrm{2}}}  \vdash  \ottnt{S_{{\mathrm{1}}}}  \ottsym{<:}  \ottnt{S_{{\mathrm{2}}}}$.
 \begin{caseanalysis}
  \case \Srule{Empty} and \Srule{Ans}: Trivial by \Srule{Empty} and \Srule{Ans},
   respectively, because $\Phi_{{\mathrm{2}}}$ is not referenced in the premises.

  \case \Srule{Embed}:
   We are given
   \begin{prooftree}
    \AxiomC{$\Gamma  \ottsym{,}  \mathit{x} \,  {:}  \, \ottnt{T}  \vdash  \Phi_{{\mathrm{2}}} \,  \tceappendop  \, \ottnt{C_{{\mathrm{21}}}}  \ottsym{<:}  \ottnt{C_{{\mathrm{22}}}}$}
    \AxiomC{$\Gamma  \ottsym{,}  \mathit{y} \,  {:}  \,  \Sigma ^\omega   \models    { \Phi_{{\mathrm{2}}} }^\nu   \ottsym{(}  \mathit{y}  \ottsym{)}    \Rightarrow     { \ottnt{C_{{\mathrm{22}}}} }^\nu (  \mathit{y}  )  $}
    \AxiomC{$\mathit{x}  \ottsym{,}  \mathit{y} \,  \not\in  \,  \mathit{fv}  (  \ottnt{C_{{\mathrm{22}}}}  )  \,  \mathbin{\cup}  \,  \mathit{fv}  (  \Phi_{{\mathrm{2}}}  ) $}
    \TrinaryInfC{$\Gamma \,  |  \, \ottnt{T} \,  \,\&\,  \, \Phi_{{\mathrm{2}}}  \vdash   \square   \ottsym{<:}   ( \forall  \mathit{x}  .  \ottnt{C_{{\mathrm{21}}}}  )  \Rightarrow   \ottnt{C_{{\mathrm{22}}}} $}
   \end{prooftree}
   for some $\mathit{x}$, $\mathit{y}$, $\ottnt{C_{{\mathrm{21}}}}$, and $\ottnt{C_{{\mathrm{22}}}}$ such that
   $\ottnt{S_{{\mathrm{2}}}} \,  =  \,  ( \forall  \mathit{x}  .  \ottnt{C_{{\mathrm{21}}}}  )  \Rightarrow   \ottnt{C_{{\mathrm{22}}}} $.
   We also have $\ottnt{S_{{\mathrm{1}}}} \,  =  \,  \square $.
   Without loss of generality, we can suppose that $\mathit{x}  \ottsym{,}  \mathit{y} \,  \not\in  \,  \mathit{fv}  (  \Phi_{{\mathrm{1}}}  )  \,  \mathbin{\cup}  \,  \mathit{dom}  (  \Gamma  ) $.
   \reflem{ts-weak-subtyping} with $\Gamma  \vdash  \Phi_{{\mathrm{1}}}  \ottsym{<:}  \Phi_{{\mathrm{2}}}$ implies
   $\Gamma  \ottsym{,}  \mathit{x} \,  {:}  \, \ottnt{T}  \vdash  \Phi_{{\mathrm{1}}}  \ottsym{<:}  \Phi_{{\mathrm{2}}}$.
   Thus, \reflem{ts-strength-preeff-subtyping} implies
   \[
    \Gamma  \ottsym{,}  \mathit{x} \,  {:}  \, \ottnt{T}  \vdash  \Phi_{{\mathrm{1}}} \,  \tceappendop  \, \ottnt{C_{{\mathrm{21}}}}  \ottsym{<:}  \ottnt{C_{{\mathrm{22}}}} ~.
   \]
   Further, $\Gamma  \vdash  \Phi_{{\mathrm{1}}}  \ottsym{<:}  \Phi_{{\mathrm{2}}}$ implies $\Gamma  \ottsym{,}  \mathit{y} \,  {:}  \,  \Sigma ^\omega   \models    { \Phi_{{\mathrm{1}}} }^\nu   \ottsym{(}  \mathit{y}  \ottsym{)}    \Rightarrow     { \Phi_{{\mathrm{2}}} }^\nu   \ottsym{(}  \mathit{y}  \ottsym{)} $.
   Then, \reflem{ts-transitivity-valid} with $\Gamma  \ottsym{,}  \mathit{y} \,  {:}  \,  \Sigma ^\omega   \models    { \Phi_{{\mathrm{2}}} }^\nu   \ottsym{(}  \mathit{y}  \ottsym{)}    \Rightarrow     { \ottnt{C_{{\mathrm{22}}}} }^\nu (  \mathit{y}  )  $
   implies
   \[
    \Gamma  \ottsym{,}  \mathit{y} \,  {:}  \,  \Sigma ^\omega   \models    { \Phi_{{\mathrm{1}}} }^\nu   \ottsym{(}  \mathit{y}  \ottsym{)}    \Rightarrow     { \ottnt{C_{{\mathrm{22}}}} }^\nu (  \mathit{y}  )   ~.
   \]
   Then, \Srule{Embed} implies the conclusion
   \[
    \Gamma \,  |  \, \ottnt{T} \,  \,\&\,  \, \Phi_{{\mathrm{1}}}  \vdash   \square   \ottsym{<:}   ( \forall  \mathit{x}  .  \ottnt{C_{{\mathrm{21}}}}  )  \Rightarrow   \ottnt{C_{{\mathrm{22}}}}  ~.
   \]
 \end{caseanalysis}
\end{proof}

\begin{lemmap}{Associativity of Pre-Effect Composition in Effect Subtyping}{ts-assoc-preeff-compose-eff-subtyping}
 For any $\Gamma$, $\Phi_{{\mathrm{1}}}$, $\Phi_{{\mathrm{2}}}$, and $\Phi_{{\mathrm{3}}}$,
 \begin{enumerate}
  \item $\Gamma  \vdash  \ottsym{(}  \Phi_{{\mathrm{1}}} \,  \tceappendop  \, \Phi_{{\mathrm{2}}}  \ottsym{)} \,  \tceappendop  \, \Phi_{{\mathrm{3}}}  \ottsym{<:}  \Phi_{{\mathrm{1}}} \,  \tceappendop  \, \ottsym{(}  \Phi_{{\mathrm{2}}} \,  \tceappendop  \, \Phi_{{\mathrm{3}}}  \ottsym{)}$ and
  \item $\Gamma  \vdash  \Phi_{{\mathrm{1}}} \,  \tceappendop  \, \ottsym{(}  \Phi_{{\mathrm{2}}} \,  \tceappendop  \, \Phi_{{\mathrm{3}}}  \ottsym{)}  \ottsym{<:}  \ottsym{(}  \Phi_{{\mathrm{1}}} \,  \tceappendop  \, \Phi_{{\mathrm{2}}}  \ottsym{)} \,  \tceappendop  \, \Phi_{{\mathrm{3}}}$.
 \end{enumerate}
\end{lemmap}
\begin{proof}
 \noindent
 \begin{enumerate}
  \item By the following.
        \begin{itemize}
         \item We show that
               \[
                \Gamma  \ottsym{,}  \mathit{x} \,  {:}  \,  \Sigma ^\ast   \models    { \ottsym{(}  \ottsym{(}  \Phi_{{\mathrm{1}}} \,  \tceappendop  \, \Phi_{{\mathrm{2}}}  \ottsym{)} \,  \tceappendop  \, \Phi_{{\mathrm{3}}}  \ottsym{)} }^\mu   \ottsym{(}  \mathit{x}  \ottsym{)}    \Rightarrow     { \ottsym{(}  \Phi_{{\mathrm{1}}} \,  \tceappendop  \, \ottsym{(}  \Phi_{{\mathrm{2}}} \,  \tceappendop  \, \Phi_{{\mathrm{3}}}  \ottsym{)}  \ottsym{)} }^\mu   \ottsym{(}  \mathit{x}  \ottsym{)} 
               \]
               for $\mathit{x} \,  \not\in  \,  \mathit{fv}  (  \ottsym{(}  \Phi_{{\mathrm{1}}} \,  \tceappendop  \, \Phi_{{\mathrm{2}}}  \ottsym{)} \,  \tceappendop  \, \Phi_{{\mathrm{3}}}  )  \,  \mathbin{\cup}  \,  \mathit{fv}  (  \Phi_{{\mathrm{1}}} \,  \tceappendop  \, \ottsym{(}  \Phi_{{\mathrm{2}}} \,  \tceappendop  \, \Phi_{{\mathrm{3}}}  \ottsym{)}  ) $.
               We have
               \[\begin{array}{ll} &
                 { \ottsym{(}  \ottsym{(}  \Phi_{{\mathrm{1}}} \,  \tceappendop  \, \Phi_{{\mathrm{2}}}  \ottsym{)} \,  \tceappendop  \, \Phi_{{\mathrm{3}}}  \ottsym{)} }^\mu   \ottsym{(}  \mathit{x}  \ottsym{)} \\ =&
                   \exists  \,  \mathit{x_{{\mathrm{12}}}}  \mathrel{  \in  }   \Sigma ^\ast   . \    \exists  \,  \mathit{x_{{\mathrm{3}}}}  \mathrel{  \in  }   \Sigma ^\ast   . \     \mathit{x}    =    \mathit{x_{{\mathrm{12}}}} \,  \tceappendop  \, \mathit{x_{{\mathrm{3}}}}     \wedge     { \ottsym{(}  \Phi_{{\mathrm{1}}} \,  \tceappendop  \, \Phi_{{\mathrm{2}}}  \ottsym{)} }^\mu   \ottsym{(}  \mathit{x_{{\mathrm{12}}}}  \ottsym{)}     \wedge     { \Phi_{{\mathrm{3}}} }^\mu   \ottsym{(}  \mathit{x_{{\mathrm{3}}}}  \ottsym{)}    \\ =&
                   \exists  \,  \mathit{x_{{\mathrm{12}}}}  \mathrel{  \in  }   \Sigma ^\ast   . \    \exists  \,  \mathit{x_{{\mathrm{3}}}}  \mathrel{  \in  }   \Sigma ^\ast   . \     \mathit{x}    =    \mathit{x_{{\mathrm{12}}}} \,  \tceappendop  \, \mathit{x_{{\mathrm{3}}}}     \wedge    \ottsym{(}    \exists  \,  \mathit{x_{{\mathrm{1}}}}  \mathrel{  \in  }   \Sigma ^\ast   . \    \exists  \,  \mathit{x_{{\mathrm{2}}}}  \mathrel{  \in  }   \Sigma ^\ast   . \     \mathit{x_{{\mathrm{12}}}}    =    \mathit{x_{{\mathrm{1}}}} \,  \tceappendop  \, \mathit{x_{{\mathrm{2}}}}     \wedge     { \Phi_{{\mathrm{1}}} }^\mu   \ottsym{(}  \mathit{x_{{\mathrm{1}}}}  \ottsym{)}     \wedge     { \Phi_{{\mathrm{2}}} }^\mu   \ottsym{(}  \mathit{x_{{\mathrm{2}}}}  \ottsym{)}     \ottsym{)}     \wedge     { \Phi_{{\mathrm{3}}} }^\mu   \ottsym{(}  \mathit{x_{{\mathrm{3}}}}  \ottsym{)}    ~.
                 \end{array}
               \]
               Then, it implies
               \[\begin{array}{ll} &
                  \exists  \,  \mathit{x_{{\mathrm{1}}}}  \mathrel{  \in  }   \Sigma ^\ast   . \    \exists  \,  \mathit{x_{{\mathrm{23}}}}  \mathrel{  \in  }   \Sigma ^\ast   . \     \mathit{x}    =    \mathit{x_{{\mathrm{1}}}} \,  \tceappendop  \, \mathit{x_{{\mathrm{23}}}}     \wedge     { \Phi_{{\mathrm{1}}} }^\mu   \ottsym{(}  \mathit{x_{{\mathrm{1}}}}  \ottsym{)}     \wedge    \ottsym{(}    \exists  \,  \mathit{x_{{\mathrm{2}}}}  \mathrel{  \in  }   \Sigma ^\ast   . \    \exists  \,  \mathit{x_{{\mathrm{3}}}}  \mathrel{  \in  }   \Sigma ^\ast   . \     \mathit{x_{{\mathrm{23}}}}    =    \mathit{x_{{\mathrm{2}}}} \,  \tceappendop  \, \mathit{x_{{\mathrm{3}}}}     \wedge     { \Phi_{{\mathrm{2}}} }^\mu   \ottsym{(}  \mathit{x_{{\mathrm{2}}}}  \ottsym{)}     \wedge     { \Phi_{{\mathrm{3}}} }^\mu   \ottsym{(}  \mathit{x_{{\mathrm{3}}}}  \ottsym{)}     \ottsym{)}    \\ =&
                  \exists  \,  \mathit{x_{{\mathrm{1}}}}  \mathrel{  \in  }   \Sigma ^\ast   . \    \exists  \,  \mathit{x_{{\mathrm{23}}}}  \mathrel{  \in  }   \Sigma ^\ast   . \     \mathit{x}    =    \mathit{x_{{\mathrm{1}}}} \,  \tceappendop  \, \mathit{x_{{\mathrm{23}}}}     \wedge     { \Phi_{{\mathrm{1}}} }^\mu   \ottsym{(}  \mathit{x_{{\mathrm{1}}}}  \ottsym{)}     \wedge     { \ottsym{(}  \Phi_{{\mathrm{2}}} \,  \tceappendop  \, \Phi_{{\mathrm{3}}}  \ottsym{)} }^\mu   \ottsym{(}  \mathit{x_{{\mathrm{23}}}}  \ottsym{)}    \\ =&
                 { \ottsym{(}  \Phi_{{\mathrm{1}}} \,  \tceappendop  \, \ottsym{(}  \Phi_{{\mathrm{2}}} \,  \tceappendop  \, \Phi_{{\mathrm{3}}}  \ottsym{)}  \ottsym{)} }^\mu   \ottsym{(}  \mathit{x}  \ottsym{)} ~.
                 \end{array}
               \]

         \item We show that
               \[
                \Gamma  \ottsym{,}  \mathit{y} \,  {:}  \,  \Sigma ^\omega   \models    { \ottsym{(}  \ottsym{(}  \Phi_{{\mathrm{1}}} \,  \tceappendop  \, \Phi_{{\mathrm{2}}}  \ottsym{)} \,  \tceappendop  \, \Phi_{{\mathrm{3}}}  \ottsym{)} }^\nu   \ottsym{(}  \mathit{y}  \ottsym{)}    \Rightarrow     { \ottsym{(}  \Phi_{{\mathrm{1}}} \,  \tceappendop  \, \ottsym{(}  \Phi_{{\mathrm{2}}} \,  \tceappendop  \, \Phi_{{\mathrm{3}}}  \ottsym{)}  \ottsym{)} }^\nu   \ottsym{(}  \mathit{y}  \ottsym{)} 
               \]
               for $\mathit{y} \,  \not\in  \,  \mathit{fv}  (  \ottsym{(}  \Phi_{{\mathrm{1}}} \,  \tceappendop  \, \Phi_{{\mathrm{2}}}  \ottsym{)} \,  \tceappendop  \, \Phi_{{\mathrm{3}}}  )  \,  \mathbin{\cup}  \,  \mathit{fv}  (  \Phi_{{\mathrm{1}}} \,  \tceappendop  \, \ottsym{(}  \Phi_{{\mathrm{2}}} \,  \tceappendop  \, \Phi_{{\mathrm{3}}}  \ottsym{)}  ) $.
               Let
               \[
                \phi_{{\mathrm{3}}} \,  =  \,   \exists  \,  \mathit{x_{{\mathrm{12}}}}  \mathrel{  \in  }   \Sigma ^\ast   . \    \exists  \,  \mathit{y_{{\mathrm{3}}}}  \mathrel{  \in  }   \Sigma ^\omega   . \     \mathit{y}    =    \mathit{x_{{\mathrm{12}}}} \,  \tceappendop  \, \mathit{y_{{\mathrm{3}}}}     \wedge     { \ottsym{(}  \Phi_{{\mathrm{1}}} \,  \tceappendop  \, \Phi_{{\mathrm{2}}}  \ottsym{)} }^\mu   \ottsym{(}  \mathit{x_{{\mathrm{12}}}}  \ottsym{)}     \wedge     { \Phi_{{\mathrm{3}}} }^\nu   \ottsym{(}  \mathit{y_{{\mathrm{3}}}}  \ottsym{)}    ~.
               \]
               Then,
               \[\begin{array}{ll} &
                 { \ottsym{(}  \ottsym{(}  \Phi_{{\mathrm{1}}} \,  \tceappendop  \, \Phi_{{\mathrm{2}}}  \ottsym{)} \,  \tceappendop  \, \Phi_{{\mathrm{3}}}  \ottsym{)} }^\nu   \ottsym{(}  \mathit{y}  \ottsym{)} \\ =&
                   { \ottsym{(}  \Phi_{{\mathrm{1}}} \,  \tceappendop  \, \Phi_{{\mathrm{2}}}  \ottsym{)} }^\nu   \ottsym{(}  \mathit{y}  \ottsym{)}    \vee    \phi_{{\mathrm{3}}}  \\ =&
                  \ottsym{(}      { \Phi_{{\mathrm{1}}} }^\nu   \ottsym{(}  \mathit{y}  \ottsym{)}    \vee      \exists  \,  \mathit{x_{{\mathrm{1}}}}  \mathrel{  \in  }   \Sigma ^\ast   . \    \exists  \,  \mathit{y_{{\mathrm{2}}}}  \mathrel{  \in  }   \Sigma ^\omega   . \   \mathit{y}    =    \mathit{x_{{\mathrm{1}}}} \,  \tceappendop  \, \mathit{y_{{\mathrm{2}}}}        \wedge     { \Phi_{{\mathrm{1}}} }^\mu   \ottsym{(}  \mathit{x_{{\mathrm{1}}}}  \ottsym{)}     \wedge     { \Phi_{{\mathrm{2}}} }^\nu   \ottsym{(}  \mathit{y}  \ottsym{)}   \ottsym{)}    \vee    \phi_{{\mathrm{3}}} 
                 \end{array}
               \]
               Then, it implies
               \[
                    { \Phi_{{\mathrm{1}}} }^\nu   \ottsym{(}  \mathit{y}  \ottsym{)}    \vee      \exists  \,  \mathit{x_{{\mathrm{1}}}}  \mathrel{  \in  }   \Sigma ^\ast   . \    \exists  \,  \mathit{y_{{\mathrm{23}}}}  \mathrel{  \in  }   \Sigma ^\omega   . \   \mathit{y}    =    \mathit{x_{{\mathrm{1}}}} \,  \tceappendop  \, \mathit{y_{{\mathrm{23}}}}        \wedge     { \Phi_{{\mathrm{1}}} }^\mu   \ottsym{(}  \mathit{x_{{\mathrm{1}}}}  \ottsym{)}     \wedge    \phi_{{\mathrm{23}}} 
               \]
               where
               \[
                 \phi_{{\mathrm{23}}} \,  =  \,     { \Phi_{{\mathrm{2}}} }^\nu   \ottsym{(}  \mathit{y_{{\mathrm{23}}}}  \ottsym{)}    \vee      \exists  \,  \mathit{x_{{\mathrm{2}}}}  \mathrel{  \in  }   \Sigma ^\ast   . \    \exists  \,  \mathit{y_{{\mathrm{3}}}}  \mathrel{  \in  }   \Sigma ^\omega   . \   \mathit{y_{{\mathrm{23}}}}    =    \mathit{x_{{\mathrm{2}}}} \,  \tceappendop  \, \mathit{y_{{\mathrm{3}}}}        \wedge     { \Phi_{{\mathrm{2}}} }^\mu   \ottsym{(}  \mathit{x_{{\mathrm{2}}}}  \ottsym{)}     \wedge     { \Phi_{{\mathrm{3}}} }^\nu   \ottsym{(}  \mathit{y_{{\mathrm{3}}}}  \ottsym{)}  ~.
               \]
               Because $\phi_{{\mathrm{23}}}$ is equivalent to $ { \ottsym{(}  \Phi_{{\mathrm{2}}} \,  \tceappendop  \, \Phi_{{\mathrm{3}}}  \ottsym{)} }^\nu   \ottsym{(}  \mathit{y_{{\mathrm{23}}}}  \ottsym{)}$, we can find
               $ { \ottsym{(}  \ottsym{(}  \Phi_{{\mathrm{1}}} \,  \tceappendop  \, \Phi_{{\mathrm{2}}}  \ottsym{)} \,  \tceappendop  \, \Phi_{{\mathrm{3}}}  \ottsym{)} }^\nu   \ottsym{(}  \mathit{y}  \ottsym{)}$ implies $ { \ottsym{(}  \Phi_{{\mathrm{1}}} \,  \tceappendop  \, \ottsym{(}  \Phi_{{\mathrm{2}}} \,  \tceappendop  \, \Phi_{{\mathrm{3}}}  \ottsym{)}  \ottsym{)} }^\nu $.
        \end{itemize}

  \item By converse of the above case.
 \end{enumerate}
\end{proof}

\begin{lemmap}{Associativity of Pre-Effect Composition in Subtyping: Left to Right}{ts-assoc-preeff-compose-subtyping-left-to-right}
 \begin{enumerate}
  \item If $\Gamma  \vdash  \Phi_{{\mathrm{1}}} \,  \tceappendop  \, \ottsym{(}  \Phi_{{\mathrm{2}}} \,  \tceappendop  \, \ottnt{C_{{\mathrm{1}}}}  \ottsym{)}  \ottsym{<:}  \ottnt{C_{{\mathrm{2}}}}$,
        then $\Gamma  \vdash  \ottsym{(}  \Phi_{{\mathrm{1}}} \,  \tceappendop  \, \Phi_{{\mathrm{2}}}  \ottsym{)} \,  \tceappendop  \, \ottnt{C_{{\mathrm{1}}}}  \ottsym{<:}  \ottnt{C_{{\mathrm{2}}}}$.
  \item If $\Gamma \,  |  \, \ottnt{T} \,  \,\&\,  \, \Phi_{{\mathrm{1}}} \,  \tceappendop  \, \ottsym{(}  \Phi_{{\mathrm{2}}} \,  \tceappendop  \, \Phi  \ottsym{)}  \vdash  \Phi_{{\mathrm{1}}} \,  \tceappendop  \, \ottsym{(}  \Phi_{{\mathrm{2}}} \,  \tceappendop  \, \ottnt{S_{{\mathrm{1}}}}  \ottsym{)}  \ottsym{<:}  \ottnt{S_{{\mathrm{2}}}}$,
        then $\Gamma \,  |  \, \ottnt{T} \,  \,\&\,  \, \ottsym{(}  \Phi_{{\mathrm{1}}} \,  \tceappendop  \, \Phi_{{\mathrm{2}}}  \ottsym{)} \,  \tceappendop  \, \Phi  \vdash  \ottsym{(}  \Phi_{{\mathrm{1}}} \,  \tceappendop  \, \Phi_{{\mathrm{2}}}  \ottsym{)} \,  \tceappendop  \, \ottnt{S_{{\mathrm{1}}}}  \ottsym{<:}  \ottnt{S_{{\mathrm{2}}}}$.
 \end{enumerate}
\end{lemmap}
\begin{proof}
 By simultaneous induction on the derivations.
 \begin{caseanalysis}
  \case \Srule{Comp}:
   Because $\Phi_{{\mathrm{1}}} \,  \tceappendop  \, \ottsym{(}  \Phi_{{\mathrm{2}}} \,  \tceappendop  \, \ottnt{C_{{\mathrm{1}}}}  \ottsym{)} \,  =  \,   \ottnt{C_{{\mathrm{1}}}}  . \ottnt{T}   \,  \,\&\,  \,  \Phi_{{\mathrm{1}}} \,  \tceappendop  \, \ottsym{(}  \Phi_{{\mathrm{2}}} \,  \tceappendop  \, \ottsym{(}   \ottnt{C_{{\mathrm{1}}}}  . \Phi   \ottsym{)}  \ottsym{)}  \, / \,  \Phi_{{\mathrm{1}}} \,  \tceappendop  \, \ottsym{(}  \Phi_{{\mathrm{2}}} \,  \tceappendop  \, \ottsym{(}   \ottnt{C_{{\mathrm{1}}}}  . \ottnt{S}   \ottsym{)}  \ottsym{)} $,
   we are given
   \begin{prooftree}
    \AxiomC{$\Gamma  \vdash   \ottnt{C_{{\mathrm{1}}}}  . \ottnt{T}   \ottsym{<:}   \ottnt{C_{{\mathrm{2}}}}  . \ottnt{T} $}
    \AxiomC{$\Gamma  \vdash  \Phi_{{\mathrm{1}}} \,  \tceappendop  \, \ottsym{(}  \Phi_{{\mathrm{2}}} \,  \tceappendop  \, \ottsym{(}   \ottnt{C_{{\mathrm{1}}}}  . \Phi   \ottsym{)}  \ottsym{)}  \ottsym{<:}   \ottnt{C_{{\mathrm{2}}}}  . \Phi $}
    \AxiomC{$\Gamma \,  |  \,  \ottnt{C_{{\mathrm{1}}}}  . \ottnt{T}  \,  \,\&\,  \, \Phi_{{\mathrm{1}}} \,  \tceappendop  \, \ottsym{(}  \Phi_{{\mathrm{2}}} \,  \tceappendop  \, \ottsym{(}   \ottnt{C_{{\mathrm{1}}}}  . \Phi   \ottsym{)}  \ottsym{)}  \vdash  \Phi_{{\mathrm{1}}} \,  \tceappendop  \, \ottsym{(}  \Phi_{{\mathrm{2}}} \,  \tceappendop  \, \ottsym{(}   \ottnt{C_{{\mathrm{1}}}}  . \ottnt{S}   \ottsym{)}  \ottsym{)}  \ottsym{<:}   \ottnt{C_{{\mathrm{2}}}}  . \ottnt{S} $}
    \TrinaryInfC{$\Gamma  \vdash  \Phi_{{\mathrm{1}}} \,  \tceappendop  \, \ottsym{(}  \Phi_{{\mathrm{2}}} \,  \tceappendop  \, \ottnt{C_{{\mathrm{1}}}}  \ottsym{)}  \ottsym{<:}  \ottnt{C_{{\mathrm{2}}}}$}
   \end{prooftree}
   By the IH,
   \[
    \Gamma \,  |  \,  \ottnt{C_{{\mathrm{1}}}}  . \ottnt{T}  \,  \,\&\,  \, \ottsym{(}  \Phi_{{\mathrm{1}}} \,  \tceappendop  \, \Phi_{{\mathrm{2}}}  \ottsym{)} \,  \tceappendop  \, \ottsym{(}   \ottnt{C_{{\mathrm{1}}}}  . \Phi   \ottsym{)}  \vdash  \ottsym{(}  \Phi_{{\mathrm{1}}} \,  \tceappendop  \, \Phi_{{\mathrm{2}}}  \ottsym{)} \,  \tceappendop  \, \ottsym{(}   \ottnt{C_{{\mathrm{1}}}}  . \ottnt{S}   \ottsym{)}  \ottsym{<:}   \ottnt{C_{{\mathrm{2}}}}  . \ottnt{S}  ~.
   \]
   Thus, \Srule{Comp} indicates that it suffices to show that
   \[
    \Gamma  \vdash  \ottsym{(}  \Phi_{{\mathrm{1}}} \,  \tceappendop  \, \Phi_{{\mathrm{2}}}  \ottsym{)} \,  \tceappendop  \, \ottsym{(}   \ottnt{C_{{\mathrm{1}}}}  . \Phi   \ottsym{)}  \ottsym{<:}   \ottnt{C_{{\mathrm{2}}}}  . \Phi  ~.
   \]
   \reflem{ts-assoc-preeff-compose-eff-subtyping} implies
   $\Gamma  \vdash  \ottsym{(}  \Phi_{{\mathrm{1}}} \,  \tceappendop  \, \Phi_{{\mathrm{2}}}  \ottsym{)} \,  \tceappendop  \, \ottsym{(}   \ottnt{C_{{\mathrm{1}}}}  . \Phi   \ottsym{)}  \ottsym{<:}  \Phi_{{\mathrm{1}}} \,  \tceappendop  \, \ottsym{(}  \Phi_{{\mathrm{2}}} \,  \tceappendop  \, \ottsym{(}   \ottnt{C_{{\mathrm{1}}}}  . \Phi   \ottsym{)}  \ottsym{)}$.
   Thus, \reflem{ts-transitivity-eff-subtyping} with
   $\Gamma  \vdash  \Phi_{{\mathrm{1}}} \,  \tceappendop  \, \ottsym{(}  \Phi_{{\mathrm{2}}} \,  \tceappendop  \, \ottsym{(}   \ottnt{C_{{\mathrm{1}}}}  . \Phi   \ottsym{)}  \ottsym{)}  \ottsym{<:}   \ottnt{C_{{\mathrm{2}}}}  . \Phi $ implies
   the conclusion $\Gamma  \vdash  \ottsym{(}  \Phi_{{\mathrm{1}}} \,  \tceappendop  \, \Phi_{{\mathrm{2}}}  \ottsym{)} \,  \tceappendop  \, \ottsym{(}   \ottnt{C_{{\mathrm{1}}}}  . \Phi   \ottsym{)}  \ottsym{<:}   \ottnt{C_{{\mathrm{2}}}}  . \Phi $.

  \case \Srule{Empty}:
   Trivial by \Srule{Empty}.

  \case \Srule{Ans}:
   We are given
   \begin{prooftree}
    \AxiomC{$\Gamma  \ottsym{,}  \mathit{x} \,  {:}  \, \ottnt{T}  \vdash  \ottnt{C_{{\mathrm{21}}}}  \ottsym{<:}  \ottnt{C_{{\mathrm{11}}}}$}
    \AxiomC{$\Gamma  \vdash  \Phi_{{\mathrm{1}}} \,  \tceappendop  \, \ottsym{(}  \Phi_{{\mathrm{2}}} \,  \tceappendop  \, \ottnt{C_{{\mathrm{12}}}}  \ottsym{)}  \ottsym{<:}  \ottnt{C_{{\mathrm{22}}}}$}
    \BinaryInfC{$\Gamma \,  |  \, \ottnt{T} \,  \,\&\,  \, \Phi_{{\mathrm{1}}} \,  \tceappendop  \, \ottsym{(}  \Phi_{{\mathrm{2}}} \,  \tceappendop  \, \Phi  \ottsym{)}  \vdash   ( \forall  \mathit{x}  .  \ottnt{C_{{\mathrm{11}}}}  )  \Rightarrow   \Phi_{{\mathrm{1}}} \,  \tceappendop  \, \ottsym{(}  \Phi_{{\mathrm{2}}} \,  \tceappendop  \, \ottnt{C_{{\mathrm{12}}}}  \ottsym{)}   \ottsym{<:}   ( \forall  \mathit{x}  .  \ottnt{C_{{\mathrm{21}}}}  )  \Rightarrow   \ottnt{C_{{\mathrm{22}}}} $}
   \end{prooftree}
   for some $\mathit{x}$, $\ottnt{C_{{\mathrm{11}}}}$, $\ottnt{C_{{\mathrm{12}}}}$, $\ottnt{C_{{\mathrm{21}}}}$, and $\ottnt{C_{{\mathrm{22}}}}$ such that
   $\ottnt{S_{{\mathrm{1}}}} \,  =  \,  ( \forall  \mathit{x}  .  \ottnt{C_{{\mathrm{11}}}}  )  \Rightarrow   \ottnt{C_{{\mathrm{12}}}} $ and $\ottnt{S_{{\mathrm{2}}}} \,  =  \,  ( \forall  \mathit{x}  .  \ottnt{C_{{\mathrm{21}}}}  )  \Rightarrow   \ottnt{C_{{\mathrm{22}}}} $.
   By the IH, $\Gamma  \vdash  \ottsym{(}  \Phi_{{\mathrm{1}}} \,  \tceappendop  \, \Phi_{{\mathrm{2}}}  \ottsym{)} \,  \tceappendop  \, \ottnt{C_{{\mathrm{12}}}}  \ottsym{<:}  \ottnt{C_{{\mathrm{22}}}}$.
   Thus, \Srule{Ans} implies the conclusion
   \[
    \Gamma \,  |  \, \ottnt{T} \,  \,\&\,  \, \ottsym{(}  \Phi_{{\mathrm{1}}} \,  \tceappendop  \, \Phi_{{\mathrm{2}}}  \ottsym{)} \,  \tceappendop  \, \Phi  \vdash   ( \forall  \mathit{x}  .  \ottnt{C_{{\mathrm{11}}}}  )  \Rightarrow   \ottsym{(}  \Phi_{{\mathrm{1}}} \,  \tceappendop  \, \Phi_{{\mathrm{2}}}  \ottsym{)} \,  \tceappendop  \, \ottnt{C_{{\mathrm{12}}}}   \ottsym{<:}   ( \forall  \mathit{x}  .  \ottnt{C_{{\mathrm{21}}}}  )  \Rightarrow   \ottnt{C_{{\mathrm{22}}}} 
   \]
   (note that $ ( \forall  \mathit{x}  .  \ottnt{C_{{\mathrm{11}}}}  )  \Rightarrow   \ottsym{(}  \Phi_{{\mathrm{1}}} \,  \tceappendop  \, \Phi_{{\mathrm{2}}}  \ottsym{)} \,  \tceappendop  \, \ottnt{C_{{\mathrm{12}}}}  \,  =  \, \ottsym{(}  \Phi_{{\mathrm{1}}} \,  \tceappendop  \, \Phi_{{\mathrm{2}}}  \ottsym{)} \,  \tceappendop  \, \ottnt{S_{{\mathrm{1}}}}$).

  \case \Srule{Embed}:
   We are given
   \begin{prooftree}
    \AxiomC{$\Gamma  \ottsym{,}  \mathit{x} \,  {:}  \, \ottnt{T}  \vdash  \ottsym{(}  \Phi_{{\mathrm{1}}} \,  \tceappendop  \, \ottsym{(}  \Phi_{{\mathrm{2}}} \,  \tceappendop  \, \Phi  \ottsym{)}  \ottsym{)} \,  \tceappendop  \, \ottnt{C_{{\mathrm{21}}}}  \ottsym{<:}  \ottnt{C_{{\mathrm{22}}}}$}
    \AxiomC{$\Gamma  \ottsym{,}  \mathit{y} \,  {:}  \,  \Sigma ^\omega   \models    { \ottsym{(}  \Phi_{{\mathrm{1}}} \,  \tceappendop  \, \ottsym{(}  \Phi_{{\mathrm{2}}} \,  \tceappendop  \, \Phi  \ottsym{)}  \ottsym{)} }^\nu   \ottsym{(}  \mathit{y}  \ottsym{)}    \Rightarrow     { \ottnt{C_{{\mathrm{22}}}} }^\nu (  \mathit{y}  )  $}
    \AxiomC{$\mathit{x}  \ottsym{,}  \mathit{y} \,  \not\in  \,  \mathit{fv}  (  \ottnt{C_{{\mathrm{22}}}}  )  \,  \mathbin{\cup}  \,  \mathit{fv}  (  \Phi_{{\mathrm{1}}} \,  \tceappendop  \, \ottsym{(}  \Phi_{{\mathrm{2}}} \,  \tceappendop  \, \Phi  \ottsym{)}  ) $}
    \TrinaryInfC{$\Gamma \,  |  \, \ottnt{T} \,  \,\&\,  \, \Phi_{{\mathrm{1}}} \,  \tceappendop  \, \ottsym{(}  \Phi_{{\mathrm{2}}} \,  \tceappendop  \, \Phi  \ottsym{)}  \vdash   \square   \ottsym{<:}   ( \forall  \mathit{x}  .  \ottnt{C_{{\mathrm{21}}}}  )  \Rightarrow   \ottnt{C_{{\mathrm{22}}}} $}
   \end{prooftree}
   for some $\mathit{x}$, $\mathit{y}$, $\ottnt{C_{{\mathrm{21}}}}$, and $\ottnt{C_{{\mathrm{22}}}}$ such that
   $\ottnt{S_{{\mathrm{2}}}} \,  =  \,  ( \forall  \mathit{x}  .  \ottnt{C_{{\mathrm{21}}}}  )  \Rightarrow   \ottnt{C_{{\mathrm{22}}}} $.
   We also have $\ottnt{S_{{\mathrm{1}}}} \,  =  \,  \square $.
   Because \reflem{ts-assoc-preeff-compose-eff-subtyping} implies
   $\Gamma  \ottsym{,}  \mathit{x} \,  {:}  \, \ottnt{T}  \vdash  \ottsym{(}  \Phi_{{\mathrm{1}}} \,  \tceappendop  \, \Phi_{{\mathrm{2}}}  \ottsym{)} \,  \tceappendop  \, \Phi  \ottsym{<:}  \Phi_{{\mathrm{1}}} \,  \tceappendop  \, \ottsym{(}  \Phi_{{\mathrm{2}}} \,  \tceappendop  \, \Phi  \ottsym{)}$,
   \reflem{ts-strength-preeff-subtyping} with $\Gamma  \ottsym{,}  \mathit{x} \,  {:}  \, \ottnt{T}  \vdash  \ottsym{(}  \Phi_{{\mathrm{1}}} \,  \tceappendop  \, \ottsym{(}  \Phi_{{\mathrm{2}}} \,  \tceappendop  \, \Phi  \ottsym{)}  \ottsym{)} \,  \tceappendop  \, \ottnt{C_{{\mathrm{21}}}}  \ottsym{<:}  \ottnt{C_{{\mathrm{22}}}}$ implies
   \[
    \Gamma  \ottsym{,}  \mathit{x} \,  {:}  \, \ottnt{T}  \vdash  \ottsym{(}  \ottsym{(}  \Phi_{{\mathrm{1}}} \,  \tceappendop  \, \Phi_{{\mathrm{2}}}  \ottsym{)} \,  \tceappendop  \, \Phi  \ottsym{)} \,  \tceappendop  \, \ottnt{C_{{\mathrm{21}}}}  \ottsym{<:}  \ottnt{C_{{\mathrm{22}}}} ~.
   \]
   Further, because $\Gamma  \vdash  \ottsym{(}  \Phi_{{\mathrm{1}}} \,  \tceappendop  \, \Phi_{{\mathrm{2}}}  \ottsym{)} \,  \tceappendop  \, \Phi  \ottsym{<:}  \Phi_{{\mathrm{1}}} \,  \tceappendop  \, \ottsym{(}  \Phi_{{\mathrm{2}}} \,  \tceappendop  \, \Phi  \ottsym{)}$ by \reflem{ts-assoc-preeff-compose-eff-subtyping},
   we have $\Gamma  \ottsym{,}  \mathit{y} \,  {:}  \,  \Sigma ^\omega   \models    { \ottsym{(}  \ottsym{(}  \Phi_{{\mathrm{1}}} \,  \tceappendop  \, \Phi_{{\mathrm{2}}}  \ottsym{)} \,  \tceappendop  \, \Phi  \ottsym{)} }^\nu   \ottsym{(}  \mathit{y}  \ottsym{)}    \Rightarrow     { \ottsym{(}  \Phi_{{\mathrm{1}}} \,  \tceappendop  \, \ottsym{(}  \Phi_{{\mathrm{2}}} \,  \tceappendop  \, \Phi  \ottsym{)}  \ottsym{)} }^\nu   \ottsym{(}  \mathit{y}  \ottsym{)} $.
   Then, \reflem{ts-transitivity-valid} with $\Gamma  \ottsym{,}  \mathit{y} \,  {:}  \,  \Sigma ^\omega   \models    { \ottsym{(}  \Phi_{{\mathrm{1}}} \,  \tceappendop  \, \ottsym{(}  \Phi_{{\mathrm{2}}} \,  \tceappendop  \, \Phi  \ottsym{)}  \ottsym{)} }^\nu   \ottsym{(}  \mathit{y}  \ottsym{)}    \Rightarrow     { \ottnt{C_{{\mathrm{22}}}} }^\nu (  \mathit{y}  )  $
   implies
   \[
    \Gamma  \ottsym{,}  \mathit{y} \,  {:}  \,  \Sigma ^\omega   \models    { \ottsym{(}  \ottsym{(}  \Phi_{{\mathrm{1}}} \,  \tceappendop  \, \Phi_{{\mathrm{2}}}  \ottsym{)} \,  \tceappendop  \, \Phi  \ottsym{)} }^\nu   \ottsym{(}  \mathit{y}  \ottsym{)}    \Rightarrow     { \ottnt{C_{{\mathrm{22}}}} }^\nu (  \mathit{y}  )   ~.
   \]
   Because $ \mathit{fv}  (  \Phi_{{\mathrm{1}}} \,  \tceappendop  \, \ottsym{(}  \Phi_{{\mathrm{2}}} \,  \tceappendop  \, \Phi  \ottsym{)}  )  \,  =  \,  \mathit{fv}  (  \ottsym{(}  \Phi_{{\mathrm{1}}} \,  \tceappendop  \, \Phi_{{\mathrm{2}}}  \ottsym{)} \,  \tceappendop  \, \Phi  ) $,
   \Srule{Embed} implies the conclusion
   \[
    \Gamma \,  |  \, \ottnt{T} \,  \,\&\,  \, \ottsym{(}  \Phi_{{\mathrm{1}}} \,  \tceappendop  \, \Phi_{{\mathrm{2}}}  \ottsym{)} \,  \tceappendop  \, \Phi  \vdash   \square   \ottsym{<:}   ( \forall  \mathit{x}  .  \ottnt{C_{{\mathrm{21}}}}  )  \Rightarrow   \ottnt{C_{{\mathrm{22}}}}  ~.
   \]
 \end{caseanalysis}
\end{proof}

\begin{lemmap}{Associativity of Pre-Effect Composition in Subtyping: Right to Left}{ts-assoc-preeff-compose-subtyping-right-to-left}
 \begin{enumerate}
  \item If $\Gamma  \vdash  \ottsym{(}  \Phi_{{\mathrm{1}}} \,  \tceappendop  \, \Phi_{{\mathrm{2}}}  \ottsym{)} \,  \tceappendop  \, \ottnt{C_{{\mathrm{1}}}}  \ottsym{<:}  \ottnt{C_{{\mathrm{2}}}}$,
        then $\Gamma  \vdash  \Phi_{{\mathrm{1}}} \,  \tceappendop  \, \ottsym{(}  \Phi_{{\mathrm{2}}} \,  \tceappendop  \, \ottnt{C_{{\mathrm{1}}}}  \ottsym{)}  \ottsym{<:}  \ottnt{C_{{\mathrm{2}}}}$.
  \item If $\Gamma \,  |  \, \ottnt{T} \,  \,\&\,  \, \ottsym{(}  \Phi_{{\mathrm{1}}} \,  \tceappendop  \, \Phi_{{\mathrm{2}}}  \ottsym{)} \,  \tceappendop  \, \Phi  \vdash  \ottsym{(}  \Phi_{{\mathrm{1}}} \,  \tceappendop  \, \Phi_{{\mathrm{2}}}  \ottsym{)} \,  \tceappendop  \, \ottnt{S_{{\mathrm{1}}}}  \ottsym{<:}  \ottnt{S_{{\mathrm{2}}}}$,
        then $\Gamma \,  |  \, \ottnt{T} \,  \,\&\,  \, \Phi_{{\mathrm{1}}} \,  \tceappendop  \, \ottsym{(}  \Phi_{{\mathrm{2}}} \,  \tceappendop  \, \Phi  \ottsym{)}  \vdash  \Phi_{{\mathrm{1}}} \,  \tceappendop  \, \ottsym{(}  \Phi_{{\mathrm{2}}} \,  \tceappendop  \, \ottnt{S_{{\mathrm{1}}}}  \ottsym{)}  \ottsym{<:}  \ottnt{S_{{\mathrm{2}}}}$.
 \end{enumerate}
\end{lemmap}
\begin{proof}
 By simultaneous induction on the derivations.
 \begin{caseanalysis}
  \case \Srule{Comp}:
   We are given
   \begin{prooftree}
    \AxiomC{$\Gamma  \vdash   \ottnt{C_{{\mathrm{1}}}}  . \ottnt{T}   \ottsym{<:}   \ottnt{C_{{\mathrm{2}}}}  . \ottnt{T} $}
    \AxiomC{$\Gamma  \vdash  \ottsym{(}  \Phi_{{\mathrm{1}}} \,  \tceappendop  \, \Phi_{{\mathrm{2}}}  \ottsym{)} \,  \tceappendop  \, \ottsym{(}   \ottnt{C_{{\mathrm{1}}}}  . \Phi   \ottsym{)}  \ottsym{<:}   \ottnt{C_{{\mathrm{2}}}}  . \Phi $}
    \AxiomC{$\Gamma \,  |  \,  \ottnt{C_{{\mathrm{1}}}}  . \ottnt{T}  \,  \,\&\,  \, \ottsym{(}  \Phi_{{\mathrm{1}}} \,  \tceappendop  \, \Phi_{{\mathrm{2}}}  \ottsym{)} \,  \tceappendop  \, \ottsym{(}   \ottnt{C_{{\mathrm{1}}}}  . \Phi   \ottsym{)}  \vdash  \ottsym{(}  \Phi_{{\mathrm{1}}} \,  \tceappendop  \, \Phi_{{\mathrm{2}}}  \ottsym{)} \,  \tceappendop  \, \ottsym{(}   \ottnt{C_{{\mathrm{1}}}}  . \ottnt{S}   \ottsym{)}  \ottsym{<:}   \ottnt{C_{{\mathrm{2}}}}  . \ottnt{S} $}
    \TrinaryInfC{$\Gamma  \vdash  \ottsym{(}  \Phi_{{\mathrm{1}}} \,  \tceappendop  \, \Phi_{{\mathrm{2}}}  \ottsym{)} \,  \tceappendop  \, \ottnt{C_{{\mathrm{1}}}}  \ottsym{<:}  \ottnt{C_{{\mathrm{2}}}}$}
   \end{prooftree}
   By the IH,
   \[
    \Gamma \,  |  \,  \ottnt{C_{{\mathrm{1}}}}  . \ottnt{T}  \,  \,\&\,  \, \Phi_{{\mathrm{1}}} \,  \tceappendop  \, \ottsym{(}  \Phi_{{\mathrm{2}}} \,  \tceappendop  \, \ottsym{(}   \ottnt{C_{{\mathrm{1}}}}  . \Phi   \ottsym{)}  \ottsym{)}  \vdash  \Phi_{{\mathrm{1}}} \,  \tceappendop  \, \ottsym{(}  \Phi_{{\mathrm{2}}} \,  \tceappendop  \, \ottsym{(}   \ottnt{C_{{\mathrm{1}}}}  . \ottnt{S}   \ottsym{)}  \ottsym{)}  \ottsym{<:}   \ottnt{C_{{\mathrm{2}}}}  . \ottnt{S}  ~.
   \]
   Thus, \Srule{Comp} indicates that it suffices to show that
   \[
    \Gamma  \vdash  \Phi_{{\mathrm{1}}} \,  \tceappendop  \, \ottsym{(}  \Phi_{{\mathrm{2}}} \,  \tceappendop  \, \ottsym{(}   \ottnt{C_{{\mathrm{1}}}}  . \Phi   \ottsym{)}  \ottsym{)}  \ottsym{<:}   \ottnt{C_{{\mathrm{2}}}}  . \Phi  ~.
   \]
   \reflem{ts-assoc-preeff-compose-eff-subtyping} implies
   $\Gamma  \vdash  \Phi_{{\mathrm{1}}} \,  \tceappendop  \, \ottsym{(}  \Phi_{{\mathrm{2}}} \,  \tceappendop  \, \ottsym{(}   \ottnt{C_{{\mathrm{1}}}}  . \Phi   \ottsym{)}  \ottsym{)}  \ottsym{<:}  \ottsym{(}  \Phi_{{\mathrm{1}}} \,  \tceappendop  \, \Phi_{{\mathrm{2}}}  \ottsym{)} \,  \tceappendop  \, \ottsym{(}   \ottnt{C_{{\mathrm{1}}}}  . \Phi   \ottsym{)}$.
   Thus, \reflem{ts-transitivity-eff-subtyping} with
   $\Gamma  \vdash  \ottsym{(}  \Phi_{{\mathrm{1}}} \,  \tceappendop  \, \Phi_{{\mathrm{2}}}  \ottsym{)} \,  \tceappendop  \, \ottsym{(}   \ottnt{C_{{\mathrm{1}}}}  . \Phi   \ottsym{)}  \ottsym{<:}   \ottnt{C_{{\mathrm{2}}}}  . \Phi $ implies
   the conclusion $\Gamma  \vdash  \Phi_{{\mathrm{1}}} \,  \tceappendop  \, \ottsym{(}  \Phi_{{\mathrm{2}}} \,  \tceappendop  \, \ottsym{(}   \ottnt{C_{{\mathrm{1}}}}  . \Phi   \ottsym{)}  \ottsym{)}  \ottsym{<:}   \ottnt{C_{{\mathrm{2}}}}  . \Phi $.

  \case \Srule{Empty}:
   Trivial by \Srule{Empty}.

  \case \Srule{Ans}:
   We are given
   \begin{prooftree}
    \AxiomC{$\Gamma  \ottsym{,}  \mathit{x} \,  {:}  \, \ottnt{T}  \vdash  \ottnt{C_{{\mathrm{21}}}}  \ottsym{<:}  \ottnt{C_{{\mathrm{11}}}}$}
    \AxiomC{$\Gamma  \vdash  \ottsym{(}  \Phi_{{\mathrm{1}}} \,  \tceappendop  \, \Phi_{{\mathrm{2}}}  \ottsym{)} \,  \tceappendop  \, \ottnt{C_{{\mathrm{12}}}}  \ottsym{<:}  \ottnt{C_{{\mathrm{22}}}}$}
    \BinaryInfC{$\Gamma \,  |  \, \ottnt{T} \,  \,\&\,  \, \ottsym{(}  \Phi_{{\mathrm{1}}} \,  \tceappendop  \, \Phi_{{\mathrm{2}}}  \ottsym{)} \,  \tceappendop  \, \Phi  \vdash   ( \forall  \mathit{x}  .  \ottnt{C_{{\mathrm{11}}}}  )  \Rightarrow   \ottsym{(}  \Phi_{{\mathrm{1}}} \,  \tceappendop  \, \Phi_{{\mathrm{2}}}  \ottsym{)} \,  \tceappendop  \, \ottnt{C_{{\mathrm{12}}}}   \ottsym{<:}   ( \forall  \mathit{x}  .  \ottnt{C_{{\mathrm{21}}}}  )  \Rightarrow   \ottnt{C_{{\mathrm{22}}}} $}
   \end{prooftree}
   for some $\mathit{x}$, $\ottnt{C_{{\mathrm{11}}}}$, $\ottnt{C_{{\mathrm{12}}}}$, $\ottnt{C_{{\mathrm{21}}}}$, and $\ottnt{C_{{\mathrm{22}}}}$ such that
   $\ottnt{S_{{\mathrm{1}}}} \,  =  \,  ( \forall  \mathit{x}  .  \ottnt{C_{{\mathrm{11}}}}  )  \Rightarrow   \ottnt{C_{{\mathrm{12}}}} $ and $\ottnt{S_{{\mathrm{2}}}} \,  =  \,  ( \forall  \mathit{x}  .  \ottnt{C_{{\mathrm{21}}}}  )  \Rightarrow   \ottnt{C_{{\mathrm{22}}}} $.
   By the IH, $\Gamma  \vdash  \Phi_{{\mathrm{1}}} \,  \tceappendop  \, \ottsym{(}  \Phi_{{\mathrm{2}}} \,  \tceappendop  \, \ottnt{C_{{\mathrm{12}}}}  \ottsym{)}  \ottsym{<:}  \ottnt{C_{{\mathrm{22}}}}$.
   Thus, \Srule{Ans} implies the conclusion
   \[
    \Gamma \,  |  \, \ottnt{T} \,  \,\&\,  \, \Phi_{{\mathrm{1}}} \,  \tceappendop  \, \ottsym{(}  \Phi_{{\mathrm{2}}} \,  \tceappendop  \, \Phi  \ottsym{)}  \vdash   ( \forall  \mathit{x}  .  \ottnt{C_{{\mathrm{11}}}}  )  \Rightarrow   \Phi_{{\mathrm{1}}} \,  \tceappendop  \, \ottsym{(}  \Phi_{{\mathrm{2}}} \,  \tceappendop  \, \ottnt{C_{{\mathrm{12}}}}  \ottsym{)}   \ottsym{<:}   ( \forall  \mathit{x}  .  \ottnt{C_{{\mathrm{21}}}}  )  \Rightarrow   \ottnt{C_{{\mathrm{22}}}} 
   \]
   (note that $ ( \forall  \mathit{x}  .  \ottnt{C_{{\mathrm{11}}}}  )  \Rightarrow   \Phi_{{\mathrm{1}}} \,  \tceappendop  \, \ottsym{(}  \Phi_{{\mathrm{2}}} \,  \tceappendop  \, \ottnt{C_{{\mathrm{12}}}}  \ottsym{)}  \,  =  \, \Phi_{{\mathrm{1}}} \,  \tceappendop  \, \ottsym{(}  \Phi_{{\mathrm{2}}} \,  \tceappendop  \, \ottnt{S_{{\mathrm{1}}}}  \ottsym{)}$).

  \case \Srule{Embed}:
   We are given
   \begin{prooftree}
    \AxiomC{$\Gamma  \ottsym{,}  \mathit{x} \,  {:}  \, \ottnt{T}  \vdash  \ottsym{(}  \ottsym{(}  \Phi_{{\mathrm{1}}} \,  \tceappendop  \, \Phi_{{\mathrm{2}}}  \ottsym{)} \,  \tceappendop  \, \Phi  \ottsym{)} \,  \tceappendop  \, \ottnt{C_{{\mathrm{21}}}}  \ottsym{<:}  \ottnt{C_{{\mathrm{22}}}}$}
    \AxiomC{$\Gamma  \ottsym{,}  \mathit{y} \,  {:}  \,  \Sigma ^\omega   \models    { \ottsym{(}  \ottsym{(}  \Phi_{{\mathrm{1}}} \,  \tceappendop  \, \Phi_{{\mathrm{2}}}  \ottsym{)} \,  \tceappendop  \, \Phi  \ottsym{)} }^\nu   \ottsym{(}  \mathit{y}  \ottsym{)}    \Rightarrow     { \ottnt{C_{{\mathrm{22}}}} }^\nu (  \mathit{y}  )  $}
    \AxiomC{$\mathit{x}  \ottsym{,}  \mathit{y} \,  \not\in  \,  \mathit{fv}  (  \ottnt{C_{{\mathrm{22}}}}  )  \,  \mathbin{\cup}  \,  \mathit{fv}  (  \ottsym{(}  \Phi_{{\mathrm{1}}} \,  \tceappendop  \, \Phi_{{\mathrm{2}}}  \ottsym{)} \,  \tceappendop  \, \Phi  ) $}
    \TrinaryInfC{$\Gamma \,  |  \, \ottnt{T} \,  \,\&\,  \, \ottsym{(}  \Phi_{{\mathrm{1}}} \,  \tceappendop  \, \Phi_{{\mathrm{2}}}  \ottsym{)} \,  \tceappendop  \, \Phi  \vdash   \square   \ottsym{<:}   ( \forall  \mathit{x}  .  \ottnt{C_{{\mathrm{21}}}}  )  \Rightarrow   \ottnt{C_{{\mathrm{22}}}} $}
   \end{prooftree}
   for some $\mathit{x}$, $\mathit{y}$, $\ottnt{C_{{\mathrm{21}}}}$, and $\ottnt{C_{{\mathrm{22}}}}$ such that
   $\ottnt{S_{{\mathrm{2}}}} \,  =  \,  ( \forall  \mathit{x}  .  \ottnt{C_{{\mathrm{21}}}}  )  \Rightarrow   \ottnt{C_{{\mathrm{22}}}} $.
   We also have $\ottnt{S_{{\mathrm{1}}}} \,  =  \,  \square $.
   Because \reflem{ts-assoc-preeff-compose-eff-subtyping} implies
   $\Gamma  \ottsym{,}  \mathit{x} \,  {:}  \, \ottnt{T}  \vdash  \Phi_{{\mathrm{1}}} \,  \tceappendop  \, \ottsym{(}  \Phi_{{\mathrm{2}}} \,  \tceappendop  \, \Phi  \ottsym{)}  \ottsym{<:}  \ottsym{(}  \Phi_{{\mathrm{1}}} \,  \tceappendop  \, \Phi_{{\mathrm{2}}}  \ottsym{)} \,  \tceappendop  \, \Phi$,
   \reflem{ts-strength-preeff-subtyping} with $\Gamma  \ottsym{,}  \mathit{x} \,  {:}  \, \ottnt{T}  \vdash  \ottsym{(}  \ottsym{(}  \Phi_{{\mathrm{1}}} \,  \tceappendop  \, \Phi_{{\mathrm{2}}}  \ottsym{)} \,  \tceappendop  \, \Phi  \ottsym{)} \,  \tceappendop  \, \ottnt{C_{{\mathrm{21}}}}  \ottsym{<:}  \ottnt{C_{{\mathrm{22}}}}$ implies
   \[
    \Gamma  \ottsym{,}  \mathit{x} \,  {:}  \, \ottnt{T}  \vdash  \ottsym{(}  \Phi_{{\mathrm{1}}} \,  \tceappendop  \, \ottsym{(}  \Phi_{{\mathrm{2}}} \,  \tceappendop  \, \Phi  \ottsym{)}  \ottsym{)} \,  \tceappendop  \, \ottnt{C_{{\mathrm{21}}}}  \ottsym{<:}  \ottnt{C_{{\mathrm{22}}}} ~.
   \]
   Further, because $\Gamma  \vdash  \Phi_{{\mathrm{1}}} \,  \tceappendop  \, \ottsym{(}  \Phi_{{\mathrm{2}}} \,  \tceappendop  \, \Phi  \ottsym{)}  \ottsym{<:}  \ottsym{(}  \Phi_{{\mathrm{1}}} \,  \tceappendop  \, \Phi_{{\mathrm{2}}}  \ottsym{)} \,  \tceappendop  \, \Phi$ by \reflem{ts-assoc-preeff-compose-eff-subtyping},
   we have $\Gamma  \ottsym{,}  \mathit{y} \,  {:}  \,  \Sigma ^\omega   \models    { \ottsym{(}  \Phi_{{\mathrm{1}}} \,  \tceappendop  \, \ottsym{(}  \Phi_{{\mathrm{2}}} \,  \tceappendop  \, \Phi  \ottsym{)}  \ottsym{)} }^\nu   \ottsym{(}  \mathit{y}  \ottsym{)}    \Rightarrow     { \ottsym{(}  \ottsym{(}  \Phi_{{\mathrm{1}}} \,  \tceappendop  \, \Phi_{{\mathrm{2}}}  \ottsym{)} \,  \tceappendop  \, \Phi  \ottsym{)} }^\nu   \ottsym{(}  \mathit{y}  \ottsym{)} $.
   Then, \reflem{ts-transitivity-valid} with $\Gamma  \ottsym{,}  \mathit{y} \,  {:}  \,  \Sigma ^\omega   \models    { \ottsym{(}  \ottsym{(}  \Phi_{{\mathrm{1}}} \,  \tceappendop  \, \Phi_{{\mathrm{2}}}  \ottsym{)} \,  \tceappendop  \, \Phi  \ottsym{)} }^\nu   \ottsym{(}  \mathit{y}  \ottsym{)}    \Rightarrow     { \ottnt{C_{{\mathrm{22}}}} }^\nu (  \mathit{y}  )  $
   implies
   \[
    \Gamma  \ottsym{,}  \mathit{y} \,  {:}  \,  \Sigma ^\omega   \models    { \ottsym{(}  \Phi_{{\mathrm{1}}} \,  \tceappendop  \, \ottsym{(}  \Phi_{{\mathrm{2}}} \,  \tceappendop  \, \Phi  \ottsym{)}  \ottsym{)} }^\nu   \ottsym{(}  \mathit{y}  \ottsym{)}    \Rightarrow     { \ottnt{C_{{\mathrm{22}}}} }^\nu (  \mathit{y}  )   ~.
   \]
   Because $ \mathit{fv}  (  \Phi_{{\mathrm{1}}} \,  \tceappendop  \, \ottsym{(}  \Phi_{{\mathrm{2}}} \,  \tceappendop  \, \Phi  \ottsym{)}  )  \,  =  \,  \mathit{fv}  (  \ottsym{(}  \Phi_{{\mathrm{1}}} \,  \tceappendop  \, \Phi_{{\mathrm{2}}}  \ottsym{)} \,  \tceappendop  \, \Phi  ) $,
   \Srule{Embed} implies the conclusion
   \[
    \Gamma \,  |  \, \ottnt{T} \,  \,\&\,  \, \Phi_{{\mathrm{1}}} \,  \tceappendop  \, \ottsym{(}  \Phi_{{\mathrm{2}}} \,  \tceappendop  \, \Phi  \ottsym{)}  \vdash   \square   \ottsym{<:}   ( \forall  \mathit{x}  .  \ottnt{C_{{\mathrm{21}}}}  )  \Rightarrow   \ottnt{C_{{\mathrm{22}}}}  ~.
   \]
 \end{caseanalysis}
\end{proof}

\begin{lemmap}{Associativity of Pre-Effect Composition in Subtyping}{ts-assoc-preeff-compose-subtyping}
 For any $\Gamma$, $\Phi_{{\mathrm{1}}}$, $\Phi_{{\mathrm{2}}}$, $\ottnt{C_{{\mathrm{1}}}}$, and $\ottnt{C_{{\mathrm{2}}}}$,
 $\Gamma  \vdash  \Phi_{{\mathrm{1}}} \,  \tceappendop  \, \ottsym{(}  \Phi_{{\mathrm{2}}} \,  \tceappendop  \, \ottnt{C_{{\mathrm{1}}}}  \ottsym{)}  \ottsym{<:}  \ottnt{C_{{\mathrm{2}}}}$ if and only if $\Gamma  \vdash  \ottsym{(}  \Phi_{{\mathrm{1}}} \,  \tceappendop  \, \Phi_{{\mathrm{2}}}  \ottsym{)} \,  \tceappendop  \, \ottnt{C_{{\mathrm{1}}}}  \ottsym{<:}  \ottnt{C_{{\mathrm{2}}}}$.
\end{lemmap}
\begin{proof}
 By Lemmas~\ref{lem:ts-assoc-preeff-compose-subtyping-left-to-right} and \ref{lem:ts-assoc-preeff-compose-subtyping-right-to-left}.
\end{proof}

\begin{lemma}{ts-add-preeff-eff-subtyping-aux}
 If
 $\Gamma  \ottsym{,}  \mathit{x} \,  {:}  \, \ottnt{s}  \models   \phi_{{\mathrm{1}}}    \Rightarrow    \phi_{{\mathrm{2}}} $,
 then
 \[
  \Gamma  \models   \ottsym{(}    \exists  \,  \mathit{x'}  \mathrel{  \in  }  \ottnt{s'}  . \    \exists  \,  \mathit{x}  \mathrel{  \in  }  \ottnt{s}  . \    \phi    \wedge    \phi'     \wedge    \phi_{{\mathrm{1}}}     \ottsym{)}    \Rightarrow    \ottsym{(}    \exists  \,  \mathit{x'}  \mathrel{  \in  }  \ottnt{s'}  . \    \exists  \,  \mathit{x}  \mathrel{  \in  }  \ottnt{s}  . \    \phi    \wedge    \phi'     \wedge    \phi_{{\mathrm{2}}}     \ottsym{)}  ~.
 \]
\end{lemma}
\begin{proof}
 \TS{Validity}
 Obvious by \reflem{ts-logic-closing-dsubs}.
\end{proof}

\begin{lemmap}{Adding Pre-Effects in Effect Subtyping}{ts-add-preeff-eff-subtyping}
 If $\Gamma  \vdash  \Phi_{{\mathrm{1}}}  \ottsym{<:}  \Phi_{{\mathrm{2}}}$, then $\Gamma  \vdash  \Phi \,  \tceappendop  \, \Phi_{{\mathrm{1}}}  \ottsym{<:}  \Phi \,  \tceappendop  \, \Phi_{{\mathrm{2}}}$ for any $\Phi$.
\end{lemmap}
\begin{proof}
 Let $\mathit{x} \,  \not\in  \,  \mathit{fv}  (  \Phi \,  \tceappendop  \, \Phi_{{\mathrm{1}}}  )  \,  \mathbin{\cup}  \,  \mathit{fv}  (  \Phi \,  \tceappendop  \, \Phi_{{\mathrm{2}}}  ) $.
 \begin{itemize}
  \item We show that
        \[
         \Gamma  \ottsym{,}  \mathit{x} \,  {:}  \,  \Sigma ^\ast   \models    { \ottsym{(}  \Phi \,  \tceappendop  \, \Phi_{{\mathrm{1}}}  \ottsym{)} }^\mu   \ottsym{(}  \mathit{x}  \ottsym{)}    \Rightarrow     { \ottsym{(}  \Phi \,  \tceappendop  \, \Phi_{{\mathrm{2}}}  \ottsym{)} }^\mu   \ottsym{(}  \mathit{x}  \ottsym{)}  ~.
        \]
        Because
        \begin{itemize}
         \item $ { \ottsym{(}  \Phi \,  \tceappendop  \, \Phi_{{\mathrm{1}}}  \ottsym{)} }^\mu   \ottsym{(}  \mathit{x}  \ottsym{)} \,  =  \,   \exists  \,  \mathit{x_{{\mathrm{1}}}}  \mathrel{  \in  }   \Sigma ^\ast   . \    \exists  \,  \mathit{x_{{\mathrm{2}}}}  \mathrel{  \in  }   \Sigma ^\ast   . \     \mathit{x}    =    \mathit{x_{{\mathrm{1}}}} \,  \tceappendop  \, \mathit{x_{{\mathrm{2}}}}     \wedge     { \Phi }^\mu   \ottsym{(}  \mathit{x_{{\mathrm{1}}}}  \ottsym{)}     \wedge     { \Phi_{{\mathrm{1}}} }^\mu   \ottsym{(}  \mathit{x_{{\mathrm{2}}}}  \ottsym{)}   $,
         \item $ { \ottsym{(}  \Phi \,  \tceappendop  \, \Phi_{{\mathrm{2}}}  \ottsym{)} }^\mu   \ottsym{(}  \mathit{x}  \ottsym{)} \,  =  \,   \exists  \,  \mathit{x_{{\mathrm{1}}}}  \mathrel{  \in  }   \Sigma ^\ast   . \    \exists  \,  \mathit{x_{{\mathrm{2}}}}  \mathrel{  \in  }   \Sigma ^\ast   . \     \mathit{x}    =    \mathit{x_{{\mathrm{1}}}} \,  \tceappendop  \, \mathit{x_{{\mathrm{2}}}}     \wedge     { \Phi }^\mu   \ottsym{(}  \mathit{x_{{\mathrm{1}}}}  \ottsym{)}     \wedge     { \Phi_{{\mathrm{2}}} }^\mu   \ottsym{(}  \mathit{x_{{\mathrm{2}}}}  \ottsym{)}   $, and
         \item $\Gamma  \ottsym{,}  \mathit{x} \,  {:}  \,  \Sigma ^\ast   \ottsym{,}  \mathit{x_{{\mathrm{2}}}} \,  {:}  \,  \Sigma ^\ast   \models    { \Phi_{{\mathrm{1}}} }^\mu   \ottsym{(}  \mathit{x_{{\mathrm{2}}}}  \ottsym{)}    \Rightarrow     { \Phi_{{\mathrm{2}}} }^\mu   \ottsym{(}  \mathit{x_{{\mathrm{2}}}}  \ottsym{)} $ by
               \reflem{ts-weak-valid} and $\Gamma  \vdash  \Phi_{{\mathrm{1}}}  \ottsym{<:}  \Phi_{{\mathrm{2}}}$,
        \end{itemize}
        \reflem{ts-add-preeff-eff-subtyping-aux} implies the conclusion.

  \item We show that
        \[
         \Gamma  \ottsym{,}  \mathit{x} \,  {:}  \,  \Sigma ^\omega   \models    { \ottsym{(}  \Phi \,  \tceappendop  \, \Phi_{{\mathrm{1}}}  \ottsym{)} }^\nu   \ottsym{(}  \mathit{x}  \ottsym{)}    \Rightarrow     { \ottsym{(}  \Phi \,  \tceappendop  \, \Phi_{{\mathrm{2}}}  \ottsym{)} }^\nu   \ottsym{(}  \mathit{x}  \ottsym{)}  ~.
        \]
        Because
        \begin{itemize}
         \item $ { \ottsym{(}  \Phi \,  \tceappendop  \, \Phi_{{\mathrm{1}}}  \ottsym{)} }^\nu   \ottsym{(}  \mathit{x}  \ottsym{)} \,  =  \,     { \Phi }^\nu   \ottsym{(}  \mathit{x}  \ottsym{)}    \vee      \exists  \,  \mathit{x'}  \mathrel{  \in  }   \Sigma ^\ast   . \    \exists  \,  \mathit{y'}  \mathrel{  \in  }   \Sigma ^\omega   . \   \mathit{x}    =    \mathit{x'} \,  \tceappendop  \, \mathit{y'}        \wedge     { \Phi }^\mu   \ottsym{(}  \mathit{x'}  \ottsym{)}     \wedge     { \Phi_{{\mathrm{1}}} }^\nu   \ottsym{(}  \mathit{y'}  \ottsym{)} $,
         \item $ { \ottsym{(}  \Phi \,  \tceappendop  \, \Phi_{{\mathrm{2}}}  \ottsym{)} }^\nu   \ottsym{(}  \mathit{x}  \ottsym{)} \,  =  \,     { \Phi }^\nu   \ottsym{(}  \mathit{x}  \ottsym{)}    \vee      \exists  \,  \mathit{x'}  \mathrel{  \in  }   \Sigma ^\ast   . \    \exists  \,  \mathit{y'}  \mathrel{  \in  }   \Sigma ^\omega   . \   \mathit{x}    =    \mathit{x'} \,  \tceappendop  \, \mathit{y'}        \wedge     { \Phi }^\mu   \ottsym{(}  \mathit{x'}  \ottsym{)}     \wedge     { \Phi_{{\mathrm{2}}} }^\nu   \ottsym{(}  \mathit{y'}  \ottsym{)} $, and
         \item $\Gamma  \ottsym{,}  \mathit{x} \,  {:}  \,  \Sigma ^\omega   \ottsym{,}  \mathit{y'} \,  {:}  \,  \Sigma ^\omega   \models    { \Phi_{{\mathrm{1}}} }^\nu   \ottsym{(}  \mathit{y'}  \ottsym{)}    \Rightarrow     { \Phi_{{\mathrm{2}}} }^\nu   \ottsym{(}  \mathit{y'}  \ottsym{)} $ by
               \reflem{ts-weak-valid} and $\Gamma  \vdash  \Phi_{{\mathrm{1}}}  \ottsym{<:}  \Phi_{{\mathrm{2}}}$,
        \end{itemize}
        Lemmas~\ref{lem:ts-imp-or} and \ref{lem:ts-add-preeff-eff-subtyping-aux} imply the conclusion
        $\Gamma  \ottsym{,}  \mathit{x} \,  {:}  \,  \Sigma ^\omega   \models    { \ottsym{(}  \Phi \,  \tceappendop  \, \Phi_{{\mathrm{1}}}  \ottsym{)} }^\nu   \ottsym{(}  \mathit{x}  \ottsym{)}    \Rightarrow     { \ottsym{(}  \Phi \,  \tceappendop  \, \Phi_{{\mathrm{2}}}  \ottsym{)} }^\nu   \ottsym{(}  \mathit{x}  \ottsym{)} $.
 \end{itemize}
\end{proof}

\begin{lemmap}{Introduction of Intersection Implication}{ts-valid-intro-impl-intersect}
 If $\Gamma  \models   \phi    \Rightarrow    \phi_{{\mathrm{1}}} $ and $\Gamma  \models   \phi    \Rightarrow    \phi_{{\mathrm{2}}} $,
 then $\Gamma  \models   \phi    \Rightarrow    \ottsym{(}   \phi_{{\mathrm{1}}}    \wedge    \phi_{{\mathrm{2}}}   \ottsym{)} $.
\end{lemmap}
\begin{proof}
 \TS{Validity}
 Obvious by \reflem{ts-logic-closing-dsubs}.
\end{proof}

\begin{lemmap}{Implication of $ \top $}{ts-valid-intro-impl-top}
 For any $\Gamma$ and $\phi$, $\Gamma  \models   \phi    \Rightarrow     \top  $ holds.
\end{lemmap}
\begin{proof}
 \TS{Validity}
 Obvious by \reflem{ts-logic-closing-dsubs}.
\end{proof}

\begin{lemmap}{Elimination of Intersection Implication}{ts-valid-elim-impl-intersect}
 If $\Gamma  \models   \phi    \Rightarrow    \ottsym{(}   \phi_{{\mathrm{1}}}    \wedge    \phi_{{\mathrm{2}}}   \ottsym{)} $,
 then $\Gamma  \models   \phi    \Rightarrow    \phi_{{\mathrm{1}}} $ and $\Gamma  \models   \phi    \Rightarrow    \phi_{{\mathrm{2}}} $.
\end{lemmap}
\begin{proof}
 \TS{Validity}
 Obvious by \reflem{ts-logic-closing-dsubs}.
\end{proof}

\begin{lemmap}{Adding Pre-Effects to $ \nu $-Effect Implication Validity}{ts-add-preeff-nu-eff-valid}
 If $\Gamma  \ottsym{,}  \mathit{y} \,  {:}  \,  \Sigma ^\omega   \models    { \Phi }^\nu   \ottsym{(}  \mathit{y}  \ottsym{)}    \Rightarrow     { \ottnt{C} }^\nu (  \mathit{y}  )  $ and $\mathit{y} \,  \not\in  \,  \mathit{fv}  (  \Phi  )  \,  \mathbin{\cup}  \,  \mathit{fv}  (  \ottnt{C}  ) $,
 then $\Gamma  \ottsym{,}  \mathit{y} \,  {:}  \,  \Sigma ^\omega   \models    { \ottsym{(}  \Phi' \,  \tceappendop  \, \Phi  \ottsym{)} }^\nu   \ottsym{(}  \mathit{y}  \ottsym{)}    \Rightarrow     { \ottsym{(}  \Phi' \,  \tceappendop  \, \ottnt{C}  \ottsym{)} }^\nu (  \mathit{y}  )  $ for any $\Phi'$.
\end{lemmap}
\begin{proof}
 By induction on $\ottnt{C}$.
 Because $ { \ottsym{(}  \Phi' \,  \tceappendop  \, \ottnt{C}  \ottsym{)} }^\nu (  \mathit{y}  )  \,  =  \,   { \ottsym{(}  \Phi' \,  \tceappendop  \, \ottsym{(}   \ottnt{C}  . \Phi   \ottsym{)}  \ottsym{)} }^\nu   \ottsym{(}  \mathit{y}  \ottsym{)}    \wedge     { \ottsym{(}  \Phi' \,  \tceappendop  \, \ottsym{(}   \ottnt{C}  . \ottnt{S}   \ottsym{)}  \ottsym{)} }^\nu (  \mathit{y}  )  $,
 \reflem{ts-valid-intro-impl-intersect} implies that it suffices to show the following.
 \begin{itemize}
  \item We show that
        \[
         \Gamma  \ottsym{,}  \mathit{y} \,  {:}  \,  \Sigma ^\omega   \models    { \ottsym{(}  \Phi' \,  \tceappendop  \, \Phi  \ottsym{)} }^\nu   \ottsym{(}  \mathit{y}  \ottsym{)}    \Rightarrow     { \ottsym{(}  \Phi' \,  \tceappendop  \, \ottsym{(}   \ottnt{C}  . \Phi   \ottsym{)}  \ottsym{)} }^\nu   \ottsym{(}  \mathit{y}  \ottsym{)}  ~.
        \]
        Because
        $\Gamma  \ottsym{,}  \mathit{y} \,  {:}  \,  \Sigma ^\omega   \models    { \Phi }^\nu   \ottsym{(}  \mathit{y}  \ottsym{)}    \Rightarrow     { \ottnt{C} }^\nu (  \mathit{y}  )  $ and
        $ { \ottnt{C} }^\nu (  \mathit{y}  )  \,  =  \,   { \ottsym{(}   \ottnt{C}  . \Phi   \ottsym{)} }^\nu   \ottsym{(}  \mathit{y}  \ottsym{)}    \wedge     { \ottsym{(}   \ottnt{C}  . \ottnt{S}   \ottsym{)} }^\nu (  \mathit{y}  )  $,
        \reflem{ts-valid-elim-impl-intersect} implies
        $\Gamma  \ottsym{,}  \mathit{y} \,  {:}  \,  \Sigma ^\omega   \models    { \Phi }^\nu   \ottsym{(}  \mathit{y}  \ottsym{)}    \Rightarrow     { \ottsym{(}   \ottnt{C}  . \Phi   \ottsym{)} }^\nu   \ottsym{(}  \mathit{y}  \ottsym{)} $.
        Therefore,
        \[
         \Gamma  \ottsym{,}  \mathit{y} \,  {:}  \,  \Sigma ^\omega   \ottsym{,}  \mathit{y'} \,  {:}  \,  \Sigma ^\omega   \models    { \Phi }^\nu   \ottsym{(}  \mathit{y'}  \ottsym{)}    \Rightarrow     { \ottsym{(}   \ottnt{C}  . \Phi   \ottsym{)} }^\nu   \ottsym{(}  \mathit{y'}  \ottsym{)} 
        \]
        by renaming $\mathit{y}$ to $\mathit{y'}$ and \reflem{ts-weak-valid}.
        Further,
        \begin{itemize}
         \item $ { \ottsym{(}  \Phi' \,  \tceappendop  \, \Phi  \ottsym{)} }^\nu   \ottsym{(}  \mathit{y}  \ottsym{)} \,  =  \,     { \Phi' }^\nu   \ottsym{(}  \mathit{y}  \ottsym{)}    \vee      \exists  \,  \mathit{x'}  \mathrel{  \in  }   \Sigma ^\ast   . \    \exists  \,  \mathit{y'}  \mathrel{  \in  }   \Sigma ^\omega   . \   \mathit{y}    =    \mathit{x'} \,  \tceappendop  \, \mathit{y'}        \wedge     { \Phi' }^\mu   \ottsym{(}  \mathit{x'}  \ottsym{)}     \wedge     { \Phi }^\nu   \ottsym{(}  \mathit{y'}  \ottsym{)} $ and
         \item $ { \ottsym{(}  \Phi' \,  \tceappendop  \, \ottsym{(}   \ottnt{C}  . \Phi   \ottsym{)}  \ottsym{)} }^\nu   \ottsym{(}  \mathit{y}  \ottsym{)} \,  =  \,     { \Phi' }^\nu   \ottsym{(}  \mathit{y}  \ottsym{)}    \vee      \exists  \,  \mathit{x'}  \mathrel{  \in  }   \Sigma ^\ast   . \    \exists  \,  \mathit{y'}  \mathrel{  \in  }   \Sigma ^\omega   . \   \mathit{y}    =    \mathit{x'} \,  \tceappendop  \, \mathit{y'}        \wedge     { \Phi' }^\mu   \ottsym{(}  \mathit{x'}  \ottsym{)}     \wedge     { \ottsym{(}   \ottnt{C}  . \Phi   \ottsym{)} }^\nu   \ottsym{(}  \mathit{y'}  \ottsym{)} $,.
        \end{itemize}
        Lemmas~\ref{lem:ts-imp-or} and \ref{lem:ts-add-preeff-eff-subtyping-aux} imply
        the conclusion.

  \item We show that
        \[
         \Gamma  \ottsym{,}  \mathit{y} \,  {:}  \,  \Sigma ^\omega   \models    { \ottsym{(}  \Phi' \,  \tceappendop  \, \Phi  \ottsym{)} }^\nu   \ottsym{(}  \mathit{y}  \ottsym{)}    \Rightarrow     { \ottsym{(}  \Phi' \,  \tceappendop  \, \ottsym{(}   \ottnt{C}  . \ottnt{S}   \ottsym{)}  \ottsym{)} }^\nu (  \mathit{y}  )  
        \]
        by case analysis on $ \ottnt{C}  . \ottnt{S} $.
        \begin{caseanalysis}
         \case $ \ottnt{C}  . \ottnt{S}  \,  =  \,  \square $:
          Because $ { \ottsym{(}  \Phi' \,  \tceappendop  \, \ottsym{(}   \ottnt{C}  . \ottnt{S}   \ottsym{)}  \ottsym{)} }^\nu (  \mathit{y}  )  =  {  \square  }^\nu (  \mathit{y}  )  =  \top $,
          we must show that
          \[
           \Gamma  \ottsym{,}  \mathit{y} \,  {:}  \,  \Sigma ^\omega   \models    { \ottsym{(}  \Phi' \,  \tceappendop  \, \Phi  \ottsym{)} }^\nu   \ottsym{(}  \mathit{y}  \ottsym{)}    \Rightarrow     \top   ~,
          \]
          which is implied by \reflem{ts-valid-intro-impl-top}.

         \case $  \exists  \, { \mathit{x}  \ottsym{,}  \ottnt{C_{{\mathrm{1}}}}  \ottsym{,}  \ottnt{C_{{\mathrm{2}}}} } . \   \ottnt{C}  . \ottnt{S}  \,  =  \,  ( \forall  \mathit{x}  .  \ottnt{C_{{\mathrm{1}}}}  )  \Rightarrow   \ottnt{C_{{\mathrm{2}}}}  $:
          Because $ { \ottsym{(}  \Phi' \,  \tceappendop  \, \ottsym{(}   \ottnt{C}  . \ottnt{S}   \ottsym{)}  \ottsym{)} }^\nu (  \mathit{y}  )  =  { \ottsym{(}  \Phi' \,  \tceappendop  \, \ottnt{C_{{\mathrm{2}}}}  \ottsym{)} }^\nu (  \mathit{y}  ) $,
          we must show that
          \[
           \Gamma  \ottsym{,}  \mathit{y} \,  {:}  \,  \Sigma ^\omega   \models    { \ottsym{(}  \Phi' \,  \tceappendop  \, \Phi  \ottsym{)} }^\nu   \ottsym{(}  \mathit{y}  \ottsym{)}    \Rightarrow     { \ottsym{(}  \Phi' \,  \tceappendop  \, \ottnt{C_{{\mathrm{2}}}}  \ottsym{)} }^\nu (  \mathit{y}  )   ~.
          \]
          Because
          $\Gamma  \ottsym{,}  \mathit{y} \,  {:}  \,  \Sigma ^\omega   \models    { \Phi }^\nu   \ottsym{(}  \mathit{y}  \ottsym{)}    \Rightarrow     { \ottnt{C} }^\nu (  \mathit{y}  )  $ and
          $ { \ottnt{C} }^\nu (  \mathit{y}  )  \,  =  \,   { \ottsym{(}   \ottnt{C}  . \Phi   \ottsym{)} }^\nu   \ottsym{(}  \mathit{y}  \ottsym{)}    \wedge     { \ottnt{C_{{\mathrm{2}}}} }^\nu (  \mathit{y}  )  $,
          \reflem{ts-valid-elim-impl-intersect} implies
          $\Gamma  \ottsym{,}  \mathit{y} \,  {:}  \,  \Sigma ^\omega   \models    { \Phi }^\nu   \ottsym{(}  \mathit{y}  \ottsym{)}    \Rightarrow     { \ottnt{C_{{\mathrm{2}}}} }^\nu (  \mathit{y}  )  $.
          Therefore, by the IH, we have the conclusion.
        \end{caseanalysis}
 \end{itemize}
\end{proof}

\begin{lemmap}{Adding Pre-Effects in Subtyping}{ts-add-preeff-subtyping}
 \begin{enumerate}
  \item If $\Gamma  \vdash  \ottnt{C_{{\mathrm{1}}}}  \ottsym{<:}  \ottnt{C_{{\mathrm{2}}}}$, then $\Gamma  \vdash  \Phi \,  \tceappendop  \, \ottnt{C_{{\mathrm{1}}}}  \ottsym{<:}  \Phi \,  \tceappendop  \, \ottnt{C_{{\mathrm{2}}}}$ for any $\Phi$.
  \item If $\Gamma \,  |  \, \ottnt{T_{{\mathrm{0}}}} \,  \,\&\,  \, \Phi_{{\mathrm{0}}}  \vdash  \ottnt{S_{{\mathrm{1}}}}  \ottsym{<:}  \ottnt{S_{{\mathrm{2}}}}$, then $\Gamma \,  |  \, \ottnt{T_{{\mathrm{0}}}} \,  \,\&\,  \, \Phi \,  \tceappendop  \, \Phi_{{\mathrm{0}}}  \vdash  \Phi \,  \tceappendop  \, \ottnt{S_{{\mathrm{1}}}}  \ottsym{<:}  \Phi \,  \tceappendop  \, \ottnt{S_{{\mathrm{2}}}}$ for any $\Phi$.
 \end{enumerate}
\end{lemmap}
\begin{proof}
 By simultaneous induction on the derivations.
 \begin{caseanalysis}
  \case \Srule{Comp}:
   We are given
   \begin{itemize}
    \item $\Gamma  \vdash   \ottnt{C_{{\mathrm{1}}}}  . \ottnt{T}   \ottsym{<:}   \ottnt{C_{{\mathrm{2}}}}  . \ottnt{T} $,
    \item $\Gamma  \vdash   \ottnt{C_{{\mathrm{1}}}}  . \Phi   \ottsym{<:}   \ottnt{C_{{\mathrm{2}}}}  . \Phi $, and
    \item $\Gamma \,  |  \,  \ottnt{C_{{\mathrm{1}}}}  . \ottnt{T}  \,  \,\&\,  \,  \ottnt{C_{{\mathrm{1}}}}  . \Phi   \vdash   \ottnt{C_{{\mathrm{1}}}}  . \ottnt{S}   \ottsym{<:}   \ottnt{C_{{\mathrm{2}}}}  . \ottnt{S} $.
   \end{itemize}
   \reflem{ts-add-preeff-eff-subtyping} implies
   \[
    \Gamma  \vdash  \Phi \,  \tceappendop  \, \ottsym{(}   \ottnt{C_{{\mathrm{1}}}}  . \Phi   \ottsym{)}  \ottsym{<:}  \Phi \,  \tceappendop  \, \ottsym{(}   \ottnt{C_{{\mathrm{2}}}}  . \Phi   \ottsym{)} ~.
   \]
   The IH implies
   \[
    \Gamma \,  |  \,  \ottnt{C_{{\mathrm{1}}}}  . \ottnt{T}  \,  \,\&\,  \, \Phi \,  \tceappendop  \, \ottsym{(}   \ottnt{C_{{\mathrm{1}}}}  . \Phi   \ottsym{)}  \vdash  \Phi \,  \tceappendop  \, \ottsym{(}   \ottnt{C_{{\mathrm{1}}}}  . \ottnt{S}   \ottsym{)}  \ottsym{<:}  \Phi \,  \tceappendop  \, \ottsym{(}   \ottnt{C_{{\mathrm{2}}}}  . \ottnt{S}   \ottsym{)} ~.
   \]
   Thus, \Srule{Comp} implies the conclusion.

  \case \Srule{Empty}:
   Trivial by \Srule{Empty} because \Srule{Empty} has no premise and
   $\ottnt{S_{{\mathrm{1}}}} = \ottnt{S_{{\mathrm{2}}}} = \Phi \,  \tceappendop  \, \ottnt{S_{{\mathrm{1}}}} = \Phi \,  \tceappendop  \, \ottnt{S_{{\mathrm{2}}}} =  \square $.

  \case \Srule{Ans}:
   We are given
   \begin{prooftree}
    \AxiomC{$\Gamma  \ottsym{,}  \mathit{x} \,  {:}  \, \ottnt{T_{{\mathrm{0}}}}  \vdash  \ottnt{C_{{\mathrm{21}}}}  \ottsym{<:}  \ottnt{C_{{\mathrm{11}}}}$}
    \AxiomC{$\Gamma  \vdash  \ottnt{C_{{\mathrm{12}}}}  \ottsym{<:}  \ottnt{C_{{\mathrm{22}}}}$}
    \BinaryInfC{$\Gamma \,  |  \, \ottnt{T_{{\mathrm{0}}}} \,  \,\&\,  \, \Phi_{{\mathrm{0}}}  \vdash   ( \forall  \mathit{x}  .  \ottnt{C_{{\mathrm{11}}}}  )  \Rightarrow   \ottnt{C_{{\mathrm{12}}}}   \ottsym{<:}   ( \forall  \mathit{x}  .  \ottnt{C_{{\mathrm{21}}}}  )  \Rightarrow   \ottnt{C_{{\mathrm{22}}}} $}
   \end{prooftree}
   for some $\mathit{x}$, $\ottnt{C_{{\mathrm{11}}}}$, $\ottnt{C_{{\mathrm{12}}}}$, $\ottnt{C_{{\mathrm{21}}}}$, and $\ottnt{C_{{\mathrm{22}}}}$ such that
   $\ottnt{S_{{\mathrm{1}}}} \,  =  \,  ( \forall  \mathit{x}  .  \ottnt{C_{{\mathrm{11}}}}  )  \Rightarrow   \ottnt{C_{{\mathrm{12}}}} $ and $\ottnt{S_{{\mathrm{2}}}} \,  =  \,  ( \forall  \mathit{x}  .  \ottnt{C_{{\mathrm{21}}}}  )  \Rightarrow   \ottnt{C_{{\mathrm{22}}}} $.
   By the IH, $\Gamma  \vdash  \Phi \,  \tceappendop  \, \ottnt{C_{{\mathrm{12}}}}  \ottsym{<:}  \Phi \,  \tceappendop  \, \ottnt{C_{{\mathrm{22}}}}$.
   Thus, \Srule{Ans} implies the conclusion
   \[
    \Gamma \,  |  \, \ottnt{T_{{\mathrm{0}}}} \,  \,\&\,  \, \Phi \,  \tceappendop  \, \Phi_{{\mathrm{0}}}  \vdash   ( \forall  \mathit{x}  .  \ottnt{C_{{\mathrm{11}}}}  )  \Rightarrow   \Phi \,  \tceappendop  \, \ottnt{C_{{\mathrm{12}}}}   \ottsym{<:}   ( \forall  \mathit{x}  .  \ottnt{C_{{\mathrm{21}}}}  )  \Rightarrow   \Phi \,  \tceappendop  \, \ottnt{C_{{\mathrm{22}}}}  ~.
   \]

  \case \Srule{Embed}:
   We are given
   \begin{prooftree}
    \AxiomC{$\Gamma  \ottsym{,}  \mathit{x} \,  {:}  \, \ottnt{T_{{\mathrm{0}}}}  \vdash  \Phi_{{\mathrm{0}}} \,  \tceappendop  \, \ottnt{C_{{\mathrm{21}}}}  \ottsym{<:}  \ottnt{C_{{\mathrm{22}}}}$}
    \AxiomC{$\Gamma  \ottsym{,}  \mathit{y} \,  {:}  \,  \Sigma ^\omega   \models    { \Phi_{{\mathrm{0}}} }^\nu   \ottsym{(}  \mathit{y}  \ottsym{)}    \Rightarrow     { \ottnt{C_{{\mathrm{22}}}} }^\nu (  \mathit{y}  )  $}
    \AxiomC{$\mathit{x}  \ottsym{,}  \mathit{y} \,  \not\in  \,  \mathit{fv}  (  \ottnt{C_{{\mathrm{22}}}}  )  \,  \mathbin{\cup}  \,  \mathit{fv}  (  \Phi_{{\mathrm{0}}}  ) $}
    \TrinaryInfC{$\Gamma \,  |  \, \ottnt{T_{{\mathrm{0}}}} \,  \,\&\,  \, \Phi_{{\mathrm{0}}}  \vdash   \square   \ottsym{<:}   ( \forall  \mathit{x}  .  \ottnt{C_{{\mathrm{21}}}}  )  \Rightarrow   \ottnt{C_{{\mathrm{22}}}} $}
   \end{prooftree}
   for some $\mathit{x}$, $\mathit{y}$, $\ottnt{C_{{\mathrm{21}}}}$, and $\ottnt{C_{{\mathrm{22}}}}$ such that
   $\ottnt{S_{{\mathrm{2}}}} \,  =  \,  ( \forall  \mathit{x}  .  \ottnt{C_{{\mathrm{21}}}}  )  \Rightarrow   \ottnt{C_{{\mathrm{22}}}} $.
   We also have $\ottnt{S_{{\mathrm{1}}}} \,  =  \,  \square $.
   Without loss of generality, we can suppose that $\mathit{x}  \ottsym{,}  \mathit{y} \,  \not\in  \,  \mathit{fv}  (  \Phi  ) $.
   By the IH,
   \[
    \Gamma  \ottsym{,}  \mathit{x} \,  {:}  \, \ottnt{T_{{\mathrm{0}}}}  \vdash  \Phi \,  \tceappendop  \, \ottsym{(}  \Phi_{{\mathrm{0}}} \,  \tceappendop  \, \ottnt{C_{{\mathrm{21}}}}  \ottsym{)}  \ottsym{<:}  \Phi \,  \tceappendop  \, \ottnt{C_{{\mathrm{22}}}} ~.
   \]
   \reflem{ts-assoc-preeff-compose-subtyping} implies
   \[
    \Gamma  \ottsym{,}  \mathit{x} \,  {:}  \, \ottnt{T_{{\mathrm{0}}}}  \vdash  \ottsym{(}  \Phi \,  \tceappendop  \, \Phi_{{\mathrm{0}}}  \ottsym{)} \,  \tceappendop  \, \ottnt{C_{{\mathrm{21}}}}  \ottsym{<:}  \Phi \,  \tceappendop  \, \ottnt{C_{{\mathrm{22}}}} ~.
   \]
   Further, because $\Gamma  \ottsym{,}  \mathit{y} \,  {:}  \,  \Sigma ^\omega   \models    { \Phi_{{\mathrm{0}}} }^\nu   \ottsym{(}  \mathit{y}  \ottsym{)}    \Rightarrow     { \ottnt{C_{{\mathrm{22}}}} }^\nu (  \mathit{y}  )  $,
   \reflem{ts-add-preeff-nu-eff-valid} implies
   \[
    \Gamma  \ottsym{,}  \mathit{y} \,  {:}  \,  \Sigma ^\omega   \models    { \ottsym{(}  \Phi \,  \tceappendop  \, \Phi_{{\mathrm{0}}}  \ottsym{)} }^\nu   \ottsym{(}  \mathit{y}  \ottsym{)}    \Rightarrow     { \ottsym{(}  \Phi \,  \tceappendop  \, \ottnt{C_{{\mathrm{22}}}}  \ottsym{)} }^\nu (  \mathit{y}  )   ~.
   \]
   Because $\mathit{x} \,  \not\in  \,  \mathit{fv}  (  \Phi \,  \tceappendop  \, \ottnt{C_{{\mathrm{22}}}}  )  \,  \mathbin{\cup}  \,  \mathit{fv}  (  \Phi \,  \tceappendop  \, \Phi_{{\mathrm{0}}}  ) $,
   \Srule{Embed} implies the conclusion
   \[
    \Gamma \,  |  \, \ottnt{T_{{\mathrm{0}}}} \,  \,\&\,  \, \Phi \,  \tceappendop  \, \Phi_{{\mathrm{0}}}  \vdash   \square   \ottsym{<:}   ( \forall  \mathit{x}  .  \ottnt{C_{{\mathrm{21}}}}  )  \Rightarrow   \Phi \,  \tceappendop  \, \ottnt{C_{{\mathrm{22}}}}  ~.
   \]
 \end{caseanalysis}
\end{proof}

\begin{lemmap}{Subtyping Preserves $\nu$-Effects}{lr-subtyping-nu}
 If $\Gamma  \vdash  \ottnt{C_{{\mathrm{1}}}}  \ottsym{<:}  \ottnt{C_{{\mathrm{2}}}}$ and $\Gamma  \ottsym{,}  \mathit{y} \,  {:}  \,  \Sigma ^\omega   \models   \phi    \Rightarrow     { \ottnt{C_{{\mathrm{1}}}} }^\nu (  \mathit{y}  )  $ and $\mathit{y} \,  \not\in  \,  \mathit{fv}  (  \ottnt{C_{{\mathrm{1}}}}  )  \,  \mathbin{\cup}  \,  \mathit{fv}  (  \ottnt{C_{{\mathrm{2}}}}  ) $,
 then $\Gamma  \ottsym{,}  \mathit{y} \,  {:}  \,  \Sigma ^\omega   \models   \phi    \Rightarrow     { \ottnt{C_{{\mathrm{2}}}} }^\nu (  \mathit{y}  )  $.
\end{lemmap}
\begin{proof}
 By induction on the subtyping derivation.
 Because $\Gamma  \vdash  \ottnt{C_{{\mathrm{1}}}}  \ottsym{<:}  \ottnt{C_{{\mathrm{2}}}}$ can be derived only by \Srule{Comp},
 we have
 \begin{itemize}
  \item $\Gamma  \vdash   \ottnt{C_{{\mathrm{1}}}}  . \Phi   \ottsym{<:}   \ottnt{C_{{\mathrm{2}}}}  . \Phi $ and
  \item $\Gamma \,  |  \,  \ottnt{C_{{\mathrm{1}}}}  . \ottnt{T}  \,  \,\&\,  \,  \ottnt{C_{{\mathrm{1}}}}  . \Phi   \vdash   \ottnt{C_{{\mathrm{1}}}}  . \ottnt{S}   \ottsym{<:}   \ottnt{C_{{\mathrm{2}}}}  . \ottnt{S} $.
 \end{itemize}
 Further, because $ { \ottnt{C_{{\mathrm{1}}}} }^\nu (  \mathit{y}  )  \,  =  \,   { \ottsym{(}   \ottnt{C_{{\mathrm{1}}}}  . \Phi   \ottsym{)} }^\nu   \ottsym{(}  \mathit{y}  \ottsym{)}    \wedge     { \ottsym{(}   \ottnt{C_{{\mathrm{1}}}}  . \ottnt{S}   \ottsym{)} }^\nu (  \mathit{y}  )  $,
 we have $\Gamma  \ottsym{,}  \mathit{y} \,  {:}  \,  \Sigma ^\omega   \models   \phi    \Rightarrow    \ottsym{(}    { \ottsym{(}   \ottnt{C_{{\mathrm{1}}}}  . \Phi   \ottsym{)} }^\nu   \ottsym{(}  \mathit{y}  \ottsym{)}    \wedge     { \ottsym{(}   \ottnt{C_{{\mathrm{1}}}}  . \ottnt{S}   \ottsym{)} }^\nu (  \mathit{y}  )    \ottsym{)} $.
 Then, \reflem{ts-valid-elim-impl-intersect} implies
 \begin{itemize}
  \item $\Gamma  \ottsym{,}  \mathit{y} \,  {:}  \,  \Sigma ^\omega   \models   \phi    \Rightarrow     { \ottsym{(}   \ottnt{C_{{\mathrm{1}}}}  . \Phi   \ottsym{)} }^\nu   \ottsym{(}  \mathit{y}  \ottsym{)} $ and
  \item $\Gamma  \ottsym{,}  \mathit{y} \,  {:}  \,  \Sigma ^\omega   \models   \phi    \Rightarrow     { \ottsym{(}   \ottnt{C_{{\mathrm{1}}}}  . \ottnt{S}   \ottsym{)} }^\nu (  \mathit{y}  )  $.
 \end{itemize}

 Because $ { \ottnt{C_{{\mathrm{2}}}} }^\nu (  \mathit{y}  )  \,  =  \,   { \ottsym{(}   \ottnt{C_{{\mathrm{2}}}}  . \Phi   \ottsym{)} }^\nu   \ottsym{(}  \mathit{y}  \ottsym{)}    \wedge     { \ottsym{(}   \ottnt{C_{{\mathrm{2}}}}  . \ottnt{S}   \ottsym{)} }^\nu (  \mathit{y}  )  $,
 \reflem{ts-valid-intro-impl-intersect} implies that it suffices to show the following.
 \begin{itemize}
  \item We show that $\Gamma  \ottsym{,}  \mathit{y} \,  {:}  \,  \Sigma ^\omega   \models   \phi    \Rightarrow     { \ottsym{(}   \ottnt{C_{{\mathrm{2}}}}  . \Phi   \ottsym{)} }^\nu   \ottsym{(}  \mathit{y}  \ottsym{)} $.
        By \reflem{ts-transitivity-valid}
        with $\Gamma  \ottsym{,}  \mathit{y} \,  {:}  \,  \Sigma ^\omega   \models   \phi    \Rightarrow     { \ottsym{(}   \ottnt{C_{{\mathrm{1}}}}  . \Phi   \ottsym{)} }^\nu   \ottsym{(}  \mathit{y}  \ottsym{)} $,
        it suffices to show that
        \[
         \Gamma  \ottsym{,}  \mathit{y} \,  {:}  \,  \Sigma ^\omega   \models    { \ottsym{(}   \ottnt{C_{{\mathrm{1}}}}  . \Phi   \ottsym{)} }^\nu   \ottsym{(}  \mathit{y}  \ottsym{)}    \Rightarrow     { \ottsym{(}   \ottnt{C_{{\mathrm{2}}}}  . \Phi   \ottsym{)} }^\nu   \ottsym{(}  \mathit{y}  \ottsym{)}  ~,
        \]
        which is implied by $\Gamma  \vdash   \ottnt{C_{{\mathrm{1}}}}  . \Phi   \ottsym{<:}   \ottnt{C_{{\mathrm{2}}}}  . \Phi $.

  \item We show that $\Gamma  \ottsym{,}  \mathit{y} \,  {:}  \,  \Sigma ^\omega   \models   \phi    \Rightarrow     { \ottsym{(}   \ottnt{C_{{\mathrm{2}}}}  . \ottnt{S}   \ottsym{)} }^\nu (  \mathit{y}  )  $.
        By case analysis on the subtyping rule applied last to derive
        $\Gamma \,  |  \,  \ottnt{C_{{\mathrm{1}}}}  . \ottnt{T}  \,  \,\&\,  \,  \ottnt{C_{{\mathrm{1}}}}  . \Phi   \vdash   \ottnt{C_{{\mathrm{1}}}}  . \ottnt{S}   \ottsym{<:}   \ottnt{C_{{\mathrm{2}}}}  . \ottnt{S} $.
        \begin{caseanalysis}
         \case \Srule{Empty}:
          We are given $ \ottnt{C_{{\mathrm{1}}}}  . \ottnt{S}  =  \ottnt{C_{{\mathrm{2}}}}  . \ottnt{S}  =  \square $.
          Because $ { \ottsym{(}   \ottnt{C_{{\mathrm{2}}}}  . \ottnt{S}   \ottsym{)} }^\nu (  \mathit{y}  )  \,  =  \,  \top $,
          it suffices to show that $\Gamma  \ottsym{,}  \mathit{y} \,  {:}  \,  \Sigma ^\omega   \models   \phi    \Rightarrow     \top  $,
          which is implied by \reflem{ts-valid-intro-impl-top}.

         \case \Srule{Ans}:
          We are given
          \begin{prooftree}
           \AxiomC{$\Gamma  \ottsym{,}  \mathit{x} \,  {:}  \,  \ottnt{C_{{\mathrm{1}}}}  . \ottnt{T}   \vdash  \ottnt{C_{{\mathrm{21}}}}  \ottsym{<:}  \ottnt{C_{{\mathrm{11}}}}$}
           \AxiomC{$\Gamma  \vdash  \ottnt{C_{{\mathrm{12}}}}  \ottsym{<:}  \ottnt{C_{{\mathrm{22}}}}$}
           \BinaryInfC{$\Gamma \,  |  \,  \ottnt{C_{{\mathrm{1}}}}  . \ottnt{T}  \,  \,\&\,  \,  \ottnt{C_{{\mathrm{1}}}}  . \Phi   \vdash   ( \forall  \mathit{x}  .  \ottnt{C_{{\mathrm{11}}}}  )  \Rightarrow   \ottnt{C_{{\mathrm{12}}}}   \ottsym{<:}   ( \forall  \mathit{x}  .  \ottnt{C_{{\mathrm{21}}}}  )  \Rightarrow   \ottnt{C_{{\mathrm{22}}}} $}
          \end{prooftree}
          where
          $ \ottnt{C_{{\mathrm{1}}}}  . \ottnt{S}  \,  =  \,  ( \forall  \mathit{x}  .  \ottnt{C_{{\mathrm{11}}}}  )  \Rightarrow   \ottnt{C_{{\mathrm{12}}}} $ and
          $ \ottnt{C_{{\mathrm{2}}}}  . \ottnt{S}  \,  =  \,  ( \forall  \mathit{x}  .  \ottnt{C_{{\mathrm{21}}}}  )  \Rightarrow   \ottnt{C_{{\mathrm{22}}}} $
          for some $\mathit{x}$, $\ottnt{C_{{\mathrm{11}}}}$, $\ottnt{C_{{\mathrm{12}}}}$, $\ottnt{C_{{\mathrm{21}}}}$, and $\ottnt{C_{{\mathrm{22}}}}$.
          Because $ { \ottsym{(}   \ottnt{C_{{\mathrm{1}}}}  . \ottnt{S}   \ottsym{)} }^\nu (  \mathit{y}  )  \,  =  \,  { \ottnt{C_{{\mathrm{12}}}} }^\nu (  \mathit{y}  ) $,
          $\Gamma  \ottsym{,}  \mathit{y} \,  {:}  \,  \Sigma ^\omega   \models   \phi    \Rightarrow     { \ottsym{(}   \ottnt{C_{{\mathrm{1}}}}  . \ottnt{S}   \ottsym{)} }^\nu (  \mathit{y}  )  $ implies
          $\Gamma  \ottsym{,}  \mathit{y} \,  {:}  \,  \Sigma ^\omega   \models   \phi    \Rightarrow     { \ottnt{C_{{\mathrm{12}}}} }^\nu (  \mathit{y}  )  $.
          Because $ { \ottsym{(}   \ottnt{C_{{\mathrm{2}}}}  . \ottnt{S}   \ottsym{)} }^\nu (  \mathit{y}  )  \,  =  \,  { \ottnt{C_{{\mathrm{22}}}} }^\nu (  \mathit{y}  ) $,
          we must show that $\Gamma  \ottsym{,}  \mathit{y} \,  {:}  \,  \Sigma ^\omega   \models   \phi    \Rightarrow     { \ottnt{C_{{\mathrm{22}}}} }^\nu (  \mathit{y}  )  $,
          which is proven by the IH on $\Gamma  \vdash  \ottnt{C_{{\mathrm{12}}}}  \ottsym{<:}  \ottnt{C_{{\mathrm{22}}}}$.

         \case \Srule{Embed}:
          We are given
          \begin{prooftree}
           \AxiomC{$\Gamma  \ottsym{,}  \mathit{x} \,  {:}  \,  \ottnt{C_{{\mathrm{1}}}}  . \ottnt{T}   \vdash  \ottsym{(}   \ottnt{C_{{\mathrm{1}}}}  . \Phi   \ottsym{)} \,  \tceappendop  \, \ottnt{C_{{\mathrm{21}}}}  \ottsym{<:}  \ottnt{C_{{\mathrm{22}}}}$}
           \AxiomC{$\Gamma  \ottsym{,}  \mathit{y} \,  {:}  \,  \Sigma ^\omega   \models    { \ottsym{(}   \ottnt{C_{{\mathrm{1}}}}  . \Phi   \ottsym{)} }^\nu   \ottsym{(}  \mathit{y}  \ottsym{)}    \Rightarrow     { \ottnt{C_{{\mathrm{22}}}} }^\nu (  \mathit{y}  )  $}
           \AxiomC{$\mathit{x}  \ottsym{,}  \mathit{y} \,  \not\in  \,  \mathit{fv}  (   \ottnt{C_{{\mathrm{1}}}}  . \Phi   )  \,  \mathbin{\cup}  \,  \mathit{fv}  (  \ottnt{C_{{\mathrm{22}}}}  ) $}
           \TrinaryInfC{$\Gamma \,  |  \,  \ottnt{C_{{\mathrm{1}}}}  . \ottnt{T}  \,  \,\&\,  \,  \ottnt{C_{{\mathrm{1}}}}  . \Phi   \vdash   \square   \ottsym{<:}   ( \forall  \mathit{x}  .  \ottnt{C_{{\mathrm{21}}}}  )  \Rightarrow   \ottnt{C_{{\mathrm{22}}}} $}
          \end{prooftree}
          where
          $ \ottnt{C_{{\mathrm{1}}}}  . \ottnt{S}  \,  =  \,  \square $ and
          $ \ottnt{C_{{\mathrm{2}}}}  . \ottnt{S}  \,  =  \,  ( \forall  \mathit{x}  .  \ottnt{C_{{\mathrm{21}}}}  )  \Rightarrow   \ottnt{C_{{\mathrm{22}}}} $
          for some $\mathit{x}$, $\ottnt{C_{{\mathrm{21}}}}$, and $\ottnt{C_{{\mathrm{22}}}}$.
          Because
          \begin{itemize}
           \item $\Gamma  \ottsym{,}  \mathit{y} \,  {:}  \,  \Sigma ^\omega   \models   \phi    \Rightarrow     { \ottsym{(}   \ottnt{C_{{\mathrm{1}}}}  . \Phi   \ottsym{)} }^\nu   \ottsym{(}  \mathit{y}  \ottsym{)} $,
           \item $\Gamma  \ottsym{,}  \mathit{y} \,  {:}  \,  \Sigma ^\omega   \models    { \ottsym{(}   \ottnt{C_{{\mathrm{1}}}}  . \Phi   \ottsym{)} }^\nu   \ottsym{(}  \mathit{y}  \ottsym{)}    \Rightarrow     { \ottnt{C_{{\mathrm{22}}}} }^\nu (  \mathit{y}  )  $, and
           \item $ { \ottnt{C_{{\mathrm{22}}}} }^\nu (  \mathit{y}  )  \,  =  \,  { \ottsym{(}   \ottnt{C_{{\mathrm{2}}}}  . \ottnt{S}   \ottsym{)} }^\nu (  \mathit{y}  ) $,
          \end{itemize}
          \reflem{ts-transitivity-valid} implies the conclusion
          $\Gamma  \ottsym{,}  \mathit{y} \,  {:}  \,  \Sigma ^\omega   \vdash   \phi    \Rightarrow     { \ottsym{(}   \ottnt{C_{{\mathrm{2}}}}  . \ottnt{S}   \ottsym{)} }^\nu (  \mathit{y}  )  $.
        \end{caseanalysis}
 \end{itemize}
\end{proof}

\begin{lemmap}{Transitivity of Subtyping}{ts-transitivity-subtyping}
 \begin{enumerate}
  \item If $\Gamma  \vdash  \ottnt{T_{{\mathrm{1}}}}  \ottsym{<:}  \ottnt{T_{{\mathrm{2}}}}$ and $\Gamma  \vdash  \ottnt{T_{{\mathrm{2}}}}  \ottsym{<:}  \ottnt{T_{{\mathrm{3}}}}$,
        then $\Gamma  \vdash  \ottnt{T_{{\mathrm{1}}}}  \ottsym{<:}  \ottnt{T_{{\mathrm{3}}}}$.

  \item \label{lem:ts-transitivity-subtyping:cty}
        If $\Gamma  \vdash  \ottnt{C_{{\mathrm{1}}}}  \ottsym{<:}  \ottnt{C_{{\mathrm{2}}}}$ and $\Gamma  \vdash  \ottnt{C_{{\mathrm{2}}}}  \ottsym{<:}  \ottnt{C_{{\mathrm{3}}}}$,
        then $\Gamma  \vdash  \ottnt{C_{{\mathrm{1}}}}  \ottsym{<:}  \ottnt{C_{{\mathrm{3}}}}$.

  \item \label{lem:ts-transitivity-subtyping:sty}
        If
        $\Gamma \,  |  \, \ottnt{T} \,  \,\&\,  \, \Phi  \vdash  \ottnt{S_{{\mathrm{1}}}}  \ottsym{<:}  \ottnt{S_{{\mathrm{2}}}}$ and
        $\Gamma \,  |  \, \ottnt{T} \,  \,\&\,  \, \Phi  \vdash  \ottnt{S_{{\mathrm{2}}}}  \ottsym{<:}  \ottnt{S_{{\mathrm{3}}}}$,
        then $\Gamma \,  |  \, \ottnt{T} \,  \,\&\,  \, \Phi  \vdash  \ottnt{S_{{\mathrm{1}}}}  \ottsym{<:}  \ottnt{S_{{\mathrm{3}}}}$.

  \item \label{lem:ts-transitivity-subtyping:eff}
        If $\Gamma  \vdash  \Phi_{{\mathrm{1}}}  \ottsym{<:}  \Phi_{{\mathrm{2}}}$ and $\Gamma  \vdash  \Phi_{{\mathrm{2}}}  \ottsym{<:}  \Phi_{{\mathrm{3}}}$,
        then $\Gamma  \vdash  \Phi_{{\mathrm{1}}}  \ottsym{<:}  \Phi_{{\mathrm{3}}}$.
 \end{enumerate}
\end{lemmap}
\begin{proof}
 The case (\ref{lem:ts-transitivity-subtyping:eff}) is shown by
 \reflem{ts-transitivity-eff-subtyping}.
 We prove the other cases by induction on the numbers of the outermost type
 constructors of $\ottnt{T_{{\mathrm{2}}}}$, $\ottnt{C_{{\mathrm{2}}}}$, and $\ottnt{S_{{\mathrm{2}}}}$.
 \begin{enumerate}
  \item By case analysis on the subtyping rule applied last to derive
        $\Gamma  \vdash  \ottnt{T_{{\mathrm{1}}}}  \ottsym{<:}  \ottnt{T_{{\mathrm{2}}}}$.
        \begin{caseanalysis}
         \case \Srule{Refine}:
          We are given
          \begin{prooftree}
           \AxiomC{$\Gamma  \ottsym{,}  \mathit{x} \,  {:}  \, \iota  \vdash   \phi_{{\mathrm{1}}}    \Rightarrow    \phi_{{\mathrm{2}}} $}
           \UnaryInfC{$\Gamma  \vdash  \ottsym{\{}  \mathit{x} \,  {:}  \, \iota \,  |  \, \phi_{{\mathrm{1}}}  \ottsym{\}}  \ottsym{<:}  \ottsym{\{}  \mathit{x} \,  {:}  \, \iota \,  |  \, \phi_{{\mathrm{2}}}  \ottsym{\}}$}
          \end{prooftree}
          for some $\mathit{x}$, $\iota$, $\phi_{{\mathrm{1}}}$, and $\phi_{{\mathrm{2}}}$ such that
          $\ottnt{T_{{\mathrm{1}}}} \,  =  \, \ottsym{\{}  \mathit{x} \,  {:}  \, \iota \,  |  \, \phi_{{\mathrm{1}}}  \ottsym{\}}$ and
          $\ottnt{T_{{\mathrm{2}}}} \,  =  \, \ottsym{\{}  \mathit{x} \,  {:}  \, \iota \,  |  \, \phi_{{\mathrm{2}}}  \ottsym{\}}$.
          Because $\Gamma  \vdash  \ottnt{T_{{\mathrm{2}}}}  \ottsym{<:}  \ottnt{T_{{\mathrm{3}}}}$ (i.e., $\Gamma  \vdash  \ottsym{\{}  \mathit{x} \,  {:}  \, \iota \,  |  \, \phi_{{\mathrm{2}}}  \ottsym{\}}  \ottsym{<:}  \ottnt{T_{{\mathrm{3}}}}$)
          can be derived only by \Srule{Refine},
          we can find that
          \begin{prooftree}
           \AxiomC{$\Gamma  \ottsym{,}  \mathit{x} \,  {:}  \, \iota  \vdash   \phi_{{\mathrm{2}}}    \Rightarrow    \phi_{{\mathrm{3}}} $}
           \UnaryInfC{$\Gamma  \vdash  \ottsym{\{}  \mathit{x} \,  {:}  \, \iota \,  |  \, \phi_{{\mathrm{2}}}  \ottsym{\}}  \ottsym{<:}  \ottsym{\{}  \mathit{x} \,  {:}  \, \iota \,  |  \, \phi_{{\mathrm{3}}}  \ottsym{\}}$}
          \end{prooftree}
          for some $\phi_{{\mathrm{3}}}$ such that $\ottnt{T_{{\mathrm{3}}}} \,  =  \, \ottsym{\{}  \mathit{x} \,  {:}  \, \iota \,  |  \, \phi_{{\mathrm{3}}}  \ottsym{\}}$.
          By \reflem{ts-transitivity-valid},
          \[
           \Gamma  \ottsym{,}  \mathit{x} \,  {:}  \, \iota  \vdash   \phi_{{\mathrm{1}}}    \Rightarrow    \phi_{{\mathrm{3}}}  ~.
          \]
          Thus, by \Srule{Refine},
          \[
           \Gamma  \vdash  \ottsym{\{}  \mathit{x} \,  {:}  \, \iota \,  |  \, \phi_{{\mathrm{1}}}  \ottsym{\}}  \ottsym{<:}  \ottsym{\{}  \mathit{x} \,  {:}  \, \iota \,  |  \, \phi_{{\mathrm{3}}}  \ottsym{\}} ~,
          \]
          which we must show.

         \case \Srule{Fun}:
          We are given
          \begin{prooftree}
           \AxiomC{$\Gamma  \vdash  \ottnt{T'_{{\mathrm{2}}}}  \ottsym{<:}  \ottnt{T'_{{\mathrm{1}}}}$}
           \AxiomC{$\Gamma  \ottsym{,}  \mathit{x} \,  {:}  \, \ottnt{T'_{{\mathrm{2}}}}  \vdash  \ottnt{C'_{{\mathrm{1}}}}  \ottsym{<:}  \ottnt{C'_{{\mathrm{2}}}}$}
           \BinaryInfC{$\Gamma  \vdash  \ottsym{(}  \mathit{x} \,  {:}  \, \ottnt{T'_{{\mathrm{1}}}}  \ottsym{)}  \rightarrow  \ottnt{C'_{{\mathrm{1}}}}  \ottsym{<:}  \ottsym{(}  \mathit{x} \,  {:}  \, \ottnt{T'_{{\mathrm{2}}}}  \ottsym{)}  \rightarrow  \ottnt{C'_{{\mathrm{2}}}}$}
          \end{prooftree}
          for some $\mathit{x}$, $\ottnt{T'_{{\mathrm{1}}}}$, $\ottnt{T'_{{\mathrm{2}}}}$, $\ottnt{C'_{{\mathrm{1}}}}$, and $\ottnt{C'_{{\mathrm{2}}}}$
          such that $\ottnt{T_{{\mathrm{1}}}} \,  =  \, \ottsym{(}  \mathit{x} \,  {:}  \, \ottnt{T'_{{\mathrm{1}}}}  \ottsym{)}  \rightarrow  \ottnt{C'_{{\mathrm{1}}}}$ and $\ottnt{T_{{\mathrm{2}}}} \,  =  \, \ottsym{(}  \mathit{x} \,  {:}  \, \ottnt{T'_{{\mathrm{2}}}}  \ottsym{)}  \rightarrow  \ottnt{C'_{{\mathrm{2}}}}$.
          Because $\Gamma  \vdash  \ottnt{T_{{\mathrm{2}}}}  \ottsym{<:}  \ottnt{T_{{\mathrm{3}}}}$ (i.e., $\Gamma  \vdash  \ottsym{(}  \mathit{x} \,  {:}  \, \ottnt{T'_{{\mathrm{2}}}}  \ottsym{)}  \rightarrow  \ottnt{C'_{{\mathrm{2}}}}  \ottsym{<:}  \ottnt{T_{{\mathrm{3}}}}$)
          can be derived only by \Srule{Fun},
          we can find that
          \begin{prooftree}
           \AxiomC{$\Gamma  \vdash  \ottnt{T'_{{\mathrm{3}}}}  \ottsym{<:}  \ottnt{T'_{{\mathrm{2}}}}$}
           \AxiomC{$\Gamma  \ottsym{,}  \mathit{x} \,  {:}  \, \ottnt{T'_{{\mathrm{3}}}}  \vdash  \ottnt{C'_{{\mathrm{2}}}}  \ottsym{<:}  \ottnt{C'_{{\mathrm{3}}}}$}
           \BinaryInfC{$\Gamma  \vdash  \ottsym{(}  \mathit{x} \,  {:}  \, \ottnt{T'_{{\mathrm{2}}}}  \ottsym{)}  \rightarrow  \ottnt{C'_{{\mathrm{2}}}}  \ottsym{<:}  \ottsym{(}  \mathit{x} \,  {:}  \, \ottnt{T'_{{\mathrm{3}}}}  \ottsym{)}  \rightarrow  \ottnt{C'_{{\mathrm{3}}}}$}
          \end{prooftree}
          for some $\ottnt{T'_{{\mathrm{3}}}}$ and $\ottnt{C'_{{\mathrm{3}}}}$ such that $\ottnt{T_{{\mathrm{3}}}} \,  =  \, \ottsym{(}  \mathit{x} \,  {:}  \, \ottnt{T'_{{\mathrm{3}}}}  \ottsym{)}  \rightarrow  \ottnt{C'_{{\mathrm{3}}}}$.
          By the IH on $\ottnt{T'_{{\mathrm{2}}}}$ in $\Gamma  \vdash  \ottnt{T'_{{\mathrm{3}}}}  \ottsym{<:}  \ottnt{T'_{{\mathrm{2}}}}$ and $\Gamma  \vdash  \ottnt{T'_{{\mathrm{2}}}}  \ottsym{<:}  \ottnt{T'_{{\mathrm{1}}}}$,
          we have
          \begin{equation}
           \Gamma  \vdash  \ottnt{T'_{{\mathrm{3}}}}  \ottsym{<:}  \ottnt{T'_{{\mathrm{1}}}} ~.
            \label{lem:ts-transitivity-subtyping:funtyp:arg}
          \end{equation}
          \reflem{ts-strength-ty-hypo-subtyping} with
          $\Gamma  \ottsym{,}  \mathit{x} \,  {:}  \, \ottnt{T'_{{\mathrm{2}}}}  \vdash  \ottnt{C'_{{\mathrm{1}}}}  \ottsym{<:}  \ottnt{C'_{{\mathrm{2}}}}$ and $\Gamma  \vdash  \ottnt{T'_{{\mathrm{3}}}}  \ottsym{<:}  \ottnt{T'_{{\mathrm{2}}}}$
          implies
          \[
           \Gamma  \ottsym{,}  \mathit{x} \,  {:}  \, \ottnt{T'_{{\mathrm{3}}}}  \vdash  \ottnt{C'_{{\mathrm{1}}}}  \ottsym{<:}  \ottnt{C'_{{\mathrm{2}}}} ~.
          \]
          Therefore, by the IH on $\ottnt{C'_{{\mathrm{2}}}}$
          in $\Gamma  \ottsym{,}  \mathit{x} \,  {:}  \, \ottnt{T'_{{\mathrm{3}}}}  \vdash  \ottnt{C'_{{\mathrm{1}}}}  \ottsym{<:}  \ottnt{C'_{{\mathrm{2}}}}$ and $\Gamma  \ottsym{,}  \mathit{x} \,  {:}  \, \ottnt{T'_{{\mathrm{3}}}}  \vdash  \ottnt{C'_{{\mathrm{2}}}}  \ottsym{<:}  \ottnt{C'_{{\mathrm{3}}}}$, we have
          \begin{equation}
           \Gamma  \ottsym{,}  \mathit{x} \,  {:}  \, \ottnt{T'_{{\mathrm{3}}}}  \vdash  \ottnt{C'_{{\mathrm{1}}}}  \ottsym{<:}  \ottnt{C'_{{\mathrm{3}}}} ~.
            \label{lem:ts-transitivity-subtyping:funtyp:ret}
          \end{equation}
          \Srule{Fun} with
          (\ref{lem:ts-transitivity-subtyping:funtyp:arg}) and
          (\ref{lem:ts-transitivity-subtyping:funtyp:ret})
          implies
          \[
           \Gamma  \vdash  \ottsym{(}  \mathit{x} \,  {:}  \, \ottnt{T'_{{\mathrm{1}}}}  \ottsym{)}  \rightarrow  \ottnt{C'_{{\mathrm{1}}}}  \ottsym{<:}  \ottsym{(}  \mathit{x} \,  {:}  \, \ottnt{T'_{{\mathrm{3}}}}  \ottsym{)}  \rightarrow  \ottnt{C'_{{\mathrm{3}}}} ~,
          \]
          which we must show.

         \case \Srule{Forall}:
          We are given
          \begin{prooftree}
           \AxiomC{$\Gamma  \ottsym{,}  \mathit{X} \,  {:}  \,  \tceseq{ \ottnt{s} }   \vdash  \ottnt{C_{{\mathrm{1}}}}  \ottsym{<:}  \ottnt{C_{{\mathrm{2}}}}$}
           \UnaryInfC{$\Gamma  \vdash   \forall   \ottsym{(}  \mathit{X} \,  {:}  \,  \tceseq{ \ottnt{s} }   \ottsym{)}  \ottsym{.}  \ottnt{C_{{\mathrm{1}}}}  \ottsym{<:}   \forall   \ottsym{(}  \mathit{X} \,  {:}  \,  \tceseq{ \ottnt{s} }   \ottsym{)}  \ottsym{.}  \ottnt{C_{{\mathrm{2}}}}$}
          \end{prooftree}
          for some $\mathit{X}$, $ \tceseq{ \ottnt{s} } $, $\ottnt{C_{{\mathrm{1}}}}$, and $\ottnt{C_{{\mathrm{2}}}}$ such that
          $\ottnt{T_{{\mathrm{1}}}} \,  =  \,  \forall   \ottsym{(}  \mathit{X} \,  {:}  \,  \tceseq{ \ottnt{s} }   \ottsym{)}  \ottsym{.}  \ottnt{C_{{\mathrm{1}}}}$ and $\ottnt{T_{{\mathrm{2}}}} \,  =  \,  \forall   \ottsym{(}  \mathit{X} \,  {:}  \,  \tceseq{ \ottnt{s} }   \ottsym{)}  \ottsym{.}  \ottnt{C_{{\mathrm{2}}}}$.
          Because $\Gamma  \vdash  \ottnt{T_{{\mathrm{2}}}}  \ottsym{<:}  \ottnt{T_{{\mathrm{3}}}}$ (i.e., $\Gamma  \vdash   \forall   \ottsym{(}  \mathit{X} \,  {:}  \,  \tceseq{ \ottnt{s} }   \ottsym{)}  \ottsym{.}  \ottnt{C_{{\mathrm{2}}}}  \ottsym{<:}  \ottnt{T_{{\mathrm{3}}}}$)
          can be derived only by \Srule{Forall},
          we can find that
          \begin{prooftree}
           \AxiomC{$\Gamma  \ottsym{,}  \mathit{X} \,  {:}  \,  \tceseq{ \ottnt{s} }   \vdash  \ottnt{C_{{\mathrm{2}}}}  \ottsym{<:}  \ottnt{C_{{\mathrm{3}}}}$}
           \UnaryInfC{$\Gamma  \vdash   \forall   \ottsym{(}  \mathit{X} \,  {:}  \,  \tceseq{ \ottnt{s} }   \ottsym{)}  \ottsym{.}  \ottnt{C_{{\mathrm{2}}}}  \ottsym{<:}   \forall   \ottsym{(}  \mathit{X} \,  {:}  \,  \tceseq{ \ottnt{s} }   \ottsym{)}  \ottsym{.}  \ottnt{C_{{\mathrm{3}}}}$}
          \end{prooftree}
          for some $\ottnt{C_{{\mathrm{3}}}}$ such that $\ottnt{T_{{\mathrm{3}}}} \,  =  \,  \forall   \ottsym{(}  \mathit{X} \,  {:}  \,  \tceseq{ \ottnt{s} }   \ottsym{)}  \ottsym{.}  \ottnt{C_{{\mathrm{3}}}}$.
          The IH on $\ottnt{C_{{\mathrm{2}}}}$ implies $\Gamma  \ottsym{,}  \mathit{X} \,  {:}  \,  \tceseq{ \ottnt{s} }   \vdash  \ottnt{C_{{\mathrm{1}}}}  \ottsym{<:}  \ottnt{C_{{\mathrm{3}}}}$.
          By \Srule{Forall},
          \[
           \Gamma  \vdash   \forall   \ottsym{(}  \mathit{X} \,  {:}  \,  \tceseq{ \ottnt{s} }   \ottsym{)}  \ottsym{.}  \ottnt{C_{{\mathrm{1}}}}  \ottsym{<:}   \forall   \ottsym{(}  \mathit{X} \,  {:}  \,  \tceseq{ \ottnt{s} }   \ottsym{)}  \ottsym{.}  \ottnt{C_{{\mathrm{3}}}} ~,
          \]
          which we must show.
        \end{caseanalysis}

  \item Because $\Gamma  \vdash  \ottnt{C_{{\mathrm{1}}}}  \ottsym{<:}  \ottnt{C_{{\mathrm{2}}}}$ can be derived only by \Srule{Comp},
        its inversion implies
        \begin{itemize}
         \item $\Gamma  \vdash   \ottnt{C_{{\mathrm{1}}}}  . \ottnt{T}   \ottsym{<:}   \ottnt{C_{{\mathrm{2}}}}  . \ottnt{T} $,
         \item $\Gamma  \vdash   \ottnt{C_{{\mathrm{1}}}}  . \Phi   \ottsym{<:}   \ottnt{C_{{\mathrm{2}}}}  . \Phi $, and
         \item $\Gamma \,  |  \,  \ottnt{C_{{\mathrm{1}}}}  . \ottnt{T}  \,  \,\&\,  \,  \ottnt{C_{{\mathrm{1}}}}  . \Phi   \vdash   \ottnt{C_{{\mathrm{1}}}}  . \ottnt{S}   \ottsym{<:}   \ottnt{C_{{\mathrm{2}}}}  . \ottnt{S} $.
        \end{itemize}
        Similarly, the inversion of $\Gamma  \vdash  \ottnt{C_{{\mathrm{2}}}}  \ottsym{<:}  \ottnt{C_{{\mathrm{3}}}}$ implies
        \begin{itemize}
         \item $\Gamma  \vdash   \ottnt{C_{{\mathrm{2}}}}  . \ottnt{T}   \ottsym{<:}   \ottnt{C_{{\mathrm{3}}}}  . \ottnt{T} $,
         \item $\Gamma  \vdash   \ottnt{C_{{\mathrm{2}}}}  . \Phi   \ottsym{<:}   \ottnt{C_{{\mathrm{3}}}}  . \Phi $, and
         \item $\Gamma \,  |  \,  \ottnt{C_{{\mathrm{2}}}}  . \ottnt{T}  \,  \,\&\,  \,  \ottnt{C_{{\mathrm{2}}}}  . \Phi   \vdash   \ottnt{C_{{\mathrm{2}}}}  . \ottnt{S}   \ottsym{<:}   \ottnt{C_{{\mathrm{3}}}}  . \ottnt{S} $.
        \end{itemize}
        %
        By the IH on $ \ottnt{C_{{\mathrm{2}}}}  . \ottnt{T} $ in $\Gamma  \vdash   \ottnt{C_{{\mathrm{1}}}}  . \ottnt{T}   \ottsym{<:}   \ottnt{C_{{\mathrm{2}}}}  . \ottnt{T} $ and $\Gamma  \vdash   \ottnt{C_{{\mathrm{2}}}}  . \ottnt{T}   \ottsym{<:}   \ottnt{C_{{\mathrm{3}}}}  . \ottnt{T} $,
        we have
        \begin{equation}
         \Gamma  \vdash   \ottnt{C_{{\mathrm{1}}}}  . \ottnt{T}   \ottsym{<:}   \ottnt{C_{{\mathrm{3}}}}  . \ottnt{T}  ~.
          \label{lem:ts-transitivity-subtyping:cty:ty}
        \end{equation}
        By the case (\ref{lem:ts-transitivity-subtyping:eff})
        with $\Gamma  \vdash   \ottnt{C_{{\mathrm{1}}}}  . \Phi   \ottsym{<:}   \ottnt{C_{{\mathrm{2}}}}  . \Phi $ and $\Gamma  \vdash   \ottnt{C_{{\mathrm{2}}}}  . \Phi   \ottsym{<:}   \ottnt{C_{{\mathrm{3}}}}  . \Phi $,
        we have
        \begin{equation}
         \Gamma  \vdash   \ottnt{C_{{\mathrm{1}}}}  . \Phi   \ottsym{<:}   \ottnt{C_{{\mathrm{3}}}}  . \Phi  ~.
          \label{lem:ts-transitivity-subtyping:cty:tp}
        \end{equation}
        Because
        $\Gamma  \vdash   \ottnt{C_{{\mathrm{1}}}}  . \ottnt{T}   \ottsym{<:}   \ottnt{C_{{\mathrm{2}}}}  . \ottnt{T} $ and
        $\Gamma  \vdash   \ottnt{C_{{\mathrm{1}}}}  . \Phi   \ottsym{<:}   \ottnt{C_{{\mathrm{2}}}}  . \Phi $,
        we have
        \[
         \Gamma \,  |  \,  \ottnt{C_{{\mathrm{1}}}}  . \ottnt{T}  \,  \,\&\,  \,  \ottnt{C_{{\mathrm{1}}}}  . \Phi   \vdash   \ottnt{C_{{\mathrm{2}}}}  . \ottnt{S}   \ottsym{<:}   \ottnt{C_{{\mathrm{3}}}}  . \ottnt{S} 
        \]
        by Lemmas~\ref{lem:ts-strength-ty-hypo-stack-subtyping} and
        \ref{lem:ts-strength-preeff-stack-subtyping}
        with
        $\Gamma \,  |  \,  \ottnt{C_{{\mathrm{2}}}}  . \ottnt{T}  \,  \,\&\,  \,  \ottnt{C_{{\mathrm{2}}}}  . \Phi   \vdash   \ottnt{C_{{\mathrm{2}}}}  . \ottnt{S}   \ottsym{<:}   \ottnt{C_{{\mathrm{3}}}}  . \ottnt{S} $.
        Therefore, the IH on $ \ottnt{C_{{\mathrm{2}}}}  . \ottnt{S} $
        implies
        \begin{equation}
         \Gamma \,  |  \,  \ottnt{C_{{\mathrm{1}}}}  . \ottnt{T}  \,  \,\&\,  \,  \ottnt{C_{{\mathrm{1}}}}  . \Phi   \vdash   \ottnt{C_{{\mathrm{1}}}}  . \ottnt{S}   \ottsym{<:}   \ottnt{C_{{\mathrm{3}}}}  . \ottnt{S} 
          \label{lem:ts-transitivity-subtyping:cty:sty} ~.
        \end{equation}
        Then, \Srule{Comp} with
        (\ref{lem:ts-transitivity-subtyping:cty:ty}),
        (\ref{lem:ts-transitivity-subtyping:cty:tp}), and
        (\ref{lem:ts-transitivity-subtyping:cty:sty}) implies
        the conclusion
        \[
         \Gamma  \vdash  \ottnt{C_{{\mathrm{1}}}}  \ottsym{<:}  \ottnt{C_{{\mathrm{3}}}} ~.
        \]

  \item By case analysis on the subtyping rule applied last to derive
        $\Gamma \,  |  \, \ottnt{T} \,  \,\&\,  \, \Phi  \vdash  \ottnt{S_{{\mathrm{1}}}}  \ottsym{<:}  \ottnt{S_{{\mathrm{2}}}}$.
        \begin{caseanalysis}
         \case \Srule{Empty}:
          We are given
          \begin{itemize}
           \item $\ottnt{S_{{\mathrm{1}}}} \,  =  \,  \square $ and
           \item $\ottnt{S_{{\mathrm{2}}}} \,  =  \,  \square $
          \end{itemize}
          (i.e., $\Gamma \,  |  \, \ottnt{T} \,  \,\&\,  \, \Phi  \vdash   \square   \ottsym{<:}   \square $).
          Thus, $\Gamma \,  |  \, \ottnt{T} \,  \,\&\,  \, \Phi  \vdash  \ottnt{S_{{\mathrm{2}}}}  \ottsym{<:}  \ottnt{S_{{\mathrm{3}}}}$ implies
          $\Gamma \,  |  \, \ottnt{T} \,  \,\&\,  \, \Phi  \vdash   \square   \ottsym{<:}  \ottnt{S_{{\mathrm{3}}}}$, which we must show.

         \case \Srule{Ans}:
          We are given
          \begin{prooftree}
           \AxiomC{$\Gamma  \ottsym{,}  \mathit{x} \,  {:}  \, \ottnt{T}  \vdash  \ottnt{C_{{\mathrm{21}}}}  \ottsym{<:}  \ottnt{C_{{\mathrm{11}}}}$}
           \AxiomC{$\Gamma  \vdash  \ottnt{C_{{\mathrm{12}}}}  \ottsym{<:}  \ottnt{C_{{\mathrm{22}}}}$}
           \BinaryInfC{$\Gamma \,  |  \, \ottnt{T} \,  \,\&\,  \, \Phi  \vdash   ( \forall  \mathit{x}  .  \ottnt{C_{{\mathrm{11}}}}  )  \Rightarrow   \ottnt{C_{{\mathrm{12}}}}   \ottsym{<:}   ( \forall  \mathit{x}  .  \ottnt{C_{{\mathrm{21}}}}  )  \Rightarrow   \ottnt{C_{{\mathrm{22}}}} $}
          \end{prooftree}
          for some $\mathit{x}$, $\ottnt{C_{{\mathrm{11}}}}$, $\ottnt{C_{{\mathrm{12}}}}$, $\ottnt{C_{{\mathrm{21}}}}$, and $\ottnt{C_{{\mathrm{22}}}}$ such that
          $\ottnt{S_{{\mathrm{1}}}} \,  =  \,  ( \forall  \mathit{x}  .  \ottnt{C_{{\mathrm{11}}}}  )  \Rightarrow   \ottnt{C_{{\mathrm{12}}}} $ and
          $\ottnt{S_{{\mathrm{2}}}} \,  =  \,  ( \forall  \mathit{x}  .  \ottnt{C_{{\mathrm{21}}}}  )  \Rightarrow   \ottnt{C_{{\mathrm{22}}}} $.
          Because $\Gamma \,  |  \, \ottnt{T} \,  \,\&\,  \, \Phi  \vdash  \ottnt{S_{{\mathrm{2}}}}  \ottsym{<:}  \ottnt{S_{{\mathrm{3}}}}$
          (i.e., $\Gamma \,  |  \, \ottnt{T} \,  \,\&\,  \, \Phi  \vdash   ( \forall  \mathit{x}  .  \ottnt{C_{{\mathrm{21}}}}  )  \Rightarrow   \ottnt{C_{{\mathrm{22}}}}   \ottsym{<:}  \ottnt{S_{{\mathrm{3}}}}$)
          can be derived only by \Srule{Ans},
          we can find
          \begin{prooftree}
           \AxiomC{$\Gamma  \ottsym{,}  \mathit{x} \,  {:}  \, \ottnt{T}  \vdash  \ottnt{C_{{\mathrm{31}}}}  \ottsym{<:}  \ottnt{C_{{\mathrm{21}}}}$}
           \AxiomC{$\Gamma  \vdash  \ottnt{C_{{\mathrm{22}}}}  \ottsym{<:}  \ottnt{C_{{\mathrm{32}}}}$}
           \BinaryInfC{$\Gamma \,  |  \, \ottnt{T} \,  \,\&\,  \, \Phi  \vdash   ( \forall  \mathit{x}  .  \ottnt{C_{{\mathrm{21}}}}  )  \Rightarrow   \ottnt{C_{{\mathrm{22}}}}   \ottsym{<:}   ( \forall  \mathit{x}  .  \ottnt{C_{{\mathrm{31}}}}  )  \Rightarrow   \ottnt{C_{{\mathrm{32}}}} $}
          \end{prooftree}
          for some $\ottnt{C_{{\mathrm{31}}}}$ and $\ottnt{C_{{\mathrm{32}}}}$ such that $\ottnt{S_{{\mathrm{3}}}} \,  =  \,  ( \forall  \mathit{x}  .  \ottnt{C_{{\mathrm{31}}}}  )  \Rightarrow   \ottnt{C_{{\mathrm{32}}}} $.

          The IH on $\ottnt{C_{{\mathrm{21}}}}$ in $\Gamma  \ottsym{,}  \mathit{x} \,  {:}  \, \ottnt{T}  \vdash  \ottnt{C_{{\mathrm{31}}}}  \ottsym{<:}  \ottnt{C_{{\mathrm{21}}}}$ and $\Gamma  \ottsym{,}  \mathit{x} \,  {:}  \, \ottnt{T}  \vdash  \ottnt{C_{{\mathrm{21}}}}  \ottsym{<:}  \ottnt{C_{{\mathrm{11}}}}$ implies
          \[
           \Gamma  \ottsym{,}  \mathit{x} \,  {:}  \, \ottnt{T}  \vdash  \ottnt{C_{{\mathrm{31}}}}  \ottsym{<:}  \ottnt{C_{{\mathrm{11}}}} ~.
          \]
          The IH on $\ottnt{C_{{\mathrm{22}}}}$ in $\Gamma  \vdash  \ottnt{C_{{\mathrm{12}}}}  \ottsym{<:}  \ottnt{C_{{\mathrm{22}}}}$ and $\Gamma  \vdash  \ottnt{C_{{\mathrm{22}}}}  \ottsym{<:}  \ottnt{C_{{\mathrm{32}}}}$
          implies
          \[
           \Gamma  \vdash  \ottnt{C_{{\mathrm{12}}}}  \ottsym{<:}  \ottnt{C_{{\mathrm{32}}}} ~.
          \]
          Thus, by \Srule{Ans}, we have the conclusion
          \[
           \Gamma \,  |  \, \ottnt{T} \,  \,\&\,  \, \Phi  \vdash   ( \forall  \mathit{x}  .  \ottnt{C_{{\mathrm{11}}}}  )  \Rightarrow   \ottnt{C_{{\mathrm{12}}}}   \ottsym{<:}   ( \forall  \mathit{x}  .  \ottnt{C_{{\mathrm{31}}}}  )  \Rightarrow   \ottnt{C_{{\mathrm{32}}}}  ~.
          \]

         \case \Srule{Embed}:
          We are given
          \begin{prooftree}
           \AxiomC{$\Gamma  \ottsym{,}  \mathit{x} \,  {:}  \, \ottnt{T}  \vdash  \Phi \,  \tceappendop  \, \ottnt{C_{{\mathrm{21}}}}  \ottsym{<:}  \ottnt{C_{{\mathrm{22}}}}$}
           \AxiomC{$\Gamma  \ottsym{,}  \mathit{y} \,  {:}  \,  \Sigma ^\omega   \models    { \Phi }^\nu   \ottsym{(}  \mathit{y}  \ottsym{)}    \Rightarrow     { \ottnt{C_{{\mathrm{22}}}} }^\nu (  \mathit{y}  )  $}
           \AxiomC{$\mathit{x}  \ottsym{,}  \mathit{y} \,  \not\in  \,  \mathit{fv}  (  \ottnt{C_{{\mathrm{22}}}}  )  \,  \mathbin{\cup}  \,  \mathit{fv}  (  \Phi  ) $}
           \TrinaryInfC{$\Gamma \,  |  \, \ottnt{T} \,  \,\&\,  \, \Phi  \vdash   \square   \ottsym{<:}   ( \forall  \mathit{x}  .  \ottnt{C_{{\mathrm{21}}}}  )  \Rightarrow   \ottnt{C_{{\mathrm{22}}}} $}
          \end{prooftree}
          for some $\mathit{x}$, $\mathit{y}$, $\ottnt{C_{{\mathrm{21}}}}$, and $\ottnt{C_{{\mathrm{22}}}}$ such that
          $\ottnt{S_{{\mathrm{2}}}} \,  =  \,  ( \forall  \mathit{x}  .  \ottnt{C_{{\mathrm{21}}}}  )  \Rightarrow   \ottnt{C_{{\mathrm{22}}}} $.
          We also have $\ottnt{S_{{\mathrm{1}}}} \,  =  \,  \square $.
          Without loss of generality, we can suppose that
          $\mathit{x}  \ottsym{,}  \mathit{y} \,  \not\in  \,  \mathit{dom}  (  \Gamma  )  \,  \mathbin{\cup}  \,  \mathit{fv}  (  \ottnt{S_{{\mathrm{3}}}}  ) $.
          Because $\Gamma \,  |  \, \ottnt{T} \,  \,\&\,  \, \Phi  \vdash  \ottnt{S_{{\mathrm{2}}}}  \ottsym{<:}  \ottnt{S_{{\mathrm{3}}}}$
          (i.e., $\Gamma \,  |  \, \ottnt{T} \,  \,\&\,  \, \Phi  \vdash   ( \forall  \mathit{x}  .  \ottnt{C_{{\mathrm{21}}}}  )  \Rightarrow   \ottnt{C_{{\mathrm{22}}}}   \ottsym{<:}  \ottnt{S_{{\mathrm{3}}}}$)
          can be derived only by \Srule{Ans},
          we can find that
          \begin{prooftree}
           \AxiomC{$\Gamma  \ottsym{,}  \mathit{x} \,  {:}  \, \ottnt{T}  \vdash  \ottnt{C_{{\mathrm{31}}}}  \ottsym{<:}  \ottnt{C_{{\mathrm{21}}}}$}
           \AxiomC{$\Gamma  \vdash  \ottnt{C_{{\mathrm{22}}}}  \ottsym{<:}  \ottnt{C_{{\mathrm{32}}}}$}
           \BinaryInfC{$\Gamma \,  |  \, \ottnt{T} \,  \,\&\,  \, \Phi  \vdash   ( \forall  \mathit{x}  .  \ottnt{C_{{\mathrm{21}}}}  )  \Rightarrow   \ottnt{C_{{\mathrm{22}}}}   \ottsym{<:}   ( \forall  \mathit{x}  .  \ottnt{C_{{\mathrm{31}}}}  )  \Rightarrow   \ottnt{C_{{\mathrm{32}}}} $}
          \end{prooftree}
          for some $\ottnt{C_{{\mathrm{31}}}}$ and $\ottnt{C_{{\mathrm{32}}}}$ such that $\ottnt{S_{{\mathrm{3}}}} \,  =  \,  ( \forall  \mathit{x}  .  \ottnt{C_{{\mathrm{31}}}}  )  \Rightarrow   \ottnt{C_{{\mathrm{32}}}} $.

          \reflem{ts-weak-subtyping} with $\Gamma  \vdash  \ottnt{C_{{\mathrm{22}}}}  \ottsym{<:}  \ottnt{C_{{\mathrm{32}}}}$ implies
          $\Gamma  \ottsym{,}  \mathit{x} \,  {:}  \, \ottnt{T}  \vdash  \ottnt{C_{{\mathrm{22}}}}  \ottsym{<:}  \ottnt{C_{{\mathrm{32}}}}$.
          Then, the IH on $\ottnt{C_{{\mathrm{22}}}}$ in $\Gamma  \ottsym{,}  \mathit{x} \,  {:}  \, \ottnt{T}  \vdash  \Phi \,  \tceappendop  \, \ottnt{C_{{\mathrm{21}}}}  \ottsym{<:}  \ottnt{C_{{\mathrm{22}}}}$ and $\Gamma  \ottsym{,}  \mathit{x} \,  {:}  \, \ottnt{T}  \vdash  \ottnt{C_{{\mathrm{22}}}}  \ottsym{<:}  \ottnt{C_{{\mathrm{32}}}}$ implies
          \[
           \Gamma  \ottsym{,}  \mathit{x} \,  {:}  \, \ottnt{T}  \vdash  \Phi \,  \tceappendop  \, \ottnt{C_{{\mathrm{21}}}}  \ottsym{<:}  \ottnt{C_{{\mathrm{32}}}} ~.
          \]
          Because \reflem{ts-add-preeff-subtyping} with $\Gamma  \ottsym{,}  \mathit{x} \,  {:}  \, \ottnt{T}  \vdash  \ottnt{C_{{\mathrm{31}}}}  \ottsym{<:}  \ottnt{C_{{\mathrm{21}}}}$
          implies $\Gamma  \ottsym{,}  \mathit{x} \,  {:}  \, \ottnt{T}  \vdash  \Phi \,  \tceappendop  \, \ottnt{C_{{\mathrm{31}}}}  \ottsym{<:}  \Phi \,  \tceappendop  \, \ottnt{C_{{\mathrm{21}}}}$,
          the IH on $\ottnt{C_{{\mathrm{21}}}}$ (note that the number of outermost type constructors of $\ottnt{C_{{\mathrm{21}}}}$ is the same as that of $\Phi \,  \tceappendop  \, \ottnt{C_{{\mathrm{21}}}}$) implies
          \[
           \Gamma  \ottsym{,}  \mathit{x} \,  {:}  \, \ottnt{T}  \vdash  \Phi \,  \tceappendop  \, \ottnt{C_{{\mathrm{31}}}}  \ottsym{<:}  \ottnt{C_{{\mathrm{32}}}} ~.
          \]
          Further, \reflem{lr-subtyping-nu} with
          $\Gamma  \vdash  \ottnt{C_{{\mathrm{22}}}}  \ottsym{<:}  \ottnt{C_{{\mathrm{32}}}}$ and
          $\Gamma  \ottsym{,}  \mathit{y} \,  {:}  \,  \Sigma ^\omega   \models    { \Phi }^\nu   \ottsym{(}  \mathit{y}  \ottsym{)}    \Rightarrow     { \ottnt{C_{{\mathrm{22}}}} }^\nu (  \mathit{y}  )  $ and
          $\mathit{y} \,  \not\in  \,  \mathit{fv}  (  \ottnt{C_{{\mathrm{22}}}}  )  \,  \mathbin{\cup}  \,  \mathit{fv}  (  \ottnt{C_{{\mathrm{32}}}}  ) $
          implies
          \[
           \Gamma  \ottsym{,}  \mathit{y} \,  {:}  \,  \Sigma ^\omega   \models    { \Phi }^\nu   \ottsym{(}  \mathit{y}  \ottsym{)}    \Rightarrow     { \ottnt{C_{{\mathrm{32}}}} }^\nu (  \mathit{y}  )   ~.
          \]
          Then, \Srule{Embed} with $\mathit{x}  \ottsym{,}  \mathit{y} \,  \not\in  \,  \mathit{fv}  (  \ottnt{C_{{\mathrm{32}}}}  )  \,  \mathbin{\cup}  \,  \mathit{fv}  (  \Phi  ) $ implies the conclusion
          \[
           \Gamma \,  |  \, \ottnt{T} \,  \,\&\,  \, \Phi  \vdash   \square   \ottsym{<:}   ( \forall  \mathit{x}  .  \ottnt{C_{{\mathrm{31}}}}  )  \Rightarrow   \ottnt{C_{{\mathrm{32}}}}  ~.
          \]
        \end{caseanalysis}
 \end{enumerate}
\end{proof}

\begin{lemmap}{Well-Formedness of Typing Contexts in Well-Formedness and Formula Typing}{ts-wf-tctx-wf-ftyping}
 \begin{enumerate}
  \item If $\Gamma  \vdash  \ottnt{T}$, then $\vdash  \Gamma$.
  \item If $\Gamma  \vdash  \ottnt{C}$, then $\vdash  \Gamma$.
  \item If $\Gamma \,  |  \, \ottnt{T}  \vdash  \ottnt{S}$, then $\vdash  \Gamma$.
  \item If $\Gamma  \vdash  \Phi$, then $\vdash  \Gamma$.
  \item If $\Gamma  \vdash  \ottnt{t}  \ottsym{:}  \ottnt{s}$, then $\vdash  \Gamma$.
  \item If $\Gamma  \vdash  \ottnt{A}  \ottsym{:}   \tceseq{ \ottnt{s} } $, then $\vdash  \Gamma$.
  \item If $\Gamma  \vdash  \phi$, then $\vdash  \Gamma$.
 \end{enumerate}
\end{lemmap}
\begin{proof}
 Straightforward by simultaneous induction on the derivations.
\end{proof}

\begin{lemmap}{Reflexivity of Subtyping}{ts-refl-subtyping}
 \begin{enumerate}
  \item If $\Gamma  \vdash  \ottnt{T}$, then $\Gamma  \vdash  \ottnt{T}  \ottsym{<:}  \ottnt{T}$.
  \item If $\Gamma  \vdash  \ottnt{C}$, then $\Gamma  \vdash  \ottnt{C}  \ottsym{<:}  \ottnt{C}$.
  \item If $\Gamma \,  |  \, \ottnt{T}  \vdash  \ottnt{S}$, then $\Gamma \,  |  \, \ottnt{T} \,  \,\&\,  \, \Phi  \vdash  \ottnt{S}  \ottsym{<:}  \ottnt{S}$
        for any $\Phi$.
  \item \label{lem:ts-refl-subtyping:eff}
        If $\Gamma  \vdash  \Phi$, then $\Gamma  \vdash  \Phi  \ottsym{<:}  \Phi$.
 \end{enumerate}
\end{lemmap}
\begin{proof}
 The case (\ref{lem:ts-refl-subtyping:eff}) is shown by \reflem{ts-refl-valid}.

 We prove the other cases
 by simultaneous induction on the sizes of
 $\ottnt{T}$, $\ottnt{C}$, and $\ottnt{S}$.
 \begin{caseanalysis}
  \case \WFT{Refine}: By \reflem{ts-refl-valid} and \Srule{Refine}.
  \case \WFT{Fun}: We are given
   \begin{prooftree}
    \AxiomC{$\Gamma  \ottsym{,}  \mathit{x} \,  {:}  \, \ottnt{T'}  \vdash  \ottnt{C'}$}
    \UnaryInfC{$\Gamma  \vdash  \ottsym{(}  \mathit{x} \,  {:}  \, \ottnt{T'}  \ottsym{)}  \rightarrow  \ottnt{C'}$}
   \end{prooftree}
   for some $\mathit{x}$, $\ottnt{T'}$, and $\ottnt{C'}$ such that $\ottnt{T} \,  =  \, \ottsym{(}  \mathit{x} \,  {:}  \, \ottnt{T'}  \ottsym{)}  \rightarrow  \ottnt{C'}$.
   By \reflem{ts-wf-tctx-wf-ftyping},
   $\vdash  \Gamma  \ottsym{,}  \mathit{x} \,  {:}  \, \ottnt{T'}$, which implies $\Gamma  \vdash  \ottnt{T'}$.
   By the IHs on $\ottnt{T'}$ and on $\ottnt{C'}$,
   we have $\Gamma  \vdash  \ottnt{T'}  \ottsym{<:}  \ottnt{T'}$ and $\Gamma  \ottsym{,}  \mathit{x} \,  {:}  \, \ottnt{T'}  \vdash  \ottnt{C'}  \ottsym{<:}  \ottnt{C'}$.
   By \Srule{Fun}, we have the conclusion
   $\Gamma  \vdash  \ottsym{(}  \mathit{x} \,  {:}  \, \ottnt{T'}  \ottsym{)}  \rightarrow  \ottnt{C'}  \ottsym{<:}  \ottsym{(}  \mathit{x} \,  {:}  \, \ottnt{T'}  \ottsym{)}  \rightarrow  \ottnt{C'}$.

  \case \WFT{Forall}: By the IH and \Srule{Forall}.
  \case \WFT{Comp}: By the IHs, the case (\ref{lem:ts-refl-subtyping:eff}), and \Srule{Comp}.
  \case \WFT{Empty}: By \Srule{Empty}.
  \case \WFT{Ans}: By the IHs and \Srule{Ans}.
 \end{caseanalysis}
\end{proof}

\begin{lemmap}{Well-Formedness of $ \Phi_\mathsf{val} $}{ts-wf-TP_val}
 \begin{enumerate}
  \item \label{lem:ts-wf-TP_val:eff}
        If $\vdash  \Gamma$, then $\Gamma  \vdash   \Phi_\mathsf{val} $.
  \item If $\Gamma  \vdash  \ottnt{T}$, then $\Gamma  \vdash   \ottnt{T}  \,  \,\&\,  \,   \Phi_\mathsf{val}   \, / \,   \square  $.
 \end{enumerate}
\end{lemmap}
\begin{proof}
 \noindent
 \begin{enumerate}
  \item Note that $ \Phi_\mathsf{val}  \,  =  \,  (   \lambda_\mu\!   \,  \mathit{x}  .\,   \mathit{x}    =     \epsilon    ,   \lambda_\nu\!   \,  \mathit{y}  .\,   \bot   ) $.
        Let $\mathit{x}, \mathit{y}$ be fresh variables.
        We have
        \[
         \Gamma  \ottsym{,}  \mathit{x} \,  {:}  \,  \Sigma ^\ast   \vdash   \mathit{x}    =     \epsilon  
        \]
        by \Tf{Pred}, \Tf{Prim}, \Tf{VarS}, and \Tf{Term} with $\vdash  \Gamma  \ottsym{,}  \mathit{x} \,  {:}  \,  \Sigma ^\ast $.
        We also have
        \[
         \Gamma  \ottsym{,}  \mathit{y} \,  {:}  \,  \Sigma ^\omega   \vdash   \bot 
        \]
        by \Tf{Bot} with $\vdash  \Gamma  \ottsym{,}  \mathit{y} \,  {:}  \,  \Sigma ^\omega $.

  \item By \reflem{ts-wf-tctx-wf-ftyping} with the assumption $\Gamma  \vdash  \ottnt{T}$
        implies $\vdash  \Gamma$.
        Thus, the case (\ref{lem:ts-wf-TP_val:eff}) implies $\Gamma  \vdash   \Phi_\mathsf{val} $,
        and \WFT{Empty} implies $\Gamma \,  |  \, \ottnt{T}  \vdash   \square $.
        Therefore, \WF{Comp} with the assumption $\Gamma  \vdash  \ottnt{T}$ implies
        the conclusion $\Gamma  \vdash   \ottnt{T}  \,  \,\&\,  \,   \Phi_\mathsf{val}   \, / \,   \square  $.
 \end{enumerate}
\end{proof}

\begin{lemmap}{Eliminating Unused Bindings in Well-Formedness and Formula Typing}{ts-elim-unused-binding-wf-ftyping}
 \begin{enumerate}
  \item If $\vdash  \Gamma_{{\mathrm{1}}}  \ottsym{,}  \Gamma_{{\mathrm{2}}}  \ottsym{,}  \Gamma_{{\mathrm{3}}}$ and $  \mathit{dom}  (  \Gamma_{{\mathrm{2}}}  )   \, \ottsym{\#} \,   \mathit{afv}  (  \Gamma_{{\mathrm{3}}}  )  $,
        then $\vdash  \Gamma_{{\mathrm{1}}}  \ottsym{,}  \Gamma_{{\mathrm{3}}}$.
  \item If $\Gamma_{{\mathrm{1}}}  \ottsym{,}  \Gamma_{{\mathrm{2}}}  \ottsym{,}  \Gamma_{{\mathrm{3}}}  \vdash  \ottnt{T}$ and $  \mathit{dom}  (  \Gamma_{{\mathrm{2}}}  )   \, \ottsym{\#} \,  \ottsym{(}   \mathit{afv}  (  \Gamma_{{\mathrm{3}}}  )  \,  \mathbin{\cup}  \,  \mathit{afv}  (  \ottnt{T}  )   \ottsym{)} $,
        then $\Gamma_{{\mathrm{1}}}  \ottsym{,}  \Gamma_{{\mathrm{3}}}  \vdash  \ottnt{T}$.
  \item If $\Gamma_{{\mathrm{1}}}  \ottsym{,}  \Gamma_{{\mathrm{2}}}  \ottsym{,}  \Gamma_{{\mathrm{3}}}  \vdash  \ottnt{C}$ and $  \mathit{dom}  (  \Gamma_{{\mathrm{2}}}  )   \, \ottsym{\#} \,  \ottsym{(}   \mathit{afv}  (  \Gamma_{{\mathrm{3}}}  )  \,  \mathbin{\cup}  \,  \mathit{afv}  (  \ottnt{C}  )   \ottsym{)} $,
        then $\Gamma_{{\mathrm{1}}}  \ottsym{,}  \Gamma_{{\mathrm{3}}}  \vdash  \ottnt{C}$.
  \item If $\Gamma_{{\mathrm{1}}}  \ottsym{,}  \Gamma_{{\mathrm{2}}}  \ottsym{,}  \Gamma_{{\mathrm{3}}} \,  |  \, \ottnt{T}  \vdash  \ottnt{S}$ and $  \mathit{dom}  (  \Gamma_{{\mathrm{2}}}  )   \, \ottsym{\#} \,  \ottsym{(}   \mathit{afv}  (  \Gamma_{{\mathrm{3}}}  )  \,  \mathbin{\cup}  \,  \mathit{afv}  (  \ottnt{T}  )  \,  \mathbin{\cup}  \,  \mathit{afv}  (  \ottnt{S}  )   \ottsym{)} $,
        then $\Gamma_{{\mathrm{1}}}  \ottsym{,}  \Gamma_{{\mathrm{3}}} \,  |  \, \ottnt{T}  \vdash  \ottnt{S}$.
  \item If $\Gamma_{{\mathrm{1}}}  \ottsym{,}  \Gamma_{{\mathrm{2}}}  \ottsym{,}  \Gamma_{{\mathrm{3}}}  \vdash  \Phi$ and $  \mathit{dom}  (  \Gamma_{{\mathrm{2}}}  )   \, \ottsym{\#} \,  \ottsym{(}   \mathit{afv}  (  \Gamma_{{\mathrm{3}}}  )  \,  \mathbin{\cup}  \,  \mathit{afv}  (  \Phi  )   \ottsym{)} $,
        then $\Gamma_{{\mathrm{1}}}  \ottsym{,}  \Gamma_{{\mathrm{3}}}  \vdash  \Phi$.
  \item If $\Gamma_{{\mathrm{1}}}  \ottsym{,}  \Gamma_{{\mathrm{2}}}  \ottsym{,}  \Gamma_{{\mathrm{3}}}  \vdash  \ottnt{t}  \ottsym{:}  \ottnt{s}$ and $  \mathit{dom}  (  \Gamma_{{\mathrm{2}}}  )   \, \ottsym{\#} \,  \ottsym{(}   \mathit{afv}  (  \Gamma_{{\mathrm{3}}}  )  \,  \mathbin{\cup}  \,  \mathit{afv}  (  \ottnt{t}  )   \ottsym{)} $,
        then $\Gamma_{{\mathrm{1}}}  \ottsym{,}  \Gamma_{{\mathrm{3}}}  \vdash  \ottnt{t}  \ottsym{:}  \ottnt{s}$.
  \item If $\Gamma_{{\mathrm{1}}}  \ottsym{,}  \Gamma_{{\mathrm{2}}}  \ottsym{,}  \Gamma_{{\mathrm{3}}}  \vdash  \ottnt{A}  \ottsym{:}   \tceseq{ \ottnt{s} } $ and $  \mathit{dom}  (  \Gamma_{{\mathrm{2}}}  )   \, \ottsym{\#} \,  \ottsym{(}   \mathit{afv}  (  \Gamma_{{\mathrm{3}}}  )  \,  \mathbin{\cup}  \,  \mathit{afv}  (  \ottnt{A}  )   \ottsym{)} $,
        then $\Gamma_{{\mathrm{1}}}  \ottsym{,}  \Gamma_{{\mathrm{3}}}  \vdash  \ottnt{A}  \ottsym{:}   \tceseq{ \ottnt{s} } $.
  \item If $\Gamma_{{\mathrm{1}}}  \ottsym{,}  \Gamma_{{\mathrm{2}}}  \ottsym{,}  \Gamma_{{\mathrm{3}}}  \vdash  \phi$ and $  \mathit{dom}  (  \Gamma_{{\mathrm{2}}}  )   \, \ottsym{\#} \,  \ottsym{(}   \mathit{afv}  (  \Gamma_{{\mathrm{3}}}  )  \,  \mathbin{\cup}  \,  \mathit{afv}  (  \phi  )   \ottsym{)} $,
        then $\Gamma_{{\mathrm{1}}}  \ottsym{,}  \Gamma_{{\mathrm{3}}}  \vdash  \phi$.
 \end{enumerate}
\end{lemmap}
\begin{proof}
 Straightforward by simultaneous induction on the derivations.
 Note that it is ensured that bound (tern and predicate) variables in types and
 formulas (e.g., as in dependent function types) are not in $ \mathit{dom}  (  \Gamma_{{\mathrm{1}}}  \ottsym{,}  \Gamma_{{\mathrm{2}}}  \ottsym{,}  \Gamma_{{\mathrm{3}}}  ) $
 by the inversion of the derivations for typing context well-formedness
 judgments obtained by \reflem{ts-wf-tctx-wf-ftyping}.
\end{proof}

\begin{lemmap}{Variables of Non-Refinement Types are Unused in Typing Contexts, Types, and Formulas}{ts-unuse-non-refine-vars}
 Suppose that $\ottnt{T}$ is not a refinement type.
 \begin{enumerate}
  \item If $\vdash  \Gamma_{{\mathrm{1}}}  \ottsym{,}  \mathit{x} \,  {:}  \, \ottnt{T}  \ottsym{,}  \Gamma_{{\mathrm{2}}}$, then $\mathit{x} \,  \not\in  \,  \mathit{fv}  (  \Gamma_{{\mathrm{2}}}  ) $.
  \item If $\Gamma_{{\mathrm{1}}}  \ottsym{,}  \mathit{x} \,  {:}  \, \ottnt{T}  \ottsym{,}  \Gamma_{{\mathrm{2}}}  \vdash  \ottnt{T_{{\mathrm{0}}}}$, then $\mathit{x} \,  \not\in  \,  \mathit{fv}  (  \Gamma_{{\mathrm{2}}}  )  \,  \mathbin{\cup}  \,  \mathit{fv}  (  \ottnt{T_{{\mathrm{0}}}}  ) $.
  \item If $\Gamma_{{\mathrm{1}}}  \ottsym{,}  \mathit{x} \,  {:}  \, \ottnt{T}  \ottsym{,}  \Gamma_{{\mathrm{2}}}  \vdash  \ottnt{C}$, then $\mathit{x} \,  \not\in  \,  \mathit{fv}  (  \Gamma_{{\mathrm{2}}}  )  \,  \mathbin{\cup}  \,  \mathit{fv}  (  \ottnt{C}  ) $.
  \item If $\Gamma_{{\mathrm{1}}}  \ottsym{,}  \mathit{x} \,  {:}  \, \ottnt{T}  \ottsym{,}  \Gamma_{{\mathrm{2}}} \,  |  \, \ottnt{T_{{\mathrm{0}}}}  \vdash  \ottnt{S}$, then $\mathit{x} \,  \not\in  \,  \mathit{fv}  (  \Gamma_{{\mathrm{2}}}  )  \,  \mathbin{\cup}  \,  \mathit{fv}  (  \ottnt{S}  ) $.
  \item If $\Gamma_{{\mathrm{1}}}  \ottsym{,}  \mathit{x} \,  {:}  \, \ottnt{T}  \ottsym{,}  \Gamma_{{\mathrm{2}}}  \vdash  \Phi$, then $\mathit{x} \,  \not\in  \,  \mathit{fv}  (  \Gamma_{{\mathrm{2}}}  )  \,  \mathbin{\cup}  \,  \mathit{fv}  (  \Phi  ) $.
  \item If $\Gamma_{{\mathrm{1}}}  \ottsym{,}  \mathit{x} \,  {:}  \, \ottnt{T}  \ottsym{,}  \Gamma_{{\mathrm{2}}}  \vdash  \ottnt{t}  \ottsym{:}  \ottnt{s}$, then $\mathit{x} \,  \not\in  \,  \mathit{fv}  (  \Gamma_{{\mathrm{2}}}  )  \,  \mathbin{\cup}  \,  \mathit{fv}  (  \ottnt{t}  ) $.
  \item If $\Gamma_{{\mathrm{1}}}  \ottsym{,}  \mathit{x} \,  {:}  \, \ottnt{T}  \ottsym{,}  \Gamma_{{\mathrm{2}}}  \vdash  \ottnt{A}  \ottsym{:}   \tceseq{ \ottnt{s} } $, then $\mathit{x} \,  \not\in  \,  \mathit{fv}  (  \Gamma_{{\mathrm{2}}}  )  \,  \mathbin{\cup}  \,  \mathit{fv}  (  \ottnt{A}  ) $.
  \item If $\Gamma_{{\mathrm{1}}}  \ottsym{,}  \mathit{x} \,  {:}  \, \ottnt{T}  \ottsym{,}  \Gamma_{{\mathrm{2}}}  \vdash  \phi$, then $\mathit{x} \,  \not\in  \,  \mathit{fv}  (  \Gamma_{{\mathrm{2}}}  )  \,  \mathbin{\cup}  \,  \mathit{fv}  (  \phi  ) $.
 \end{enumerate}
\end{lemmap}
\begin{proof}
 Straightforward by simultaneous induction on the derivations.
\end{proof}

\begin{lemmap}{Pulling Back Sorted Variables}{ts-pull-back-sorted-vars}
 Suppose that $\Gamma \,  =  \, \mathit{x} \,  {:}  \, \ottnt{s}$ or $\Gamma \,  =  \, \mathit{X} \,  {:}  \,  \tceseq{ \ottnt{s} } $.
 \begin{enumerate}
  \item \label{lem:ts-pull-back-sorted-vars:tctx}
        If $\vdash  \Gamma_{{\mathrm{1}}}  \ottsym{,}  \Gamma_{{\mathrm{2}}}  \ottsym{,}  \Gamma  \ottsym{,}  \Gamma_{{\mathrm{3}}}$, then $\vdash  \Gamma_{{\mathrm{1}}}  \ottsym{,}  \Gamma  \ottsym{,}  \Gamma_{{\mathrm{2}}}  \ottsym{,}  \Gamma_{{\mathrm{3}}}$.
  \item If $\Gamma_{{\mathrm{1}}}  \ottsym{,}  \Gamma_{{\mathrm{2}}}  \ottsym{,}  \Gamma  \ottsym{,}  \Gamma_{{\mathrm{3}}}  \vdash  \ottnt{T}$, then $\Gamma_{{\mathrm{1}}}  \ottsym{,}  \Gamma  \ottsym{,}  \Gamma_{{\mathrm{2}}}  \ottsym{,}  \Gamma_{{\mathrm{3}}}  \vdash  \ottnt{T}$.
  \item If $\Gamma_{{\mathrm{1}}}  \ottsym{,}  \Gamma_{{\mathrm{2}}}  \ottsym{,}  \Gamma  \ottsym{,}  \Gamma_{{\mathrm{3}}}  \vdash  \ottnt{C}$, then $\Gamma_{{\mathrm{1}}}  \ottsym{,}  \Gamma  \ottsym{,}  \Gamma_{{\mathrm{2}}}  \ottsym{,}  \Gamma_{{\mathrm{3}}}  \vdash  \ottnt{C}$.
  \item If $\Gamma_{{\mathrm{1}}}  \ottsym{,}  \Gamma_{{\mathrm{2}}}  \ottsym{,}  \Gamma  \ottsym{,}  \Gamma_{{\mathrm{3}}} \,  |  \, \ottnt{T}  \vdash  \ottnt{S}$, then $\Gamma_{{\mathrm{1}}}  \ottsym{,}  \Gamma  \ottsym{,}  \Gamma_{{\mathrm{2}}}  \ottsym{,}  \Gamma_{{\mathrm{3}}} \,  |  \, \ottnt{T}  \vdash  \ottnt{S}$.
  \item If $\Gamma_{{\mathrm{1}}}  \ottsym{,}  \Gamma_{{\mathrm{2}}}  \ottsym{,}  \Gamma  \ottsym{,}  \Gamma_{{\mathrm{3}}}  \vdash  \Phi$, then $\Gamma_{{\mathrm{1}}}  \ottsym{,}  \Gamma  \ottsym{,}  \Gamma_{{\mathrm{2}}}  \ottsym{,}  \Gamma_{{\mathrm{3}}}  \vdash  \Phi$.
  \item If $\Gamma_{{\mathrm{1}}}  \ottsym{,}  \Gamma_{{\mathrm{2}}}  \ottsym{,}  \Gamma  \ottsym{,}  \Gamma_{{\mathrm{3}}}  \vdash  \ottnt{t}  \ottsym{:}  \ottnt{s}$, then $\Gamma_{{\mathrm{1}}}  \ottsym{,}  \Gamma  \ottsym{,}  \Gamma_{{\mathrm{2}}}  \ottsym{,}  \Gamma_{{\mathrm{3}}}  \vdash  \ottnt{t}  \ottsym{:}  \ottnt{s}$.
  \item If $\Gamma_{{\mathrm{1}}}  \ottsym{,}  \Gamma_{{\mathrm{2}}}  \ottsym{,}  \Gamma  \ottsym{,}  \Gamma_{{\mathrm{3}}}  \vdash  \ottnt{A}  \ottsym{:}   \tceseq{ \ottnt{s} } $, then $\Gamma_{{\mathrm{1}}}  \ottsym{,}  \Gamma  \ottsym{,}  \Gamma_{{\mathrm{2}}}  \ottsym{,}  \Gamma_{{\mathrm{3}}}  \vdash  \ottnt{A}  \ottsym{:}   \tceseq{ \ottnt{s} } $.
  \item If $\Gamma_{{\mathrm{1}}}  \ottsym{,}  \Gamma_{{\mathrm{2}}}  \ottsym{,}  \Gamma  \ottsym{,}  \Gamma_{{\mathrm{3}}}  \vdash  \phi$, then $\Gamma_{{\mathrm{1}}}  \ottsym{,}  \Gamma  \ottsym{,}  \Gamma_{{\mathrm{2}}}  \ottsym{,}  \Gamma_{{\mathrm{3}}}  \vdash  \phi$.
 \end{enumerate}
\end{lemmap}
\begin{proof}
 Straightforward by simultaneous induction on the derivations.
 The case for $\Gamma_{{\mathrm{3}}} \,  =  \,  \emptyset $ in the case (\ref{lem:ts-pull-back-sorted-vars:tctx})
 can be proven by \reflem{ts-weak-wf-ftyping}.
\end{proof}

\begin{lemmap}{Un-refining Refinement Types}{ts-unrefine-refine-ty}
 \begin{enumerate}
  \item If $\vdash  \Gamma_{{\mathrm{1}}}  \ottsym{,}  \mathit{x} \,  {:}  \, \ottsym{\{}  \mathit{y} \,  {:}  \, \iota \,  |  \, \phi  \ottsym{\}}  \ottsym{,}  \Gamma_{{\mathrm{2}}}$, then $\vdash  \Gamma_{{\mathrm{1}}}  \ottsym{,}  \mathit{x} \,  {:}  \, \iota  \ottsym{,}  \Gamma_{{\mathrm{2}}}$.
  \item If $\Gamma_{{\mathrm{1}}}  \ottsym{,}  \mathit{x} \,  {:}  \, \ottsym{\{}  \mathit{y} \,  {:}  \, \iota \,  |  \, \phi  \ottsym{\}}  \ottsym{,}  \Gamma_{{\mathrm{2}}}  \vdash  \ottnt{T}$, then $\Gamma_{{\mathrm{1}}}  \ottsym{,}  \mathit{x} \,  {:}  \, \iota  \ottsym{,}  \Gamma_{{\mathrm{2}}}  \vdash  \ottnt{T}$.
  \item If $\Gamma_{{\mathrm{1}}}  \ottsym{,}  \mathit{x} \,  {:}  \, \ottsym{\{}  \mathit{y} \,  {:}  \, \iota \,  |  \, \phi  \ottsym{\}}  \ottsym{,}  \Gamma_{{\mathrm{2}}}  \vdash  \ottnt{C}$, then $\Gamma_{{\mathrm{1}}}  \ottsym{,}  \mathit{x} \,  {:}  \, \iota  \ottsym{,}  \Gamma_{{\mathrm{2}}}  \vdash  \ottnt{C}$.
  \item If $\Gamma_{{\mathrm{1}}}  \ottsym{,}  \mathit{x} \,  {:}  \, \ottsym{\{}  \mathit{y} \,  {:}  \, \iota \,  |  \, \phi  \ottsym{\}}  \ottsym{,}  \Gamma_{{\mathrm{2}}} \,  |  \, \ottnt{T}  \vdash  \ottnt{S}$, then $\Gamma_{{\mathrm{1}}}  \ottsym{,}  \mathit{x} \,  {:}  \, \iota  \ottsym{,}  \Gamma_{{\mathrm{2}}} \,  |  \, \ottnt{T}  \vdash  \ottnt{S}$.
  \item If $\Gamma_{{\mathrm{1}}}  \ottsym{,}  \mathit{x} \,  {:}  \, \ottsym{\{}  \mathit{y} \,  {:}  \, \iota \,  |  \, \phi  \ottsym{\}}  \ottsym{,}  \Gamma_{{\mathrm{2}}}  \vdash  \Phi$, then $\Gamma_{{\mathrm{1}}}  \ottsym{,}  \mathit{x} \,  {:}  \, \iota  \ottsym{,}  \Gamma_{{\mathrm{2}}}  \vdash  \Phi$.
  \item \label{lem:ts-unrefine-refine-ty:term}
        If $\Gamma_{{\mathrm{1}}}  \ottsym{,}  \mathit{x} \,  {:}  \, \ottsym{\{}  \mathit{y} \,  {:}  \, \iota \,  |  \, \phi  \ottsym{\}}  \ottsym{,}  \Gamma_{{\mathrm{2}}}  \vdash  \ottnt{t}  \ottsym{:}  \ottnt{s}$, then $\Gamma_{{\mathrm{1}}}  \ottsym{,}  \mathit{x} \,  {:}  \, \iota  \ottsym{,}  \Gamma_{{\mathrm{2}}}  \vdash  \ottnt{t}  \ottsym{:}  \ottnt{s}$.
  \item If $\Gamma_{{\mathrm{1}}}  \ottsym{,}  \mathit{x} \,  {:}  \, \ottsym{\{}  \mathit{y} \,  {:}  \, \iota \,  |  \, \phi  \ottsym{\}}  \ottsym{,}  \Gamma_{{\mathrm{2}}}  \vdash  \ottnt{A}  \ottsym{:}   \tceseq{ \ottnt{s} } $, then $\Gamma_{{\mathrm{1}}}  \ottsym{,}  \mathit{x} \,  {:}  \, \iota  \ottsym{,}  \Gamma_{{\mathrm{2}}}  \vdash  \ottnt{A}  \ottsym{:}   \tceseq{ \ottnt{s} } $.
  \item If $\Gamma_{{\mathrm{1}}}  \ottsym{,}  \mathit{x} \,  {:}  \, \ottsym{\{}  \mathit{y} \,  {:}  \, \iota \,  |  \, \phi  \ottsym{\}}  \ottsym{,}  \Gamma_{{\mathrm{2}}}  \vdash  \phi_{{\mathrm{0}}}$, then $\Gamma_{{\mathrm{1}}}  \ottsym{,}  \mathit{x} \,  {:}  \, \iota  \ottsym{,}  \Gamma_{{\mathrm{2}}}  \vdash  \phi_{{\mathrm{0}}}$.
 \end{enumerate}
\end{lemmap}
\begin{proof}
 Straightforward by simultaneous induction on the derivations.
 The interesting case is only for \Tf{VarT} in the case (\ref{lem:ts-unrefine-refine-ty:term}).

 Consider the case for \Tf{VarT}.  We are given
 \begin{prooftree}
  \AxiomC{$\vdash  \Gamma_{{\mathrm{1}}}  \ottsym{,}  \mathit{x} \,  {:}  \, \ottsym{\{}  \mathit{y} \,  {:}  \, \iota \,  |  \, \phi  \ottsym{\}}  \ottsym{,}  \Gamma_{{\mathrm{2}}}$}
  \AxiomC{$\ottsym{(}  \Gamma_{{\mathrm{1}}}  \ottsym{,}  \mathit{x} \,  {:}  \, \ottsym{\{}  \mathit{y} \,  {:}  \, \iota \,  |  \, \phi  \ottsym{\}}  \ottsym{,}  \Gamma_{{\mathrm{2}}}  \ottsym{)}  \ottsym{(}  \mathit{z}  \ottsym{)} \,  =  \, \ottsym{\{}  \mathit{y_{{\mathrm{0}}}} \,  {:}  \, \iota_{{\mathrm{0}}} \,  |  \, \phi_{{\mathrm{0}}}  \ottsym{\}}$}
  \BinaryInfC{$\Gamma_{{\mathrm{1}}}  \ottsym{,}  \mathit{x} \,  {:}  \, \ottsym{\{}  \mathit{y} \,  {:}  \, \iota \,  |  \, \phi  \ottsym{\}}  \ottsym{,}  \Gamma_{{\mathrm{2}}}  \vdash  \mathit{z}  \ottsym{:}  \iota_{{\mathrm{0}}}$}
 \end{prooftree}
 for some $\mathit{z}$, $\mathit{y_{{\mathrm{0}}}}$, $\iota_{{\mathrm{0}}}$, and $\phi_{{\mathrm{0}}}$
 such that $ \ottnt{t}    =    \mathit{z} $ and $\ottnt{s} \,  =  \, \iota_{{\mathrm{0}}}$.
 By the IH, $\vdash  \Gamma_{{\mathrm{1}}}  \ottsym{,}  \mathit{x} \,  {:}  \, \iota  \ottsym{,}  \Gamma_{{\mathrm{2}}}$.
 By case analysis on whether $ \mathit{z}    =    \mathit{x} $.
 \begin{caseanalysis}
  \case $ \mathit{z}    =    \mathit{x} $:
   We have $\ottsym{\{}  \mathit{y} \,  {:}  \, \iota \,  |  \, \phi  \ottsym{\}} = \ottsym{(}  \Gamma_{{\mathrm{1}}}  \ottsym{,}  \mathit{x} \,  {:}  \, \ottsym{\{}  \mathit{y} \,  {:}  \, \iota \,  |  \, \phi  \ottsym{\}}  \ottsym{,}  \Gamma_{{\mathrm{2}}}  \ottsym{)}  \ottsym{(}  \mathit{z}  \ottsym{)} = \ottsym{\{}  \mathit{y_{{\mathrm{0}}}} \,  {:}  \, \iota_{{\mathrm{0}}} \,  |  \, \phi_{{\mathrm{0}}}  \ottsym{\}}$, so
   $\iota \,  =  \, \iota_{{\mathrm{0}}}$.
   Then, by \Tf{VarS} with $\vdash  \Gamma_{{\mathrm{1}}}  \ottsym{,}  \mathit{x} \,  {:}  \, \iota  \ottsym{,}  \Gamma_{{\mathrm{2}}}$ and
   the fact that $\ottsym{(}  \Gamma_{{\mathrm{1}}}  \ottsym{,}  \mathit{x} \,  {:}  \, \iota  \ottsym{,}  \Gamma_{{\mathrm{2}}}  \ottsym{)}  \ottsym{(}  \mathit{z}  \ottsym{)} = \iota = \iota_{{\mathrm{0}}}$,
   we have the conclusion $\Gamma_{{\mathrm{1}}}  \ottsym{,}  \mathit{x} \,  {:}  \, \iota  \ottsym{,}  \Gamma_{{\mathrm{2}}}  \vdash  \mathit{z}  \ottsym{:}  \iota_{{\mathrm{0}}}$.

  \case $ \mathit{z}    \not=    \mathit{x} $:
   We have $\ottsym{(}  \Gamma_{{\mathrm{1}}}  \ottsym{,}  \mathit{x} \,  {:}  \, \iota  \ottsym{,}  \Gamma_{{\mathrm{2}}}  \ottsym{)}  \ottsym{(}  \mathit{z}  \ottsym{)} = \ottsym{(}  \Gamma_{{\mathrm{1}}}  \ottsym{,}  \mathit{x} \,  {:}  \, \ottsym{\{}  \mathit{y} \,  {:}  \, \iota \,  |  \, \phi  \ottsym{\}}  \ottsym{,}  \Gamma_{{\mathrm{2}}}  \ottsym{)}  \ottsym{(}  \mathit{z}  \ottsym{)} = \ottsym{\{}  \mathit{y_{{\mathrm{0}}}} \,  {:}  \, \iota_{{\mathrm{0}}} \,  |  \, \phi_{{\mathrm{0}}}  \ottsym{\}}$, so
   \Tf{VarT} implies the conclusion $\Gamma_{{\mathrm{1}}}  \ottsym{,}  \mathit{x} \,  {:}  \, \iota  \ottsym{,}  \Gamma_{{\mathrm{2}}}  \vdash  \mathit{z}  \ottsym{:}  \iota_{{\mathrm{0}}}$.
 \end{caseanalysis}
\end{proof}

\begin{lemmap}{Refining Base Sorts}{ts-refine-base-sorts}
 Suppose that $\Gamma_{{\mathrm{1}}}  \ottsym{,}  \mathit{y} \,  {:}  \, \iota  \vdash  \phi$.
 \begin{enumerate}
  \item If $\vdash  \Gamma_{{\mathrm{1}}}  \ottsym{,}  \mathit{x} \,  {:}  \, \iota  \ottsym{,}  \Gamma_{{\mathrm{2}}}$, then $\vdash  \Gamma_{{\mathrm{1}}}  \ottsym{,}  \mathit{x} \,  {:}  \, \ottsym{\{}  \mathit{y} \,  {:}  \, \iota \,  |  \, \phi  \ottsym{\}}  \ottsym{,}  \Gamma_{{\mathrm{2}}}$.
  \item If $\Gamma_{{\mathrm{1}}}  \ottsym{,}  \mathit{x} \,  {:}  \, \iota  \ottsym{,}  \Gamma_{{\mathrm{2}}}  \vdash  \ottnt{T}$, then $\Gamma_{{\mathrm{1}}}  \ottsym{,}  \mathit{x} \,  {:}  \, \ottsym{\{}  \mathit{y} \,  {:}  \, \iota \,  |  \, \phi  \ottsym{\}}  \ottsym{,}  \Gamma_{{\mathrm{2}}}  \vdash  \ottnt{T}$.
  \item If $\Gamma_{{\mathrm{1}}}  \ottsym{,}  \mathit{x} \,  {:}  \, \iota  \ottsym{,}  \Gamma_{{\mathrm{2}}}  \vdash  \ottnt{C}$, then $\Gamma_{{\mathrm{1}}}  \ottsym{,}  \mathit{x} \,  {:}  \, \ottsym{\{}  \mathit{y} \,  {:}  \, \iota \,  |  \, \phi  \ottsym{\}}  \ottsym{,}  \Gamma_{{\mathrm{2}}}  \vdash  \ottnt{C}$.
  \item If $\Gamma_{{\mathrm{1}}}  \ottsym{,}  \mathit{x} \,  {:}  \, \iota  \ottsym{,}  \Gamma_{{\mathrm{2}}} \,  |  \, \ottnt{T}  \vdash  \ottnt{S}$, then $\Gamma_{{\mathrm{1}}}  \ottsym{,}  \mathit{x} \,  {:}  \, \ottsym{\{}  \mathit{y} \,  {:}  \, \iota \,  |  \, \phi  \ottsym{\}}  \ottsym{,}  \Gamma_{{\mathrm{2}}} \,  |  \, \ottnt{T}  \vdash  \ottnt{S}$.
  \item If $\Gamma_{{\mathrm{1}}}  \ottsym{,}  \mathit{x} \,  {:}  \, \iota  \ottsym{,}  \Gamma_{{\mathrm{2}}}  \vdash  \Phi$, then $\Gamma_{{\mathrm{1}}}  \ottsym{,}  \mathit{x} \,  {:}  \, \ottsym{\{}  \mathit{y} \,  {:}  \, \iota \,  |  \, \phi  \ottsym{\}}  \ottsym{,}  \Gamma_{{\mathrm{2}}}  \vdash  \Phi$.
  \item If $\Gamma_{{\mathrm{1}}}  \ottsym{,}  \mathit{x} \,  {:}  \, \iota  \ottsym{,}  \Gamma_{{\mathrm{2}}}  \vdash  \ottnt{t}  \ottsym{:}  \ottnt{s}$, then $\Gamma_{{\mathrm{1}}}  \ottsym{,}  \mathit{x} \,  {:}  \, \ottsym{\{}  \mathit{y} \,  {:}  \, \iota \,  |  \, \phi  \ottsym{\}}  \ottsym{,}  \Gamma_{{\mathrm{2}}}  \vdash  \ottnt{t}  \ottsym{:}  \ottnt{s}$.
  \item If $\Gamma_{{\mathrm{1}}}  \ottsym{,}  \mathit{x} \,  {:}  \, \iota  \ottsym{,}  \Gamma_{{\mathrm{2}}}  \vdash  \ottnt{A}  \ottsym{:}   \tceseq{ \ottnt{s} } $, then $\Gamma_{{\mathrm{1}}}  \ottsym{,}  \mathit{x} \,  {:}  \, \ottsym{\{}  \mathit{y} \,  {:}  \, \iota \,  |  \, \phi  \ottsym{\}}  \ottsym{,}  \Gamma_{{\mathrm{2}}}  \vdash  \ottnt{A}  \ottsym{:}   \tceseq{ \ottnt{s} } $.
  \item If $\Gamma_{{\mathrm{1}}}  \ottsym{,}  \mathit{x} \,  {:}  \, \iota  \ottsym{,}  \Gamma_{{\mathrm{2}}}  \vdash  \phi_{{\mathrm{0}}}$, then $\Gamma_{{\mathrm{1}}}  \ottsym{,}  \mathit{x} \,  {:}  \, \ottsym{\{}  \mathit{y} \,  {:}  \, \iota \,  |  \, \phi  \ottsym{\}}  \ottsym{,}  \Gamma_{{\mathrm{2}}}  \vdash  \phi_{{\mathrm{0}}}$.
 \end{enumerate}
\end{lemmap}
\begin{proof}
 Straightforward by simultaneous induction on the derivations.
\end{proof}

\begin{lemmap}{Weakening and Strengthening Type Hypothesis in Well-Formedness and Formula Typing}{ts-weak-strength-ty-hypo-wf-ftyping}
 Suppose that $\Gamma_{{\mathrm{1}}}  \vdash  \ottnt{T_{{\mathrm{01}}}}$ and that $\Gamma_{{\mathrm{1}}}  \vdash  \ottnt{T_{{\mathrm{01}}}}  \ottsym{<:}  \ottnt{T_{{\mathrm{02}}}}$ or $\Gamma_{{\mathrm{1}}}  \vdash  \ottnt{T_{{\mathrm{02}}}}  \ottsym{<:}  \ottnt{T_{{\mathrm{01}}}}$.
 \begin{enumerate}
  \item If $\vdash  \Gamma_{{\mathrm{1}}}  \ottsym{,}  \mathit{x} \,  {:}  \, \ottnt{T_{{\mathrm{02}}}}  \ottsym{,}  \Gamma_{{\mathrm{2}}}$, then $\vdash  \Gamma_{{\mathrm{1}}}  \ottsym{,}  \mathit{x} \,  {:}  \, \ottnt{T_{{\mathrm{01}}}}  \ottsym{,}  \Gamma_{{\mathrm{2}}}$.
  \item If $\Gamma_{{\mathrm{1}}}  \ottsym{,}  \mathit{x} \,  {:}  \, \ottnt{T_{{\mathrm{02}}}}  \ottsym{,}  \Gamma_{{\mathrm{2}}}  \vdash  \ottnt{T}$, then $\Gamma_{{\mathrm{1}}}  \ottsym{,}  \mathit{x} \,  {:}  \, \ottnt{T_{{\mathrm{01}}}}  \ottsym{,}  \Gamma_{{\mathrm{2}}}  \vdash  \ottnt{T}$.
  \item If $\Gamma_{{\mathrm{1}}}  \ottsym{,}  \mathit{x} \,  {:}  \, \ottnt{T_{{\mathrm{02}}}}  \ottsym{,}  \Gamma_{{\mathrm{2}}}  \vdash  \ottnt{C}$, then $\Gamma_{{\mathrm{1}}}  \ottsym{,}  \mathit{x} \,  {:}  \, \ottnt{T_{{\mathrm{01}}}}  \ottsym{,}  \Gamma_{{\mathrm{2}}}  \vdash  \ottnt{C}$.
  \item If $\Gamma_{{\mathrm{1}}}  \ottsym{,}  \mathit{x} \,  {:}  \, \ottnt{T_{{\mathrm{02}}}}  \ottsym{,}  \Gamma_{{\mathrm{2}}}  \vdash  \Phi$, then $\Gamma_{{\mathrm{1}}}  \ottsym{,}  \mathit{x} \,  {:}  \, \ottnt{T_{{\mathrm{01}}}}  \ottsym{,}  \Gamma_{{\mathrm{2}}}  \vdash  \Phi$.
  \item If $\Gamma_{{\mathrm{1}}}  \ottsym{,}  \mathit{x} \,  {:}  \, \ottnt{T_{{\mathrm{02}}}}  \ottsym{,}  \Gamma_{{\mathrm{2}}}  \vdash  \ottnt{t}  \ottsym{:}  \ottnt{s}$, then $\Gamma_{{\mathrm{1}}}  \ottsym{,}  \mathit{x} \,  {:}  \, \ottnt{T_{{\mathrm{01}}}}  \ottsym{,}  \Gamma_{{\mathrm{2}}}  \vdash  \ottnt{t}  \ottsym{:}  \ottnt{s}$.
  \item If $\Gamma_{{\mathrm{1}}}  \ottsym{,}  \mathit{x} \,  {:}  \, \ottnt{T_{{\mathrm{02}}}}  \ottsym{,}  \Gamma_{{\mathrm{2}}}  \vdash  \ottnt{A}  \ottsym{:}   \tceseq{ \ottnt{s} } $, then $\Gamma_{{\mathrm{1}}}  \ottsym{,}  \mathit{x} \,  {:}  \, \ottnt{T_{{\mathrm{01}}}}  \ottsym{,}  \Gamma_{{\mathrm{2}}}  \vdash  \ottnt{A}  \ottsym{:}   \tceseq{ \ottnt{s} } $.
  \item If $\Gamma_{{\mathrm{1}}}  \ottsym{,}  \mathit{x} \,  {:}  \, \ottnt{T_{{\mathrm{02}}}}  \ottsym{,}  \Gamma_{{\mathrm{2}}}  \vdash  \phi$, then $\Gamma_{{\mathrm{1}}}  \ottsym{,}  \mathit{x} \,  {:}  \, \ottnt{T_{{\mathrm{01}}}}  \ottsym{,}  \Gamma_{{\mathrm{2}}}  \vdash  \phi$.
 \end{enumerate}
\end{lemmap}
\begin{proof}
 If $\ottnt{T_{{\mathrm{02}}}}$ is not a refinement type,
 then we have the conclusion
 by Lemmas~\ref{lem:ts-unuse-non-refine-vars}, \ref{lem:ts-elim-unused-binding-wf-ftyping}, and \ref{lem:ts-weak-wf-ftyping}.

 If $\ottnt{T_{{\mathrm{02}}}}$ is a refinement type $\ottsym{\{}  \mathit{y} \,  {:}  \, \iota \,  |  \, \phi_{{\mathrm{2}}}  \ottsym{\}}$
 for some $\mathit{y}$, $\iota$, and $\phi_{{\mathrm{2}}}$,
 then the assumption $\Gamma_{{\mathrm{1}}}  \vdash  \ottnt{T_{{\mathrm{01}}}}  \ottsym{<:}  \ottnt{T_{{\mathrm{02}}}}$ or $\Gamma_{{\mathrm{1}}}  \vdash  \ottnt{T_{{\mathrm{02}}}}  \ottsym{<:}  \ottnt{T_{{\mathrm{01}}}}$
 implies $\ottnt{T_{{\mathrm{01}}}} \,  =  \, \ottsym{\{}  \mathit{y} \,  {:}  \, \iota \,  |  \, \phi_{{\mathrm{1}}}  \ottsym{\}}$ for some $\phi_{{\mathrm{1}}}$ by inversion.
 Because $\Gamma_{{\mathrm{1}}}  \vdash  \ottnt{T_{{\mathrm{01}}}}$, we have $\Gamma_{{\mathrm{1}}}  \ottsym{,}  \mathit{y} \,  {:}  \, \iota  \vdash  \phi_{{\mathrm{1}}}$ by inversion.
 Therefore, we have the conclusions
 by Lemmas~\ref{lem:ts-unrefine-refine-ty} and \ref{lem:ts-refine-base-sorts}.
\end{proof}

\begin{lemmap}{Effect Composition Preserves Well-Formedness}{ts-wf-eff-compose}
 If $\Gamma  \vdash  \Phi_{{\mathrm{1}}}$ and $\Gamma  \vdash  \Phi_{{\mathrm{2}}}$, then $\Gamma  \vdash  \Phi_{{\mathrm{1}}} \,  \tceappendop  \, \Phi_{{\mathrm{2}}}$.
\end{lemmap}
\begin{proof}
 Let $\mathit{x}, \mathit{x_{{\mathrm{1}}}}, \mathit{x_{{\mathrm{2}}}}$ be fresh variables, and
 $\Gamma'$ be typing context $\Gamma  \ottsym{,}  \mathit{x} \,  {:}  \,  \Sigma ^\ast   \ottsym{,}  \mathit{x_{{\mathrm{1}}}} \,  {:}  \,  \Sigma ^\ast   \ottsym{,}  \mathit{x_{{\mathrm{2}}}} \,  {:}  \,  \Sigma ^\ast $.
 We have
 \[
  \Gamma  \ottsym{,}  \mathit{x} \,  {:}  \,  \Sigma ^\ast   \vdash   { \ottsym{(}  \Phi_{{\mathrm{1}}} \,  \tceappendop  \, \Phi_{{\mathrm{2}}}  \ottsym{)} }^\mu   \ottsym{(}  \mathit{x}  \ottsym{)}
 \]
 because:
 \begin{prooftree}
  \AxiomC{{$\Gamma'  \vdash   \mathit{x}    =    \mathit{x_{{\mathrm{1}}}} \,  \tceappendop  \, \mathit{x_{{\mathrm{2}}}} $}}
  \AxiomC{$\Gamma'  \vdash   { \Phi_{{\mathrm{1}}} }^\mu   \ottsym{(}  \mathit{x_{{\mathrm{1}}}}  \ottsym{)}$}
  \AxiomC{$\Gamma'  \vdash   { \Phi_{{\mathrm{2}}} }^\mu   \ottsym{(}  \mathit{x_{{\mathrm{2}}}}  \ottsym{)}$}
  \RightLabel{\Tf{BinOp}}
  \BinaryInfC{{$\Gamma'  \vdash    { \Phi_{{\mathrm{1}}} }^\mu   \ottsym{(}  \mathit{x_{{\mathrm{1}}}}  \ottsym{)}    \wedge     { \Phi_{{\mathrm{2}}} }^\mu   \ottsym{(}  \mathit{x_{{\mathrm{2}}}}  \ottsym{)} $}}
  \RightLabel{\Tf{BinOp}}
  \BinaryInfC{$\Gamma'  \vdash     \mathit{x}    =    \mathit{x_{{\mathrm{1}}}} \,  \tceappendop  \, \mathit{x_{{\mathrm{2}}}}     \wedge     { \Phi_{{\mathrm{1}}} }^\mu   \ottsym{(}  \mathit{x_{{\mathrm{1}}}}  \ottsym{)}     \wedge     { \Phi_{{\mathrm{2}}} }^\mu   \ottsym{(}  \mathit{x_{{\mathrm{2}}}}  \ottsym{)} $}
  \RightLabel{\Tf{ExistS}}
  \UnaryInfC{$\Gamma  \ottsym{,}  \mathit{x} \,  {:}  \,  \Sigma ^\ast   \vdash    \exists  \,  \mathit{x_{{\mathrm{1}}}}  \mathrel{  \in  }   \Sigma ^\ast   . \    \exists  \,  \mathit{x_{{\mathrm{2}}}}  \mathrel{  \in  }   \Sigma ^\ast   . \     \mathit{x}    =    \mathit{x_{{\mathrm{1}}}} \,  \tceappendop  \, \mathit{x_{{\mathrm{2}}}}     \wedge     { \Phi_{{\mathrm{1}}} }^\mu   \ottsym{(}  \mathit{x_{{\mathrm{1}}}}  \ottsym{)}     \wedge     { \Phi_{{\mathrm{2}}} }^\mu   \ottsym{(}  \mathit{x_{{\mathrm{2}}}}  \ottsym{)}   $}
 \end{prooftree}
 where
 \begin{itemize}
  \item $\Gamma'  \vdash   \mathit{x}    =    \mathit{x_{{\mathrm{1}}}} \,  \tceappendop  \, \mathit{x_{{\mathrm{2}}}} $ is derived
        by \Tf{Pred}, \Tf{VarS}, and \Tf{Fun},
  \item $\Gamma'  \vdash   { \Phi_{{\mathrm{1}}} }^\mu   \ottsym{(}  \mathit{x_{{\mathrm{1}}}}  \ottsym{)}$ is derived
        by \reflem{ts-weak-wf-ftyping} with $\Gamma  \vdash  \Phi_{{\mathrm{1}}}$, and
  \item $\Gamma'  \vdash   { \Phi_{{\mathrm{2}}} }^\mu   \ottsym{(}  \mathit{x_{{\mathrm{2}}}}  \ottsym{)}$ is derived
        by \reflem{ts-weak-wf-ftyping} with $\Gamma  \vdash  \Phi_{{\mathrm{2}}}$.
 \end{itemize}

 Let $\mathit{y}, \mathit{x'}, \mathit{y'}$ be fresh variables, and
 $\Gamma'$ be typing context $\Gamma  \ottsym{,}  \mathit{y} \,  {:}  \,  \Sigma ^\omega   \ottsym{,}  \mathit{x'} \,  {:}  \,  \Sigma ^\ast   \ottsym{,}  \mathit{y'} \,  {:}  \,  \Sigma ^\omega $.
 We have
 \[
  \Gamma  \ottsym{,}  \mathit{y} \,  {:}  \,  \Sigma ^\omega   \vdash   { \ottsym{(}  \Phi_{{\mathrm{1}}} \,  \tceappendop  \, \Phi_{{\mathrm{2}}}  \ottsym{)} }^\nu   \ottsym{(}  \mathit{y}  \ottsym{)}
 \]
 because:
 \begin{prooftree}
  \AxiomC{$\Gamma  \ottsym{,}  \mathit{y} \,  {:}  \,  \Sigma ^\omega   \vdash   { \Phi_{{\mathrm{1}}} }^\nu   \ottsym{(}  \mathit{y}  \ottsym{)}$}
  \AxiomC{$\Gamma'  \vdash   \mathit{y}    =    \mathit{x'} \,  \tceappendop  \, \mathit{y'} $}
  \AxiomC{$\Gamma'  \vdash   { \Phi_{{\mathrm{1}}} }^\mu   \ottsym{(}  \mathit{x'}  \ottsym{)}$}
  \AxiomC{$\Gamma'  \vdash   { \Phi_{{\mathrm{2}}} }^\nu   \ottsym{(}  \mathit{y'}  \ottsym{)}$}
  \BinaryInfC{$\Gamma'  \vdash    { \Phi_{{\mathrm{1}}} }^\mu   \ottsym{(}  \mathit{x'}  \ottsym{)}    \wedge     { \Phi_{{\mathrm{2}}} }^\nu   \ottsym{(}  \mathit{y'}  \ottsym{)} $}
  \BinaryInfC{$\Gamma'  \vdash     \mathit{y}    =    \mathit{x'} \,  \tceappendop  \, \mathit{y'}     \wedge     { \Phi_{{\mathrm{1}}} }^\mu   \ottsym{(}  \mathit{x'}  \ottsym{)}     \wedge     { \Phi_{{\mathrm{2}}} }^\nu   \ottsym{(}  \mathit{y'}  \ottsym{)} $}
  \RightLabel{\Tf{ExistS}}
  \UnaryInfC{$\Gamma  \ottsym{,}  \mathit{y} \,  {:}  \,  \Sigma ^\omega   \vdash    \exists  \,  \mathit{x'}  \mathrel{  \in  }   \Sigma ^\ast   . \    \exists  \,  \mathit{y'}  \mathrel{  \in  }   \Sigma ^\omega   . \     \mathit{y}    =    \mathit{x'} \,  \tceappendop  \, \mathit{y'}     \wedge     { \Phi_{{\mathrm{1}}} }^\mu   \ottsym{(}  \mathit{x'}  \ottsym{)}     \wedge     { \Phi_{{\mathrm{2}}} }^\nu   \ottsym{(}  \mathit{y'}  \ottsym{)}   $}
  \RightLabel{\Tf{BinOp}}
  \BinaryInfC{$\Gamma  \ottsym{,}  \mathit{y} \,  {:}  \,  \Sigma ^\omega   \vdash    { \Phi_{{\mathrm{1}}} }^\nu   \ottsym{(}  \mathit{y}  \ottsym{)}    \vee    \ottsym{(}    \exists  \,  \mathit{x'}  \mathrel{  \in  }   \Sigma ^\ast   . \    \exists  \,  \mathit{y'}  \mathrel{  \in  }   \Sigma ^\omega   . \     \mathit{y}    =    \mathit{x'} \,  \tceappendop  \, \mathit{y'}     \wedge     { \Phi_{{\mathrm{1}}} }^\mu   \ottsym{(}  \mathit{x'}  \ottsym{)}     \wedge     { \Phi_{{\mathrm{2}}} }^\nu   \ottsym{(}  \mathit{y'}  \ottsym{)}     \ottsym{)} $}
 \end{prooftree}
 where
 \begin{itemize}
  \item $\Gamma  \ottsym{,}  \mathit{y} \,  {:}  \,  \Sigma ^\omega   \vdash   { \Phi_{{\mathrm{1}}} }^\nu   \ottsym{(}  \mathit{y}  \ottsym{)}$ is derived
        by $\Gamma  \vdash  \Phi_{{\mathrm{1}}}$,
  \item $\Gamma'  \vdash   \mathit{y}    =    \mathit{x'} \,  \tceappendop  \, \mathit{y'} $ is derived
        by \Tf{Pred}, \Tf{VarS}, and \Tf{Fun},
  \item $\Gamma'  \vdash   { \Phi_{{\mathrm{1}}} }^\mu   \ottsym{(}  \mathit{x'}  \ottsym{)}$ is derived
        by \reflem{ts-weak-wf-ftyping} with $\Gamma  \vdash  \Phi_{{\mathrm{1}}}$, and
  \item $\Gamma'  \vdash   { \Phi_{{\mathrm{2}}} }^\nu   \ottsym{(}  \mathit{y'}  \ottsym{)}$ is derived
        by \reflem{ts-weak-wf-ftyping} with $\Gamma  \vdash  \Phi_{{\mathrm{2}}}$.
 \end{itemize}
\end{proof}

\begin{lemmap}{Well-Formedness of $ \mu $-Effects of Recursive Functions}{lr-mu-eff-rec-wf-XXX}
 Assume that
 \begin{itemize}
  \item $\Gamma  \ottsym{,}  \ottsym{(}   \tceseq{ \mathit{X'} }   \ottsym{,}  \mathit{X_{{\mathrm{0}}}}  \ottsym{,}   \tceseq{ \mathit{X} }   \ottsym{)} \,  {:}  \,  (   \Sigma ^\ast   ,   \tceseq{ \iota_{{\mathrm{0}}} }   )   \ottsym{,}  \ottsym{(}   \tceseq{ \mathit{Y'} }   \ottsym{,}  \mathit{Y_{{\mathrm{0}}}}  \ottsym{,}   \tceseq{ \mathit{Y} }   \ottsym{)} \,  {:}  \,  (   \Sigma ^\omega   ,   \tceseq{ \iota_{{\mathrm{0}}} }   )   \ottsym{,}   \tceseq{ \mathit{z_{{\mathrm{0}}}}  \,  {:}  \,  \iota_{{\mathrm{0}}} }   \vdash  \Phi$,
  \item $  \{   \tceseq{ \mathit{X} }   \ottsym{,}   \tceseq{ \mathit{Y} }   \}   \, \ottsym{\#} \,   \mathit{fpv}  (  \Phi  )  $, and
  \item $\mathit{Y_{{\mathrm{0}}}} \,  \not\in  \,  \mathit{fpv}  (   { \Phi }^\mu   ) $.
 \end{itemize}
 Let $\Gamma' \,  =  \,  \tceseq{ \mathit{X'} }  \,  {:}  \,  (   \Sigma ^\ast   ,   \tceseq{ \iota_{{\mathrm{0}}} }   )   \ottsym{,}   \tceseq{ \mathit{Y'} }  \,  {:}  \,  (   \Sigma ^\omega   ,   \tceseq{ \iota_{{\mathrm{0}}} }   ) $.
 Then, $\Gamma  \ottsym{,}  \Gamma'  \vdash    \mu\!   \, (  \mathit{X_{{\mathrm{0}}}} ,  \mathit{Y_{{\mathrm{0}}}} ,   \tceseq{ \mathit{z_{{\mathrm{0}}}}  \,  {:}  \,  \iota_{{\mathrm{0}}} }  ,  \Phi  )   \ottsym{:}   (   \Sigma ^\ast   ,   \tceseq{ \iota_{{\mathrm{0}}} }   ) $.
\end{lemmap}
\begin{proof}
 \TS{CHECKED: 2022/07/06}
 By \Tf{Mu}, it suffices to show that
 \[
  \Gamma  \ottsym{,}  \Gamma'  \ottsym{,}  \mathit{X_{{\mathrm{0}}}} \,  {:}  \,  (   \Sigma ^\ast   ,   \tceseq{ \iota_{{\mathrm{0}}} }   )   \ottsym{,}  \mathit{x} \,  {:}  \,  \Sigma ^\ast   \ottsym{,}   \tceseq{ \mathit{z_{{\mathrm{0}}}}  \,  {:}  \,  \iota_{{\mathrm{0}}} }   \vdash   { \Phi }^\mu   \ottsym{(}  \mathit{x}  \ottsym{)}
 \]
 for some fresh $\mathit{x}$.
 Because
 $\Gamma  \ottsym{,}  \ottsym{(}   \tceseq{ \mathit{X'} }   \ottsym{,}  \mathit{X_{{\mathrm{0}}}}  \ottsym{,}   \tceseq{ \mathit{X} }   \ottsym{)} \,  {:}  \,  (   \Sigma ^\ast   ,   \tceseq{ \iota_{{\mathrm{0}}} }   )   \ottsym{,}  \ottsym{(}   \tceseq{ \mathit{Y'} }   \ottsym{,}  \mathit{Y_{{\mathrm{0}}}}  \ottsym{,}   \tceseq{ \mathit{Y} }   \ottsym{)} \,  {:}  \,  (   \Sigma ^\omega   ,   \tceseq{ \iota_{{\mathrm{0}}} }   )   \ottsym{,}   \tceseq{ \mathit{z_{{\mathrm{0}}}}  \,  {:}  \,  \iota_{{\mathrm{0}}} }   \vdash  \Phi$
 by the assumption,
 we have
 \[
  \Gamma  \ottsym{,}  \ottsym{(}   \tceseq{ \mathit{X'} }   \ottsym{,}  \mathit{X_{{\mathrm{0}}}}  \ottsym{,}   \tceseq{ \mathit{X} }   \ottsym{)} \,  {:}  \,  (   \Sigma ^\ast   ,   \tceseq{ \iota_{{\mathrm{0}}} }   )   \ottsym{,}  \ottsym{(}   \tceseq{ \mathit{Y'} }   \ottsym{,}  \mathit{Y_{{\mathrm{0}}}}  \ottsym{,}   \tceseq{ \mathit{Y} }   \ottsym{)} \,  {:}  \,  (   \Sigma ^\omega   ,   \tceseq{ \iota_{{\mathrm{0}}} }   )   \ottsym{,}   \tceseq{ \mathit{z_{{\mathrm{0}}}}  \,  {:}  \,  \iota_{{\mathrm{0}}} }   \ottsym{,}  \mathit{x} \,  {:}  \,  \Sigma ^\ast   \vdash   { \Phi }^\mu   \ottsym{(}  \mathit{x}  \ottsym{)} ~.
 \]
 Because
 $  \{   \tceseq{ \mathit{X} }   \ottsym{,}   \tceseq{ \mathit{Y} }   \}   \, \ottsym{\#} \,   \mathit{fpv}  (  \Phi  )  $ and
 $\mathit{Y_{{\mathrm{0}}}} \,  \not\in  \,  \mathit{fpv}  (   { \Phi }^\mu   ) $,
 we have the conclusion
 by Lemmas~\ref{lem:ts-pull-back-sorted-vars} and \ref{lem:ts-elim-unused-binding-wf-ftyping}.
\end{proof}

\begin{lemmap}{Well-Formedness of $ \nu $-Effects of Recursive Functions}{lr-nu-eff-rec-wf-XXX}
 Assume that
 \begin{itemize}
  \item $\Gamma  \ottsym{,}  \ottsym{(}   \tceseq{ \mathit{X'} }   \ottsym{,}  \mathit{X_{{\mathrm{0}}}}  \ottsym{,}   \tceseq{ \mathit{X} }   \ottsym{)} \,  {:}  \,  (   \Sigma ^\ast   ,   \tceseq{ \iota_{{\mathrm{0}}} }   )   \ottsym{,}  \ottsym{(}   \tceseq{ \mathit{Y'} }   \ottsym{,}  \mathit{Y_{{\mathrm{0}}}}  \ottsym{,}   \tceseq{ \mathit{Y} }   \ottsym{)} \,  {:}  \,  (   \Sigma ^\omega   ,   \tceseq{ \iota_{{\mathrm{0}}} }   )   \ottsym{,}   \tceseq{ \mathit{z_{{\mathrm{0}}}}  \,  {:}  \,  \iota_{{\mathrm{0}}} }   \vdash  \Phi$,
  \item $  \{   \tceseq{ \mathit{X} }   \ottsym{,}   \tceseq{ \mathit{Y} }   \}   \, \ottsym{\#} \,   \mathit{fpv}  (  \Phi  )  $, and
  \item $\mathit{Y_{{\mathrm{0}}}} \,  \not\in  \,  \mathit{fpv}  (   { \Phi }^\mu   ) $.
 \end{itemize}
 Let $\Gamma' \,  =  \,  \tceseq{ \mathit{X'} }  \,  {:}  \,  (   \Sigma ^\ast   ,   \tceseq{ \iota_{{\mathrm{0}}} }   )   \ottsym{,}   \tceseq{ \mathit{Y'} }  \,  {:}  \,  (   \Sigma ^\omega   ,   \tceseq{ \iota_{{\mathrm{0}}} }   ) $.
 Then, $\Gamma  \ottsym{,}  \Gamma'  \vdash    \nu\!   \, (  \mathit{X_{{\mathrm{0}}}} ,  \mathit{Y_{{\mathrm{0}}}} ,   \tceseq{ \mathit{z_{{\mathrm{0}}}}  \,  {:}  \,  \iota_{{\mathrm{0}}} }  ,  \Phi  )   \ottsym{:}   (   \Sigma ^\omega   ,   \tceseq{ \iota_{{\mathrm{0}}} }   ) $.
\end{lemmap}
\begin{proof}
 \TS{CHECKED: 2022/07/07}
 By \Tf{Nu}, it suffices to show that
 \[
  \Gamma  \ottsym{,}  \Gamma'  \ottsym{,}  \mathit{Y_{{\mathrm{0}}}} \,  {:}  \,  (   \Sigma ^\omega   ,   \tceseq{ \iota_{{\mathrm{0}}} }   )   \ottsym{,}  \mathit{y} \,  {:}  \,  \Sigma ^\omega   \ottsym{,}   \tceseq{ \mathit{z_{{\mathrm{0}}}}  \,  {:}  \,  \iota_{{\mathrm{0}}} }   \vdash    { \Phi }^\nu   \ottsym{(}  \mathit{y}  \ottsym{)}    [    \mu\!   \, (  \mathit{X_{{\mathrm{0}}}} ,  \mathit{Y_{{\mathrm{0}}}} ,   \tceseq{ \mathit{z_{{\mathrm{0}}}}  \,  {:}  \,  \iota_{{\mathrm{0}}} }  ,  \Phi  )   /  \mathit{X_{{\mathrm{0}}}}  ]  
 \]
 for some fresh $\mathit{y}$.
 Because
 $\Gamma  \ottsym{,}  \ottsym{(}   \tceseq{ \mathit{X'} }   \ottsym{,}  \mathit{X_{{\mathrm{0}}}}  \ottsym{,}   \tceseq{ \mathit{X} }   \ottsym{)} \,  {:}  \,  (   \Sigma ^\ast   ,   \tceseq{ \iota_{{\mathrm{0}}} }   )   \ottsym{,}  \ottsym{(}   \tceseq{ \mathit{Y'} }   \ottsym{,}  \mathit{Y_{{\mathrm{0}}}}  \ottsym{,}   \tceseq{ \mathit{Y} }   \ottsym{)} \,  {:}  \,  (   \Sigma ^\omega   ,   \tceseq{ \iota_{{\mathrm{0}}} }   )   \ottsym{,}   \tceseq{ \mathit{z_{{\mathrm{0}}}}  \,  {:}  \,  \iota_{{\mathrm{0}}} }   \vdash  \Phi$
 by the assumption,
 we have
 \[
  \Gamma  \ottsym{,}  \ottsym{(}   \tceseq{ \mathit{X'} }   \ottsym{,}  \mathit{X_{{\mathrm{0}}}}  \ottsym{,}   \tceseq{ \mathit{X} }   \ottsym{)} \,  {:}  \,  (   \Sigma ^\ast   ,   \tceseq{ \iota_{{\mathrm{0}}} }   )   \ottsym{,}  \ottsym{(}   \tceseq{ \mathit{Y'} }   \ottsym{,}  \mathit{Y_{{\mathrm{0}}}}  \ottsym{,}   \tceseq{ \mathit{Y} }   \ottsym{)} \,  {:}  \,  (   \Sigma ^\omega   ,   \tceseq{ \iota_{{\mathrm{0}}} }   )   \ottsym{,}   \tceseq{ \mathit{z_{{\mathrm{0}}}}  \,  {:}  \,  \iota_{{\mathrm{0}}} }   \ottsym{,}  \mathit{y} \,  {:}  \,  \Sigma ^\omega   \vdash   { \Phi }^\nu   \ottsym{(}  \mathit{y}  \ottsym{)} ~.
 \]
 By \reflem{lr-mu-eff-rec-wf-XXX},
 $\Gamma  \ottsym{,}  \Gamma'  \vdash    \mu\!   \, (  \mathit{X_{{\mathrm{0}}}} ,  \mathit{Y_{{\mathrm{0}}}} ,   \tceseq{ \mathit{z_{{\mathrm{0}}}}  \,  {:}  \,  \iota_{{\mathrm{0}}} }  ,  \Phi  )   \ottsym{:}   (   \Sigma ^\ast   ,   \tceseq{ \iota_{{\mathrm{0}}} }   ) $.
 Therefore, by Lemmas~\ref{lem:ts-pred-subst-wf-ftyping}, \ref{lem:ts-pull-back-sorted-vars}, and \ref{lem:ts-elim-unused-binding-wf-ftyping},
 we have the conclusion.
\end{proof}

\begin{lemmap}{Well-Formedness of Effects of Recursive Functions}{lr-eff-rec-wf-XXX}
 Assume that
 \begin{itemize}
  \item $\Gamma  \ottsym{,}  \ottsym{(}   \tceseq{ \mathit{X'} }   \ottsym{,}   \tceseq{ \mathit{X} }   \ottsym{)} \,  {:}  \,  (   \Sigma ^\ast   ,   \tceseq{ \iota_{{\mathrm{0}}} }   )   \ottsym{,}  \ottsym{(}   \tceseq{ \mathit{Y'} }   \ottsym{,}   \tceseq{ \mathit{Y} }   \ottsym{)} \,  {:}  \,  (   \Sigma ^\omega   ,   \tceseq{ \iota_{{\mathrm{0}}} }   )   \ottsym{,}   \tceseq{ \mathit{z_{{\mathrm{0}}}}  \,  {:}  \,  \iota_{{\mathrm{0}}} }   \vdash  \ottnt{C}$,
  \item $ \tceseq{ \ottsym{(}  \mathit{X}  \ottsym{,}  \mathit{Y}  \ottsym{)} }   \ottsym{;}   \tceseq{ \mathit{z_{{\mathrm{0}}}}  \,  {:}  \,  \iota_{{\mathrm{0}}} }  \,  |  \, \ottnt{C}  \vdash  \sigma$, and
  \item $ \tceseq{ \ottsym{(}  \mathit{X'}  \ottsym{,}  \mathit{Y'}  \ottsym{)} }   \ottsym{;}   \tceseq{ \ottsym{(}  \mathit{X}  \ottsym{,}  \mathit{Y}  \ottsym{)} }  \,  \vdash ^{\circ}  \, \ottnt{C}$.
 \end{itemize}
 Let $\Gamma' \,  =  \,  \tceseq{ \mathit{X'} }  \,  {:}  \,  (   \Sigma ^\ast   ,   \tceseq{ \iota_{{\mathrm{0}}} }   )   \ottsym{,}   \tceseq{ \mathit{Y'} }  \,  {:}  \,  (   \Sigma ^\omega   ,   \tceseq{ \iota_{{\mathrm{0}}} }   ) $.
 Then,
 \[
   {   \forall  \, { \mathit{Y_{{\mathrm{0}}}} \,  \in  \,  \{   \tceseq{ \mathit{Y} }   \}  } . \  \Gamma  \ottsym{,}  \Gamma'  \vdash  \sigma  \ottsym{(}  \mathit{Y_{{\mathrm{0}}}}  \ottsym{)}  \ottsym{:}   (   \Sigma ^\omega   ,   \tceseq{ \iota_{{\mathrm{0}}} }   )   } \,\mathrel{\wedge}\, {   \forall  \, { \mathit{X_{{\mathrm{0}}}} \,  \in  \,  \{   \tceseq{ \mathit{X} }   \}  } . \  \Gamma  \ottsym{,}  \Gamma'  \vdash  \sigma  \ottsym{(}  \mathit{X_{{\mathrm{0}}}}  \ottsym{)}  \ottsym{:}   (   \Sigma ^\ast   ,   \tceseq{ \iota_{{\mathrm{0}}} }   )   }  ~.
 \]
\end{lemmap}
\begin{proof}
 \TS{CHECKED: 2022/07/07}
 Straightforward by induction on the derivation of $ \tceseq{ \ottsym{(}  \mathit{X}  \ottsym{,}  \mathit{Y}  \ottsym{)} }   \ottsym{;}   \tceseq{ \mathit{z_{{\mathrm{0}}}}  \,  {:}  \,  \iota_{{\mathrm{0}}} }  \,  |  \, \ottnt{C}  \vdash  \sigma$
 with Lemmas~\ref{lem:lr-mu-eff-rec-wf-XXX}, \ref{lem:lr-nu-eff-rec-wf-XXX}, and \ref{lem:ts-pred-subst-wf-ftyping}.
\end{proof}

\begin{lemmap}{Well-Formedness of Types and Typing Contexts in Typing}{ts-wf-ty-tctx-typing}
 If $\Gamma  \vdash  \ottnt{e}  \ottsym{:}  \ottnt{C}$, then $\Gamma  \vdash  \ottnt{C}$.
\end{lemmap}
\begin{proof}
 By induction on the typing derivation.
 Note that, for any $\ottnt{T}$, if $\ottnt{C} \,  =  \,  \ottnt{T}  \,  \,\&\,  \,   \Phi_\mathsf{val}   \, / \,   \square  $, then it suffices to show that
 $\Gamma  \vdash  \ottnt{T}$ by \reflem{ts-wf-TP_val}.
 By case analysis on the typing rule applied last to derive $\Gamma  \vdash  \ottnt{e}  \ottsym{:}  \ottnt{C}$.
   \begin{caseanalysis}
    \case \T{CVar}:
     We are given
     \begin{prooftree}
      \AxiomC{$\vdash  \Gamma$}
      \AxiomC{$\Gamma  \ottsym{(}  \mathit{x}  \ottsym{)} \,  =  \, \ottsym{\{}  \mathit{y} \,  {:}  \, \iota \,  |  \, \phi  \ottsym{\}}$}
      \BinaryInfC{$\Gamma  \vdash  \mathit{x}  \ottsym{:}  \ottsym{\{}  \mathit{y} \,  {:}  \, \iota \,  |  \,  \mathit{x}    =    \mathit{y}   \ottsym{\}}$}
     \end{prooftree}
     for some $\mathit{x}$, $\mathit{y}$, $\iota$, and $\phi$
     such that $\ottnt{e} \,  =  \, \mathit{x}$ and $\ottnt{C} \,  =  \,  \ottsym{\{}  \mathit{y} \,  {:}  \, \iota \,  |  \,  \mathit{x}    =    \mathit{y}   \ottsym{\}}  \,  \,\&\,  \,   \Phi_\mathsf{val}   \, / \,   \square  $.
     Without loss of generality, we can suppose that $\mathit{y} \,  \not\in  \,  \mathit{dom}  (  \Gamma  ) $.
     Then, we have $\vdash  \Gamma  \ottsym{,}  \mathit{y} \,  {:}  \, \iota$ by \WFC{SVar}.
     Thus, we have the conclusion $\Gamma  \vdash  \ottsym{\{}  \mathit{y} \,  {:}  \, \iota \,  |  \,  \mathit{x}    =    \mathit{y}   \ottsym{\}}$ by the following derivation:
     \begin{prooftree}
      \AxiomC{$\vdash  \Gamma  \ottsym{,}  \mathit{y} \,  {:}  \, \iota$}
      \RightLabel{\Tf{Prim}}
      \UnaryInfC{$\Gamma  \ottsym{,}  \mathit{y} \,  {:}  \, \iota  \vdash   (=)   \ottsym{:}   (  \iota  ,  \iota  ) $}
      \AxiomC{$\Gamma  \ottsym{,}  \mathit{y} \,  {:}  \, \iota  \vdash  \mathit{x}  \ottsym{:}  \iota$}
      \AxiomC{$\Gamma  \ottsym{,}  \mathit{y} \,  {:}  \, \iota  \vdash  \mathit{y}  \ottsym{:}  \iota$}
      \RightLabel{\Tf{Pred}}
      \TrinaryInfC{$\Gamma  \ottsym{,}  \mathit{y} \,  {:}  \, \iota  \vdash   \mathit{x}    =    \mathit{y} $}
      \RightLabel{\WFT{Refine}}
      \UnaryInfC{$\Gamma  \vdash  \ottsym{\{}  \mathit{y} \,  {:}  \, \iota \,  |  \,  \mathit{x}    =    \mathit{y}   \ottsym{\}}$}
     \end{prooftree}
     where $\Gamma  \ottsym{,}  \mathit{y} \,  {:}  \, \iota  \vdash  \mathit{x}  \ottsym{:}  \iota$ and $\Gamma  \ottsym{,}  \mathit{y} \,  {:}  \, \iota  \vdash  \mathit{y}  \ottsym{:}  \iota$ are derived
     by \Tf{VarT} and \Tf{VarS}, respectively.

    \case \T{Var}: By inversion and \reflem{ts-weak-wf-ftyping}.

    \case \T{Const}:
     We are given
     \begin{prooftree}
      \AxiomC{$\vdash  \Gamma$}
      \AxiomC{$ \mathit{ty}  (  \ottnt{c}  )  \,  =  \, \iota$}
      \BinaryInfC{$\Gamma  \vdash  \ottnt{c}  \ottsym{:}  \ottsym{\{}  \mathit{x} \,  {:}  \, \iota \,  |  \,  \ottnt{c}    =    \mathit{x}   \ottsym{\}}$}
     \end{prooftree}
     for some $\mathit{x}$, $\ottnt{c}$, and $\iota$
     such that $\ottnt{e} \,  =  \, \ottnt{c}$ and $\ottnt{C} \,  =  \,  \ottsym{\{}  \mathit{x} \,  {:}  \, \iota \,  |  \,  \ottnt{c}    =    \mathit{x}   \ottsym{\}}  \,  \,\&\,  \,   \Phi_\mathsf{val}   \, / \,   \square  $.
     Without loss of generality, we can suppose that $\mathit{x} \,  \not\in  \,  \mathit{dom}  (  \Gamma  ) $.
     Then, we have $\vdash  \Gamma  \ottsym{,}  \mathit{x} \,  {:}  \, \iota$ by \WFC{SVar}.
     Thus, we have the conclusion $\Gamma  \vdash  \ottsym{\{}  \mathit{x} \,  {:}  \, \iota \,  |  \,  \ottnt{c}    =    \mathit{x}   \ottsym{\}}$ by the following derivation:
     \begin{prooftree}
      \AxiomC{$\vdash  \Gamma  \ottsym{,}  \mathit{x} \,  {:}  \, \iota$}
      \RightLabel{\Tf{Prim}}
      \UnaryInfC{$\Gamma  \ottsym{,}  \mathit{x} \,  {:}  \, \iota  \vdash   (=)   \ottsym{:}   (  \iota  ,  \iota  ) $}
      \AxiomC{$\vdash  \Gamma  \ottsym{,}  \mathit{x} \,  {:}  \, \iota$}
      \RightLabel{\Tf{Fun}}
      \UnaryInfC{$\Gamma  \ottsym{,}  \mathit{x} \,  {:}  \, \iota  \vdash  \ottnt{c}  \ottsym{:}  \iota$}
      \AxiomC{$\vdash  \Gamma  \ottsym{,}  \mathit{x} \,  {:}  \, \iota$}
      \AxiomC{$\ottsym{(}  \Gamma  \ottsym{,}  \mathit{x} \,  {:}  \, \iota  \ottsym{)}  \ottsym{(}  \mathit{x}  \ottsym{)} \,  =  \, \iota$}
      \RightLabel{\Tf{VarS}}
      \BinaryInfC{$\Gamma  \ottsym{,}  \mathit{x} \,  {:}  \, \iota  \vdash  \mathit{x}  \ottsym{:}  \iota$}
      \RightLabel{\Tf{Pred}}
      \TrinaryInfC{$\Gamma  \ottsym{,}  \mathit{x} \,  {:}  \, \iota  \vdash   \ottnt{c}    =    \mathit{x} $}
      \RightLabel{\WFT{Refine}}
      \UnaryInfC{$\Gamma  \vdash  \ottsym{\{}  \mathit{x} \,  {:}  \, \iota \,  |  \,  \ottnt{c}    =    \mathit{x}   \ottsym{\}}$}
     \end{prooftree}
     where we assume that there exists some $\mathit{F}$ that corresponds to $\ottnt{c}$
     (\refasm{termfun}) in the application of \Tf{Fun}.

    \case \T{PAbs}: By the IH and \WFT{Forall}.

    \case \T{Abs}: By the IH, \WFT{Fun}, and \reflem{ts-wf-TP_val}.

    \case \T{Fun}:
     We are given
     \begin{prooftree}
      \AxiomC{$\Gamma  \ottsym{,}   \tceseq{ \mathit{X} }  \,  {:}  \,  (   \Sigma ^\ast   ,   \tceseq{ \iota_{{\mathrm{0}}} }   )   \ottsym{,}   \tceseq{ \mathit{Y} }  \,  {:}  \,  (   \Sigma ^\omega   ,   \tceseq{ \iota_{{\mathrm{0}}} }   )   \ottsym{,}  \mathit{f} \,  {:}  \, \ottsym{(}   \tceseq{ \mathit{z}  \,  {:}  \,  \ottnt{T_{{\mathrm{1}}}} }   \ottsym{)}  \rightarrow  \ottnt{C_{{\mathrm{0}}}}  \ottsym{,}   \tceseq{ \mathit{z}  \,  {:}  \,  \ottnt{T_{{\mathrm{1}}}} }   \vdash  \ottnt{e'}  \ottsym{:}  \ottnt{C}$}
      \UnaryInfC{$\Gamma  \vdash   \mathsf{rec}  \, (  \tceseq{  \mathit{X} ^{ \sigma  \ottsym{(}  \mathit{X}  \ottsym{)} },  \mathit{Y} ^{ \sigma  \ottsym{(}  \mathit{Y}  \ottsym{)} }  }  ,  \mathit{f} ,  \mathit{z} ) . \,  \ottnt{e'}   \ottsym{:}  \ottsym{(}   \tceseq{ \mathit{z}  \,  {:}  \,  \ottnt{T_{{\mathrm{1}}}} }   \ottsym{)}  \rightarrow  \sigma  \ottsym{(}  \ottnt{C_{{\mathrm{0}}}}  \ottsym{)}$}
     \end{prooftree}
     for some $ \tceseq{ \mathit{X} } $, $ \tceseq{ \mathit{Y} } $, $ \tceseq{ \mathit{z} } $, $ \tceseq{ \mathit{z_{{\mathrm{0}}}} } $, $\mathit{f}$, $\ottnt{e'}$, $ \tceseq{ \ottnt{T_{{\mathrm{1}}}} } $, $\ottnt{C_{{\mathrm{0}}}}$, $\ottnt{C}$, $\sigma$, and $ \tceseq{ \iota_{{\mathrm{0}}} } $ such that
     \begin{itemize}
      \item $\ottnt{e} \,  =  \,  \mathsf{rec}  \, (  \tceseq{  \mathit{X} ^{ \sigma  \ottsym{(}  \mathit{X}  \ottsym{)} },  \mathit{Y} ^{ \sigma  \ottsym{(}  \mathit{Y}  \ottsym{)} }  }  ,  \mathit{f} ,  \mathit{z} ) . \,  \ottnt{e'} $,
      \item $\ottnt{C} \,  =  \,  \ottsym{(}   \tceseq{ \mathit{z}  \,  {:}  \,  \ottnt{T_{{\mathrm{1}}}} }   \ottsym{)}  \rightarrow  \sigma  \ottsym{(}  \ottnt{C_{{\mathrm{0}}}}  \ottsym{)}  \,  \,\&\,  \,   \Phi_\mathsf{val}   \, / \,   \square  $,
      \item $ \tceseq{ \mathit{z_{{\mathrm{0}}}}  \,  {:}  \,  \iota_{{\mathrm{0}}} }  \,  =  \,  \gamma  (   \tceseq{ \mathit{z}  \,  {:}  \,  \ottnt{T_{{\mathrm{1}}}} }   ) $,
      \item $ \{   \tceseq{ \mathit{X} }   \ottsym{,}   \tceseq{ \mathit{Y} }   \}  \,  \mathbin{\cap}  \,  \mathit{fpv}  (   \tceseq{ \ottnt{T_{{\mathrm{1}}}} }   )  \,  =  \,  \emptyset $,
      \item $ \tceseq{ \ottsym{(}  \mathit{X}  \ottsym{,}  \mathit{Y}  \ottsym{)} }   \ottsym{;}   \tceseq{ \mathit{z_{{\mathrm{0}}}}  \,  {:}  \,  \iota_{{\mathrm{0}}} }   \vdash  \ottnt{C_{{\mathrm{0}}}}  \stackrel{\succ}{\circ}  \ottnt{C}$, and
      \item $ \tceseq{ \ottsym{(}  \mathit{X}  \ottsym{,}  \mathit{Y}  \ottsym{)} }   \ottsym{;}   \tceseq{ \mathit{z_{{\mathrm{0}}}}  \,  {:}  \,  \iota_{{\mathrm{0}}} }  \,  |  \, \ottnt{C}  \vdash  \sigma$.
     \end{itemize}
     It suffices to show that
     \[
      \Gamma  \vdash  \ottsym{(}   \tceseq{ \mathit{z}  \,  {:}  \,  \ottnt{T_{{\mathrm{1}}}} }   \ottsym{)}  \rightarrow  \sigma  \ottsym{(}  \ottnt{C_{{\mathrm{0}}}}  \ottsym{)} ~.
     \]
     By the IH,
     \[
      \Gamma  \ottsym{,}   \tceseq{ \mathit{X} }  \,  {:}  \,  (   \Sigma ^\ast   ,   \tceseq{ \iota_{{\mathrm{0}}} }   )   \ottsym{,}   \tceseq{ \mathit{Y} }  \,  {:}  \,  (   \Sigma ^\omega   ,   \tceseq{ \iota_{{\mathrm{0}}} }   )   \ottsym{,}  \mathit{f} \,  {:}  \, \ottsym{(}   \tceseq{ \mathit{z}  \,  {:}  \,  \ottnt{T_{{\mathrm{1}}}} }   \ottsym{)}  \rightarrow  \ottnt{C_{{\mathrm{0}}}}  \ottsym{,}   \tceseq{ \mathit{z}  \,  {:}  \,  \ottnt{T_{{\mathrm{1}}}} }   \vdash  \ottnt{C} ~.
     \]
     Then, \reflem{ts-wf-tctx-wf-ftyping} implies
     \[
      \Gamma  \ottsym{,}   \tceseq{ \mathit{X} }  \,  {:}  \,  (   \Sigma ^\ast   ,   \tceseq{ \iota_{{\mathrm{0}}} }   )   \ottsym{,}   \tceseq{ \mathit{Y} }  \,  {:}  \,  (   \Sigma ^\omega   ,   \tceseq{ \iota_{{\mathrm{0}}} }   )   \vdash  \ottsym{(}   \tceseq{ \mathit{z}  \,  {:}  \,  \ottnt{T_{{\mathrm{1}}}} }   \ottsym{)}  \rightarrow  \ottnt{C_{{\mathrm{0}}}} ~.
     \]
     Its inversion implies that
     \[
      \Gamma  \ottsym{,}   \tceseq{ \mathit{X} }  \,  {:}  \,  (   \Sigma ^\ast   ,   \tceseq{ \iota_{{\mathrm{0}}} }   )   \ottsym{,}   \tceseq{ \mathit{Y} }  \,  {:}  \,  (   \Sigma ^\omega   ,   \tceseq{ \iota_{{\mathrm{0}}} }   )   \ottsym{,}   \tceseq{ \mathit{z}  \,  {:}  \,  \ottnt{T_{{\mathrm{1}}}} }   \vdash  \ottnt{C_{{\mathrm{0}}}} ~.
     \]
     Then, by Lemmas~\ref{lem:lr-eff-rec-wf-XXX} and \ref{lem:ts-pred-subst-wf-ftyping}, and \WFT{Fun}, we have the conclusion.
    \case \T{Sub}: By inversion.

    \case \T{Op}:
     We are given
     \begin{prooftree}
      \AxiomC{$\vdash  \Gamma$}
      \AxiomC{$ \tceseq{ \Gamma  \vdash  \ottnt{v}  \ottsym{:}  \ottnt{T} } $}
      \AxiomC{$\mathit{ty} \, \ottsym{(}  \ottnt{o}  \ottsym{)} \,  =  \,  \tceseq{(  \mathit{x}  \,  {:}  \,  \ottnt{T}  )}  \rightarrow   \ottnt{T_{{\mathrm{0}}}} $}
      \TrinaryInfC{$\Gamma  \vdash  \ottnt{o}  \ottsym{(}   \tceseq{ \ottnt{v} }   \ottsym{)}  \ottsym{:}   \ottnt{T_{{\mathrm{0}}}}    [ \tceseq{ \ottnt{v}  /  \mathit{x} } ]  $}
     \end{prooftree}
     for some $\ottnt{o}$, $ \tceseq{ \ottnt{v} } $, $ \tceseq{ \mathit{x} } $, $ \tceseq{ \ottnt{T} } $, and $\ottnt{T_{{\mathrm{0}}}}$
     such that
     $\ottnt{e} \,  =  \, \ottnt{o}  \ottsym{(}   \tceseq{ \ottnt{v} }   \ottsym{)}$ and
     $\ottnt{C} \,  =  \,   \ottnt{T_{{\mathrm{0}}}}    [ \tceseq{ \ottnt{v}  /  \mathit{x} } ]    \,  \,\&\,  \,   \Phi_\mathsf{val}   \, / \,   \square  $.
     Without loss of generality, we can suppose that $ \tceseq{ \mathit{x} } $ do not occur in $\Gamma$.
     \refasm{const} states that $\mathit{ty} \, \ottsym{(}  \ottnt{o}  \ottsym{)}$ is closed and well-formed.
     Thus, \reflem{ts-weak-wf-ftyping} with $\vdash  \Gamma$ implies
     $\Gamma  \ottsym{,}   \tceseq{ \mathit{x}  \,  {:}  \,  \ottnt{T} }   \vdash  \ottnt{T_{{\mathrm{0}}}}$.
     Therefore, \reflem{ts-val-subst-wf-ftyping} with $ \tceseq{ \Gamma  \vdash  \ottnt{v}  \ottsym{:}  \ottnt{T} } $ implies
     the conclusion $\Gamma  \vdash   \ottnt{T_{{\mathrm{0}}}}    [ \tceseq{ \ottnt{v}  /  \mathit{x} } ]  $.

    \case \T{App}:
     We are given
     \begin{prooftree}
      \AxiomC{$\Gamma  \vdash  \ottnt{v_{{\mathrm{1}}}}  \ottsym{:}  \ottsym{(}  \mathit{x} \,  {:}  \, \ottnt{T'}  \ottsym{)}  \rightarrow  \ottnt{C'}$}
      \AxiomC{$\Gamma  \vdash  \ottnt{v_{{\mathrm{2}}}}  \ottsym{:}  \ottnt{T'}$}
      \BinaryInfC{$\Gamma  \vdash  \ottnt{v_{{\mathrm{1}}}} \, \ottnt{v_{{\mathrm{2}}}}  \ottsym{:}   \ottnt{C'}    [  \ottnt{v_{{\mathrm{2}}}}  /  \mathit{x}  ]  $}
     \end{prooftree}
     for some $\ottnt{v_{{\mathrm{1}}}}$, $\ottnt{v_{{\mathrm{2}}}}$, $\mathit{x}$, $\ottnt{T'}$, and $\ottnt{C'}$
     such that $\ottnt{e} \,  =  \, \ottnt{v_{{\mathrm{1}}}} \, \ottnt{v_{{\mathrm{2}}}}$ and $\ottnt{C} \,  =  \,  \ottnt{C'}    [  \ottnt{v_{{\mathrm{2}}}}  /  \mathit{x}  ]  $.
     By the IH, $\Gamma  \vdash  \ottsym{(}  \mathit{x} \,  {:}  \, \ottnt{T'}  \ottsym{)}  \rightarrow  \ottnt{C'}$, which implies
     $\Gamma  \ottsym{,}  \mathit{x} \,  {:}  \, \ottnt{T'}  \vdash  \ottnt{C'}$ by inversion.
     Thus, \reflem{ts-val-subst-wf-ftyping} with $\Gamma  \vdash  \ottnt{v_{{\mathrm{2}}}}  \ottsym{:}  \ottnt{T'}$
     implies the conclusion $\Gamma  \vdash   \ottnt{C'}    [  \ottnt{v_{{\mathrm{2}}}}  /  \mathit{x}  ]  $.

    \case \T{PApp}:
     We are given
     \begin{prooftree}
      \AxiomC{$\Gamma  \vdash  \ottnt{v}  \ottsym{:}   \forall   \ottsym{(}  \mathit{X} \,  {:}  \,  \tceseq{ \ottnt{s} }   \ottsym{)}  \ottsym{.}  \ottnt{C'}$}
      \AxiomC{$\Gamma  \vdash  \ottnt{A}  \ottsym{:}   \tceseq{ \ottnt{s} } $}
      \BinaryInfC{$\Gamma  \vdash  \ottnt{v} \, \ottnt{A}  \ottsym{:}   \ottnt{C'}    [  \ottnt{A}  /  \mathit{X}  ]  $}
     \end{prooftree}
     for some $\ottnt{v}$, $\ottnt{A}$, $\ottnt{C'}$, $\mathit{X}$, and $ \tceseq{ \ottnt{s} } $
     such that $\ottnt{e} \,  =  \, \ottnt{v} \, \ottnt{A}$ and $\ottnt{C} \,  =  \,  \ottnt{C'}    [  \ottnt{A}  /  \mathit{X}  ]  $.
     By the IH, $\Gamma  \vdash   \forall   \ottsym{(}  \mathit{X} \,  {:}  \,  \tceseq{ \ottnt{s} }   \ottsym{)}  \ottsym{.}  \ottnt{C'}$, which implies
     $\Gamma  \ottsym{,}  \mathit{X} \,  {:}  \,  \tceseq{ \ottnt{s} }   \vdash  \ottnt{C'}$ by inversion.
     Thus, \reflem{ts-pred-subst-wf-ftyping} with $\Gamma  \vdash  \ottnt{A}  \ottsym{:}   \tceseq{ \ottnt{s} } $
     implies the conclusion $\Gamma  \vdash   \ottnt{C'}    [  \ottnt{A}  /  \mathit{X}  ]  $.

    \case \T{If}:
     We are given
     \begin{prooftree}
      \AxiomC{$\Gamma  \vdash  \ottnt{v}  \ottsym{:}  \ottsym{\{}  \mathit{x} \,  {:}  \,  \mathsf{bool}  \,  |  \, \phi  \ottsym{\}}$}
      \AxiomC{$\Gamma  \ottsym{,}    \ottnt{v}     =     \mathsf{true}    \vdash  \ottnt{e_{{\mathrm{1}}}}  \ottsym{:}  \ottnt{C}$}
      \AxiomC{$\Gamma  \ottsym{,}    \ottnt{v}     =     \mathsf{false}    \vdash  \ottnt{e_{{\mathrm{2}}}}  \ottsym{:}  \ottnt{C}$}
      \TrinaryInfC{$\Gamma  \vdash  \mathsf{if} \, \ottnt{v} \, \mathsf{then} \, \ottnt{e_{{\mathrm{1}}}} \, \mathsf{else} \, \ottnt{e_{{\mathrm{2}}}}  \ottsym{:}  \ottnt{C}$}
     \end{prooftree}
     for some $\ottnt{v}$, $\ottnt{e_{{\mathrm{1}}}}$, $\ottnt{e_{{\mathrm{2}}}}$, $\mathit{x}$, and $\phi$
     such that $\ottnt{e} \,  =  \, \mathsf{if} \, \ottnt{v} \, \mathsf{then} \, \ottnt{e_{{\mathrm{1}}}} \, \mathsf{else} \, \ottnt{e_{{\mathrm{2}}}}$.
     By the IH, $\Gamma  \ottsym{,}    \ottnt{v}     =     \mathsf{true}    \vdash  \ottnt{C}$.
     Then, \reflem{ts-elim-unused-binding-wf-ftyping} implies the conclusion
     $\Gamma  \vdash  \ottnt{C}$.

    \case \T{Let}:
     We are given
     \begin{prooftree}
      \AxiomC{$\Gamma  \vdash  \ottnt{e_{{\mathrm{1}}}}  \ottsym{:}   \ottnt{T_{{\mathrm{1}}}}  \,  \,\&\,  \,  \Phi_{{\mathrm{1}}}  \, / \,  \ottnt{S_{{\mathrm{1}}}} $}
      \AxiomC{$\Gamma  \ottsym{,}  \mathit{x} \,  {:}  \, \ottnt{T_{{\mathrm{1}}}}  \vdash  \ottnt{e_{{\mathrm{2}}}}  \ottsym{:}   \ottnt{T_{{\mathrm{2}}}}  \,  \,\&\,  \,  \Phi_{{\mathrm{2}}}  \, / \,  \ottnt{S_{{\mathrm{2}}}} $}
      \AxiomC{$\mathit{x_{{\mathrm{1}}}} \,  \not\in  \,  \mathit{fv}  (  \ottnt{T_{{\mathrm{2}}}}  )  \,  \mathbin{\cup}  \,  \mathit{fv}  (  \Phi_{{\mathrm{2}}}  ) $}
      \TrinaryInfC{$\Gamma  \vdash  \mathsf{let} \, \mathit{x_{{\mathrm{1}}}}  \ottsym{=}  \ottnt{e_{{\mathrm{1}}}} \,  \mathsf{in}  \, \ottnt{e_{{\mathrm{2}}}}  \ottsym{:}   \ottnt{T_{{\mathrm{2}}}}  \,  \,\&\,  \,  \Phi_{{\mathrm{1}}} \,  \tceappendop  \, \Phi_{{\mathrm{2}}}  \, / \,  \ottsym{(}  \ottnt{S_{{\mathrm{1}}}}  \gg\!\!=  \ottsym{(}   \lambda\!  \, \mathit{x_{{\mathrm{1}}}}  \ottsym{.}  \ottnt{S_{{\mathrm{2}}}}  \ottsym{)}  \ottsym{)} $}
     \end{prooftree}
     for some $\mathit{x_{{\mathrm{1}}}}$, $\ottnt{e_{{\mathrm{1}}}}$, $\ottnt{e_{{\mathrm{2}}}}$, $\ottnt{T_{{\mathrm{1}}}}$, $\ottnt{T_{{\mathrm{2}}}}$, $\Phi_{{\mathrm{1}}}$, $\Phi_{{\mathrm{2}}}$, $\ottnt{S_{{\mathrm{1}}}}$, and $\ottnt{S_{{\mathrm{2}}}}$
     such that
     \begin{itemize}
      \item $\ottnt{e} \,  =  \, \ottsym{(}  \mathsf{let} \, \mathit{x_{{\mathrm{1}}}}  \ottsym{=}  \ottnt{e_{{\mathrm{1}}}} \,  \mathsf{in}  \, \ottnt{e_{{\mathrm{2}}}}  \ottsym{)}$ and
      \item $\ottnt{C} \,  =  \,  \ottnt{T_{{\mathrm{2}}}}  \,  \,\&\,  \,  \Phi_{{\mathrm{1}}} \,  \tceappendop  \, \Phi_{{\mathrm{2}}}  \, / \,  \ottsym{(}  \ottnt{S_{{\mathrm{1}}}}  \gg\!\!=  \ottsym{(}   \lambda\!  \, \mathit{x_{{\mathrm{1}}}}  \ottsym{.}  \ottnt{S_{{\mathrm{2}}}}  \ottsym{)}  \ottsym{)} $.
     \end{itemize}

     By the IHs,
     \begin{equation}
      \Gamma  \vdash   \ottnt{T_{{\mathrm{1}}}}  \,  \,\&\,  \,  \Phi_{{\mathrm{1}}}  \, / \,  \ottnt{S_{{\mathrm{1}}}} 
       \label{lem:ts-wf-ty-tctx-typing:let:IH1}
     \end{equation}
     and
     \begin{equation}
      \Gamma  \ottsym{,}  \mathit{x_{{\mathrm{1}}}} \,  {:}  \, \ottnt{T_{{\mathrm{1}}}}  \vdash   \ottnt{T_{{\mathrm{2}}}}  \,  \,\&\,  \,  \Phi_{{\mathrm{2}}}  \, / \,  \ottnt{S_{{\mathrm{2}}}} 
       \label{lem:ts-wf-ty-tctx-typing:let:IH2} ~.
     \end{equation}
     By inversion of (\ref{lem:ts-wf-ty-tctx-typing:let:IH2}), and
     \reflem{ts-elim-unused-binding-wf-ftyping} with $\mathit{x_{{\mathrm{1}}}} \,  \not\in  \,  \mathit{fv}  (  \ottnt{T_{{\mathrm{2}}}}  )  \,  \mathbin{\cup}  \,  \mathit{fv}  (  \Phi_{{\mathrm{2}}}  ) $,
     we have $\Gamma  \vdash  \ottnt{T_{{\mathrm{2}}}}$ and $\Gamma  \vdash  \Phi_{{\mathrm{2}}}$.
     Further, by inversion of (\ref{lem:ts-wf-ty-tctx-typing:let:IH1}),
     we have $\Gamma  \vdash  \Phi_{{\mathrm{1}}}$.
     Thus, \reflem{ts-wf-eff-compose} implies $\Gamma  \vdash  \Phi_{{\mathrm{1}}} \,  \tceappendop  \, \Phi_{{\mathrm{2}}}$.
     Then, by \WFT{Comp}, it suffices to show that
     \[
      \Gamma \,  |  \, \ottnt{T_{{\mathrm{2}}}}  \vdash  \ottnt{S_{{\mathrm{1}}}}  \gg\!\!=  \ottsym{(}   \lambda\!  \, \mathit{x_{{\mathrm{1}}}}  \ottsym{.}  \ottnt{S_{{\mathrm{2}}}}  \ottsym{)} ~.
     \]
     By case analysis on $\ottnt{S_{{\mathrm{1}}}}$.
     \begin{caseanalysis}
      \case $\ottnt{S_{{\mathrm{1}}}} \,  =  \,  \square $:
       The definition of $\ottnt{S_{{\mathrm{1}}}}  \gg\!\!=  \ottsym{(}   \lambda\!  \, \mathit{x_{{\mathrm{1}}}}  \ottsym{.}  \ottnt{S_{{\mathrm{2}}}}  \ottsym{)}$ implies
       $\ottnt{S_{{\mathrm{2}}}} \,  =  \,  \square $ and $\ottnt{S_{{\mathrm{1}}}}  \gg\!\!=  \ottsym{(}   \lambda\!  \, \mathit{x_{{\mathrm{1}}}}  \ottsym{.}  \ottnt{S_{{\mathrm{2}}}}  \ottsym{)} \,  =  \,  \square $.
       Thus, it suffices to show that
       \[
        \Gamma \,  |  \, \ottnt{T_{{\mathrm{2}}}}  \vdash   \square  ~.
       \]
       It is derived by \WFT{Empty} with $\vdash  \Gamma$, which is implied
       by \reflem{ts-wf-tctx-wf-ftyping} with (\ref{lem:ts-wf-ty-tctx-typing:let:IH1}).

      \case $  \exists  \, { \ottnt{C'_{{\mathrm{1}}}}  \ottsym{,}  \ottnt{C'} } . \  \ottnt{S_{{\mathrm{1}}}} \,  =  \,  ( \forall  \mathit{x_{{\mathrm{1}}}}  .  \ottnt{C'_{{\mathrm{1}}}}  )  \Rightarrow   \ottnt{C'}  $:
       The definition of $\ottnt{S_{{\mathrm{1}}}}  \gg\!\!=  \ottsym{(}   \lambda\!  \, \mathit{x_{{\mathrm{1}}}}  \ottsym{.}  \ottnt{S_{{\mathrm{2}}}}  \ottsym{)}$ implies
       \begin{itemize}
        \item $\ottnt{S_{{\mathrm{2}}}} \,  =  \,  ( \forall  \mathit{x_{{\mathrm{2}}}}  .  \ottnt{C'_{{\mathrm{2}}}}  )  \Rightarrow   \ottnt{C'_{{\mathrm{1}}}} $ and
        \item $\ottnt{S_{{\mathrm{1}}}}  \gg\!\!=  \ottsym{(}   \lambda\!  \, \mathit{x_{{\mathrm{1}}}}  \ottsym{.}  \ottnt{S_{{\mathrm{2}}}}  \ottsym{)} \,  =  \,  ( \forall  \mathit{x_{{\mathrm{2}}}}  .  \ottnt{C'_{{\mathrm{2}}}}  )  \Rightarrow   \ottnt{C'} $
       \end{itemize}
       for some $\mathit{x_{{\mathrm{2}}}}$ and $\ottnt{C'_{{\mathrm{2}}}}$ such that
       $\mathit{x_{{\mathrm{1}}}} \,  \not\in  \,  \mathit{fv}  (  \ottnt{C'_{{\mathrm{2}}}}  )  \,  \mathbin{\backslash}  \,  \{  \mathit{x_{{\mathrm{2}}}}  \} $.
       Without loss of generality, we can suppose that
       $ \mathit{x_{{\mathrm{1}}}}    \not=    \mathit{x_{{\mathrm{2}}}} $.  Thus, $\mathit{x_{{\mathrm{1}}}} \,  \not\in  \,  \mathit{fv}  (  \ottnt{C'_{{\mathrm{2}}}}  ) $.
       Then, it suffices to show that
       \[
        \Gamma \,  |  \, \ottnt{T_{{\mathrm{2}}}}  \vdash   ( \forall  \mathit{x_{{\mathrm{2}}}}  .  \ottnt{C'_{{\mathrm{2}}}}  )  \Rightarrow   \ottnt{C'}  ~,
       \]
       which is proven by \WFT{Ans} with the following.
       \begin{itemize}
        \item We show that $\Gamma  \ottsym{,}  \mathit{x_{{\mathrm{2}}}} \,  {:}  \, \ottnt{T_{{\mathrm{2}}}}  \vdash  \ottnt{C'_{{\mathrm{2}}}}$.
              By inversion of (\ref{lem:ts-wf-ty-tctx-typing:let:IH2}),
              we have $\Gamma  \ottsym{,}  \mathit{x_{{\mathrm{1}}}} \,  {:}  \, \ottnt{T_{{\mathrm{1}}}} \,  |  \, \ottnt{T_{{\mathrm{2}}}}  \vdash  \ottnt{S_{{\mathrm{2}}}}$, i.e.,
              $\Gamma  \ottsym{,}  \mathit{x_{{\mathrm{1}}}} \,  {:}  \, \ottnt{T_{{\mathrm{1}}}} \,  |  \, \ottnt{T_{{\mathrm{2}}}}  \vdash   ( \forall  \mathit{x_{{\mathrm{2}}}}  .  \ottnt{C'_{{\mathrm{2}}}}  )  \Rightarrow   \ottnt{C'_{{\mathrm{1}}}} $.
              By its inversion, we have $\Gamma  \ottsym{,}  \mathit{x_{{\mathrm{1}}}} \,  {:}  \, \ottnt{T_{{\mathrm{1}}}}  \ottsym{,}  \mathit{x_{{\mathrm{2}}}} \,  {:}  \, \ottnt{T_{{\mathrm{2}}}}  \vdash  \ottnt{C'_{{\mathrm{2}}}}$.
              Because $\mathit{x_{{\mathrm{1}}}} \,  \not\in  \,  \mathit{fv}  (  \ottnt{T_{{\mathrm{2}}}}  )  \,  \mathbin{\cup}  \,  \mathit{fv}  (  \ottnt{C'_{{\mathrm{2}}}}  ) $,
              \reflem{ts-elim-unused-binding-wf-ftyping} implies $\Gamma  \ottsym{,}  \mathit{x_{{\mathrm{2}}}} \,  {:}  \, \ottnt{T_{{\mathrm{2}}}}  \vdash  \ottnt{C'_{{\mathrm{2}}}}$.
        \item We show that $\Gamma  \vdash  \ottnt{C'}$.
              By inversion of (\ref{lem:ts-wf-ty-tctx-typing:let:IH1}),
              we have $\Gamma \,  |  \, \ottnt{T_{{\mathrm{1}}}}  \vdash  \ottnt{S_{{\mathrm{1}}}}$, i.e.,
              $\Gamma \,  |  \, \ottnt{T_{{\mathrm{1}}}}  \vdash   ( \forall  \mathit{x_{{\mathrm{1}}}}  .  \ottnt{C'_{{\mathrm{1}}}}  )  \Rightarrow   \ottnt{C'} $.
              By its inversion, $\Gamma  \vdash  \ottnt{C'}$.
       \end{itemize}
     \end{caseanalysis}

    \case \T{LetV}: By the IH and \reflem{ts-val-subst-wf-ftyping}.

    \case \T{Shift}:
     We are given
     \begin{prooftree}
      \AxiomC{$\Gamma  \ottsym{,}  \mathit{x} \,  {:}  \, \ottsym{(}  \mathit{y} \,  {:}  \, \ottnt{T}  \ottsym{)}  \rightarrow  \ottnt{C_{{\mathrm{1}}}}  \vdash  \ottnt{e'}  \ottsym{:}  \ottnt{C_{{\mathrm{2}}}}$}
      \UnaryInfC{$\Gamma  \vdash  \mathcal{S}_0 \, \mathit{x}  \ottsym{.}  \ottnt{e'}  \ottsym{:}   \ottnt{T}  \,  \,\&\,  \,   \Phi_\mathsf{val}   \, / \,   ( \forall  \mathit{y}  .  \ottnt{C_{{\mathrm{1}}}}  )  \Rightarrow   \ottnt{C_{{\mathrm{2}}}}  $}
     \end{prooftree}
     for some $\mathit{x}$, $\mathit{y}$, $\ottnt{e'}$, $\ottnt{T}$, $\ottnt{C_{{\mathrm{1}}}}$, and $\ottnt{C_{{\mathrm{2}}}}$
     such that
     \begin{itemize}
      \item $\ottnt{e} \,  =  \, \mathcal{S}_0 \, \mathit{x}  \ottsym{.}  \ottnt{e'}$ and
      \item $\ottnt{C} \,  =  \,  \ottnt{T}  \,  \,\&\,  \,   \Phi_\mathsf{val}   \, / \,   ( \forall  \mathit{y}  .  \ottnt{C_{{\mathrm{1}}}}  )  \Rightarrow   \ottnt{C_{{\mathrm{2}}}}  $.
     \end{itemize}

     By the IH, $\Gamma  \ottsym{,}  \mathit{x} \,  {:}  \, \ottsym{(}  \mathit{y} \,  {:}  \, \ottnt{T}  \ottsym{)}  \rightarrow  \ottnt{C_{{\mathrm{1}}}}  \vdash  \ottnt{C_{{\mathrm{2}}}}$.
     By \reflem{ts-wf-tctx-wf-ftyping}, $\vdash  \Gamma  \ottsym{,}  \mathit{x} \,  {:}  \, \ottsym{(}  \mathit{y} \,  {:}  \, \ottnt{T}  \ottsym{)}  \rightarrow  \ottnt{C_{{\mathrm{1}}}}$, which implies
     $\Gamma  \vdash  \ottsym{(}  \mathit{y} \,  {:}  \, \ottnt{T}  \ottsym{)}  \rightarrow  \ottnt{C_{{\mathrm{1}}}}$.  By its inversion, $\Gamma  \ottsym{,}  \mathit{y} \,  {:}  \, \ottnt{T}  \vdash  \ottnt{C_{{\mathrm{1}}}}$.
     Again, by \reflem{ts-wf-tctx-wf-ftyping}, $\vdash  \Gamma  \ottsym{,}  \mathit{y} \,  {:}  \, \ottnt{T}$, so
     $\vdash  \Gamma$ and $\Gamma  \vdash  \ottnt{T}$.
     Thus, \reflem{ts-wf-TP_val} implies $\Gamma  \vdash   \Phi_\mathsf{val} $.
     Then, by \WFT{Comp}, it suffices to show that
     \[
      \Gamma \,  |  \, \ottnt{T}  \vdash   ( \forall  \mathit{y}  .  \ottnt{C_{{\mathrm{1}}}}  )  \Rightarrow   \ottnt{C_{{\mathrm{2}}}}  ~,
     \]
     which is proven by \WFT{Ans} with the following.
     \begin{itemize}
      \item $\Gamma  \ottsym{,}  \mathit{y} \,  {:}  \, \ottnt{T}  \vdash  \ottnt{C_{{\mathrm{1}}}}$, which has been obtained.
      \item $\Gamma  \vdash  \ottnt{C_{{\mathrm{2}}}}$, which is proven by
            Lemmas~\ref{lem:ts-unuse-non-refine-vars} and \ref{lem:ts-elim-unused-binding-wf-ftyping}
            with $\Gamma  \ottsym{,}  \mathit{x} \,  {:}  \, \ottsym{(}  \mathit{y} \,  {:}  \, \ottnt{T}  \ottsym{)}  \rightarrow  \ottnt{C_{{\mathrm{1}}}}  \vdash  \ottnt{C_{{\mathrm{2}}}}$ (which is the result of applying the IH).
     \end{itemize}

    \case \T{ShiftD}: By inversion.

    \case \T{Reset}:
     We are given
     \begin{prooftree}
      \AxiomC{$\Gamma  \vdash  \ottnt{e'}  \ottsym{:}   \ottnt{T}  \,  \,\&\,  \,  \Phi  \, / \,   ( \forall  \mathit{x}  .   \ottnt{T}  \,  \,\&\,  \,   \Phi_\mathsf{val}   \, / \,   \square    )  \Rightarrow   \ottnt{C}  $}
      \AxiomC{$\mathit{x} \,  \not\in  \,  \mathit{fv}  (  \ottnt{T}  ) $}
      \BinaryInfC{$\Gamma  \vdash   \resetcnstr{ \ottnt{e'} }   \ottsym{:}  \ottnt{C}$}
     \end{prooftree}
     for some $\mathit{x}$, $\ottnt{e'}$, $\ottnt{T}$, and $\Phi$
     such that $\ottnt{e} \,  =  \,  \resetcnstr{ \ottnt{e'} } $.
     By the IH, $\Gamma  \vdash   \ottnt{T}  \,  \,\&\,  \,  \Phi  \, / \,   ( \forall  \mathit{x}  .   \ottnt{T}  \,  \,\&\,  \,   \Phi_\mathsf{val}   \, / \,   \square    )  \Rightarrow   \ottnt{C}  $,
     which implies $\Gamma \,  |  \, \ottnt{T}  \vdash   ( \forall  \mathit{x}  .   \ottnt{T}  \,  \,\&\,  \,   \Phi_\mathsf{val}   \, / \,   \square    )  \Rightarrow   \ottnt{C} $.
     Further, its inversion implies the conclusion $\Gamma  \vdash  \ottnt{C}$.

    \case \T{Event}:
     We are given
     \begin{prooftree}
      \AxiomC{$\vdash  \Gamma$}
      \UnaryInfC{$\Gamma  \vdash   \mathsf{ev}  [  \mathbf{a}  ]   \ottsym{:}   \ottsym{\{}  \mathit{z} \,  {:}  \,  \mathsf{unit}  \,  |  \,  \top   \ottsym{\}}  \,  \,\&\,  \,   (   \lambda_\mu\!   \,  \mathit{x}  .\,   \mathit{x}    =    \mathbf{a}   ,   \lambda_\nu\!   \,  \mathit{y}  .\,   \bot   )   \, / \,   \square  $}
     \end{prooftree}
     for some $\mathit{x}$, $\mathit{y}$, $\mathit{z}$, and $\mathbf{a}$
     such that $\ottnt{e} \,  =  \,  \mathsf{ev}  [  \mathbf{a}  ] $ and $\ottnt{C} \,  =  \,  \ottsym{\{}  \mathit{z} \,  {:}  \,  \mathsf{unit}  \,  |  \,  \top   \ottsym{\}}  \,  \,\&\,  \,   (   \lambda_\mu\!   \,  \mathit{x}  .\,   \mathit{x}    =    \mathbf{a}   ,   \lambda_\nu\!   \,  \mathit{y}  .\,   \bot   )   \, / \,   \square  $.
     Because $ \emptyset   \vdash  \ottnt{C}$ and $\vdash  \Gamma$, we have the conclusion $\Gamma  \vdash  \ottnt{C}$
     by \reflem{ts-weak-wf-ftyping}.

    \case \T{Div}: By inversion.
   \end{caseanalysis}
\end{proof}

\begin{lemmap}{Subtyping of Pure Effects}{ts-refl-subtyping-pure-eff}
 If $\vdash  \Gamma$ and $\Gamma  \vdash  \ottnt{T_{{\mathrm{1}}}}  \ottsym{<:}  \ottnt{T_{{\mathrm{2}}}}$,
 then $\Gamma  \vdash   \ottnt{T_{{\mathrm{1}}}}  \,  \,\&\,  \,   \Phi_\mathsf{val}   \, / \,   \square    \ottsym{<:}   \ottnt{T_{{\mathrm{2}}}}  \,  \,\&\,  \,   \Phi_\mathsf{val}   \, / \,   \square  $.
\end{lemmap}
\begin{proof}
 \reflem{ts-wf-TP_val} with the assumption $\vdash  \Gamma$ implies
 $\Gamma  \vdash   \Phi_\mathsf{val} $.
 Thus, \reflem{ts-refl-subtyping} implies $\Gamma  \vdash   \Phi_\mathsf{val}   \ottsym{<:}   \Phi_\mathsf{val} $.
 Then, we can derive the conclusion as follows:
 \begin{prooftree}
  \AxiomC{$\Gamma  \vdash  \ottnt{T_{{\mathrm{1}}}}  \ottsym{<:}  \ottnt{T_{{\mathrm{2}}}}$}
  \AxiomC{$\Gamma  \vdash   \Phi_\mathsf{val}   \ottsym{<:}   \Phi_\mathsf{val} $}
  \AxiomC{}
  \RightLabel{\Srule{Empty}}
  \UnaryInfC{$\Gamma \,  |  \, \ottnt{T_{{\mathrm{1}}}} \,  \,\&\,  \,  \Phi_\mathsf{val}   \vdash   \square   \ottsym{<:}   \square $}
  \TrinaryInfC{$\Gamma  \vdash   \ottnt{T_{{\mathrm{1}}}}  \,  \,\&\,  \,   \Phi_\mathsf{val}   \, / \,   \square    \ottsym{<:}   \ottnt{T_{{\mathrm{2}}}}  \,  \,\&\,  \,   \Phi_\mathsf{val}   \, / \,   \square  $}
 \end{prooftree}
\end{proof}


\begin{lemmap}{Extracting Substituted Values in Validity}{ts-extract-subst-val-valid}
 \begin{enumerate}
  \item If $ { \models  \,   \phi    [  \ottnt{t}  /  \mathit{x}  ]   } $ and $ \emptyset   \vdash  \ottnt{t}  \ottsym{:}  \ottnt{s}$,
        then $\mathit{x} \,  {:}  \, \ottnt{s}  \models   \ottsym{(}   \ottnt{t}    =    \mathit{x}   \ottsym{)}    \Rightarrow    \phi $.
  \item If $ { \models  \,   \phi    [  \ottnt{c}  /  \mathit{x}  ]   } $ and $ \mathit{ty}  (  \ottnt{c}  )  \,  =  \, \iota$,
        then $\mathit{x} \,  {:}  \, \iota  \models   \ottsym{(}   \ottnt{c}    =    \mathit{x}   \ottsym{)}    \Rightarrow    \phi $.
 \end{enumerate}
\end{lemmap}
\begin{proof}
 \noindent
 \begin{enumerate}
  \item \TS{Validity}
        Let $\ottnt{d} \,  \in  \,  \mathcal{D}_{ \ottnt{s} } $.
        It suffices to show that $ { [\![   \ottsym{(}   \ottnt{t}    =    \mathit{x}   \ottsym{)}    \Rightarrow    \phi   ]\!]_{  \{  \mathit{x}  \mapsto  \ottnt{d}  \}  } }  \,  =  \,  \top $.
        If $ { [\![  \ottnt{t}  ]\!]_{  \{  \mathit{x}  \mapsto  \ottnt{d}  \}  } }  \,  \not=  \, \ottnt{d}$,
        then it holds obviously.
        Otherwise, suppose that $ { [\![  \ottnt{t}  ]\!]_{  \{  \mathit{x}  \mapsto  \ottnt{d}  \}  } }  \,  =  \, \ottnt{d}$.
        We must show that $ { [\![  \phi  ]\!]_{  \{  \mathit{x}  \mapsto  \ottnt{d}  \}  } }  \,  =  \,  \top $.
        Because $ { \models  \,   \phi    [  \ottnt{t}  /  \mathit{x}  ]   } $,
        \reflem{ts-logic-val-subst} implies
        $ { [\![  \phi  ]\!]_{  \{  \mathit{x}  \mapsto   { [\![  \ottnt{t}  ]\!]_{  \emptyset  } }   \}  } }  \,  =  \,  \top $.
        Because $ \emptyset   \vdash  \ottnt{t}  \ottsym{:}  \ottnt{s}$ implies that $\mathit{x}$ does not occur free in $\ottnt{t}$,
        $ { [\![  \ottnt{t}  ]\!]_{  \{  \mathit{x}  \mapsto  \ottnt{d}  \}  } }  \,  =  \, \ottnt{d}$ implies $ { [\![  \ottnt{t}  ]\!]_{  \emptyset  } }  \,  =  \, \ottnt{d}$.
        Therefore, we have the conclusion $ { [\![  \phi  ]\!]_{  \{  \mathit{x}  \mapsto  \ottnt{d}  \}  } }  \,  =  \,  \top $.
  \item By the first case; note that
        $ \emptyset   \vdash  \ottnt{c}  \ottsym{:}  \iota$ if $ \mathit{ty}  (  \ottnt{c}  )  \,  =  \, \iota$ by \refasm{termfun}.
 \end{enumerate}
\end{proof}

\begin{lemmap}{Value Inversion of Refinement Types}{ts-val-inv-refine-ty}
 If $ \emptyset   \vdash  \ottnt{v}  \ottsym{:}   \ottsym{\{}  \mathit{x} \,  {:}  \, \iota \,  |  \, \phi  \ottsym{\}}  \,  \,\&\,  \,  \Phi  \, / \,  \ottnt{S} $,
 then there exists some $\ottnt{c}$ such that
 \begin{itemize}
  \item $\ottnt{v} \,  =  \, \ottnt{c}$
  \item $ \mathit{ty}  (  \ottnt{c}  )  \,  =  \, \iota$,
  \item $ { \models  \,   \phi    [  \ottnt{c}  /  \mathit{x}  ]   } $,
  \item $ \emptyset   \vdash  \ottnt{v}  \ottsym{:}  \ottsym{\{}  \mathit{x} \,  {:}  \, \iota \,  |  \,  \ottnt{c}    =    \mathit{x}   \ottsym{\}}$, and
  \item $ \emptyset   \vdash   \ottsym{\{}  \mathit{x} \,  {:}  \, \iota \,  |  \,  \ottnt{c}    =    \mathit{x}   \ottsym{\}}  \,  \,\&\,  \,   \Phi_\mathsf{val}   \, / \,   \square    \ottsym{<:}   \ottsym{\{}  \mathit{x} \,  {:}  \, \iota \,  |  \, \phi  \ottsym{\}}  \,  \,\&\,  \,  \Phi  \, / \,  \ottnt{S} $.
 \end{itemize}
\end{lemmap}
\begin{proof}
 By induction on the typing derivation.
 \begin{caseanalysis}
  \case \T{Const}:
   We are given
   \begin{prooftree}
    \AxiomC{$\vdash   \emptyset $}
    \AxiomC{$ \mathit{ty}  (  \ottnt{c}  )  \,  =  \, \iota$}
    \BinaryInfC{$ \emptyset   \vdash  \ottnt{c}  \ottsym{:}  \ottsym{\{}  \mathit{x} \,  {:}  \, \iota \,  |  \,  \ottnt{c}    =    \mathit{x}   \ottsym{\}}$}
   \end{prooftree}
   for some $\ottnt{c}$ such that $\ottnt{v} \,  =  \, \ottnt{c}$ and $\phi \,  =  \, \ottsym{(}   \ottnt{c}    =    \mathit{x}   \ottsym{)}$.
   Because $ { \models  \,   \ottnt{c}    =    \ottnt{c}  } $ by \reflem{ts-refl-const-eq},
   we have $ { \models  \,   \phi    [  \ottnt{c}  /  \mathit{x}  ]   } $.
   Now, it suffices to show that
   \[
     \emptyset   \vdash   \ottsym{\{}  \mathit{x} \,  {:}  \, \iota \,  |  \,  \ottnt{c}    =    \mathit{x}   \ottsym{\}}  \,  \,\&\,  \,   \Phi_\mathsf{val}   \, / \,   \square    \ottsym{<:}   \ottsym{\{}  \mathit{x} \,  {:}  \, \iota \,  |  \,  \ottnt{c}    =    \mathit{x}   \ottsym{\}}  \,  \,\&\,  \,   \Phi_\mathsf{val}   \, / \,   \square   ~.
   \]
   Because $ \emptyset   \vdash   \ottsym{\{}  \mathit{x} \,  {:}  \, \iota \,  |  \,  \ottnt{c}    =    \mathit{x}   \ottsym{\}}  \,  \,\&\,  \,   \Phi_\mathsf{val}   \, / \,   \square  $ by \reflem{ts-wf-TP_val},
   we have the conclusion by \reflem{ts-refl-subtyping}.

  \case \T{Sub}:
   We are given
   \begin{prooftree}
    \AxiomC{$ \emptyset   \vdash  \ottnt{v}  \ottsym{:}  \ottnt{C}$}
    \AxiomC{$ \emptyset   \vdash  \ottnt{C}  \ottsym{<:}   \ottsym{\{}  \mathit{x} \,  {:}  \, \iota \,  |  \, \phi  \ottsym{\}}  \,  \,\&\,  \,  \Phi  \, / \,  \ottnt{S} $}
    \AxiomC{$ \emptyset   \vdash   \ottsym{\{}  \mathit{x} \,  {:}  \, \iota \,  |  \, \phi  \ottsym{\}}  \,  \,\&\,  \,  \Phi  \, / \,  \ottnt{S} $}
    \TrinaryInfC{$ \emptyset   \vdash  \ottnt{v}  \ottsym{:}   \ottsym{\{}  \mathit{x} \,  {:}  \, \iota \,  |  \, \phi  \ottsym{\}}  \,  \,\&\,  \,  \Phi  \, / \,  \ottnt{S} $}
   \end{prooftree}
   for some $\ottnt{C}$.
   Because $ \emptyset   \vdash  \ottnt{C}  \ottsym{<:}   \ottsym{\{}  \mathit{x} \,  {:}  \, \iota \,  |  \, \phi  \ottsym{\}}  \,  \,\&\,  \,  \Phi  \, / \,  \ottnt{S} $ can be derived only by
   \Srule{Comp}, and $ \emptyset   \vdash   \ottnt{C}  . \ottnt{T}   \ottsym{<:}  \ottsym{\{}  \mathit{x} \,  {:}  \, \iota \,  |  \, \phi  \ottsym{\}}$ can be only by \Srule{Refine},
   its inversion implies $ \ottnt{C}  . \ottnt{T}  \,  =  \, \ottsym{\{}  \mathit{x} \,  {:}  \, \iota \,  |  \, \phi'  \ottsym{\}}$ for some $\phi'$ such that
   $\mathit{x} \,  {:}  \, \iota  \models   \phi'    \Rightarrow    \phi $.
   Because $ \emptyset   \vdash  \ottnt{v}  \ottsym{:}  \ottnt{C}$ means $ \emptyset   \vdash  \ottnt{v}  \ottsym{:}   \ottsym{\{}  \mathit{x} \,  {:}  \, \iota \,  |  \, \phi'  \ottsym{\}}  \,  \,\&\,  \,   \ottnt{C}  . \Phi   \, / \,   \ottnt{C}  . \ottnt{S}  $,
   the IH implies
   \begin{itemize}
    \item $\ottnt{v} \,  =  \, \ottnt{c}$,
    \item $ \mathit{ty}  (  \ottnt{c}  )  \,  =  \, \iota$,
    \item $ { \models  \,   \phi'    [  \ottnt{c}  /  \mathit{x}  ]   } $,
    \item $ \emptyset   \vdash  \ottnt{v}  \ottsym{:}  \ottsym{\{}  \mathit{x} \,  {:}  \, \iota \,  |  \,  \ottnt{c}    =    \mathit{x}   \ottsym{\}}$, and
    \item $ \emptyset   \vdash   \ottsym{\{}  \mathit{x} \,  {:}  \, \iota \,  |  \,  \ottnt{c}    =    \mathit{x}   \ottsym{\}}  \,  \,\&\,  \,   \Phi_\mathsf{val}   \, / \,   \square    \ottsym{<:}   \ottsym{\{}  \mathit{x} \,  {:}  \, \iota \,  |  \, \phi'  \ottsym{\}}  \,  \,\&\,  \,   \ottnt{C}  . \Phi   \, / \,   \ottnt{C}  . \ottnt{S}  $
   \end{itemize}
   for some $\ottnt{c}$.
   Then, it suffices to show the following.
   \begin{itemize}
    \item We show that $ { \models  \,   \phi    [  \ottnt{c}  /  \mathit{x}  ]   } $.
          Because $ \mathit{ty}  (  \ottnt{c}  )  \,  =  \, \iota$ implies $ \emptyset   \vdash  \ottnt{c}  \ottsym{:}  \iota$ by \refasm{termfun},
          \reflem{ts-term-subst-valid} with $\mathit{x} \,  {:}  \, \iota  \models   \phi'    \Rightarrow    \phi $
          implies $ { \models  \,     \phi'    [  \ottnt{c}  /  \mathit{x}  ]      \Rightarrow    \phi     [  \ottnt{c}  /  \mathit{x}  ]   } $.
          Because $ { \models  \,   \phi'    [  \ottnt{c}  /  \mathit{x}  ]   } $,
          \reflem{ts-discharge-assumption-valid} implies
          $ { \models  \,   \phi    [  \ottnt{c}  /  \mathit{x}  ]   } $.

    \item We have $ \emptyset   \vdash   \ottsym{\{}  \mathit{x} \,  {:}  \, \iota \,  |  \,  \ottnt{c}    =    \mathit{x}   \ottsym{\}}  \,  \,\&\,  \,   \Phi_\mathsf{val}   \, / \,   \square    \ottsym{<:}   \ottsym{\{}  \mathit{x} \,  {:}  \, \iota \,  |  \, \phi  \ottsym{\}}  \,  \,\&\,  \,  \Phi  \, / \,  \ottnt{S} $
          by \reflem{ts-transitivity-subtyping} with
          \begin{itemize}
           \item $ \emptyset   \vdash   \ottsym{\{}  \mathit{x} \,  {:}  \, \iota \,  |  \,  \ottnt{c}    =    \mathit{x}   \ottsym{\}}  \,  \,\&\,  \,   \Phi_\mathsf{val}   \, / \,   \square    \ottsym{<:}   \ottsym{\{}  \mathit{x} \,  {:}  \, \iota \,  |  \, \phi'  \ottsym{\}}  \,  \,\&\,  \,   \ottnt{C}  . \Phi   \, / \,   \ottnt{C}  . \ottnt{S}  $,
           \item $ \emptyset   \vdash  \ottnt{C}  \ottsym{<:}   \ottsym{\{}  \mathit{x} \,  {:}  \, \iota \,  |  \, \phi  \ottsym{\}}  \,  \,\&\,  \,  \Phi  \, / \,  \ottnt{S} $, and
           \item $\ottnt{C} \,  =  \,  \ottsym{\{}  \mathit{x} \,  {:}  \, \iota \,  |  \, \phi'  \ottsym{\}}  \,  \,\&\,  \,   \ottnt{C}  . \Phi   \, / \,   \ottnt{C}  . \ottnt{S}  $.
          \end{itemize}
   \end{itemize}

  \case \T{App}: Contradictory by \reflem{ts-val-applied-fun}.

  \otherwise: Contradictory.
 \end{caseanalysis}
\end{proof}

\begin{lemmap}{Value Inversion of Function Types}{ts-val-inv-fun-ty}
 If $ \emptyset   \vdash  \ottnt{v}  \ottsym{:}   \ottsym{(}  \ottsym{(}  \mathit{x} \,  {:}  \, \ottnt{T}  \ottsym{)}  \rightarrow  \ottnt{C}  \ottsym{)}  \,  \,\&\,  \,  \Phi  \, / \,  \ottnt{S} $, then
 there exist some $\ottnt{v_{{\mathrm{1}}}}$, $ \tceseq{ \ottnt{v_{{\mathrm{2}}}} } $, and $ \tceseq{ \mathit{z} } $ such that:
 \begin{itemize}
  \item $\ottnt{v} \,  =  \, \ottnt{v_{{\mathrm{1}}}} \,  \tceseq{ \ottnt{v_{{\mathrm{2}}}} } $;
  \item $  \exists  \, { \ottnt{e} } . \  \ottnt{v_{{\mathrm{1}}}} \,  =  \,  \lambda\!  \,  \tceseq{ \mathit{z} }   \ottsym{.}  \ottnt{e} $ or $  \exists  \, {  \tceseq{ \mathit{X} }   \ottsym{,}   \tceseq{ \mathit{Y} }   \ottsym{,}   \tceseq{  \ottnt{A} _{  \mu  }  }   \ottsym{,}   \tceseq{  \ottnt{A} _{  \nu  }  }   \ottsym{,}  \mathit{f}  \ottsym{,}  \ottnt{e} } . \  \ottnt{v_{{\mathrm{1}}}} \,  =  \,  \mathsf{rec}  \, (  \tceseq{  \mathit{X} ^{  \ottnt{A} _{  \mu  }  },  \mathit{Y} ^{  \ottnt{A} _{  \nu  }  }  }  ,  \mathit{f} ,   \tceseq{ \mathit{z} }  ) . \,  \ottnt{e}  $; and
  \item $\ottsym{\mbox{$\mid$}}   \tceseq{ \ottnt{v_{{\mathrm{2}}}} }   \ottsym{\mbox{$\mid$}} \,  <  \, \ottsym{\mbox{$\mid$}}   \tceseq{ \mathit{z} }   \ottsym{\mbox{$\mid$}}$.
 \end{itemize}
\end{lemmap}
\begin{proof}
 By induction on the typing derivation.
 \begin{caseanalysis}
  \case \T{Abs} and \T{Fun}: Obvious.
  \case \T{App}: Obvious by definition.
  \case \T{Sub}:
   We are given
   \begin{prooftree}
    \AxiomC{$ \emptyset   \vdash  \ottnt{v}  \ottsym{:}  \ottnt{C'}$}
    \AxiomC{$ \emptyset   \vdash  \ottnt{C'}  \ottsym{<:}   \ottsym{(}  \ottsym{(}  \mathit{x} \,  {:}  \, \ottnt{T}  \ottsym{)}  \rightarrow  \ottnt{C}  \ottsym{)}  \,  \,\&\,  \,  \Phi  \, / \,  \ottnt{S} $}
    \AxiomC{$ \emptyset   \vdash   \ottsym{(}  \ottsym{(}  \mathit{x} \,  {:}  \, \ottnt{T}  \ottsym{)}  \rightarrow  \ottnt{C}  \ottsym{)}  \,  \,\&\,  \,  \Phi  \, / \,  \ottnt{S} $}
    \TrinaryInfC{$ \emptyset   \vdash  \ottnt{v}  \ottsym{:}   \ottsym{(}  \ottsym{(}  \mathit{x} \,  {:}  \, \ottnt{T}  \ottsym{)}  \rightarrow  \ottnt{C}  \ottsym{)}  \,  \,\&\,  \,  \Phi  \, / \,  \ottnt{S} $}
   \end{prooftree}
   for some $\ottnt{C'}$.
   Because $ \emptyset   \vdash  \ottnt{C'}  \ottsym{<:}   \ottsym{(}  \ottsym{(}  \mathit{x} \,  {:}  \, \ottnt{T}  \ottsym{)}  \rightarrow  \ottnt{C}  \ottsym{)}  \,  \,\&\,  \,  \Phi  \, / \,  \ottnt{S} $ can be derived only by \Srule{Comp},
   its inversion implies that $ \emptyset   \vdash   \ottnt{C'}  . \ottnt{T}   \ottsym{<:}  \ottsym{(}  \mathit{x} \,  {:}  \, \ottnt{T}  \ottsym{)}  \rightarrow  \ottnt{C}$.
   Further, it can be derived only by \Srule{Fun},
   its inversion implies that $ \ottnt{C'}  . \ottnt{T}  \,  =  \, \ottsym{(}  \mathit{x} \,  {:}  \, \ottnt{T''}  \ottsym{)}  \rightarrow  \ottnt{C''}$ for some
   $\ottnt{T''}$ and $\ottnt{C''}$.
   Therefore, $ \emptyset   \vdash  \ottnt{v}  \ottsym{:}  \ottnt{C'}$ means $ \emptyset   \vdash  \ottnt{v}  \ottsym{:}   \ottsym{(}  \ottsym{(}  \mathit{x} \,  {:}  \, \ottnt{T''}  \ottsym{)}  \rightarrow  \ottnt{C''}  \ottsym{)}  \,  \,\&\,  \,   \ottnt{C'}  . \Phi   \, / \,   \ottnt{C'}  . \ottnt{S}  $.
   Then, the IH implies the conclusion.

  \otherwise: Contradictory.
 \end{caseanalysis}
\end{proof}

\begin{lemmap}{Value Inversion of Lambda Abstractions}{ts-val-inv-abs}
 If $\Gamma  \vdash  \ottsym{(}   \lambda\!  \,  \tceseq{ \mathit{z} }   \ottsym{.}  \ottnt{e}  \ottsym{)} \,  \tceseq{ \ottnt{v_{{\mathrm{2}}}} }   \ottsym{:}  \ottnt{C}$ and $\ottsym{\mbox{$\mid$}}   \tceseq{ \ottnt{v_{{\mathrm{2}}}} }   \ottsym{\mbox{$\mid$}} \,  <  \, \ottsym{\mbox{$\mid$}}   \tceseq{ \mathit{z} }   \ottsym{\mbox{$\mid$}}$, then
 there exist $ \tceseq{ \mathit{z_{{\mathrm{1}}}} } $, $ \tceseq{ \mathit{z_{{\mathrm{2}}}} } $, $ \tceseq{ \ottnt{T_{{\mathrm{1}}}} } $, $ \tceseq{ \ottnt{T_{{\mathrm{2}}}} } $, and $\ottnt{C_{{\mathrm{0}}}}$ such that
 \begin{itemize}
  \item $ \tceseq{ \mathit{z} }  \,  =  \,  \tceseq{ \mathit{z_{{\mathrm{2}}}} }   \ottsym{,}   \tceseq{ \mathit{z_{{\mathrm{1}}}} } $,
  \item $\Gamma  \ottsym{,}   \tceseq{ \mathit{z_{{\mathrm{2}}}}  \,  {:}  \,  \ottnt{T_{{\mathrm{2}}}} }   \ottsym{,}   \tceseq{ \mathit{z_{{\mathrm{1}}}}  \,  {:}  \,  \ottnt{T_{{\mathrm{1}}}} }   \vdash  \ottnt{e}  \ottsym{:}  \ottnt{C_{{\mathrm{0}}}}$,
  \item $ \tceseq{ \Gamma  \vdash  \ottnt{v_{{\mathrm{2}}}}  \ottsym{:}   \ottnt{T_{{\mathrm{2}}}}    [ \tceseq{ \ottnt{v_{{\mathrm{2}}}}  /  \mathit{z_{{\mathrm{2}}}} } ]   } $, and
  \item $\Gamma  \vdash    \ottsym{(}  \ottsym{(}   \tceseq{ \mathit{z_{{\mathrm{1}}}}  \,  {:}  \,  \ottnt{T_{{\mathrm{1}}}} }   \ottsym{)}  \rightarrow  \ottnt{C_{{\mathrm{0}}}}  \ottsym{)}    [ \tceseq{ \ottnt{v_{{\mathrm{2}}}}  /  \mathit{z_{{\mathrm{2}}}} } ]    \,  \,\&\,  \,   \Phi_\mathsf{val}   \, / \,   \square    \ottsym{<:}  \ottnt{C}$
 \end{itemize}
\end{lemmap}
\begin{proof}
 By induction on the derivation of $\Gamma  \vdash  \ottsym{(}   \lambda\!  \,  \tceseq{ \mathit{z} }   \ottsym{.}  \ottnt{e}  \ottsym{)} \,  \tceseq{ \ottnt{v_{{\mathrm{2}}}} }   \ottsym{:}  \ottnt{C}$.
 \begin{caseanalysis}
  \case \T{Abs}:
   By inversion and Lemmas~\ref{lem:ts-wf-ty-tctx-typing} and \ref{lem:ts-refl-subtyping}.

  \case \T{App}:
   We are given
   \begin{prooftree}
    \AxiomC{$\Gamma  \vdash  \ottsym{(}   \lambda\!  \,  \tceseq{ \mathit{z} }   \ottsym{.}  \ottnt{e}  \ottsym{)} \,  \tceseq{ \ottnt{v_{{\mathrm{21}}}} }   \ottsym{:}  \ottsym{(}  \mathit{x} \,  {:}  \, \ottnt{T'}  \ottsym{)}  \rightarrow  \ottnt{C'}$}
    \AxiomC{$\Gamma  \vdash  \ottnt{v_{{\mathrm{22}}}}  \ottsym{:}  \ottnt{T'}$}
    \BinaryInfC{$\Gamma  \vdash  \ottsym{(}   \lambda\!  \,  \tceseq{ \mathit{z} }   \ottsym{.}  \ottnt{e}  \ottsym{)} \,  \tceseq{ \ottnt{v_{{\mathrm{21}}}} }  \, \ottnt{v_{{\mathrm{22}}}}  \ottsym{:}   \ottnt{C'}    [  \ottnt{v_{{\mathrm{22}}}}  /  \mathit{x}  ]  $}
   \end{prooftree}
   for some $ \tceseq{ \ottnt{v_{{\mathrm{21}}}} } $, $\ottnt{v_{{\mathrm{22}}}}$, $\mathit{x}$, $\ottnt{T'}$, and $\ottnt{C'}$ such that
   $ \tceseq{ \ottnt{v_{{\mathrm{2}}}} }  \,  =  \,  \tceseq{ \ottnt{v_{{\mathrm{21}}}} }   \ottsym{,}  \ottnt{v_{{\mathrm{22}}}}$ and $\ottnt{C} \,  =  \,  \ottnt{C'}    [  \ottnt{v_{{\mathrm{22}}}}  /  \mathit{x}  ]  $.
   Because $\ottsym{\mbox{$\mid$}}   \tceseq{ \ottnt{v_{{\mathrm{21}}}} }   \ottsym{\mbox{$\mid$}} \,  <  \, \ottsym{\mbox{$\mid$}}   \tceseq{ \ottnt{v_{{\mathrm{21}}}} }   \ottsym{\mbox{$\mid$}}  \ottsym{+}  \ottsym{1} = \ottsym{\mbox{$\mid$}}   \tceseq{ \ottnt{v_{{\mathrm{2}}}} }   \ottsym{\mbox{$\mid$}} \,  <  \, \ottsym{\mbox{$\mid$}}   \tceseq{ \mathit{z} }   \ottsym{\mbox{$\mid$}}$, the IH implies
   \begin{itemize}
    \item $ \tceseq{ \mathit{z} }  \,  =  \,  \tceseq{ \mathit{z_{{\mathrm{21}}}} }   \ottsym{,}   \tceseq{ \mathit{z_{{\mathrm{01}}}} } $,
    \item $\Gamma  \ottsym{,}   \tceseq{ \mathit{z_{{\mathrm{21}}}}  \,  {:}  \,  \ottnt{T_{{\mathrm{21}}}} }   \ottsym{,}   \tceseq{ \mathit{z_{{\mathrm{01}}}}  \,  {:}  \,  \ottnt{T_{{\mathrm{01}}}} }   \vdash  \ottnt{e}  \ottsym{:}  \ottnt{C_{{\mathrm{0}}}}$,
    \item $ \tceseq{ \Gamma  \vdash  \ottnt{v_{{\mathrm{21}}}}  \ottsym{:}   \ottnt{T_{{\mathrm{21}}}}    [ \tceseq{ \ottnt{v_{{\mathrm{21}}}}  /  \mathit{z_{{\mathrm{21}}}} } ]   } $, and
    \item $\Gamma  \vdash    \ottsym{(}  \ottsym{(}   \tceseq{ \mathit{z_{{\mathrm{01}}}}  \,  {:}  \,  \ottnt{T_{{\mathrm{01}}}} }   \ottsym{)}  \rightarrow  \ottnt{C_{{\mathrm{0}}}}  \ottsym{)}    [ \tceseq{ \ottnt{v_{{\mathrm{21}}}}  /  \mathit{z_{{\mathrm{21}}}} } ]    \,  \,\&\,  \,   \Phi_\mathsf{val}   \, / \,   \square    \ottsym{<:}   \ottsym{(}  \ottsym{(}  \mathit{x} \,  {:}  \, \ottnt{T'}  \ottsym{)}  \rightarrow  \ottnt{C'}  \ottsym{)}  \,  \,\&\,  \,   \Phi_\mathsf{val}   \, / \,   \square  $
   \end{itemize}
   for some $ \tceseq{ \mathit{z_{{\mathrm{01}}}} } $, $ \tceseq{ \mathit{z_{{\mathrm{21}}}} } $, $ \tceseq{ \ottnt{T_{{\mathrm{01}}}} } $, $ \tceseq{ \ottnt{T_{{\mathrm{21}}}} } $, and $\ottnt{C_{{\mathrm{0}}}}$.
   Because
   \[
    \ottsym{\mbox{$\mid$}}   \tceseq{ \ottnt{v_{{\mathrm{21}}}} }   \ottsym{\mbox{$\mid$}}  \ottsym{+}  \ottsym{1} \,  <  \, \ottsym{\mbox{$\mid$}}   \tceseq{ \mathit{z} }   \ottsym{\mbox{$\mid$}} = \ottsym{\mbox{$\mid$}}   \tceseq{ \mathit{z_{{\mathrm{21}}}} }   \ottsym{,}   \tceseq{ \mathit{z_{{\mathrm{01}}}} }   \ottsym{\mbox{$\mid$}} = \ottsym{\mbox{$\mid$}}   \tceseq{ \mathit{z_{{\mathrm{21}}}} }   \ottsym{\mbox{$\mid$}}  \ottsym{+}  \ottsym{\mbox{$\mid$}}   \tceseq{ \mathit{z_{{\mathrm{01}}}} }   \ottsym{\mbox{$\mid$}} \,  =  \, \ottsym{\mbox{$\mid$}}   \tceseq{ \ottnt{v_{{\mathrm{21}}}} }   \ottsym{\mbox{$\mid$}}  \ottsym{+}  \ottsym{\mbox{$\mid$}}   \tceseq{ \mathit{z_{{\mathrm{01}}}} }   \ottsym{\mbox{$\mid$}} ~,
   \]
   we have $\ottsym{1} \,  <  \, \ottsym{\mbox{$\mid$}}   \tceseq{ \mathit{z_{{\mathrm{01}}}} }   \ottsym{\mbox{$\mid$}}$.
   Therefore, there exist some $\mathit{z_{{\mathrm{11}}}}$, $ \tceseq{ \mathit{z_{{\mathrm{1}}}} } $, $\ottnt{T_{{\mathrm{11}}}}$, and $ \tceseq{ \ottnt{T_{{\mathrm{1}}}} } $ such that
   \begin{itemize}
    \item $ \tceseq{ \mathit{z_{{\mathrm{01}}}} }  \,  =  \, \mathit{z_{{\mathrm{11}}}}  \ottsym{,}   \tceseq{ \mathit{z_{{\mathrm{1}}}} } $ and
    \item $ \tceseq{ \ottnt{T_{{\mathrm{01}}}} }  \,  =  \, \ottnt{T_{{\mathrm{11}}}}  \ottsym{,}   \tceseq{ \ottnt{T_{{\mathrm{1}}}} } $.
   \end{itemize}

   Let $ \tceseq{ \mathit{z_{{\mathrm{2}}}} }  \,  =  \,  \tceseq{ \mathit{z_{{\mathrm{21}}}} }   \ottsym{,}  \mathit{z_{{\mathrm{11}}}}$ and $ \tceseq{ \ottnt{T_{{\mathrm{2}}}} }  \,  =  \,  \tceseq{ \ottnt{T_{{\mathrm{21}}}} }   \ottsym{,}  \ottnt{T_{{\mathrm{11}}}}$.
   Now, we have the conclusion by the following.
   \begin{itemize}
    \item We have $ \tceseq{ \mathit{z} }  =  \tceseq{ \mathit{z_{{\mathrm{21}}}} }   \ottsym{,}   \tceseq{ \mathit{z_{{\mathrm{01}}}} }  =  \tceseq{ \mathit{z_{{\mathrm{21}}}} }   \ottsym{,}  \mathit{z_{{\mathrm{11}}}}  \ottsym{,}   \tceseq{ \mathit{z_{{\mathrm{1}}}} }  =  \tceseq{ \mathit{z_{{\mathrm{2}}}} }   \ottsym{,}   \tceseq{ \mathit{z_{{\mathrm{1}}}} } $.

    \item We have $\Gamma  \ottsym{,}   \tceseq{ \mathit{z_{{\mathrm{2}}}}  \,  {:}  \,  \ottnt{T_{{\mathrm{2}}}} }   \ottsym{,}   \tceseq{ \mathit{z_{{\mathrm{1}}}}  \,  {:}  \,  \ottnt{T_{{\mathrm{1}}}} }   \vdash  \ottnt{e}  \ottsym{:}  \ottnt{C_{{\mathrm{0}}}}$
          because $\Gamma  \ottsym{,}   \tceseq{ \mathit{z_{{\mathrm{21}}}}  \,  {:}  \,  \ottnt{T_{{\mathrm{21}}}} }   \ottsym{,}   \tceseq{ \mathit{z_{{\mathrm{01}}}}  \,  {:}  \,  \ottnt{T_{{\mathrm{01}}}} }   \vdash  \ottnt{e}  \ottsym{:}  \ottnt{C_{{\mathrm{0}}}}$ and
          \[
            \tceseq{ \mathit{z_{{\mathrm{21}}}}  \,  {:}  \,  \ottnt{T_{{\mathrm{21}}}} }   \ottsym{,}   \tceseq{ \mathit{z_{{\mathrm{01}}}}  \,  {:}  \,  \ottnt{T_{{\mathrm{01}}}} }  =  \tceseq{ \mathit{z_{{\mathrm{21}}}}  \,  {:}  \,  \ottnt{T_{{\mathrm{21}}}} }   \ottsym{,}  \mathit{z_{{\mathrm{11}}}} \,  {:}  \, \ottnt{T_{{\mathrm{11}}}}  \ottsym{,}   \tceseq{ \mathit{z_{{\mathrm{1}}}}  \,  {:}  \,  \ottnt{T_{{\mathrm{1}}}} }  =  \tceseq{ \mathit{z_{{\mathrm{2}}}}  \,  {:}  \,  \ottnt{T_{{\mathrm{2}}}} }   \ottsym{,}   \tceseq{ \mathit{z_{{\mathrm{1}}}}  \,  {:}  \,  \ottnt{T_{{\mathrm{1}}}} }  ~.
          \]

    \item We show that $ \tceseq{ \Gamma  \vdash  \ottnt{v_{{\mathrm{2}}}}  \ottsym{:}   \ottnt{T_{{\mathrm{2}}}}    [  \ottnt{v_{{\mathrm{2}}}}  /  \mathit{z_{{\mathrm{2}}}}  ]   } $.
          Because
          \begin{itemize}
           \item $ \tceseq{ \ottnt{v_{{\mathrm{2}}}} }  \,  =  \,  \tceseq{ \ottnt{v_{{\mathrm{21}}}} }   \ottsym{,}  \ottnt{v_{{\mathrm{22}}}}$,
           \item $ \tceseq{ \ottnt{T_{{\mathrm{2}}}} }  \,  =  \,  \tceseq{ \ottnt{T_{{\mathrm{21}}}} }   \ottsym{,}  \ottnt{T_{{\mathrm{11}}}}$,
           \item $ \tceseq{ \mathit{z_{{\mathrm{2}}}} }  \,  =  \,  \tceseq{ \mathit{z_{{\mathrm{21}}}} }   \ottsym{,}  \mathit{z_{{\mathrm{11}}}}$,
           \item $ \tceseq{ \Gamma  \vdash  \ottnt{v_{{\mathrm{21}}}}  \ottsym{:}   \ottnt{T_{{\mathrm{21}}}}    [ \tceseq{ \ottnt{v_{{\mathrm{21}}}}  /  \mathit{z_{{\mathrm{21}}}} } ]   } $, and
           \item $\mathit{z_{{\mathrm{11}}}}$ does not occur in $ \tceseq{ \ottnt{T_{{\mathrm{21}}}} } $ and $\ottnt{T_{{\mathrm{11}}}}$,
          \end{itemize}
          it suffices to show that
          \[
           \Gamma  \vdash  \ottnt{v_{{\mathrm{22}}}}  \ottsym{:}   \ottnt{T_{{\mathrm{11}}}}    [ \tceseq{ \ottnt{v_{{\mathrm{21}}}}  /  \mathit{z_{{\mathrm{21}}}} } ]   ~.
          \]
          Because
          \begin{itemize}
           \item $\Gamma  \vdash    \ottsym{(}  \ottsym{(}   \tceseq{ \mathit{z_{{\mathrm{01}}}}  \,  {:}  \,  \ottnt{T_{{\mathrm{01}}}} }   \ottsym{)}  \rightarrow  \ottnt{C_{{\mathrm{0}}}}  \ottsym{)}    [ \tceseq{ \ottnt{v_{{\mathrm{21}}}}  /  \mathit{z_{{\mathrm{21}}}} } ]    \,  \,\&\,  \,   \Phi_\mathsf{val}   \, / \,   \square    \ottsym{<:}   \ottsym{(}  \ottsym{(}  \mathit{x} \,  {:}  \, \ottnt{T'}  \ottsym{)}  \rightarrow  \ottnt{C'}  \ottsym{)}  \,  \,\&\,  \,   \Phi_\mathsf{val}   \, / \,   \square  $ and
           \item $ \tceseq{ \ottnt{T_{{\mathrm{01}}}} }  \,  =  \, \ottnt{T_{{\mathrm{11}}}}  \ottsym{,}   \tceseq{ \ottnt{T_{{\mathrm{1}}}} } $,
          \end{itemize}
          we have $\Gamma  \vdash  \ottnt{T'}  \ottsym{<:}   \ottnt{T_{{\mathrm{11}}}}    [ \tceseq{ \ottnt{v_{{\mathrm{21}}}}  /  \mathit{z_{{\mathrm{21}}}} } ]  $.
          Because $\Gamma  \vdash  \ottnt{v_{{\mathrm{22}}}}  \ottsym{:}  \ottnt{T'}$, \T{Sub} implies that
          it suffices to show that
          \[
           \Gamma  \vdash   \ottnt{T_{{\mathrm{11}}}}    [ \tceseq{ \ottnt{v_{{\mathrm{21}}}}  /  \mathit{z_{{\mathrm{21}}}} } ]   ~.
          \]
          Because $ \tceseq{ \mathit{z_{{\mathrm{01}}}}  \,  {:}  \,  \ottnt{T_{{\mathrm{01}}}} }  \,  =  \, \mathit{z_{{\mathrm{11}}}} \,  {:}  \, \ottnt{T_{{\mathrm{11}}}}  \ottsym{,}   \tceseq{ \mathit{z_{{\mathrm{1}}}}  \,  {:}  \,  \ottnt{T_{{\mathrm{1}}}} } $,
          Lemmas~\ref{lem:ts-wf-ty-tctx-typing} and \ref{lem:ts-wf-tctx-wf-ftyping}
          with $\Gamma  \ottsym{,}   \tceseq{ \mathit{z_{{\mathrm{21}}}}  \,  {:}  \,  \ottnt{T_{{\mathrm{21}}}} }   \ottsym{,}   \tceseq{ \mathit{z_{{\mathrm{01}}}}  \,  {:}  \,  \ottnt{T_{{\mathrm{01}}}} }   \vdash  \ottnt{e}  \ottsym{:}  \ottnt{C_{{\mathrm{0}}}}$
          imply
          \[
            \Gamma  \ottsym{,}   \tceseq{ \mathit{z_{{\mathrm{21}}}}  \,  {:}  \,  \ottnt{T_{{\mathrm{21}}}} }   \vdash  \ottnt{T_{{\mathrm{11}}}} ~.
          \]
          Because $ \tceseq{ \Gamma  \vdash  \ottnt{v_{{\mathrm{21}}}}  \ottsym{:}   \ottnt{T_{{\mathrm{21}}}}    [ \tceseq{ \ottnt{v_{{\mathrm{21}}}}  /  \mathit{z_{{\mathrm{21}}}} } ]   } $,
          \reflem{ts-val-subst-wf-ftyping} implies the conclusion.

    \item We show that
          \[
           \Gamma  \vdash    \ottsym{(}  \ottsym{(}   \tceseq{ \mathit{z_{{\mathrm{1}}}}  \,  {:}  \,  \ottnt{T_{{\mathrm{1}}}} }   \ottsym{)}  \rightarrow  \ottnt{C_{{\mathrm{0}}}}  \ottsym{)}    [ \tceseq{ \ottnt{v_{{\mathrm{2}}}}  /  \mathit{z_{{\mathrm{2}}}} } ]    \,  \,\&\,  \,   \Phi_\mathsf{val}   \, / \,   \square    \ottsym{<:}  \ottnt{C} ~.
          \]
          Because
          \begin{itemize}
           \item $\ottnt{C} \,  =  \,  \ottnt{C'}    [  \ottnt{v_{{\mathrm{22}}}}  /  \mathit{x}  ]  $,
           \item $ \tceseq{ \ottnt{v_{{\mathrm{2}}}} }  \,  =  \,  \tceseq{ \ottnt{v_{{\mathrm{21}}}} }   \ottsym{,}  \ottnt{v_{{\mathrm{22}}}}$,
           \item $ \tceseq{ \mathit{z_{{\mathrm{2}}}} }  \,  =  \,  \tceseq{ \mathit{z_{{\mathrm{21}}}} }   \ottsym{,}  \mathit{z_{{\mathrm{11}}}}$, and
           \item $\mathit{z_{{\mathrm{11}}}}$ does not occur in $ \tceseq{ \ottnt{v_{{\mathrm{21}}}} } $,
          \end{itemize}
          it suffices to show that
          \[
           \Gamma  \vdash     \ottsym{(}  \ottsym{(}   \tceseq{ \mathit{z_{{\mathrm{1}}}}  \,  {:}  \,  \ottnt{T_{{\mathrm{1}}}} }   \ottsym{)}  \rightarrow  \ottnt{C_{{\mathrm{0}}}}  \ottsym{)}    [ \tceseq{ \ottnt{v_{{\mathrm{21}}}}  /  \mathit{z_{{\mathrm{21}}}} } ]      [  \ottnt{v_{{\mathrm{22}}}}  /  \mathit{z_{{\mathrm{11}}}}  ]    \,  \,\&\,  \,   \Phi_\mathsf{val}   \, / \,   \square    \ottsym{<:}   \ottnt{C'}    [  \ottnt{v_{{\mathrm{22}}}}  /  \mathit{x}  ]   ~.
          \]
          Because
          \begin{itemize}
           \item $\Gamma  \vdash    \ottsym{(}  \ottsym{(}   \tceseq{ \mathit{z_{{\mathrm{01}}}}  \,  {:}  \,  \ottnt{T_{{\mathrm{01}}}} }   \ottsym{)}  \rightarrow  \ottnt{C_{{\mathrm{0}}}}  \ottsym{)}    [ \tceseq{ \ottnt{v_{{\mathrm{21}}}}  /  \mathit{z_{{\mathrm{21}}}} } ]    \,  \,\&\,  \,   \Phi_\mathsf{val}   \, / \,   \square    \ottsym{<:}   \ottsym{(}  \ottsym{(}  \mathit{x} \,  {:}  \, \ottnt{T'}  \ottsym{)}  \rightarrow  \ottnt{C'}  \ottsym{)}  \,  \,\&\,  \,   \Phi_\mathsf{val}   \, / \,   \square  $,
           \item $ \tceseq{ \mathit{z_{{\mathrm{01}}}} }  \,  =  \, \mathit{z_{{\mathrm{11}}}}  \ottsym{,}   \tceseq{ \mathit{z_{{\mathrm{1}}}} } $, and
           \item $ \tceseq{ \ottnt{T_{{\mathrm{01}}}} }  \,  =  \, \ottnt{T_{{\mathrm{11}}}}  \ottsym{,}   \tceseq{ \ottnt{T_{{\mathrm{1}}}} } $,
          \end{itemize}
          we have
          \begin{itemize}
           \item $ \mathit{x}    =    \mathit{z_{{\mathrm{11}}}} $ and
           \item $\Gamma  \ottsym{,}  \mathit{z_{{\mathrm{11}}}} \,  {:}  \, \ottnt{T'}  \vdash    \ottsym{(}  \ottsym{(}   \tceseq{ \mathit{z_{{\mathrm{1}}}}  \,  {:}  \,  \ottnt{T_{{\mathrm{1}}}} }   \ottsym{)}  \rightarrow  \ottnt{C_{{\mathrm{0}}}}  \ottsym{)}    [ \tceseq{ \ottnt{v_{{\mathrm{21}}}}  /  \mathit{z_{{\mathrm{21}}}} } ]    \,  \,\&\,  \,   \Phi_\mathsf{val}   \, / \,   \square    \ottsym{<:}  \ottnt{C'}$.
          \end{itemize}
          Because $\Gamma  \vdash  \ottnt{v_{{\mathrm{22}}}}  \ottsym{:}  \ottnt{T'}$,
          we have the conclusion by \reflem{ts-val-subst-subtyping}.
   \end{itemize}

  \case \T{Sub}: By the IH and \reflem{ts-transitivity-subtyping}.
  \otherwise: Contradictory.
 \end{caseanalysis}
\end{proof}

\begin{lemmap}{Value Inversion of Recursive Functions}{ts-val-inv-rec}
 If $\Gamma  \vdash  \ottsym{(}   \mathsf{rec}  \, (  \tceseq{  \mathit{X} ^{  \ottnt{A} _{  \mu  }  },  \mathit{Y} ^{  \ottnt{A} _{  \nu  }  }  }  ,  \mathit{f} ,   \tceseq{ \mathit{z} }  ) . \,  \ottnt{e}   \ottsym{)} \,  \tceseq{ \ottnt{v_{{\mathrm{2}}}} }   \ottsym{:}  \ottnt{C}$ and $\ottsym{\mbox{$\mid$}}   \tceseq{ \ottnt{v_{{\mathrm{2}}}} }   \ottsym{\mbox{$\mid$}} \,  <  \, \ottsym{\mbox{$\mid$}}   \tceseq{ \mathit{z} }   \ottsym{\mbox{$\mid$}}$,
 then there exist some $ \tceseq{ \mathit{z_{{\mathrm{1}}}} } $, $ \tceseq{ \mathit{z_{{\mathrm{2}}}} } $, $ \tceseq{ \ottnt{T_{{\mathrm{11}}}} } $, $ \tceseq{ \ottnt{T_{{\mathrm{12}}}} } $, $\ottnt{C_{{\mathrm{0}}}}$, $\ottnt{C}$, $ \tceseq{ \mathit{z_{{\mathrm{0}}}} } $, $ \tceseq{ \iota_{{\mathrm{0}}} } $, and $\sigma$ such that
 \begin{itemize}
  \item $ \tceseq{ \mathit{z} }  \,  =  \,  \tceseq{ \mathit{z_{{\mathrm{2}}}} }   \ottsym{,}   \tceseq{ \mathit{z_{{\mathrm{1}}}} } $.
  \item $\Gamma  \vdash   \mathsf{rec}  \, (  \tceseq{  \mathit{X} ^{  \ottnt{A} _{  \mu  }  },  \mathit{Y} ^{  \ottnt{A} _{  \nu  }  }  }  ,  \mathit{f} ,   \tceseq{ \mathit{z} }  ) . \,  \ottnt{e}   \ottsym{:}  \ottsym{(}   \tceseq{ \mathit{z_{{\mathrm{2}}}}  \,  {:}  \,  \ottnt{T_{{\mathrm{12}}}} }   \ottsym{,}   \tceseq{ \mathit{z_{{\mathrm{1}}}}  \,  {:}  \,  \ottnt{T_{{\mathrm{11}}}} }   \ottsym{)}  \rightarrow  \sigma  \ottsym{(}  \ottnt{C_{{\mathrm{0}}}}  \ottsym{)}$,
  \item $\Gamma  \vdash    \ottsym{(}  \ottsym{(}   \tceseq{ \mathit{z_{{\mathrm{1}}}}  \,  {:}  \,  \ottnt{T_{{\mathrm{11}}}} }   \ottsym{)}  \rightarrow  \sigma  \ottsym{(}  \ottnt{C_{{\mathrm{0}}}}  \ottsym{)}  \ottsym{)}    [ \tceseq{ \ottnt{v_{{\mathrm{2}}}}  /  \mathit{z_{{\mathrm{2}}}} } ]    \,  \,\&\,  \,   \Phi_\mathsf{val}   \, / \,   \square    \ottsym{<:}  \ottnt{C}$,
  \item $\Gamma  \ottsym{,}   \tceseq{ \mathit{X} }  \,  {:}  \,  (   \Sigma ^\ast   ,   \tceseq{ \iota_{{\mathrm{0}}} }   )   \ottsym{,}   \tceseq{ \mathit{Y} }  \,  {:}  \,  (   \Sigma ^\omega   ,   \tceseq{ \iota_{{\mathrm{0}}} }   )   \ottsym{,}  \mathit{f} \,  {:}  \, \ottsym{(}   \tceseq{ \mathit{z_{{\mathrm{2}}}}  \,  {:}  \,  \ottnt{T_{{\mathrm{12}}}} }   \ottsym{,}   \tceseq{ \mathit{z_{{\mathrm{1}}}}  \,  {:}  \,  \ottnt{T_{{\mathrm{11}}}} }   \ottsym{)}  \rightarrow  \ottnt{C_{{\mathrm{0}}}}  \ottsym{,}   \tceseq{ \mathit{z_{{\mathrm{2}}}}  \,  {:}  \,  \ottnt{T_{{\mathrm{12}}}} }   \ottsym{,}   \tceseq{ \mathit{z_{{\mathrm{1}}}}  \,  {:}  \,  \ottnt{T_{{\mathrm{11}}}} }   \vdash  \ottnt{e}  \ottsym{:}  \ottnt{C}$,
  \item $ \tceseq{ \mathit{z_{{\mathrm{0}}}}  \,  {:}  \,  \iota_{{\mathrm{0}}} }  \,  =  \,  \gamma  (   \tceseq{ \mathit{z_{{\mathrm{2}}}}  \,  {:}  \,  \ottnt{T_{{\mathrm{12}}}} }   \ottsym{,}   \tceseq{ \mathit{z_{{\mathrm{1}}}}  \,  {:}  \,  \ottnt{T_{{\mathrm{11}}}} }   ) $,
  \item $ \tceseq{  \ottnt{A} _{  \mu  }  \,  =  \, \sigma  \ottsym{(}  \mathit{X}  \ottsym{)} } $,
  \item $ \tceseq{  \ottnt{A} _{  \nu  }  \,  =  \, \sigma  \ottsym{(}  \mathit{Y}  \ottsym{)} } $,
  \item $  \{   \tceseq{ \mathit{X} }   \ottsym{,}   \tceseq{ \mathit{Y} }   \}   \, \ottsym{\#} \,  \ottsym{(}   \mathit{fpv}  (   \tceseq{ \ottnt{T_{{\mathrm{12}}}} }   )  \,  \mathbin{\cup}  \,  \mathit{fpv}  (   \tceseq{ \ottnt{T_{{\mathrm{11}}}} }   )   \ottsym{)} $,
  \item $ \tceseq{ \ottsym{(}  \mathit{X}  \ottsym{,}  \mathit{Y}  \ottsym{)} }   \ottsym{;}   \tceseq{ \mathit{z_{{\mathrm{0}}}}  \,  {:}  \,  \iota_{{\mathrm{0}}} }   \vdash  \ottnt{C_{{\mathrm{0}}}}  \stackrel{\succ}{\circ}  \ottnt{C}$,
  \item $ \tceseq{ \ottsym{(}  \mathit{X}  \ottsym{,}  \mathit{Y}  \ottsym{)} }   \ottsym{;}   \tceseq{ \mathit{z_{{\mathrm{0}}}}  \,  {:}  \,  \iota_{{\mathrm{0}}} }  \,  |  \, \ottnt{C}  \vdash  \sigma$, and
  \item $ \tceseq{ \Gamma  \vdash  \ottnt{v_{{\mathrm{2}}}}  \ottsym{:}   \ottnt{T_{{\mathrm{12}}}}    [ \tceseq{ \ottnt{v_{{\mathrm{2}}}}  /  \mathit{z_{{\mathrm{2}}}} } ]   } $.
 \end{itemize}
\end{lemmap}
\begin{proof}
 By induction on the typing derivation.
 Only \T{Fun}, \T{Sub}, and \T{App} are applicable.
 \begin{caseanalysis}
  \case \T{Fun}: By inversion, and Lemmas~\ref{lem:ts-wf-ty-tctx-typing} and \ref{lem:ts-refl-subtyping} for
   \[
    \Gamma  \vdash    \ottsym{(}   \tceseq{ \mathit{z_{{\mathrm{1}}}}  \,  {:}  \,  \ottnt{T_{{\mathrm{11}}}} }   \ottsym{)}  \rightarrow  \sigma  \ottsym{(}  \ottnt{C_{{\mathrm{0}}}}  \ottsym{)}    [ \tceseq{ \ottnt{v_{{\mathrm{2}}}}  /  \mathit{z_{{\mathrm{2}}}} } ]    \,  \,\&\,  \,   \Phi_\mathsf{val}   \, / \,   \square    \ottsym{<:}  \ottnt{C} ~.
   \]

  \case \T{Sub}:
   By the IH and \reflem{ts-transitivity-subtyping}.

  \case \T{App}: Similarly to the case of \T{App} in \reflem{ts-val-inv-abs}.
 \end{caseanalysis}
\end{proof}

\begin{lemmap}{Value Inversion of Universal Types}{ts-val-inv-univ-ty}
 If $ \emptyset   \vdash  \ottnt{v}  \ottsym{:}    \forall   \ottsym{(}  \mathit{X} \,  {:}  \,  \tceseq{ \ottnt{s} }   \ottsym{)}  \ottsym{.}  \ottnt{C}  \,  \,\&\,  \,  \Phi  \, / \,  \ottnt{S} $,
 then there exist some $\ottnt{e}$ and $\ottnt{C_{{\mathrm{0}}}}$ such that
 \begin{itemize}
  \item $\ottnt{v} \,  =  \,  \Lambda   \ottsym{(}  \mathit{X} \,  {:}  \,  \tceseq{ \ottnt{s} }   \ottsym{)}  \ottsym{.}  \ottnt{e}$,
  \item $\mathit{X} \,  {:}  \,  \tceseq{ \ottnt{s} }   \vdash  \ottnt{e}  \ottsym{:}  \ottnt{C_{{\mathrm{0}}}}$, and
  \item $ \emptyset   \vdash    \forall   \ottsym{(}  \mathit{X} \,  {:}  \,  \tceseq{ \ottnt{s} }   \ottsym{)}  \ottsym{.}  \ottnt{C_{{\mathrm{0}}}}  \,  \,\&\,  \,   \Phi_\mathsf{val}   \, / \,   \square    \ottsym{<:}    \forall   \ottsym{(}  \mathit{X} \,  {:}  \,  \tceseq{ \ottnt{s} }   \ottsym{)}  \ottsym{.}  \ottnt{C}  \,  \,\&\,  \,  \Phi  \, / \,  \ottnt{S} $.
 \end{itemize}
\end{lemmap}
\begin{proof}
 By induction on the typing derivation.
 \begin{caseanalysis}
  \case \T{PAbs}:
   By inversion and Lemmas~\ref{lem:ts-wf-ty-tctx-typing} and \ref{lem:ts-refl-subtyping}.

  \case \T{Sub}:
   We are given
   \begin{prooftree}
    \AxiomC{$ \emptyset   \vdash  \ottnt{v}  \ottsym{:}  \ottnt{C'}$}
    \AxiomC{$ \emptyset   \vdash  \ottnt{C'}  \ottsym{<:}    \forall   \ottsym{(}  \mathit{X} \,  {:}  \,  \tceseq{ \ottnt{s} }   \ottsym{)}  \ottsym{.}  \ottnt{C}  \,  \,\&\,  \,  \Phi  \, / \,  \ottnt{S} $}
    \AxiomC{$ \emptyset   \vdash    \forall   \ottsym{(}  \mathit{X} \,  {:}  \,  \tceseq{ \ottnt{s} }   \ottsym{)}  \ottsym{.}  \ottnt{C}  \,  \,\&\,  \,  \Phi  \, / \,  \ottnt{S} $}
    \TrinaryInfC{$ \emptyset   \vdash  \ottnt{v}  \ottsym{:}    \forall   \ottsym{(}  \mathit{X} \,  {:}  \,  \tceseq{ \ottnt{s} }   \ottsym{)}  \ottsym{.}  \ottnt{C}  \,  \,\&\,  \,  \Phi  \, / \,  \ottnt{S} $}
   \end{prooftree}
   for some $\ottnt{C'}$.
   Because $ \emptyset   \vdash  \ottnt{C'}  \ottsym{<:}    \forall   \ottsym{(}  \mathit{X} \,  {:}  \,  \tceseq{ \ottnt{s} }   \ottsym{)}  \ottsym{.}  \ottnt{C}  \,  \,\&\,  \,  \Phi  \, / \,  \ottnt{S} $ can be derived only by \Srule{Comp}, and
   $ \emptyset   \vdash   \ottnt{C'}  . \ottnt{T}   \ottsym{<:}   \forall   \ottsym{(}  \mathit{X} \,  {:}  \,  \tceseq{ \ottnt{s} }   \ottsym{)}  \ottsym{.}  \ottnt{C}$ can be only by \Srule{Forall},
   its inversion implies $ \ottnt{C'}  . \ottnt{T}  \,  =  \,  \forall   \ottsym{(}  \mathit{X} \,  {:}  \,  \tceseq{ \ottnt{s} }   \ottsym{)}  \ottsym{.}  \ottnt{C''}$ for some $\ottnt{C''}$.
   Therefore, $ \emptyset   \vdash  \ottnt{v}  \ottsym{:}  \ottnt{C'}$ means $ \emptyset   \vdash  \ottnt{v}  \ottsym{:}    \forall   \ottsym{(}  \mathit{X} \,  {:}  \,  \tceseq{ \ottnt{s} }   \ottsym{)}  \ottsym{.}  \ottnt{C''}  \,  \,\&\,  \,   \ottnt{C'}  . \Phi   \, / \,   \ottnt{C'}  . \ottnt{S}  $.
   Then, we have the conclusion by the IH and \reflem{ts-transitivity-subtyping}
   with $ \emptyset   \vdash  \ottnt{C'}  \ottsym{<:}    \forall   \ottsym{(}  \mathit{X} \,  {:}  \,  \tceseq{ \ottnt{s} }   \ottsym{)}  \ottsym{.}  \ottnt{C}  \,  \,\&\,  \,  \Phi  \, / \,  \ottnt{S} $.

  \case \T{App}: Contradictory by \reflem{ts-val-applied-fun}.
  \otherwise: Contradictory.
 \end{caseanalysis}
\end{proof}

\begin{lemmap}{Folding and unfolding $ \mu $- and $ \nu $-Predicates in Validity}{ts-mu-nu-valid}
 Let $\Gamma$ be a typing context.  Then, the following holds.
 \begin{itemize}
  \item $\Gamma  \models     \ottsym{(}   \mu\!  \, \mathit{X}  \ottsym{(}   \tceseq{ \mathit{x}  \,  {:}  \,  \ottnt{s} }   \ottsym{)}  \ottsym{.}  \phi  \ottsym{)}  \ottsym{(}   \tceseq{ \ottnt{t} }   \ottsym{)}    \Rightarrow    \phi     [  \ottsym{(}   \mu\!  \, \mathit{X}  \ottsym{(}   \tceseq{ \mathit{x}  \,  {:}  \,  \ottnt{s} }   \ottsym{)}  \ottsym{.}  \phi  \ottsym{)}  /  \mathit{X}  ]      [ \tceseq{ \ottnt{t}  /  \mathit{x} } ]  $.
  \item $\Gamma  \models     \ottsym{(}   \nu\!  \, \mathit{X}  \ottsym{(}   \tceseq{ \mathit{x}  \,  {:}  \,  \ottnt{s} }   \ottsym{)}  \ottsym{.}  \phi  \ottsym{)}  \ottsym{(}   \tceseq{ \ottnt{t} }   \ottsym{)}    \Rightarrow    \phi     [  \ottsym{(}   \nu\!  \, \mathit{X}  \ottsym{(}   \tceseq{ \mathit{x}  \,  {:}  \,  \ottnt{s} }   \ottsym{)}  \ottsym{.}  \phi  \ottsym{)}  /  \mathit{X}  ]      [ \tceseq{ \ottnt{t}  /  \mathit{x} } ]  $.
  \item $\Gamma  \models     \phi    [  \ottsym{(}   \mu\!  \, \mathit{X}  \ottsym{(}   \tceseq{ \mathit{x}  \,  {:}  \,  \ottnt{s} }   \ottsym{)}  \ottsym{.}  \phi  \ottsym{)}  /  \mathit{X}  ]      [ \tceseq{ \ottnt{t}  /  \mathit{x} } ]      \Rightarrow    \ottsym{(}   \mu\!  \, \mathit{X}  \ottsym{(}   \tceseq{ \mathit{x}  \,  {:}  \,  \ottnt{s} }   \ottsym{)}  \ottsym{.}  \phi  \ottsym{)}  \ottsym{(}   \tceseq{ \ottnt{t} }   \ottsym{)} $.
  \item $\Gamma  \models     \phi    [  \ottsym{(}   \nu\!  \, \mathit{X}  \ottsym{(}   \tceseq{ \mathit{x}  \,  {:}  \,  \ottnt{s} }   \ottsym{)}  \ottsym{.}  \phi  \ottsym{)}  /  \mathit{X}  ]      [ \tceseq{ \ottnt{t}  /  \mathit{x} } ]      \Rightarrow    \ottsym{(}   \nu\!  \, \mathit{X}  \ottsym{(}   \tceseq{ \mathit{x}  \,  {:}  \,  \ottnt{s} }   \ottsym{)}  \ottsym{.}  \phi  \ottsym{)}  \ottsym{(}   \tceseq{ \ottnt{t} }   \ottsym{)} $.
 \end{itemize}
\end{lemmap}
\begin{proof}
 \TS{Validity}
 Obvious by Lemmas~\ref{lem:ts-logic-closing-dsubs},
 \ref{lem:ts-logic-val-subst}, and \ref{lem:ts-logic-pred-subst} and the
 definitions of the least and greatest fixpoint operators.
\end{proof}

\begin{lemmap}{Value Inversion}{ts-inv-val}
 If $ \emptyset   \vdash  \ottnt{v}  \ottsym{:}  \ottnt{C}$,
 then $ \emptyset   \vdash  \ottnt{v}  \ottsym{:}  \ottnt{T}$ and $ \emptyset   \vdash   \ottnt{T}  \,  \,\&\,  \,   \Phi_\mathsf{val}   \, / \,   \square    \ottsym{<:}  \ottnt{C}$ for some $\ottnt{T}$.
\end{lemmap}
\begin{proof}
 By case analysis on $ \ottnt{C}  . \ottnt{T} $.
 \begin{caseanalysis}
  \case $  \exists  \, { \mathit{x}  \ottsym{,}  \iota  \ottsym{,}  \phi } . \   \ottnt{C}  . \ottnt{T}  \,  =  \, \ottsym{\{}  \mathit{x} \,  {:}  \, \iota \,  |  \, \phi  \ottsym{\}} $:
   By \reflem{ts-val-inv-refine-ty}.

  \case $  \exists  \, { \mathit{x}  \ottsym{,}  \ottnt{T'}  \ottsym{,}  \ottnt{C'} } . \   \ottnt{C}  . \ottnt{T}  \,  =  \, \ottsym{(}  \mathit{x} \,  {:}  \, \ottnt{T'}  \ottsym{)}  \rightarrow  \ottnt{C'} $:
   We have $ \emptyset   \vdash  \ottnt{v}  \ottsym{:}   \ottsym{(}  \ottsym{(}  \mathit{x} \,  {:}  \, \ottnt{T'}  \ottsym{)}  \rightarrow  \ottnt{C'}  \ottsym{)}  \,  \,\&\,  \,   \ottnt{C}  . \Phi   \, / \,   \ottnt{C}  . \ottnt{S}  $.
   By case analysis on the result of applying \reflem{ts-val-inv-fun-ty}.
   \begin{caseanalysis}
    \case $ {   \exists  \, {  \tceseq{ \mathit{z} }   \ottsym{,}  \ottnt{e}  \ottsym{,}   \tceseq{ \ottnt{v_{{\mathrm{2}}}} }  } . \  \ottnt{v} \,  =  \, \ottsym{(}   \lambda\!  \,  \tceseq{ \mathit{z} }   \ottsym{.}  \ottnt{e}  \ottsym{)} \,  \tceseq{ \ottnt{v_{{\mathrm{2}}}} }   } \,\mathrel{\wedge}\, { \ottsym{\mbox{$\mid$}}   \tceseq{ \ottnt{v_{{\mathrm{2}}}} }   \ottsym{\mbox{$\mid$}} \,  <  \, \ottsym{\mbox{$\mid$}}   \tceseq{ \mathit{z} }   \ottsym{\mbox{$\mid$}} } $:
     Because $ \emptyset   \vdash  \ottsym{(}   \lambda\!  \,  \tceseq{ \mathit{z} }   \ottsym{.}  \ottnt{e}  \ottsym{)} \,  \tceseq{ \ottnt{v_{{\mathrm{2}}}} }   \ottsym{:}  \ottnt{C}$,
     \reflem{ts-val-inv-abs} implies that
     \begin{itemize}
      \item $ \tceseq{ \mathit{z} }  \,  =  \,  \tceseq{ \mathit{z_{{\mathrm{2}}}} }   \ottsym{,}   \tceseq{ \mathit{z_{{\mathrm{1}}}} } $,
      \item $ \tceseq{ \mathit{z_{{\mathrm{2}}}}  \,  {:}  \,  \ottnt{T_{{\mathrm{2}}}} }   \ottsym{,}   \tceseq{ \mathit{z_{{\mathrm{1}}}}  \,  {:}  \,  \ottnt{T_{{\mathrm{1}}}} }   \vdash  \ottnt{e}  \ottsym{:}  \ottnt{C_{{\mathrm{0}}}}$,
      \item $ \tceseq{  \emptyset   \vdash  \ottnt{v_{{\mathrm{2}}}}  \ottsym{:}   \ottnt{T_{{\mathrm{2}}}}    [ \tceseq{ \ottnt{v_{{\mathrm{2}}}}  /  \mathit{z_{{\mathrm{2}}}} } ]   } $, and
      \item $ \emptyset   \vdash    \ottsym{(}  \ottsym{(}   \tceseq{ \mathit{z_{{\mathrm{1}}}}  \,  {:}  \,  \ottnt{T_{{\mathrm{1}}}} }   \ottsym{)}  \rightarrow  \ottnt{C_{{\mathrm{0}}}}  \ottsym{)}    [ \tceseq{ \ottnt{v_{{\mathrm{2}}}}  /  \mathit{z_{{\mathrm{2}}}} } ]    \,  \,\&\,  \,   \Phi_\mathsf{val}   \, / \,   \square    \ottsym{<:}  \ottnt{C}$
     \end{itemize}
     for some $ \tceseq{ \mathit{z_{{\mathrm{1}}}} } $, $ \tceseq{ \mathit{z_{{\mathrm{2}}}} } $, $ \tceseq{ \ottnt{T_{{\mathrm{1}}}} } $, $ \tceseq{ \ottnt{T_{{\mathrm{2}}}} } $, and $\ottnt{C_{{\mathrm{0}}}}$.
     By \T{Abs} and \T{App},
     \[
       \emptyset   \vdash  \ottsym{(}   \lambda\!  \,  \tceseq{ \mathit{z} }   \ottsym{.}  \ottnt{e}  \ottsym{)} \,  \tceseq{ \ottnt{v_{{\mathrm{2}}}} }   \ottsym{:}   \ottsym{(}  \ottsym{(}   \tceseq{ \mathit{z_{{\mathrm{1}}}}  \,  {:}  \,  \ottnt{T_{{\mathrm{1}}}} }   \ottsym{)}  \rightarrow  \ottnt{C_{{\mathrm{0}}}}  \ottsym{)}    [ \tceseq{ \ottnt{v_{{\mathrm{2}}}}  /  \mathit{z_{{\mathrm{2}}}} } ]   ~.
     \]
     Therefore, we have the conclusion by letting $\ottnt{T} \,  =  \,  \ottsym{(}  \ottsym{(}   \tceseq{ \mathit{z_{{\mathrm{1}}}}  \,  {:}  \,  \ottnt{T_{{\mathrm{1}}}} }   \ottsym{)}  \rightarrow  \ottnt{C_{{\mathrm{0}}}}  \ottsym{)}    [ \tceseq{ \ottnt{v_{{\mathrm{2}}}}  /  \mathit{z_{{\mathrm{2}}}} } ]  $.

    \case $\exists\, \mathit{X}, \ottnt{A} _{  \mu  } ,\mathit{Y}, \ottnt{A} _{  \nu  } ,\mathit{f}, \tceseq{ \mathit{z} } ,\ottnt{e}, \tceseq{ \ottnt{v_{{\mathrm{2}}}} } .\,  { \ottnt{v} \,  =  \, \ottsym{(}   \mathsf{rec}  \, (  \tceseq{  \mathit{X} ^{  \ottnt{A} _{  \mu  }  },  \mathit{Y} ^{  \ottnt{A} _{  \nu  }  }  }  ,  \mathit{f} ,   \tceseq{ \mathit{z} }  ) . \,  \ottnt{e}   \ottsym{)} \,  \tceseq{ \ottnt{v_{{\mathrm{2}}}} }  } \,\mathrel{\wedge}\, { \ottsym{\mbox{$\mid$}}   \tceseq{ \ottnt{v_{{\mathrm{2}}}} }   \ottsym{\mbox{$\mid$}} \,  <  \, \ottsym{\mbox{$\mid$}}   \tceseq{ \mathit{z} }   \ottsym{\mbox{$\mid$}} } $:
     By \reflem{ts-val-inv-rec} and \T{App}.
   \end{caseanalysis}

  \case $  \exists  \, { \mathit{X}  \ottsym{,}   \tceseq{ \ottnt{s} }   \ottsym{,}  \ottnt{C'} } . \   \ottnt{C}  . \ottnt{T}  \,  =  \,  \forall   \ottsym{(}  \mathit{X} \,  {:}  \,  \tceseq{ \ottnt{s} }   \ottsym{)}  \ottsym{.}  \ottnt{C'} $:
   By \reflem{ts-val-inv-univ-ty} and \T{PAbs}.
 \end{caseanalysis}
\end{proof}

\begin{lemmap}{Inversion of Variables}{ts-inv-var}
 If $\Gamma  \vdash  \mathit{x}  \ottsym{:}  \ottnt{C}$,
 then $\Gamma  \vdash  \mathit{x}  \ottsym{:}  \ottnt{T}$ and $\Gamma  \vdash   \ottnt{T}  \,  \,\&\,  \,   \Phi_\mathsf{val}   \, / \,   \square    \ottsym{<:}  \ottnt{C}$ for some $\ottnt{T}$.
\end{lemmap}
\begin{proof}
 By induction on the typing derivation.
 \begin{caseanalysis}
  \case \T{CVar} and \T{Var}:
   Let $\ottnt{T} \,  =  \,  \ottnt{C}  . \ottnt{T} $.
   Then, because $\ottnt{C} \,  =  \,  \ottnt{T}  \,  \,\&\,  \,   \Phi_\mathsf{val}   \, / \,   \square  $, we have $\Gamma  \vdash  \mathit{x}  \ottsym{:}  \ottnt{T}$.
   Further, we also have $\Gamma  \vdash   \ottnt{T}  \,  \,\&\,  \,   \Phi_\mathsf{val}   \, / \,   \square    \ottsym{<:}  \ottnt{C}$ by
   Lemmas~\ref{lem:ts-wf-ty-tctx-typing} and \ref{lem:ts-refl-subtyping}.

  \case \T{Sub}: By the IH and \reflem{ts-transitivity-subtyping}.

  \otherwise: Contradictory.
 \end{caseanalysis}
\end{proof}

\begin{lemmap}{Stack Types of Shift Expressions}{ts-inv-shift-sty}
 If $\Gamma  \vdash   \ottnt{K}  [  \mathcal{S}_0 \, \mathit{x}  \ottsym{.}  \ottnt{e}  ]   \ottsym{:}  \ottnt{C}$,
 then $ \ottnt{C}  . \ottnt{S}  \,  =  \,  ( \forall  \mathit{y}  .  \ottnt{C_{{\mathrm{1}}}}  )  \Rightarrow   \ottnt{C_{{\mathrm{2}}}} $
 for some $\mathit{y}$, $\ottnt{C_{{\mathrm{1}}}}$, and $\ottnt{C_{{\mathrm{2}}}}$.
\end{lemmap}
\begin{proof}
 Straightforward by induction on the typing derivation with the facts that:
 \[
     \forall  \, { \ottnt{T}  \ottsym{,}  \Phi  \ottsym{,}  \mathit{y}  \ottsym{,}  \ottnt{C_{{\mathrm{1}}}}  \ottsym{,}  \ottnt{C_{{\mathrm{2}}}}  \ottsym{,}  \ottnt{S} } . \  \Gamma \,  |  \, \ottnt{T} \,  \,\&\,  \, \Phi  \vdash   ( \forall  \mathit{y}  .  \ottnt{C_{{\mathrm{1}}}}  )  \Rightarrow   \ottnt{C_{{\mathrm{2}}}}   \ottsym{<:}  \ottnt{S}   \mathrel{\Longrightarrow}    \exists  \, { \ottnt{C'_{{\mathrm{1}}}}  \ottsym{,}  \ottnt{C'_{{\mathrm{2}}}} } . \  \ottnt{S} \,  =  \,  ( \forall  \mathit{y}  .  \ottnt{C'_{{\mathrm{1}}}}  )  \Rightarrow   \ottnt{C'_{{\mathrm{2}}}}   
 \]
 (which is used in the case for \T{Sub}); and
 \[
    \forall  \, { \mathit{y}  \ottsym{,}  \ottnt{C_{{\mathrm{1}}}}  \ottsym{,}  \ottnt{C_{{\mathrm{2}}}}  \ottsym{,}  \ottnt{S} } . \    \exists  \, { \mathit{z}  \ottsym{,}  \ottnt{C'_{{\mathrm{1}}}} } . \   ( \forall  \mathit{y}  .  \ottnt{C_{{\mathrm{1}}}}  )  \Rightarrow   \ottnt{C_{{\mathrm{2}}}}   \gg\!\!=  \ottsym{(}   \lambda\!  \, \mathit{y}  \ottsym{.}  \ottnt{S}  \ottsym{)} \,  =  \,  ( \forall  \mathit{z}  .  \ottnt{C'_{{\mathrm{1}}}}  )  \Rightarrow   \ottnt{C_{{\mathrm{2}}}}   
 \]
 (which is used in the case for \T{Let}).
 Note that \T{LetV} cannot be applied.
\end{proof}

\begin{lemmap}{Weakening and Strengthening Effects by $ \Phi_\mathsf{val} $}{ts-weak-strength-TP_val}
 \begin{enumerate}
  \item For any $\Gamma$ and $\Phi$,
        \begin{itemize}
         \item $\Gamma  \vdash   \Phi_\mathsf{val}  \,  \tceappendop  \, \Phi  \ottsym{<:}  \Phi$,
         \item $\Gamma  \vdash  \Phi  \ottsym{<:}   \Phi_\mathsf{val}  \,  \tceappendop  \, \Phi$,
         \item $\Gamma  \vdash  \Phi  \ottsym{<:}  \Phi \,  \tceappendop  \,  \Phi_\mathsf{val} $, and
         \item $\Gamma  \vdash  \Phi \,  \tceappendop  \,  \Phi_\mathsf{val}   \ottsym{<:}  \Phi$.
        \end{itemize}

  \item For any $\Gamma$, $\Phi_{{\mathrm{1}}}$, and $\Phi_{{\mathrm{2}}}$,
        $\Gamma  \vdash  \Phi_{{\mathrm{1}}}  \ottsym{<:}  \Phi_{{\mathrm{2}}}$ if and only if $\Gamma  \vdash   \Phi_\mathsf{val}  \,  \tceappendop  \, \Phi_{{\mathrm{1}}}  \ottsym{<:}  \Phi_{{\mathrm{2}}}$.
 \end{enumerate}
\end{lemmap}
\begin{proof}
 \begin{enumerate}
  \item We consider each case.
        In what follows, we suppose that $\mathit{x} \,  \not\in  \,  \mathit{fv}  (  \Phi  ) $.
        \begin{itemize}
         \item We show that $\Gamma  \vdash   \Phi_\mathsf{val}  \,  \tceappendop  \, \Phi  \ottsym{<:}  \Phi$.
               \begin{itemize}
                \item We show that $\Gamma  \ottsym{,}  \mathit{x} \,  {:}  \,  \Sigma ^\ast   \models    { \ottsym{(}   \Phi_\mathsf{val}  \,  \tceappendop  \, \Phi  \ottsym{)} }^\mu   \ottsym{(}  \mathit{x}  \ottsym{)}    \Rightarrow     { \Phi }^\mu   \ottsym{(}  \mathit{x}  \ottsym{)} $.
                      By definition, it suffices to show that
                      \[
                       \Gamma  \ottsym{,}  \mathit{x} \,  {:}  \,  \Sigma ^\ast   \models   \ottsym{(}    \exists  \,  \mathit{x_{{\mathrm{1}}}}  \mathrel{  \in  }   \Sigma ^\ast   . \    \exists  \,  \mathit{x_{{\mathrm{2}}}}  \mathrel{  \in  }   \Sigma ^\ast   . \     \mathit{x}    =    \mathit{x_{{\mathrm{1}}}} \,  \tceappendop  \, \mathit{x_{{\mathrm{2}}}}     \wedge     \mathit{x_{{\mathrm{1}}}}    =     \epsilon       \wedge     { \Phi }^\mu   \ottsym{(}  \mathit{x_{{\mathrm{2}}}}  \ottsym{)}     \ottsym{)}    \Rightarrow     { \Phi }^\mu   \ottsym{(}  \mathit{x}  \ottsym{)}  ~,
                      \]
                      which is true.
                      \TS{OK informal}

                \item We show that $\Gamma  \ottsym{,}  \mathit{x} \,  {:}  \,  \Sigma ^\omega   \models    { \ottsym{(}   \Phi_\mathsf{val}  \,  \tceappendop  \, \Phi  \ottsym{)} }^\nu   \ottsym{(}  \mathit{x}  \ottsym{)}    \Rightarrow     { \Phi }^\nu   \ottsym{(}  \mathit{x}  \ottsym{)} $.
                      By definition, it suffices to show that
                      \[
                       \Gamma  \ottsym{,}  \mathit{x} \,  {:}  \,  \Sigma ^\omega   \models   \ottsym{(}    \bot     \vee    \ottsym{(}    \exists  \,  \mathit{x'}  \mathrel{  \in  }   \Sigma ^\ast   . \    \exists  \,  \mathit{y'}  \mathrel{  \in  }   \Sigma ^\omega   . \     \mathit{x}    =    \mathit{x'} \,  \tceappendop  \, \mathit{y'}     \wedge     \mathit{x'}    =     \epsilon       \wedge     { \Phi }^\nu   \ottsym{(}  \mathit{y'}  \ottsym{)}     \ottsym{)}   \ottsym{)}    \Rightarrow     { \Phi }^\nu   \ottsym{(}  \mathit{x}  \ottsym{)}  ~,
                      \]
                      which is true.
                      \TS{OK informal}
               \end{itemize}

         \item We show that $\Gamma  \vdash  \Phi  \ottsym{<:}   \Phi_\mathsf{val}  \,  \tceappendop  \, \Phi$.
               \begin{itemize}
                \item We show that $\Gamma  \ottsym{,}  \mathit{x} \,  {:}  \,  \Sigma ^\ast   \models    { \Phi }^\mu   \ottsym{(}  \mathit{x}  \ottsym{)}    \Rightarrow     { \ottsym{(}   \Phi_\mathsf{val}  \,  \tceappendop  \, \Phi  \ottsym{)} }^\mu   \ottsym{(}  \mathit{x}  \ottsym{)} $.
                      By definition, it suffices to show that
                      \[
                       \Gamma  \ottsym{,}  \mathit{x} \,  {:}  \,  \Sigma ^\ast   \models    { \Phi }^\mu   \ottsym{(}  \mathit{x}  \ottsym{)}    \Rightarrow    \ottsym{(}    \exists  \,  \mathit{x_{{\mathrm{1}}}}  \mathrel{  \in  }   \Sigma ^\ast   . \    \exists  \,  \mathit{x_{{\mathrm{2}}}}  \mathrel{  \in  }   \Sigma ^\ast   . \     \mathit{x}    =    \mathit{x_{{\mathrm{1}}}} \,  \tceappendop  \, \mathit{x_{{\mathrm{2}}}}     \wedge     \mathit{x_{{\mathrm{1}}}}    =     \epsilon       \wedge     { \Phi }^\mu   \ottsym{(}  \mathit{x_{{\mathrm{2}}}}  \ottsym{)}     \ottsym{)}  ~,
                      \]
                      which it true.
                      \TS{OK informal}

                \item We show that $\Gamma  \ottsym{,}  \mathit{x} \,  {:}  \,  \Sigma ^\omega   \models    { \Phi }^\nu   \ottsym{(}  \mathit{x}  \ottsym{)}    \Rightarrow     { \ottsym{(}   \Phi_\mathsf{val}  \,  \tceappendop  \, \Phi  \ottsym{)} }^\nu   \ottsym{(}  \mathit{x}  \ottsym{)} $.
                      By definition, it suffices to show that
                      \[
                       \Gamma  \ottsym{,}  \mathit{x} \,  {:}  \,  \Sigma ^\omega   \models    { \Phi }^\nu   \ottsym{(}  \mathit{x}  \ottsym{)}    \Rightarrow    \ottsym{(}    \bot     \vee    \ottsym{(}    \exists  \,  \mathit{x'}  \mathrel{  \in  }   \Sigma ^\ast   . \    \exists  \,  \mathit{y'}  \mathrel{  \in  }   \Sigma ^\omega   . \     \mathit{x}    =    \mathit{x'} \,  \tceappendop  \, \mathit{y'}     \wedge     \mathit{x'}    =     \epsilon       \wedge     { \Phi }^\nu   \ottsym{(}  \mathit{y'}  \ottsym{)}     \ottsym{)}   \ottsym{)}  ~,
                      \]
                      which is true.
                      \TS{OK informal}
               \end{itemize}

         \item We show that $\Gamma  \vdash  \Phi  \ottsym{<:}  \Phi \,  \tceappendop  \,  \Phi_\mathsf{val} $.
               \begin{itemize}
                \item We show that $\Gamma  \ottsym{,}  \mathit{x} \,  {:}  \,  \Sigma ^\ast   \models    { \Phi }^\mu   \ottsym{(}  \mathit{x}  \ottsym{)}    \Rightarrow     { \ottsym{(}  \Phi \,  \tceappendop  \,  \Phi_\mathsf{val}   \ottsym{)} }^\mu   \ottsym{(}  \mathit{x}  \ottsym{)} $.
                      By definition, it suffices to show that
                      \[
                       \Gamma  \ottsym{,}  \mathit{x} \,  {:}  \,  \Sigma ^\ast   \models    { \Phi }^\mu   \ottsym{(}  \mathit{x}  \ottsym{)}    \Rightarrow    \ottsym{(}    \exists  \,  \mathit{x_{{\mathrm{1}}}}  \mathrel{  \in  }   \Sigma ^\ast   . \    \exists  \,  \mathit{x_{{\mathrm{2}}}}  \mathrel{  \in  }   \Sigma ^\ast   . \     \mathit{x}    =    \mathit{x_{{\mathrm{1}}}} \,  \tceappendop  \, \mathit{x_{{\mathrm{2}}}}     \wedge     { \Phi }^\mu   \ottsym{(}  \mathit{x_{{\mathrm{1}}}}  \ottsym{)}     \wedge     \mathit{x_{{\mathrm{2}}}}    =     \epsilon       \ottsym{)}  ~,
                      \]
                      which is true.
                      \TS{OK informal}

                \item We show that $\Gamma  \ottsym{,}  \mathit{x} \,  {:}  \,  \Sigma ^\omega   \models    { \Phi }^\nu   \ottsym{(}  \mathit{x}  \ottsym{)}    \Rightarrow     { \ottsym{(}  \Phi \,  \tceappendop  \,  \Phi_\mathsf{val}   \ottsym{)} }^\nu   \ottsym{(}  \mathit{x}  \ottsym{)} $.
                      By definition, it suffices to show that
                      \[
                       \Gamma  \ottsym{,}  \mathit{x} \,  {:}  \,  \Sigma ^\omega   \models    { \Phi }^\nu   \ottsym{(}  \mathit{x}  \ottsym{)}    \Rightarrow    \ottsym{(}    { \Phi }^\nu   \ottsym{(}  \mathit{x}  \ottsym{)}    \vee    \ottsym{(}    \exists  \,  \mathit{x'}  \mathrel{  \in  }   \Sigma ^\ast   . \    \exists  \,  \mathit{y'}  \mathrel{  \in  }   \Sigma ^\omega   . \     \mathit{x}    =    \mathit{x'} \,  \tceappendop  \, \mathit{y'}     \wedge     { \Phi }^\mu   \ottsym{(}  \mathit{x'}  \ottsym{)}     \wedge     \bot      \ottsym{)}   \ottsym{)}  ~,
                      \]
                      which is true.
                      \TS{OK informal}
               \end{itemize}

         \item We show that $\Gamma  \vdash  \Phi \,  \tceappendop  \,  \Phi_\mathsf{val}   \ottsym{<:}  \Phi$.
               \begin{itemize}
                \item We show that $\Gamma  \ottsym{,}  \mathit{x} \,  {:}  \,  \Sigma ^\ast   \models    { \ottsym{(}  \Phi \,  \tceappendop  \,  \Phi_\mathsf{val}   \ottsym{)} }^\mu   \ottsym{(}  \mathit{x}  \ottsym{)}    \Rightarrow     { \Phi }^\mu   \ottsym{(}  \mathit{x}  \ottsym{)} $.
                      By definition, it suffices to show that
                      \[
                       \Gamma  \ottsym{,}  \mathit{x} \,  {:}  \,  \Sigma ^\ast   \models   \ottsym{(}    \exists  \,  \mathit{x_{{\mathrm{1}}}}  \mathrel{  \in  }   \Sigma ^\ast   . \    \exists  \,  \mathit{x_{{\mathrm{2}}}}  \mathrel{  \in  }   \Sigma ^\ast   . \     \mathit{x}    =    \mathit{x_{{\mathrm{1}}}} \,  \tceappendop  \, \mathit{x_{{\mathrm{2}}}}     \wedge     { \Phi }^\mu   \ottsym{(}  \mathit{x_{{\mathrm{1}}}}  \ottsym{)}     \wedge     \mathit{x_{{\mathrm{2}}}}    =     \epsilon       \ottsym{)}    \Rightarrow     { \Phi }^\mu   \ottsym{(}  \mathit{x}  \ottsym{)}  ~,
                      \]
                      which is true.
                      \TS{OK informal}

                \item We show that $\Gamma  \ottsym{,}  \mathit{x} \,  {:}  \,  \Sigma ^\omega   \models    { \ottsym{(}  \Phi \,  \tceappendop  \,  \Phi_\mathsf{val}   \ottsym{)} }^\nu   \ottsym{(}  \mathit{x}  \ottsym{)}    \Rightarrow     { \Phi }^\nu   \ottsym{(}  \mathit{x}  \ottsym{)} $.
                      By definition, it suffices to show that
                      \[
                       \Gamma  \ottsym{,}  \mathit{x} \,  {:}  \,  \Sigma ^\omega   \models   \ottsym{(}    { \Phi }^\nu   \ottsym{(}  \mathit{x}  \ottsym{)}    \vee    \ottsym{(}    \exists  \,  \mathit{x'}  \mathrel{  \in  }   \Sigma ^\ast   . \    \exists  \,  \mathit{y'}  \mathrel{  \in  }   \Sigma ^\omega   . \     \mathit{x}    =    \mathit{x'} \,  \tceappendop  \, \mathit{y'}     \wedge     { \Phi }^\mu   \ottsym{(}  \mathit{x'}  \ottsym{)}     \wedge     \bot      \ottsym{)}   \ottsym{)}    \Rightarrow     { \Phi }^\nu   \ottsym{(}  \mathit{x}  \ottsym{)}  ~,
                      \]
                      which is true.
                      \TS{OK informal}
               \end{itemize}
        \end{itemize}

  \item  We show each direction.
         \begin{itemize}
          \item Consider the left-to-right direction.
                Suppose that $\Gamma  \vdash  \Phi_{{\mathrm{1}}}  \ottsym{<:}  \Phi_{{\mathrm{2}}}$.
                From the first part, we have $\Gamma  \vdash   \Phi_\mathsf{val}  \,  \tceappendop  \, \Phi_{{\mathrm{1}}}  \ottsym{<:}  \Phi_{{\mathrm{1}}}$.
                Then, \reflem{ts-transitivity-subtyping} with $\Gamma  \vdash  \Phi_{{\mathrm{1}}}  \ottsym{<:}  \Phi_{{\mathrm{2}}}$ implies
                the conclusion $\Gamma  \vdash   \Phi_\mathsf{val}  \,  \tceappendop  \, \Phi_{{\mathrm{1}}}  \ottsym{<:}  \Phi_{{\mathrm{2}}}$.
          \item Consider the right-to-left direction.
                Suppose that $\Gamma  \vdash   \Phi_\mathsf{val}  \,  \tceappendop  \, \Phi_{{\mathrm{1}}}  \ottsym{<:}  \Phi_{{\mathrm{2}}}$.
                From the first part, we have $\Gamma  \vdash  \Phi_{{\mathrm{1}}}  \ottsym{<:}   \Phi_\mathsf{val}  \,  \tceappendop  \, \Phi_{{\mathrm{1}}}$.
                Then, \reflem{ts-transitivity-subtyping} with $\Gamma  \vdash   \Phi_\mathsf{val}  \,  \tceappendop  \, \Phi_{{\mathrm{1}}}  \ottsym{<:}  \Phi_{{\mathrm{2}}}$ implies
                the conclusion $\Gamma  \vdash  \Phi_{{\mathrm{1}}}  \ottsym{<:}  \Phi_{{\mathrm{2}}}$.
         \end{itemize}
 \end{enumerate}
\end{proof}

\begin{lemmap}{Adding $ \Phi_\mathsf{val} $ as Pre-Effects to Subtypes}{ts-add-TP_val-subty}
 \begin{enumerate}
  \item If $\Gamma  \vdash  \ottnt{C_{{\mathrm{1}}}}  \ottsym{<:}  \ottnt{C_{{\mathrm{2}}}}$, then $\Gamma  \vdash   \Phi_\mathsf{val}  \,  \tceappendop  \, \ottnt{C_{{\mathrm{1}}}}  \ottsym{<:}  \ottnt{C_{{\mathrm{2}}}}$.
  \item If $\Gamma \,  |  \, \ottnt{T} \,  \,\&\,  \, \Phi  \vdash  \ottnt{S_{{\mathrm{1}}}}  \ottsym{<:}  \ottnt{S_{{\mathrm{2}}}}$, then $\Gamma \,  |  \, \ottnt{T} \,  \,\&\,  \,  \Phi_\mathsf{val}  \,  \tceappendop  \, \Phi  \vdash   \Phi_\mathsf{val}  \,  \tceappendop  \, \ottnt{S_{{\mathrm{1}}}}  \ottsym{<:}  \ottnt{S_{{\mathrm{2}}}}$
 \end{enumerate}
\end{lemmap}
\begin{proof}
 By simultaneous induction on the derivations.
 \begin{caseanalysis}
  \case \Srule{Comp}:
   We are given
   \begin{prooftree}
    \AxiomC{$\Gamma  \vdash   \ottnt{C_{{\mathrm{1}}}}  . \ottnt{T}   \ottsym{<:}   \ottnt{C_{{\mathrm{2}}}}  . \ottnt{T} $}
    \AxiomC{$\Gamma  \vdash   \ottnt{C_{{\mathrm{1}}}}  . \Phi   \ottsym{<:}   \ottnt{C_{{\mathrm{2}}}}  . \Phi $}
    \AxiomC{$\Gamma \,  |  \,  \ottnt{C_{{\mathrm{1}}}}  . \ottnt{T}  \,  \,\&\,  \,  \ottnt{C_{{\mathrm{1}}}}  . \Phi   \vdash   \ottnt{C_{{\mathrm{1}}}}  . \ottnt{S}   \ottsym{<:}   \ottnt{C_{{\mathrm{2}}}}  . \ottnt{S} $}
    \TrinaryInfC{$\Gamma  \vdash  \ottnt{C_{{\mathrm{1}}}}  \ottsym{<:}  \ottnt{C_{{\mathrm{2}}}}$}
   \end{prooftree}
   We have the conclusion by the IH, \reflem{ts-weak-strength-TP_val} with $\Gamma  \vdash   \ottnt{C_{{\mathrm{1}}}}  . \Phi   \ottsym{<:}   \ottnt{C_{{\mathrm{2}}}}  . \Phi $, and \Srule{Comp}.

  \case \Srule{Empty}: Trivial by \Srule{Empty}
   because $\ottnt{S_{{\mathrm{1}}}} = \ottnt{S_{{\mathrm{2}}}} =  \Phi_\mathsf{val}  \,  \tceappendop  \, \ottnt{S_{{\mathrm{1}}}} =  \square $.

  \case \Srule{Ans}: By the IH and \Srule{Ans}.
  \case \Srule{Embed}:
   We are given
   \begin{prooftree}
    \AxiomC{$\Gamma  \ottsym{,}  \mathit{x} \,  {:}  \, \ottnt{T}  \vdash  \Phi \,  \tceappendop  \, \ottnt{C_{{\mathrm{21}}}}  \ottsym{<:}  \ottnt{C_{{\mathrm{22}}}}$}
    \AxiomC{$\Gamma  \ottsym{,}  \mathit{y} \,  {:}  \,  \Sigma ^\omega   \models    { \Phi }^\nu   \ottsym{(}  \mathit{y}  \ottsym{)}    \Rightarrow     { \ottnt{C_{{\mathrm{22}}}} }^\nu (  \mathit{y}  )  $}
    \AxiomC{$\mathit{x}  \ottsym{,}  \mathit{y} \,  \not\in  \,  \mathit{fv}  (  \ottnt{C_{{\mathrm{22}}}}  )  \,  \mathbin{\cup}  \,  \mathit{fv}  (  \Phi  ) $}
    \TrinaryInfC{$\Gamma \,  |  \, \ottnt{T} \,  \,\&\,  \, \Phi  \vdash   \square   \ottsym{<:}   ( \forall  \mathit{x}  .  \ottnt{C_{{\mathrm{21}}}}  )  \Rightarrow   \ottnt{C_{{\mathrm{22}}}} $}
   \end{prooftree}
   for some $\mathit{x}$, $\ottnt{C_{{\mathrm{21}}}}$, and $\ottnt{C_{{\mathrm{22}}}}$ such that
   $\ottnt{S_{{\mathrm{2}}}} \,  =  \,  ( \forall  \mathit{x}  .  \ottnt{C_{{\mathrm{21}}}}  )  \Rightarrow   \ottnt{C_{{\mathrm{22}}}} $.
   We also have $\ottnt{S_{{\mathrm{1}}}} \,  =  \,  \square $.
   We have the conclusion by \Srule{Embed} with the following.
   \begin{itemize}
    \item We show that
          \[
           \Gamma  \ottsym{,}  \mathit{x} \,  {:}  \, \ottnt{T}  \vdash  \ottsym{(}   \Phi_\mathsf{val}  \,  \tceappendop  \, \Phi  \ottsym{)} \,  \tceappendop  \, \ottnt{C_{{\mathrm{21}}}}  \ottsym{<:}  \ottnt{C_{{\mathrm{22}}}} ~.
          \]
          By the IH,
          \[
           \Gamma  \ottsym{,}  \mathit{x} \,  {:}  \, \ottnt{T}  \vdash   \Phi_\mathsf{val}  \,  \tceappendop  \, \ottsym{(}  \Phi \,  \tceappendop  \, \ottnt{C_{{\mathrm{21}}}}  \ottsym{)}  \ottsym{<:}  \ottnt{C_{{\mathrm{22}}}} ~.
          \]
          Then, \reflem{ts-assoc-preeff-compose-subtyping} implies the conclusion.

    \item We show that
          \[
           \Gamma  \ottsym{,}  \mathit{y} \,  {:}  \,  \Sigma ^\omega   \models    { \ottsym{(}   \Phi_\mathsf{val}  \,  \tceappendop  \, \Phi  \ottsym{)} }^\nu   \ottsym{(}  \mathit{y}  \ottsym{)}    \Rightarrow     { \ottnt{C_{{\mathrm{22}}}} }^\nu (  \mathit{y}  )   ~.
          \]
          By \reflem{ts-weak-strength-TP_val}, $\Gamma  \vdash   \Phi_\mathsf{val}  \,  \tceappendop  \, \Phi  \ottsym{<:}  \Phi$.
          Therefore, $\Gamma  \ottsym{,}  \mathit{y} \,  {:}  \,  \Sigma ^\omega   \models    { \ottsym{(}   \Phi_\mathsf{val}  \,  \tceappendop  \, \Phi  \ottsym{)} }^\nu   \ottsym{(}  \mathit{y}  \ottsym{)}    \Rightarrow     { \Phi }^\nu   \ottsym{(}  \mathit{y}  \ottsym{)} $.
          Because $\Gamma  \ottsym{,}  \mathit{y} \,  {:}  \,  \Sigma ^\omega   \models    { \Phi }^\nu   \ottsym{(}  \mathit{y}  \ottsym{)}    \Rightarrow     { \ottnt{C_{{\mathrm{22}}}} }^\nu (  \mathit{y}  )  $,
          \reflem{ts-transitivity-valid} implies the conclusion.
   \end{itemize}
 \end{caseanalysis}
\end{proof}

\begin{lemmap}{Typing Variables}{ts-typing-vars}
 If $\vdash  \Gamma$ and $\Gamma  \ottsym{(}  \mathit{x}  \ottsym{)} \,  =  \, \ottnt{T}$,
 then $\Gamma  \vdash  \mathit{x}  \ottsym{:}  \ottnt{T}$.
\end{lemmap}
\begin{proof}
 If $\ottnt{T}$ is not a refinement type, it can be derived by \T{Var}.

 Suppose that $\ottnt{T}$ is a refinement type, that is, $\ottnt{T} \,  =  \, \ottsym{\{}  \mathit{y} \,  {:}  \, \iota \,  |  \, \phi  \ottsym{\}}$
 for some $\mathit{y}$, $\iota$, and $\phi$.
 By \T{CVar},
 \[
  \Gamma  \vdash  \mathit{x}  \ottsym{:}  \ottsym{\{}  \mathit{y} \,  {:}  \, \iota \,  |  \,  \mathit{x}    =    \mathit{y}   \ottsym{\}} ~.
 \]
 Because $\Gamma  \ottsym{(}  \mathit{x}  \ottsym{)} \,  =  \, \ottnt{T} = \ottsym{\{}  \mathit{y} \,  {:}  \, \iota \,  |  \, \phi  \ottsym{\}}$, we can suppose that $\Gamma \,  =  \, \Gamma_{{\mathrm{1}}}  \ottsym{,}  \mathit{x} \,  {:}  \, \ottsym{\{}  \mathit{y} \,  {:}  \, \iota \,  |  \, \phi  \ottsym{\}}  \ottsym{,}  \Gamma_{{\mathrm{2}}}$
 for some $\Gamma_{{\mathrm{1}}}$ and $\Gamma_{{\mathrm{2}}}$.
 We have
 \[
  \Gamma  \ottsym{,}  \mathit{y} \,  {:}  \, \iota  \models   \ottsym{(}   \mathit{x}    =    \mathit{y}   \ottsym{)}    \Rightarrow    \phi 
 \]
 by Lemmas~\ref{lem:ts-logic-closing-dsubs} and \ref{lem:ts-logic-val-subst}.
 \TS{Validity}
 %
 Hence, $\Gamma  \vdash  \ottsym{\{}  \mathit{y} \,  {:}  \, \iota \,  |  \,  \mathit{x}    =    \mathit{y}   \ottsym{\}}  \ottsym{<:}  \ottsym{\{}  \mathit{y} \,  {:}  \, \iota \,  |  \, \phi  \ottsym{\}} = \ottnt{T}$ by \Srule{Refine}.
 Further, $\vdash  \Gamma$ and $\Gamma \,  =  \, \Gamma_{{\mathrm{1}}}  \ottsym{,}  \mathit{x} \,  {:}  \, \ottnt{T}  \ottsym{,}  \Gamma_{{\mathrm{2}}}$ imply $\Gamma_{{\mathrm{1}}}  \vdash  \ottnt{T}$.
 Therefore, $\Gamma_{{\mathrm{1}}}  \ottsym{,}  \mathit{x} \,  {:}  \, \ottnt{T}  \ottsym{,}  \Gamma_{{\mathrm{2}}}  \vdash  \ottnt{T}$ by \reflem{ts-weak-wf-ftyping}.
 Then, Lemmas~\ref{lem:ts-refl-subtyping-pure-eff} and \ref{lem:ts-wf-TP_val}, and \T{Sub} imply
 the conclusion $\Gamma  \vdash  \mathit{x}  \ottsym{:}  \ottnt{T}$.
\end{proof}

\begin{lemmap}{Assuming $ \bot $ in Validity}{ts-valid-intro-bot-impl}
 For any $\Gamma$ and $\phi$, $\Gamma  \models    \bot     \Rightarrow    \phi $ holds.
\end{lemmap}
\begin{proof}
 \TS{Validity}
 Obvious by \reflem{ts-logic-closing-dsubs}.
\end{proof}

\begin{lemmap}{Inversion of Shift Expressions}{ts-inv-shift}
 If $ \emptyset   \vdash  \mathcal{S}_0 \, \mathit{x}  \ottsym{.}  \ottnt{e}  \ottsym{:}   \ottnt{T}  \,  \,\&\,  \,  \Phi  \, / \,   ( \forall  \mathit{y}  .  \ottnt{C_{{\mathrm{1}}}}  )  \Rightarrow   \ottnt{C_{{\mathrm{2}}}}  $,
 then either of the following holds:
 \begin{itemize}
  \item there exist some $\mathit{z}$, $\ottnt{T_{{\mathrm{0}}}}$, and $\ottnt{C_{{\mathrm{0}}}}$ such that
        \begin{itemize}
         \item $\mathit{x} \,  {:}  \, \ottsym{(}  \mathit{z} \,  {:}  \, \ottnt{T_{{\mathrm{0}}}}  \ottsym{)}  \rightarrow  \ottnt{C_{{\mathrm{0}}}}  \vdash  \ottnt{e}  \ottsym{:}  \ottnt{C_{{\mathrm{2}}}}$,
         \item $\mathit{z} \,  {:}  \, \ottnt{T_{{\mathrm{0}}}}  \vdash  \mathit{z}  \ottsym{:}   \ottnt{T}  \,  \,\&\,  \,  \Phi  \, / \,   ( \forall  \mathit{y}  .   \ottnt{C_{{\mathrm{1}}}}    [  \mathit{z}  /  \mathit{y}  ]    )  \Rightarrow   \ottnt{C_{{\mathrm{0}}}}  $,
         \item $ \emptyset   \vdash  \ottnt{T_{{\mathrm{0}}}}  \ottsym{<:}  \ottnt{T}$, and
         \item $ \mathit{z}    \not=    \mathit{y} $;
        \end{itemize}
        or
  \item $ \emptyset   \vdash  \ottnt{e}  \ottsym{:}  \ottnt{C_{{\mathrm{2}}}}$.
 \end{itemize}
\end{lemmap}
\begin{proof}
 By induction on the typing derivation.
 \begin{caseanalysis}
  \case \T{Sub}:
   We are given
   \begin{prooftree}
    \AxiomC{$ \emptyset   \vdash  \mathcal{S}_0 \, \mathit{x}  \ottsym{.}  \ottnt{e}  \ottsym{:}  \ottnt{C'}$}
    \AxiomC{$ \emptyset   \vdash  \ottnt{C'}  \ottsym{<:}   \ottnt{T}  \,  \,\&\,  \,  \Phi  \, / \,   ( \forall  \mathit{y}  .  \ottnt{C_{{\mathrm{1}}}}  )  \Rightarrow   \ottnt{C_{{\mathrm{2}}}}  $}
    \AxiomC{$ \emptyset   \vdash   \ottnt{T}  \,  \,\&\,  \,  \Phi  \, / \,   ( \forall  \mathit{y}  .  \ottnt{C_{{\mathrm{1}}}}  )  \Rightarrow   \ottnt{C_{{\mathrm{2}}}}  $}
    \TrinaryInfC{$ \emptyset   \vdash  \mathcal{S}_0 \, \mathit{x}  \ottsym{.}  \ottnt{e}  \ottsym{:}   \ottnt{T}  \,  \,\&\,  \,  \Phi  \, / \,   ( \forall  \mathit{y}  .  \ottnt{C_{{\mathrm{1}}}}  )  \Rightarrow   \ottnt{C_{{\mathrm{2}}}}  $}
   \end{prooftree}
   for some $\ottnt{C'}$.
   \reflem{ts-inv-shift-sty} with $ \emptyset   \vdash  \mathcal{S}_0 \, \mathit{x}  \ottsym{.}  \ottnt{e}  \ottsym{:}  \ottnt{C'}$
   implies
   $ \ottnt{C'}  . \ottnt{S}  \,  =  \,  ( \forall  \mathit{y}  .  \ottnt{C'_{{\mathrm{1}}}}  )  \Rightarrow   \ottnt{C'_{{\mathrm{2}}}} $ for some $\ottnt{C'_{{\mathrm{1}}}}$ and $\ottnt{C'_{{\mathrm{2}}}}$.
   Thus, we have
   \[
     \emptyset   \vdash  \mathcal{S}_0 \, \mathit{x}  \ottsym{.}  \ottnt{e}  \ottsym{:}    \ottnt{C'}  . \ottnt{T}   \,  \,\&\,  \,   \ottnt{C'}  . \Phi   \, / \,   ( \forall  \mathit{y}  .  \ottnt{C'_{{\mathrm{1}}}}  )  \Rightarrow   \ottnt{C'_{{\mathrm{2}}}}   ~.
   \]

   Assume that $ \emptyset   \vdash  \ottnt{e}  \ottsym{:}  \ottnt{C'_{{\mathrm{2}}}}$.
   Because
   $ \emptyset   \vdash  \ottnt{C'}  \ottsym{<:}   \ottnt{T}  \,  \,\&\,  \,  \Phi  \, / \,   ( \forall  \mathit{y}  .  \ottnt{C_{{\mathrm{1}}}}  )  \Rightarrow   \ottnt{C_{{\mathrm{2}}}}  $ and
   $ \ottnt{C'}  . \ottnt{S}  \,  =  \,  ( \forall  \mathit{y}  .  \ottnt{C'_{{\mathrm{1}}}}  )  \Rightarrow   \ottnt{C'_{{\mathrm{2}}}} $
   imply
   $ \emptyset   \vdash  \ottnt{C'_{{\mathrm{2}}}}  \ottsym{<:}  \ottnt{C_{{\mathrm{2}}}}$, and
   $ \emptyset   \vdash   \ottnt{T}  \,  \,\&\,  \,  \Phi  \, / \,   ( \forall  \mathit{y}  .  \ottnt{C_{{\mathrm{1}}}}  )  \Rightarrow   \ottnt{C_{{\mathrm{2}}}}  $ implies $ \emptyset   \vdash  \ottnt{C_{{\mathrm{2}}}}$,
   \T{Sub} implies $ \emptyset   \vdash  \ottnt{e}  \ottsym{:}  \ottnt{C_{{\mathrm{2}}}}$.
   Therefore, we have the conclusion.

   In what follows,
   we assume that $ \emptyset   \vdash  \ottnt{e}  \ottsym{:}  \ottnt{C'_{{\mathrm{2}}}}$ does not hold.
   Then, by the IH, there exit some $\mathit{z}$, $\ottnt{T_{{\mathrm{0}}}}$, and $\ottnt{C_{{\mathrm{0}}}}$ such that
   \begin{itemize}
    \item $\mathit{x} \,  {:}  \, \ottsym{(}  \mathit{z} \,  {:}  \, \ottnt{T_{{\mathrm{0}}}}  \ottsym{)}  \rightarrow  \ottnt{C_{{\mathrm{0}}}}  \vdash  \ottnt{e}  \ottsym{:}  \ottnt{C'_{{\mathrm{2}}}}$,
    \item $\mathit{z} \,  {:}  \, \ottnt{T_{{\mathrm{0}}}}  \vdash  \mathit{z}  \ottsym{:}    \ottnt{C'}  . \ottnt{T}   \,  \,\&\,  \,   \ottnt{C'}  . \Phi   \, / \,   ( \forall  \mathit{y}  .   \ottnt{C'_{{\mathrm{1}}}}    [  \mathit{z}  /  \mathit{y}  ]    )  \Rightarrow   \ottnt{C_{{\mathrm{0}}}}  $,
    \item $ \emptyset   \vdash  \ottnt{T_{{\mathrm{0}}}}  \ottsym{<:}   \ottnt{C'}  . \ottnt{T} $, and
    \item $ \mathit{z}    \not=    \mathit{y} $.
   \end{itemize}

   Because $ \emptyset   \vdash  \ottnt{C'}  \ottsym{<:}   \ottnt{T}  \,  \,\&\,  \,  \Phi  \, / \,   ( \forall  \mathit{y}  .  \ottnt{C_{{\mathrm{1}}}}  )  \Rightarrow   \ottnt{C_{{\mathrm{2}}}}  $ can be derived only by \Srule{Comp},
   its inversion implies
   \begin{itemize}
    \item $ \emptyset   \vdash   \ottnt{C'}  . \ottnt{T}   \ottsym{<:}  \ottnt{T}$,
    \item $ \emptyset   \vdash   \ottnt{C'}  . \Phi   \ottsym{<:}  \Phi$, and
    \item $ \emptyset  \,  |  \,  \ottnt{C'}  . \ottnt{T}  \,  \,\&\,  \,  \ottnt{C'}  . \Phi   \vdash   ( \forall  \mathit{y}  .  \ottnt{C'_{{\mathrm{1}}}}  )  \Rightarrow   \ottnt{C'_{{\mathrm{2}}}}   \ottsym{<:}   ( \forall  \mathit{y}  .  \ottnt{C_{{\mathrm{1}}}}  )  \Rightarrow   \ottnt{C_{{\mathrm{2}}}} $.
   \end{itemize}
   Further, because the last subtyping judgment can be derived only by \Srule{Ans},
   the inversion implies
   \begin{itemize}
    \item $\mathit{y} \,  {:}  \,  \ottnt{C'}  . \ottnt{T}   \vdash  \ottnt{C_{{\mathrm{1}}}}  \ottsym{<:}  \ottnt{C'_{{\mathrm{1}}}}$ and
    \item $ \emptyset   \vdash  \ottnt{C'_{{\mathrm{2}}}}  \ottsym{<:}  \ottnt{C_{{\mathrm{2}}}}$.
   \end{itemize}

   We first show that
   \[
    \mathit{x} \,  {:}  \, \ottsym{(}  \mathit{z} \,  {:}  \, \ottnt{T_{{\mathrm{0}}}}  \ottsym{)}  \rightarrow  \ottnt{C_{{\mathrm{0}}}}  \vdash  \ottnt{e}  \ottsym{:}  \ottnt{C_{{\mathrm{2}}}} ~,
   \]
   which is derived by \T{Sub} with $\mathit{x} \,  {:}  \, \ottsym{(}  \mathit{z} \,  {:}  \, \ottnt{T_{{\mathrm{0}}}}  \ottsym{)}  \rightarrow  \ottnt{C_{{\mathrm{0}}}}  \vdash  \ottnt{e}  \ottsym{:}  \ottnt{C'_{{\mathrm{2}}}}$ and the following.
   \begin{itemize}
    \item $\mathit{x} \,  {:}  \, \ottsym{(}  \mathit{z} \,  {:}  \, \ottnt{T_{{\mathrm{0}}}}  \ottsym{)}  \rightarrow  \ottnt{C_{{\mathrm{0}}}}  \vdash  \ottnt{C'_{{\mathrm{2}}}}  \ottsym{<:}  \ottnt{C_{{\mathrm{2}}}}$ by
          \reflem{ts-weak-subtyping} with $ \emptyset   \vdash  \ottnt{C'_{{\mathrm{2}}}}  \ottsym{<:}  \ottnt{C_{{\mathrm{2}}}}$.
    \item $\mathit{x} \,  {:}  \, \ottsym{(}  \mathit{z} \,  {:}  \, \ottnt{T_{{\mathrm{0}}}}  \ottsym{)}  \rightarrow  \ottnt{C_{{\mathrm{0}}}}  \vdash  \ottnt{C_{{\mathrm{2}}}}$, which is proven as follows.
          Because \reflem{ts-wf-ty-tctx-typing} with
          $ \emptyset   \vdash  \mathcal{S}_0 \, \mathit{x}  \ottsym{.}  \ottnt{e}  \ottsym{:}   \ottnt{T}  \,  \,\&\,  \,  \Phi  \, / \,   ( \forall  \mathit{y}  .  \ottnt{C_{{\mathrm{1}}}}  )  \Rightarrow   \ottnt{C_{{\mathrm{2}}}}  $ implies
          $ \emptyset   \vdash   \ottnt{T}  \,  \,\&\,  \,  \Phi  \, / \,   ( \forall  \mathit{y}  .  \ottnt{C_{{\mathrm{1}}}}  )  \Rightarrow   \ottnt{C_{{\mathrm{2}}}}  $,
          its inversion implies $ \emptyset   \vdash  \ottnt{C_{{\mathrm{2}}}}$.
          Because \reflem{ts-wf-ty-tctx-typing} with $\mathit{x} \,  {:}  \, \ottsym{(}  \mathit{z} \,  {:}  \, \ottnt{T_{{\mathrm{0}}}}  \ottsym{)}  \rightarrow  \ottnt{C_{{\mathrm{0}}}}  \vdash  \ottnt{e}  \ottsym{:}  \ottnt{C'_{{\mathrm{2}}}}$ implies
          $\mathit{x} \,  {:}  \, \ottsym{(}  \mathit{z} \,  {:}  \, \ottnt{T_{{\mathrm{0}}}}  \ottsym{)}  \rightarrow  \ottnt{C_{{\mathrm{0}}}}  \vdash  \ottnt{C'_{{\mathrm{2}}}}$,
          we have $\vdash   \emptyset   \ottsym{,}  \mathit{x} \,  {:}  \, \ottsym{(}  \mathit{z} \,  {:}  \, \ottnt{T_{{\mathrm{0}}}}  \ottsym{)}  \rightarrow  \ottnt{C_{{\mathrm{0}}}}$ by \reflem{ts-wf-tctx-wf-ftyping}.
          Thus, we have
          $\mathit{x} \,  {:}  \, \ottsym{(}  \mathit{z} \,  {:}  \, \ottnt{T_{{\mathrm{0}}}}  \ottsym{)}  \rightarrow  \ottnt{C_{{\mathrm{0}}}}  \vdash  \ottnt{C_{{\mathrm{2}}}}$ by \reflem{ts-weak-wf-ftyping}.
   \end{itemize}

   Next, we show that
   \[
     \emptyset   \vdash  \ottnt{T_{{\mathrm{0}}}}  \ottsym{<:}  \ottnt{T} ~.
   \]
   It is proven by \reflem{ts-transitivity-subtyping} with
   $ \emptyset   \vdash  \ottnt{T_{{\mathrm{0}}}}  \ottsym{<:}   \ottnt{C'}  . \ottnt{T} $ and $ \emptyset   \vdash   \ottnt{C'}  . \ottnt{T}   \ottsym{<:}  \ottnt{T}$.

   Finally, we show that
   \[
    \mathit{z} \,  {:}  \, \ottnt{T_{{\mathrm{0}}}}  \vdash  \mathit{z}  \ottsym{:}   \ottnt{T}  \,  \,\&\,  \,  \Phi  \, / \,   ( \forall  \mathit{y}  .   \ottnt{C_{{\mathrm{1}}}}    [  \mathit{z}  /  \mathit{y}  ]    )  \Rightarrow   \ottnt{C_{{\mathrm{0}}}}   ~.
   \]
   By \T{Sub} with $\mathit{z} \,  {:}  \, \ottnt{T_{{\mathrm{0}}}}  \vdash  \mathit{z}  \ottsym{:}    \ottnt{C'}  . \ottnt{T}   \,  \,\&\,  \,   \ottnt{C'}  . \Phi   \, / \,   ( \forall  \mathit{y}  .   \ottnt{C'_{{\mathrm{1}}}}    [  \mathit{z}  /  \mathit{y}  ]    )  \Rightarrow   \ottnt{C_{{\mathrm{0}}}}  $,
   it suffices to show the following.
   \begin{itemize}
    \item We show that
          \[
           \mathit{z} \,  {:}  \, \ottnt{T_{{\mathrm{0}}}}  \vdash    \ottnt{C'}  . \ottnt{T}   \,  \,\&\,  \,   \ottnt{C'}  . \Phi   \, / \,   ( \forall  \mathit{y}  .   \ottnt{C'_{{\mathrm{1}}}}    [  \mathit{z}  /  \mathit{y}  ]    )  \Rightarrow   \ottnt{C_{{\mathrm{0}}}}    \ottsym{<:}   \ottnt{T}  \,  \,\&\,  \,  \Phi  \, / \,   ( \forall  \mathit{y}  .   \ottnt{C_{{\mathrm{1}}}}    [  \mathit{z}  /  \mathit{y}  ]    )  \Rightarrow   \ottnt{C_{{\mathrm{0}}}}   ~.
          \]
          Because $ \emptyset   \vdash   \ottnt{C'}  . \ottnt{T}   \ottsym{<:}  \ottnt{T}$ and $ \emptyset   \vdash   \ottnt{C'}  . \Phi   \ottsym{<:}  \Phi$,
          \reflem{ts-weak-subtyping} implies
          $\mathit{z} \,  {:}  \, \ottnt{T_{{\mathrm{0}}}}  \vdash   \ottnt{C'}  . \ottnt{T}   \ottsym{<:}  \ottnt{T}$ and $\mathit{z} \,  {:}  \, \ottnt{T_{{\mathrm{0}}}}  \vdash   \ottnt{C'}  . \Phi   \ottsym{<:}  \Phi$.
          Thus, by \Srule{Comp}, it suffices to show that
          \[
           \mathit{z} \,  {:}  \, \ottnt{T_{{\mathrm{0}}}} \,  |  \,  \ottnt{C'}  . \ottnt{T}  \,  \,\&\,  \,  \ottnt{C'}  . \Phi   \vdash   ( \forall  \mathit{y}  .   \ottnt{C'_{{\mathrm{1}}}}    [  \mathit{z}  /  \mathit{y}  ]    )  \Rightarrow   \ottnt{C_{{\mathrm{0}}}}   \ottsym{<:}   ( \forall  \mathit{y}  .   \ottnt{C_{{\mathrm{1}}}}    [  \mathit{z}  /  \mathit{y}  ]    )  \Rightarrow   \ottnt{C_{{\mathrm{0}}}}  ~.
          \]
          Because $\vdash   \emptyset   \ottsym{,}  \mathit{x} \,  {:}  \, \ottsym{(}  \mathit{z} \,  {:}  \, \ottnt{T_{{\mathrm{0}}}}  \ottsym{)}  \rightarrow  \ottnt{C_{{\mathrm{0}}}}$, which has been proven above,
          its inversion implies $\mathit{z} \,  {:}  \, \ottnt{T_{{\mathrm{0}}}}  \vdash  \ottnt{C_{{\mathrm{0}}}}$.
          Thus, \reflem{ts-refl-subtyping} implies $\mathit{z} \,  {:}  \, \ottnt{T_{{\mathrm{0}}}}  \vdash  \ottnt{C_{{\mathrm{0}}}}  \ottsym{<:}  \ottnt{C_{{\mathrm{0}}}}$.
          Thus, by \Srule{Ans}, it suffices to show that
          \[
           \mathit{z} \,  {:}  \, \ottnt{T_{{\mathrm{0}}}}  \ottsym{,}  \mathit{y} \,  {:}  \,  \ottnt{C'}  . \ottnt{T}   \vdash   \ottnt{C_{{\mathrm{1}}}}    [  \mathit{z}  /  \mathit{y}  ]    \ottsym{<:}   \ottnt{C'_{{\mathrm{1}}}}    [  \mathit{z}  /  \mathit{y}  ]   ~.
          \]
          Because
          \begin{itemize}
           \item $\mathit{y} \,  {:}  \,  \ottnt{C'}  . \ottnt{T}   \vdash  \ottnt{C_{{\mathrm{1}}}}  \ottsym{<:}  \ottnt{C'_{{\mathrm{1}}}}$ and
           \item $\vdash   \emptyset   \ottsym{,}  \mathit{z} \,  {:}  \,  \ottnt{C'}  . \ottnt{T} $
                 by Lemmas~\ref{lem:ts-wf-ty-tctx-typing} and \ref{lem:ts-wf-tctx-wf-ftyping} with $ \emptyset   \vdash  \mathcal{S}_0 \, \mathit{x}  \ottsym{.}  \ottnt{e}  \ottsym{:}  \ottnt{C'}$,
          \end{itemize}
          we have $\mathit{z} \,  {:}  \,  \ottnt{C'}  . \ottnt{T}   \vdash   \ottnt{C_{{\mathrm{1}}}}    [  \mathit{z}  /  \mathit{y}  ]    \ottsym{<:}   \ottnt{C'_{{\mathrm{1}}}}    [  \mathit{z}  /  \mathit{y}  ]  $
          by Lemmas~\ref{lem:ts-weak-subtyping}, \ref{lem:ts-typing-vars}, and \ref{lem:ts-val-subst-subtyping}.
          Then, by \reflem{ts-strength-ty-hypo-subtyping} with
          $ \emptyset   \vdash  \ottnt{T_{{\mathrm{0}}}}  \ottsym{<:}   \ottnt{C'}  . \ottnt{T} $,
          we have $\mathit{z} \,  {:}  \, \ottnt{T_{{\mathrm{0}}}}  \vdash   \ottnt{C_{{\mathrm{1}}}}    [  \mathit{z}  /  \mathit{y}  ]    \ottsym{<:}   \ottnt{C'_{{\mathrm{1}}}}    [  \mathit{z}  /  \mathit{y}  ]  $.
          By \reflem{ts-weak-subtyping}, we have the conclusion.

    \item We show that
          \[
           \mathit{z} \,  {:}  \, \ottnt{T_{{\mathrm{0}}}}  \vdash   \ottnt{T}  \,  \,\&\,  \,  \Phi  \, / \,   ( \forall  \mathit{y}  .   \ottnt{C_{{\mathrm{1}}}}    [  \mathit{z}  /  \mathit{y}  ]    )  \Rightarrow   \ottnt{C_{{\mathrm{0}}}}   ~.
          \]
          We already have $\mathit{z} \,  {:}  \, \ottnt{T_{{\mathrm{0}}}}  \vdash  \ottnt{C_{{\mathrm{0}}}}$.
          By \reflem{ts-wf-tctx-wf-ftyping}, $\vdash   \emptyset   \ottsym{,}  \mathit{z} \,  {:}  \, \ottnt{T_{{\mathrm{0}}}}$, so $ \emptyset   \vdash  \ottnt{T_{{\mathrm{0}}}}$.
          Because we already have $ \emptyset   \vdash   \ottnt{T}  \,  \,\&\,  \,  \Phi  \, / \,   ( \forall  \mathit{y}  .  \ottnt{C_{{\mathrm{1}}}}  )  \Rightarrow   \ottnt{C_{{\mathrm{2}}}}  $,
          its inversion implies
          $ \emptyset   \vdash  \ottnt{T}$ and $ \emptyset   \vdash  \Phi$ and $\mathit{y} \,  {:}  \, \ottnt{T}  \vdash  \ottnt{C_{{\mathrm{1}}}}$.
          By \reflem{ts-weak-wf-ftyping} with $ \emptyset   \vdash  \ottnt{T_{{\mathrm{0}}}}$,
          $\mathit{z} \,  {:}  \, \ottnt{T_{{\mathrm{0}}}}  \vdash  \ottnt{T}$ and $\mathit{z} \,  {:}  \, \ottnt{T_{{\mathrm{0}}}}  \vdash  \Phi$.
          Thus, by \Srule{Comp}, it suffices to show that
          \[
           \mathit{z} \,  {:}  \, \ottnt{T_{{\mathrm{0}}}}  \ottsym{,}  \mathit{y} \,  {:}  \, \ottnt{T}  \vdash   \ottnt{C_{{\mathrm{1}}}}    [  \mathit{z}  /  \mathit{y}  ]   ~.
          \]
          \reflem{ts-weak-strength-ty-hypo-wf-ftyping} with $ \emptyset   \vdash  \ottnt{T_{{\mathrm{0}}}}$ and $ \emptyset   \vdash  \ottnt{T_{{\mathrm{0}}}}  \ottsym{<:}  \ottnt{T}$ and $\mathit{y} \,  {:}  \, \ottnt{T}  \vdash  \ottnt{C_{{\mathrm{1}}}}$ implies $\mathit{y} \,  {:}  \, \ottnt{T_{{\mathrm{0}}}}  \vdash  \ottnt{C_{{\mathrm{1}}}}$.
          Then, by Lemmas~\ref{lem:ts-weak-subtyping}, \ref{lem:ts-typing-vars}, and \ref{lem:ts-val-subst-subtyping},
          $\mathit{z} \,  {:}  \, \ottnt{T_{{\mathrm{0}}}}  \vdash   \ottnt{C_{{\mathrm{1}}}}    [  \mathit{z}  /  \mathit{y}  ]  $, and then \reflem{ts-weak-wf-ftyping} implies the conclusion.
   \end{itemize}

  \case \T{Shift}:
   We are given
   \begin{prooftree}
    \AxiomC{$\mathit{x} \,  {:}  \, \ottsym{(}  \mathit{y} \,  {:}  \, \ottnt{T}  \ottsym{)}  \rightarrow  \ottnt{C_{{\mathrm{1}}}}  \vdash  \ottnt{e}  \ottsym{:}  \ottnt{C_{{\mathrm{2}}}}$}
    \UnaryInfC{$ \emptyset   \vdash  \mathcal{S}_0 \, \mathit{x}  \ottsym{.}  \ottnt{e}  \ottsym{:}   \ottnt{T}  \,  \,\&\,  \,   \Phi_\mathsf{val}   \, / \,   ( \forall  \mathit{y}  .  \ottnt{C_{{\mathrm{1}}}}  )  \Rightarrow   \ottnt{C_{{\mathrm{2}}}}  $}
   \end{prooftree}
   where $\Phi \,  =  \,  \Phi_\mathsf{val} $.
   Let $\mathit{z}$ be a variable such that $ \mathit{z}    \not=    \mathit{y} $.
   Further, let $\ottnt{T_{{\mathrm{0}}}} \,  =  \, \ottnt{T}$, and $\ottnt{C_{{\mathrm{0}}}} \,  =  \,  \ottnt{C_{{\mathrm{1}}}}    [  \mathit{z}  /  \mathit{y}  ]  $.
   Because $\mathit{x} \,  {:}  \, \ottsym{(}  \mathit{y} \,  {:}  \, \ottnt{T}  \ottsym{)}  \rightarrow  \ottnt{C_{{\mathrm{1}}}}  \vdash  \ottnt{e}  \ottsym{:}  \ottnt{C_{{\mathrm{2}}}}$,
   we have
   \[
     \mathit{x} \,  {:}  \, \ottsym{(}  \mathit{z} \,  {:}  \, \ottnt{T}  \ottsym{)}  \rightarrow  \ottnt{C_{{\mathrm{1}}}}    [  \mathit{z}  /  \mathit{y}  ]    \vdash  \ottnt{e}  \ottsym{:}  \ottnt{C_{{\mathrm{2}}}} ~,
   \]
   which is equivalent to $\mathit{x} \,  {:}  \, \ottsym{(}  \mathit{z} \,  {:}  \, \ottnt{T_{{\mathrm{0}}}}  \ottsym{)}  \rightarrow  \ottnt{C_{{\mathrm{0}}}}  \vdash  \ottnt{e}  \ottsym{:}  \ottnt{C_{{\mathrm{2}}}}$.

   Next, we show that
   \[
     \emptyset   \vdash  \ottnt{T_{{\mathrm{0}}}}  \ottsym{<:}  \ottnt{T} ~,
   \]
   i.e.,
   $ \emptyset   \vdash  \ottnt{T}  \ottsym{<:}  \ottnt{T}$.
   By \reflem{ts-refl-subtyping}, it suffices to show that $ \emptyset   \vdash  \ottnt{T}$.
   \reflem{ts-wf-ty-tctx-typing} with $ \emptyset   \vdash  \mathcal{S}_0 \, \mathit{x}  \ottsym{.}  \ottnt{e}  \ottsym{:}   \ottnt{T}  \,  \,\&\,  \,   \Phi_\mathsf{val}   \, / \,   ( \forall  \mathit{y}  .  \ottnt{C_{{\mathrm{1}}}}  )  \Rightarrow   \ottnt{C_{{\mathrm{2}}}}  $
   implies $ \emptyset   \vdash   \ottnt{T}  \,  \,\&\,  \,   \Phi_\mathsf{val}   \, / \,   ( \forall  \mathit{y}  .  \ottnt{C_{{\mathrm{1}}}}  )  \Rightarrow   \ottnt{C_{{\mathrm{2}}}}  $, which further implies
   $ \emptyset   \vdash  \ottnt{T}$ by inversion.

   Finally, we show that $\mathit{z} \,  {:}  \, \ottnt{T_{{\mathrm{0}}}}  \vdash  \mathit{z}  \ottsym{:}   \ottnt{T}  \,  \,\&\,  \,   \Phi_\mathsf{val}   \, / \,   ( \forall  \mathit{y}  .   \ottnt{C_{{\mathrm{1}}}}    [  \mathit{z}  /  \mathit{y}  ]    )  \Rightarrow   \ottnt{C_{{\mathrm{0}}}}  $, i.e.,
   \[
    \mathit{z} \,  {:}  \, \ottnt{T}  \vdash  \mathit{z}  \ottsym{:}    \ottnt{T}  \,  \,\&\,  \,   \Phi_\mathsf{val}   \, / \,   ( \forall  \mathit{y}  .   \ottnt{C_{{\mathrm{1}}}}    [  \mathit{z}  /  \mathit{y}  ]    )  \Rightarrow   \ottnt{C_{{\mathrm{1}}}}      [  \mathit{z}  /  \mathit{y}  ]   ~.
   \]
   By \T{Sub} with the following, we have the conclusion.
   \begin{itemize}
    \item We have $\mathit{z} \,  {:}  \, \ottnt{T}  \vdash  \mathit{z}  \ottsym{:}   \ottnt{T}  \,  \,\&\,  \,   \Phi_\mathsf{val}   \, / \,   \square  $
          by \reflem{ts-typing-vars} with $\vdash   \emptyset   \ottsym{,}  \mathit{z} \,  {:}  \, \ottnt{T}$, which is proven
          by \WFC{Var} with $ \emptyset   \vdash  \ottnt{T}$ shown above.

    \item We show that
          \[
           \mathit{z} \,  {:}  \, \ottnt{T}  \vdash   \ottnt{T}  \,  \,\&\,  \,   \Phi_\mathsf{val}   \, / \,   \square    \ottsym{<:}    \ottnt{T}  \,  \,\&\,  \,   \Phi_\mathsf{val}   \, / \,   ( \forall  \mathit{y}  .   \ottnt{C_{{\mathrm{1}}}}    [  \mathit{z}  /  \mathit{y}  ]    )  \Rightarrow   \ottnt{C_{{\mathrm{1}}}}      [  \mathit{z}  /  \mathit{y}  ]   ~.
          \]
          Because $ \emptyset   \vdash  \ottnt{T}$,
          we have $\mathit{z} \,  {:}  \, \ottnt{T}  \vdash  \ottnt{T}  \ottsym{<:}  \ottnt{T}$
          by Lemmas~\ref{lem:ts-weak-wf-ftyping} and \ref{lem:ts-refl-subtyping},
          and $\mathit{z} \,  {:}  \, \ottnt{T}  \vdash   \Phi_\mathsf{val}   \ottsym{<:}   \Phi_\mathsf{val} $ by Lemmas~\ref{lem:ts-wf-TP_val} and \ref{lem:ts-refl-subtyping}.
          Then, by \Srule{Comp}, it suffices to show that
          \[
           \mathit{z} \,  {:}  \, \ottnt{T} \,  |  \, \ottnt{T} \,  \,\&\,  \,  \Phi_\mathsf{val}   \vdash   \square   \ottsym{<:}    ( \forall  \mathit{y}  .   \ottnt{C_{{\mathrm{1}}}}    [  \mathit{z}  /  \mathit{y}  ]    )  \Rightarrow   \ottnt{C_{{\mathrm{1}}}}     [  \mathit{z}  /  \mathit{y}  ]   ~,
          \]
          which is implied by \Srule{Embed} with the following.
          \begin{itemize}
           \item We show that
                 \[
                  \mathit{z} \,  {:}  \, \ottnt{T}  \ottsym{,}  \mathit{y} \,  {:}  \, \ottnt{T}  \vdash    \Phi_\mathsf{val}  \,  \tceappendop  \, \ottnt{C_{{\mathrm{1}}}}    [  \mathit{z}  /  \mathit{y}  ]    \ottsym{<:}   \ottnt{C_{{\mathrm{1}}}}    [  \mathit{z}  /  \mathit{y}  ]    ~.
                 \]
                 By \reflem{ts-add-TP_val-subty}, it suffices to show that
                 \[
                  \mathit{z} \,  {:}  \, \ottnt{T}  \ottsym{,}  \mathit{y} \,  {:}  \, \ottnt{T}  \vdash   \ottnt{C_{{\mathrm{1}}}}    [  \mathit{z}  /  \mathit{y}  ]    \ottsym{<:}   \ottnt{C_{{\mathrm{1}}}}    [  \mathit{z}  /  \mathit{y}  ]   ~.
                 \]
                 By \reflem{ts-refl-subtyping}, it suffices to show that
                 \[
                  \mathit{z} \,  {:}  \, \ottnt{T}  \ottsym{,}  \mathit{y} \,  {:}  \, \ottnt{T}  \vdash   \ottnt{C_{{\mathrm{1}}}}    [  \mathit{z}  /  \mathit{y}  ]   ~.
                 \]
                 \reflem{ts-wf-ty-tctx-typing} with $ \emptyset   \vdash  \mathcal{S}_0 \, \mathit{x}  \ottsym{.}  \ottnt{e}  \ottsym{:}   \ottnt{T}  \,  \,\&\,  \,  \Phi  \, / \,   ( \forall  \mathit{y}  .  \ottnt{C_{{\mathrm{1}}}}  )  \Rightarrow   \ottnt{C_{{\mathrm{2}}}}  $ implies
                 $ \emptyset   \vdash   \ottnt{T}  \,  \,\&\,  \,  \Phi  \, / \,   ( \forall  \mathit{y}  .  \ottnt{C_{{\mathrm{1}}}}  )  \Rightarrow   \ottnt{C_{{\mathrm{2}}}}  $, which further implies
                 $\mathit{y} \,  {:}  \, \ottnt{T}  \vdash  \ottnt{C_{{\mathrm{1}}}}$ by inversion.
                 By Lemmas~\ref{lem:ts-weak-wf-ftyping}, \ref{lem:ts-typing-vars}, and \ref{lem:ts-val-subst-wf-ftyping},
                 $\mathit{z} \,  {:}  \, \ottnt{T}  \vdash   \ottnt{C_{{\mathrm{1}}}}    [  \mathit{z}  /  \mathit{y}  ]  $.
                 Therefore, we have the conclusion $\mathit{z} \,  {:}  \, \ottnt{T}  \ottsym{,}  \mathit{y} \,  {:}  \, \ottnt{T}  \vdash   \ottnt{C_{{\mathrm{1}}}}    [  \mathit{z}  /  \mathit{y}  ]  $
                 by \reflem{ts-weak-wf-ftyping}.

           \item We show that
                 \[
                  \mathit{z} \,  {:}  \, \ottnt{T}  \ottsym{,}  \mathit{y} \,  {:}  \,  \Sigma ^\omega   \models    { \ottsym{(}   \Phi_\mathsf{val}   \ottsym{)} }^\nu   \ottsym{(}  \mathit{y}  \ottsym{)}    \Rightarrow     { \ottsym{(}   \ottnt{C_{{\mathrm{1}}}}    [  \mathit{z}  /  \mathit{y}  ]    \ottsym{)} }^\nu (  \mathit{y}  )   ~.
                 \]
                 Because $ { \ottsym{(}   \Phi_\mathsf{val}   \ottsym{)} }^\nu   \ottsym{(}  \mathit{y}  \ottsym{)} \,  =  \,  \bot $, it suffices to show that
                 \[
                  \mathit{z} \,  {:}  \, \ottnt{T}  \ottsym{,}  \mathit{y} \,  {:}  \,  \Sigma ^\omega   \models    \bot     \Rightarrow     { \ottsym{(}   \ottnt{C_{{\mathrm{1}}}}    [  \mathit{z}  /  \mathit{y}  ]    \ottsym{)} }^\nu (  \mathit{y}  )   ~,
                 \]
                 which is implied by \reflem{ts-valid-intro-bot-impl}.
          \end{itemize}

    \item We show that $\mathit{z} \,  {:}  \, \ottnt{T}  \vdash    \ottnt{T}  \,  \,\&\,  \,   \Phi_\mathsf{val}   \, / \,   ( \forall  \mathit{y}  .   \ottnt{C_{{\mathrm{1}}}}    [  \mathit{z}  /  \mathit{y}  ]    )  \Rightarrow   \ottnt{C_{{\mathrm{1}}}}      [  \mathit{z}  /  \mathit{y}  ]  $,
          which is proven by \WFT{Comp}, \WFT{Ans}, and the following.
          \begin{itemize}
           \item $\mathit{z} \,  {:}  \, \ottnt{T}  \vdash  \ottnt{T}$ by \reflem{ts-weak-wf-ftyping} with $ \emptyset   \vdash  \ottnt{T}$
                 which has been proven.
           \item $\mathit{z} \,  {:}  \, \ottnt{T}  \vdash   \Phi_\mathsf{val} $ by \reflem{ts-wf-TP_val} with $\vdash   \emptyset   \ottsym{,}  \mathit{z} \,  {:}  \, \ottnt{T}$.
           \item $\mathit{z} \,  {:}  \, \ottnt{T}  \vdash   \ottnt{C_{{\mathrm{1}}}}    [  \mathit{z}  /  \mathit{y}  ]  $, which has been proven.
           \item $\mathit{z} \,  {:}  \, \ottnt{T}  \ottsym{,}  \mathit{y} \,  {:}  \, \ottnt{T}  \vdash   \ottnt{C_{{\mathrm{1}}}}    [  \mathit{z}  /  \mathit{y}  ]  $, which has been proven.
          \end{itemize}
   \end{itemize}

  \case \T{ShiftD}: By inversion.

  \otherwise: Contradictory.
 \end{caseanalysis}
\end{proof}

\begin{lemmap}{Removing Pre-Effect $ \Phi_\mathsf{val} $ from Subtypes}{ts-remove-TP_val-subty}
 \begin{enumerate}
  \item If $\Gamma  \vdash   \Phi_\mathsf{val}  \,  \tceappendop  \, \ottnt{C_{{\mathrm{1}}}}  \ottsym{<:}  \ottnt{C_{{\mathrm{2}}}}$, then $\Gamma  \vdash  \ottnt{C_{{\mathrm{1}}}}  \ottsym{<:}  \ottnt{C_{{\mathrm{2}}}}$.
  \item If $\Gamma \,  |  \, \ottnt{T} \,  \,\&\,  \,  \Phi_\mathsf{val}  \,  \tceappendop  \, \Phi  \vdash   \Phi_\mathsf{val}  \,  \tceappendop  \, \ottnt{S_{{\mathrm{1}}}}  \ottsym{<:}  \ottnt{S_{{\mathrm{2}}}}$, then $\Gamma \,  |  \, \ottnt{T} \,  \,\&\,  \, \Phi  \vdash  \ottnt{S_{{\mathrm{1}}}}  \ottsym{<:}  \ottnt{S_{{\mathrm{2}}}}$
 \end{enumerate}
\end{lemmap}
\begin{proof}
 By simultaneous induction on the total numbers of outermost type constructors of $(\ottnt{C_{{\mathrm{1}}}},\ottnt{C_{{\mathrm{2}}}})$ and those of $(\ottnt{S_{{\mathrm{1}}}},\ottnt{S_{{\mathrm{2}}}})$.
 \begin{caseanalysis}
  \case \Srule{Comp}:
   We are given
   \begin{prooftree}
    \AxiomC{$\Gamma  \vdash   \ottnt{C_{{\mathrm{1}}}}  . \ottnt{T}   \ottsym{<:}   \ottnt{C_{{\mathrm{2}}}}  . \ottnt{T} $}
    \AxiomC{$\Gamma  \vdash   \Phi_\mathsf{val}  \,  \tceappendop  \, \ottsym{(}   \ottnt{C_{{\mathrm{1}}}}  . \Phi   \ottsym{)}  \ottsym{<:}   \ottnt{C_{{\mathrm{2}}}}  . \Phi $}
    \AxiomC{$\Gamma \,  |  \,  \ottnt{C_{{\mathrm{1}}}}  . \ottnt{T}  \,  \,\&\,  \,  \Phi_\mathsf{val}  \,  \tceappendop  \, \ottsym{(}   \ottnt{C_{{\mathrm{1}}}}  . \Phi   \ottsym{)}  \vdash   \Phi_\mathsf{val}  \,  \tceappendop  \, \ottsym{(}   \ottnt{C_{{\mathrm{1}}}}  . \ottnt{S}   \ottsym{)}  \ottsym{<:}   \ottnt{C_{{\mathrm{2}}}}  . \ottnt{S} $}
    \TrinaryInfC{$\Gamma  \vdash   \Phi_\mathsf{val}  \,  \tceappendop  \, \ottnt{C_{{\mathrm{1}}}}  \ottsym{<:}  \ottnt{C_{{\mathrm{2}}}}$}
   \end{prooftree}
   We have the conclusion by the IH, Lemmas~\ref{lem:ts-weak-strength-TP_val} and \ref{lem:ts-transitivity-subtyping} with $\Gamma  \vdash   \Phi_\mathsf{val}  \,  \tceappendop  \, \ottsym{(}   \ottnt{C_{{\mathrm{1}}}}  . \Phi   \ottsym{)}  \ottsym{<:}   \ottnt{C_{{\mathrm{2}}}}  . \Phi $, and \Srule{Comp}.

  \case \Srule{Empty}: Trivial by \Srule{Empty}
   because $\ottnt{S_{{\mathrm{1}}}} = \ottnt{S_{{\mathrm{2}}}} =  \Phi_\mathsf{val}  \,  \tceappendop  \, \ottnt{S_{{\mathrm{1}}}} =  \square $.

  \case \Srule{Ans}: By the IH and \Srule{Ans}.

  \case \Srule{Embed}:
   We are given
   \begin{prooftree}
    \AxiomC{$\Gamma  \ottsym{,}  \mathit{x} \,  {:}  \, \ottnt{T}  \vdash  \ottsym{(}   \Phi_\mathsf{val}  \,  \tceappendop  \, \Phi  \ottsym{)} \,  \tceappendop  \, \ottnt{C_{{\mathrm{21}}}}  \ottsym{<:}  \ottnt{C_{{\mathrm{22}}}}$}
    \AxiomC{$\Gamma  \ottsym{,}  \mathit{y} \,  {:}  \,  \Sigma ^\omega   \models    { \ottsym{(}   \Phi_\mathsf{val}  \,  \tceappendop  \, \Phi  \ottsym{)} }^\nu   \ottsym{(}  \mathit{y}  \ottsym{)}    \Rightarrow     { \ottnt{C_{{\mathrm{22}}}} }^\nu (  \mathit{y}  )  $}
    \AxiomC{$\mathit{x}  \ottsym{,}  \mathit{y} \,  \not\in  \,  \mathit{fv}  (  \ottnt{C_{{\mathrm{22}}}}  )  \,  \mathbin{\cup}  \,  \mathit{fv}  (   \Phi_\mathsf{val}  \,  \tceappendop  \, \Phi  ) $}
    \TrinaryInfC{$\Gamma \,  |  \, \ottnt{T} \,  \,\&\,  \,  \Phi_\mathsf{val}  \,  \tceappendop  \, \Phi  \vdash   \square   \ottsym{<:}   ( \forall  \mathit{x}  .  \ottnt{C_{{\mathrm{21}}}}  )  \Rightarrow   \ottnt{C_{{\mathrm{22}}}} $}
   \end{prooftree}
   for some $\mathit{x}$, $\mathit{y}$, $\ottnt{C_{{\mathrm{21}}}}$, and $\ottnt{C_{{\mathrm{22}}}}$ such that
   $\ottnt{S_{{\mathrm{2}}}} \,  =  \,  ( \forall  \mathit{x}  .  \ottnt{C_{{\mathrm{21}}}}  )  \Rightarrow   \ottnt{C_{{\mathrm{22}}}} $.
   We also have $\ottnt{S_{{\mathrm{1}}}} \,  =  \,  \square $.
   By \Srule{Embed}, it suffices to show the following.
   \begin{itemize}
    \item We show that
          \[
           \Gamma  \ottsym{,}  \mathit{x} \,  {:}  \, \ottnt{T}  \vdash  \Phi \,  \tceappendop  \, \ottnt{C_{{\mathrm{21}}}}  \ottsym{<:}  \ottnt{C_{{\mathrm{22}}}} ~.
          \]
          By \reflem{ts-assoc-preeff-compose-subtyping} with $\Gamma  \ottsym{,}  \mathit{x} \,  {:}  \, \ottnt{T}  \vdash  \ottsym{(}   \Phi_\mathsf{val}  \,  \tceappendop  \, \Phi  \ottsym{)} \,  \tceappendop  \, \ottnt{C_{{\mathrm{21}}}}  \ottsym{<:}  \ottnt{C_{{\mathrm{22}}}}$, we have
          \[
           \Gamma  \ottsym{,}  \mathit{x} \,  {:}  \, \ottnt{T}  \vdash   \Phi_\mathsf{val}  \,  \tceappendop  \, \ottsym{(}  \Phi \,  \tceappendop  \, \ottnt{C_{{\mathrm{21}}}}  \ottsym{)}  \ottsym{<:}  \ottnt{C_{{\mathrm{22}}}} ~.
          \]
          Then, the IH implies the conclusion.
          Note that the total number of outermost type constructors of
          $(\Phi \,  \tceappendop  \, \ottnt{C_{{\mathrm{21}}}},\ottnt{C_{{\mathrm{22}}}})$ is smaller than that of
          $( \square ,  ( \forall  \mathit{x}  .  \ottnt{C_{{\mathrm{21}}}}  )  \Rightarrow   \ottnt{C_{{\mathrm{22}}}} )$.

    \item We show that
          \[
           \Gamma  \ottsym{,}  \mathit{y} \,  {:}  \,  \Sigma ^\omega   \models    { \Phi }^\nu   \ottsym{(}  \mathit{y}  \ottsym{)}    \Rightarrow     { \ottnt{C_{{\mathrm{22}}}} }^\nu (  \mathit{y}  )   ~.
          \]
          Because $\Gamma  \vdash  \Phi  \ottsym{<:}   \Phi_\mathsf{val}  \,  \tceappendop  \, \Phi$ by \reflem{ts-weak-strength-TP_val},
          we have $\Gamma  \ottsym{,}  \mathit{y} \,  {:}  \,  \Sigma ^\omega   \models    { \Phi }^\nu   \ottsym{(}  \mathit{y}  \ottsym{)}    \Rightarrow     { \ottsym{(}   \Phi_\mathsf{val}  \,  \tceappendop  \, \Phi  \ottsym{)} }^\nu   \ottsym{(}  \mathit{y}  \ottsym{)} $.
          Because $\Gamma  \ottsym{,}  \mathit{y} \,  {:}  \,  \Sigma ^\omega   \models    { \ottsym{(}   \Phi_\mathsf{val}  \,  \tceappendop  \, \Phi  \ottsym{)} }^\nu   \ottsym{(}  \mathit{y}  \ottsym{)}    \Rightarrow     { \ottnt{C_{{\mathrm{22}}}} }^\nu (  \mathit{y}  )  $,
          we have the conclusion by \reflem{ts-transitivity-valid}.
   \end{itemize}
 \end{caseanalysis}
\end{proof}

\begin{lemmap}{Inversion of Shift Expressions Under Evaluation Contexts}{ts-inv-shift-ectx}
 If $ \emptyset   \vdash   \ottnt{K}  [  \mathcal{S}_0 \, \mathit{x}  \ottsym{.}  \ottnt{e}  ]   \ottsym{:}   \ottnt{T}  \,  \,\&\,  \,  \Phi  \, / \,   ( \forall  \mathit{y}  .  \ottnt{C_{{\mathrm{1}}}}  )  \Rightarrow   \ottnt{C_{{\mathrm{2}}}}  $,
 then either of the following holds:
 \begin{itemize}
  \item there exist some $\mathit{z}$, $\ottnt{T_{{\mathrm{0}}}}$, and $\ottnt{C_{{\mathrm{0}}}}$ such that
        \begin{itemize}
         \item $\mathit{x} \,  {:}  \, \ottsym{(}  \mathit{z} \,  {:}  \, \ottnt{T_{{\mathrm{0}}}}  \ottsym{)}  \rightarrow  \ottnt{C_{{\mathrm{0}}}}  \vdash  \ottnt{e}  \ottsym{:}  \ottnt{C_{{\mathrm{2}}}}$ and
         \item $\mathit{z} \,  {:}  \, \ottnt{T_{{\mathrm{0}}}}  \vdash   \ottnt{K}  [  \mathit{z}  ]   \ottsym{:}   \ottnt{T}  \,  \,\&\,  \,  \Phi  \, / \,   ( \forall  \mathit{y}  .  \ottnt{C_{{\mathrm{1}}}}  )  \Rightarrow   \ottnt{C_{{\mathrm{0}}}}  $;
        \end{itemize}
        or
  \item $ \emptyset   \vdash  \ottnt{e}  \ottsym{:}  \ottnt{C_{{\mathrm{2}}}}$.
 \end{itemize}
\end{lemmap}
\begin{proof}
 By induction on the typing derivation of $ \emptyset   \vdash   \ottnt{K}  [  \mathcal{S}_0 \, \mathit{x}  \ottsym{.}  \ottnt{e}  ]   \ottsym{:}   \ottnt{T}  \,  \,\&\,  \,  \Phi  \, / \,   ( \forall  \mathit{y}  .  \ottnt{C_{{\mathrm{1}}}}  )  \Rightarrow   \ottnt{C_{{\mathrm{2}}}}  $.
 \begin{caseanalysis}
  \case \T{Sub}:
   We are given
   \begin{prooftree}
    \AxiomC{$ \emptyset   \vdash   \ottnt{K}  [  \mathcal{S}_0 \, \mathit{x}  \ottsym{.}  \ottnt{e}  ]   \ottsym{:}  \ottnt{C'}$}
    \AxiomC{$ \emptyset   \vdash  \ottnt{C'}  \ottsym{<:}   \ottnt{T}  \,  \,\&\,  \,  \Phi  \, / \,   ( \forall  \mathit{y}  .  \ottnt{C_{{\mathrm{1}}}}  )  \Rightarrow   \ottnt{C_{{\mathrm{2}}}}  $}
    \AxiomC{$ \emptyset   \vdash   \ottnt{T}  \,  \,\&\,  \,  \Phi  \, / \,   ( \forall  \mathit{y}  .  \ottnt{C_{{\mathrm{1}}}}  )  \Rightarrow   \ottnt{C_{{\mathrm{2}}}}  $}
    \TrinaryInfC{$ \emptyset   \vdash   \ottnt{K}  [  \mathcal{S}_0 \, \mathit{x}  \ottsym{.}  \ottnt{e}  ]   \ottsym{:}   \ottnt{T}  \,  \,\&\,  \,  \Phi  \, / \,   ( \forall  \mathit{y}  .  \ottnt{C_{{\mathrm{1}}}}  )  \Rightarrow   \ottnt{C_{{\mathrm{2}}}}  $}
   \end{prooftree}
   for some $\ottnt{C'}$.
   \reflem{ts-inv-shift-sty} with $ \emptyset   \vdash   \ottnt{K}  [  \mathcal{S}_0 \, \mathit{x}  \ottsym{.}  \ottnt{e}  ]   \ottsym{:}  \ottnt{C'}$
   implies
   $ \ottnt{C'}  . \ottnt{S}  \,  =  \,  ( \forall  \mathit{y}  .  \ottnt{C'_{{\mathrm{1}}}}  )  \Rightarrow   \ottnt{C'_{{\mathrm{2}}}} $ for some $\ottnt{C'_{{\mathrm{1}}}}$ and $\ottnt{C'_{{\mathrm{2}}}}$.
   Thus, $ \emptyset   \vdash   \ottnt{K}  [  \mathcal{S}_0 \, \mathit{x}  \ottsym{.}  \ottnt{e}  ]   \ottsym{:}    \ottnt{C'}  . \ottnt{T}   \,  \,\&\,  \,   \ottnt{C'}  . \Phi   \, / \,   ( \forall  \mathit{y}  .  \ottnt{C'_{{\mathrm{1}}}}  )  \Rightarrow   \ottnt{C'_{{\mathrm{2}}}}  $.

   Assume that $ \emptyset   \vdash  \ottnt{e}  \ottsym{:}  \ottnt{C'_{{\mathrm{2}}}}$.
   Then, because
   $ \emptyset   \vdash  \ottnt{C'}  \ottsym{<:}   \ottnt{T}  \,  \,\&\,  \,  \Phi  \, / \,   ( \forall  \mathit{y}  .  \ottnt{C_{{\mathrm{1}}}}  )  \Rightarrow   \ottnt{C_{{\mathrm{2}}}}  $ and $ \ottnt{C'}  . \ottnt{S}  \,  =  \,  ( \forall  \mathit{y}  .  \ottnt{C'_{{\mathrm{1}}}}  )  \Rightarrow   \ottnt{C'_{{\mathrm{2}}}} $
   imply $ \emptyset   \vdash  \ottnt{C'_{{\mathrm{2}}}}  \ottsym{<:}  \ottnt{C_{{\mathrm{2}}}}$, and
   $ \emptyset   \vdash   \ottnt{T}  \,  \,\&\,  \,  \Phi  \, / \,   ( \forall  \mathit{y}  .  \ottnt{C_{{\mathrm{1}}}}  )  \Rightarrow   \ottnt{C_{{\mathrm{2}}}}  $ implies $ \emptyset   \vdash  \ottnt{C_{{\mathrm{2}}}}$,
   \T{Sub} implies $ \emptyset   \vdash  \ottnt{e}  \ottsym{:}  \ottnt{C_{{\mathrm{2}}}}$.
   Therefore, we have the conclusion.

   In what follows, we assume that
   $ \emptyset   \vdash  \ottnt{e}  \ottsym{:}  \ottnt{C_{{\mathrm{2}}}}$ does not hold.
   Then, by the IH, there exit some $\mathit{z}$, $\ottnt{T_{{\mathrm{0}}}}$, and $\ottnt{C_{{\mathrm{0}}}}$ such that
   \begin{itemize}
    \item $\mathit{x} \,  {:}  \, \ottsym{(}  \mathit{z} \,  {:}  \, \ottnt{T_{{\mathrm{0}}}}  \ottsym{)}  \rightarrow  \ottnt{C_{{\mathrm{0}}}}  \vdash  \ottnt{e}  \ottsym{:}  \ottnt{C'_{{\mathrm{2}}}}$ and
    \item $\mathit{z} \,  {:}  \, \ottnt{T_{{\mathrm{0}}}}  \vdash   \ottnt{K}  [  \mathit{z}  ]   \ottsym{:}    \ottnt{C'}  . \ottnt{T}   \,  \,\&\,  \,   \ottnt{C'}  . \Phi   \, / \,   ( \forall  \mathit{y}  .  \ottnt{C'_{{\mathrm{1}}}}  )  \Rightarrow   \ottnt{C_{{\mathrm{0}}}}  $.
   \end{itemize}

   Because $ \emptyset   \vdash  \ottnt{C'}  \ottsym{<:}   \ottnt{T}  \,  \,\&\,  \,  \Phi  \, / \,   ( \forall  \mathit{y}  .  \ottnt{C_{{\mathrm{1}}}}  )  \Rightarrow   \ottnt{C_{{\mathrm{2}}}}  $ can be derived only by \Srule{Comp},
   its inversion implies
   \begin{itemize}
    \item $ \emptyset   \vdash   \ottnt{C'}  . \ottnt{T}   \ottsym{<:}  \ottnt{T}$,
    \item $ \emptyset   \vdash   \ottnt{C'}  . \Phi   \ottsym{<:}  \Phi$, and
    \item $ \emptyset  \,  |  \,  \ottnt{C'}  . \ottnt{T}  \,  \,\&\,  \,  \ottnt{C'}  . \Phi   \vdash   ( \forall  \mathit{y}  .  \ottnt{C'_{{\mathrm{1}}}}  )  \Rightarrow   \ottnt{C'_{{\mathrm{2}}}}   \ottsym{<:}   ( \forall  \mathit{y}  .  \ottnt{C_{{\mathrm{1}}}}  )  \Rightarrow   \ottnt{C_{{\mathrm{2}}}} $.
   \end{itemize}
   Furthermore, because the last subtyping judgment can be derived only by \Srule{Ans},
   we have
   \begin{itemize}
    \item $\mathit{y} \,  {:}  \,  \ottnt{C'}  . \ottnt{T}   \vdash  \ottnt{C_{{\mathrm{1}}}}  \ottsym{<:}  \ottnt{C'_{{\mathrm{1}}}}$ and
    \item $ \emptyset   \vdash  \ottnt{C'_{{\mathrm{2}}}}  \ottsym{<:}  \ottnt{C_{{\mathrm{2}}}}$.
   \end{itemize}

   We first show that
   \[
    \mathit{x} \,  {:}  \, \ottsym{(}  \mathit{z} \,  {:}  \, \ottnt{T_{{\mathrm{0}}}}  \ottsym{)}  \rightarrow  \ottnt{C_{{\mathrm{0}}}}  \vdash  \ottnt{e}  \ottsym{:}  \ottnt{C_{{\mathrm{2}}}} ~,
   \]
   which is implied by \T{Sub} with $\mathit{x} \,  {:}  \, \ottsym{(}  \mathit{z} \,  {:}  \, \ottnt{T_{{\mathrm{0}}}}  \ottsym{)}  \rightarrow  \ottnt{C_{{\mathrm{0}}}}  \vdash  \ottnt{e}  \ottsym{:}  \ottnt{C'_{{\mathrm{2}}}}$ and the following.
   \begin{itemize}
    \item $\mathit{x} \,  {:}  \, \ottsym{(}  \mathit{z} \,  {:}  \, \ottnt{T_{{\mathrm{0}}}}  \ottsym{)}  \rightarrow  \ottnt{C_{{\mathrm{0}}}}  \vdash  \ottnt{C'_{{\mathrm{2}}}}  \ottsym{<:}  \ottnt{C_{{\mathrm{2}}}}$
          by \reflem{ts-weak-subtyping} with $ \emptyset   \vdash  \ottnt{C'_{{\mathrm{2}}}}  \ottsym{<:}  \ottnt{C_{{\mathrm{2}}}}$.
    \item $\mathit{x} \,  {:}  \, \ottsym{(}  \mathit{z} \,  {:}  \, \ottnt{T_{{\mathrm{0}}}}  \ottsym{)}  \rightarrow  \ottnt{C_{{\mathrm{0}}}}  \vdash  \ottnt{C_{{\mathrm{2}}}}$, which is proven as follows.
          Because \reflem{ts-wf-ty-tctx-typing} with
          $ \emptyset   \vdash   \ottnt{K}  [  \mathcal{S}_0 \, \mathit{x}  \ottsym{.}  \ottnt{e}  ]   \ottsym{:}   \ottnt{T}  \,  \,\&\,  \,  \Phi  \, / \,   ( \forall  \mathit{y}  .  \ottnt{C_{{\mathrm{1}}}}  )  \Rightarrow   \ottnt{C_{{\mathrm{2}}}}  $ implies
          $ \emptyset   \vdash   \ottnt{T}  \,  \,\&\,  \,  \Phi  \, / \,   ( \forall  \mathit{y}  .  \ottnt{C_{{\mathrm{1}}}}  )  \Rightarrow   \ottnt{C_{{\mathrm{2}}}}  $,
          its inversion implies $ \emptyset   \vdash  \ottnt{C_{{\mathrm{2}}}}$.
          Because \reflem{ts-wf-ty-tctx-typing} with $\mathit{x} \,  {:}  \, \ottsym{(}  \mathit{z} \,  {:}  \, \ottnt{T_{{\mathrm{0}}}}  \ottsym{)}  \rightarrow  \ottnt{C_{{\mathrm{0}}}}  \vdash  \ottnt{e}  \ottsym{:}  \ottnt{C'_{{\mathrm{2}}}}$ implies
          $\mathit{x} \,  {:}  \, \ottsym{(}  \mathit{z_{{\mathrm{0}}}} \,  {:}  \, \ottnt{T_{{\mathrm{0}}}}  \ottsym{)}  \rightarrow  \ottnt{C_{{\mathrm{0}}}}  \vdash  \ottnt{C'_{{\mathrm{2}}}}$, we have
          $\vdash   \emptyset   \ottsym{,}  \mathit{x} \,  {:}  \, \ottsym{(}  \mathit{z_{{\mathrm{0}}}} \,  {:}  \, \ottnt{T_{{\mathrm{0}}}}  \ottsym{)}  \rightarrow  \ottnt{C_{{\mathrm{0}}}}$ by \reflem{ts-wf-tctx-wf-ftyping}.
          Thus, we have
          $\mathit{x} \,  {:}  \, \ottsym{(}  \mathit{z} \,  {:}  \, \ottnt{T_{{\mathrm{0}}}}  \ottsym{)}  \rightarrow  \ottnt{C_{{\mathrm{0}}}}  \vdash  \ottnt{C_{{\mathrm{2}}}}$ by \reflem{ts-weak-wf-ftyping}.
   \end{itemize}

   Next, we show that
   \[
    \mathit{z} \,  {:}  \, \ottnt{T_{{\mathrm{0}}}}  \vdash   \ottnt{K}  [  \mathit{z}  ]   \ottsym{:}   \ottnt{T}  \,  \,\&\,  \,  \Phi  \, / \,   ( \forall  \mathit{y}  .  \ottnt{C_{{\mathrm{1}}}}  )  \Rightarrow   \ottnt{C_{{\mathrm{0}}}}   ~.
   \]
   By \T{Sub} with $\mathit{z} \,  {:}  \, \ottnt{T_{{\mathrm{0}}}}  \vdash   \ottnt{K}  [  \mathit{z}  ]   \ottsym{:}    \ottnt{C'}  . \ottnt{T}   \,  \,\&\,  \,   \ottnt{C'}  . \Phi   \, / \,   ( \forall  \mathit{y}  .  \ottnt{C'_{{\mathrm{1}}}}  )  \Rightarrow   \ottnt{C_{{\mathrm{0}}}}  $,
   it suffices to show the following.
   \begin{itemize}
    \item We show that
          \[
           \mathit{z} \,  {:}  \, \ottnt{T_{{\mathrm{0}}}}  \vdash    \ottnt{C'}  . \ottnt{T}   \,  \,\&\,  \,   \ottnt{C'}  . \Phi   \, / \,   ( \forall  \mathit{y}  .  \ottnt{C'_{{\mathrm{1}}}}  )  \Rightarrow   \ottnt{C_{{\mathrm{0}}}}    \ottsym{<:}   \ottnt{T}  \,  \,\&\,  \,  \Phi  \, / \,   ( \forall  \mathit{y}  .  \ottnt{C_{{\mathrm{1}}}}  )  \Rightarrow   \ottnt{C_{{\mathrm{0}}}}   ~.
          \]
          Because $ \emptyset   \vdash   \ottnt{C'}  . \ottnt{T}   \ottsym{<:}  \ottnt{T}$ and $ \emptyset   \vdash   \ottnt{C'}  . \Phi   \ottsym{<:}  \Phi$,
          \reflem{ts-weak-subtyping} implies
          $\mathit{z} \,  {:}  \, \ottnt{T_{{\mathrm{0}}}}  \vdash   \ottnt{C'}  . \ottnt{T}   \ottsym{<:}  \ottnt{T}$ and $\mathit{z} \,  {:}  \, \ottnt{T_{{\mathrm{0}}}}  \vdash   \ottnt{C'}  . \Phi   \ottsym{<:}  \Phi$.
          Thus, by \Srule{Comp}, it suffices to show that
          \[
           \mathit{z} \,  {:}  \, \ottnt{T_{{\mathrm{0}}}} \,  |  \,  \ottnt{C'}  . \ottnt{T}  \,  \,\&\,  \,  \ottnt{C'}  . \Phi   \vdash   ( \forall  \mathit{y}  .  \ottnt{C'_{{\mathrm{1}}}}  )  \Rightarrow   \ottnt{C_{{\mathrm{0}}}}   \ottsym{<:}   ( \forall  \mathit{y}  .  \ottnt{C_{{\mathrm{1}}}}  )  \Rightarrow   \ottnt{C_{{\mathrm{0}}}}  ~.
          \]
          Because $\vdash   \emptyset   \ottsym{,}  \mathit{x} \,  {:}  \, \ottsym{(}  \mathit{z_{{\mathrm{0}}}} \,  {:}  \, \ottnt{T_{{\mathrm{0}}}}  \ottsym{)}  \rightarrow  \ottnt{C_{{\mathrm{0}}}}$ as shown above,
          its inversion implies $\mathit{z_{{\mathrm{0}}}} \,  {:}  \, \ottnt{T_{{\mathrm{0}}}}  \vdash  \ottnt{C_{{\mathrm{0}}}}$.
          Thus, \reflem{ts-refl-subtyping} implies $\mathit{z_{{\mathrm{0}}}} \,  {:}  \, \ottnt{T_{{\mathrm{0}}}}  \vdash  \ottnt{C_{{\mathrm{0}}}}  \ottsym{<:}  \ottnt{C_{{\mathrm{0}}}}$.
          Thus, by \Srule{Ans}, it suffices to show that
          \[
           \mathit{z} \,  {:}  \, \ottnt{T_{{\mathrm{0}}}}  \ottsym{,}  \mathit{y} \,  {:}  \,  \ottnt{C'}  . \ottnt{T}   \vdash  \ottnt{C_{{\mathrm{1}}}}  \ottsym{<:}  \ottnt{C'_{{\mathrm{1}}}} ~.
          \]
          Because $\mathit{y} \,  {:}  \,  \ottnt{C'}  . \ottnt{T}   \vdash  \ottnt{C_{{\mathrm{1}}}}  \ottsym{<:}  \ottnt{C'_{{\mathrm{1}}}}$,
          \reflem{ts-weak-subtyping} implies the conclusion.

    \item We show that
          \[
           \mathit{z} \,  {:}  \, \ottnt{T_{{\mathrm{0}}}}  \vdash   \ottnt{T}  \,  \,\&\,  \,  \Phi  \, / \,   ( \forall  \mathit{y}  .  \ottnt{C_{{\mathrm{1}}}}  )  \Rightarrow   \ottnt{C_{{\mathrm{0}}}}   ~.
          \]
          Because we have $ \emptyset   \vdash   \ottnt{T}  \,  \,\&\,  \,  \Phi  \, / \,   ( \forall  \mathit{y}  .  \ottnt{C_{{\mathrm{1}}}}  )  \Rightarrow   \ottnt{C_{{\mathrm{2}}}}  $ already and
          it can be derived only \WFT{Comp}, its inversion implies
          \begin{itemize}
           \item $ \emptyset   \vdash  \ottnt{T}$,
           \item $ \emptyset   \vdash  \Phi$, and
           \item $\mathit{y} \,  {:}  \, \ottnt{T}  \vdash  \ottnt{C_{{\mathrm{1}}}}$.
          \end{itemize}
          Lemmas~\ref{lem:ts-wf-ty-tctx-typing} and \ref{lem:ts-wf-tctx-wf-ftyping} with
          $\mathit{z} \,  {:}  \, \ottnt{T_{{\mathrm{0}}}}  \vdash   \ottnt{K}  [  \mathit{z}  ]   \ottsym{:}    \ottnt{C'}  . \ottnt{T}   \,  \,\&\,  \,   \ottnt{C'}  . \Phi   \, / \,   ( \forall  \mathit{y}  .  \ottnt{C'_{{\mathrm{1}}}}  )  \Rightarrow   \ottnt{C_{{\mathrm{0}}}}  $
          imply $ \emptyset   \vdash  \ottnt{T_{{\mathrm{0}}}}$.
          Therefore, by \reflem{ts-weak-wf-ftyping},
          \begin{itemize}
           \item $\mathit{z} \,  {:}  \, \ottnt{T_{{\mathrm{0}}}}  \vdash  \ottnt{T}$,
           \item $\mathit{z} \,  {:}  \, \ottnt{T_{{\mathrm{0}}}}  \vdash  \Phi$, and
           \item $\mathit{z} \,  {:}  \, \ottnt{T_{{\mathrm{0}}}}  \ottsym{,}  \mathit{y} \,  {:}  \, \ottnt{T}  \vdash  \ottnt{C_{{\mathrm{1}}}}$.
          \end{itemize}
          Then, by \WF{Ans} and \WFT{Comp}, it suffices to show that
          $\mathit{z} \,  {:}  \, \ottnt{T_{{\mathrm{0}}}}  \vdash  \ottnt{C_{{\mathrm{0}}}}$, which has been proven above.
   \end{itemize}

  \case \T{Let}:
   We are given
   \begin{prooftree}
    \AxiomC{$ \emptyset   \vdash   \ottnt{K'}  [  \mathcal{S}_0 \, \mathit{x}  \ottsym{.}  \ottnt{e}  ]   \ottsym{:}   \ottnt{T_{{\mathrm{1}}}}  \,  \,\&\,  \,  \Phi_{{\mathrm{1}}}  \, / \,   ( \forall  \mathit{x_{{\mathrm{1}}}}  .  \ottnt{C}  )  \Rightarrow   \ottnt{C_{{\mathrm{2}}}}  $}
    \AxiomC{$\mathit{x_{{\mathrm{1}}}} \,  {:}  \, \ottnt{T_{{\mathrm{1}}}}  \vdash  \ottnt{e_{{\mathrm{2}}}}  \ottsym{:}   \ottnt{T}  \,  \,\&\,  \,  \Phi_{{\mathrm{2}}}  \, / \,   ( \forall  \mathit{y}  .  \ottnt{C_{{\mathrm{1}}}}  )  \Rightarrow   \ottnt{C}  $}
    \BinaryInfC{$ \emptyset   \vdash  \mathsf{let} \, \mathit{x_{{\mathrm{1}}}}  \ottsym{=}   \ottnt{K'}  [  \mathcal{S}_0 \, \mathit{x}  \ottsym{.}  \ottnt{e}  ]  \,  \mathsf{in}  \, \ottnt{e_{{\mathrm{2}}}}  \ottsym{:}   \ottnt{T}  \,  \,\&\,  \,  \Phi_{{\mathrm{1}}} \,  \tceappendop  \, \Phi_{{\mathrm{2}}}  \, / \,   ( \forall  \mathit{y}  .  \ottnt{C_{{\mathrm{1}}}}  )  \Rightarrow   \ottnt{C_{{\mathrm{2}}}}  $}
   \end{prooftree}
   for some $\ottnt{K'}$, $\mathit{x_{{\mathrm{1}}}}$, $\ottnt{e_{{\mathrm{2}}}}$, $\ottnt{T_{{\mathrm{1}}}}$, $\Phi_{{\mathrm{1}}}$, $\Phi_{{\mathrm{2}}}$, $\ottnt{C}$
   such that
   \begin{itemize}
    \item $\ottnt{K} \,  =  \, \ottsym{(}  \mathsf{let} \, \mathit{x_{{\mathrm{1}}}}  \ottsym{=}  \ottnt{K'} \,  \mathsf{in}  \, \ottnt{e_{{\mathrm{2}}}}  \ottsym{)}$,
    \item $\Phi \,  =  \, \Phi_{{\mathrm{1}}} \,  \tceappendop  \, \Phi_{{\mathrm{2}}}$, and
    \item $\mathit{x_{{\mathrm{1}}}} \,  \not\in  \,  \mathit{fv}  (  \ottnt{T}  )  \,  \mathbin{\cup}  \,  \mathit{fv}  (  \Phi_{{\mathrm{2}}}  )  \,  \mathbin{\cup}  \, \ottsym{(}   \mathit{fv}  (  \ottnt{C_{{\mathrm{1}}}}  )  \,  \mathbin{\backslash}  \,  \{  \mathit{y}  \}   \ottsym{)}$.
   \end{itemize}
   In what follows, we assume that $ \emptyset   \vdash  \ottnt{e}  \ottsym{:}  \ottnt{C_{{\mathrm{2}}}}$ does not hold
   (if it holds, then we have the conclusion).
   We proceed by analysis on $\ottnt{K'}$.
   \begin{caseanalysis}
    \case $\ottnt{K'} \,  =  \, \,[\,]\,$:
     We have
     $ \emptyset   \vdash  \mathcal{S}_0 \, \mathit{x}  \ottsym{.}  \ottnt{e}  \ottsym{:}   \ottnt{T_{{\mathrm{1}}}}  \,  \,\&\,  \,  \Phi_{{\mathrm{1}}}  \, / \,   ( \forall  \mathit{x_{{\mathrm{1}}}}  .  \ottnt{C}  )  \Rightarrow   \ottnt{C_{{\mathrm{2}}}}  $.
     By \reflem{ts-inv-shift} and the assumption that $ \emptyset   \vdash  \ottnt{e}  \ottsym{:}  \ottnt{C_{{\mathrm{2}}}}$ does not hold,
     there exist some $\mathit{z}$, $\ottnt{T_{{\mathrm{0}}}}$, and $\ottnt{C_{{\mathrm{0}}}}$ such that
     \begin{itemize}
      \item $\mathit{x} \,  {:}  \, \ottsym{(}  \mathit{z} \,  {:}  \, \ottnt{T_{{\mathrm{0}}}}  \ottsym{)}  \rightarrow  \ottnt{C_{{\mathrm{0}}}}  \vdash  \ottnt{e}  \ottsym{:}  \ottnt{C_{{\mathrm{2}}}}$,
      \item $\mathit{z} \,  {:}  \, \ottnt{T_{{\mathrm{0}}}}  \vdash  \mathit{z}  \ottsym{:}   \ottnt{T_{{\mathrm{1}}}}  \,  \,\&\,  \,  \Phi_{{\mathrm{1}}}  \, / \,   ( \forall  \mathit{x_{{\mathrm{1}}}}  .   \ottnt{C}    [  \mathit{z}  /  \mathit{x_{{\mathrm{1}}}}  ]    )  \Rightarrow   \ottnt{C_{{\mathrm{0}}}}  $,
      \item $ \emptyset   \vdash  \ottnt{T_{{\mathrm{0}}}}  \ottsym{<:}  \ottnt{T_{{\mathrm{1}}}}$, and
      \item $ \mathit{z}    \not=    \mathit{x_{{\mathrm{1}}}} $.
     \end{itemize}
     Without loss of generality, we can suppose that $ \mathit{z}    \not=    \mathit{y} $.
     Because $ \ottnt{K}  [  \mathit{z}  ]  \,  =  \, \ottsym{(}  \mathsf{let} \, \mathit{x_{{\mathrm{1}}}}  \ottsym{=}  \mathit{z} \,  \mathsf{in}  \, \ottnt{e_{{\mathrm{2}}}}  \ottsym{)}$ by definition,
     it suffices to show that
     \[
      \mathit{z} \,  {:}  \, \ottnt{T_{{\mathrm{0}}}}  \vdash  \mathsf{let} \, \mathit{x_{{\mathrm{1}}}}  \ottsym{=}  \mathit{z} \,  \mathsf{in}  \, \ottnt{e_{{\mathrm{2}}}}  \ottsym{:}   \ottnt{T}  \,  \,\&\,  \,  \Phi_{{\mathrm{1}}} \,  \tceappendop  \, \Phi_{{\mathrm{2}}}  \, / \,   ( \forall  \mathit{y}  .  \ottnt{C_{{\mathrm{1}}}}  )  \Rightarrow   \ottnt{C_{{\mathrm{0}}}}   ~,
     \]
     which is derived by \T{Sub} with the following.
     \begin{itemize}
      \item We show that
            \[
             \mathit{z} \,  {:}  \, \ottnt{T_{{\mathrm{0}}}}  \vdash  \mathsf{let} \, \mathit{x_{{\mathrm{1}}}}  \ottsym{=}  \mathit{z} \,  \mathsf{in}  \, \ottnt{e_{{\mathrm{2}}}}  \ottsym{:}    \ottnt{T}  \,  \,\&\,  \,  \Phi_{{\mathrm{2}}}  \, / \,   ( \forall  \mathit{y}  .  \ottnt{C_{{\mathrm{1}}}}  )  \Rightarrow   \ottnt{C}      [  \mathit{z}  /  \mathit{x_{{\mathrm{1}}}}  ]   ~.
            \]
            By \reflem{ts-inv-var} with $\mathit{z} \,  {:}  \, \ottnt{T_{{\mathrm{0}}}}  \vdash  \mathit{z}  \ottsym{:}   \ottnt{T_{{\mathrm{1}}}}  \,  \,\&\,  \,  \Phi_{{\mathrm{1}}}  \, / \,   ( \forall  \mathit{x_{{\mathrm{1}}}}  .   \ottnt{C}    [  \mathit{z}  /  \mathit{x_{{\mathrm{1}}}}  ]    )  \Rightarrow   \ottnt{C_{{\mathrm{0}}}}  $,
            there exists some $\ottnt{T'_{{\mathrm{1}}}}$ such that
            \begin{itemize}
             \item $\mathit{z} \,  {:}  \, \ottnt{T_{{\mathrm{0}}}}  \vdash  \mathit{z}  \ottsym{:}  \ottnt{T'_{{\mathrm{1}}}}$ and
             \item $\mathit{z} \,  {:}  \, \ottnt{T_{{\mathrm{0}}}}  \vdash   \ottnt{T'_{{\mathrm{1}}}}  \,  \,\&\,  \,   \Phi_\mathsf{val}   \, / \,   \square    \ottsym{<:}   \ottnt{T_{{\mathrm{1}}}}  \,  \,\&\,  \,  \Phi_{{\mathrm{1}}}  \, / \,   ( \forall  \mathit{x_{{\mathrm{1}}}}  .   \ottnt{C}    [  \mathit{z}  /  \mathit{x_{{\mathrm{1}}}}  ]    )  \Rightarrow   \ottnt{C_{{\mathrm{0}}}}  $, which
                   implies $\mathit{z} \,  {:}  \, \ottnt{T_{{\mathrm{0}}}}  \vdash  \ottnt{T'_{{\mathrm{1}}}}  \ottsym{<:}  \ottnt{T_{{\mathrm{1}}}}$.
            \end{itemize}
            Because $\vdash   \emptyset   \ottsym{,}  \mathit{z} \,  {:}  \, \ottnt{T_{{\mathrm{0}}}}$ by Lemmas~\ref{lem:ts-wf-ty-tctx-typing} and \ref{lem:ts-wf-tctx-wf-ftyping} with $\mathit{z} \,  {:}  \, \ottnt{T_{{\mathrm{0}}}}  \vdash  \mathit{z}  \ottsym{:}  \ottnt{T'_{{\mathrm{1}}}}$,
            \reflem{ts-refl-subtyping-pure-eff} with $\mathit{z} \,  {:}  \, \ottnt{T_{{\mathrm{0}}}}  \vdash  \ottnt{T'_{{\mathrm{1}}}}  \ottsym{<:}  \ottnt{T_{{\mathrm{1}}}}$ implies
            \[
             \mathit{z} \,  {:}  \, \ottnt{T_{{\mathrm{0}}}}  \vdash   \ottnt{T'_{{\mathrm{1}}}}  \,  \,\&\,  \,   \Phi_\mathsf{val}   \, / \,   \square    \ottsym{<:}   \ottnt{T_{{\mathrm{1}}}}  \,  \,\&\,  \,   \Phi_\mathsf{val}   \, / \,   \square   ~.
            \]
            Furthermore, because $\mathit{z} \,  {:}  \, \ottnt{T_{{\mathrm{0}}}}  \vdash  \ottnt{T_{{\mathrm{1}}}}$ by \reflem{ts-wf-ty-tctx-typing} with $\mathit{z} \,  {:}  \, \ottnt{T_{{\mathrm{0}}}}  \vdash  \mathit{z}  \ottsym{:}   \ottnt{T_{{\mathrm{1}}}}  \,  \,\&\,  \,  \Phi_{{\mathrm{1}}}  \, / \,   ( \forall  \mathit{x_{{\mathrm{1}}}}  .   \ottnt{C}    [  \mathit{z}  /  \mathit{x_{{\mathrm{1}}}}  ]    )  \Rightarrow   \ottnt{C_{{\mathrm{0}}}}  $, \reflem{ts-wf-TP_val} implies
            \[
             \mathit{z} \,  {:}  \, \ottnt{T_{{\mathrm{0}}}}  \vdash   \ottnt{T_{{\mathrm{1}}}}  \,  \,\&\,  \,   \Phi_\mathsf{val}   \, / \,   \square   ~.
            \]
            Therefore, \T{Sub} with $\mathit{z} \,  {:}  \, \ottnt{T_{{\mathrm{0}}}}  \vdash  \mathit{z}  \ottsym{:}  \ottnt{T'_{{\mathrm{1}}}}$ implies
            \[
             \mathit{z} \,  {:}  \, \ottnt{T_{{\mathrm{0}}}}  \vdash  \mathit{z}  \ottsym{:}  \ottnt{T_{{\mathrm{1}}}} ~.
            \]
            Because
            \begin{itemize}
             \item $\mathit{x_{{\mathrm{1}}}} \,  {:}  \, \ottnt{T_{{\mathrm{1}}}}  \vdash  \ottnt{e_{{\mathrm{2}}}}  \ottsym{:}   \ottnt{T}  \,  \,\&\,  \,  \Phi_{{\mathrm{2}}}  \, / \,   ( \forall  \mathit{y}  .  \ottnt{C_{{\mathrm{1}}}}  )  \Rightarrow   \ottnt{C}  $ and
             \item $\vdash   \emptyset   \ottsym{,}  \mathit{z} \,  {:}  \, \ottnt{T_{{\mathrm{0}}}}$,
            \end{itemize}
            \reflem{ts-weak-typing} implies
            \[
             \mathit{z} \,  {:}  \, \ottnt{T_{{\mathrm{0}}}}  \ottsym{,}  \mathit{x_{{\mathrm{1}}}} \,  {:}  \, \ottnt{T_{{\mathrm{1}}}}  \vdash  \ottnt{e_{{\mathrm{2}}}}  \ottsym{:}   \ottnt{T}  \,  \,\&\,  \,  \Phi_{{\mathrm{2}}}  \, / \,   ( \forall  \mathit{y}  .  \ottnt{C_{{\mathrm{1}}}}  )  \Rightarrow   \ottnt{C}   ~.
            \]
            Then, \T{LetV} implies the conclusion.
            Note that $\mathit{x_{{\mathrm{1}}}} \,  \not\in  \,  \mathit{fv}  (  \ottnt{T}  )  \,  \mathbin{\cup}  \,  \mathit{fv}  (  \Phi_{{\mathrm{2}}}  )  \,  \mathbin{\cup}  \, \ottsym{(}   \mathit{fv}  (  \ottnt{C_{{\mathrm{1}}}}  )  \,  \mathbin{\backslash}  \,  \{  \mathit{y}  \}   \ottsym{)}$.

      \item We show that
            \[
             \mathit{z} \,  {:}  \, \ottnt{T_{{\mathrm{0}}}}  \vdash    \ottnt{T}  \,  \,\&\,  \,  \Phi_{{\mathrm{2}}}  \, / \,   ( \forall  \mathit{y}  .  \ottnt{C_{{\mathrm{1}}}}  )  \Rightarrow   \ottnt{C}      [  \mathit{z}  /  \mathit{x_{{\mathrm{1}}}}  ]    \ottsym{<:}   \ottnt{T}  \,  \,\&\,  \,  \Phi_{{\mathrm{1}}} \,  \tceappendop  \, \Phi_{{\mathrm{2}}}  \, / \,   ( \forall  \mathit{y}  .  \ottnt{C_{{\mathrm{1}}}}  )  \Rightarrow   \ottnt{C_{{\mathrm{0}}}}   ~.
            \]
            \reflem{ts-inv-var} with $\mathit{z} \,  {:}  \, \ottnt{T_{{\mathrm{0}}}}  \vdash  \mathit{z}  \ottsym{:}   \ottnt{T_{{\mathrm{1}}}}  \,  \,\&\,  \,  \Phi_{{\mathrm{1}}}  \, / \,   ( \forall  \mathit{x_{{\mathrm{1}}}}  .   \ottnt{C}    [  \mathit{z}  /  \mathit{x_{{\mathrm{1}}}}  ]    )  \Rightarrow   \ottnt{C_{{\mathrm{0}}}}  $
            implies that there exists some $\ottnt{T'_{{\mathrm{1}}}}$ such that
            \begin{itemize}
             \item $\mathit{z} \,  {:}  \, \ottnt{T_{{\mathrm{0}}}}  \vdash  \mathit{z}  \ottsym{:}  \ottnt{T'_{{\mathrm{1}}}}$ and
             \item $\mathit{z} \,  {:}  \, \ottnt{T_{{\mathrm{0}}}}  \vdash   \ottnt{T'_{{\mathrm{1}}}}  \,  \,\&\,  \,   \Phi_\mathsf{val}   \, / \,   \square    \ottsym{<:}   \ottnt{T_{{\mathrm{1}}}}  \,  \,\&\,  \,  \Phi_{{\mathrm{1}}}  \, / \,   ( \forall  \mathit{x_{{\mathrm{1}}}}  .   \ottnt{C}    [  \mathit{z}  /  \mathit{x_{{\mathrm{1}}}}  ]    )  \Rightarrow   \ottnt{C_{{\mathrm{0}}}}  $,
                   which implies that
                   \begin{itemize}
                    \item $\mathit{z} \,  {:}  \, \ottnt{T_{{\mathrm{0}}}}  \vdash  \ottnt{T'_{{\mathrm{1}}}}  \ottsym{<:}  \ottnt{T_{{\mathrm{1}}}}$,
                    \item $\mathit{z} \,  {:}  \, \ottnt{T_{{\mathrm{0}}}}  \vdash   \Phi_\mathsf{val}   \ottsym{<:}  \Phi_{{\mathrm{1}}}$, and
                    \item $\mathit{z} \,  {:}  \, \ottnt{T_{{\mathrm{0}}}} \,  |  \, \ottnt{T'_{{\mathrm{1}}}} \,  \,\&\,  \,  \Phi_\mathsf{val}   \vdash   \square   \ottsym{<:}   ( \forall  \mathit{x_{{\mathrm{1}}}}  .   \ottnt{C}    [  \mathit{z}  /  \mathit{x_{{\mathrm{1}}}}  ]    )  \Rightarrow   \ottnt{C_{{\mathrm{0}}}} $.
                   \end{itemize}
            \end{itemize}
            Then, we have the conclusion by \Srule{Comp} and \Srule{Ans} with the following.
            \begin{itemize}
             \item We show that
                   \[
                    \mathit{z} \,  {:}  \, \ottnt{T_{{\mathrm{0}}}}  \vdash  \ottnt{T}  \ottsym{<:}  \ottnt{T} ~.
                   \]
                   \reflem{ts-wf-ty-tctx-typing} with $ \emptyset   \vdash   \ottnt{K}  [  \mathcal{S}_0 \, \mathit{x}  \ottsym{.}  \ottnt{e}  ]   \ottsym{:}   \ottnt{T}  \,  \,\&\,  \,  \Phi  \, / \,   ( \forall  \mathit{y}  .  \ottnt{C_{{\mathrm{1}}}}  )  \Rightarrow   \ottnt{C_{{\mathrm{2}}}}  $ implies
                   $ \emptyset   \vdash   \ottnt{T}  \,  \,\&\,  \,  \Phi  \, / \,   ( \forall  \mathit{y}  .  \ottnt{C_{{\mathrm{1}}}}  )  \Rightarrow   \ottnt{C_{{\mathrm{2}}}}  $, which further implies
                   $ \emptyset   \vdash  \ottnt{T}$ by inversion.
                   Then, Lemmas~\ref{lem:ts-refl-subtyping} and \ref{lem:ts-weak-subtyping}
                   imply $\mathit{z} \,  {:}  \, \ottnt{T_{{\mathrm{0}}}}  \vdash  \ottnt{T}  \ottsym{<:}  \ottnt{T}$.

             \item We show that
                   \[
                    \mathit{z} \,  {:}  \, \ottnt{T_{{\mathrm{0}}}}  \vdash  \Phi_{{\mathrm{2}}}  \ottsym{<:}  \Phi_{{\mathrm{1}}} \,  \tceappendop  \, \Phi_{{\mathrm{2}}} ~.
                   \]
                   \reflem{ts-weak-strength-TP_val} implies
                   $\mathit{z} \,  {:}  \, \ottnt{T_{{\mathrm{0}}}}  \vdash  \Phi_{{\mathrm{2}}}  \ottsym{<:}   \Phi_\mathsf{val}  \,  \tceappendop  \, \Phi_{{\mathrm{2}}}$, and
                   \reflem{ts-mono-eff-compose-preeff} with $\mathit{z} \,  {:}  \, \ottnt{T_{{\mathrm{0}}}}  \vdash   \Phi_\mathsf{val}   \ottsym{<:}  \Phi_{{\mathrm{1}}}$ implies
                   $\mathit{z} \,  {:}  \, \ottnt{T_{{\mathrm{0}}}}  \vdash   \Phi_\mathsf{val}  \,  \tceappendop  \, \Phi_{{\mathrm{2}}}  \ottsym{<:}  \Phi_{{\mathrm{1}}} \,  \tceappendop  \, \Phi_{{\mathrm{2}}}$.
                   Therefore, \reflem{ts-transitivity-subtyping} implies
                   $\mathit{z} \,  {:}  \, \ottnt{T_{{\mathrm{0}}}}  \vdash  \Phi_{{\mathrm{2}}}  \ottsym{<:}  \Phi_{{\mathrm{1}}} \,  \tceappendop  \, \Phi_{{\mathrm{2}}}$.

             \item We show that
                   \[
                    \mathit{z} \,  {:}  \, \ottnt{T_{{\mathrm{0}}}}  \ottsym{,}  \mathit{y} \,  {:}  \, \ottnt{T}  \vdash  \ottnt{C_{{\mathrm{1}}}}  \ottsym{<:}  \ottnt{C_{{\mathrm{1}}}} ~.
                   \]
                   Because $ \emptyset   \vdash   \ottnt{T}  \,  \,\&\,  \,  \Phi  \, / \,   ( \forall  \mathit{y}  .  \ottnt{C_{{\mathrm{1}}}}  )  \Rightarrow   \ottnt{C_{{\mathrm{2}}}}  $ as shown above, its inversion implies
                   $\mathit{y} \,  {:}  \, \ottnt{T}  \vdash  \ottnt{C_{{\mathrm{1}}}}$.
                   Then, Lemmas~\ref{lem:ts-refl-subtyping} and \ref{lem:ts-weak-subtyping}
                   imply $\mathit{z} \,  {:}  \, \ottnt{T_{{\mathrm{0}}}}  \ottsym{,}  \mathit{y} \,  {:}  \, \ottnt{T}  \vdash  \ottnt{C_{{\mathrm{1}}}}  \ottsym{<:}  \ottnt{C_{{\mathrm{1}}}}$.

             \item We show that
                   \[
                    \mathit{z} \,  {:}  \, \ottnt{T_{{\mathrm{0}}}}  \vdash   \ottnt{C}    [  \mathit{z}  /  \mathit{x_{{\mathrm{1}}}}  ]    \ottsym{<:}  \ottnt{C_{{\mathrm{0}}}} ~.
                   \]
                   Because $\mathit{z} \,  {:}  \, \ottnt{T_{{\mathrm{0}}}} \,  |  \, \ottnt{T'_{{\mathrm{1}}}} \,  \,\&\,  \,  \Phi_\mathsf{val}   \vdash   \square   \ottsym{<:}   ( \forall  \mathit{x_{{\mathrm{1}}}}  .   \ottnt{C}    [  \mathit{z}  /  \mathit{x_{{\mathrm{1}}}}  ]    )  \Rightarrow   \ottnt{C_{{\mathrm{0}}}} $
                   can be derived only by \Srule{Embed}, its inversion implies that
                   \begin{itemize}
                    \item $\mathit{x_{{\mathrm{1}}}} \,  \not\in  \,  \mathit{fv}  (  \ottnt{C_{{\mathrm{0}}}}  ) $ and
                    \item $\mathit{z} \,  {:}  \, \ottnt{T_{{\mathrm{0}}}}  \ottsym{,}  \mathit{x_{{\mathrm{1}}}} \,  {:}  \, \ottnt{T'_{{\mathrm{1}}}}  \vdash    \Phi_\mathsf{val}  \,  \tceappendop  \, \ottnt{C}    [  \mathit{z}  /  \mathit{x_{{\mathrm{1}}}}  ]    \ottsym{<:}  \ottnt{C_{{\mathrm{0}}}}$.
                   \end{itemize}
                   By \reflem{ts-val-subst-subtyping} with $\mathit{z} \,  {:}  \, \ottnt{T_{{\mathrm{0}}}}  \vdash  \mathit{z}  \ottsym{:}  \ottnt{T'_{{\mathrm{1}}}}$,
                   we have $\mathit{z} \,  {:}  \, \ottnt{T_{{\mathrm{0}}}}  \vdash    \Phi_\mathsf{val}  \,  \tceappendop  \, \ottnt{C}    [  \mathit{z}  /  \mathit{x_{{\mathrm{1}}}}  ]    \ottsym{<:}  \ottnt{C_{{\mathrm{0}}}}$
                   (note that $\mathit{x_{{\mathrm{1}}}}$ does not occur free in $ \Phi_\mathsf{val} $, $ \ottnt{C}    [  \mathit{z}  /  \mathit{x_{{\mathrm{1}}}}  ]  $, nor $\ottnt{C_{{\mathrm{0}}}}$).
                   Then, \reflem{ts-remove-TP_val-subty} implies the conclusion
                   $\mathit{z} \,  {:}  \, \ottnt{T_{{\mathrm{0}}}}  \vdash   \ottnt{C}    [  \mathit{z}  /  \mathit{x_{{\mathrm{1}}}}  ]    \ottsym{<:}  \ottnt{C_{{\mathrm{0}}}}$.
            \end{itemize}

      \item We show that
            \[
             \mathit{z} \,  {:}  \, \ottnt{T_{{\mathrm{0}}}}  \vdash   \ottnt{T}  \,  \,\&\,  \,  \Phi_{{\mathrm{1}}} \,  \tceappendop  \, \Phi_{{\mathrm{2}}}  \, / \,   ( \forall  \mathit{y}  .  \ottnt{C_{{\mathrm{1}}}}  )  \Rightarrow   \ottnt{C_{{\mathrm{0}}}}   ~.
            \]
            Because $ \emptyset   \vdash   \ottnt{T}  \,  \,\&\,  \,  \Phi  \, / \,   ( \forall  \mathit{y}  .  \ottnt{C_{{\mathrm{1}}}}  )  \Rightarrow   \ottnt{C_{{\mathrm{2}}}}  $ and $\Phi \,  =  \, \Phi_{{\mathrm{1}}} \,  \tceappendop  \, \Phi_{{\mathrm{2}}}$ and $\vdash   \emptyset   \ottsym{,}  \mathit{z} \,  {:}  \, \ottnt{T_{{\mathrm{0}}}}$ as shown above,
            \reflem{ts-weak-wf-ftyping} implies
            $\mathit{z} \,  {:}  \, \ottnt{T_{{\mathrm{0}}}}  \vdash  \ottnt{T}$ and $\mathit{z} \,  {:}  \, \ottnt{T_{{\mathrm{0}}}}  \vdash  \Phi_{{\mathrm{1}}} \,  \tceappendop  \, \Phi_{{\mathrm{2}}}$ and $\mathit{z} \,  {:}  \, \ottnt{T_{{\mathrm{0}}}}  \ottsym{,}  \mathit{y} \,  {:}  \, \ottnt{T}  \vdash  \ottnt{C_{{\mathrm{1}}}}$.
            Furthermore, \reflem{ts-wf-ty-tctx-typing} with $\mathit{z} \,  {:}  \, \ottnt{T_{{\mathrm{0}}}}  \vdash  \mathit{z}  \ottsym{:}   \ottnt{T_{{\mathrm{1}}}}  \,  \,\&\,  \,  \Phi_{{\mathrm{1}}}  \, / \,   ( \forall  \mathit{x_{{\mathrm{1}}}}  .   \ottnt{C}    [  \mathit{z}  /  \mathit{x_{{\mathrm{1}}}}  ]    )  \Rightarrow   \ottnt{C_{{\mathrm{0}}}}  $
            implies $\mathit{z} \,  {:}  \, \ottnt{T_{{\mathrm{0}}}}  \vdash  \ottnt{C_{{\mathrm{0}}}}$.
            Therefore, \WFT{Comp} implies the conclusion.
     \end{itemize}

    \case $\ottnt{K'} \,  \not=  \, \,[\,]\,$:
     By the IH and the assumption that $ \emptyset   \vdash  \ottnt{e}  \ottsym{:}  \ottnt{C_{{\mathrm{2}}}}$ does not hold,
     there exist some $\mathit{z}$, $\ottnt{T_{{\mathrm{0}}}}$, and $\ottnt{C_{{\mathrm{0}}}}$ such that
     \begin{itemize}
      \item $\mathit{x} \,  {:}  \, \ottsym{(}  \mathit{z} \,  {:}  \, \ottnt{T_{{\mathrm{0}}}}  \ottsym{)}  \rightarrow  \ottnt{C_{{\mathrm{0}}}}  \vdash  \ottnt{e}  \ottsym{:}  \ottnt{C_{{\mathrm{2}}}}$ and
      \item $\mathit{z} \,  {:}  \, \ottnt{T_{{\mathrm{0}}}}  \vdash   \ottnt{K'}  [  \mathit{z}  ]   \ottsym{:}   \ottnt{T_{{\mathrm{1}}}}  \,  \,\&\,  \,  \Phi_{{\mathrm{1}}}  \, / \,   ( \forall  \mathit{x_{{\mathrm{1}}}}  .  \ottnt{C}  )  \Rightarrow   \ottnt{C_{{\mathrm{0}}}}  $.
     \end{itemize}
     Because \reflem{ts-wf-ty-tctx-typing} implies $\mathit{z} \,  {:}  \, \ottnt{T_{{\mathrm{0}}}}  \vdash   \ottnt{T_{{\mathrm{1}}}}  \,  \,\&\,  \,  \Phi_{{\mathrm{1}}}  \, / \,   ( \forall  \mathit{x_{{\mathrm{1}}}}  .  \ottnt{C}  )  \Rightarrow   \ottnt{C_{{\mathrm{0}}}}  $,
     we have $\vdash   \emptyset   \ottsym{,}  \mathit{z} \,  {:}  \, \ottnt{T_{{\mathrm{0}}}}$ by \reflem{ts-wf-tctx-wf-ftyping}.
     Therefore, \reflem{ts-weak-typing} with $\mathit{x_{{\mathrm{1}}}} \,  {:}  \, \ottnt{T_{{\mathrm{1}}}}  \vdash  \ottnt{e_{{\mathrm{2}}}}  \ottsym{:}   \ottnt{T}  \,  \,\&\,  \,  \Phi_{{\mathrm{2}}}  \, / \,   ( \forall  \mathit{y}  .  \ottnt{C_{{\mathrm{1}}}}  )  \Rightarrow   \ottnt{C}  $
     implies
     \[
      \mathit{z} \,  {:}  \, \ottnt{T_{{\mathrm{0}}}}  \ottsym{,}  \mathit{x_{{\mathrm{1}}}} \,  {:}  \, \ottnt{T_{{\mathrm{1}}}}  \vdash  \ottnt{e_{{\mathrm{2}}}}  \ottsym{:}   \ottnt{T}  \,  \,\&\,  \,  \Phi_{{\mathrm{2}}}  \, / \,   ( \forall  \mathit{y}  .  \ottnt{C_{{\mathrm{1}}}}  )  \Rightarrow   \ottnt{C}   ~.
     \]
     Then, \T{Let} implies the conclusion
     \[
      \mathit{z} \,  {:}  \, \ottnt{T_{{\mathrm{0}}}}  \vdash  \mathsf{let} \, \mathit{x_{{\mathrm{1}}}}  \ottsym{=}   \ottnt{K'}  [  \mathit{z}  ]  \,  \mathsf{in}  \, \ottnt{e_{{\mathrm{2}}}}  \ottsym{:}   \ottnt{T}  \,  \,\&\,  \,  \Phi_{{\mathrm{1}}} \,  \tceappendop  \, \Phi_{{\mathrm{2}}}  \, / \,   ( \forall  \mathit{y}  .  \ottnt{C_{{\mathrm{1}}}}  )  \Rightarrow   \ottnt{C_{{\mathrm{0}}}}   ~.
     \]
   \end{caseanalysis}

  \T{LetV}: Contradictory because $ \ottnt{K'}  [  \mathcal{S}_0 \, \mathit{x}  \ottsym{.}  \ottnt{e}  ] $ is not a value for any $\ottnt{K'}$.

  \otherwise: Contradictory.
 \end{caseanalysis}
\end{proof}

\begin{lemmap}{Subject Reduction of Capturing Continuations}{ts-subject-red-shift}
 If $ \emptyset   \vdash   \ottnt{K}  [  \mathcal{S}_0 \, \mathit{x}  \ottsym{.}  \ottnt{e}  ]   \ottsym{:}   \ottnt{T}  \,  \,\&\,  \,  \Phi  \, / \,   ( \forall  \mathit{y}  .   \ottnt{T}  \,  \,\&\,  \,   \Phi_\mathsf{val}   \, / \,   \square    )  \Rightarrow   \ottnt{C}  $,
 then $ \emptyset   \vdash   \ottnt{e}    [   \lambda\!  \, \mathit{x}  \ottsym{.}   \resetcnstr{  \ottnt{K}  [  \mathit{x}  ]  }   /  \mathit{x}  ]    \ottsym{:}  \ottnt{C}$.
\end{lemmap}
\begin{proof}
 If $ \emptyset   \vdash  \ottnt{e}  \ottsym{:}  \ottnt{C}$, then we have the conclusion.
 Note that $ \emptyset   \vdash  \ottnt{e}  \ottsym{:}  \ottnt{C}$ implies $\mathit{x} \,  \not\in  \,  \mathit{fv}  (  \ottnt{e}  ) $.
 In what follows, we assume that $ \emptyset   \vdash  \ottnt{e}  \ottsym{:}  \ottnt{C}$ does not hold.

 Then, \reflem{ts-inv-shift-ectx} with $ \emptyset   \vdash   \ottnt{K}  [  \mathcal{S}_0 \, \mathit{x}  \ottsym{.}  \ottnt{e}  ]   \ottsym{:}   \ottnt{T}  \,  \,\&\,  \,  \Phi  \, / \,   ( \forall  \mathit{y}  .   \ottnt{T}  \,  \,\&\,  \,   \Phi_\mathsf{val}   \, / \,   \square    )  \Rightarrow   \ottnt{C}  $ implies
 \begin{itemize}
  \item $\mathit{x} \,  {:}  \, \ottsym{(}  \mathit{z} \,  {:}  \, \ottnt{T_{{\mathrm{0}}}}  \ottsym{)}  \rightarrow  \ottnt{C_{{\mathrm{0}}}}  \vdash  \ottnt{e}  \ottsym{:}  \ottnt{C}$ and
  \item $\mathit{z} \,  {:}  \, \ottnt{T_{{\mathrm{0}}}}  \vdash   \ottnt{K}  [  \mathit{z}  ]   \ottsym{:}   \ottnt{T}  \,  \,\&\,  \,  \Phi  \, / \,   ( \forall  \mathit{y}  .   \ottnt{T}  \,  \,\&\,  \,   \Phi_\mathsf{val}   \, / \,   \square    )  \Rightarrow   \ottnt{C_{{\mathrm{0}}}}  $
 \end{itemize}
 for some $\mathit{z}$, $\ottnt{T_{{\mathrm{0}}}}$, and $\ottnt{C_{{\mathrm{0}}}}$.
 We have $\mathit{y} \,  \not\in  \,  \mathit{fv}  (  \ottnt{T}  ) $ because
 $ \emptyset   \vdash  \ottnt{T}$ is implied by \reflem{ts-wf-ty-tctx-typing}.
 Then, by \T{Reset} with $\mathit{z} \,  {:}  \, \ottnt{T_{{\mathrm{0}}}}  \vdash   \ottnt{K}  [  \mathit{z}  ]   \ottsym{:}   \ottnt{T}  \,  \,\&\,  \,  \Phi  \, / \,   ( \forall  \mathit{y}  .   \ottnt{T}  \,  \,\&\,  \,   \Phi_\mathsf{val}   \, / \,   \square    )  \Rightarrow   \ottnt{C_{{\mathrm{0}}}}  $,
 \[
  \mathit{z} \,  {:}  \, \ottnt{T_{{\mathrm{0}}}}  \vdash   \resetcnstr{  \ottnt{K}  [  \mathit{z}  ]  }   \ottsym{:}  \ottnt{C_{{\mathrm{0}}}} ~.
 \]
 Therefore, \T{Abs} implies
 \[
   \emptyset   \vdash   \lambda\!  \, \mathit{z}  \ottsym{.}   \resetcnstr{  \ottnt{K}  [  \mathit{z}  ]  }   \ottsym{:}  \ottsym{(}  \mathit{z} \,  {:}  \, \ottnt{T_{{\mathrm{0}}}}  \ottsym{)}  \rightarrow  \ottnt{C_{{\mathrm{0}}}} ~.
 \]
 Then, \reflem{ts-val-subst-typing} with $\mathit{x} \,  {:}  \, \ottsym{(}  \mathit{z} \,  {:}  \, \ottnt{T_{{\mathrm{0}}}}  \ottsym{)}  \rightarrow  \ottnt{C_{{\mathrm{0}}}}  \vdash  \ottnt{e}  \ottsym{:}  \ottnt{C}$
 implies the conclusion
 \[
   \emptyset   \vdash   \ottnt{e}    [   \lambda\!  \, \mathit{z}  \ottsym{.}   \resetcnstr{  \ottnt{K}  [  \mathit{z}  ]  }   /  \mathit{x}  ]    \ottsym{:}  \ottnt{C}
 \]
 where note that $\mathit{x}$ does not occur free in $\ottnt{C}$ because
 $ \emptyset   \vdash  \ottnt{C}$, which is implied by
 \reflem{ts-wf-ty-tctx-typing} with $ \emptyset   \vdash   \ottnt{K}  [  \mathcal{S}_0 \, \mathit{x}  \ottsym{.}  \ottnt{e}  ]   \ottsym{:}   \ottnt{T}  \,  \,\&\,  \,  \Phi  \, / \,   ( \forall  \mathit{y}  .   \ottnt{T}  \,  \,\&\,  \,   \Phi_\mathsf{val}   \, / \,   \square    )  \Rightarrow   \ottnt{C}  $.
\end{proof}

\begin{lemmap}{Substitution of Equivalent Predicates in Validity}{lr-iff-valid}
 Suppose that $\ottnt{A_{{\mathrm{1}}}} \, \Leftrightarrow \, \ottnt{A_{{\mathrm{2}}}}  \ottsym{:}   \tceseq{ \ottnt{s} } $.
 If $\Gamma  \models  \phi$ and $ \Gamma  \mathrel{ \Leftrightarrow _{( \ottnt{A_{{\mathrm{1}}}} , \ottnt{A_{{\mathrm{2}}}} ) /  \mathit{X} } }  \Gamma' $ and $ \phi  \mathrel{ \Leftrightarrow _{( \ottnt{A_{{\mathrm{1}}}} , \ottnt{A_{{\mathrm{2}}}} ) /  \mathit{X} } }  \phi' $,
 then $\Gamma'  \models  \phi'$.
\end{lemmap}
\begin{proof}
 First, we show that
 $\ottnt{A_{{\mathrm{1}}}} \, \Leftrightarrow \, \ottnt{A_{{\mathrm{2}}}}  \ottsym{:}   \tceseq{ \ottnt{s} } $ implies
 $ { [\![  \ottnt{A_{{\mathrm{1}}}}  ]\!]_{  \emptyset  } }  \,  =  \,  { [\![  \ottnt{A_{{\mathrm{2}}}}  ]\!]_{  \emptyset  } } $.
 Let $ \tceseq{ \ottnt{d} \,  \in  \,  \mathcal{D}_{ \ottnt{s} }  } $.
 It suffices to show that $ { [\![  \ottnt{A_{{\mathrm{1}}}}  ]\!]_{  \emptyset  } }   \ottsym{(}   \tceseq{ \ottnt{d} }   \ottsym{)} \,  =  \,  { [\![  \ottnt{A_{{\mathrm{2}}}}  ]\!]_{  \emptyset  } }   \ottsym{(}   \tceseq{ \ottnt{d} }   \ottsym{)}$.
 Because $\ottnt{A_{{\mathrm{1}}}} \, \Leftrightarrow \, \ottnt{A_{{\mathrm{2}}}}  \ottsym{:}   \tceseq{ \ottnt{s} } $,
 we have $ \tceseq{ \mathit{x}  \,  {:}  \,  \ottnt{s} }   \models   \ottnt{A_{{\mathrm{1}}}}  \ottsym{(}   \tceseq{ \mathit{x} }   \ottsym{)}    \Rightarrow    \ottnt{A_{{\mathrm{2}}}}  \ottsym{(}   \tceseq{ \mathit{x} }   \ottsym{)} $ and $ \tceseq{ \mathit{x}  \,  {:}  \,  \ottnt{s} }   \models   \ottnt{A_{{\mathrm{2}}}}  \ottsym{(}   \tceseq{ \mathit{x} }   \ottsym{)}    \Rightarrow    \ottnt{A_{{\mathrm{1}}}}  \ottsym{(}   \tceseq{ \mathit{x} }   \ottsym{)} $.
 By definition,
 $ { [\![  \ottnt{A_{{\mathrm{1}}}}  ]\!]_{  \{ \tceseq{ \mathit{x}  \mapsto  \ottnt{d} } \}  } }   \ottsym{(}   \tceseq{ \ottnt{d} }   \ottsym{)} \,  \sqsubseteq_{  \mathbb{B}  }  \,  { [\![  \ottnt{A_{{\mathrm{2}}}}  ]\!]_{  \{ \tceseq{ \mathit{x}  \mapsto  \ottnt{d} } \}  } }   \ottsym{(}   \tceseq{ \ottnt{d} }   \ottsym{)}$ and
 $ { [\![  \ottnt{A_{{\mathrm{2}}}}  ]\!]_{  \{ \tceseq{ \mathit{x}  \mapsto  \ottnt{d} } \}  } }   \ottsym{(}   \tceseq{ \ottnt{d} }   \ottsym{)} \,  \sqsubseteq_{  \mathbb{B}  }  \,  { [\![  \ottnt{A_{{\mathrm{1}}}}  ]\!]_{  \{ \tceseq{ \mathit{x}  \mapsto  \ottnt{d} } \}  } }   \ottsym{(}   \tceseq{ \ottnt{d} }   \ottsym{)}$,
 which jointly imply
 $ { [\![  \ottnt{A_{{\mathrm{1}}}}  ]\!]_{  \{ \tceseq{ \mathit{x}  \mapsto  \ottnt{d} } \}  } }   \ottsym{(}   \tceseq{ \ottnt{d} }   \ottsym{)} \,  =  \,  { [\![  \ottnt{A_{{\mathrm{2}}}}  ]\!]_{  \{ \tceseq{ \mathit{x}  \mapsto  \ottnt{d} } \}  } }   \ottsym{(}   \tceseq{ \ottnt{d} }   \ottsym{)}$.
 Because $\ottnt{A_{{\mathrm{1}}}}$ and $\ottnt{A_{{\mathrm{2}}}}$ are closed,
 we have the conclusion $ { [\![  \ottnt{A_{{\mathrm{1}}}}  ]\!]_{  \emptyset  } }   \ottsym{(}   \tceseq{ \ottnt{d} }   \ottsym{)} \,  =  \,  { [\![  \ottnt{A_{{\mathrm{2}}}}  ]\!]_{  \emptyset  } }   \ottsym{(}   \tceseq{ \ottnt{d} }   \ottsym{)}$.

 Let $\theta$ be any substitution such that $\Gamma'  \vdash  \theta$.
 By \reflem{ts-logic-closing-dsubs}, it suffices to show that
 $ { [\![  \phi'  ]\!]_{ \theta } }  \,  =  \,  \top $.
 Because
 \begin{itemize}
  \item $ \Gamma  \mathrel{ \Leftrightarrow _{( \ottnt{A_{{\mathrm{1}}}} , \ottnt{A_{{\mathrm{2}}}} ) /  \mathit{X} } }  \Gamma' $,
  \item $\Gamma'  \vdash  \theta$, and
  \item $ { [\![  \ottnt{A_{{\mathrm{1}}}}  ]\!]_{  \emptyset  } }  \,  =  \,  { [\![  \ottnt{A_{{\mathrm{2}}}}  ]\!]_{  \emptyset  } } $,
 \end{itemize}
 we have
 $\Gamma  \vdash  \theta$
 with \reflem{ts-logic-pred-subst}.
 By \reflem{ts-logic-closing-dsubs} with $\Gamma  \models  \phi$,
 we have $ { [\![  \phi  ]\!]_{ \theta } }  \,  =  \,  \top $.
 Because
 \begin{itemize}
  \item $ \phi  \mathrel{ \Leftrightarrow _{( \ottnt{A_{{\mathrm{1}}}} , \ottnt{A_{{\mathrm{2}}}} ) /  \mathit{X} } }  \phi' $,
  \item $ { [\![  \phi  ]\!]_{ \theta } }  \,  =  \,  \top $, and
  \item $ { [\![  \ottnt{A_{{\mathrm{1}}}}  ]\!]_{  \emptyset  } }  \,  =  \,  { [\![  \ottnt{A_{{\mathrm{2}}}}  ]\!]_{  \emptyset  } } $,
 \end{itemize}
 we have the conclusion $ { [\![  \phi'  ]\!]_{ \theta } }  \,  =  \,  \top $
 with \reflem{ts-logic-pred-subst}.
\end{proof}

\begin{lemmap}{Substitution of Equivalent Predicates in Implication Validity}{lr-iff-impl-valid}
 Suppose that $\ottnt{A_{{\mathrm{1}}}} \, \Leftrightarrow \, \ottnt{A_{{\mathrm{2}}}}  \ottsym{:}   \tceseq{ \ottnt{s} } $.
 If $\Gamma  \models   \phi_{{\mathrm{1}}}    \Rightarrow    \phi_{{\mathrm{2}}} $ and $ \Gamma  \mathrel{ \Leftrightarrow _{( \ottnt{A_{{\mathrm{1}}}} , \ottnt{A_{{\mathrm{2}}}} ) /  \mathit{X} } }  \Gamma' $ and
 $ \phi_{{\mathrm{1}}}  \mathrel{ \Leftrightarrow _{( \ottnt{A_{{\mathrm{1}}}} , \ottnt{A_{{\mathrm{2}}}} ) /  \mathit{X} } }  \phi'_{{\mathrm{1}}} $ and $ \phi_{{\mathrm{2}}}  \mathrel{ \Leftrightarrow _{( \ottnt{A_{{\mathrm{1}}}} , \ottnt{A_{{\mathrm{2}}}} ) /  \mathit{X} } }  \phi'_{{\mathrm{2}}} $,
 then $\Gamma'  \models   \phi'_{{\mathrm{1}}}    \Rightarrow    \phi'_{{\mathrm{2}}} $.
\end{lemmap}
\begin{proof}
 \TS{Validity}
 By \reflem{lr-iff-valid}.
\end{proof}

\begin{lemmap}{Substitution of Equivalent Predicates in Subtyping}{lr-iff-subtyping}
 Suppose that $\ottnt{A_{{\mathrm{1}}}} \, \Leftrightarrow \, \ottnt{A_{{\mathrm{2}}}}  \ottsym{:}   \tceseq{ \ottnt{s} } $.
 \begin{enumerate}
  \item If $\Gamma  \vdash  \sigma_{{\mathrm{1}}}  \ottsym{(}  \ottnt{T}  \ottsym{)}  \ottsym{<:}  \sigma_{{\mathrm{2}}}  \ottsym{(}  \ottnt{T}  \ottsym{)}$ and $ \Gamma  \mathrel{ \Leftrightarrow _{( \ottnt{A_{{\mathrm{1}}}} , \ottnt{A_{{\mathrm{2}}}} ) /  \mathit{X} } }  \Gamma' $ and
        $ \sigma_{{\mathrm{1}}}  \ottsym{(}  \ottnt{T}  \ottsym{)}  \mathrel{ \Leftrightarrow _{( \ottnt{A_{{\mathrm{1}}}} , \ottnt{A_{{\mathrm{2}}}} ) /  \mathit{X} } }  \ottnt{T'_{{\mathrm{1}}}} $ and $ \sigma_{{\mathrm{2}}}  \ottsym{(}  \ottnt{T}  \ottsym{)}  \mathrel{ \Leftrightarrow _{( \ottnt{A_{{\mathrm{1}}}} , \ottnt{A_{{\mathrm{2}}}} ) /  \mathit{X} } }  \ottnt{T'_{{\mathrm{2}}}} $,
        then $\Gamma'  \vdash  \ottnt{T'_{{\mathrm{1}}}}  \ottsym{<:}  \ottnt{T'_{{\mathrm{2}}}}$.

  \item If $\Gamma  \vdash  \sigma_{{\mathrm{1}}}  \ottsym{(}  \ottnt{C}  \ottsym{)}  \ottsym{<:}  \sigma_{{\mathrm{2}}}  \ottsym{(}  \ottnt{C}  \ottsym{)}$ and $ \Gamma  \mathrel{ \Leftrightarrow _{( \ottnt{A_{{\mathrm{1}}}} , \ottnt{A_{{\mathrm{2}}}} ) /  \mathit{X} } }  \Gamma' $ and
        $ \sigma_{{\mathrm{1}}}  \ottsym{(}  \ottnt{C}  \ottsym{)}  \mathrel{ \Leftrightarrow _{( \ottnt{A_{{\mathrm{1}}}} , \ottnt{A_{{\mathrm{2}}}} ) /  \mathit{X} } }  \ottnt{C'_{{\mathrm{1}}}} $ and $ \sigma_{{\mathrm{2}}}  \ottsym{(}  \ottnt{C}  \ottsym{)}  \mathrel{ \Leftrightarrow _{( \ottnt{A_{{\mathrm{1}}}} , \ottnt{A_{{\mathrm{2}}}} ) /  \mathit{X} } }  \ottnt{C'_{{\mathrm{2}}}} $,
        then $\Gamma'  \vdash  \ottnt{C'_{{\mathrm{1}}}}  \ottsym{<:}  \ottnt{C'_{{\mathrm{2}}}}$.

  \item If
        $\Gamma \,  |  \, \ottnt{T} \,  \,\&\,  \, \Phi  \vdash  \sigma_{{\mathrm{1}}}  \ottsym{(}  \ottnt{S}  \ottsym{)}  \ottsym{<:}  \sigma_{{\mathrm{2}}}  \ottsym{(}  \ottnt{S}  \ottsym{)}$ and
        $ \Gamma  \mathrel{ \Leftrightarrow _{( \ottnt{A_{{\mathrm{1}}}} , \ottnt{A_{{\mathrm{2}}}} ) /  \mathit{X} } }  \Gamma' $ and
        $ \ottnt{T}  \mathrel{ \Leftrightarrow _{( \ottnt{A_{{\mathrm{1}}}} , \ottnt{A_{{\mathrm{2}}}} ) /  \mathit{X} } }  \ottnt{T'} $ and
        $ \sigma_{{\mathrm{1}}}  \ottsym{(}  \ottnt{S}  \ottsym{)}  \mathrel{ \Leftrightarrow _{( \ottnt{A_{{\mathrm{1}}}} , \ottnt{A_{{\mathrm{2}}}} ) /  \mathit{X} } }  \ottnt{S'_{{\mathrm{1}}}} $ and
        $ \sigma_{{\mathrm{2}}}  \ottsym{(}  \ottnt{S}  \ottsym{)}  \mathrel{ \Leftrightarrow _{( \ottnt{A_{{\mathrm{1}}}} , \ottnt{A_{{\mathrm{2}}}} ) /  \mathit{X} } }  \ottnt{S'_{{\mathrm{2}}}} $,
        then $  \forall  \, { \Phi' } . \  \Gamma' \,  |  \, \ottnt{T'} \,  \,\&\,  \, \Phi'  \vdash  \ottnt{S'_{{\mathrm{1}}}}  \ottsym{<:}  \ottnt{S'_{{\mathrm{2}}}} $.

  \item If $\Gamma  \vdash  \sigma_{{\mathrm{1}}}  \ottsym{(}  \Phi  \ottsym{)}  \ottsym{<:}  \sigma_{{\mathrm{2}}}  \ottsym{(}  \Phi  \ottsym{)}$ and $ \Gamma  \mathrel{ \Leftrightarrow _{( \ottnt{A_{{\mathrm{1}}}} , \ottnt{A_{{\mathrm{2}}}} ) /  \mathit{X} } }  \Gamma' $ and
        $ \sigma_{{\mathrm{1}}}  \ottsym{(}  \Phi  \ottsym{)}  \mathrel{ \Leftrightarrow _{( \ottnt{A_{{\mathrm{1}}}} , \ottnt{A_{{\mathrm{2}}}} ) /  \mathit{X} } }  \Phi'_{{\mathrm{1}}} $ and $ \sigma_{{\mathrm{2}}}  \ottsym{(}  \Phi  \ottsym{)}  \mathrel{ \Leftrightarrow _{( \ottnt{A_{{\mathrm{1}}}} , \ottnt{A_{{\mathrm{2}}}} ) /  \mathit{X} } }  \Phi'_{{\mathrm{2}}} $,
        then $\Gamma'  \vdash  \Phi'_{{\mathrm{1}}}  \ottsym{<:}  \Phi'_{{\mathrm{2}}}$.
 \end{enumerate}
\end{lemmap}
\begin{proof}
 \TS{OK: 2022/01/25}
 By simultaneous induction on the subtyping derivations.
 \begin{caseanalysis}
  \case \Srule{Refine}:
   We are given
   \begin{prooftree}
    \AxiomC{$\Gamma  \ottsym{,}  \mathit{x} \,  {:}  \, \iota  \models   \sigma_{{\mathrm{1}}}  \ottsym{(}  \phi  \ottsym{)}    \Rightarrow    \sigma_{{\mathrm{2}}}  \ottsym{(}  \phi  \ottsym{)} $}
    \UnaryInfC{$\Gamma  \vdash  \ottsym{\{}  \mathit{x} \,  {:}  \, \iota \,  |  \, \sigma_{{\mathrm{1}}}  \ottsym{(}  \phi  \ottsym{)}  \ottsym{\}}  \ottsym{<:}  \ottsym{\{}  \mathit{x} \,  {:}  \, \iota \,  |  \, \sigma_{{\mathrm{2}}}  \ottsym{(}  \phi  \ottsym{)}  \ottsym{\}}$}
   \end{prooftree}
   for some $\mathit{x}$, $\iota$, and $\phi$ such that
   $\ottnt{T} \,  =  \, \ottsym{\{}  \mathit{x} \,  {:}  \, \iota \,  |  \, \phi  \ottsym{\}}$ and $\mathit{x}$ does not occur in $\sigma_{{\mathrm{1}}}$ nor $\sigma_{{\mathrm{2}}}$.
   Because $ \sigma_{{\mathrm{1}}}  \ottsym{(}  \ottnt{T}  \ottsym{)}  \mathrel{ \Leftrightarrow _{( \ottnt{A_{{\mathrm{1}}}} , \ottnt{A_{{\mathrm{2}}}} ) /  \mathit{X} } }  \ottnt{T'_{{\mathrm{1}}}} $ and $ \sigma_{{\mathrm{2}}}  \ottsym{(}  \ottnt{T}  \ottsym{)}  \mathrel{ \Leftrightarrow _{( \ottnt{A_{{\mathrm{1}}}} , \ottnt{A_{{\mathrm{2}}}} ) /  \mathit{X} } }  \ottnt{T'_{{\mathrm{2}}}} $,
   there exist some $\phi''_{{\mathrm{1}}}$, $\phi''_{{\mathrm{2}}}$, and $ { \ottmv{i} }_{ \ottsym{1} } ,  { \ottmv{i} }_{ \ottsym{2} }  \in \{1,2\}$ such that
   \begin{itemize}
    \item $\sigma_{{\mathrm{1}}}  \ottsym{(}  \phi  \ottsym{)} \,  =  \,  \phi''_{{\mathrm{1}}}    [   \ottnt{A} _{  { \ottmv{i} }_{ \ottsym{1} }  }   /  \mathit{X}  ]  $,
    \item $\ottnt{T'_{{\mathrm{1}}}} \,  =  \, \ottsym{\{}  \mathit{x} \,  {:}  \, \iota \,  |  \,  \phi''_{{\mathrm{1}}}    [   \ottnt{A} _{  { \ottsym{3}  \ottsym{-}  \ottmv{i} }_{ \ottsym{1} }  }   /  \mathit{X}  ]    \ottsym{\}}$,
    \item $\sigma_{{\mathrm{2}}}  \ottsym{(}  \phi  \ottsym{)} \,  =  \,  \phi''_{{\mathrm{2}}}    [   \ottnt{A} _{  { \ottmv{i} }_{ \ottsym{2} }  }   /  \mathit{X}  ]  $, and
    \item $\ottnt{T'_{{\mathrm{2}}}} \,  =  \, \ottsym{\{}  \mathit{x} \,  {:}  \, \iota \,  |  \,  \phi''_{{\mathrm{2}}}    [   \ottnt{A} _{  { \ottsym{3}  \ottsym{-}  \ottmv{i} }_{ \ottsym{2} }  }   /  \mathit{X}  ]    \ottsym{\}}$.
   \end{itemize}
   Because
   \begin{itemize}
    \item $\Gamma  \ottsym{,}  \mathit{x} \,  {:}  \, \iota  \models   \sigma_{{\mathrm{1}}}  \ottsym{(}  \phi  \ottsym{)}    \Rightarrow    \sigma_{{\mathrm{2}}}  \ottsym{(}  \phi  \ottsym{)} $ means
          $\Gamma  \ottsym{,}  \mathit{x} \,  {:}  \, \iota  \models     \phi''_{{\mathrm{1}}}    [   \ottnt{A} _{  { \ottmv{i} }_{ \ottsym{1} }  }   /  \mathit{X}  ]      \Rightarrow    \phi''_{{\mathrm{2}}}     [   \ottnt{A} _{  { \ottmv{i} }_{ \ottsym{2} }  }   /  \mathit{X}  ]  $,
    \item $ \Gamma  \ottsym{,}  \mathit{x} \,  {:}  \, \iota  \mathrel{ \Leftrightarrow _{( \ottnt{A_{{\mathrm{1}}}} , \ottnt{A_{{\mathrm{2}}}} ) /  \mathit{X} } }  \Gamma'  \ottsym{,}  \mathit{x} \,  {:}  \, \iota $,
    \item $  \phi''_{{\mathrm{1}}}    [   \ottnt{A} _{  { \ottmv{i} }_{ \ottsym{1} }  }   /  \mathit{X}  ]    \mathrel{ \Leftrightarrow _{( \ottnt{A_{{\mathrm{1}}}} , \ottnt{A_{{\mathrm{2}}}} ) /  \mathit{X} } }   \phi''_{{\mathrm{1}}}    [   \ottnt{A} _{  { \ottsym{3}  \ottsym{-}  \ottmv{i} }_{ \ottsym{1} }  }   /  \mathit{X}  ]   $, and
    \item $  \phi''_{{\mathrm{2}}}    [   \ottnt{A} _{  { \ottmv{i} }_{ \ottsym{2} }  }   /  \mathit{X}  ]    \mathrel{ \Leftrightarrow _{( \ottnt{A_{{\mathrm{1}}}} , \ottnt{A_{{\mathrm{2}}}} ) /  \mathit{X} } }   \phi''_{{\mathrm{2}}}    [   \ottnt{A} _{  { \ottsym{3}  \ottsym{-}  \ottmv{i} }_{ \ottsym{2} }  }   /  \mathit{X}  ]   $
   \end{itemize}
   we have
   \[
    \Gamma'  \ottsym{,}  \mathit{x} \,  {:}  \, \iota  \models     \phi''_{{\mathrm{1}}}    [   \ottnt{A} _{  { \ottsym{3}  \ottsym{-}  \ottmv{i} }_{ \ottsym{1} }  }   /  \mathit{X}  ]      \Rightarrow    \phi''_{{\mathrm{2}}}     [   \ottnt{A} _{  { \ottsym{3}  \ottsym{-}  \ottmv{i} }_{ \ottsym{2} }  }   /  \mathit{X}  ]  
   \]
   by \reflem{lr-iff-impl-valid}.
   Then, \Srule{Refine} implies the conclusion $\Gamma'  \vdash  \ottnt{T'_{{\mathrm{1}}}}  \ottsym{<:}  \ottnt{T'_{{\mathrm{2}}}}$.

  \case \Srule{Fun}:
   We are given
   \begin{prooftree}
    \AxiomC{$\Gamma  \vdash  \sigma_{{\mathrm{2}}}  \ottsym{(}  \ottnt{T_{{\mathrm{0}}}}  \ottsym{)}  \ottsym{<:}  \sigma_{{\mathrm{1}}}  \ottsym{(}  \ottnt{T_{{\mathrm{0}}}}  \ottsym{)}$}
    \AxiomC{$\Gamma  \ottsym{,}  \mathit{x} \,  {:}  \, \sigma_{{\mathrm{2}}}  \ottsym{(}  \ottnt{T_{{\mathrm{0}}}}  \ottsym{)}  \vdash  \sigma_{{\mathrm{1}}}  \ottsym{(}  \ottnt{C_{{\mathrm{0}}}}  \ottsym{)}  \ottsym{<:}  \sigma_{{\mathrm{2}}}  \ottsym{(}  \ottnt{C_{{\mathrm{0}}}}  \ottsym{)}$}
    \BinaryInfC{$\Gamma  \vdash  \ottsym{(}  \mathit{x} \,  {:}  \, \sigma_{{\mathrm{1}}}  \ottsym{(}  \ottnt{T_{{\mathrm{0}}}}  \ottsym{)}  \ottsym{)}  \rightarrow  \sigma_{{\mathrm{1}}}  \ottsym{(}  \ottnt{C_{{\mathrm{0}}}}  \ottsym{)}  \ottsym{<:}  \ottsym{(}  \mathit{x} \,  {:}  \, \sigma_{{\mathrm{2}}}  \ottsym{(}  \ottnt{T_{{\mathrm{0}}}}  \ottsym{)}  \ottsym{)}  \rightarrow  \sigma_{{\mathrm{2}}}  \ottsym{(}  \ottnt{C_{{\mathrm{0}}}}  \ottsym{)}$}
   \end{prooftree}
   for some $\mathit{x}$, $\ottnt{T_{{\mathrm{0}}}}$, and $\ottnt{C_{{\mathrm{0}}}}$ such that
   $\ottnt{T} \,  =  \, \ottsym{(}  \mathit{x} \,  {:}  \, \ottnt{T_{{\mathrm{0}}}}  \ottsym{)}  \rightarrow  \ottnt{C_{{\mathrm{0}}}}$ and $\mathit{x}$ does not occur in $\sigma_{{\mathrm{1}}}$ nor $\sigma_{{\mathrm{2}}}$.
   Because $ \sigma_{{\mathrm{1}}}  \ottsym{(}  \ottnt{T}  \ottsym{)}  \mathrel{ \Leftrightarrow _{( \ottnt{A_{{\mathrm{1}}}} , \ottnt{A_{{\mathrm{2}}}} ) /  \mathit{X} } }  \ottnt{T'_{{\mathrm{1}}}} $ and $ \sigma_{{\mathrm{2}}}  \ottsym{(}  \ottnt{T}  \ottsym{)}  \mathrel{ \Leftrightarrow _{( \ottnt{A_{{\mathrm{1}}}} , \ottnt{A_{{\mathrm{2}}}} ) /  \mathit{X} } }  \ottnt{T'_{{\mathrm{2}}}} $,
   there exist $\ottnt{T''_{{\mathrm{01}}}}$, $\ottnt{C''_{{\mathrm{01}}}}$, $\ottnt{T''_{{\mathrm{02}}}}$, $\ottnt{C''_{{\mathrm{02}}}}$, and
   $ { \ottmv{i} }_{ \ottsym{1} } ,  { \ottmv{i} }_{ \ottsym{2} }  \in \{1,2\}$ such that
   \begin{itemize}
    \item $\sigma_{{\mathrm{1}}}  \ottsym{(}  \ottnt{T_{{\mathrm{0}}}}  \ottsym{)} \,  =  \,  \ottnt{T''_{{\mathrm{01}}}}    [   \ottnt{A} _{  { \ottmv{i} }_{ \ottsym{1} }  }   /  \mathit{X}  ]  $,
    \item $\sigma_{{\mathrm{1}}}  \ottsym{(}  \ottnt{C_{{\mathrm{0}}}}  \ottsym{)} \,  =  \,  \ottnt{C''_{{\mathrm{01}}}}    [   \ottnt{A} _{  { \ottmv{i} }_{ \ottsym{1} }  }   /  \mathit{X}  ]  $,
    \item $\ottnt{T'_{{\mathrm{1}}}} \,  =  \,  \ottsym{(}  \ottsym{(}  \mathit{x} \,  {:}  \, \ottnt{T''_{{\mathrm{01}}}}  \ottsym{)}  \rightarrow  \ottnt{C''_{{\mathrm{01}}}}  \ottsym{)}    [   \ottnt{A} _{  { \ottsym{3}  \ottsym{-}  \ottmv{i} }_{ \ottsym{1} }  }   /  \mathit{X}  ]  $,
    \item $\sigma_{{\mathrm{2}}}  \ottsym{(}  \ottnt{T_{{\mathrm{0}}}}  \ottsym{)} \,  =  \,  \ottnt{T''_{{\mathrm{02}}}}    [   \ottnt{A} _{  { \ottmv{i} }_{ \ottsym{2} }  }   /  \mathit{X}  ]  $,
    \item $\sigma_{{\mathrm{2}}}  \ottsym{(}  \ottnt{C_{{\mathrm{0}}}}  \ottsym{)} \,  =  \,  \ottnt{C''_{{\mathrm{02}}}}    [   \ottnt{A} _{  { \ottmv{i} }_{ \ottsym{2} }  }   /  \mathit{X}  ]  $, and
    \item $\ottnt{T'_{{\mathrm{2}}}} \,  =  \,  \ottsym{(}  \ottsym{(}  \mathit{x} \,  {:}  \, \ottnt{T''_{{\mathrm{02}}}}  \ottsym{)}  \rightarrow  \ottnt{C''_{{\mathrm{02}}}}  \ottsym{)}    [   \ottnt{A} _{  { \ottsym{3}  \ottsym{-}  \ottmv{i} }_{ \ottsym{2} }  }   /  \mathit{X}  ]  $.
   \end{itemize}
   Because
   \begin{itemize}
    \item $\Gamma  \vdash  \sigma_{{\mathrm{2}}}  \ottsym{(}  \ottnt{T_{{\mathrm{0}}}}  \ottsym{)}  \ottsym{<:}  \sigma_{{\mathrm{1}}}  \ottsym{(}  \ottnt{T_{{\mathrm{0}}}}  \ottsym{)}$,
    \item $ \Gamma  \mathrel{ \Leftrightarrow _{( \ottnt{A_{{\mathrm{1}}}} , \ottnt{A_{{\mathrm{2}}}} ) /  \mathit{X} } }  \Gamma' $,
    \item $\sigma_{{\mathrm{2}}}  \ottsym{(}  \ottnt{T_{{\mathrm{0}}}}  \ottsym{)} =   \ottnt{T''_{{\mathrm{02}}}}    [   \ottnt{A} _{  { \ottmv{i} }_{ \ottsym{2} }  }   /  \mathit{X}  ]    \mathrel{ \Leftrightarrow _{( \ottnt{A_{{\mathrm{1}}}} , \ottnt{A_{{\mathrm{2}}}} ) /  \mathit{X} } }   \ottnt{T''_{{\mathrm{02}}}}    [   \ottnt{A} _{  { \ottsym{3}  \ottsym{-}  \ottmv{i} }_{ \ottsym{2} }  }   /  \mathit{X}  ]   $, and
    \item $\sigma_{{\mathrm{1}}}  \ottsym{(}  \ottnt{T_{{\mathrm{0}}}}  \ottsym{)} =   \ottnt{T''_{{\mathrm{01}}}}    [   \ottnt{A} _{  { \ottmv{i} }_{ \ottsym{1} }  }   /  \mathit{X}  ]    \mathrel{ \Leftrightarrow _{( \ottnt{A_{{\mathrm{1}}}} , \ottnt{A_{{\mathrm{2}}}} ) /  \mathit{X} } }   \ottnt{T''_{{\mathrm{01}}}}    [   \ottnt{A} _{  { \ottsym{3}  \ottsym{-}  \ottmv{i} }_{ \ottsym{1} }  }   /  \mathit{X}  ]   $,
   \end{itemize}
   we have
   \[
    \Gamma'  \vdash   \ottnt{T''_{{\mathrm{02}}}}    [   \ottnt{A} _{  { \ottsym{3}  \ottsym{-}  \ottmv{i} }_{ \ottsym{2} }  }   /  \mathit{X}  ]    \ottsym{<:}   \ottnt{T''_{{\mathrm{01}}}}    [   \ottnt{A} _{  { \ottsym{3}  \ottsym{-}  \ottmv{i} }_{ \ottsym{1} }  }   /  \mathit{X}  ]  
   \]
   by the IH.
   Further, because
   \begin{itemize}
    \item $\Gamma  \ottsym{,}  \mathit{x} \,  {:}  \, \sigma_{{\mathrm{2}}}  \ottsym{(}  \ottnt{T_{{\mathrm{0}}}}  \ottsym{)}  \vdash  \sigma_{{\mathrm{1}}}  \ottsym{(}  \ottnt{C_{{\mathrm{0}}}}  \ottsym{)}  \ottsym{<:}  \sigma_{{\mathrm{2}}}  \ottsym{(}  \ottnt{C_{{\mathrm{0}}}}  \ottsym{)}$,
    \item $ \Gamma  \ottsym{,}  \mathit{x} \,  {:}  \, \sigma_{{\mathrm{2}}}  \ottsym{(}  \ottnt{T_{{\mathrm{0}}}}  \ottsym{)}  \mathrel{ \Leftrightarrow _{( \ottnt{A_{{\mathrm{1}}}} , \ottnt{A_{{\mathrm{2}}}} ) /  \mathit{X} } }   \Gamma'  \ottsym{,}  \mathit{x} \,  {:}  \, \ottnt{T''_{{\mathrm{02}}}}    [   \ottnt{A} _{  { \ottsym{3}  \ottsym{-}  \ottmv{i} }_{ \ottsym{2} }  }   /  \mathit{X}  ]   $,
    \item $\sigma_{{\mathrm{1}}}  \ottsym{(}  \ottnt{C_{{\mathrm{0}}}}  \ottsym{)} =   \ottnt{C''_{{\mathrm{01}}}}    [   \ottnt{A} _{  { \ottmv{i} }_{ \ottsym{1} }  }   /  \mathit{X}  ]    \mathrel{ \Leftrightarrow _{( \ottnt{A_{{\mathrm{1}}}} , \ottnt{A_{{\mathrm{2}}}} ) /  \mathit{X} } }   \ottnt{C''_{{\mathrm{01}}}}    [   \ottnt{A} _{  { \ottsym{3}  \ottsym{-}  \ottmv{i} }_{ \ottsym{1} }  }   /  \mathit{X}  ]   $, and
    \item $\sigma_{{\mathrm{2}}}  \ottsym{(}  \ottnt{C_{{\mathrm{0}}}}  \ottsym{)} =   \ottnt{C''_{{\mathrm{02}}}}    [   \ottnt{A} _{  { \ottmv{i} }_{ \ottsym{2} }  }   /  \mathit{X}  ]    \mathrel{ \Leftrightarrow _{( \ottnt{A_{{\mathrm{1}}}} , \ottnt{A_{{\mathrm{2}}}} ) /  \mathit{X} } }   \ottnt{C''_{{\mathrm{02}}}}    [   \ottnt{A} _{  { \ottsym{3}  \ottsym{-}  \ottmv{i} }_{ \ottsym{2} }  }   /  \mathit{X}  ]   $,
   \end{itemize}
   we have
   \[
     \Gamma'  \ottsym{,}  \mathit{x} \,  {:}  \, \ottnt{T''_{{\mathrm{02}}}}    [   \ottnt{A} _{  { \ottsym{3}  \ottsym{-}  \ottmv{i} }_{ \ottsym{2} }  }   /  \mathit{X}  ]    \vdash   \ottnt{C''_{{\mathrm{01}}}}    [   \ottnt{A} _{  { \ottsym{3}  \ottsym{-}  \ottmv{i} }_{ \ottsym{1} }  }   /  \mathit{X}  ]    \ottsym{<:}   \ottnt{C''_{{\mathrm{02}}}}    [   \ottnt{A} _{  { \ottsym{3}  \ottsym{-}  \ottmv{i} }_{ \ottsym{2} }  }   /  \mathit{X}  ]  
   \]
   by the IH.
   By \Srule{Fun}, we have
   \[
    \Gamma'  \vdash   \ottsym{(}  \ottsym{(}  \mathit{x} \,  {:}  \, \ottnt{T''_{{\mathrm{01}}}}  \ottsym{)}  \rightarrow  \ottnt{C''_{{\mathrm{01}}}}  \ottsym{)}    [   \ottnt{A} _{  { \ottsym{3}  \ottsym{-}  \ottmv{i} }_{ \ottsym{1} }  }   /  \mathit{X}  ]    \ottsym{<:}   \ottsym{(}  \ottsym{(}  \mathit{x} \,  {:}  \, \ottnt{T''_{{\mathrm{02}}}}  \ottsym{)}  \rightarrow  \ottnt{C''_{{\mathrm{02}}}}  \ottsym{)}    [   \ottnt{A} _{  { \ottsym{3}  \ottsym{-}  \ottmv{i} }_{ \ottsym{2} }  }   /  \mathit{X}  ]  
   \]
   by \Srule{Fun}.
   Therefore, we have the conclusion $\Gamma'  \vdash  \ottnt{T'_{{\mathrm{1}}}}  \ottsym{<:}  \ottnt{T'_{{\mathrm{2}}}}$.

  \case \Srule{Forall}:
   We are given
   \begin{prooftree}
    \AxiomC{$\Gamma  \ottsym{,}  \mathit{Y} \,  {:}  \,  \tceseq{ \ottnt{s} }   \vdash  \sigma_{{\mathrm{1}}}  \ottsym{(}  \ottnt{C_{{\mathrm{0}}}}  \ottsym{)}  \ottsym{<:}  \sigma_{{\mathrm{2}}}  \ottsym{(}  \ottnt{C_{{\mathrm{0}}}}  \ottsym{)}$}
    \UnaryInfC{$\Gamma  \vdash   \forall   \ottsym{(}  \mathit{Y} \,  {:}  \,  \tceseq{ \ottnt{s} }   \ottsym{)}  \ottsym{.}  \sigma_{{\mathrm{1}}}  \ottsym{(}  \ottnt{C_{{\mathrm{0}}}}  \ottsym{)}  \ottsym{<:}   \forall   \ottsym{(}  \mathit{Y} \,  {:}  \,  \tceseq{ \ottnt{s} }   \ottsym{)}  \ottsym{.}  \sigma_{{\mathrm{2}}}  \ottsym{(}  \ottnt{C_{{\mathrm{0}}}}  \ottsym{)}$}
   \end{prooftree}
   for some $\mathit{Y}$, $ \tceseq{ \ottnt{s} } $, and $\ottnt{C_{{\mathrm{0}}}}$
   such that $\ottnt{T} \,  =  \,  \forall   \ottsym{(}  \mathit{Y} \,  {:}  \,  \tceseq{ \ottnt{s} }   \ottsym{)}  \ottsym{.}  \ottnt{C_{{\mathrm{0}}}}$ and $\mathit{Y}$ does not occur in $\sigma_{{\mathrm{1}}}$ nor $\sigma_{{\mathrm{2}}}$.
   Without loss of generality, we can suppose that
   $\mathit{Y} \,  \not=  \, \mathit{X}$.
   Because $ \sigma_{{\mathrm{1}}}  \ottsym{(}  \ottnt{T}  \ottsym{)}  \mathrel{ \Leftrightarrow _{( \ottnt{A_{{\mathrm{1}}}} , \ottnt{A_{{\mathrm{2}}}} ) /  \mathit{X} } }  \ottnt{T'_{{\mathrm{1}}}} $ and $ \sigma_{{\mathrm{2}}}  \ottsym{(}  \ottnt{T}  \ottsym{)}  \mathrel{ \Leftrightarrow _{( \ottnt{A_{{\mathrm{1}}}} , \ottnt{A_{{\mathrm{2}}}} ) /  \mathit{X} } }  \ottnt{T'_{{\mathrm{2}}}} $,
   there exist $\ottnt{C''_{{\mathrm{01}}}}$, $\ottnt{C''_{{\mathrm{02}}}}$, and $ { \ottmv{i} }_{ \ottsym{1} } ,  { \ottmv{i} }_{ \ottsym{2} }  \in \{1,2\}$ such that
   \begin{itemize}
    \item $\sigma_{{\mathrm{1}}}  \ottsym{(}  \ottnt{C_{{\mathrm{0}}}}  \ottsym{)} \,  =  \,  \ottnt{C''_{{\mathrm{01}}}}    [   \ottnt{A} _{  { \ottmv{i} }_{ \ottsym{1} }  }   /  \mathit{X}  ]  $,
    \item $\ottnt{T'_{{\mathrm{1}}}} \,  =  \,  \ottsym{(}   \forall   \ottsym{(}  \mathit{Y} \,  {:}  \,  \tceseq{ \ottnt{s} }   \ottsym{)}  \ottsym{.}  \ottnt{C''_{{\mathrm{01}}}}  \ottsym{)}    [   \ottnt{A} _{  { \ottsym{3}  \ottsym{-}  \ottmv{i} }_{ \ottsym{1} }  }   /  \mathit{X}  ]  $,
    \item $\sigma_{{\mathrm{2}}}  \ottsym{(}  \ottnt{C_{{\mathrm{0}}}}  \ottsym{)} \,  =  \,  \ottnt{C''_{{\mathrm{02}}}}    [   \ottnt{A} _{  { \ottmv{i} }_{ \ottsym{2} }  }   /  \mathit{X}  ]  $, and
    \item $\ottnt{T'_{{\mathrm{2}}}} \,  =  \,  \ottsym{(}   \forall   \ottsym{(}  \mathit{Y} \,  {:}  \,  \tceseq{ \ottnt{s} }   \ottsym{)}  \ottsym{.}  \ottnt{C''_{{\mathrm{02}}}}  \ottsym{)}    [   \ottnt{A} _{  { \ottsym{3}  \ottsym{-}  \ottmv{i} }_{ \ottsym{2} }  }   /  \mathit{X}  ]  $.
   \end{itemize}
   Because
   \begin{itemize}
    \item $\Gamma  \ottsym{,}  \mathit{Y} \,  {:}  \,  \tceseq{ \ottnt{s} }   \vdash  \sigma_{{\mathrm{1}}}  \ottsym{(}  \ottnt{C_{{\mathrm{0}}}}  \ottsym{)}  \ottsym{<:}  \sigma_{{\mathrm{2}}}  \ottsym{(}  \ottnt{C_{{\mathrm{0}}}}  \ottsym{)}$,
    \item $ \Gamma  \ottsym{,}  \mathit{Y} \,  {:}  \,  \tceseq{ \ottnt{s} }   \mathrel{ \Leftrightarrow _{( \ottnt{A_{{\mathrm{1}}}} , \ottnt{A_{{\mathrm{2}}}} ) /  \mathit{X} } }  \Gamma'  \ottsym{,}  \mathit{Y} \,  {:}  \,  \tceseq{ \ottnt{s} }  $,
    \item $\sigma_{{\mathrm{1}}}  \ottsym{(}  \ottnt{C_{{\mathrm{0}}}}  \ottsym{)} =   \ottnt{C''_{{\mathrm{01}}}}    [   \ottnt{A} _{  { \ottmv{i} }_{ \ottsym{1} }  }   /  \mathit{X}  ]    \mathrel{ \Leftrightarrow _{( \ottnt{A_{{\mathrm{1}}}} , \ottnt{A_{{\mathrm{2}}}} ) /  \mathit{X} } }   \ottnt{C''_{{\mathrm{01}}}}    [   \ottnt{A} _{  { \ottsym{3}  \ottsym{-}  \ottmv{i} }_{ \ottsym{1} }  }   /  \mathit{X}  ]   $, and
    \item $\sigma_{{\mathrm{2}}}  \ottsym{(}  \ottnt{C_{{\mathrm{0}}}}  \ottsym{)} =   \ottnt{C''_{{\mathrm{02}}}}    [   \ottnt{A} _{  { \ottmv{i} }_{ \ottsym{1} }  }   /  \mathit{X}  ]    \mathrel{ \Leftrightarrow _{( \ottnt{A_{{\mathrm{1}}}} , \ottnt{A_{{\mathrm{2}}}} ) /  \mathit{X} } }   \ottnt{C''_{{\mathrm{02}}}}    [   \ottnt{A} _{  { \ottsym{3}  \ottsym{-}  \ottmv{i} }_{ \ottsym{1} }  }   /  \mathit{X}  ]   $,
   \end{itemize}
   we have
   \[
    \Gamma'  \ottsym{,}  \mathit{Y} \,  {:}  \,  \tceseq{ \ottnt{s} }   \vdash   \ottnt{C''_{{\mathrm{01}}}}    [   \ottnt{A} _{  { \ottsym{3}  \ottsym{-}  \ottmv{i} }_{ \ottsym{1} }  }   /  \mathit{X}  ]    \ottsym{<:}   \ottnt{C''_{{\mathrm{02}}}}    [   \ottnt{A} _{  { \ottsym{3}  \ottsym{-}  \ottmv{i} }_{ \ottsym{2} }  }   /  \mathit{X}  ]  
   \]
   by the IH.
   By \Srule{Forall}, we have
   \[
    \Gamma'  \vdash   \ottsym{(}   \forall   \ottsym{(}  \mathit{Y} \,  {:}  \,  \tceseq{ \ottnt{s} }   \ottsym{)}  \ottsym{.}  \ottnt{C''_{{\mathrm{01}}}}  \ottsym{)}    [   \ottnt{A} _{  { \ottsym{3}  \ottsym{-}  \ottmv{i} }_{ \ottsym{1} }  }   /  \mathit{X}  ]    \ottsym{<:}   \ottsym{(}   \forall   \ottsym{(}  \mathit{Y} \,  {:}  \,  \tceseq{ \ottnt{s} }   \ottsym{)}  \ottsym{.}  \ottnt{C''_{{\mathrm{02}}}}  \ottsym{)}    [   \ottnt{A} _{  { \ottsym{3}  \ottsym{-}  \ottmv{i} }_{ \ottsym{1} }  }   /  \mathit{X}  ]   ~.
   \]
   Therefore, we have the conclusion $\Gamma'  \vdash  \ottnt{T'_{{\mathrm{1}}}}  \ottsym{<:}  \ottnt{T'_{{\mathrm{2}}}}$.

  \case \Srule{Comp}:
   We are given
   \begin{prooftree}
    \AxiomC{$\Gamma  \vdash  \sigma_{{\mathrm{1}}}  \ottsym{(}  \ottnt{T}  \ottsym{)}  \ottsym{<:}  \sigma_{{\mathrm{2}}}  \ottsym{(}  \ottnt{T}  \ottsym{)}$}
    \AxiomC{$\Gamma  \vdash  \sigma_{{\mathrm{1}}}  \ottsym{(}  \Phi  \ottsym{)}  \ottsym{<:}  \sigma_{{\mathrm{2}}}  \ottsym{(}  \Phi  \ottsym{)}$}
    \AxiomC{$\Gamma \,  |  \, \sigma_{{\mathrm{1}}}  \ottsym{(}  \ottnt{T}  \ottsym{)} \,  \,\&\,  \, \sigma_{{\mathrm{1}}}  \ottsym{(}  \Phi  \ottsym{)}  \vdash  \sigma_{{\mathrm{1}}}  \ottsym{(}  \ottnt{S}  \ottsym{)}  \ottsym{<:}  \sigma_{{\mathrm{2}}}  \ottsym{(}  \ottnt{S}  \ottsym{)}$}
    \TrinaryInfC{$\Gamma  \vdash   \sigma_{{\mathrm{1}}}  \ottsym{(}  \ottnt{T}  \ottsym{)}  \,  \,\&\,  \,  \sigma_{{\mathrm{1}}}  \ottsym{(}  \Phi  \ottsym{)}  \, / \,  \sigma_{{\mathrm{1}}}  \ottsym{(}  \ottnt{S}  \ottsym{)}   \ottsym{<:}   \sigma_{{\mathrm{2}}}  \ottsym{(}  \ottnt{T}  \ottsym{)}  \,  \,\&\,  \,  \sigma_{{\mathrm{2}}}  \ottsym{(}  \Phi  \ottsym{)}  \, / \,  \sigma_{{\mathrm{2}}}  \ottsym{(}  \ottnt{S}  \ottsym{)} $}
   \end{prooftree}
   for some $\ottnt{T}$, $\Phi$, and $\ottnt{S}$
   such that $\ottnt{C} \,  =  \,  \ottnt{T}  \,  \,\&\,  \,  \Phi  \, / \,  \ottnt{S} $.
   Because $ \sigma_{{\mathrm{1}}}  \ottsym{(}  \ottnt{C}  \ottsym{)}  \mathrel{ \Leftrightarrow _{( \ottnt{A_{{\mathrm{1}}}} , \ottnt{A_{{\mathrm{2}}}} ) /  \mathit{X} } }  \ottnt{C'_{{\mathrm{1}}}} $ and $ \sigma_{{\mathrm{2}}}  \ottsym{(}  \ottnt{C}  \ottsym{)}  \mathrel{ \Leftrightarrow _{( \ottnt{A_{{\mathrm{1}}}} , \ottnt{A_{{\mathrm{2}}}} ) /  \mathit{X} } }  \ottnt{C'_{{\mathrm{2}}}} $,
   there exist some $\ottnt{T''_{{\mathrm{1}}}}$, $\Phi''_{{\mathrm{1}}}$, $\ottnt{S''_{{\mathrm{1}}}}$, $\ottnt{T''_{{\mathrm{2}}}}$, $\Phi''_{{\mathrm{2}}}$, $\ottnt{S''_{{\mathrm{2}}}}$, and $ { \ottmv{i} }_{ \ottsym{1} } ,  { \ottmv{i} }_{ \ottsym{2} }  \in \{1,2\}$ such that
   \begin{itemize}
    \item $\sigma_{{\mathrm{1}}}  \ottsym{(}  \ottnt{T}  \ottsym{)} \,  =  \,  \ottnt{T''_{{\mathrm{1}}}}    [   \ottnt{A} _{  { \ottmv{i} }_{ \ottsym{1} }  }   /  \mathit{X}  ]  $,
    \item $\sigma_{{\mathrm{1}}}  \ottsym{(}  \Phi  \ottsym{)} \,  =  \,  \Phi''_{{\mathrm{1}}}    [   \ottnt{A} _{  { \ottmv{i} }_{ \ottsym{1} }  }   /  \mathit{X}  ]  $,
    \item $\sigma_{{\mathrm{1}}}  \ottsym{(}  \ottnt{S}  \ottsym{)} \,  =  \,  \ottnt{S''_{{\mathrm{1}}}}    [   \ottnt{A} _{  { \ottmv{i} }_{ \ottsym{1} }  }   /  \mathit{X}  ]  $,
    \item $\ottnt{C'_{{\mathrm{1}}}} \,  =  \,  \ottsym{(}   \ottnt{T''_{{\mathrm{1}}}}  \,  \,\&\,  \,  \Phi''_{{\mathrm{1}}}  \, / \,  \ottnt{S''_{{\mathrm{1}}}}   \ottsym{)}    [   \ottnt{A} _{  { \ottsym{3}  \ottsym{-}  \ottmv{i} }_{ \ottsym{1} }  }   /  \mathit{X}  ]  $,
    \item $\sigma_{{\mathrm{2}}}  \ottsym{(}  \ottnt{T}  \ottsym{)} \,  =  \,  \ottnt{T''_{{\mathrm{2}}}}    [   \ottnt{A} _{  { \ottmv{i} }_{ \ottsym{2} }  }   /  \mathit{X}  ]  $,
    \item $\sigma_{{\mathrm{2}}}  \ottsym{(}  \Phi  \ottsym{)} \,  =  \,  \Phi''_{{\mathrm{2}}}    [   \ottnt{A} _{  { \ottmv{i} }_{ \ottsym{2} }  }   /  \mathit{X}  ]  $,
    \item $\sigma_{{\mathrm{2}}}  \ottsym{(}  \ottnt{S}  \ottsym{)} \,  =  \,  \ottnt{S''_{{\mathrm{2}}}}    [   \ottnt{A} _{  { \ottmv{i} }_{ \ottsym{2} }  }   /  \mathit{X}  ]  $, and
    \item $\ottnt{C'_{{\mathrm{2}}}} \,  =  \,  \ottsym{(}   \ottnt{T''_{{\mathrm{2}}}}  \,  \,\&\,  \,  \Phi''_{{\mathrm{2}}}  \, / \,  \ottnt{S''_{{\mathrm{2}}}}   \ottsym{)}    [   \ottnt{A} _{  { \ottsym{3}  \ottsym{-}  \ottmv{i} }_{ \ottsym{2} }  }   /  \mathit{X}  ]  $.
   \end{itemize}
   Because
   \begin{itemize}
    \item $\Gamma  \vdash  \sigma_{{\mathrm{1}}}  \ottsym{(}  \ottnt{T}  \ottsym{)}  \ottsym{<:}  \sigma_{{\mathrm{2}}}  \ottsym{(}  \ottnt{T}  \ottsym{)}$,
    \item $ \Gamma  \mathrel{ \Leftrightarrow _{( \ottnt{A_{{\mathrm{1}}}} , \ottnt{A_{{\mathrm{2}}}} ) /  \mathit{X} } }  \Gamma' $,
    \item $\sigma_{{\mathrm{1}}}  \ottsym{(}  \ottnt{T}  \ottsym{)} =   \ottnt{T''_{{\mathrm{1}}}}    [   \ottnt{A} _{  { \ottmv{i} }_{ \ottsym{1} }  }   /  \mathit{X}  ]    \mathrel{ \Leftrightarrow _{( \ottnt{A_{{\mathrm{1}}}} , \ottnt{A_{{\mathrm{2}}}} ) /  \mathit{X} } }   \ottnt{T''_{{\mathrm{1}}}}    [   \ottnt{A} _{  { \ottsym{3}  \ottsym{-}  \ottmv{i} }_{ \ottsym{1} }  }   /  \mathit{X}  ]   $, and
    \item $\sigma_{{\mathrm{2}}}  \ottsym{(}  \ottnt{T}  \ottsym{)} =   \ottnt{T''_{{\mathrm{2}}}}    [   \ottnt{A} _{  { \ottmv{i} }_{ \ottsym{2} }  }   /  \mathit{X}  ]    \mathrel{ \Leftrightarrow _{( \ottnt{A_{{\mathrm{1}}}} , \ottnt{A_{{\mathrm{2}}}} ) /  \mathit{X} } }   \ottnt{T''_{{\mathrm{2}}}}    [   \ottnt{A} _{  { \ottsym{3}  \ottsym{-}  \ottmv{i} }_{ \ottsym{2} }  }   /  \mathit{X}  ]   $,
   \end{itemize}
   we have
   \[
    \Gamma'  \vdash   \ottnt{T''_{{\mathrm{1}}}}    [   \ottnt{A} _{  { \ottsym{3}  \ottsym{-}  \ottmv{i} }_{ \ottsym{1} }  }   /  \mathit{X}  ]    \ottsym{<:}   \ottnt{T''_{{\mathrm{2}}}}    [   \ottnt{A} _{  { \ottsym{3}  \ottsym{-}  \ottmv{i} }_{ \ottsym{2} }  }   /  \mathit{X}  ]  
   \]
   by the IH.
   Because
   \begin{itemize}
    \item $\Gamma  \vdash  \sigma_{{\mathrm{1}}}  \ottsym{(}  \Phi  \ottsym{)}  \ottsym{<:}  \sigma_{{\mathrm{2}}}  \ottsym{(}  \Phi  \ottsym{)}$,
    \item $ \Gamma  \mathrel{ \Leftrightarrow _{( \ottnt{A_{{\mathrm{1}}}} , \ottnt{A_{{\mathrm{2}}}} ) /  \mathit{X} } }  \Gamma' $,
    \item $\sigma_{{\mathrm{1}}}  \ottsym{(}  \Phi  \ottsym{)} =   \Phi''_{{\mathrm{1}}}    [   \ottnt{A} _{  { \ottmv{i} }_{ \ottsym{1} }  }   /  \mathit{X}  ]    \mathrel{ \Leftrightarrow _{( \ottnt{A_{{\mathrm{1}}}} , \ottnt{A_{{\mathrm{2}}}} ) /  \mathit{X} } }   \Phi''_{{\mathrm{1}}}    [   \ottnt{A} _{  { \ottsym{3}  \ottsym{-}  \ottmv{i} }_{ \ottsym{1} }  }   /  \mathit{X}  ]   $, and
    \item $\sigma_{{\mathrm{2}}}  \ottsym{(}  \Phi  \ottsym{)} =   \Phi''_{{\mathrm{2}}}    [   \ottnt{A} _{  { \ottmv{i} }_{ \ottsym{2} }  }   /  \mathit{X}  ]    \mathrel{ \Leftrightarrow _{( \ottnt{A_{{\mathrm{1}}}} , \ottnt{A_{{\mathrm{2}}}} ) /  \mathit{X} } }   \Phi''_{{\mathrm{2}}}    [   \ottnt{A} _{  { \ottsym{3}  \ottsym{-}  \ottmv{i} }_{ \ottsym{2} }  }   /  \mathit{X}  ]   $,
   \end{itemize}
   we have
   \[
    \Gamma'  \vdash   \Phi''_{{\mathrm{1}}}    [   \ottnt{A} _{  { \ottsym{3}  \ottsym{-}  \ottmv{i} }_{ \ottsym{1} }  }   /  \mathit{X}  ]    \ottsym{<:}   \Phi''_{{\mathrm{2}}}    [   \ottnt{A} _{  { \ottsym{3}  \ottsym{-}  \ottmv{i} }_{ \ottsym{2} }  }   /  \mathit{X}  ]  
   \]
   by the IH.
   Because
   \begin{itemize}
    \item $\Gamma \,  |  \, \sigma_{{\mathrm{1}}}  \ottsym{(}  \ottnt{T}  \ottsym{)} \,  \,\&\,  \, \sigma_{{\mathrm{1}}}  \ottsym{(}  \Phi  \ottsym{)}  \vdash  \sigma_{{\mathrm{1}}}  \ottsym{(}  \ottnt{S}  \ottsym{)}  \ottsym{<:}  \sigma_{{\mathrm{2}}}  \ottsym{(}  \ottnt{S}  \ottsym{)}$,
    \item $ \Gamma  \mathrel{ \Leftrightarrow _{( \ottnt{A_{{\mathrm{1}}}} , \ottnt{A_{{\mathrm{2}}}} ) /  \mathit{X} } }  \Gamma' $,
    \item $\sigma_{{\mathrm{1}}}  \ottsym{(}  \ottnt{T}  \ottsym{)} =   \ottnt{T''_{{\mathrm{1}}}}    [   \ottnt{A} _{  { \ottmv{i} }_{ \ottsym{1} }  }   /  \mathit{X}  ]    \mathrel{ \Leftrightarrow _{( \ottnt{A_{{\mathrm{1}}}} , \ottnt{A_{{\mathrm{2}}}} ) /  \mathit{X} } }   \ottnt{T''_{{\mathrm{1}}}}    [   \ottnt{A} _{  { \ottsym{3}  \ottsym{-}  \ottmv{i} }_{ \ottsym{1} }  }   /  \mathit{X}  ]   $,
    \item $\sigma_{{\mathrm{1}}}  \ottsym{(}  \ottnt{S}  \ottsym{)} =   \ottnt{S''_{{\mathrm{1}}}}    [   \ottnt{A} _{  { \ottmv{i} }_{ \ottsym{1} }  }   /  \mathit{X}  ]    \mathrel{ \Leftrightarrow _{( \ottnt{A_{{\mathrm{1}}}} , \ottnt{A_{{\mathrm{2}}}} ) /  \mathit{X} } }   \ottnt{S''_{{\mathrm{1}}}}    [   \ottnt{A} _{  { \ottsym{3}  \ottsym{-}  \ottmv{i} }_{ \ottsym{1} }  }   /  \mathit{X}  ]   $, and
    \item $\sigma_{{\mathrm{2}}}  \ottsym{(}  \ottnt{S}  \ottsym{)} =   \ottnt{S''_{{\mathrm{2}}}}    [   \ottnt{A} _{  { \ottmv{i} }_{ \ottsym{2} }  }   /  \mathit{X}  ]    \mathrel{ \Leftrightarrow _{( \ottnt{A_{{\mathrm{1}}}} , \ottnt{A_{{\mathrm{2}}}} ) /  \mathit{X} } }   \ottnt{S''_{{\mathrm{2}}}}    [   \ottnt{A} _{  { \ottsym{3}  \ottsym{-}  \ottmv{i} }_{ \ottsym{2} }  }   /  \mathit{X}  ]   $,
   \end{itemize}
   we have
   \[
    \Gamma' \,  |  \,  \ottnt{T''_{{\mathrm{1}}}}    [   \ottnt{A} _{  { \ottsym{3}  \ottsym{-}  \ottmv{i} }_{ \ottsym{1} }  }   /  \mathit{X}  ]   \,  \,\&\,  \,  \Phi''_{{\mathrm{1}}}    [   \ottnt{A} _{  { \ottsym{3}  \ottsym{-}  \ottmv{i} }_{ \ottsym{1} }  }   /  \mathit{X}  ]    \vdash   \ottnt{S''_{{\mathrm{1}}}}    [   \ottnt{A} _{  { \ottsym{3}  \ottsym{-}  \ottmv{i} }_{ \ottsym{1} }  }   /  \mathit{X}  ]    \ottsym{<:}   \ottnt{S''_{{\mathrm{2}}}}    [   \ottnt{A} _{  { \ottsym{3}  \ottsym{-}  \ottmv{i} }_{ \ottsym{2} }  }   /  \mathit{X}  ]  
   \]
   by the IH.
   Therefore, by \Srule{Comp}, we have
   \[
    \Gamma'  \vdash   \ottsym{(}   \ottnt{T''_{{\mathrm{1}}}}  \,  \,\&\,  \,  \Phi''_{{\mathrm{1}}}  \, / \,  \ottnt{S''_{{\mathrm{1}}}}   \ottsym{)}    [   \ottnt{A} _{  { \ottsym{3}  \ottsym{-}  \ottmv{i} }_{ \ottsym{1} }  }   /  \mathit{X}  ]    \ottsym{<:}   \ottsym{(}   \ottnt{T''_{{\mathrm{2}}}}  \,  \,\&\,  \,  \Phi''_{{\mathrm{2}}}  \, / \,  \ottnt{S''_{{\mathrm{2}}}}   \ottsym{)}    [   \ottnt{A} _{  { \ottsym{3}  \ottsym{-}  \ottmv{i} }_{ \ottsym{2} }  }   /  \mathit{X}  ]   ~.
   \]
   Therefore, we have the conclusion
   $\Gamma'  \vdash  \ottnt{C'_{{\mathrm{1}}}}  \ottsym{<:}  \ottnt{C'_{{\mathrm{2}}}}$.

  \case \Srule{Empty}:
   We are given $\ottnt{S} \,  =  \,  \square $.
   Because $ \sigma_{{\mathrm{1}}}  \ottsym{(}  \ottnt{S}  \ottsym{)}  \mathrel{ \Leftrightarrow _{( \ottnt{A_{{\mathrm{1}}}} , \ottnt{A_{{\mathrm{2}}}} ) /  \mathit{X} } }  \ottnt{S'_{{\mathrm{1}}}} $ and $ \sigma_{{\mathrm{2}}}  \ottsym{(}  \ottnt{S}  \ottsym{)}  \mathrel{ \Leftrightarrow _{( \ottnt{A_{{\mathrm{1}}}} , \ottnt{A_{{\mathrm{2}}}} ) /  \mathit{X} } }  \ottnt{S'_{{\mathrm{2}}}} $,
   we have $\ottnt{S'_{{\mathrm{1}}}} = \ottnt{S'_{{\mathrm{2}}}} =  \square $.
   Then, by \Srule{Empty},
   we have the conclusion $\Gamma' \,  |  \, \ottnt{T'} \,  \,\&\,  \, \Phi'  \vdash  \ottnt{S'_{{\mathrm{1}}}}  \ottsym{<:}  \ottnt{S'_{{\mathrm{2}}}}$.

  \case \Srule{Ans}:
   We are given
   \begin{prooftree}
    \AxiomC{$\Gamma  \ottsym{,}  \mathit{x} \,  {:}  \, \ottnt{T}  \vdash  \sigma_{{\mathrm{2}}}  \ottsym{(}  \ottnt{C_{{\mathrm{1}}}}  \ottsym{)}  \ottsym{<:}  \sigma_{{\mathrm{1}}}  \ottsym{(}  \ottnt{C_{{\mathrm{1}}}}  \ottsym{)}$}
    \AxiomC{$\Gamma  \vdash  \sigma_{{\mathrm{1}}}  \ottsym{(}  \ottnt{C_{{\mathrm{2}}}}  \ottsym{)}  \ottsym{<:}  \sigma_{{\mathrm{2}}}  \ottsym{(}  \ottnt{C_{{\mathrm{2}}}}  \ottsym{)}$}
    \BinaryInfC{$\Gamma \,  |  \, \ottnt{T} \,  \,\&\,  \, \Phi  \vdash   ( \forall  \mathit{x}  .  \sigma_{{\mathrm{1}}}  \ottsym{(}  \ottnt{C_{{\mathrm{1}}}}  \ottsym{)}  )  \Rightarrow   \sigma_{{\mathrm{1}}}  \ottsym{(}  \ottnt{C_{{\mathrm{2}}}}  \ottsym{)}   \ottsym{<:}   ( \forall  \mathit{x}  .  \sigma_{{\mathrm{2}}}  \ottsym{(}  \ottnt{C_{{\mathrm{1}}}}  \ottsym{)}  )  \Rightarrow   \sigma_{{\mathrm{2}}}  \ottsym{(}  \ottnt{C_{{\mathrm{2}}}}  \ottsym{)} $}
   \end{prooftree}
   for some $\mathit{x}$, $\ottnt{C_{{\mathrm{1}}}}$, and $\ottnt{C_{{\mathrm{2}}}}$ such that
   $\ottnt{S} \,  =  \,  ( \forall  \mathit{x}  .  \ottnt{C_{{\mathrm{1}}}}  )  \Rightarrow   \ottnt{C_{{\mathrm{2}}}} $ and $\mathit{x}$ does not occur in $\sigma_{{\mathrm{1}}}$ nor $\sigma_{{\mathrm{2}}}$.
   Because $ \ottnt{T}  \mathrel{ \Leftrightarrow _{( \ottnt{A_{{\mathrm{1}}}} , \ottnt{A_{{\mathrm{2}}}} ) /  \mathit{X} } }  \ottnt{T'} $ and $ \sigma_{{\mathrm{1}}}  \ottsym{(}  \ottnt{S}  \ottsym{)}  \mathrel{ \Leftrightarrow _{( \ottnt{A_{{\mathrm{1}}}} , \ottnt{A_{{\mathrm{2}}}} ) /  \mathit{X} } }  \ottnt{S'_{{\mathrm{1}}}} $ and $ \sigma_{{\mathrm{2}}}  \ottsym{(}  \ottnt{S}  \ottsym{)}  \mathrel{ \Leftrightarrow _{( \ottnt{A_{{\mathrm{1}}}} , \ottnt{A_{{\mathrm{2}}}} ) /  \mathit{X} } }  \ottnt{S'_{{\mathrm{2}}}} $,
   there exist $\ottnt{T''}$, $\ottnt{C''_{{\mathrm{11}}}}$, $\ottnt{C''_{{\mathrm{12}}}}$, $\ottnt{C''_{{\mathrm{21}}}}$, $\ottnt{C''_{{\mathrm{22}}}}$, and
   $ { \ottmv{i} }_{ \ottsym{0} } ,  { \ottmv{i} }_{ \ottsym{1} } ,  { \ottmv{i} }_{ \ottsym{2} }  \in \{1,2\}$ such that
   \begin{itemize}
    \item $\ottnt{T} \,  =  \,  \ottnt{T''}    [   \ottnt{A} _{  { \ottmv{i} }_{ \ottsym{0} }  }   /  \mathit{X}  ]  $,
    \item $\ottnt{T'} \,  =  \,  \ottnt{T''}    [   \ottnt{A} _{  { \ottsym{3}  \ottsym{-}  \ottmv{i} }_{ \ottsym{0} }  }   /  \mathit{X}  ]  $,
    \item $\sigma_{{\mathrm{1}}}  \ottsym{(}  \ottnt{C_{{\mathrm{1}}}}  \ottsym{)} \,  =  \,  \ottnt{C''_{{\mathrm{11}}}}    [   \ottnt{A} _{  { \ottmv{i} }_{ \ottsym{1} }  }   /  \mathit{X}  ]  $,
    \item $\sigma_{{\mathrm{1}}}  \ottsym{(}  \ottnt{C_{{\mathrm{2}}}}  \ottsym{)} \,  =  \,  \ottnt{C''_{{\mathrm{12}}}}    [   \ottnt{A} _{  { \ottmv{i} }_{ \ottsym{1} }  }   /  \mathit{X}  ]  $,
    \item $\ottnt{S'_{{\mathrm{1}}}} \,  =  \,  \ottsym{(}   ( \forall  \mathit{x}  .  \ottnt{C''_{{\mathrm{11}}}}  )  \Rightarrow   \ottnt{C''_{{\mathrm{12}}}}   \ottsym{)}    [   \ottnt{A} _{  { \ottsym{3}  \ottsym{-}  \ottmv{i} }_{ \ottsym{1} }  }   /  \mathit{X}  ]  $,
    \item $\sigma_{{\mathrm{2}}}  \ottsym{(}  \ottnt{C_{{\mathrm{1}}}}  \ottsym{)} \,  =  \,  \ottnt{C''_{{\mathrm{21}}}}    [   \ottnt{A} _{  { \ottmv{i} }_{ \ottsym{2} }  }   /  \mathit{X}  ]  $,
    \item $\sigma_{{\mathrm{2}}}  \ottsym{(}  \ottnt{C_{{\mathrm{2}}}}  \ottsym{)} \,  =  \,  \ottnt{C''_{{\mathrm{22}}}}    [   \ottnt{A} _{  { \ottmv{i} }_{ \ottsym{2} }  }   /  \mathit{X}  ]  $, and
    \item $\ottnt{S'_{{\mathrm{2}}}} \,  =  \,  \ottsym{(}   ( \forall  \mathit{x}  .  \ottnt{C''_{{\mathrm{21}}}}  )  \Rightarrow   \ottnt{C''_{{\mathrm{22}}}}   \ottsym{)}    [   \ottnt{A} _{  { \ottsym{3}  \ottsym{-}  \ottmv{i} }_{ \ottsym{2} }  }   /  \mathit{X}  ]  $.
   \end{itemize}
   Because
   \begin{itemize}
    \item $\Gamma  \ottsym{,}  \mathit{x} \,  {:}  \, \ottnt{T}  \vdash  \sigma_{{\mathrm{2}}}  \ottsym{(}  \ottnt{C_{{\mathrm{1}}}}  \ottsym{)}  \ottsym{<:}  \sigma_{{\mathrm{1}}}  \ottsym{(}  \ottnt{C_{{\mathrm{1}}}}  \ottsym{)}$ means
          $ \Gamma  \ottsym{,}  \mathit{x} \,  {:}  \, \ottnt{T''}    [   \ottnt{A} _{  { \ottmv{i} }_{ \ottsym{0} }  }   /  \mathit{X}  ]    \vdash  \sigma_{{\mathrm{2}}}  \ottsym{(}  \ottnt{C_{{\mathrm{1}}}}  \ottsym{)}  \ottsym{<:}  \sigma_{{\mathrm{1}}}  \ottsym{(}  \ottnt{C_{{\mathrm{1}}}}  \ottsym{)}$,
    \item $  \Gamma  \ottsym{,}  \mathit{x} \,  {:}  \, \ottnt{T''}    [   \ottnt{A} _{  { \ottmv{i} }_{ \ottsym{0} }  }   /  \mathit{X}  ]    \mathrel{ \Leftrightarrow _{( \ottnt{A_{{\mathrm{1}}}} , \ottnt{A_{{\mathrm{2}}}} ) /  \mathit{X} } }   \Gamma'  \ottsym{,}  \mathit{x} \,  {:}  \, \ottnt{T''}    [   \ottnt{A} _{  { \ottsym{3}  \ottsym{-}  \ottmv{i} }_{ \ottsym{0} }  }   /  \mathit{X}  ]   $,
    \item $\sigma_{{\mathrm{2}}}  \ottsym{(}  \ottnt{C_{{\mathrm{1}}}}  \ottsym{)} =   \ottnt{C''_{{\mathrm{21}}}}    [   \ottnt{A} _{  { \ottmv{i} }_{ \ottsym{2} }  }   /  \mathit{X}  ]    \mathrel{ \Leftrightarrow _{( \ottnt{A_{{\mathrm{1}}}} , \ottnt{A_{{\mathrm{2}}}} ) /  \mathit{X} } }   \ottnt{C''_{{\mathrm{21}}}}    [   \ottnt{A} _{  { \ottsym{3}  \ottsym{-}  \ottmv{i} }_{ \ottsym{2} }  }   /  \mathit{X}  ]   $, and
    \item $\sigma_{{\mathrm{1}}}  \ottsym{(}  \ottnt{C_{{\mathrm{1}}}}  \ottsym{)} =   \ottnt{C''_{{\mathrm{11}}}}    [   \ottnt{A} _{  { \ottmv{i} }_{ \ottsym{1} }  }   /  \mathit{X}  ]    \mathrel{ \Leftrightarrow _{( \ottnt{A_{{\mathrm{1}}}} , \ottnt{A_{{\mathrm{2}}}} ) /  \mathit{X} } }   \ottnt{C''_{{\mathrm{11}}}}    [   \ottnt{A} _{  { \ottsym{3}  \ottsym{-}  \ottmv{i} }_{ \ottsym{1} }  }   /  \mathit{X}  ]   $,
   \end{itemize}
   we have
   \[
     \Gamma'  \ottsym{,}  \mathit{x} \,  {:}  \, \ottnt{T''}    [   \ottnt{A} _{  { \ottsym{3}  \ottsym{-}  \ottmv{i} }_{ \ottsym{0} }  }   /  \mathit{X}  ]    \vdash   \ottnt{C''_{{\mathrm{21}}}}    [   \ottnt{A} _{  { \ottsym{3}  \ottsym{-}  \ottmv{i} }_{ \ottsym{2} }  }   /  \mathit{X}  ]    \ottsym{<:}   \ottnt{C''_{{\mathrm{11}}}}    [   \ottnt{A} _{  { \ottsym{3}  \ottsym{-}  \ottmv{i} }_{ \ottsym{1} }  }   /  \mathit{X}  ]  
   \]
   by the IH.
   Because
   \begin{itemize}
    \item $\Gamma  \vdash  \sigma_{{\mathrm{1}}}  \ottsym{(}  \ottnt{C_{{\mathrm{2}}}}  \ottsym{)}  \ottsym{<:}  \sigma_{{\mathrm{2}}}  \ottsym{(}  \ottnt{C_{{\mathrm{2}}}}  \ottsym{)}$,
    \item $ \Gamma  \mathrel{ \Leftrightarrow _{( \ottnt{A_{{\mathrm{1}}}} , \ottnt{A_{{\mathrm{2}}}} ) /  \mathit{X} } }  \Gamma' $,
    \item $\sigma_{{\mathrm{1}}}  \ottsym{(}  \ottnt{C_{{\mathrm{2}}}}  \ottsym{)} =   \ottnt{C''_{{\mathrm{12}}}}    [   \ottnt{A} _{  { \ottmv{i} }_{ \ottsym{1} }  }   /  \mathit{X}  ]    \mathrel{ \Leftrightarrow _{( \ottnt{A_{{\mathrm{1}}}} , \ottnt{A_{{\mathrm{2}}}} ) /  \mathit{X} } }   \ottnt{C''_{{\mathrm{12}}}}    [   \ottnt{A} _{  { \ottsym{3}  \ottsym{-}  \ottmv{i} }_{ \ottsym{1} }  }   /  \mathit{X}  ]   $, and
    \item $\sigma_{{\mathrm{2}}}  \ottsym{(}  \ottnt{C_{{\mathrm{2}}}}  \ottsym{)} =   \ottnt{C''_{{\mathrm{22}}}}    [   \ottnt{A} _{  { \ottmv{i} }_{ \ottsym{2} }  }   /  \mathit{X}  ]    \mathrel{ \Leftrightarrow _{( \ottnt{A_{{\mathrm{1}}}} , \ottnt{A_{{\mathrm{2}}}} ) /  \mathit{X} } }   \ottnt{C''_{{\mathrm{22}}}}    [   \ottnt{A} _{  { \ottsym{3}  \ottsym{-}  \ottmv{i} }_{ \ottsym{2} }  }   /  \mathit{X}  ]   $,
   \end{itemize}
   we have
   \[
    \Gamma'  \vdash   \ottnt{C''_{{\mathrm{12}}}}    [   \ottnt{A} _{  { \ottsym{3}  \ottsym{-}  \ottmv{i} }_{ \ottsym{1} }  }   /  \mathit{X}  ]    \ottsym{<:}   \ottnt{C''_{{\mathrm{22}}}}    [   \ottnt{A} _{  { \ottsym{3}  \ottsym{-}  \ottmv{i} }_{ \ottsym{2} }  }   /  \mathit{X}  ]  
   \]
   by the IH.
   By \Srule{Ans}, we have
   \[
      \forall  \, { \Phi' } . \  \Gamma' \,  |  \,  \ottnt{T''}    [   \ottnt{A} _{  { \ottsym{3}  \ottsym{-}  \ottmv{i} }_{ \ottsym{0} }  }   /  \mathit{X}  ]   \,  \,\&\,  \, \Phi'  \vdash   \ottsym{(}   ( \forall  \mathit{x}  .  \ottnt{C''_{{\mathrm{11}}}}  )  \Rightarrow   \ottnt{C''_{{\mathrm{12}}}}   \ottsym{)}    [   \ottnt{A} _{  { \ottsym{3}  \ottsym{-}  \ottmv{i} }_{ \ottsym{1} }  }   /  \mathit{X}  ]    \ottsym{<:}   \ottsym{(}   ( \forall  \mathit{x}  .  \ottnt{C''_{{\mathrm{21}}}}  )  \Rightarrow   \ottnt{C''_{{\mathrm{22}}}}   \ottsym{)}    [   \ottnt{A} _{  { \ottsym{3}  \ottsym{-}  \ottmv{i} }_{ \ottsym{2} }  }   /  \mathit{X}  ]    ~.
   \]
   Therefore, we have the conclusion
   $  \forall  \, { \Phi' } . \  \Gamma' \,  |  \, \ottnt{T'} \,  \,\&\,  \, \Phi'  \vdash  \ottnt{S'_{{\mathrm{1}}}}  \ottsym{<:}  \ottnt{S'_{{\mathrm{2}}}} $.

  \case \Srule{Embed}:
   We are given
   $\sigma_{{\mathrm{1}}}  \ottsym{(}  \ottnt{S}  \ottsym{)} \,  =  \,  \square $ and
   $\sigma_{{\mathrm{2}}}  \ottsym{(}  \ottnt{S}  \ottsym{)} \,  =  \,  ( \forall  \mathit{x}  .  \ottnt{C_{{\mathrm{21}}}}  )  \Rightarrow   \ottnt{C_{{\mathrm{22}}}} $ for some $\mathit{x}$, $\ottnt{C_{{\mathrm{21}}}}$, and $\ottnt{C_{{\mathrm{22}}}}$,
   but it is contradictory.

  \case \Srule{Eff}:
   We are given
   \begin{prooftree}
    \AxiomC{$\Gamma  \ottsym{,}  \mathit{x} \,  {:}  \,  \Sigma ^\ast   \models    { \ottsym{(}  \sigma_{{\mathrm{1}}}  \ottsym{(}  \Phi  \ottsym{)}  \ottsym{)} }^\mu   \ottsym{(}  \mathit{x}  \ottsym{)}    \Rightarrow     { \ottsym{(}  \sigma_{{\mathrm{2}}}  \ottsym{(}  \Phi  \ottsym{)}  \ottsym{)} }^\mu   \ottsym{(}  \mathit{x}  \ottsym{)} $}
    \AxiomC{$\Gamma  \ottsym{,}  \mathit{y} \,  {:}  \,  \Sigma ^\omega   \models    { \ottsym{(}  \sigma_{{\mathrm{1}}}  \ottsym{(}  \Phi  \ottsym{)}  \ottsym{)} }^\nu   \ottsym{(}  \mathit{y}  \ottsym{)}    \Rightarrow     { \ottsym{(}  \sigma_{{\mathrm{2}}}  \ottsym{(}  \Phi  \ottsym{)}  \ottsym{)} }^\nu   \ottsym{(}  \mathit{y}  \ottsym{)} $}
    \BinaryInfC{$\Gamma  \vdash  \sigma_{{\mathrm{1}}}  \ottsym{(}  \Phi  \ottsym{)}  \ottsym{<:}  \sigma_{{\mathrm{2}}}  \ottsym{(}  \Phi  \ottsym{)}$}
   \end{prooftree}
   for some $\mathit{x}$ and $\mathit{y}$ such that $\mathit{x}  \ottsym{,}  \mathit{y} \,  \not\in  \,  \mathit{fv}  (  \sigma_{{\mathrm{1}}}  \ottsym{(}  \Phi  \ottsym{)}  )  \,  \mathbin{\cup}  \,  \mathit{fv}  (  \sigma_{{\mathrm{2}}}  \ottsym{(}  \Phi  \ottsym{)}  ) $.
   Because $ \sigma_{{\mathrm{1}}}  \ottsym{(}  \Phi  \ottsym{)}  \mathrel{ \Leftrightarrow _{( \ottnt{A_{{\mathrm{1}}}} , \ottnt{A_{{\mathrm{2}}}} ) /  \mathit{X} } }  \Phi'_{{\mathrm{1}}} $ and $ \sigma_{{\mathrm{2}}}  \ottsym{(}  \Phi  \ottsym{)}  \mathrel{ \Leftrightarrow _{( \ottnt{A_{{\mathrm{1}}}} , \ottnt{A_{{\mathrm{2}}}} ) /  \mathit{X} } }  \Phi'_{{\mathrm{2}}} $,
   there exist some $\Phi''_{{\mathrm{1}}}$, $\Phi''_{{\mathrm{2}}}$, and $ { \ottmv{i} }_{ \ottsym{1} } ,  { \ottmv{i} }_{ \ottsym{2} }  \in \{1,2\}$ such that
   \begin{itemize}
    \item $\sigma_{{\mathrm{1}}}  \ottsym{(}  \Phi  \ottsym{)} \,  =  \,  \Phi''_{{\mathrm{1}}}    [   \ottnt{A} _{  { \ottmv{i} }_{ \ottsym{1} }  }   /  \mathit{X}  ]  $,
    \item $\Phi'_{{\mathrm{1}}} \,  =  \,  \Phi''_{{\mathrm{1}}}    [   \ottnt{A} _{  { \ottsym{3}  \ottsym{-}  \ottmv{i} }_{ \ottsym{1} }  }   /  \mathit{X}  ]  $,
    \item $\sigma_{{\mathrm{2}}}  \ottsym{(}  \Phi  \ottsym{)} \,  =  \,  \Phi''_{{\mathrm{2}}}    [   \ottnt{A} _{  { \ottmv{i} }_{ \ottsym{2} }  }   /  \mathit{X}  ]  $, and
    \item $\Phi'_{{\mathrm{2}}} \,  =  \,  \Phi''_{{\mathrm{2}}}    [   \ottnt{A} _{  { \ottsym{3}  \ottsym{-}  \ottmv{i} }_{ \ottsym{2} }  }   /  \mathit{X}  ]  $.
   \end{itemize}
   Because
   \begin{itemize}
    \item $\Gamma  \ottsym{,}  \mathit{x} \,  {:}  \,  \Sigma ^\ast   \models    { \ottsym{(}  \sigma_{{\mathrm{1}}}  \ottsym{(}  \Phi  \ottsym{)}  \ottsym{)} }^\mu   \ottsym{(}  \mathit{x}  \ottsym{)}    \Rightarrow     { \ottsym{(}  \sigma_{{\mathrm{2}}}  \ottsym{(}  \Phi  \ottsym{)}  \ottsym{)} }^\mu   \ottsym{(}  \mathit{x}  \ottsym{)} $ means
          $\Gamma  \ottsym{,}  \mathit{x} \,  {:}  \,  \Sigma ^\ast   \models    { \ottsym{(}   \Phi''_{{\mathrm{1}}}    [   \ottnt{A} _{  { \ottmv{i} }_{ \ottsym{1} }  }   /  \mathit{X}  ]    \ottsym{)} }^\mu   \ottsym{(}  \mathit{x}  \ottsym{)}    \Rightarrow     { \ottsym{(}   \Phi''_{{\mathrm{2}}}    [   \ottnt{A} _{  { \ottmv{i} }_{ \ottsym{2} }  }   /  \mathit{X}  ]    \ottsym{)} }^\mu   \ottsym{(}  \mathit{x}  \ottsym{)} $,
    \item $ \Gamma  \ottsym{,}  \mathit{x} \,  {:}  \,  \Sigma ^\ast   \mathrel{ \Leftrightarrow _{( \ottnt{A_{{\mathrm{1}}}} , \ottnt{A_{{\mathrm{2}}}} ) /  \mathit{X} } }  \Gamma'  \ottsym{,}  \mathit{x} \,  {:}  \,  \Sigma ^\ast  $,
    \item $  { \ottsym{(}   \Phi''_{{\mathrm{1}}}    [   \ottnt{A} _{  { \ottmv{i} }_{ \ottsym{1} }  }   /  \mathit{X}  ]    \ottsym{)} }^\mu   \ottsym{(}  \mathit{x}  \ottsym{)}  \mathrel{ \Leftrightarrow _{( \ottnt{A_{{\mathrm{1}}}} , \ottnt{A_{{\mathrm{2}}}} ) /  \mathit{X} } }   { \ottsym{(}   \Phi''_{{\mathrm{1}}}    [   \ottnt{A} _{  { \ottsym{3}  \ottsym{-}  \ottmv{i} }_{ \ottsym{1} }  }   /  \mathit{X}  ]    \ottsym{)} }^\mu   \ottsym{(}  \mathit{x}  \ottsym{)} $, and
    \item $  { \ottsym{(}   \Phi''_{{\mathrm{2}}}    [   \ottnt{A} _{  { \ottmv{i} }_{ \ottsym{2} }  }   /  \mathit{X}  ]    \ottsym{)} }^\mu   \ottsym{(}  \mathit{x}  \ottsym{)}  \mathrel{ \Leftrightarrow _{( \ottnt{A_{{\mathrm{1}}}} , \ottnt{A_{{\mathrm{2}}}} ) /  \mathit{X} } }   { \ottsym{(}   \Phi''_{{\mathrm{2}}}    [   \ottnt{A} _{  { \ottsym{3}  \ottsym{-}  \ottmv{i} }_{ \ottsym{2} }  }   /  \mathit{X}  ]    \ottsym{)} }^\mu   \ottsym{(}  \mathit{x}  \ottsym{)} $,
   \end{itemize}
   we have
   \[
    \Gamma'  \ottsym{,}  \mathit{x} \,  {:}  \,  \Sigma ^\ast   \models    { \ottsym{(}   \Phi''_{{\mathrm{1}}}    [   \ottnt{A} _{  { \ottsym{3}  \ottsym{-}  \ottmv{i} }_{ \ottsym{1} }  }   /  \mathit{X}  ]    \ottsym{)} }^\mu   \ottsym{(}  \mathit{x}  \ottsym{)}    \Rightarrow     { \ottsym{(}   \Phi''_{{\mathrm{2}}}    [   \ottnt{A} _{  { \ottsym{3}  \ottsym{-}  \ottmv{i} }_{ \ottsym{2} }  }   /  \mathit{X}  ]    \ottsym{)} }^\mu   \ottsym{(}  \mathit{x}  \ottsym{)} 
   \]
   by \reflem{lr-iff-impl-valid}.
   Because
   \begin{itemize}
    \item $\Gamma  \ottsym{,}  \mathit{y} \,  {:}  \,  \Sigma ^\omega   \models    { \ottsym{(}  \sigma_{{\mathrm{1}}}  \ottsym{(}  \Phi  \ottsym{)}  \ottsym{)} }^\nu   \ottsym{(}  \mathit{y}  \ottsym{)}    \Rightarrow     { \ottsym{(}  \sigma_{{\mathrm{2}}}  \ottsym{(}  \Phi  \ottsym{)}  \ottsym{)} }^\nu   \ottsym{(}  \mathit{y}  \ottsym{)} $ means
          $\Gamma  \ottsym{,}  \mathit{y} \,  {:}  \,  \Sigma ^\omega   \models    { \ottsym{(}   \Phi''_{{\mathrm{1}}}    [   \ottnt{A} _{  { \ottmv{i} }_{ \ottsym{1} }  }   /  \mathit{X}  ]    \ottsym{)} }^\nu   \ottsym{(}  \mathit{y}  \ottsym{)}    \Rightarrow     { \ottsym{(}   \Phi''_{{\mathrm{2}}}    [   \ottnt{A} _{  { \ottmv{i} }_{ \ottsym{2} }  }   /  \mathit{X}  ]    \ottsym{)} }^\nu   \ottsym{(}  \mathit{y}  \ottsym{)} $,
    \item $ \Gamma  \ottsym{,}  \mathit{y} \,  {:}  \,  \Sigma ^\omega   \mathrel{ \Leftrightarrow _{( \ottnt{A_{{\mathrm{1}}}} , \ottnt{A_{{\mathrm{2}}}} ) /  \mathit{X} } }  \Gamma'  \ottsym{,}  \mathit{y} \,  {:}  \,  \Sigma ^\omega  $,
    \item $  { \ottsym{(}   \Phi''_{{\mathrm{1}}}    [   \ottnt{A} _{  { \ottmv{i} }_{ \ottsym{1} }  }   /  \mathit{X}  ]    \ottsym{)} }^\nu   \ottsym{(}  \mathit{y}  \ottsym{)}  \mathrel{ \Leftrightarrow _{( \ottnt{A_{{\mathrm{1}}}} , \ottnt{A_{{\mathrm{2}}}} ) /  \mathit{X} } }   { \ottsym{(}   \Phi''_{{\mathrm{1}}}    [   \ottnt{A} _{  { \ottsym{3}  \ottsym{-}  \ottmv{i} }_{ \ottsym{1} }  }   /  \mathit{X}  ]    \ottsym{)} }^\nu   \ottsym{(}  \mathit{y}  \ottsym{)} $, and
    \item $  { \ottsym{(}   \Phi''_{{\mathrm{2}}}    [   \ottnt{A} _{  { \ottmv{i} }_{ \ottsym{2} }  }   /  \mathit{X}  ]    \ottsym{)} }^\nu   \ottsym{(}  \mathit{y}  \ottsym{)}  \mathrel{ \Leftrightarrow _{( \ottnt{A_{{\mathrm{1}}}} , \ottnt{A_{{\mathrm{2}}}} ) /  \mathit{X} } }   { \ottsym{(}   \Phi''_{{\mathrm{2}}}    [   \ottnt{A} _{  { \ottsym{3}  \ottsym{-}  \ottmv{i} }_{ \ottsym{2} }  }   /  \mathit{X}  ]    \ottsym{)} }^\nu   \ottsym{(}  \mathit{y}  \ottsym{)} $,
   \end{itemize}
   we have
   \[
    \Gamma'  \ottsym{,}  \mathit{y} \,  {:}  \,  \Sigma ^\omega   \models    { \ottsym{(}   \Phi''_{{\mathrm{1}}}    [   \ottnt{A} _{  { \ottsym{3}  \ottsym{-}  \ottmv{i} }_{ \ottsym{1} }  }   /  \mathit{X}  ]    \ottsym{)} }^\nu   \ottsym{(}  \mathit{y}  \ottsym{)}    \Rightarrow     { \ottsym{(}   \Phi''_{{\mathrm{2}}}    [   \ottnt{A} _{  { \ottsym{3}  \ottsym{-}  \ottmv{i} }_{ \ottsym{2} }  }   /  \mathit{X}  ]    \ottsym{)} }^\nu   \ottsym{(}  \mathit{y}  \ottsym{)} 
   \]
   by \reflem{lr-iff-impl-valid}.
   By \Srule{Eff}, we have
   \[
    \Gamma'  \vdash   \Phi''_{{\mathrm{1}}}    [   \ottnt{A} _{  { \ottsym{3}  \ottsym{-}  \ottmv{i} }_{ \ottsym{1} }  }   /  \mathit{X}  ]    \ottsym{<:}   \Phi''_{{\mathrm{2}}}    [   \ottnt{A} _{  { \ottsym{3}  \ottsym{-}  \ottmv{i} }_{ \ottsym{2} }  }   /  \mathit{X}  ]   ~.
   \]
   Therefore, we have the conclusion $\Gamma'  \vdash  \Phi'_{{\mathrm{1}}}  \ottsym{<:}  \Phi'_{{\mathrm{2}}}$.
 \end{caseanalysis}
\end{proof}

\begin{lemmap}{Substitution of Predicate Fixpoints}{lr-subtype-subst-fixpoint}
 Assume that
 \begin{itemize}
  \item $\Gamma  \ottsym{,}  \ottsym{(}   \tceseq{ \mathit{X'} }   \ottsym{,}   \tceseq{ \mathit{X} }   \ottsym{)} \,  {:}  \,  (   \Sigma ^\ast   ,   \tceseq{ \iota }   )   \ottsym{,}  \ottsym{(}   \tceseq{ \mathit{Y'} }   \ottsym{,}   \tceseq{ \mathit{Y} }   \ottsym{)} \,  {:}  \,  (   \Sigma ^\omega   ,   \tceseq{ \iota }   )   \ottsym{,}   \tceseq{ \mathit{z_{{\mathrm{0}}}}  \,  {:}  \,  \iota_{{\mathrm{0}}} }   \vdash  \ottnt{C}$,
  \item $ \tceseq{ \ottsym{(}  \mathit{X}  \ottsym{,}  \mathit{Y}  \ottsym{)} }   \ottsym{;}   \tceseq{ \mathit{z_{{\mathrm{0}}}}  \,  {:}  \,  \iota_{{\mathrm{0}}} }   \vdash  \ottnt{C_{{\mathrm{0}}}}  \succ  \ottnt{C}$,
  \item $ \tceseq{ \ottsym{(}  \mathit{X'}  \ottsym{,}  \mathit{Y'}  \ottsym{)} }   \ottsym{;}   \tceseq{ \ottsym{(}  \mathit{X}  \ottsym{,}  \mathit{Y}  \ottsym{)} }  \,  \vdash ^{\circ}  \, \ottnt{C}$, and
  \item $ \tceseq{ \ottsym{(}  \mathit{X}  \ottsym{,}  \mathit{Y}  \ottsym{)} }   \ottsym{;}   \tceseq{ \mathit{z_{{\mathrm{0}}}}  \,  {:}  \,  \iota_{{\mathrm{0}}} }  \,  |  \, \ottnt{C}  \vdash  \sigma$.
 \end{itemize}
 Then,
 $\Gamma  \ottsym{,}   \tceseq{ \mathit{X'} }  \,  {:}  \,  (   \Sigma ^\ast   ,   \tceseq{ \iota }   )   \ottsym{,}   \tceseq{ \mathit{Y'} }  \,  {:}  \,  (   \Sigma ^\omega   ,   \tceseq{ \iota }   )   \ottsym{,}   \tceseq{ \mathit{z_{{\mathrm{0}}}}  \,  {:}  \,  \iota_{{\mathrm{0}}} }   \vdash  \sigma  \ottsym{(}  \ottnt{C}  \ottsym{)}  \ottsym{<:}  \sigma  \ottsym{(}  \ottnt{C_{{\mathrm{0}}}}  \ottsym{)}$.
\end{lemmap}
\begin{proof}
 By induction on $\ottnt{C}$.
 It suffices to show the following with \Srule{Comp}.
 Let $\Gamma' \,  =  \,  \tceseq{ \mathit{X'} }  \,  {:}  \,  (   \Sigma ^\ast   ,   \tceseq{ \iota }   )   \ottsym{,}   \tceseq{ \mathit{Y'} }  \,  {:}  \,  (   \Sigma ^\omega   ,   \tceseq{ \iota }   )   \ottsym{,}   \tceseq{ \mathit{z_{{\mathrm{0}}}}  \,  {:}  \,  \iota_{{\mathrm{0}}} } $
 in what follows.
 \begin{itemize}
  \item We show that $\Gamma  \ottsym{,}  \Gamma'  \vdash  \sigma  \ottsym{(}   \ottnt{C}  . \ottnt{T}   \ottsym{)}  \ottsym{<:}  \sigma  \ottsym{(}   \ottnt{C_{{\mathrm{0}}}}  . \ottnt{T}   \ottsym{)}$.
        $ \tceseq{ \ottsym{(}  \mathit{X}  \ottsym{,}  \mathit{Y}  \ottsym{)} }   \ottsym{;}   \tceseq{ \mathit{z_{{\mathrm{0}}}}  \,  {:}  \,  \iota_{{\mathrm{0}}} }   \vdash  \ottnt{C_{{\mathrm{0}}}}  \succ  \ottnt{C}$ implies $ \ottnt{C_{{\mathrm{0}}}}  . \ottnt{T}  \,  =  \,  \ottnt{C}  . \ottnt{T} $.
        $ \tceseq{ \ottsym{(}  \mathit{X'}  \ottsym{,}  \mathit{Y'}  \ottsym{)} }   \ottsym{;}   \tceseq{ \ottsym{(}  \mathit{X}  \ottsym{,}  \mathit{Y}  \ottsym{)} }  \,  \vdash ^{\circ}  \, \ottnt{C}$ implies $  \{   \tceseq{ \mathit{X} }   \ottsym{,}   \tceseq{ \mathit{X'} }   \ottsym{,}   \tceseq{ \mathit{Y} }   \ottsym{,}   \tceseq{ \mathit{Y'} }   \}   \, \ottsym{\#} \,   \mathit{fpv}  (   \ottnt{C}  . \ottnt{T}   )  $.
        Therefore, it suffices to show that $\Gamma  \ottsym{,}  \Gamma'  \vdash   \ottnt{C}  . \ottnt{T}   \ottsym{<:}   \ottnt{C}  . \ottnt{T} $,
        which is implied by
        Lemmas~\ref{lem:ts-elim-unused-binding-wf-ftyping} and \ref{lem:ts-refl-subtyping}
        with the assumption
        $\Gamma  \ottsym{,}  \ottsym{(}   \tceseq{ \mathit{X'} }   \ottsym{,}   \tceseq{ \mathit{X} }   \ottsym{)} \,  {:}  \,  (   \Sigma ^\ast   ,   \tceseq{ \iota }   )   \ottsym{,}  \ottsym{(}   \tceseq{ \mathit{Y'} }   \ottsym{,}   \tceseq{ \mathit{Y} }   \ottsym{)} \,  {:}  \,  (   \Sigma ^\omega   ,   \tceseq{ \iota }   )   \ottsym{,}   \tceseq{ \mathit{z_{{\mathrm{0}}}}  \,  {:}  \,  \iota_{{\mathrm{0}}} }   \vdash  \ottnt{C}$.

  \item We show that $\Gamma  \ottsym{,}  \Gamma'  \vdash  \sigma  \ottsym{(}   \ottnt{C}  . \Phi   \ottsym{)}  \ottsym{<:}  \sigma  \ottsym{(}   \ottnt{C_{{\mathrm{0}}}}  . \Phi   \ottsym{)}$.
        $ \tceseq{ \ottsym{(}  \mathit{X}  \ottsym{,}  \mathit{Y}  \ottsym{)} }   \ottsym{;}   \tceseq{ \mathit{z_{{\mathrm{0}}}}  \,  {:}  \,  \iota_{{\mathrm{0}}} }   \vdash  \ottnt{C_{{\mathrm{0}}}}  \succ  \ottnt{C}$ implies
        \begin{itemize}
         \item $ \tceseq{ \ottsym{(}  \mathit{X}  \ottsym{,}  \mathit{Y}  \ottsym{)} }  \,  =  \, \ottsym{(}  \mathit{X_{{\mathrm{0}}}}  \ottsym{,}  \mathit{Y_{{\mathrm{0}}}}  \ottsym{)}  \ottsym{,}   \tceseq{ \ottsym{(}  \mathit{X_{{\mathrm{1}}}}  \ottsym{,}  \mathit{Y_{{\mathrm{1}}}}  \ottsym{)} } $ and
         \item $ \ottnt{C_{{\mathrm{0}}}}  . \Phi  \,  =  \,  (   \lambda_\mu\!   \,  \mathit{x}  .\,  \mathit{X_{{\mathrm{0}}}}  \ottsym{(}  \mathit{x}  \ottsym{,}   \tceseq{ \mathit{z_{{\mathrm{0}}}} }   \ottsym{)}  ,   \lambda_\nu\!   \,  \mathit{y}  .\,  \mathit{Y_{{\mathrm{0}}}}  \ottsym{(}  \mathit{y}  \ottsym{,}   \tceseq{ \mathit{z_{{\mathrm{0}}}} }   \ottsym{)}  ) $
        \end{itemize}
        for some $\mathit{X_{{\mathrm{0}}}}$, $\mathit{Y_{{\mathrm{0}}}}$, $ \tceseq{ \mathit{X_{{\mathrm{1}}}} } $, $ \tceseq{ \mathit{Y_{{\mathrm{1}}}} } $, $\mathit{x}$, and $\mathit{y}$.
        Without loss of generality, we can assume that $\mathit{x}$ and $\mathit{y}$ are fresh.
        By $ \tceseq{ \ottsym{(}  \mathit{X}  \ottsym{,}  \mathit{Y}  \ottsym{)} }   \ottsym{;}   \tceseq{ \mathit{z_{{\mathrm{0}}}}  \,  {:}  \,  \iota_{{\mathrm{0}}} }  \,  |  \, \ottnt{C}  \vdash  \sigma$, we have
        \[
         \sigma  \ottsym{(}   \ottnt{C_{{\mathrm{0}}}}  . \Phi   \ottsym{)} \,  =  \,  (   \lambda_\mu\!   \,  \mathit{x}  .\,    \mu\!   \, (  \mathit{X_{{\mathrm{0}}}} ,  \mathit{Y_{{\mathrm{0}}}} ,   \tceseq{ \mathit{z_{{\mathrm{0}}}}  \,  {:}  \,  \iota }  ,   \ottnt{C}  . \Phi   )   \ottsym{(}  \mathit{x}  \ottsym{,}   \tceseq{ \mathit{z_{{\mathrm{0}}}} }   \ottsym{)}  ,   \lambda_\nu\!   \,  \mathit{y}  .\,    \nu\!   \, (  \mathit{X_{{\mathrm{0}}}} ,  \mathit{Y_{{\mathrm{0}}}} ,   \tceseq{ \mathit{z_{{\mathrm{0}}}}  \,  {:}  \,  \iota_{{\mathrm{0}}} }  ,   \ottnt{C}  . \Phi   )   \ottsym{(}  \mathit{y}  \ottsym{,}   \tceseq{ \mathit{z_{{\mathrm{0}}}} }   \ottsym{)}  )  ~.
        \]
        We show the conclusion as follows.
         \begin{itemize}
          \item Assume that $ { \ottsym{(}   \ottnt{C}  . \Phi   \ottsym{)} }^\mu   \ottsym{(}  \mathit{x}  \ottsym{)} \,  =  \, \phi_{{\mathrm{1}}}$.
                Then, we show that
                \[
                 \Gamma  \ottsym{,}  \Gamma'  \ottsym{,}  \mathit{x} \,  {:}  \,  \Sigma ^\ast   \models   \sigma  \ottsym{(}  \phi_{{\mathrm{1}}}  \ottsym{)}    \Rightarrow    \ottsym{(}   \mu\!  \, \mathit{X_{{\mathrm{0}}}}  \ottsym{(}  \mathit{x} \,  {:}  \,  \Sigma ^\ast   \ottsym{,}   \tceseq{ \mathit{z_{{\mathrm{0}}}}  \,  {:}  \,  \iota_{{\mathrm{0}}} }   \ottsym{)}  \ottsym{.}  \phi_{{\mathrm{1}}}  \ottsym{)}  \ottsym{(}  \mathit{x}  \ottsym{,}   \tceseq{ \mathit{z_{{\mathrm{0}}}} }   \ottsym{)}  ~.
                \]
                $ \tceseq{ \ottsym{(}  \mathit{X'}  \ottsym{,}  \mathit{Y'}  \ottsym{)} }   \ottsym{;}   \tceseq{ \ottsym{(}  \mathit{X}  \ottsym{,}  \mathit{Y}  \ottsym{)} }  \,  \vdash ^{\circ}  \, \ottnt{C}$ implies $  \{   \tceseq{ \mathit{X_{{\mathrm{1}}}} }   \ottsym{,}   \tceseq{ \mathit{Y_{{\mathrm{1}}}} }   \ottsym{,}  \mathit{Y_{{\mathrm{0}}}}  \}   \, \ottsym{\#} \,   \mathit{fpv}  (  \phi_{{\mathrm{1}}}  )  $.
                Therefore, it suffices to show that
                \[
                 \Gamma  \ottsym{,}  \Gamma'  \ottsym{,}  \mathit{x} \,  {:}  \,  \Sigma ^\ast   \models    \phi_{{\mathrm{1}}}    [    \mu\!   \, (  \mathit{X_{{\mathrm{0}}}} ,  \mathit{Y_{{\mathrm{0}}}} ,   \tceseq{ \mathit{z_{{\mathrm{0}}}}  \,  {:}  \,  \iota }  ,   \ottnt{C}  . \Phi   )   /  \mathit{X_{{\mathrm{0}}}}  ]      \Rightarrow    \ottsym{(}   \mu\!  \, \mathit{X_{{\mathrm{0}}}}  \ottsym{(}  \mathit{x} \,  {:}  \,  \Sigma ^\ast   \ottsym{,}   \tceseq{ \mathit{z_{{\mathrm{0}}}}  \,  {:}  \,  \iota_{{\mathrm{0}}} }   \ottsym{)}  \ottsym{.}  \phi_{{\mathrm{1}}}  \ottsym{)}  \ottsym{(}  \mathit{x}  \ottsym{,}   \tceseq{ \mathit{z_{{\mathrm{0}}}} }   \ottsym{)}  ~,
                \]
                which is implied by \reflem{ts-mu-nu-valid}.
          \item Assume that $ { \ottsym{(}   \ottnt{C}  . \Phi   \ottsym{)} }^\nu   \ottsym{(}  \mathit{y}  \ottsym{)} \,  =  \, \phi_{{\mathrm{2}}}$.
                Then, we show that
                \[
                 \Gamma  \ottsym{,}  \Gamma'  \ottsym{,}  \mathit{y} \,  {:}  \,  \Sigma ^\omega   \models   \sigma  \ottsym{(}  \phi_{{\mathrm{2}}}  \ottsym{)}    \Rightarrow    \ottsym{(}   \nu\!  \, \mathit{Y_{{\mathrm{0}}}}  \ottsym{(}  \mathit{x} \,  {:}  \,  \Sigma ^\omega   \ottsym{,}   \tceseq{ \mathit{z_{{\mathrm{0}}}}  \,  {:}  \,  \iota_{{\mathrm{0}}} }   \ottsym{)}  \ottsym{.}   \phi_{{\mathrm{2}}}    [    \mu\!   \, (  \mathit{X_{{\mathrm{0}}}} ,  \mathit{Y_{{\mathrm{0}}}} ,   \tceseq{ \mathit{z_{{\mathrm{0}}}}  \,  {:}  \,  \iota }  ,   \ottnt{C}  . \Phi   )   /  \mathit{X_{{\mathrm{0}}}}  ]    \ottsym{)}  \ottsym{(}  \mathit{y}  \ottsym{,}   \tceseq{ \mathit{z_{{\mathrm{0}}}} }   \ottsym{)}  ~.
                \]
                $ \tceseq{ \ottsym{(}  \mathit{X'}  \ottsym{,}  \mathit{Y'}  \ottsym{)} }   \ottsym{;}   \tceseq{ \ottsym{(}  \mathit{X}  \ottsym{,}  \mathit{Y}  \ottsym{)} }  \,  \vdash ^{\circ}  \, \ottnt{C}$ implies $  \{   \tceseq{ \mathit{X_{{\mathrm{1}}}} }   \ottsym{,}   \tceseq{ \mathit{Y_{{\mathrm{1}}}} }   \}   \, \ottsym{\#} \,   \mathit{fpv}  (  \phi_{{\mathrm{2}}}  )  $.
                Therefore, it suffices to show that
                \[
                 \Gamma  \ottsym{,}  \Gamma'  \ottsym{,}  \mathit{y} \,  {:}  \,  \Sigma ^\omega   \models   \phi'_{{\mathrm{2}}}    \Rightarrow    \ottsym{(}   \nu\!  \, \mathit{Y_{{\mathrm{0}}}}  \ottsym{(}  \mathit{y} \,  {:}  \,  \Sigma ^\ast   \ottsym{,}   \tceseq{ \mathit{z_{{\mathrm{0}}}}  \,  {:}  \,  \iota_{{\mathrm{0}}} }   \ottsym{)}  \ottsym{.}   \phi_{{\mathrm{2}}}    [    \mu\!   \, (  \mathit{X_{{\mathrm{0}}}} ,  \mathit{Y_{{\mathrm{0}}}} ,   \tceseq{ \mathit{z_{{\mathrm{0}}}}  \,  {:}  \,  \iota }  ,   \ottnt{C}  . \Phi   )   /  \mathit{X_{{\mathrm{0}}}}  ]    \ottsym{)}  \ottsym{(}  \mathit{y}  \ottsym{,}   \tceseq{ \mathit{z_{{\mathrm{0}}}} }   \ottsym{)} 
                \]
                where $\phi'_{{\mathrm{2}}} \,  =  \,   \phi_{{\mathrm{2}}}    [    \mu\!   \, (  \mathit{X_{{\mathrm{0}}}} ,  \mathit{Y_{{\mathrm{0}}}} ,   \tceseq{ \mathit{z_{{\mathrm{0}}}}  \,  {:}  \,  \iota }  ,   \ottnt{C}  . \Phi   )   /  \mathit{X_{{\mathrm{0}}}}  ]      [    \nu\!   \, (  \mathit{X_{{\mathrm{0}}}} ,  \mathit{Y_{{\mathrm{0}}}} ,   \tceseq{ \mathit{z_{{\mathrm{0}}}}  \,  {:}  \,  \iota }  ,   \ottnt{C}  . \Phi   )   /  \mathit{Y_{{\mathrm{0}}}}  ]  $.
                This is implied by \reflem{ts-mu-nu-valid}.
         \end{itemize}

  \item We show that $\Gamma  \ottsym{,}  \Gamma' \,  |  \,  \ottnt{C}  . \ottnt{T}  \,  \,\&\,  \, \sigma  \ottsym{(}   \ottnt{C}  . \Phi   \ottsym{)}  \vdash  \sigma  \ottsym{(}   \ottnt{C}  . \ottnt{S}   \ottsym{)}  \ottsym{<:}  \sigma  \ottsym{(}   \ottnt{C_{{\mathrm{0}}}}  . \ottnt{S}   \ottsym{)}$.
        Note that $\sigma  \ottsym{(}   \ottnt{C}  . \ottnt{T}   \ottsym{)} \,  =  \,  \ottnt{C}  . \ottnt{T} $ as shown above.
        By case analysis on $ \ottnt{C}  . \ottnt{S} $.
        \begin{caseanalysis}
         \case $ \ottnt{C}  . \ottnt{S}  \,  =  \,  \square $:
          Because $ \tceseq{ \ottsym{(}  \mathit{X}  \ottsym{,}  \mathit{Y}  \ottsym{)} }   \ottsym{;}   \tceseq{ \mathit{z_{{\mathrm{0}}}}  \,  {:}  \,  \iota_{{\mathrm{0}}} }   \vdash  \ottnt{C_{{\mathrm{0}}}}  \succ  \ottnt{C}$ implies $ \ottnt{C_{{\mathrm{0}}}}  . \ottnt{S}  \,  =  \,  \square $,
          we have the conclusion by \Srule{Empty}.

         \case $  \exists  \, { \mathit{x}  \ottsym{,}  \ottnt{C_{{\mathrm{1}}}}  \ottsym{,}  \ottnt{C_{{\mathrm{2}}}} } . \   \ottnt{C}  . \ottnt{S}  \,  =  \,  ( \forall  \mathit{x}  .  \ottnt{C_{{\mathrm{1}}}}  )  \Rightarrow   \ottnt{C_{{\mathrm{2}}}}  $:
          Because $ \tceseq{ \ottsym{(}  \mathit{X}  \ottsym{,}  \mathit{Y}  \ottsym{)} }   \ottsym{;}   \tceseq{ \mathit{z_{{\mathrm{0}}}}  \,  {:}  \,  \iota_{{\mathrm{0}}} }   \vdash  \ottnt{C_{{\mathrm{0}}}}  \succ  \ottnt{C}$ implies $ \ottnt{C_{{\mathrm{0}}}}  . \ottnt{S}  \,  =  \,  ( \forall  \mathit{x}  .  \ottnt{C_{{\mathrm{1}}}}  )  \Rightarrow   \ottnt{C_{{\mathrm{02}}}} $
          for some $\ottnt{C_{{\mathrm{02}}}}$, it suffices to show the following with \Srule{Ans}.
          \begin{itemize}
           \item We show that $\Gamma  \ottsym{,}  \Gamma'  \ottsym{,}  \mathit{x} \,  {:}  \,  \ottnt{C}  . \ottnt{T}   \vdash  \sigma  \ottsym{(}  \ottnt{C_{{\mathrm{1}}}}  \ottsym{)}  \ottsym{<:}  \sigma  \ottsym{(}  \ottnt{C_{{\mathrm{1}}}}  \ottsym{)}$.
                 Because $ \tceseq{ \ottsym{(}  \mathit{X'}  \ottsym{,}  \mathit{Y'}  \ottsym{)} }   \ottsym{;}   \tceseq{ \ottsym{(}  \mathit{X}  \ottsym{,}  \mathit{Y}  \ottsym{)} }  \,  \vdash ^{\circ}  \, \ottnt{C}$ implies $  \{   \tceseq{ \mathit{X} }   \ottsym{,}   \tceseq{ \mathit{X'} }   \ottsym{,}   \tceseq{ \mathit{Y} }   \ottsym{,}   \tceseq{ \mathit{Y'} }   \}   \, \ottsym{\#} \,   \mathit{fpv}  (  \ottnt{C_{{\mathrm{1}}}}  )  $,
                 we have the conclusion by
                 Lemmas~\ref{lem:ts-elim-unused-binding-wf-ftyping} and \ref{lem:ts-refl-subtyping}
                 with the assumption
                 $\Gamma  \ottsym{,}  \ottsym{(}   \tceseq{ \mathit{X'} }   \ottsym{,}   \tceseq{ \mathit{X} }   \ottsym{)} \,  {:}  \,  (   \Sigma ^\ast   ,   \tceseq{ \iota }   )   \ottsym{,}  \ottsym{(}   \tceseq{ \mathit{Y'} }   \ottsym{,}   \tceseq{ \mathit{Y} }   \ottsym{)} \,  {:}  \,  (   \Sigma ^\omega   ,   \tceseq{ \iota }   )   \ottsym{,}   \tceseq{ \mathit{z_{{\mathrm{0}}}}  \,  {:}  \,  \iota_{{\mathrm{0}}} }   \vdash  \ottnt{C}$.

           \item We show that $\Gamma  \ottsym{,}  \Gamma'  \vdash  \sigma  \ottsym{(}  \ottnt{C_{{\mathrm{2}}}}  \ottsym{)}  \ottsym{<:}  \sigma  \ottsym{(}  \ottnt{C_{{\mathrm{02}}}}  \ottsym{)}$.
                 We have the following.
                 \begin{itemize}
                  \item $\Gamma  \ottsym{,}  \ottsym{(}   \tceseq{ \mathit{X'} }   \ottsym{,}  \mathit{X_{{\mathrm{0}}}}  \ottsym{,}   \tceseq{ \mathit{X_{{\mathrm{1}}}} }   \ottsym{)} \,  {:}  \,  (   \Sigma ^\ast   ,   \tceseq{ \iota }   )   \ottsym{,}  \ottsym{(}   \tceseq{ \mathit{Y'} }   \ottsym{,}  \mathit{Y_{{\mathrm{0}}}}  \ottsym{,}   \tceseq{ \mathit{Y_{{\mathrm{1}}}} }   \ottsym{)} \,  {:}  \,  (   \Sigma ^\omega   ,   \tceseq{ \iota }   )   \ottsym{,}   \tceseq{ \mathit{z_{{\mathrm{0}}}}  \,  {:}  \,  \iota_{{\mathrm{0}}} }   \vdash  \ottnt{C_{{\mathrm{2}}}}$
                        by
                        $\Gamma  \ottsym{,}  \ottsym{(}   \tceseq{ \mathit{X'} }   \ottsym{,}   \tceseq{ \mathit{X} }   \ottsym{)} \,  {:}  \,  (   \Sigma ^\ast   ,   \tceseq{ \iota }   )   \ottsym{,}  \ottsym{(}   \tceseq{ \mathit{Y'} }   \ottsym{,}   \tceseq{ \mathit{Y} }   \ottsym{)} \,  {:}  \,  (   \Sigma ^\omega   ,   \tceseq{ \iota }   )   \ottsym{,}   \tceseq{ \mathit{z_{{\mathrm{0}}}}  \,  {:}  \,  \iota_{{\mathrm{0}}} }   \vdash  \ottnt{C}$.
                  \item $ \tceseq{ \ottsym{(}  \mathit{X_{{\mathrm{2}}}}  \ottsym{,}  \mathit{Y_{{\mathrm{2}}}}  \ottsym{)} }   \ottsym{;}   \tceseq{ \mathit{z_{{\mathrm{0}}}}  \,  {:}  \,  \iota_{{\mathrm{0}}} }   \vdash  \ottnt{C_{{\mathrm{02}}}}  \succ  \ottnt{C_{{\mathrm{2}}}}$ by $ \tceseq{ \ottsym{(}  \mathit{X}  \ottsym{,}  \mathit{Y}  \ottsym{)} }   \ottsym{;}   \tceseq{ \mathit{z_{{\mathrm{0}}}}  \,  {:}  \,  \iota_{{\mathrm{0}}} }   \vdash  \ottnt{C_{{\mathrm{0}}}}  \succ  \ottnt{C}$.
                  \item $ \tceseq{ \ottsym{(}  \mathit{X'}  \ottsym{,}  \mathit{Y'}  \ottsym{)} }   \ottsym{,}  \ottsym{(}  \mathit{X_{{\mathrm{0}}}}  \ottsym{,}  \mathit{Y_{{\mathrm{0}}}}  \ottsym{)}  \ottsym{;}   \tceseq{ \ottsym{(}  \mathit{X_{{\mathrm{1}}}}  \ottsym{,}  \mathit{Y_{{\mathrm{1}}}}  \ottsym{)} }  \,  \vdash ^{\circ}  \, \ottnt{C_{{\mathrm{2}}}}$ by $ \tceseq{ \ottsym{(}  \mathit{X'}  \ottsym{,}  \mathit{Y'}  \ottsym{)} }   \ottsym{;}   \tceseq{ \ottsym{(}  \mathit{X}  \ottsym{,}  \mathit{Y}  \ottsym{)} }  \,  \vdash ^{\circ}  \, \ottnt{C}$.
                  \item $ \tceseq{ \ottsym{(}  \mathit{X_{{\mathrm{1}}}}  \ottsym{,}  \mathit{Y_{{\mathrm{1}}}}  \ottsym{)} }   \ottsym{;}   \tceseq{ \mathit{z_{{\mathrm{0}}}}  \,  {:}  \,  \iota_{{\mathrm{0}}} }  \,  |  \, \ottnt{C_{{\mathrm{2}}}}  \vdash  \sigma'$ for some $\sigma'$ such that
                        $\sigma \,  =  \, \sigma''  \ottsym{(}  \sigma'  \ottsym{)}  \mathbin{\uplus}  \sigma''$ where
                        $\sigma'' \,  =  \,  [    \mu\!   \, (  \mathit{X_{{\mathrm{0}}}} ,  \mathit{Y_{{\mathrm{0}}}} ,   \tceseq{ \mathit{z_{{\mathrm{0}}}}  \,  {:}  \,  \iota_{{\mathrm{0}}} }  ,   \ottnt{C}  . \Phi   )   /  \mathit{X_{{\mathrm{0}}}}  ]   \mathbin{\uplus}   [    \nu\!   \, (  \mathit{X_{{\mathrm{0}}}} ,  \mathit{Y_{{\mathrm{0}}}} ,   \tceseq{ \mathit{z_{{\mathrm{0}}}}  \,  {:}  \,  \iota_{{\mathrm{0}}} }  ,   \ottnt{C}  . \Phi   )   /  \mathit{Y_{{\mathrm{0}}}}  ] $.
                 \end{itemize}
                 Therefore, by the IH,
                 \[
                  \Gamma  \ottsym{,}  \ottsym{(}   \tceseq{ \mathit{X'} }   \ottsym{,}  \mathit{X_{{\mathrm{0}}}}  \ottsym{)} \,  {:}  \,  (   \Sigma ^\ast   ,   \tceseq{ \iota }   )   \ottsym{,}  \ottsym{(}   \tceseq{ \mathit{Y'} }   \ottsym{,}  \mathit{Y_{{\mathrm{0}}}}  \ottsym{)} \,  {:}  \,  (   \Sigma ^\omega   ,   \tceseq{ \iota }   )   \ottsym{,}   \tceseq{ \mathit{z_{{\mathrm{0}}}}  \,  {:}  \,  \iota_{{\mathrm{0}}} }   \vdash  \sigma'  \ottsym{(}  \ottnt{C_{{\mathrm{2}}}}  \ottsym{)}  \ottsym{<:}  \sigma'  \ottsym{(}  \ottnt{C_{{\mathrm{02}}}}  \ottsym{)} ~.
                 \]
                 By Lemmas~\ref{lem:lr-eff-rec-wf-XXX} and \ref{lem:ts-pred-subst-subtyping},
                 \[
                  \Gamma  \ottsym{,}   \tceseq{ \mathit{X'} }  \,  {:}  \,  (   \Sigma ^\ast   ,   \tceseq{ \iota }   )   \ottsym{,}   \tceseq{ \mathit{Y'} }  \,  {:}  \,  (   \Sigma ^\omega   ,   \tceseq{ \iota }   )   \ottsym{,}   \tceseq{ \mathit{z_{{\mathrm{0}}}}  \,  {:}  \,  \iota_{{\mathrm{0}}} }   \vdash  \sigma''  \ottsym{(}  \sigma'  \ottsym{(}  \ottnt{C_{{\mathrm{2}}}}  \ottsym{)}  \ottsym{)}  \ottsym{<:}  \sigma''  \ottsym{(}  \sigma'  \ottsym{(}  \ottnt{C_{{\mathrm{02}}}}  \ottsym{)}  \ottsym{)} ~.
                 \]
                 Therefore, we have the conclusion.
          \end{itemize}
        \end{caseanalysis}
 \end{itemize}
\end{proof}

\begin{lemmap}{Subject Reduction: Reduction}{ts-subject-red-red}
 If $ \emptyset   \vdash  \ottnt{e}  \ottsym{:}  \ottnt{C}$ and $\ottnt{e} \,  \rightsquigarrow  \, \ottnt{e'}$,
 then $ \emptyset   \vdash  \ottnt{e'}  \ottsym{:}  \ottnt{C}$.
\end{lemmap}
\begin{proof}
 By induction on the typing derivation of $ \emptyset   \vdash  \ottnt{e}  \ottsym{:}  \ottnt{C}$.
 \begin{caseanalysis}
  \case \T{CVar}, \T{Var}, \T{Const}, \T{PAbs}, \T{Abs}, \T{Fun}, \T{Shift}, \T{ShiftD}, \T{Event}, and \T{Div}: \mbox{}
   Contradictory because there is no reduction rule applicable to $\ottnt{e}$.

  \case \T{Sub}: By the IH and \T{Sub}.

  \case \T{Op}:
   We are given
   \begin{prooftree}
    \AxiomC{$\vdash   \emptyset $}
    \AxiomC{$ \tceseq{  \emptyset   \vdash  \ottnt{v}  \ottsym{:}  \ottnt{T} } $}
    \AxiomC{$\mathit{ty} \, \ottsym{(}  \ottnt{o}  \ottsym{)} \,  =  \,  \tceseq{(  \mathit{x}  \,  {:}  \,  \ottnt{T}  )}  \rightarrow   \ottnt{T_{{\mathrm{0}}}} $}
    \TrinaryInfC{$ \emptyset   \vdash  \ottnt{o}  \ottsym{(}   \tceseq{ \ottnt{v} }   \ottsym{)}  \ottsym{:}   \ottnt{T_{{\mathrm{0}}}}    [ \tceseq{ \ottnt{v}  /  \mathit{x} } ]  $}
   \end{prooftree}
   for some $\ottnt{o}$, $ \tceseq{ \ottnt{v} } $, $ \tceseq{ \mathit{x} } $, $ \tceseq{ \ottnt{T} } $, and $\ottnt{T_{{\mathrm{0}}}}$
   such that
   $\ottnt{e} \,  =  \, \ottnt{o}  \ottsym{(}   \tceseq{ \ottnt{v} }   \ottsym{)}$ and
   $\ottnt{C} \,  =  \,   \ottnt{T_{{\mathrm{0}}}}    [ \tceseq{ \ottnt{v}  /  \mathit{x} } ]    \,  \,\&\,  \,   \Phi_\mathsf{val}   \, / \,   \square  $.
   The reduction rule applicable to $\ottnt{o}  \ottsym{(}   \tceseq{ \ottnt{v} }   \ottsym{)}$ is only \R{Const}.
   Therefore, there exists some $ \tceseq{ \ottnt{c} } $ such that
   $ \tceseq{ \ottnt{v} \,  =  \, \ottnt{c} } $ and $\ottnt{e'} \,  =  \,  \zeta  (  \ottnt{o}  ,   \tceseq{ \ottnt{c} }   ) $; namely,
   \[
    \ottnt{e} ~=~ \ottnt{o}  \ottsym{(}   \tceseq{ \ottnt{c} }   \ottsym{)} \,  \rightsquigarrow  \,  \zeta  (  \ottnt{o}  ,   \tceseq{ \ottnt{c} }   )  ~=~ \ottnt{e'} ~.
   \]

   \refasm{const} states that $ \tceseq{(  \mathit{x}  \,  {:}  \,  \ottnt{T}  )}  \rightarrow   \ottnt{T_{{\mathrm{0}}}}  =  \tceseq{(  \mathit{x}  \,  {:}  \,  \ottsym{\{}  \mathit{x} \,  {:}  \, \iota \,  |  \, \phi  \ottsym{\}}  )}  \rightarrow   \ottsym{\{}  \mathit{x_{{\mathrm{0}}}} \,  {:}  \, \iota_{{\mathrm{0}}} \,  |  \, \phi_{{\mathrm{0}}}  \ottsym{\}} $
   for some $ \tceseq{ \ottsym{\{}  \mathit{x} \,  {:}  \, \iota \,  |  \, \phi  \ottsym{\}} } $, $\mathit{x_{{\mathrm{0}}}}$, $\iota_{{\mathrm{0}}}$, and $\phi_{{\mathrm{0}}}$.
   Because $ \tceseq{  \emptyset   \vdash  \ottnt{v}  \ottsym{:}  \ottnt{T} } $ implies $ \tceseq{  \emptyset   \vdash  \ottnt{c}  \ottsym{:}  \ottsym{\{}  \mathit{x} \,  {:}  \, \iota \,  |  \, \phi  \ottsym{\}} } $,
   \reflem{ts-val-inv-refine-ty} implies
   \begin{itemize}
    \item $ \tceseq{  \mathit{ty}  (  \ottnt{c}  )  \,  =  \, \iota } $, and
    \item $ \tceseq{  { \models  \,   \phi    [  \ottnt{c}  /  \mathit{x}  ]   }  } $.
   \end{itemize}
   Thus, \refasm{const} implies $ \mathit{ty}  (   \zeta  (  \ottnt{o}  ,   \tceseq{ \ottnt{c} }   )   )  \,  =  \, \iota_{{\mathrm{0}}}$ and
   $ { \models  \,    \phi_{{\mathrm{0}}}    [ \tceseq{ \ottnt{c}  /  \mathit{x} } ]      [   \zeta  (  \ottnt{o}  ,   \tceseq{ \ottnt{c} }   )   /  \mathit{x_{{\mathrm{0}}}}  ]   } $.
   Then, \reflem{ts-extract-subst-val-valid} implies
   $\mathit{x_{{\mathrm{0}}}} \,  {:}  \, \iota_{{\mathrm{0}}}  \models    \ottsym{(}    \zeta  (  \ottnt{o}  ,   \tceseq{ \ottnt{c} }   )     =    \mathit{x_{{\mathrm{0}}}}   \ottsym{)}    \Rightarrow    \phi_{{\mathrm{0}}}     [ \tceseq{ \ottnt{c}  /  \mathit{x} } ]  $.
   Thus, \Srule{Refine} implies $ \emptyset   \vdash  \ottsym{\{}  \mathit{x_{{\mathrm{0}}}} \,  {:}  \, \iota_{{\mathrm{0}}} \,  |  \,   \zeta  (  \ottnt{o}  ,   \tceseq{ \ottnt{c} }   )     =    \mathit{x_{{\mathrm{0}}}}   \ottsym{\}}  \ottsym{<:}  \ottsym{\{}  \mathit{x_{{\mathrm{0}}}} \,  {:}  \, \iota_{{\mathrm{0}}} \,  |  \,  \phi_{{\mathrm{0}}}    [ \tceseq{ \ottnt{c}  /  \mathit{x} } ]    \ottsym{\}}$.
   Thus, noting $\vdash   \emptyset $ (which is derived by \WFC{Empty}), we have
   \[
     \emptyset   \vdash   \ottsym{\{}  \mathit{x_{{\mathrm{0}}}} \,  {:}  \, \iota_{{\mathrm{0}}} \,  |  \,   \zeta  (  \ottnt{o}  ,   \tceseq{ \ottnt{c} }   )     =    \mathit{x_{{\mathrm{0}}}}   \ottsym{\}}  \,  \,\&\,  \,   \Phi_\mathsf{val}   \, / \,   \square    \ottsym{<:}   \ottsym{\{}  \mathit{x_{{\mathrm{0}}}} \,  {:}  \, \iota_{{\mathrm{0}}} \,  |  \,  \phi_{{\mathrm{0}}}    [ \tceseq{ \ottnt{c}  /  \mathit{x} } ]    \ottsym{\}}  \,  \,\&\,  \,   \Phi_\mathsf{val}   \, / \,   \square  
   \]
   by \reflem{ts-refl-subtyping-pure-eff}.
   We also have
   \[
     \emptyset   \vdash   \zeta  (  \ottnt{o}  ,   \tceseq{ \ottnt{c} }   )   \ottsym{:}  \ottsym{\{}  \mathit{x_{{\mathrm{0}}}} \,  {:}  \, \iota_{{\mathrm{0}}} \,  |  \,   \zeta  (  \ottnt{o}  ,   \tceseq{ \ottnt{c} }   )     =    \mathit{x_{{\mathrm{0}}}}   \ottsym{\}}
   \]
   by \T{Const}, and
   \[
     \emptyset   \vdash   \ottsym{\{}  \mathit{x_{{\mathrm{0}}}} \,  {:}  \, \iota_{{\mathrm{0}}} \,  |  \,  \phi_{{\mathrm{0}}}    [ \tceseq{ \ottnt{c}  /  \mathit{x} } ]    \ottsym{\}}  \,  \,\&\,  \,   \Phi_\mathsf{val}   \, / \,   \square  
   \]
   by \reflem{ts-wf-ty-tctx-typing}
   with $ \emptyset   \vdash  \ottnt{o}  \ottsym{(}   \tceseq{ \ottnt{v} }   \ottsym{)}  \ottsym{:}   \ottnt{T_{{\mathrm{0}}}}    [ \tceseq{ \ottnt{v}  /  \mathit{x} } ]   = \ottsym{\{}  \mathit{x_{{\mathrm{0}}}} \,  {:}  \, \iota_{{\mathrm{0}}} \,  |  \,  \phi_{{\mathrm{0}}}    [ \tceseq{ \ottnt{c}  /  \mathit{x} } ]    \ottsym{\}}$
   (note that we can suppose that
   $\mathit{x_{{\mathrm{0}}}}$ is not included in $ \tceseq{ \mathit{x} } $ without loss of generality).
   Therefore, \T{Sub} implies
   \[
     \emptyset   \vdash   \zeta  (  \ottnt{o}  ,   \tceseq{ \ottnt{c} }   )   \ottsym{:}  \ottsym{\{}  \mathit{x_{{\mathrm{0}}}} \,  {:}  \, \iota_{{\mathrm{0}}} \,  |  \,  \phi_{{\mathrm{0}}}    [ \tceseq{ \ottnt{c}  /  \mathit{x} } ]    \ottsym{\}} ~,
   \]
   which we must show.

  \case \T{App}:
   We are given
   \begin{prooftree}
    \AxiomC{$ \emptyset   \vdash  \ottnt{v_{{\mathrm{1}}}}  \ottsym{:}  \ottsym{(}  \mathit{x} \,  {:}  \, \ottnt{T'}  \ottsym{)}  \rightarrow  \ottnt{C'}$}
    \AxiomC{$ \emptyset   \vdash  \ottnt{v_{{\mathrm{2}}}}  \ottsym{:}  \ottnt{T'}$}
    \BinaryInfC{$ \emptyset   \vdash  \ottnt{v_{{\mathrm{1}}}} \, \ottnt{v_{{\mathrm{2}}}}  \ottsym{:}   \ottnt{C'}    [  \ottnt{v_{{\mathrm{2}}}}  /  \mathit{x}  ]  $}
   \end{prooftree}
   for some $\ottnt{v_{{\mathrm{1}}}}$, $\ottnt{v_{{\mathrm{2}}}}$, $\mathit{x}$, $\ottnt{T'}$, and $\ottnt{C'}$
   such that $\ottnt{e} \,  =  \, \ottnt{v_{{\mathrm{1}}}} \, \ottnt{v_{{\mathrm{2}}}}$ and $\ottnt{C} \,  =  \,  \ottnt{C'}    [  \ottnt{v_{{\mathrm{2}}}}  /  \mathit{x}  ]  $.
   We perform case analysis on the reduction rule applied to reduce $\ottnt{v_{{\mathrm{1}}}} \, \ottnt{v_{{\mathrm{2}}}}$;
   it is either \R{Beta} or \R{RBeta}.
   \begin{caseanalysis}
    \case \R{Beta}:
     We are given
     $\ottnt{v_{{\mathrm{1}}}} \,  =  \, \ottsym{(}   \lambda\!  \,  \tceseq{ \mathit{z_{{\mathrm{0}}}} }   \ottsym{,}  \mathit{z_{{\mathrm{2}}}}  \ottsym{.}  \ottnt{e}  \ottsym{)} \,  \tceseq{ \ottnt{v_{{\mathrm{0}}}} } $ and $\ottnt{e'} \,  =  \,   \ottnt{e}    [ \tceseq{ \ottnt{v_{{\mathrm{0}}}}  /  \mathit{z_{{\mathrm{0}}}} } ]      [  \ottnt{v_{{\mathrm{2}}}}  /  \mathit{z_{{\mathrm{2}}}}  ]  $
     for some $ \tceseq{ \mathit{z_{{\mathrm{0}}}} } $, $\mathit{z_{{\mathrm{2}}}}$, $\ottnt{e}$, and $ \tceseq{ \ottnt{v_{{\mathrm{0}}}} } $ such that
     $\ottsym{\mbox{$\mid$}}   \tceseq{ \mathit{z_{{\mathrm{0}}}} }   \ottsym{\mbox{$\mid$}} \,  =  \, \ottsym{\mbox{$\mid$}}   \tceseq{ \ottnt{v_{{\mathrm{0}}}} }   \ottsym{\mbox{$\mid$}}$; namely,
     \[
      \ottnt{e} = \ottsym{(}   \lambda\!  \,  \tceseq{ \mathit{z_{{\mathrm{0}}}} }   \ottsym{,}  \mathit{z_{{\mathrm{2}}}}  \ottsym{.}  \ottnt{e}  \ottsym{)} \,  \tceseq{ \ottnt{v_{{\mathrm{0}}}} }  \, \ottnt{v_{{\mathrm{2}}}} \,  \rightsquigarrow  \,   \ottnt{e}    [ \tceseq{ \ottnt{v_{{\mathrm{0}}}}  /  \mathit{z_{{\mathrm{0}}}} } ]      [  \ottnt{v_{{\mathrm{2}}}}  /  \mathit{z_{{\mathrm{2}}}}  ]   = \ottnt{e'} ~.
     \]
     Therefore, it suffices to show that
     \[
       \emptyset   \vdash    \ottnt{e}    [ \tceseq{ \ottnt{v_{{\mathrm{0}}}}  /  \mathit{z_{{\mathrm{0}}}} } ]      [  \ottnt{v_{{\mathrm{2}}}}  /  \mathit{z_{{\mathrm{2}}}}  ]    \ottsym{:}   \ottnt{C'}    [  \ottnt{v_{{\mathrm{2}}}}  /  \mathit{x}  ]   ~.
     \]

     Because $ \emptyset   \vdash  \ottnt{v_{{\mathrm{1}}}}  \ottsym{:}  \ottsym{(}  \mathit{x} \,  {:}  \, \ottnt{T'}  \ottsym{)}  \rightarrow  \ottnt{C'}$ means $ \emptyset   \vdash  \ottsym{(}   \lambda\!  \,  \tceseq{ \mathit{z_{{\mathrm{0}}}} }   \ottsym{,}  \mathit{z_{{\mathrm{2}}}}  \ottsym{.}  \ottnt{e}  \ottsym{)} \,  \tceseq{ \ottnt{v_{{\mathrm{0}}}} }   \ottsym{:}  \ottsym{(}  \mathit{x} \,  {:}  \, \ottnt{T'}  \ottsym{)}  \rightarrow  \ottnt{C'}$, and
     $\ottsym{\mbox{$\mid$}}   \tceseq{ \mathit{z_{{\mathrm{0}}}} }   \ottsym{\mbox{$\mid$}} \,  =  \, \ottsym{\mbox{$\mid$}}   \tceseq{ \ottnt{v_{{\mathrm{0}}}} }   \ottsym{\mbox{$\mid$}}$,
     \reflem{ts-val-inv-abs} implies that there exist some $ \tceseq{ \ottnt{T_{{\mathrm{0}}}} } $, $\ottnt{T_{{\mathrm{2}}}}$, and $\ottnt{C_{{\mathrm{0}}}}$ such that
     \begin{itemize}
      \item $ \tceseq{ \mathit{z_{{\mathrm{0}}}}  \,  {:}  \,  \ottnt{T_{{\mathrm{0}}}} }   \ottsym{,}  \mathit{z_{{\mathrm{2}}}} \,  {:}  \, \ottnt{T_{{\mathrm{2}}}}  \vdash  \ottnt{e}  \ottsym{:}  \ottnt{C_{{\mathrm{0}}}}$,
      \item $ \tceseq{  \emptyset   \vdash  \ottnt{v_{{\mathrm{0}}}}  \ottsym{:}   \ottnt{T_{{\mathrm{0}}}}    [ \tceseq{ \ottnt{v_{{\mathrm{0}}}}  /  \mathit{z_{{\mathrm{0}}}} } ]   } $, and
      \item $ \emptyset   \vdash    \ottsym{(}  \ottsym{(}  \mathit{z_{{\mathrm{2}}}} \,  {:}  \, \ottnt{T_{{\mathrm{2}}}}  \ottsym{)}  \rightarrow  \ottnt{C_{{\mathrm{0}}}}  \ottsym{)}    [ \tceseq{ \ottnt{v_{{\mathrm{0}}}}  /  \mathit{z_{{\mathrm{0}}}} } ]    \,  \,\&\,  \,   \Phi_\mathsf{val}   \, / \,   \square    \ottsym{<:}   \ottsym{(}  \mathit{x} \,  {:}  \, \ottnt{T'}  \ottsym{)}  \rightarrow  \ottnt{C'}  \,  \,\&\,  \,   \Phi_\mathsf{val}   \, / \,   \square  $.
     \end{itemize}
     By \reflem{ts-val-subst-typing},
     \begin{equation}
       \mathit{z_{{\mathrm{2}}}} \,  {:}  \, \ottnt{T_{{\mathrm{2}}}}    [ \tceseq{ \ottnt{v_{{\mathrm{0}}}}  /  \mathit{z_{{\mathrm{0}}}} } ]    \vdash   \ottnt{e}    [ \tceseq{ \ottnt{v_{{\mathrm{0}}}}  /  \mathit{z_{{\mathrm{0}}}} } ]    \ottsym{:}   \ottnt{C_{{\mathrm{0}}}}    [ \tceseq{ \ottnt{v_{{\mathrm{0}}}}  /  \mathit{z_{{\mathrm{0}}}} } ]   ~.
       \label{lem:ts-subject-red:app:beta':body}
     \end{equation}
     Then, Lemmas~\ref{lem:ts-wf-ty-tctx-typing} and \ref{lem:ts-wf-tctx-wf-ftyping}
     imply $ \emptyset   \vdash   \ottnt{T_{{\mathrm{2}}}}    [ \tceseq{ \ottnt{v_{{\mathrm{0}}}}  /  \mathit{z_{{\mathrm{0}}}} } ]  $.
     Because $ \emptyset   \vdash    \ottsym{(}  \ottsym{(}  \mathit{z_{{\mathrm{2}}}} \,  {:}  \, \ottnt{T_{{\mathrm{2}}}}  \ottsym{)}  \rightarrow  \ottnt{C_{{\mathrm{0}}}}  \ottsym{)}    [ \tceseq{ \ottnt{v_{{\mathrm{0}}}}  /  \mathit{z_{{\mathrm{0}}}} } ]    \,  \,\&\,  \,   \Phi_\mathsf{val}   \, / \,   \square    \ottsym{<:}   \ottsym{(}  \mathit{x} \,  {:}  \, \ottnt{T'}  \ottsym{)}  \rightarrow  \ottnt{C'}  \,  \,\&\,  \,   \Phi_\mathsf{val}   \, / \,   \square  $, its inversion implies that
     \begin{itemize}
      \item $ \mathit{z_{{\mathrm{2}}}}    =    \mathit{x} $,
      \item $ \emptyset   \vdash  \ottnt{T'}  \ottsym{<:}   \ottnt{T_{{\mathrm{2}}}}    [ \tceseq{ \ottnt{v_{{\mathrm{0}}}}  /  \mathit{z_{{\mathrm{0}}}} } ]  $, and
      \item $\mathit{z_{{\mathrm{2}}}} \,  {:}  \, \ottnt{T'}  \vdash   \ottnt{C_{{\mathrm{0}}}}    [ \tceseq{ \ottnt{v_{{\mathrm{0}}}}  /  \mathit{z_{{\mathrm{0}}}} } ]    \ottsym{<:}  \ottnt{C'}$.
     \end{itemize}
     Because
     \begin{itemize}
      \item $ \emptyset   \vdash  \ottnt{v_{{\mathrm{2}}}}  \ottsym{:}  \ottnt{T'}$,
      \item $ \emptyset   \vdash   \ottnt{T'}  \,  \,\&\,  \,   \Phi_\mathsf{val}   \, / \,   \square    \ottsym{<:}    \ottnt{T_{{\mathrm{2}}}}    [ \tceseq{ \ottnt{v_{{\mathrm{0}}}}  /  \mathit{z_{{\mathrm{0}}}} } ]    \,  \,\&\,  \,   \Phi_\mathsf{val}   \, / \,   \square  $
            by \reflem{ts-refl-subtyping-pure-eff} with
            $ \emptyset   \vdash  \ottnt{T'}  \ottsym{<:}   \ottnt{T_{{\mathrm{2}}}}    [ \tceseq{ \ottnt{v_{{\mathrm{0}}}}  /  \mathit{z_{{\mathrm{0}}}} } ]  $, and
      \item $ \emptyset   \vdash    \ottnt{T_{{\mathrm{2}}}}    [ \tceseq{ \ottnt{v_{{\mathrm{0}}}}  /  \mathit{z_{{\mathrm{0}}}} } ]    \,  \,\&\,  \,   \Phi_\mathsf{val}   \, / \,   \square  $
            by \reflem{ts-wf-TP_val} with
            $ \emptyset   \vdash   \ottnt{T_{{\mathrm{2}}}}    [ \tceseq{ \ottnt{v_{{\mathrm{0}}}}  /  \mathit{z_{{\mathrm{0}}}} } ]  $,
     \end{itemize}
     \T{Sub} implies
     \begin{equation}
       \emptyset   \vdash  \ottnt{v_{{\mathrm{2}}}}  \ottsym{:}   \ottnt{T_{{\mathrm{2}}}}    [ \tceseq{ \ottnt{v_{{\mathrm{0}}}}  /  \mathit{z_{{\mathrm{0}}}} } ]   ~.
       \label{lem:ts-subject-red:app:beta':arg}
     \end{equation}
     Then, \reflem{ts-val-subst-typing} with
     (\ref{lem:ts-subject-red:app:beta':body}) and (\ref{lem:ts-subject-red:app:beta':arg})
     implies
     \[
       \emptyset   \vdash    \ottnt{e}    [ \tceseq{ \ottnt{v_{{\mathrm{0}}}}  /  \mathit{z_{{\mathrm{0}}}} } ]      [  \ottnt{v_{{\mathrm{2}}}}  /  \mathit{z_{{\mathrm{2}}}}  ]    \ottsym{:}    \ottnt{C_{{\mathrm{0}}}}    [ \tceseq{ \ottnt{v_{{\mathrm{0}}}}  /  \mathit{z_{{\mathrm{0}}}} } ]      [  \ottnt{v_{{\mathrm{2}}}}  /  \mathit{z_{{\mathrm{2}}}}  ]   ~.
     \]
     Then, we have the conclusion by \T{Sub} with the following,
     where note that $ \mathit{z_{{\mathrm{2}}}}    =    \mathit{x} $.
     \begin{itemize}
      \item We show that
            \[
              \emptyset   \vdash    \ottnt{C_{{\mathrm{0}}}}    [ \tceseq{ \ottnt{v_{{\mathrm{0}}}}  /  \mathit{z_{{\mathrm{0}}}} } ]      [  \ottnt{v_{{\mathrm{2}}}}  /  \mathit{z_{{\mathrm{2}}}}  ]    \ottsym{<:}   \ottnt{C'}    [  \ottnt{v_{{\mathrm{2}}}}  /  \mathit{z_{{\mathrm{2}}}}  ]   ~,
            \]
            which is implied by
            \reflem{ts-val-subst-subtyping} with
            $\mathit{z_{{\mathrm{2}}}} \,  {:}  \, \ottnt{T'}  \vdash   \ottnt{C_{{\mathrm{0}}}}    [ \tceseq{ \ottnt{v_{{\mathrm{0}}}}  /  \mathit{z_{{\mathrm{0}}}} } ]    \ottsym{<:}  \ottnt{C'}$ and
            $ \emptyset   \vdash  \ottnt{v_{{\mathrm{2}}}}  \ottsym{:}  \ottnt{T'}$.

      \item We show that
            \[
              \emptyset   \vdash   \ottnt{C'}    [  \ottnt{v_{{\mathrm{2}}}}  /  \mathit{z_{{\mathrm{2}}}}  ]   ~,
            \]
            which is implied by \reflem{ts-wf-ty-tctx-typing}
            with $ \emptyset   \vdash  \ottnt{v_{{\mathrm{1}}}} \, \ottnt{v_{{\mathrm{2}}}}  \ottsym{:}   \ottnt{C'}    [  \ottnt{v_{{\mathrm{2}}}}  /  \mathit{x}  ]  $.
     \end{itemize}

    \case \R{RBeta}:
     We are given
     $\ottnt{v_{{\mathrm{1}}}} \,  =  \, \ottnt{v} \,  \tceseq{ \ottnt{v_{{\mathrm{0}}}} } $ and $\ottnt{e'} \,  =  \,      \ottnt{e_{{\mathrm{0}}}}    [ \tceseq{  \ottnt{A} _{  \mu  }   /  \mathit{X} } ]      [ \tceseq{  \ottnt{A} _{  \nu  }   /  \mathit{Y} } ]      [  \ottnt{v}  /  \mathit{f}  ]      [ \tceseq{ \ottnt{v_{{\mathrm{0}}}}  /  \mathit{z_{{\mathrm{0}}}} } ]      [  \ottnt{v_{{\mathrm{2}}}}  /  \mathit{z_{{\mathrm{2}}}}  ]  $
     for some $\ottnt{v}$, $ \tceseq{ \ottnt{v_{{\mathrm{0}}}} } $, $ \tceseq{ \mathit{X} } $, $ \tceseq{ \mathit{Y} } $, $ \tceseq{  \ottnt{A} _{  \mu  }  } $, $ \tceseq{  \ottnt{A} _{  \nu  }  } $, $\mathit{f}$, $ \tceseq{ \mathit{z_{{\mathrm{0}}}} } $, $\mathit{z_{{\mathrm{2}}}}$, and $\ottnt{e_{{\mathrm{0}}}}$ such that
     $\ottsym{\mbox{$\mid$}}   \tceseq{ \ottnt{v_{{\mathrm{0}}}} }   \ottsym{\mbox{$\mid$}} \,  =  \, \ottsym{\mbox{$\mid$}}   \tceseq{ \mathit{z_{{\mathrm{0}}}} }   \ottsym{\mbox{$\mid$}}$ and
     \[
      \ottnt{v} \,  =  \,  \mathsf{rec}  \, (  \tceseq{  \mathit{X} ^{  \ottnt{A} _{  \mu  }  },  \mathit{Y} ^{  \ottnt{A} _{  \nu  }  }  }  ,  \mathit{f} ,  \ottsym{(}   \tceseq{ \mathit{z_{{\mathrm{0}}}} }   \ottsym{,}  \mathit{z_{{\mathrm{2}}}}  \ottsym{)} ) . \,  \ottnt{e_{{\mathrm{0}}}}  ~.
     \]
     Namely, we have
     \[
      \ottnt{e} = \ottsym{(}   \mathsf{rec}  \, (  \tceseq{  \mathit{X} ^{  \ottnt{A} _{  \mu  }  },  \mathit{Y} ^{  \ottnt{A} _{  \nu  }  }  }  ,  \mathit{f} ,  \ottsym{(}   \tceseq{ \mathit{z_{{\mathrm{0}}}} }   \ottsym{,}  \mathit{z_{{\mathrm{2}}}}  \ottsym{)} ) . \,  \ottnt{e_{{\mathrm{0}}}}   \ottsym{)} \,  \tceseq{ \ottnt{v_{{\mathrm{0}}}} }  \, \ottnt{v_{{\mathrm{2}}}} \,  \rightsquigarrow  \,      \ottnt{e_{{\mathrm{0}}}}    [ \tceseq{  \ottnt{A} _{  \mu  }   /  \mathit{X} } ]      [ \tceseq{  \ottnt{A} _{  \nu  }   /  \mathit{Y} } ]      [  \ottnt{v}  /  \mathit{f}  ]      [ \tceseq{ \ottnt{v_{{\mathrm{0}}}}  /  \mathit{z_{{\mathrm{0}}}} } ]      [  \ottnt{v_{{\mathrm{2}}}}  /  \mathit{z_{{\mathrm{2}}}}  ]   = \ottnt{e'} ~.
     \]
     Therefore, it suffices to show that
     \[
       \emptyset   \vdash       \ottnt{e_{{\mathrm{0}}}}    [ \tceseq{  \ottnt{A} _{  \mu  }   /  \mathit{X} } ]      [ \tceseq{  \ottnt{A} _{  \nu  }   /  \mathit{Y} } ]      [  \ottnt{v}  /  \mathit{f}  ]      [ \tceseq{ \ottnt{v_{{\mathrm{0}}}}  /  \mathit{z_{{\mathrm{0}}}} } ]      [  \ottnt{v_{{\mathrm{2}}}}  /  \mathit{z_{{\mathrm{2}}}}  ]    \ottsym{:}   \ottnt{C'}    [  \ottnt{v_{{\mathrm{2}}}}  /  \mathit{x}  ]   ~.
     \]

     Because $ \emptyset   \vdash  \ottnt{v_{{\mathrm{1}}}}  \ottsym{:}  \ottsym{(}  \mathit{x} \,  {:}  \, \ottnt{T'}  \ottsym{)}  \rightarrow  \ottnt{C'}$ means
     $ \emptyset   \vdash  \ottsym{(}   \mathsf{rec}  \, (  \tceseq{  \mathit{X} ^{  \ottnt{A} _{  \mu  }  },  \mathit{Y} ^{  \ottnt{A} _{  \nu  }  }  }  ,  \mathit{f} ,  \ottsym{(}   \tceseq{ \mathit{z_{{\mathrm{0}}}} }   \ottsym{,}  \mathit{z_{{\mathrm{2}}}}  \ottsym{)} ) . \,  \ottnt{e_{{\mathrm{0}}}}   \ottsym{)} \,  \tceseq{ \ottnt{v_{{\mathrm{0}}}} }   \ottsym{:}  \ottsym{(}  \mathit{x} \,  {:}  \, \ottnt{T'}  \ottsym{)}  \rightarrow  \ottnt{C'}$,
     \reflem{ts-val-inv-rec} implies that
     there exist some $ \tceseq{ \ottnt{T_{{\mathrm{10}}}} } $, $\ottnt{T_{{\mathrm{12}}}}$, $\ottnt{T_{{\mathrm{2}}}}$, $\Phi_{{\mathrm{0}}}$, $\ottnt{S_{{\mathrm{0}}}}$, $ \tceseq{ \mathit{z} } $, $ \tceseq{ \iota } $, $\Phi$, $\ottnt{S}$, $\mathit{x}$, $\mathit{y}$ such that
     \begin{itemize}
      \item $ \emptyset   \vdash  \ottnt{v}  \ottsym{:}  \ottsym{(}   \tceseq{ \mathit{z_{{\mathrm{0}}}}  \,  {:}  \,  \ottnt{T_{{\mathrm{10}}}} }   \ottsym{,}  \mathit{z_{{\mathrm{2}}}} \,  {:}  \, \ottnt{T_{{\mathrm{12}}}}  \ottsym{)}  \rightarrow  \sigma  \ottsym{(}  \ottnt{C_{{\mathrm{0}}}}  \ottsym{)}$,
      \item $ \emptyset   \vdash    \ottsym{(}  \ottsym{(}  \mathit{z_{{\mathrm{2}}}} \,  {:}  \, \ottnt{T_{{\mathrm{12}}}}  \ottsym{)}  \rightarrow  \sigma  \ottsym{(}  \ottnt{C_{{\mathrm{0}}}}  \ottsym{)}  \ottsym{)}    [ \tceseq{ \ottnt{v_{{\mathrm{0}}}}  /  \mathit{z_{{\mathrm{0}}}} } ]    \,  \,\&\,  \,   \Phi_\mathsf{val}   \, / \,   \square    \ottsym{<:}   \ottsym{(}  \mathit{x} \,  {:}  \, \ottnt{T'}  \ottsym{)}  \rightarrow  \ottnt{C'}  \,  \,\&\,  \,   \Phi_\mathsf{val}   \, / \,   \square  $,
      \item $ \tceseq{ \mathit{X} }  \,  {:}  \,  (   \Sigma ^\ast   ,   \tceseq{ \iota }   )   \ottsym{,}   \tceseq{ \mathit{Y} }  \,  {:}  \,  (   \Sigma ^\omega   ,   \tceseq{ \iota }   )   \ottsym{,}  \mathit{f} \,  {:}  \, \ottsym{(}   \tceseq{ \mathit{z_{{\mathrm{0}}}}  \,  {:}  \,  \ottnt{T_{{\mathrm{10}}}} }   \ottsym{,}  \mathit{z_{{\mathrm{2}}}} \,  {:}  \, \ottnt{T_{{\mathrm{12}}}}  \ottsym{)}  \rightarrow   \ottnt{T_{{\mathrm{2}}}}  \,  \,\&\,  \,  \Phi_{{\mathrm{0}}}  \, / \,  \ottnt{S_{{\mathrm{0}}}}   \ottsym{,}   \tceseq{ \mathit{z_{{\mathrm{0}}}}  \,  {:}  \,  \ottnt{T_{{\mathrm{10}}}} }   \ottsym{,}  \mathit{z_{{\mathrm{2}}}} \,  {:}  \, \ottnt{T_{{\mathrm{12}}}}  \vdash  \ottnt{e_{{\mathrm{0}}}}  \ottsym{:}  \ottnt{C}$,
      \item $ \tceseq{ \mathit{z}  \,  {:}  \,  \iota }  \,  =  \,  \gamma  (   \tceseq{ \mathit{z_{{\mathrm{0}}}}  \,  {:}  \,  \ottnt{T_{{\mathrm{10}}}} }   \ottsym{,}  \mathit{z_{{\mathrm{2}}}} \,  {:}  \, \ottnt{T_{{\mathrm{12}}}}  ) $,
      \item $ \tceseq{  \ottnt{A} _{  \mu  }  \,  =  \, \sigma  \ottsym{(}  \mathit{X}  \ottsym{)} } $,
      \item $ \tceseq{  \ottnt{A} _{  \nu  }  \,  =  \, \sigma  \ottsym{(}  \mathit{Y}  \ottsym{)} } $,
      \item $  \{   \tceseq{ \mathit{X} }   \ottsym{,}   \tceseq{ \mathit{Y} }   \}   \, \ottsym{\#} \,  \ottsym{(}   \mathit{fpv}  (   \tceseq{ \ottnt{T_{{\mathrm{10}}}} }   )  \,  \mathbin{\cup}  \,  \mathit{fpv}  (  \ottnt{T_{{\mathrm{12}}}}  )   \ottsym{)} $,
      \item $ \tceseq{ \ottsym{(}  \mathit{X}  \ottsym{,}  \mathit{Y}  \ottsym{)} }   \ottsym{;}   \tceseq{ \mathit{z}  \,  {:}  \,  \iota }   \vdash  \ottnt{C_{{\mathrm{0}}}}  \stackrel{\succ}{\circ}  \ottnt{C}$,
      \item $ \tceseq{ \ottsym{(}  \mathit{X}  \ottsym{,}  \mathit{Y}  \ottsym{)} }   \ottsym{;}   \tceseq{ \mathit{z}  \,  {:}  \,  \iota }  \,  |  \, \ottnt{C}  \vdash  \sigma$, and
      \item $ \tceseq{  \emptyset   \vdash  \ottnt{v_{{\mathrm{0}}}}  \ottsym{:}   \ottnt{T_{{\mathrm{10}}}}    [ \tceseq{ \ottnt{v_{{\mathrm{0}}}}  /  \mathit{z_{{\mathrm{0}}}} } ]   } $.
     \end{itemize}
     Then, Lemmas~\ref{lem:ts-unuse-non-refine-vars}, \ref{lem:ts-elim-unused-binding-wf-ftyping}, \ref{lem:lr-eff-rec-wf-XXX}, \ref{lem:ts-pred-subst-typing}, and \ref{lem:ts-val-subst-typing} imply
     \begin{equation}
       \mathit{z_{{\mathrm{2}}}} \,  {:}  \, \ottnt{T_{{\mathrm{12}}}}    [ \tceseq{ \ottnt{v_{{\mathrm{0}}}}  /  \mathit{z_{{\mathrm{0}}}} } ]    \vdash    \sigma  \ottsym{(}  \ottnt{e_{{\mathrm{0}}}}  \ottsym{)}    [  \ottnt{v}  /  \mathit{f}  ]      [ \tceseq{ \ottnt{v_{{\mathrm{0}}}}  /  \mathit{z_{{\mathrm{0}}}} } ]    \ottsym{:}   \sigma  \ottsym{(}  \ottnt{C}  \ottsym{)}    [ \tceseq{ \ottnt{v_{{\mathrm{0}}}}  /  \mathit{z_{{\mathrm{0}}}} } ]   ~.
       \label{lem:ts-subject-red:app:beta:body}
     \end{equation}
     Because
     \[
       \emptyset   \vdash    \ottsym{(}  \ottsym{(}  \mathit{z_{{\mathrm{2}}}} \,  {:}  \, \ottnt{T_{{\mathrm{12}}}}  \ottsym{)}  \rightarrow  \sigma  \ottsym{(}  \ottnt{C_{{\mathrm{0}}}}  \ottsym{)}  \ottsym{)}    [ \tceseq{ \ottnt{v_{{\mathrm{0}}}}  /  \mathit{z_{{\mathrm{0}}}} } ]    \,  \,\&\,  \,   \Phi_\mathsf{val}   \, / \,   \square    \ottsym{<:}   \ottsym{(}  \mathit{x} \,  {:}  \, \ottnt{T'}  \ottsym{)}  \rightarrow  \ottnt{C'}  \,  \,\&\,  \,   \Phi_\mathsf{val}   \, / \,   \square   ~,
     \]
     we have
     \begin{itemize}
      \item $ \mathit{z_{{\mathrm{2}}}}    =    \mathit{x} $,
      \item $ \emptyset   \vdash  \ottnt{T'}  \ottsym{<:}   \ottnt{T_{{\mathrm{12}}}}    [ \tceseq{ \ottnt{v_{{\mathrm{0}}}}  /  \mathit{z_{{\mathrm{0}}}} } ]  $, and
      \item $\mathit{z_{{\mathrm{2}}}} \,  {:}  \, \ottnt{T'}  \vdash   \sigma  \ottsym{(}  \ottnt{C_{{\mathrm{0}}}}  \ottsym{)}    [ \tceseq{ \ottnt{v_{{\mathrm{0}}}}  /  \mathit{z_{{\mathrm{0}}}} } ]    \ottsym{<:}  \ottnt{C'}$.
     \end{itemize}
     Because $ \emptyset   \vdash  \ottnt{v_{{\mathrm{2}}}}  \ottsym{:}  \ottnt{T'}$, and $ \emptyset   \vdash   \ottnt{T_{{\mathrm{12}}}}    [ \tceseq{ \ottnt{v_{{\mathrm{0}}}}  /  \mathit{z_{{\mathrm{0}}}} } ]  $
     is implied by Lemmas~\ref{lem:ts-wf-ty-tctx-typing} and \ref{lem:ts-wf-tctx-wf-ftyping} with (\ref{lem:ts-subject-red:app:beta:body}),
     we have $ \emptyset   \vdash  \ottnt{v_{{\mathrm{2}}}}  \ottsym{:}   \ottnt{T_{{\mathrm{12}}}}    [ \tceseq{ \ottnt{v_{{\mathrm{0}}}}  /  \mathit{z_{{\mathrm{0}}}} } ]  $
     by Lemmas~\ref{lem:ts-refl-subtyping-pure-eff} and \ref{lem:ts-wf-TP_val}, and \T{Sub}.
     Then, by \reflem{ts-val-subst-typing} with (\ref{lem:ts-subject-red:app:beta:body}),
     \[
       \emptyset   \vdash     \sigma  \ottsym{(}  \ottnt{e_{{\mathrm{0}}}}  \ottsym{)}    [  \ottnt{v}  /  \mathit{f}  ]      [ \tceseq{ \ottnt{v_{{\mathrm{0}}}}  /  \mathit{z_{{\mathrm{0}}}} } ]      [  \ottnt{v_{{\mathrm{2}}}}  /  \mathit{z_{{\mathrm{2}}}}  ]    \ottsym{:}    \sigma  \ottsym{(}  \ottnt{C}  \ottsym{)}    [ \tceseq{ \ottnt{v_{{\mathrm{0}}}}  /  \mathit{z_{{\mathrm{0}}}} } ]      [  \ottnt{v_{{\mathrm{2}}}}  /  \mathit{z_{{\mathrm{2}}}}  ]   ~.
     \]
     We also have $ \emptyset   \vdash   \ottnt{C'}    [  \ottnt{v_{{\mathrm{2}}}}  /  \mathit{x}  ]  $ by \reflem{ts-wf-ty-tctx-typing} with $ \emptyset   \vdash  \ottnt{v_{{\mathrm{1}}}} \, \ottnt{v_{{\mathrm{2}}}}  \ottsym{:}   \ottnt{C'}    [  \ottnt{v_{{\mathrm{2}}}}  /  \mathit{x}  ]  $.
     Therefore, \T{Sub} implies that it suffices to show
     \[
       \emptyset   \vdash    \sigma  \ottsym{(}  \ottnt{C}  \ottsym{)}    [ \tceseq{ \ottnt{v_{{\mathrm{0}}}}  /  \mathit{z_{{\mathrm{0}}}} } ]      [  \ottnt{v_{{\mathrm{2}}}}  /  \mathit{z_{{\mathrm{2}}}}  ]    \ottsym{<:}   \ottnt{C'}    [  \ottnt{v_{{\mathrm{2}}}}  /  \mathit{z_{{\mathrm{2}}}}  ]  
     \]
     where note that $ \mathit{z_{{\mathrm{2}}}}    =    \mathit{x} $.
     Because
     $\mathit{z_{{\mathrm{2}}}} \,  {:}  \, \ottnt{T'}  \vdash   \sigma  \ottsym{(}  \ottnt{C_{{\mathrm{0}}}}  \ottsym{)}    [ \tceseq{ \ottnt{v_{{\mathrm{0}}}}  /  \mathit{z_{{\mathrm{0}}}} } ]    \ottsym{<:}  \ottnt{C'}$ and
     $ \emptyset   \vdash  \ottnt{v_{{\mathrm{2}}}}  \ottsym{:}  \ottnt{T'}$,
     \reflem{ts-val-subst-subtyping} implies
     \[
       \emptyset   \vdash    \sigma  \ottsym{(}  \ottnt{C_{{\mathrm{0}}}}  \ottsym{)}    [ \tceseq{ \ottnt{v_{{\mathrm{0}}}}  /  \mathit{z_{{\mathrm{0}}}} } ]      [  \ottnt{v_{{\mathrm{2}}}}  /  \mathit{z_{{\mathrm{2}}}}  ]    \ottsym{<:}   \ottnt{C'}    [  \ottnt{v_{{\mathrm{2}}}}  /  \mathit{z_{{\mathrm{2}}}}  ]   ~.
     \]
     By \reflem{ts-transitivity-subtyping}, it suffices to show that
     \[
       \emptyset   \vdash    \sigma  \ottsym{(}  \ottnt{C}  \ottsym{)}    [ \tceseq{ \ottnt{v_{{\mathrm{0}}}}  /  \mathit{z_{{\mathrm{0}}}} } ]      [  \ottnt{v_{{\mathrm{2}}}}  /  \mathit{z_{{\mathrm{2}}}}  ]    \ottsym{<:}    \sigma  \ottsym{(}  \ottnt{C_{{\mathrm{0}}}}  \ottsym{)}    [ \tceseq{ \ottnt{v_{{\mathrm{0}}}}  /  \mathit{z_{{\mathrm{0}}}} } ]      [  \ottnt{v_{{\mathrm{2}}}}  /  \mathit{z_{{\mathrm{2}}}}  ]   ~.
     \]
     By \reflem{ts-val-subst-subtyping}, it suffices to show that
     \[
       \tceseq{ \mathit{z_{{\mathrm{0}}}}  \,  {:}  \,  \ottnt{T_{{\mathrm{10}}}} }   \ottsym{,}  \mathit{z_{{\mathrm{2}}}} \,  {:}  \, \ottnt{T_{{\mathrm{12}}}}  \vdash  \sigma  \ottsym{(}  \ottnt{C}  \ottsym{)}  \ottsym{<:}  \sigma  \ottsym{(}  \ottnt{C_{{\mathrm{0}}}}  \ottsym{)} ~,
     \]
     which is implied by \Srule{Comp} and \Srule{Eff} with the following;
     in what follows, we use
     \[
       \tceseq{ \mathit{X} }  \,  {:}  \,  (   \Sigma ^\ast   ,   \tceseq{ \iota }   )   \ottsym{,}   \tceseq{ \mathit{Y} }  \,  {:}  \,  (   \Sigma ^\omega   ,   \tceseq{ \iota }   )   \ottsym{,}   \tceseq{ \mathit{z_{{\mathrm{0}}}}  \,  {:}  \,  \ottnt{T_{{\mathrm{10}}}} }   \ottsym{,}  \mathit{z_{{\mathrm{2}}}} \,  {:}  \, \ottnt{T_{{\mathrm{12}}}}  \vdash  \ottnt{C} ~,
     \]
     which is implied by Lemmas~\ref{lem:ts-wf-ty-tctx-typing}, \ref{lem:ts-unuse-non-refine-vars}, \ref{lem:ts-elim-unused-binding-wf-ftyping}, and
     \ref{lem:lr-subtype-subst-fixpoint}.
   \end{caseanalysis}

  \case \T{PApp}:
   We are given
   \begin{prooftree}
    \AxiomC{$ \emptyset   \vdash  \ottnt{v}  \ottsym{:}   \forall   \ottsym{(}  \mathit{X} \,  {:}  \,  \tceseq{ \ottnt{s} }   \ottsym{)}  \ottsym{.}  \ottnt{C'}$}
    \AxiomC{$ \emptyset   \vdash  \ottnt{A}  \ottsym{:}   \tceseq{ \ottnt{s} } $}
    \BinaryInfC{$ \emptyset   \vdash  \ottnt{v} \, \ottnt{A}  \ottsym{:}   \ottnt{C'}    [  \ottnt{A}  /  \mathit{X}  ]  $}
   \end{prooftree}
   for some $\ottnt{v}$, $\ottnt{A}$, $\ottnt{C'}$, $\mathit{X}$, and $ \tceseq{ \ottnt{s} } $
   such that $\ottnt{e} \,  =  \, \ottnt{v} \, \ottnt{A}$ and $\ottnt{C} \,  =  \,  \ottnt{C'}    [  \ottnt{A}  /  \mathit{X}  ]  $.
   The reduction rule applicable to $\ottnt{v} \, \ottnt{A}$ is only \R{PBeta}.
   Therefore, there exists $\ottnt{e_{{\mathrm{0}}}}$ such that
   $\ottnt{v} \,  =  \,  \Lambda   \ottsym{(}  \mathit{X} \,  {:}  \,  \tceseq{ \ottnt{s} }   \ottsym{)}  \ottsym{.}  \ottnt{e_{{\mathrm{0}}}}$ and $\ottnt{e'} \,  =  \,  \ottnt{e_{{\mathrm{0}}}}    [  \ottnt{A}  /  \mathit{X}  ]  $; namely,
   \[
    \ottnt{e} ~=~ \ottsym{(}   \Lambda   \ottsym{(}  \mathit{X} \,  {:}  \,  \tceseq{ \ottnt{s} }   \ottsym{)}  \ottsym{.}  \ottnt{e_{{\mathrm{0}}}}  \ottsym{)} \, \ottnt{A} \,  \rightsquigarrow  \,  \ottnt{e_{{\mathrm{0}}}}    [  \ottnt{A}  /  \mathit{X}  ]   ~=~ \ottnt{e'} ~.
   \]
   Therefore, it suffices to show that
   \[
     \emptyset   \vdash   \ottnt{e_{{\mathrm{0}}}}    [  \ottnt{A}  /  \mathit{X}  ]    \ottsym{:}   \ottnt{C'}    [  \ottnt{A}  /  \mathit{X}  ]   ~.
   \]

   \reflem{ts-val-inv-univ-ty} with $ \emptyset   \vdash  \ottnt{v}  \ottsym{:}   \forall   \ottsym{(}  \mathit{X} \,  {:}  \,  \tceseq{ \ottnt{s} }   \ottsym{)}  \ottsym{.}  \ottnt{C'}$
   implies that there exists some $\ottnt{C_{{\mathrm{0}}}}$ such that
   \begin{itemize}
    \item $\mathit{X} \,  {:}  \,  \tceseq{ \ottnt{s} }   \vdash  \ottnt{e_{{\mathrm{0}}}}  \ottsym{:}  \ottnt{C_{{\mathrm{0}}}}$, and
    \item $ \emptyset   \vdash    \forall   \ottsym{(}  \mathit{X} \,  {:}  \,  \tceseq{ \ottnt{s} }   \ottsym{)}  \ottsym{.}  \ottnt{C_{{\mathrm{0}}}}  \,  \,\&\,  \,   \Phi_\mathsf{val}   \, / \,   \square    \ottsym{<:}    \forall   \ottsym{(}  \mathit{X} \,  {:}  \,  \tceseq{ \ottnt{s} }   \ottsym{)}  \ottsym{.}  \ottnt{C'}  \,  \,\&\,  \,   \Phi_\mathsf{val}   \, / \,   \square  $,
          which implies $\mathit{X} \,  {:}  \,  \tceseq{ \ottnt{s} }   \vdash  \ottnt{C_{{\mathrm{0}}}}  \ottsym{<:}  \ottnt{C'}$.
   \end{itemize}
   Because $ \emptyset   \vdash  \ottnt{A}  \ottsym{:}   \tceseq{ \ottnt{s} } $,
   \reflem{ts-pred-subst-typing} implies
   $ \emptyset   \vdash   \ottnt{e_{{\mathrm{0}}}}    [  \ottnt{A}  /  \mathit{X}  ]    \ottsym{:}   \ottnt{C_{{\mathrm{0}}}}    [  \ottnt{A}  /  \mathit{X}  ]  $, and
   \reflem{ts-pred-subst-subtyping} implies
   $ \emptyset   \vdash   \ottnt{C_{{\mathrm{0}}}}    [  \ottnt{A}  /  \mathit{X}  ]    \ottsym{<:}   \ottnt{C'}    [  \ottnt{A}  /  \mathit{X}  ]  $.
   Because $ \emptyset   \vdash   \ottnt{C'}    [  \ottnt{A}  /  \mathit{X}  ]  $ by \reflem{ts-wf-ty-tctx-typing},
   \T{Sub} implies the conclusion.

  \case \T{If}:
   We are given
   \begin{prooftree}
    \AxiomC{$ \emptyset   \vdash  \ottnt{v}  \ottsym{:}  \ottsym{\{}  \mathit{x} \,  {:}  \,  \mathsf{bool}  \,  |  \, \phi  \ottsym{\}}$}
    \AxiomC{$  \ottnt{v}     =     \mathsf{true}    \vdash  \ottnt{e_{{\mathrm{1}}}}  \ottsym{:}  \ottnt{C}$}
    \AxiomC{$  \ottnt{v}     =     \mathsf{false}    \vdash  \ottnt{e_{{\mathrm{2}}}}  \ottsym{:}  \ottnt{C}$}
    \TrinaryInfC{$ \emptyset   \vdash  \mathsf{if} \, \ottnt{v} \, \mathsf{then} \, \ottnt{e_{{\mathrm{1}}}} \, \mathsf{else} \, \ottnt{e_{{\mathrm{2}}}}  \ottsym{:}  \ottnt{C}$}
   \end{prooftree}
   for some $\ottnt{v}$, $\ottnt{e_{{\mathrm{1}}}}$, $\ottnt{e_{{\mathrm{2}}}}$, $\mathit{x}$, and $\phi$
   such that $\ottnt{e} \,  =  \, \mathsf{if} \, \ottnt{v} \, \mathsf{then} \, \ottnt{e_{{\mathrm{1}}}} \, \mathsf{else} \, \ottnt{e_{{\mathrm{2}}}}$.
   The reduction rules applicable to $\ottnt{e}$ are only \R{IfT} and \R{IfF}.
   By case analysis on the reduction rule applied to $\ottnt{e}$.
   \begin{caseanalysis}
    \case \R{IfT}:
     We are given $\ottnt{v} \,  =  \,  \mathsf{true} $, and
     \[
      \ottnt{e} ~=~ \mathsf{if} \,  \mathsf{true}  \, \mathsf{then} \, \ottnt{e_{{\mathrm{1}}}} \, \mathsf{else} \, \ottnt{e_{{\mathrm{2}}}} \,  \rightsquigarrow  \, \ottnt{e_{{\mathrm{1}}}} ~=~ \ottnt{e'} ~.
     \]
     Further, $  \ottnt{v}     =     \mathsf{true}    \vdash  \ottnt{e_{{\mathrm{1}}}}  \ottsym{:}  \ottnt{C}$ implies
     $  \mathsf{true}     =     \mathsf{true}    \vdash  \ottnt{e_{{\mathrm{1}}}}  \ottsym{:}  \ottnt{C}$.
     Because
     \begin{itemize}
      \item $ \emptyset   \vdash   ()   \ottsym{:}  \ottsym{\{}  \mathit{x} \,  {:}  \,  \mathsf{unit}  \,  |  \,   ()     =    \mathit{x}   \ottsym{\}}$ by \T{Const},
      \item $ \emptyset   \vdash  \ottsym{\{}  \mathit{x} \,  {:}  \,  \mathsf{unit}  \,  |  \,   ()     =    \mathit{x}   \ottsym{\}}  \ottsym{<:}  \ottsym{\{}  \mathit{x} \,  {:}  \,  \mathsf{unit}  \,  |  \,   \mathsf{true}     =     \mathsf{true}    \ottsym{\}}$
            by Lemmas~\ref{lem:ts-refl-const-eq} and \ref{lem:ts-extract-subst-val-valid}, and \Srule{Refine}, and
      \item $ \emptyset   \vdash  \ottsym{\{}  \mathit{x} \,  {:}  \,  \mathsf{unit}  \,  |  \,   \mathsf{true}     =     \mathsf{true}    \ottsym{\}}$,
     \end{itemize}
     we have $ \emptyset   \vdash   ()   \ottsym{:}  \ottsym{\{}  \mathit{x} \,  {:}  \,  \mathsf{unit}  \,  |  \,   \mathsf{true}     =     \mathsf{true}    \ottsym{\}}$.
     Thus, \reflem{ts-val-subst-typing} implies $ \emptyset   \vdash  \ottnt{e_{{\mathrm{1}}}}  \ottsym{:}  \ottnt{C}$, which we must show.

    \case \R{IfF}:
     We are given $\ottnt{v} \,  =  \,  \mathsf{false} $, and
     \[
      \ottnt{e} ~=~ \mathsf{if} \,  \mathsf{false}  \, \mathsf{then} \, \ottnt{e_{{\mathrm{1}}}} \, \mathsf{else} \, \ottnt{e_{{\mathrm{2}}}} \,  \rightsquigarrow  \, \ottnt{e_{{\mathrm{2}}}} ~=~ \ottnt{e'} ~.
     \]
     Further, $  \ottnt{v}     =     \mathsf{false}    \vdash  \ottnt{e_{{\mathrm{2}}}}  \ottsym{:}  \ottnt{C}$ implies
     $  \mathsf{false}     =     \mathsf{false}    \vdash  \ottnt{e_{{\mathrm{2}}}}  \ottsym{:}  \ottnt{C}$.
     Because $ \emptyset   \vdash   ()   \ottsym{:}  \ottsym{\{}  \mathit{x} \,  {:}  \,  \mathsf{unit}  \,  |  \,   \mathsf{false}     =     \mathsf{false}    \ottsym{\}}$ by \T{Const}, \T{Sub}, Lemmas~\ref{lem:ts-refl-const-eq} and \ref{lem:ts-extract-subst-val-valid}, and \Srule{Refine},
     \reflem{ts-val-subst-typing} implies $ \emptyset   \vdash  \ottnt{e_{{\mathrm{2}}}}  \ottsym{:}  \ottnt{C}$, which we must show.
   \end{caseanalysis}

  \case \T{Let}:
   We are given
   \begin{prooftree}
    \AxiomC{$ \emptyset   \vdash  \ottnt{e_{{\mathrm{1}}}}  \ottsym{:}   \ottnt{T_{{\mathrm{1}}}}  \,  \,\&\,  \,  \Phi_{{\mathrm{1}}}  \, / \,  \ottnt{S_{{\mathrm{1}}}} $}
    \AxiomC{$\mathit{x_{{\mathrm{1}}}} \,  {:}  \, \ottnt{T_{{\mathrm{1}}}}  \vdash  \ottnt{e_{{\mathrm{2}}}}  \ottsym{:}   \ottnt{T_{{\mathrm{2}}}}  \,  \,\&\,  \,  \Phi_{{\mathrm{2}}}  \, / \,  \ottnt{S_{{\mathrm{2}}}} $}
    \AxiomC{$\mathit{x_{{\mathrm{1}}}} \,  \not\in  \,  \mathit{fv}  (  \ottnt{T_{{\mathrm{2}}}}  )  \,  \mathbin{\cup}  \,  \mathit{fv}  (  \Phi_{{\mathrm{2}}}  ) $}
    \TrinaryInfC{$ \emptyset   \vdash  \mathsf{let} \, \mathit{x_{{\mathrm{1}}}}  \ottsym{=}  \ottnt{e_{{\mathrm{1}}}} \,  \mathsf{in}  \, \ottnt{e_{{\mathrm{2}}}}  \ottsym{:}   \ottnt{T_{{\mathrm{2}}}}  \,  \,\&\,  \,  \Phi_{{\mathrm{1}}} \,  \tceappendop  \, \Phi_{{\mathrm{2}}}  \, / \,  \ottnt{S_{{\mathrm{1}}}}  \gg\!\!=  \ottsym{(}   \lambda\!  \, \mathit{x_{{\mathrm{1}}}}  \ottsym{.}  \ottnt{S_{{\mathrm{2}}}}  \ottsym{)} $}
   \end{prooftree}
   for some $\mathit{x_{{\mathrm{1}}}}$, $\ottnt{e_{{\mathrm{1}}}}$, $\ottnt{e_{{\mathrm{2}}}}$, $\ottnt{T_{{\mathrm{1}}}}$, $\ottnt{T_{{\mathrm{2}}}}$, $\Phi_{{\mathrm{1}}}$, $\Phi_{{\mathrm{2}}}$, $\ottnt{S_{{\mathrm{1}}}}$, and $\ottnt{S_{{\mathrm{2}}}}$ such that
   \begin{itemize}
    \item $\ottnt{e} \,  =  \, \ottsym{(}  \mathsf{let} \, \mathit{x_{{\mathrm{1}}}}  \ottsym{=}  \ottnt{e_{{\mathrm{1}}}} \,  \mathsf{in}  \, \ottnt{e_{{\mathrm{2}}}}  \ottsym{)}$ and
    \item $\ottnt{C} \,  =  \,  \ottnt{T_{{\mathrm{2}}}}  \,  \,\&\,  \,  \Phi_{{\mathrm{1}}} \,  \tceappendop  \, \Phi_{{\mathrm{2}}}  \, / \,  \ottnt{S_{{\mathrm{1}}}}  \gg\!\!=  \ottsym{(}   \lambda\!  \, \mathit{x_{{\mathrm{1}}}}  \ottsym{.}  \ottnt{S_{{\mathrm{2}}}}  \ottsym{)} $.
   \end{itemize}
   The reduction rule applicable to $\ottnt{e}$ is only \R{Let}.
   Therefore, there exists some $\ottnt{v_{{\mathrm{1}}}}$ such that $\ottnt{e_{{\mathrm{1}}}} \,  =  \, \ottnt{v_{{\mathrm{1}}}}$ and $\ottnt{e'} \,  =  \,  \ottnt{e_{{\mathrm{2}}}}    [  \ottnt{v_{{\mathrm{1}}}}  /  \mathit{x_{{\mathrm{1}}}}  ]  $;
   namely,
   \[
    \ottnt{e} = \ottsym{(}  \mathsf{let} \, \mathit{x_{{\mathrm{1}}}}  \ottsym{=}  \ottnt{v_{{\mathrm{1}}}} \,  \mathsf{in}  \, \ottnt{e_{{\mathrm{2}}}}  \ottsym{)} \,  \rightsquigarrow  \,  \ottnt{e_{{\mathrm{2}}}}    [  \ottnt{v_{{\mathrm{1}}}}  /  \mathit{x_{{\mathrm{1}}}}  ]   = \ottnt{e'} ~.
   \]
   Therefore, it suffices to show that
   \[
     \emptyset   \vdash   \ottnt{e_{{\mathrm{2}}}}    [  \ottnt{v_{{\mathrm{1}}}}  /  \mathit{x_{{\mathrm{1}}}}  ]    \ottsym{:}   \ottnt{T_{{\mathrm{2}}}}  \,  \,\&\,  \,  \Phi_{{\mathrm{1}}} \,  \tceappendop  \, \Phi_{{\mathrm{2}}}  \, / \,  \ottnt{S_{{\mathrm{1}}}}  \gg\!\!=  \ottsym{(}   \lambda\!  \, \mathit{x_{{\mathrm{1}}}}  \ottsym{.}  \ottnt{S_{{\mathrm{2}}}}  \ottsym{)}  ~.
   \]

   Because $ \emptyset   \vdash  \ottnt{v_{{\mathrm{1}}}}  \ottsym{:}   \ottnt{T_{{\mathrm{1}}}}  \,  \,\&\,  \,  \Phi_{{\mathrm{1}}}  \, / \,  \ottnt{S_{{\mathrm{1}}}} $,
   \reflem{ts-inv-val} implies that there exists some $\ottnt{T'_{{\mathrm{1}}}}$ such that
   \begin{itemize}
    \item $ \emptyset   \vdash  \ottnt{v_{{\mathrm{1}}}}  \ottsym{:}  \ottnt{T'_{{\mathrm{1}}}}$ and
    \item $ \emptyset   \vdash   \ottnt{T'_{{\mathrm{1}}}}  \,  \,\&\,  \,   \Phi_\mathsf{val}   \, / \,   \square    \ottsym{<:}   \ottnt{T_{{\mathrm{1}}}}  \,  \,\&\,  \,  \Phi_{{\mathrm{1}}}  \, / \,  \ottnt{S_{{\mathrm{1}}}} $, which implies
          \begin{itemize}
           \item $ \emptyset   \vdash  \ottnt{T'_{{\mathrm{1}}}}  \ottsym{<:}  \ottnt{T_{{\mathrm{1}}}}$,
           \item $ \emptyset   \vdash   \Phi_\mathsf{val}   \ottsym{<:}  \Phi_{{\mathrm{1}}}$, and
           \item $ \emptyset  \,  |  \, \ottnt{T'_{{\mathrm{1}}}} \,  \,\&\,  \,  \Phi_\mathsf{val}   \vdash   \square   \ottsym{<:}  \ottnt{S_{{\mathrm{1}}}}$.
          \end{itemize}
   \end{itemize}
   Because
   \begin{itemize}
    \item $ \emptyset   \vdash  \ottnt{v_{{\mathrm{1}}}}  \ottsym{:}  \ottnt{T'_{{\mathrm{1}}}}$,
    \item $ \emptyset   \vdash  \ottnt{T'_{{\mathrm{1}}}}  \ottsym{<:}  \ottnt{T_{{\mathrm{1}}}}$, and
    \item $ \emptyset   \vdash  \ottnt{T_{{\mathrm{1}}}}$ by \reflem{ts-wf-ty-tctx-typing} with $ \emptyset   \vdash  \ottnt{v_{{\mathrm{1}}}}  \ottsym{:}   \ottnt{T_{{\mathrm{1}}}}  \,  \,\&\,  \,  \Phi_{{\mathrm{1}}}  \, / \,  \ottnt{S_{{\mathrm{1}}}} $,
   \end{itemize}
   we have $ \emptyset   \vdash  \ottnt{v_{{\mathrm{1}}}}  \ottsym{:}  \ottnt{T_{{\mathrm{1}}}}$
   by Lemmas~\ref{lem:ts-refl-subtyping-pure-eff} and \ref{lem:ts-wf-TP_val}, and \T{Sub}.
   Then, by \reflem{ts-val-subst-typing} with $\mathit{x_{{\mathrm{1}}}} \,  {:}  \, \ottnt{T_{{\mathrm{1}}}}  \vdash  \ottnt{e_{{\mathrm{2}}}}  \ottsym{:}   \ottnt{T_{{\mathrm{2}}}}  \,  \,\&\,  \,  \Phi_{{\mathrm{2}}}  \, / \,  \ottnt{S_{{\mathrm{2}}}} $,
   we have
   \[
     \emptyset   \vdash   \ottnt{e_{{\mathrm{2}}}}    [  \ottnt{v_{{\mathrm{1}}}}  /  \mathit{x_{{\mathrm{1}}}}  ]    \ottsym{:}   \ottnt{T_{{\mathrm{2}}}}  \,  \,\&\,  \,  \Phi_{{\mathrm{2}}}  \, / \,   \ottnt{S_{{\mathrm{2}}}}    [  \ottnt{v_{{\mathrm{1}}}}  /  \mathit{x_{{\mathrm{1}}}}  ]    ~.
   \]
   Note that $\mathit{x_{{\mathrm{1}}}} \,  \not\in  \,  \mathit{fv}  (  \ottnt{T_{{\mathrm{2}}}}  )  \,  \mathbin{\cup}  \,  \mathit{fv}  (  \Phi_{{\mathrm{2}}}  ) $.
   Because $ \emptyset   \vdash   \ottnt{T_{{\mathrm{2}}}}  \,  \,\&\,  \,  \Phi_{{\mathrm{1}}} \,  \tceappendop  \, \Phi_{{\mathrm{2}}}  \, / \,  \ottnt{S_{{\mathrm{1}}}}  \gg\!\!=  \ottsym{(}   \lambda\!  \, \mathit{x_{{\mathrm{1}}}}  \ottsym{.}  \ottnt{S_{{\mathrm{2}}}}  \ottsym{)} $ by \reflem{ts-wf-tctx-wf-ftyping}
   with $ \emptyset   \vdash  \mathsf{let} \, \mathit{x_{{\mathrm{1}}}}  \ottsym{=}  \ottnt{v_{{\mathrm{1}}}} \,  \mathsf{in}  \, \ottnt{e_{{\mathrm{2}}}}  \ottsym{:}   \ottnt{T_{{\mathrm{2}}}}  \,  \,\&\,  \,  \Phi_{{\mathrm{1}}} \,  \tceappendop  \, \Phi_{{\mathrm{2}}}  \, / \,  \ottnt{S_{{\mathrm{1}}}}  \gg\!\!=  \ottsym{(}   \lambda\!  \, \mathit{x_{{\mathrm{1}}}}  \ottsym{.}  \ottnt{S_{{\mathrm{2}}}}  \ottsym{)} $,
   \T{Sub} implies that it suffices to show that
   \[
     \emptyset   \vdash    \ottnt{T_{{\mathrm{2}}}}  \,  \,\&\,  \,  \Phi_{{\mathrm{2}}}  \, / \,  \ottnt{S_{{\mathrm{2}}}}     [  \ottnt{v_{{\mathrm{1}}}}  /  \mathit{x_{{\mathrm{1}}}}  ]    \ottsym{<:}   \ottnt{T_{{\mathrm{2}}}}  \,  \,\&\,  \,  \Phi_{{\mathrm{1}}} \,  \tceappendop  \, \Phi_{{\mathrm{2}}}  \, / \,  \ottnt{S_{{\mathrm{1}}}}  \gg\!\!=  \ottsym{(}   \lambda\!  \, \mathit{x_{{\mathrm{1}}}}  \ottsym{.}  \ottnt{S_{{\mathrm{2}}}}  \ottsym{)}  ~,
   \]
   which is implied by \Srule{Comp} with the following.
   \begin{itemize}
    \item $ \emptyset   \vdash  \ottnt{T_{{\mathrm{2}}}}  \ottsym{<:}  \ottnt{T_{{\mathrm{2}}}}$ by \reflem{ts-refl-subtyping}
          with $ \emptyset   \vdash  \ottnt{T_{{\mathrm{2}}}}$, which is implied by
          Lemmas~\ref{lem:ts-wf-ty-tctx-typing} and \ref{lem:ts-elim-unused-binding-wf-ftyping} with
          $\mathit{x_{{\mathrm{1}}}} \,  {:}  \, \ottnt{T_{{\mathrm{1}}}}  \vdash  \ottnt{e_{{\mathrm{2}}}}  \ottsym{:}   \ottnt{T_{{\mathrm{2}}}}  \,  \,\&\,  \,  \Phi_{{\mathrm{2}}}  \, / \,  \ottnt{S_{{\mathrm{2}}}} $ and $\mathit{x_{{\mathrm{1}}}} \,  \not\in  \,  \mathit{fv}  (  \ottnt{T_{{\mathrm{2}}}}  ) $.

    \item $ \emptyset   \vdash  \Phi_{{\mathrm{2}}}  \ottsym{<:}  \Phi_{{\mathrm{1}}} \,  \tceappendop  \, \Phi_{{\mathrm{2}}}$ by
          Lemmas~\ref{lem:ts-transitivity-subtyping} with
          \begin{itemize}
           \item $ \emptyset   \vdash  \Phi_{{\mathrm{2}}}  \ottsym{<:}   \Phi_\mathsf{val}  \,  \tceappendop  \, \Phi_{{\mathrm{2}}}$ by \reflem{ts-weak-strength-TP_val}, and
           \item $ \emptyset   \vdash   \Phi_\mathsf{val}  \,  \tceappendop  \, \Phi_{{\mathrm{2}}}  \ottsym{<:}  \Phi_{{\mathrm{1}}} \,  \tceappendop  \, \Phi_{{\mathrm{2}}}$ by \reflem{ts-mono-eff-compose-preeff}
                 and $ \emptyset   \vdash   \Phi_\mathsf{val}   \ottsym{<:}  \Phi_{{\mathrm{1}}}$.
          \end{itemize}

    \item We show that
          \[
            \emptyset  \,  |  \, \ottnt{T_{{\mathrm{2}}}} \,  \,\&\,  \, \Phi_{{\mathrm{2}}}  \vdash   \ottnt{S_{{\mathrm{2}}}}    [  \ottnt{v_{{\mathrm{1}}}}  /  \mathit{x_{{\mathrm{1}}}}  ]    \ottsym{<:}  \ottnt{S_{{\mathrm{1}}}}  \gg\!\!=  \ottsym{(}   \lambda\!  \, \mathit{x_{{\mathrm{1}}}}  \ottsym{.}  \ottnt{S_{{\mathrm{2}}}}  \ottsym{)} ~.
          \]
          By case analysis on $\ottnt{S_{{\mathrm{1}}}}  \gg\!\!=  \ottsym{(}   \lambda\!  \, \mathit{x_{{\mathrm{1}}}}  \ottsym{.}  \ottnt{S_{{\mathrm{2}}}}  \ottsym{)}$.
          \begin{caseanalysis}
           \case $\ottnt{S_{{\mathrm{1}}}}  \gg\!\!=  \ottsym{(}   \lambda\!  \, \mathit{x_{{\mathrm{1}}}}  \ottsym{.}  \ottnt{S_{{\mathrm{2}}}}  \ottsym{)} \,  =  \,  \square $:
            We have $\ottnt{S_{{\mathrm{1}}}} = \ottnt{S_{{\mathrm{2}}}} =  \square $.
            Therefore, we have the conclusion by \Srule{Empty}.

           \case $  \exists  \, { \mathit{y}  \ottsym{,}  \ottnt{C_{{\mathrm{1}}}}  \ottsym{,}  \ottnt{C_{{\mathrm{2}}}} } . \  \ottsym{(}  \ottnt{S_{{\mathrm{1}}}}  \gg\!\!=  \ottsym{(}   \lambda\!  \, \mathit{x_{{\mathrm{1}}}}  \ottsym{.}  \ottnt{S_{{\mathrm{2}}}}  \ottsym{)}  \ottsym{)} \,  =  \,  ( \forall  \mathit{y}  .  \ottnt{C_{{\mathrm{2}}}}  )  \Rightarrow   \ottnt{C_{{\mathrm{1}}}}  $:
            We have
            $\ottnt{S_{{\mathrm{1}}}} \,  =  \,  ( \forall  \mathit{x_{{\mathrm{1}}}}  .  \ottnt{C}  )  \Rightarrow   \ottnt{C_{{\mathrm{1}}}} $ and
            $\ottnt{S_{{\mathrm{2}}}} \,  =  \,  ( \forall  \mathit{y}  .  \ottnt{C_{{\mathrm{2}}}}  )  \Rightarrow   \ottnt{C} $ and
            $\mathit{x_{{\mathrm{1}}}} \,  \not\in  \,  \mathit{fv}  (  \ottnt{C_{{\mathrm{2}}}}  )  \,  \mathbin{\backslash}  \,  \{  \mathit{y}  \} $
            for some $\ottnt{C}$.
            Without loss of generality, we can suppose that $ \mathit{x_{{\mathrm{1}}}}    \not=    \mathit{y} $.
            Then, we must show that
            \[
              \emptyset  \,  |  \, \ottnt{T_{{\mathrm{2}}}} \,  \,\&\,  \, \Phi_{{\mathrm{2}}}  \vdash    ( \forall  \mathit{y}  .  \ottnt{C_{{\mathrm{2}}}}  )  \Rightarrow   \ottnt{C}     [  \ottnt{v_{{\mathrm{1}}}}  /  \mathit{x_{{\mathrm{1}}}}  ]    \ottsym{<:}   ( \forall  \mathit{y}  .  \ottnt{C_{{\mathrm{2}}}}  )  \Rightarrow   \ottnt{C_{{\mathrm{1}}}}  ~,
            \]
            which is implied by \Srule{Ans} with the following.
            \begin{itemize}
             \item We show that
                   \[
                    \mathit{y} \,  {:}  \, \ottnt{T_{{\mathrm{2}}}}  \vdash  \ottnt{C_{{\mathrm{2}}}}  \ottsym{<:}  \ottnt{C_{{\mathrm{2}}}} ~.
                   \]
                   By \reflem{ts-refl-subtyping}, it suffices to show that
                   $\mathit{y} \,  {:}  \, \ottnt{T_{{\mathrm{2}}}}  \vdash  \ottnt{C_{{\mathrm{2}}}}$.
                   Because $\mathit{x_{{\mathrm{1}}}} \,  {:}  \, \ottnt{T_{{\mathrm{1}}}}  \vdash  \ottnt{e_{{\mathrm{2}}}}  \ottsym{:}   \ottnt{T_{{\mathrm{2}}}}  \,  \,\&\,  \,  \Phi_{{\mathrm{2}}}  \, / \,  \ottnt{S_{{\mathrm{2}}}} $ and $\ottnt{S_{{\mathrm{2}}}} \,  =  \,  ( \forall  \mathit{y}  .  \ottnt{C_{{\mathrm{2}}}}  )  \Rightarrow   \ottnt{C} $,
                   \reflem{ts-wf-ty-tctx-typing} implies
                   $\mathit{x_{{\mathrm{1}}}} \,  {:}  \, \ottnt{T_{{\mathrm{1}}}}  \ottsym{,}  \mathit{y} \,  {:}  \, \ottnt{T_{{\mathrm{2}}}}  \vdash  \ottnt{C_{{\mathrm{2}}}}$.
                   Because $\mathit{x} \,  \not\in  \,  \mathit{fv}  (  \ottnt{C_{{\mathrm{2}}}}  ) $,
                   \reflem{ts-elim-unused-binding-wf-ftyping} implies the conclusion.

             \item We show that
                   \[
                     \emptyset   \vdash   \ottnt{C}    [  \ottnt{v_{{\mathrm{1}}}}  /  \mathit{x_{{\mathrm{1}}}}  ]    \ottsym{<:}  \ottnt{C_{{\mathrm{1}}}} ~.
                   \]
                   Because
                   $ \emptyset  \,  |  \, \ottnt{T'_{{\mathrm{1}}}} \,  \,\&\,  \,  \Phi_\mathsf{val}   \vdash   \square   \ottsym{<:}  \ottnt{S_{{\mathrm{1}}}}$ and
                   $\ottnt{S_{{\mathrm{1}}}} \,  =  \,  ( \forall  \mathit{x_{{\mathrm{1}}}}  .  \ottnt{C}  )  \Rightarrow   \ottnt{C_{{\mathrm{1}}}} $,
                   we have
                   $ \emptyset  \,  |  \, \ottnt{T'_{{\mathrm{1}}}} \,  \,\&\,  \,  \Phi_\mathsf{val}   \vdash   \square   \ottsym{<:}   ( \forall  \mathit{x_{{\mathrm{1}}}}  .  \ottnt{C}  )  \Rightarrow   \ottnt{C_{{\mathrm{1}}}} $.
                   Then, its inversion implies
                   $\mathit{x_{{\mathrm{1}}}} \,  {:}  \, \ottnt{T'_{{\mathrm{1}}}}  \vdash   \Phi_\mathsf{val}  \,  \tceappendop  \, \ottnt{C}  \ottsym{<:}  \ottnt{C_{{\mathrm{1}}}}$ and
                   $\mathit{x_{{\mathrm{1}}}} \,  \not\in  \,  \mathit{fv}  (  \ottnt{C_{{\mathrm{1}}}}  ) $.
                   By \reflem{ts-remove-TP_val-subty},
                   $\mathit{x_{{\mathrm{1}}}} \,  {:}  \, \ottnt{T'_{{\mathrm{1}}}}  \vdash  \ottnt{C}  \ottsym{<:}  \ottnt{C_{{\mathrm{1}}}}$.
                   Because $ \emptyset   \vdash  \ottnt{v_{{\mathrm{1}}}}  \ottsym{:}  \ottnt{T'_{{\mathrm{1}}}}$,
                   we have $ \emptyset   \vdash   \ottnt{C}    [  \ottnt{v_{{\mathrm{1}}}}  /  \mathit{x_{{\mathrm{1}}}}  ]    \ottsym{<:}  \ottnt{C_{{\mathrm{1}}}}$
                   by \reflem{ts-val-subst-subtyping}.
            \end{itemize}
          \end{caseanalysis}
   \end{itemize}

  \case \T{LetV}:
   We are given
   \begin{prooftree}
    \AxiomC{$ \emptyset   \vdash  \ottnt{v_{{\mathrm{1}}}}  \ottsym{:}  \ottnt{T_{{\mathrm{1}}}}$}
    \AxiomC{$\mathit{x_{{\mathrm{1}}}} \,  {:}  \, \ottnt{T_{{\mathrm{1}}}}  \vdash  \ottnt{e_{{\mathrm{2}}}}  \ottsym{:}  \ottnt{C_{{\mathrm{2}}}}$}
    \BinaryInfC{$ \emptyset   \vdash  \mathsf{let} \, \mathit{x_{{\mathrm{1}}}}  \ottsym{=}  \ottnt{v_{{\mathrm{1}}}} \,  \mathsf{in}  \, \ottnt{e_{{\mathrm{2}}}}  \ottsym{:}   \ottnt{C_{{\mathrm{2}}}}    [  \ottnt{v_{{\mathrm{1}}}}  /  \mathit{x_{{\mathrm{1}}}}  ]  $}
   \end{prooftree}
   for some $\mathit{x_{{\mathrm{1}}}}$, $\ottnt{v_{{\mathrm{1}}}}$, $\ottnt{e_{{\mathrm{2}}}}$, $\ottnt{T_{{\mathrm{1}}}}$, and $\ottnt{C_{{\mathrm{2}}}}$ such that
   \begin{itemize}
    \item $\ottnt{e} \,  =  \, \ottsym{(}  \mathsf{let} \, \mathit{x_{{\mathrm{1}}}}  \ottsym{=}  \ottnt{v_{{\mathrm{1}}}} \,  \mathsf{in}  \, \ottnt{e_{{\mathrm{2}}}}  \ottsym{)}$ and
    \item $\ottnt{C} \,  =  \,  \ottnt{C_{{\mathrm{2}}}}    [  \ottnt{v_{{\mathrm{1}}}}  /  \mathit{x_{{\mathrm{1}}}}  ]  $.
   \end{itemize}
   The reduction rule applicable to $\ottnt{e}$ is only \R{Let}.
   Therefore, $\ottnt{e'} \,  =  \,  \ottnt{e_{{\mathrm{2}}}}    [  \ottnt{v_{{\mathrm{1}}}}  /  \mathit{x_{{\mathrm{1}}}}  ]  $;
   namely,
   \[
    \ottnt{e} = \ottsym{(}  \mathsf{let} \, \mathit{x_{{\mathrm{1}}}}  \ottsym{=}  \ottnt{v_{{\mathrm{1}}}} \,  \mathsf{in}  \, \ottnt{e_{{\mathrm{2}}}}  \ottsym{)} \,  \rightsquigarrow  \,  \ottnt{e_{{\mathrm{2}}}}    [  \ottnt{v_{{\mathrm{1}}}}  /  \mathit{x_{{\mathrm{1}}}}  ]   = \ottnt{e'} ~.
   \]
   Therefore, it suffices to show that
   \[
     \emptyset   \vdash   \ottnt{e_{{\mathrm{2}}}}    [  \ottnt{v_{{\mathrm{1}}}}  /  \mathit{x_{{\mathrm{1}}}}  ]    \ottsym{:}   \ottnt{C_{{\mathrm{2}}}}    [  \ottnt{v_{{\mathrm{1}}}}  /  \mathit{x_{{\mathrm{1}}}}  ]   ~,
   \]
   which is implied by \reflem{ts-val-subst-typing} with
   $ \emptyset   \vdash  \ottnt{v_{{\mathrm{1}}}}  \ottsym{:}  \ottnt{T_{{\mathrm{1}}}}$ and $\mathit{x_{{\mathrm{1}}}} \,  {:}  \, \ottnt{T_{{\mathrm{1}}}}  \vdash  \ottnt{e_{{\mathrm{2}}}}  \ottsym{:}  \ottnt{C_{{\mathrm{2}}}}$.

  \case \T{Reset}:
   We are given
   \begin{prooftree}
    \AxiomC{$ \emptyset   \vdash  \ottnt{e_{{\mathrm{0}}}}  \ottsym{:}   \ottnt{T}  \,  \,\&\,  \,  \Phi  \, / \,   ( \forall  \mathit{y}  .   \ottnt{T}  \,  \,\&\,  \,   \Phi_\mathsf{val}   \, / \,   \square    )  \Rightarrow   \ottnt{C}  $}
    \AxiomC{$\mathit{y} \,  \not\in  \,  \mathit{fv}  (  \ottnt{T}  ) $}
    \BinaryInfC{$ \emptyset   \vdash   \resetcnstr{ \ottnt{e_{{\mathrm{0}}}} }   \ottsym{:}  \ottnt{C}$}
   \end{prooftree}
   for some $\ottnt{e_{{\mathrm{0}}}}$, $\ottnt{T}$, $\Phi$, and $\mathit{y}$ such that
   $\ottnt{e} \,  =  \,  \resetcnstr{ \ottnt{e_{{\mathrm{0}}}} } $.
   The reduction rules applicable to $\ottnt{e}$ are only \R{Shift} and \R{Reset}.
   By case analysis on the reduction rule applied to $\ottnt{e}$.
   \begin{caseanalysis}
    \case \R{Shift}:
     We are given
     $\ottnt{e_{{\mathrm{0}}}} \,  =  \,  \ottnt{K}  [  \mathcal{S}_0 \, \mathit{x}  \ottsym{.}  \ottnt{e'_{{\mathrm{0}}}}  ] $ and
     $\ottnt{e'} \,  =  \,  \ottnt{e'_{{\mathrm{0}}}}    [   \lambda\!  \, \mathit{x}  \ottsym{.}   \resetcnstr{  \ottnt{K}  [  \mathit{x}  ]  }   /  \mathit{x}  ]  $
     for some $\ottnt{K}$, $\mathit{x}$, and $\ottnt{e'_{{\mathrm{0}}}}$; namely,
     \[
      \ottnt{e} =  \resetcnstr{  \ottnt{K}  [  \mathcal{S}_0 \, \mathit{x}  \ottsym{.}  \ottnt{e'_{{\mathrm{0}}}}  ]  }  \,  \rightsquigarrow  \,  \ottnt{e'_{{\mathrm{0}}}}    [   \lambda\!  \, \mathit{x}  \ottsym{.}   \resetcnstr{  \ottnt{K}  [  \mathit{x}  ]  }   /  \mathit{x}  ]   = \ottnt{e'} ~.
     \]
     Therefore, it suffices to show that
     \[
       \emptyset   \vdash   \ottnt{e'_{{\mathrm{0}}}}    [   \lambda\!  \, \mathit{x}  \ottsym{.}   \resetcnstr{  \ottnt{K}  [  \mathit{x}  ]  }   /  \mathit{x}  ]    \ottsym{:}  \ottnt{C} ~,
     \]
     which is implied by \reflem{ts-subject-red-shift}
     with $ \emptyset   \vdash   \ottnt{K}  [  \mathcal{S}_0 \, \mathit{x}  \ottsym{.}  \ottnt{e'_{{\mathrm{0}}}}  ]   \ottsym{:}   \ottnt{T}  \,  \,\&\,  \,  \Phi  \, / \,   ( \forall  \mathit{y}  .   \ottnt{T}  \,  \,\&\,  \,   \Phi_\mathsf{val}   \, / \,   \square    )  \Rightarrow   \ottnt{C}  $.

    \case \R{Reset}:
     We are given
     $\ottnt{e_{{\mathrm{0}}}} = \ottnt{e'} = \ottnt{v}$ for some $\ottnt{v}$; namely,
     \[
      \ottnt{e} =  \resetcnstr{ \ottnt{v} }  \,  \rightsquigarrow  \, \ottnt{v} = \ottnt{e'} ~.
     \]
     Therefore, it suffices to show that
     \[
       \emptyset   \vdash  \ottnt{v}  \ottsym{:}  \ottnt{C} ~.
     \]

     \reflem{ts-inv-val} with $ \emptyset   \vdash  \ottnt{v}  \ottsym{:}   \ottnt{T}  \,  \,\&\,  \,  \Phi  \, / \,   ( \forall  \mathit{y}  .   \ottnt{T}  \,  \,\&\,  \,   \Phi_\mathsf{val}   \, / \,   \square    )  \Rightarrow   \ottnt{C}  $
     implies that there exists some $\ottnt{T'}$ such that
     \begin{itemize}
      \item $ \emptyset   \vdash  \ottnt{v}  \ottsym{:}  \ottnt{T'}$ and
      \item $ \emptyset   \vdash   \ottnt{T'}  \,  \,\&\,  \,   \Phi_\mathsf{val}   \, / \,   \square    \ottsym{<:}   \ottnt{T}  \,  \,\&\,  \,  \Phi  \, / \,   ( \forall  \mathit{y}  .   \ottnt{T}  \,  \,\&\,  \,   \Phi_\mathsf{val}   \, / \,   \square    )  \Rightarrow   \ottnt{C}  $.
     \end{itemize}
     The last subtyping derivation implies
     \begin{itemize}
      \item $ \emptyset   \vdash  \ottnt{T'}  \ottsym{<:}  \ottnt{T}$,
      \item $ \emptyset  \,  |  \, \ottnt{T'} \,  \,\&\,  \,  \Phi_\mathsf{val}   \vdash   \square   \ottsym{<:}   ( \forall  \mathit{y}  .   \ottnt{T}  \,  \,\&\,  \,   \Phi_\mathsf{val}   \, / \,   \square    )  \Rightarrow   \ottnt{C} $, and
      \item $\mathit{y} \,  {:}  \, \ottnt{T'}  \vdash   \ottnt{T}  \,  \,\&\,  \,   \Phi_\mathsf{val}  \,  \tceappendop  \,  \Phi_\mathsf{val}   \, / \,   \square    \ottsym{<:}  \ottnt{C}$
            where $\mathit{y} \,  \not\in  \,  \mathit{fv}  (  \ottnt{C}  ) $.
     \end{itemize}
     Because $ \emptyset   \vdash  \ottnt{v}  \ottsym{:}  \ottnt{T'}$ and $ \emptyset   \vdash  \ottnt{T'}  \ottsym{<:}  \ottnt{T}$ and $ \emptyset   \vdash  \ottnt{T}$ (which is implied by
     \reflem{ts-wf-ty-tctx-typing} with $ \emptyset   \vdash  \ottnt{e_{{\mathrm{0}}}}  \ottsym{:}   \ottnt{T}  \,  \,\&\,  \,  \Phi  \, / \,   ( \forall  \mathit{y}  .   \ottnt{T}  \,  \,\&\,  \,   \Phi_\mathsf{val}   \, / \,   \square    )  \Rightarrow   \ottnt{C}  $),
     we have $ \emptyset   \vdash  \ottnt{v}  \ottsym{:}  \ottnt{T}$ by Lemmas~\ref{lem:ts-refl-subtyping-pure-eff} and \ref{lem:ts-wf-TP_val}, and \T{Sub}.
     Furthermore, we have
     \begin{itemize}
      \item $ \emptyset   \vdash  \ottnt{C}$ by \reflem{ts-wf-ty-tctx-typing} with $ \emptyset   \vdash  \ottnt{e}  \ottsym{:}  \ottnt{C}$ and
      \item $ \emptyset   \vdash   \ottnt{T}  \,  \,\&\,  \,   \Phi_\mathsf{val}   \, / \,   \square    \ottsym{<:}  \ottnt{C}$ by
            Lemmas~\ref{lem:ts-val-subst-subtyping} and
            \ref{lem:ts-remove-TP_val-subty}
            with
            $\mathit{y} \,  {:}  \, \ottnt{T'}  \vdash   \ottnt{T}  \,  \,\&\,  \,   \Phi_\mathsf{val}  \,  \tceappendop  \,  \Phi_\mathsf{val}   \, / \,   \square    \ottsym{<:}  \ottnt{C}$ and
            $ \emptyset   \vdash  \ottnt{v}  \ottsym{:}  \ottnt{T'}$ (note that $\mathit{y} \,  \not\in  \,  \mathit{fv}  (  \ottnt{T}  )  \,  \mathbin{\cup}  \,  \mathit{fv}  (  \ottnt{C}  ) $).
     \end{itemize}
     Therefore, \T{Sub} implies the conclusion.

   \end{caseanalysis}
 \end{caseanalysis}
\end{proof}

\begin{lemmap}{Pre-Effect Composition to Pre-Effect Quotient in Effect Subtyping}{ts-preeff-comp-to-preeff-quot-eff-subtyping}
 Suppose that $\mathit{x} \,  \not\in  \,  \mathit{fv}  (  \Phi_{{\mathrm{1}}}  )  \,  \mathbin{\cup}  \,  \mathit{fv}  (  \Phi_{{\mathrm{2}}}  ) $.
 \begin{enumerate}
  \item If $\Gamma  \ottsym{,}  \mathit{x} \,  {:}  \,  \Sigma ^\ast   \models    { \ottsym{(}   (  \varpi  , \bot )  \,  \tceappendop  \, \Phi_{{\mathrm{1}}}  \ottsym{)} }^\mu   \ottsym{(}  \mathit{x}  \ottsym{)}    \Rightarrow     { \Phi_{{\mathrm{2}}} }^\mu   \ottsym{(}  \mathit{x}  \ottsym{)} $,
        then $\Gamma  \ottsym{,}  \mathit{x} \,  {:}  \,  \Sigma ^\ast   \models    { \Phi_{{\mathrm{1}}} }^\mu   \ottsym{(}  \mathit{x}  \ottsym{)}    \Rightarrow     { \ottsym{(}   { \varpi }^{-1}  \,  \tceappendop  \, \Phi_{{\mathrm{2}}}  \ottsym{)} }^\mu   \ottsym{(}  \mathit{x}  \ottsym{)} $.
  \item If $\Gamma  \ottsym{,}  \mathit{x} \,  {:}  \,  \Sigma ^\omega   \models    { \ottsym{(}   (  \varpi  , \bot )  \,  \tceappendop  \, \Phi_{{\mathrm{1}}}  \ottsym{)} }^\nu   \ottsym{(}  \mathit{x}  \ottsym{)}    \Rightarrow     { \Phi_{{\mathrm{2}}} }^\nu   \ottsym{(}  \mathit{x}  \ottsym{)} $,
        then $\Gamma  \ottsym{,}  \mathit{x} \,  {:}  \,  \Sigma ^\omega   \models    { \Phi_{{\mathrm{1}}} }^\nu   \ottsym{(}  \mathit{x}  \ottsym{)}    \Rightarrow     { \ottsym{(}   { \varpi }^{-1}  \,  \tceappendop  \, \Phi_{{\mathrm{2}}}  \ottsym{)} }^\nu   \ottsym{(}  \mathit{x}  \ottsym{)} $.
  \item If $\Gamma  \vdash   (  \varpi  , \bot )  \,  \tceappendop  \, \Phi_{{\mathrm{1}}}  \ottsym{<:}  \Phi_{{\mathrm{2}}}$, then $\Gamma  \vdash  \Phi_{{\mathrm{1}}}  \ottsym{<:}   { \varpi }^{-1}  \,  \tceappendop  \, \Phi_{{\mathrm{2}}}$.
 \end{enumerate}
\end{lemmap}
\begin{proof}
 \noindent
 \begin{enumerate}
  \item Because $ { \ottsym{(}   { \varpi }^{-1}  \,  \tceappendop  \, \Phi_{{\mathrm{2}}}  \ottsym{)} }^\mu   \ottsym{(}  \mathit{x}  \ottsym{)} \,  =  \,  { \Phi_{{\mathrm{2}}} }^\mu   \ottsym{(}  \varpi \,  \tceappendop  \, \mathit{x}  \ottsym{)}$,
        it suffices to show that
        \[
         \Gamma  \ottsym{,}  \mathit{x} \,  {:}  \,  \Sigma ^\ast   \models    { \Phi_{{\mathrm{1}}} }^\mu   \ottsym{(}  \mathit{x}  \ottsym{)}    \Rightarrow     { \Phi_{{\mathrm{2}}} }^\mu   \ottsym{(}  \varpi \,  \tceappendop  \, \mathit{x}  \ottsym{)}  ~.
        \]
        Because
        \[\begin{array}{lll}
          { \ottsym{(}   (  \varpi  , \bot )  \,  \tceappendop  \, \Phi_{{\mathrm{1}}}  \ottsym{)} }^\mu   \ottsym{(}  \mathit{x}  \ottsym{)} &=&
           \exists  \,  \mathit{x_{{\mathrm{1}}}}  \mathrel{  \in  }   \Sigma ^\ast   . \    \exists  \,  \mathit{x_{{\mathrm{2}}}}  \mathrel{  \in  }   \Sigma ^\ast   . \     \mathit{x}    =    \mathit{x_{{\mathrm{1}}}} \,  \tceappendop  \, \mathit{x_{{\mathrm{2}}}}     \wedge     { \ottsym{(}   (  \varpi  , \bot )   \ottsym{)} }^\mu   \ottsym{(}  \mathit{x_{{\mathrm{1}}}}  \ottsym{)}     \wedge     { \Phi_{{\mathrm{1}}} }^\mu   \ottsym{(}  \mathit{x_{{\mathrm{2}}}}  \ottsym{)}    \\ &=&
           \exists  \,  \mathit{x_{{\mathrm{1}}}}  \mathrel{  \in  }   \Sigma ^\ast   . \    \exists  \,  \mathit{x_{{\mathrm{2}}}}  \mathrel{  \in  }   \Sigma ^\ast   . \     \mathit{x}    =    \mathit{x_{{\mathrm{1}}}} \,  \tceappendop  \, \mathit{x_{{\mathrm{2}}}}     \wedge     \mathit{x_{{\mathrm{1}}}}    =    \varpi      \wedge     { \Phi_{{\mathrm{1}}} }^\mu   \ottsym{(}  \mathit{x_{{\mathrm{2}}}}  \ottsym{)}    ~,
          \end{array}
        \]
        the assumption $\Gamma  \ottsym{,}  \mathit{x} \,  {:}  \,  \Sigma ^\ast   \models    { \ottsym{(}   (  \varpi  , \bot )  \,  \tceappendop  \, \Phi_{{\mathrm{1}}}  \ottsym{)} }^\mu   \ottsym{(}  \mathit{x}  \ottsym{)}    \Rightarrow     { \Phi_{{\mathrm{2}}} }^\mu   \ottsym{(}  \mathit{x}  \ottsym{)} $ implies
        \[
         \Gamma  \ottsym{,}  \mathit{x} \,  {:}  \,  \Sigma ^\ast   \models   \ottsym{(}    \exists  \,  \mathit{x_{{\mathrm{1}}}}  \mathrel{  \in  }   \Sigma ^\ast   . \    \exists  \,  \mathit{x_{{\mathrm{2}}}}  \mathrel{  \in  }   \Sigma ^\ast   . \     \mathit{x}    =    \mathit{x_{{\mathrm{1}}}} \,  \tceappendop  \, \mathit{x_{{\mathrm{2}}}}     \wedge     \mathit{x_{{\mathrm{1}}}}    =    \varpi      \wedge     { \Phi_{{\mathrm{1}}} }^\mu   \ottsym{(}  \mathit{x_{{\mathrm{2}}}}  \ottsym{)}     \ottsym{)}    \Rightarrow     { \Phi_{{\mathrm{2}}} }^\mu   \ottsym{(}  \mathit{x}  \ottsym{)}  ~.
        \]
        It means $ { \Phi_{{\mathrm{2}}} }^\mu   \ottsym{(}  \varpi \,  \tceappendop  \, \mathit{x}  \ottsym{)}$ is true if $ { \Phi_{{\mathrm{1}}} }^\mu   \ottsym{(}  \mathit{x}  \ottsym{)}$ is true.
        Therefore, we have the conclusion.
        \TS{OK informal}

  \item Because $ { \ottsym{(}   { \varpi }^{-1}  \,  \tceappendop  \, \Phi_{{\mathrm{2}}}  \ottsym{)} }^\nu   \ottsym{(}  \mathit{x}  \ottsym{)} \,  =  \,  { \Phi_{{\mathrm{2}}} }^\nu   \ottsym{(}  \varpi \,  \tceappendop  \, \mathit{x}  \ottsym{)}$,
        it suffices to show that
        \[
         \Gamma  \ottsym{,}  \mathit{x} \,  {:}  \,  \Sigma ^\omega   \models    { \Phi_{{\mathrm{1}}} }^\nu   \ottsym{(}  \mathit{x}  \ottsym{)}    \Rightarrow     { \Phi_{{\mathrm{2}}} }^\nu   \ottsym{(}  \varpi \,  \tceappendop  \, \mathit{x}  \ottsym{)}  ~.
        \]
        Because
        \[\begin{array}{lll}
          { \ottsym{(}   (  \varpi  , \bot )  \,  \tceappendop  \, \Phi_{{\mathrm{1}}}  \ottsym{)} }^\nu   \ottsym{(}  \mathit{x}  \ottsym{)} &=&
           { \ottsym{(}   (  \varpi  , \bot )   \ottsym{)} }^\nu   \ottsym{(}  \mathit{x}  \ottsym{)}    \vee    \ottsym{(}    \exists  \,  \mathit{x'}  \mathrel{  \in  }   \Sigma ^\ast   . \    \exists  \,  \mathit{y'}  \mathrel{  \in  }   \Sigma ^\omega   . \     \mathit{x}    =    \mathit{x'} \,  \tceappendop  \, \mathit{y'}     \wedge     { \ottsym{(}   (  \varpi  , \bot )   \ottsym{)} }^\mu   \ottsym{(}  \mathit{x'}  \ottsym{)}     \wedge     { \Phi_{{\mathrm{1}}} }^\nu   \ottsym{(}  \mathit{y'}  \ottsym{)}     \ottsym{)}  \\ &=&
           \bot     \vee    \ottsym{(}    \exists  \,  \mathit{x'}  \mathrel{  \in  }   \Sigma ^\ast   . \    \exists  \,  \mathit{y'}  \mathrel{  \in  }   \Sigma ^\omega   . \     \mathit{x}    =    \mathit{x'} \,  \tceappendop  \, \mathit{y'}     \wedge     \mathit{x'}    =    \varpi      \wedge     { \Phi_{{\mathrm{1}}} }^\nu   \ottsym{(}  \mathit{y'}  \ottsym{)}     \ottsym{)}  ~,
          \end{array}
        \]
        the assumption $\Gamma  \ottsym{,}  \mathit{x} \,  {:}  \,  \Sigma ^\omega   \models    { \ottsym{(}   (  \varpi  , \bot )  \,  \tceappendop  \, \Phi_{{\mathrm{1}}}  \ottsym{)} }^\nu   \ottsym{(}  \mathit{x}  \ottsym{)}    \Rightarrow     { \Phi_{{\mathrm{2}}} }^\nu   \ottsym{(}  \mathit{x}  \ottsym{)} $ implies
        \[
         \Gamma  \ottsym{,}  \mathit{x} \,  {:}  \,  \Sigma ^\omega   \models   \ottsym{(}    \bot     \vee    \ottsym{(}    \exists  \,  \mathit{x'}  \mathrel{  \in  }   \Sigma ^\ast   . \    \exists  \,  \mathit{y'}  \mathrel{  \in  }   \Sigma ^\omega   . \     \mathit{x}    =    \mathit{x'} \,  \tceappendop  \, \mathit{y'}     \wedge     \mathit{x'}    =    \varpi      \wedge     { \Phi_{{\mathrm{1}}} }^\nu   \ottsym{(}  \mathit{y'}  \ottsym{)}     \ottsym{)}   \ottsym{)}    \Rightarrow     { \Phi_{{\mathrm{2}}} }^\nu   \ottsym{(}  \mathit{x}  \ottsym{)}  ~.
        \]
        It means $ { \Phi_{{\mathrm{2}}} }^\nu   \ottsym{(}  \varpi \,  \tceappendop  \, \mathit{x}  \ottsym{)}$ is true if $ { \Phi_{{\mathrm{1}}} }^\nu   \ottsym{(}  \mathit{x}  \ottsym{)}$ is true.
        Therefore, we have the conclusion.
        \TS{OK informal}

  \item By \Srule{Eff} with the first and second cases.
 \end{enumerate}
\end{proof}

\begin{lemmap}{Pre-Effect Composition to Pre-Effect Quotient in $ \nu $-Effects}{ts-preeff-comp-to-preeff-quot-nu}
 If $\vdash  \Gamma$ and $\Gamma  \ottsym{,}  \mathit{y} \,  {:}  \,  \Sigma ^\omega   \models    { \ottsym{(}   (  \varpi  , \bot )  \,  \tceappendop  \, \Phi  \ottsym{)} }^\nu   \ottsym{(}  \mathit{y}  \ottsym{)}    \Rightarrow     { \ottnt{C} }^\nu (  \mathit{y}  )  $,
 then $\Gamma  \ottsym{,}  \mathit{y} \,  {:}  \,  \Sigma ^\omega   \models    { \Phi }^\nu   \ottsym{(}  \mathit{y}  \ottsym{)}    \Rightarrow     { \ottsym{(}   { \varpi }^{-1}  \,  \tceappendop  \, \ottnt{C}  \ottsym{)} }^\nu (  \mathit{y}  )  $.
\end{lemmap}
\begin{proof}
 By induction on $\ottnt{C}$.
 Because $ { \ottnt{C} }^\nu (  \mathit{y}  )  \,  =  \,   { \ottsym{(}   \ottnt{C}  . \Phi   \ottsym{)} }^\nu   \ottsym{(}  \mathit{y}  \ottsym{)}    \wedge     { \ottsym{(}   \ottnt{C}  . \ottnt{S}   \ottsym{)} }^\nu (  \mathit{y}  )  $,
 \reflem{ts-valid-elim-impl-intersect} with $\Gamma  \ottsym{,}  \mathit{y} \,  {:}  \,  \Sigma ^\omega   \models    { \ottsym{(}   (  \varpi  , \bot )  \,  \tceappendop  \, \Phi  \ottsym{)} }^\nu   \ottsym{(}  \mathit{y}  \ottsym{)}    \Rightarrow     { \ottnt{C} }^\nu (  \mathit{y}  )  $ implies
 \begin{itemize}
  \item $\Gamma  \ottsym{,}  \mathit{y} \,  {:}  \,  \Sigma ^\omega   \models    { \ottsym{(}   (  \varpi  , \bot )  \,  \tceappendop  \, \Phi  \ottsym{)} }^\nu   \ottsym{(}  \mathit{y}  \ottsym{)}    \Rightarrow     { \ottsym{(}   \ottnt{C}  . \Phi   \ottsym{)} }^\nu   \ottsym{(}  \mathit{y}  \ottsym{)} $ and
  \item $\Gamma  \ottsym{,}  \mathit{y} \,  {:}  \,  \Sigma ^\omega   \models    { \ottsym{(}   (  \varpi  , \bot )  \,  \tceappendop  \, \Phi  \ottsym{)} }^\nu   \ottsym{(}  \mathit{y}  \ottsym{)}    \Rightarrow     { \ottsym{(}   \ottnt{C}  . \ottnt{S}   \ottsym{)} }^\nu (  \mathit{y}  )  $.
 \end{itemize}
 Because $ { \ottsym{(}   { \varpi }^{-1}  \,  \tceappendop  \, \ottnt{C}  \ottsym{)} }^\nu (  \mathit{y}  )  \,  =  \,   { \ottsym{(}   { \varpi }^{-1}  \,  \tceappendop  \, \ottsym{(}   \ottnt{C}  . \Phi   \ottsym{)}  \ottsym{)} }^\nu   \ottsym{(}  \mathit{y}  \ottsym{)}    \wedge     { \ottsym{(}   { \varpi }^{-1}  \,  \tceappendop  \, \ottsym{(}   \ottnt{C}  . \ottnt{S}   \ottsym{)}  \ottsym{)} }^\nu (  \mathit{y}  )  $,
 \reflem{ts-valid-intro-impl-intersect} implies that it suffices to show the following.
 \begin{itemize}
  \item We show that
        \[
         \Gamma  \ottsym{,}  \mathit{y} \,  {:}  \,  \Sigma ^\omega   \models    { \Phi }^\nu   \ottsym{(}  \mathit{y}  \ottsym{)}    \Rightarrow     { \ottsym{(}   { \varpi }^{-1}  \,  \tceappendop  \, \ottsym{(}   \ottnt{C}  . \Phi   \ottsym{)}  \ottsym{)} }^\nu   \ottsym{(}  \mathit{y}  \ottsym{)}  ~.
        \]
        Because $ { \ottsym{(}   { \varpi }^{-1}  \,  \tceappendop  \, \ottsym{(}   \ottnt{C}  . \Phi   \ottsym{)}  \ottsym{)} }^\nu   \ottsym{(}  \mathit{y}  \ottsym{)} \,  =  \,  { \ottsym{(}   \ottnt{C}  . \Phi   \ottsym{)} }^\nu   \ottsym{(}  \varpi \,  \tceappendop  \, \mathit{y}  \ottsym{)}$,
        it suffices to show that
        \[
         \Gamma  \ottsym{,}  \mathit{y} \,  {:}  \,  \Sigma ^\omega   \models    { \Phi }^\nu   \ottsym{(}  \mathit{y}  \ottsym{)}    \Rightarrow     { \ottsym{(}   \ottnt{C}  . \Phi   \ottsym{)} }^\nu   \ottsym{(}  \varpi \,  \tceappendop  \, \mathit{y}  \ottsym{)}  ~.
        \]
        Because
        $\Gamma  \ottsym{,}  \mathit{y} \,  {:}  \,  \Sigma ^\omega   \models    { \ottsym{(}   (  \varpi  , \bot )  \,  \tceappendop  \, \Phi  \ottsym{)} }^\nu   \ottsym{(}  \mathit{y}  \ottsym{)}    \Rightarrow     { \ottsym{(}   \ottnt{C}  . \Phi   \ottsym{)} }^\nu   \ottsym{(}  \mathit{y}  \ottsym{)} $ and
        $\Gamma  \ottsym{,}  \mathit{y} \,  {:}  \,  \Sigma ^\omega   \vdash  \varpi \,  \tceappendop  \, \mathit{y}  \ottsym{:}   \Sigma ^\omega $,
        we have
        \[
         \Gamma  \ottsym{,}  \mathit{y} \,  {:}  \,  \Sigma ^\omega   \models    { \ottsym{(}   (  \varpi  , \bot )  \,  \tceappendop  \, \Phi  \ottsym{)} }^\nu   \ottsym{(}  \varpi \,  \tceappendop  \, \mathit{y}  \ottsym{)}    \Rightarrow     { \ottsym{(}   \ottnt{C}  . \Phi   \ottsym{)} }^\nu   \ottsym{(}  \varpi \,  \tceappendop  \, \mathit{y}  \ottsym{)} 
        \]
        by Lemmas~\ref{lem:ts-weak-valid} and \ref{lem:ts-term-subst-valid}.
        Then, by \reflem{ts-transitivity-valid}, it suffices to show that
        \[
         \Gamma  \ottsym{,}  \mathit{y} \,  {:}  \,  \Sigma ^\omega   \models    { \Phi }^\nu   \ottsym{(}  \mathit{y}  \ottsym{)}    \Rightarrow     { \ottsym{(}   (  \varpi  , \bot )  \,  \tceappendop  \, \Phi  \ottsym{)} }^\nu   \ottsym{(}  \varpi \,  \tceappendop  \, \mathit{y}  \ottsym{)}  ~.
        \]
        Because
        \[\begin{array}{lll}
          { \ottsym{(}   (  \varpi  , \bot )  \,  \tceappendop  \, \Phi  \ottsym{)} }^\nu   \ottsym{(}  \varpi \,  \tceappendop  \, \mathit{y}  \ottsym{)} &=&
           { \ottsym{(}   (  \varpi  , \bot )   \ottsym{)} }^\nu   \ottsym{(}  \varpi \,  \tceappendop  \, \mathit{y}  \ottsym{)}    \vee    \ottsym{(}    \exists  \,  \mathit{x'}  \mathrel{  \in  }   \Sigma ^\ast   . \    \exists  \,  \mathit{y'}  \mathrel{  \in  }   \Sigma ^\omega   . \     \varpi \,  \tceappendop  \, \mathit{y}    =    \mathit{x'} \,  \tceappendop  \, \mathit{y'}     \wedge     { \ottsym{(}   (  \varpi  , \bot )   \ottsym{)} }^\mu   \ottsym{(}  \mathit{x'}  \ottsym{)}     \wedge     { \Phi }^\nu   \ottsym{(}  \mathit{y'}  \ottsym{)}     \ottsym{)}  \\ &=&
           \bot     \vee    \ottsym{(}    \exists  \,  \mathit{x'}  \mathrel{  \in  }   \Sigma ^\ast   . \    \exists  \,  \mathit{y'}  \mathrel{  \in  }   \Sigma ^\omega   . \     \varpi \,  \tceappendop  \, \mathit{y}    =    \mathit{x'} \,  \tceappendop  \, \mathit{y'}     \wedge     \mathit{x'}    =    \varpi      \wedge     { \Phi }^\nu   \ottsym{(}  \mathit{y'}  \ottsym{)}     \ottsym{)}  ~,
          \end{array}
        \]
        we have the conclusion by letting $ \mathit{x'}    =    \varpi $ and $ \mathit{y'}    =    \mathit{y} $.
        \TS{OK informal}

  \item We show that
        \[
         \Gamma  \ottsym{,}  \mathit{y} \,  {:}  \,  \Sigma ^\omega   \models    { \Phi }^\nu   \ottsym{(}  \mathit{y}  \ottsym{)}    \Rightarrow     { \ottsym{(}   { \varpi }^{-1}  \,  \tceappendop  \, \ottsym{(}   \ottnt{C}  . \ottnt{S}   \ottsym{)}  \ottsym{)} }^\nu (  \mathit{y}  )   ~.
        \]
        By case analysis on $ \ottnt{C}  . \ottnt{S} $.
        \begin{caseanalysis}
         \case $ \ottnt{C}  . \ottnt{S}  \,  =  \,  \square $: By \reflem{ts-valid-intro-impl-top}
          because $ { \ottsym{(}   { \varpi }^{-1}  \,  \tceappendop  \, \ottsym{(}   \ottnt{C}  . \ottnt{S}   \ottsym{)}  \ottsym{)} }^\nu (  \mathit{y}  )  =  {  \square  }^\nu (  \mathit{y}  )  =  \top $.
         \case $  \exists  \, { \mathit{x}  \ottsym{,}  \ottnt{C_{{\mathrm{1}}}}  \ottsym{,}  \ottnt{C_{{\mathrm{2}}}} } . \   \ottnt{C}  . \ottnt{S}  \,  =  \,  ( \forall  \mathit{x}  .  \ottnt{C_{{\mathrm{1}}}}  )  \Rightarrow   \ottnt{C_{{\mathrm{2}}}}  $:
          Because $ { \ottsym{(}   { \varpi }^{-1}  \,  \tceappendop  \, \ottsym{(}   \ottnt{C}  . \ottnt{S}   \ottsym{)}  \ottsym{)} }^\nu (  \mathit{y}  )  \,  =  \,  { \ottsym{(}   { \varpi }^{-1}  \,  \tceappendop  \, \ottnt{C_{{\mathrm{2}}}}  \ottsym{)} }^\nu (  \mathit{y}  ) $,
          we have the conclusion by the IH with
          $\Gamma  \ottsym{,}  \mathit{y} \,  {:}  \,  \Sigma ^\omega   \models    { \ottsym{(}   (  \varpi  , \bot )  \,  \tceappendop  \, \Phi  \ottsym{)} }^\nu   \ottsym{(}  \mathit{y}  \ottsym{)}    \Rightarrow     { \ottnt{C_{{\mathrm{2}}}} }^\nu (  \mathit{y}  )  $,
          which is implied by $\Gamma  \ottsym{,}  \mathit{y} \,  {:}  \,  \Sigma ^\omega   \models    { \ottsym{(}   (  \varpi  , \bot )  \,  \tceappendop  \, \Phi  \ottsym{)} }^\nu   \ottsym{(}  \mathit{y}  \ottsym{)}    \Rightarrow     { \ottsym{(}   \ottnt{C}  . \ottnt{S}   \ottsym{)} }^\nu (  \mathit{y}  )  $.
        \end{caseanalysis}
 \end{itemize}
\end{proof}

\begin{lemmap}{Pre-Effect Composition to Pre-Effect Quotient in Subtyping}{ts-preeff-comp-to-preeff-quot}
 Suppose that $\vdash  \Gamma$.
 \begin{enumerate}
  \item If $\Gamma  \vdash   (  \varpi  , \bot )  \,  \tceappendop  \, \ottnt{C_{{\mathrm{1}}}}  \ottsym{<:}  \ottnt{C_{{\mathrm{2}}}}$, then $\Gamma  \vdash  \ottnt{C_{{\mathrm{1}}}}  \ottsym{<:}   { \varpi }^{-1}  \,  \tceappendop  \, \ottnt{C_{{\mathrm{2}}}}$.
  \item If $\Gamma \,  |  \, \ottnt{T} \,  \,\&\,  \,  (  \varpi  , \bot )  \,  \tceappendop  \, \Phi  \vdash   (  \varpi  , \bot )  \,  \tceappendop  \, \ottnt{S_{{\mathrm{1}}}}  \ottsym{<:}  \ottnt{S_{{\mathrm{2}}}}$,
        then $\Gamma \,  |  \, \ottnt{T} \,  \,\&\,  \, \Phi  \vdash  \ottnt{S_{{\mathrm{1}}}}  \ottsym{<:}   { \varpi }^{-1}  \,  \tceappendop  \, \ottnt{S_{{\mathrm{2}}}}$.
 \end{enumerate}
\end{lemmap}
\begin{proof}
 By simultaneous induction on the total numbers of
 type constructors in $(\ottnt{C_{{\mathrm{1}}}}, \ottnt{C_{{\mathrm{2}}}})$ and $(\ottnt{S_{{\mathrm{1}}}}, \ottnt{S_{{\mathrm{2}}}})$.
 \begin{caseanalysis}
  \case \Srule{Comp}:
   We are given the following derivation.
   \begin{prooftree}
    \AxiomC{$\Gamma  \vdash   \ottnt{C_{{\mathrm{1}}}}  . \ottnt{T}   \ottsym{<:}   \ottnt{C_{{\mathrm{2}}}}  . \ottnt{T} $}
    \AxiomC{$\Gamma  \vdash   (  \varpi  , \bot )  \,  \tceappendop  \, \ottsym{(}   \ottnt{C_{{\mathrm{1}}}}  . \Phi   \ottsym{)}  \ottsym{<:}   \ottnt{C_{{\mathrm{2}}}}  . \Phi $}
    \AxiomC{$\Gamma \,  |  \,  \ottnt{C_{{\mathrm{1}}}}  . \ottnt{T}  \,  \,\&\,  \,  (  \varpi  , \bot )  \,  \tceappendop  \, \ottsym{(}   \ottnt{C_{{\mathrm{1}}}}  . \Phi   \ottsym{)}  \vdash   (  \varpi  , \bot )  \,  \tceappendop  \, \ottsym{(}   \ottnt{C_{{\mathrm{1}}}}  . \ottnt{S}   \ottsym{)}  \ottsym{<:}   \ottnt{C_{{\mathrm{2}}}}  . \ottnt{S} $}
    \TrinaryInfC{$\Gamma  \vdash   (  \varpi  , \bot )  \,  \tceappendop  \, \ottnt{C_{{\mathrm{1}}}}  \ottsym{<:}  \ottnt{C_{{\mathrm{2}}}}$}
   \end{prooftree}
   We have the conclusion by \Srule{Comp} with
   \begin{itemize}
    \item $\Gamma  \vdash   \ottnt{C_{{\mathrm{1}}}}  . \ottnt{T}   \ottsym{<:}   \ottnt{C_{{\mathrm{2}}}}  . \ottnt{T} $,
    \item $\Gamma  \vdash   \ottnt{C_{{\mathrm{1}}}}  . \Phi   \ottsym{<:}   { \varpi }^{-1}  \,  \tceappendop  \, \ottsym{(}   \ottnt{C_{{\mathrm{2}}}}  . \Phi   \ottsym{)}$
          by \reflem{ts-preeff-comp-to-preeff-quot-eff-subtyping}
          with $\Gamma  \vdash   (  \varpi  , \bot )  \,  \tceappendop  \, \ottsym{(}   \ottnt{C_{{\mathrm{1}}}}  . \Phi   \ottsym{)}  \ottsym{<:}   \ottnt{C_{{\mathrm{2}}}}  . \Phi $, and
    \item $\Gamma \,  |  \,  \ottnt{C_{{\mathrm{1}}}}  . \ottnt{T}  \,  \,\&\,  \,  \ottnt{C_{{\mathrm{1}}}}  . \Phi   \vdash   \ottnt{C_{{\mathrm{1}}}}  . \ottnt{S}   \ottsym{<:}   { \varpi }^{-1}  \,  \tceappendop  \, \ottsym{(}   \ottnt{C_{{\mathrm{2}}}}  . \ottnt{S}   \ottsym{)}$
          by the IH.
   \end{itemize}

  \case \Srule{Empty}:
   We are given
   \begin{prooftree}
    \AxiomC{$\Gamma \,  |  \, \ottnt{T} \,  \,\&\,  \,  (  \varpi  , \bot )  \,  \tceappendop  \, \Phi  \vdash   \square   \ottsym{<:}   \square $}
   \end{prooftree}
   where $\ottnt{S_{{\mathrm{1}}}} = \ottnt{S_{{\mathrm{2}}}} =  \square $.
   Because $ { \varpi }^{-1}  \,  \tceappendop  \,  \square  \,  =  \,  \square $,
   we have the conclusion $\Gamma \,  |  \, \ottnt{T} \,  \,\&\,  \, \Phi  \vdash   \square   \ottsym{<:}   \square $ by \Srule{Empty}.

  \case \Srule{Ans}:
   We are given
   \begin{prooftree}
    \AxiomC{$\Gamma  \ottsym{,}  \mathit{x} \,  {:}  \, \ottnt{T}  \vdash  \ottnt{C_{{\mathrm{21}}}}  \ottsym{<:}  \ottnt{C_{{\mathrm{11}}}}$}
    \AxiomC{$\Gamma  \vdash   (  \varpi  , \bot )  \,  \tceappendop  \, \ottnt{C_{{\mathrm{12}}}}  \ottsym{<:}  \ottnt{C_{{\mathrm{22}}}}$}
    \BinaryInfC{$\Gamma \,  |  \, \ottnt{T} \,  \,\&\,  \,  (  \varpi  , \bot )  \,  \tceappendop  \, \Phi  \vdash   ( \forall  \mathit{x}  .  \ottnt{C_{{\mathrm{11}}}}  )  \Rightarrow    (  \varpi  , \bot )  \,  \tceappendop  \, \ottnt{C_{{\mathrm{12}}}}   \ottsym{<:}   ( \forall  \mathit{x}  .  \ottnt{C_{{\mathrm{21}}}}  )  \Rightarrow   \ottnt{C_{{\mathrm{22}}}} $}
   \end{prooftree}
   for some $\mathit{x}$, $\ottnt{C_{{\mathrm{11}}}}$, $\ottnt{C_{{\mathrm{12}}}}$, $\ottnt{C_{{\mathrm{21}}}}$, and $\ottnt{C_{{\mathrm{22}}}}$ such that
   $\ottnt{S_{{\mathrm{1}}}} \,  =  \,  ( \forall  \mathit{x}  .  \ottnt{C_{{\mathrm{11}}}}  )  \Rightarrow   \ottnt{C_{{\mathrm{12}}}} $ and $\ottnt{S_{{\mathrm{2}}}} \,  =  \,  ( \forall  \mathit{x}  .  \ottnt{C_{{\mathrm{21}}}}  )  \Rightarrow   \ottnt{C_{{\mathrm{22}}}} $.
   By the IH, $\Gamma  \vdash  \ottnt{C_{{\mathrm{12}}}}  \ottsym{<:}   { \varpi }^{-1}  \,  \tceappendop  \, \ottnt{C_{{\mathrm{22}}}}$.
   Therefore, by \Srule{Ans}, we have the conclusion
   \[
    \Gamma \,  |  \, \ottnt{T} \,  \,\&\,  \, \Phi  \vdash   ( \forall  \mathit{x}  .  \ottnt{C_{{\mathrm{11}}}}  )  \Rightarrow   \ottnt{C_{{\mathrm{12}}}}   \ottsym{<:}   ( \forall  \mathit{x}  .  \ottnt{C_{{\mathrm{21}}}}  )  \Rightarrow    { \varpi }^{-1}  \,  \tceappendop  \, \ottnt{C_{{\mathrm{22}}}}  ~.
   \]

  \case \Srule{Embed}:
   We are given
   \begin{prooftree}
    \AxiomC{$\Gamma  \ottsym{,}  \mathit{x} \,  {:}  \, \ottnt{T}  \vdash  \ottsym{(}   (  \varpi  , \bot )  \,  \tceappendop  \, \Phi  \ottsym{)} \,  \tceappendop  \, \ottnt{C_{{\mathrm{21}}}}  \ottsym{<:}  \ottnt{C_{{\mathrm{22}}}}$}
    \AxiomC{$\Gamma  \ottsym{,}  \mathit{y} \,  {:}  \,  \Sigma ^\omega   \models    { \ottsym{(}   (  \varpi  , \bot )  \,  \tceappendop  \, \Phi  \ottsym{)} }^\nu   \ottsym{(}  \mathit{y}  \ottsym{)}    \Rightarrow     { \ottnt{C_{{\mathrm{22}}}} }^\nu (  \mathit{y}  )  $}
    \AxiomC{$\mathit{x}  \ottsym{,}  \mathit{y} \,  \not\in  \,  \mathit{fv}  (   (  \varpi  , \bot )  \,  \tceappendop  \, \Phi  )  \,  \mathbin{\cup}  \,  \mathit{fv}  (  \ottnt{C_{{\mathrm{22}}}}  ) $}
    \TrinaryInfC{$\Gamma \,  |  \, \ottnt{T} \,  \,\&\,  \,  (  \varpi  , \bot )  \,  \tceappendop  \, \Phi  \vdash   \square   \ottsym{<:}   ( \forall  \mathit{x}  .  \ottnt{C_{{\mathrm{21}}}}  )  \Rightarrow   \ottnt{C_{{\mathrm{22}}}} $}
   \end{prooftree}
   for some $\mathit{x}$, $\ottnt{C_{{\mathrm{21}}}}$, and $\ottnt{C_{{\mathrm{22}}}}$ such that $\ottnt{S_{{\mathrm{2}}}} \,  =  \,  ( \forall  \mathit{x}  .  \ottnt{C_{{\mathrm{21}}}}  )  \Rightarrow   \ottnt{C_{{\mathrm{22}}}} $.
   We also have $\ottnt{S_{{\mathrm{1}}}} \,  =  \,  \square $.
   \reflem{ts-assoc-preeff-compose-subtyping} with $\Gamma  \ottsym{,}  \mathit{x} \,  {:}  \, \ottnt{T}  \vdash  \ottsym{(}   (  \varpi  , \bot )  \,  \tceappendop  \, \Phi  \ottsym{)} \,  \tceappendop  \, \ottnt{C_{{\mathrm{21}}}}  \ottsym{<:}  \ottnt{C_{{\mathrm{22}}}}$
   implies
   \[
    \Gamma  \ottsym{,}  \mathit{x} \,  {:}  \, \ottnt{T}  \vdash   (  \varpi  , \bot )  \,  \tceappendop  \, \ottsym{(}  \Phi \,  \tceappendop  \, \ottnt{C_{{\mathrm{21}}}}  \ottsym{)}  \ottsym{<:}  \ottnt{C_{{\mathrm{22}}}} ~.
   \]
   Because the total number of type constructors in $( (  \varpi  , \bot )  \,  \tceappendop  \, \ottsym{(}  \Phi \,  \tceappendop  \, \ottnt{C_{{\mathrm{21}}}}  \ottsym{)}, \ottnt{C_{{\mathrm{22}}}})$
   is smaller than that of $(\ottnt{S_{{\mathrm{1}}}}, \ottnt{S_{{\mathrm{2}}}})$ (i.e., $( \square ,  ( \forall  \mathit{x}  .  \ottnt{C_{{\mathrm{21}}}}  )  \Rightarrow   \ottnt{C_{{\mathrm{22}}}} )$),
   the IH implies
   \[
    \Gamma  \ottsym{,}  \mathit{x} \,  {:}  \, \ottnt{T}  \vdash  \Phi \,  \tceappendop  \, \ottnt{C_{{\mathrm{21}}}}  \ottsym{<:}   { \varpi }^{-1}  \,  \tceappendop  \, \ottnt{C_{{\mathrm{22}}}} ~.
   \]
   Further, \reflem{ts-preeff-comp-to-preeff-quot-nu} with
   $\vdash  \Gamma$ and $\Gamma  \ottsym{,}  \mathit{y} \,  {:}  \,  \Sigma ^\omega   \models    { \ottsym{(}   (  \varpi  , \bot )  \,  \tceappendop  \, \Phi  \ottsym{)} }^\nu   \ottsym{(}  \mathit{y}  \ottsym{)}    \Rightarrow     { \ottnt{C_{{\mathrm{22}}}} }^\nu (  \mathit{y}  )  $
   implies
   \[
    \Gamma  \ottsym{,}  \mathit{y} \,  {:}  \,  \Sigma ^\omega   \models    { \Phi }^\nu   \ottsym{(}  \mathit{y}  \ottsym{)}    \Rightarrow     { \ottsym{(}   { \varpi }^{-1}  \,  \tceappendop  \, \ottnt{C_{{\mathrm{22}}}}  \ottsym{)} }^\nu (  \mathit{y}  )   ~.
   \]
   Therefore, \Srule{Empty} implies the conclusion
   \[
    \Gamma \,  |  \, \ottnt{T} \,  \,\&\,  \, \Phi  \vdash   \square   \ottsym{<:}   ( \forall  \mathit{x}  .  \ottnt{C_{{\mathrm{21}}}}  )  \Rightarrow    { \varpi }^{-1}  \,  \tceappendop  \, \ottnt{C_{{\mathrm{22}}}}  ~.
   \]
 \end{caseanalysis}
\end{proof}

\begin{lemmap}{Term Substitution in Well-Formedness and Formula Typing}{ts-term-subst-wf-ftyping}
 Suppose that $\Gamma_{{\mathrm{1}}}  \vdash  \ottnt{t}  \ottsym{:}  \ottnt{s}$.
 \begin{enumerate}
  \item If $\vdash  \Gamma_{{\mathrm{1}}}  \ottsym{,}  \mathit{x} \,  {:}  \, \ottnt{s}  \ottsym{,}  \Gamma_{{\mathrm{2}}}$, then $\vdash   \Gamma_{{\mathrm{1}}}  \ottsym{,}  \Gamma_{{\mathrm{2}}}    [  \ottnt{t}  /  \mathit{x}  ]  $.
  \item If $\Gamma_{{\mathrm{1}}}  \ottsym{,}  \mathit{x} \,  {:}  \, \ottnt{s}  \ottsym{,}  \Gamma_{{\mathrm{2}}}  \vdash  \ottnt{T}$, then $ \Gamma_{{\mathrm{1}}}  \ottsym{,}  \Gamma_{{\mathrm{2}}}    [  \ottnt{t}  /  \mathit{x}  ]    \vdash   \ottnt{T}    [  \ottnt{t}  /  \mathit{x}  ]  $.
  \item If $\Gamma_{{\mathrm{1}}}  \ottsym{,}  \mathit{x} \,  {:}  \, \ottnt{s}  \ottsym{,}  \Gamma_{{\mathrm{2}}}  \vdash  \ottnt{C}$, then $ \Gamma_{{\mathrm{1}}}  \ottsym{,}  \Gamma_{{\mathrm{2}}}    [  \ottnt{t}  /  \mathit{x}  ]    \vdash   \ottnt{C}    [  \ottnt{t}  /  \mathit{x}  ]  $.
  \item If $\Gamma_{{\mathrm{1}}}  \ottsym{,}  \mathit{x} \,  {:}  \, \ottnt{s}  \ottsym{,}  \Gamma_{{\mathrm{2}}} \,  |  \, \ottnt{T}  \vdash  \ottnt{S}$, then $ \Gamma_{{\mathrm{1}}}  \ottsym{,}  \Gamma_{{\mathrm{2}}}    [  \ottnt{t}  /  \mathit{x}  ]   \,  |  \,  \ottnt{T}    [  \ottnt{t}  /  \mathit{x}  ]    \vdash   \ottnt{S}    [  \ottnt{t}  /  \mathit{x}  ]  $.
  \item If $\Gamma_{{\mathrm{1}}}  \ottsym{,}  \mathit{x} \,  {:}  \, \ottnt{s}  \ottsym{,}  \Gamma_{{\mathrm{2}}}  \vdash  \Phi$, then $ \Gamma_{{\mathrm{1}}}  \ottsym{,}  \Gamma_{{\mathrm{2}}}    [  \ottnt{t}  /  \mathit{x}  ]    \vdash   \Phi    [  \ottnt{t}  /  \mathit{x}  ]  $.
  \item If $\Gamma_{{\mathrm{1}}}  \ottsym{,}  \mathit{x} \,  {:}  \, \ottnt{s}  \ottsym{,}  \Gamma_{{\mathrm{2}}}  \vdash  \ottnt{t_{{\mathrm{0}}}}  \ottsym{:}  \ottnt{s_{{\mathrm{0}}}}$, then $ \Gamma_{{\mathrm{1}}}  \ottsym{,}  \Gamma_{{\mathrm{2}}}    [  \ottnt{t}  /  \mathit{x}  ]    \vdash   \ottnt{t_{{\mathrm{0}}}}    [  \ottnt{t}  /  \mathit{x}  ]    \ottsym{:}  \ottnt{s_{{\mathrm{0}}}}$.
  \item If $\Gamma_{{\mathrm{1}}}  \ottsym{,}  \mathit{x} \,  {:}  \, \ottnt{s}  \ottsym{,}  \Gamma_{{\mathrm{2}}}  \vdash  \ottnt{A}  \ottsym{:}   \tceseq{ \ottnt{s_{{\mathrm{0}}}} } $, then $ \Gamma_{{\mathrm{1}}}  \ottsym{,}  \Gamma_{{\mathrm{2}}}    [  \ottnt{t}  /  \mathit{x}  ]    \vdash   \ottnt{A}    [  \ottnt{t}  /  \mathit{x}  ]    \ottsym{:}   \tceseq{ \ottnt{s_{{\mathrm{0}}}} } $.
  \item If $\Gamma_{{\mathrm{1}}}  \ottsym{,}  \mathit{x} \,  {:}  \, \ottnt{s}  \ottsym{,}  \Gamma_{{\mathrm{2}}}  \vdash  \phi$, then $ \Gamma_{{\mathrm{1}}}  \ottsym{,}  \Gamma_{{\mathrm{2}}}    [  \ottnt{t}  /  \mathit{x}  ]    \vdash   \phi    [  \ottnt{t}  /  \mathit{x}  ]  $.
 \end{enumerate}
\end{lemmap}
\begin{proof}
 Straightforward by simultaneous induction on the derivations.
 The case for \Tf{VarS} uses \reflem{ts-weak-wf-ftyping}.
\end{proof}

\begin{lemmap}{Well-Formedness of Pre-Effect Quotient}{ts-wf-preeff-quot}
 If $\Gamma  \vdash  \ottnt{C}$, then $\Gamma  \vdash   { \varpi }^{-1}  \,  \tceappendop  \, \ottnt{C}$.
\end{lemmap}
\begin{proof}
 By induction on $\ottnt{C}$.
 Because $\Gamma  \vdash  \ottnt{C}$ can be derived only by \WFT{Comp},
 the inversion implies $\Gamma  \vdash   \ottnt{C}  . \ottnt{T} $ and $\Gamma  \vdash   \ottnt{C}  . \Phi $ and $\Gamma \,  |  \,  \ottnt{C}  . \ottnt{T}   \vdash   \ottnt{C}  . \ottnt{S} $.
 Because $ { \varpi }^{-1}  \,  \tceappendop  \, \ottnt{C} \,  =  \,   \ottnt{C}  . \ottnt{T}   \,  \,\&\,  \,   { \varpi }^{-1}  \,  \tceappendop  \, \ottsym{(}   \ottnt{C}  . \Phi   \ottsym{)}  \, / \,   { \varpi }^{-1}  \,  \tceappendop  \, \ottsym{(}   \ottnt{C}  . \ottnt{S}   \ottsym{)} $,
 we have the conclusion by \WFT{Comp} with $\Gamma  \vdash   \ottnt{C}  . \ottnt{T} $ and the following.
 \begin{itemize}
  \item We show that $\Gamma  \vdash   { \varpi }^{-1}  \,  \tceappendop  \, \ottsym{(}   \ottnt{C}  . \Phi   \ottsym{)}$,
        which is derived by \WFT{Eff} with the following, where we suppose $\mathit{x} \,  \not\in  \,  \mathit{fv}  (   \ottnt{C}  . \Phi   ) $.
        \begin{itemize}
         \item We show that
               \[
                \Gamma  \ottsym{,}  \mathit{x} \,  {:}  \,  \Sigma ^\ast   \vdash   { \ottsym{(}   { \varpi }^{-1}  \,  \tceappendop  \, \ottsym{(}   \ottnt{C}  . \Phi   \ottsym{)}  \ottsym{)} }^\mu   \ottsym{(}  \mathit{x}  \ottsym{)} ~.
               \]
               Because $ { \ottsym{(}   { \varpi }^{-1}  \,  \tceappendop  \, \ottsym{(}   \ottnt{C}  . \Phi   \ottsym{)}  \ottsym{)} }^\mu   \ottsym{(}  \mathit{x}  \ottsym{)} \,  =  \,  { \ottsym{(}   \ottnt{C}  . \Phi   \ottsym{)} }^\mu   \ottsym{(}  \varpi \,  \tceappendop  \, \mathit{x}  \ottsym{)}$,
               it suffices to show that $\Gamma  \ottsym{,}  \mathit{x} \,  {:}  \,  \Sigma ^\ast   \vdash   { \ottsym{(}   \ottnt{C}  . \Phi   \ottsym{)} }^\mu   \ottsym{(}  \varpi \,  \tceappendop  \, \mathit{x}  \ottsym{)}$.
               Because $\Gamma  \vdash   \ottnt{C}  . \Phi $, we have $\Gamma  \ottsym{,}  \mathit{x} \,  {:}  \,  \Sigma ^\ast   \vdash   { \ottsym{(}   \ottnt{C}  . \Phi   \ottsym{)} }^\mu   \ottsym{(}  \mathit{x}  \ottsym{)}$.
               Because $\vdash  \Gamma$ by \reflem{ts-wf-tctx-wf-ftyping} with $\Gamma  \vdash  \ottnt{C}$,
               we have $\Gamma  \ottsym{,}  \mathit{x} \,  {:}  \,  \Sigma ^\ast   \vdash  \varpi \,  \tceappendop  \, \mathit{x}  \ottsym{:}   \Sigma ^\ast $.
               Therefore, we have the conclusion by Lemmas~\ref{lem:ts-weak-wf-ftyping} and \ref{lem:ts-term-subst-wf-ftyping}.

         \item We show that
               \[
                \Gamma  \ottsym{,}  \mathit{x} \,  {:}  \,  \Sigma ^\omega   \vdash   { \ottsym{(}   { \varpi }^{-1}  \,  \tceappendop  \, \ottsym{(}   \ottnt{C}  . \Phi   \ottsym{)}  \ottsym{)} }^\nu   \ottsym{(}  \mathit{x}  \ottsym{)} ~.
               \]
               Because $ { \ottsym{(}   { \varpi }^{-1}  \,  \tceappendop  \, \ottsym{(}   \ottnt{C}  . \Phi   \ottsym{)}  \ottsym{)} }^\nu   \ottsym{(}  \mathit{x}  \ottsym{)} \,  =  \,  { \ottsym{(}   \ottnt{C}  . \Phi   \ottsym{)} }^\nu   \ottsym{(}  \varpi \,  \tceappendop  \, \mathit{x}  \ottsym{)}$,
               it suffices to show that $\Gamma  \ottsym{,}  \mathit{x} \,  {:}  \,  \Sigma ^\omega   \vdash   { \ottsym{(}   \ottnt{C}  . \Phi   \ottsym{)} }^\nu   \ottsym{(}  \varpi \,  \tceappendop  \, \mathit{x}  \ottsym{)}$.
               Because $\Gamma  \vdash   \ottnt{C}  . \Phi $, we have $\Gamma  \ottsym{,}  \mathit{x} \,  {:}  \,  \Sigma ^\omega   \vdash   { \ottsym{(}   \ottnt{C}  . \Phi   \ottsym{)} }^\nu   \ottsym{(}  \mathit{x}  \ottsym{)}$.
               Because $\vdash  \Gamma$ by \reflem{ts-wf-tctx-wf-ftyping} with $\Gamma  \vdash  \ottnt{C}$,
               we have $\Gamma  \ottsym{,}  \mathit{x} \,  {:}  \,  \Sigma ^\omega   \vdash  \varpi \,  \tceappendop  \, \mathit{x}  \ottsym{:}   \Sigma ^\omega $.
               Therefore, we have the conclusion by Lemmas~\ref{lem:ts-weak-wf-ftyping} and \ref{lem:ts-term-subst-wf-ftyping}.
        \end{itemize}

  \item We show that $\Gamma \,  |  \,  \ottnt{C}  . \ottnt{T}   \vdash   { \varpi }^{-1}  \,  \tceappendop  \, \ottsym{(}   \ottnt{C}  . \ottnt{S}   \ottsym{)}$.
        By case analysis on $ \ottnt{C}  . \ottnt{S} $.
        \begin{caseanalysis}
         \case $ \ottnt{C}  . \ottnt{S}  \,  =  \,  \square $:
          By \WFT{Empty} with $\vdash  \Gamma$, which is derived by \reflem{ts-wf-tctx-wf-ftyping} with $\Gamma  \vdash  \ottnt{C}$.

         \case $  \exists  \, { \mathit{x}  \ottsym{,}  \ottnt{C_{{\mathrm{1}}}}  \ottsym{,}  \ottnt{C_{{\mathrm{2}}}} } . \   \ottnt{C}  . \ottnt{S}  \,  =  \,  ( \forall  \mathit{x}  .  \ottnt{C_{{\mathrm{1}}}}  )  \Rightarrow   \ottnt{C_{{\mathrm{2}}}}  $:
          Because $ { \varpi }^{-1}  \,  \tceappendop  \, \ottsym{(}   \ottnt{C}  . \ottnt{S}   \ottsym{)} \,  =  \,  ( \forall  \mathit{x}  .  \ottnt{C_{{\mathrm{1}}}}  )  \Rightarrow    { \varpi }^{-1}  \,  \tceappendop  \, \ottnt{C_{{\mathrm{2}}}} $,
          we have the conclusion by \WFT{Ans} with the following.
          \begin{itemize}
           \item $\Gamma  \ottsym{,}  \mathit{x} \,  {:}  \,  \ottnt{C}  . \ottnt{T}   \vdash  \ottnt{C_{{\mathrm{1}}}}$, which is implied by $\Gamma \,  |  \,  \ottnt{C}  . \ottnt{T}   \vdash   \ottnt{C}  . \ottnt{S} $.
           \item $\Gamma  \vdash   { \varpi }^{-1}  \,  \tceappendop  \, \ottnt{C_{{\mathrm{2}}}}$, which is derived by the IH with $\Gamma  \vdash  \ottnt{C_{{\mathrm{2}}}}$
                 implied by $\Gamma \,  |  \,  \ottnt{C}  . \ottnt{T}   \vdash   \ottnt{C}  . \ottnt{S} $.
          \end{itemize}
        \end{caseanalysis}
 \end{itemize}
\end{proof}

\begin{lemmap}{Monadic Binding of Stack Types is Consistent with Pre-Effect Quotient}{ts-sty-preeff-quot}
 If $\ottnt{S_{{\mathrm{1}}}}  \gg\!\!=  \ottsym{(}   \lambda\!  \, \mathit{x}  \ottsym{.}  \ottnt{S_{{\mathrm{2}}}}  \ottsym{)}$ is well defined,
 then $\ottsym{(}   { \varpi }^{-1}  \,  \tceappendop  \, \ottnt{S_{{\mathrm{1}}}}  \ottsym{)}  \gg\!\!=  \ottsym{(}   \lambda\!  \, \mathit{x}  \ottsym{.}  \ottnt{S_{{\mathrm{2}}}}  \ottsym{)} \,  =  \,  { \varpi }^{-1}  \,  \tceappendop  \, \ottsym{(}  \ottnt{S_{{\mathrm{1}}}}  \gg\!\!=  \ottsym{(}   \lambda\!  \, \mathit{x}  \ottsym{.}  \ottnt{S_{{\mathrm{2}}}}  \ottsym{)}  \ottsym{)}$.
\end{lemmap}
\begin{proof}
 By case analysis on $\ottnt{S_{{\mathrm{1}}}}$, $\ottnt{S_{{\mathrm{2}}}}$, and $\ottnt{S_{{\mathrm{1}}}}  \gg\!\!=  \ottsym{(}   \lambda\!  \, \mathit{x}  \ottsym{.}  \ottnt{S_{{\mathrm{2}}}}  \ottsym{)}$.
 \begin{caseanalysis}
  \case $\ottnt{S_{{\mathrm{1}}}} = \ottnt{S_{{\mathrm{2}}}} = \ottsym{(}  \ottnt{S_{{\mathrm{1}}}}  \gg\!\!=  \ottsym{(}   \lambda\!  \, \mathit{x}  \ottsym{.}  \ottnt{S_{{\mathrm{2}}}}  \ottsym{)}  \ottsym{)} =  \square $:
   Obvious because $ { \varpi }^{-1}  \,  \tceappendop  \,  \square  \,  =  \,  \square $.

  \case $ {  {  {   \exists  \, { \mathit{y}  \ottsym{,}  \ottnt{C}  \ottsym{,}  \ottnt{C_{{\mathrm{1}}}}  \ottsym{,}  \ottnt{C_{{\mathrm{2}}}} } . \  \ottnt{S_{{\mathrm{1}}}} \,  =  \,  ( \forall  \mathit{x}  .  \ottnt{C}  )  \Rightarrow   \ottnt{C_{{\mathrm{1}}}}   } \,\mathrel{\wedge}\, { \ottnt{S_{{\mathrm{2}}}} \,  =  \,  ( \forall  \mathit{y}  .  \ottnt{C_{{\mathrm{2}}}}  )  \Rightarrow   \ottnt{C}  }  } \,\mathrel{\wedge}\, { \ottnt{S_{{\mathrm{1}}}}  \gg\!\!=  \ottsym{(}   \lambda\!  \, \mathit{x}  \ottsym{.}  \ottnt{S_{{\mathrm{2}}}}  \ottsym{)} \,  =  \,  ( \forall  \mathit{y}  .  \ottnt{C_{{\mathrm{2}}}}  )  \Rightarrow   \ottnt{C_{{\mathrm{1}}}}  }  } \,\mathrel{\wedge}\, { \mathit{x} \,  \not\in  \,  \mathit{fv}  (  \ottnt{C_{{\mathrm{2}}}}  )  \,  \mathbin{\backslash}  \,  \{  \mathit{y}  \}  } $:
   Because $ { \varpi }^{-1}  \,  \tceappendop  \, \ottnt{S_{{\mathrm{1}}}} \,  =  \,  ( \forall  \mathit{x}  .  \ottnt{C}  )  \Rightarrow    { \varpi }^{-1}  \,  \tceappendop  \, \ottnt{C_{{\mathrm{1}}}} $, we have
   \[\begin{array}{lll}
    \ottsym{(}   { \varpi }^{-1}  \,  \tceappendop  \, \ottnt{S_{{\mathrm{1}}}}  \ottsym{)}  \gg\!\!=  \ottsym{(}   \lambda\!  \, \mathit{x}  \ottsym{.}  \ottnt{S_{{\mathrm{2}}}}  \ottsym{)} &=&
    \ottsym{(}   ( \forall  \mathit{x}  .  \ottnt{C}  )  \Rightarrow    { \varpi }^{-1}  \,  \tceappendop  \, \ottnt{C_{{\mathrm{1}}}}   \ottsym{)}  \gg\!\!=  \ottsym{(}   \lambda\!  \, \mathit{x}  \ottsym{.}   ( \forall  \mathit{y}  .  \ottnt{C_{{\mathrm{2}}}}  )  \Rightarrow   \ottnt{C}   \ottsym{)} \\ &=&
     ( \forall  \mathit{y}  .  \ottnt{C_{{\mathrm{2}}}}  )  \Rightarrow    { \varpi }^{-1}  \,  \tceappendop  \, \ottnt{C_{{\mathrm{1}}}}  \\ &=&
     { \varpi }^{-1}  \,  \tceappendop  \, \ottsym{(}  \ottnt{S_{{\mathrm{1}}}}  \gg\!\!=  \ottsym{(}   \lambda\!  \, \mathit{x}  \ottsym{.}  \ottnt{S_{{\mathrm{2}}}}  \ottsym{)}  \ottsym{)} ~.
     \end{array}
   \]
 \end{caseanalysis}
\end{proof}

\begin{lemmap}{Associativity of Pre-Effect Quotient in Effect Subtyping}{ts-assoc-preeff-quot-eff-subtyping}
 $\Gamma  \vdash  \ottsym{(}   { \varpi }^{-1}  \,  \tceappendop  \, \Phi_{{\mathrm{1}}}  \ottsym{)} \,  \tceappendop  \, \Phi_{{\mathrm{2}}}  \ottsym{<:}   { \varpi }^{-1}  \,  \tceappendop  \, \ottsym{(}  \Phi_{{\mathrm{1}}} \,  \tceappendop  \, \Phi_{{\mathrm{2}}}  \ottsym{)}$.
\end{lemmap}
\begin{proof}
 By \Srule{Eff}, it suffices to show the following.
 \begin{itemize}
  \item We show that
        \[
         \Gamma  \ottsym{,}  \mathit{x} \,  {:}  \,  \Sigma ^\ast   \models    { \ottsym{(}  \ottsym{(}   { \varpi }^{-1}  \,  \tceappendop  \, \Phi_{{\mathrm{1}}}  \ottsym{)} \,  \tceappendop  \, \Phi_{{\mathrm{2}}}  \ottsym{)} }^\mu   \ottsym{(}  \mathit{x}  \ottsym{)}    \Rightarrow     { \ottsym{(}   { \varpi }^{-1}  \,  \tceappendop  \, \ottsym{(}  \Phi_{{\mathrm{1}}} \,  \tceappendop  \, \Phi_{{\mathrm{2}}}  \ottsym{)}  \ottsym{)} }^\mu   \ottsym{(}  \mathit{x}  \ottsym{)} 
        \]
        for $\mathit{x} \,  \not\in  \,  \mathit{fv}  (  \Phi_{{\mathrm{1}}}  )  \,  \mathbin{\cup}  \,  \mathit{fv}  (  \Phi_{{\mathrm{2}}}  ) $.
        It suffices to show that
        \[
         \Gamma  \ottsym{,}  \mathit{x} \,  {:}  \,  \Sigma ^\ast   \models   \phi_{{\mathrm{1}}}    \Rightarrow    \phi_{{\mathrm{2}}} 
        \]
        where
        \begin{itemize}
         \item $\phi_{{\mathrm{1}}} \,  =  \,   \exists  \,  \mathit{x_{{\mathrm{11}}}}  \mathrel{  \in  }   \Sigma ^\ast   . \    \exists  \,  \mathit{x_{{\mathrm{12}}}}  \mathrel{  \in  }   \Sigma ^\ast   . \     \mathit{x}    =    \mathit{x_{{\mathrm{11}}}} \,  \tceappendop  \, \mathit{x_{{\mathrm{12}}}}     \wedge     { \Phi_{{\mathrm{1}}} }^\mu   \ottsym{(}  \varpi \,  \tceappendop  \, \mathit{x_{{\mathrm{11}}}}  \ottsym{)}     \wedge     { \Phi_{{\mathrm{2}}} }^\mu   \ottsym{(}  \mathit{x_{{\mathrm{12}}}}  \ottsym{)}   $
               because
               \[\begin{array}{ll} &
                 { \ottsym{(}  \ottsym{(}   { \varpi }^{-1}  \,  \tceappendop  \, \Phi_{{\mathrm{1}}}  \ottsym{)} \,  \tceappendop  \, \Phi_{{\mathrm{2}}}  \ottsym{)} }^\mu   \ottsym{(}  \mathit{x}  \ottsym{)} \\ = &
                  \exists  \,  \mathit{x_{{\mathrm{11}}}}  \mathrel{  \in  }   \Sigma ^\ast   . \    \exists  \,  \mathit{x_{{\mathrm{12}}}}  \mathrel{  \in  }   \Sigma ^\ast   . \     \mathit{x}    =    \mathit{x_{{\mathrm{11}}}} \,  \tceappendop  \, \mathit{x_{{\mathrm{12}}}}     \wedge     { \ottsym{(}   { \varpi }^{-1}  \,  \tceappendop  \, \Phi_{{\mathrm{1}}}  \ottsym{)} }^\mu   \ottsym{(}  \mathit{x_{{\mathrm{11}}}}  \ottsym{)}     \wedge     { \Phi_{{\mathrm{2}}} }^\mu   \ottsym{(}  \mathit{x_{{\mathrm{12}}}}  \ottsym{)}    \\ = &
                \phi_{{\mathrm{1}}} ~,
                 \end{array}
               \]
               and
         \item $\phi_{{\mathrm{2}}} \,  =  \,   \exists  \,  \mathit{x_{{\mathrm{21}}}}  \mathrel{  \in  }   \Sigma ^\ast   . \    \exists  \,  \mathit{x_{{\mathrm{22}}}}  \mathrel{  \in  }   \Sigma ^\ast   . \     \varpi \,  \tceappendop  \, \mathit{x}    =    \mathit{x_{{\mathrm{21}}}} \,  \tceappendop  \, \mathit{x_{{\mathrm{22}}}}     \wedge     { \Phi_{{\mathrm{1}}} }^\mu   \ottsym{(}  \mathit{x_{{\mathrm{21}}}}  \ottsym{)}     \wedge     { \Phi_{{\mathrm{2}}} }^\mu   \ottsym{(}  \mathit{x_{{\mathrm{22}}}}  \ottsym{)}   $
               because
               \[
                 { \ottsym{(}   { \varpi }^{-1}  \,  \tceappendop  \, \ottsym{(}  \Phi_{{\mathrm{1}}} \,  \tceappendop  \, \Phi_{{\mathrm{2}}}  \ottsym{)}  \ottsym{)} }^\mu   \ottsym{(}  \mathit{x}  \ottsym{)} =  { \ottsym{(}  \Phi_{{\mathrm{1}}} \,  \tceappendop  \, \Phi_{{\mathrm{2}}}  \ottsym{)} }^\mu   \ottsym{(}  \varpi \,  \tceappendop  \, \mathit{x}  \ottsym{)} = \phi_{{\mathrm{2}}} ~.
               \]
        \end{itemize}
        We can find it holds by letting $ \mathit{x_{{\mathrm{21}}}}    =    \varpi \,  \tceappendop  \, \mathit{x_{{\mathrm{11}}}} $ and $ \mathit{x_{{\mathrm{22}}}}    =    \mathit{x_{{\mathrm{12}}}} $.
        \TS{OK informal}

  \item We show that
        \[
         \Gamma  \ottsym{,}  \mathit{x} \,  {:}  \,  \Sigma ^\omega   \models    { \ottsym{(}  \ottsym{(}   { \varpi }^{-1}  \,  \tceappendop  \, \Phi_{{\mathrm{1}}}  \ottsym{)} \,  \tceappendop  \, \Phi_{{\mathrm{2}}}  \ottsym{)} }^\nu   \ottsym{(}  \mathit{x}  \ottsym{)}    \Rightarrow     { \ottsym{(}   { \varpi }^{-1}  \,  \tceappendop  \, \ottsym{(}  \Phi_{{\mathrm{1}}} \,  \tceappendop  \, \Phi_{{\mathrm{2}}}  \ottsym{)}  \ottsym{)} }^\nu   \ottsym{(}  \mathit{x}  \ottsym{)} 
        \]
        for $\mathit{x} \,  \not\in  \,  \mathit{fv}  (  \Phi_{{\mathrm{1}}}  )  \,  \mathbin{\cup}  \,  \mathit{fv}  (  \Phi_{{\mathrm{2}}}  ) $.
        It suffices to show that
        \[
         \Gamma  \ottsym{,}  \mathit{x} \,  {:}  \,  \Sigma ^\omega   \models   \phi_{{\mathrm{1}}}    \Rightarrow    \phi_{{\mathrm{2}}} 
        \]
        where
        \begin{itemize}
         \item $\phi_{{\mathrm{1}}} \,  =  \,   { \Phi_{{\mathrm{1}}} }^\nu   \ottsym{(}  \varpi \,  \tceappendop  \, \mathit{x}  \ottsym{)}    \vee    \ottsym{(}    \exists  \,  \mathit{x'_{{\mathrm{1}}}}  \mathrel{  \in  }   \Sigma ^\ast   . \    \exists  \,  \mathit{y'_{{\mathrm{2}}}}  \mathrel{  \in  }   \Sigma ^\omega   . \     \mathit{x}    =    \mathit{x'_{{\mathrm{1}}}} \,  \tceappendop  \, \mathit{y'_{{\mathrm{1}}}}     \wedge     { \Phi_{{\mathrm{1}}} }^\mu   \ottsym{(}  \varpi \,  \tceappendop  \, \mathit{x'_{{\mathrm{1}}}}  \ottsym{)}     \wedge     { \Phi_{{\mathrm{2}}} }^\nu   \ottsym{(}  \mathit{y'_{{\mathrm{1}}}}  \ottsym{)}     \ottsym{)} $
               because
               \[\begin{array}{ll} &
                 { \ottsym{(}  \ottsym{(}   { \varpi }^{-1}  \,  \tceappendop  \, \Phi_{{\mathrm{1}}}  \ottsym{)} \,  \tceappendop  \, \Phi_{{\mathrm{2}}}  \ottsym{)} }^\nu   \ottsym{(}  \mathit{x}  \ottsym{)} \\ = &
                  { \ottsym{(}   { \varpi }^{-1}  \,  \tceappendop  \, \Phi_{{\mathrm{1}}}  \ottsym{)} }^\nu   \ottsym{(}  \mathit{x}  \ottsym{)}    \vee    \ottsym{(}    \exists  \,  \mathit{x'_{{\mathrm{1}}}}  \mathrel{  \in  }   \Sigma ^\ast   . \    \exists  \,  \mathit{y'_{{\mathrm{2}}}}  \mathrel{  \in  }   \Sigma ^\omega   . \     \mathit{x}    =    \mathit{x'_{{\mathrm{1}}}} \,  \tceappendop  \, \mathit{y'_{{\mathrm{1}}}}     \wedge     { \ottsym{(}   { \varpi }^{-1}  \,  \tceappendop  \, \Phi_{{\mathrm{1}}}  \ottsym{)} }^\mu   \ottsym{(}  \mathit{x'_{{\mathrm{1}}}}  \ottsym{)}     \wedge     { \Phi_{{\mathrm{2}}} }^\nu   \ottsym{(}  \mathit{y'_{{\mathrm{1}}}}  \ottsym{)}     \ottsym{)}  \\ = &
                \phi_{{\mathrm{1}}} ~,
                 \end{array}
               \]
               and
         \item $\phi_{{\mathrm{2}}} \,  =  \,   { \Phi_{{\mathrm{1}}} }^\nu   \ottsym{(}  \varpi \,  \tceappendop  \, \mathit{x}  \ottsym{)}    \vee    \ottsym{(}    \exists  \,  \mathit{x'_{{\mathrm{2}}}}  \mathrel{  \in  }   \Sigma ^\ast   . \    \exists  \,  \mathit{y'_{{\mathrm{2}}}}  \mathrel{  \in  }   \Sigma ^\omega   . \     \varpi \,  \tceappendop  \, \mathit{x}    =    \mathit{x'_{{\mathrm{2}}}} \,  \tceappendop  \, \mathit{y'_{{\mathrm{2}}}}     \wedge     { \Phi_{{\mathrm{1}}} }^\mu   \ottsym{(}  \mathit{x'_{{\mathrm{2}}}}  \ottsym{)}     \wedge     { \Phi_{{\mathrm{2}}} }^\nu   \ottsym{(}  \mathit{y'_{{\mathrm{2}}}}  \ottsym{)}     \ottsym{)} $
               because
               \[
                 { \ottsym{(}   { \varpi }^{-1}  \,  \tceappendop  \, \ottsym{(}  \Phi_{{\mathrm{1}}} \,  \tceappendop  \, \Phi_{{\mathrm{2}}}  \ottsym{)}  \ottsym{)} }^\nu   \ottsym{(}  \mathit{x}  \ottsym{)} =  { \ottsym{(}  \Phi_{{\mathrm{1}}} \,  \tceappendop  \, \Phi_{{\mathrm{2}}}  \ottsym{)} }^\nu   \ottsym{(}  \varpi \,  \tceappendop  \, \mathit{x}  \ottsym{)} = \phi_{{\mathrm{2}}} ~.
               \]
        \end{itemize}
        We can find it holds by letting $ \mathit{x'_{{\mathrm{2}}}}    =    \varpi \,  \tceappendop  \, \mathit{x'_{{\mathrm{1}}}} $ and $ \mathit{y'_{{\mathrm{2}}}}    =    \mathit{y'_{{\mathrm{1}}}} $
        if $ { \Phi_{{\mathrm{1}}} }^\nu   \ottsym{(}  \varpi \,  \tceappendop  \, \mathit{x}  \ottsym{)}$ is not true.
        \TS{OK informal}
 \end{itemize}
\end{proof}

\begin{lemmap}{Multiplicative Inverse in Subtype Side}{ts-preeff-comp-quot-minv-subty}
 If $\Gamma  \vdash  \ottnt{C}$, then $\Gamma  \vdash   (  \varpi  , \bot )  \,  \tceappendop  \, \ottsym{(}   { \varpi }^{-1}  \,  \tceappendop  \, \ottnt{C}  \ottsym{)}  \ottsym{<:}  \ottnt{C}$.
\end{lemmap}
\begin{proof}
 By induction on $\ottnt{C}$.
 Because $ (  \varpi  , \bot )  \,  \tceappendop  \, \ottsym{(}   { \varpi }^{-1}  \,  \tceappendop  \, \ottnt{C}  \ottsym{)} \,  =  \,   \ottnt{C}  . \ottnt{T}   \,  \,\&\,  \,   (  \varpi  , \bot )  \,  \tceappendop  \, \ottsym{(}   { \varpi }^{-1}  \,  \tceappendop  \, \ottsym{(}   \ottnt{C}  . \Phi   \ottsym{)}  \ottsym{)}  \, / \,   (  \varpi  , \bot )  \,  \tceappendop  \, \ottsym{(}   { \varpi }^{-1}  \,  \tceappendop  \, \ottsym{(}   \ottnt{C}  . \ottnt{S}   \ottsym{)}  \ottsym{)} $,
 we have the conclusion by \Srule{Comp} with the following.
 \begin{itemize}
  \item We have $\Gamma  \vdash   \ottnt{C}  . \ottnt{T}   \ottsym{<:}   \ottnt{C}  . \ottnt{T} $
        by \reflem{ts-refl-subtyping} with $\Gamma  \vdash   \ottnt{C}  . \ottnt{T} $,
        which is implied by $\Gamma  \vdash  \ottnt{C}$.

  \item We show that
        \[
         \Gamma  \vdash   (  \varpi  , \bot )  \,  \tceappendop  \, \ottsym{(}   { \varpi }^{-1}  \,  \tceappendop  \, \ottsym{(}   \ottnt{C}  . \Phi   \ottsym{)}  \ottsym{)}  \ottsym{<:}   \ottnt{C}  . \Phi  ~,
        \]
        which is derived by \Srule{Eff} with the following.
        \begin{itemize}
         \item We show that
               \begin{equation}
                \Gamma  \ottsym{,}  \mathit{x} \,  {:}  \,  \Sigma ^\ast   \models    { \ottsym{(}   (  \varpi  , \bot )  \,  \tceappendop  \, \ottsym{(}   { \varpi }^{-1}  \,  \tceappendop  \, \ottsym{(}   \ottnt{C}  . \Phi   \ottsym{)}  \ottsym{)}  \ottsym{)} }^\mu   \ottsym{(}  \mathit{x}  \ottsym{)}    \Rightarrow     { \ottsym{(}   \ottnt{C}  . \Phi   \ottsym{)} }^\mu   \ottsym{(}  \mathit{x}  \ottsym{)} 
                 \label{lem:ts-preeff-comp-quot-minv-subty:eff:mu}
               \end{equation}
               for $\mathit{x} \,  \not\in  \,  \mathit{fv}  (   \ottnt{C}  . \Phi   ) $.
               Because
               \[\begin{array}{lll}
                 { \ottsym{(}   (  \varpi  , \bot )  \,  \tceappendop  \, \ottsym{(}   { \varpi }^{-1}  \,  \tceappendop  \, \ottsym{(}   \ottnt{C}  . \Phi   \ottsym{)}  \ottsym{)}  \ottsym{)} }^\mu   \ottsym{(}  \mathit{x}  \ottsym{)} &=&
                  \exists  \,  \mathit{x_{{\mathrm{1}}}}  \mathrel{  \in  }   \Sigma ^\ast   . \    \exists  \,  \mathit{x_{{\mathrm{2}}}}  \mathrel{  \in  }   \Sigma ^\ast   . \     \mathit{x}    =    \mathit{x_{{\mathrm{1}}}} \,  \tceappendop  \, \mathit{x_{{\mathrm{2}}}}     \wedge     { \ottsym{(}   (  \varpi  , \bot )   \ottsym{)} }^\mu   \ottsym{(}  \mathit{x_{{\mathrm{1}}}}  \ottsym{)}     \wedge     { \ottsym{(}   { \varpi }^{-1}  \,  \tceappendop  \, \ottsym{(}   \ottnt{C}  . \Phi   \ottsym{)}  \ottsym{)} }^\mu   \ottsym{(}  \mathit{x_{{\mathrm{2}}}}  \ottsym{)}    \\ &=&
                  \exists  \,  \mathit{x_{{\mathrm{1}}}}  \mathrel{  \in  }   \Sigma ^\ast   . \    \exists  \,  \mathit{x_{{\mathrm{2}}}}  \mathrel{  \in  }   \Sigma ^\ast   . \     \mathit{x}    =    \mathit{x_{{\mathrm{1}}}} \,  \tceappendop  \, \mathit{x_{{\mathrm{2}}}}     \wedge     \mathit{x_{{\mathrm{1}}}}    =    \varpi      \wedge     { \ottsym{(}   \ottnt{C}  . \Phi   \ottsym{)} }^\mu   \ottsym{(}  \varpi \,  \tceappendop  \, \mathit{x_{{\mathrm{2}}}}  \ottsym{)}    ~,
                 \end{array}
               \]
               the variable $\mathit{x}$ in the conclusion formula
               $ { \ottsym{(}   \ottnt{C}  . \Phi   \ottsym{)} }^\mu   \ottsym{(}  \mathit{x}  \ottsym{)}$ of the validity judgment (\ref{lem:ts-preeff-comp-quot-minv-subty:eff:mu}) must be $\varpi \,  \tceappendop  \, \mathit{x_{{\mathrm{2}}}}$.
               Because $ { \ottsym{(}   \ottnt{C}  . \Phi   \ottsym{)} }^\mu   \ottsym{(}  \varpi \,  \tceappendop  \, \mathit{x_{{\mathrm{2}}}}  \ottsym{)}$ is supposed to be true, $ { \ottsym{(}   \ottnt{C}  . \Phi   \ottsym{)} }^\mu   \ottsym{(}  \mathit{x}  \ottsym{)}$ is true.
               Therefore, we have the conclusion.
               \TS{OK informal}

         \item We show that
               \begin{equation}
                \Gamma  \ottsym{,}  \mathit{y} \,  {:}  \,  \Sigma ^\omega   \models    { \ottsym{(}   (  \varpi  , \bot )  \,  \tceappendop  \, \ottsym{(}   { \varpi }^{-1}  \,  \tceappendop  \, \ottsym{(}   \ottnt{C}  . \Phi   \ottsym{)}  \ottsym{)}  \ottsym{)} }^\nu   \ottsym{(}  \mathit{y}  \ottsym{)}    \Rightarrow     { \ottsym{(}   \ottnt{C}  . \Phi   \ottsym{)} }^\nu   \ottsym{(}  \mathit{y}  \ottsym{)} 
                 \label{lem:ts-preeff-comp-quot-minv-subty:eff:nu}
               \end{equation}
               for $\mathit{y} \,  \not\in  \,  \mathit{fv}  (   \ottnt{C}  . \Phi   ) $.
               Because
               \[\begin{array}{l@{\ }l@{\ }l}
                 { \ottsym{(}   (  \varpi  , \bot )  \,  \tceappendop  \, \ottsym{(}   { \varpi }^{-1}  \,  \tceappendop  \, \ottsym{(}   \ottnt{C}  . \Phi   \ottsym{)}  \ottsym{)}  \ottsym{)} }^\nu   \ottsym{(}  \mathit{y}  \ottsym{)} &=&
                  { \ottsym{(}   (  \varpi  , \bot )   \ottsym{)} }^\nu   \ottsym{(}  \mathit{y}  \ottsym{)}    \vee    \ottsym{(}    \exists  \,  \mathit{x'}  \mathrel{  \in  }   \Sigma ^\ast   . \    \exists  \,  \mathit{y'}  \mathrel{  \in  }   \Sigma ^\omega   . \     \mathit{y}    =    \mathit{x'} \,  \tceappendop  \, \mathit{y'}     \wedge     { \ottsym{(}   (  \varpi  , \bot )   \ottsym{)} }^\mu   \ottsym{(}  \mathit{x'}  \ottsym{)}     \wedge     { \ottsym{(}   { \varpi }^{-1}  \,  \tceappendop  \, \ottsym{(}   \ottnt{C}  . \Phi   \ottsym{)}  \ottsym{)} }^\nu   \ottsym{(}  \mathit{y'}  \ottsym{)}     \ottsym{)}  \\ &=&
                  \bot     \vee    \ottsym{(}    \exists  \,  \mathit{x'}  \mathrel{  \in  }   \Sigma ^\ast   . \    \exists  \,  \mathit{y'}  \mathrel{  \in  }   \Sigma ^\omega   . \     \mathit{y}    =    \mathit{x'} \,  \tceappendop  \, \mathit{y'}     \wedge     \mathit{x'}    =    \varpi      \wedge     { \ottsym{(}   \ottnt{C}  . \Phi   \ottsym{)} }^\nu   \ottsym{(}  \varpi \,  \tceappendop  \, \mathit{y'}  \ottsym{)}     \ottsym{)}  ~,
                 \end{array}
               \]
               the variable $\mathit{y}$ in the conclusion formula
               $ { \ottsym{(}   \ottnt{C}  . \Phi   \ottsym{)} }^\nu   \ottsym{(}  \mathit{y}  \ottsym{)}$ of the validity formula (\ref{lem:ts-preeff-comp-quot-minv-subty:eff:nu}) must be $\varpi \,  \tceappendop  \, \mathit{y'}$.
               Because $ { \ottsym{(}   \ottnt{C}  . \Phi   \ottsym{)} }^\nu   \ottsym{(}  \varpi \,  \tceappendop  \, \mathit{y'}  \ottsym{)}$ is supposed to be true, $ { \ottsym{(}   \ottnt{C}  . \Phi   \ottsym{)} }^\nu   \ottsym{(}  \mathit{y}  \ottsym{)}$ is true.
               Therefore, we have the conclusion.
               \TS{OK informal}
        \end{itemize}

  \item We show that
        \[
         \Gamma \,  |  \,  \ottnt{C}  . \ottnt{T}  \,  \,\&\,  \,  (  \varpi  , \bot )  \,  \tceappendop  \, \ottsym{(}   { \varpi }^{-1}  \,  \tceappendop  \, \ottsym{(}   \ottnt{C}  . \Phi   \ottsym{)}  \ottsym{)}  \vdash   (  \varpi  , \bot )  \,  \tceappendop  \, \ottsym{(}   { \varpi }^{-1}  \,  \tceappendop  \, \ottsym{(}   \ottnt{C}  . \ottnt{S}   \ottsym{)}  \ottsym{)}  \ottsym{<:}   \ottnt{C}  . \ottnt{S} 
        \]
        by case analysis on $ \ottnt{C}  . \ottnt{S} $.
        \begin{caseanalysis}
         \case $ \ottnt{C}  . \ottnt{S}  \,  =  \,  \square $:
          Because $ (  \varpi  , \bot )  \,  \tceappendop  \, \ottsym{(}   { \varpi }^{-1}  \,  \tceappendop  \, \ottsym{(}   \ottnt{C}  . \ottnt{S}   \ottsym{)}  \ottsym{)} \,  =  \,  \square $,
          it suffices to show that
          \[
           \Gamma \,  |  \,  \ottnt{C}  . \ottnt{T}  \,  \,\&\,  \,  (  \varpi  , \bot )  \,  \tceappendop  \, \ottsym{(}   { \varpi }^{-1}  \,  \tceappendop  \, \ottsym{(}   \ottnt{C}  . \Phi   \ottsym{)}  \ottsym{)}  \vdash   \square   \ottsym{<:}   \square  ~,
          \]
          which is derived by \Srule{Empty}.

         \case $  \exists  \, { \mathit{x}  \ottsym{,}  \ottnt{C_{{\mathrm{1}}}}  \ottsym{,}  \ottnt{C_{{\mathrm{2}}}} } . \   \ottnt{C}  . \ottnt{S}  \,  =  \,  ( \forall  \mathit{x}  .  \ottnt{C_{{\mathrm{1}}}}  )  \Rightarrow   \ottnt{C_{{\mathrm{2}}}}  $:
          Because $ (  \varpi  , \bot )  \,  \tceappendop  \, \ottsym{(}   { \varpi }^{-1}  \,  \tceappendop  \, \ottsym{(}   \ottnt{C}  . \ottnt{S}   \ottsym{)}  \ottsym{)} \,  =  \,  ( \forall  \mathit{x}  .  \ottnt{C_{{\mathrm{1}}}}  )  \Rightarrow    (  \varpi  , \bot )  \,  \tceappendop  \, \ottsym{(}   { \varpi }^{-1}  \,  \tceappendop  \, \ottnt{C_{{\mathrm{2}}}}  \ottsym{)} $,
          it suffices to show that
          \[
           \Gamma \,  |  \,  \ottnt{C}  . \ottnt{T}  \,  \,\&\,  \,  (  \varpi  , \bot )  \,  \tceappendop  \, \ottsym{(}   { \varpi }^{-1}  \,  \tceappendop  \, \ottsym{(}   \ottnt{C}  . \Phi   \ottsym{)}  \ottsym{)}  \vdash   ( \forall  \mathit{x}  .  \ottnt{C_{{\mathrm{1}}}}  )  \Rightarrow    (  \varpi  , \bot )  \,  \tceappendop  \, \ottsym{(}   { \varpi }^{-1}  \,  \tceappendop  \, \ottnt{C_{{\mathrm{2}}}}  \ottsym{)}   \ottsym{<:}   ( \forall  \mathit{x}  .  \ottnt{C_{{\mathrm{1}}}}  )  \Rightarrow   \ottnt{C_{{\mathrm{2}}}}  ~,
          \]
          which is derived by \Srule{Ans} with the following.
          \begin{itemize}
           \item We show that $\Gamma  \ottsym{,}  \mathit{x} \,  {:}  \,  \ottnt{C}  . \ottnt{T}   \vdash  \ottnt{C_{{\mathrm{1}}}}  \ottsym{<:}  \ottnt{C_{{\mathrm{1}}}}$.
                 Because $\Gamma  \vdash  \ottnt{C}$ implies $\Gamma \,  |  \,  \ottnt{C}  . \ottnt{T}   \vdash   ( \forall  \mathit{x}  .  \ottnt{C_{{\mathrm{1}}}}  )  \Rightarrow   \ottnt{C_{{\mathrm{2}}}} $,
                 the inversion implies $\Gamma  \ottsym{,}  \mathit{x} \,  {:}  \,  \ottnt{C}  . \ottnt{T}   \vdash  \ottnt{C_{{\mathrm{1}}}}$.
                 Therefore, we have the conclusion by \reflem{ts-refl-subtyping}.
           \item We show that $\Gamma  \vdash   (  \varpi  , \bot )  \,  \tceappendop  \, \ottsym{(}   { \varpi }^{-1}  \,  \tceappendop  \, \ottnt{C_{{\mathrm{2}}}}  \ottsym{)}  \ottsym{<:}  \ottnt{C_{{\mathrm{2}}}}$.
                 Because $\Gamma  \vdash  \ottnt{C}$ implies $\Gamma \,  |  \,  \ottnt{C}  . \ottnt{T}   \vdash   ( \forall  \mathit{x}  .  \ottnt{C_{{\mathrm{1}}}}  )  \Rightarrow   \ottnt{C_{{\mathrm{2}}}} $,
                 the inversion implies $\Gamma  \vdash  \ottnt{C_{{\mathrm{2}}}}$.
                 Therefore, we have the conclusion by the IH.
          \end{itemize}
        \end{caseanalysis}
 \end{itemize}
\end{proof}

\begin{lemmap}{Subject Reduction: Evaluation}{ts-subject-red-eval}
 If $ \emptyset   \vdash  \ottnt{e}  \ottsym{:}  \ottnt{C}$ and $\ottnt{e} \,  \mathrel{  \longrightarrow  }  \, \ottnt{e'} \,  \,\&\,  \, \varpi$,
 then $ \emptyset   \vdash  \ottnt{e'}  \ottsym{:}   { \varpi }^{-1}  \,  \tceappendop  \, \ottnt{C}$.
\end{lemmap}
\begin{proof}
 By induction on the typing derivation of $ \emptyset   \vdash  \ottnt{e}  \ottsym{:}  \ottnt{C}$.
 If $\ottnt{e} \,  \mathrel{  \longrightarrow  }  \, \ottnt{e'} \,  \,\&\,  \, \varpi$ is derived by \E{Red},
 then $\ottnt{e} \,  \rightsquigarrow  \, \ottnt{e'}$ and $ \varpi    =     \epsilon  $, and it suffices to show that
 $ \emptyset   \vdash  \ottnt{e'}  \ottsym{:}   {  \epsilon  }^{-1}  \,  \tceappendop  \, \ottnt{C}$,
 which is derived by \T{Sub} with the following:
 \begin{itemize}
  \item $ \emptyset   \vdash  \ottnt{e'}  \ottsym{:}  \ottnt{C}$ by \reflem{ts-subject-red-red};
  \item $ \emptyset   \vdash  \ottnt{C}  \ottsym{<:}   {  \epsilon  }^{-1}  \,  \tceappendop  \, \ottnt{C}$
        by \reflem{ts-preeff-comp-to-preeff-quot} with $ \emptyset   \vdash   \Phi_\mathsf{val}  \,  \tceappendop  \, \ottnt{C}  \ottsym{<:}  \ottnt{C}$,
        which is implied by Lemmas~\ref{lem:ts-wf-ty-tctx-typing}, \ref{lem:ts-refl-subtyping}, and \ref{lem:ts-add-TP_val-subty} with $ \emptyset   \vdash  \ottnt{e}  \ottsym{:}  \ottnt{C}$; and
  \item $ \emptyset   \vdash   {  \epsilon  }^{-1}  \,  \tceappendop  \, \ottnt{C}$ by \reflem{ts-wf-preeff-quot} with $ \emptyset   \vdash  \ottnt{C}$,
        which is implied by \reflem{ts-wf-ty-tctx-typing} with $ \emptyset   \vdash  \ottnt{e}  \ottsym{:}  \ottnt{C}$.
 \end{itemize}

 In what follows, we suppose that the evaluation rule applied last to derive
 $\ottnt{e} \,  \mathrel{  \longrightarrow  }  \, \ottnt{e'} \,  \,\&\,  \, \varpi$ is not \E{Red}.
 \begin{caseanalysis}
  \case \T{CVar}, \T{Var}, \T{Const}, \T{PAbs}, \T{Abs}, \T{Fun}, \T{Shift}, \T{ShiftD}, and \T{Div}: \mbox{}
   Contradictory because there is no evaluation rule applicable to $\ottnt{e}$.

  \case \T{Sub}:
   We are given
   \begin{prooftree}
    \AxiomC{$ \emptyset   \vdash  \ottnt{e}  \ottsym{:}  \ottnt{C'}$}
    \AxiomC{$ \emptyset   \vdash  \ottnt{C'}  \ottsym{<:}  \ottnt{C}$}
    \AxiomC{$ \emptyset   \vdash  \ottnt{C}$}
    \TrinaryInfC{$ \emptyset   \vdash  \ottnt{e}  \ottsym{:}  \ottnt{C}$}
   \end{prooftree}
   for some $\ottnt{C'}$.
   By the IH, $ \emptyset   \vdash  \ottnt{e'}  \ottsym{:}   { \varpi }^{-1}  \,  \tceappendop  \, \ottnt{C'}$.
   Because $ \emptyset   \vdash   { \varpi }^{-1}  \,  \tceappendop  \, \ottnt{C}$ by \reflem{ts-wf-preeff-quot} with $ \emptyset   \vdash  \ottnt{C}$,
   \T{Sub} implies that it suffices to show that
   \[
     \emptyset   \vdash   { \varpi }^{-1}  \,  \tceappendop  \, \ottnt{C'}  \ottsym{<:}   { \varpi }^{-1}  \,  \tceappendop  \, \ottnt{C} ~.
   \]
   By \reflem{ts-preeff-comp-to-preeff-quot}, it suffices to show that
   \[
     \emptyset   \vdash   (  \varpi  , \bot )  \,  \tceappendop  \, \ottsym{(}   { \varpi }^{-1}  \,  \tceappendop  \, \ottnt{C'}  \ottsym{)}  \ottsym{<:}  \ottnt{C} ~.
   \]
   By \reflem{ts-transitivity-subtyping} with $ \emptyset   \vdash  \ottnt{C'}  \ottsym{<:}  \ottnt{C}$,
   it suffices to show that
   \[
     \emptyset   \vdash   (  \varpi  , \bot )  \,  \tceappendop  \, \ottsym{(}   { \varpi }^{-1}  \,  \tceappendop  \, \ottnt{C'}  \ottsym{)}  \ottsym{<:}  \ottnt{C'} ~.
   \]
   By \reflem{ts-preeff-comp-quot-minv-subty}, it suffices to show that $ \emptyset   \vdash  \ottnt{C'}$,
   which is implied by \reflem{ts-wf-ty-tctx-typing} with $ \emptyset   \vdash  \ottnt{e}  \ottsym{:}  \ottnt{C'}$.

  \case \T{Op}, \T{App}, \T{PApp}, and \T{If}:
   Contradictory because $\ottnt{e}$ can be evaluated only by \E{Red}.

  \case \T{Let}:
   We are given
   \begin{prooftree}
    \AxiomC{$ \emptyset   \vdash  \ottnt{e_{{\mathrm{1}}}}  \ottsym{:}   \ottnt{T_{{\mathrm{1}}}}  \,  \,\&\,  \,  \Phi_{{\mathrm{1}}}  \, / \,  \ottnt{S_{{\mathrm{1}}}} $}
    \AxiomC{$\mathit{x_{{\mathrm{1}}}} \,  {:}  \, \ottnt{T_{{\mathrm{1}}}}  \vdash  \ottnt{e_{{\mathrm{2}}}}  \ottsym{:}   \ottnt{T_{{\mathrm{2}}}}  \,  \,\&\,  \,  \Phi_{{\mathrm{2}}}  \, / \,  \ottnt{S_{{\mathrm{2}}}} $}
    \AxiomC{$\mathit{x_{{\mathrm{1}}}} \,  \not\in  \,  \mathit{fv}  (  \ottnt{T_{{\mathrm{2}}}}  )  \,  \mathbin{\cup}  \,  \mathit{fv}  (  \Phi_{{\mathrm{2}}}  ) $}
    \TrinaryInfC{$ \emptyset   \vdash  \mathsf{let} \, \mathit{x_{{\mathrm{1}}}}  \ottsym{=}  \ottnt{e_{{\mathrm{1}}}} \,  \mathsf{in}  \, \ottnt{e_{{\mathrm{2}}}}  \ottsym{:}   \ottnt{T_{{\mathrm{2}}}}  \,  \,\&\,  \,  \Phi_{{\mathrm{1}}} \,  \tceappendop  \, \Phi_{{\mathrm{2}}}  \, / \,  \ottnt{S_{{\mathrm{1}}}}  \gg\!\!=  \ottsym{(}   \lambda\!  \, \mathit{x_{{\mathrm{1}}}}  \ottsym{.}  \ottnt{S_{{\mathrm{2}}}}  \ottsym{)} $}
   \end{prooftree}
   for some $\mathit{x_{{\mathrm{1}}}}$, $\ottnt{e_{{\mathrm{1}}}}$, $\ottnt{e_{{\mathrm{2}}}}$, $\ottnt{T_{{\mathrm{1}}}}$, $\ottnt{T_{{\mathrm{2}}}}$,
   $\Phi_{{\mathrm{1}}}$, $\Phi_{{\mathrm{2}}}$, $\ottnt{S_{{\mathrm{1}}}}$, and $\ottnt{S_{{\mathrm{2}}}}$ such that
   $\ottnt{e} \,  =  \, \ottsym{(}  \mathsf{let} \, \mathit{x_{{\mathrm{1}}}}  \ottsym{=}  \ottnt{e_{{\mathrm{1}}}} \,  \mathsf{in}  \, \ottnt{e_{{\mathrm{2}}}}  \ottsym{)}$ and $\ottnt{C} \,  =  \,  \ottnt{T_{{\mathrm{2}}}}  \,  \,\&\,  \,  \Phi_{{\mathrm{1}}} \,  \tceappendop  \, \Phi_{{\mathrm{2}}}  \, / \,  \ottnt{S_{{\mathrm{1}}}}  \gg\!\!=  \ottsym{(}   \lambda\!  \, \mathit{x_{{\mathrm{1}}}}  \ottsym{.}  \ottnt{S_{{\mathrm{2}}}}  \ottsym{)} $.
   Because the evaluation rule applicable to $\ottnt{e}$ is only \E{Let} (except for \E{Red}),
   we have
   \begin{prooftree}
    \AxiomC{$\ottnt{e_{{\mathrm{1}}}} \,  \mathrel{  \longrightarrow  }  \, \ottnt{e'_{{\mathrm{1}}}} \,  \,\&\,  \, \varpi$}
    \UnaryInfC{$\mathsf{let} \, \mathit{x_{{\mathrm{1}}}}  \ottsym{=}  \ottnt{e_{{\mathrm{1}}}} \,  \mathsf{in}  \, \ottnt{e_{{\mathrm{2}}}} \,  \mathrel{  \longrightarrow  }  \, \mathsf{let} \, \mathit{x_{{\mathrm{1}}}}  \ottsym{=}  \ottnt{e'_{{\mathrm{1}}}} \,  \mathsf{in}  \, \ottnt{e_{{\mathrm{2}}}} \,  \,\&\,  \, \varpi$}
   \end{prooftree}
   for some $\ottnt{e'_{{\mathrm{1}}}}$ such that $\ottnt{e'} \,  =  \, \ottsym{(}  \mathsf{let} \, \mathit{x_{{\mathrm{1}}}}  \ottsym{=}  \ottnt{e'_{{\mathrm{1}}}} \,  \mathsf{in}  \, \ottnt{e_{{\mathrm{2}}}}  \ottsym{)}$.
   By the IH, $ \emptyset   \vdash  \ottnt{e'_{{\mathrm{1}}}}  \ottsym{:}   \ottnt{T_{{\mathrm{1}}}}  \,  \,\&\,  \,   { \varpi }^{-1}  \,  \tceappendop  \, \Phi_{{\mathrm{1}}}  \, / \,   { \varpi }^{-1}  \,  \tceappendop  \, \ottnt{S_{{\mathrm{1}}}} $.
   Because $\ottnt{S_{{\mathrm{1}}}}  \gg\!\!=  \ottsym{(}   \lambda\!  \, \mathit{x_{{\mathrm{1}}}}  \ottsym{.}  \ottnt{S_{{\mathrm{2}}}}  \ottsym{)}$ is well defined,
   \reflem{ts-sty-preeff-quot} implies that $\ottsym{(}   { \varpi }^{-1}  \,  \tceappendop  \, \ottnt{S_{{\mathrm{1}}}}  \ottsym{)}  \gg\!\!=  \ottsym{(}   \lambda\!  \, \mathit{x_{{\mathrm{1}}}}  \ottsym{.}  \ottnt{S_{{\mathrm{2}}}}  \ottsym{)}$ is well defined, and
   $\ottsym{(}   { \varpi }^{-1}  \,  \tceappendop  \, \ottnt{S_{{\mathrm{1}}}}  \ottsym{)}  \gg\!\!=  \ottsym{(}   \lambda\!  \, \mathit{x_{{\mathrm{1}}}}  \ottsym{.}  \ottnt{S_{{\mathrm{2}}}}  \ottsym{)} \,  =  \,  { \varpi }^{-1}  \,  \tceappendop  \, \ottsym{(}  \ottnt{S_{{\mathrm{1}}}}  \gg\!\!=  \ottsym{(}   \lambda\!  \, \mathit{x_{{\mathrm{1}}}}  \ottsym{.}  \ottnt{S_{{\mathrm{2}}}}  \ottsym{)}  \ottsym{)}$.
   Therefore, by \T{Let}, we have
   \begin{equation}
     \emptyset   \vdash  \mathsf{let} \, \mathit{x_{{\mathrm{1}}}}  \ottsym{=}  \ottnt{e'_{{\mathrm{1}}}} \,  \mathsf{in}  \, \ottnt{e_{{\mathrm{2}}}}  \ottsym{:}   \ottnt{T_{{\mathrm{2}}}}  \,  \,\&\,  \,  \ottsym{(}   { \varpi }^{-1}  \,  \tceappendop  \, \Phi_{{\mathrm{1}}}  \ottsym{)} \,  \tceappendop  \, \Phi_{{\mathrm{2}}}  \, / \,   { \varpi }^{-1}  \,  \tceappendop  \, \ottsym{(}  \ottnt{S_{{\mathrm{1}}}}  \gg\!\!=  \ottsym{(}   \lambda\!  \, \mathit{x_{{\mathrm{1}}}}  \ottsym{.}  \ottnt{S_{{\mathrm{2}}}}  \ottsym{)}  \ottsym{)}  ~.
     \label{lem:ts-subject-red-eval:let:reduced-e}
   \end{equation}
   Because $ \emptyset   \vdash   \ottnt{T_{{\mathrm{2}}}}  \,  \,\&\,  \,  \Phi_{{\mathrm{1}}} \,  \tceappendop  \, \Phi_{{\mathrm{2}}}  \, / \,  \ottnt{S_{{\mathrm{1}}}}  \gg\!\!=  \ottsym{(}   \lambda\!  \, \mathit{x_{{\mathrm{1}}}}  \ottsym{.}  \ottnt{S_{{\mathrm{2}}}}  \ottsym{)} $
   by \reflem{ts-wf-ty-tctx-typing} with $ \emptyset   \vdash  \ottnt{e}  \ottsym{:}   \ottnt{T_{{\mathrm{2}}}}  \,  \,\&\,  \,  \Phi_{{\mathrm{1}}} \,  \tceappendop  \, \Phi_{{\mathrm{2}}}  \, / \,  \ottnt{S_{{\mathrm{1}}}}  \gg\!\!=  \ottsym{(}   \lambda\!  \, \mathit{x_{{\mathrm{1}}}}  \ottsym{.}  \ottnt{S_{{\mathrm{2}}}}  \ottsym{)} $,
   we have $ \emptyset   \vdash   \ottnt{T_{{\mathrm{2}}}}  \,  \,\&\,  \,   { \varpi }^{-1}  \,  \tceappendop  \, \ottsym{(}  \Phi_{{\mathrm{1}}} \,  \tceappendop  \, \Phi_{{\mathrm{2}}}  \ottsym{)}  \, / \,   { \varpi }^{-1}  \,  \tceappendop  \, \ottsym{(}  \ottnt{S_{{\mathrm{1}}}}  \gg\!\!=  \ottsym{(}   \lambda\!  \, \mathit{x_{{\mathrm{1}}}}  \ottsym{.}  \ottnt{S_{{\mathrm{2}}}}  \ottsym{)}  \ottsym{)} $
   by \reflem{ts-wf-preeff-quot}.
   Therefore, \T{Sub} with (\ref{lem:ts-subject-red-eval:let:reduced-e}) implies that it suffices to show that
   \[
     \emptyset   \vdash   \ottnt{T_{{\mathrm{2}}}}  \,  \,\&\,  \,  \ottsym{(}   { \varpi }^{-1}  \,  \tceappendop  \, \Phi_{{\mathrm{1}}}  \ottsym{)} \,  \tceappendop  \, \Phi_{{\mathrm{2}}}  \, / \,   { \varpi }^{-1}  \,  \tceappendop  \, \ottsym{(}  \ottnt{S_{{\mathrm{1}}}}  \gg\!\!=  \ottsym{(}   \lambda\!  \, \mathit{x_{{\mathrm{1}}}}  \ottsym{.}  \ottnt{S_{{\mathrm{2}}}}  \ottsym{)}  \ottsym{)}   \ottsym{<:}   \ottnt{T_{{\mathrm{2}}}}  \,  \,\&\,  \,   { \varpi }^{-1}  \,  \tceappendop  \, \ottsym{(}  \Phi_{{\mathrm{1}}} \,  \tceappendop  \, \Phi_{{\mathrm{2}}}  \ottsym{)}  \, / \,   { \varpi }^{-1}  \,  \tceappendop  \, \ottsym{(}  \ottnt{S_{{\mathrm{1}}}}  \gg\!\!=  \ottsym{(}   \lambda\!  \, \mathit{x_{{\mathrm{1}}}}  \ottsym{.}  \ottnt{S_{{\mathrm{2}}}}  \ottsym{)}  \ottsym{)}  ~,
   \]
   which is implied by \Srule{Comp} with the following.
   \begin{itemize}
    \item $ \emptyset   \vdash  \ottnt{T_{{\mathrm{2}}}}  \ottsym{<:}  \ottnt{T_{{\mathrm{2}}}}$
          by \reflem{ts-refl-subtyping} with $ \emptyset   \vdash  \ottnt{T_{{\mathrm{2}}}}$,
          which is implied by $ \emptyset   \vdash   \ottnt{T_{{\mathrm{2}}}}  \,  \,\&\,  \,  \Phi_{{\mathrm{1}}} \,  \tceappendop  \, \Phi_{{\mathrm{2}}}  \, / \,  \ottnt{S_{{\mathrm{1}}}}  \gg\!\!=  \ottsym{(}   \lambda\!  \, \mathit{x_{{\mathrm{1}}}}  \ottsym{.}  \ottnt{S_{{\mathrm{2}}}}  \ottsym{)} $ shown above.

    \item $ \emptyset   \vdash  \ottsym{(}   { \varpi }^{-1}  \,  \tceappendop  \, \Phi_{{\mathrm{1}}}  \ottsym{)} \,  \tceappendop  \, \Phi_{{\mathrm{2}}}  \ottsym{<:}   { \varpi }^{-1}  \,  \tceappendop  \, \ottsym{(}  \Phi_{{\mathrm{1}}} \,  \tceappendop  \, \Phi_{{\mathrm{2}}}  \ottsym{)}$
          by \reflem{ts-assoc-preeff-quot-eff-subtyping}.

    \item $ \emptyset  \,  |  \, \ottnt{T_{{\mathrm{2}}}} \,  \,\&\,  \, \ottsym{(}   { \varpi }^{-1}  \,  \tceappendop  \, \Phi_{{\mathrm{1}}}  \ottsym{)} \,  \tceappendop  \, \Phi_{{\mathrm{2}}}  \vdash   { \varpi }^{-1}  \,  \tceappendop  \, \ottsym{(}  \ottnt{S_{{\mathrm{1}}}}  \gg\!\!=  \ottsym{(}   \lambda\!  \, \mathit{x_{{\mathrm{1}}}}  \ottsym{.}  \ottnt{S_{{\mathrm{2}}}}  \ottsym{)}  \ottsym{)}  \ottsym{<:}   { \varpi }^{-1}  \,  \tceappendop  \, \ottsym{(}  \ottnt{S_{{\mathrm{1}}}}  \gg\!\!=  \ottsym{(}   \lambda\!  \, \mathit{x_{{\mathrm{1}}}}  \ottsym{.}  \ottnt{S_{{\mathrm{2}}}}  \ottsym{)}  \ottsym{)}$
          by \reflem{ts-refl-subtyping} with $ \emptyset  \,  |  \, \ottnt{T_{{\mathrm{2}}}}  \vdash   { \varpi }^{-1}  \,  \tceappendop  \, \ottsym{(}  \ottnt{S_{{\mathrm{1}}}}  \gg\!\!=  \ottsym{(}   \lambda\!  \, \mathit{x_{{\mathrm{1}}}}  \ottsym{.}  \ottnt{S_{{\mathrm{2}}}}  \ottsym{)}  \ottsym{)}$,
          which is implied by \reflem{ts-wf-ty-tctx-typing} with (\ref{lem:ts-subject-red-eval:let:reduced-e}).
   \end{itemize}

  \case \T{LetV}:
   Contradictory because no value can be evaluated.

  \case \T{Reset}:
   We are given
   \begin{prooftree}
    \AxiomC{$ \emptyset   \vdash  \ottnt{e_{{\mathrm{0}}}}  \ottsym{:}   \ottnt{T}  \,  \,\&\,  \,  \Phi  \, / \,   ( \forall  \mathit{x}  .   \ottnt{T}  \,  \,\&\,  \,   \Phi_\mathsf{val}   \, / \,   \square    )  \Rightarrow   \ottnt{C}  $}
    \AxiomC{$\mathit{x} \,  \not\in  \,  \mathit{fv}  (  \ottnt{T}  ) $}
    \BinaryInfC{$ \emptyset   \vdash   \resetcnstr{ \ottnt{e_{{\mathrm{0}}}} }   \ottsym{:}  \ottnt{C}$}
   \end{prooftree}
   for some $\ottnt{e_{{\mathrm{0}}}}$, $\ottnt{T}$, $\Phi$, and $\mathit{x}$ such that
   $\ottnt{e} \,  =  \,  \resetcnstr{ \ottnt{e_{{\mathrm{0}}}} } $.
   Because the evaluation rule applicable to $\ottnt{e}$ is only \E{Reset} (except for \E{Red}),
   we have
   \begin{prooftree}
    \AxiomC{$\ottnt{e_{{\mathrm{0}}}} \,  \mathrel{  \longrightarrow  }  \, \ottnt{e'_{{\mathrm{0}}}} \,  \,\&\,  \, \varpi$}
    \UnaryInfC{$ \resetcnstr{ \ottnt{e_{{\mathrm{0}}}} }  \,  \mathrel{  \longrightarrow  }  \,  \resetcnstr{ \ottnt{e'_{{\mathrm{0}}}} }  \,  \,\&\,  \, \varpi$}
   \end{prooftree}
   for some $\ottnt{e'_{{\mathrm{0}}}}$ such that $\ottnt{e'} \,  =  \,  \resetcnstr{ \ottnt{e'_{{\mathrm{0}}}} } $.
   By the IH,
   \[
     \emptyset   \vdash  \ottnt{e'_{{\mathrm{0}}}}  \ottsym{:}   \ottnt{T}  \,  \,\&\,  \,   { \varpi }^{-1}  \,  \tceappendop  \, \Phi  \, / \,   ( \forall  \mathit{x}  .   \ottnt{T}  \,  \,\&\,  \,   \Phi_\mathsf{val}   \, / \,   \square    )  \Rightarrow    { \varpi }^{-1}  \,  \tceappendop  \, \ottnt{C}   ~.
   \]
   Then, \T{Reset} implies the conclusion
   \[
     \emptyset   \vdash   \resetcnstr{ \ottnt{e'_{{\mathrm{0}}}} }   \ottsym{:}   { \varpi }^{-1}  \,  \tceappendop  \, \ottnt{C} ~.
   \]

  \case \T{Event}:
   We are given
   \begin{prooftree}
    \AxiomC{$\vdash   \emptyset $}
    \UnaryInfC{$ \emptyset   \vdash   \mathsf{ev}  [  \mathbf{a}  ]   \ottsym{:}   \ottsym{\{}  \mathit{x} \,  {:}  \,  \mathsf{unit}  \,  |  \,  \top   \ottsym{\}}  \,  \,\&\,  \,   (  \mathbf{a}  , \bot )   \, / \,   \square  $}
   \end{prooftree}
   for some $\mathbf{a}$ and $\mathit{x}$.
   Because $ \mathsf{ev}  [  \mathbf{a}  ] $ can be evaluated only by \E{Ev}, we have
   $\ottnt{e'} \,  =  \,  () $ and $ \varpi    =    \mathbf{a} $.
   We must show that
   \[
     \emptyset   \vdash   ()   \ottsym{:}   \ottsym{\{}  \mathit{x} \,  {:}  \,  \mathsf{unit}  \,  |  \,  \top   \ottsym{\}}  \,  \,\&\,  \,   { \mathbf{a} }^{-1}  \,  \tceappendop  \,  (  \mathbf{a}  , \bot )   \, / \,   \square   ~,
   \]
   which is derived by \T{Sub} with the following.
   \begin{itemize}
    \item $ \emptyset   \vdash   ()   \ottsym{:}  \ottsym{\{}  \mathit{x} \,  {:}  \,  \mathsf{unit}  \,  |  \,   ()     =    \mathit{x}   \ottsym{\}}$ by \T{Const}.
    \item $ \emptyset   \vdash   \ottsym{\{}  \mathit{x} \,  {:}  \,  \mathsf{unit}  \,  |  \,   ()     =    \mathit{x}   \ottsym{\}}  \,  \,\&\,  \,   \Phi_\mathsf{val}   \, / \,   \square    \ottsym{<:}   \ottsym{\{}  \mathit{x} \,  {:}  \,  \mathsf{unit}  \,  |  \,  \top   \ottsym{\}}  \,  \,\&\,  \,   { \mathbf{a} }^{-1}  \,  \tceappendop  \,  (  \mathbf{a}  , \bot )   \, / \,   \square  $ by \Srule{Comp} with
          \begin{itemize}
           \item $ \emptyset   \vdash  \ottsym{\{}  \mathit{x} \,  {:}  \,  \mathsf{unit}  \,  |  \,   ()     =    \mathit{x}   \ottsym{\}}  \ottsym{<:}  \ottsym{\{}  \mathit{x} \,  {:}  \,  \mathsf{unit}  \,  |  \,  \top   \ottsym{\}}$
                 by \Srule{Refine} and \reflem{ts-valid-intro-impl-top},
           \item $ \emptyset   \vdash   \Phi_\mathsf{val}   \ottsym{<:}   { \mathbf{a} }^{-1}  \,  \tceappendop  \,  (  \mathbf{a}  , \bot ) $
                 by \reflem{ts-preeff-comp-to-preeff-quot} with $ \emptyset   \vdash   (  \mathbf{a}  , \bot )  \,  \tceappendop  \,  \Phi_\mathsf{val}   \ottsym{<:}   (  \mathbf{a}  , \bot ) $,
                 which is derived by \reflem{ts-weak-strength-TP_val}, and
           \item $ \emptyset  \,  |  \, \ottsym{\{}  \mathit{x} \,  {:}  \,  \mathsf{unit}  \,  |  \,   ()     =    \mathit{x}   \ottsym{\}} \,  \,\&\,  \,  \Phi_\mathsf{val}   \vdash   \square   \ottsym{<:}   \square $
                 by \Srule{Empty}.
          \end{itemize}
    \item $ \emptyset   \vdash   \ottsym{\{}  \mathit{x} \,  {:}  \,  \mathsf{unit}  \,  |  \,  \top   \ottsym{\}}  \,  \,\&\,  \,   { \mathbf{a} }^{-1}  \,  \tceappendop  \,  (  \mathbf{a}  , \bot )   \, / \,   \square  $
          by \WFT{Comp} with
          \begin{itemize}
           \item $ \emptyset   \vdash  \ottsym{\{}  \mathit{x} \,  {:}  \,  \mathsf{unit}  \,  |  \,  \top   \ottsym{\}}$ by \WFT{Refine} and \Tf{Top},
           \item $ \emptyset   \vdash   { \mathbf{a} }^{-1}  \,  \tceappendop  \,  (  \mathbf{a}  , \bot ) $ by \WFT{Eff}, and
           \item $ \emptyset  \,  |  \, \ottsym{\{}  \mathit{x} \,  {:}  \,  \mathsf{unit}  \,  |  \,  \top   \ottsym{\}}  \vdash   \square $ by \WFT{Empty}.
          \end{itemize}
   \end{itemize}
 \end{caseanalysis}
\end{proof}

\begin{lemmap}{Composition of Pre-Effect Quotient in Subtyping}{ts-comp-preeff-quot-subtyping}
 If $\Gamma  \vdash  \ottnt{C}$, then $\Gamma  \vdash   { \varpi_{{\mathrm{2}}} }^{-1}  \,  \tceappendop  \, \ottsym{(}   { \varpi_{{\mathrm{1}}} }^{-1}  \,  \tceappendop  \, \ottnt{C}  \ottsym{)}  \ottsym{<:}   { \ottsym{(}  \varpi_{{\mathrm{1}}} \,  \tceappendop  \, \varpi_{{\mathrm{2}}}  \ottsym{)} }^{-1}  \,  \tceappendop  \, \ottnt{C}$.
\end{lemmap}
\begin{proof}
 By induction on $\ottnt{C}$.
 Because
 \begin{itemize}
  \item $ { \varpi_{{\mathrm{2}}} }^{-1}  \,  \tceappendop  \, \ottsym{(}   { \varpi_{{\mathrm{1}}} }^{-1}  \,  \tceappendop  \, \ottnt{C}  \ottsym{)} \,  =  \,   \ottnt{C}  . \ottnt{T}   \,  \,\&\,  \,   { \varpi_{{\mathrm{2}}} }^{-1}  \,  \tceappendop  \, \ottsym{(}   { \varpi_{{\mathrm{1}}} }^{-1}  \,  \tceappendop  \, \ottsym{(}   \ottnt{C}  . \Phi   \ottsym{)}  \ottsym{)}  \, / \,   { \varpi_{{\mathrm{2}}} }^{-1}  \,  \tceappendop  \, \ottsym{(}   { \varpi_{{\mathrm{1}}} }^{-1}  \,  \tceappendop  \, \ottsym{(}   \ottnt{C}  . \ottnt{S}   \ottsym{)}  \ottsym{)} $ and
  \item $ { \ottsym{(}  \varpi_{{\mathrm{1}}} \,  \tceappendop  \, \varpi_{{\mathrm{2}}}  \ottsym{)} }^{-1}  \,  \tceappendop  \, \ottnt{C} \,  =  \,   \ottnt{C}  . \ottnt{T}   \,  \,\&\,  \,   { \ottsym{(}  \varpi_{{\mathrm{1}}} \,  \tceappendop  \, \varpi_{{\mathrm{2}}}  \ottsym{)} }^{-1}  \,  \tceappendop  \, \ottsym{(}   \ottnt{C}  . \Phi   \ottsym{)}  \, / \,   { \ottsym{(}  \varpi_{{\mathrm{1}}} \,  \tceappendop  \, \varpi_{{\mathrm{2}}}  \ottsym{)} }^{-1}  \,  \tceappendop  \, \ottsym{(}   \ottnt{C}  . \ottnt{S}   \ottsym{)} $,
 \end{itemize}
 we have the conclusion by \Srule{Comp} with the following.
 \begin{itemize}
  \item $\Gamma  \vdash   \ottnt{C}  . \ottnt{T}   \ottsym{<:}   \ottnt{C}  . \ottnt{T} $ by \reflem{ts-refl-subtyping} with $\Gamma  \vdash   \ottnt{C}  . \ottnt{T} $,
        which is implied by $\Gamma  \vdash  \ottnt{C}$.

  \item We show that
        \[
         \Gamma  \vdash   { \varpi_{{\mathrm{2}}} }^{-1}  \,  \tceappendop  \, \ottsym{(}   { \varpi_{{\mathrm{1}}} }^{-1}  \,  \tceappendop  \, \ottsym{(}   \ottnt{C}  . \Phi   \ottsym{)}  \ottsym{)}  \ottsym{<:}   { \ottsym{(}  \varpi_{{\mathrm{1}}} \,  \tceappendop  \, \varpi_{{\mathrm{2}}}  \ottsym{)} }^{-1}  \,  \tceappendop  \, \ottsym{(}   \ottnt{C}  . \Phi   \ottsym{)}
        \]
        by \Srule{Eff} with the following.
        \begin{itemize}
         \item We show that
               \[
                \Gamma  \ottsym{,}  \mathit{x} \,  {:}  \,  \Sigma ^\ast   \models    { \ottsym{(}   { \varpi_{{\mathrm{2}}} }^{-1}  \,  \tceappendop  \, \ottsym{(}   { \varpi_{{\mathrm{1}}} }^{-1}  \,  \tceappendop  \, \ottsym{(}   \ottnt{C}  . \Phi   \ottsym{)}  \ottsym{)}  \ottsym{)} }^\mu   \ottsym{(}  \mathit{x}  \ottsym{)}    \Rightarrow     { \ottsym{(}   { \ottsym{(}  \varpi_{{\mathrm{1}}} \,  \tceappendop  \, \varpi_{{\mathrm{2}}}  \ottsym{)} }^{-1}  \,  \tceappendop  \, \ottsym{(}   \ottnt{C}  . \Phi   \ottsym{)}  \ottsym{)} }^\mu   \ottsym{(}  \mathit{x}  \ottsym{)} 
               \]
               for $\mathit{x} \,  \not\in  \,  \mathit{fv}  (   \ottnt{C}  . \Phi   ) $.
               Because
               \[
                 { \ottsym{(}   { \varpi_{{\mathrm{2}}} }^{-1}  \,  \tceappendop  \, \ottsym{(}   { \varpi_{{\mathrm{1}}} }^{-1}  \,  \tceappendop  \, \ottsym{(}   \ottnt{C}  . \Phi   \ottsym{)}  \ottsym{)}  \ottsym{)} }^\mu   \ottsym{(}  \mathit{x}  \ottsym{)} =
                 { \ottsym{(}   { \varpi_{{\mathrm{1}}} }^{-1}  \,  \tceappendop  \, \ottsym{(}   \ottnt{C}  . \Phi   \ottsym{)}  \ottsym{)} }^\mu   \ottsym{(}  \varpi_{{\mathrm{2}}} \,  \tceappendop  \, \mathit{x}  \ottsym{)} =
                 { \ottsym{(}   \ottnt{C}  . \Phi   \ottsym{)} }^\mu   \ottsym{(}  \varpi_{{\mathrm{1}}} \,  \tceappendop  \, \ottsym{(}  \varpi_{{\mathrm{2}}} \,  \tceappendop  \, \mathit{x}  \ottsym{)}  \ottsym{)}
               \]
               and
               \[
                 { \ottsym{(}   { \ottsym{(}  \varpi_{{\mathrm{1}}} \,  \tceappendop  \, \varpi_{{\mathrm{2}}}  \ottsym{)} }^{-1}  \,  \tceappendop  \, \ottsym{(}   \ottnt{C}  . \Phi   \ottsym{)}  \ottsym{)} }^\mu   \ottsym{(}  \mathit{x}  \ottsym{)} \,  =  \,  { \ottsym{(}   \ottnt{C}  . \Phi   \ottsym{)} }^\mu   \ottsym{(}  \ottsym{(}  \varpi_{{\mathrm{1}}} \,  \tceappendop  \, \varpi_{{\mathrm{2}}}  \ottsym{)} \,  \tceappendop  \, \mathit{x}  \ottsym{)} ~,
               \]
               it suffices to show that
               \[
                \Gamma  \ottsym{,}  \mathit{x} \,  {:}  \,  \Sigma ^\ast   \models    { \ottsym{(}   \ottnt{C}  . \Phi   \ottsym{)} }^\mu   \ottsym{(}  \varpi_{{\mathrm{1}}} \,  \tceappendop  \, \ottsym{(}  \varpi_{{\mathrm{2}}} \,  \tceappendop  \, \mathit{x}  \ottsym{)}  \ottsym{)}    \Rightarrow     { \ottsym{(}   \ottnt{C}  . \Phi   \ottsym{)} }^\mu   \ottsym{(}  \ottsym{(}  \varpi_{{\mathrm{1}}} \,  \tceappendop  \, \varpi_{{\mathrm{2}}}  \ottsym{)} \,  \tceappendop  \, \mathit{x}  \ottsym{)}  ~,
               \]
               which is true because, for any $\varpi$, $ \varpi_{{\mathrm{1}}} \,  \tceappendop  \, \ottsym{(}  \varpi_{{\mathrm{2}}} \,  \tceappendop  \, \varpi  \ottsym{)}    =    \ottsym{(}  \varpi_{{\mathrm{1}}} \,  \tceappendop  \, \varpi_{{\mathrm{2}}}  \ottsym{)} \,  \tceappendop  \, \varpi $. \TS{OK informal}

         \item We show that
               \[
                \Gamma  \ottsym{,}  \mathit{x} \,  {:}  \,  \Sigma ^\omega   \models    { \ottsym{(}   { \varpi_{{\mathrm{2}}} }^{-1}  \,  \tceappendop  \, \ottsym{(}   { \varpi_{{\mathrm{1}}} }^{-1}  \,  \tceappendop  \, \ottsym{(}   \ottnt{C}  . \Phi   \ottsym{)}  \ottsym{)}  \ottsym{)} }^\nu   \ottsym{(}  \mathit{x}  \ottsym{)}    \Rightarrow     { \ottsym{(}   { \ottsym{(}  \varpi_{{\mathrm{1}}} \,  \tceappendop  \, \varpi_{{\mathrm{2}}}  \ottsym{)} }^{-1}  \,  \tceappendop  \, \ottsym{(}   \ottnt{C}  . \Phi   \ottsym{)}  \ottsym{)} }^\nu   \ottsym{(}  \mathit{x}  \ottsym{)} 
               \]
               for $\mathit{x} \,  \not\in  \,  \mathit{fv}  (   \ottnt{C}  . \Phi   ) $.
               Because
               \[
                 { \ottsym{(}   { \varpi_{{\mathrm{2}}} }^{-1}  \,  \tceappendop  \, \ottsym{(}   { \varpi_{{\mathrm{1}}} }^{-1}  \,  \tceappendop  \, \ottsym{(}   \ottnt{C}  . \Phi   \ottsym{)}  \ottsym{)}  \ottsym{)} }^\nu   \ottsym{(}  \mathit{x}  \ottsym{)} =  { \ottsym{(}   { \varpi_{{\mathrm{1}}} }^{-1}  \,  \tceappendop  \, \ottsym{(}   \ottnt{C}  . \Phi   \ottsym{)}  \ottsym{)} }^\nu   \ottsym{(}  \varpi_{{\mathrm{2}}} \,  \tceappendop  \, \mathit{x}  \ottsym{)} =  { \ottsym{(}   \ottnt{C}  . \Phi   \ottsym{)} }^\nu   \ottsym{(}  \varpi_{{\mathrm{1}}} \,  \tceappendop  \, \ottsym{(}  \varpi_{{\mathrm{2}}} \,  \tceappendop  \, \mathit{x}  \ottsym{)}  \ottsym{)}
               \]
               and
               \[
                 { \ottsym{(}   { \ottsym{(}  \varpi_{{\mathrm{1}}} \,  \tceappendop  \, \varpi_{{\mathrm{2}}}  \ottsym{)} }^{-1}  \,  \tceappendop  \, \ottsym{(}   \ottnt{C}  . \Phi   \ottsym{)}  \ottsym{)} }^\nu   \ottsym{(}  \mathit{x}  \ottsym{)} \,  =  \,  { \ottsym{(}   \ottnt{C}  . \Phi   \ottsym{)} }^\nu   \ottsym{(}  \ottsym{(}  \varpi_{{\mathrm{1}}} \,  \tceappendop  \, \varpi_{{\mathrm{2}}}  \ottsym{)} \,  \tceappendop  \, \mathit{x}  \ottsym{)} ~,
               \]
               it suffices to show that
               \[
                \Gamma  \ottsym{,}  \mathit{x} \,  {:}  \,  \Sigma ^\omega   \models    { \ottsym{(}   \ottnt{C}  . \Phi   \ottsym{)} }^\nu   \ottsym{(}  \varpi_{{\mathrm{1}}} \,  \tceappendop  \, \ottsym{(}  \varpi_{{\mathrm{2}}} \,  \tceappendop  \, \mathit{x}  \ottsym{)}  \ottsym{)}    \Rightarrow     { \ottsym{(}   \ottnt{C}  . \Phi   \ottsym{)} }^\nu   \ottsym{(}  \ottsym{(}  \varpi_{{\mathrm{1}}} \,  \tceappendop  \, \varpi_{{\mathrm{2}}}  \ottsym{)} \,  \tceappendop  \, \mathit{x}  \ottsym{)}  ~,
               \]
               which is true because, for any $\pi$, $ \varpi_{{\mathrm{1}}} \,  \tceappendop  \, \ottsym{(}  \varpi_{{\mathrm{2}}} \,  \tceappendop  \, \pi  \ottsym{)}    =    \ottsym{(}  \varpi_{{\mathrm{1}}} \,  \tceappendop  \, \varpi_{{\mathrm{2}}}  \ottsym{)} \,  \tceappendop  \, \pi $. \TS{OK informal}
        \end{itemize}

  \item We show that
        \[
         \Gamma \,  |  \,  \ottnt{C}  . \ottnt{T}  \,  \,\&\,  \,  { \varpi_{{\mathrm{2}}} }^{-1}  \,  \tceappendop  \, \ottsym{(}   { \varpi_{{\mathrm{1}}} }^{-1}  \,  \tceappendop  \, \ottsym{(}   \ottnt{C}  . \Phi   \ottsym{)}  \ottsym{)}  \vdash   { \varpi_{{\mathrm{2}}} }^{-1}  \,  \tceappendop  \, \ottsym{(}   { \varpi_{{\mathrm{1}}} }^{-1}  \,  \tceappendop  \, \ottsym{(}   \ottnt{C}  . \ottnt{S}   \ottsym{)}  \ottsym{)}  \ottsym{<:}   { \ottsym{(}  \varpi_{{\mathrm{1}}} \,  \tceappendop  \, \varpi_{{\mathrm{2}}}  \ottsym{)} }^{-1}  \,  \tceappendop  \, \ottsym{(}   \ottnt{C}  . \ottnt{S}   \ottsym{)}
        \]
        by case analysis on $ \ottnt{C}  . \ottnt{S} $.
        \begin{caseanalysis}
         \case $ \ottnt{C}  . \ottnt{S}  \,  =  \,  \square $:
          Because $ { \varpi_{{\mathrm{2}}} }^{-1}  \,  \tceappendop  \, \ottsym{(}   { \varpi_{{\mathrm{1}}} }^{-1}  \,  \tceappendop  \, \ottsym{(}   \ottnt{C}  . \ottnt{S}   \ottsym{)}  \ottsym{)} =  { \ottsym{(}  \varpi_{{\mathrm{1}}} \,  \tceappendop  \, \varpi_{{\mathrm{2}}}  \ottsym{)} }^{-1}  \,  \tceappendop  \, \ottsym{(}   \ottnt{C}  . \ottnt{S}   \ottsym{)} =  \square $,
          we have the conclusion by \Srule{Empty}.

         \case $  \exists  \, { \mathit{x}  \ottsym{,}  \ottnt{C_{{\mathrm{1}}}}  \ottsym{,}  \ottnt{C_{{\mathrm{2}}}} } . \   \ottnt{C}  . \ottnt{S}  \,  =  \,  ( \forall  \mathit{x}  .  \ottnt{C_{{\mathrm{1}}}}  )  \Rightarrow   \ottnt{C_{{\mathrm{2}}}}  $:
          Because
          $ { \varpi_{{\mathrm{2}}} }^{-1}  \,  \tceappendop  \, \ottsym{(}   { \varpi_{{\mathrm{1}}} }^{-1}  \,  \tceappendop  \, \ottsym{(}   \ottnt{C}  . \ottnt{S}   \ottsym{)}  \ottsym{)} \,  =  \,  ( \forall  \mathit{x}  .  \ottnt{C_{{\mathrm{1}}}}  )  \Rightarrow    { \varpi_{{\mathrm{2}}} }^{-1}  \,  \tceappendop  \, \ottsym{(}   { \varpi_{{\mathrm{1}}} }^{-1}  \,  \tceappendop  \, \ottnt{C_{{\mathrm{2}}}}  \ottsym{)} $ and
          $ { \ottsym{(}  \varpi_{{\mathrm{1}}} \,  \tceappendop  \, \varpi_{{\mathrm{2}}}  \ottsym{)} }^{-1}  \,  \tceappendop  \, \ottsym{(}   \ottnt{C}  . \ottnt{S}   \ottsym{)} \,  =  \,  ( \forall  \mathit{x}  .  \ottnt{C_{{\mathrm{1}}}}  )  \Rightarrow    { \ottsym{(}  \varpi_{{\mathrm{1}}} \,  \tceappendop  \, \varpi_{{\mathrm{2}}}  \ottsym{)} }^{-1}  \,  \tceappendop  \, \ottnt{C_{{\mathrm{2}}}} $,
          \Srule{Ans} implies that it suffices to show the following.
          \begin{itemize}
           \item We show that $\Gamma  \ottsym{,}  \mathit{x} \,  {:}  \,  \ottnt{C}  . \ottnt{T}   \vdash  \ottnt{C_{{\mathrm{1}}}}  \ottsym{<:}  \ottnt{C_{{\mathrm{1}}}}$,
                 which is implied by \reflem{ts-refl-subtyping} with $\Gamma  \vdash  \ottnt{C}$.
           \item We show that $\Gamma  \vdash   { \varpi_{{\mathrm{2}}} }^{-1}  \,  \tceappendop  \, \ottsym{(}   { \varpi_{{\mathrm{1}}} }^{-1}  \,  \tceappendop  \, \ottnt{C_{{\mathrm{2}}}}  \ottsym{)}  \ottsym{<:}   { \ottsym{(}  \varpi_{{\mathrm{1}}} \,  \tceappendop  \, \varpi_{{\mathrm{2}}}  \ottsym{)} }^{-1}  \,  \tceappendop  \, \ottnt{C_{{\mathrm{2}}}}$,
                 which is implied by the IH with $\Gamma  \vdash  \ottnt{C_{{\mathrm{2}}}}$.
          \end{itemize}
        \end{caseanalysis}
 \end{itemize}
\end{proof}

\begin{lemmap}{Subject Reduction: Multi-Step Evaluation}{ts-subject-red-meval}
 If $ \emptyset   \vdash  \ottnt{e}  \ottsym{:}  \ottnt{C}$ and $\ottnt{e} \,  \mathrel{  \longrightarrow  }^{ \ottmv{n} }  \, \ottnt{e'} \,  \,\&\,  \, \varpi$,
 then $ \emptyset   \vdash  \ottnt{e'}  \ottsym{:}   { \varpi }^{-1}  \,  \tceappendop  \, \ottnt{C}$.
\end{lemmap}
\begin{proof}
 By induction on $\ottmv{n}$.
 \begin{caseanalysis}
  \case $\ottmv{n} \,  =  \, \ottsym{0}$:
   We are given $\ottnt{e} \,  =  \, \ottnt{e'}$ and $ \varpi    =     \epsilon  $.
   It suffices to show that $ \emptyset   \vdash  \ottnt{e}  \ottsym{:}   {  \epsilon  }^{-1}  \,  \tceappendop  \, \ottnt{C}$,
   which is derived by \T{Sub} with the following.
   \begin{itemize}
    \item $ \emptyset   \vdash  \ottnt{e}  \ottsym{:}  \ottnt{C}$.
    \item $ \emptyset   \vdash  \ottnt{C}  \ottsym{<:}   {  \epsilon  }^{-1}  \,  \tceappendop  \, \ottnt{C}$ by \reflem{ts-preeff-comp-to-preeff-quot}
          with $ \emptyset   \vdash   \Phi_\mathsf{val}  \,  \tceappendop  \, \ottnt{C}  \ottsym{<:}  \ottnt{C}$,
          which is shown by Lemmas~\ref{lem:ts-wf-ty-tctx-typing}, \ref{lem:ts-refl-subtyping}, and \reflem{ts-add-TP_val-subty}
          with $ \emptyset   \vdash  \ottnt{e}  \ottsym{:}  \ottnt{C}$.
    \item $ \emptyset   \vdash   {  \epsilon  }^{-1}  \,  \tceappendop  \, \ottnt{C}$ by \reflem{ts-wf-preeff-quot}
          with $ \emptyset   \vdash  \ottnt{C}$, which is implied by \reflem{ts-wf-ty-tctx-typing} with $ \emptyset   \vdash  \ottnt{e}  \ottsym{:}  \ottnt{C}$.
   \end{itemize}

  \case $\ottmv{n} \,  >  \, \ottsym{0}$:
   We are given
   $\ottnt{e} \,  \mathrel{  \longrightarrow  }  \, \ottnt{e''} \,  \,\&\,  \, \varpi_{{\mathrm{1}}}$ and $\ottnt{e''} \,  \mathrel{  \longrightarrow  }^{ \ottmv{n}  \ottsym{-}  \ottsym{1} }  \, \ottnt{e'} \,  \,\&\,  \, \varpi_{{\mathrm{2}}}$
   for some $\ottnt{e''}$, $\varpi_{{\mathrm{1}}}$, and $\varpi_{{\mathrm{2}}}$ such that
   $ \varpi    =    \varpi_{{\mathrm{1}}} \,  \tceappendop  \, \varpi_{{\mathrm{2}}} $.
   By \reflem{ts-subject-red-eval}, $ \emptyset   \vdash  \ottnt{e''}  \ottsym{:}   { \varpi_{{\mathrm{1}}} }^{-1}  \,  \tceappendop  \, \ottnt{C}$.
   By the IH, $ \emptyset   \vdash  \ottnt{e'}  \ottsym{:}   { \varpi_{{\mathrm{2}}} }^{-1}  \,  \tceappendop  \, \ottsym{(}   { \varpi_{{\mathrm{1}}} }^{-1}  \,  \tceappendop  \, \ottnt{C}  \ottsym{)}$.
   By Lemmas~\ref{lem:ts-wf-ty-tctx-typing}, \ref{lem:ts-comp-preeff-quot-subtyping}, and \ref{lem:ts-wf-preeff-quot}, and \T{Sub},
   we have the conclusion $ \emptyset   \vdash  \ottnt{e'}  \ottsym{:}   { \ottsym{(}  \varpi_{{\mathrm{1}}} \,  \tceappendop  \, \varpi_{{\mathrm{2}}}  \ottsym{)} }^{-1}  \,  \tceappendop  \, \ottnt{C}$.
 \end{caseanalysis}
\end{proof}

\begin{lemmap}{Progress}{ts-progress}
 If $ \emptyset   \vdash  \ottnt{e}  \ottsym{:}  \ottnt{C}$,
 then one of the following holds:
 \begin{itemize}
  \item $  \exists  \, { \ottnt{v} } . \  \ottnt{e} \,  =  \, \ottnt{v} $;
  \item $  \exists  \, { \ottnt{K}  \ottsym{,}  \mathit{x}  \ottsym{,}  \ottnt{e'} } . \  \ottnt{e} \,  =  \,  \ottnt{K}  [  \mathcal{S}_0 \, \mathit{x}  \ottsym{.}  \ottnt{e'}  ]  $;
  \item $  \exists  \, { \ottnt{E}  \ottsym{,}  \pi } . \  \ottnt{e} \,  =  \,  \ottnt{E}  [   \bullet ^{ \pi }   ]  $; or
  \item $  \exists  \, { \ottnt{e'}  \ottsym{,}  \varpi } . \  \ottnt{e} \,  \mathrel{  \longrightarrow  }  \, \ottnt{e'} \,  \,\&\,  \, \varpi $.
 \end{itemize}
\end{lemmap}
\begin{proof}
 By induction on the typing derivation of $ \emptyset   \vdash  \ottnt{e}  \ottsym{:}  \ottnt{C}$.
 \begin{caseanalysis}
  \case \T{CVar} and \T{Var}: Contradictory.
  \case \T{Const}, \T{PAbs}, \T{Abs}, \T{Fun}, \T{Shift}, \T{ShiftD}, and \T{Div}: Obvious.
  \case \T{Sub}: By the IH.
  \case \T{Op}:
   We are given
   \begin{prooftree}
    \AxiomC{$\vdash   \emptyset $}
    \AxiomC{$ \tceseq{  \emptyset   \vdash  \ottnt{v_{{\mathrm{0}}}}  \ottsym{:}  \ottnt{T} } $}
    \AxiomC{$\mathit{ty} \, \ottsym{(}  \ottnt{o}  \ottsym{)} \,  =  \,  \tceseq{(  \mathit{x}  \,  {:}  \,  \ottnt{T}  )}  \rightarrow   \ottnt{T_{{\mathrm{0}}}} $}
    \TrinaryInfC{$ \emptyset   \vdash  \ottnt{o}  \ottsym{(}   \tceseq{ \ottnt{v_{{\mathrm{0}}}} }   \ottsym{)}  \ottsym{:}   \ottnt{T_{{\mathrm{0}}}}    [ \tceseq{ \ottnt{v_{{\mathrm{0}}}}  /  \mathit{x} } ]  $}
   \end{prooftree}
   for some $\ottnt{o}$, $ \tceseq{ \ottnt{v_{{\mathrm{0}}}} } $, $ \tceseq{ \mathit{x} } $, $ \tceseq{ \ottnt{T} } $, and $\ottnt{T_{{\mathrm{0}}}}$
   such that
   $\ottnt{e} \,  =  \, \ottnt{o}  \ottsym{(}   \tceseq{ \ottnt{v_{{\mathrm{0}}}} }   \ottsym{)}$ and
   $\ottnt{C} \,  =  \,   \ottnt{T_{{\mathrm{0}}}}    [ \tceseq{ \ottnt{v_{{\mathrm{0}}}}  /  \mathit{x} } ]    \,  \,\&\,  \,   \Phi_\mathsf{val}   \, / \,   \square  $.
  By \refasm{const},
  $\mathit{ty} \, \ottsym{(}  \ottnt{o}  \ottsym{)} \,  =  \,  \tceseq{(  \mathit{x}  \,  {:}  \,  \ottsym{\{}  \mathit{x} \,  {:}  \, \iota \,  |  \, \phi  \ottsym{\}}  )}  \rightarrow   \ottsym{\{}  \mathit{x_{{\mathrm{0}}}} \,  {:}  \, \iota_{{\mathrm{0}}} \,  |  \, \phi_{{\mathrm{0}}}  \ottsym{\}} $
  for some $ \tceseq{ \ottsym{\{}  \mathit{x} \,  {:}  \, \iota \,  |  \, \phi  \ottsym{\}} } $, $\mathit{x_{{\mathrm{0}}}}$, $\iota_{{\mathrm{0}}}$, and $\phi_{{\mathrm{0}}}$.
  Because $ \tceseq{  \emptyset   \vdash  \ottnt{v_{{\mathrm{0}}}}  \ottsym{:}  \ottsym{\{}  \mathit{x} \,  {:}  \, \iota \,  |  \, \phi  \ottsym{\}} } $,
  \reflem{ts-val-inv-refine-ty} implies that
  there exist some $ \tceseq{ \ottnt{c_{{\mathrm{0}}}} } $ such that
  \begin{itemize}
   \item $ \tceseq{ \ottnt{v_{{\mathrm{0}}}} \,  =  \, \ottnt{c_{{\mathrm{0}}}} } $,
   \item $ \tceseq{  \mathit{ty}  (  \ottnt{c_{{\mathrm{0}}}}  )  \,  =  \, \iota } $, and
   \item $ \tceseq{  { \models  \,   \phi    [  \ottnt{c_{{\mathrm{0}}}}  /  \mathit{x}  ]   }  } $.
  \end{itemize}
  Then, \refasm{const} implies that $ \zeta  (  \ottnt{o}  ,   \tceseq{ \ottnt{c_{{\mathrm{0}}}} }   ) $ is well defiend.
  Therefore, we have $\ottnt{o}  \ottsym{(}   \tceseq{ \ottnt{v_{{\mathrm{0}}}} }   \ottsym{)} \,  \mathrel{  \longrightarrow  }  \,  \zeta  (  \ottnt{o}  ,   \tceseq{ \ottnt{c_{{\mathrm{0}}}} }   )  \,  \,\&\,  \,  \epsilon $ by \E{Red}/\R{Const}.

  \case \T{App}:
   We are given
   \begin{prooftree}
    \AxiomC{$ \emptyset   \vdash  \ottnt{v_{{\mathrm{1}}}}  \ottsym{:}  \ottsym{(}  \mathit{x} \,  {:}  \, \ottnt{T'}  \ottsym{)}  \rightarrow  \ottnt{C'}$}
    \AxiomC{$ \emptyset   \vdash  \ottnt{v_{{\mathrm{2}}}}  \ottsym{:}  \ottnt{T'}$}
    \BinaryInfC{$ \emptyset   \vdash  \ottnt{v_{{\mathrm{1}}}} \, \ottnt{v_{{\mathrm{2}}}}  \ottsym{:}   \ottnt{C'}    [  \ottnt{v_{{\mathrm{2}}}}  /  \mathit{x}  ]  $}
   \end{prooftree}
   for some $\ottnt{v_{{\mathrm{1}}}}$, $\ottnt{v_{{\mathrm{2}}}}$, $\mathit{x}$, $\ottnt{T'}$, and $\ottnt{C'}$
   such that $\ottnt{e} \,  =  \, \ottnt{v_{{\mathrm{1}}}} \, \ottnt{v_{{\mathrm{2}}}}$ and $\ottnt{C} \,  =  \,  \ottnt{C'}    [  \ottnt{v_{{\mathrm{2}}}}  /  \mathit{x}  ]  $.
   By \reflem{ts-val-inv-fun-ty} with $ \emptyset   \vdash  \ottnt{v_{{\mathrm{1}}}}  \ottsym{:}  \ottsym{(}  \mathit{x} \,  {:}  \, \ottnt{T'}  \ottsym{)}  \rightarrow  \ottnt{C'}$,
   there exist some $\ottnt{v_{{\mathrm{11}}}}$, $ \tceseq{ \ottnt{v_{{\mathrm{12}}}} } $, and $ \tceseq{ \mathit{z} } $ such that
   \begin{itemize}
    \item $\ottnt{v_{{\mathrm{1}}}} \,  =  \, \ottnt{v_{{\mathrm{11}}}} \,  \tceseq{ \ottnt{v_{{\mathrm{12}}}} } $,
    \item $  \exists  \, { \ottnt{e_{{\mathrm{0}}}} } . \  \ottnt{v_{{\mathrm{11}}}} \,  =  \,  \lambda\!  \,  \tceseq{ \mathit{z} }   \ottsym{.}  \ottnt{e_{{\mathrm{0}}}} $ or
          $  \exists  \, { \mathit{X}  \ottsym{,}  \mathit{Y}  \ottsym{,}   \ottnt{A} _{  \mu  }   \ottsym{,}   \ottnt{A} _{  \nu  }   \ottsym{,}  \mathit{f}  \ottsym{,}  \ottnt{e_{{\mathrm{0}}}} } . \  \ottnt{v_{{\mathrm{11}}}} \,  =  \,  \mathsf{rec}  \, (  \tceseq{  \mathit{X} ^{  \ottnt{A} _{  \mu  }  },  \mathit{Y} ^{  \ottnt{A} _{  \nu  }  }  }  ,  \mathit{f} ,   \tceseq{ \mathit{z} }  ) . \,  \ottnt{e_{{\mathrm{0}}}}  $, and
    \item $\ottsym{\mbox{$\mid$}}   \tceseq{ \ottnt{v_{{\mathrm{12}}}} }   \ottsym{\mbox{$\mid$}} \,  <  \, \ottsym{\mbox{$\mid$}}   \tceseq{ \mathit{z} }   \ottsym{\mbox{$\mid$}}$.
   \end{itemize}
   If $\ottsym{\mbox{$\mid$}}   \tceseq{ \ottnt{v_{{\mathrm{12}}}} }   \ottsym{,}  \ottnt{v_{{\mathrm{2}}}}  \ottsym{\mbox{$\mid$}} \,  =  \, \ottsym{\mbox{$\mid$}}   \tceseq{ \mathit{z} }   \ottsym{\mbox{$\mid$}}$,
   then we have the conclusion by \E{Red}/\R{Beta} or \E{Red}/\R{RBeta}.
   Otherwise, if $\ottsym{\mbox{$\mid$}}   \tceseq{ \ottnt{v_{{\mathrm{12}}}} }   \ottsym{,}  \ottnt{v_{{\mathrm{2}}}}  \ottsym{\mbox{$\mid$}} \,  <  \, \ottsym{\mbox{$\mid$}}   \tceseq{ \mathit{z} }   \ottsym{\mbox{$\mid$}}$,
   then $\ottnt{v_{{\mathrm{1}}}} \, \ottnt{v_{{\mathrm{2}}}} \,  =  \, \ottnt{v_{{\mathrm{11}}}} \,  \tceseq{ \ottnt{v_{{\mathrm{12}}}} }  \, \ottnt{v_{{\mathrm{2}}}}$ is a value.

  \case \T{PApp}:
   We are given
   \begin{prooftree}
    \AxiomC{$ \emptyset   \vdash  \ottnt{v}  \ottsym{:}   \forall   \ottsym{(}  \mathit{X} \,  {:}  \,  \tceseq{ \ottnt{s} }   \ottsym{)}  \ottsym{.}  \ottnt{C'}$}
    \AxiomC{$ \emptyset   \vdash  \ottnt{A}  \ottsym{:}   \tceseq{ \ottnt{s} } $}
    \BinaryInfC{$ \emptyset   \vdash  \ottnt{v} \, \ottnt{A}  \ottsym{:}   \ottnt{C'}    [  \ottnt{A}  /  \mathit{X}  ]  $}
   \end{prooftree}
   for some $\ottnt{v}$, $\ottnt{A}$, $\ottnt{C'}$, $\mathit{X}$, and $ \tceseq{ \ottnt{s} } $
   such that $\ottnt{e} \,  =  \, \ottnt{v} \, \ottnt{A}$ and $\ottnt{C} \,  =  \,  \ottnt{C'}    [  \ottnt{A}  /  \mathit{X}  ]  $.
   By \reflem{ts-val-inv-univ-ty} with $ \emptyset   \vdash  \ottnt{v}  \ottsym{:}   \forall   \ottsym{(}  \mathit{X} \,  {:}  \,  \tceseq{ \ottnt{s} }   \ottsym{)}  \ottsym{.}  \ottnt{C'}$,
   there exists some $\ottnt{e_{{\mathrm{0}}}}$ such that
   $\ottnt{v} \,  =  \,  \Lambda   \ottsym{(}  \mathit{X} \,  {:}  \,  \tceseq{ \ottnt{s} }   \ottsym{)}  \ottsym{.}  \ottnt{e_{{\mathrm{0}}}}$.
   Therefore, we have $\ottnt{v} \, \ottnt{A} \,  \mathrel{  \longrightarrow  }  \,  \ottnt{e_{{\mathrm{0}}}}    [  \ottnt{A}  /  \mathit{X}  ]   \,  \,\&\,  \,  \epsilon $ by \E{Red}/\R{PBeta}.

  \case \T{If}:
   We are given
   \begin{prooftree}
    \AxiomC{$ \emptyset   \vdash  \ottnt{v}  \ottsym{:}  \ottsym{\{}  \mathit{x} \,  {:}  \,  \mathsf{bool}  \,  |  \, \phi  \ottsym{\}}$}
    \AxiomC{$  \ottnt{v}     =     \mathsf{true}    \vdash  \ottnt{e_{{\mathrm{1}}}}  \ottsym{:}  \ottnt{C}$}
    \AxiomC{$  \ottnt{v}     =     \mathsf{false}    \vdash  \ottnt{e_{{\mathrm{2}}}}  \ottsym{:}  \ottnt{C}$}
    \TrinaryInfC{$ \emptyset   \vdash  \mathsf{if} \, \ottnt{v} \, \mathsf{then} \, \ottnt{e_{{\mathrm{1}}}} \, \mathsf{else} \, \ottnt{e_{{\mathrm{2}}}}  \ottsym{:}  \ottnt{C}$}
   \end{prooftree}
   for some $\ottnt{v}$, $\ottnt{e_{{\mathrm{1}}}}$, $\ottnt{e_{{\mathrm{2}}}}$, $\mathit{x}$, and $\phi$
   such that $\ottnt{e} \,  =  \, \mathsf{if} \, \ottnt{v} \, \mathsf{then} \, \ottnt{e_{{\mathrm{1}}}} \, \mathsf{else} \, \ottnt{e_{{\mathrm{2}}}}$.
   By \reflem{ts-val-inv-refine-ty} with $ \emptyset   \vdash  \ottnt{v}  \ottsym{:}  \ottsym{\{}  \mathit{x} \,  {:}  \,  \mathsf{bool}  \,  |  \, \phi  \ottsym{\}}$, and
   \refasm{const},
   $\ottnt{v} \,  =  \,  \mathsf{true} $ or $\ottnt{v} \,  =  \,  \mathsf{false} $.
   If $\ottnt{v} \,  =  \,  \mathsf{true} $, then $\ottnt{e} \,  \mathrel{  \longrightarrow  }  \, \ottnt{e_{{\mathrm{1}}}} \,  \,\&\,  \,  \epsilon $ by \E{Red}/\R{IfT};
   otherwise, if $\ottnt{v} \,  =  \,  \mathsf{false} $, then $\ottnt{e} \,  \mathrel{  \longrightarrow  }  \, \ottnt{e_{{\mathrm{2}}}} \,  \,\&\,  \,  \epsilon $ by \E{Red}/\R{IfF}.

  \case \T{Let}:
   We are given
   \begin{prooftree}
    \AxiomC{$ \emptyset   \vdash  \ottnt{e_{{\mathrm{1}}}}  \ottsym{:}   \ottnt{T_{{\mathrm{1}}}}  \,  \,\&\,  \,  \Phi_{{\mathrm{1}}}  \, / \,  \ottnt{S_{{\mathrm{1}}}} $}
    \AxiomC{$\mathit{x_{{\mathrm{1}}}} \,  {:}  \, \ottnt{T_{{\mathrm{1}}}}  \vdash  \ottnt{e_{{\mathrm{2}}}}  \ottsym{:}   \ottnt{T_{{\mathrm{2}}}}  \,  \,\&\,  \,  \Phi_{{\mathrm{2}}}  \, / \,  \ottnt{S_{{\mathrm{2}}}} $}
    \AxiomC{$\mathit{x_{{\mathrm{1}}}} \,  \not\in  \,  \mathit{fv}  (  \ottnt{T_{{\mathrm{2}}}}  )  \,  \mathbin{\cup}  \,  \mathit{fv}  (  \Phi_{{\mathrm{2}}}  ) $}
    \TrinaryInfC{$ \emptyset   \vdash  \mathsf{let} \, \mathit{x_{{\mathrm{1}}}}  \ottsym{=}  \ottnt{e_{{\mathrm{1}}}} \,  \mathsf{in}  \, \ottnt{e_{{\mathrm{2}}}}  \ottsym{:}   \ottnt{T_{{\mathrm{2}}}}  \,  \,\&\,  \,  \Phi_{{\mathrm{1}}} \,  \tceappendop  \, \Phi_{{\mathrm{2}}}  \, / \,  \ottnt{S_{{\mathrm{1}}}}  \gg\!\!=  \ottsym{(}   \lambda\!  \, \mathit{x_{{\mathrm{1}}}}  \ottsym{.}  \ottnt{S_{{\mathrm{2}}}}  \ottsym{)} $}
   \end{prooftree}
   for some $\mathit{x_{{\mathrm{1}}}}$, $\ottnt{e_{{\mathrm{1}}}}$, $\ottnt{e_{{\mathrm{2}}}}$, $\ottnt{T_{{\mathrm{1}}}}$, $\ottnt{T_{{\mathrm{2}}}}$,
   $\Phi_{{\mathrm{1}}}$, $\Phi_{{\mathrm{2}}}$, $\ottnt{S_{{\mathrm{1}}}}$, and $\ottnt{S_{{\mathrm{2}}}}$ such that
   $\ottnt{e} \,  =  \, \ottsym{(}  \mathsf{let} \, \mathit{x_{{\mathrm{1}}}}  \ottsym{=}  \ottnt{e_{{\mathrm{1}}}} \,  \mathsf{in}  \, \ottnt{e_{{\mathrm{2}}}}  \ottsym{)}$ and $\ottnt{C} \,  =  \,  \ottnt{T_{{\mathrm{2}}}}  \,  \,\&\,  \,  \Phi_{{\mathrm{1}}} \,  \tceappendop  \, \Phi_{{\mathrm{2}}}  \, / \,  \ottnt{S_{{\mathrm{1}}}}  \gg\!\!=  \ottsym{(}   \lambda\!  \, \mathit{x_{{\mathrm{1}}}}  \ottsym{.}  \ottnt{S_{{\mathrm{2}}}}  \ottsym{)} $.
   By case analysis on the result of the IH on $ \emptyset   \vdash  \ottnt{e_{{\mathrm{1}}}}  \ottsym{:}   \ottnt{T_{{\mathrm{1}}}}  \,  \,\&\,  \,  \Phi_{{\mathrm{1}}}  \, / \,  \ottnt{S_{{\mathrm{1}}}} $.
   \begin{caseanalysis}
    \case $  \exists  \, { \ottnt{v_{{\mathrm{1}}}} } . \  \ottnt{e_{{\mathrm{1}}}} \,  =  \, \ottnt{v_{{\mathrm{1}}}} $: By \E{Red}/\R{Let}.
    \case $  \exists  \, { \ottnt{K_{{\mathrm{1}}}}  \ottsym{,}  \mathit{x}  \ottsym{,}  \ottnt{e'} } . \  \ottnt{e_{{\mathrm{1}}}} \,  =  \,  \ottnt{K_{{\mathrm{1}}}}  [  \mathcal{S}_0 \, \mathit{x}  \ottsym{.}  \ottnt{e'}  ]  $: By letting $\ottnt{K} \,  =  \, \ottsym{(}  \mathsf{let} \, \mathit{x_{{\mathrm{1}}}}  \ottsym{=}  \ottnt{K_{{\mathrm{1}}}} \,  \mathsf{in}  \, \ottnt{e_{{\mathrm{2}}}}  \ottsym{)}$ in the second case of the conclusion.
    \case $  \exists  \, { \ottnt{E_{{\mathrm{1}}}}  \ottsym{,}  \pi } . \  \ottnt{e_{{\mathrm{1}}}} \,  =  \,  \ottnt{E_{{\mathrm{1}}}}  [   \bullet ^{ \pi }   ]  $: By letting $\ottnt{E} \,  =  \, \ottsym{(}  \mathsf{let} \, \mathit{x_{{\mathrm{1}}}}  \ottsym{=}  \ottnt{E_{{\mathrm{1}}}} \,  \mathsf{in}  \, \ottnt{e_{{\mathrm{2}}}}  \ottsym{)}$
     in the third case of the conclusion.
    \case $  \exists  \, { \ottnt{e'_{{\mathrm{1}}}}  \ottsym{,}  \varpi } . \  \ottnt{e_{{\mathrm{1}}}} \,  \mathrel{  \longrightarrow  }  \, \ottnt{e'_{{\mathrm{1}}}} \,  \,\&\,  \, \varpi $: By \E{Let}.
   \end{caseanalysis}

  \case \T{LetV}: By \E{Red}/\R{Let}.

  \case \T{Reset}:
   We are given
   \begin{prooftree}
    \AxiomC{$ \emptyset   \vdash  \ottnt{e_{{\mathrm{0}}}}  \ottsym{:}   \ottnt{T}  \,  \,\&\,  \,  \Phi  \, / \,   ( \forall  \mathit{x}  .   \ottnt{T}  \,  \,\&\,  \,   \Phi_\mathsf{val}   \, / \,   \square    )  \Rightarrow   \ottnt{C}  $}
    \AxiomC{$\mathit{x} \,  \not\in  \,  \mathit{fv}  (  \ottnt{T}  ) $}
    \BinaryInfC{$ \emptyset   \vdash   \resetcnstr{ \ottnt{e_{{\mathrm{0}}}} }   \ottsym{:}  \ottnt{C}$}
   \end{prooftree}
   for some $\ottnt{e_{{\mathrm{0}}}}$, $\ottnt{T}$, $\Phi$, and $\mathit{x}$ such that
   $\ottnt{e} \,  =  \,  \resetcnstr{ \ottnt{e_{{\mathrm{0}}}} } $.
   By case analysis on the result of the IH on $ \emptyset   \vdash  \ottnt{e_{{\mathrm{0}}}}  \ottsym{:}   \ottnt{T}  \,  \,\&\,  \,  \Phi  \, / \,   ( \forall  \mathit{x}  .   \ottnt{T}  \,  \,\&\,  \,   \Phi_\mathsf{val}   \, / \,   \square    )  \Rightarrow   \ottnt{C}  $.
   \begin{caseanalysis}
    \case $  \exists  \, { \ottnt{v_{{\mathrm{0}}}} } . \  \ottnt{e_{{\mathrm{0}}}} \,  =  \, \ottnt{v_{{\mathrm{0}}}} $: By \E{Red}/\R{Reset}.
    \case $  \exists  \, { \ottnt{K_{{\mathrm{0}}}}  \ottsym{,}  \mathit{x}  \ottsym{,}  \ottnt{e'_{{\mathrm{0}}}} } . \  \ottnt{e_{{\mathrm{0}}}} \,  =  \,  \ottnt{K_{{\mathrm{0}}}}  [  \mathcal{S}_0 \, \mathit{x}  \ottsym{.}  \ottnt{e'_{{\mathrm{0}}}}  ]  $: By \E{Red}/\R{Shift}.
    \case $  \exists  \, { \ottnt{E_{{\mathrm{0}}}}  \ottsym{,}  \pi } . \  \ottnt{e_{{\mathrm{0}}}} \,  =  \,  \ottnt{E_{{\mathrm{0}}}}  [   \bullet ^{ \pi }   ]  $: By letting $\ottnt{E} \,  =  \,  \resetcnstr{ \ottnt{E_{{\mathrm{0}}}} } $
     in the third case of the conclusion.
    \case $  \exists  \, { \ottnt{e'_{{\mathrm{0}}}}  \ottsym{,}  \varpi } . \  \ottnt{e_{{\mathrm{0}}}} \,  \mathrel{  \longrightarrow  }  \, \ottnt{e'_{{\mathrm{0}}}} \,  \,\&\,  \, \varpi $: By \E{Reset}.
   \end{caseanalysis}

  \case \T{Event}: By \E{Ev}.
 \end{caseanalysis}
\end{proof}

\begin{lemmap}{Type Safety}{ts-type-sound-terminating}
 If $ \emptyset   \vdash  \ottnt{e}  \ottsym{:}  \ottnt{C}$ and $\ottnt{e} \,  \mathrel{  \longrightarrow  ^*}  \, \ottnt{e'} \,  \,\&\,  \, \varpi$ and $\ottnt{e'}$ cannot be evaluated further,
 then $ \emptyset   \vdash  \ottnt{e'}  \ottsym{:}   { \varpi }^{-1}  \,  \tceappendop  \, \ottnt{C}$, and either of the following holds:
 \begin{enumerate}
  \item $ {  {   \exists  \, { \ottnt{v}  \ottsym{,}  \ottnt{T} } . \  \ottnt{e'} \,  =  \, \ottnt{v}  } \,\mathrel{\wedge}\, {  \emptyset   \vdash  \ottnt{v}  \ottsym{:}  \ottnt{T} }  } \,\mathrel{\wedge}\, {  \emptyset   \vdash   \ottnt{T}  \,  \,\&\,  \,   (  \varpi  , \bot )   \, / \,   \square    \ottsym{<:}  \ottnt{C} } $.
  \item $ {   \exists  \, { \ottnt{K}  \ottsym{,}  \mathit{x}  \ottsym{,}  \ottnt{e_{{\mathrm{0}}}} } . \  \ottnt{e'} \,  =  \,  \ottnt{K}  [  \mathcal{S}_0 \, \mathit{x}  \ottsym{.}  \ottnt{e_{{\mathrm{0}}}}  ]   } \,\mathrel{\wedge}\, {  \ottnt{C}  . \ottnt{S}  \,  \not=  \,  \square  } $.
  \item $  \exists  \, { \ottnt{E}  \ottsym{,}  \pi } . \  \ottnt{e'} \,  =  \,  \ottnt{E}  [   \bullet ^{ \pi }   ]  $.
 \end{enumerate}
\end{lemmap}
\begin{proof}
 By \reflem{ts-subject-red-meval}, $ \emptyset   \vdash  \ottnt{e'}  \ottsym{:}   { \varpi }^{-1}  \,  \tceappendop  \, \ottnt{C}$.
 We proceed by case analysis on the result of applying \reflem{ts-progress}
 to $ \emptyset   \vdash  \ottnt{e'}  \ottsym{:}   { \varpi }^{-1}  \,  \tceappendop  \, \ottnt{C}$
 \begin{caseanalysis}
  \case $  \exists  \, { \ottnt{v} } . \  \ottnt{e'} \,  =  \, \ottnt{v} $:
   Because $ \emptyset   \vdash  \ottnt{e'}  \ottsym{:}   { \varpi }^{-1}  \,  \tceappendop  \, \ottnt{C}$ implies $ \emptyset   \vdash  \ottnt{v}  \ottsym{:}   { \varpi }^{-1}  \,  \tceappendop  \, \ottnt{C}$,
   \reflem{ts-inv-val} implies
   $ \emptyset   \vdash  \ottnt{v}  \ottsym{:}  \ottnt{T}$ and $ \emptyset   \vdash   \ottnt{T}  \,  \,\&\,  \,   \Phi_\mathsf{val}   \, / \,   \square    \ottsym{<:}   { \varpi }^{-1}  \,  \tceappendop  \, \ottnt{C}$
   for some $\ottnt{T}$.
   Now, it suffices to show that
   \[
     \emptyset   \vdash   \ottnt{T}  \,  \,\&\,  \,   (  \varpi  , \bot )   \, / \,   \square    \ottsym{<:}  \ottnt{C} ~,
   \]
   which is implied by \reflem{ts-transitivity-subtyping} with the following.
   \begin{itemize}
    \item We show that
          \[
            \emptyset   \vdash   \ottnt{T}  \,  \,\&\,  \,   (  \varpi  , \bot )   \, / \,   \square    \ottsym{<:}   \ottnt{T}  \,  \,\&\,  \,   (  \varpi  , \bot )  \,  \tceappendop  \,  \Phi_\mathsf{val}   \, / \,   \square   ~,
          \]
          which is implied by \Srule{Comp} with the following.
          \begin{itemize}
           \item $ \emptyset   \vdash  \ottnt{T}  \ottsym{<:}  \ottnt{T}$
                 by Lemmas~\ref{lem:ts-wf-ty-tctx-typing} and \ref{lem:ts-refl-subtyping}
                 with $ \emptyset   \vdash  \ottnt{v}  \ottsym{:}  \ottnt{T}$.
           \item $ \emptyset   \vdash   (  \varpi  , \bot )   \ottsym{<:}   (  \varpi  , \bot )  \,  \tceappendop  \,  \Phi_\mathsf{val} $
                 by \reflem{ts-weak-strength-TP_val}.
           \item $ \emptyset  \,  |  \, \ottnt{T} \,  \,\&\,  \,  (  \varpi  , \bot )   \vdash   \square   \ottsym{<:}   \square $ by \Srule{Empty}.
          \end{itemize}

    \item We show that
          \[
            \emptyset   \vdash   \ottnt{T}  \,  \,\&\,  \,   (  \varpi  , \bot )  \,  \tceappendop  \,  \Phi_\mathsf{val}   \, / \,   \square    \ottsym{<:}   (  \varpi  , \bot )  \,  \tceappendop  \, \ottsym{(}   { \varpi }^{-1}  \,  \tceappendop  \, \ottnt{C}  \ottsym{)} ~,
          \]
          which is implied by \reflem{ts-add-preeff-subtyping} with $ \emptyset   \vdash   \ottnt{T}  \,  \,\&\,  \,   \Phi_\mathsf{val}   \, / \,   \square    \ottsym{<:}   { \varpi }^{-1}  \,  \tceappendop  \, \ottnt{C}$.

    \item We show that
          \[
            \emptyset   \vdash   (  \varpi  , \bot )  \,  \tceappendop  \, \ottsym{(}   { \varpi }^{-1}  \,  \tceappendop  \, \ottnt{C}  \ottsym{)}  \ottsym{<:}  \ottnt{C} ~,
          \]
          which is implied by \reflem{ts-preeff-comp-quot-minv-subty}
          with $ \emptyset   \vdash  \ottnt{C}$, which is implied by \reflem{ts-wf-ty-tctx-typing} with $ \emptyset   \vdash  \ottnt{e}  \ottsym{:}  \ottnt{C}$.
   \end{itemize}

  \case $  \exists  \, { \ottnt{K}  \ottsym{,}  \mathit{x}  \ottsym{,}  \ottnt{e_{{\mathrm{0}}}} } . \  \ottnt{e'} \,  =  \,  \ottnt{K}  [  \mathcal{S}_0 \, \mathit{x}  \ottsym{.}  \ottnt{e_{{\mathrm{0}}}}  ]  $:
   It suffices to show that $ \ottnt{C}  . \ottnt{S}  \,  \not=  \,  \square $.
   Because $ \emptyset   \vdash  \ottnt{e'}  \ottsym{:}   { \varpi }^{-1}  \,  \tceappendop  \, \ottnt{C}$ implies $ \emptyset   \vdash   \ottnt{K}  [  \mathcal{S}_0 \, \mathit{x}  \ottsym{.}  \ottnt{e_{{\mathrm{0}}}}  ]   \ottsym{:}   { \varpi }^{-1}  \,  \tceappendop  \, \ottnt{C}$,
   \reflem{ts-inv-shift-sty} implies that $ \ottsym{(}   { \varpi }^{-1}  \,  \tceappendop  \, \ottnt{C}  \ottsym{)}  . \ottnt{S}  \,  \not=  \,  \square $,
   which then implies the conclusion $ \ottnt{C}  . \ottnt{S}  \,  \not=  \,  \square $.

  \case $  \exists  \, { \ottnt{E}  \ottsym{,}  \pi } . \  \ottnt{e'} \,  =  \,  \ottnt{E}  [   \bullet ^{ \pi }   ]  $: Nothing to prove further.
  \case $  \exists  \, { \ottnt{e''}  \ottsym{,}  \varpi'' } . \  \ottnt{e'} \,  \mathrel{  \longrightarrow  }  \, \ottnt{e''} \,  \,\&\,  \, \varpi'' $:
   Contradictory with the assumption that $\ottnt{e'}$ cannot be evaluated.
 \end{caseanalysis}
\end{proof}

%% file: proof/lr.tex
\subsection{Soundness of Logical Relation}

\begin{lemmap}{Partially Ordered Domain Sets are Complete Lattices}{ts-logic-domain-lattice}
 A partially ordered set $( \mathcal{D}_{ \mathcal{T} } ,  \sqsubseteq_{ \mathcal{T} } )$ is a complete lattice.
 Furthermore, the following holds.
 \begin{itemize}
  \item If $\mathcal{T} \,  =  \, \ottnt{s}  \rightarrow  \mathcal{T}'$ for some $\ottnt{s}$ and $\mathcal{T}'$,
        the meet and join of $( \mathcal{D}_{ \mathcal{T} } ,  \sqsubseteq_{ \mathcal{T} } )$ for $\mathit{D} \subseteq  \mathcal{D}_{ \ottnt{s}  \rightarrow  \mathcal{T}' } $
        are defined by
        $ \lambda\!  \, \mathit{x} \,  {:}  \, \ottnt{s}  \ottsym{.}    \bigsqcap\nolimits_{ \mathcal{T}' }     \{ \,  \mathbf{\mathit{f} }  \ottsym{(}  \mathit{x}  \ottsym{)}  \mathrel{  \in  }   \mathcal{D}_{ \mathcal{T}' }   \mid  \mathbf{\mathit{f} }  \mathrel{  \in  }  \mathit{D}  \, \}  $ and
        $ \lambda\!  \, \mathit{x} \,  {:}  \, \ottnt{s}  \ottsym{.}    \bigsqcup\nolimits_{ \mathcal{T}' }     \{ \,  \mathbf{\mathit{f} }  \ottsym{(}  \mathit{x}  \ottsym{)}  \mathrel{  \in  }   \mathcal{D}_{ \mathcal{T}' }   \mid  \mathbf{\mathit{f} }  \mathrel{  \in  }  \mathit{D}  \, \}  $,
        respectively.
  \item If $\mathcal{T} \,  =  \,  \mathcal{T}_{{\mathrm{1}}}  \times  \mathcal{T}_{{\mathrm{2}}} $ for some $\phi_{{\mathrm{1}}}$ and $\mathcal{T}_{{\mathrm{2}}}$,
        the meet and join of $( \mathcal{D}_{ \mathcal{T} } ,  \sqsubseteq_{ \mathcal{T} } )$ for $\mathit{D} \subseteq  \mathcal{D}_{  \mathcal{T}_{{\mathrm{1}}}  \times  \mathcal{T}_{{\mathrm{2}}}  } $
        are defined by
        $\ottsym{(}    \bigsqcap\nolimits_{ \mathcal{T}_{{\mathrm{1}}} }     \{ \,  \mathbf{\mathit{f} }_{{\mathrm{1}}}  \mid  \ottsym{(}  \mathbf{\mathit{f} }_{{\mathrm{1}}}  \ottsym{,}  \mathbf{\mathit{f} }_{{\mathrm{2}}}  \ottsym{)}  \mathrel{  \in  }  \mathit{D}  \, \}    \ottsym{,}    \bigsqcap\nolimits_{ \mathcal{T}_{{\mathrm{2}}} }     \{ \,  \mathbf{\mathit{f} }_{{\mathrm{2}}}  \mid  \ottsym{(}  \mathbf{\mathit{f} }_{{\mathrm{1}}}  \ottsym{,}  \mathbf{\mathit{f} }_{{\mathrm{2}}}  \ottsym{)}  \mathrel{  \in  }  \mathit{D}  \, \}    \ottsym{)}$ and
        $\ottsym{(}    \bigsqcup\nolimits_{ \mathcal{T}_{{\mathrm{1}}} }     \{ \,  \mathbf{\mathit{f} }_{{\mathrm{1}}}  \mid  \ottsym{(}  \mathbf{\mathit{f} }_{{\mathrm{1}}}  \ottsym{,}  \mathbf{\mathit{f} }_{{\mathrm{2}}}  \ottsym{)}  \mathrel{  \in  }  \mathit{D}  \, \}    \ottsym{,}    \bigsqcup\nolimits_{ \mathcal{T}_{{\mathrm{2}}} }     \{ \,  \mathbf{\mathit{f} }_{{\mathrm{2}}}  \mid  \ottsym{(}  \mathbf{\mathit{f} }_{{\mathrm{1}}}  \ottsym{,}  \mathbf{\mathit{f} }_{{\mathrm{2}}}  \ottsym{)}  \mathrel{  \in  }  \mathit{D}  \, \}    \ottsym{)}$,
        respectively.
 \end{itemize}
\end{lemmap}
\begin{proof}
 By induction on $\mathcal{T}$.
 \begin{caseanalysis}
  \case $\mathcal{T} \,  =  \,  \mathbb{B} $:
   Obvious by definition.
   The meet is the conjunction and the join is the disjunction.

  \case $  \exists  \, { \ottnt{s}  \ottsym{,}  \mathcal{T}' } . \  \mathcal{T} \,  =  \, \ottnt{s}  \rightarrow  \mathcal{T}' $:
   Let $\mathit{D}$ be a subset of $ \mathcal{D}_{ \ottnt{s}  \rightarrow  \mathcal{T}' } $.
   By the IH, $( \mathcal{D}_{ \mathcal{T}' } ,  \sqsubseteq_{ \mathcal{T}' } )$ is a complete lattice.
   Therefore, there exists the following functions $\mathbf{\mathit{f} }_{{\mathrm{1}}}$ and $\mathbf{\mathit{f} }_{{\mathrm{2}}}$:
   \begin{itemize}
    \item $\mathbf{\mathit{f} }_{{\mathrm{1}}} \,  =  \,  \lambda\!  \, \mathit{x} \,  {:}  \, \ottnt{s}  \ottsym{.}    \bigsqcap\nolimits_{ \mathcal{T}' }     \{ \,  \mathbf{\mathit{f} }  \ottsym{(}  \mathit{x}  \ottsym{)}  \mathrel{  \in  }   \mathcal{D}_{ \mathcal{T}' }   \mid  \mathbf{\mathit{f} }  \mathrel{  \in  }  \mathit{D}  \, \}  $ and
    \item $\mathbf{\mathit{f} }_{{\mathrm{2}}} \,  =  \,  \lambda\!  \, \mathit{x} \,  {:}  \, \ottnt{s}  \ottsym{.}    \bigsqcup\nolimits_{ \mathcal{T}' }     \{ \,  \mathbf{\mathit{f} }  \ottsym{(}  \mathit{x}  \ottsym{)}  \mathrel{  \in  }   \mathcal{D}_{ \mathcal{T}' }   \mid  \mathbf{\mathit{f} }  \mathrel{  \in  }  \mathit{D}  \, \}  $.
   \end{itemize}
   We show
   that $\mathbf{\mathit{f} }_{{\mathrm{1}}}$ is the greatest lower bound of $\mathit{D}$ and
   that $\mathbf{\mathit{f} }_{{\mathrm{2}}}$ is the least upper bound of $\mathit{D}$.

   \begin{itemize}
    \item We show that $\mathbf{\mathit{f} }_{{\mathrm{1}}}$ is the greatest lower bound of $\mathit{D}$.

          Let $\mathbf{\mathit{f} } \,  \in  \, \mathit{D}$ and $\ottnt{d} \,  \in  \,  \mathcal{D}_{ \ottnt{s} } $.
          By definition,
          $\mathbf{\mathit{f} }_{{\mathrm{1}}}  \ottsym{(}  \ottnt{d}  \ottsym{)} \,  \sqsubseteq_{ \mathcal{T}' }  \, \mathbf{\mathit{f} }  \ottsym{(}  \ottnt{d}  \ottsym{)}$.
          Therefore, $\mathbf{\mathit{f} }_{{\mathrm{1}}}  \ottsym{(}  \ottnt{d}  \ottsym{)}$ is a lower bound of $\mathit{D}$.

          Let $\mathbf{\mathit{f} }' \,  \in  \,  \mathcal{D}_{ \ottnt{s}  \rightarrow  \mathcal{T}' } $ be a lower bound of $\mathit{D}$.
          Then,
          \[
             \forall  \, { \mathbf{\mathit{f} } \,  \in  \, \mathit{D} } . \    \forall  \, { \ottnt{d} \,  \in  \,  \mathcal{D}_{ \ottnt{s} }  } . \  \mathbf{\mathit{f} }'  \ottsym{(}  \ottnt{d}  \ottsym{)} \,  \sqsubseteq_{ \mathcal{T}' }  \, \mathbf{\mathit{f} }  \ottsym{(}  \ottnt{d}  \ottsym{)}   ~.
          \]
          Therefore,
          \[
             \forall  \, { \ottnt{d} \,  \in  \,  \mathcal{D}_{ \ottnt{s} }  } . \  \mathbf{\mathit{f} }'  \ottsym{(}  \ottnt{d}  \ottsym{)} \,  \sqsubseteq_{ \mathcal{T}' }  \,   \bigsqcap\nolimits_{ \mathcal{T}' }     \{ \,  \mathbf{\mathit{f} }  \ottsym{(}  \ottnt{d}  \ottsym{)}  \mathrel{  \in  }   \mathcal{D}_{ \mathcal{T}' }   \mid  \mathbf{\mathit{f} }  \mathrel{  \in  }  \mathit{D}  \, \}   
           = \mathbf{\mathit{f} }_{{\mathrm{1}}}  \ottsym{(}  \ottnt{d}  \ottsym{)} ~.
          \]
          Hence, $\mathbf{\mathit{f} }_{{\mathrm{1}}}$ is the greatest lower bound of $\mathit{D}$.

    \item We show that $\mathbf{\mathit{f} }_{{\mathrm{2}}}$ is the least upper bound of $\mathit{D}$.

          Let $\mathbf{\mathit{f} } \,  \in  \, \mathit{D}$ and $\ottnt{d} \,  \in  \,  \mathcal{D}_{ \ottnt{s} } $.
          By definition, $\mathbf{\mathit{f} }  \ottsym{(}  \ottnt{d}  \ottsym{)} \,  \sqsubseteq_{ \mathcal{T}' }  \, \mathbf{\mathit{f} }_{{\mathrm{2}}}  \ottsym{(}  \ottnt{d}  \ottsym{)}$.
          Therefore, $\mathbf{\mathit{f} }_{{\mathrm{2}}}$ is an upper bound of $\mathit{D}$.

          Let $\mathbf{\mathit{f} }' \,  \in  \,  \mathcal{D}_{ \ottnt{s}  \rightarrow  \mathcal{T}' } $ be a upper bound of $\mathit{D}$.
          Then,
          \[
             \forall  \, { \mathbf{\mathit{f} } \,  \in  \, \mathit{D} } . \    \forall  \, { \ottnt{d} \,  \in  \,  \mathcal{D}_{ \ottnt{s} }  } . \  \mathbf{\mathit{f} }  \ottsym{(}  \ottnt{d}  \ottsym{)} \,  \sqsubseteq_{ \mathcal{T}' }  \, \mathbf{\mathit{f} }'  \ottsym{(}  \ottnt{d}  \ottsym{)}   ~.
          \]
          Therefore,
          \[
             \forall  \, { \ottnt{d} \,  \in  \,  \mathcal{D}_{ \ottnt{s} }  } . \    \bigsqcup\nolimits_{ \mathcal{T}' }     \{ \,  \mathbf{\mathit{f} }  \ottsym{(}  \ottnt{d}  \ottsym{)}  \mathrel{  \in  }   \mathcal{D}_{ \mathcal{T}' }   \mid  \mathbf{\mathit{f} }  \mathrel{  \in  }  \mathit{D}  \, \}   \,  \sqsubseteq_{ \mathcal{T}' }  \, \mathbf{\mathit{f} }'  \ottsym{(}  \ottnt{d}  \ottsym{)}  ~.
          \]
          Because the left-hand expression is $\mathbf{\mathit{f} }_{{\mathrm{2}}}  \ottsym{(}  \ottnt{d}  \ottsym{)}$,
          $\mathbf{\mathit{f} }_{{\mathrm{2}}}$ is the least upper bound of $\mathit{D}$.
   \end{itemize}

  \case $  \exists  \, { \mathcal{T}_{{\mathrm{1}}}  \ottsym{,}  \mathcal{T}_{{\mathrm{2}}} } . \  \mathcal{T} \,  =  \,  \mathcal{T}_{{\mathrm{1}}}  \times  \mathcal{T}_{{\mathrm{2}}}  $:
   Let $\mathit{D}$ be a subset of $ \mathcal{D}_{  \mathcal{T}_{{\mathrm{1}}}  \times  \mathcal{T}_{{\mathrm{2}}}  } $.
   By the IH, $( \mathcal{D}_{ \mathcal{T}_{{\mathrm{1}}} } ,  \sqsubseteq_{ \mathcal{T}_{{\mathrm{1}}} } )$ and $( \mathcal{D}_{ \mathcal{T}_{{\mathrm{2}}} } ,  \sqsubseteq_{ \mathcal{T}_{{\mathrm{2}}} } )$ are complete lattices.
   It is easy to show the meet and joint given in the statements are the greatest lower bound of $\mathit{D}$ and the least upper bound of $\mathit{D}$, respectively.
 \end{caseanalysis}
\end{proof}

\begin{lemmap}{Term Substitution in Type Composition}{ts-term-subst-ty-compose}
 \begin{enumerate}
  \item $   \forall  \, { \ottnt{S_{{\mathrm{1}}}}  \ottsym{,}  \ottnt{S_{{\mathrm{2}}}}  \ottsym{,}  \ottnt{t}  \ottsym{,}  \mathit{x}  \ottsym{,}  \mathit{y} } . \  \mathit{y} \,  \not\in  \,  \mathit{fv}  (  \ottnt{t}  )  \,  \mathbin{\cup}  \,  \{  \mathit{x}  \}    \mathrel{\Longrightarrow}  \ottsym{(}   \ottnt{S_{{\mathrm{1}}}}    [  \ottnt{t}  /  \mathit{x}  ]    \gg\!\!=  \ottsym{(}   \lambda\!  \, \mathit{y}  \ottsym{.}   \ottnt{S_{{\mathrm{2}}}}    [  \ottnt{t}  /  \mathit{x}  ]    \ottsym{)}  \ottsym{)} \,  =  \,  \ottsym{(}  \ottnt{S_{{\mathrm{1}}}}  \gg\!\!=  \ottsym{(}   \lambda\!  \, \mathit{y}  \ottsym{.}  \ottnt{S_{{\mathrm{2}}}}  \ottsym{)}  \ottsym{)}    [  \ottnt{t}  /  \mathit{x}  ]   $.
  \item $  \forall  \, { \Phi_{{\mathrm{1}}}  \ottsym{,}  \Phi_{{\mathrm{2}}}  \ottsym{,}  \ottnt{t}  \ottsym{,}  \mathit{x} } . \    \Phi_{{\mathrm{1}}}    [  \ottnt{t}  /  \mathit{x}  ]   \,  \tceappendop  \, \Phi_{{\mathrm{2}}}    [  \ottnt{t}  /  \mathit{x}  ]   \,  =  \,  \ottsym{(}  \Phi_{{\mathrm{1}}} \,  \tceappendop  \, \Phi_{{\mathrm{2}}}  \ottsym{)}    [  \ottnt{t}  /  \mathit{x}  ]   $.
  \item $  \forall  \, { \Phi  \ottsym{,}  \ottnt{C}  \ottsym{,}  \ottnt{t}  \ottsym{,}  \mathit{x} } . \    \Phi    [  \ottnt{t}  /  \mathit{x}  ]   \,  \tceappendop  \, \ottnt{C}    [  \ottnt{t}  /  \mathit{x}  ]   \,  =  \,  \ottsym{(}  \Phi \,  \tceappendop  \, \ottnt{C}  \ottsym{)}    [  \ottnt{t}  /  \mathit{x}  ]   $.
  \item $  \forall  \, { \Phi  \ottsym{,}  \ottnt{S}  \ottsym{,}  \ottnt{t}  \ottsym{,}  \mathit{x} } . \    \Phi    [  \ottnt{t}  /  \mathit{x}  ]   \,  \tceappendop  \, \ottnt{S}    [  \ottnt{t}  /  \mathit{x}  ]   \,  =  \,  \ottsym{(}  \Phi \,  \tceappendop  \, \ottnt{S}  \ottsym{)}    [  \ottnt{t}  /  \mathit{x}  ]   $.
 \end{enumerate}
\end{lemmap}
\begin{proof}
 The first and second cases are by definition.
 The third and fourth cases are straightforward by simultaneous induction on $\ottnt{C}$ and $\ottnt{S}$ with the second case.
\end{proof}

\begin{lemmap}{Term Substitution in Subtyping}{ts-term-subst-subtyping}
 Suppose that $\Gamma_{{\mathrm{1}}}  \vdash  \ottnt{t}  \ottsym{:}  \ottnt{s}$.
 \begin{enumerate}
  \item If $\Gamma_{{\mathrm{1}}}  \ottsym{,}  \mathit{x} \,  {:}  \, \ottnt{s}  \ottsym{,}  \Gamma_{{\mathrm{2}}}  \vdash  \ottnt{T_{{\mathrm{1}}}}  \ottsym{<:}  \ottnt{T_{{\mathrm{2}}}}$, then $ \Gamma_{{\mathrm{1}}}  \ottsym{,}  \Gamma_{{\mathrm{2}}}    [  \ottnt{t}  /  \mathit{x}  ]    \vdash   \ottnt{T_{{\mathrm{1}}}}    [  \ottnt{t}  /  \mathit{x}  ]    \ottsym{<:}   \ottnt{T_{{\mathrm{2}}}}    [  \ottnt{t}  /  \mathit{x}  ]  $.
  \item If $\Gamma_{{\mathrm{1}}}  \ottsym{,}  \mathit{x} \,  {:}  \, \ottnt{s}  \ottsym{,}  \Gamma_{{\mathrm{2}}}  \vdash  \ottnt{C_{{\mathrm{1}}}}  \ottsym{<:}  \ottnt{C_{{\mathrm{2}}}}$, then $ \Gamma_{{\mathrm{1}}}  \ottsym{,}  \Gamma_{{\mathrm{2}}}    [  \ottnt{t}  /  \mathit{x}  ]    \vdash   \ottnt{C_{{\mathrm{1}}}}    [  \ottnt{t}  /  \mathit{x}  ]    \ottsym{<:}   \ottnt{C_{{\mathrm{2}}}}    [  \ottnt{t}  /  \mathit{x}  ]  $.
  \item If $\Gamma_{{\mathrm{1}}}  \ottsym{,}  \mathit{x} \,  {:}  \, \ottnt{s}  \ottsym{,}  \Gamma_{{\mathrm{2}}} \,  |  \, \ottnt{T} \,  \,\&\,  \, \Phi  \vdash  \ottnt{S_{{\mathrm{1}}}}  \ottsym{<:}  \ottnt{S_{{\mathrm{2}}}}$, then $ \Gamma_{{\mathrm{1}}}  \ottsym{,}  \Gamma_{{\mathrm{2}}}    [  \ottnt{t}  /  \mathit{x}  ]   \,  |  \,  \ottnt{T}    [  \ottnt{t}  /  \mathit{x}  ]   \,  \,\&\,  \,  \Phi    [  \ottnt{t}  /  \mathit{x}  ]    \vdash   \ottnt{S_{{\mathrm{1}}}}    [  \ottnt{t}  /  \mathit{x}  ]    \ottsym{<:}   \ottnt{S_{{\mathrm{2}}}}    [  \ottnt{t}  /  \mathit{x}  ]  $.
  \item If $\Gamma_{{\mathrm{1}}}  \ottsym{,}  \mathit{x} \,  {:}  \, \ottnt{s}  \ottsym{,}  \Gamma_{{\mathrm{2}}}  \vdash  \Phi_{{\mathrm{1}}}  \ottsym{<:}  \Phi_{{\mathrm{2}}}$, then $ \Gamma_{{\mathrm{1}}}  \ottsym{,}  \Gamma_{{\mathrm{2}}}    [  \ottnt{t}  /  \mathit{x}  ]    \vdash   \Phi_{{\mathrm{1}}}    [  \ottnt{t}  /  \mathit{x}  ]    \ottsym{<:}   \Phi_{{\mathrm{2}}}    [  \ottnt{t}  /  \mathit{x}  ]  $.
 \end{enumerate}
\end{lemmap}
\begin{proof}
 Straightforward by simultaneous induction on the derivations.
 The cases for \Srule{Refine}, \Srule{Eff}, and \Srule{Embed} use \reflem{ts-term-subst-valid}, and
 the case for \Srule{Embed} also uses Lemmas~\ref{lem:ts-subst-nu-cty} and \ref{lem:ts-term-subst-ty-compose}.
\end{proof}

\begin{lemmap}{Term Substitution in Typing}{ts-term-subst-typing}
 Suppose that $\Gamma_{{\mathrm{1}}}  \vdash  \ottnt{t}  \ottsym{:}  \ottnt{s}$.
 If $\Gamma_{{\mathrm{1}}}  \ottsym{,}  \mathit{x} \,  {:}  \, \ottnt{s}  \ottsym{,}  \Gamma_{{\mathrm{2}}}  \vdash  \ottnt{e}  \ottsym{:}  \ottnt{C}$, then $ \Gamma_{{\mathrm{1}}}  \ottsym{,}  \Gamma_{{\mathrm{2}}}    [  \ottnt{t}  /  \mathit{x}  ]    \vdash   \ottnt{e}    [  \ottnt{t}  /  \mathit{x}  ]    \ottsym{:}   \ottnt{C}    [  \ottnt{t}  /  \mathit{x}  ]  $.
 Especially, if $\Gamma_{{\mathrm{1}}}  \ottsym{,}  \mathit{x} \,  {:}  \, \ottnt{s}  \ottsym{,}  \Gamma_{{\mathrm{2}}}  \vdash  \ottnt{e}  \ottsym{:}  \ottnt{T}$, then $ \Gamma_{{\mathrm{1}}}  \ottsym{,}  \Gamma_{{\mathrm{2}}}    [  \ottnt{t}  /  \mathit{x}  ]    \vdash   \ottnt{e}    [  \ottnt{t}  /  \mathit{x}  ]    \ottsym{:}   \ottnt{T}    [  \ottnt{t}  /  \mathit{x}  ]  $.
\end{lemmap}
\begin{proof}
 By induction on the typing derivation with \reflem{ts-term-subst-wf-ftyping}.
 We mention only the interesting cases.
 \begin{caseanalysis}
  \case \T{Fun}: By the IH and Lemmas~\ref{lem:lr-subst-occur}, \ref{lem:lr-subst-consist}, and \ref{lem:lr-subst-eff-subst}.

  \case \T{Sub}: By the IH and Lemmas~\ref{lem:ts-term-subst-wf-ftyping} and \ref{lem:ts-term-subst-subtyping}.

  \case \T{Op}:  By the IHs and \reflem{ts-term-subst-wf-ftyping}.
   Note that $\mathit{ty} \, \ottsym{(}  \ottnt{o}  \ottsym{)}$ is closed (see \refasm{const}).

  \case \T{Let}: By the IHs and \reflem{ts-term-subst-ty-compose}.

  \case \T{Div}: By Lemmas~\ref{lem:ts-term-subst-wf-ftyping}, \ref{lem:ts-term-subst-valid}, and \ref{lem:ts-subst-nu-cty}.
 \end{caseanalysis}
\end{proof}

\begin{lemmap}{Substitution Lemmas with Related Substitution}{lr-typeability}
 Let $\sigma \,  \in  \,  \LRRel{ \mathcal{G} }{ \Gamma } $.
 \begin{itemize}
  \item If $\Gamma  \ottsym{,}  \Gamma'  \vdash  \ottnt{A}  \ottsym{:}   \tceseq{ \ottnt{s} } $, then $\sigma  \ottsym{(}  \Gamma'  \ottsym{)}  \vdash  \sigma  \ottsym{(}  \ottnt{A}  \ottsym{)}  \ottsym{:}   \tceseq{ \ottnt{s} } $.
  \item If $\Gamma  \ottsym{,}  \Gamma'  \vdash  \phi$, then $\sigma  \ottsym{(}  \Gamma'  \ottsym{)}  \vdash  \sigma  \ottsym{(}  \phi  \ottsym{)}$.
  \item If $\Gamma  \ottsym{,}  \Gamma'  \vdash  \ottnt{C}$, then $\sigma  \ottsym{(}  \Gamma'  \ottsym{)}  \vdash  \sigma  \ottsym{(}  \ottnt{C}  \ottsym{)}$.
  \item If $\Gamma  \ottsym{,}  \Gamma'  \vdash  \ottnt{C_{{\mathrm{1}}}}  \ottsym{<:}  \ottnt{C_{{\mathrm{2}}}}$, then $\sigma  \ottsym{(}  \Gamma'  \ottsym{)}  \vdash  \sigma  \ottsym{(}  \ottnt{C_{{\mathrm{1}}}}  \ottsym{)}  \ottsym{<:}  \sigma  \ottsym{(}  \ottnt{C_{{\mathrm{2}}}}  \ottsym{)}$.
  \item If $\Gamma  \ottsym{,}  \Gamma'  \vdash  \ottnt{e}  \ottsym{:}  \ottnt{C}$, then $\sigma  \ottsym{(}  \Gamma'  \ottsym{)}  \vdash  \sigma  \ottsym{(}  \ottnt{e}  \ottsym{)}  \ottsym{:}  \sigma  \ottsym{(}  \ottnt{C}  \ottsym{)}$.
  \item If $\Gamma  \ottsym{,}  \Gamma'  \models  \phi$, then $\sigma  \ottsym{(}  \Gamma'  \ottsym{)}  \models  \sigma  \ottsym{(}  \phi  \ottsym{)}$.
 \end{itemize}
\end{lemmap}
\begin{proof}
 Straightforward by induction on $\Gamma$ with:
 \begin{itemize}
  \item Lemmas~\ref{lem:ts-val-subst-wf-ftyping}, \ref{lem:ts-pred-subst-wf-ftyping}, and \ref{lem:ts-term-subst-wf-ftyping} for $\sigma  \ottsym{(}  \Gamma'  \ottsym{)}  \vdash  \sigma  \ottsym{(}  \ottnt{A}  \ottsym{)}  \ottsym{:}   \tceseq{ \ottnt{s} } $, $\sigma  \ottsym{(}  \Gamma'  \ottsym{)}  \vdash  \sigma  \ottsym{(}  \phi  \ottsym{)}$, and $\sigma  \ottsym{(}  \Gamma'  \ottsym{)}  \vdash  \sigma  \ottsym{(}  \ottnt{C}  \ottsym{)}$;
  \item Lemmas~\ref{lem:ts-val-subst-subtyping}, \ref{lem:ts-pred-subst-subtyping}, and \ref{lem:ts-term-subst-subtyping} for $\sigma  \ottsym{(}  \Gamma'  \ottsym{)}  \vdash  \sigma  \ottsym{(}  \ottnt{C_{{\mathrm{1}}}}  \ottsym{)}  \ottsym{<:}  \sigma  \ottsym{(}  \ottnt{C_{{\mathrm{2}}}}  \ottsym{)}$;
  \item Lemmas~\ref{lem:ts-val-subst-typing}, \ref{lem:ts-pred-subst-typing}, and \ref{lem:ts-term-subst-typing} for $\sigma  \ottsym{(}  \Gamma'  \ottsym{)}  \vdash  \sigma  \ottsym{(}  \ottnt{e}  \ottsym{)}  \ottsym{:}  \sigma  \ottsym{(}  \ottnt{C}  \ottsym{)}$;
        and
  \item Lemmas~\ref{lem:ts-val-subst-valid}, \ref{lem:ts-pred-subst-valid}, and \ref{lem:ts-term-subst-valid} for $\sigma  \ottsym{(}  \Gamma'  \ottsym{)}  \models  \sigma  \ottsym{(}  \phi  \ottsym{)}$.
 \end{itemize}
\end{proof}

\begin{lemmap}{Well-Formedness of Effects in Continuation Typing}{lr-typing-cont-wf-eff}
 If $\ottnt{K}  \ottsym{:}    \ottnt{T}  \, / \, ( \forall  \mathit{x}  .  \ottnt{C}  )   \mathrel{ \Rrightarrow }  \ottnt{T_{{\mathrm{0}}}}  \,  \,\&\,  \,  \Phi_{{\mathrm{0}}}  \, / \, ( \forall  \mathit{x_{{\mathrm{0}}}}  .  \ottnt{C_{{\mathrm{0}}}}  ) $,
 then $ \emptyset   \vdash  \Phi_{{\mathrm{0}}}$.
\end{lemmap}
\begin{proof}
 By induction on the typing derivation.
 \begin{caseanalysis}
  \case \TK{Empty}: By \reflem{ts-wf-TP_val} with $\vdash   \emptyset $.
  \case \TK{Sub}: By the IH.
  \case \TK{Let}:
   We are given
   \begin{prooftree}
    \AxiomC{$\ottnt{K'}  \ottsym{:}    \ottnt{T}  \, / \, ( \forall  \mathit{x}  .  \ottnt{C}  )   \mathrel{ \Rrightarrow }  \ottnt{T_{{\mathrm{1}}}}  \,  \,\&\,  \,  \Phi_{{\mathrm{1}}}  \, / \, ( \forall  \mathit{x_{{\mathrm{1}}}}  .  \ottnt{C'_{{\mathrm{0}}}}  ) $}
    \AxiomC{$\mathit{x_{{\mathrm{1}}}} \,  {:}  \, \ottnt{T_{{\mathrm{1}}}}  \vdash  \ottnt{e_{{\mathrm{2}}}}  \ottsym{:}   \ottnt{T_{{\mathrm{0}}}}  \,  \,\&\,  \,  \Phi_{{\mathrm{2}}}  \, / \,   ( \forall  \mathit{x_{{\mathrm{0}}}}  .  \ottnt{C_{{\mathrm{0}}}}  )  \Rightarrow   \ottnt{C'_{{\mathrm{0}}}}  $}
    \BinaryInfC{$\mathsf{let} \, \mathit{x_{{\mathrm{1}}}}  \ottsym{=}  \ottnt{K'} \,  \mathsf{in}  \, \ottnt{e_{{\mathrm{2}}}}  \ottsym{:}    \ottnt{T}  \, / \, ( \forall  \mathit{x}  .  \ottnt{C}  )   \mathrel{ \Rrightarrow }  \ottnt{T_{{\mathrm{0}}}}  \,  \,\&\,  \,  \Phi_{{\mathrm{1}}} \,  \tceappendop  \, \Phi_{{\mathrm{2}}}  \, / \, ( \forall  \mathit{x_{{\mathrm{0}}}}  .  \ottnt{C_{{\mathrm{0}}}}  ) $}
   \end{prooftree}
   for some $\ottnt{K'}$, $\ottnt{T_{{\mathrm{1}}}}$, $\Phi_{{\mathrm{1}}}$, $\Phi_{{\mathrm{2}}}$, $\mathit{x_{{\mathrm{1}}}}$, $\ottnt{e_{{\mathrm{2}}}}$, and $\ottnt{C'_{{\mathrm{0}}}}$ such that
   \begin{itemize}
    \item $\ottnt{K} \,  =  \, \ottsym{(}  \mathsf{let} \, \mathit{x_{{\mathrm{1}}}}  \ottsym{=}  \ottnt{K'} \,  \mathsf{in}  \, \ottnt{e_{{\mathrm{2}}}}  \ottsym{)}$,
    \item $\Phi_{{\mathrm{0}}} \,  =  \, \Phi_{{\mathrm{1}}} \,  \tceappendop  \, \Phi_{{\mathrm{2}}}$, and
    \item $\mathit{x_{{\mathrm{1}}}} \,  \not\in  \,  \mathit{fv}  (  \ottnt{T_{{\mathrm{0}}}}  )  \,  \mathbin{\cup}  \,  \mathit{fv}  (  \Phi_{{\mathrm{2}}}  )  \,  \mathbin{\cup}  \, \ottsym{(}   \mathit{fv}  (  \ottnt{C_{{\mathrm{0}}}}  )  \,  \mathbin{\backslash}  \,  \{  \mathit{x_{{\mathrm{0}}}}  \}   \ottsym{)}$.
   \end{itemize}
   By the IH, $ \emptyset   \vdash  \Phi_{{\mathrm{1}}}$.
   \reflem{ts-wf-ty-tctx-typing} with $\mathit{x_{{\mathrm{1}}}} \,  {:}  \, \ottnt{T_{{\mathrm{1}}}}  \vdash  \ottnt{e_{{\mathrm{2}}}}  \ottsym{:}   \ottnt{T_{{\mathrm{0}}}}  \,  \,\&\,  \,  \Phi_{{\mathrm{2}}}  \, / \,   ( \forall  \mathit{x_{{\mathrm{0}}}}  .  \ottnt{C_{{\mathrm{0}}}}  )  \Rightarrow   \ottnt{C'_{{\mathrm{0}}}}  $
   implies $\mathit{x_{{\mathrm{1}}}} \,  {:}  \, \ottnt{T_{{\mathrm{1}}}}  \vdash  \Phi_{{\mathrm{2}}}$.
   Because $\mathit{x_{{\mathrm{1}}}} \,  \not\in  \,  \mathit{fv}  (  \Phi_{{\mathrm{2}}}  ) $,
   \reflem{ts-elim-unused-binding-wf-ftyping} implies $ \emptyset   \vdash  \Phi_{{\mathrm{2}}}$.
   Then, \reflem{ts-wf-eff-compose} implies the conclusion $ \emptyset   \vdash  \Phi_{{\mathrm{1}}} \,  \tceappendop  \, \Phi_{{\mathrm{2}}}$.
 \end{caseanalysis}
\end{proof}

\begin{lemmap}{Filling Well-Typed Continuations with Well-Typed Expressions}{lr-typing-cont-exp}
 \begin{enumerate}
  \item \label{lem:lr-typing-pure-ectx-exp:cont}
        If
        $\ottnt{K}  \ottsym{:}    \ottnt{T}  \, / \, ( \forall  \mathit{x}  .  \ottnt{C_{{\mathrm{1}}}}  )   \mathrel{ \Rrightarrow }  \ottnt{T_{{\mathrm{0}}}}  \,  \,\&\,  \,  \Phi_{{\mathrm{0}}}  \, / \, ( \forall  \mathit{x_{{\mathrm{0}}}}  .  \ottnt{C_{{\mathrm{0}}}}  ) $ and
        $\Gamma  \vdash  \ottnt{e}  \ottsym{:}   \ottnt{T}  \,  \,\&\,  \,  \Phi  \, / \,   ( \forall  \mathit{x}  .  \ottnt{C_{{\mathrm{1}}}}  )  \Rightarrow   \ottnt{C_{{\mathrm{2}}}}  $,
        then
        $\Gamma  \vdash   \ottnt{K}  [  \ottnt{e}  ]   \ottsym{:}   \ottnt{T_{{\mathrm{0}}}}  \,  \,\&\,  \,  \Phi \,  \tceappendop  \, \Phi_{{\mathrm{0}}}  \, / \,   ( \forall  \mathit{x_{{\mathrm{0}}}}  .  \ottnt{C_{{\mathrm{0}}}}  )  \Rightarrow   \ottnt{C_{{\mathrm{2}}}}  $.
  \item If
        $\ottnt{K}  \ottsym{:}   \ottnt{T}  \, / \, ( \forall  \mathit{x}  .  \ottnt{C_{{\mathrm{1}}}}  ) $ and
        $\Gamma  \vdash  \ottnt{e}  \ottsym{:}   \ottnt{T}  \,  \,\&\,  \,  \Phi  \, / \,   ( \forall  \mathit{x}  .  \ottnt{C_{{\mathrm{1}}}}  )  \Rightarrow   \ottnt{C_{{\mathrm{2}}}}  $,
        then $\Gamma  \vdash   \resetcnstr{  \ottnt{K}  [  \ottnt{e}  ]  }   \ottsym{:}  \ottnt{C_{{\mathrm{2}}}}$
 \end{enumerate}
\end{lemmap}
\begin{proof}
 \noindent
 \begin{enumerate}
  \item By induction on the typing derivation of
        $\ottnt{K}  \ottsym{:}    \ottnt{T}  \, / \, ( \forall  \mathit{x}  .  \ottnt{C_{{\mathrm{1}}}}  )   \mathrel{ \Rrightarrow }  \ottnt{T_{{\mathrm{0}}}}  \,  \,\&\,  \,  \Phi_{{\mathrm{0}}}  \, / \, ( \forall  \mathit{x_{{\mathrm{0}}}}  .  \ottnt{C_{{\mathrm{0}}}}  ) $.
        \begin{caseanalysis}
         \case \TK{Empty}:
          We are given
          \begin{prooftree}
           \AxiomC{$\,[\,]\,  \ottsym{:}    \ottnt{T}  \, / \, ( \forall  \mathit{x}  .  \ottnt{C_{{\mathrm{1}}}}  )   \mathrel{ \Rrightarrow }  \ottnt{T}  \,  \,\&\,  \,   \Phi_\mathsf{val}   \, / \, ( \forall  \mathit{x}  .  \ottnt{C_{{\mathrm{1}}}}  ) $}
          \end{prooftree}
          where
          $\ottnt{K} \,  =  \, \,[\,]\,$ and $\ottnt{T_{{\mathrm{0}}}} \,  =  \, \ottnt{T}$ and $\Phi_{{\mathrm{0}}} \,  =  \,  \Phi_\mathsf{val} $ and
          $ \mathit{x_{{\mathrm{0}}}}    =    \mathit{x} $ and $\ottnt{C_{{\mathrm{0}}}} \,  =  \, \ottnt{C_{{\mathrm{1}}}}$.
          We must show that
          \[
           \Gamma  \vdash  \ottnt{e}  \ottsym{:}   \ottnt{T}  \,  \,\&\,  \,  \Phi \,  \tceappendop  \,  \Phi_\mathsf{val}   \, / \,   ( \forall  \mathit{x}  .  \ottnt{C_{{\mathrm{1}}}}  )  \Rightarrow   \ottnt{C_{{\mathrm{2}}}}   ~.
          \]
          Because $\Gamma  \vdash  \ottnt{e}  \ottsym{:}   \ottnt{T}  \,  \,\&\,  \,  \Phi  \, / \,   ( \forall  \mathit{x}  .  \ottnt{C_{{\mathrm{1}}}}  )  \Rightarrow   \ottnt{C_{{\mathrm{2}}}}  $,
          the conclusion is implied by
          \T{Sub} with the following.
          \begin{itemize}
           \item We show that
                 \[
                  \Gamma  \vdash   \ottnt{T}  \,  \,\&\,  \,  \Phi  \, / \,   ( \forall  \mathit{x}  .  \ottnt{C_{{\mathrm{1}}}}  )  \Rightarrow   \ottnt{C_{{\mathrm{2}}}}    \ottsym{<:}   \ottnt{T}  \,  \,\&\,  \,  \Phi \,  \tceappendop  \,  \Phi_\mathsf{val}   \, / \,   ( \forall  \mathit{x}  .  \ottnt{C_{{\mathrm{1}}}}  )  \Rightarrow   \ottnt{C_{{\mathrm{2}}}}   ~.
                 \]
                 \reflem{ts-wf-ty-tctx-typing} with $\Gamma  \vdash  \ottnt{e}  \ottsym{:}   \ottnt{T}  \,  \,\&\,  \,  \Phi  \, / \,   ( \forall  \mathit{x}  .  \ottnt{C_{{\mathrm{1}}}}  )  \Rightarrow   \ottnt{C_{{\mathrm{2}}}}  $
                 implies
                 \begin{itemize}
                  \item $\Gamma  \vdash  \ottnt{T}$,
                  \item $\Gamma  \vdash  \Phi$,
                  \item $\Gamma  \ottsym{,}  \mathit{x} \,  {:}  \, \ottnt{T}  \vdash  \ottnt{C_{{\mathrm{1}}}}$, and
                  \item $\Gamma  \vdash  \ottnt{C_{{\mathrm{2}}}}$.
                 \end{itemize}
                 Therefore, by \reflem{ts-refl-subtyping}, we have
                 \begin{itemize}
                  \item $\Gamma  \vdash  \ottnt{T}  \ottsym{<:}  \ottnt{T}$,
                  \item $\Gamma  \ottsym{,}  \mathit{x} \,  {:}  \, \ottnt{T}  \vdash  \ottnt{C_{{\mathrm{1}}}}  \ottsym{<:}  \ottnt{C_{{\mathrm{1}}}}$, and
                  \item $\Gamma  \vdash  \ottnt{C_{{\mathrm{2}}}}  \ottsym{<:}  \ottnt{C_{{\mathrm{2}}}}$.
                 \end{itemize}
                 Further, \reflem{ts-weak-strength-TP_val} implies
                 $\Gamma  \vdash  \Phi  \ottsym{<:}  \Phi \,  \tceappendop  \,  \Phi_\mathsf{val} $.
                 Thus, \Srule{Ans} and \Srule{Comp} imply the conclusion.

           \item We show that
                 \[
                  \Gamma  \vdash   \ottnt{T}  \,  \,\&\,  \,  \Phi \,  \tceappendop  \,  \Phi_\mathsf{val}   \, / \,   ( \forall  \mathit{x}  .  \ottnt{C_{{\mathrm{1}}}}  )  \Rightarrow   \ottnt{C_{{\mathrm{2}}}}   ~.
                 \]
                 It is proven by \WFT{Ans} and \WFT{Comp} with the following.
                 \begin{itemize}
                  \item $\Gamma  \vdash  \ottnt{T}$, which is shown above.
                  \item $\Gamma  \vdash  \Phi \,  \tceappendop  \,  \Phi_\mathsf{val} $, which is shown by
                        \reflem{ts-wf-eff-compose} with
                        $\Gamma  \vdash  \Phi$ (shown above) and
                        $\Gamma  \vdash   \Phi_\mathsf{val} $
                        (shown by \reflem{ts-wf-TP_val} with $\vdash  \Gamma$ that is implied by \reflem{ts-wf-tctx-wf-ftyping} with $\Gamma  \vdash  \ottnt{T}$).
                  \item $\Gamma  \ottsym{,}  \mathit{x} \,  {:}  \, \ottnt{T}  \vdash  \ottnt{C_{{\mathrm{1}}}}$, which is shown above.
                  \item $\Gamma  \vdash  \ottnt{C_{{\mathrm{2}}}}$, which is shown above.
                 \end{itemize}
          \end{itemize}

         \case \TK{Sub}:
          We are given
          \begin{prooftree}
           \AxiomC{$\ottnt{K}  \ottsym{:}    \ottnt{T}  \, / \, ( \forall  \mathit{x}  .  \ottnt{C_{{\mathrm{1}}}}  )   \mathrel{ \Rrightarrow }  \ottnt{T_{{\mathrm{0}}}}  \,  \,\&\,  \,  \Phi_{{\mathrm{0}}}  \, / \, ( \forall  \mathit{x_{{\mathrm{0}}}}  .  \ottnt{C'_{{\mathrm{0}}}}  ) $}
           \AxiomC{$\mathit{x_{{\mathrm{0}}}} \,  {:}  \, \ottnt{T_{{\mathrm{0}}}}  \vdash  \ottnt{C_{{\mathrm{0}}}}  \ottsym{<:}  \ottnt{C'_{{\mathrm{0}}}}$}
           \AxiomC{$\mathit{x_{{\mathrm{0}}}} \,  {:}  \, \ottnt{T_{{\mathrm{0}}}}  \vdash  \ottnt{C_{{\mathrm{0}}}}$}
           \TrinaryInfC{$\ottnt{K}  \ottsym{:}    \ottnt{T}  \, / \, ( \forall  \mathit{x}  .  \ottnt{C_{{\mathrm{1}}}}  )   \mathrel{ \Rrightarrow }  \ottnt{T_{{\mathrm{0}}}}  \,  \,\&\,  \,  \Phi_{{\mathrm{0}}}  \, / \, ( \forall  \mathit{x_{{\mathrm{0}}}}  .  \ottnt{C_{{\mathrm{0}}}}  ) $}
          \end{prooftree}
          for some $\ottnt{C'_{{\mathrm{0}}}}$.
          Without loss of generality, we can suppose that
          $\mathit{x_{{\mathrm{0}}}} \,  \not\in  \,  \mathit{dom}  (  \Gamma  ) $.
          By the IH,
          \begin{equation}
           \Gamma  \vdash   \ottnt{K}  [  \ottnt{e}  ]   \ottsym{:}   \ottnt{T_{{\mathrm{0}}}}  \,  \,\&\,  \,  \Phi \,  \tceappendop  \, \Phi_{{\mathrm{0}}}  \, / \,   ( \forall  \mathit{x_{{\mathrm{0}}}}  .  \ottnt{C'_{{\mathrm{0}}}}  )  \Rightarrow   \ottnt{C_{{\mathrm{2}}}}   ~.
           \label{lem:lr-typing-pure-ectx-exp:sub:IH}
          \end{equation}
          Then, the conclusion is implied by \T{Sub} with the following.
          \begin{itemize}
           \item We show that
                 \[
                  \Gamma  \vdash   \ottnt{T_{{\mathrm{0}}}}  \,  \,\&\,  \,  \Phi \,  \tceappendop  \, \Phi_{{\mathrm{0}}}  \, / \,   ( \forall  \mathit{x_{{\mathrm{0}}}}  .  \ottnt{C'_{{\mathrm{0}}}}  )  \Rightarrow   \ottnt{C_{{\mathrm{2}}}}    \ottsym{<:}   \ottnt{T_{{\mathrm{0}}}}  \,  \,\&\,  \,  \Phi \,  \tceappendop  \, \Phi_{{\mathrm{0}}}  \, / \,   ( \forall  \mathit{x_{{\mathrm{0}}}}  .  \ottnt{C_{{\mathrm{0}}}}  )  \Rightarrow   \ottnt{C_{{\mathrm{2}}}}   ~.
                 \]
                 By \reflem{ts-wf-ty-tctx-typing} with (\ref{lem:lr-typing-pure-ectx-exp:sub:IH}),
                 we have
                 \begin{itemize}
                  \item $\Gamma  \vdash  \ottnt{T_{{\mathrm{0}}}}$,
                  \item $\Gamma  \vdash  \Phi \,  \tceappendop  \, \Phi_{{\mathrm{0}}}$, and
                  \item $\Gamma  \vdash  \ottnt{C_{{\mathrm{2}}}}$.
                 \end{itemize}
                 By \reflem{ts-refl-subtyping},
                 \begin{itemize}
                  \item $\Gamma  \vdash  \ottnt{T_{{\mathrm{0}}}}  \ottsym{<:}  \ottnt{T_{{\mathrm{0}}}}$,
                  \item $\Gamma  \vdash  \Phi \,  \tceappendop  \, \Phi_{{\mathrm{0}}}  \ottsym{<:}  \Phi \,  \tceappendop  \, \Phi_{{\mathrm{0}}}$,
                  \item $\Gamma  \vdash  \ottnt{C_{{\mathrm{2}}}}  \ottsym{<:}  \ottnt{C_{{\mathrm{2}}}}$.
                 \end{itemize}
                 Further, we also have $\Gamma  \ottsym{,}  \mathit{x_{{\mathrm{0}}}} \,  {:}  \, \ottnt{T_{{\mathrm{0}}}}  \vdash  \ottnt{C_{{\mathrm{0}}}}  \ottsym{<:}  \ottnt{C'_{{\mathrm{0}}}}$
                 by \reflem{ts-weak-subtyping} with $\mathit{x_{{\mathrm{0}}}} \,  {:}  \, \ottnt{T_{{\mathrm{0}}}}  \vdash  \ottnt{C_{{\mathrm{0}}}}  \ottsym{<:}  \ottnt{C'_{{\mathrm{0}}}}$.
                 Therefore, \Srule{Ans} and \Srule{Comp} imply
                 the conclusion.

           \item We show that
                 \[
                  \Gamma  \vdash   \ottnt{T_{{\mathrm{0}}}}  \,  \,\&\,  \,  \Phi \,  \tceappendop  \, \Phi_{{\mathrm{0}}}  \, / \,   ( \forall  \mathit{x_{{\mathrm{0}}}}  .  \ottnt{C_{{\mathrm{0}}}}  )  \Rightarrow   \ottnt{C_{{\mathrm{2}}}}   ~.
                 \]
                 By \WFT{Ans} and \WFT{Comp}, it suffices to show the following.
                 \begin{itemize}
                  \item $\Gamma  \vdash  \ottnt{T_{{\mathrm{0}}}}$, which is shown above.
                  \item $\Gamma  \vdash  \Phi \,  \tceappendop  \, \Phi_{{\mathrm{0}}}$, which is shown above.
                  \item $\Gamma  \ottsym{,}  \mathit{x_{{\mathrm{0}}}} \,  {:}  \, \ottnt{T_{{\mathrm{0}}}}  \vdash  \ottnt{C_{{\mathrm{0}}}}$, which is implied by
                        \reflem{ts-weak-wf-ftyping} with
                        $\vdash  \Gamma$ (shown by \reflem{ts-wf-tctx-wf-ftyping} with $\Gamma  \vdash  \ottnt{T_{{\mathrm{0}}}}$) and
                        $\mathit{x_{{\mathrm{0}}}} \,  {:}  \, \ottnt{T_{{\mathrm{0}}}}  \vdash  \ottnt{C_{{\mathrm{0}}}}$.
                  \item $\Gamma  \vdash  \ottnt{C_{{\mathrm{2}}}}$, which is shown above.
                 \end{itemize}
          \end{itemize}

         \case \TK{Let}:
          We are given
          \begin{prooftree}
           \AxiomC{$\ottnt{K'}  \ottsym{:}    \ottnt{T}  \, / \, ( \forall  \mathit{x}  .  \ottnt{C_{{\mathrm{1}}}}  )   \mathrel{ \Rrightarrow }  \ottnt{T_{{\mathrm{1}}}}  \,  \,\&\,  \,  \Phi_{{\mathrm{1}}}  \, / \, ( \forall  \mathit{x_{{\mathrm{1}}}}  .  \ottnt{C'_{{\mathrm{0}}}}  ) $}
           \AxiomC{$\mathit{x_{{\mathrm{1}}}} \,  {:}  \, \ottnt{T_{{\mathrm{1}}}}  \vdash  \ottnt{e_{{\mathrm{2}}}}  \ottsym{:}   \ottnt{T_{{\mathrm{0}}}}  \,  \,\&\,  \,  \Phi_{{\mathrm{2}}}  \, / \,   ( \forall  \mathit{x_{{\mathrm{0}}}}  .  \ottnt{C_{{\mathrm{0}}}}  )  \Rightarrow   \ottnt{C'_{{\mathrm{0}}}}  $}
           \BinaryInfC{$\mathsf{let} \, \mathit{x_{{\mathrm{1}}}}  \ottsym{=}  \ottnt{K'} \,  \mathsf{in}  \, \ottnt{e_{{\mathrm{2}}}}  \ottsym{:}    \ottnt{T}  \, / \, ( \forall  \mathit{x}  .  \ottnt{C_{{\mathrm{1}}}}  )   \mathrel{ \Rrightarrow }  \ottnt{T_{{\mathrm{0}}}}  \,  \,\&\,  \,  \Phi_{{\mathrm{1}}} \,  \tceappendop  \, \Phi_{{\mathrm{2}}}  \, / \, ( \forall  \mathit{x_{{\mathrm{0}}}}  .  \ottnt{C_{{\mathrm{0}}}}  ) $}
          \end{prooftree}
          for some $\ottnt{K'}$, $\ottnt{T_{{\mathrm{1}}}}$, $\Phi_{{\mathrm{1}}}$, $\Phi_{{\mathrm{2}}}$, $\mathit{x_{{\mathrm{1}}}}$, $\ottnt{e_{{\mathrm{2}}}}$, and $\ottnt{C'_{{\mathrm{0}}}}$ such that
          \begin{itemize}
           \item $\ottnt{K} \,  =  \, \ottsym{(}  \mathsf{let} \, \mathit{x_{{\mathrm{1}}}}  \ottsym{=}  \ottnt{K'} \,  \mathsf{in}  \, \ottnt{e_{{\mathrm{2}}}}  \ottsym{)}$,
           \item $\Phi_{{\mathrm{0}}} \,  =  \, \Phi_{{\mathrm{1}}} \,  \tceappendop  \, \Phi_{{\mathrm{2}}}$, and
           \item $\mathit{x_{{\mathrm{1}}}} \,  \not\in  \,  \mathit{fv}  (  \ottnt{T_{{\mathrm{0}}}}  )  \,  \mathbin{\cup}  \,  \mathit{fv}  (  \Phi_{{\mathrm{2}}}  )  \,  \mathbin{\cup}  \, \ottsym{(}   \mathit{fv}  (  \ottnt{C_{{\mathrm{0}}}}  )  \,  \mathbin{\backslash}  \,  \{  \mathit{x_{{\mathrm{0}}}}  \}   \ottsym{)}$.
          \end{itemize}
          By the IH,
          \[
           \Gamma  \vdash   \ottnt{K'}  [  \ottnt{e}  ]   \ottsym{:}   \ottnt{T_{{\mathrm{1}}}}  \,  \,\&\,  \,  \Phi \,  \tceappendop  \, \Phi_{{\mathrm{1}}}  \, / \,   ( \forall  \mathit{x_{{\mathrm{1}}}}  .  \ottnt{C'_{{\mathrm{0}}}}  )  \Rightarrow   \ottnt{C_{{\mathrm{2}}}}   ~.
          \]
          Without loss of generality, we can suppose that $\mathit{x_{{\mathrm{1}}}} \,  \not\in  \,  \mathit{dom}  (  \Gamma  ) $.
          Because Lemmas~\ref{lem:ts-wf-ty-tctx-typing} and \ref{lem:ts-wf-tctx-wf-ftyping}
          imply $\vdash  \Gamma$,
          we have
          \[
           \Gamma  \ottsym{,}  \mathit{x_{{\mathrm{1}}}} \,  {:}  \, \ottnt{T_{{\mathrm{1}}}}  \vdash  \ottnt{e_{{\mathrm{2}}}}  \ottsym{:}   \ottnt{T_{{\mathrm{0}}}}  \,  \,\&\,  \,  \Phi_{{\mathrm{2}}}  \, / \,   ( \forall  \mathit{x_{{\mathrm{0}}}}  .  \ottnt{C_{{\mathrm{0}}}}  )  \Rightarrow   \ottnt{C'_{{\mathrm{0}}}}  
          \]
          by \reflem{ts-weak-typing}.
          Then, by \T{Let},
          \begin{equation}
           \Gamma  \vdash  \mathsf{let} \, \mathit{x_{{\mathrm{1}}}}  \ottsym{=}   \ottnt{K'}  [  \ottnt{e}  ]  \,  \mathsf{in}  \, \ottnt{e_{{\mathrm{2}}}}  \ottsym{:}   \ottnt{T_{{\mathrm{0}}}}  \,  \,\&\,  \,  \ottsym{(}  \Phi \,  \tceappendop  \, \Phi_{{\mathrm{1}}}  \ottsym{)} \,  \tceappendop  \, \Phi_{{\mathrm{2}}}  \, / \,   ( \forall  \mathit{x_{{\mathrm{0}}}}  .  \ottnt{C_{{\mathrm{0}}}}  )  \Rightarrow   \ottnt{C_{{\mathrm{2}}}}   ~.
            \label{lem:lr-typing-pure-ectx-exp:let:let}
          \end{equation}
          Then, the conclusion
          $\Gamma  \vdash  \mathsf{let} \, \mathit{x_{{\mathrm{1}}}}  \ottsym{=}   \ottnt{K'}  [  \ottnt{e}  ]  \,  \mathsf{in}  \, \ottnt{e_{{\mathrm{2}}}}  \ottsym{:}   \ottnt{T_{{\mathrm{0}}}}  \,  \,\&\,  \,  \Phi \,  \tceappendop  \, \ottsym{(}  \Phi_{{\mathrm{1}}} \,  \tceappendop  \, \Phi_{{\mathrm{2}}}  \ottsym{)}  \, / \,   ( \forall  \mathit{x_{{\mathrm{0}}}}  .  \ottnt{C_{{\mathrm{0}}}}  )  \Rightarrow   \ottnt{C_{{\mathrm{2}}}}  $
          is proven by \T{Sub} with the following.
          \begin{itemize}
           \item We show that
                 \[
                  \Gamma  \vdash   \ottnt{T_{{\mathrm{0}}}}  \,  \,\&\,  \,  \ottsym{(}  \Phi \,  \tceappendop  \, \Phi_{{\mathrm{1}}}  \ottsym{)} \,  \tceappendop  \, \Phi_{{\mathrm{2}}}  \, / \,   ( \forall  \mathit{x_{{\mathrm{0}}}}  .  \ottnt{C_{{\mathrm{0}}}}  )  \Rightarrow   \ottnt{C_{{\mathrm{2}}}}    \ottsym{<:}   \ottnt{T_{{\mathrm{0}}}}  \,  \,\&\,  \,  \Phi \,  \tceappendop  \, \ottsym{(}  \Phi_{{\mathrm{1}}} \,  \tceappendop  \, \Phi_{{\mathrm{2}}}  \ottsym{)}  \, / \,   ( \forall  \mathit{x_{{\mathrm{0}}}}  .  \ottnt{C_{{\mathrm{0}}}}  )  \Rightarrow   \ottnt{C_{{\mathrm{2}}}}   ~.
                 \]
                 \reflem{ts-wf-ty-tctx-typing} with (\ref{lem:lr-typing-pure-ectx-exp:let:let}) implies
                 \begin{itemize}
                  \item $\Gamma  \vdash  \ottnt{T_{{\mathrm{0}}}}$,
                  \item $\Gamma  \ottsym{,}  \mathit{x_{{\mathrm{0}}}} \,  {:}  \, \ottnt{T_{{\mathrm{0}}}}  \vdash  \ottnt{C_{{\mathrm{0}}}}$, and
                  \item $\Gamma  \vdash  \ottnt{C_{{\mathrm{2}}}}$.
                 \end{itemize}
                 Therefore, \reflem{ts-refl-subtyping} implies
                 \begin{itemize}
                  \item $\Gamma  \vdash  \ottnt{T_{{\mathrm{0}}}}  \ottsym{<:}  \ottnt{T_{{\mathrm{0}}}}$,
                  \item $\Gamma  \ottsym{,}  \mathit{x_{{\mathrm{0}}}} \,  {:}  \, \ottnt{T_{{\mathrm{0}}}}  \vdash  \ottnt{C_{{\mathrm{0}}}}  \ottsym{<:}  \ottnt{C_{{\mathrm{0}}}}$, and
                  \item $\Gamma  \vdash  \ottnt{C_{{\mathrm{2}}}}  \ottsym{<:}  \ottnt{C_{{\mathrm{2}}}}$.
                 \end{itemize}
                 Further, $\Gamma  \vdash  \ottsym{(}  \Phi \,  \tceappendop  \, \Phi_{{\mathrm{1}}}  \ottsym{)} \,  \tceappendop  \, \Phi_{{\mathrm{2}}}  \ottsym{<:}  \Phi \,  \tceappendop  \, \ottsym{(}  \Phi_{{\mathrm{1}}} \,  \tceappendop  \, \Phi_{{\mathrm{2}}}  \ottsym{)}$
                 by \reflem{ts-assoc-preeff-compose-eff-subtyping}.
                 Thus, we have the conclusion by \Srule{Ans} and \Srule{Comp}.

           \item We show that
                 \[
                  \Gamma  \vdash   \ottnt{T_{{\mathrm{0}}}}  \,  \,\&\,  \,  \Phi \,  \tceappendop  \, \ottsym{(}  \Phi_{{\mathrm{1}}} \,  \tceappendop  \, \Phi_{{\mathrm{2}}}  \ottsym{)}  \, / \,   ( \forall  \mathit{x_{{\mathrm{0}}}}  .  \ottnt{C_{{\mathrm{0}}}}  )  \Rightarrow   \ottnt{C_{{\mathrm{2}}}}   ~.
                 \]
                 By \WFT{Ans} and \WFT{Comp}, it suffices to show the following.
                 \begin{itemize}
                  \item $\Gamma  \vdash  \ottnt{T_{{\mathrm{0}}}}$, which is shown above.
                  \item We show that $\Gamma  \vdash  \Phi \,  \tceappendop  \, \ottsym{(}  \Phi_{{\mathrm{1}}} \,  \tceappendop  \, \Phi_{{\mathrm{2}}}  \ottsym{)}$.
                        By \reflem{ts-wf-eff-compose},
                        it suffices to show that
                        $\Gamma  \vdash  \Phi$ and $\Gamma  \vdash  \Phi_{{\mathrm{1}}}$ and
                        $\Gamma  \vdash  \Phi_{{\mathrm{2}}}$.
                        \reflem{ts-wf-ty-tctx-typing} with
                        $\Gamma  \vdash  \ottnt{e}  \ottsym{:}   \ottnt{T}  \,  \,\&\,  \,  \Phi  \, / \,   ( \forall  \mathit{x}  .  \ottnt{C_{{\mathrm{1}}}}  )  \Rightarrow   \ottnt{C_{{\mathrm{2}}}}  $
                        implies $\Gamma  \vdash  \Phi$.
                        \reflem{ts-wf-ty-tctx-typing} with
                        $\Gamma  \ottsym{,}  \mathit{x_{{\mathrm{1}}}} \,  {:}  \, \ottnt{T_{{\mathrm{1}}}}  \vdash  \ottnt{e_{{\mathrm{2}}}}  \ottsym{:}   \ottnt{T_{{\mathrm{0}}}}  \,  \,\&\,  \,  \Phi_{{\mathrm{2}}}  \, / \,   ( \forall  \mathit{x_{{\mathrm{0}}}}  .  \ottnt{C_{{\mathrm{0}}}}  )  \Rightarrow   \ottnt{C'_{{\mathrm{0}}}}  $
                        implies $\Gamma  \ottsym{,}  \mathit{x_{{\mathrm{1}}}} \,  {:}  \, \ottnt{T_{{\mathrm{1}}}}  \vdash  \Phi_{{\mathrm{2}}}$.
                        Because $\mathit{x_{{\mathrm{1}}}} \,  \not\in  \,  \mathit{fv}  (  \Phi_{{\mathrm{2}}}  ) $,
                        \reflem{ts-elim-unused-binding-wf-ftyping}
                        implies $\Gamma  \vdash  \Phi_{{\mathrm{2}}}$.
                        Finally, \reflem{lr-typing-cont-wf-eff} with
                        $\ottnt{K'}  \ottsym{:}    \ottnt{T}  \, / \, ( \forall  \mathit{x}  .  \ottnt{C_{{\mathrm{1}}}}  )   \mathrel{ \Rrightarrow }  \ottnt{T_{{\mathrm{1}}}}  \,  \,\&\,  \,  \Phi_{{\mathrm{1}}}  \, / \, ( \forall  \mathit{x_{{\mathrm{1}}}}  .  \ottnt{C'_{{\mathrm{0}}}}  ) $
                        implies $ \emptyset   \vdash  \Phi_{{\mathrm{1}}}$, so
                        $\Gamma  \vdash  \Phi_{{\mathrm{1}}}$ by \reflem{ts-weak-wf-ftyping} with $\vdash  \Gamma$.
                  \item $\Gamma  \ottsym{,}  \mathit{x_{{\mathrm{0}}}} \,  {:}  \, \ottnt{T_{{\mathrm{0}}}}  \vdash  \ottnt{C_{{\mathrm{0}}}}$, which is shown above.
                  \item $\Gamma  \vdash  \ottnt{C_{{\mathrm{2}}}}$, which is shown above.
                 \end{itemize}
          \end{itemize}
        \end{caseanalysis}

  \item Because $\ottnt{K}  \ottsym{:}   \ottnt{T}  \, / \, ( \forall  \mathit{x}  .  \ottnt{C_{{\mathrm{1}}}}  ) $,
        we have $\ottnt{K}  \ottsym{:}    \ottnt{T}  \, / \, ( \forall  \mathit{x}  .  \ottnt{C_{{\mathrm{1}}}}  )   \mathrel{ \Rrightarrow }  \ottnt{T_{{\mathrm{0}}}}  \,  \,\&\,  \,  \Phi_{{\mathrm{0}}}  \, / \, ( \forall  \mathit{x_{{\mathrm{0}}}}  .   \ottnt{T_{{\mathrm{0}}}}  \,  \,\&\,  \,   \Phi_\mathsf{val}   \, / \,   \square    ) $
        for some $\ottnt{T_{{\mathrm{0}}}}$, $\Phi_{{\mathrm{0}}}$, and $\mathit{x_{{\mathrm{0}}}} \,  \not\in  \,  \mathit{fv}  (  \ottnt{T_{{\mathrm{0}}}}  ) $.
        By the case (\ref{lem:lr-typing-pure-ectx-exp:cont}),
        $\Gamma  \vdash   \ottnt{K}  [  \ottnt{e}  ]   \ottsym{:}   \ottnt{T_{{\mathrm{0}}}}  \,  \,\&\,  \,  \Phi \,  \tceappendop  \, \Phi_{{\mathrm{0}}}  \, / \,   ( \forall  \mathit{x_{{\mathrm{0}}}}  .   \ottnt{T_{{\mathrm{0}}}}  \,  \,\&\,  \,   \Phi_\mathsf{val}   \, / \,   \square    )  \Rightarrow   \ottnt{C_{{\mathrm{2}}}}  $.
        Therefore, \T{Reset} implies the conclusion
        $\Gamma  \vdash   \resetcnstr{  \ottnt{K}  [  \ottnt{e}  ]  }   \ottsym{:}  \ottnt{C_{{\mathrm{2}}}}$.
 \end{enumerate}
\end{proof}

\begin{lemmap}{Dependence of Expression Relations on Observational Behavior}{lr-exp-rel-obs-res}
 Let $\ottnt{e}$ and $\ottnt{e'}$ be expressions such that
 \begin{itemize}
  \item $   \forall  \, { \ottnt{E}  \ottsym{,}  \ottnt{r}  \ottsym{,}  \varpi } . \   \ottnt{E}  [  \ottnt{e}  ]  \,  \Downarrow  \, \ottnt{r} \,  \,\&\,  \, \varpi   \mathrel{\Longrightarrow}   \ottnt{E}  [  \ottnt{e'}  ]  \,  \Downarrow  \, \ottnt{r} \,  \,\&\,  \, \varpi $ and
  \item $   \forall  \, { \ottnt{E}  \ottsym{,}  \pi } . \   \ottnt{E}  [  \ottnt{e}  ]  \,  \Uparrow  \, \pi   \mathrel{\Longrightarrow}   \ottnt{E}  [  \ottnt{e'}  ]  \,  \Uparrow  \, \pi $.
 \end{itemize}
 If $\ottnt{e} \,  \in  \,  \LRRel{ \mathrm{Atom} }{ \ottnt{C} } $ and $\ottnt{e'} \,  \in  \,  \LRRel{ \mathcal{E} }{ \ottnt{C} } $, then $\ottnt{e} \,  \in  \,  \LRRel{ \mathcal{E} }{ \ottnt{C} } $.
\end{lemmap}
\begin{proof}
 By induction on $\ottnt{C}$.
 We consider each case on the result of evaluating $\ottnt{e}$.
 \begin{caseanalysis}
  \case $  \exists  \, { \ottnt{v}  \ottsym{,}  \varpi } . \  \ottnt{e} \,  \Downarrow  \, \ottnt{v} \,  \,\&\,  \, \varpi $:
   In this case, we must show that $\ottnt{v} \,  \in  \,  \LRRel{ \mathcal{V} }{  \ottnt{C}  . \ottnt{T}  } $.
   By the assumption, $\ottnt{e'} \,  \Downarrow  \, \ottnt{v} \,  \,\&\,  \, \varpi$.
   Because $\ottnt{e'} \,  \in  \,  \LRRel{ \mathcal{E} }{ \ottnt{C} } $, we have the conclusion.

  \case $  \exists  \, { \ottnt{K_{{\mathrm{0}}}}  \ottsym{,}  \mathit{x_{{\mathrm{0}}}}  \ottsym{,}  \ottnt{e_{{\mathrm{0}}}}  \ottsym{,}  \varpi } . \  \ottnt{e} \,  \Downarrow  \,  \ottnt{K_{{\mathrm{0}}}}  [  \mathcal{S}_0 \, \mathit{x_{{\mathrm{0}}}}  \ottsym{.}  \ottnt{e_{{\mathrm{0}}}}  ]  \,  \,\&\,  \, \varpi $:
   In this case, we must show that
   \begin{itemize}
    \item $ \ottnt{C}  . \ottnt{S}  \,  =  \,  ( \forall  \mathit{x}  .  \ottnt{C_{{\mathrm{1}}}}  )  \Rightarrow   \ottnt{C_{{\mathrm{2}}}} $,
    \item $  \forall  \, { \ottnt{K} \,  \in  \,  \LRRel{ \mathcal{K} }{   \ottnt{C}  . \ottnt{T}   \, / \, ( \forall  \mathit{x}  .  \ottnt{C_{{\mathrm{1}}}}  )  }  } . \   \resetcnstr{  \ottnt{K}  [  \ottnt{e}  ]  }  \,  \in  \,  \LRRel{ \mathcal{E} }{ \ottnt{C_{{\mathrm{2}}}} }  $
   \end{itemize}
   for some $\mathit{x}$, $\ottnt{C_{{\mathrm{1}}}}$, and $\ottnt{C_{{\mathrm{2}}}}$.
   By the assumption, $\ottnt{e'} \,  \Downarrow  \,  \ottnt{K_{{\mathrm{0}}}}  [  \mathcal{S}_0 \, \mathit{x_{{\mathrm{0}}}}  \ottsym{.}  \ottnt{e_{{\mathrm{0}}}}  ]  \,  \,\&\,  \, \varpi$.
   Because $\ottnt{e'} \,  \in  \,  \LRRel{ \mathcal{E} }{ \ottnt{C} } $, we have
   \begin{itemize}
    \item $ \ottnt{C}  . \ottnt{S}  \,  =  \,  ( \forall  \mathit{x}  .  \ottnt{C_{{\mathrm{1}}}}  )  \Rightarrow   \ottnt{C_{{\mathrm{2}}}} $,
    \item $  \forall  \, { \ottnt{K} \,  \in  \,  \LRRel{ \mathcal{K} }{   \ottnt{C}  . \ottnt{T}   \, / \, ( \forall  \mathit{x}  .  \ottnt{C_{{\mathrm{1}}}}  )  }  } . \   \resetcnstr{  \ottnt{K}  [  \ottnt{e'}  ]  }  \,  \in  \,  \LRRel{ \mathcal{E} }{ \ottnt{C_{{\mathrm{2}}}} }  $
   \end{itemize}
   for some $\mathit{x}$, $\ottnt{C_{{\mathrm{1}}}}$, and $\ottnt{C_{{\mathrm{2}}}}$.

   Let $\ottnt{K} \,  \in  \,  \LRRel{ \mathcal{K} }{   \ottnt{C}  . \ottnt{T}   \, / \, ( \forall  \mathit{x}  .  \ottnt{C_{{\mathrm{1}}}}  )  } $.
   It suffices to show that $ \resetcnstr{  \ottnt{K}  [  \ottnt{e}  ]  }  \,  \in  \,  \LRRel{ \mathcal{E} }{ \ottnt{C_{{\mathrm{2}}}} } $.
   By the assumption,
   \begin{itemize}
    \item $   \forall  \, { \ottnt{E}  \ottsym{,}  \ottnt{r}  \ottsym{,}  \varpi } . \   \ottnt{E}  [   \resetcnstr{  \ottnt{K}  [  \ottnt{e}  ]  }   ]  \,  \Downarrow  \, \ottnt{r} \,  \,\&\,  \, \varpi   \mathrel{\Longrightarrow}   \ottnt{E}  [   \resetcnstr{  \ottnt{K}  [  \ottnt{e'}  ]  }   ]  \,  \Downarrow  \, \ottnt{r} \,  \,\&\,  \, \varpi $ and
   \item $   \forall  \, { \ottnt{E}  \ottsym{,}  \pi } . \   \ottnt{E}  [   \resetcnstr{  \ottnt{K}  [  \ottnt{e}  ]  }   ]  \,  \Uparrow  \, \pi   \mathrel{\Longrightarrow}   \ottnt{E}  [   \resetcnstr{  \ottnt{K}  [  \ottnt{e'}  ]  }   ]  \,  \Uparrow  \, \pi $.
   \end{itemize}
   Because
   $\ottnt{K} \,  \in  \,  \LRRel{ \mathcal{K} }{   \ottnt{C}  . \ottnt{T}   \, / \, ( \forall  \mathit{x}  .  \ottnt{C_{{\mathrm{1}}}}  )  } $ implies $\ottnt{K}  \ottsym{:}    \ottnt{C}  . \ottnt{T}   \, / \, ( \forall  \mathit{x}  .  \ottnt{C_{{\mathrm{1}}}}  ) $, and
   $\ottnt{e} \,  \in  \,  \LRRel{ \mathrm{Atom} }{ \ottnt{C} } $ implies $ \emptyset   \vdash  \ottnt{e}  \ottsym{:}  \ottnt{C}$,
   \reflem{lr-typing-cont-exp} implies
   $ \emptyset   \vdash   \resetcnstr{  \ottnt{K}  [  \ottnt{e}  ]  }   \ottsym{:}  \ottnt{C_{{\mathrm{2}}}}$, i.e., $ \resetcnstr{  \ottnt{K}  [  \ottnt{e}  ]  }  \,  \in  \,  \LRRel{ \mathrm{Atom} }{ \ottnt{C_{{\mathrm{2}}}} } $.
   We also have $ \resetcnstr{  \ottnt{K}  [  \ottnt{e'}  ]  }  \,  \in  \,  \LRRel{ \mathcal{E} }{ \ottnt{C_{{\mathrm{2}}}} } $.
   Therefore, by the IH,
   we have the conclusion $ \resetcnstr{  \ottnt{K}  [  \ottnt{e}  ]  }  \,  \in  \,  \LRRel{ \mathcal{E} }{ \ottnt{C_{{\mathrm{2}}}} } $.

  \case $  \exists  \, { \pi } . \  \ottnt{e} \,  \Uparrow  \, \pi $:
   In this case, we must show that $ { \models  \,   { \ottnt{C} }^\nu (  \pi  )  } $.
   By the assumption, $\ottnt{e'} \,  \Uparrow  \, \pi$.
   Because $\ottnt{e'} \,  \in  \,  \LRRel{ \mathcal{E} }{ \ottnt{C} } $, we have the conclusion.
 \end{caseanalysis}
\end{proof}

\begin{lemmap}{Impossibility of Evaluating Shift}{lr-red-shift}
 \begin{enumerate}
  \item For any $\ottnt{K}$, $\mathit{x}$, $\ottnt{e}$, $\ottnt{e'}$, and $\varpi$,
        $ \ottnt{K}  [  \mathcal{S}_0 \, \mathit{x}  \ottsym{.}  \ottnt{e}  ]  \,  \mathrel{  \longrightarrow  }  \, \ottnt{e'} \,  \,\&\,  \, \varpi$ cannot be derived.
  \item For any $\ottnt{K}$, $\mathit{x}$, $\ottnt{e}$, and $\pi$,
        $ \ottnt{K}  [  \mathcal{S}_0 \, \mathit{x}  \ottsym{.}  \ottnt{e}  ]  \,  \Uparrow  \, \pi$ cannot be derived.
 \end{enumerate}
\end{lemmap}
\begin{proof}
 \noindent
 \begin{enumerate}
  \item Straightforward by induction on $\ottnt{K}$.
  \item Suppose that $ \ottnt{K}  [  \mathcal{S}_0 \, \mathit{x}  \ottsym{.}  \ottnt{e}  ]  \,  \Uparrow  \, \pi$.
        Because the first case implies that $ \ottnt{K}  [  \mathcal{S}_0 \, \mathit{x}  \ottsym{.}  \ottnt{e}  ] $ cannot be evaluated,
        $ \ottnt{K}  [  \mathcal{S}_0 \, \mathit{x}  \ottsym{.}  \ottnt{e}  ]  \,  =  \,  \ottnt{E}  [   \bullet ^{ \pi }   ] $ for some $\ottnt{E}$, but
        it does not hold.
 \end{enumerate}
\end{proof}

\begin{lemmap}{Impossibility of Evaluating Values}{lr-red-val}
 \begin{enumerate}
  \item If $\ottnt{e} \,  \mathrel{  \longrightarrow  }  \, \ottnt{e'} \,  \,\&\,  \, \varpi$, then $\ottnt{e}$ is not a value.
  \item If $\ottnt{e} \,  \Uparrow  \, \pi$, then $\ottnt{e}$ is not a value.
 \end{enumerate}
\end{lemmap}
\begin{proof}
 \noindent
 \begin{enumerate}
  \item Straightforward by case analysis on the derivation of $\ottnt{e} \,  \mathrel{  \longrightarrow  }  \, \ottnt{e'} \,  \,\&\,  \, \varpi$.
  \item Suppose that $\ottnt{v} \,  \Uparrow  \, \pi$ for some $\ottnt{v}$.
        Because the first case implies that $\ottnt{v}$ cannot be derived,
        $\ottnt{v} \,  =  \,  \ottnt{E}  [   \bullet ^{ \pi }   ] $ for some $\ottnt{E}$, but it does not hold.
 \end{enumerate}
 
\end{proof}

\begin{lemmap}{Determinism}{lr-eval-det}
 If $\ottnt{e} \,  \mathrel{  \longrightarrow  }  \, \ottnt{e_{{\mathrm{1}}}} \,  \,\&\,  \, \varpi_{{\mathrm{1}}}$ and $\ottnt{e} \,  \mathrel{  \longrightarrow  }  \, \ottnt{e_{{\mathrm{2}}}} \,  \,\&\,  \, \varpi_{{\mathrm{2}}}$,
 then $\ottnt{e_{{\mathrm{1}}}} \,  =  \, \ottnt{e_{{\mathrm{2}}}}$ and $ \varpi_{{\mathrm{1}}}    =    \varpi_{{\mathrm{2}}} $.
\end{lemmap}
\begin{proof}
 By induction on the derivation of $\ottnt{e} \,  \mathrel{  \longrightarrow  }  \, \ottnt{e_{{\mathrm{1}}}} \,  \,\&\,  \, \varpi_{{\mathrm{1}}}$.
 \begin{caseanalysis}
  \case \E{Red} and \E{Ev}:
   The evaluation rule applicable to $\ottnt{e}$ is only the applied evaluation/reduction rule.
   For the cases of \E{Red}/\R{Let} and \E{Red}/\R{Reset}, it is found
   because values cannot be evaluated by \reflem{lr-red-val}; and
   for the case of \E{Red}/\R{Shift}, it is found
   because $ \ottnt{K}  [  \mathcal{S}_0 \, \mathit{x}  \ottsym{.}  \ottnt{e_{{\mathrm{0}}}}  ] $ cannot be evaluated by \reflem{lr-red-shift}.

  \case \E{Let}:
   By \reflem{lr-red-val} and the IH.

  \case \E{Reset}:
   By Lemmas~\ref{lem:lr-red-shift} and \ref{lem:lr-red-val} and the IH.
 \end{caseanalysis}
\end{proof}

\begin{lemmap}{Evaluation of Subterms}{lr-red-subterm}
 \begin{enumerate}
  \item If $\ottnt{e} \,  \mathrel{  \longrightarrow  }  \, \ottnt{e'} \,  \,\&\,  \, \varpi$, then $ \ottnt{E}  [  \ottnt{e}  ]  \,  \mathrel{  \longrightarrow  }  \,  \ottnt{E}  [  \ottnt{e'}  ]  \,  \,\&\,  \, \varpi$.
  \item If $\ottnt{e} \,  \mathrel{  \longrightarrow  }^{ \ottmv{n} }  \, \ottnt{e'} \,  \,\&\,  \, \varpi$, then $ \ottnt{E}  [  \ottnt{e}  ]  \,  \mathrel{  \longrightarrow  }^{ \ottmv{n} }  \,  \ottnt{E}  [  \ottnt{e'}  ]  \,  \,\&\,  \, \varpi$.
 \end{enumerate}
\end{lemmap}
\begin{proof}
 \noindent
 \begin{enumerate}
  \item Straightforward by induction on $\ottnt{E}$.
  \item Straightforward by induction on $\ottmv{n}$ with the first case.
 \end{enumerate}
\end{proof}

\begin{lemmap}{Impossibility of Evaluating Diverging Expressions}{lr-red-div}
 For any $\ottnt{E}$, $\pi$, $\ottnt{e}$, and $\varpi$,
 $ \ottnt{E}  [   \bullet ^{ \pi }   ]  \,  \mathrel{  \longrightarrow  }  \, \ottnt{e} \,  \,\&\,  \, \varpi$ cannot be derived.
\end{lemmap}
\begin{proof}
 Straightforward by induction on $\ottnt{E}$.
\end{proof}

\begin{lemmap}{Pure Computation of Redexes Preserves Behavior}{lr-red-subterm-obs}
 Suppose that $\ottnt{e} \,  \mathrel{  \longrightarrow  }  \, \ottnt{e'} \,  \,\&\,  \,  \epsilon $.
 \begin{enumerate}
  \item If $ \ottnt{E}  [  \ottnt{e}  ]  \,  \Downarrow  \, \ottnt{r} \,  \,\&\,  \, \varpi$, then $ \ottnt{E}  [  \ottnt{e'}  ]  \,  \Downarrow  \, \ottnt{r} \,  \,\&\,  \, \varpi$.
  \item If $ \ottnt{E}  [  \ottnt{e}  ]  \,  \Uparrow  \, \pi$, then $ \ottnt{E}  [  \ottnt{e'}  ]  \,  \Uparrow  \, \pi$.
 \end{enumerate}
\end{lemmap}
\begin{proof}
 \noindent
 \begin{enumerate}
  \item By \reflem{lr-red-subterm}, $ \ottnt{E}  [  \ottnt{e}  ]  \,  \mathrel{  \longrightarrow  }  \,  \ottnt{E}  [  \ottnt{e'}  ]  \,  \,\&\,  \,  \epsilon $.
        Therefore, by Lemmas~\ref{lem:lr-red-val} and \ref{lem:lr-red-shift},
        $ \ottnt{E}  [  \ottnt{e}  ]  \,  \not=  \, \ottnt{r}$.
        Hence, $ \ottnt{E}  [  \ottnt{e}  ]  \,  \Downarrow  \, \ottnt{r} \,  \,\&\,  \, \varpi$ implies
        $ \ottnt{E}  [  \ottnt{e}  ]  \,  \mathrel{  \longrightarrow  }  \, \ottnt{e_{{\mathrm{0}}}} \,  \,\&\,  \, \varpi_{{\mathrm{1}}}$ and $\ottnt{e_{{\mathrm{0}}}} \,  \mathrel{  \longrightarrow  ^*}  \, \ottnt{r} \,  \,\&\,  \, \varpi_{{\mathrm{2}}}$
        for some $\ottnt{e_{{\mathrm{0}}}}$, $\varpi_{{\mathrm{1}}}$, and $\varpi_{{\mathrm{2}}}$ such that $ \varpi    =    \varpi_{{\mathrm{1}}} \,  \tceappendop  \, \varpi_{{\mathrm{2}}} $.
         By \reflem{lr-eval-det}, $\ottnt{e_{{\mathrm{0}}}} \,  =  \,  \ottnt{E}  [  \ottnt{e'}  ] $ and $ \varpi_{{\mathrm{1}}}    =     \epsilon  $.
         Therefore, $ \ottnt{E}  [  \ottnt{e'}  ]  \,  \mathrel{  \longrightarrow  ^*}  \, \ottnt{r} \,  \,\&\,  \, \varpi$, that is, $ \ottnt{E}  [  \ottnt{e'}  ]  \,  \Downarrow  \, \ottnt{r} \,  \,\&\,  \, \varpi$.

  \item By case analysis on the definition of $ \ottnt{E}  [  \ottnt{e}  ]  \,  \Uparrow  \, \pi$.
        \begin{caseanalysis}
         \case $   \forall  \, { \varpi'  \ottsym{,}  \pi' } . \  \pi \,  =  \, \varpi' \,  \tceappendop  \, \pi'   \mathrel{\Longrightarrow}    \exists  \, { \ottnt{e_{{\mathrm{0}}}} } . \   \ottnt{E}  [  \ottnt{e}  ]  \,  \mathrel{  \longrightarrow  ^*}  \, \ottnt{e_{{\mathrm{0}}}} \,  \,\&\,  \, \varpi'  $:
          Let $\varpi'$ be any finite trace and $\pi'$ be any infinite trace
          such that $ \pi    =    \varpi' \,  \tceappendop  \, \pi' $.
          Then, it suffices to show that
          \[
             \exists  \, { \ottnt{e_{{\mathrm{0}}}} } . \   \ottnt{E}  [  \ottnt{e'}  ]  \,  \mathrel{  \longrightarrow  ^*}  \, \ottnt{e_{{\mathrm{0}}}} \,  \,\&\,  \, \varpi'  ~.
          \]

          If $ \varpi'    =     \epsilon  $,
          then we have the conclusion by letting $\ottnt{e_{{\mathrm{0}}}} \,  =  \,  \ottnt{E}  [  \ottnt{e'}  ] $.
          Otherwise, suppose that $ \varpi'    \not=     \epsilon  $.
          By the definition of $ \ottnt{E}  [  \ottnt{e}  ]  \,  \Uparrow  \, \pi$,
          $ \ottnt{E}  [  \ottnt{e}  ]  \,  \mathrel{  \longrightarrow  }^{ \ottmv{n} }  \, \ottnt{e_{{\mathrm{0}}}} \,  \,\&\,  \, \varpi'$ for some $\ottmv{n}$ and $\ottnt{e_{{\mathrm{0}}}}$.
          Because $ \varpi'    \not=     \epsilon  $, $\ottmv{n} \,  >  \, \ottsym{0}$.
          Therefore, $ \ottnt{E}  [  \ottnt{e}  ]  \,  \mathrel{  \longrightarrow  }  \, \ottnt{e_{{\mathrm{1}}}} \,  \,\&\,  \, \varpi'_{{\mathrm{1}}}$ and $\ottnt{e_{{\mathrm{1}}}} \,  \mathrel{  \longrightarrow  }^{ \ottmv{n}  \ottsym{-}  \ottsym{1} }  \, \ottnt{e_{{\mathrm{0}}}} \,  \,\&\,  \, \varpi'_{{\mathrm{2}}}$
          for some $\ottnt{e_{{\mathrm{1}}}}$, $\varpi'_{{\mathrm{1}}}$, and $\varpi'_{{\mathrm{2}}}$ such that
          $ \varpi'    =    \varpi'_{{\mathrm{1}}} \,  \tceappendop  \, \varpi'_{{\mathrm{2}}} $.
          By \reflem{lr-red-subterm} with $\ottnt{e} \,  \mathrel{  \longrightarrow  }  \, \ottnt{e'} \,  \,\&\,  \,  \epsilon $,
          $ \ottnt{E}  [  \ottnt{e}  ]  \,  \mathrel{  \longrightarrow  }  \,  \ottnt{E}  [  \ottnt{e'}  ]  \,  \,\&\,  \,  \epsilon $.
          By \reflem{lr-eval-det}, $\ottnt{e_{{\mathrm{1}}}} \,  =  \,  \ottnt{E}  [  \ottnt{e'}  ] $ and $ \varpi'_{{\mathrm{1}}}    =     \epsilon  $.
          Therefore, $ \ottnt{E}  [  \ottnt{e'}  ]  \,  \mathrel{  \longrightarrow  }^{ \ottmv{n}  \ottsym{-}  \ottsym{1} }  \, \ottnt{e_{{\mathrm{0}}}} \,  \,\&\,  \, \varpi'$.

         \case $ {   \exists  \, { \varpi'  \ottsym{,}  \pi'  \ottsym{,}  \ottnt{E'} } . \  \pi \,  =  \, \varpi' \,  \tceappendop  \, \pi'  } \,\mathrel{\wedge}\, {  \ottnt{E}  [  \ottnt{e}  ]  \,  \mathrel{  \longrightarrow  ^*}  \,  \ottnt{E'}  [   \bullet ^{ \pi' }   ]  \,  \,\&\,  \, \varpi' } $:
          By \reflem{lr-red-subterm}, $ \ottnt{E}  [  \ottnt{e}  ]  \,  \mathrel{  \longrightarrow  }  \,  \ottnt{E}  [  \ottnt{e'}  ]  \,  \,\&\,  \,  \epsilon $.

          If $ \ottnt{E}  [  \ottnt{e}  ]  \,  \mathrel{  \longrightarrow  }^{ \ottsym{0} }  \,  \ottnt{E'}  [   \bullet ^{ \pi' }   ]  \,  \,\&\,  \, \varpi'$,
          then $ \ottnt{E}  [  \ottnt{e}  ]  \,  =  \,  \ottnt{E'}  [   \bullet ^{ \pi' }   ] $, but then
          $ \ottnt{E}  [  \ottnt{e}  ] $ cannot be evaluated by \reflem{lr-red-div}.
          It is contradictory with $ \ottnt{E}  [  \ottnt{e}  ]  \,  \mathrel{  \longrightarrow  }  \,  \ottnt{E}  [  \ottnt{e'}  ]  \,  \,\&\,  \,  \epsilon $.

          Otherwise, suppose that $ \ottnt{E}  [  \ottnt{e}  ]  \,  \mathrel{  \longrightarrow  }^{ \ottmv{n} }  \,  \ottnt{E'}  [   \bullet ^{ \pi' }   ]  \,  \,\&\,  \, \varpi'$
          for some $\ottmv{n} \,  >  \, \ottsym{0}$.
          Then, $ \ottnt{E}  [  \ottnt{e}  ]  \,  \mathrel{  \longrightarrow  }  \, \ottnt{e_{{\mathrm{1}}}} \,  \,\&\,  \, \varpi'_{{\mathrm{1}}}$ and $\ottnt{e_{{\mathrm{1}}}} \,  \mathrel{  \longrightarrow  }^{ \ottmv{n}  \ottsym{-}  \ottsym{1} }  \,  \ottnt{E'}  [   \bullet ^{ \pi' }   ]  \,  \,\&\,  \, \varpi'_{{\mathrm{2}}}$
          for some $\ottnt{e_{{\mathrm{1}}}}$, $\varpi'_{{\mathrm{1}}}$, and $\varpi'_{{\mathrm{2}}}$ such that
          $ \varpi'    =    \varpi'_{{\mathrm{1}}} \,  \tceappendop  \, \varpi'_{{\mathrm{2}}} $.
          By \reflem{lr-eval-det}, $\ottnt{e_{{\mathrm{1}}}} \,  =  \,  \ottnt{E}  [  \ottnt{e'}  ] $ and $ \varpi'_{{\mathrm{1}}}    =     \epsilon  $.
          Therefore, $ \ottnt{E}  [  \ottnt{e'}  ]  \,  \mathrel{  \longrightarrow  }^{ \ottmv{n}  \ottsym{-}  \ottsym{1} }  \,  \ottnt{E'}  [   \bullet ^{ \pi' }   ]  \,  \,\&\,  \, \varpi'$, so
          $ \ottnt{E}  [  \ottnt{e'}  ]  \,  \Uparrow  \, \pi$.
        \end{caseanalysis}
 \end{enumerate}
\end{proof}

\begin{lemmap}{Anti-Reduction of Logical Relation}{lr-anti-red}
 If $ \ottnt{E}  [  \ottnt{e}  ]  \,  \in  \,  \LRRel{ \mathrm{Atom} }{ \ottnt{C} } $ and $\ottnt{e} \,  \mathrel{  \longrightarrow  }  \, \ottnt{e'} \,  \,\&\,  \,  \epsilon $ and $ \ottnt{E}  [  \ottnt{e'}  ]  \,  \in  \,  \LRRel{ \mathcal{E} }{ \ottnt{C} } $,
 then $ \ottnt{E}  [  \ottnt{e}  ]  \,  \in  \,  \LRRel{ \mathcal{E} }{ \ottnt{C} } $.
\end{lemmap}
\begin{proof}
 By Lemmas~\ref{lem:lr-exp-rel-obs-res} and \ref{lem:lr-red-subterm-obs}.
\end{proof}

\begin{lemmap}{Value Relation}{lr-val}
 If $\ottnt{v} \,  \in  \,  \LRRel{ \mathcal{V} }{ \ottnt{T} } $,
 then $\ottnt{v} \,  \in  \,  \LRRel{ \mathcal{E} }{  \ottnt{T}  \,  \,\&\,  \,   \Phi_\mathsf{val}   \, / \,   \square   } $.
\end{lemmap}
\begin{proof}
 Because $\ottnt{v} \,  \in  \,  \LRRel{ \mathcal{V} }{ \ottnt{T} } $ implies $ \emptyset   \vdash  \ottnt{v}  \ottsym{:}  \ottnt{T}$,
 we have $\ottnt{v} \,  \in  \,  \LRRel{ \mathrm{Atom} }{  \ottnt{T}  \,  \,\&\,  \,   \Phi_\mathsf{val}   \, / \,   \square   } $.
 Because \reflem{lr-red-val} implies
 \begin{itemize}
  \item $ {    \forall  \, { \ottnt{r}  \ottsym{,}  \varpi } . \  \ottnt{v} \,  \Downarrow  \, \ottnt{r} \,  \,\&\,  \, \varpi   \mathrel{\Longrightarrow}  \ottnt{r} \,  =  \, \ottnt{v}  } \,\mathrel{\wedge}\, { \varpi \,  =  \,  \epsilon  } $ and
  \item $\ottnt{v} \,  \Uparrow  \, \pi$ cannot be derived for any $\pi$,
 \end{itemize}
 it suffices to show that $\ottnt{v} \,  \in  \,  \LRRel{ \mathcal{V} }{ \ottnt{T} } $, which is the assumption.
\end{proof}

\begin{lemmap}{Compatibility: Variables of Refinement Types}{lr-compat-cvar}
 If $\vdash  \Gamma$ and $\Gamma  \ottsym{(}  \mathit{x}  \ottsym{)} \,  =  \, \ottsym{\{}  \mathit{y} \,  {:}  \, \iota \,  |  \, \phi  \ottsym{\}}$,
 then $\mathit{x} \,  \in  \,  \LRRel{ \mathcal{O} }{ \Gamma  \, \vdash \,  \ottsym{\{}  \mathit{y} \,  {:}  \, \iota \,  |  \,  \mathit{x}    =    \mathit{y}   \ottsym{\}} } $.
\end{lemmap}
\begin{proof}
 By \T{CVar}, $\Gamma  \vdash  \mathit{x}  \ottsym{:}  \ottsym{\{}  \mathit{y} \,  {:}  \, \iota \,  |  \,  \mathit{x}    =    \mathit{y}   \ottsym{\}}$.
 Let $\sigma \,  \in  \,  \LRRel{ \mathcal{G} }{ \Gamma } $.
 We must show that
 \[
  \sigma  \ottsym{(}  \mathit{x}  \ottsym{)} \,  \in  \,  \LRRel{ \mathcal{E} }{  \sigma  \ottsym{(}  \ottsym{\{}  \mathit{y} \,  {:}  \, \iota \,  |  \,  \mathit{x}    =    \mathit{y}   \ottsym{\}}  \ottsym{)}  \,  \,\&\,  \,   \Phi_\mathsf{val}   \, / \,   \square   }  ~.
 \]
 Because $\Gamma  \ottsym{(}  \mathit{x}  \ottsym{)} \,  =  \, \ottsym{\{}  \mathit{y} \,  {:}  \, \iota \,  |  \, \phi  \ottsym{\}}$ and $\sigma \,  \in  \,  \LRRel{ \mathcal{G} }{ \Gamma } $,
 there exists some $\ottnt{v}$ such that
 $\sigma  \ottsym{(}  \mathit{x}  \ottsym{)} \,  =  \, \ottnt{v}$ and $\ottnt{v} \,  \in  \,  \LRRel{ \mathcal{V} }{ \sigma  \ottsym{(}  \ottsym{\{}  \mathit{y} \,  {:}  \, \iota \,  |  \, \phi  \ottsym{\}}  \ottsym{)} } $.
 By \reflem{lr-val}, it suffices to show that
 \[
  \ottnt{v} \,  \in  \,  \LRRel{ \mathcal{V} }{ \sigma  \ottsym{(}  \ottsym{\{}  \mathit{y} \,  {:}  \, \iota \,  |  \,  \mathit{x}    =    \mathit{y}   \ottsym{\}}  \ottsym{)} }  ~.
 \]
 By definition, it suffices to show that
 $ \emptyset   \vdash  \ottnt{v}  \ottsym{:}  \ottsym{\{}  \mathit{y} \,  {:}  \, \iota \,  |  \,   \ottnt{v}     =    \mathit{y}   \ottsym{\}}$.
 Because $\ottnt{v} \,  \in  \,  \LRRel{ \mathcal{V} }{ \sigma  \ottsym{(}  \ottsym{\{}  \mathit{y} \,  {:}  \, \iota \,  |  \, \phi  \ottsym{\}}  \ottsym{)} } $ implies $ \emptyset   \vdash  \ottnt{v}  \ottsym{:}  \sigma  \ottsym{(}  \ottsym{\{}  \mathit{y} \,  {:}  \, \iota \,  |  \, \phi  \ottsym{\}}  \ottsym{)}$,
 \reflem{ts-val-inv-refine-ty} implies $\ottnt{v} \,  =  \, \ottnt{c}$
 for some $\ottnt{c}$ such that $ \mathit{ty}  (  \ottnt{c}  )  \,  =  \, \iota$.
 Therefore, it suffices to show that
 $ \emptyset   \vdash  \ottnt{c}  \ottsym{:}  \ottsym{\{}  \mathit{y} \,  {:}  \, \iota \,  |  \,  \ottnt{c}    =    \mathit{y}   \ottsym{\}}$, which is derived by \T{Const}.
\end{proof}

\begin{lemmap}{Compatibility: Variables of Non-Refinement Types}{lr-compat-var}
 Suppose that $\Gamma  \ottsym{(}  \mathit{x}  \ottsym{)}$ is a value type but not a refinement type.
 If $\vdash  \Gamma$, then $\mathit{x} \,  \in  \,  \LRRel{ \mathcal{O} }{ \Gamma  \, \vdash \,  \Gamma  \ottsym{(}  \mathit{x}  \ottsym{)} } $.
\end{lemmap}
\begin{proof}
 By \T{Var}, $\Gamma  \vdash  \mathit{x}  \ottsym{:}  \Gamma  \ottsym{(}  \mathit{x}  \ottsym{)}$.
 Let $\sigma \,  \in  \,  \LRRel{ \mathcal{G} }{ \Gamma } $.
 We must show that
 \[
  \sigma  \ottsym{(}  \mathit{x}  \ottsym{)} \,  \in  \,  \LRRel{ \mathcal{E} }{  \sigma  \ottsym{(}  \Gamma  \ottsym{(}  \mathit{x}  \ottsym{)}  \ottsym{)}  \,  \,\&\,  \,   \Phi_\mathsf{val}   \, / \,   \square   }  ~.
 \]
 Because $\Gamma  \ottsym{(}  \mathit{x}  \ottsym{)}$ is a value type and $\sigma \,  \in  \,  \LRRel{ \mathcal{G} }{ \Gamma } $,
 there exists some $\ottnt{v}$ such that
 $\sigma  \ottsym{(}  \mathit{x}  \ottsym{)} \,  =  \, \ottnt{v}$ and $\ottnt{v} \,  \in  \,  \LRRel{ \mathcal{V} }{ \sigma  \ottsym{(}  \Gamma  \ottsym{(}  \mathit{x}  \ottsym{)}  \ottsym{)} } $.
 By \reflem{lr-val}, it suffices to show that
 \[
  \ottnt{v} \,  \in  \,  \LRRel{ \mathcal{V} }{ \sigma  \ottsym{(}  \Gamma  \ottsym{(}  \mathit{x}  \ottsym{)}  \ottsym{)} }  ~,
 \]
 which has been obtained already.
\end{proof}

\begin{lemmap}{Compatibility: Constants}{lr-compat-const}
 If $\vdash  \Gamma$ and $ \mathit{ty}  (  \ottnt{c}  )  \,  =  \, \iota$,
 then $\ottnt{c} \,  \in  \,  \LRRel{ \mathcal{O} }{ \Gamma  \, \vdash \,  \ottsym{\{}  \mathit{x} \,  {:}  \, \iota \,  |  \,  \ottnt{c}    =    \mathit{x}   \ottsym{\}} } $.
\end{lemmap}
\begin{proof}
 By \T{Const}, $\Gamma  \vdash  \ottnt{c}  \ottsym{:}  \ottsym{\{}  \mathit{x} \,  {:}  \, \iota \,  |  \,  \ottnt{c}    =    \mathit{x}   \ottsym{\}}$.
 Then, it suffices to show that
 \[
  \ottnt{c} \,  \in  \,  \LRRel{ \mathcal{E} }{  \ottsym{\{}  \mathit{x} \,  {:}  \, \iota \,  |  \,  \ottnt{c}    =    \mathit{x}   \ottsym{\}}  \,  \,\&\,  \,   \Phi_\mathsf{val}   \, / \,   \square   }  ~.
 \]
 By \reflem{lr-val}, it suffices to show that
 \[
  \ottnt{c} \,  \in  \,  \LRRel{ \mathcal{V} }{ \ottsym{\{}  \mathit{x} \,  {:}  \, \iota \,  |  \,  \ottnt{c}    =    \mathit{x}   \ottsym{\}} }  ~.
 \]
 By definition, it suffices to show that $ \emptyset   \vdash  \ottnt{c}  \ottsym{:}  \ottsym{\{}  \mathit{x} \,  {:}  \, \iota \,  |  \,  \ottnt{c}    =    \mathit{x}   \ottsym{\}}$,
 which is derived by \T{Const}.
\end{proof}

\begin{lemmap}{Compatibility: Predicate Abstractions}{lr-compat-pabs}
 If $\ottnt{e} \,  \in  \,  \LRRel{ \mathcal{O} }{ \Gamma  \ottsym{,}  \mathit{X} \,  {:}  \,  \tceseq{ \ottnt{s} }   \, \vdash \,  \ottnt{C} } $, then $ \Lambda   \ottsym{(}  \mathit{X} \,  {:}  \,  \tceseq{ \ottnt{s} }   \ottsym{)}  \ottsym{.}  \ottnt{e} \,  \in  \,  \LRRel{ \mathcal{O} }{ \Gamma  \, \vdash \,   \forall   \ottsym{(}  \mathit{X} \,  {:}  \,  \tceseq{ \ottnt{s} }   \ottsym{)}  \ottsym{.}  \ottnt{C} } $.
\end{lemmap}
\begin{proof}
 Because $\ottnt{e} \,  \in  \,  \LRRel{ \mathcal{O} }{ \Gamma  \ottsym{,}  \mathit{X} \,  {:}  \,  \tceseq{ \ottnt{s} }   \, \vdash \,  \ottnt{C} } $ implies $\Gamma  \ottsym{,}  \mathit{X} \,  {:}  \,  \tceseq{ \ottnt{s} }   \vdash  \ottnt{e}  \ottsym{:}  \ottnt{C}$,
 we have $\Gamma  \vdash   \Lambda   \ottsym{(}  \mathit{X} \,  {:}  \,  \tceseq{ \ottnt{s} }   \ottsym{)}  \ottsym{.}  \ottnt{e}  \ottsym{:}   \forall   \ottsym{(}  \mathit{X} \,  {:}  \,  \tceseq{ \ottnt{s} }   \ottsym{)}  \ottsym{.}  \ottnt{C}$ by \T{PAbs}.
 Let $\sigma \,  \in  \,  \LRRel{ \mathcal{G} }{ \Gamma } $.
 We must show that
 \[
  \sigma  \ottsym{(}   \Lambda   \ottsym{(}  \mathit{X} \,  {:}  \,  \tceseq{ \ottnt{s} }   \ottsym{)}  \ottsym{.}  \ottnt{e}  \ottsym{)} \,  \in  \,  \LRRel{ \mathcal{E} }{  \sigma  \ottsym{(}   \forall   \ottsym{(}  \mathit{X} \,  {:}  \,  \tceseq{ \ottnt{s} }   \ottsym{)}  \ottsym{.}  \ottnt{C}  \ottsym{)}  \,  \,\&\,  \,   \Phi_\mathsf{val}   \, / \,   \square   }  ~.
 \]
 By \reflem{lr-val}, it suffices to show that
 \[
  \sigma  \ottsym{(}   \Lambda   \ottsym{(}  \mathit{X} \,  {:}  \,  \tceseq{ \ottnt{s} }   \ottsym{)}  \ottsym{.}  \ottnt{e}  \ottsym{)} \,  \in  \,  \LRRel{ \mathcal{V} }{ \sigma  \ottsym{(}   \forall   \ottsym{(}  \mathit{X} \,  {:}  \,  \tceseq{ \ottnt{s} }   \ottsym{)}  \ottsym{.}  \ottnt{C}  \ottsym{)} }  ~.
 \]

 Because \reflem{lr-typeability} with $\Gamma  \vdash   \Lambda   \ottsym{(}  \mathit{X} \,  {:}  \,  \tceseq{ \ottnt{s} }   \ottsym{)}  \ottsym{.}  \ottnt{e}  \ottsym{:}   \forall   \ottsym{(}  \mathit{X} \,  {:}  \,  \tceseq{ \ottnt{s} }   \ottsym{)}  \ottsym{.}  \ottnt{C}$
 implies $ \emptyset   \vdash  \sigma  \ottsym{(}   \Lambda   \ottsym{(}  \mathit{X} \,  {:}  \,  \tceseq{ \ottnt{s} }   \ottsym{)}  \ottsym{.}  \ottnt{e}  \ottsym{)}  \ottsym{:}  \sigma  \ottsym{(}   \forall   \ottsym{(}  \mathit{X} \,  {:}  \,  \tceseq{ \ottnt{s} }   \ottsym{)}  \ottsym{.}  \ottnt{C}  \ottsym{)}$,
 we have
 \[
  \sigma  \ottsym{(}   \Lambda   \ottsym{(}  \mathit{X} \,  {:}  \,  \tceseq{ \ottnt{s} }   \ottsym{)}  \ottsym{.}  \ottnt{e}  \ottsym{)} \,  \in  \,  \LRRel{ \mathrm{Atom} }{ \sigma  \ottsym{(}   \forall   \ottsym{(}  \mathit{X} \,  {:}  \,  \tceseq{ \ottnt{s} }   \ottsym{)}  \ottsym{.}  \ottnt{C}  \ottsym{)} }  ~.
 \]

 Let $\ottnt{A}$ be a predicate such that $ \emptyset   \vdash  \ottnt{A}  \ottsym{:}   \tceseq{ \ottnt{s} } $.
 Without loss of generality, we can suppose that
 $\mathit{X}$ does not occur in $\sigma$.
 Then, it suffices to show that
 \[
  \sigma  \ottsym{(}   \Lambda   \ottsym{(}  \mathit{X} \,  {:}  \,  \tceseq{ \ottnt{s} }   \ottsym{)}  \ottsym{.}  \ottnt{e}  \ottsym{)} \, \ottnt{A} \,  \in  \,  \LRRel{ \mathcal{E} }{  \sigma  \ottsym{(}  \ottnt{C}  \ottsym{)}    [  \ottnt{A}  /  \mathit{X}  ]   }  ~.
 \]
 Because $ \emptyset   \vdash  \sigma  \ottsym{(}   \Lambda   \ottsym{(}  \mathit{X} \,  {:}  \,  \tceseq{ \ottnt{s} }   \ottsym{)}  \ottsym{.}  \ottnt{e}  \ottsym{)} \, \ottnt{A}  \ottsym{:}   \sigma  \ottsym{(}  \ottnt{C}  \ottsym{)}    [  \ottnt{A}  /  \mathit{X}  ]  $ by \T{PApp},
 we have
 \[
  \sigma  \ottsym{(}   \Lambda   \ottsym{(}  \mathit{X} \,  {:}  \,  \tceseq{ \ottnt{s} }   \ottsym{)}  \ottsym{.}  \ottnt{e}  \ottsym{)} \, \ottnt{A} \,  \in  \,  \LRRel{ \mathrm{Atom} }{  \sigma  \ottsym{(}  \ottnt{C}  \ottsym{)}    [  \ottnt{A}  /  \mathit{X}  ]   }  ~.
 \]
 Further,
 \[
  \sigma  \ottsym{(}   \Lambda   \ottsym{(}  \mathit{X} \,  {:}  \,  \tceseq{ \ottnt{s} }   \ottsym{)}  \ottsym{.}  \ottnt{e}  \ottsym{)} \, \ottnt{A} \,  \mathrel{  \longrightarrow  }  \,  \sigma  \ottsym{(}  \ottnt{e}  \ottsym{)}    [  \ottnt{A}  /  \mathit{X}  ]   \,  \,\&\,  \,  \epsilon 
 \]
 by \E{Red}/\R{PBeta}.
 Therefore, \reflem{lr-anti-red} implies that
 it suffices to show that
 \[
   \sigma  \ottsym{(}  \ottnt{e}  \ottsym{)}    [  \ottnt{A}  /  \mathit{X}  ]   \,  \in  \,  \LRRel{ \mathcal{E} }{  \sigma  \ottsym{(}  \ottnt{C}  \ottsym{)}    [  \ottnt{A}  /  \mathit{X}  ]   }  ~.
 \]
 Because $\ottnt{e} \,  \in  \,  \LRRel{ \mathcal{O} }{ \Gamma  \ottsym{,}  \mathit{X} \,  {:}  \,  \tceseq{ \ottnt{s} }   \, \vdash \,  \ottnt{C} } $, and
 $\sigma \,  \in  \,  \LRRel{ \mathcal{G} }{ \Gamma } $ and $ \emptyset   \vdash  \ottnt{A}  \ottsym{:}   \tceseq{ \ottnt{s} } $ imply
 $\sigma  \mathbin{\uplus}   [  \ottnt{A}  /  \mathit{X}  ]  \,  \in  \,  \LRRel{ \mathcal{G} }{ \Gamma  \ottsym{,}  \mathit{X} \,  {:}  \,  \tceseq{ \ottnt{s} }  } $,
 we have the conclusion.
\end{proof}

\begin{lemmap}{Compatibility: Lambda Abstractions}{lr-compat-abs}
 If $\ottnt{e} \,  \in  \,  \LRRel{ \mathcal{O} }{ \Gamma  \ottsym{,}   \tceseq{ \mathit{x}  \,  {:}  \,  \ottnt{T} }   \, \vdash \,  \ottnt{C} } $, then $ \lambda\!  \,  \tceseq{ \mathit{x} }   \ottsym{.}  \ottnt{e} \,  \in  \,  \LRRel{ \mathcal{O} }{ \Gamma  \, \vdash \,  \ottsym{(}   \tceseq{ \mathit{x}  \,  {:}  \,  \ottnt{T} }   \ottsym{)}  \rightarrow  \ottnt{C} } $.
\end{lemmap}
\begin{proof}
 By \T{Abs}, $\Gamma  \vdash   \lambda\!  \,  \tceseq{ \mathit{x} }   \ottsym{.}  \ottnt{e}  \ottsym{:}  \ottsym{(}   \tceseq{ \mathit{x}  \,  {:}  \,  \ottnt{T} }   \ottsym{)}  \rightarrow  \ottnt{C}$.
 Let $\sigma \,  \in  \,  \LRRel{ \mathcal{G} }{ \Gamma } $.
 Then, we must show that
 \[
  \sigma  \ottsym{(}   \lambda\!  \,  \tceseq{ \mathit{x} }   \ottsym{.}  \ottnt{e}  \ottsym{)} \,  \in  \,  \LRRel{ \mathcal{E} }{  \sigma  \ottsym{(}  \ottsym{(}   \tceseq{ \mathit{x}  \,  {:}  \,  \ottnt{T} }   \ottsym{)}  \rightarrow  \ottnt{C}  \ottsym{)}  \,  \,\&\,  \,   \Phi_\mathsf{val}   \, / \,   \square   }  ~.
 \]
 By \reflem{lr-val}, it suffices to show that
 \[
  \sigma  \ottsym{(}   \lambda\!  \,  \tceseq{ \mathit{x} }   \ottsym{.}  \ottnt{e}  \ottsym{)} \,  \in  \,  \LRRel{ \mathcal{V} }{ \sigma  \ottsym{(}  \ottsym{(}   \tceseq{ \mathit{x}  \,  {:}  \,  \ottnt{T} }   \ottsym{)}  \rightarrow  \ottnt{C}  \ottsym{)} }  ~.
 \]
 By \reflem{lr-typeability}, $\sigma  \ottsym{(}   \lambda\!  \,  \tceseq{ \mathit{x} }   \ottsym{.}  \ottnt{e}  \ottsym{)} \,  \in  \,  \LRRel{ \mathrm{Atom} }{ \sigma  \ottsym{(}  \ottsym{(}   \tceseq{ \mathit{x}  \,  {:}  \,  \ottnt{T} }   \ottsym{)}  \rightarrow  \ottnt{C}  \ottsym{)} } $.
 Without loss of generality, we can suppose that
 $ \tceseq{ \mathit{x} } $ do not occur in $\sigma$.

 Let $ \tceseq{ \ottnt{v} } $ be values such that $ \tceseq{ \ottnt{v} \,  \in  \,  \LRRel{ \mathcal{V} }{  \sigma  \ottsym{(}  \ottnt{T}  \ottsym{)}    [ \tceseq{ \ottnt{v}  /  \mathit{x} } ]   }  } $.
 Then, by definition and \reflem{lr-red-val}, it suffices to show that
 \[
  \sigma  \ottsym{(}   \lambda\!  \,  \tceseq{ \mathit{x} }   \ottsym{.}  \ottnt{e}  \ottsym{)} \,  \tceseq{ \ottnt{v} }  \,  \in  \,  \LRRel{ \mathcal{E} }{  \sigma  \ottsym{(}  \ottnt{C}  \ottsym{)}    [ \tceseq{ \ottnt{v}  /  \mathit{x} } ]   }  ~,
 \]
 which is implied by \reflem{lr-anti-red} with the following.
 \begin{itemize}
  \item $ \emptyset   \vdash  \sigma  \ottsym{(}   \lambda\!  \,  \tceseq{ \mathit{x} }   \ottsym{.}  \ottnt{e}  \ottsym{)} \,  \tceseq{ \ottnt{v} }   \ottsym{:}   \sigma  \ottsym{(}  \ottnt{C}  \ottsym{)}    [ \tceseq{ \ottnt{v}  /  \mathit{x} } ]  $ by \T{App}.
  \item $\sigma  \ottsym{(}   \lambda\!  \,  \tceseq{ \mathit{x} }   \ottsym{.}  \ottnt{e}  \ottsym{)} \,  \tceseq{ \ottnt{v} }  \,  \mathrel{  \longrightarrow  }  \,  \sigma  \ottsym{(}  \ottnt{e}  \ottsym{)}    [ \tceseq{ \ottnt{v}  /  \mathit{x} } ]   \,  \,\&\,  \,  \epsilon $
        by \E{Red}/\R{Beta}.
  \item $ \sigma  \ottsym{(}  \ottnt{e}  \ottsym{)}    [ \tceseq{ \ottnt{v}  /  \mathit{x} } ]   \,  \in  \,  \LRRel{ \mathcal{E} }{  \sigma  \ottsym{(}  \ottnt{C}  \ottsym{)}    [ \tceseq{ \ottnt{v}  /  \mathit{x} } ]   } $
        by $\ottnt{e} \,  \in  \,  \LRRel{ \mathcal{O} }{ \Gamma  \ottsym{,}   \tceseq{ \mathit{x}  \,  {:}  \,  \ottnt{T} }   \, \vdash \,  \ottnt{C} } $ and $\sigma  \mathbin{\uplus}   [ \tceseq{ \ottnt{v}  /  \mathit{x} } ]  \,  \in  \,  \LRRel{ \mathcal{G} }{ \Gamma  \ottsym{,}   \tceseq{ \mathit{x}  \,  {:}  \,  \ottnt{T} }  } $.
 \end{itemize}
\end{proof}

\begin{lemmap}{Strengthening Type Hypothesis in Typing}{ts-strength-ty-hypo-typing}
 If $\Gamma_{{\mathrm{1}}}  \vdash  \ottnt{T_{{\mathrm{01}}}}$ and $\Gamma_{{\mathrm{1}}}  \vdash  \ottnt{T_{{\mathrm{01}}}}  \ottsym{<:}  \ottnt{T_{{\mathrm{02}}}}$ and $\Gamma_{{\mathrm{1}}}  \ottsym{,}  \mathit{x} \,  {:}  \, \ottnt{T_{{\mathrm{02}}}}  \ottsym{,}  \Gamma_{{\mathrm{2}}}  \vdash  \ottnt{e}  \ottsym{:}  \ottnt{C}$, then $\Gamma_{{\mathrm{1}}}  \ottsym{,}  \mathit{x} \,  {:}  \, \ottnt{T_{{\mathrm{01}}}}  \ottsym{,}  \Gamma_{{\mathrm{2}}}  \vdash  \ottnt{e}  \ottsym{:}  \ottnt{C}$.
\end{lemmap}
\begin{proof}
 By induction on the derivation of $\Gamma_{{\mathrm{1}}}  \ottsym{,}  \mathit{x} \,  {:}  \, \ottnt{T_{{\mathrm{02}}}}  \ottsym{,}  \Gamma_{{\mathrm{2}}}  \vdash  \ottnt{e}  \ottsym{:}  \ottnt{C}$
 with \reflem{ts-weak-strength-ty-hypo-wf-ftyping}.
 We mention only the interesting cases.
 \begin{caseanalysis}
  \case \T{Fun}: By the IH and \reflem{ts-strength-ty-hypo-valid}.
  \case \T{Sub}: By the IH and Lemmas~\ref{lem:ts-strength-ty-hypo-subtyping} and \ref{lem:ts-weak-strength-ty-hypo-wf-ftyping}.
  \case \T{Div}: By Lemmas~\ref{lem:ts-weak-strength-ty-hypo-wf-ftyping} and \ref{lem:ts-strength-ty-hypo-valid}.
 \end{caseanalysis}
\end{proof}

\begin{lemmap}{Subsumption for Continuations}{lr-cont-subsump}
 Suppose that
 $\mathit{x} \,  {:}  \, \ottnt{T}  \vdash  \ottnt{C}$ and
 $\mathit{x} \,  {:}  \, \ottnt{T'}  \vdash  \ottnt{C'}$ and
 $ \emptyset   \vdash  \ottnt{T'}  \ottsym{<:}  \ottnt{T}$ and
 $\mathit{x} \,  {:}  \, \ottnt{T'}  \vdash  \ottnt{C}  \ottsym{<:}  \ottnt{C'}$.

 \begin{enumerate}
  \item \label{lem:lr-cont-subsump:cont-typing}
        If
        $\ottnt{K}  \ottsym{:}    \ottnt{T}  \, / \, ( \forall  \mathit{x}  .  \ottnt{C}  )   \mathrel{ \Rrightarrow }  \ottnt{T_{{\mathrm{0}}}}  \,  \,\&\,  \,  \Phi_{{\mathrm{0}}}  \, / \, ( \forall  \mathit{x_{{\mathrm{0}}}}  .  \ottnt{C_{{\mathrm{0}}}}  ) $,
        then
        $\ottnt{K}  \ottsym{:}    \ottnt{T'}  \, / \, ( \forall  \mathit{x}  .  \ottnt{C'}  )   \mathrel{ \Rrightarrow }  \ottnt{T'_{{\mathrm{0}}}}  \,  \,\&\,  \,  \Phi_{{\mathrm{0}}}  \, / \, ( \forall  \mathit{x_{{\mathrm{0}}}}  .  \ottnt{C_{{\mathrm{0}}}}  ) $
        for some $\ottnt{T'_{{\mathrm{0}}}}$ such that
        $ \emptyset   \vdash  \ottnt{T'_{{\mathrm{0}}}}  \ottsym{<:}  \ottnt{T_{{\mathrm{0}}}}$ and $ \emptyset   \vdash  \ottnt{T'_{{\mathrm{0}}}}$.

  \item If $\ottnt{K}  \ottsym{:}   \ottnt{T}  \, / \, ( \forall  \mathit{x}  .  \ottnt{C}  ) $,
        then $\ottnt{K}  \ottsym{:}   \ottnt{T'}  \, / \, ( \forall  \mathit{x}  .  \ottnt{C'}  ) $.
 \end{enumerate}
\end{lemmap}
\begin{proof}
 \noindent
 \begin{enumerate}
  \item By induction on the derivation of $\ottnt{K}  \ottsym{:}    \ottnt{T}  \, / \, ( \forall  \mathit{x}  .  \ottnt{C}  )   \mathrel{ \Rrightarrow }  \ottnt{T_{{\mathrm{0}}}}  \,  \,\&\,  \,  \Phi_{{\mathrm{0}}}  \, / \, ( \forall  \mathit{x_{{\mathrm{0}}}}  .  \ottnt{C_{{\mathrm{0}}}}  ) $.
        In what follows, we use the fact that $ \emptyset   \vdash  \ottnt{T'}$,
        which is derived by \reflem{ts-wf-tctx-wf-ftyping} with $\mathit{x} \,  {:}  \, \ottnt{T'}  \vdash  \ottnt{C'}$.
        \begin{caseanalysis}
         \case \TK{Empty}:
          We are given
          \[
           \,[\,]\,  \ottsym{:}    \ottnt{T}  \, / \, ( \forall  \mathit{x}  .  \ottnt{C}  )   \mathrel{ \Rrightarrow }  \ottnt{T}  \,  \,\&\,  \,   \Phi_\mathsf{val}   \, / \, ( \forall  \mathit{x}  .  \ottnt{C}  ) 
          \]
          where
          \begin{itemize}
           \item $\ottnt{K} \,  =  \, \,[\,]\,$,
           \item $\ottnt{T_{{\mathrm{0}}}} \,  =  \, \ottnt{T}$,
           \item $\Phi_{{\mathrm{0}}} \,  =  \,  \Phi_\mathsf{val} $,
           \item $ \mathit{x_{{\mathrm{0}}}}    =    \mathit{x} $, and
           \item $\ottnt{C_{{\mathrm{0}}}} \,  =  \, \ottnt{C}$.
          \end{itemize}
          By \TK{Empty}, we have
          \[
           \,[\,]\,  \ottsym{:}    \ottnt{T'}  \, / \, ( \forall  \mathit{x}  .  \ottnt{C'}  )   \mathrel{ \Rrightarrow }  \ottnt{T'}  \,  \,\&\,  \,   \Phi_\mathsf{val}   \, / \, ( \forall  \mathit{x}  .  \ottnt{C'}  )  ~.
          \]
          Let $\ottnt{T'_{{\mathrm{0}}}} \,  =  \, \ottnt{T'}$.
          Because $ \emptyset   \vdash  \ottnt{T'}  \ottsym{<:}  \ottnt{T}$ and $ \emptyset   \vdash  \ottnt{T'}$,
          we have $ \emptyset   \vdash  \ottnt{T'_{{\mathrm{0}}}}  \ottsym{<:}  \ottnt{T_{{\mathrm{0}}}}$ and $ \emptyset   \vdash  \ottnt{T'_{{\mathrm{0}}}}$.
          We must show that
          \[
           \,[\,]\,  \ottsym{:}    \ottnt{T'}  \, / \, ( \forall  \mathit{x}  .  \ottnt{C'}  )   \mathrel{ \Rrightarrow }  \ottnt{T'}  \,  \,\&\,  \,   \Phi_\mathsf{val}   \, / \, ( \forall  \mathit{x}  .  \ottnt{C}  )  ~.
          \]
          By \TK{Sub}, it suffices to show the following.
          \begin{itemize}
           \item $\mathit{x} \,  {:}  \, \ottnt{T'}  \vdash  \ottnt{C}  \ottsym{<:}  \ottnt{C'}$, which is assumed.
           \item $\mathit{x} \,  {:}  \, \ottnt{T'}  \vdash  \ottnt{C}$, which is shown by
                 \reflem{ts-weak-strength-ty-hypo-wf-ftyping}
                 with $\mathit{x} \,  {:}  \, \ottnt{T}  \vdash  \ottnt{C}$ and $ \emptyset   \vdash  \ottnt{T'}  \ottsym{<:}  \ottnt{T}$ and $ \emptyset   \vdash  \ottnt{T'}$.
          \end{itemize}

         \case \TK{Sub}:
          We are given
          \begin{prooftree}
           \AxiomC{$\ottnt{K}  \ottsym{:}    \ottnt{T}  \, / \, ( \forall  \mathit{x}  .  \ottnt{C}  )   \mathrel{ \Rrightarrow }  \ottnt{T_{{\mathrm{0}}}}  \,  \,\&\,  \,  \Phi_{{\mathrm{0}}}  \, / \, ( \forall  \mathit{x_{{\mathrm{0}}}}  .  \ottnt{C'_{{\mathrm{0}}}}  ) $}
           \AxiomC{$\mathit{x_{{\mathrm{0}}}} \,  {:}  \, \ottnt{T_{{\mathrm{0}}}}  \vdash  \ottnt{C_{{\mathrm{0}}}}  \ottsym{<:}  \ottnt{C'_{{\mathrm{0}}}}$}
           \AxiomC{$\mathit{x_{{\mathrm{0}}}} \,  {:}  \, \ottnt{T_{{\mathrm{0}}}}  \vdash  \ottnt{C_{{\mathrm{0}}}}$}
           \TrinaryInfC{$\ottnt{K}  \ottsym{:}    \ottnt{T}  \, / \, ( \forall  \mathit{x}  .  \ottnt{C}  )   \mathrel{ \Rrightarrow }  \ottnt{T_{{\mathrm{0}}}}  \,  \,\&\,  \,  \Phi_{{\mathrm{0}}}  \, / \, ( \forall  \mathit{x_{{\mathrm{0}}}}  .  \ottnt{C_{{\mathrm{0}}}}  ) $}
          \end{prooftree}
          for some $\ottnt{C'_{{\mathrm{0}}}}$.
          By the IH,
          \[
           \ottnt{K}  \ottsym{:}    \ottnt{T'}  \, / \, ( \forall  \mathit{x}  .  \ottnt{C'}  )   \mathrel{ \Rrightarrow }  \ottnt{T'_{{\mathrm{0}}}}  \,  \,\&\,  \,  \Phi_{{\mathrm{0}}}  \, / \, ( \forall  \mathit{x_{{\mathrm{0}}}}  .  \ottnt{C'_{{\mathrm{0}}}}  ) 
          \]
          for some $\ottnt{T'_{{\mathrm{0}}}}$ such that $ \emptyset   \vdash  \ottnt{T'_{{\mathrm{0}}}}  \ottsym{<:}  \ottnt{T_{{\mathrm{0}}}}$ and $ \emptyset   \vdash  \ottnt{T'_{{\mathrm{0}}}}$.
          \reflem{ts-strength-ty-hypo-subtyping} with $\mathit{x_{{\mathrm{0}}}} \,  {:}  \, \ottnt{T_{{\mathrm{0}}}}  \vdash  \ottnt{C_{{\mathrm{0}}}}  \ottsym{<:}  \ottnt{C'_{{\mathrm{0}}}}$ implies
          \[
           \mathit{x_{{\mathrm{0}}}} \,  {:}  \, \ottnt{T'_{{\mathrm{0}}}}  \vdash  \ottnt{C_{{\mathrm{0}}}}  \ottsym{<:}  \ottnt{C'_{{\mathrm{0}}}} ~.
          \]
          \reflem{ts-weak-strength-ty-hypo-wf-ftyping} with
          $\mathit{x_{{\mathrm{0}}}} \,  {:}  \, \ottnt{T_{{\mathrm{0}}}}  \vdash  \ottnt{C_{{\mathrm{0}}}}$ and $ \emptyset   \vdash  \ottnt{T'_{{\mathrm{0}}}}$ implies
          \[
           \mathit{x_{{\mathrm{0}}}} \,  {:}  \, \ottnt{T'_{{\mathrm{0}}}}  \vdash  \ottnt{C_{{\mathrm{0}}}} ~.
          \]
          Therefore, \TK{Sub} implies the conclusion
          $\ottnt{K}  \ottsym{:}    \ottnt{T'}  \, / \, ( \forall  \mathit{x}  .  \ottnt{C'}  )   \mathrel{ \Rrightarrow }  \ottnt{T'_{{\mathrm{0}}}}  \,  \,\&\,  \,  \Phi_{{\mathrm{0}}}  \, / \, ( \forall  \mathit{x_{{\mathrm{0}}}}  .  \ottnt{C_{{\mathrm{0}}}}  ) $.

         \case \TK{Let}:
          We are given
          \begin{prooftree}
           \AxiomC{$\ottnt{K'}  \ottsym{:}    \ottnt{T}  \, / \, ( \forall  \mathit{x}  .  \ottnt{C}  )   \mathrel{ \Rrightarrow }  \ottnt{T_{{\mathrm{1}}}}  \,  \,\&\,  \,  \Phi_{{\mathrm{1}}}  \, / \, ( \forall  \mathit{x_{{\mathrm{1}}}}  .  \ottnt{C'_{{\mathrm{0}}}}  ) $}
           \AxiomC{$\mathit{x_{{\mathrm{1}}}} \,  {:}  \, \ottnt{T_{{\mathrm{1}}}}  \vdash  \ottnt{e_{{\mathrm{2}}}}  \ottsym{:}   \ottnt{T_{{\mathrm{0}}}}  \,  \,\&\,  \,  \Phi_{{\mathrm{2}}}  \, / \,   ( \forall  \mathit{x_{{\mathrm{0}}}}  .  \ottnt{C_{{\mathrm{0}}}}  )  \Rightarrow   \ottnt{C'_{{\mathrm{0}}}}  $}
           \BinaryInfC{$\mathsf{let} \, \mathit{x_{{\mathrm{1}}}}  \ottsym{=}  \ottnt{K'} \,  \mathsf{in}  \, \ottnt{e_{{\mathrm{2}}}}  \ottsym{:}    \ottnt{T}  \, / \, ( \forall  \mathit{x}  .  \ottnt{C}  )   \mathrel{ \Rrightarrow }  \ottnt{T_{{\mathrm{0}}}}  \,  \,\&\,  \,  \Phi_{{\mathrm{1}}} \,  \tceappendop  \, \Phi_{{\mathrm{2}}}  \, / \, ( \forall  \mathit{x_{{\mathrm{0}}}}  .  \ottnt{C_{{\mathrm{0}}}}  ) $}
          \end{prooftree}
          for some $\ottnt{K'}$, $\mathit{x_{{\mathrm{1}}}}$, $\ottnt{e_{{\mathrm{2}}}}$, $\ottnt{T_{{\mathrm{1}}}}$, $\Phi_{{\mathrm{1}}}$, $\Phi_{{\mathrm{2}}}$, and $\ottnt{C'_{{\mathrm{0}}}}$
          such that
          \begin{itemize}
           \item $\ottnt{K} \,  =  \, \ottsym{(}  \mathsf{let} \, \mathit{x_{{\mathrm{1}}}}  \ottsym{=}  \ottnt{K'} \,  \mathsf{in}  \, \ottnt{e_{{\mathrm{2}}}}  \ottsym{)}$,
           \item $\Phi_{{\mathrm{0}}} \,  =  \, \Phi_{{\mathrm{1}}} \,  \tceappendop  \, \Phi_{{\mathrm{2}}}$, and
           \item $\mathit{x_{{\mathrm{1}}}} \,  \not\in  \,  \mathit{fv}  (  \ottnt{T_{{\mathrm{0}}}}  )  \,  \mathbin{\cup}  \,  \mathit{fv}  (  \Phi_{{\mathrm{2}}}  )  \,  \mathbin{\cup}  \, \ottsym{(}   \mathit{fv}  (  \ottnt{C_{{\mathrm{0}}}}  )  \,  \mathbin{\backslash}  \,  \{  \mathit{x_{{\mathrm{0}}}}  \}   \ottsym{)}$.
          \end{itemize}
          By the IH,
          \[
           \ottnt{K'}  \ottsym{:}    \ottnt{T'}  \, / \, ( \forall  \mathit{x}  .  \ottnt{C'}  )   \mathrel{ \Rrightarrow }  \ottnt{T'_{{\mathrm{1}}}}  \,  \,\&\,  \,  \Phi_{{\mathrm{1}}}  \, / \, ( \forall  \mathit{x_{{\mathrm{1}}}}  .  \ottnt{C'_{{\mathrm{0}}}}  ) 
          \]
          for some $\ottnt{T'_{{\mathrm{1}}}}$ such that $ \emptyset   \vdash  \ottnt{T'_{{\mathrm{1}}}}  \ottsym{<:}  \ottnt{T_{{\mathrm{1}}}}$ and $ \emptyset   \vdash  \ottnt{T'_{{\mathrm{1}}}}$.
          By \reflem{ts-strength-ty-hypo-typing} with
          $ \emptyset   \vdash  \ottnt{T'_{{\mathrm{1}}}}  \ottsym{<:}  \ottnt{T_{{\mathrm{1}}}}$ and
          $\mathit{x_{{\mathrm{1}}}} \,  {:}  \, \ottnt{T_{{\mathrm{1}}}}  \vdash  \ottnt{e_{{\mathrm{2}}}}  \ottsym{:}   \ottnt{T_{{\mathrm{0}}}}  \,  \,\&\,  \,  \Phi_{{\mathrm{2}}}  \, / \,   ( \forall  \mathit{x_{{\mathrm{0}}}}  .  \ottnt{C_{{\mathrm{0}}}}  )  \Rightarrow   \ottnt{C'_{{\mathrm{0}}}}  $,
          we have
          \[
           \mathit{x_{{\mathrm{1}}}} \,  {:}  \, \ottnt{T'_{{\mathrm{1}}}}  \vdash  \ottnt{e_{{\mathrm{2}}}}  \ottsym{:}   \ottnt{T_{{\mathrm{0}}}}  \,  \,\&\,  \,  \Phi_{{\mathrm{2}}}  \, / \,   ( \forall  \mathit{x_{{\mathrm{0}}}}  .  \ottnt{C_{{\mathrm{0}}}}  )  \Rightarrow   \ottnt{C'_{{\mathrm{0}}}}   ~.
          \]
          Therefore, \TK{Let} implies
          \[
           \mathsf{let} \, \mathit{x_{{\mathrm{1}}}}  \ottsym{=}  \ottnt{K'} \,  \mathsf{in}  \, \ottnt{e_{{\mathrm{2}}}}  \ottsym{:}    \ottnt{T'}  \, / \, ( \forall  \mathit{x}  .  \ottnt{C'}  )   \mathrel{ \Rrightarrow }  \ottnt{T_{{\mathrm{0}}}}  \,  \,\&\,  \,  \Phi_{{\mathrm{1}}} \,  \tceappendop  \, \Phi_{{\mathrm{2}}}  \, / \, ( \forall  \mathit{x_{{\mathrm{0}}}}  .  \ottnt{C_{{\mathrm{0}}}}  )  ~.
          \]
          By letting $\ottnt{T'_{{\mathrm{0}}}} \,  =  \, \ottnt{T_{{\mathrm{0}}}}$, finally we must show that
          $ \emptyset   \vdash  \ottnt{T_{{\mathrm{0}}}}  \ottsym{<:}  \ottnt{T_{{\mathrm{0}}}}$ and $ \emptyset   \vdash  \ottnt{T_{{\mathrm{0}}}}$.
          By \reflem{ts-refl-subtyping}, it suffices to show only that $ \emptyset   \vdash  \ottnt{T_{{\mathrm{0}}}}$.
          \reflem{ts-wf-ty-tctx-typing} with $\mathit{x_{{\mathrm{1}}}} \,  {:}  \, \ottnt{T_{{\mathrm{1}}}}  \vdash  \ottnt{e_{{\mathrm{2}}}}  \ottsym{:}   \ottnt{T_{{\mathrm{0}}}}  \,  \,\&\,  \,  \Phi_{{\mathrm{2}}}  \, / \,   ( \forall  \mathit{x_{{\mathrm{0}}}}  .  \ottnt{C_{{\mathrm{0}}}}  )  \Rightarrow   \ottnt{C'_{{\mathrm{0}}}}  $
          implies $\mathit{x_{{\mathrm{1}}}} \,  {:}  \, \ottnt{T_{{\mathrm{1}}}}  \vdash  \ottnt{T_{{\mathrm{0}}}}$.
          Because $\mathit{x_{{\mathrm{1}}}} \,  \not\in  \,  \mathit{fv}  (  \ottnt{T_{{\mathrm{0}}}}  ) $,
          \reflem{ts-elim-unused-binding-wf-ftyping} implies
          $ \emptyset   \vdash  \ottnt{T_{{\mathrm{0}}}}$.
        \end{caseanalysis}

  \item Because $\ottnt{K}  \ottsym{:}   \ottnt{T}  \, / \, ( \forall  \mathit{x}  .  \ottnt{C}  ) $, we have
        $\ottnt{K}  \ottsym{:}    \ottnt{T}  \, / \, ( \forall  \mathit{x}  .  \ottnt{C}  )   \mathrel{ \Rrightarrow }  \ottnt{T_{{\mathrm{0}}}}  \,  \,\&\,  \,  \Phi_{{\mathrm{0}}}  \, / \, ( \forall  \mathit{x_{{\mathrm{0}}}}  .   \ottnt{T_{{\mathrm{0}}}}  \,  \,\&\,  \,   \Phi_\mathsf{val}   \, / \,   \square    ) $
        for some $\ottnt{T_{{\mathrm{0}}}}$, $\Phi_{{\mathrm{0}}}$, and $\mathit{x_{{\mathrm{0}}}} \,  \not\in  \,  \mathit{fv}  (  \ottnt{T_{{\mathrm{0}}}}  ) $.
        By the case (\ref{lem:lr-cont-subsump:cont-typing}),
        \[
         \ottnt{K}  \ottsym{:}    \ottnt{T'}  \, / \, ( \forall  \mathit{x}  .  \ottnt{C'}  )   \mathrel{ \Rrightarrow }  \ottnt{T'_{{\mathrm{0}}}}  \,  \,\&\,  \,  \Phi_{{\mathrm{0}}}  \, / \, ( \forall  \mathit{x_{{\mathrm{0}}}}  .   \ottnt{T_{{\mathrm{0}}}}  \,  \,\&\,  \,   \Phi_\mathsf{val}   \, / \,   \square    ) 
        \]
        for some $\ottnt{T'_{{\mathrm{0}}}}$ such that $ \emptyset   \vdash  \ottnt{T'_{{\mathrm{0}}}}  \ottsym{<:}  \ottnt{T_{{\mathrm{0}}}}$ and $ \emptyset   \vdash  \ottnt{T'_{{\mathrm{0}}}}$.
        Then, by \TK{Sub}, it suffices to show the following.
        \begin{itemize}
         \item $\mathit{x_{{\mathrm{0}}}} \,  {:}  \, \ottnt{T'_{{\mathrm{0}}}}  \vdash   \ottnt{T'_{{\mathrm{0}}}}  \,  \,\&\,  \,   \Phi_\mathsf{val}   \, / \,   \square    \ottsym{<:}   \ottnt{T_{{\mathrm{0}}}}  \,  \,\&\,  \,   \Phi_\mathsf{val}   \, / \,   \square  $,
               which is shown by \reflem{ts-refl-subtyping-pure-eff}
               with
               \begin{itemize}
                \item $\vdash   \emptyset   \ottsym{,}  \mathit{x_{{\mathrm{0}}}} \,  {:}  \, \ottnt{T'_{{\mathrm{0}}}}$, which is derived by $ \emptyset   \vdash  \ottnt{T'_{{\mathrm{0}}}}$, and
                \item $\mathit{x_{{\mathrm{0}}}} \,  {:}  \, \ottnt{T'_{{\mathrm{0}}}}  \vdash  \ottnt{T'_{{\mathrm{0}}}}  \ottsym{<:}  \ottnt{T_{{\mathrm{0}}}}$, which is shown by
                      \reflem{ts-weak-subtyping} with $ \emptyset   \vdash  \ottnt{T'_{{\mathrm{0}}}}  \ottsym{<:}  \ottnt{T_{{\mathrm{0}}}}$.
               \end{itemize}

         \item $\mathit{x_{{\mathrm{0}}}} \,  {:}  \, \ottnt{T'_{{\mathrm{0}}}}  \vdash   \ottnt{T'_{{\mathrm{0}}}}  \,  \,\&\,  \,   \Phi_\mathsf{val}   \, / \,   \square  $,
               which is shown by \reflem{ts-wf-TP_val} with
               $\mathit{x_{{\mathrm{0}}}} \,  {:}  \, \ottnt{T'_{{\mathrm{0}}}}  \vdash  \ottnt{T'_{{\mathrm{0}}}}$ that is proven
               by \reflem{ts-weak-wf-ftyping} with $ \emptyset   \vdash  \ottnt{T'_{{\mathrm{0}}}}$.
        \end{itemize}
 \end{enumerate}
\end{proof}

\begin{lemmap}{Logical Relation is Consistent with Subtyping}{lr-subtyping}
 \begin{enumerate}
  \item If $ \emptyset   \vdash  \ottnt{C_{{\mathrm{1}}}}$ and $ \emptyset   \vdash  \ottnt{C_{{\mathrm{2}}}}$ and $ \emptyset   \vdash  \ottnt{C_{{\mathrm{1}}}}  \ottsym{<:}  \ottnt{C_{{\mathrm{2}}}}$,
        then $ \LRRel{ \mathcal{E} }{ \ottnt{C_{{\mathrm{1}}}} }  \subseteq  \LRRel{ \mathcal{E} }{ \ottnt{C_{{\mathrm{2}}}} } $.
  \item If $ \emptyset   \vdash  \ottnt{T_{{\mathrm{1}}}}$ and $ \emptyset   \vdash  \ottnt{T_{{\mathrm{2}}}}$ and $ \emptyset   \vdash  \ottnt{T_{{\mathrm{1}}}}  \ottsym{<:}  \ottnt{T_{{\mathrm{2}}}}$,
        then $ \LRRel{ \mathcal{V} }{ \ottnt{T_{{\mathrm{1}}}} }  \subseteq  \LRRel{ \mathcal{V} }{ \ottnt{T_{{\mathrm{2}}}} } $.
 \end{enumerate}
\end{lemmap}
\begin{proof}
 By simultaneous induction on the total numbers of type constructors in
 the pairs $(\ottnt{C_{{\mathrm{1}}}},\ottnt{C_{{\mathrm{2}}}})$ and $(\ottnt{T_{{\mathrm{1}}}},\ottnt{T_{{\mathrm{2}}}})$.
 \begin{enumerate}
  \item Let $\ottnt{e} \,  \in  \,  \LRRel{ \mathcal{E} }{ \ottnt{C_{{\mathrm{1}}}} } $.
        Then, we must show that
        \[
         \ottnt{e} \,  \in  \,  \LRRel{ \mathcal{E} }{ \ottnt{C_{{\mathrm{2}}}} }  ~.
        \]
        Because $\ottnt{e} \,  \in  \,  \LRRel{ \mathcal{E} }{ \ottnt{C_{{\mathrm{1}}}} } $ implies $ \emptyset   \vdash  \ottnt{e}  \ottsym{:}  \ottnt{C_{{\mathrm{1}}}}$,
        \T{Sub} with $ \emptyset   \vdash  \ottnt{C_{{\mathrm{1}}}}  \ottsym{<:}  \ottnt{C_{{\mathrm{2}}}}$ and $ \emptyset   \vdash  \ottnt{C_{{\mathrm{2}}}}$
        implies $ \emptyset   \vdash  \ottnt{e}  \ottsym{:}  \ottnt{C_{{\mathrm{2}}}}$, i.e., $\ottnt{e} \,  \in  \,  \LRRel{ \mathrm{Atom} }{ \ottnt{C_{{\mathrm{2}}}} } $.

        In what follows, we consider each case on the result of evaluating
        $\ottnt{e}$.
        \begin{caseanalysis}
         \case $  \exists  \, { \ottnt{v}  \ottsym{,}  \varpi } . \  \ottnt{e} \,  \Downarrow  \, \ottnt{v} \,  \,\&\,  \, \varpi $:
          We must show that
          \[
           \ottnt{v} \,  \in  \,  \LRRel{ \mathcal{V} }{  \ottnt{C_{{\mathrm{2}}}}  . \ottnt{T}  }  ~.
          \]
          Because $\ottnt{e} \,  \in  \,  \LRRel{ \mathcal{E} }{ \ottnt{C_{{\mathrm{1}}}} } $, we have $\ottnt{v} \,  \in  \,  \LRRel{ \mathcal{V} }{  \ottnt{C_{{\mathrm{1}}}}  . \ottnt{T}  } $.
          Because $ \emptyset   \vdash  \ottnt{C_{{\mathrm{1}}}}$ and $ \emptyset   \vdash  \ottnt{C_{{\mathrm{2}}}}$ imply
          $ \emptyset   \vdash   \ottnt{C_{{\mathrm{1}}}}  . \ottnt{T} $ and $ \emptyset   \vdash   \ottnt{C_{{\mathrm{2}}}}  . \ottnt{T} $, respectively, and
          $ \emptyset   \vdash  \ottnt{C_{{\mathrm{1}}}}  \ottsym{<:}  \ottnt{C_{{\mathrm{2}}}}$ implies $ \emptyset   \vdash   \ottnt{C_{{\mathrm{1}}}}  . \ottnt{T}   \ottsym{<:}   \ottnt{C_{{\mathrm{2}}}}  . \ottnt{T} $,
          we have the conclusion by the IH on $( \ottnt{C_{{\mathrm{1}}}}  . \ottnt{T} , \ottnt{C_{{\mathrm{2}}}}  . \ottnt{T} )$ with
          $\ottnt{v} \,  \in  \,  \LRRel{ \mathcal{V} }{  \ottnt{C_{{\mathrm{1}}}}  . \ottnt{T}  } $.

         \case $  \exists  \, { \ottnt{K_{{\mathrm{0}}}}  \ottsym{,}  \mathit{x_{{\mathrm{0}}}}  \ottsym{,}  \ottnt{e_{{\mathrm{0}}}}  \ottsym{,}  \varpi } . \  \ottnt{e} \,  \Downarrow  \,  \ottnt{K_{{\mathrm{0}}}}  [  \mathcal{S}_0 \, \mathit{x_{{\mathrm{0}}}}  \ottsym{.}  \ottnt{e_{{\mathrm{0}}}}  ]  \,  \,\&\,  \, \varpi $:
               We must show that
               \begin{itemize}
                \item $ \ottnt{C_{{\mathrm{2}}}}  . \ottnt{S}  \,  =  \,  ( \forall  \mathit{x}  .  \ottnt{C_{{\mathrm{21}}}}  )  \Rightarrow   \ottnt{C_{{\mathrm{22}}}} $ and
                \item $  \forall  \, { \ottnt{K} \,  \in  \,  \LRRel{ \mathcal{K} }{   \ottnt{C_{{\mathrm{2}}}}  . \ottnt{T}   \, / \, ( \forall  \mathit{x}  .  \ottnt{C_{{\mathrm{21}}}}  )  }  } . \   \resetcnstr{  \ottnt{K}  [  \ottnt{e}  ]  }  \,  \in  \,  \LRRel{ \mathcal{E} }{ \ottnt{C_{{\mathrm{22}}}} }  $
               \end{itemize}
               for some $\mathit{x}$, $\ottnt{C_{{\mathrm{21}}}}$, and $\ottnt{C_{{\mathrm{22}}}}$.

               Because $\ottnt{e} \,  \in  \,  \LRRel{ \mathcal{E} }{ \ottnt{C_{{\mathrm{1}}}} } $ and $\ottnt{e} \,  \Downarrow  \,  \ottnt{K_{{\mathrm{0}}}}  [  \mathcal{S}_0 \, \mathit{x_{{\mathrm{0}}}}  \ottsym{.}  \ottnt{e_{{\mathrm{0}}}}  ]  \,  \,\&\,  \, \varpi$,
               there exist some $\mathit{x}$, $\ottnt{C_{{\mathrm{11}}}}$, and $\ottnt{C_{{\mathrm{12}}}}$ such that
               \begin{itemize}
                \item $ \ottnt{C_{{\mathrm{1}}}}  . \ottnt{S}  \,  =  \,  ( \forall  \mathit{x}  .  \ottnt{C_{{\mathrm{11}}}}  )  \Rightarrow   \ottnt{C_{{\mathrm{12}}}} $ and
                \item $  \forall  \, { \ottnt{K} \,  \in  \,  \LRRel{ \mathcal{K} }{   \ottnt{C_{{\mathrm{1}}}}  . \ottnt{T}   \, / \, ( \forall  \mathit{x}  .  \ottnt{C_{{\mathrm{11}}}}  )  }  } . \   \resetcnstr{  \ottnt{K}  [  \ottnt{e}  ]  }  \,  \in  \,  \LRRel{ \mathcal{E} }{ \ottnt{C_{{\mathrm{12}}}} }  $.
               \end{itemize}
               Because $ \emptyset   \vdash  \ottnt{C_{{\mathrm{1}}}}  \ottsym{<:}  \ottnt{C_{{\mathrm{2}}}}$ and $ \ottnt{C_{{\mathrm{1}}}}  . \ottnt{S}  \,  =  \,  ( \forall  \mathit{x}  .  \ottnt{C_{{\mathrm{11}}}}  )  \Rightarrow   \ottnt{C_{{\mathrm{12}}}} $,
               there exist some $\ottnt{C_{{\mathrm{21}}}}$ and $\ottnt{C_{{\mathrm{22}}}}$ such that
               \begin{itemize}
                \item $ \emptyset   \vdash   \ottnt{C_{{\mathrm{1}}}}  . \ottnt{T}   \ottsym{<:}   \ottnt{C_{{\mathrm{2}}}}  . \ottnt{T} $,
                \item $ \ottnt{C_{{\mathrm{2}}}}  . \ottnt{S}  \,  =  \,  ( \forall  \mathit{x}  .  \ottnt{C_{{\mathrm{21}}}}  )  \Rightarrow   \ottnt{C_{{\mathrm{22}}}} $,
                \item $\mathit{x} \,  {:}  \,  \ottnt{C_{{\mathrm{1}}}}  . \ottnt{T}   \vdash  \ottnt{C_{{\mathrm{21}}}}  \ottsym{<:}  \ottnt{C_{{\mathrm{11}}}}$, and
                \item $ \emptyset   \vdash  \ottnt{C_{{\mathrm{12}}}}  \ottsym{<:}  \ottnt{C_{{\mathrm{22}}}}$.
               \end{itemize}
               Further, $ \emptyset   \vdash  \ottnt{C_{{\mathrm{1}}}}$ and $ \emptyset   \vdash  \ottnt{C_{{\mathrm{2}}}}$ imply
               \begin{itemize}
                \item $ \emptyset   \vdash   \ottnt{C_{{\mathrm{1}}}}  . \ottnt{T} $,
                \item $\mathit{x} \,  {:}  \,  \ottnt{C_{{\mathrm{1}}}}  . \ottnt{T}   \vdash  \ottnt{C_{{\mathrm{11}}}}$,
                \item $ \emptyset   \vdash  \ottnt{C_{{\mathrm{12}}}}$,
                \item $ \emptyset   \vdash   \ottnt{C_{{\mathrm{2}}}}  . \ottnt{T} $,
                \item $\mathit{x} \,  {:}  \,  \ottnt{C_{{\mathrm{2}}}}  . \ottnt{T}   \vdash  \ottnt{C_{{\mathrm{21}}}}$, and
                \item $ \emptyset   \vdash  \ottnt{C_{{\mathrm{22}}}}$.
               \end{itemize}

               Let $\ottnt{K} \,  \in  \,  \LRRel{ \mathcal{K} }{   \ottnt{C_{{\mathrm{2}}}}  . \ottnt{T}   \, / \, ( \forall  \mathit{x}  .  \ottnt{C_{{\mathrm{21}}}}  )  } $.
               We must show that $ \resetcnstr{  \ottnt{K}  [  \ottnt{e}  ]  }  \,  \in  \,  \LRRel{ \mathcal{E} }{ \ottnt{C_{{\mathrm{22}}}} } $.

               We first show that
               \[
                \ottnt{K} \,  \in  \,  \LRRel{ \mathcal{K} }{   \ottnt{C_{{\mathrm{1}}}}  . \ottnt{T}   \, / \, ( \forall  \mathit{x}  .  \ottnt{C_{{\mathrm{11}}}}  )  }  ~,
               \]
               which is proven by the following.
               \begin{itemize}
                \item We have $\ottnt{K}  \ottsym{:}    \ottnt{C_{{\mathrm{1}}}}  . \ottnt{T}   \, / \, ( \forall  \mathit{x}  .  \ottnt{C_{{\mathrm{11}}}}  ) $ by
                      \reflem{lr-cont-subsump} with the following:
                      \begin{itemize}
                       \item $\mathit{x} \,  {:}  \,  \ottnt{C_{{\mathrm{2}}}}  . \ottnt{T}   \vdash  \ottnt{C_{{\mathrm{21}}}}$;
                       \item $\mathit{x} \,  {:}  \,  \ottnt{C_{{\mathrm{1}}}}  . \ottnt{T}   \vdash  \ottnt{C_{{\mathrm{11}}}}$;
                       \item $ \emptyset   \vdash   \ottnt{C_{{\mathrm{1}}}}  . \ottnt{T}   \ottsym{<:}   \ottnt{C_{{\mathrm{2}}}}  . \ottnt{T} $;
                       \item $\mathit{x} \,  {:}  \,  \ottnt{C_{{\mathrm{1}}}}  . \ottnt{T}   \vdash  \ottnt{C_{{\mathrm{21}}}}  \ottsym{<:}  \ottnt{C_{{\mathrm{11}}}}$; and
                       \item $\ottnt{K}  \ottsym{:}    \ottnt{C_{{\mathrm{2}}}}  . \ottnt{T}   \, / \, ( \forall  \mathit{x}  .  \ottnt{C_{{\mathrm{21}}}}  ) $
                             implied by $\ottnt{K} \,  \in  \,  \LRRel{ \mathcal{K} }{   \ottnt{C_{{\mathrm{2}}}}  . \ottnt{T}   \, / \, ( \forall  \mathit{x}  .  \ottnt{C_{{\mathrm{21}}}}  )  } $.
                      \end{itemize}

                \item We have $ \lambda\!  \, \mathit{x}  \ottsym{.}   \resetcnstr{  \ottnt{K}  [  \mathit{x}  ]  }  \,  \in  \,  \LRRel{ \mathcal{V} }{ \ottsym{(}  \mathit{x} \,  {:}  \,  \ottnt{C_{{\mathrm{1}}}}  . \ottnt{T}   \ottsym{)}  \rightarrow  \ottnt{C_{{\mathrm{11}}}} } $
                      by the IH with the following:
                      \begin{itemize}
                       \item $ \emptyset   \vdash  \ottsym{(}  \mathit{x} \,  {:}  \,  \ottnt{C_{{\mathrm{2}}}}  . \ottnt{T}   \ottsym{)}  \rightarrow  \ottnt{C_{{\mathrm{21}}}}$ implied by
                             \WFT{Fun} with $\mathit{x} \,  {:}  \,  \ottnt{C_{{\mathrm{2}}}}  . \ottnt{T}   \vdash  \ottnt{C_{{\mathrm{21}}}}$;
                       \item $ \emptyset   \vdash  \ottsym{(}  \mathit{x} \,  {:}  \,  \ottnt{C_{{\mathrm{1}}}}  . \ottnt{T}   \ottsym{)}  \rightarrow  \ottnt{C_{{\mathrm{11}}}}$ implied by
                             \WFT{Fun} with $\mathit{x} \,  {:}  \,  \ottnt{C_{{\mathrm{1}}}}  . \ottnt{T}   \vdash  \ottnt{C_{{\mathrm{11}}}}$;
                       \item $ \emptyset   \vdash  \ottsym{(}  \mathit{x} \,  {:}  \,  \ottnt{C_{{\mathrm{2}}}}  . \ottnt{T}   \ottsym{)}  \rightarrow  \ottnt{C_{{\mathrm{21}}}}  \ottsym{<:}  \ottsym{(}  \mathit{x} \,  {:}  \,  \ottnt{C_{{\mathrm{1}}}}  . \ottnt{T}   \ottsym{)}  \rightarrow  \ottnt{C_{{\mathrm{11}}}}$
                             implied by \Srule{Fun} with
                             $ \emptyset   \vdash   \ottnt{C_{{\mathrm{1}}}}  . \ottnt{T}   \ottsym{<:}   \ottnt{C_{{\mathrm{2}}}}  . \ottnt{T} $ and
                             $\mathit{x} \,  {:}  \,  \ottnt{C_{{\mathrm{1}}}}  . \ottnt{T}   \vdash  \ottnt{C_{{\mathrm{21}}}}  \ottsym{<:}  \ottnt{C_{{\mathrm{11}}}}$; and
                       \item $ \lambda\!  \, \mathit{x}  \ottsym{.}   \resetcnstr{  \ottnt{K}  [  \mathit{x}  ]  }  \,  \in  \,  \LRRel{ \mathcal{V} }{ \ottsym{(}  \mathit{x} \,  {:}  \,  \ottnt{C_{{\mathrm{2}}}}  . \ottnt{T}   \ottsym{)}  \rightarrow  \ottnt{C_{{\mathrm{21}}}} } $
                             implied by $\ottnt{K} \,  \in  \,  \LRRel{ \mathcal{K} }{   \ottnt{C_{{\mathrm{2}}}}  . \ottnt{T}   \, / \, ( \forall  \mathit{x}  .  \ottnt{C_{{\mathrm{21}}}}  )  } $.
                      \end{itemize}
               \end{itemize}

               Then, we show that $ \resetcnstr{  \ottnt{K}  [  \ottnt{e}  ]  }  \,  \in  \,  \LRRel{ \mathcal{E} }{ \ottnt{C_{{\mathrm{22}}}} } $.
               %
               %
               Because of $\ottnt{K} \,  \in  \,  \LRRel{ \mathcal{K} }{   \ottnt{C_{{\mathrm{1}}}}  . \ottnt{T}   \, / \, ( \forall  \mathit{x}  .  \ottnt{C_{{\mathrm{11}}}}  )  } $ and
               the obtained property
               \[
                  \forall  \, { \ottnt{K} \,  \in  \,  \LRRel{ \mathcal{K} }{   \ottnt{C_{{\mathrm{1}}}}  . \ottnt{T}   \, / \, ( \forall  \mathit{x}  .  \ottnt{C_{{\mathrm{11}}}}  )  }  } . \   \resetcnstr{  \ottnt{K}  [  \ottnt{e}  ]  }  \,  \in  \,  \LRRel{ \mathcal{E} }{ \ottnt{C_{{\mathrm{12}}}} }   ~,
               \]
               we have $ \resetcnstr{  \ottnt{K}  [  \ottnt{e}  ]  }  \,  \in  \,  \LRRel{ \mathcal{E} }{ \ottnt{C_{{\mathrm{12}}}} } $.
               Then, the IH on $(\ottnt{C_{{\mathrm{12}}}},\ottnt{C_{{\mathrm{22}}}})$ with
               $ \emptyset   \vdash  \ottnt{C_{{\mathrm{12}}}}  \ottsym{<:}  \ottnt{C_{{\mathrm{22}}}}$ and
               $ \emptyset   \vdash  \ottnt{C_{{\mathrm{12}}}}$ and $ \emptyset   \vdash  \ottnt{C_{{\mathrm{22}}}}$
               implies the conclusion $ \resetcnstr{  \ottnt{K}  [  \ottnt{e}  ]  }  \,  \in  \,  \LRRel{ \mathcal{E} }{ \ottnt{C_{{\mathrm{22}}}} } $.

         \case $  \exists  \, { \pi } . \  \ottnt{e} \,  \Uparrow  \, \pi $:
               We must show that $ { \models  \,   { \ottnt{C_{{\mathrm{2}}}} }^\nu (  \pi  )  } $.
               Because $\ottnt{e} \,  \in  \,  \LRRel{ \mathcal{E} }{ \ottnt{C_{{\mathrm{1}}}} } $, we have $ { \models  \,   { \ottnt{C_{{\mathrm{1}}}} }^\nu (  \pi  )  } $.
               Therefore, \reflem{ts-extract-subst-val-valid} implies
               $\mathit{y} \,  {:}  \,  \Sigma ^\omega   \models   \ottsym{(}   \pi    =    \mathit{y}   \ottsym{)}    \Rightarrow     { \ottnt{C_{{\mathrm{1}}}} }^\nu (  \mathit{y}  )  $
               for $\mathit{y} \,  \not\in  \,  \mathit{fv}  (  \ottnt{C_{{\mathrm{1}}}}  )  \,  \mathbin{\cup}  \,  \mathit{fv}  (  \ottnt{C_{{\mathrm{2}}}}  ) $.
               Then, \reflem{lr-subtyping-nu} with $ \emptyset   \vdash  \ottnt{C_{{\mathrm{1}}}}  \ottsym{<:}  \ottnt{C_{{\mathrm{2}}}}$ implies
               $\mathit{y} \,  {:}  \,  \Sigma ^\omega   \models   \ottsym{(}   \pi    =    \mathit{y}   \ottsym{)}    \Rightarrow     { \ottnt{C_{{\mathrm{2}}}} }^\nu (  \mathit{y}  )  $.
               By \reflem{ts-term-subst-valid}, we have
               $ { \models  \,   \ottsym{(}   \pi    =    \pi   \ottsym{)}    \Rightarrow     { \ottnt{C_{{\mathrm{2}}}} }^\nu (  \pi  )   } $.
               Because $ { \models  \,   \pi    =    \pi  } $ by \reflem{ts-refl-const-eq},
               \reflem{ts-discharge-assumption-valid} implies the conclusion
               $ { \models  \,   { \ottnt{C_{{\mathrm{2}}}} }^\nu (  \pi  )  } $.
        \end{caseanalysis}

  \item Let $\ottnt{v} \,  \in  \,  \LRRel{ \mathcal{V} }{ \ottnt{T_{{\mathrm{1}}}} } $.
        We must show that $\ottnt{v} \,  \in  \,  \LRRel{ \mathcal{V} }{ \ottnt{T_{{\mathrm{2}}}} } $.
        Because
        \begin{itemize}
         \item $\ottnt{v} \,  \in  \,  \LRRel{ \mathcal{V} }{ \ottnt{T_{{\mathrm{1}}}} } $ implies $ \emptyset   \vdash  \ottnt{v}  \ottsym{:}  \ottnt{T_{{\mathrm{1}}}}$,
         \item \reflem{ts-refl-subtyping-pure-eff} with $\vdash   \emptyset $ and $ \emptyset   \vdash  \ottnt{T_{{\mathrm{1}}}}  \ottsym{<:}  \ottnt{T_{{\mathrm{2}}}}$
               implies $ \emptyset   \vdash   \ottnt{T_{{\mathrm{1}}}}  \,  \,\&\,  \,   \Phi_\mathsf{val}   \, / \,   \square    \ottsym{<:}   \ottnt{T_{{\mathrm{2}}}}  \,  \,\&\,  \,   \Phi_\mathsf{val}   \, / \,   \square  $, and
         \item \reflem{ts-wf-TP_val} with $ \emptyset   \vdash  \ottnt{T_{{\mathrm{2}}}}$ implies $ \emptyset   \vdash   \ottnt{T_{{\mathrm{2}}}}  \,  \,\&\,  \,   \Phi_\mathsf{val}   \, / \,   \square  $,
        \end{itemize}
        \T{Sub} implies $ \emptyset   \vdash  \ottnt{v}  \ottsym{:}  \ottnt{T_{{\mathrm{2}}}}$, i.e., $\ottnt{v} \,  \in  \,  \LRRel{ \mathrm{Atom} }{ \ottnt{T_{{\mathrm{2}}}} } $.
        The remaining proceeds by case analysis on $\ottnt{T_{{\mathrm{2}}}}$.
        \begin{caseanalysis}
         \case $  \exists  \, { \mathit{x}  \ottsym{,}  \iota  \ottsym{,}  \phi_{{\mathrm{2}}} } . \  \ottnt{T_{{\mathrm{2}}}} \,  =  \, \ottsym{\{}  \mathit{x} \,  {:}  \, \iota \,  |  \, \phi_{{\mathrm{2}}}  \ottsym{\}} $:
          It suffices to show that $\ottnt{v} \,  \in  \,  \LRRel{ \mathrm{Atom} }{ \ottnt{T_{{\mathrm{2}}}} } $,
          which has been proven.

         \case $  \exists  \, { \mathit{x}  \ottsym{,}  \ottnt{T'_{{\mathrm{2}}}}  \ottsym{,}  \ottnt{C'_{{\mathrm{2}}}} } . \  \ottnt{T_{{\mathrm{2}}}} \,  =  \, \ottsym{(}  \mathit{x} \,  {:}  \, \ottnt{T'_{{\mathrm{2}}}}  \ottsym{)}  \rightarrow  \ottnt{C'_{{\mathrm{2}}}} $:
          Let $\ottnt{v'} \,  \in  \,  \LRRel{ \mathcal{V} }{ \ottnt{T'_{{\mathrm{2}}}} } $.
          Then, we must show that
          \[
           \ottnt{v} \, \ottnt{v'} \,  \in  \,  \LRRel{ \mathcal{E} }{  \ottnt{C'_{{\mathrm{2}}}}    [  \ottnt{v'}  /  \mathit{x}  ]   }  ~.
          \]
          Because $ \emptyset   \vdash  \ottnt{T_{{\mathrm{1}}}}  \ottsym{<:}  \ottnt{T_{{\mathrm{2}}}}$ can be derived only by \Srule{Fun},
          there exist some $\ottnt{T'_{{\mathrm{1}}}}$ and $\ottnt{C'_{{\mathrm{1}}}}$ such that
          \begin{itemize}
           \item $\ottnt{T_{{\mathrm{1}}}} \,  =  \, \ottsym{(}  \mathit{x} \,  {:}  \, \ottnt{T'_{{\mathrm{1}}}}  \ottsym{)}  \rightarrow  \ottnt{C'_{{\mathrm{1}}}}$,
           \item $ \emptyset   \vdash  \ottnt{T'_{{\mathrm{2}}}}  \ottsym{<:}  \ottnt{T'_{{\mathrm{1}}}}$, and
           \item $\mathit{x} \,  {:}  \, \ottnt{T'_{{\mathrm{2}}}}  \vdash  \ottnt{C'_{{\mathrm{1}}}}  \ottsym{<:}  \ottnt{C'_{{\mathrm{2}}}}$.
          \end{itemize}
          Because $ \emptyset   \vdash  \ottnt{T_{{\mathrm{1}}}}$ and $ \emptyset   \vdash  \ottnt{T_{{\mathrm{2}}}}$ imply
          \begin{itemize}
           \item $\mathit{x} \,  {:}  \, \ottnt{T'_{{\mathrm{1}}}}  \vdash  \ottnt{C'_{{\mathrm{1}}}}$ and
           \item $\mathit{x} \,  {:}  \, \ottnt{T'_{{\mathrm{2}}}}  \vdash  \ottnt{C'_{{\mathrm{2}}}}$,
          \end{itemize}
          \reflem{ts-wf-tctx-wf-ftyping} implies
          \begin{itemize}
           \item $ \emptyset   \vdash  \ottnt{T'_{{\mathrm{1}}}}$ and
           \item $ \emptyset   \vdash  \ottnt{T'_{{\mathrm{2}}}}$.
          \end{itemize}
          By the IH on $(\ottnt{T'_{{\mathrm{2}}}}, \ottnt{T'_{{\mathrm{1}}}})$, we have $\ottnt{v'} \,  \in  \,  \LRRel{ \mathcal{V} }{ \ottnt{T'_{{\mathrm{1}}}} } $.
          Because $\ottnt{v} \,  \in  \,  \LRRel{ \mathcal{V} }{ \ottsym{(}  \mathit{x} \,  {:}  \, \ottnt{T'_{{\mathrm{1}}}}  \ottsym{)}  \rightarrow  \ottnt{C'_{{\mathrm{1}}}} } $,
          we have
          \[
           \ottnt{v} \, \ottnt{v'} \,  \in  \,  \LRRel{ \mathcal{E} }{  \ottnt{C'_{{\mathrm{1}}}}    [  \ottnt{v'}  /  \mathit{x}  ]   }  ~.
          \]
          Because $\ottnt{v'} \,  \in  \,  \LRRel{ \mathcal{V} }{ \ottnt{T'_{{\mathrm{1}}}} } $ and $\ottnt{v'} \,  \in  \,  \LRRel{ \mathcal{V} }{ \ottnt{T'_{{\mathrm{2}}}} } $
          imply $ \emptyset   \vdash  \ottnt{v'}  \ottsym{:}  \ottnt{T'_{{\mathrm{1}}}}$ and $ \emptyset   \vdash  \ottnt{v'}  \ottsym{:}  \ottnt{T'_{{\mathrm{2}}}}$, respectively,
          Lemmas~\ref{lem:ts-val-subst-wf-ftyping} and \ref{lem:ts-val-subst-subtyping} imply
          \begin{itemize}
           \item $ \emptyset   \vdash   \ottnt{C'_{{\mathrm{1}}}}    [  \ottnt{v'}  /  \mathit{x}  ]  $,
           \item $ \emptyset   \vdash   \ottnt{C'_{{\mathrm{2}}}}    [  \ottnt{v'}  /  \mathit{x}  ]  $, and
           \item $ \emptyset   \vdash   \ottnt{C'_{{\mathrm{1}}}}    [  \ottnt{v'}  /  \mathit{x}  ]    \ottsym{<:}   \ottnt{C'_{{\mathrm{2}}}}    [  \ottnt{v'}  /  \mathit{x}  ]  $.
          \end{itemize}
          Because the value substitution does not change the numbers of type constructors,
          we can apply the IH on $( \ottnt{C'_{{\mathrm{1}}}}    [  \ottnt{v'}  /  \mathit{x}  ]  , \ottnt{C'_{{\mathrm{2}}}}    [  \ottnt{v'}  /  \mathit{x}  ]  )$, which implies
          the conclusion $\ottnt{v} \, \ottnt{v'} \,  \in  \,  \LRRel{ \mathcal{E} }{  \ottnt{C'_{{\mathrm{2}}}}    [  \ottnt{v'}  /  \mathit{x}  ]   } $.

         \case $  \exists  \, { \mathit{X}  \ottsym{,}   \tceseq{ \ottnt{s} }   \ottsym{,}  \ottnt{C_{{\mathrm{2}}}} } . \  \ottnt{T_{{\mathrm{2}}}} \,  =  \,  \forall   \ottsym{(}  \mathit{X} \,  {:}  \,  \tceseq{ \ottnt{s} }   \ottsym{)}  \ottsym{.}  \ottnt{C_{{\mathrm{2}}}} $:
          Let $\ottnt{A}$ be a predicate such that $ \emptyset   \vdash  \ottnt{A}  \ottsym{:}   \tceseq{ \ottnt{s} } $.
          We must show that $\ottnt{v} \, \ottnt{A} \,  \in  \,  \LRRel{ \mathcal{E} }{  \ottnt{C_{{\mathrm{2}}}}    [  \ottnt{A}  /  \mathit{X}  ]   } $.
          Because $ \emptyset   \vdash  \ottnt{T_{{\mathrm{1}}}}  \ottsym{<:}  \ottnt{T_{{\mathrm{2}}}}$ can be derived only by \Srule{Forall},
          its inversion implies
          \begin{itemize}
           \item $\ottnt{T_{{\mathrm{1}}}} \,  =  \,  \forall   \ottsym{(}  \mathit{X} \,  {:}  \,  \tceseq{ \ottnt{s} }   \ottsym{)}  \ottsym{.}  \ottnt{C_{{\mathrm{1}}}}$ and
           \item $\mathit{X} \,  {:}  \,  \tceseq{ \ottnt{s} }   \vdash  \ottnt{C_{{\mathrm{1}}}}  \ottsym{<:}  \ottnt{C_{{\mathrm{2}}}}$
          \end{itemize}
          for some $\ottnt{C_{{\mathrm{1}}}}$.
          Because $\ottnt{v} \,  \in  \,  \LRRel{ \mathcal{V} }{ \ottnt{T_{{\mathrm{1}}}} } $,
          we have $\ottnt{v} \, \ottnt{A} \,  \in  \,  \LRRel{ \mathcal{E} }{  \ottnt{C_{{\mathrm{1}}}}    [  \ottnt{A}  /  \mathit{X}  ]   } $.
          $ \emptyset   \vdash  \ottnt{T_{{\mathrm{1}}}}$ and $ \emptyset   \vdash  \ottnt{T_{{\mathrm{2}}}}$ imply
          $\mathit{X} \,  {:}  \,  \tceseq{ \ottnt{s} }   \vdash  \ottnt{C_{{\mathrm{1}}}}$ and $\mathit{X} \,  {:}  \,  \tceseq{ \ottnt{s} }   \vdash  \ottnt{C_{{\mathrm{2}}}}$, respectively.
          Then, Lemmas~\ref{lem:ts-pred-subst-wf-ftyping} and \ref{lem:ts-pred-subst-subtyping} with $ \emptyset   \vdash  \ottnt{A}  \ottsym{:}   \tceseq{ \ottnt{s} } $ imply
          \begin{itemize}
           \item $ \emptyset   \vdash   \ottnt{C_{{\mathrm{1}}}}    [  \ottnt{A}  /  \mathit{X}  ]  $,
           \item $ \emptyset   \vdash   \ottnt{C_{{\mathrm{2}}}}    [  \ottnt{A}  /  \mathit{X}  ]  $, and
           \item $ \emptyset   \vdash   \ottnt{C_{{\mathrm{1}}}}    [  \ottnt{A}  /  \mathit{X}  ]    \ottsym{<:}   \ottnt{C_{{\mathrm{2}}}}    [  \ottnt{A}  /  \mathit{X}  ]  $.
          \end{itemize}
          Because the predicate substitution does not change the numbers of type constructors,
          we can apply the IH on $( \ottnt{C_{{\mathrm{1}}}}    [  \ottnt{A}  /  \mathit{X}  ]  , \ottnt{C_{{\mathrm{2}}}}    [  \ottnt{A}  /  \mathit{X}  ]  )$, which implies
          the conclusion $\ottnt{v} \, \ottnt{A} \,  \in  \,  \LRRel{ \mathcal{E} }{  \ottnt{C_{{\mathrm{2}}}}    [  \ottnt{A}  /  \mathit{X}  ]   } $.
        \end{caseanalysis}
 \end{enumerate}
\end{proof}

\begin{lemmap}{Compatibility: Subsumption for Subtyping}{lr-compat-subtyping}
 If $\ottnt{e} \,  \in  \,  \LRRel{ \mathcal{O} }{ \Gamma  \, \vdash \,  \ottnt{C} } $ and $\Gamma  \vdash  \ottnt{C}  \ottsym{<:}  \ottnt{C'}$ and $\Gamma  \vdash  \ottnt{C'}$,
 then $\ottnt{e} \,  \in  \,  \LRRel{ \mathcal{O} }{ \Gamma  \, \vdash \,  \ottnt{C'} } $.
\end{lemmap}
\begin{proof}
 Because $\ottnt{e} \,  \in  \,  \LRRel{ \mathcal{O} }{ \Gamma  \, \vdash \,  \ottnt{C} } $ implies $\Gamma  \vdash  \ottnt{e}  \ottsym{:}  \ottnt{C}$,
 we have $\Gamma  \vdash  \ottnt{e}  \ottsym{:}  \ottnt{C'}$ by \T{Sub} with the assumptions $\Gamma  \vdash  \ottnt{C}  \ottsym{<:}  \ottnt{C'}$ and $\Gamma  \vdash  \ottnt{C'}$.
 Let $\sigma \,  \in  \,  \LRRel{ \mathcal{G} }{ \Gamma } $.
 We must show that
 \[
  \sigma  \ottsym{(}  \ottnt{e}  \ottsym{)} \,  \in  \,  \LRRel{ \mathcal{E} }{ \sigma  \ottsym{(}  \ottnt{C'}  \ottsym{)} }  ~.
 \]
 Now, we can find that
 \begin{itemize}
  \item $\ottnt{e} \,  \in  \,  \LRRel{ \mathcal{O} }{ \Gamma  \, \vdash \,  \ottnt{C} } $ implies $\sigma  \ottsym{(}  \ottnt{e}  \ottsym{)} \,  \in  \,  \LRRel{ \mathcal{E} }{ \sigma  \ottsym{(}  \ottnt{C}  \ottsym{)} } $,
  \item \reflem{lr-typeability} with $\Gamma  \vdash  \ottnt{C}  \ottsym{<:}  \ottnt{C'}$ implies
        $ \emptyset   \vdash  \sigma  \ottsym{(}  \ottnt{C}  \ottsym{)}  \ottsym{<:}  \sigma  \ottsym{(}  \ottnt{C'}  \ottsym{)}$, and
  \item \reflem{lr-typeability} with $\Gamma  \vdash  \ottnt{C'}$ implies
        $ \emptyset   \vdash  \sigma  \ottsym{(}  \ottnt{C'}  \ottsym{)}$.
 \end{itemize}
 Further, because $\sigma  \ottsym{(}  \ottnt{e}  \ottsym{)} \,  \in  \,  \LRRel{ \mathcal{E} }{ \sigma  \ottsym{(}  \ottnt{C}  \ottsym{)} } $ implies $ \emptyset   \vdash  \sigma  \ottsym{(}  \ottnt{e}  \ottsym{)}  \ottsym{:}  \sigma  \ottsym{(}  \ottnt{C}  \ottsym{)}$,
 we have $ \emptyset   \vdash  \sigma  \ottsym{(}  \ottnt{C}  \ottsym{)}$ by \reflem{ts-wf-ty-tctx-typing}.
 Then, \reflem{lr-subtyping} implies the conclusion $\sigma  \ottsym{(}  \ottnt{e}  \ottsym{)} \,  \in  \,  \LRRel{ \mathcal{E} }{ \sigma  \ottsym{(}  \ottnt{C'}  \ottsym{)} } $.
\end{proof}

\begin{lemmap}{Compatibility: Shift}{lr-compat-shift}
 If $\ottnt{e} \,  \in  \,  \LRRel{ \mathcal{O} }{ \Gamma  \ottsym{,}  \mathit{x} \,  {:}  \, \ottsym{(}  \mathit{y} \,  {:}  \, \ottnt{T}  \ottsym{)}  \rightarrow  \ottnt{C_{{\mathrm{1}}}}  \, \vdash \,  \ottnt{C_{{\mathrm{2}}}} } $,
 then $\mathcal{S}_0 \, \mathit{x}  \ottsym{.}  \ottnt{e} \,  \in  \,  \LRRel{ \mathcal{O} }{ \Gamma  \, \vdash \,   \ottnt{T}  \,  \,\&\,  \,   \Phi_\mathsf{val}   \, / \,   ( \forall  \mathit{y}  .  \ottnt{C_{{\mathrm{1}}}}  )  \Rightarrow   \ottnt{C_{{\mathrm{2}}}}   } $.
\end{lemmap}
\begin{proof}
 Let $\ottnt{C} \,  =  \,  \ottnt{T}  \,  \,\&\,  \,   \Phi_\mathsf{val}   \, / \,   ( \forall  \mathit{y}  .  \ottnt{C_{{\mathrm{1}}}}  )  \Rightarrow   \ottnt{C_{{\mathrm{2}}}}  $.
 Because $\ottnt{e} \,  \in  \,  \LRRel{ \mathcal{O} }{ \Gamma  \ottsym{,}  \mathit{x} \,  {:}  \, \ottsym{(}  \mathit{y} \,  {:}  \, \ottnt{T}  \ottsym{)}  \rightarrow  \ottnt{C_{{\mathrm{1}}}}  \, \vdash \,  \ottnt{C_{{\mathrm{2}}}} } $ implies
 $\Gamma  \ottsym{,}  \mathit{x} \,  {:}  \, \ottsym{(}  \mathit{y} \,  {:}  \, \ottnt{T}  \ottsym{)}  \rightarrow  \ottnt{C_{{\mathrm{1}}}}  \vdash  \ottnt{e}  \ottsym{:}  \ottnt{C_{{\mathrm{2}}}}$,
 \T{Shift} implies $\Gamma  \vdash  \mathcal{S}_0 \, \mathit{x}  \ottsym{.}  \ottnt{e}  \ottsym{:}  \ottnt{C}$.
 Let $\sigma \,  \in  \,  \LRRel{ \mathcal{G} }{ \Gamma } $.
 Without loss of generality, we can suppose that $\mathit{x}$ and $\mathit{y}$ do not occur in $\sigma$.
 We must show that
 \[
  \mathcal{S}_0 \, \mathit{x}  \ottsym{.}  \sigma  \ottsym{(}  \ottnt{e}  \ottsym{)} \,  \in  \,  \LRRel{ \mathcal{E} }{ \sigma  \ottsym{(}  \ottnt{C}  \ottsym{)} }  ~.
 \]

 Because $\Gamma  \vdash  \mathcal{S}_0 \, \mathit{x}  \ottsym{.}  \ottnt{e}  \ottsym{:}  \ottnt{C}$ and $\sigma \,  \in  \,  \LRRel{ \mathcal{G} }{ \Gamma } $,
 \reflem{lr-typeability} implies $ \emptyset   \vdash  \mathcal{S}_0 \, \mathit{x}  \ottsym{.}  \sigma  \ottsym{(}  \ottnt{e}  \ottsym{)}  \ottsym{:}  \sigma  \ottsym{(}  \ottnt{C}  \ottsym{)}$.
 Therefore, we have $\mathcal{S}_0 \, \mathit{x}  \ottsym{.}  \sigma  \ottsym{(}  \ottnt{e}  \ottsym{)} \,  \in  \,  \LRRel{ \mathrm{Atom} }{ \sigma  \ottsym{(}  \ottnt{C}  \ottsym{)} } $.

 Because \reflem{lr-red-shift} implies
 \begin{itemize}
  \item $ {    \forall  \, { \ottnt{r}  \ottsym{,}  \varpi } . \  \mathcal{S}_0 \, \mathit{x}  \ottsym{.}  \sigma  \ottsym{(}  \ottnt{e}  \ottsym{)} \,  \Downarrow  \, \ottnt{r} \,  \,\&\,  \, \varpi   \mathrel{\Longrightarrow}  \ottnt{r} \,  =  \, \mathcal{S}_0 \, \mathit{x}  \ottsym{.}  \sigma  \ottsym{(}  \ottnt{e}  \ottsym{)}  } \,\mathrel{\wedge}\, { \varpi \,  =  \,  \epsilon  } $ and
  \item $\mathcal{S}_0 \, \mathit{x}  \ottsym{.}  \sigma  \ottsym{(}  \ottnt{e}  \ottsym{)} \,  \Uparrow  \, \pi$ cannot be derived for any $\pi$,
 \end{itemize}
 it suffices to show that
 \[
    \forall  \, { \ottnt{K} \,  \in  \,  \LRRel{ \mathcal{K} }{  \sigma  \ottsym{(}  \ottnt{T}  \ottsym{)}  \, / \, ( \forall  \mathit{y}  .  \sigma  \ottsym{(}  \ottnt{C_{{\mathrm{1}}}}  \ottsym{)}  )  }  } . \   \resetcnstr{  \ottnt{K}  [  \mathcal{S}_0 \, \mathit{x}  \ottsym{.}  \sigma  \ottsym{(}  \ottnt{e}  \ottsym{)}  ]  }  \,  \in  \,  \LRRel{ \mathcal{E} }{ \sigma  \ottsym{(}  \ottnt{C_{{\mathrm{2}}}}  \ottsym{)} }   ~.
 \]

 Let $\ottnt{K} \,  \in  \,  \LRRel{ \mathcal{K} }{  \sigma  \ottsym{(}  \ottnt{T}  \ottsym{)}  \, / \, ( \forall  \mathit{y}  .  \sigma  \ottsym{(}  \ottnt{C_{{\mathrm{1}}}}  \ottsym{)}  )  } $.
 We must show that
 \[
   \resetcnstr{  \ottnt{K}  [  \mathcal{S}_0 \, \mathit{x}  \ottsym{.}  \sigma  \ottsym{(}  \ottnt{e}  \ottsym{)}  ]  }  \,  \in  \,  \LRRel{ \mathcal{E} }{ \sigma  \ottsym{(}  \ottnt{C_{{\mathrm{2}}}}  \ottsym{)} }  ~,
 \]
 which is implied by \reflem{lr-anti-red}
 with the following.
 \begin{itemize}
  \item We show that $ \resetcnstr{  \ottnt{K}  [  \mathcal{S}_0 \, \mathit{x}  \ottsym{.}  \sigma  \ottsym{(}  \ottnt{e}  \ottsym{)}  ]  }  \,  \in  \,  \LRRel{ \mathrm{Atom} }{ \sigma  \ottsym{(}  \ottnt{C_{{\mathrm{2}}}}  \ottsym{)} } $.
        It suffices to show that
        $ \emptyset   \vdash   \resetcnstr{  \ottnt{K}  [  \mathcal{S}_0 \, \mathit{x}  \ottsym{.}  \sigma  \ottsym{(}  \ottnt{e}  \ottsym{)}  ]  }   \ottsym{:}  \sigma  \ottsym{(}  \ottnt{C_{{\mathrm{2}}}}  \ottsym{)}$,
        which is implied by \reflem{lr-typing-cont-exp}
        with the following:
        \begin{itemize}
         \item $\ottnt{K}  \ottsym{:}   \sigma  \ottsym{(}  \ottnt{T}  \ottsym{)}  \, / \, ( \forall  \mathit{y}  .  \sigma  \ottsym{(}  \ottnt{C_{{\mathrm{1}}}}  \ottsym{)}  ) $ implied by
               $\ottnt{K} \,  \in  \,  \LRRel{ \mathcal{K} }{  \sigma  \ottsym{(}  \ottnt{T}  \ottsym{)}  \, / \, ( \forall  \mathit{y}  .  \sigma  \ottsym{(}  \ottnt{C_{{\mathrm{1}}}}  \ottsym{)}  )  } $;
         \item $ \emptyset   \vdash  \mathcal{S}_0 \, \mathit{x}  \ottsym{.}  \sigma  \ottsym{(}  \ottnt{e}  \ottsym{)}  \ottsym{:}  \sigma  \ottsym{(}  \ottnt{C}  \ottsym{)}$; and
         \item $\sigma  \ottsym{(}  \ottnt{C}  \ottsym{)} \,  =  \,  \sigma  \ottsym{(}  \ottnt{T}  \ottsym{)}  \,  \,\&\,  \,   \Phi_\mathsf{val}   \, / \,   ( \forall  \mathit{y}  .  \sigma  \ottsym{(}  \ottnt{C_{{\mathrm{1}}}}  \ottsym{)}  )  \Rightarrow   \sigma  \ottsym{(}  \ottnt{C_{{\mathrm{2}}}}  \ottsym{)}  $.
        \end{itemize}

  \item We have $ \resetcnstr{  \ottnt{K}  [  \mathcal{S}_0 \, \mathit{x}  \ottsym{.}  \sigma  \ottsym{(}  \ottnt{e}  \ottsym{)}  ]  }  \,  \mathrel{  \longrightarrow  }  \,  \sigma  \ottsym{(}  \ottnt{e}  \ottsym{)}    [   \lambda\!  \, \mathit{x}  \ottsym{.}   \resetcnstr{  \ottnt{K}  [  \mathit{x}  ]  }   /  \mathit{x}  ]   \,  \,\&\,  \,  \epsilon $
        by \E{Red}/\R{Shift}.

  \item We show that $ \sigma  \ottsym{(}  \ottnt{e}  \ottsym{)}    [   \lambda\!  \, \mathit{x}  \ottsym{.}   \resetcnstr{  \ottnt{K}  [  \mathit{x}  ]  }   /  \mathit{x}  ]   \,  \in  \,  \LRRel{ \mathcal{E} }{ \sigma  \ottsym{(}  \ottnt{C_{{\mathrm{2}}}}  \ottsym{)} } $.
        Because $\ottnt{K} \,  \in  \,  \LRRel{ \mathcal{K} }{  \sigma  \ottsym{(}  \ottnt{T}  \ottsym{)}  \, / \, ( \forall  \mathit{y}  .  \sigma  \ottsym{(}  \ottnt{C_{{\mathrm{1}}}}  \ottsym{)}  )  } $,
        we have
        \[
          \lambda\!  \, \mathit{x}  \ottsym{.}   \resetcnstr{  \ottnt{K}  [  \mathit{x}  ]  }  \,  \in  \,  \LRRel{ \mathcal{V} }{ \ottsym{(}  \mathit{y} \,  {:}  \, \sigma  \ottsym{(}  \ottnt{T}  \ottsym{)}  \ottsym{)}  \rightarrow  \sigma  \ottsym{(}  \ottnt{C_{{\mathrm{1}}}}  \ottsym{)} }  ~.
        \]
        Because $\sigma \,  \in  \,  \LRRel{ \mathcal{G} }{ \Gamma } $, we have
        \[
         \sigma  \mathbin{\uplus}   [   \lambda\!  \, \mathit{x}  \ottsym{.}   \resetcnstr{  \ottnt{K}  [  \mathit{x}  ]  }   /  \mathit{x}  ]  \,  \in  \,  \LRRel{ \mathcal{G} }{ \Gamma  \ottsym{,}  \mathit{x} \,  {:}  \, \ottsym{(}  \mathit{y} \,  {:}  \, \ottnt{T}  \ottsym{)}  \rightarrow  \ottnt{C_{{\mathrm{1}}}} }  ~.
        \]
        Because $\ottnt{e} \,  \in  \,  \LRRel{ \mathcal{O} }{ \Gamma  \ottsym{,}  \mathit{x} \,  {:}  \, \ottsym{(}  \mathit{y} \,  {:}  \, \ottnt{T}  \ottsym{)}  \rightarrow  \ottnt{C_{{\mathrm{1}}}}  \, \vdash \,  \ottnt{C_{{\mathrm{2}}}} } $,
        we have the conclusion $ \sigma  \ottsym{(}  \ottnt{e}  \ottsym{)}    [   \lambda\!  \, \mathit{x}  \ottsym{.}   \resetcnstr{  \ottnt{K}  [  \mathit{x}  ]  }   /  \mathit{x}  ]   \,  \in  \,  \LRRel{ \mathcal{E} }{ \sigma  \ottsym{(}  \ottnt{C_{{\mathrm{2}}}}  \ottsym{)} } $.
        Note that $\mathit{x} \,  \not\in  \,  \mathit{fv}  (  \ottnt{C_{{\mathrm{2}}}}  ) $ by \reflem{ts-unuse-non-refine-vars}.
 \end{itemize}
\end{proof}

\begin{lemmap}{Compatibility: Shift Discarding Continuations}{lr-compat-shift-discard-cont}
 If $\ottnt{e} \,  \in  \,  \LRRel{ \mathcal{O} }{ \Gamma  \, \vdash \,  \ottnt{C_{{\mathrm{2}}}} } $ and $\Gamma  \vdash   \ottnt{T}  \,  \,\&\,  \,  \Phi  \, / \,   ( \forall  \mathit{y}  .  \ottnt{C_{{\mathrm{1}}}}  )  \Rightarrow   \ottnt{C_{{\mathrm{2}}}}  $,
 then $\mathcal{S}_0 \, \mathit{x}  \ottsym{.}  \ottnt{e} \,  \in  \,  \LRRel{ \mathcal{O} }{ \Gamma  \, \vdash \,   \ottnt{T}  \,  \,\&\,  \,  \Phi  \, / \,   ( \forall  \mathit{y}  .  \ottnt{C_{{\mathrm{1}}}}  )  \Rightarrow   \ottnt{C_{{\mathrm{2}}}}   } $.
\end{lemmap}
\begin{proof}
 The proof is similar to that of \reflem{lr-compat-shift}.

 Let $\ottnt{C} \,  =  \,  \ottnt{T}  \,  \,\&\,  \,  \Phi  \, / \,   ( \forall  \mathit{y}  .  \ottnt{C_{{\mathrm{1}}}}  )  \Rightarrow   \ottnt{C_{{\mathrm{2}}}}  $.
 Because $\ottnt{e} \,  \in  \,  \LRRel{ \mathcal{O} }{ \Gamma  \, \vdash \,  \ottnt{C_{{\mathrm{2}}}} } $ implies $\Gamma  \vdash  \ottnt{e}  \ottsym{:}  \ottnt{C_{{\mathrm{2}}}}$,
 \T{ShiftD} with $\Gamma  \vdash  \ottnt{C}$ implies $\Gamma  \vdash  \mathcal{S}_0 \, \mathit{x}  \ottsym{.}  \ottnt{e}  \ottsym{:}  \ottnt{C}$.
 Let $\sigma \,  \in  \,  \LRRel{ \mathcal{G} }{ \Gamma } $.
 Without loss of generality, we can suppose that $\mathit{x}$ does not occur in $\sigma$.
 We must show that
 \[
  \mathcal{S}_0 \, \mathit{x}  \ottsym{.}  \sigma  \ottsym{(}  \ottnt{e}  \ottsym{)} \,  \in  \,  \LRRel{ \mathcal{E} }{ \sigma  \ottsym{(}  \ottnt{C}  \ottsym{)} }  ~.
 \]

 Because $\Gamma  \vdash  \mathcal{S}_0 \, \mathit{x}  \ottsym{.}  \ottnt{e}  \ottsym{:}  \ottnt{C}$ and $\sigma \,  \in  \,  \LRRel{ \mathcal{G} }{ \Gamma } $,
 \reflem{lr-typeability} implies $ \emptyset   \vdash  \mathcal{S}_0 \, \mathit{x}  \ottsym{.}  \sigma  \ottsym{(}  \ottnt{e}  \ottsym{)}  \ottsym{:}  \sigma  \ottsym{(}  \ottnt{C}  \ottsym{)}$.
 Therefore, we have $\mathcal{S}_0 \, \mathit{x}  \ottsym{.}  \sigma  \ottsym{(}  \ottnt{e}  \ottsym{)} \,  \in  \,  \LRRel{ \mathrm{Atom} }{ \sigma  \ottsym{(}  \ottnt{C}  \ottsym{)} } $.

 Because \reflem{lr-red-shift} implies
 \begin{itemize}
  \item $ {    \forall  \, { \ottnt{r}  \ottsym{,}  \varpi } . \  \mathcal{S}_0 \, \mathit{x}  \ottsym{.}  \sigma  \ottsym{(}  \ottnt{e}  \ottsym{)} \,  \Downarrow  \, \ottnt{r} \,  \,\&\,  \, \varpi   \mathrel{\Longrightarrow}  \ottnt{r} \,  =  \, \mathcal{S}_0 \, \mathit{x}  \ottsym{.}  \sigma  \ottsym{(}  \ottnt{e}  \ottsym{)}  } \,\mathrel{\wedge}\, { \varpi \,  =  \,  \epsilon  } $ and
  \item $\mathcal{S}_0 \, \mathit{x}  \ottsym{.}  \sigma  \ottsym{(}  \ottnt{e}  \ottsym{)} \,  \Uparrow  \, \pi$ cannot be derived for any $\pi$,
 \end{itemize}
 it suffices to show that
 \[
    \forall  \, { \ottnt{K} \,  \in  \,  \LRRel{ \mathcal{K} }{  \sigma  \ottsym{(}  \ottnt{T}  \ottsym{)}  \, / \, ( \forall  \mathit{y}  .  \sigma  \ottsym{(}  \ottnt{C_{{\mathrm{1}}}}  \ottsym{)}  )  }  } . \   \resetcnstr{  \ottnt{K}  [  \mathcal{S}_0 \, \mathit{x}  \ottsym{.}  \sigma  \ottsym{(}  \ottnt{e}  \ottsym{)}  ]  }  \,  \in  \,  \LRRel{ \mathcal{E} }{ \sigma  \ottsym{(}  \ottnt{C_{{\mathrm{2}}}}  \ottsym{)} }   ~.
 \]

 Let $\ottnt{K} \,  \in  \,  \LRRel{ \mathcal{K} }{  \sigma  \ottsym{(}  \ottnt{T}  \ottsym{)}  \, / \, ( \forall  \mathit{y}  .  \sigma  \ottsym{(}  \ottnt{C_{{\mathrm{1}}}}  \ottsym{)}  )  } $.
 We must show that
 \[
   \resetcnstr{  \ottnt{K}  [  \mathcal{S}_0 \, \mathit{x}  \ottsym{.}  \sigma  \ottsym{(}  \ottnt{e}  \ottsym{)}  ]  }  \,  \in  \,  \LRRel{ \mathcal{E} }{ \sigma  \ottsym{(}  \ottnt{C_{{\mathrm{2}}}}  \ottsym{)} }  ~,
 \]
 which is implied by \reflem{lr-anti-red}
 with the following.
 \begin{itemize}
  \item We show that $ \resetcnstr{  \ottnt{K}  [  \mathcal{S}_0 \, \mathit{x}  \ottsym{.}  \sigma  \ottsym{(}  \ottnt{e}  \ottsym{)}  ]  }  \,  \in  \,  \LRRel{ \mathrm{Atom} }{ \sigma  \ottsym{(}  \ottnt{C_{{\mathrm{2}}}}  \ottsym{)} } $.
        It suffices to show that
        $ \emptyset   \vdash   \resetcnstr{  \ottnt{K}  [  \mathcal{S}_0 \, \mathit{x}  \ottsym{.}  \sigma  \ottsym{(}  \ottnt{e}  \ottsym{)}  ]  }   \ottsym{:}  \sigma  \ottsym{(}  \ottnt{C_{{\mathrm{2}}}}  \ottsym{)}$,
        which is implied by \reflem{lr-typing-cont-exp}
        with the following:
        \begin{itemize}
         \item $\ottnt{K}  \ottsym{:}   \sigma  \ottsym{(}  \ottnt{T}  \ottsym{)}  \, / \, ( \forall  \mathit{y}  .  \sigma  \ottsym{(}  \ottnt{C_{{\mathrm{1}}}}  \ottsym{)}  ) $ implied by
               $\ottnt{K} \,  \in  \,  \LRRel{ \mathcal{K} }{  \sigma  \ottsym{(}  \ottnt{T}  \ottsym{)}  \, / \, ( \forall  \mathit{y}  .  \sigma  \ottsym{(}  \ottnt{C_{{\mathrm{1}}}}  \ottsym{)}  )  } $;
         \item $ \emptyset   \vdash  \mathcal{S}_0 \, \mathit{x}  \ottsym{.}  \sigma  \ottsym{(}  \ottnt{e}  \ottsym{)}  \ottsym{:}  \sigma  \ottsym{(}  \ottnt{C}  \ottsym{)}$; and
         \item $\sigma  \ottsym{(}  \ottnt{C}  \ottsym{)} \,  =  \,  \sigma  \ottsym{(}  \ottnt{T}  \ottsym{)}  \,  \,\&\,  \,  \sigma  \ottsym{(}  \Phi  \ottsym{)}  \, / \,   ( \forall  \mathit{y}  .  \sigma  \ottsym{(}  \ottnt{C_{{\mathrm{1}}}}  \ottsym{)}  )  \Rightarrow   \sigma  \ottsym{(}  \ottnt{C_{{\mathrm{2}}}}  \ottsym{)}  $.
        \end{itemize}

  \item We have $ \resetcnstr{  \ottnt{K}  [  \mathcal{S}_0 \, \mathit{x}  \ottsym{.}  \sigma  \ottsym{(}  \ottnt{e}  \ottsym{)}  ]  }  \,  \mathrel{  \longrightarrow  }  \, \sigma  \ottsym{(}  \ottnt{e}  \ottsym{)} \,  \,\&\,  \,  \epsilon $
        because:
        $ \resetcnstr{  \ottnt{K}  [  \mathcal{S}_0 \, \mathit{x}  \ottsym{.}  \sigma  \ottsym{(}  \ottnt{e}  \ottsym{)}  ]  }  \,  \mathrel{  \longrightarrow  }  \,  \sigma  \ottsym{(}  \ottnt{e}  \ottsym{)}    [   \lambda\!  \, \mathit{x}  \ottsym{.}   \resetcnstr{  \ottnt{K}  [  \mathit{x}  ]  }   /  \mathit{x}  ]   \,  \,\&\,  \,  \epsilon $
        by \E{Red}/\R{Shift}; and
        $ \sigma  \ottsym{(}  \ottnt{e}  \ottsym{)}    [   \lambda\!  \, \mathit{x}  \ottsym{.}   \resetcnstr{  \ottnt{K}  [  \mathit{x}  ]  }   /  \mathit{x}  ]   \,  =  \, \sigma  \ottsym{(}  \ottnt{e}  \ottsym{)}$,
        which is implied by $\mathit{x} \,  \not\in  \,  \mathit{fv}  (  \ottnt{e}  ) $ and $\sigma$ is a closing substitution.

  \item We show that $\sigma  \ottsym{(}  \ottnt{e}  \ottsym{)} \,  \in  \,  \LRRel{ \mathcal{E} }{ \sigma  \ottsym{(}  \ottnt{C_{{\mathrm{2}}}}  \ottsym{)} } $,
        which is implied by $\ottnt{e} \,  \in  \,  \LRRel{ \mathcal{O} }{ \Gamma  \, \vdash \,  \ottnt{C_{{\mathrm{2}}}} } $ and $\sigma \,  \in  \,  \LRRel{ \mathcal{G} }{ \Gamma } $.
 \end{itemize}
\end{proof}

\begin{lemmap}{Elimination of Intersection}{ts-valid-elim-intersect}
 If $\Gamma  \models   \phi_{{\mathrm{1}}}    \wedge    \phi_{{\mathrm{2}}} $,
 then $\Gamma  \models  \phi_{{\mathrm{1}}}$ and $\Gamma  \models  \phi_{{\mathrm{2}}}$.
\end{lemmap}
\begin{proof}
 \TS{Validity}
 Obvious by \reflem{ts-logic-closing-dsubs}.
\end{proof}

\begin{lemmap}{Empty Evaluation Context in the Logical Relation}{lr-rel-emp-ectx}
 If $ \emptyset   \vdash  \ottnt{T}$,
 then $\,[\,]\, \,  \in  \,  \LRRel{ \mathcal{K} }{  \ottnt{T}  \, / \, ( \forall  \mathit{x}  .   \ottnt{T}  \,  \,\&\,  \,   \Phi_\mathsf{val}   \, / \,   \square    )  } $.
\end{lemmap}
\begin{proof}
 Let $\ottnt{C} \,  =  \,  \ottnt{T}  \,  \,\&\,  \,   \Phi_\mathsf{val}   \, / \,   \square  $.
 The conclusion $\,[\,]\, \,  \in  \,  \LRRel{ \mathcal{K} }{  \ottnt{T}  \, / \, ( \forall  \mathit{x}  .  \ottnt{C}  )  } $ is implied by the following.
 \begin{itemize}
  \item We show that $\,[\,]\,  \ottsym{:}   \ottnt{T}  \, / \, ( \forall  \mathit{x}  .  \ottnt{C}  ) $.
        It suffices to show that
        \[
         \,[\,]\,  \ottsym{:}    \ottnt{T}  \, / \, ( \forall  \mathit{x}  .  \ottnt{C}  )   \mathrel{ \Rrightarrow }  \ottnt{T}  \,  \,\&\,  \,   \Phi_\mathsf{val}   \, / \, ( \forall  \mathit{x}  .  \ottnt{C}  )  ~,
        \]
        which is derivable by \TK{Empty}.

  \item We show that $ \lambda\!  \, \mathit{x}  \ottsym{.}   \resetcnstr{ \mathit{x} }  \,  \in  \,  \LRRel{ \mathcal{V} }{ \ottsym{(}  \mathit{x} \,  {:}  \, \ottnt{T}  \ottsym{)}  \rightarrow  \ottnt{C} } $.

        Let $\mathit{y}$ be a fresh variable.
        We first show that
        \[
         \mathit{x} \,  {:}  \, \ottnt{T}  \vdash  \mathit{x}  \ottsym{:}   \ottnt{T}  \,  \,\&\,  \,   \Phi_\mathsf{val}   \, / \,   ( \forall  \mathit{y}  .  \ottnt{C}  )  \Rightarrow   \ottnt{C}   ~,
        \]
        which is derived by \T{Sub} with the following.
        \begin{itemize}
         \item We show that $\mathit{x} \,  {:}  \, \ottnt{T}  \vdash  \mathit{x}  \ottsym{:}   \ottnt{T}  \,  \,\&\,  \,   \Phi_\mathsf{val}   \, / \,   \square  $,
               which is implied by \reflem{ts-typing-vars}
               because $ \emptyset   \vdash  \ottnt{T}$ implies $\vdash   \emptyset   \ottsym{,}  \mathit{x} \,  {:}  \, \ottnt{T}$.
         \item We show that
               $\mathit{x} \,  {:}  \, \ottnt{T}  \vdash   \ottnt{T}  \,  \,\&\,  \,   \Phi_\mathsf{val}   \, / \,   \square    \ottsym{<:}   \ottnt{T}  \,  \,\&\,  \,   \Phi_\mathsf{val}   \, / \,   ( \forall  \mathit{y}  .  \ottnt{C}  )  \Rightarrow   \ottnt{C}  $,
               which is proven by \Srule{Comp} with the following.
               \begin{itemize}
                \item $\mathit{x} \,  {:}  \, \ottnt{T}  \vdash  \ottnt{T}  \ottsym{<:}  \ottnt{T}$, which is implied by
                      Lemmas~\ref{lem:ts-refl-subtyping} and \ref{lem:ts-weak-subtyping}
                      with $ \emptyset   \vdash  \ottnt{T}$.
                \item $\mathit{x} \,  {:}  \, \ottnt{T}  \vdash   \Phi_\mathsf{val}   \ottsym{<:}   \Phi_\mathsf{val} $, which is implied
                      by Lemmas~\ref{lem:ts-wf-TP_val} and \ref{lem:ts-refl-subtyping}
                      with $\vdash   \emptyset   \ottsym{,}  \mathit{x} \,  {:}  \, \ottnt{T}$.
                \item We show that $\mathit{x} \,  {:}  \, \ottnt{T} \,  |  \, \ottnt{T} \,  \,\&\,  \,  \Phi_\mathsf{val}   \vdash   \square   \ottsym{<:}   ( \forall  \mathit{y}  .  \ottnt{C}  )  \Rightarrow   \ottnt{C} $.
                      Because $\mathit{y} \,  \not\in  \,  \mathit{fv}  (   \Phi_\mathsf{val}   )  \,  \mathbin{\cup}  \,  \mathit{fv}  (  \ottnt{C}  ) $,
                      \Srule{Embed} implies that it suffices to show the following.
                      \begin{itemize}
                       \item We show that $\mathit{x} \,  {:}  \, \ottnt{T}  \ottsym{,}  \mathit{y} \,  {:}  \, \ottnt{T}  \vdash   \Phi_\mathsf{val}  \,  \tceappendop  \, \ottnt{C}  \ottsym{<:}  \ottnt{C}$.
                             By \reflem{ts-add-TP_val-subty}, it suffices to show that
                             $\mathit{x} \,  {:}  \, \ottnt{T}  \ottsym{,}  \mathit{y} \,  {:}  \, \ottnt{T}  \vdash  \ottnt{C}  \ottsym{<:}  \ottnt{C}$.
                             Because \reflem{ts-wf-TP_val} with $ \emptyset   \vdash  \ottnt{T}$
                             implies $ \emptyset   \vdash   \ottnt{T}  \,  \,\&\,  \,   \Phi_\mathsf{val}   \, / \,   \square  $, i.e., $ \emptyset   \vdash  \ottnt{C}$,
                             we have the conclusion by Lemmas~\ref{lem:ts-refl-subtyping} and \ref{lem:ts-weak-subtyping}.

                       \item We show that $\mathit{x} \,  {:}  \, \ottnt{T}  \ottsym{,}  \mathit{y} \,  {:}  \,  \Sigma ^\omega   \models    {  \Phi_\mathsf{val}  }^\nu   \ottsym{(}  \mathit{y}  \ottsym{)}    \Rightarrow     { \ottnt{C} }^\nu (  \mathit{y}  )  $.
                             Because $ {  \Phi_\mathsf{val}  }^\nu   \ottsym{(}  \mathit{y}  \ottsym{)} \,  =  \,  \bot $,
                             we have the conclusion by \reflem{ts-valid-intro-bot-impl}.
                      \end{itemize}
               \end{itemize}
         \item We show that
               $\mathit{x} \,  {:}  \, \ottnt{T}  \vdash   \ottnt{T}  \,  \,\&\,  \,   \Phi_\mathsf{val}   \, / \,   ( \forall  \mathit{y}  .  \ottnt{C}  )  \Rightarrow   \ottnt{C}  $.
               By \WFT{Ans} and \WFT{Comp}, it suffices to show the following.
               \begin{itemize}
                \item $\mathit{x} \,  {:}  \, \ottnt{T}  \vdash  \ottnt{T}$, which is implied by
                      \reflem{ts-weak-wf-ftyping} with $ \emptyset   \vdash  \ottnt{T}$ and $\vdash   \emptyset   \ottsym{,}  \mathit{x} \,  {:}  \, \ottnt{T}$.
                \item $\mathit{x} \,  {:}  \, \ottnt{T}  \vdash   \Phi_\mathsf{val} $, which is implied by
                      \reflem{ts-wf-TP_val} with $\vdash   \emptyset   \ottsym{,}  \mathit{x} \,  {:}  \, \ottnt{T}$.
                \item $\mathit{x} \,  {:}  \, \ottnt{T}  \ottsym{,}  \mathit{y} \,  {:}  \, \ottnt{T}  \vdash  \ottnt{C}$ and $\mathit{x} \,  {:}  \, \ottnt{T}  \vdash  \ottnt{C}$, which are implied by
                      \reflem{ts-weak-wf-ftyping} with $ \emptyset   \vdash  \ottnt{C}$ shown above.
               \end{itemize}
        \end{itemize}
        Then, we have the following typing derivation.
        \begin{prooftree}
         \AxiomC{$\mathit{x} \,  {:}  \, \ottnt{T}  \vdash  \mathit{x}  \ottsym{:}   \ottnt{T}  \,  \,\&\,  \,   \Phi_\mathsf{val}   \, / \,   ( \forall  \mathit{y}  .  \ottnt{C}  )  \Rightarrow   \ottnt{C}  $}
         \RightLabel{\T{Reset}}
         \UnaryInfC{$\mathit{x} \,  {:}  \, \ottnt{T}  \vdash   \resetcnstr{ \mathit{x} }   \ottsym{:}  \ottnt{C}$}
         \RightLabel{\T{Abs}}
         \UnaryInfC{$ \emptyset   \vdash   \lambda\!  \, \mathit{x}  \ottsym{.}   \resetcnstr{ \mathit{x} }   \ottsym{:}  \ottsym{(}  \mathit{x} \,  {:}  \, \ottnt{T}  \ottsym{)}  \rightarrow  \ottnt{C}$}
        \end{prooftree}
        Therefore, we have $ \lambda\!  \, \mathit{x}  \ottsym{.}   \resetcnstr{ \mathit{x} }  \,  \in  \,  \LRRel{ \mathrm{Atom} }{ \ottsym{(}  \mathit{x} \,  {:}  \, \ottnt{T}  \ottsym{)}  \rightarrow  \ottnt{C} } $.

        Next, we show that
        \[
           \forall  \, { \ottnt{v} \,  \in  \,  \LRRel{ \mathcal{V} }{ \ottnt{T} }  } . \  \ottsym{(}   \lambda\!  \, \mathit{x}  \ottsym{.}   \resetcnstr{ \mathit{x} }   \ottsym{)} \, \ottnt{v} \,  \in  \,  \LRRel{ \mathcal{E} }{  \ottnt{C}    [  \ottnt{v}  /  \mathit{x}  ]   }   =  \LRRel{ \mathcal{E} }{  \ottnt{T}  \,  \,\&\,  \,   \Phi_\mathsf{val}   \, / \,   \square   } 
        \]
        (note that $\mathit{x}$ does not occur free in $\ottnt{T}$).
        Let $\ottnt{v} \,  \in  \,  \LRRel{ \mathcal{V} }{ \ottnt{T} } $.
        Because
        \begin{itemize}
         \item $ {    \forall  \, { \ottnt{r}  \ottsym{,}  \varpi } . \  \ottsym{(}   \lambda\!  \, \mathit{x}  \ottsym{.}   \resetcnstr{ \mathit{x} }   \ottsym{)} \, \ottnt{v} \,  \Downarrow  \, \ottnt{r} \,  \,\&\,  \, \varpi   \mathrel{\Longrightarrow}  \ottnt{r} \,  =  \, \ottnt{v}  } \,\mathrel{\wedge}\, { \varpi \,  =  \,  \epsilon  } $
               because $\ottsym{(}   \lambda\!  \, \mathit{x}  \ottsym{.}   \resetcnstr{ \mathit{x} }   \ottsym{)} \, \ottnt{v} \,  \mathrel{  \longrightarrow  ^*}  \, \ottnt{v} \,  \,\&\,  \,  \epsilon $ by \E{Red}/\R{Beta} and \E{Red}/\R{Reset}, and
               the semantics is deterministic (\reflem{lr-eval-det}), and
         \item $\ottsym{(}   \lambda\!  \, \mathit{x}  \ottsym{.}   \resetcnstr{ \mathit{x} }   \ottsym{)} \, \ottnt{v} \,  \Uparrow  \, \pi$ cannot be derived for any $\pi$ by \reflem{lr-red-val},
        \end{itemize}
        it suffices to show that $\ottnt{v} \,  \in  \,  \LRRel{ \mathcal{V} }{ \ottnt{T} } $, which has been assumed.

        Therefore, we have the conclusion $ \lambda\!  \, \mathit{x}  \ottsym{.}   \resetcnstr{ \mathit{x} }  \,  \in  \,  \LRRel{ \mathcal{V} }{ \ottsym{(}  \mathit{x} \,  {:}  \, \ottnt{T}  \ottsym{)}  \rightarrow  \ottnt{C} } $.
 \end{itemize}
\end{proof}

\begin{lemmap}{Terminating Evaluation of Reset}{lr-red-reset-term}
 If $ \resetcnstr{ \ottnt{e_{{\mathrm{1}}}} }  \,  \mathrel{  \longrightarrow  }^{ \ottmv{n} }  \, \ottnt{e_{{\mathrm{2}}}} \,  \,\&\,  \, \varpi$,
 then one of the following holds.
 \begin{itemize}
  \item $ {   \exists  \, { \ottnt{v} } . \  \ottnt{e_{{\mathrm{1}}}} \,  \mathrel{  \longrightarrow  ^*}  \, \ottnt{v} \,  \,\&\,  \, \varpi  } \,\mathrel{\wedge}\, { \ottnt{e_{{\mathrm{2}}}} \,  =  \, \ottnt{v} } $;
  \item $ {  {   \exists  \, { \ottnt{K_{{\mathrm{0}}}}  \ottsym{,}  \mathit{x}  \ottsym{,}  \ottnt{e_{{\mathrm{0}}}}  \ottsym{,}  \varpi_{{\mathrm{1}}}  \ottsym{,}  \varpi_{{\mathrm{2}}} } . \  \ottnt{e_{{\mathrm{1}}}} \,  \mathrel{  \longrightarrow  ^*}  \,  \ottnt{K_{{\mathrm{0}}}}  [  \mathcal{S}_0 \, \mathit{x}  \ottsym{.}  \ottnt{e_{{\mathrm{0}}}}  ]  \,  \,\&\,  \, \varpi_{{\mathrm{1}}}  } \,\mathrel{\wedge}\, {  \ottnt{e_{{\mathrm{0}}}}    [   \lambda\!  \, \mathit{x}  \ottsym{.}   \resetcnstr{  \ottnt{K_{{\mathrm{0}}}}  [  \mathit{x}  ]  }   /  \mathit{x}  ]   \,  \mathrel{  \longrightarrow  ^*}  \, \ottnt{e_{{\mathrm{2}}}} \,  \,\&\,  \, \varpi_{{\mathrm{2}}} }  } \,\mathrel{\wedge}\, { \varpi \,  =  \, \varpi_{{\mathrm{1}}} \,  \tceappendop  \, \varpi_{{\mathrm{2}}} } $; or
  \item $ {   \exists  \, { \ottnt{e'_{{\mathrm{2}}}} } . \  \ottnt{e_{{\mathrm{1}}}} \,  \mathrel{  \longrightarrow  ^*}  \, \ottnt{e'_{{\mathrm{2}}}} \,  \,\&\,  \, \varpi  } \,\mathrel{\wedge}\, { \ottnt{e_{{\mathrm{2}}}} \,  =  \,  \resetcnstr{ \ottnt{e'_{{\mathrm{2}}}} }  } $.
 \end{itemize}
\end{lemmap}
\begin{proof}
 By induction on $\ottmv{n}$.
 \begin{caseanalysis}
  \case $\ottmv{n} \,  =  \, \ottsym{0}$:
   We are given $ \resetcnstr{ \ottnt{e_{{\mathrm{1}}}} }  \,  =  \, \ottnt{e_{{\mathrm{2}}}}$ and $ \varpi    =     \epsilon  $.
   Then, we have the conclusion (the third case) by letting $\ottnt{e'_{{\mathrm{2}}}} \,  =  \, \ottnt{e_{{\mathrm{1}}}}$.

  \case $\ottmv{n} \,  >  \, \ottsym{0}$:
   We are given $ \resetcnstr{ \ottnt{e_{{\mathrm{1}}}} }  \,  \mathrel{  \longrightarrow  }  \, \ottnt{e_{{\mathrm{3}}}} \,  \,\&\,  \, \varpi_{{\mathrm{1}}}$ and $\ottnt{e_{{\mathrm{3}}}} \,  \mathrel{  \longrightarrow  }^{ \ottmv{n}  \ottsym{-}  \ottsym{1} }  \, \ottnt{e_{{\mathrm{2}}}} \,  \,\&\,  \, \varpi_{{\mathrm{2}}}$
   for some $\ottnt{e_{{\mathrm{3}}}}$, $\varpi_{{\mathrm{1}}}$, and $\varpi_{{\mathrm{2}}}$ such that $ \varpi    =    \varpi_{{\mathrm{1}}} \,  \tceappendop  \, \varpi_{{\mathrm{2}}} $.
   By case analysis on the derivation of $ \resetcnstr{ \ottnt{e_{{\mathrm{1}}}} }  \,  \mathrel{  \longrightarrow  }  \, \ottnt{e_{{\mathrm{3}}}} \,  \,\&\,  \, \varpi_{{\mathrm{1}}}$.
   \begin{caseanalysis}
    \case \E{Red}/\R{Shift}:
     We are given
     \begin{prooftree}
      \AxiomC{$ \resetcnstr{  \ottnt{K_{{\mathrm{0}}}}  [  \mathcal{S}_0 \, \mathit{x}  \ottsym{.}  \ottnt{e_{{\mathrm{0}}}}  ]  }  \,  \mathrel{  \longrightarrow  }  \,  \ottnt{e_{{\mathrm{0}}}}    [   \lambda\!  \, \mathit{x}  \ottsym{.}   \resetcnstr{  \ottnt{K_{{\mathrm{0}}}}  [  \mathit{x}  ]  }   /  \mathit{x}  ]   \,  \,\&\,  \,  \epsilon $}
     \end{prooftree}
     for some $\ottnt{K_{{\mathrm{0}}}}$, $\mathit{x}$, and $\ottnt{e_{{\mathrm{0}}}}$ such that $\ottnt{e_{{\mathrm{1}}}} \,  =  \,  \ottnt{K_{{\mathrm{0}}}}  [  \mathcal{S}_0 \, \mathit{x}  \ottsym{.}  \ottnt{e_{{\mathrm{0}}}}  ] $
     and $\ottnt{e_{{\mathrm{3}}}} \,  =  \,  \ottnt{e_{{\mathrm{0}}}}    [   \lambda\!  \, \mathit{x}  \ottsym{.}   \resetcnstr{  \ottnt{K_{{\mathrm{0}}}}  [  \mathit{x}  ]  }   /  \mathit{x}  ]  $.
     We also have $ \varpi_{{\mathrm{1}}}    =     \epsilon  $.
     Then, we have the conclusion (the second case) because
     \begin{itemize}
      \item $\ottnt{e_{{\mathrm{1}}}} \,  \mathrel{  \longrightarrow  }^{ \ottsym{0} }  \,  \ottnt{K_{{\mathrm{0}}}}  [  \mathcal{S}_0 \, \mathit{x}  \ottsym{.}  \ottnt{e_{{\mathrm{0}}}}  ]  \,  \,\&\,  \,  \epsilon $,
      \item $ \ottnt{e_{{\mathrm{0}}}}    [   \lambda\!  \, \mathit{x}  \ottsym{.}   \resetcnstr{  \ottnt{K_{{\mathrm{0}}}}  [  \mathit{x}  ]  }   /  \mathit{x}  ]   = \ottnt{e_{{\mathrm{3}}}} \,  \mathrel{  \longrightarrow  }^{ \ottmv{n}  \ottsym{-}  \ottsym{1} }  \, \ottnt{e_{{\mathrm{2}}}} \,  \,\&\,  \, \varpi_{{\mathrm{2}}}$, and
      \item $ \varpi    =     \epsilon  \,  \tceappendop  \, \varpi_{{\mathrm{2}}} $.
     \end{itemize}

    \case \E{Red}/\R{Reset}:
     We are given
     \begin{prooftree}
      \AxiomC{$ \resetcnstr{ \ottnt{v} }  \,  \mathrel{  \longrightarrow  }  \, \ottnt{v} \,  \,\&\,  \,  \epsilon $}
     \end{prooftree}
     for some $\ottnt{v}$ such that $\ottnt{e_{{\mathrm{1}}}} = \ottnt{e_{{\mathrm{3}}}} = \ottnt{v}$.
     We also have $ \varpi_{{\mathrm{1}}}    =     \epsilon  $.
     By \reflem{lr-red-val} with $\ottnt{e_{{\mathrm{3}}}} \,  \mathrel{  \longrightarrow  }^{ \ottmv{n}  \ottsym{-}  \ottsym{1} }  \, \ottnt{e_{{\mathrm{2}}}} \,  \,\&\,  \, \varpi_{{\mathrm{2}}}$,
     $\ottnt{e_{{\mathrm{2}}}} \,  =  \, \ottnt{v}$ and $ \varpi_{{\mathrm{2}}}    =     \epsilon  $.
     Then, we have the conclusion (the first case) because
     $\ottnt{e_{{\mathrm{1}}}} = \ottnt{v} \,  \mathrel{  \longrightarrow  ^*}  \, \ottnt{v} \,  \,\&\,  \,  \epsilon $ and
     $\varpi = \varpi_{{\mathrm{1}}} \,  \tceappendop  \, \varpi_{{\mathrm{2}}} =  \epsilon $.

    \case \E{Reset}:
     We are given
     \begin{prooftree}
      \AxiomC{$\ottnt{e_{{\mathrm{1}}}} \,  \mathrel{  \longrightarrow  }  \, \ottnt{e'_{{\mathrm{3}}}} \,  \,\&\,  \, \varpi_{{\mathrm{1}}}$}
      \UnaryInfC{$ \resetcnstr{ \ottnt{e_{{\mathrm{1}}}} }  \,  \mathrel{  \longrightarrow  }  \,  \resetcnstr{ \ottnt{e'_{{\mathrm{3}}}} }  \,  \,\&\,  \, \varpi_{{\mathrm{1}}}$}
     \end{prooftree}
     for some $\ottnt{e'_{{\mathrm{3}}}}$ such that $\ottnt{e_{{\mathrm{3}}}} \,  =  \,  \resetcnstr{ \ottnt{e'_{{\mathrm{3}}}} } $.
     Then, we have $ \resetcnstr{ \ottnt{e'_{{\mathrm{3}}}} }  \,  \mathrel{  \longrightarrow  }^{ \ottmv{n}  \ottsym{-}  \ottsym{1} }  \, \ottnt{e_{{\mathrm{2}}}} \,  \,\&\,  \, \varpi_{{\mathrm{2}}}$.
     We perform case analysis on the result of the IH.
     \begin{caseanalysis}
      \case $ {   \exists  \, { \ottnt{v} } . \  \ottnt{e'_{{\mathrm{3}}}} \,  \mathrel{  \longrightarrow  ^*}  \, \ottnt{v} \,  \,\&\,  \, \varpi_{{\mathrm{2}}}  } \,\mathrel{\wedge}\, { \ottnt{e_{{\mathrm{2}}}} \,  =  \, \ottnt{v} } $:
       Because $\ottnt{e_{{\mathrm{1}}}} \,  \mathrel{  \longrightarrow  }  \, \ottnt{e'_{{\mathrm{3}}}} \,  \,\&\,  \, \varpi_{{\mathrm{1}}}$ and $\ottnt{e'_{{\mathrm{3}}}} \,  \mathrel{  \longrightarrow  ^*}  \, \ottnt{v} \,  \,\&\,  \, \varpi_{{\mathrm{2}}}$,
       we have $\ottnt{e_{{\mathrm{1}}}} \,  \mathrel{  \longrightarrow  ^*}  \, \ottnt{v} \,  \,\&\,  \, \varpi_{{\mathrm{1}}} \,  \tceappendop  \, \varpi_{{\mathrm{2}}}$.
       Because $ \varpi    =    \varpi_{{\mathrm{1}}} \,  \tceappendop  \, \varpi_{{\mathrm{2}}} $, we have the conclusion (the first case).

      \case $ {  {   \exists  \, { \ottnt{K_{{\mathrm{0}}}}  \ottsym{,}  \mathit{x}  \ottsym{,}  \ottnt{e_{{\mathrm{0}}}}  \ottsym{,}  \varpi_{{\mathrm{21}}}  \ottsym{,}  \varpi_{{\mathrm{22}}} } . \  \ottnt{e'_{{\mathrm{3}}}} \,  \mathrel{  \longrightarrow  ^*}  \,  \ottnt{K_{{\mathrm{0}}}}  [  \mathcal{S}_0 \, \mathit{x}  \ottsym{.}  \ottnt{e_{{\mathrm{0}}}}  ]  \,  \,\&\,  \, \varpi_{{\mathrm{21}}}  } \,\mathrel{\wedge}\, {  \ottnt{e_{{\mathrm{0}}}}    [   \lambda\!  \, \mathit{x}  \ottsym{.}   \resetcnstr{  \ottnt{K_{{\mathrm{0}}}}  [  \mathit{x}  ]  }   /  \mathit{x}  ]   \,  \mathrel{  \longrightarrow  ^*}  \, \ottnt{e_{{\mathrm{2}}}} \,  \,\&\,  \, \varpi_{{\mathrm{22}}} }  } \,\mathrel{\wedge}\, { \varpi_{{\mathrm{2}}} \,  =  \, \varpi_{{\mathrm{21}}} \,  \tceappendop  \, \varpi_{{\mathrm{22}}} } $:
       Because $\ottnt{e_{{\mathrm{1}}}} \,  \mathrel{  \longrightarrow  }  \, \ottnt{e'_{{\mathrm{3}}}} \,  \,\&\,  \, \varpi_{{\mathrm{1}}}$ and $\ottnt{e'_{{\mathrm{3}}}} \,  \mathrel{  \longrightarrow  ^*}  \,  \ottnt{K_{{\mathrm{0}}}}  [  \mathcal{S}_0 \, \mathit{x}  \ottsym{.}  \ottnt{e_{{\mathrm{0}}}}  ]  \,  \,\&\,  \, \varpi_{{\mathrm{21}}}$,
       we have $\ottnt{e_{{\mathrm{1}}}} \,  \mathrel{  \longrightarrow  ^*}  \,  \ottnt{K_{{\mathrm{0}}}}  [  \mathcal{S}_0 \, \mathit{x}  \ottsym{.}  \ottnt{e_{{\mathrm{0}}}}  ]  \,  \,\&\,  \, \varpi_{{\mathrm{1}}} \,  \tceappendop  \, \varpi_{{\mathrm{21}}}$.
       Because $ \ottsym{(}  \varpi_{{\mathrm{1}}} \,  \tceappendop  \, \varpi_{{\mathrm{21}}}  \ottsym{)} \,  \tceappendop  \, \varpi_{{\mathrm{22}}}    =    \varpi $, we have the conclusion (the second case).

      \case $ {   \exists  \, { \ottnt{e'_{{\mathrm{2}}}} } . \  \ottnt{e'_{{\mathrm{3}}}} \,  \mathrel{  \longrightarrow  ^*}  \, \ottnt{e'_{{\mathrm{2}}}} \,  \,\&\,  \, \varpi_{{\mathrm{2}}}  } \,\mathrel{\wedge}\, { \ottnt{e_{{\mathrm{2}}}} \,  =  \,  \resetcnstr{ \ottnt{e'_{{\mathrm{2}}}} }  } $:
       Because $\ottnt{e_{{\mathrm{1}}}} \,  \mathrel{  \longrightarrow  }  \, \ottnt{e'_{{\mathrm{3}}}} \,  \,\&\,  \, \varpi_{{\mathrm{1}}}$ and $\ottnt{e'_{{\mathrm{3}}}} \,  \mathrel{  \longrightarrow  ^*}  \, \ottnt{e'_{{\mathrm{2}}}} \,  \,\&\,  \, \varpi_{{\mathrm{2}}}$,
       we have $\ottnt{e_{{\mathrm{1}}}} \,  \mathrel{  \longrightarrow  ^*}  \, \ottnt{e'_{{\mathrm{2}}}} \,  \,\&\,  \, \varpi_{{\mathrm{1}}} \,  \tceappendop  \, \varpi_{{\mathrm{2}}}$.
       Then, because $ \varpi    =    \varpi_{{\mathrm{1}}} \,  \tceappendop  \, \varpi_{{\mathrm{2}}} $, we have the conclusion (the third case).
     \end{caseanalysis}
   \end{caseanalysis}
 \end{caseanalysis}
\end{proof}

\begin{lemmap}{Decomposition of Evaluation}{lr-eval-decomp}
 If $\ottnt{e} \,  \mathrel{  \longrightarrow  }^{ \ottmv{n} }  \, \ottnt{e'} \,  \,\&\,  \, \varpi_{{\mathrm{1}}} \,  \tceappendop  \, \varpi_{{\mathrm{2}}}$,
 then $\ottnt{e} \,  \mathrel{  \longrightarrow  ^*}  \, \ottnt{e''} \,  \,\&\,  \, \varpi_{{\mathrm{1}}}$ and $\ottnt{e''} \,  \mathrel{  \longrightarrow  ^*}  \, \ottnt{e'} \,  \,\&\,  \, \varpi_{{\mathrm{2}}}$
 for some $\ottnt{e''}$.
\end{lemmap}
\begin{proof}
 By induction on $\ottmv{n}$.
 \begin{caseanalysis}
  \case $\ottmv{n} \,  =  \, \ottsym{0}$:
   Because $\ottnt{e} \,  =  \, \ottnt{e'}$ and $\varpi_{{\mathrm{1}}} = \varpi_{{\mathrm{2}}} =  \epsilon $,
   we have the conclusion by letting $\ottnt{e''} \,  =  \, \ottnt{e}$.

  \case $\ottmv{n} \,  >  \, \ottsym{0}$:
   We are given $\ottnt{e} \,  \mathrel{  \longrightarrow  }  \, \ottnt{e_{{\mathrm{0}}}} \,  \,\&\,  \, \varpi'_{{\mathrm{1}}}$ and $\ottnt{e_{{\mathrm{0}}}} \,  \mathrel{  \longrightarrow  }^{ \ottmv{n}  \ottsym{-}  \ottsym{1} }  \, \ottnt{e'} \,  \,\&\,  \, \varpi'_{{\mathrm{2}}}$
   for some $\ottnt{e_{{\mathrm{0}}}}$, $\varpi'_{{\mathrm{1}}}$, and $\varpi'_{{\mathrm{2}}}$ such that $ \varpi_{{\mathrm{1}}} \,  \tceappendop  \, \varpi_{{\mathrm{2}}}    =    \varpi'_{{\mathrm{1}}} \,  \tceappendop  \, \varpi'_{{\mathrm{2}}} $.
   By case analysis on $\varpi_{{\mathrm{1}}}$.
   \begin{caseanalysis}
    \case $ \varpi_{{\mathrm{1}}}    =     \epsilon  $:
     We have the conclusion by letting $\ottnt{e''} \,  =  \, \ottnt{e}$.

    \case $ \varpi_{{\mathrm{1}}}    \not=     \epsilon  $:
     Because the length of $\varpi'_{{\mathrm{1}}}$ is at most one, and $ \varpi_{{\mathrm{1}}}    \not=     \epsilon  $,
     $ \varpi_{{\mathrm{1}}} \,  \tceappendop  \, \varpi_{{\mathrm{2}}}    =    \varpi'_{{\mathrm{1}}} \,  \tceappendop  \, \varpi'_{{\mathrm{2}}} $ implies that
     $\varpi'_{{\mathrm{1}}}$ is a prefix of $\varpi_{{\mathrm{1}}}$, that is,
     $ \varpi_{{\mathrm{1}}}    =    \varpi'_{{\mathrm{1}}} \,  \tceappendop  \, \varpi''_{{\mathrm{1}}} $ for some $\varpi''_{{\mathrm{1}}}$.
     Then, because $ \varpi_{{\mathrm{1}}} \,  \tceappendop  \, \varpi_{{\mathrm{2}}}    =    \varpi'_{{\mathrm{1}}} \,  \tceappendop  \, \varpi'_{{\mathrm{2}}} $ implies $ \varpi'_{{\mathrm{1}}} \,  \tceappendop  \, \varpi''_{{\mathrm{1}}} \,  \tceappendop  \, \varpi_{{\mathrm{2}}}    =    \varpi'_{{\mathrm{1}}} \,  \tceappendop  \, \varpi'_{{\mathrm{2}}} $,
     we have $ \varpi'_{{\mathrm{2}}}    =    \varpi''_{{\mathrm{1}}} \,  \tceappendop  \, \varpi_{{\mathrm{2}}} $.
     Because $\ottnt{e_{{\mathrm{0}}}} \,  \mathrel{  \longrightarrow  }^{ \ottmv{n}  \ottsym{-}  \ottsym{1} }  \, \ottnt{e'} \,  \,\&\,  \, \varpi'_{{\mathrm{2}}}$ implies $\ottnt{e_{{\mathrm{0}}}} \,  \mathrel{  \longrightarrow  }^{ \ottmv{n}  \ottsym{-}  \ottsym{1} }  \, \ottnt{e'} \,  \,\&\,  \, \varpi''_{{\mathrm{1}}} \,  \tceappendop  \, \varpi_{{\mathrm{2}}}$,
     the IH implies that there exists some $\ottnt{e''}$ such that
     $\ottnt{e_{{\mathrm{0}}}} \,  \mathrel{  \longrightarrow  ^*}  \, \ottnt{e''} \,  \,\&\,  \, \varpi''_{{\mathrm{1}}}$ and $\ottnt{e''} \,  \mathrel{  \longrightarrow  ^*}  \, \ottnt{e'} \,  \,\&\,  \, \varpi_{{\mathrm{2}}}$.
     Because $\ottnt{e} \,  \mathrel{  \longrightarrow  }  \, \ottnt{e_{{\mathrm{0}}}} \,  \,\&\,  \, \varpi'_{{\mathrm{1}}}$ and $\ottnt{e_{{\mathrm{0}}}} \,  \mathrel{  \longrightarrow  ^*}  \, \ottnt{e''} \,  \,\&\,  \, \varpi''_{{\mathrm{1}}}$,
     we have $\ottnt{e} \,  \mathrel{  \longrightarrow  ^*}  \, \ottnt{e''} \,  \,\&\,  \, \varpi'_{{\mathrm{1}}} \,  \tceappendop  \, \varpi''_{{\mathrm{1}}}$, that is, $\ottnt{e} \,  \mathrel{  \longrightarrow  ^*}  \, \ottnt{e''} \,  \,\&\,  \, \varpi_{{\mathrm{1}}}$.
   \end{caseanalysis}
 \end{caseanalysis}
\end{proof}

\begin{lemmap}{Composition of Evaluation and Divergence}{lr-comp-eval-div}
 If $\ottnt{e} \,  \mathrel{  \longrightarrow  }^{ \ottmv{n} }  \, \ottnt{e'} \,  \,\&\,  \, \varpi$ and $\ottnt{e'} \,  \Uparrow  \, \pi$,
 then $\ottnt{e} \,  \Uparrow  \, \varpi \,  \tceappendop  \, \pi$.
\end{lemmap}
\begin{proof}
 By induction on $\ottmv{n}$.
 \begin{caseanalysis}
  \case $\ottmv{n} \,  =  \, \ottsym{0}$: Obvious because $\ottnt{e} \,  =  \, \ottnt{e'}$ and $ \varpi    =     \epsilon  $.
  \case $\ottmv{n} \,  >  \, \ottsym{0}$:
   By case analysis on the definition of $\ottnt{e'} \,  \Uparrow  \, \pi$.
   \begin{caseanalysis}
    \case $   \forall  \, { \varpi'  \ottsym{,}  \pi' } . \  \pi \,  =  \, \varpi' \,  \tceappendop  \, \pi'   \mathrel{\Longrightarrow}    \exists  \, { \ottnt{e''} } . \  \ottnt{e'} \,  \mathrel{  \longrightarrow  ^*}  \, \ottnt{e''} \,  \,\&\,  \, \varpi'  $:
     Let $\varpi'$ be any finite trace and $\pi'$ be any infinite trace
     such that $ \varpi \,  \tceappendop  \, \pi    =    \varpi' \,  \tceappendop  \, \pi' $.
     Then, it suffices to show that
     \[
        \exists  \, { \ottnt{e''} } . \  \ottnt{e} \,  \mathrel{  \longrightarrow  ^*}  \, \ottnt{e''} \,  \,\&\,  \, \varpi'  ~.
     \]
     Because $ \varpi \,  \tceappendop  \, \pi    =    \varpi' \,  \tceappendop  \, \pi' $,
     either of the following holds:
     $\varpi$ is a prefix of $\varpi'$; or
     $\varpi'$ is a prefix of $\varpi$.
     \begin{caseanalysis}
      \case $  \exists  \, { \varpi'' } . \  \varpi' \,  =  \, \varpi \,  \tceappendop  \, \varpi'' $:
       Because $ \varpi \,  \tceappendop  \, \pi    =    \varpi' \,  \tceappendop  \, \pi' $,
       we have $ \varpi \,  \tceappendop  \, \pi    =    \varpi \,  \tceappendop  \, \varpi'' \,  \tceappendop  \, \pi' $.
       Therefore, $ \pi    =    \varpi'' \,  \tceappendop  \, \pi' $.
       Then, $\ottnt{e'} \,  \Uparrow  \, \pi$ implies that there exists some $\ottnt{e''}$ such that
       $\ottnt{e'} \,  \mathrel{  \longrightarrow  ^*}  \, \ottnt{e''} \,  \,\&\,  \, \varpi''$.
       Because $\ottnt{e} \,  \mathrel{  \longrightarrow  }^{ \ottmv{n} }  \, \ottnt{e'} \,  \,\&\,  \, \varpi$ and $\ottnt{e'} \,  \mathrel{  \longrightarrow  ^*}  \, \ottnt{e''} \,  \,\&\,  \, \varpi''$,
       we have $\ottnt{e} \,  \mathrel{  \longrightarrow  ^*}  \, \ottnt{e''} \,  \,\&\,  \, \varpi \,  \tceappendop  \, \varpi''$, that is,
       $\ottnt{e} \,  \mathrel{  \longrightarrow  ^*}  \, \ottnt{e''} \,  \,\&\,  \, \varpi'$.

      \case $  \exists  \, { \varpi'' } . \  \varpi \,  =  \, \varpi' \,  \tceappendop  \, \varpi'' $:
       Because $\ottnt{e} \,  \mathrel{  \longrightarrow  }^{ \ottmv{n} }  \, \ottnt{e'} \,  \,\&\,  \, \varpi$ implies $\ottnt{e} \,  \mathrel{  \longrightarrow  }^{ \ottmv{n} }  \, \ottnt{e'} \,  \,\&\,  \, \varpi' \,  \tceappendop  \, \varpi''$,
       \reflem{lr-eval-decomp} implies that
       there exists some $\ottnt{e''}$ such that
       $\ottnt{e} \,  \mathrel{  \longrightarrow  ^*}  \, \ottnt{e''} \,  \,\&\,  \, \varpi'$ and $\ottnt{e''} \,  \mathrel{  \longrightarrow  ^*}  \, \ottnt{e'} \,  \,\&\,  \, \varpi''$.
       Therefore, we have the conclusion.
     \end{caseanalysis}

    \case $ {   \exists  \, { \varpi'  \ottsym{,}  \pi'  \ottsym{,}  \ottnt{E} } . \  \pi \,  =  \, \varpi' \,  \tceappendop  \, \pi'  } \,\mathrel{\wedge}\, { \ottnt{e'} \,  \mathrel{  \longrightarrow  ^*}  \,  \ottnt{E}  [   \bullet ^{ \pi' }   ]  \,  \,\&\,  \, \varpi' } $:
     Because $\ottnt{e} \,  \mathrel{  \longrightarrow  }^{ \ottmv{n} }  \, \ottnt{e'} \,  \,\&\,  \, \varpi$ and $\ottnt{e'} \,  \mathrel{  \longrightarrow  ^*}  \,  \ottnt{E}  [   \bullet ^{ \pi' }   ]  \,  \,\&\,  \, \varpi'$,
     we have $\ottnt{e} \,  \mathrel{  \longrightarrow  ^*}  \,  \ottnt{E}  [   \bullet ^{ \pi' }   ]  \,  \,\&\,  \, \varpi \,  \tceappendop  \, \varpi'$.
     Therefore, $\ottnt{e} \,  \Uparrow  \, \varpi \,  \tceappendop  \, \varpi' \,  \tceappendop  \, \pi'$, that is, $\ottnt{e} \,  \Uparrow  \, \varpi \,  \tceappendop  \, \pi$.
  \end{caseanalysis}
 \end{caseanalysis}
\end{proof}

\begin{lemmap}{Diverging Evaluation of Reset}{lr-red-reset-div}
 If $ \resetcnstr{ \ottnt{e} }  \,  \Uparrow  \, \pi$,
 then either of the following holds.
 \begin{itemize}
  \item $\ottnt{e} \,  \Uparrow  \, \pi$; or
  \item $ {  {   \exists  \, { \ottnt{K_{{\mathrm{0}}}}  \ottsym{,}  \mathit{x}  \ottsym{,}  \ottnt{e_{{\mathrm{0}}}}  \ottsym{,}  \varpi_{{\mathrm{1}}}  \ottsym{,}  \pi_{{\mathrm{2}}} } . \  \ottnt{e} \,  \Downarrow  \,  \ottnt{K_{{\mathrm{0}}}}  [  \mathcal{S}_0 \, \mathit{x}  \ottsym{.}  \ottnt{e_{{\mathrm{0}}}}  ]  \,  \,\&\,  \, \varpi_{{\mathrm{1}}}  } \,\mathrel{\wedge}\, {  \ottnt{e_{{\mathrm{0}}}}    [   \lambda\!  \, \mathit{x}  \ottsym{.}   \resetcnstr{  \ottnt{K_{{\mathrm{0}}}}  [  \mathit{x}  ]  }   /  \mathit{x}  ]   \,  \Uparrow  \, \pi_{{\mathrm{2}}} }  } \,\mathrel{\wedge}\, { \pi \,  =  \, \varpi_{{\mathrm{1}}} \,  \tceappendop  \, \pi_{{\mathrm{2}}} } $.
 \end{itemize}
\end{lemmap}
\begin{proof}
 By case analysis on the definition of $ \resetcnstr{ \ottnt{e} }  \,  \Uparrow  \, \pi$.
 \begin{caseanalysis}
  \case $   \forall  \, { \varpi'  \ottsym{,}  \pi' } . \  \pi \,  =  \, \varpi' \,  \tceappendop  \, \pi'   \mathrel{\Longrightarrow}    \exists  \, { \ottnt{e'} } . \   \resetcnstr{ \ottnt{e} }  \,  \mathrel{  \longrightarrow  ^*}  \, \ottnt{e'} \,  \,\&\,  \, \varpi'  $:
   By case analysis on whether there exist some $\ottnt{K_{{\mathrm{0}}}}$, $\mathit{x}$, $\ottnt{e_{{\mathrm{0}}}}$, $\varpi_{{\mathrm{1}}}$, and $\pi_{{\mathrm{2}}}$ such that
  $ { \ottnt{e} \,  \Downarrow  \,  \ottnt{K_{{\mathrm{0}}}}  [  \mathcal{S}_0 \, \mathit{x}  \ottsym{.}  \ottnt{e_{{\mathrm{0}}}}  ]  \,  \,\&\,  \, \varpi_{{\mathrm{1}}} } \,\mathrel{\wedge}\, { \pi \,  =  \, \varpi_{{\mathrm{1}}} \,  \tceappendop  \, \pi_{{\mathrm{2}}} } $.
   \begin{caseanalysis}
    \case $ {   \exists  \, { \ottnt{K_{{\mathrm{0}}}}  \ottsym{,}  \mathit{x}  \ottsym{,}  \ottnt{e_{{\mathrm{0}}}}  \ottsym{,}  \varpi_{{\mathrm{1}}}  \ottsym{,}  \pi_{{\mathrm{2}}} } . \  \ottnt{e} \,  \Downarrow  \,  \ottnt{K_{{\mathrm{0}}}}  [  \mathcal{S}_0 \, \mathit{x}  \ottsym{.}  \ottnt{e_{{\mathrm{0}}}}  ]  \,  \,\&\,  \, \varpi_{{\mathrm{1}}}  } \,\mathrel{\wedge}\, { \pi \,  =  \, \varpi_{{\mathrm{1}}} \,  \tceappendop  \, \pi_{{\mathrm{2}}} } $:
     It suffices to show that
     \[
       \ottnt{e_{{\mathrm{0}}}}    [   \lambda\!  \, \mathit{x}  \ottsym{.}   \resetcnstr{  \ottnt{K_{{\mathrm{0}}}}  [  \mathit{x}  ]  }   /  \mathit{x}  ]   \,  \Uparrow  \, \pi_{{\mathrm{2}}} ~.
     \]
     Let $\varpi'_{{\mathrm{2}}}$ be any finite trace and $\pi'_{{\mathrm{2}}}$ be any infinite trace
     such that $ \pi_{{\mathrm{2}}}    =    \varpi'_{{\mathrm{2}}} \,  \tceappendop  \, \pi'_{{\mathrm{2}}} $.
     Then, it suffices to show that
     \[
        \exists  \, { \ottnt{e'} } . \   \ottnt{e_{{\mathrm{0}}}}    [   \lambda\!  \, \mathit{x}  \ottsym{.}   \resetcnstr{  \ottnt{K_{{\mathrm{0}}}}  [  \mathit{x}  ]  }   /  \mathit{x}  ]   \,  \mathrel{  \longrightarrow  ^*}  \, \ottnt{e'} \,  \,\&\,  \, \varpi'_{{\mathrm{2}}}  ~.
     \]

     If $ \varpi'_{{\mathrm{2}}}    =     \epsilon  $, then we have the conclusion
     by letting $\ottnt{e'} \,  =  \,  \ottnt{e_{{\mathrm{0}}}}    [   \lambda\!  \, \mathit{x}  \ottsym{.}   \resetcnstr{  \ottnt{K_{{\mathrm{0}}}}  [  \mathit{x}  ]  }   /  \mathit{x}  ]  $.

     Otherwise, suppose that $ \varpi'_{{\mathrm{2}}}    \not=     \epsilon  $.
     Because $\pi = \varpi_{{\mathrm{1}}} \,  \tceappendop  \, \pi_{{\mathrm{2}}} = \varpi_{{\mathrm{1}}} \,  \tceappendop  \, \varpi'_{{\mathrm{2}}} \,  \tceappendop  \, \pi'_{{\mathrm{2}}}$,
     $ \resetcnstr{ \ottnt{e} }  \,  \Uparrow  \, \pi$ implies that there exists some $\ottnt{e'}$ such that
     $ \resetcnstr{ \ottnt{e} }  \,  \mathrel{  \longrightarrow  ^*}  \, \ottnt{e'} \,  \,\&\,  \, \varpi_{{\mathrm{1}}} \,  \tceappendop  \, \varpi'_{{\mathrm{2}}}$.
     Because $\ottnt{e} \,  \Downarrow  \,  \ottnt{K_{{\mathrm{0}}}}  [  \mathcal{S}_0 \, \mathit{x}  \ottsym{.}  \ottnt{e_{{\mathrm{0}}}}  ]  \,  \,\&\,  \, \varpi_{{\mathrm{1}}}$,
     \reflem{lr-red-subterm} and \E{Red}/\R{Shift} imply
     $ \resetcnstr{ \ottnt{e} }  \,  \mathrel{  \longrightarrow  ^*}  \,  \ottnt{e_{{\mathrm{0}}}}    [   \lambda\!  \, \mathit{x}  \ottsym{.}   \resetcnstr{  \ottnt{K_{{\mathrm{0}}}}  [  \mathit{x}  ]  }   /  \mathit{x}  ]   \,  \,\&\,  \, \varpi_{{\mathrm{1}}}$.
     Then, \reflem{lr-eval-det} implies that
     $ \ottnt{e_{{\mathrm{0}}}}    [   \lambda\!  \, \mathit{x}  \ottsym{.}   \resetcnstr{  \ottnt{K_{{\mathrm{0}}}}  [  \mathit{x}  ]  }   /  \mathit{x}  ]   \,  \mathrel{  \longrightarrow  ^*}  \, \ottnt{e'} \,  \,\&\,  \, \varpi'_{{\mathrm{2}}}$.
     Therefore, we have the conclusion.

    \case $ \neg  \ottsym{(}   {   \exists  \, { \ottnt{K_{{\mathrm{0}}}}  \ottsym{,}  \mathit{x}  \ottsym{,}  \ottnt{e_{{\mathrm{0}}}}  \ottsym{,}  \varpi_{{\mathrm{1}}}  \ottsym{,}  \pi_{{\mathrm{2}}} } . \  \ottnt{e} \,  \Downarrow  \,  \ottnt{K_{{\mathrm{0}}}}  [  \mathcal{S}_0 \, \mathit{x}  \ottsym{.}  \ottnt{e_{{\mathrm{0}}}}  ]  \,  \,\&\,  \, \varpi_{{\mathrm{1}}}  } \,\mathrel{\wedge}\, { \pi \,  =  \, \varpi_{{\mathrm{1}}} \,  \tceappendop  \, \pi_{{\mathrm{2}}} }   \ottsym{)} $:
     We show that $\ottnt{e} \,  \Uparrow  \, \pi$.
     Let $\varpi'$ be any finite trace and $\pi'$ be any infinite trace
     such that $ \pi    =    \varpi' \,  \tceappendop  \, \pi' $.
     Then, it suffices to show that
     \[
        \exists  \, { \ottnt{e''} } . \  \ottnt{e} \,  \mathrel{  \longrightarrow  ^*}  \, \ottnt{e''} \,  \,\&\,  \, \varpi'  ~.
     \]
     $ \resetcnstr{ \ottnt{e} }  \,  \Uparrow  \, \pi$ implies $ \resetcnstr{ \ottnt{e} }  \,  \mathrel{  \longrightarrow  ^*}  \, \ottnt{e'} \,  \,\&\,  \, \varpi'$ for some $\ottnt{e'}$.
     By case analysis on the result of \reflem{lr-red-reset-term} with
     $ \resetcnstr{ \ottnt{e} }  \,  \mathrel{  \longrightarrow  ^*}  \, \ottnt{e'} \,  \,\&\,  \, \varpi'$.
     \begin{caseanalysis}
      \case $ {   \exists  \, { \ottnt{v} } . \  \ottnt{e} \,  \mathrel{  \longrightarrow  ^*}  \, \ottnt{v} \,  \,\&\,  \, \varpi'  } \,\mathrel{\wedge}\, { \ottnt{e'} \,  =  \, \ottnt{v} } $:
       By \reflem{lr-red-subterm} and \E{Red}/\R{Reset},
       $ \resetcnstr{ \ottnt{e} }  \,  \mathrel{  \longrightarrow  ^*}  \, \ottnt{v} \,  \,\&\,  \, \varpi'$.
       However, it is contradictory with $ \resetcnstr{ \ottnt{e} }  \,  \Uparrow  \, \pi$:
       $ \resetcnstr{ \ottnt{e} }  \,  \Uparrow  \, \pi$ ensures that, for any non-empty prefix $\varpi'_{{\mathrm{0}}}$ of $\pi'$,
       there exists some $\ottnt{e'_{{\mathrm{0}}}}$ such that $ \resetcnstr{ \ottnt{e} }  \,  \mathrel{  \longrightarrow  ^*}  \, \ottnt{e'_{{\mathrm{0}}}} \,  \,\&\,  \, \varpi' \,  \tceappendop  \, \varpi'_{{\mathrm{0}}}$;
       then \reflem{lr-eval-det} implies that $\ottnt{v} \,  \mathrel{  \longrightarrow  ^*}  \, \ottnt{e'_{{\mathrm{0}}}} \,  \,\&\,  \, \varpi'_{{\mathrm{0}}}$;
       however, it cannot be derived by \reflem{lr-red-val}.

      \case $ {  {   \exists  \, { \ottnt{K_{{\mathrm{0}}}}  \ottsym{,}  \mathit{x}  \ottsym{,}  \ottnt{e_{{\mathrm{0}}}}  \ottsym{,}  \varpi_{{\mathrm{1}}}  \ottsym{,}  \varpi_{{\mathrm{2}}} } . \  \ottnt{e} \,  \mathrel{  \longrightarrow  ^*}  \,  \ottnt{K_{{\mathrm{0}}}}  [  \mathcal{S}_0 \, \mathit{x}  \ottsym{.}  \ottnt{e_{{\mathrm{0}}}}  ]  \,  \,\&\,  \, \varpi_{{\mathrm{1}}}  } \,\mathrel{\wedge}\, {  \ottnt{e_{{\mathrm{0}}}}    [   \lambda\!  \, \mathit{x}  \ottsym{.}   \resetcnstr{  \ottnt{K_{{\mathrm{0}}}}  [  \mathit{x}  ]  }   /  \mathit{x}  ]   \,  \mathrel{  \longrightarrow  ^*}  \, \ottnt{e'} \,  \,\&\,  \, \varpi_{{\mathrm{2}}} }  } \,\mathrel{\wedge}\, { \varpi' \,  =  \, \varpi_{{\mathrm{1}}} \,  \tceappendop  \, \varpi_{{\mathrm{2}}} } $: \mbox{}
      Contradictory because $\pi = \varpi' \,  \tceappendop  \, \pi' = \varpi_{{\mathrm{1}}} \,  \tceappendop  \, \varpi_{{\mathrm{2}}} \,  \tceappendop  \, \pi'$.

      \case $ {   \exists  \, { \ottnt{e''} } . \  \ottnt{e} \,  \mathrel{  \longrightarrow  ^*}  \, \ottnt{e''} \,  \,\&\,  \, \varpi'  } \,\mathrel{\wedge}\, { \ottnt{e'} \,  =  \,  \resetcnstr{ \ottnt{e''} }  } $:
       We have the conclusion.
     \end{caseanalysis}
   \end{caseanalysis}

  \case $ {   \exists  \, { \varpi'  \ottsym{,}  \pi'  \ottsym{,}  \ottnt{E} } . \  \pi \,  =  \, \varpi' \,  \tceappendop  \, \pi'  } \,\mathrel{\wedge}\, {  \resetcnstr{ \ottnt{e} }  \,  \mathrel{  \longrightarrow  ^*}  \,  \ottnt{E}  [   \bullet ^{ \pi' }   ]  \,  \,\&\,  \, \varpi' } $:
   By case analysis on the result of \reflem{lr-red-reset-term}
   with $ \resetcnstr{ \ottnt{e} }  \,  \mathrel{  \longrightarrow  ^*}  \,  \ottnt{E}  [   \bullet ^{ \pi' }   ]  \,  \,\&\,  \, \varpi'$.
   \begin{caseanalysis}
    \case $ {   \exists  \, { \ottnt{v} } . \  \ottnt{e} \,  \mathrel{  \longrightarrow  ^*}  \, \ottnt{v} \,  \,\&\,  \, \varpi'  } \,\mathrel{\wedge}\, {  \ottnt{E}  [   \bullet ^{ \pi' }   ]  \,  =  \, \ottnt{v} } $:
     Contradictory.

    \case $ {  {   \exists  \, { \ottnt{K_{{\mathrm{0}}}}  \ottsym{,}  \mathit{x}  \ottsym{,}  \ottnt{e_{{\mathrm{0}}}}  \ottsym{,}  \varpi_{{\mathrm{1}}}  \ottsym{,}  \varpi_{{\mathrm{2}}} } . \  \ottnt{e} \,  \mathrel{  \longrightarrow  ^*}  \,  \ottnt{K_{{\mathrm{0}}}}  [  \mathcal{S}_0 \, \mathit{x}  \ottsym{.}  \ottnt{e_{{\mathrm{0}}}}  ]  \,  \,\&\,  \, \varpi_{{\mathrm{1}}}  } \,\mathrel{\wedge}\, {  \ottnt{e_{{\mathrm{0}}}}    [   \lambda\!  \, \mathit{x}  \ottsym{.}   \resetcnstr{  \ottnt{K_{{\mathrm{0}}}}  [  \mathit{x}  ]  }   /  \mathit{x}  ]   \,  \mathrel{  \longrightarrow  ^*}  \,  \ottnt{E}  [   \bullet ^{ \pi' }   ]  \,  \,\&\,  \, \varpi_{{\mathrm{2}}} }  } \,\mathrel{\wedge}\, { \varpi' \,  =  \, \varpi_{{\mathrm{1}}} \,  \tceappendop  \, \varpi_{{\mathrm{2}}} } $: \mbox{}
     Because $\pi = \varpi' \,  \tceappendop  \, \pi' = \varpi_{{\mathrm{1}}} \,  \tceappendop  \, \varpi_{{\mathrm{2}}} \,  \tceappendop  \, \pi'$,
     it suffices to show that
     \[
       \ottnt{e_{{\mathrm{0}}}}    [   \lambda\!  \, \mathit{x}  \ottsym{.}   \resetcnstr{  \ottnt{K_{{\mathrm{0}}}}  [  \mathit{x}  ]  }   /  \mathit{x}  ]   \,  \Uparrow  \, \varpi_{{\mathrm{2}}} \,  \tceappendop  \, \pi' ~,
     \]
     which is implied by $ \ottnt{e_{{\mathrm{0}}}}    [   \lambda\!  \, \mathit{x}  \ottsym{.}   \resetcnstr{  \ottnt{K_{{\mathrm{0}}}}  [  \mathit{x}  ]  }   /  \mathit{x}  ]   \,  \mathrel{  \longrightarrow  ^*}  \,  \ottnt{E}  [   \bullet ^{ \pi' }   ]  \,  \,\&\,  \, \varpi_{{\mathrm{2}}}$.

    \case $ {   \exists  \, { \ottnt{e'} } . \  \ottnt{e} \,  \mathrel{  \longrightarrow  ^*}  \, \ottnt{e'} \,  \,\&\,  \, \varpi'  } \,\mathrel{\wedge}\, {  \ottnt{E}  [   \bullet ^{ \pi' }   ]  \,  =  \,  \resetcnstr{ \ottnt{e'} }  } $:
     Because $ \ottnt{E}  [   \bullet ^{ \pi' }   ]  \,  =  \,  \resetcnstr{ \ottnt{e'} } $,
     there exists some $\ottnt{E'}$ such that
     $\ottnt{E} \,  =  \,  \resetcnstr{ \ottnt{E'} } $ and $\ottnt{e'} \,  =  \,  \ottnt{E'}  [   \bullet ^{ \pi' }   ] $.
     Because $\ottnt{e} \,  \mathrel{  \longrightarrow  ^*}  \,  \ottnt{E'}  [   \bullet ^{ \pi' }   ]  \,  \,\&\,  \, \varpi'$ and $ \pi    =    \varpi' \,  \tceappendop  \, \pi' $,
     we have the conclusion $\ottnt{e} \,  \Uparrow  \, \pi$.
   \end{caseanalysis}
 \end{caseanalysis}
\end{proof}

\begin{lemmap}{Compatibility: Reset}{lr-compat-reset}
 If $\ottnt{e} \,  \in  \,  \LRRel{ \mathcal{O} }{ \Gamma  \, \vdash \,   \ottnt{T}  \,  \,\&\,  \,  \Phi  \, / \,   ( \forall  \mathit{x}  .   \ottnt{T}  \,  \,\&\,  \,   \Phi_\mathsf{val}   \, / \,   \square    )  \Rightarrow   \ottnt{C}   } $ and $\mathit{x} \,  \not\in  \,  \mathit{fv}  (  \ottnt{T}  ) $,
 then $ \resetcnstr{ \ottnt{e} }  \,  \in  \,  \LRRel{ \mathcal{O} }{ \Gamma  \, \vdash \,  \ottnt{C} } $.
\end{lemmap}
\begin{proof}
 Let $\ottnt{C'} \,  =  \,  \ottnt{T}  \,  \,\&\,  \,  \Phi  \, / \,   ( \forall  \mathit{x}  .   \ottnt{T}  \,  \,\&\,  \,   \Phi_\mathsf{val}   \, / \,   \square    )  \Rightarrow   \ottnt{C}  $.
 Because $\ottnt{e} \,  \in  \,  \LRRel{ \mathcal{O} }{ \Gamma  \, \vdash \,  \ottnt{C'} } $ implies $\Gamma  \vdash  \ottnt{e}  \ottsym{:}  \ottnt{C'}$,
 we have $\Gamma  \vdash   \resetcnstr{ \ottnt{e} }   \ottsym{:}  \ottnt{C}$ by \T{Reset} with $\mathit{x} \,  \not\in  \,  \mathit{fv}  (  \ottnt{T}  ) $.
 Let $\sigma \,  \in  \,  \LRRel{ \mathcal{G} }{ \Gamma } $.
 Without loss of generality, we can suppose that $\mathit{x}$ does not occur in $\sigma$.
 We must show that
 \[
   \resetcnstr{ \sigma  \ottsym{(}  \ottnt{e}  \ottsym{)} }  \,  \in  \,  \LRRel{ \mathcal{E} }{ \sigma  \ottsym{(}  \ottnt{C}  \ottsym{)} }  ~.
 \]
 Note that $ \resetcnstr{ \sigma  \ottsym{(}  \ottnt{e}  \ottsym{)} }  \,  \in  \,  \LRRel{ \mathrm{Atom} }{ \sigma  \ottsym{(}  \ottnt{C}  \ottsym{)} } $
 by \reflem{lr-typeability} with $\Gamma  \vdash   \resetcnstr{ \ottnt{e} }   \ottsym{:}  \ottnt{C}$ and $\sigma \,  \in  \,  \LRRel{ \mathcal{G} }{ \Gamma } $.

 We first show that
 \[
     \forall  \, { \ottnt{K_{{\mathrm{0}}}}  \ottsym{,}  \mathit{x_{{\mathrm{0}}}}  \ottsym{,}  \ottnt{e_{{\mathrm{0}}}}  \ottsym{,}  \varpi } . \  \sigma  \ottsym{(}  \ottnt{e}  \ottsym{)} \,  \Downarrow  \,  \ottnt{K_{{\mathrm{0}}}}  [  \mathcal{S}_0 \, \mathit{x_{{\mathrm{0}}}}  \ottsym{.}  \ottnt{e_{{\mathrm{0}}}}  ]  \,  \,\&\,  \, \varpi   \mathrel{\Longrightarrow}   \resetcnstr{ \sigma  \ottsym{(}  \ottnt{e}  \ottsym{)} }  \,  \in  \,  \LRRel{ \mathcal{E} }{ \sigma  \ottsym{(}  \ottnt{C}  \ottsym{)} }   ~.
 \]
 Suppose that $\sigma  \ottsym{(}  \ottnt{e}  \ottsym{)} \,  \Downarrow  \,  \ottnt{K_{{\mathrm{0}}}}  [  \mathcal{S}_0 \, \mathit{x_{{\mathrm{0}}}}  \ottsym{.}  \ottnt{e_{{\mathrm{0}}}}  ]  \,  \,\&\,  \, \varpi$ for some
 $\ottnt{K_{{\mathrm{0}}}}$, $\mathit{x_{{\mathrm{0}}}}$, $\ottnt{e_{{\mathrm{0}}}}$, and $\varpi$.
 We have $\sigma  \ottsym{(}  \ottnt{e}  \ottsym{)} \,  \in  \,  \LRRel{ \mathcal{E} }{ \sigma  \ottsym{(}  \ottnt{C'}  \ottsym{)} } $ because $\ottnt{e} \,  \in  \,  \LRRel{ \mathcal{O} }{ \Gamma  \, \vdash \,  \ottnt{C'} } $.
 Because $ \emptyset   \vdash  \sigma  \ottsym{(}  \ottnt{e}  \ottsym{)}  \ottsym{:}  \sigma  \ottsym{(}  \ottnt{C'}  \ottsym{)}$,
 \reflem{ts-wf-ty-tctx-typing} implies $ \emptyset   \vdash  \sigma  \ottsym{(}  \ottnt{T}  \ottsym{)}$.
 Then,
 $\,[\,]\, \,  \in  \,  \LRRel{ \mathcal{K} }{  \sigma  \ottsym{(}  \ottnt{T}  \ottsym{)}  \, / \, ( \forall  \mathit{x}  .   \sigma  \ottsym{(}  \ottnt{T}  \ottsym{)}  \,  \,\&\,  \,   \Phi_\mathsf{val}   \, / \,   \square    )  } $ by \reflem{lr-rel-emp-ectx}.
 Therefore, $\sigma  \ottsym{(}  \ottnt{e}  \ottsym{)} \,  \in  \,  \LRRel{ \mathcal{E} }{ \sigma  \ottsym{(}  \ottnt{C'}  \ottsym{)} }  =  \LRRel{ \mathcal{E} }{  \sigma  \ottsym{(}  \ottnt{T}  \ottsym{)}  \,  \,\&\,  \,  \sigma  \ottsym{(}  \Phi  \ottsym{)}  \, / \,   ( \forall  \mathit{x}  .   \sigma  \ottsym{(}  \ottnt{T}  \ottsym{)}  \,  \,\&\,  \,   \Phi_\mathsf{val}   \, / \,   \square    )  \Rightarrow   \sigma  \ottsym{(}  \ottnt{C}  \ottsym{)}   } $ and $\sigma  \ottsym{(}  \ottnt{e}  \ottsym{)} \,  \Downarrow  \,  \ottnt{K_{{\mathrm{0}}}}  [  \mathcal{S}_0 \, \mathit{x_{{\mathrm{0}}}}  \ottsym{.}  \ottnt{e_{{\mathrm{0}}}}  ]  \,  \,\&\,  \, \varpi$
 imply the conclusion $ \resetcnstr{ \sigma  \ottsym{(}  \ottnt{e}  \ottsym{)} }  \,  \in  \,  \LRRel{ \mathcal{E} }{ \sigma  \ottsym{(}  \ottnt{C}  \ottsym{)} } $.

 In what follows, we suppose that
 \begin{equation}
   \neg    \exists  \, { \ottnt{K_{{\mathrm{0}}}}  \ottsym{,}  \mathit{x_{{\mathrm{0}}}}  \ottsym{,}  \ottnt{e_{{\mathrm{0}}}}  \ottsym{,}  \varpi } . \  \sigma  \ottsym{(}  \ottnt{e}  \ottsym{)} \,  \Downarrow  \,  \ottnt{K_{{\mathrm{0}}}}  [  \mathcal{S}_0 \, \mathit{x_{{\mathrm{0}}}}  \ottsym{.}  \ottnt{e_{{\mathrm{0}}}}  ]  \,  \,\&\,  \, \varpi   ~.
   \label{lem:lr-compat-reset:shift}
 \end{equation}

 We consider each case on the result of of evaluating $ \resetcnstr{ \sigma  \ottsym{(}  \ottnt{e}  \ottsym{)} } $.
 \begin{caseanalysis}
  \case $  \exists  \, { \ottnt{v}  \ottsym{,}  \varpi } . \   \resetcnstr{ \sigma  \ottsym{(}  \ottnt{e}  \ottsym{)} }  \,  \Downarrow  \, \ottnt{v} \,  \,\&\,  \, \varpi $:
   We must show that $\ottnt{v} \,  \in  \,  \LRRel{ \mathcal{V} }{ \sigma  \ottsym{(}   \ottnt{C}  . \ottnt{T}   \ottsym{)} } $.
   By case analysis on the result of \reflem{lr-red-reset-term} with $ \resetcnstr{ \sigma  \ottsym{(}  \ottnt{e}  \ottsym{)} }  \,  \Downarrow  \, \ottnt{v} \,  \,\&\,  \, \varpi$.
   \begin{caseanalysis}
    \case $\sigma  \ottsym{(}  \ottnt{e}  \ottsym{)} \,  \mathrel{  \longrightarrow  ^*}  \, \ottnt{v} \,  \,\&\,  \, \varpi$:
     Because $\sigma  \ottsym{(}  \ottnt{e}  \ottsym{)} \,  \in  \,  \LRRel{ \mathcal{E} }{ \sigma  \ottsym{(}  \ottnt{C'}  \ottsym{)} } $, we have $\ottnt{v} \,  \in  \,  \LRRel{ \mathcal{V} }{ \sigma  \ottsym{(}  \ottnt{T}  \ottsym{)} } $.
     Because $ \emptyset   \vdash  \sigma  \ottsym{(}  \ottnt{e}  \ottsym{)}  \ottsym{:}  \sigma  \ottsym{(}  \ottnt{C'}  \ottsym{)}$,
     Lemmas~\ref{lem:ts-type-sound-terminating} and \ref{lem:lr-red-val} imply that
     there exists some $\ottnt{T''}$ such that
     $ \emptyset   \vdash  \ottnt{v}  \ottsym{:}  \ottnt{T''}$ and $ \emptyset   \vdash   \ottnt{T''}  \,  \,\&\,  \,   (  \varpi  , \bot )   \, / \,   \square    \ottsym{<:}  \sigma  \ottsym{(}  \ottnt{C'}  \ottsym{)}$.
     Because $ \emptyset   \vdash   \ottnt{T''}  \,  \,\&\,  \,   (  \varpi  , \bot )   \, / \,   \square    \ottsym{<:}  \sigma  \ottsym{(}  \ottnt{C'}  \ottsym{)}$ means
     \[
       \emptyset   \vdash   \ottnt{T''}  \,  \,\&\,  \,   (  \varpi  , \bot )   \, / \,   \square    \ottsym{<:}   \sigma  \ottsym{(}  \ottnt{T}  \ottsym{)}  \,  \,\&\,  \,  \sigma  \ottsym{(}  \Phi  \ottsym{)}  \, / \,   ( \forall  \mathit{x}  .   \sigma  \ottsym{(}  \ottnt{T}  \ottsym{)}  \,  \,\&\,  \,   \Phi_\mathsf{val}   \, / \,   \square    )  \Rightarrow   \sigma  \ottsym{(}  \ottnt{C}  \ottsym{)}   ~,
     \]
     its inversion implies
     \begin{itemize}
      \item $ \emptyset   \vdash  \ottnt{T''}  \ottsym{<:}  \sigma  \ottsym{(}  \ottnt{T}  \ottsym{)}$ and
      \item $\mathit{x} \,  {:}  \, \ottnt{T''}  \vdash   \sigma  \ottsym{(}  \ottnt{T}  \ottsym{)}  \,  \,\&\,  \,   (  \varpi  , \bot )  \,  \tceappendop  \,  \Phi_\mathsf{val}   \, / \,   \square    \ottsym{<:}  \sigma  \ottsym{(}  \ottnt{C}  \ottsym{)}$
            where $\mathit{x} \,  \not\in  \,  \mathit{fv}  (  \sigma  \ottsym{(}  \ottnt{C}  \ottsym{)}  ) $.
     \end{itemize}
     Further, the last subtyping derivation implies
     \[
      \mathit{x} \,  {:}  \, \ottnt{T''}  \vdash  \sigma  \ottsym{(}  \ottnt{T}  \ottsym{)}  \ottsym{<:}  \sigma  \ottsym{(}   \ottnt{C}  . \ottnt{T}   \ottsym{)} ~.
     \]
     Because $ \emptyset   \vdash  \ottnt{v}  \ottsym{:}  \ottnt{T''}$ and $\mathit{x} \,  \not\in  \,  \mathit{fv}  (  \sigma  \ottsym{(}  \ottnt{T}  \ottsym{)}  )  \,  \mathbin{\cup}  \,  \mathit{fv}  (  \sigma  \ottsym{(}  \ottnt{C}  \ottsym{)}  ) $,
     \reflem{ts-val-subst-subtyping} implies
     \[
       \emptyset   \vdash  \sigma  \ottsym{(}  \ottnt{T}  \ottsym{)}  \ottsym{<:}  \sigma  \ottsym{(}   \ottnt{C}  . \ottnt{T}   \ottsym{)} ~.
     \]
     Then, because
     \begin{itemize}
      \item $\ottnt{v} \,  \in  \,  \LRRel{ \mathcal{V} }{ \sigma  \ottsym{(}  \ottnt{T}  \ottsym{)} } $,
      \item $ \emptyset   \vdash  \sigma  \ottsym{(}  \ottnt{T}  \ottsym{)}$
            by Lemmas~\ref{lem:lr-typeability} and \ref{lem:ts-wf-ty-tctx-typing}
            with $\Gamma  \vdash  \ottnt{e}  \ottsym{:}  \ottnt{C'}$ and $\sigma \,  \in  \,  \LRRel{ \mathcal{G} }{ \Gamma } $, and
      \item $ \emptyset   \vdash  \sigma  \ottsym{(}   \ottnt{C}  . \ottnt{T}   \ottsym{)}$
            by Lemmas~\ref{lem:lr-typeability} and \ref{lem:ts-wf-ty-tctx-typing}
            with $\Gamma  \vdash   \resetcnstr{ \ottnt{e} }   \ottsym{:}  \ottnt{C}$ and $\sigma \,  \in  \,  \LRRel{ \mathcal{G} }{ \Gamma } $,
     \end{itemize}
     \reflem{lr-subtyping} implies the conclusion $\ottnt{v} \,  \in  \,  \LRRel{ \mathcal{V} }{ \sigma  \ottsym{(}   \ottnt{C}  . \ottnt{T}   \ottsym{)} } $.

    \case $ {  {   \exists  \, { \ottnt{K_{{\mathrm{0}}}}  \ottsym{,}  \mathit{x_{{\mathrm{0}}}}  \ottsym{,}  \ottnt{e_{{\mathrm{0}}}}  \ottsym{,}  \varpi_{{\mathrm{1}}} \,  \tceappendop  \, \varpi_{{\mathrm{2}}} } . \  \sigma  \ottsym{(}  \ottnt{e}  \ottsym{)} \,  \mathrel{  \longrightarrow  ^*}  \,  \ottnt{K_{{\mathrm{0}}}}  [  \mathcal{S}_0 \, \mathit{x_{{\mathrm{0}}}}  \ottsym{.}  \ottnt{e_{{\mathrm{0}}}}  ]  \,  \,\&\,  \, \varpi_{{\mathrm{1}}}  } \,\mathrel{\wedge}\, {  \ottnt{e_{{\mathrm{0}}}}    [   \lambda\!  \, \mathit{x_{{\mathrm{0}}}}  \ottsym{.}   \resetcnstr{  \ottnt{K_{{\mathrm{0}}}}  [  \mathit{x_{{\mathrm{0}}}}  ]  }   /  \mathit{x_{{\mathrm{0}}}}  ]   \,  \mathrel{  \longrightarrow  ^*}  \, \ottnt{v} \,  \,\&\,  \, \varpi_{{\mathrm{2}}} }  } \,\mathrel{\wedge}\, { \varpi \,  =  \, \varpi_{{\mathrm{1}}} \,  \tceappendop  \, \varpi_{{\mathrm{2}}} } $: \mbox{}
     Contradictory with the assumption (\ref{lem:lr-compat-reset:shift}).

    \case $ {   \exists  \, { \ottnt{e'} } . \  \sigma  \ottsym{(}  \ottnt{e}  \ottsym{)} \,  \mathrel{  \longrightarrow  ^*}  \, \ottnt{e'} \,  \,\&\,  \, \varpi  } \,\mathrel{\wedge}\, { \ottnt{v} \,  =  \,  \resetcnstr{ \ottnt{e'} }  } $:
     Contradictory.
   \end{caseanalysis}

  \case $  \exists  \, { \ottnt{K_{{\mathrm{0}}}}  \ottsym{,}  \mathit{x_{{\mathrm{0}}}}  \ottsym{,}  \ottnt{e_{{\mathrm{0}}}}  \ottsym{,}  \varpi } . \   \resetcnstr{ \sigma  \ottsym{(}  \ottnt{e}  \ottsym{)} }  \,  \Downarrow  \,  \ottnt{K_{{\mathrm{0}}}}  [  \mathcal{S}_0 \, \mathit{x_{{\mathrm{0}}}}  \ottsym{.}  \ottnt{e_{{\mathrm{0}}}}  ]  \,  \,\&\,  \, \varpi $:
   Contradictory with the assumption (\ref{lem:lr-compat-reset:shift})
   by \reflem{lr-red-reset-term}.

  \case $  \exists  \, { \pi } . \   \resetcnstr{ \sigma  \ottsym{(}  \ottnt{e}  \ottsym{)} }  \,  \Uparrow  \, \pi $:
   We must show that $ { \models  \,   { \sigma  \ottsym{(}  \ottnt{C}  \ottsym{)} }^\nu (  \pi  )  } $.
   By case analysis on the result of \reflem{lr-red-reset-div}
   with $ \resetcnstr{ \sigma  \ottsym{(}  \ottnt{e}  \ottsym{)} }  \,  \Uparrow  \, \pi$.
   \begin{caseanalysis}
    \case $\sigma  \ottsym{(}  \ottnt{e}  \ottsym{)} \,  \Uparrow  \, \pi$:
     Because $\sigma  \ottsym{(}  \ottnt{e}  \ottsym{)} \,  \in  \,  \LRRel{ \mathcal{E} }{ \sigma  \ottsym{(}  \ottnt{C'}  \ottsym{)} } $ and $\sigma  \ottsym{(}  \ottnt{e}  \ottsym{)} \,  \Uparrow  \, \pi$,
     we have $ { \models  \,   { \sigma  \ottsym{(}  \ottnt{C'}  \ottsym{)} }^\nu (  \pi  )  } $.
     Because $ { \sigma  \ottsym{(}  \ottnt{C'}  \ottsym{)} }^\nu (  \pi  )  \,  =  \,   { \sigma  \ottsym{(}  \Phi  \ottsym{)} }^\nu   \ottsym{(}  \pi  \ottsym{)}    \wedge     { \sigma  \ottsym{(}  \ottnt{C}  \ottsym{)} }^\nu (  \pi  )  $
     by definition,
     we have the conclusion $ { \models  \,   { \sigma  \ottsym{(}  \ottnt{C}  \ottsym{)} }^\nu (  \pi  )  } $
     by \reflem{ts-valid-elim-intersect}.

    \case $ {  {   \exists  \, { \ottnt{K_{{\mathrm{0}}}}  \ottsym{,}  \mathit{x_{{\mathrm{0}}}}  \ottsym{,}  \ottnt{e_{{\mathrm{0}}}}  \ottsym{,}  \varpi_{{\mathrm{0}}}  \ottsym{,}  \pi_{{\mathrm{0}}} } . \  \sigma  \ottsym{(}  \ottnt{e}  \ottsym{)} \,  \Downarrow  \,  \ottnt{K_{{\mathrm{0}}}}  [  \mathcal{S}_0 \, \mathit{x_{{\mathrm{0}}}}  \ottsym{.}  \ottnt{e_{{\mathrm{0}}}}  ]  \,  \,\&\,  \, \varpi_{{\mathrm{0}}}  } \,\mathrel{\wedge}\, {  \ottnt{e_{{\mathrm{0}}}}    [   \lambda\!  \, \mathit{x_{{\mathrm{0}}}}  \ottsym{.}   \resetcnstr{  \ottnt{K_{{\mathrm{0}}}}  [  \mathit{x_{{\mathrm{0}}}}  ]  }   /  \mathit{x_{{\mathrm{0}}}}  ]   \,  \Uparrow  \, \pi_{{\mathrm{0}}} }  } \,\mathrel{\wedge}\, { \pi \,  =  \, \varpi_{{\mathrm{0}}} \,  \tceappendop  \, \pi_{{\mathrm{0}}} } $:
     This case is contradictory with the assumption (\ref{lem:lr-compat-reset:shift}).

   \end{caseanalysis}
 \end{caseanalysis}
\end{proof}

\begin{lemmap}{Related Substitution in Type Composition}{lr-rel-subst-ty-compose}
 Let $\sigma \,  \in  \,  \LRRel{ \mathcal{G} }{ \Gamma } $.
 \begin{enumerate}
  \item If $\mathit{x}$ does not occur in $\sigma$,
        then $\sigma  \ottsym{(}  \ottnt{S_{{\mathrm{1}}}}  \gg\!\!=  \ottsym{(}   \lambda\!  \, \mathit{x}  \ottsym{.}  \ottnt{S_{{\mathrm{2}}}}  \ottsym{)}  \ottsym{)} \,  =  \, \sigma  \ottsym{(}  \ottnt{S_{{\mathrm{1}}}}  \ottsym{)}  \gg\!\!=  \ottsym{(}   \lambda\!  \, \mathit{x}  \ottsym{.}  \sigma  \ottsym{(}  \ottnt{S_{{\mathrm{2}}}}  \ottsym{)}  \ottsym{)}$.
  \item $\sigma  \ottsym{(}  \Phi_{{\mathrm{1}}} \,  \tceappendop  \, \Phi_{{\mathrm{2}}}  \ottsym{)} \,  =  \, \sigma  \ottsym{(}  \Phi_{{\mathrm{1}}}  \ottsym{)} \,  \tceappendop  \, \sigma  \ottsym{(}  \Phi_{{\mathrm{2}}}  \ottsym{)}$ for any $\Phi_{{\mathrm{1}}}$ and $\Phi_{{\mathrm{2}}}$.
 \end{enumerate}
\end{lemmap}
\begin{proof}
 By Lemmas~\ref{lem:ts-val-subst-ty-compose}, \ref{lem:ts-pred-subst-ty-compose}, and \ref{lem:ts-term-subst-ty-compose}.
\end{proof}

\begin{lemmap}{Continuation Composition with Let}{lr-cont-compose-let}
 Suppose that $\mathit{x_{{\mathrm{1}}}} \,  {:}  \, \ottnt{T_{{\mathrm{1}}}}  \vdash  \ottnt{e_{{\mathrm{2}}}}  \ottsym{:}   \ottnt{T_{{\mathrm{2}}}}  \,  \,\&\,  \,  \Phi_{{\mathrm{2}}}  \, / \,   ( \forall  \mathit{y}  .  \ottnt{C_{{\mathrm{1}}}}  )  \Rightarrow   \ottnt{C_{{\mathrm{2}}}}  $ and
 $\mathit{x_{{\mathrm{1}}}} \,  \not\in  \,  \mathit{fv}  (  \ottnt{T_{{\mathrm{2}}}}  )  \,  \mathbin{\cup}  \,  \mathit{fv}  (  \Phi_{{\mathrm{2}}}  )  \,  \mathbin{\cup}  \, \ottsym{(}   \mathit{fv}  (  \ottnt{C_{{\mathrm{1}}}}  )  \,  \mathbin{\backslash}  \,  \{  \mathit{y}  \}   \ottsym{)}$.
 \begin{enumerate}
  \item \label{lem:lr-cont-compse-let:cont}
        If $\ottnt{K}  \ottsym{:}    \ottnt{T_{{\mathrm{2}}}}  \, / \, ( \forall  \mathit{y}  .  \ottnt{C_{{\mathrm{1}}}}  )   \mathrel{ \Rrightarrow }  \ottnt{T_{{\mathrm{0}}}}  \,  \,\&\,  \,  \Phi_{{\mathrm{0}}}  \, / \, ( \forall  \mathit{z}  .  \ottnt{C_{{\mathrm{0}}}}  ) $,
        then $ \ottnt{K}  [  \mathsf{let} \, \mathit{x_{{\mathrm{1}}}}  \ottsym{=}  \,[\,]\, \,  \mathsf{in}  \, \ottnt{e_{{\mathrm{2}}}}  ]   \ottsym{:}    \ottnt{T_{{\mathrm{1}}}}  \, / \, ( \forall  \mathit{x_{{\mathrm{1}}}}  .  \ottnt{C_{{\mathrm{2}}}}  )   \mathrel{ \Rrightarrow }  \ottnt{T_{{\mathrm{0}}}}  \,  \,\&\,  \,  \Phi  \, / \, ( \forall  \mathit{z}  .  \ottnt{C_{{\mathrm{0}}}}  ) $
        for some $\Phi$.
  \item $\ottnt{K}  \ottsym{:}   \ottnt{T_{{\mathrm{2}}}}  \, / \, ( \forall  \mathit{y}  .  \ottnt{C_{{\mathrm{1}}}}  ) $,
        then $ \ottnt{K}  [  \mathsf{let} \, \mathit{x_{{\mathrm{1}}}}  \ottsym{=}  \,[\,]\, \,  \mathsf{in}  \, \ottnt{e_{{\mathrm{2}}}}  ]   \ottsym{:}   \ottnt{T_{{\mathrm{1}}}}  \, / \, ( \forall  \mathit{x_{{\mathrm{1}}}}  .  \ottnt{C_{{\mathrm{2}}}}  ) $.
 \end{enumerate}
\end{lemmap}
\begin{proof}
 \noindent
 \begin{enumerate}
  \item By induction on the typing derivation of
        $\ottnt{K}  \ottsym{:}    \ottnt{T_{{\mathrm{2}}}}  \, / \, ( \forall  \mathit{y}  .  \ottnt{C_{{\mathrm{1}}}}  )   \mathrel{ \Rrightarrow }  \ottnt{T_{{\mathrm{0}}}}  \,  \,\&\,  \,  \Phi_{{\mathrm{0}}}  \, / \, ( \forall  \mathit{z}  .  \ottnt{C_{{\mathrm{0}}}}  ) $.
        \begin{caseanalysis}
         \case \TK{Empty}:
          We are given
          \begin{itemize}
           \item $\ottnt{K} \,  =  \, \,[\,]\,$,
           \item $\ottnt{T_{{\mathrm{2}}}} \,  =  \, \ottnt{T_{{\mathrm{0}}}}$,
           \item $\Phi_{{\mathrm{0}}} \,  =  \,  \Phi_\mathsf{val} $,
           \item $ \mathit{y}    =    \mathit{z} $, and
           \item $\ottnt{C_{{\mathrm{1}}}} \,  =  \, \ottnt{C_{{\mathrm{0}}}}$.
          \end{itemize}
          It suffices to show that
          \[
           \mathsf{let} \, \mathit{x_{{\mathrm{1}}}}  \ottsym{=}  \,[\,]\, \,  \mathsf{in}  \, \ottnt{e_{{\mathrm{2}}}}  \ottsym{:}    \ottnt{T_{{\mathrm{1}}}}  \, / \, ( \forall  \mathit{x_{{\mathrm{1}}}}  .  \ottnt{C_{{\mathrm{2}}}}  )   \mathrel{ \Rrightarrow }  \ottnt{T_{{\mathrm{2}}}}  \,  \,\&\,  \,   \Phi_\mathsf{val}  \,  \tceappendop  \, \Phi_{{\mathrm{2}}}  \, / \, ( \forall  \mathit{y}  .  \ottnt{C_{{\mathrm{1}}}}  )  ~
          \]
          where we let $\Phi \,  =  \,  \Phi_\mathsf{val}  \,  \tceappendop  \, \Phi_{{\mathrm{2}}}$.
          It is derivable as follows.
          \begin{prooftree}
           \AxiomC{}
           \RightLabel{\TK{Empty}}
           \UnaryInfC{$\,[\,]\,  \ottsym{:}    \ottnt{T_{{\mathrm{1}}}}  \, / \, ( \forall  \mathit{x_{{\mathrm{1}}}}  .  \ottnt{C_{{\mathrm{2}}}}  )   \mathrel{ \Rrightarrow }  \ottnt{T_{{\mathrm{1}}}}  \,  \,\&\,  \,   \Phi_\mathsf{val}   \, / \, ( \forall  \mathit{x_{{\mathrm{1}}}}  .  \ottnt{C_{{\mathrm{2}}}}  ) $}
           \AxiomC{$\mathit{x_{{\mathrm{1}}}} \,  {:}  \, \ottnt{T_{{\mathrm{1}}}}  \vdash  \ottnt{e_{{\mathrm{2}}}}  \ottsym{:}   \ottnt{T_{{\mathrm{2}}}}  \,  \,\&\,  \,  \Phi_{{\mathrm{2}}}  \, / \,   ( \forall  \mathit{y}  .  \ottnt{C_{{\mathrm{1}}}}  )  \Rightarrow   \ottnt{C_{{\mathrm{2}}}}  $}
           \RightLabel{\TK{Let}}
           \BinaryInfC{$\mathsf{let} \, \mathit{x_{{\mathrm{1}}}}  \ottsym{=}  \,[\,]\, \,  \mathsf{in}  \, \ottnt{e_{{\mathrm{2}}}}  \ottsym{:}    \ottnt{T_{{\mathrm{1}}}}  \, / \, ( \forall  \mathit{x_{{\mathrm{1}}}}  .  \ottnt{C_{{\mathrm{2}}}}  )   \mathrel{ \Rrightarrow }  \ottnt{T_{{\mathrm{2}}}}  \,  \,\&\,  \,   \Phi_\mathsf{val}  \,  \tceappendop  \, \Phi_{{\mathrm{2}}}  \, / \, ( \forall  \mathit{y}  .  \ottnt{C_{{\mathrm{1}}}}  ) $}
          \end{prooftree}
          Note that $\mathit{x_{{\mathrm{1}}}} \,  \not\in  \,  \mathit{fv}  (  \ottnt{T_{{\mathrm{2}}}}  )  \,  \mathbin{\cup}  \,  \mathit{fv}  (  \Phi_{{\mathrm{2}}}  )  \,  \mathbin{\cup}  \, \ottsym{(}   \mathit{fv}  (  \ottnt{C_{{\mathrm{1}}}}  )  \,  \mathbin{\backslash}  \,  \{  \mathit{y}  \}   \ottsym{)}$
          by the assumption.

         \case \TK{Sub}:
          We are given
          \begin{prooftree}
           \AxiomC{$\ottnt{K}  \ottsym{:}    \ottnt{T_{{\mathrm{2}}}}  \, / \, ( \forall  \mathit{y}  .  \ottnt{C_{{\mathrm{1}}}}  )   \mathrel{ \Rrightarrow }  \ottnt{T_{{\mathrm{0}}}}  \,  \,\&\,  \,  \Phi_{{\mathrm{0}}}  \, / \, ( \forall  \mathit{z}  .  \ottnt{C'_{{\mathrm{0}}}}  ) $}
           \AxiomC{$\mathit{z} \,  {:}  \, \ottnt{T_{{\mathrm{0}}}}  \vdash  \ottnt{C_{{\mathrm{0}}}}  \ottsym{<:}  \ottnt{C'_{{\mathrm{0}}}}$}
           \AxiomC{$\mathit{z} \,  {:}  \, \ottnt{T_{{\mathrm{0}}}}  \vdash  \ottnt{C_{{\mathrm{0}}}}$}
           \TrinaryInfC{$\ottnt{K}  \ottsym{:}    \ottnt{T_{{\mathrm{2}}}}  \, / \, ( \forall  \mathit{y}  .  \ottnt{C_{{\mathrm{1}}}}  )   \mathrel{ \Rrightarrow }  \ottnt{T_{{\mathrm{0}}}}  \,  \,\&\,  \,  \Phi_{{\mathrm{0}}}  \, / \, ( \forall  \mathit{z}  .  \ottnt{C_{{\mathrm{0}}}}  ) $}
          \end{prooftree}
          for some $\ottnt{C'_{{\mathrm{0}}}}$.
          By the IH,
          \[
            \ottnt{K}  [  \mathsf{let} \, \mathit{x_{{\mathrm{1}}}}  \ottsym{=}  \,[\,]\, \,  \mathsf{in}  \, \ottnt{e_{{\mathrm{2}}}}  ]   \ottsym{:}    \ottnt{T_{{\mathrm{1}}}}  \, / \, ( \forall  \mathit{x_{{\mathrm{1}}}}  .  \ottnt{C_{{\mathrm{2}}}}  )   \mathrel{ \Rrightarrow }  \ottnt{T_{{\mathrm{0}}}}  \,  \,\&\,  \,  \Phi  \, / \, ( \forall  \mathit{z}  .  \ottnt{C'_{{\mathrm{0}}}}  ) 
          \]
          for some $\Phi$.
          By \TK{Sub}, we have the conclusion
          \[
            \ottnt{K}  [  \mathsf{let} \, \mathit{x_{{\mathrm{1}}}}  \ottsym{=}  \,[\,]\, \,  \mathsf{in}  \, \ottnt{e_{{\mathrm{2}}}}  ]   \ottsym{:}    \ottnt{T_{{\mathrm{1}}}}  \, / \, ( \forall  \mathit{x_{{\mathrm{1}}}}  .  \ottnt{C_{{\mathrm{2}}}}  )   \mathrel{ \Rrightarrow }  \ottnt{T_{{\mathrm{0}}}}  \,  \,\&\,  \,  \Phi  \, / \, ( \forall  \mathit{z}  .  \ottnt{C_{{\mathrm{0}}}}  )  ~.
          \]

         \case \TK{Let}:
          We are given
          \begin{prooftree}
           \AxiomC{$\ottnt{K'}  \ottsym{:}    \ottnt{T_{{\mathrm{2}}}}  \, / \, ( \forall  \mathit{y}  .  \ottnt{C_{{\mathrm{1}}}}  )   \mathrel{ \Rrightarrow }  \ottnt{T'_{{\mathrm{1}}}}  \,  \,\&\,  \,  \Phi'_{{\mathrm{1}}}  \, / \, ( \forall  \mathit{x'_{{\mathrm{1}}}}  .  \ottnt{C'_{{\mathrm{1}}}}  ) $}
           \AxiomC{$\mathit{x'_{{\mathrm{1}}}} \,  {:}  \, \ottnt{T'_{{\mathrm{1}}}}  \vdash  \ottnt{e'_{{\mathrm{2}}}}  \ottsym{:}   \ottnt{T_{{\mathrm{0}}}}  \,  \,\&\,  \,  \Phi'_{{\mathrm{2}}}  \, / \,   ( \forall  \mathit{z}  .  \ottnt{C_{{\mathrm{0}}}}  )  \Rightarrow   \ottnt{C'_{{\mathrm{1}}}}  $}
           \BinaryInfC{$\mathsf{let} \, \mathit{x'_{{\mathrm{1}}}}  \ottsym{=}  \ottnt{K'} \,  \mathsf{in}  \, \ottnt{e'_{{\mathrm{2}}}}  \ottsym{:}    \ottnt{T_{{\mathrm{2}}}}  \, / \, ( \forall  \mathit{y}  .  \ottnt{C_{{\mathrm{1}}}}  )   \mathrel{ \Rrightarrow }  \ottnt{T_{{\mathrm{0}}}}  \,  \,\&\,  \,  \Phi'_{{\mathrm{1}}} \,  \tceappendop  \, \Phi'_{{\mathrm{2}}}  \, / \, ( \forall  \mathit{z}  .  \ottnt{C_{{\mathrm{0}}}}  ) $}
          \end{prooftree}
          for some $\mathit{x'_{{\mathrm{1}}}}$, $\ottnt{K'}$, $\ottnt{e'_{{\mathrm{2}}}}$, $\Phi'_{{\mathrm{1}}}$, $\Phi'_{{\mathrm{2}}}$, $\ottnt{T'_{{\mathrm{1}}}}$, and $\ottnt{C'_{{\mathrm{1}}}}$ such that
          \begin{itemize}
           \item $\ottnt{K} \,  =  \, \ottsym{(}  \mathsf{let} \, \mathit{x'_{{\mathrm{1}}}}  \ottsym{=}  \ottnt{K'} \,  \mathsf{in}  \, \ottnt{e'_{{\mathrm{2}}}}  \ottsym{)}$,
           \item $\Phi_{{\mathrm{0}}} \,  =  \, \Phi'_{{\mathrm{1}}} \,  \tceappendop  \, \Phi'_{{\mathrm{2}}}$, and
           \item $\mathit{x'_{{\mathrm{1}}}} \,  \not\in  \,  \mathit{fv}  (  \ottnt{T_{{\mathrm{0}}}}  )  \,  \mathbin{\cup}  \,  \mathit{fv}  (  \Phi'_{{\mathrm{2}}}  )  \,  \mathbin{\cup}  \, \ottsym{(}   \mathit{fv}  (  \ottnt{C_{{\mathrm{0}}}}  )  \,  \mathbin{\backslash}  \,  \{  \mathit{z}  \}   \ottsym{)}$.
          \end{itemize}
          By the IH,
          \[
            \ottnt{K'}  [  \mathsf{let} \, \mathit{x_{{\mathrm{1}}}}  \ottsym{=}  \,[\,]\, \,  \mathsf{in}  \, \ottnt{e_{{\mathrm{2}}}}  ]   \ottsym{:}    \ottnt{T_{{\mathrm{1}}}}  \, / \, ( \forall  \mathit{x_{{\mathrm{1}}}}  .  \ottnt{C_{{\mathrm{2}}}}  )   \mathrel{ \Rrightarrow }  \ottnt{T'_{{\mathrm{1}}}}  \,  \,\&\,  \,  \Phi'  \, / \, ( \forall  \mathit{x'_{{\mathrm{1}}}}  .  \ottnt{C'_{{\mathrm{1}}}}  ) 
          \]
          for some $\Phi'$.
          By \TK{Let}, we have the conclusion
          \[
           \mathsf{let} \, \mathit{x'_{{\mathrm{1}}}}  \ottsym{=}   \ottnt{K'}  [  \mathsf{let} \, \mathit{x_{{\mathrm{1}}}}  \ottsym{=}  \,[\,]\, \,  \mathsf{in}  \, \ottnt{e_{{\mathrm{2}}}}  ]  \,  \mathsf{in}  \, \ottnt{e'_{{\mathrm{2}}}}  \ottsym{:}    \ottnt{T_{{\mathrm{1}}}}  \, / \, ( \forall  \mathit{x_{{\mathrm{1}}}}  .  \ottnt{C_{{\mathrm{2}}}}  )   \mathrel{ \Rrightarrow }  \ottnt{T_{{\mathrm{0}}}}  \,  \,\&\,  \,  \Phi' \,  \tceappendop  \, \Phi'_{{\mathrm{2}}}  \, / \, ( \forall  \mathit{z}  .  \ottnt{C_{{\mathrm{0}}}}  )  ~.
          \]
          where we let $\Phi \,  =  \, \Phi' \,  \tceappendop  \, \Phi'_{{\mathrm{2}}}$.
        \end{caseanalysis}

  \item Because $\ottnt{K}  \ottsym{:}   \ottnt{T_{{\mathrm{2}}}}  \, / \, ( \forall  \mathit{y}  .  \ottnt{C_{{\mathrm{1}}}}  ) $ implies
        $\ottnt{K}  \ottsym{:}    \ottnt{T_{{\mathrm{2}}}}  \, / \, ( \forall  \mathit{y}  .  \ottnt{C_{{\mathrm{1}}}}  )   \mathrel{ \Rrightarrow }  \ottnt{T_{{\mathrm{0}}}}  \,  \,\&\,  \,  \Phi_{{\mathrm{0}}}  \, / \, ( \forall  \mathit{x_{{\mathrm{0}}}}  .   \ottnt{T_{{\mathrm{0}}}}  \,  \,\&\,  \,   \Phi_\mathsf{val}   \, / \,   \square    ) $
        for some $\ottnt{T_{{\mathrm{0}}}}$, $\Phi_{{\mathrm{0}}}$, and $\mathit{x_{{\mathrm{0}}}} \,  \not\in  \,  \mathit{fv}  (  \ottnt{T_{{\mathrm{0}}}}  ) $,
        the case (\ref{lem:lr-cont-compse-let:cont}) implies
        $ \ottnt{K}  [  \mathsf{let} \, \mathit{x_{{\mathrm{1}}}}  \ottsym{=}  \,[\,]\, \,  \mathsf{in}  \, \ottnt{e_{{\mathrm{2}}}}  ]   \ottsym{:}    \ottnt{T_{{\mathrm{1}}}}  \, / \, ( \forall  \mathit{x_{{\mathrm{1}}}}  .  \ottnt{C_{{\mathrm{2}}}}  )   \mathrel{ \Rrightarrow }  \ottnt{T_{{\mathrm{0}}}}  \,  \,\&\,  \,  \Phi  \, / \, ( \forall  \mathit{x_{{\mathrm{0}}}}  .   \ottnt{T_{{\mathrm{0}}}}  \,  \,\&\,  \,   \Phi_\mathsf{val}   \, / \,   \square    ) $
        for some $\Phi$.
        Therefore, we have
        $ \ottnt{K}  [  \mathsf{let} \, \mathit{x_{{\mathrm{1}}}}  \ottsym{=}  \,[\,]\, \,  \mathsf{in}  \, \ottnt{e_{{\mathrm{2}}}}  ]   \ottsym{:}   \ottnt{T_{{\mathrm{1}}}}  \, / \, ( \forall  \mathit{x_{{\mathrm{1}}}}  .  \ottnt{C_{{\mathrm{2}}}}  ) $.
 \end{enumerate}
\end{proof}

\begin{lemmap}{Well-Typed Evaluation Contexts are Closed}{lr-typing-ectx-closed}
 If $\ottnt{K}  \ottsym{:}   \ottnt{T}  \, / \, ( \forall  \mathit{x}  .  \ottnt{C}  ) $, then no free variable occurs in $\ottnt{K}$.
\end{lemmap}
\begin{proof}
 It suffices to show that,
 if $\ottnt{K}  \ottsym{:}    \ottnt{T}  \, / \, ( \forall  \mathit{x}  .  \ottnt{C}  )   \mathrel{ \Rrightarrow }  \ottnt{T_{{\mathrm{0}}}}  \,  \,\&\,  \,  \Phi_{{\mathrm{0}}}  \, / \, ( \forall  \mathit{y}  .  \ottnt{C_{{\mathrm{0}}}}  ) $,
 then no free variable occurs in $\ottnt{K}$.
 It is proven straightforwardly by induction on the typing derivaiton
 of $\ottnt{K}  \ottsym{:}    \ottnt{T}  \, / \, ( \forall  \mathit{x}  .  \ottnt{C}  )   \mathrel{ \Rrightarrow }  \ottnt{T_{{\mathrm{0}}}}  \,  \,\&\,  \,  \Phi_{{\mathrm{0}}}  \, / \, ( \forall  \mathit{y}  .  \ottnt{C_{{\mathrm{0}}}}  ) $.
\end{proof}

\begin{lemmap}{One-Step Reduction Under Continuations}{lr-red-under-cont-one}
 If $ \ottnt{K}  [  \ottnt{e_{{\mathrm{1}}}}  ]  \,  \mathrel{  \longrightarrow  }  \, \ottnt{e_{{\mathrm{2}}}} \,  \,\&\,  \, \varpi$,
 then either of the following holds:
 \begin{itemize}
  \item $ {   \exists  \, { \ottnt{e'_{{\mathrm{2}}}} } . \  \ottnt{e_{{\mathrm{1}}}} \,  \mathrel{  \longrightarrow  }  \, \ottnt{e'_{{\mathrm{2}}}} \,  \,\&\,  \, \varpi  } \,\mathrel{\wedge}\, { \ottnt{e_{{\mathrm{2}}}} \,  =  \,  \ottnt{K}  [  \ottnt{e'_{{\mathrm{2}}}}  ]  } $; or
  \item $ {  {  {   \exists  \, { \ottnt{v_{{\mathrm{1}}}}  \ottsym{,}  \ottnt{K'}  \ottsym{,}  \mathit{x_{{\mathrm{1}}}}  \ottsym{,}  \ottnt{e'} } . \  \ottnt{e_{{\mathrm{1}}}} \,  =  \, \ottnt{v_{{\mathrm{1}}}}  } \,\mathrel{\wedge}\, { \ottnt{K} \,  =  \,  \ottnt{K'}  [  \mathsf{let} \, \mathit{x_{{\mathrm{1}}}}  \ottsym{=}  \,[\,]\, \,  \mathsf{in}  \, \ottnt{e'}  ]  }  } \,\mathrel{\wedge}\, { \ottnt{e_{{\mathrm{2}}}} \,  =  \,  \ottnt{K'}  [   \ottnt{e'}    [  \ottnt{v_{{\mathrm{1}}}}  /  \mathit{x_{{\mathrm{1}}}}  ]    ]  }  } \,\mathrel{\wedge}\, { \varpi \,  =  \,  \epsilon  } $.
 \end{itemize}
\end{lemmap}
\begin{proof}
 By induction on $\ottnt{K}$.
 \begin{caseanalysis}
  \case $\ottnt{K} \,  =  \, \,[\,]\,$:
   We are given $ \ottnt{K}  [  \ottnt{e_{{\mathrm{1}}}}  ]  \,  =  \, \ottnt{e_{{\mathrm{1}}}}$.
   Therefore, $\ottnt{e_{{\mathrm{1}}}} \,  \mathrel{  \longrightarrow  }  \, \ottnt{e_{{\mathrm{2}}}} \,  \,\&\,  \, \varpi$.
   Then, we have the conclusion (the first case) by letting $\ottnt{e'_{{\mathrm{2}}}} \,  =  \, \ottnt{e_{{\mathrm{2}}}}$.

  \case $  \exists  \, { \mathit{x_{{\mathrm{0}}}}  \ottsym{,}  \ottnt{K_{{\mathrm{0}}}}  \ottsym{,}  \ottnt{e_{{\mathrm{0}}}} } . \  \ottnt{K} \,  =  \, \ottsym{(}  \mathsf{let} \, \mathit{x_{{\mathrm{0}}}}  \ottsym{=}  \ottnt{K_{{\mathrm{0}}}} \,  \mathsf{in}  \, \ottnt{e_{{\mathrm{0}}}}  \ottsym{)} $:
   We are given $\mathsf{let} \, \mathit{x_{{\mathrm{0}}}}  \ottsym{=}   \ottnt{K_{{\mathrm{0}}}}  [  \ottnt{e_{{\mathrm{1}}}}  ]  \,  \mathsf{in}  \, \ottnt{e_{{\mathrm{0}}}} \,  \mathrel{  \longrightarrow  }  \, \ottnt{e_{{\mathrm{2}}}} \,  \,\&\,  \, \varpi$.
   By case analysis on the evaluation rule applied to derive $\mathsf{let} \, \mathit{x_{{\mathrm{0}}}}  \ottsym{=}   \ottnt{K_{{\mathrm{0}}}}  [  \ottnt{e_{{\mathrm{1}}}}  ]  \,  \mathsf{in}  \, \ottnt{e_{{\mathrm{0}}}} \,  \mathrel{  \longrightarrow  }  \, \ottnt{e_{{\mathrm{2}}}} \,  \,\&\,  \, \varpi$.
   \begin{caseanalysis}
    \case \E{Red}/\R{Let}:
     We are given
     \begin{prooftree}
      \AxiomC{$\mathsf{let} \, \mathit{x_{{\mathrm{0}}}}  \ottsym{=}  \ottnt{v_{{\mathrm{1}}}} \,  \mathsf{in}  \, \ottnt{e_{{\mathrm{0}}}} \,  \mathrel{  \longrightarrow  }  \,  \ottnt{e_{{\mathrm{0}}}}    [  \ottnt{v_{{\mathrm{1}}}}  /  \mathit{x_{{\mathrm{0}}}}  ]   \,  \,\&\,  \,  \epsilon $}
     \end{prooftree}
     for some $\ottnt{v_{{\mathrm{1}}}}$ such that $ \ottnt{K_{{\mathrm{0}}}}  [  \ottnt{e_{{\mathrm{1}}}}  ]  \,  =  \, \ottnt{v_{{\mathrm{1}}}}$ and $\ottnt{e_{{\mathrm{2}}}} \,  =  \,  \ottnt{e_{{\mathrm{0}}}}    [  \ottnt{v_{{\mathrm{1}}}}  /  \mathit{x_{{\mathrm{0}}}}  ]  $.
     We also have $ \varpi    =     \epsilon  $.
     Because $ \ottnt{K_{{\mathrm{0}}}}  [  \ottnt{e_{{\mathrm{1}}}}  ]  \,  =  \, \ottnt{v_{{\mathrm{1}}}}$, we have $\ottnt{K_{{\mathrm{0}}}} \,  =  \, \,[\,]\,$ and $\ottnt{e_{{\mathrm{1}}}} \,  =  \, \ottnt{v_{{\mathrm{1}}}}$.
     Therefore, we have the conclusion (the second case)
     by letting $\ottnt{K'} \,  =  \, \,[\,]\,$, $ \mathit{x_{{\mathrm{1}}}}    =    \mathit{x_{{\mathrm{0}}}} $, and $\ottnt{e'} \,  =  \, \ottnt{e_{{\mathrm{0}}}}$.

    \case \E{Let}:
     We are given
     \begin{prooftree}
      \AxiomC{$ \ottnt{K_{{\mathrm{0}}}}  [  \ottnt{e_{{\mathrm{1}}}}  ]  \,  \mathrel{  \longrightarrow  }  \, \ottnt{e_{{\mathrm{02}}}} \,  \,\&\,  \, \varpi$}
      \UnaryInfC{$\mathsf{let} \, \mathit{x_{{\mathrm{0}}}}  \ottsym{=}   \ottnt{K_{{\mathrm{0}}}}  [  \ottnt{e_{{\mathrm{1}}}}  ]  \,  \mathsf{in}  \, \ottnt{e_{{\mathrm{0}}}} \,  \mathrel{  \longrightarrow  }  \, \mathsf{let} \, \mathit{x_{{\mathrm{0}}}}  \ottsym{=}  \ottnt{e_{{\mathrm{02}}}} \,  \mathsf{in}  \, \ottnt{e_{{\mathrm{0}}}} \,  \,\&\,  \, \varpi$}
     \end{prooftree}
     for some $\ottnt{e_{{\mathrm{02}}}}$ such that $\ottnt{e_{{\mathrm{2}}}} \,  =  \, \ottsym{(}  \mathsf{let} \, \mathit{x_{{\mathrm{0}}}}  \ottsym{=}  \ottnt{e_{{\mathrm{02}}}} \,  \mathsf{in}  \, \ottnt{e_{{\mathrm{0}}}}  \ottsym{)}$.
     By case analysis on the result of the IH with $ \ottnt{K_{{\mathrm{0}}}}  [  \ottnt{e_{{\mathrm{1}}}}  ]  \,  \mathrel{  \longrightarrow  }  \, \ottnt{e_{{\mathrm{02}}}} \,  \,\&\,  \, \varpi$.
     \begin{caseanalysis}
      \case $ {   \exists  \, { \ottnt{e'_{{\mathrm{2}}}} } . \  \ottnt{e_{{\mathrm{1}}}} \,  \mathrel{  \longrightarrow  }  \, \ottnt{e'_{{\mathrm{2}}}} \,  \,\&\,  \, \varpi  } \,\mathrel{\wedge}\, { \ottnt{e_{{\mathrm{02}}}} \,  =  \,  \ottnt{K_{{\mathrm{0}}}}  [  \ottnt{e'_{{\mathrm{2}}}}  ]  } $:
       Because $\ottnt{e_{{\mathrm{2}}}} = \ottsym{(}  \mathsf{let} \, \mathit{x_{{\mathrm{0}}}}  \ottsym{=}  \ottnt{e_{{\mathrm{02}}}} \,  \mathsf{in}  \, \ottnt{e_{{\mathrm{0}}}}  \ottsym{)} = \ottsym{(}  \mathsf{let} \, \mathit{x_{{\mathrm{0}}}}  \ottsym{=}   \ottnt{K_{{\mathrm{0}}}}  [  \ottnt{e'_{{\mathrm{2}}}}  ]  \,  \mathsf{in}  \, \ottnt{e_{{\mathrm{0}}}}  \ottsym{)} =  \ottnt{K}  [  \ottnt{e'_{{\mathrm{2}}}}  ] $,
       we have the conclusion (the first case).

      \case $ {  {  {   \exists  \, { \ottnt{v_{{\mathrm{1}}}}  \ottsym{,}  \ottnt{K''}  \ottsym{,}  \mathit{x_{{\mathrm{1}}}}  \ottsym{,}  \ottnt{e'} } . \  \ottnt{e_{{\mathrm{1}}}} \,  =  \, \ottnt{v_{{\mathrm{1}}}}  } \,\mathrel{\wedge}\, { \ottnt{K_{{\mathrm{0}}}} \,  =  \,  \ottnt{K''}  [  \mathsf{let} \, \mathit{x_{{\mathrm{1}}}}  \ottsym{=}  \,[\,]\, \,  \mathsf{in}  \, \ottnt{e'}  ]  }  } \,\mathrel{\wedge}\, { \ottnt{e_{{\mathrm{02}}}} \,  =  \,  \ottnt{K''}  [   \ottnt{e'}    [  \ottnt{v_{{\mathrm{1}}}}  /  \mathit{x_{{\mathrm{1}}}}  ]    ]  }  } \,\mathrel{\wedge}\, { \varpi \,  =  \,  \epsilon  } $:
       Because
       $\ottnt{K} = \ottsym{(}  \mathsf{let} \, \mathit{x_{{\mathrm{0}}}}  \ottsym{=}  \ottnt{K_{{\mathrm{0}}}} \,  \mathsf{in}  \, \ottnt{e_{{\mathrm{0}}}}  \ottsym{)} = \ottsym{(}  \mathsf{let} \, \mathit{x_{{\mathrm{0}}}}  \ottsym{=}   \ottnt{K''}  [  \mathsf{let} \, \mathit{x_{{\mathrm{1}}}}  \ottsym{=}  \,[\,]\, \,  \mathsf{in}  \, \ottnt{e'}  ]  \,  \mathsf{in}  \, \ottnt{e_{{\mathrm{0}}}}  \ottsym{)}$ and
       $\ottnt{e_{{\mathrm{2}}}} = \ottsym{(}  \mathsf{let} \, \mathit{x_{{\mathrm{0}}}}  \ottsym{=}  \ottnt{e_{{\mathrm{02}}}} \,  \mathsf{in}  \, \ottnt{e_{{\mathrm{0}}}}  \ottsym{)} = \ottsym{(}  \mathsf{let} \, \mathit{x_{{\mathrm{0}}}}  \ottsym{=}   \ottnt{K''}  [   \ottnt{e'}    [  \ottnt{v_{{\mathrm{1}}}}  /  \mathit{x_{{\mathrm{1}}}}  ]    ]  \,  \mathsf{in}  \, \ottnt{e_{{\mathrm{0}}}}  \ottsym{)}$,
       we have the conclusion (the second case) by letting $\ottnt{K'} \,  =  \, \ottsym{(}  \mathsf{let} \, \mathit{x_{{\mathrm{0}}}}  \ottsym{=}  \ottnt{K''} \,  \mathsf{in}  \, \ottnt{e_{{\mathrm{0}}}}  \ottsym{)}$.
     \end{caseanalysis}
   \end{caseanalysis}
 \end{caseanalysis}
\end{proof}

\begin{lemmap}{Evaluation Under Continuations}{lr-red-under-cont-term}
 If $ \ottnt{K}  [  \ottnt{e_{{\mathrm{1}}}}  ]  \,  \mathrel{  \longrightarrow  }^{ \ottmv{n} }  \, \ottnt{e_{{\mathrm{2}}}} \,  \,\&\,  \, \varpi$,
 then one of the following hods.
 \begin{itemize}
  \item $ {  {   \exists  \, { \ottnt{v}  \ottsym{,}  \varpi_{{\mathrm{1}}}  \ottsym{,}  \varpi_{{\mathrm{2}}} } . \  \ottnt{e_{{\mathrm{1}}}} \,  \mathrel{  \longrightarrow  ^*}  \, \ottnt{v} \,  \,\&\,  \, \varpi_{{\mathrm{1}}}  } \,\mathrel{\wedge}\, {  \ottnt{K}  [  \ottnt{v}  ]  \,  \mathrel{  \longrightarrow  ^*}  \, \ottnt{e_{{\mathrm{2}}}} \,  \,\&\,  \, \varpi_{{\mathrm{2}}} }  } \,\mathrel{\wedge}\, { \varpi \,  =  \, \varpi_{{\mathrm{1}}} \,  \tceappendop  \, \varpi_{{\mathrm{2}}} } $;
  \item $ {   \exists  \, { \ottnt{K_{{\mathrm{0}}}}  \ottsym{,}  \mathit{x_{{\mathrm{0}}}}  \ottsym{,}  \ottnt{e_{{\mathrm{0}}}} } . \  \ottnt{e_{{\mathrm{1}}}} \,  \mathrel{  \longrightarrow  ^*}  \,  \ottnt{K_{{\mathrm{0}}}}  [  \mathcal{S}_0 \, \mathit{x_{{\mathrm{0}}}}  \ottsym{.}  \ottnt{e_{{\mathrm{0}}}}  ]  \,  \,\&\,  \, \varpi  } \,\mathrel{\wedge}\, { \ottnt{e_{{\mathrm{2}}}} \,  =  \,  \ottnt{K}  [   \ottnt{K_{{\mathrm{0}}}}  [  \mathcal{S}_0 \, \mathit{x_{{\mathrm{0}}}}  \ottsym{.}  \ottnt{e_{{\mathrm{0}}}}  ]   ]  } $; or
  \item $ {   \exists  \, { \ottnt{e'_{{\mathrm{2}}}} } . \  \ottnt{e_{{\mathrm{1}}}} \,  \mathrel{  \longrightarrow  ^*}  \, \ottnt{e'_{{\mathrm{2}}}} \,  \,\&\,  \, \varpi  } \,\mathrel{\wedge}\, { \ottnt{e_{{\mathrm{2}}}} \,  =  \,  \ottnt{K}  [  \ottnt{e'_{{\mathrm{2}}}}  ]  } $.
 \end{itemize}
\end{lemmap}
\begin{proof}
 By induction on $\ottmv{n}$.
 \begin{caseanalysis}
  \case $\ottmv{n} \,  =  \, \ottsym{0}$:
   We are given $ \ottnt{K}  [  \ottnt{e_{{\mathrm{1}}}}  ]  \,  =  \, \ottnt{e_{{\mathrm{2}}}}$ and $ \varpi    =     \epsilon  $.
   Therefore, we have the conclusion (the third case) by letting $\ottnt{e'_{{\mathrm{2}}}} \,  =  \, \ottnt{e_{{\mathrm{1}}}}$.

  \case $\ottmv{n} \,  >  \, \ottsym{0}$:
   We are given $ \ottnt{K}  [  \ottnt{e_{{\mathrm{1}}}}  ]  \,  \mathrel{  \longrightarrow  }  \, \ottnt{e_{{\mathrm{3}}}} \,  \,\&\,  \, \varpi'_{{\mathrm{1}}}$ and $\ottnt{e_{{\mathrm{3}}}} \,  \mathrel{  \longrightarrow  }^{ \ottmv{n}  \ottsym{-}  \ottsym{1} }  \, \ottnt{e_{{\mathrm{2}}}} \,  \,\&\,  \, \varpi'_{{\mathrm{2}}}$
   for some $\ottnt{e_{{\mathrm{3}}}}$, $\varpi'_{{\mathrm{1}}}$, and $\varpi'_{{\mathrm{2}}}$ such that $ \varpi    =    \varpi'_{{\mathrm{1}}} \,  \tceappendop  \, \varpi'_{{\mathrm{2}}} $.
   By case analysis on the result of \reflem{lr-red-under-cont-one} with $ \ottnt{K}  [  \ottnt{e_{{\mathrm{1}}}}  ]  \,  \mathrel{  \longrightarrow  }  \, \ottnt{e_{{\mathrm{3}}}} \,  \,\&\,  \, \varpi'_{{\mathrm{1}}}$.
   \begin{caseanalysis}
    \case $ {   \exists  \, { \ottnt{e'_{{\mathrm{3}}}} } . \  \ottnt{e_{{\mathrm{1}}}} \,  \mathrel{  \longrightarrow  }  \, \ottnt{e'_{{\mathrm{3}}}} \,  \,\&\,  \, \varpi'_{{\mathrm{1}}}  } \,\mathrel{\wedge}\, { \ottnt{e_{{\mathrm{3}}}} \,  =  \,  \ottnt{K}  [  \ottnt{e'_{{\mathrm{3}}}}  ]  } $:
     Because $ \ottnt{K}  [  \ottnt{e'_{{\mathrm{3}}}}  ]  = \ottnt{e_{{\mathrm{3}}}} \,  \mathrel{  \longrightarrow  }^{ \ottmv{n}  \ottsym{-}  \ottsym{1} }  \, \ottnt{e_{{\mathrm{2}}}} \,  \,\&\,  \, \varpi'_{{\mathrm{2}}}$,
     the IH implies one of the following.
     \begin{caseanalysis}
      \case $ {  {   \exists  \, { \ottnt{v}  \ottsym{,}  \varpi'_{{\mathrm{21}}}  \ottsym{,}  \varpi'_{{\mathrm{22}}} } . \  \ottnt{e'_{{\mathrm{3}}}} \,  \mathrel{  \longrightarrow  ^*}  \, \ottnt{v} \,  \,\&\,  \, \varpi'_{{\mathrm{21}}}  } \,\mathrel{\wedge}\, {  \ottnt{K}  [  \ottnt{v}  ]  \,  \mathrel{  \longrightarrow  ^*}  \, \ottnt{e_{{\mathrm{2}}}} \,  \,\&\,  \, \varpi'_{{\mathrm{22}}} }  } \,\mathrel{\wedge}\, { \varpi'_{{\mathrm{2}}} \,  =  \, \varpi'_{{\mathrm{21}}} \,  \tceappendop  \, \varpi'_{{\mathrm{22}}} } $:
       Because $\ottnt{e_{{\mathrm{1}}}} \,  \mathrel{  \longrightarrow  }  \, \ottnt{e'_{{\mathrm{3}}}} \,  \,\&\,  \, \varpi'_{{\mathrm{1}}}$, we have $\ottnt{e_{{\mathrm{1}}}} \,  \mathrel{  \longrightarrow  ^*}  \, \ottnt{v} \,  \,\&\,  \, \varpi'_{{\mathrm{1}}} \,  \tceappendop  \, \varpi'_{{\mathrm{21}}}$.
       Then, because $\varpi = \varpi'_{{\mathrm{1}}} \,  \tceappendop  \, \varpi'_{{\mathrm{2}}} = \varpi'_{{\mathrm{1}}} \,  \tceappendop  \, \varpi'_{{\mathrm{21}}} \,  \tceappendop  \, \varpi'_{{\mathrm{22}}}$, we have the conclusion (the first case).

      \case $ {   \exists  \, { \ottnt{K_{{\mathrm{0}}}}  \ottsym{,}  \mathit{x_{{\mathrm{0}}}}  \ottsym{,}  \ottnt{e_{{\mathrm{0}}}} } . \  \ottnt{e'_{{\mathrm{3}}}} \,  \mathrel{  \longrightarrow  ^*}  \,  \ottnt{K_{{\mathrm{0}}}}  [  \mathcal{S}_0 \, \mathit{x_{{\mathrm{0}}}}  \ottsym{.}  \ottnt{e_{{\mathrm{0}}}}  ]  \,  \,\&\,  \, \varpi'_{{\mathrm{2}}}  } \,\mathrel{\wedge}\, { \ottnt{e_{{\mathrm{2}}}} \,  =  \,  \ottnt{K}  [   \ottnt{K_{{\mathrm{0}}}}  [  \mathcal{S}_0 \, \mathit{x_{{\mathrm{0}}}}  \ottsym{.}  \ottnt{e_{{\mathrm{0}}}}  ]   ]  } $:
       Because $\ottnt{e_{{\mathrm{1}}}} \,  \mathrel{  \longrightarrow  }  \, \ottnt{e'_{{\mathrm{3}}}} \,  \,\&\,  \, \varpi'_{{\mathrm{1}}}$, we have $\ottnt{e_{{\mathrm{1}}}} \,  \mathrel{  \longrightarrow  ^*}  \,  \ottnt{K_{{\mathrm{0}}}}  [  \mathcal{S}_0 \, \mathit{x_{{\mathrm{0}}}}  \ottsym{.}  \ottnt{e_{{\mathrm{0}}}}  ]  \,  \,\&\,  \, \varpi'_{{\mathrm{1}}} \,  \tceappendop  \, \varpi'_{{\mathrm{2}}}$.
       Because $ \varpi    =    \varpi'_{{\mathrm{1}}} \,  \tceappendop  \, \varpi'_{{\mathrm{2}}} $, we have the conclusion (the second case).

      \case $ {   \exists  \, { \ottnt{e'_{{\mathrm{2}}}} } . \  \ottnt{e'_{{\mathrm{3}}}} \,  \mathrel{  \longrightarrow  ^*}  \, \ottnt{e'_{{\mathrm{2}}}} \,  \,\&\,  \, \varpi'_{{\mathrm{2}}}  } \,\mathrel{\wedge}\, { \ottnt{e_{{\mathrm{2}}}} \,  =  \,  \ottnt{K}  [  \ottnt{e'_{{\mathrm{2}}}}  ]  } $:
       Because $\ottnt{e_{{\mathrm{1}}}} \,  \mathrel{  \longrightarrow  }  \, \ottnt{e'_{{\mathrm{3}}}} \,  \,\&\,  \, \varpi'_{{\mathrm{1}}}$, we have $\ottnt{e_{{\mathrm{1}}}} \,  \mathrel{  \longrightarrow  ^*}  \, \ottnt{e'_{{\mathrm{2}}}} \,  \,\&\,  \, \varpi'_{{\mathrm{1}}} \,  \tceappendop  \, \varpi'_{{\mathrm{2}}}$.
       Because $ \varpi    =    \varpi'_{{\mathrm{1}}} \,  \tceappendop  \, \varpi'_{{\mathrm{2}}} $, we have the conclusion (the third case).
     \end{caseanalysis}

    \case $ {  {  {   \exists  \, { \ottnt{v}  \ottsym{,}  \ottnt{K'}  \ottsym{,}  \mathit{x_{{\mathrm{1}}}}  \ottsym{,}  \ottnt{e'} } . \  \ottnt{e_{{\mathrm{1}}}} \,  =  \, \ottnt{v}  } \,\mathrel{\wedge}\, { \ottnt{K} \,  =  \,  \ottnt{K'}  [  \mathsf{let} \, \mathit{x_{{\mathrm{1}}}}  \ottsym{=}  \,[\,]\, \,  \mathsf{in}  \, \ottnt{e'}  ]  }  } \,\mathrel{\wedge}\, { \ottnt{e_{{\mathrm{3}}}} \,  =  \,  \ottnt{K'}  [   \ottnt{e'}    [  \ottnt{v}  /  \mathit{x_{{\mathrm{1}}}}  ]    ]  }  } \,\mathrel{\wedge}\, { \varpi'_{{\mathrm{1}}} \,  =  \,  \epsilon  } $:
     We have the conclusion (the first case) because
     \begin{itemize}
      \item $\ottnt{e_{{\mathrm{1}}}} = \ottnt{v} \,  \mathrel{  \longrightarrow  ^*}  \, \ottnt{v} \,  \,\&\,  \,  \epsilon $,
      \item $ \ottnt{K}  [  \ottnt{v}  ]  \,  \mathrel{  \longrightarrow  ^*}  \, \ottnt{e_{{\mathrm{2}}}} \,  \,\&\,  \, \varpi'_{{\mathrm{2}}}$ by
            \begin{itemize}
             \item $ \ottnt{K}  [  \ottnt{v}  ]  =  \ottnt{K'}  [  \mathsf{let} \, \mathit{x_{{\mathrm{1}}}}  \ottsym{=}  \ottnt{v} \,  \mathsf{in}  \, \ottnt{e'}  ]  \,  \mathrel{  \longrightarrow  }  \,  \ottnt{K'}  [   \ottnt{e'}    [  \ottnt{v}  /  \mathit{x_{{\mathrm{1}}}}  ]    ]  \,  \,\&\,  \,  \epsilon $
                   by \reflem{lr-red-subterm} and \E{Red}/\R{Let}, and
             \item $ \ottnt{K'}  [   \ottnt{e'}    [  \ottnt{v}  /  \mathit{x_{{\mathrm{1}}}}  ]    ] = \ottnt{e_{{\mathrm{3}}}} \,  \mathrel{  \longrightarrow  }^{ \ottmv{n}  \ottsym{-}  \ottsym{1} }  \, \ottnt{e_{{\mathrm{2}}}} \,  \,\&\,  \, \varpi'_{{\mathrm{2}}}$,
            \end{itemize}
            and
      \item $\varpi = \varpi'_{{\mathrm{1}}} \,  \tceappendop  \, \varpi'_{{\mathrm{2}}} =  \epsilon  \,  \tceappendop  \, \varpi'_{{\mathrm{2}}}$.
     \end{itemize}
   \end{caseanalysis}
 \end{caseanalysis}
\end{proof}

\begin{lemmap}{Divergence Under Continuations}{lr-red-under-cont-div}
 If $ \ottnt{K}  [  \ottnt{e}  ]  \,  \Uparrow  \, \pi$, then either of the following holds.
 \begin{itemize}
  \item $ {  {   \exists  \, { \ottnt{v}  \ottsym{,}  \varpi_{{\mathrm{1}}}  \ottsym{,}  \pi_{{\mathrm{2}}} } . \  \ottnt{e} \,  \Downarrow  \, \ottnt{v} \,  \,\&\,  \, \varpi_{{\mathrm{1}}}  } \,\mathrel{\wedge}\, {  \ottnt{K}  [  \ottnt{v}  ]  \,  \Uparrow  \, \pi_{{\mathrm{2}}} }  } \,\mathrel{\wedge}\, { \pi \,  =  \, \varpi_{{\mathrm{1}}} \,  \tceappendop  \, \pi_{{\mathrm{2}}} } $; or
  \item $\ottnt{e} \,  \Uparrow  \, \pi$.
 \end{itemize}
\end{lemmap}
\begin{proof}
 By case analysis on the definition of $ \ottnt{K}  [  \ottnt{e}  ]  \,  \Uparrow  \, \pi$.
 \begin{caseanalysis}
  \case $   \forall  \, { \varpi'  \ottsym{,}  \pi' } . \  \pi \,  =  \, \varpi' \,  \tceappendop  \, \pi'   \mathrel{\Longrightarrow}    \exists  \, { \ottnt{e'} } . \   \ottnt{K}  [  \ottnt{e}  ]  \,  \mathrel{  \longrightarrow  ^*}  \, \ottnt{e'} \,  \,\&\,  \, \varpi'  $:
   By case analysis on whether there exist some $\ottnt{v}$, $\varpi_{{\mathrm{1}}}$, and $\pi_{{\mathrm{2}}}$
   such that $ { \ottnt{e} \,  \Downarrow  \, \ottnt{v} \,  \,\&\,  \, \varpi_{{\mathrm{1}}} } \,\mathrel{\wedge}\, { \pi \,  =  \, \varpi_{{\mathrm{1}}} \,  \tceappendop  \, \pi_{{\mathrm{2}}} } $.
   \begin{caseanalysis}
    \case $ {   \exists  \, { \ottnt{v}  \ottsym{,}  \varpi_{{\mathrm{1}}}  \ottsym{,}  \pi_{{\mathrm{2}}} } . \  \ottnt{e} \,  \Downarrow  \, \ottnt{v} \,  \,\&\,  \, \varpi_{{\mathrm{1}}}  } \,\mathrel{\wedge}\, { \pi \,  =  \, \varpi_{{\mathrm{1}}} \,  \tceappendop  \, \pi_{{\mathrm{2}}} } $:
     It suffices to show that $ \ottnt{K}  [  \ottnt{v}  ]  \,  \Uparrow  \, \pi_{{\mathrm{2}}}$.
     Let $\varpi'_{{\mathrm{2}}}$ be any finite trace and $\pi'_{{\mathrm{2}}}$ be any infinite trace such that
     $ \pi_{{\mathrm{2}}}    =    \varpi'_{{\mathrm{2}}} \,  \tceappendop  \, \pi'_{{\mathrm{2}}} $.
     Then, it suffices to show that
     \[
        \exists  \, { \ottnt{e'} } . \   \ottnt{K}  [  \ottnt{v}  ]  \,  \mathrel{  \longrightarrow  ^*}  \, \ottnt{e'} \,  \,\&\,  \, \varpi'_{{\mathrm{2}}}  ~.
     \]

     If $ \varpi'_{{\mathrm{2}}}    =     \epsilon  $, then we have the conclusion by letting $\ottnt{e'} \,  =  \,  \ottnt{K}  [  \ottnt{v}  ] $.

     Otherwise, suppose that $ \varpi'_{{\mathrm{2}}}    \not=     \epsilon  $.
     Because $\pi = \varpi_{{\mathrm{1}}} \,  \tceappendop  \, \pi_{{\mathrm{2}}} = \varpi_{{\mathrm{1}}} \,  \tceappendop  \, \varpi'_{{\mathrm{2}}} \,  \tceappendop  \, \pi'_{{\mathrm{2}}}$,
     $ \ottnt{K}  [  \ottnt{e}  ]  \,  \Uparrow  \, \pi$ implies that there exists some $\ottnt{e'}$ such that
     $ \ottnt{K}  [  \ottnt{e}  ]  \,  \mathrel{  \longrightarrow  ^*}  \, \ottnt{e'} \,  \,\&\,  \, \varpi_{{\mathrm{1}}} \,  \tceappendop  \, \varpi'_{{\mathrm{2}}}$.
     Because $\ottnt{e} \,  \Downarrow  \, \ottnt{v} \,  \,\&\,  \, \varpi_{{\mathrm{1}}}$ implies $ \ottnt{K}  [  \ottnt{e}  ]  \,  \mathrel{  \longrightarrow  ^*}  \,  \ottnt{K}  [  \ottnt{v}  ]  \,  \,\&\,  \, \varpi_{{\mathrm{1}}}$
     by \reflem{lr-red-subterm},
     \reflem{lr-eval-det} implies that the conclusion
     $ \ottnt{K}  [  \ottnt{v}  ]  \,  \mathrel{  \longrightarrow  ^*}  \, \ottnt{e'} \,  \,\&\,  \, \varpi'_{{\mathrm{2}}}$.

    \case $ \neg  \ottsym{(}   {   \exists  \, { \ottnt{v}  \ottsym{,}  \varpi_{{\mathrm{1}}}  \ottsym{,}  \pi_{{\mathrm{2}}} } . \  \ottnt{e} \,  \Downarrow  \, \ottnt{v} \,  \,\&\,  \, \varpi_{{\mathrm{1}}}  } \,\mathrel{\wedge}\, { \pi \,  =  \, \varpi_{{\mathrm{1}}} \,  \tceappendop  \, \pi_{{\mathrm{2}}} }   \ottsym{)} $:
     We show that $\ottnt{e} \,  \Uparrow  \, \pi$.
     Let $\varpi'$ be any finite trace and $\pi'$ be any infinite trace
     such that $ \pi    =    \varpi' \,  \tceappendop  \, \pi' $.
     Then, it suffices to show that
     \[
        \exists  \, { \ottnt{e''} } . \  \ottnt{e} \,  \mathrel{  \longrightarrow  ^*}  \, \ottnt{e''} \,  \,\&\,  \, \varpi'  ~.
     \]
     $ \ottnt{K}  [  \ottnt{e}  ]  \,  \Uparrow  \, \pi$ implies that there exists some $\ottnt{e'}$ such that
     $ \ottnt{K}  [  \ottnt{e}  ]  \,  \mathrel{  \longrightarrow  ^*}  \, \ottnt{e'} \,  \,\&\,  \, \varpi'$.
     By case analysis on the result of \reflem{lr-red-under-cont-term}
     with $ \ottnt{K}  [  \ottnt{e}  ]  \,  \mathrel{  \longrightarrow  ^*}  \, \ottnt{e'} \,  \,\&\,  \, \varpi'$.
     \begin{caseanalysis}
      \case $ {  {   \exists  \, { \ottnt{v}  \ottsym{,}  \varpi'_{{\mathrm{1}}}  \ottsym{,}  \varpi'_{{\mathrm{2}}} } . \  \ottnt{e} \,  \mathrel{  \longrightarrow  ^*}  \, \ottnt{v} \,  \,\&\,  \, \varpi'_{{\mathrm{1}}}  } \,\mathrel{\wedge}\, {  \ottnt{K}  [  \ottnt{v}  ]  \,  \mathrel{  \longrightarrow  ^*}  \, \ottnt{e'} \,  \,\&\,  \, \varpi'_{{\mathrm{2}}} }  } \,\mathrel{\wedge}\, { \varpi' \,  =  \, \varpi'_{{\mathrm{1}}} \,  \tceappendop  \, \varpi'_{{\mathrm{2}}} } $:
       Contradictory because $\pi = \varpi' \,  \tceappendop  \, \pi' = \varpi'_{{\mathrm{1}}} \,  \tceappendop  \, \varpi'_{{\mathrm{2}}} \,  \tceappendop  \, \pi'$ (i.e., $\varpi'_{{\mathrm{1}}}$ is a prefix of $\pi$).

      \case $ {   \exists  \, { \ottnt{K_{{\mathrm{0}}}}  \ottsym{,}  \mathit{x_{{\mathrm{0}}}}  \ottsym{,}  \ottnt{e_{{\mathrm{0}}}} } . \  \ottnt{e} \,  \mathrel{  \longrightarrow  ^*}  \,  \ottnt{K_{{\mathrm{0}}}}  [  \mathcal{S}_0 \, \mathit{x_{{\mathrm{0}}}}  \ottsym{.}  \ottnt{e_{{\mathrm{0}}}}  ]  \,  \,\&\,  \, \varpi'  } \,\mathrel{\wedge}\, { \ottnt{e'} \,  =  \,  \ottnt{K}  [   \ottnt{K_{{\mathrm{0}}}}  [  \mathcal{S}_0 \, \mathit{x_{{\mathrm{0}}}}  \ottsym{.}  \ottnt{e_{{\mathrm{0}}}}  ]   ]  } $:
       We have the conclusion by letting $\ottnt{e''} \,  =  \,  \ottnt{K_{{\mathrm{0}}}}  [  \mathcal{S}_0 \, \mathit{x_{{\mathrm{0}}}}  \ottsym{.}  \ottnt{e_{{\mathrm{0}}}}  ] $.

      \case $ {   \exists  \, { \ottnt{e''} } . \  \ottnt{e} \,  \mathrel{  \longrightarrow  ^*}  \, \ottnt{e''} \,  \,\&\,  \, \varpi'  } \,\mathrel{\wedge}\, { \ottnt{e'} \,  =  \,  \ottnt{K}  [  \ottnt{e''}  ]  } $:
       We have the conclusion.
     \end{caseanalysis}

    \case $ {   \exists  \, { \varpi'  \ottsym{,}  \pi'  \ottsym{,}  \ottnt{E} } . \  \pi \,  =  \, \varpi' \,  \tceappendop  \, \pi'  } \,\mathrel{\wedge}\, {  \ottnt{K}  [  \ottnt{e}  ]  \,  \mathrel{  \longrightarrow  ^*}  \,  \ottnt{E}  [   \bullet ^{ \pi' }   ]  \,  \,\&\,  \, \varpi' } $:
     By case analysis on the result of \reflem{lr-red-under-cont-term}
     with $ \ottnt{K}  [  \ottnt{e}  ]  \,  \mathrel{  \longrightarrow  ^*}  \,  \ottnt{E}  [   \bullet ^{ \pi' }   ]  \,  \,\&\,  \, \varpi'$.
     \begin{caseanalysis}
      \case $ {  {   \exists  \, { \ottnt{v}  \ottsym{,}  \varpi_{{\mathrm{1}}}  \ottsym{,}  \varpi_{{\mathrm{2}}} } . \  \ottnt{e} \,  \mathrel{  \longrightarrow  ^*}  \, \ottnt{v} \,  \,\&\,  \, \varpi_{{\mathrm{1}}}  } \,\mathrel{\wedge}\, {  \ottnt{K}  [  \ottnt{v}  ]  \,  \mathrel{  \longrightarrow  ^*}  \,  \ottnt{E}  [   \bullet ^{ \pi' }   ]  \,  \,\&\,  \, \varpi_{{\mathrm{2}}} }  } \,\mathrel{\wedge}\, { \varpi' \,  =  \, \varpi_{{\mathrm{1}}} \,  \tceappendop  \, \varpi_{{\mathrm{2}}} } $:
       Because $\pi = \varpi' \,  \tceappendop  \, \pi' = \varpi_{{\mathrm{1}}} \,  \tceappendop  \, \varpi_{{\mathrm{2}}} \,  \tceappendop  \, \pi'$,
       it suffices to show that
       \[
         \ottnt{K}  [  \ottnt{v}  ]  \,  \Uparrow  \, \varpi_{{\mathrm{2}}} \,  \tceappendop  \, \pi' ~,
       \]
       which is implied by $ \ottnt{K}  [  \ottnt{v}  ]  \,  \mathrel{  \longrightarrow  ^*}  \,  \ottnt{E}  [   \bullet ^{ \pi' }   ]  \,  \,\&\,  \, \varpi_{{\mathrm{2}}}$.

      \case $ {   \exists  \, { \ottnt{K_{{\mathrm{0}}}}  \ottsym{,}  \mathit{x_{{\mathrm{0}}}}  \ottsym{,}  \ottnt{e_{{\mathrm{0}}}} } . \  \ottnt{e} \,  \mathrel{  \longrightarrow  ^*}  \,  \ottnt{K_{{\mathrm{0}}}}  [  \mathcal{S}_0 \, \mathit{x_{{\mathrm{0}}}}  \ottsym{.}  \ottnt{e_{{\mathrm{0}}}}  ]  \,  \,\&\,  \, \varpi'  } \,\mathrel{\wedge}\, {  \ottnt{E}  [   \bullet ^{ \pi' }   ]  \,  =  \,  \ottnt{K}  [   \ottnt{K_{{\mathrm{0}}}}  [  \mathcal{S}_0 \, \mathit{x_{{\mathrm{0}}}}  \ottsym{.}  \ottnt{e_{{\mathrm{0}}}}  ]   ]  } $:
       $ \ottnt{E}  [   \bullet ^{ \pi' }   ]  \,  =  \,  \ottnt{K}  [   \ottnt{K_{{\mathrm{0}}}}  [  \mathcal{S}_0 \, \mathit{x_{{\mathrm{0}}}}  \ottsym{.}  \ottnt{e_{{\mathrm{0}}}}  ]   ] $ is contradictory.

      \case $ {   \exists  \, { \ottnt{e''} } . \  \ottnt{e} \,  \mathrel{  \longrightarrow  ^*}  \, \ottnt{e''} \,  \,\&\,  \, \varpi'  } \,\mathrel{\wedge}\, {  \ottnt{E}  [   \bullet ^{ \pi' }   ]  \,  =  \,  \ottnt{K}  [  \ottnt{e''}  ]  } $:
       Because $ \ottnt{E}  [   \bullet ^{ \pi' }   ]  \,  =  \,  \ottnt{K}  [  \ottnt{e''}  ] $, we have
       $\ottnt{e''} \,  =  \,  \ottnt{E''}  [   \bullet ^{ \pi' }   ] $ for some $\ottnt{E''}$.
       Therefore, $\ottnt{e} \,  \mathrel{  \longrightarrow  ^*}  \,  \ottnt{E''}  [   \bullet ^{ \pi' }   ]  \,  \,\&\,  \, \varpi'$.
       Because $ \pi    =    \varpi' \,  \tceappendop  \, \pi' $, we have $\ottnt{e} \,  \Uparrow  \, \pi$.
     \end{caseanalysis}
   \end{caseanalysis}
  \end{caseanalysis}
\end{proof}

\begin{lemmap}{Divergence of Subterms}{lr-red-subterm-div}
 If $\ottnt{e} \,  \Uparrow  \, \pi$, then $ \ottnt{E}  [  \ottnt{e}  ]  \,  \Uparrow  \, \pi$ for any $\ottnt{E}$.
\end{lemmap}
\begin{proof}
 By case analysis on $\ottnt{e} \,  \Uparrow  \, \pi$.
 \begin{caseanalysis}
  \case $   \forall  \, { \varpi'  \ottsym{,}  \pi' } . \  \pi \,  =  \, \varpi' \,  \tceappendop  \, \pi'   \mathrel{\Longrightarrow}    \exists  \, { \ottnt{e'} } . \  \ottnt{e} \,  \mathrel{  \longrightarrow  ^*}  \, \ottnt{e'} \,  \,\&\,  \, \varpi'  $:
   Let $\varpi'$ be any finite trace and $\pi'$ be any infinite trace
   such that $ \pi    =    \varpi' \,  \tceappendop  \, \pi' $.  Then, it suffices to show that
   \[
      \exists  \, { \ottnt{e''} } . \   \ottnt{E}  [  \ottnt{e}  ]  \,  \mathrel{  \longrightarrow  ^*}  \, \ottnt{e''} \,  \,\&\,  \, \varpi'  ~.
   \]
   $\ottnt{e} \,  \Uparrow  \, \pi$ implies $\ottnt{e} \,  \mathrel{  \longrightarrow  ^*}  \, \ottnt{e'} \,  \,\&\,  \, \varpi'$.
   By \reflem{lr-red-subterm}, $ \ottnt{E}  [  \ottnt{e}  ]  \,  \mathrel{  \longrightarrow  ^*}  \,  \ottnt{E'}  [  \ottnt{e'}  ]  \,  \,\&\,  \, \varpi'$.
   Therefore, we have the conclusion.

  \case $ {   \exists  \, { \varpi'  \ottsym{,}  \pi'  \ottsym{,}  \ottnt{E'} } . \  \pi \,  =  \, \varpi' \,  \tceappendop  \, \pi'  } \,\mathrel{\wedge}\, { \ottnt{e} \,  \mathrel{  \longrightarrow  ^*}  \,  \ottnt{E'}  [   \bullet ^{ \pi' }   ]  \,  \,\&\,  \, \varpi' } $:
   By \reflem{lr-red-subterm}, $ \ottnt{E}  [  \ottnt{e}  ]  \,  \mathrel{  \longrightarrow  ^*}  \,  \ottnt{E}  [   \ottnt{E'}  [   \bullet ^{ \pi' }   ]   ]  \,  \,\&\,  \, \varpi'$.
   Because $ \pi    =    \varpi' \,  \tceappendop  \, \pi' $, we have the conclusion $ \ottnt{E}  [  \ottnt{e}  ]  \,  \Uparrow  \, \pi$.
 \end{caseanalysis}
\end{proof}

\begin{lemmap}{Decomposition of Embedding Expressions into Evaluation Contexts}{lr-decompose-ectx-embed-exp}
 If $\ottnt{e_{{\mathrm{1}}}}$ and $\ottnt{e_{{\mathrm{2}}}}$ are not values, let-expressions, nor reset expressions,
 then $ \ottnt{E_{{\mathrm{1}}}}  [  \ottnt{e_{{\mathrm{1}}}}  ]  \,  =  \,  \ottnt{E_{{\mathrm{2}}}}  [  \ottnt{e_{{\mathrm{2}}}}  ] $ implies $\ottnt{E_{{\mathrm{1}}}} \,  =  \, \ottnt{E_{{\mathrm{2}}}}$ and $\ottnt{e_{{\mathrm{1}}}} \,  =  \, \ottnt{e_{{\mathrm{2}}}}$.
\end{lemmap}
\begin{proof}
 \TS{OK: 2022/01/20}
 By induction on $\ottnt{E_{{\mathrm{1}}}}$.
 \begin{caseanalysis}
  \case $\ottnt{E_{{\mathrm{1}}}} \,  =  \, \,[\,]\,$:
   Because $\ottnt{e_{{\mathrm{1}}}}$ is neither of a let-expression and reset expression,
   $ \ottnt{E_{{\mathrm{1}}}}  [  \ottnt{e_{{\mathrm{1}}}}  ]  = \ottnt{e_{{\mathrm{1}}}} \,  =  \,  \ottnt{E_{{\mathrm{2}}}}  [  \ottnt{e_{{\mathrm{2}}}}  ] $ implies $\ottnt{E_{{\mathrm{2}}}} \,  =  \, \,[\,]\,$.
   Therefore, $\ottnt{E_{{\mathrm{1}}}} \,  =  \, \ottnt{E_{{\mathrm{2}}}}$ and $\ottnt{e_{{\mathrm{1}}}} \,  =  \, \ottnt{e_{{\mathrm{2}}}}$.

  \case $  \exists  \, { \mathit{x_{{\mathrm{1}}}}  \ottsym{,}  \ottnt{E'_{{\mathrm{1}}}}  \ottsym{,}  \ottnt{e_{{\mathrm{12}}}} } . \  \ottnt{E_{{\mathrm{1}}}} \,  =  \, \ottsym{(}  \mathsf{let} \, \mathit{x_{{\mathrm{1}}}}  \ottsym{=}  \ottnt{E'_{{\mathrm{1}}}} \,  \mathsf{in}  \, \ottnt{e_{{\mathrm{12}}}}  \ottsym{)} $:
   We have $ \ottnt{E_{{\mathrm{1}}}}  [  \ottnt{e_{{\mathrm{1}}}}  ]  = \ottsym{(}  \mathsf{let} \, \mathit{x_{{\mathrm{1}}}}  \ottsym{=}   \ottnt{E'_{{\mathrm{1}}}}  [  \ottnt{e_{{\mathrm{1}}}}  ]  \,  \mathsf{in}  \, \ottnt{e_{{\mathrm{12}}}}  \ottsym{)} \,  =  \,  \ottnt{E_{{\mathrm{2}}}}  [  \ottnt{e_{{\mathrm{2}}}}  ] $.
   Because $\ottnt{e_{{\mathrm{2}}}}$ is not a let-expression, $\ottnt{E_{{\mathrm{2}}}} \,  =  \, \ottsym{(}  \mathsf{let} \, \mathit{x_{{\mathrm{1}}}}  \ottsym{=}  \ottnt{E'_{{\mathrm{2}}}} \,  \mathsf{in}  \, \ottnt{e_{{\mathrm{22}}}}  \ottsym{)}$
   for some $\ottnt{E'_{{\mathrm{2}}}}$ and $\ottnt{e_{{\mathrm{22}}}}$.
   Because $\ottsym{(}  \mathsf{let} \, \mathit{x_{{\mathrm{1}}}}  \ottsym{=}   \ottnt{E'_{{\mathrm{1}}}}  [  \ottnt{e_{{\mathrm{1}}}}  ]  \,  \mathsf{in}  \, \ottnt{e_{{\mathrm{12}}}}  \ottsym{)} \,  =  \, \ottsym{(}  \mathsf{let} \, \mathit{x_{{\mathrm{1}}}}  \ottsym{=}   \ottnt{E'_{{\mathrm{2}}}}  [  \ottnt{e_{{\mathrm{2}}}}  ]  \,  \mathsf{in}  \, \ottnt{e_{{\mathrm{22}}}}  \ottsym{)}$,
   we have $ \ottnt{E'_{{\mathrm{1}}}}  [  \ottnt{e_{{\mathrm{1}}}}  ]  \,  =  \,  \ottnt{E'_{{\mathrm{2}}}}  [  \ottnt{e_{{\mathrm{2}}}}  ] $ and $\ottnt{e_{{\mathrm{12}}}} \,  =  \, \ottnt{e_{{\mathrm{22}}}}$.
   By the IH, $\ottnt{E'_{{\mathrm{1}}}} \,  =  \, \ottnt{E'_{{\mathrm{2}}}}$ and $\ottnt{e_{{\mathrm{1}}}} \,  =  \, \ottnt{e_{{\mathrm{2}}}}$.
   Therefore, $\ottnt{E_{{\mathrm{1}}}} = \ottsym{(}  \mathsf{let} \, \mathit{x_{{\mathrm{1}}}}  \ottsym{=}  \ottnt{E'_{{\mathrm{1}}}} \,  \mathsf{in}  \, \ottnt{e_{{\mathrm{12}}}}  \ottsym{)} = \ottsym{(}  \mathsf{let} \, \mathit{x_{{\mathrm{1}}}}  \ottsym{=}  \ottnt{E'_{{\mathrm{2}}}} \,  \mathsf{in}  \, \ottnt{e_{{\mathrm{22}}}}  \ottsym{)} = \ottnt{E_{{\mathrm{2}}}}$.

  \case $  \exists  \, { \ottnt{E'_{{\mathrm{1}}}} } . \  \ottnt{E_{{\mathrm{1}}}} \,  =  \,  \resetcnstr{ \ottnt{E'_{{\mathrm{1}}}} }  $:
   We have $ \ottnt{E_{{\mathrm{1}}}}  [  \ottnt{e_{{\mathrm{1}}}}  ]  =  \resetcnstr{  \ottnt{E'_{{\mathrm{1}}}}  [  \ottnt{e_{{\mathrm{1}}}}  ]  }  =  \ottnt{E_{{\mathrm{2}}}}  [  \ottnt{e_{{\mathrm{2}}}}  ] $.
   Because $\ottnt{e_{{\mathrm{2}}}}$ is not a reset expression, $\ottnt{E_{{\mathrm{2}}}} \,  =  \,  \resetcnstr{ \ottnt{E'_{{\mathrm{2}}}} } $ for some $\ottnt{E'_{{\mathrm{2}}}}$.
   Because $ \resetcnstr{  \ottnt{E'_{{\mathrm{1}}}}  [  \ottnt{e_{{\mathrm{1}}}}  ]  }  \,  =  \,  \resetcnstr{  \ottnt{E'_{{\mathrm{2}}}}  [  \ottnt{e_{{\mathrm{2}}}}  ]  } $,
   we have $ \ottnt{E'_{{\mathrm{1}}}}  [  \ottnt{e_{{\mathrm{1}}}}  ]  \,  =  \,  \ottnt{E'_{{\mathrm{2}}}}  [  \ottnt{e_{{\mathrm{2}}}}  ] $.
   By the IH, $\ottnt{E'_{{\mathrm{1}}}} \,  =  \, \ottnt{E'_{{\mathrm{2}}}}$ and $\ottnt{e_{{\mathrm{1}}}} \,  =  \, \ottnt{e_{{\mathrm{2}}}}$.
   Therefore, $\ottnt{E_{{\mathrm{1}}}} =  \resetcnstr{ \ottnt{E'_{{\mathrm{1}}}} }  =  \resetcnstr{ \ottnt{E'_{{\mathrm{2}}}} }  = \ottnt{E_{{\mathrm{2}}}}$.
 \end{caseanalysis}
\end{proof}

\begin{lemmap}{Decomposition of Embedding Expressions into Continuations}{lr-decompose-cont-embed-exp}
 Suppose that $\ottnt{e_{{\mathrm{1}}}}$ is not a let-expression.
 Then, $ \ottnt{K_{{\mathrm{1}}}}  [  \ottnt{e_{{\mathrm{1}}}}  ]  \,  =  \,  \ottnt{K_{{\mathrm{2}}}}  [  \ottnt{e_{{\mathrm{2}}}}  ] $ implies that $\ottnt{e_{{\mathrm{2}}}} \,  =  \,  \ottnt{K'_{{\mathrm{2}}}}  [  \ottnt{e_{{\mathrm{1}}}}  ] $ for some $\ottnt{K'_{{\mathrm{2}}}}$
 such that $\ottnt{K_{{\mathrm{1}}}} \,  =  \,  \ottnt{K_{{\mathrm{2}}}}  [  \ottnt{K'_{{\mathrm{2}}}}  ] $.
\end{lemmap}
\begin{proof}
 By induction on $\ottnt{K_{{\mathrm{2}}}}$.
 \begin{caseanalysis}
  \case $\ottnt{K_{{\mathrm{2}}}} \,  =  \, \,[\,]\,$: Because $ \ottnt{K_{{\mathrm{1}}}}  [  \ottnt{e_{{\mathrm{1}}}}  ]  \,  =  \, \ottnt{e_{{\mathrm{2}}}}$,
   we have the conclusion by letting $\ottnt{K'_{{\mathrm{2}}}} \,  =  \, \ottnt{K_{{\mathrm{1}}}}$.

  \case $  \exists  \, { \mathit{x'_{{\mathrm{1}}}}  \ottsym{,}  \ottnt{K_{{\mathrm{02}}}}  \ottsym{,}  \ottnt{e'_{{\mathrm{2}}}} } . \  \ottnt{K_{{\mathrm{2}}}} \,  =  \, \ottsym{(}  \mathsf{let} \, \mathit{x'_{{\mathrm{1}}}}  \ottsym{=}  \ottnt{K_{{\mathrm{02}}}} \,  \mathsf{in}  \, \ottnt{e'_{{\mathrm{2}}}}  \ottsym{)} $:
   Because $ \ottnt{K_{{\mathrm{1}}}}  [  \ottnt{e_{{\mathrm{1}}}}  ]  \,  =  \,  \ottnt{K_{{\mathrm{2}}}}  [  \ottnt{e_{{\mathrm{2}}}}  ] $,
   we have $ \ottnt{K_{{\mathrm{1}}}}  [  \ottnt{e_{{\mathrm{1}}}}  ]  \,  =  \, \ottsym{(}  \mathsf{let} \, \mathit{x'_{{\mathrm{1}}}}  \ottsym{=}   \ottnt{K_{{\mathrm{02}}}}  [  \ottnt{e_{{\mathrm{2}}}}  ]  \,  \mathsf{in}  \, \ottnt{e'_{{\mathrm{2}}}}  \ottsym{)}$.
   Because $\ottnt{e_{{\mathrm{1}}}}$ is not a let-expression, $\ottnt{K_{{\mathrm{1}}}} \,  \not=  \, \,[\,]\,$.
   Therefore, $\ottnt{K_{{\mathrm{1}}}} \,  =  \, \ottsym{(}  \mathsf{let} \, \mathit{x'_{{\mathrm{1}}}}  \ottsym{=}  \ottnt{K_{{\mathrm{01}}}} \,  \mathsf{in}  \, \ottnt{e'_{{\mathrm{2}}}}  \ottsym{)}$ for some $\ottnt{K_{{\mathrm{01}}}}$ such that
   $ \ottnt{K_{{\mathrm{01}}}}  [  \ottnt{e_{{\mathrm{1}}}}  ]  \,  =  \,  \ottnt{K_{{\mathrm{02}}}}  [  \ottnt{e_{{\mathrm{2}}}}  ] $.
   By the IH, $\ottnt{e_{{\mathrm{2}}}} \,  =  \,  \ottnt{K'_{{\mathrm{2}}}}  [  \ottnt{e_{{\mathrm{1}}}}  ] $ and $\ottnt{K_{{\mathrm{01}}}} \,  =  \,  \ottnt{K_{{\mathrm{02}}}}  [  \ottnt{K'_{{\mathrm{2}}}}  ] $ for some $\ottnt{K'_{{\mathrm{2}}}}$.
   It suffices to show that $\ottnt{K_{{\mathrm{1}}}} \,  =  \,  \ottnt{K_{{\mathrm{2}}}}  [  \ottnt{K'_{{\mathrm{2}}}}  ] $, which is implied by
   \[\begin{array}{lll}
    \ottnt{K_{{\mathrm{1}}}} &=&
    \ottsym{(}  \mathsf{let} \, \mathit{x'_{{\mathrm{1}}}}  \ottsym{=}  \ottnt{K_{{\mathrm{01}}}} \,  \mathsf{in}  \, \ottnt{e'_{{\mathrm{2}}}}  \ottsym{)} \\ &=&
    \ottsym{(}  \mathsf{let} \, \mathit{x'_{{\mathrm{1}}}}  \ottsym{=}   \ottnt{K_{{\mathrm{02}}}}  [  \ottnt{K'_{{\mathrm{2}}}}  ]  \,  \mathsf{in}  \, \ottnt{e'_{{\mathrm{2}}}}  \ottsym{)} \\ &=&
     \ottnt{K_{{\mathrm{2}}}}  [  \ottnt{K'_{{\mathrm{2}}}}  ]  ~.
     \end{array}
   \]
 \end{caseanalysis}
\end{proof}

\begin{lemmap}{Evaluation and Divergence under Evaluation Contexts}{lr-red-under-ectx}
 \begin{enumerate}
  \item \label{lem:lr-red-under-ectx:TM}
        If $ \ottnt{E}  [  \ottnt{e}  ]  \,  \Downarrow  \, \ottnt{r} \,  \,\&\,  \, \varpi$,
        then $ {  {   \exists  \, { \ottnt{r'}  \ottsym{,}  \varpi_{{\mathrm{1}}}  \ottsym{,}  \varpi_{{\mathrm{2}}} } . \  \ottnt{e} \,  \Downarrow  \, \ottnt{r'} \,  \,\&\,  \, \varpi_{{\mathrm{1}}}  } \,\mathrel{\wedge}\, {  \ottnt{E}  [  \ottnt{r'}  ]  \,  \Downarrow  \, \ottnt{r} \,  \,\&\,  \, \varpi_{{\mathrm{2}}} }  } \,\mathrel{\wedge}\, { \varpi \,  =  \, \varpi_{{\mathrm{1}}} \,  \tceappendop  \, \varpi_{{\mathrm{2}}} } $

  \item If $ \ottnt{E}  [  \ottnt{e}  ]  \,  \Uparrow  \, \pi$,
        then one of the following holds:
        \begin{itemize}
         \item $ {  {   \exists  \, { \ottnt{r}  \ottsym{,}  \varpi_{{\mathrm{1}}}  \ottsym{,}  \pi_{{\mathrm{2}}} } . \  \ottnt{e} \,  \Downarrow  \, \ottnt{r} \,  \,\&\,  \, \varpi_{{\mathrm{1}}}  } \,\mathrel{\wedge}\, {  \ottnt{E}  [  \ottnt{r}  ]  \,  \Uparrow  \, \pi_{{\mathrm{2}}} }  } \,\mathrel{\wedge}\, { \pi \,  =  \, \varpi_{{\mathrm{1}}} \,  \tceappendop  \, \pi_{{\mathrm{2}}} } $; or
         \item $\ottnt{e} \,  \Uparrow  \, \pi$.
        \end{itemize}
 \end{enumerate}
\end{lemmap}
\begin{proof}
 \noindent
 \begin{enumerate}
  \item By induction on $\ottnt{E}$.
        \begin{caseanalysis}
         \case $\ottnt{E} \,  =  \, \,[\,]\,$:
          Because $ \ottnt{E}  [  \ottnt{e}  ]  = \ottnt{e} \,  \Downarrow  \, \ottnt{r} \,  \,\&\,  \, \varpi$,
          we have the conclusion by letting $\ottnt{r'} \,  =  \, \ottnt{r}$, $ \varpi_{{\mathrm{1}}}    =    \varpi $, and $ \varpi_{{\mathrm{2}}}    =     \epsilon  $.

         \case $  \exists  \, { \mathit{x_{{\mathrm{1}}}}  \ottsym{,}  \ottnt{E'}  \ottsym{,}  \ottnt{e_{{\mathrm{2}}}} } . \  \ottnt{E} \,  =  \, \ottsym{(}  \mathsf{let} \, \mathit{x_{{\mathrm{1}}}}  \ottsym{=}  \ottnt{E'} \,  \mathsf{in}  \, \ottnt{e_{{\mathrm{2}}}}  \ottsym{)} $:
          We are given $ \ottnt{E}  [  \ottnt{e}  ]  = \ottsym{(}  \mathsf{let} \, \mathit{x_{{\mathrm{1}}}}  \ottsym{=}   \ottnt{E'}  [  \ottnt{e}  ]  \,  \mathsf{in}  \, \ottnt{e_{{\mathrm{2}}}}  \ottsym{)} \,  \Downarrow  \, \ottnt{r} \,  \,\&\,  \, \varpi$.
          By case analysis on the result of \reflem{lr-red-under-cont-term}.
          \begin{caseanalysis}
           \case $ {  {   \exists  \, { \ottnt{v}  \ottsym{,}  \varpi_{{\mathrm{1}}}  \ottsym{,}  \varpi_{{\mathrm{2}}} } . \   \ottnt{E'}  [  \ottnt{e}  ]  \,  \mathrel{  \longrightarrow  ^*}  \, \ottnt{v} \,  \,\&\,  \, \varpi_{{\mathrm{1}}}  } \,\mathrel{\wedge}\, { \ottsym{(}  \mathsf{let} \, \mathit{x_{{\mathrm{1}}}}  \ottsym{=}  \ottnt{v} \,  \mathsf{in}  \, \ottnt{e_{{\mathrm{2}}}}  \ottsym{)} \,  \mathrel{  \longrightarrow  ^*}  \, \ottnt{r} \,  \,\&\,  \, \varpi_{{\mathrm{2}}} }  } \,\mathrel{\wedge}\, { \varpi \,  =  \, \varpi_{{\mathrm{1}}} \,  \tceappendop  \, \varpi_{{\mathrm{2}}} } $:
            The IH with $ \ottnt{E'}  [  \ottnt{e}  ]  \,  \mathrel{  \longrightarrow  ^*}  \, \ottnt{v} \,  \,\&\,  \, \varpi_{{\mathrm{1}}}$ implies
            $\ottnt{e} \,  \Downarrow  \, \ottnt{r'} \,  \,\&\,  \, \varpi_{{\mathrm{11}}}$ and $ \ottnt{E'}  [  \ottnt{r'}  ]  \,  \Downarrow  \, \ottnt{v} \,  \,\&\,  \, \varpi_{{\mathrm{12}}}$ and $ \varpi_{{\mathrm{1}}}    =    \varpi_{{\mathrm{11}}} \,  \tceappendop  \, \varpi_{{\mathrm{12}}} $
            for some $\ottnt{r'}$, $\varpi_{{\mathrm{11}}}$, and $\varpi_{{\mathrm{12}}}$.
            Because $\varpi = \varpi_{{\mathrm{1}}} \,  \tceappendop  \, \varpi_{{\mathrm{2}}} = \varpi_{{\mathrm{11}}} \,  \tceappendop  \, \varpi_{{\mathrm{12}}} \,  \tceappendop  \, \varpi_{{\mathrm{2}}}$,
            it suffices to show that $ \ottnt{E}  [  \ottnt{r'}  ]  \,  \Downarrow  \, \ottnt{r} \,  \,\&\,  \, \varpi_{{\mathrm{12}}} \,  \tceappendop  \, \varpi_{{\mathrm{2}}}$.
            Because $ \ottnt{E}  [  \ottnt{r'}  ]  = \ottsym{(}  \mathsf{let} \, \mathit{x_{{\mathrm{1}}}}  \ottsym{=}   \ottnt{E'}  [  \ottnt{r'}  ]  \,  \mathsf{in}  \, \ottnt{e_{{\mathrm{2}}}}  \ottsym{)}$,
            \reflem{lr-red-subterm} implies $ \ottnt{E}  [  \ottnt{r'}  ]  \,  \mathrel{  \longrightarrow  ^*}  \, \ottsym{(}  \mathsf{let} \, \mathit{x_{{\mathrm{1}}}}  \ottsym{=}  \ottnt{v} \,  \mathsf{in}  \, \ottnt{e_{{\mathrm{2}}}}  \ottsym{)} \,  \,\&\,  \, \varpi_{{\mathrm{12}}}$.
            Because $\ottsym{(}  \mathsf{let} \, \mathit{x_{{\mathrm{1}}}}  \ottsym{=}  \ottnt{v} \,  \mathsf{in}  \, \ottnt{e_{{\mathrm{2}}}}  \ottsym{)} \,  \mathrel{  \longrightarrow  ^*}  \, \ottnt{r} \,  \,\&\,  \, \varpi_{{\mathrm{2}}}$,
            we have $ \ottnt{E}  [  \ottnt{r'}  ]  \,  \Downarrow  \, \ottnt{r} \,  \,\&\,  \, \varpi_{{\mathrm{12}}} \,  \tceappendop  \, \varpi$.

           \case $ {   \exists  \, { \ottnt{K_{{\mathrm{0}}}}  \ottsym{,}  \mathit{x_{{\mathrm{0}}}}  \ottsym{,}  \ottnt{e_{{\mathrm{0}}}} } . \   \ottnt{E'}  [  \ottnt{e}  ]  \,  \mathrel{  \longrightarrow  ^*}  \,  \ottnt{K_{{\mathrm{0}}}}  [  \mathcal{S}_0 \, \mathit{x_{{\mathrm{0}}}}  \ottsym{.}  \ottnt{e_{{\mathrm{0}}}}  ]  \,  \,\&\,  \, \varpi  } \,\mathrel{\wedge}\, { \ottnt{r} \,  =  \, \ottsym{(}  \mathsf{let} \, \mathit{x_{{\mathrm{1}}}}  \ottsym{=}   \ottnt{K_{{\mathrm{0}}}}  [  \mathcal{S}_0 \, \mathit{x_{{\mathrm{0}}}}  \ottsym{.}  \ottnt{e_{{\mathrm{0}}}}  ]  \,  \mathsf{in}  \, \ottnt{e_{{\mathrm{2}}}}  \ottsym{)} } $:
            The IH with $ \ottnt{E'}  [  \ottnt{e}  ]  \,  \mathrel{  \longrightarrow  ^*}  \,  \ottnt{K_{{\mathrm{0}}}}  [  \mathcal{S}_0 \, \mathit{x_{{\mathrm{0}}}}  \ottsym{.}  \ottnt{e_{{\mathrm{0}}}}  ]  \,  \,\&\,  \, \varpi$ implies
            $\ottnt{e} \,  \Downarrow  \, \ottnt{r'} \,  \,\&\,  \, \varpi_{{\mathrm{1}}}$ and $ \ottnt{E'}  [  \ottnt{r'}  ]  \,  \Downarrow  \,  \ottnt{K_{{\mathrm{0}}}}  [  \mathcal{S}_0 \, \mathit{x_{{\mathrm{0}}}}  \ottsym{.}  \ottnt{e_{{\mathrm{0}}}}  ]  \,  \,\&\,  \, \varpi_{{\mathrm{2}}}$ and $ \varpi    =    \varpi_{{\mathrm{1}}} \,  \tceappendop  \, \varpi_{{\mathrm{2}}} $
            for some $\ottnt{r'}$, $\varpi_{{\mathrm{1}}}$, and $\varpi_{{\mathrm{2}}}$.
            Because $\varpi = \varpi_{{\mathrm{1}}} \,  \tceappendop  \, \varpi_{{\mathrm{2}}}$,
            it suffices to show that $ \ottnt{E}  [  \ottnt{r'}  ]  \,  \Downarrow  \, \ottnt{r} \,  \,\&\,  \, \varpi_{{\mathrm{2}}}$.
            Because $ \ottnt{E}  [  \ottnt{r'}  ]  = \ottsym{(}  \mathsf{let} \, \mathit{x_{{\mathrm{1}}}}  \ottsym{=}   \ottnt{E'}  [  \ottnt{r'}  ]  \,  \mathsf{in}  \, \ottnt{e_{{\mathrm{2}}}}  \ottsym{)}$,
            \reflem{lr-red-subterm} implies $ \ottnt{E}  [  \ottnt{r'}  ]  \,  \mathrel{  \longrightarrow  ^*}  \, \ottsym{(}  \mathsf{let} \, \mathit{x_{{\mathrm{1}}}}  \ottsym{=}   \ottnt{K_{{\mathrm{0}}}}  [  \mathcal{S}_0 \, \mathit{x_{{\mathrm{0}}}}  \ottsym{.}  \ottnt{e_{{\mathrm{0}}}}  ]  \,  \mathsf{in}  \, \ottnt{e_{{\mathrm{2}}}}  \ottsym{)} \,  \,\&\,  \, \varpi_{{\mathrm{2}}}$.
            Because $\ottnt{r} \,  =  \, \ottsym{(}  \mathsf{let} \, \mathit{x_{{\mathrm{1}}}}  \ottsym{=}   \ottnt{K_{{\mathrm{0}}}}  [  \mathcal{S}_0 \, \mathit{x_{{\mathrm{0}}}}  \ottsym{.}  \ottnt{e_{{\mathrm{0}}}}  ]  \,  \mathsf{in}  \, \ottnt{e_{{\mathrm{2}}}}  \ottsym{)}$,
            we have $ \ottnt{E}  [  \ottnt{r'}  ]  \,  \Downarrow  \, \ottnt{r} \,  \,\&\,  \, \varpi_{{\mathrm{2}}}$.

           \case $ {   \exists  \, { \ottnt{e'} } . \   \ottnt{E'}  [  \ottnt{e}  ]  \,  \mathrel{  \longrightarrow  ^*}  \, \ottnt{e'} \,  \,\&\,  \, \varpi  } \,\mathrel{\wedge}\, { \ottnt{r} \,  =  \, \ottsym{(}  \mathsf{let} \, \mathit{x_{{\mathrm{1}}}}  \ottsym{=}  \ottnt{e'} \,  \mathsf{in}  \, \ottnt{e_{{\mathrm{2}}}}  \ottsym{)} } $:
            $\ottnt{r} \,  =  \, \ottsym{(}  \mathsf{let} \, \mathit{x_{{\mathrm{1}}}}  \ottsym{=}  \ottnt{e'} \,  \mathsf{in}  \, \ottnt{e_{{\mathrm{2}}}}  \ottsym{)}$ implies
            $\ottnt{e'} \,  =  \,  \ottnt{K_{{\mathrm{0}}}}  [  \mathcal{S}_0 \, \mathit{x_{{\mathrm{0}}}}  \ottsym{.}  \ottnt{e_{{\mathrm{0}}}}  ] $ for some $\ottnt{K_{{\mathrm{0}}}}$, $\mathit{x_{{\mathrm{0}}}}$, and $\ottnt{e_{{\mathrm{0}}}}$.
            The IH with $ \ottnt{E'}  [  \ottnt{e}  ]  \,  \mathrel{  \longrightarrow  ^*}  \,  \ottnt{K_{{\mathrm{0}}}}  [  \mathcal{S}_0 \, \mathit{x_{{\mathrm{0}}}}  \ottsym{.}  \ottnt{e_{{\mathrm{0}}}}  ]  \,  \,\&\,  \, \varpi$ implies
            $\ottnt{e} \,  \Downarrow  \, \ottnt{r'} \,  \,\&\,  \, \varpi_{{\mathrm{1}}}$ and $ \ottnt{E'}  [  \ottnt{r'}  ]  \,  \Downarrow  \,  \ottnt{K_{{\mathrm{0}}}}  [  \mathcal{S}_0 \, \mathit{x_{{\mathrm{0}}}}  \ottsym{.}  \ottnt{e_{{\mathrm{0}}}}  ]  \,  \,\&\,  \, \varpi_{{\mathrm{2}}}$ and $ \varpi    =    \varpi_{{\mathrm{1}}} \,  \tceappendop  \, \varpi_{{\mathrm{2}}} $
            for some $\ottnt{r'}$, $\varpi_{{\mathrm{1}}}$, and $\varpi_{{\mathrm{2}}}$.
            Then, it suffices to show that $ \ottnt{E}  [  \ottnt{r'}  ]  \,  \Downarrow  \, \ottnt{r} \,  \,\&\,  \, \varpi_{{\mathrm{2}}}$.
            Because $ \ottnt{E}  [  \ottnt{r'}  ]  = \ottsym{(}  \mathsf{let} \, \mathit{x_{{\mathrm{1}}}}  \ottsym{=}   \ottnt{E'}  [  \ottnt{r'}  ]  \,  \mathsf{in}  \, \ottnt{e_{{\mathrm{2}}}}  \ottsym{)}$,
            we have $ \ottnt{E}  [  \ottnt{r'}  ]  \,  \mathrel{  \longrightarrow  ^*}  \, \ottsym{(}  \mathsf{let} \, \mathit{x_{{\mathrm{1}}}}  \ottsym{=}   \ottnt{K_{{\mathrm{0}}}}  [  \mathcal{S}_0 \, \mathit{x_{{\mathrm{0}}}}  \ottsym{.}  \ottnt{e_{{\mathrm{0}}}}  ]  \,  \mathsf{in}  \, \ottnt{e_{{\mathrm{2}}}}  \ottsym{)} \,  \,\&\,  \, \varpi_{{\mathrm{2}}}$.
            Because $\ottnt{r} \,  =  \, \ottsym{(}  \mathsf{let} \, \mathit{x_{{\mathrm{1}}}}  \ottsym{=}   \ottnt{K_{{\mathrm{0}}}}  [  \mathcal{S}_0 \, \mathit{x_{{\mathrm{0}}}}  \ottsym{.}  \ottnt{e_{{\mathrm{0}}}}  ]  \,  \mathsf{in}  \, \ottnt{e_{{\mathrm{2}}}}  \ottsym{)}$,
            we have the conclusion.
          \end{caseanalysis}

         \case $  \exists  \, { \ottnt{E'} } . \  \ottnt{E} \,  =  \,  \resetcnstr{ \ottnt{E'} }  $:
          We are given $ \ottnt{E}  [  \ottnt{e}  ]  =  \resetcnstr{  \ottnt{E'}  [  \ottnt{e}  ]  }  \,  \Downarrow  \, \ottnt{r} \,  \,\&\,  \, \varpi$.
          By case analysis on the result of \reflem{lr-red-reset-term}.
          \begin{caseanalysis}
           \case $ {   \exists  \, { \ottnt{v} } . \   \ottnt{E'}  [  \ottnt{e}  ]  \,  \mathrel{  \longrightarrow  ^*}  \, \ottnt{v} \,  \,\&\,  \, \varpi  } \,\mathrel{\wedge}\, { \ottnt{r} \,  =  \, \ottnt{v} } $:
            The IH with $ \ottnt{E'}  [  \ottnt{e}  ]  \,  \mathrel{  \longrightarrow  ^*}  \, \ottnt{v} \,  \,\&\,  \, \varpi$ implies
            $\ottnt{e} \,  \Downarrow  \, \ottnt{r'} \,  \,\&\,  \, \varpi_{{\mathrm{1}}}$ and $ \ottnt{E'}  [  \ottnt{r'}  ]  \,  \Downarrow  \, \ottnt{v} \,  \,\&\,  \, \varpi_{{\mathrm{2}}}$ and $ \varpi    =    \varpi_{{\mathrm{1}}} \,  \tceappendop  \, \varpi_{{\mathrm{2}}} $
            for some $\ottnt{r'}$, $\varpi_{{\mathrm{1}}}$, and $\varpi_{{\mathrm{2}}}$.
            It suffices to show that $ \ottnt{E}  [  \ottnt{r'}  ]  =  \resetcnstr{  \ottnt{E'}  [  \ottnt{r'}  ]  }  \,  \Downarrow  \, \ottnt{r} \,  \,\&\,  \, \varpi_{{\mathrm{2}}}$, which is implied by
            \begin{itemize}
             \item $ \resetcnstr{  \ottnt{E'}  [  \ottnt{r'}  ]  }  \,  \mathrel{  \longrightarrow  ^*}  \,  \resetcnstr{ \ottnt{v} }  \,  \,\&\,  \, \varpi_{{\mathrm{2}}}$ by \reflem{lr-red-subterm} with $ \ottnt{E'}  [  \ottnt{r'}  ]  \,  \Downarrow  \, \ottnt{v} \,  \,\&\,  \, \varpi_{{\mathrm{2}}}$,
             \item $ \resetcnstr{ \ottnt{v} }  \,  \mathrel{  \longrightarrow  }  \, \ottnt{v} \,  \,\&\,  \,  \epsilon $ by \E{Red}/\R{Reset}, and
             \item $\ottnt{r} \,  =  \, \ottnt{v}$.
            \end{itemize}

           \case $ {  {   \exists  \, { \ottnt{K_{{\mathrm{0}}}}  \ottsym{,}  \mathit{x_{{\mathrm{0}}}}  \ottsym{,}  \ottnt{e_{{\mathrm{0}}}}  \ottsym{,}  \varpi_{{\mathrm{1}}}  \ottsym{,}  \varpi_{{\mathrm{2}}} } . \   \ottnt{E'}  [  \ottnt{e}  ]  \,  \mathrel{  \longrightarrow  ^*}  \,  \ottnt{K_{{\mathrm{0}}}}  [  \mathcal{S}_0 \, \mathit{x_{{\mathrm{0}}}}  \ottsym{.}  \ottnt{e_{{\mathrm{0}}}}  ]  \,  \,\&\,  \, \varpi_{{\mathrm{1}}}  } \,\mathrel{\wedge}\, {  \ottnt{e_{{\mathrm{0}}}}    [   \lambda\!  \, \mathit{x_{{\mathrm{0}}}}  \ottsym{.}   \resetcnstr{  \ottnt{K_{{\mathrm{0}}}}  [  \mathit{x_{{\mathrm{0}}}}  ]  }   /  \mathit{x_{{\mathrm{0}}}}  ]   \,  \mathrel{  \longrightarrow  ^*}  \, \ottnt{r} \,  \,\&\,  \, \varpi_{{\mathrm{2}}} }  } \,\mathrel{\wedge}\, { \varpi \,  =  \, \varpi_{{\mathrm{1}}} \,  \tceappendop  \, \varpi_{{\mathrm{2}}} } $:
            The IH with $ \ottnt{E'}  [  \ottnt{e}  ]  \,  \mathrel{  \longrightarrow  ^*}  \,  \ottnt{K_{{\mathrm{0}}}}  [  \mathcal{S}_0 \, \mathit{x_{{\mathrm{0}}}}  \ottsym{.}  \ottnt{e_{{\mathrm{0}}}}  ]  \,  \,\&\,  \, \varpi_{{\mathrm{1}}}$ implies
            $\ottnt{e} \,  \Downarrow  \, \ottnt{r'} \,  \,\&\,  \, \varpi_{{\mathrm{11}}}$ and $ \ottnt{E'}  [  \ottnt{r'}  ]  \,  \Downarrow  \,  \ottnt{K_{{\mathrm{0}}}}  [  \mathcal{S}_0 \, \mathit{x_{{\mathrm{0}}}}  \ottsym{.}  \ottnt{e_{{\mathrm{0}}}}  ]  \,  \,\&\,  \, \varpi_{{\mathrm{12}}}$ and $ \varpi_{{\mathrm{1}}}    =    \varpi_{{\mathrm{11}}} \,  \tceappendop  \, \varpi_{{\mathrm{12}}} $
            for some $\ottnt{r'}$, $\varpi_{{\mathrm{11}}}$, and $\varpi_{{\mathrm{12}}}$.
            Because $\varpi = \varpi_{{\mathrm{1}}} \,  \tceappendop  \, \varpi_{{\mathrm{2}}} = \varpi_{{\mathrm{11}}} \,  \tceappendop  \, \varpi_{{\mathrm{12}}} \,  \tceappendop  \, \varpi_{{\mathrm{2}}}$,
            it suffices to show that $ \ottnt{E}  [  \ottnt{r'}  ]  =  \resetcnstr{  \ottnt{E'}  [  \ottnt{r'}  ]  }  \,  \Downarrow  \, \ottnt{r} \,  \,\&\,  \, \varpi_{{\mathrm{12}}} \,  \tceappendop  \, \varpi_{{\mathrm{2}}}$, which is implied by
            \begin{itemize}
             \item $ \resetcnstr{  \ottnt{E'}  [  \ottnt{r'}  ]  }  \,  \mathrel{  \longrightarrow  ^*}  \,  \resetcnstr{  \ottnt{K_{{\mathrm{0}}}}  [  \mathcal{S}_0 \, \mathit{x_{{\mathrm{0}}}}  \ottsym{.}  \ottnt{e_{{\mathrm{0}}}}  ]  }  \,  \,\&\,  \, \varpi_{{\mathrm{12}}}$
                   by \reflem{lr-red-subterm} with $ \ottnt{E'}  [  \ottnt{r'}  ]  \,  \Downarrow  \,  \ottnt{K_{{\mathrm{0}}}}  [  \mathcal{S}_0 \, \mathit{x_{{\mathrm{0}}}}  \ottsym{.}  \ottnt{e_{{\mathrm{0}}}}  ]  \,  \,\&\,  \, \varpi_{{\mathrm{12}}}$,
             \item $ \resetcnstr{  \ottnt{K_{{\mathrm{0}}}}  [  \mathcal{S}_0 \, \mathit{x_{{\mathrm{0}}}}  \ottsym{.}  \ottnt{e_{{\mathrm{0}}}}  ]  }  \,  \mathrel{  \longrightarrow  }  \,  \ottnt{e_{{\mathrm{0}}}}    [   \lambda\!  \, \mathit{x_{{\mathrm{0}}}}  \ottsym{.}   \resetcnstr{  \ottnt{K_{{\mathrm{0}}}}  [  \mathit{x_{{\mathrm{0}}}}  ]  }   /  \mathit{x_{{\mathrm{0}}}}  ]   \,  \,\&\,  \,  \epsilon $
                   by \E{Red}/\R{Shift}, and
             \item $ \ottnt{e_{{\mathrm{0}}}}    [   \lambda\!  \, \mathit{x_{{\mathrm{0}}}}  \ottsym{.}   \resetcnstr{  \ottnt{K_{{\mathrm{0}}}}  [  \mathit{x_{{\mathrm{0}}}}  ]  }   /  \mathit{x_{{\mathrm{0}}}}  ]   \,  \mathrel{  \longrightarrow  ^*}  \, \ottnt{r} \,  \,\&\,  \, \varpi_{{\mathrm{2}}}$.
            \end{itemize}

           \case $ {   \exists  \, { \ottnt{e'} } . \   \ottnt{E'}  [  \ottnt{e}  ]  \,  \mathrel{  \longrightarrow  ^*}  \, \ottnt{e'} \,  \,\&\,  \, \varpi  } \,\mathrel{\wedge}\, { \ottnt{r} \,  =  \,  \resetcnstr{ \ottnt{e'} }  } $:
            $\ottnt{r} \,  =  \,  \resetcnstr{ \ottnt{e'} } $ is contradictory.
          \end{caseanalysis}
        \end{caseanalysis}

  \item By induction on $\ottnt{E}$.
        \begin{caseanalysis}
         \case $\ottnt{E} \,  =  \, \,[\,]\,$: We have the conclusion (the second case) because $ \ottnt{E}  [  \ottnt{e}  ]  = \ottnt{e} \,  \Uparrow  \, \pi$.

         \case $  \exists  \, { \mathit{x_{{\mathrm{1}}}}  \ottsym{,}  \ottnt{E'}  \ottsym{,}  \ottnt{e_{{\mathrm{2}}}} } . \  \ottnt{E} \,  =  \, \ottsym{(}  \mathsf{let} \, \mathit{x_{{\mathrm{1}}}}  \ottsym{=}  \ottnt{E'} \,  \mathsf{in}  \, \ottnt{e_{{\mathrm{2}}}}  \ottsym{)} $:
          We are given
          $ \ottnt{E}  [  \ottnt{e}  ]  = \ottsym{(}  \mathsf{let} \, \mathit{x_{{\mathrm{1}}}}  \ottsym{=}   \ottnt{E'}  [  \ottnt{e}  ]  \,  \mathsf{in}  \, \ottnt{e_{{\mathrm{2}}}}  \ottsym{)} \,  \Uparrow  \, \pi$.
          By case analysis on the result of \reflem{lr-red-under-cont-div}.
          \begin{caseanalysis}
           \case $ {  {   \exists  \, { \ottnt{v}  \ottsym{,}  \varpi_{{\mathrm{1}}}  \ottsym{,}  \pi_{{\mathrm{2}}} } . \   \ottnt{E'}  [  \ottnt{e}  ]  \,  \Downarrow  \, \ottnt{v} \,  \,\&\,  \, \varpi_{{\mathrm{1}}}  } \,\mathrel{\wedge}\, { \ottsym{(}  \mathsf{let} \, \mathit{x_{{\mathrm{1}}}}  \ottsym{=}  \ottnt{v} \,  \mathsf{in}  \, \ottnt{e_{{\mathrm{2}}}}  \ottsym{)} \,  \Uparrow  \, \pi_{{\mathrm{2}}} }  } \,\mathrel{\wedge}\, { \pi \,  =  \, \varpi_{{\mathrm{1}}} \,  \tceappendop  \, \pi_{{\mathrm{2}}} } $:
            The case (\ref{lem:lr-red-under-ectx:TM}) with $ \ottnt{E'}  [  \ottnt{e}  ]  \,  \Downarrow  \, \ottnt{v} \,  \,\&\,  \, \varpi_{{\mathrm{1}}}$ implies
            $\ottnt{e} \,  \Downarrow  \, \ottnt{r} \,  \,\&\,  \, \varpi_{{\mathrm{11}}}$ and $ \ottnt{E'}  [  \ottnt{r}  ]  \,  \Downarrow  \, \ottnt{v} \,  \,\&\,  \, \varpi_{{\mathrm{12}}}$ and $ \varpi_{{\mathrm{1}}}    =    \varpi_{{\mathrm{11}}} \,  \tceappendop  \, \varpi_{{\mathrm{12}}} $
            for some $\ottnt{r}$, $\varpi_{{\mathrm{11}}}$, and $\varpi_{{\mathrm{12}}}$.
            Because $\pi = \varpi_{{\mathrm{1}}} \,  \tceappendop  \, \pi_{{\mathrm{2}}} = \varpi_{{\mathrm{11}}} \,  \tceappendop  \, \varpi_{{\mathrm{12}}} \,  \tceappendop  \, \pi_{{\mathrm{2}}}$,
            it suffices to show that $ \ottnt{E}  [  \ottnt{r}  ]  \,  \Uparrow  \, \varpi_{{\mathrm{12}}} \,  \tceappendop  \, \pi_{{\mathrm{2}}}$, which is implied by
            \reflem{lr-comp-eval-div} with
            \begin{itemize}
             \item $ \ottnt{E}  [  \ottnt{r}  ]  = \ottsym{(}  \mathsf{let} \, \mathit{x_{{\mathrm{1}}}}  \ottsym{=}   \ottnt{E'}  [  \ottnt{r}  ]  \,  \mathsf{in}  \, \ottnt{e_{{\mathrm{2}}}}  \ottsym{)} \,  \mathrel{  \longrightarrow  ^*}  \, \ottsym{(}  \mathsf{let} \, \mathit{x_{{\mathrm{1}}}}  \ottsym{=}  \ottnt{v} \,  \mathsf{in}  \, \ottnt{e_{{\mathrm{2}}}}  \ottsym{)} \,  \,\&\,  \, \varpi_{{\mathrm{12}}}$
                   by \reflem{lr-red-subterm} with $ \ottnt{E'}  [  \ottnt{r}  ]  \,  \Downarrow  \, \ottnt{v} \,  \,\&\,  \, \varpi_{{\mathrm{12}}}$, and
             \item $\ottsym{(}  \mathsf{let} \, \mathit{x_{{\mathrm{1}}}}  \ottsym{=}  \ottnt{v} \,  \mathsf{in}  \, \ottnt{e_{{\mathrm{2}}}}  \ottsym{)} \,  \Uparrow  \, \pi_{{\mathrm{2}}}$.
            \end{itemize}

           \case $ \ottnt{E'}  [  \ottnt{e}  ]  \,  \Uparrow  \, \pi$:  By case analysis on the IH.
            \begin{caseanalysis}
             \case $ {  {   \exists  \, { \ottnt{r}  \ottsym{,}  \varpi_{{\mathrm{1}}}  \ottsym{,}  \pi_{{\mathrm{2}}} } . \  \ottnt{e} \,  \Downarrow  \, \ottnt{r} \,  \,\&\,  \, \varpi_{{\mathrm{1}}}  } \,\mathrel{\wedge}\, {  \ottnt{E'}  [  \ottnt{r}  ]  \,  \Uparrow  \, \pi_{{\mathrm{2}}} }  } \,\mathrel{\wedge}\, { \pi \,  =  \, \varpi_{{\mathrm{1}}} \,  \tceappendop  \, \pi_{{\mathrm{2}}} } $:
              It suffices to show that $ \ottnt{E}  [  \ottnt{r}  ]  = \ottsym{(}  \mathsf{let} \, \mathit{x_{{\mathrm{1}}}}  \ottsym{=}   \ottnt{E'}  [  \ottnt{r}  ]  \,  \mathsf{in}  \, \ottnt{e_{{\mathrm{2}}}}  \ottsym{)} \,  \Uparrow  \, \pi_{{\mathrm{2}}}$,
              which is implied by \reflem{lr-red-subterm-div}.

             \case $\ottnt{e} \,  \Uparrow  \, \pi$: We have the conclusion (the second case).
            \end{caseanalysis}
          \end{caseanalysis}

         \case $  \exists  \, { \ottnt{E'} } . \  \ottnt{E} \,  =  \,  \resetcnstr{ \ottnt{E'} }  $:
          We are given
          $ \ottnt{E}  [  \ottnt{e}  ]  =  \resetcnstr{  \ottnt{E'}  [  \ottnt{e}  ]  }  \,  \Uparrow  \, \pi$.
          By case analysis on the result of \reflem{lr-red-reset-div}.
          \begin{caseanalysis}
           \case $ \ottnt{E'}  [  \ottnt{e}  ]  \,  \Uparrow  \, \pi$:
            By case analysis on the IH.
            \begin{caseanalysis}
             \case $ {  {   \exists  \, { \ottnt{r}  \ottsym{,}  \varpi_{{\mathrm{1}}}  \ottsym{,}  \pi_{{\mathrm{2}}} } . \  \ottnt{e} \,  \Downarrow  \, \ottnt{r} \,  \,\&\,  \, \varpi_{{\mathrm{1}}}  } \,\mathrel{\wedge}\, {  \ottnt{E'}  [  \ottnt{r}  ]  \,  \Uparrow  \, \pi_{{\mathrm{2}}} }  } \,\mathrel{\wedge}\, { \pi \,  =  \, \varpi_{{\mathrm{1}}} \,  \tceappendop  \, \pi_{{\mathrm{2}}} } $:
              It suffices to show that $ \ottnt{E}  [  \ottnt{r}  ]  =  \resetcnstr{  \ottnt{E'}  [  \ottnt{r}  ]  }  \,  \Uparrow  \, \pi_{{\mathrm{2}}}$,
              which is implied by \reflem{lr-red-subterm-div}.
             \case $\ottnt{e} \,  \Uparrow  \, \pi$: We have the conclusion (the second case).
            \end{caseanalysis}

           \case $ {  {   \exists  \, { \ottnt{K_{{\mathrm{0}}}}  \ottsym{,}  \mathit{x_{{\mathrm{0}}}}  \ottsym{,}  \ottnt{e_{{\mathrm{0}}}}  \ottsym{,}  \varpi_{{\mathrm{1}}}  \ottsym{,}  \pi_{{\mathrm{2}}} } . \   \ottnt{E'}  [  \ottnt{e}  ]  \,  \Downarrow  \,  \ottnt{K_{{\mathrm{0}}}}  [  \mathcal{S}_0 \, \mathit{x_{{\mathrm{0}}}}  \ottsym{.}  \ottnt{e_{{\mathrm{0}}}}  ]  \,  \,\&\,  \, \varpi_{{\mathrm{1}}}  } \,\mathrel{\wedge}\, {  \ottnt{e_{{\mathrm{0}}}}    [   \lambda\!  \, \mathit{x_{{\mathrm{0}}}}  \ottsym{.}   \resetcnstr{  \ottnt{K_{{\mathrm{0}}}}  [  \mathit{x_{{\mathrm{0}}}}  ]  }   /  \mathit{x_{{\mathrm{0}}}}  ]   \,  \Uparrow  \, \pi_{{\mathrm{2}}} }  } \,\mathrel{\wedge}\, { \pi \,  =  \, \varpi_{{\mathrm{1}}} \,  \tceappendop  \, \pi_{{\mathrm{2}}} } $:
            The case (\ref{lem:lr-red-under-ectx:TM}) with $ \ottnt{E'}  [  \ottnt{e}  ]  \,  \Downarrow  \,  \ottnt{K_{{\mathrm{0}}}}  [  \mathcal{S}_0 \, \mathit{x_{{\mathrm{0}}}}  \ottsym{.}  \ottnt{e_{{\mathrm{0}}}}  ]  \,  \,\&\,  \, \varpi_{{\mathrm{1}}}$ implies
            $\ottnt{e} \,  \Downarrow  \, \ottnt{r} \,  \,\&\,  \, \varpi_{{\mathrm{11}}}$ and $ \ottnt{E'}  [  \ottnt{r}  ]  \,  \Downarrow  \,  \ottnt{K_{{\mathrm{0}}}}  [  \mathcal{S}_0 \, \mathit{x_{{\mathrm{0}}}}  \ottsym{.}  \ottnt{e_{{\mathrm{0}}}}  ]  \,  \,\&\,  \, \varpi_{{\mathrm{12}}}$ and $ \varpi_{{\mathrm{1}}}    =    \varpi_{{\mathrm{11}}} \,  \tceappendop  \, \varpi_{{\mathrm{12}}} $
            for some $\ottnt{r}$, $\varpi_{{\mathrm{11}}}$, and $\varpi_{{\mathrm{12}}}$.
            Because $\pi = \varpi_{{\mathrm{1}}} \,  \tceappendop  \, \pi_{{\mathrm{2}}} = \varpi_{{\mathrm{11}}} \,  \tceappendop  \, \varpi_{{\mathrm{12}}} \,  \tceappendop  \, \pi_{{\mathrm{2}}}$,
            it suffices to show that $ \ottnt{E}  [  \ottnt{r}  ]  \,  \Uparrow  \, \varpi_{{\mathrm{12}}} \,  \tceappendop  \, \pi_{{\mathrm{2}}}$, which is implied by
            \reflem{lr-comp-eval-div} with
            \begin{itemize}
             \item $ \ottnt{E}  [  \ottnt{r}  ]  =  \resetcnstr{  \ottnt{E'}  [  \ottnt{r}  ]  }  \,  \mathrel{  \longrightarrow  ^*}  \,  \resetcnstr{  \ottnt{K_{{\mathrm{0}}}}  [  \mathcal{S}_0 \, \mathit{x_{{\mathrm{0}}}}  \ottsym{.}  \ottnt{e_{{\mathrm{0}}}}  ]  }  \,  \,\&\,  \, \varpi_{{\mathrm{12}}}$
                   by \reflem{lr-red-subterm} with $ \ottnt{E'}  [  \ottnt{r}  ]  \,  \Downarrow  \,  \ottnt{K_{{\mathrm{0}}}}  [  \mathcal{S}_0 \, \mathit{x_{{\mathrm{0}}}}  \ottsym{.}  \ottnt{e_{{\mathrm{0}}}}  ]  \,  \,\&\,  \, \varpi_{{\mathrm{12}}}$,
             \item $ \resetcnstr{  \ottnt{K_{{\mathrm{0}}}}  [  \mathcal{S}_0 \, \mathit{x_{{\mathrm{0}}}}  \ottsym{.}  \ottnt{e_{{\mathrm{0}}}}  ]  }  \,  \mathrel{  \longrightarrow  }  \,  \ottnt{e_{{\mathrm{0}}}}    [   \lambda\!  \, \mathit{x_{{\mathrm{0}}}}  \ottsym{.}   \resetcnstr{  \ottnt{K_{{\mathrm{0}}}}  [  \mathit{x_{{\mathrm{0}}}}  ]  }   /  \mathit{x_{{\mathrm{0}}}}  ]   \,  \,\&\,  \,  \epsilon $
                   by \E{Red}/\R{Shift}, and
             \item $ \ottnt{e_{{\mathrm{0}}}}    [   \lambda\!  \, \mathit{x_{{\mathrm{0}}}}  \ottsym{.}   \resetcnstr{  \ottnt{K_{{\mathrm{0}}}}  [  \mathit{x_{{\mathrm{0}}}}  ]  }   /  \mathit{x_{{\mathrm{0}}}}  ]   \,  \Uparrow  \, \pi_{{\mathrm{2}}}$.
            \end{itemize}
          \end{caseanalysis}
        \end{caseanalysis}
 \end{enumerate}
\end{proof}

\begin{lemmap}{Decomposition of Evaluation}{lr-red-decomp-eval}
 Let $\varpi_{{\mathrm{1}}}$ and $\varpi_{{\mathrm{2}}}$ be finite traces such that
 the length of $\varpi_{{\mathrm{2}}}$ is larger than that of $\varpi_{{\mathrm{1}}}$.
 If $\ottnt{e} \,  \mathrel{  \longrightarrow  }^{ \ottmv{n} }  \, \ottnt{e_{{\mathrm{1}}}} \,  \,\&\,  \, \varpi_{{\mathrm{1}}}$ and $\ottnt{e} \,  \mathrel{  \longrightarrow  ^*}  \, \ottnt{e_{{\mathrm{2}}}} \,  \,\&\,  \, \varpi_{{\mathrm{2}}}$,
 then $\ottnt{e_{{\mathrm{1}}}} \,  \mathrel{  \longrightarrow  ^*}  \, \ottnt{e_{{\mathrm{2}}}} \,  \,\&\,  \, \varpi'$ for some $\varpi'$ such that $ \varpi_{{\mathrm{2}}}    =    \varpi_{{\mathrm{1}}} \,  \tceappendop  \, \varpi' $.
\end{lemmap}
\begin{proof}
 By induction on $\ottmv{n}$.
 \begin{caseanalysis}
  \case $\ottmv{n} \,  =  \, \ottsym{0}$: We are given $\ottnt{e} \,  =  \, \ottnt{e_{{\mathrm{1}}}}$ and $ \varpi_{{\mathrm{1}}}    =     \epsilon  $.
   Then, we have the conclusion by letting $ \varpi'    =    \varpi_{{\mathrm{2}}} $ because
   \begin{itemize}
    \item $\ottnt{e_{{\mathrm{1}}}} = \ottnt{e} \,  \mathrel{  \longrightarrow  ^*}  \, \ottnt{e_{{\mathrm{2}}}} \,  \,\&\,  \, \varpi_{{\mathrm{2}}} = \ottnt{e_{{\mathrm{2}}}} \,  \,\&\,  \, \varpi'$ by the assumption, and
    \item $\varpi_{{\mathrm{2}}} =  \epsilon  \,  \tceappendop  \, \varpi_{{\mathrm{2}}} = \varpi_{{\mathrm{1}}} \,  \tceappendop  \, \varpi'$.
   \end{itemize}

  \case $\ottmv{n} \,  >  \, \ottsym{0}$: We are given $\ottnt{e} \,  \mathrel{  \longrightarrow  }  \, \ottnt{e'_{{\mathrm{1}}}} \,  \,\&\,  \, \varpi_{{\mathrm{11}}}$ and $\ottnt{e'_{{\mathrm{1}}}} \,  \mathrel{  \longrightarrow  }^{ \ottmv{n}  \ottsym{-}  \ottsym{1} }  \, \ottnt{e_{{\mathrm{1}}}} \,  \,\&\,  \, \varpi_{{\mathrm{12}}}$
   for some $\ottnt{e'_{{\mathrm{1}}}}$, $\varpi_{{\mathrm{11}}}$, and $\varpi_{{\mathrm{12}}}$ such that
   $ \varpi_{{\mathrm{1}}}    =    \varpi_{{\mathrm{11}}} \,  \tceappendop  \, \varpi_{{\mathrm{12}}} $.
   Because the length of $\varpi_{{\mathrm{2}}}$ is larger than that of $\varpi_{{\mathrm{1}}}$,
   $ \varpi_{{\mathrm{2}}}    \not=     \epsilon  $.
   Therefore, $\ottnt{e} \,  \mathrel{  \longrightarrow  ^*}  \, \ottnt{e_{{\mathrm{2}}}} \,  \,\&\,  \, \varpi_{{\mathrm{2}}}$ implies that
   there exist some $\ottnt{e'_{{\mathrm{2}}}}$, $\varpi_{{\mathrm{21}}}$, and $\varpi_{{\mathrm{22}}}$ such that
   $\ottnt{e} \,  \mathrel{  \longrightarrow  }  \, \ottnt{e'_{{\mathrm{2}}}} \,  \,\&\,  \, \varpi_{{\mathrm{21}}}$ and $\ottnt{e'_{{\mathrm{2}}}} \,  \mathrel{  \longrightarrow  ^*}  \, \ottnt{e_{{\mathrm{2}}}} \,  \,\&\,  \, \varpi_{{\mathrm{22}}}$ and $ \varpi_{{\mathrm{2}}}    =    \varpi_{{\mathrm{21}}} \,  \tceappendop  \, \varpi_{{\mathrm{22}}} $.
   \reflem{lr-eval-det} with $\ottnt{e} \,  \mathrel{  \longrightarrow  }  \, \ottnt{e'_{{\mathrm{1}}}} \,  \,\&\,  \, \varpi_{{\mathrm{11}}}$ and $\ottnt{e} \,  \mathrel{  \longrightarrow  }  \, \ottnt{e'_{{\mathrm{2}}}} \,  \,\&\,  \, \varpi_{{\mathrm{21}}}$
   implies $\ottnt{e'_{{\mathrm{1}}}} \,  =  \, \ottnt{e'_{{\mathrm{2}}}}$  and $ \varpi_{{\mathrm{11}}}    =    \varpi_{{\mathrm{21}}} $.
   Because the length of $\varpi_{{\mathrm{2}}} = \varpi_{{\mathrm{21}}} \,  \tceappendop  \, \varpi_{{\mathrm{22}}} = \varpi_{{\mathrm{11}}} \,  \tceappendop  \, \varpi_{{\mathrm{22}}}$ is
   larger than that of $ \varpi_{{\mathrm{1}}}    =    \varpi_{{\mathrm{11}}} \,  \tceappendop  \, \varpi_{{\mathrm{12}}} $, the length of $\varpi_{{\mathrm{22}}}$ is
   larger than that of $\varpi_{{\mathrm{12}}}$.
   Then, by the IH with $\ottnt{e'_{{\mathrm{1}}}} \,  \mathrel{  \longrightarrow  }^{ \ottmv{n}  \ottsym{-}  \ottsym{1} }  \, \ottnt{e_{{\mathrm{1}}}} \,  \,\&\,  \, \varpi_{{\mathrm{12}}}$ and $\ottnt{e'_{{\mathrm{1}}}} = \ottnt{e'_{{\mathrm{2}}}} \,  \mathrel{  \longrightarrow  ^*}  \, \ottnt{e_{{\mathrm{2}}}} \,  \,\&\,  \, \varpi_{{\mathrm{22}}}$,
   we have $\ottnt{e_{{\mathrm{1}}}} \,  \mathrel{  \longrightarrow  ^*}  \, \ottnt{e_{{\mathrm{2}}}} \,  \,\&\,  \, \varpi'$ for some $\varpi'$ such that
   $ \varpi_{{\mathrm{22}}}    =    \varpi_{{\mathrm{12}}} \,  \tceappendop  \, \varpi' $.
   Because
   \[
    \varpi_{{\mathrm{2}}} = \varpi_{{\mathrm{21}}} \,  \tceappendop  \, \varpi_{{\mathrm{22}}} = \varpi_{{\mathrm{11}}} \,  \tceappendop  \, \varpi_{{\mathrm{22}}} = \varpi_{{\mathrm{11}}} \,  \tceappendop  \, \varpi_{{\mathrm{12}}} \,  \tceappendop  \, \varpi' = \varpi_{{\mathrm{1}}} \,  \tceappendop  \, \varpi' ~,
   \]
   we have the conclusion.
 \end{caseanalysis}
\end{proof}

\begin{lemmap}{Decomposition of Divergence}{lr-red-decomp-div}
 \begin{enumerate}
  \item \label{lem:lr-red-decomp-div:TM}
        If $\ottnt{e} \,  \Uparrow  \, \pi$ and $\ottnt{e} \,  \mathrel{  \longrightarrow  }  \, \ottnt{e'} \,  \,\&\,  \, \varpi'$,
        then $\ottnt{e'} \,  \Uparrow  \, \pi'$ for some $\pi'$ such that $ \pi    =    \varpi' \,  \tceappendop  \, \pi' $.

  \item If $\ottnt{e} \,  \Uparrow  \, \pi$ and $\ottnt{e} \,  \mathrel{  \longrightarrow  }^{ \ottmv{n} }  \, \ottnt{e'} \,  \,\&\,  \, \varpi'$,
        then $\ottnt{e'} \,  \Uparrow  \, \pi'$ for some $\pi'$ such that $ \pi    =    \varpi' \,  \tceappendop  \, \pi' $.
 \end{enumerate}
\end{lemmap}
\begin{proof}
 \noindent
 \begin{enumerate}
  \item  By case analysis on $\ottnt{e} \,  \Uparrow  \, \pi$.
         \begin{caseanalysis}
          \case $   \forall  \, { \varpi_{{\mathrm{0}}}  \ottsym{,}  \pi_{{\mathrm{0}}} } . \  \pi \,  =  \, \varpi_{{\mathrm{0}}} \,  \tceappendop  \, \pi_{{\mathrm{0}}}   \mathrel{\Longrightarrow}    \exists  \, { \ottnt{e_{{\mathrm{0}}}} } . \  \ottnt{e} \,  \mathrel{  \longrightarrow  ^*}  \, \ottnt{e_{{\mathrm{0}}}} \,  \,\&\,  \, \varpi_{{\mathrm{0}}}  $:
           First, we show that $ \pi    =    \varpi' \,  \tceappendop  \, \pi' $ for some $\pi'$.
           Let $\varpi_{{\mathrm{0}}}$ be a finite trace and $\pi_{{\mathrm{0}}}$ be an infinite trace
           such that $ \pi    =    \varpi_{{\mathrm{0}}} \,  \tceappendop  \, \pi_{{\mathrm{0}}} $ and the length of $\varpi_{{\mathrm{0}}}$ is
           larger than that of $\varpi'$; such $\varpi_{{\mathrm{0}}}$ and $\pi_{{\mathrm{0}}}$ exist
           because $\pi$ is an infinite trace.
           Then, $\ottnt{e} \,  \Uparrow  \, \pi$ implies that there exists some $\ottnt{e_{{\mathrm{0}}}}$ such
           that $\ottnt{e} \,  \mathrel{  \longrightarrow  ^*}  \, \ottnt{e_{{\mathrm{0}}}} \,  \,\&\,  \, \varpi_{{\mathrm{0}}}$.
           \reflem{lr-red-decomp-eval} with $\ottnt{e} \,  \mathrel{  \longrightarrow  }  \, \ottnt{e'} \,  \,\&\,  \, \varpi'$ and $\ottnt{e} \,  \mathrel{  \longrightarrow  ^*}  \, \ottnt{e_{{\mathrm{0}}}} \,  \,\&\,  \, \varpi_{{\mathrm{0}}}$
           implies $\ottnt{e'} \,  \mathrel{  \longrightarrow  ^*}  \, \ottnt{e_{{\mathrm{0}}}} \,  \,\&\,  \, \varpi''$ for some $\varpi''$ such that
           $ \varpi_{{\mathrm{0}}}    =    \varpi' \,  \tceappendop  \, \varpi'' $.
           Because $\pi= \varpi_{{\mathrm{0}}} \,  \tceappendop  \, \pi_{{\mathrm{0}}} = \varpi' \,  \tceappendop  \, \varpi'' \,  \tceappendop  \, \pi_{{\mathrm{0}}}$,
           we have $ \pi    =    \varpi' \,  \tceappendop  \, \pi' $ where $ \pi'    =    \varpi'' \,  \tceappendop  \, \pi_{{\mathrm{0}}} $.

           Next, we show that $\ottnt{e'} \,  \Uparrow  \, \pi'$.
           Let $\varpi'_{{\mathrm{0}}}$ be any finite trace and $\pi'_{{\mathrm{0}}}$ be any infinite trace such that
           $ \pi'    =    \varpi'_{{\mathrm{0}}} \,  \tceappendop  \, \pi'_{{\mathrm{0}}} $.
           Then, it suffices to show that
           \[
             \exists  \, { \ottnt{e'_{{\mathrm{0}}}} } . \  \ottnt{e'} \,  \mathrel{  \longrightarrow  ^*}  \, \ottnt{e'_{{\mathrm{0}}}} \,  \,\&\,  \, \varpi'_{{\mathrm{0}}}  ~.
           \]

           If $ \varpi'_{{\mathrm{0}}}    =     \epsilon  $, then we have the conclusion by letting $\ottnt{e'_{{\mathrm{0}}}} \,  =  \, \ottnt{e'}$.

           Otherwise, we suppose that $ \varpi'_{{\mathrm{0}}}    \not=     \epsilon  $.
           Because $\pi = \varpi' \,  \tceappendop  \, \pi' = \varpi' \,  \tceappendop  \, \varpi'_{{\mathrm{0}}} \,  \tceappendop  \, \pi'_{{\mathrm{0}}}$,
           $\ottnt{e} \,  \Uparrow  \, \pi$ implies that there exists some $\ottnt{e'_{{\mathrm{0}}}}$ such that
           $\ottnt{e} \,  \mathrel{  \longrightarrow  ^*}  \, \ottnt{e'_{{\mathrm{0}}}} \,  \,\&\,  \, \varpi' \,  \tceappendop  \, \varpi'_{{\mathrm{0}}}$.
           Because $ \varpi'_{{\mathrm{0}}}    \not=     \epsilon  $, the length of $\varpi' \,  \tceappendop  \, \varpi'_{{\mathrm{0}}}$ is larger than
           that of $\varpi'$.
           Then, \reflem{lr-red-decomp-eval} with
           $\ottnt{e} \,  \mathrel{  \longrightarrow  }  \, \ottnt{e'} \,  \,\&\,  \, \varpi'$ and $\ottnt{e} \,  \mathrel{  \longrightarrow  ^*}  \, \ottnt{e'_{{\mathrm{0}}}} \,  \,\&\,  \, \varpi' \,  \tceappendop  \, \varpi'_{{\mathrm{0}}}$
           implies the conclusion $\ottnt{e'} \,  \mathrel{  \longrightarrow  ^*}  \, \ottnt{e'_{{\mathrm{0}}}} \,  \,\&\,  \, \varpi'_{{\mathrm{0}}}$.

          \case $ {   \exists  \, { \varpi_{{\mathrm{0}}}  \ottsym{,}  \pi_{{\mathrm{0}}}  \ottsym{,}  \ottnt{E} } . \  \pi \,  =  \, \varpi_{{\mathrm{0}}} \,  \tceappendop  \, \pi_{{\mathrm{0}}}  } \,\mathrel{\wedge}\, { \ottnt{e} \,  \mathrel{  \longrightarrow  ^*}  \,  \ottnt{E}  [   \bullet ^{ \pi_{{\mathrm{0}}} }   ]  \,  \,\&\,  \, \varpi_{{\mathrm{0}}} } $:
           Because $\ottnt{e} \,  \mathrel{  \longrightarrow  }  \, \ottnt{e'} \,  \,\&\,  \, \varpi'$, and $ \ottnt{E}  [   \bullet ^{ \pi_{{\mathrm{0}}} }   ] $ cannot be evaluated
           by \reflem{lr-red-div}, we have $\ottnt{e} \,  \not=  \,  \ottnt{E}  [   \bullet ^{ \pi_{{\mathrm{0}}} }   ] $.
           Therefore, $\ottnt{e} \,  \mathrel{  \longrightarrow  ^*}  \,  \ottnt{E}  [   \bullet ^{ \pi_{{\mathrm{0}}} }   ]  \,  \,\&\,  \, \varpi_{{\mathrm{0}}}$ implies
           $\ottnt{e} \,  \mathrel{  \longrightarrow  }  \, \ottnt{e''} \,  \,\&\,  \, \varpi_{{\mathrm{01}}}$ and $\ottnt{e''} \,  \mathrel{  \longrightarrow  ^*}  \,  \ottnt{E}  [   \bullet ^{ \pi_{{\mathrm{0}}} }   ]  \,  \,\&\,  \, \varpi_{{\mathrm{02}}}$
           for some $\ottnt{e''}$, $\varpi_{{\mathrm{01}}}$, and $\varpi_{{\mathrm{02}}}$ such that $ \varpi_{{\mathrm{0}}}    =    \varpi_{{\mathrm{01}}} \,  \tceappendop  \, \varpi_{{\mathrm{02}}} $.
           Then, \reflem{lr-eval-det} with $\ottnt{e} \,  \mathrel{  \longrightarrow  }  \, \ottnt{e'} \,  \,\&\,  \, \varpi'$ implies
           $\ottnt{e'} \,  =  \, \ottnt{e''}$ and $ \varpi'    =    \varpi_{{\mathrm{01}}} $.
           Therefore, we have $\ottnt{e'} = \ottnt{e''} \,  \mathrel{  \longrightarrow  ^*}  \,  \ottnt{E}  [   \bullet ^{ \pi_{{\mathrm{0}}} }   ]  \,  \,\&\,  \, \varpi_{{\mathrm{02}}}$.
           Then, we have $\ottnt{e'} \,  \Uparrow  \, \pi'$ where $ \pi'    =    \varpi_{{\mathrm{02}}} \,  \tceappendop  \, \pi_{{\mathrm{0}}} $.
           Because $\pi = \varpi_{{\mathrm{0}}} \,  \tceappendop  \, \pi_{{\mathrm{0}}} = \varpi_{{\mathrm{01}}} \,  \tceappendop  \, \varpi_{{\mathrm{02}}} \,  \tceappendop  \, \pi_{{\mathrm{0}}} = \varpi' \,  \tceappendop  \, \pi'$,
           we have the conclusion.

         \end{caseanalysis}

  \item By induction on $\ottmv{n}$.
        \begin{caseanalysis}
         \case $\ottmv{n} \,  =  \, \ottsym{0}$: Because $\ottnt{e} \,  =  \, \ottnt{e'}$ and $ \varpi'    =     \epsilon  $, we have the conclusion by letting $ \pi'    =    \pi $.
         \case $\ottmv{n} \,  >  \, \ottsym{0}$:
          We are given $\ottnt{e} \,  \mathrel{  \longrightarrow  }  \, \ottnt{e_{{\mathrm{0}}}} \,  \,\&\,  \, \varpi_{{\mathrm{0}}}$ and $\ottnt{e_{{\mathrm{0}}}} \,  \mathrel{  \longrightarrow  }^{ \ottmv{n}  \ottsym{-}  \ottsym{1} }  \, \ottnt{e'} \,  \,\&\,  \, \varpi'_{{\mathrm{0}}}$
          for some $\ottnt{e_{{\mathrm{0}}}}$, $\varpi_{{\mathrm{0}}}$, and $\varpi'_{{\mathrm{0}}}$ such that $ \varpi'    =    \varpi_{{\mathrm{0}}} \,  \tceappendop  \, \varpi'_{{\mathrm{0}}} $.
          By the case (\ref{lem:lr-red-decomp-div:TM}) with $\ottnt{e} \,  \Uparrow  \, \pi$ and $\ottnt{e} \,  \mathrel{  \longrightarrow  }  \, \ottnt{e_{{\mathrm{0}}}} \,  \,\&\,  \, \varpi_{{\mathrm{0}}}$,
          $\ottnt{e_{{\mathrm{0}}}} \,  \Uparrow  \, \pi_{{\mathrm{0}}}$ for some $\pi_{{\mathrm{0}}}$ such that $ \pi    =    \varpi_{{\mathrm{0}}} \,  \tceappendop  \, \pi_{{\mathrm{0}}} $.
          Further, the IH with $\ottnt{e_{{\mathrm{0}}}} \,  \Uparrow  \, \pi_{{\mathrm{0}}}$ and $\ottnt{e_{{\mathrm{0}}}} \,  \mathrel{  \longrightarrow  }^{ \ottmv{n}  \ottsym{-}  \ottsym{1} }  \, \ottnt{e'} \,  \,\&\,  \, \varpi'_{{\mathrm{0}}}$
          implies $\ottnt{e'} \,  \Uparrow  \, \pi'$ for some $\pi'$ such that $ \pi_{{\mathrm{0}}}    =    \varpi'_{{\mathrm{0}}} \,  \tceappendop  \, \pi' $.
          Because $\pi = \varpi_{{\mathrm{0}}} \,  \tceappendop  \, \pi_{{\mathrm{0}}} = \varpi_{{\mathrm{0}}} \,  \tceappendop  \, \varpi'_{{\mathrm{0}}} \,  \tceappendop  \, \pi' = \varpi' \,  \tceappendop  \, \pi'$, we have the conclusion.
        \end{caseanalysis}
 \end{enumerate}
\end{proof}

\begin{lemmap}{Reduction of Logical Relation}{lr-red}
 If $\ottnt{e} \,  \in  \,  \LRRel{ \mathcal{E} }{ \ottnt{C} } $ and $\ottnt{e} \,  \rightsquigarrow  \, \ottnt{e'}$,
 then $\ottnt{e'} \,  \in  \,  \LRRel{ \mathcal{E} }{ \ottnt{C} } $.
\end{lemmap}
\begin{proof}
 We have the conclusion by \reflem{lr-exp-rel-obs-res}
 with the following.
 \begin{itemize}
  \item $\ottnt{e'} \,  \in  \,  \LRRel{ \mathrm{Atom} }{ \ottnt{C} } $ by \reflem{ts-subject-red-red}.

  \item $   \forall  \, { \ottnt{E}  \ottsym{,}  \ottnt{r}  \ottsym{,}  \varpi } . \   \ottnt{E}  [  \ottnt{e'}  ]  \,  \Downarrow  \, \ottnt{r} \,  \,\&\,  \, \varpi   \mathrel{\Longrightarrow}   \ottnt{E}  [  \ottnt{e}  ]  \,  \Downarrow  \, \ottnt{r} \,  \,\&\,  \, \varpi $
        by \reflem{lr-red-subterm} with $\ottnt{e} \,  \mathrel{  \longrightarrow  }  \, \ottnt{e'} \,  \,\&\,  \,  \epsilon $
        that is derived by \E{Red} with $\ottnt{e} \,  \rightsquigarrow  \, \ottnt{e'}$.

  \item $   \forall  \, { \ottnt{E}  \ottsym{,}  \pi } . \   \ottnt{E}  [  \ottnt{e'}  ]  \,  \Uparrow  \, \pi   \mathrel{\Longrightarrow}   \ottnt{E}  [  \ottnt{e}  ]  \,  \Uparrow  \, \pi $
        by \reflem{lr-comp-eval-div} with
        \begin{itemize}
         \item $ \ottnt{E}  [  \ottnt{e}  ]  \,  \mathrel{  \longrightarrow  }  \,  \ottnt{E}  [  \ottnt{e'}  ]  \,  \,\&\,  \,  \epsilon $, which is implied by
               \reflem{lr-red-subterm} with $\ottnt{e} \,  \mathrel{  \longrightarrow  }  \, \ottnt{e'} \,  \,\&\,  \,  \epsilon $ that is
               derived by \E{Red} with $\ottnt{e} \,  \rightsquigarrow  \, \ottnt{e'}$, and
         \item $ \ottnt{E}  [  \ottnt{e'}  ]  \,  \Uparrow  \, \pi$ by the assumption.
        \end{itemize}
 \end{itemize}
\end{proof}

\begin{lemmap}{Introduction of Intersection}{ts-valid-intro-intersect}
 If $ { \models  \,  \phi_{{\mathrm{1}}} } $ and $ { \models  \,  \phi_{{\mathrm{2}}} } $, then $ { \models  \,   \phi_{{\mathrm{1}}}    \wedge    \phi_{{\mathrm{2}}}  } $.
\end{lemmap}
\begin{proof}
 \TS{Validity}
 By definition.
\end{proof}

\begin{lemmap}{Introduction of Top}{ts-valid-top}
 $ { \models  \,   \top  } $ is true.
\end{lemmap}
\begin{proof}
 \TS{Validity}
 By definition.
\end{proof}

\begin{lemmap}{Effect Composition is Consistent with Trace Composition}{ts-eff-compose-consist-trace-compose}
 \begin{enumerate}
 \item If $ { \models  \,   { \Phi_{{\mathrm{1}}} }^\mu   \ottsym{(}  \varpi_{{\mathrm{1}}}  \ottsym{)} } $ and $ { \models  \,   { \Phi_{{\mathrm{2}}} }^\mu   \ottsym{(}  \varpi_{{\mathrm{2}}}  \ottsym{)} } $, then $ { \models  \,   { \ottsym{(}  \Phi_{{\mathrm{1}}} \,  \tceappendop  \, \Phi_{{\mathrm{2}}}  \ottsym{)} }^\mu   \ottsym{(}  \varpi_{{\mathrm{1}}} \,  \tceappendop  \, \varpi_{{\mathrm{2}}}  \ottsym{)} } $.
 \item If $ { \models  \,   { \Phi_{{\mathrm{1}}} }^\mu   \ottsym{(}  \varpi_{{\mathrm{1}}}  \ottsym{)} } $ and $ { \models  \,   { \Phi_{{\mathrm{2}}} }^\nu   \ottsym{(}  \pi_{{\mathrm{2}}}  \ottsym{)} } $, then $ { \models  \,   { \ottsym{(}  \Phi_{{\mathrm{1}}} \,  \tceappendop  \, \Phi_{{\mathrm{2}}}  \ottsym{)} }^\nu   \ottsym{(}  \varpi_{{\mathrm{1}}} \,  \tceappendop  \, \pi_{{\mathrm{2}}}  \ottsym{)} } $.
 \item If $ { \models  \,   { \Phi_{{\mathrm{1}}} }^\nu   \ottsym{(}  \pi  \ottsym{)} } $, then $ { \models  \,   { \ottsym{(}  \Phi_{{\mathrm{1}}} \,  \tceappendop  \, \Phi_{{\mathrm{2}}}  \ottsym{)} }^\nu   \ottsym{(}  \pi  \ottsym{)} } $ for any $\Phi_{{\mathrm{2}}}$.
 \end{enumerate}
\end{lemmap}
\begin{proof}
 \noindent
 \begin{enumerate}
  \item By definition, it suffices to show that
        \[
          { \models  \,    \exists  \,  \mathit{x_{{\mathrm{1}}}}  \mathrel{  \in  }   \Sigma ^\ast   . \    \exists  \,  \mathit{x_{{\mathrm{2}}}}  \mathrel{  \in  }   \Sigma ^\ast  \,  \tceappendop  \, \varpi_{{\mathrm{1}}}  . \     \varpi_{{\mathrm{2}}}    =    \mathit{x_{{\mathrm{1}}}} \,  \tceappendop  \, \mathit{x_{{\mathrm{2}}}}     \wedge     { \Phi_{{\mathrm{1}}} }^\mu   \ottsym{(}  \mathit{x_{{\mathrm{1}}}}  \ottsym{)}     \wedge     { \Phi_{{\mathrm{2}}} }^\mu   \ottsym{(}  \mathit{x_{{\mathrm{2}}}}  \ottsym{)}    }  ~,
        \]
        which is true from the assumptions by taking $\varpi_{{\mathrm{1}}}$ and $\varpi_{{\mathrm{2}}}$
        as $\mathit{x_{{\mathrm{1}}}}$ and $\mathit{x_{{\mathrm{2}}}}$, respectively.
        \TS{OK informal}

  \item By definition, it suffices to show that
        \[
          { \models  \,    { \Phi_{{\mathrm{1}}} }^\nu   \ottsym{(}  \varpi_{{\mathrm{1}}} \,  \tceappendop  \, \pi_{{\mathrm{2}}}  \ottsym{)}    \vee    \ottsym{(}    \exists  \,  \mathit{x'}  \mathrel{  \in  }   \Sigma ^\ast   . \    \exists  \,  \mathit{y'}  \mathrel{  \in  }   \Sigma ^\omega   . \     \varpi_{{\mathrm{1}}} \,  \tceappendop  \, \pi_{{\mathrm{2}}}    =    \mathit{x'} \,  \tceappendop  \, \mathit{y'}     \wedge     { \Phi_{{\mathrm{1}}} }^\mu   \ottsym{(}  \mathit{x'}  \ottsym{)}     \wedge     { \Phi_{{\mathrm{2}}} }^\nu   \ottsym{(}  \mathit{y'}  \ottsym{)}     \ottsym{)}  }  ~,
        \]
        which is true from the assumptions by taking $\varpi_{{\mathrm{1}}}$ and $\pi_{{\mathrm{2}}}$
        as $\mathit{x'}$ and $\mathit{y'}$, respectively.
        \TS{OK informal}

  \item By definition, it suffices to show that
        \[
          { \models  \,    { \Phi_{{\mathrm{1}}} }^\nu   \ottsym{(}  \pi  \ottsym{)}    \vee    \ottsym{(}    \exists  \,  \mathit{x'}  \mathrel{  \in  }   \Sigma ^\ast   . \    \exists  \,  \mathit{y'}  \mathrel{  \in  }   \Sigma ^\omega   . \     \pi    =    \mathit{x'} \,  \tceappendop  \, \mathit{y'}     \wedge     { \Phi_{{\mathrm{1}}} }^\mu   \ottsym{(}  \mathit{x'}  \ottsym{)}     \wedge     { \Phi_{{\mathrm{2}}} }^\nu   \ottsym{(}  \mathit{y'}  \ottsym{)}     \ottsym{)}  }  ~,
        \]
        which is true from the assumption.
        \TS{OK informal}
 \end{enumerate}
\end{proof}

\begin{lemmap}{Composition of Effects and Computation Types is Consistent with Trace Composition}{lr-eff-cty-compose-consist-trace-compose}
 If $ { \models  \,   { \Phi }^\mu   \ottsym{(}  \varpi  \ottsym{)} } $ and $ { \models  \,   { \ottnt{C} }^\nu (  \pi  )  } $,
 then $ { \models  \,   { \ottsym{(}  \Phi \,  \tceappendop  \, \ottnt{C}  \ottsym{)} }^\nu (  \varpi \,  \tceappendop  \, \pi  )  } $.
\end{lemmap}
\begin{proof}
 By induction on $\ottnt{C}$.
 Because $\Phi \,  \tceappendop  \, \ottnt{C} \,  =  \,   \ottnt{C}  . \ottnt{T}   \,  \,\&\,  \,  \Phi \,  \tceappendop  \, \ottsym{(}   \ottnt{C}  . \Phi   \ottsym{)}  \, / \,  \Phi \,  \tceappendop  \, \ottsym{(}   \ottnt{C}  . \ottnt{S}   \ottsym{)} $,
 we must show that
 \[
   { \models  \,    { \ottsym{(}  \Phi \,  \tceappendop  \, \ottsym{(}   \ottnt{C}  . \Phi   \ottsym{)}  \ottsym{)} }^\nu   \ottsym{(}  \varpi \,  \tceappendop  \, \pi  \ottsym{)}    \wedge     { \ottsym{(}  \Phi \,  \tceappendop  \, \ottsym{(}   \ottnt{C}  . \ottnt{S}   \ottsym{)}  \ottsym{)} }^\nu (  \varpi \,  \tceappendop  \, \pi  )   }  ~.
 \]
 We prove it by \reflem{ts-valid-intro-intersect} withe the following.
 \begin{itemize}
  \item We show that $ { \models  \,   { \ottsym{(}  \Phi \,  \tceappendop  \, \ottsym{(}   \ottnt{C}  . \Phi   \ottsym{)}  \ottsym{)} }^\nu   \ottsym{(}  \varpi \,  \tceappendop  \, \pi  \ottsym{)} } $.
        Because $ { \ottnt{C} }^\nu (  \pi  )  \,  =  \,   { \ottsym{(}   \ottnt{C}  . \Phi   \ottsym{)} }^\nu   \ottsym{(}  \pi  \ottsym{)}    \wedge     { \ottsym{(}   \ottnt{C}  . \ottnt{S}   \ottsym{)} }^\nu (  \pi  )  $,
        we have $ { \models  \,    { \ottsym{(}   \ottnt{C}  . \Phi   \ottsym{)} }^\nu   \ottsym{(}  \pi  \ottsym{)}    \wedge     { \ottsym{(}   \ottnt{C}  . \ottnt{S}   \ottsym{)} }^\nu (  \pi  )   } $
        by $ { \models  \,   { \ottnt{C} }^\nu (  \pi  )  } $.
        By \reflem{ts-valid-elim-intersect}, $ { \models  \,   { \ottsym{(}   \ottnt{C}  . \Phi   \ottsym{)} }^\nu   \ottsym{(}  \pi  \ottsym{)} } $.
        Then, the conclusion is implied
        by \reflem{ts-eff-compose-consist-trace-compose}
        with $ { \models  \,   { \Phi }^\mu   \ottsym{(}  \varpi  \ottsym{)} } $ and $ { \models  \,   { \ottsym{(}   \ottnt{C}  . \Phi   \ottsym{)} }^\nu   \ottsym{(}  \pi  \ottsym{)} } $.

  \item We show that $ { \models  \,   { \ottsym{(}  \Phi \,  \tceappendop  \, \ottsym{(}   \ottnt{C}  . \ottnt{S}   \ottsym{)}  \ottsym{)} }^\nu (  \varpi \,  \tceappendop  \, \pi  )  } $
        by case analysis on $ \ottnt{C}  . \ottnt{S} $.
        \begin{caseanalysis}
         \case $ \ottnt{C}  . \ottnt{S}  \,  =  \,  \square $:
          Because $\Phi \,  \tceappendop  \,  \square  \,  =  \,  \square $ and
          $ {  \square  }^\nu (  \varpi \,  \tceappendop  \, \pi  )  \,  =  \,  \top $,
          we must show that $ { \models  \,   \top  } $, which is implied by
          \reflem{ts-valid-top}.

         \case $  \exists  \, { \mathit{x}  \ottsym{,}  \ottnt{C_{{\mathrm{1}}}}  \ottsym{,}  \ottnt{C_{{\mathrm{2}}}} } . \   \ottnt{C}  . \ottnt{S}  \,  =  \,  ( \forall  \mathit{x}  .  \ottnt{C_{{\mathrm{1}}}}  )  \Rightarrow   \ottnt{C_{{\mathrm{2}}}}  $:
          Because
          \begin{itemize}
           \item $\Phi \,  \tceappendop  \, \ottsym{(}   ( \forall  \mathit{x}  .  \ottnt{C_{{\mathrm{1}}}}  )  \Rightarrow   \ottnt{C_{{\mathrm{2}}}}   \ottsym{)} \,  =  \,  ( \forall  \mathit{x}  .  \ottnt{C_{{\mathrm{1}}}}  )  \Rightarrow   \Phi \,  \tceappendop  \, \ottnt{C_{{\mathrm{2}}}} $ and
           \item $ { \ottsym{(}   ( \forall  \mathit{x}  .  \ottnt{C_{{\mathrm{1}}}}  )  \Rightarrow   \Phi \,  \tceappendop  \, \ottnt{C_{{\mathrm{2}}}}   \ottsym{)} }^\nu (  \varpi \,  \tceappendop  \, \pi  )  \,  =  \,  { \ottsym{(}  \Phi \,  \tceappendop  \, \ottnt{C_{{\mathrm{2}}}}  \ottsym{)} }^\nu (  \varpi \,  \tceappendop  \, \pi  ) $,
          \end{itemize}
          we must show that $ { \models  \,   { \ottsym{(}  \Phi \,  \tceappendop  \, \ottnt{C_{{\mathrm{2}}}}  \ottsym{)} }^\nu (  \varpi \,  \tceappendop  \, \pi  )  } $.
          By the IH, it suffices to show that $ { \models  \,   { \ottnt{C_{{\mathrm{2}}}} }^\nu (  \pi  )  } $.
          Because $ { \models  \,   { \ottnt{C} }^\nu (  \pi  )  } $, we have
          $ { \models  \,    { \ottsym{(}   \ottnt{C}  . \Phi   \ottsym{)} }^\nu   \ottsym{(}  \pi  \ottsym{)}    \wedge     { \ottnt{C_{{\mathrm{2}}}} }^\nu (  \pi  )   } $.
          Then, the conclusion $ { \models  \,   { \ottnt{C_{{\mathrm{2}}}} }^\nu (  \pi  )  } $ is implied
          by \reflem{ts-valid-elim-intersect}.
        \end{caseanalysis}
 \end{itemize}
\end{proof}

\begin{lemmap}{Inconsistency of Evaluation and Divergence}{lr-red-inconsist-eval-div}
 If $\ottnt{e} \,  \Downarrow  \, \ottnt{r} \,  \,\&\,  \, \varpi$,
 then $\ottnt{e} \,  \Uparrow  \, \pi$ does not hold for any $\pi$.
\end{lemmap}
\begin{proof}
 Suppose that $\ottnt{e} \,  \Uparrow  \, \pi$.
 We show a contradiction.
 By case analysis on $\ottnt{e} \,  \Uparrow  \, \pi$.
 \begin{caseanalysis}
  \case $   \forall  \, { \varpi'  \ottsym{,}  \pi' } . \  \pi \,  =  \, \varpi' \,  \tceappendop  \, \pi'   \mathrel{\Longrightarrow}    \exists  \, { \ottnt{e'} } . \  \ottnt{e} \,  \mathrel{  \longrightarrow  ^*}  \, \ottnt{e'} \,  \,\&\,  \, \varpi'  $:
   Let $\varpi'$ be a finite trace and $\pi'$ be an infinite trace such that
   $ \pi    =    \varpi' \,  \tceappendop  \, \pi' $ and the length of $\varpi'$ is larger than that of $\varpi$.
   Then, $\ottnt{e} \,  \Uparrow  \, \pi$ implies $\ottnt{e} \,  \mathrel{  \longrightarrow  ^*}  \, \ottnt{e'} \,  \,\&\,  \, \varpi'$ for some $\ottnt{e'}$.
   By \reflem{lr-red-decomp-eval} with $\ottnt{e} \,  \Downarrow  \, \ottnt{r} \,  \,\&\,  \, \varpi$,
   $\ottnt{r} \,  \mathrel{  \longrightarrow  ^*}  \, \ottnt{e'} \,  \,\&\,  \, \varpi''$ for some $\varpi''$ such that $ \varpi'    =    \varpi \,  \tceappendop  \, \varpi'' $.
   Because $\varpi'$ is larger than $\varpi$, $ \varpi''    \not=     \epsilon  $.
   Therefore, $\ottnt{r} \,  \mathrel{  \longrightarrow  }  \, \ottnt{e''} \,  \,\&\,  \, \varpi''_{{\mathrm{1}}}$ for some $\ottnt{e''}$ and $\varpi''_{{\mathrm{1}}}$.
   However, it is contradictory with Lemmas~\ref{lem:lr-red-val} and \ref{lem:lr-red-shift}.

  \case $ {   \exists  \, { \varpi'  \ottsym{,}  \pi'  \ottsym{,}  \ottnt{E} } . \  \pi \,  =  \, \varpi' \,  \tceappendop  \, \pi'  } \,\mathrel{\wedge}\, { \ottnt{e} \,  \mathrel{  \longrightarrow  ^*}  \,  \ottnt{E}  [   \bullet ^{ \pi' }   ]  \,  \,\&\,  \, \varpi' } $:
   By Lemmas~\ref{lem:lr-eval-det}, \ref{lem:lr-red-val}, \ref{lem:lr-red-shift}, and \ref{lem:lr-red-div}, $\ottnt{r} \,  =  \,  \ottnt{E}  [   \bullet ^{ \pi' }   ] $, but it is contradictory:
   if $\ottnt{r}$ is a value, then a contradiction is clear;
   otherwise, if $\ottnt{r} \,  =  \,  \ottnt{K_{{\mathrm{0}}}}  [  \mathcal{S}_0 \, \mathit{x_{{\mathrm{0}}}}  \ottsym{.}  \ottnt{e_{{\mathrm{0}}}}  ] $ for some $\ottnt{K_{{\mathrm{0}}}}$, $\mathit{x_{{\mathrm{0}}}}$, and $\ottnt{r_{{\mathrm{0}}}}$,
   then \reflem{lr-decompose-ectx-embed-exp} implies $\mathcal{S}_0 \, \mathit{x_{{\mathrm{0}}}}  \ottsym{.}  \ottnt{e_{{\mathrm{0}}}} \,  =  \,  \bullet ^{ \pi' } $,
   which is a contradiction.
 \end{caseanalysis}
\end{proof}

\begin{lemmap}{Embedding Terminating Expressions into Evaluation Contexts}{lr-embed-term-exp-into-ectx}
 If
 $\ottnt{e} \,  \Downarrow  \, \ottnt{v} \,  \,\&\,  \, \varpi$ and
 $ { \models  \,   { \Phi }^\mu   \ottsym{(}  \varpi  \ottsym{)} } $ and
 $ \ottnt{E}  [  \ottnt{v}  ]  \,  \in  \,  \LRRel{ \mathcal{E} }{  \ottnt{C_{{\mathrm{1}}}}    [  \ottnt{v}  /  \mathit{x}  ]   } $ and
 $\ottnt{E} \,  \not=  \, \,[\,]\,$ and
 $ \emptyset   \vdash  \Phi \,  \tceappendop  \, \ottsym{(}   \ottnt{C_{{\mathrm{1}}}}    [  \ottnt{v}  /  \mathit{x}  ]    \ottsym{)}  \ottsym{<:}  \ottnt{C_{{\mathrm{2}}}}$ and
 $ \ottnt{E}  [  \ottnt{e}  ]  \,  \in  \,  \LRRel{ \mathrm{Atom} }{ \ottnt{C_{{\mathrm{2}}}} } $,
 then $ \ottnt{E}  [  \ottnt{e}  ]  \,  \in  \,  \LRRel{ \mathcal{E} }{ \ottnt{C_{{\mathrm{2}}}} } $.
\end{lemmap}
\begin{proof}
 By induction on $\ottnt{C_{{\mathrm{1}}}}$.
 Consider each on the result of evaluating $ \ottnt{E}  [  \ottnt{e}  ] $.
 \begin{caseanalysis}
  \case $  \exists  \, { \ottnt{v'}  \ottsym{,}  \varpi' } . \   \ottnt{E}  [  \ottnt{e}  ]  \,  \Downarrow  \, \ottnt{v'} \,  \,\&\,  \, \varpi' $:
   We must show that $\ottnt{v'} \,  \in  \,  \LRRel{ \mathcal{V} }{  \ottnt{C_{{\mathrm{2}}}}  . \ottnt{T}  } $.
   By Lemmas~\ref{lem:lr-red-under-ectx}, \ref{lem:lr-eval-det}, and \ref{lem:lr-red-val},
   $ \ottnt{E}  [  \ottnt{v}  ]  \,  \Downarrow  \, \ottnt{v'} \,  \,\&\,  \, \varpi''$ and
   $ \varpi'    =    \varpi \,  \tceappendop  \, \varpi'' $ for some $\varpi''$.
   %
   %
   Because $ \ottnt{E}  [  \ottnt{v}  ]  \,  \in  \,  \LRRel{ \mathcal{E} }{  \ottnt{C_{{\mathrm{1}}}}    [  \ottnt{v}  /  \mathit{x}  ]   } $,
   we have $\ottnt{v'} \,  \in  \,  \LRRel{ \mathcal{V} }{  \ottsym{(}   \ottnt{C_{{\mathrm{1}}}}  . \ottnt{T}   \ottsym{)}    [  \ottnt{v}  /  \mathit{x}  ]   } $.
   Now, we have the conclusion $\ottnt{v'} \,  \in  \,  \LRRel{ \mathcal{V} }{  \ottnt{C_{{\mathrm{2}}}}  . \ottnt{T}  } $
   by \reflem{lr-subtyping} with the following.
   \begin{itemize}
    \item $ \emptyset   \vdash   \ottsym{(}   \ottnt{C_{{\mathrm{1}}}}  . \ottnt{T}   \ottsym{)}    [  \ottnt{v}  /  \mathit{x}  ]  $ and $ \emptyset   \vdash   \ottnt{C_{{\mathrm{2}}}}  . \ottnt{T} $
          by \reflem{ts-wf-ty-tctx-typing},
    \item $ \emptyset   \vdash   \ottsym{(}   \ottnt{C_{{\mathrm{1}}}}  . \ottnt{T}   \ottsym{)}    [  \ottnt{v}  /  \mathit{x}  ]    \ottsym{<:}   \ottnt{C_{{\mathrm{2}}}}  . \ottnt{T} $,
          which is implied by $ \emptyset   \vdash  \Phi \,  \tceappendop  \, \ottsym{(}   \ottnt{C_{{\mathrm{1}}}}    [  \ottnt{v}  /  \mathit{x}  ]    \ottsym{)}  \ottsym{<:}  \ottnt{C_{{\mathrm{2}}}}$, and
    \item $\ottnt{v'} \,  \in  \,  \LRRel{ \mathcal{V} }{  \ottsym{(}   \ottnt{C_{{\mathrm{1}}}}  . \ottnt{T}   \ottsym{)}    [  \ottnt{v}  /  \mathit{x}  ]   } $.
   \end{itemize}

  \case $  \exists  \, { \ottnt{K_{{\mathrm{0}}}}  \ottsym{,}  \mathit{x_{{\mathrm{0}}}}  \ottsym{,}  \ottnt{e_{{\mathrm{0}}}}  \ottsym{,}  \varpi' } . \   \ottnt{E}  [  \ottnt{e}  ]  \,  \Downarrow  \,  \ottnt{K_{{\mathrm{0}}}}  [  \mathcal{S}_0 \, \mathit{x_{{\mathrm{0}}}}  \ottsym{.}  \ottnt{e_{{\mathrm{0}}}}  ]  \,  \,\&\,  \, \varpi' $:
   We must show that
   \begin{itemize}
    \item $ \ottnt{C_{{\mathrm{2}}}}  . \ottnt{S}  \,  =  \,  ( \forall  \mathit{y}  .  \ottnt{C_{{\mathrm{21}}}}  )  \Rightarrow   \ottnt{C_{{\mathrm{22}}}} $ and
    \item $  \forall  \, { \ottnt{K'} \,  \in  \,  \LRRel{ \mathcal{K} }{   \ottnt{C_{{\mathrm{2}}}}  . \ottnt{T}   \, / \, ( \forall  \mathit{y}  .  \ottnt{C_{{\mathrm{21}}}}  )  }  } . \   \resetcnstr{  \ottnt{K'}  [   \ottnt{E}  [  \ottnt{e}  ]   ]  }  \,  \in  \,  \LRRel{ \mathcal{E} }{ \ottnt{C_{{\mathrm{22}}}} }  $
   \end{itemize}
   for some $\mathit{y}$, $\ottnt{C_{{\mathrm{21}}}}$, and $\ottnt{C_{{\mathrm{22}}}}$.
   By Lemmas~\ref{lem:lr-red-under-ectx}, \ref{lem:lr-eval-det}, and \ref{lem:lr-red-val},
   $ \ottnt{E}  [  \ottnt{v}  ]  \,  \Downarrow  \,  \ottnt{K_{{\mathrm{0}}}}  [  \mathcal{S}_0 \, \mathit{x_{{\mathrm{0}}}}  \ottsym{.}  \ottnt{e_{{\mathrm{0}}}}  ]  \,  \,\&\,  \, \varpi''$ and
   $ \varpi'    =    \varpi \,  \tceappendop  \, \varpi'' $ for some $\varpi''$.
   Because $ \ottnt{E}  [  \ottnt{v}  ]  \,  \in  \,  \LRRel{ \mathcal{E} }{  \ottnt{C_{{\mathrm{1}}}}    [  \ottnt{v}  /  \mathit{x}  ]   } $,
   we have
   \begin{itemize}
    \item $ \ottnt{C_{{\mathrm{1}}}}  . \ottnt{S}  \,  =  \,  ( \forall  \mathit{y}  .  \ottnt{C_{{\mathrm{11}}}}  )  \Rightarrow   \ottnt{C_{{\mathrm{12}}}} $ and
    \item $  \forall  \, { \ottnt{K'} \,  \in  \,  \LRRel{ \mathcal{K} }{   \ottsym{(}   \ottnt{C_{{\mathrm{1}}}}  . \ottnt{T}   \ottsym{)}    [  \ottnt{v}  /  \mathit{x}  ]    \, / \, ( \forall  \mathit{y}  .   \ottnt{C_{{\mathrm{11}}}}    [  \ottnt{v}  /  \mathit{x}  ]    )  }  } . \   \resetcnstr{  \ottnt{K'}  [   \ottnt{E}  [  \ottnt{v}  ]   ]  }  \,  \in  \,  \LRRel{ \mathcal{E} }{  \ottnt{C_{{\mathrm{12}}}}    [  \ottnt{v}  /  \mathit{x}  ]   }  $
   \end{itemize}
   for some $\mathit{y} \,  \not\in  \,  \mathit{fv}  (  \ottnt{v}  )  \,  \mathbin{\cup}  \,  \{  \mathit{x}  \} $, $\ottnt{C_{{\mathrm{11}}}}$, and $\ottnt{C_{{\mathrm{12}}}}$.
   Because $ \emptyset   \vdash  \Phi \,  \tceappendop  \, \ottsym{(}   \ottnt{C_{{\mathrm{1}}}}    [  \ottnt{v}  /  \mathit{x}  ]    \ottsym{)}  \ottsym{<:}  \ottnt{C_{{\mathrm{2}}}}$, we have
   \begin{itemize}
    \item $ \emptyset   \vdash   \ottsym{(}   \ottnt{C_{{\mathrm{1}}}}  . \ottnt{T}   \ottsym{)}    [  \ottnt{v}  /  \mathit{x}  ]    \ottsym{<:}   \ottnt{C_{{\mathrm{2}}}}  . \ottnt{T} $,
    \item $ \ottnt{C_{{\mathrm{2}}}}  . \ottnt{S}  \,  =  \,  ( \forall  \mathit{y}  .  \ottnt{C_{{\mathrm{21}}}}  )  \Rightarrow   \ottnt{C_{{\mathrm{22}}}} $,
    \item $ \mathit{y} \,  {:}  \, \ottsym{(}   \ottnt{C_{{\mathrm{1}}}}  . \ottnt{T}   \ottsym{)}    [  \ottnt{v}  /  \mathit{x}  ]    \vdash  \ottnt{C_{{\mathrm{21}}}}  \ottsym{<:}   \ottnt{C_{{\mathrm{11}}}}    [  \ottnt{v}  /  \mathit{x}  ]  $, and
    \item $ \emptyset   \vdash  \Phi \,  \tceappendop  \, \ottsym{(}   \ottnt{C_{{\mathrm{12}}}}    [  \ottnt{v}  /  \mathit{x}  ]    \ottsym{)}  \ottsym{<:}  \ottnt{C_{{\mathrm{22}}}}$.
   \end{itemize}

   Now, it suffices to show that
   \[
      \forall  \, { \ottnt{K'} \,  \in  \,  \LRRel{ \mathcal{K} }{   \ottnt{C_{{\mathrm{2}}}}  . \ottnt{T}   \, / \, ( \forall  \mathit{y}  .  \ottnt{C_{{\mathrm{21}}}}  )  }  } . \   \resetcnstr{  \ottnt{K'}  [   \ottnt{E}  [  \ottnt{e}  ]   ]  }  \,  \in  \,  \LRRel{ \mathcal{E} }{ \ottnt{C_{{\mathrm{22}}}} }   ~.
   \]
   Let $\ottnt{K'} \,  \in  \,  \LRRel{ \mathcal{K} }{   \ottnt{C_{{\mathrm{2}}}}  . \ottnt{T}   \, / \, ( \forall  \mathit{y}  .  \ottnt{C_{{\mathrm{21}}}}  )  } $.
   Because
   \reflem{ts-wf-ty-tctx-typing} with $ \ottnt{E}  [  \ottnt{v}  ]  \,  \in  \,  \LRRel{ \mathcal{E} }{  \ottnt{C_{{\mathrm{1}}}}    [  \ottnt{v}  /  \mathit{x}  ]   } $
   implies $ \emptyset   \vdash   \ottnt{C_{{\mathrm{1}}}}    [  \ottnt{v}  /  \mathit{x}  ]  $, and
   \reflem{ts-wf-ty-tctx-typing} with $ \ottnt{E}  [  \ottnt{e}  ]  \,  \in  \,  \LRRel{ \mathrm{Atom} }{ \ottnt{C_{{\mathrm{2}}}} } $
   implies $ \emptyset   \vdash  \ottnt{C_{{\mathrm{2}}}}$,
   we have
   \begin{itemize}
    \item $ \emptyset   \vdash   \ottsym{(}   \ottnt{C_{{\mathrm{1}}}}  . \ottnt{T}   \ottsym{)}    [  \ottnt{v}  /  \mathit{x}  ]  $,
    \item $ \mathit{y} \,  {:}  \, \ottsym{(}   \ottnt{C_{{\mathrm{1}}}}  . \ottnt{T}   \ottsym{)}    [  \ottnt{v}  /  \mathit{x}  ]    \vdash   \ottnt{C_{{\mathrm{11}}}}    [  \ottnt{v}  /  \mathit{x}  ]  $,
    \item $ \emptyset   \vdash   \ottnt{C_{{\mathrm{12}}}}    [  \ottnt{v}  /  \mathit{x}  ]  $,
    \item $ \emptyset   \vdash   \ottnt{C_{{\mathrm{2}}}}  . \ottnt{T} $,
    \item $\mathit{y} \,  {:}  \,  \ottnt{C_{{\mathrm{2}}}}  . \ottnt{T}   \vdash  \ottnt{C_{{\mathrm{21}}}}$, and
    \item $ \emptyset   \vdash  \ottnt{C_{{\mathrm{22}}}}$.
   \end{itemize}
   Then, because
   $ \emptyset   \vdash  \ottsym{(}  \mathit{y} \,  {:}  \,  \ottnt{C_{{\mathrm{2}}}}  . \ottnt{T}   \ottsym{)}  \rightarrow  \ottnt{C_{{\mathrm{21}}}}  \ottsym{<:}   \ottsym{(}  \mathit{y} \,  {:}  \,  \ottsym{(}   \ottnt{C_{{\mathrm{1}}}}  . \ottnt{T}   \ottsym{)}    [  \ottnt{v}  /  \mathit{x}  ]    \ottsym{)}  \rightarrow  \ottnt{C_{{\mathrm{11}}}}    [  \ottnt{v}  /  \mathit{x}  ]  $ by \Srule{Fun}, and
   $ \emptyset   \vdash  \ottsym{(}  \mathit{y} \,  {:}  \,  \ottnt{C_{{\mathrm{2}}}}  . \ottnt{T}   \ottsym{)}  \rightarrow  \ottnt{C_{{\mathrm{21}}}}$ and $ \emptyset   \vdash   \ottsym{(}  \mathit{y} \,  {:}  \,  \ottsym{(}   \ottnt{C_{{\mathrm{1}}}}  . \ottnt{T}   \ottsym{)}    [  \ottnt{v}  /  \mathit{x}  ]    \ottsym{)}  \rightarrow  \ottnt{C_{{\mathrm{11}}}}    [  \ottnt{v}  /  \mathit{x}  ]  $ by \WFT{Fun},
   we have $\ottnt{K'} \,  \in  \,  \LRRel{ \mathcal{K} }{    \ottnt{C_{{\mathrm{1}}}}  . \ottnt{T}     [  \ottnt{v}  /  \mathit{x}  ]    \, / \, ( \forall  \mathit{y}  .   \ottnt{C_{{\mathrm{11}}}}    [  \ottnt{v}  /  \mathit{x}  ]    )  } $
   by Lemmas~\ref{lem:lr-cont-subsump} and \ref{lem:lr-subtyping}.
   Therefore, we have
   \[
     \resetcnstr{  \ottnt{K'}  [   \ottnt{E}  [  \ottnt{v}  ]   ]  }  \,  \in  \,  \LRRel{ \mathcal{E} }{  \ottnt{C_{{\mathrm{12}}}}    [  \ottnt{v}  /  \mathit{x}  ]   }  ~.
   \]
   %
   Because $ \resetcnstr{  \ottnt{K'}  [   \ottnt{E}  [  \ottnt{v}  ]   ]  }  =  \ottsym{(}    \resetcnstr{ \ottnt{K'} }   [  \ottnt{E}  ]   \ottsym{)}  [  \ottnt{v}  ] $,
   we have
   \[
     \ottsym{(}    \resetcnstr{ \ottnt{K'} }   [  \ottnt{E}  ]   \ottsym{)}  [  \ottnt{v}  ]  \,  \in  \,  \LRRel{ \mathcal{E} }{  \ottnt{C_{{\mathrm{12}}}}    [  \ottnt{v}  /  \mathit{x}  ]   }  ~.
   \]
   %
   Furthermore, $ \ottsym{(}    \resetcnstr{ \ottnt{K'} }   [  \ottnt{E}  ]   \ottsym{)}  [  \ottnt{e}  ]  =  \resetcnstr{  \ottnt{K'}  [   \ottnt{E}  [  \ottnt{e}  ]   ]  }  \,  \in  \,  \LRRel{ \mathrm{Atom} }{ \ottnt{C_{{\mathrm{22}}}} } $ by \reflem{lr-typing-cont-exp}.
   Then, the IH implies the conclusion $ \resetcnstr{  \ottnt{K'}  [   \ottnt{E}  [  \ottnt{e}  ]   ]  }  \,  \in  \,  \LRRel{ \mathcal{E} }{ \ottnt{C_{{\mathrm{22}}}} } $.

  \case $  \exists  \, { \pi' } . \   \ottnt{E}  [  \ottnt{e}  ]  \,  \Uparrow  \, \pi' $:
   We must show that $ { \models  \,   { \ottnt{C_{{\mathrm{2}}}} }^\nu (  \pi'  )  } $.
   By Lemmas~\ref{lem:lr-red-under-ectx}, \ref{lem:lr-red-inconsist-eval-div}, \ref{lem:lr-eval-det}, and \ref{lem:lr-red-val},
   $ \ottnt{E}  [  \ottnt{v}  ]  \,  \Uparrow  \, \pi''$ and
   $ \pi'    =    \varpi \,  \tceappendop  \, \pi'' $ for some $\pi''$.
   Because $ \ottnt{E}  [  \ottnt{v}  ]  \,  \in  \,  \LRRel{ \mathcal{E} }{  \ottnt{C_{{\mathrm{1}}}}    [  \ottnt{v}  /  \mathit{x}  ]   } $,
   we have $ { \models  \,   { \ottsym{(}   \ottnt{C_{{\mathrm{1}}}}    [  \ottnt{v}  /  \mathit{x}  ]    \ottsym{)} }^\nu (  \pi''  )  } $.
   Then, by \reflem{lr-eff-cty-compose-consist-trace-compose} with
   $ { \models  \,   { \Phi }^\mu   \ottsym{(}  \varpi  \ottsym{)} } $,
   we have $ { \models  \,   { \ottsym{(}  \Phi \,  \tceappendop  \, \ottsym{(}   \ottnt{C_{{\mathrm{1}}}}    [  \ottnt{v}  /  \mathit{x}  ]    \ottsym{)}  \ottsym{)} }^\nu (  \varpi \,  \tceappendop  \, \pi''  )  } $, i.e.,
   $ { \models  \,   { \ottsym{(}  \Phi \,  \tceappendop  \, \ottsym{(}   \ottnt{C_{{\mathrm{1}}}}    [  \ottnt{v}  /  \mathit{x}  ]    \ottsym{)}  \ottsym{)} }^\nu (  \pi'  )  } $.
   Therefore, \reflem{ts-extract-subst-val-valid} implies
   $\mathit{y} \,  {:}  \,  \Sigma ^\omega   \models   \ottsym{(}   \pi'    =    \mathit{y}   \ottsym{)}    \Rightarrow     { \ottsym{(}  \Phi \,  \tceappendop  \, \ottsym{(}   \ottnt{C_{{\mathrm{1}}}}    [  \ottnt{v}  /  \mathit{x}  ]    \ottsym{)}  \ottsym{)} }^\nu (  \mathit{y}  )  $
   for fresh variable $\mathit{y}$.
   Because $ \emptyset   \vdash  \Phi \,  \tceappendop  \, \ottsym{(}   \ottnt{C_{{\mathrm{1}}}}    [  \ottnt{v}  /  \mathit{x}  ]    \ottsym{)}  \ottsym{<:}  \ottnt{C_{{\mathrm{2}}}}$,
   \reflem{lr-subtyping-nu} implies
   $\mathit{y} \,  {:}  \,  \Sigma ^\omega   \models   \ottsym{(}   \pi'    =    \mathit{y}   \ottsym{)}    \Rightarrow     { \ottnt{C_{{\mathrm{2}}}} }^\nu (  \mathit{y}  )  $.
   By \reflem{ts-term-subst-valid}, we have
   $ { \models  \,   \ottsym{(}   \pi'    =    \pi'   \ottsym{)}    \Rightarrow     { \ottnt{C_{{\mathrm{2}}}} }^\nu (  \pi'  )   } $.
   Because $ { \models  \,   \pi'    =    \pi'  } $ by \reflem{ts-refl-const-eq},
   \reflem{ts-discharge-assumption-valid} implies the conclusion
   $ { \models  \,   { \ottnt{C_{{\mathrm{2}}}} }^\nu (  \pi'  )  } $.
 \end{caseanalysis}
\end{proof}

\begin{lemmap}{Logical Relation is Closed Under Reset for Terminating Expressions}{lr-closed-reset-terminate}
 If
 $\ottnt{e} \,  \in  \,  \LRRel{ \mathcal{E} }{  \ottnt{T}  \,  \,\&\,  \,  \Phi  \, / \,   ( \forall  \mathit{x}  .  \ottnt{C_{{\mathrm{1}}}}  )  \Rightarrow   \ottnt{C_{{\mathrm{2}}}}   } $ and
 $\ottnt{e} \,  \Downarrow  \, \ottnt{v} \,  \,\&\,  \, \varpi$ and
 $\ottnt{K} \,  \in  \,  \LRRel{ \mathcal{K} }{  \ottnt{T}  \, / \, ( \forall  \mathit{x}  .  \ottnt{C_{{\mathrm{1}}}}  )  } $,
 then $ \resetcnstr{  \ottnt{K}  [  \ottnt{e}  ]  }  \,  \in  \,  \LRRel{ \mathcal{E} }{ \ottnt{C_{{\mathrm{2}}}} } $.
\end{lemmap}
\begin{proof}
 Because $\ottnt{e} \,  \in  \,  \LRRel{ \mathcal{E} }{  \ottnt{T}  \,  \,\&\,  \,  \Phi  \, / \,   ( \forall  \mathit{x}  .  \ottnt{C_{{\mathrm{1}}}}  )  \Rightarrow   \ottnt{C_{{\mathrm{2}}}}   } $ and $\ottnt{e} \,  \Downarrow  \, \ottnt{v} \,  \,\&\,  \, \varpi$,
 we have $\ottnt{v} \,  \in  \,  \LRRel{ \mathcal{V} }{ \ottnt{T} } $.
 Furthermore, Lemmas~\ref{lem:ts-type-sound-terminating} and \ref{lem:lr-red-val}
 imply
 \begin{itemize}
  \item $ \emptyset   \vdash  \ottnt{v}  \ottsym{:}  \ottnt{T_{{\mathrm{0}}}}$,
  \item $ \emptyset   \vdash   \ottnt{T_{{\mathrm{0}}}}  \,  \,\&\,  \,   (  \varpi  , \bot )   \, / \,   \square    \ottsym{<:}   \ottnt{T}  \,  \,\&\,  \,  \Phi  \, / \,   ( \forall  \mathit{x}  .  \ottnt{C_{{\mathrm{1}}}}  )  \Rightarrow   \ottnt{C_{{\mathrm{2}}}}  $
 \end{itemize}
 for some $\ottnt{T_{{\mathrm{0}}}}$.
 The last subtyping derivation implies
 \begin{itemize}
  \item $\mathit{x} \,  {:}  \, \ottnt{T_{{\mathrm{0}}}}  \vdash   (  \varpi  , \bot )  \,  \tceappendop  \, \ottnt{C_{{\mathrm{1}}}}  \ottsym{<:}  \ottnt{C_{{\mathrm{2}}}}$ and
  \item $\mathit{x} \,  \not\in  \,  \mathit{fv}  (  \ottnt{C_{{\mathrm{2}}}}  ) $.
 \end{itemize}
 Then, \reflem{ts-val-subst-subtyping} with $ \emptyset   \vdash  \ottnt{v}  \ottsym{:}  \ottnt{T_{{\mathrm{0}}}}$, and
 \reflem{ts-val-subst-ty-compose} imply
 \[
   \emptyset   \vdash   (  \varpi  , \bot )  \,  \tceappendop  \, \ottsym{(}   \ottnt{C_{{\mathrm{1}}}}    [  \ottnt{v}  /  \mathit{x}  ]    \ottsym{)}  \ottsym{<:}  \ottnt{C_{{\mathrm{2}}}} ~.
 \]
 Because $\ottnt{K} \,  \in  \,  \LRRel{ \mathcal{K} }{  \ottnt{T}  \, / \, ( \forall  \mathit{x}  .  \ottnt{C_{{\mathrm{1}}}}  )  } $ implies $ \lambda\!  \, \mathit{x}  \ottsym{.}   \resetcnstr{  \ottnt{K}  [  \mathit{x}  ]  }  \,  \in  \,  \LRRel{ \mathcal{V} }{ \ottsym{(}  \mathit{x} \,  {:}  \, \ottnt{T}  \ottsym{)}  \rightarrow  \ottnt{C_{{\mathrm{1}}}} } $,
 we have $\ottsym{(}   \lambda\!  \, \mathit{x}  \ottsym{.}   \resetcnstr{  \ottnt{K}  [  \mathit{x}  ]  }   \ottsym{)} \, \ottnt{v} \,  \in  \,  \LRRel{ \mathcal{E} }{  \ottnt{C_{{\mathrm{1}}}}    [  \ottnt{v}  /  \mathit{x}  ]   } $.
 Because $\mathit{x} \,  \not\in  \,  \mathit{fv}  (  \ottnt{K}  ) $ by \reflem{lr-typing-ectx-closed},
 we have $\ottsym{(}   \lambda\!  \, \mathit{x}  \ottsym{.}   \resetcnstr{  \ottnt{K}  [  \mathit{x}  ]  }   \ottsym{)} \, \ottnt{v} \,  \rightsquigarrow  \,  \resetcnstr{  \ottnt{K}  [  \ottnt{v}  ]  } $ by \R{Beta}.
 Therefore, \reflem{lr-red} implies
 \[
   \resetcnstr{  \ottnt{K}  [  \ottnt{v}  ]  }  \,  \in  \,  \LRRel{ \mathcal{E} }{  \ottnt{C_{{\mathrm{1}}}}    [  \ottnt{v}  /  \mathit{x}  ]   }  ~.
 \]
 We also have
 \[
   \resetcnstr{  \ottnt{K}  [  \ottnt{e}  ]  }  \,  \in  \,  \LRRel{ \mathrm{Atom} }{ \ottnt{C_{{\mathrm{2}}}} } 
 \]
 by \reflem{lr-typing-cont-exp} with
 $\ottnt{K} \,  \in  \,  \LRRel{ \mathcal{K} }{  \ottnt{T}  \, / \, ( \forall  \mathit{x}  .  \ottnt{C_{{\mathrm{1}}}}  )  } $ and
 $\ottnt{e} \,  \in  \,  \LRRel{ \mathcal{E} }{  \ottnt{T}  \,  \,\&\,  \,  \Phi  \, / \,   ( \forall  \mathit{x}  .  \ottnt{C_{{\mathrm{1}}}}  )  \Rightarrow   \ottnt{C_{{\mathrm{2}}}}   } $.
 Furthermore, \reflem{ts-refl-const-eq} implies
 \[
   \emptyset   \models   { \ottsym{(}   (  \varpi  , \bot )   \ottsym{)} }^\mu   \ottsym{(}  \varpi  \ottsym{)} ~.
 \]
 Therefore, \reflem{lr-embed-term-exp-into-ectx} implies the conclusion.
\end{proof}

\begin{lemmap}{Logical Relation is Closed Under Reset}{lr-closed-reset}
 If $\ottnt{e} \,  \in  \,  \LRRel{ \mathcal{E} }{  \ottnt{T}  \,  \,\&\,  \,  \Phi  \, / \,   ( \forall  \mathit{x}  .  \ottnt{C_{{\mathrm{1}}}}  )  \Rightarrow   \ottnt{C_{{\mathrm{2}}}}   } $ and $\ottnt{K} \,  \in  \,  \LRRel{ \mathcal{K} }{  \ottnt{T}  \, / \, ( \forall  \mathit{x}  .  \ottnt{C_{{\mathrm{1}}}}  )  } $,
 then $ \resetcnstr{  \ottnt{K}  [  \ottnt{e}  ]  }  \,  \in  \,  \LRRel{ \mathcal{E} }{ \ottnt{C_{{\mathrm{2}}}} } $.
\end{lemmap}
\begin{proof}
 \reflem{lr-typing-cont-exp} with
 $\ottnt{e} \,  \in  \,  \LRRel{ \mathcal{E} }{  \ottnt{T}  \,  \,\&\,  \,  \Phi  \, / \,   ( \forall  \mathit{x}  .  \ottnt{C_{{\mathrm{1}}}}  )  \Rightarrow   \ottnt{C_{{\mathrm{2}}}}   } $ and
 $\ottnt{K} \,  \in  \,  \LRRel{ \mathcal{K} }{  \ottnt{T}  \, / \, ( \forall  \mathit{x}  .  \ottnt{C_{{\mathrm{1}}}}  )  } $ implies
 $ \emptyset   \vdash   \resetcnstr{  \ottnt{K}  [  \ottnt{e}  ]  }   \ottsym{:}  \ottnt{C_{{\mathrm{2}}}}$, i.e.,
 $ \resetcnstr{  \ottnt{K}  [  \ottnt{e}  ]  }  \,  \in  \,  \LRRel{ \mathrm{Atom} }{ \ottnt{C_{{\mathrm{2}}}} } $.

 Consider the case that $\ottnt{e} \,  \Downarrow  \,  \ottnt{K_{{\mathrm{0}}}}  [  \mathcal{S}_0 \, \mathit{x_{{\mathrm{0}}}}  \ottsym{.}  \ottnt{e_{{\mathrm{0}}}}  ]  \,  \,\&\,  \, \varpi$
 for some $\ottnt{K_{{\mathrm{0}}}}$, $\mathit{x_{{\mathrm{0}}}}$, $\ottnt{e_{{\mathrm{0}}}}$, and $\varpi$.
 Then, because $\ottnt{e} \,  \in  \,  \LRRel{ \mathcal{E} }{  \ottnt{T}  \,  \,\&\,  \,  \Phi  \, / \,   ( \forall  \mathit{x}  .  \ottnt{C_{{\mathrm{1}}}}  )  \Rightarrow   \ottnt{C_{{\mathrm{2}}}}   } $ and $\ottnt{K} \,  \in  \,  \LRRel{ \mathcal{K} }{  \ottnt{T}  \, / \, ( \forall  \mathit{x}  .  \ottnt{C_{{\mathrm{1}}}}  )  } $,
 we have the conclusion $ \resetcnstr{  \ottnt{K}  [  \ottnt{e}  ]  }  \,  \in  \,  \LRRel{ \mathcal{E} }{ \ottnt{C_{{\mathrm{2}}}} } $.

 In what follows, we assume
 \begin{equation}
   \neg    \exists  \, { \ottnt{K_{{\mathrm{0}}}}  \ottsym{,}  \mathit{x_{{\mathrm{0}}}}  \ottsym{,}  \ottnt{e_{{\mathrm{0}}}}  \ottsym{,}  \varpi } . \  \ottnt{e} \,  \Downarrow  \,  \ottnt{K_{{\mathrm{0}}}}  [  \mathcal{S}_0 \, \mathit{x_{{\mathrm{0}}}}  \ottsym{.}  \ottnt{e_{{\mathrm{0}}}}  ]  \,  \,\&\,  \, \varpi   ~.
   \label{lem:lr-closed-reset:shift-assump}
 \end{equation}

 Next, consider the case that $\ottnt{e} \,  \Downarrow  \, \ottnt{v} \,  \,\&\,  \, \varpi$
 for some $\ottnt{v}$ and $\varpi$.
 Then, \reflem{lr-closed-reset-terminate} implies the conclusion.

 In what follows, we assume
 \begin{equation}
   \neg    \exists  \, { \ottnt{v}  \ottsym{,}  \varpi } . \  \ottnt{e} \,  \Downarrow  \, \ottnt{v} \,  \,\&\,  \, \varpi   ~.
   \label{lem:lr-closed-reset:val-assump}
 \end{equation}

 We consider each case on the result of evaluating $ \resetcnstr{  \ottnt{K}  [  \ottnt{e}  ]  } $.
 \begin{caseanalysis}
  \case $  \exists  \, { \ottnt{v}  \ottsym{,}  \varpi } . \   \resetcnstr{  \ottnt{K}  [  \ottnt{e}  ]  }  \,  \Downarrow  \, \ottnt{v} \,  \,\&\,  \, \varpi $:
   Contradictory by \reflem{lr-red-under-ectx} and
   the assumptions (\ref{lem:lr-closed-reset:shift-assump}) and (\ref{lem:lr-closed-reset:val-assump}).

  \case $  \exists  \, { \ottnt{K_{{\mathrm{0}}}}  \ottsym{,}  \mathit{x_{{\mathrm{0}}}}  \ottsym{,}  \ottnt{e_{{\mathrm{0}}}}  \ottsym{,}  \varpi } . \   \resetcnstr{  \ottnt{K}  [  \ottnt{e}  ]  }  \,  \Downarrow  \,  \ottnt{K_{{\mathrm{0}}}}  [  \mathcal{S}_0 \, \mathit{x_{{\mathrm{0}}}}  \ottsym{.}  \ottnt{e_{{\mathrm{0}}}}  ]  \,  \,\&\,  \, \varpi $:
   Contradictory by \reflem{lr-red-under-ectx} and
 the assumptions (\ref{lem:lr-closed-reset:shift-assump}) and (\ref{lem:lr-closed-reset:val-assump}).

  \case $  \exists  \, { \pi } . \   \resetcnstr{  \ottnt{K}  [  \ottnt{e}  ]  }  \,  \Uparrow  \, \pi $:
   We must show that $ { \models  \,   { \ottnt{C_{{\mathrm{2}}}} }^\nu (  \pi  )  } $.
   By \reflem{lr-red-under-ectx} and
   the assumptions (\ref{lem:lr-closed-reset:val-assump}) and (\ref{lem:lr-closed-reset:shift-assump}),
   we can suppose that $\ottnt{e} \,  \Uparrow  \, \pi$.
   Then, because $\ottnt{e} \,  \in  \,  \LRRel{ \mathcal{E} }{  \ottnt{T}  \,  \,\&\,  \,  \Phi  \, / \,   ( \forall  \mathit{x}  .  \ottnt{C_{{\mathrm{1}}}}  )  \Rightarrow   \ottnt{C_{{\mathrm{2}}}}   } $,
   we have $ { \models  \,    { \Phi }^\nu   \ottsym{(}  \pi  \ottsym{)}    \wedge     { \ottnt{C_{{\mathrm{2}}}} }^\nu (  \pi  )   } $.
   By \reflem{ts-valid-elim-intersect}, we have the conclusion
   $ { \models  \,   { \ottnt{C_{{\mathrm{2}}}} }^\nu (  \pi  )  } $.
 \end{caseanalysis}
\end{proof}

\begin{lemmap}{Compatibility: Let}{lr-compat-let}
 If
 $\ottnt{e_{{\mathrm{1}}}} \,  \in  \,  \LRRel{ \mathcal{O} }{ \Gamma  \, \vdash \,   \ottnt{T_{{\mathrm{1}}}}  \,  \,\&\,  \,  \Phi_{{\mathrm{1}}}  \, / \,  \ottnt{S_{{\mathrm{1}}}}  } $ and
 $\ottnt{e_{{\mathrm{2}}}} \,  \in  \,  \LRRel{ \mathcal{O} }{ \Gamma  \ottsym{,}  \mathit{x_{{\mathrm{1}}}} \,  {:}  \, \ottnt{T_{{\mathrm{1}}}}  \, \vdash \,   \ottnt{T_{{\mathrm{2}}}}  \,  \,\&\,  \,  \Phi_{{\mathrm{2}}}  \, / \,  \ottnt{S_{{\mathrm{2}}}}  } $ and
 $\mathit{x_{{\mathrm{1}}}} \,  \not\in  \,  \mathit{fv}  (  \ottnt{T_{{\mathrm{2}}}}  )  \,  \mathbin{\cup}  \,  \mathit{fv}  (  \Phi_{{\mathrm{2}}}  ) $,
 then
 $\mathsf{let} \, \mathit{x_{{\mathrm{1}}}}  \ottsym{=}  \ottnt{e_{{\mathrm{1}}}} \,  \mathsf{in}  \, \ottnt{e_{{\mathrm{2}}}} \,  \in  \,  \LRRel{ \mathcal{O} }{ \Gamma  \, \vdash \,   \ottnt{T_{{\mathrm{2}}}}  \,  \,\&\,  \,  \Phi_{{\mathrm{1}}} \,  \tceappendop  \, \Phi_{{\mathrm{2}}}  \, / \,  \ottnt{S_{{\mathrm{1}}}}  \gg\!\!=  \ottsym{(}   \lambda\!  \, \mathit{x_{{\mathrm{1}}}}  \ottsym{.}  \ottnt{S_{{\mathrm{2}}}}  \ottsym{)}  } $.
\end{lemmap}
\begin{proof}
 Let $\ottnt{C} \,  =  \,  \ottnt{T_{{\mathrm{2}}}}  \,  \,\&\,  \,  \Phi_{{\mathrm{1}}} \,  \tceappendop  \, \Phi_{{\mathrm{2}}}  \, / \,  \ottnt{S_{{\mathrm{1}}}}  \gg\!\!=  \ottsym{(}   \lambda\!  \, \mathit{x_{{\mathrm{1}}}}  \ottsym{.}  \ottnt{S_{{\mathrm{2}}}}  \ottsym{)} $.
 By \T{Let}, we have
 $\Gamma  \vdash  \mathsf{let} \, \mathit{x_{{\mathrm{1}}}}  \ottsym{=}  \ottnt{e_{{\mathrm{1}}}} \,  \mathsf{in}  \, \ottnt{e_{{\mathrm{2}}}}  \ottsym{:}  \ottnt{C}$.
 Let $\sigma \,  \in  \,  \LRRel{ \mathcal{G} }{ \Gamma } $.
 Without loss of generality, we can suppose that $\mathit{x_{{\mathrm{1}}}}$ does not occur in $\sigma$.
 We must show that
 \[
  \mathsf{let} \, \mathit{x_{{\mathrm{1}}}}  \ottsym{=}  \sigma  \ottsym{(}  \ottnt{e_{{\mathrm{1}}}}  \ottsym{)} \,  \mathsf{in}  \, \sigma  \ottsym{(}  \ottnt{e_{{\mathrm{2}}}}  \ottsym{)} \,  \in  \,  \LRRel{ \mathcal{E} }{ \sigma  \ottsym{(}  \ottnt{C}  \ottsym{)} }  ~.
 \]
 We have $\mathsf{let} \, \mathit{x_{{\mathrm{1}}}}  \ottsym{=}  \sigma  \ottsym{(}  \ottnt{e_{{\mathrm{1}}}}  \ottsym{)} \,  \mathsf{in}  \, \sigma  \ottsym{(}  \ottnt{e_{{\mathrm{2}}}}  \ottsym{)} \,  \in  \,  \LRRel{ \mathrm{Atom} }{ \sigma  \ottsym{(}  \ottnt{C}  \ottsym{)} } $
 by \reflem{lr-typeability}.

 We consider each case on the result of evaluating $\mathsf{let} \, \mathit{x_{{\mathrm{1}}}}  \ottsym{=}  \sigma  \ottsym{(}  \ottnt{e_{{\mathrm{1}}}}  \ottsym{)} \,  \mathsf{in}  \, \sigma  \ottsym{(}  \ottnt{e_{{\mathrm{2}}}}  \ottsym{)}$.
 \begin{caseanalysis}
  \case $  \exists  \, { \ottnt{v}  \ottsym{,}  \varpi } . \  \ottsym{(}  \mathsf{let} \, \mathit{x_{{\mathrm{1}}}}  \ottsym{=}  \sigma  \ottsym{(}  \ottnt{e_{{\mathrm{1}}}}  \ottsym{)} \,  \mathsf{in}  \, \sigma  \ottsym{(}  \ottnt{e_{{\mathrm{2}}}}  \ottsym{)}  \ottsym{)} \,  \Downarrow  \, \ottnt{v} \,  \,\&\,  \, \varpi $:
   We must show that $\ottnt{v} \,  \in  \,  \LRRel{ \mathcal{V} }{ \sigma  \ottsym{(}  \ottnt{T_{{\mathrm{2}}}}  \ottsym{)} } $.
   By \reflem{lr-red-under-ectx} with $\ottsym{(}  \mathsf{let} \, \mathit{x_{{\mathrm{1}}}}  \ottsym{=}  \sigma  \ottsym{(}  \ottnt{e_{{\mathrm{1}}}}  \ottsym{)} \,  \mathsf{in}  \, \sigma  \ottsym{(}  \ottnt{e_{{\mathrm{2}}}}  \ottsym{)}  \ottsym{)} \,  \Downarrow  \, \ottnt{v} \,  \,\&\,  \, \varpi$,
   and Lemmas~\ref{lem:lr-red-shift},
   we can suppose that
   \begin{itemize}
    \item $\sigma  \ottsym{(}  \ottnt{e_{{\mathrm{1}}}}  \ottsym{)} \,  \mathrel{  \longrightarrow  ^*}  \, \ottnt{v_{{\mathrm{1}}}} \,  \,\&\,  \, \varpi_{{\mathrm{1}}}$,
    \item $\ottsym{(}  \mathsf{let} \, \mathit{x_{{\mathrm{1}}}}  \ottsym{=}  \ottnt{v_{{\mathrm{1}}}} \,  \mathsf{in}  \, \sigma  \ottsym{(}  \ottnt{e_{{\mathrm{2}}}}  \ottsym{)}  \ottsym{)} \,  \mathrel{  \longrightarrow  ^*}  \, \ottnt{v} \,  \,\&\,  \, \varpi_{{\mathrm{2}}}$, and
    \item $ \varpi    =    \varpi_{{\mathrm{1}}} \,  \tceappendop  \, \varpi_{{\mathrm{2}}} $
   \end{itemize}
   for some $\ottnt{v_{{\mathrm{1}}}}$, $\varpi_{{\mathrm{1}}}$, and $\varpi_{{\mathrm{2}}}$.
   Because $\ottnt{e_{{\mathrm{1}}}} \,  \in  \,  \LRRel{ \mathcal{O} }{ \Gamma  \, \vdash \,   \ottnt{T_{{\mathrm{1}}}}  \,  \,\&\,  \,  \Phi_{{\mathrm{1}}}  \, / \,  \ottnt{S_{{\mathrm{1}}}}  } $ and $\sigma \,  \in  \,  \LRRel{ \mathcal{G} }{ \Gamma } $,
   we have $\sigma  \ottsym{(}  \ottnt{e_{{\mathrm{1}}}}  \ottsym{)} \,  \in  \,  \LRRel{ \mathcal{E} }{ \sigma  \ottsym{(}   \ottnt{T_{{\mathrm{1}}}}  \,  \,\&\,  \,  \Phi_{{\mathrm{1}}}  \, / \,  \ottnt{S_{{\mathrm{1}}}}   \ottsym{)} } $.
   Because $\sigma  \ottsym{(}  \ottnt{e_{{\mathrm{1}}}}  \ottsym{)} \,  \Downarrow  \, \ottnt{v_{{\mathrm{1}}}} \,  \,\&\,  \, \varpi_{{\mathrm{1}}}$, we have $\ottnt{v_{{\mathrm{1}}}} \,  \in  \,  \LRRel{ \mathcal{V} }{ \sigma  \ottsym{(}  \ottnt{T_{{\mathrm{1}}}}  \ottsym{)} } $.
   Then, $\sigma  \mathbin{\uplus}   [  \ottnt{v_{{\mathrm{1}}}}  /  \mathit{x_{{\mathrm{1}}}}  ]  \,  \in  \,  \LRRel{ \mathcal{G} }{ \Gamma  \ottsym{,}  \mathit{x_{{\mathrm{1}}}} \,  {:}  \, \ottnt{T_{{\mathrm{1}}}} } $.
   Because $\ottnt{e_{{\mathrm{2}}}} \,  \in  \,  \LRRel{ \mathcal{O} }{ \Gamma  \ottsym{,}  \mathit{x_{{\mathrm{1}}}} \,  {:}  \, \ottnt{T_{{\mathrm{1}}}}  \, \vdash \,   \ottnt{T_{{\mathrm{2}}}}  \,  \,\&\,  \,  \Phi_{{\mathrm{2}}}  \, / \,  \ottnt{S_{{\mathrm{2}}}}  } $,
   we have $ \sigma  \ottsym{(}  \ottnt{e_{{\mathrm{2}}}}  \ottsym{)}    [  \ottnt{v_{{\mathrm{1}}}}  /  \mathit{x_{{\mathrm{1}}}}  ]   \,  \in  \,  \LRRel{ \mathcal{E} }{  \sigma  \ottsym{(}   \ottnt{T_{{\mathrm{2}}}}  \,  \,\&\,  \,  \Phi_{{\mathrm{2}}}  \, / \,  \ottnt{S_{{\mathrm{2}}}}   \ottsym{)}    [  \ottnt{v_{{\mathrm{1}}}}  /  \mathit{x_{{\mathrm{1}}}}  ]   } $.
   Because $\mathit{x_{{\mathrm{1}}}} \,  \not\in  \,  \mathit{fv}  (  \ottnt{T_{{\mathrm{2}}}}  )  \,  \mathbin{\cup}  \,  \mathit{fv}  (  \Phi_{{\mathrm{2}}}  ) $ and $\mathit{x_{{\mathrm{1}}}}$ does not occur in $\sigma$,
   we have $ \sigma  \ottsym{(}  \ottnt{e_{{\mathrm{2}}}}  \ottsym{)}    [  \ottnt{v_{{\mathrm{1}}}}  /  \mathit{x_{{\mathrm{1}}}}  ]   \,  \in  \,  \LRRel{ \mathcal{E} }{  \sigma  \ottsym{(}  \ottnt{T_{{\mathrm{2}}}}  \ottsym{)}  \,  \,\&\,  \,  \sigma  \ottsym{(}  \Phi_{{\mathrm{2}}}  \ottsym{)}  \, / \,   \sigma  \ottsym{(}  \ottnt{S_{{\mathrm{2}}}}  \ottsym{)}    [  \ottnt{v_{{\mathrm{1}}}}  /  \mathit{x_{{\mathrm{1}}}}  ]    } $.
   Because $ \sigma  \ottsym{(}  \ottnt{e_{{\mathrm{2}}}}  \ottsym{)}    [  \ottnt{v_{{\mathrm{1}}}}  /  \mathit{x_{{\mathrm{1}}}}  ]   \,  \Downarrow  \, \ottnt{v} \,  \,\&\,  \, \varpi_{{\mathrm{2}}}$ by \reflem{lr-eval-det} and \E{Red}/\R{Let},
   we have the conclusion $\ottnt{v} \,  \in  \,  \LRRel{ \mathcal{V} }{ \sigma  \ottsym{(}  \ottnt{T_{{\mathrm{2}}}}  \ottsym{)} } $.

  \case $  \exists  \, { \ottnt{K_{{\mathrm{0}}}}  \ottsym{,}  \mathit{x_{{\mathrm{0}}}}  \ottsym{,}  \ottnt{e_{{\mathrm{0}}}}  \ottsym{,}  \varpi } . \  \ottsym{(}  \mathsf{let} \, \mathit{x_{{\mathrm{1}}}}  \ottsym{=}  \sigma  \ottsym{(}  \ottnt{e_{{\mathrm{1}}}}  \ottsym{)} \,  \mathsf{in}  \, \sigma  \ottsym{(}  \ottnt{e_{{\mathrm{2}}}}  \ottsym{)}  \ottsym{)} \,  \Downarrow  \,  \ottnt{K_{{\mathrm{0}}}}  [  \mathcal{S}_0 \, \mathit{x_{{\mathrm{0}}}}  \ottsym{.}  \ottnt{e_{{\mathrm{0}}}}  ]  \,  \,\&\,  \, \varpi $:
   We must show that
   \begin{itemize}
    \item $\sigma  \ottsym{(}  \ottnt{S_{{\mathrm{1}}}}  \gg\!\!=  \ottsym{(}   \lambda\!  \, \mathit{x_{{\mathrm{1}}}}  \ottsym{.}  \ottnt{S_{{\mathrm{2}}}}  \ottsym{)}  \ottsym{)} \,  =  \,  ( \forall  \mathit{y}  .  \ottnt{C'_{{\mathrm{1}}}}  )  \Rightarrow   \ottnt{C'_{{\mathrm{2}}}} $ and
    \item $  \forall  \, { \ottnt{K} \,  \in  \,  \LRRel{ \mathcal{K} }{  \sigma  \ottsym{(}  \ottnt{T_{{\mathrm{2}}}}  \ottsym{)}  \, / \, ( \forall  \mathit{y}  .  \ottnt{C'_{{\mathrm{1}}}}  )  }  } . \   \resetcnstr{  \ottnt{K}  [  \mathsf{let} \, \mathit{x_{{\mathrm{1}}}}  \ottsym{=}  \sigma  \ottsym{(}  \ottnt{e_{{\mathrm{1}}}}  \ottsym{)} \,  \mathsf{in}  \, \sigma  \ottsym{(}  \ottnt{e_{{\mathrm{2}}}}  \ottsym{)}  ]  }  \,  \in  \,  \LRRel{ \mathcal{E} }{ \ottnt{C'_{{\mathrm{2}}}} }  $
   \end{itemize}
   for some $\mathit{y}$, $\ottnt{C'_{{\mathrm{1}}}}$, and $\ottnt{C'_{{\mathrm{2}}}}$.

   We first show that
   \begin{itemize}
    \item $\ottnt{S_{{\mathrm{1}}}} \,  =  \,  ( \forall  \mathit{x_{{\mathrm{1}}}}  .  \ottnt{C_{{\mathrm{3}}}}  )  \Rightarrow   \ottnt{C_{{\mathrm{2}}}} $,
    \item $\ottnt{S_{{\mathrm{2}}}} \,  =  \,  ( \forall  \mathit{y}  .  \ottnt{C_{{\mathrm{1}}}}  )  \Rightarrow   \ottnt{C_{{\mathrm{3}}}} $, and
    \item $\ottnt{S_{{\mathrm{1}}}}  \gg\!\!=  \ottsym{(}   \lambda\!  \, \mathit{x_{{\mathrm{1}}}}  \ottsym{.}  \ottnt{S_{{\mathrm{2}}}}  \ottsym{)} \,  =  \,  ( \forall  \mathit{y}  .  \ottnt{C_{{\mathrm{1}}}}  )  \Rightarrow   \ottnt{C_{{\mathrm{2}}}} $
   \end{itemize}
   for some $\mathit{y}$, $\ottnt{C_{{\mathrm{1}}}}$, $\ottnt{C_{{\mathrm{2}}}}$, and $\ottnt{C_{{\mathrm{3}}}}$ such that
   $\mathit{x_{{\mathrm{1}}}} \,  \not\in  \,  \mathit{fv}  (  \ottnt{C_{{\mathrm{1}}}}  )  \,  \mathbin{\backslash}  \,  \{  \mathit{y}  \} $.
   Because
   \begin{itemize}
    \item $\mathsf{let} \, \mathit{x_{{\mathrm{1}}}}  \ottsym{=}  \sigma  \ottsym{(}  \ottnt{e_{{\mathrm{1}}}}  \ottsym{)} \,  \mathsf{in}  \, \sigma  \ottsym{(}  \ottnt{e_{{\mathrm{2}}}}  \ottsym{)} \,  \in  \,  \LRRel{ \mathrm{Atom} }{ \sigma  \ottsym{(}  \ottnt{C}  \ottsym{)} } $,
    \item $\ottsym{(}  \mathsf{let} \, \mathit{x_{{\mathrm{1}}}}  \ottsym{=}  \sigma  \ottsym{(}  \ottnt{e_{{\mathrm{1}}}}  \ottsym{)} \,  \mathsf{in}  \, \sigma  \ottsym{(}  \ottnt{e_{{\mathrm{2}}}}  \ottsym{)}  \ottsym{)} \,  \Downarrow  \,  \ottnt{K_{{\mathrm{0}}}}  [  \mathcal{S}_0 \, \mathit{x_{{\mathrm{0}}}}  \ottsym{.}  \ottnt{e_{{\mathrm{0}}}}  ]  \,  \,\&\,  \, \varpi$, and
    \item $ \ottnt{K_{{\mathrm{0}}}}  [  \mathcal{S}_0 \, \mathit{x_{{\mathrm{0}}}}  \ottsym{.}  \ottnt{e_{{\mathrm{0}}}}  ] $ cannot be evaluated furthermore (\reflem{lr-red-shift}),
   \end{itemize}
   \reflem{ts-type-sound-terminating} implies
   $\sigma  \ottsym{(}  \ottnt{S_{{\mathrm{1}}}}  \gg\!\!=  \ottsym{(}   \lambda\!  \, \mathit{x_{{\mathrm{1}}}}  \ottsym{.}  \ottnt{S_{{\mathrm{2}}}}  \ottsym{)}  \ottsym{)} \,  \not=  \,  \square $;
   note that $ \ottnt{K_{{\mathrm{0}}}}  [  \mathcal{S}_0 \, \mathit{x_{{\mathrm{0}}}}  \ottsym{.}  \ottnt{e_{{\mathrm{0}}}}  ]  \,  \not=  \,  \ottnt{E}  [   \bullet ^{ \pi }   ] $
   for any $\ottnt{E}$ and $\pi$ by \reflem{lr-decompose-ectx-embed-exp}.
   Therefore,
   \begin{itemize}
    \item $\ottnt{S_{{\mathrm{1}}}} \,  =  \,  ( \forall  \mathit{x_{{\mathrm{1}}}}  .  \ottnt{C_{{\mathrm{3}}}}  )  \Rightarrow   \ottnt{C_{{\mathrm{2}}}} $,
    \item $\ottnt{S_{{\mathrm{2}}}} \,  =  \,  ( \forall  \mathit{y}  .  \ottnt{C_{{\mathrm{1}}}}  )  \Rightarrow   \ottnt{C_{{\mathrm{3}}}} $, and
    \item $\ottnt{S_{{\mathrm{1}}}}  \gg\!\!=  \ottsym{(}   \lambda\!  \, \mathit{x_{{\mathrm{1}}}}  \ottsym{.}  \ottnt{S_{{\mathrm{2}}}}  \ottsym{)} \,  =  \,  ( \forall  \mathit{y}  .  \ottnt{C_{{\mathrm{1}}}}  )  \Rightarrow   \ottnt{C_{{\mathrm{2}}}} $
   \end{itemize}
   for some $\mathit{y}$, $\ottnt{C_{{\mathrm{1}}}}$, $\ottnt{C_{{\mathrm{2}}}}$, and $\ottnt{C_{{\mathrm{3}}}}$ such that
   $\mathit{x_{{\mathrm{1}}}} \,  \not\in  \,  \mathit{fv}  (  \ottnt{C_{{\mathrm{1}}}}  )  \,  \mathbin{\backslash}  \,  \{  \mathit{y}  \} $.

   Without loss of generality,
   we can suppose that $\mathit{y}$ does not occur in $\sigma$.
   Next, we show the conclusion by letting $\ottnt{C'_{{\mathrm{1}}}} \,  =  \, \sigma  \ottsym{(}  \ottnt{C_{{\mathrm{1}}}}  \ottsym{)}$ and $\ottnt{C'_{{\mathrm{2}}}} \,  =  \, \sigma  \ottsym{(}  \ottnt{C_{{\mathrm{2}}}}  \ottsym{)}$.
   \begin{itemize}
    \item We have $\sigma  \ottsym{(}  \ottnt{S_{{\mathrm{1}}}}  \gg\!\!=  \ottsym{(}   \lambda\!  \, \mathit{x_{{\mathrm{1}}}}  \ottsym{.}  \ottnt{S_{{\mathrm{2}}}}  \ottsym{)}  \ottsym{)} \,  =  \,  ( \forall  \mathit{y}  .  \sigma  \ottsym{(}  \ottnt{C_{{\mathrm{1}}}}  \ottsym{)}  )  \Rightarrow   \sigma  \ottsym{(}  \ottnt{C_{{\mathrm{2}}}}  \ottsym{)} $.
    \item Let $\ottnt{K} \,  \in  \,  \LRRel{ \mathcal{K} }{  \sigma  \ottsym{(}  \ottnt{T_{{\mathrm{2}}}}  \ottsym{)}  \, / \, ( \forall  \mathit{y}  .  \sigma  \ottsym{(}  \ottnt{C_{{\mathrm{1}}}}  \ottsym{)}  )  } $.
          We must show that
          \[
            \resetcnstr{  \ottnt{K}  [  \mathsf{let} \, \mathit{x_{{\mathrm{1}}}}  \ottsym{=}  \sigma  \ottsym{(}  \ottnt{e_{{\mathrm{1}}}}  \ottsym{)} \,  \mathsf{in}  \, \sigma  \ottsym{(}  \ottnt{e_{{\mathrm{2}}}}  \ottsym{)}  ]  }  \,  \in  \,  \LRRel{ \mathcal{E} }{ \sigma  \ottsym{(}  \ottnt{C_{{\mathrm{2}}}}  \ottsym{)} }  ~.
          \]
          Because $\sigma  \ottsym{(}  \ottnt{e_{{\mathrm{1}}}}  \ottsym{)} \,  \in  \,  \LRRel{ \mathcal{E} }{  \sigma  \ottsym{(}  \ottnt{T_{{\mathrm{1}}}}  \ottsym{)}  \,  \,\&\,  \,  \sigma  \ottsym{(}  \Phi_{{\mathrm{1}}}  \ottsym{)}  \, / \,   ( \forall  \mathit{x_{{\mathrm{1}}}}  .  \sigma  \ottsym{(}  \ottnt{C_{{\mathrm{3}}}}  \ottsym{)}  )  \Rightarrow   \sigma  \ottsym{(}  \ottnt{C_{{\mathrm{2}}}}  \ottsym{)}   } $,
          \reflem{lr-closed-reset} implies that it suffices to show that
          \[
            \ottnt{K}  [  \mathsf{let} \, \mathit{x_{{\mathrm{1}}}}  \ottsym{=}  \,[\,]\, \,  \mathsf{in}  \, \sigma  \ottsym{(}  \ottnt{e_{{\mathrm{2}}}}  \ottsym{)}  ]  \,  \in  \,  \LRRel{ \mathcal{K} }{  \sigma  \ottsym{(}  \ottnt{T_{{\mathrm{1}}}}  \ottsym{)}  \, / \, ( \forall  \mathit{x_{{\mathrm{1}}}}  .  \sigma  \ottsym{(}  \ottnt{C_{{\mathrm{3}}}}  \ottsym{)}  )  }  ~,
          \]
          which is proven by the following.
          \begin{itemize}
           \item We show that $ \ottnt{K}  [  \mathsf{let} \, \mathit{x_{{\mathrm{1}}}}  \ottsym{=}  \,[\,]\, \,  \mathsf{in}  \, \sigma  \ottsym{(}  \ottnt{e_{{\mathrm{2}}}}  \ottsym{)}  ]   \ottsym{:}   \sigma  \ottsym{(}  \ottnt{T_{{\mathrm{1}}}}  \ottsym{)}  \, / \, ( \forall  \mathit{x_{{\mathrm{1}}}}  .  \sigma  \ottsym{(}  \ottnt{C_{{\mathrm{3}}}}  \ottsym{)}  ) $.
                 Now, we have the following:
                 \begin{itemize}
                  \item $\ottnt{K}  \ottsym{:}   \sigma  \ottsym{(}  \ottnt{T_{{\mathrm{2}}}}  \ottsym{)}  \, / \, ( \forall  \mathit{y}  .  \sigma  \ottsym{(}  \ottnt{C_{{\mathrm{1}}}}  \ottsym{)}  ) $
                        because $\ottnt{K} \,  \in  \,  \LRRel{ \mathcal{K} }{  \sigma  \ottsym{(}  \ottnt{T_{{\mathrm{2}}}}  \ottsym{)}  \, / \, ( \forall  \mathit{y}  .  \sigma  \ottsym{(}  \ottnt{C_{{\mathrm{1}}}}  \ottsym{)}  )  } $;
                  \item $\mathit{x_{{\mathrm{1}}}} \,  {:}  \, \sigma  \ottsym{(}  \ottnt{T_{{\mathrm{1}}}}  \ottsym{)}  \vdash  \sigma  \ottsym{(}  \ottnt{e_{{\mathrm{2}}}}  \ottsym{)}  \ottsym{:}   \sigma  \ottsym{(}  \ottnt{T_{{\mathrm{2}}}}  \ottsym{)}  \,  \,\&\,  \,  \sigma  \ottsym{(}  \Phi_{{\mathrm{2}}}  \ottsym{)}  \, / \,   ( \forall  \mathit{y}  .  \sigma  \ottsym{(}  \ottnt{C_{{\mathrm{1}}}}  \ottsym{)}  )  \Rightarrow   \sigma  \ottsym{(}  \ottnt{C_{{\mathrm{3}}}}  \ottsym{)}  $ by \reflem{lr-typeability}
                        with
                        $\Gamma  \ottsym{,}  \mathit{x_{{\mathrm{1}}}} \,  {:}  \, \ottnt{T_{{\mathrm{1}}}}  \vdash  \ottnt{e_{{\mathrm{2}}}}  \ottsym{:}   \ottnt{T_{{\mathrm{2}}}}  \,  \,\&\,  \,  \Phi_{{\mathrm{2}}}  \, / \,   ( \forall  \mathit{y}  .  \ottnt{C_{{\mathrm{1}}}}  )  \Rightarrow   \ottnt{C_{{\mathrm{3}}}}  $ and
                        $\sigma \,  \in  \,  \LRRel{ \mathcal{G} }{ \Gamma } $; and
                  \item $\mathit{x_{{\mathrm{1}}}} \,  \not\in  \,  \mathit{fv}  (  \sigma  \ottsym{(}  \ottnt{T_{{\mathrm{2}}}}  \ottsym{)}  )  \,  \mathbin{\cup}  \,  \mathit{fv}  (  \sigma  \ottsym{(}  \Phi_{{\mathrm{2}}}  \ottsym{)}  )  \,  \mathbin{\cup}  \, \ottsym{(}   \mathit{fv}  (  \sigma  \ottsym{(}  \ottnt{C_{{\mathrm{1}}}}  \ottsym{)}  )  \,  \mathbin{\backslash}  \,  \{  \mathit{y}  \}   \ottsym{)}$.
                 \end{itemize}
                 Therefore, \reflem{lr-cont-compose-let} implies the conclusion
                 $ \ottnt{K}  [  \mathsf{let} \, \mathit{x_{{\mathrm{1}}}}  \ottsym{=}  \,[\,]\, \,  \mathsf{in}  \, \sigma  \ottsym{(}  \ottnt{e_{{\mathrm{2}}}}  \ottsym{)}  ]   \ottsym{:}   \sigma  \ottsym{(}  \ottnt{T_{{\mathrm{1}}}}  \ottsym{)}  \, / \, ( \forall  \mathit{x_{{\mathrm{1}}}}  .  \sigma  \ottsym{(}  \ottnt{C_{{\mathrm{3}}}}  \ottsym{)}  ) $.

           \item We show that $ \lambda\!  \, \mathit{z_{{\mathrm{1}}}}  \ottsym{.}   \resetcnstr{  \ottnt{K}  [  \mathsf{let} \, \mathit{x_{{\mathrm{1}}}}  \ottsym{=}  \mathit{z_{{\mathrm{1}}}} \,  \mathsf{in}  \, \sigma  \ottsym{(}  \ottnt{e_{{\mathrm{2}}}}  \ottsym{)}  ]  }  \,  \in  \,  \LRRel{ \mathcal{V} }{  \ottsym{(}  \mathit{z_{{\mathrm{1}}}} \,  {:}  \, \sigma  \ottsym{(}  \ottnt{T_{{\mathrm{1}}}}  \ottsym{)}  \ottsym{)}  \rightarrow  \sigma  \ottsym{(}  \ottnt{C_{{\mathrm{3}}}}  \ottsym{)}    [  \mathit{z_{{\mathrm{1}}}}  /  \mathit{x_{{\mathrm{1}}}}  ]   } $
                 for some fresh variable $\mathit{z_{{\mathrm{1}}}}$;
                 note the $\alpha$-renaming between $\mathit{x_{{\mathrm{1}}}}$ and $\mathit{z_{{\mathrm{1}}}}$, which is for avoiding the conflict of variable names.
                 Because
                 \begin{itemize}
                  \item $\mathit{x_{{\mathrm{1}}}} \,  {:}  \, \sigma  \ottsym{(}  \ottnt{T_{{\mathrm{1}}}}  \ottsym{)}  \vdash  \sigma  \ottsym{(}  \ottnt{e_{{\mathrm{2}}}}  \ottsym{)}  \ottsym{:}   \sigma  \ottsym{(}  \ottnt{T_{{\mathrm{2}}}}  \ottsym{)}  \,  \,\&\,  \,  \sigma  \ottsym{(}  \Phi_{{\mathrm{2}}}  \ottsym{)}  \, / \,   ( \forall  \mathit{y}  .  \sigma  \ottsym{(}  \ottnt{C_{{\mathrm{1}}}}  \ottsym{)}  )  \Rightarrow   \sigma  \ottsym{(}  \ottnt{C_{{\mathrm{3}}}}  \ottsym{)}  $ (shown above), and
                  \item $\mathit{z_{{\mathrm{1}}}} \,  {:}  \, \sigma  \ottsym{(}  \ottnt{T_{{\mathrm{1}}}}  \ottsym{)}  \vdash  \mathit{z_{{\mathrm{1}}}}  \ottsym{:}  \sigma  \ottsym{(}  \ottnt{T_{{\mathrm{1}}}}  \ottsym{)}$ by \reflem{ts-typing-vars},
                 \end{itemize}
                 \reflem{ts-weak-typing} and \T{LetV} imply
                 \[
                  \mathit{z_{{\mathrm{1}}}} \,  {:}  \, \sigma  \ottsym{(}  \ottnt{T_{{\mathrm{1}}}}  \ottsym{)}  \vdash  \mathsf{let} \, \mathit{x_{{\mathrm{1}}}}  \ottsym{=}  \mathit{z_{{\mathrm{1}}}} \,  \mathsf{in}  \, \sigma  \ottsym{(}  \ottnt{e_{{\mathrm{2}}}}  \ottsym{)}  \ottsym{:}    \sigma  \ottsym{(}  \ottnt{T_{{\mathrm{2}}}}  \ottsym{)}  \,  \,\&\,  \,  \sigma  \ottsym{(}  \Phi_{{\mathrm{2}}}  \ottsym{)}  \, / \,   ( \forall  \mathit{y}  .  \sigma  \ottsym{(}  \ottnt{C_{{\mathrm{1}}}}  \ottsym{)}  )  \Rightarrow   \sigma  \ottsym{(}  \ottnt{C_{{\mathrm{3}}}}  \ottsym{)}      [  \mathit{z_{{\mathrm{1}}}}  /  \mathit{x_{{\mathrm{1}}}}  ]   ~.
                 \]
                 Note that $\mathit{x_{{\mathrm{1}}}} \,  \not\in  \,  \mathit{fv}  (  \sigma  \ottsym{(}  \ottnt{T_{{\mathrm{2}}}}  \ottsym{)}  )  \,  \mathbin{\cup}  \,  \mathit{fv}  (  \sigma  \ottsym{(}  \Phi_{{\mathrm{2}}}  \ottsym{)}  )  \,  \mathbin{\cup}  \, \ottsym{(}   \mathit{fv}  (  \sigma  \ottsym{(}  \ottnt{C_{{\mathrm{1}}}}  \ottsym{)}  )  \,  \mathbin{\backslash}  \,  \{  \mathit{y}  \}   \ottsym{)}$.
                 Because $\ottnt{K}  \ottsym{:}   \sigma  \ottsym{(}  \ottnt{T_{{\mathrm{2}}}}  \ottsym{)}  \, / \, ( \forall  \mathit{y}  .  \sigma  \ottsym{(}  \ottnt{C_{{\mathrm{1}}}}  \ottsym{)}  ) $,
                 \reflem{lr-typing-cont-exp} implies
                 \[
                  \mathit{z_{{\mathrm{1}}}} \,  {:}  \, \sigma  \ottsym{(}  \ottnt{T_{{\mathrm{1}}}}  \ottsym{)}  \vdash   \resetcnstr{  \ottnt{K}  [  \mathsf{let} \, \mathit{x_{{\mathrm{1}}}}  \ottsym{=}  \mathit{z_{{\mathrm{1}}}} \,  \mathsf{in}  \, \sigma  \ottsym{(}  \ottnt{e_{{\mathrm{2}}}}  \ottsym{)}  ]  }   \ottsym{:}   \sigma  \ottsym{(}  \ottnt{C_{{\mathrm{3}}}}  \ottsym{)}    [  \mathit{z_{{\mathrm{1}}}}  /  \mathit{x_{{\mathrm{1}}}}  ]   ~.
                 \]
                 Therefore, by \T{Abs},
                 \[
                   \emptyset   \vdash   \lambda\!  \, \mathit{z_{{\mathrm{1}}}}  \ottsym{.}   \resetcnstr{  \ottnt{K}  [  \mathsf{let} \, \mathit{x_{{\mathrm{1}}}}  \ottsym{=}  \mathit{z_{{\mathrm{1}}}} \,  \mathsf{in}  \, \sigma  \ottsym{(}  \ottnt{e_{{\mathrm{2}}}}  \ottsym{)}  ]  }   \ottsym{:}   \ottsym{(}  \mathit{z_{{\mathrm{1}}}} \,  {:}  \, \sigma  \ottsym{(}  \ottnt{T_{{\mathrm{1}}}}  \ottsym{)}  \ottsym{)}  \rightarrow  \sigma  \ottsym{(}  \ottnt{C_{{\mathrm{3}}}}  \ottsym{)}    [  \mathit{z_{{\mathrm{1}}}}  /  \mathit{x_{{\mathrm{1}}}}  ]   ~,
                 \]
                 i.e., $ \lambda\!  \, \mathit{z_{{\mathrm{1}}}}  \ottsym{.}   \resetcnstr{  \ottnt{K}  [  \mathsf{let} \, \mathit{x_{{\mathrm{1}}}}  \ottsym{=}  \mathit{z_{{\mathrm{1}}}} \,  \mathsf{in}  \, \sigma  \ottsym{(}  \ottnt{e_{{\mathrm{2}}}}  \ottsym{)}  ]  }  \,  \in  \,  \LRRel{ \mathrm{Atom} }{  \ottsym{(}  \mathit{z_{{\mathrm{1}}}} \,  {:}  \, \sigma  \ottsym{(}  \ottnt{T_{{\mathrm{1}}}}  \ottsym{)}  \ottsym{)}  \rightarrow  \sigma  \ottsym{(}  \ottnt{C_{{\mathrm{3}}}}  \ottsym{)}    [  \mathit{z_{{\mathrm{1}}}}  /  \mathit{x_{{\mathrm{1}}}}  ]   } $.

                 Let $\ottnt{v} \,  \in  \,  \LRRel{ \mathcal{V} }{ \sigma  \ottsym{(}  \ottnt{T_{{\mathrm{1}}}}  \ottsym{)} } $.
                 Then, we must show that
                 \[
                  \ottsym{(}   \lambda\!  \, \mathit{z_{{\mathrm{1}}}}  \ottsym{.}   \resetcnstr{  \ottnt{K}  [  \mathsf{let} \, \mathit{x_{{\mathrm{1}}}}  \ottsym{=}  \mathit{z_{{\mathrm{1}}}} \,  \mathsf{in}  \, \sigma  \ottsym{(}  \ottnt{e_{{\mathrm{2}}}}  \ottsym{)}  ]  }   \ottsym{)} \, \ottnt{v} \,  \in  \,  \LRRel{ \mathcal{E} }{  \sigma  \ottsym{(}  \ottnt{C_{{\mathrm{3}}}}  \ottsym{)}    [  \ottnt{v}  /  \mathit{x_{{\mathrm{1}}}}  ]   }  ~,
                 \]
                 which is proven by Lemmas~\ref{lem:lr-anti-red} and \ref{lem:ts-subject-red-red} with the following.
                 \begin{itemize}
                  \item $\ottsym{(}   \lambda\!  \, \mathit{z_{{\mathrm{1}}}}  \ottsym{.}   \resetcnstr{  \ottnt{K}  [  \mathsf{let} \, \mathit{x_{{\mathrm{1}}}}  \ottsym{=}  \mathit{z_{{\mathrm{1}}}} \,  \mathsf{in}  \, \sigma  \ottsym{(}  \ottnt{e_{{\mathrm{2}}}}  \ottsym{)}  ]  }   \ottsym{)} \, \ottnt{v} \,  \in  \,  \LRRel{ \mathrm{Atom} }{  \sigma  \ottsym{(}  \ottnt{C_{{\mathrm{3}}}}  \ottsym{)}    [  \ottnt{v}  /  \mathit{x_{{\mathrm{1}}}}  ]   } $,
                        which is implied by \T{App} because
                        $ \emptyset   \vdash   \lambda\!  \, \mathit{z_{{\mathrm{1}}}}  \ottsym{.}   \resetcnstr{  \ottnt{K}  [  \mathsf{let} \, \mathit{x_{{\mathrm{1}}}}  \ottsym{=}  \mathit{z_{{\mathrm{1}}}} \,  \mathsf{in}  \, \sigma  \ottsym{(}  \ottnt{e_{{\mathrm{2}}}}  \ottsym{)}  ]  }   \ottsym{:}   \ottsym{(}  \mathit{z_{{\mathrm{1}}}} \,  {:}  \, \sigma  \ottsym{(}  \ottnt{T_{{\mathrm{1}}}}  \ottsym{)}  \ottsym{)}  \rightarrow  \sigma  \ottsym{(}  \ottnt{C_{{\mathrm{3}}}}  \ottsym{)}    [  \mathit{z_{{\mathrm{1}}}}  /  \mathit{x_{{\mathrm{1}}}}  ]  $ and
                        $ \emptyset   \vdash  \ottnt{v}  \ottsym{:}  \sigma  \ottsym{(}  \ottnt{T_{{\mathrm{1}}}}  \ottsym{)}$.
                        Note that $\mathit{z_{{\mathrm{1}}}} \,  \not\in  \,  \mathit{fv}  (  \sigma  \ottsym{(}  \ottnt{C_{{\mathrm{3}}}}  \ottsym{)}  ) $.
                  \item $\ottsym{(}   \lambda\!  \, \mathit{z_{{\mathrm{1}}}}  \ottsym{.}   \resetcnstr{  \ottnt{K}  [  \mathsf{let} \, \mathit{x_{{\mathrm{1}}}}  \ottsym{=}  \mathit{z_{{\mathrm{1}}}} \,  \mathsf{in}  \, \sigma  \ottsym{(}  \ottnt{e_{{\mathrm{2}}}}  \ottsym{)}  ]  }   \ottsym{)} \, \ottnt{v} \,  \mathrel{  \longrightarrow  ^*}  \,  \resetcnstr{  \ottnt{K}  [   \sigma  \ottsym{(}  \ottnt{e_{{\mathrm{2}}}}  \ottsym{)}    [  \ottnt{v}  /  \mathit{x_{{\mathrm{1}}}}  ]    ]  }  \,  \,\&\,  \,  \epsilon $
                        by \E{Red}/\R{Beta} and \E{Red}/\R{Let}.
                        Note that $\mathit{z_{{\mathrm{1}}}} \,  \not\in  \,  \mathit{fv}  (  \ottnt{K}  ) $.
                  \item $ \resetcnstr{  \ottnt{K}  [   \sigma  \ottsym{(}  \ottnt{e_{{\mathrm{2}}}}  \ottsym{)}    [  \ottnt{v}  /  \mathit{x_{{\mathrm{1}}}}  ]    ]  }  \,  \in  \,  \LRRel{ \mathcal{E} }{  \sigma  \ottsym{(}  \ottnt{C_{{\mathrm{3}}}}  \ottsym{)}    [  \ottnt{v}  /  \mathit{x_{{\mathrm{1}}}}  ]   } $,
                        which is implied by \reflem{lr-closed-reset} with the following.
                        \begin{itemize}
                         \item $ \sigma  \ottsym{(}  \ottnt{e_{{\mathrm{2}}}}  \ottsym{)}    [  \ottnt{v}  /  \mathit{x_{{\mathrm{1}}}}  ]   \,  \in  \,  \LRRel{ \mathcal{E} }{   \sigma  \ottsym{(}  \ottnt{T_{{\mathrm{2}}}}  \ottsym{)}  \,  \,\&\,  \,  \sigma  \ottsym{(}  \Phi_{{\mathrm{2}}}  \ottsym{)}  \, / \,   ( \forall  \mathit{y}  .  \sigma  \ottsym{(}  \ottnt{C_{{\mathrm{1}}}}  \ottsym{)}  )  \Rightarrow   \sigma  \ottsym{(}  \ottnt{C_{{\mathrm{3}}}}  \ottsym{)}      [  \ottnt{v}  /  \mathit{x_{{\mathrm{1}}}}  ]   } $
                               by:
                               $\ottnt{e_{{\mathrm{2}}}} \,  \in  \,  \LRRel{ \mathcal{O} }{ \Gamma  \ottsym{,}  \mathit{x_{{\mathrm{1}}}} \,  {:}  \, \ottnt{T_{{\mathrm{1}}}}  \, \vdash \,   \ottnt{T_{{\mathrm{2}}}}  \,  \,\&\,  \,  \Phi_{{\mathrm{2}}}  \, / \,  \ottnt{S_{{\mathrm{2}}}}  } $;
                               $\sigma  \mathbin{\uplus}   [  \ottnt{v}  /  \mathit{x_{{\mathrm{1}}}}  ]  \,  \in  \,  \LRRel{ \mathcal{G} }{ \Gamma  \ottsym{,}  \mathit{x_{{\mathrm{1}}}} \,  {:}  \, \ottnt{T_{{\mathrm{1}}}} } $
                               because $\sigma \,  \in  \,  \LRRel{ \mathcal{G} }{ \Gamma } $ and
                               $\ottnt{v} \,  \in  \,  \LRRel{ \mathcal{V} }{ \sigma  \ottsym{(}  \ottnt{T_{{\mathrm{1}}}}  \ottsym{)} } $; and
                               $\mathit{x_{{\mathrm{1}}}} \,  \not\in  \,  \mathit{fv}  (  \ottnt{T_{{\mathrm{2}}}}  )  \,  \mathbin{\cup}  \,  \mathit{fv}  (  \Phi_{{\mathrm{2}}}  )  \,  \mathbin{\cup}  \, \ottsym{(}   \mathit{fv}  (  \ottnt{C_{{\mathrm{1}}}}  )  \,  \mathbin{\backslash}  \,  \{  \mathit{y}  \}   \ottsym{)}$.
                         \item $\ottnt{K} \,  \in  \,  \LRRel{ \mathcal{K} }{  \sigma  \ottsym{(}  \ottnt{T_{{\mathrm{2}}}}  \ottsym{)}  \, / \, ( \forall  \mathit{y}  .  \sigma  \ottsym{(}  \ottnt{C_{{\mathrm{1}}}}  \ottsym{)}  )  } $.
                        \end{itemize}
                 \end{itemize}
          \end{itemize}
   \end{itemize}

  \case $  \exists  \, { \pi } . \  \ottsym{(}  \mathsf{let} \, \mathit{x_{{\mathrm{1}}}}  \ottsym{=}  \sigma  \ottsym{(}  \ottnt{e_{{\mathrm{1}}}}  \ottsym{)} \,  \mathsf{in}  \, \sigma  \ottsym{(}  \ottnt{e_{{\mathrm{2}}}}  \ottsym{)}  \ottsym{)} \,  \Uparrow  \, \pi $:
   We must show that
   $ { \models  \,    { \ottsym{(}  \sigma  \ottsym{(}  \Phi_{{\mathrm{1}}} \,  \tceappendop  \, \Phi_{{\mathrm{2}}}  \ottsym{)}  \ottsym{)} }^\nu   \ottsym{(}  \pi  \ottsym{)}    \wedge     { \ottsym{(}  \sigma  \ottsym{(}  \ottnt{S_{{\mathrm{1}}}}  \gg\!\!=  \ottsym{(}   \lambda\!  \, \mathit{x_{{\mathrm{1}}}}  \ottsym{.}  \ottnt{S_{{\mathrm{2}}}}  \ottsym{)}  \ottsym{)}  \ottsym{)} }^\nu (  \pi  )   } $.
   By case analysis on the result of Lemmas~\ref{lem:lr-red-under-ectx} and \ref{lem:lr-red-shift}
   with $\ottsym{(}  \mathsf{let} \, \mathit{x_{{\mathrm{1}}}}  \ottsym{=}  \sigma  \ottsym{(}  \ottnt{e_{{\mathrm{1}}}}  \ottsym{)} \,  \mathsf{in}  \, \sigma  \ottsym{(}  \ottnt{e_{{\mathrm{2}}}}  \ottsym{)}  \ottsym{)} \,  \Uparrow  \, \pi$.
   \begin{caseanalysis}
    \case $\sigma  \ottsym{(}  \ottnt{e_{{\mathrm{1}}}}  \ottsym{)} \,  \Uparrow  \, \pi$:
     As $\sigma  \ottsym{(}  \ottnt{e_{{\mathrm{1}}}}  \ottsym{)} \,  \in  \,  \LRRel{ \mathcal{E} }{ \sigma  \ottsym{(}   \ottnt{T_{{\mathrm{1}}}}  \,  \,\&\,  \,  \Phi_{{\mathrm{1}}}  \, / \,  \ottnt{S_{{\mathrm{1}}}}   \ottsym{)} } $,
     we have $ { \models  \,    { \ottsym{(}  \sigma  \ottsym{(}  \Phi_{{\mathrm{1}}}  \ottsym{)}  \ottsym{)} }^\nu   \ottsym{(}  \pi  \ottsym{)}    \wedge     { \ottsym{(}  \sigma  \ottsym{(}  \ottnt{S_{{\mathrm{1}}}}  \ottsym{)}  \ottsym{)} }^\nu (  \pi  )   } $.
     By \reflem{ts-valid-elim-intersect},
     $ { \models  \,   { \ottsym{(}  \sigma  \ottsym{(}  \Phi_{{\mathrm{1}}}  \ottsym{)}  \ottsym{)} }^\nu   \ottsym{(}  \pi  \ottsym{)} } $ and $ { \models  \,   { \ottsym{(}  \sigma  \ottsym{(}  \ottnt{S_{{\mathrm{1}}}}  \ottsym{)}  \ottsym{)} }^\nu (  \pi  )  } $.
     Then, Lemmas~\ref{lem:lr-rel-subst-ty-compose} and \ref{lem:ts-eff-compose-consist-trace-compose} imply
     $ { \models  \,   { \ottsym{(}  \sigma  \ottsym{(}  \Phi_{{\mathrm{1}}} \,  \tceappendop  \, \Phi_{{\mathrm{2}}}  \ottsym{)}  \ottsym{)} }^\nu   \ottsym{(}  \pi  \ottsym{)} } $.
     By \reflem{ts-valid-intro-intersect},
     it suffices to show that $ { \models  \,   { \ottsym{(}  \sigma  \ottsym{(}  \ottnt{S_{{\mathrm{1}}}}  \gg\!\!=  \ottsym{(}   \lambda\!  \, \mathit{x_{{\mathrm{1}}}}  \ottsym{.}  \ottnt{S_{{\mathrm{2}}}}  \ottsym{)}  \ottsym{)}  \ottsym{)} }^\nu (  \pi  )  } $.
     By case analysis on $\ottnt{S_{{\mathrm{1}}}}$.
     \begin{caseanalysis}
      \case $\ottnt{S_{{\mathrm{1}}}} \,  =  \,  \square $:
       By definition, $\ottnt{S_{{\mathrm{2}}}} \,  =  \,  \square $ and
       $\ottnt{S_{{\mathrm{1}}}}  \gg\!\!=  \ottsym{(}   \lambda\!  \, \mathit{x_{{\mathrm{1}}}}  \ottsym{.}  \ottnt{S_{{\mathrm{2}}}}  \ottsym{)} \,  =  \,  \square $.
       Therefore, we must show that $ { \models  \,   \top  } $, which is implied
       by \reflem{ts-valid-top}.

      \case $  \exists  \, { \ottnt{C_{{\mathrm{1}}}}  \ottsym{,}  \ottnt{C_{{\mathrm{2}}}} } . \  \ottnt{S_{{\mathrm{1}}}} \,  =  \,  ( \forall  \mathit{x_{{\mathrm{1}}}}  .  \ottnt{C_{{\mathrm{1}}}}  )  \Rightarrow   \ottnt{C_{{\mathrm{2}}}}  $:
       By definition,
       \begin{itemize}
        \item $\ottnt{S_{{\mathrm{2}}}} \,  =  \,  ( \forall  \mathit{y}  .  \ottnt{C_{{\mathrm{3}}}}  )  \Rightarrow   \ottnt{C_{{\mathrm{1}}}} $,
        \item $\ottnt{S_{{\mathrm{1}}}}  \gg\!\!=  \ottsym{(}   \lambda\!  \, \mathit{x_{{\mathrm{1}}}}  \ottsym{.}  \ottnt{S_{{\mathrm{2}}}}  \ottsym{)} \,  =  \,  ( \forall  \mathit{y}  .  \ottnt{C_{{\mathrm{3}}}}  )  \Rightarrow   \ottnt{C_{{\mathrm{2}}}} $, and
        \item $\mathit{x_{{\mathrm{1}}}} \,  \not\in  \,  \mathit{fv}  (  \ottnt{C_{{\mathrm{3}}}}  )  \,  \mathbin{\backslash}  \,  \{  \mathit{y}  \} $.
       \end{itemize}
       Because $ { \ottsym{(}  \sigma  \ottsym{(}   ( \forall  \mathit{y}  .  \ottnt{C_{{\mathrm{3}}}}  )  \Rightarrow   \ottnt{C_{{\mathrm{2}}}}   \ottsym{)}  \ottsym{)} }^\nu (  \pi  )  \,  =  \,  { \ottsym{(}  \sigma  \ottsym{(}  \ottnt{C_{{\mathrm{2}}}}  \ottsym{)}  \ottsym{)} }^\nu (  \pi  ) $,
       we must show that $ { \models  \,   { \ottsym{(}  \sigma  \ottsym{(}  \ottnt{C_{{\mathrm{2}}}}  \ottsym{)}  \ottsym{)} }^\nu (  \pi  )  } $,
       which is implied by $ { \models  \,   { \ottsym{(}  \sigma  \ottsym{(}  \ottnt{S_{{\mathrm{1}}}}  \ottsym{)}  \ottsym{)} }^\nu (  \pi  )  } $
       because $ { \ottsym{(}  \sigma  \ottsym{(}  \ottnt{S_{{\mathrm{1}}}}  \ottsym{)}  \ottsym{)} }^\nu (  \pi  )  \,  =  \,  { \ottsym{(}  \sigma  \ottsym{(}  \ottnt{C_{{\mathrm{2}}}}  \ottsym{)}  \ottsym{)} }^\nu (  \pi  ) $.
     \end{caseanalysis}

    \case $ {  {   \exists  \, { \ottnt{v_{{\mathrm{1}}}}  \ottsym{,}  \varpi_{{\mathrm{1}}}  \ottsym{,}  \pi_{{\mathrm{2}}} } . \  \sigma  \ottsym{(}  \ottnt{e_{{\mathrm{1}}}}  \ottsym{)} \,  \Downarrow  \, \ottnt{v_{{\mathrm{1}}}} \,  \,\&\,  \, \varpi_{{\mathrm{1}}}  } \,\mathrel{\wedge}\, { \ottsym{(}  \mathsf{let} \, \mathit{x_{{\mathrm{1}}}}  \ottsym{=}  \ottnt{v_{{\mathrm{1}}}} \,  \mathsf{in}  \, \sigma  \ottsym{(}  \ottnt{e_{{\mathrm{2}}}}  \ottsym{)}  \ottsym{)} \,  \Uparrow  \, \pi_{{\mathrm{2}}} }  } \,\mathrel{\wedge}\, { \pi \,  =  \, \varpi_{{\mathrm{1}}} \,  \tceappendop  \, \pi_{{\mathrm{2}}} } $:
     By \reflem{lr-red-decomp-div} and \E{Red}/\R{Let},
     we have $ \sigma  \ottsym{(}  \ottnt{e_{{\mathrm{2}}}}  \ottsym{)}    [  \ottnt{v_{{\mathrm{1}}}}  /  \mathit{x_{{\mathrm{1}}}}  ]   \,  \Uparrow  \, \pi_{{\mathrm{2}}}$.
     Because $\sigma  \ottsym{(}  \ottnt{e_{{\mathrm{1}}}}  \ottsym{)} \,  \in  \,  \LRRel{ \mathcal{E} }{ \sigma  \ottsym{(}   \ottnt{T_{{\mathrm{1}}}}  \,  \,\&\,  \,  \Phi_{{\mathrm{1}}}  \, / \,  \ottnt{S_{{\mathrm{1}}}}   \ottsym{)} } $,
     $\ottnt{v_{{\mathrm{1}}}} \,  \in  \,  \LRRel{ \mathcal{V} }{ \sigma  \ottsym{(}  \ottnt{T_{{\mathrm{1}}}}  \ottsym{)} } $.
     By Lemmas~\ref{lem:ts-type-sound-terminating} and \ref{lem:lr-red-val},
     \begin{itemize}
      \item $ \emptyset   \vdash  \ottnt{v_{{\mathrm{1}}}}  \ottsym{:}  \ottnt{T'_{{\mathrm{1}}}}$,
      \item $ \emptyset   \vdash   \ottnt{T'_{{\mathrm{1}}}}  \,  \,\&\,  \,   (  \varpi_{{\mathrm{1}}}  , \bot )   \, / \,   \square    \ottsym{<:}  \sigma  \ottsym{(}   \ottnt{T_{{\mathrm{1}}}}  \,  \,\&\,  \,  \Phi_{{\mathrm{1}}}  \, / \,  \ottnt{S_{{\mathrm{1}}}}   \ottsym{)}$
     \end{itemize}
     for some $\ottnt{T'_{{\mathrm{1}}}}$.
     The last subtyping derivation implies
     \begin{itemize}
      \item $ \emptyset   \vdash  \ottnt{T'_{{\mathrm{1}}}}  \ottsym{<:}  \sigma  \ottsym{(}  \ottnt{T_{{\mathrm{1}}}}  \ottsym{)}$,
      \item $ \emptyset   \vdash   (  \varpi_{{\mathrm{1}}}  , \bot )   \ottsym{<:}  \sigma  \ottsym{(}  \Phi_{{\mathrm{1}}}  \ottsym{)}$, and
      \item $ \emptyset  \,  |  \, \ottnt{T'_{{\mathrm{1}}}} \,  \,\&\,  \,  (  \varpi_{{\mathrm{1}}}  , \bot )   \vdash   \square   \ottsym{<:}  \sigma  \ottsym{(}  \ottnt{S_{{\mathrm{1}}}}  \ottsym{)}$.
     \end{itemize}
     Because $\sigma \,  \in  \,  \LRRel{ \mathcal{G} }{ \Gamma } $ and $\ottnt{v_{{\mathrm{1}}}} \,  \in  \,  \LRRel{ \mathcal{V} }{ \sigma  \ottsym{(}  \ottnt{T_{{\mathrm{1}}}}  \ottsym{)} } $,
     we have $\sigma  \mathbin{\uplus}   [  \ottnt{v_{{\mathrm{1}}}}  /  \mathit{x_{{\mathrm{1}}}}  ]  \,  \in  \,  \LRRel{ \mathcal{G} }{ \Gamma  \ottsym{,}  \mathit{x_{{\mathrm{1}}}} \,  {:}  \, \ottnt{T_{{\mathrm{1}}}} } $.
     Because $\ottnt{e_{{\mathrm{2}}}} \,  \in  \,  \LRRel{ \mathcal{O} }{ \Gamma  \ottsym{,}  \mathit{x_{{\mathrm{1}}}} \,  {:}  \, \ottnt{T_{{\mathrm{1}}}}  \, \vdash \,   \ottnt{T_{{\mathrm{2}}}}  \,  \,\&\,  \,  \Phi_{{\mathrm{2}}}  \, / \,  \ottnt{S_{{\mathrm{2}}}}  } $ and
     $\mathit{x_{{\mathrm{1}}}} \,  \not\in  \,  \mathit{fv}  (  \ottnt{T_{{\mathrm{2}}}}  )  \,  \mathbin{\cup}  \,  \mathit{fv}  (  \Phi_{{\mathrm{2}}}  ) $,
     we have $ \sigma  \ottsym{(}  \ottnt{e_{{\mathrm{2}}}}  \ottsym{)}    [  \ottnt{v_{{\mathrm{1}}}}  /  \mathit{x_{{\mathrm{1}}}}  ]   \,  \in  \,  \LRRel{ \mathcal{E} }{  \sigma  \ottsym{(}  \ottnt{T_{{\mathrm{2}}}}  \ottsym{)}  \,  \,\&\,  \,  \sigma  \ottsym{(}  \Phi_{{\mathrm{2}}}  \ottsym{)}  \, / \,   \sigma  \ottsym{(}  \ottnt{S_{{\mathrm{2}}}}  \ottsym{)}    [  \ottnt{v_{{\mathrm{1}}}}  /  \mathit{x_{{\mathrm{1}}}}  ]    } $.
     Because $ \sigma  \ottsym{(}  \ottnt{e_{{\mathrm{2}}}}  \ottsym{)}    [  \ottnt{v_{{\mathrm{1}}}}  /  \mathit{x_{{\mathrm{1}}}}  ]   \,  \Uparrow  \, \pi_{{\mathrm{2}}}$,
     we have $ { \models  \,    { \ottsym{(}  \sigma  \ottsym{(}  \Phi_{{\mathrm{2}}}  \ottsym{)}  \ottsym{)} }^\nu   \ottsym{(}  \pi_{{\mathrm{2}}}  \ottsym{)}    \wedge     { \ottsym{(}   \sigma  \ottsym{(}  \ottnt{S_{{\mathrm{2}}}}  \ottsym{)}    [  \ottnt{v_{{\mathrm{1}}}}  /  \mathit{x_{{\mathrm{1}}}}  ]    \ottsym{)} }^\nu (  \pi_{{\mathrm{2}}}  )   } $.
     By \reflem{ts-valid-elim-intersect},
     $ { \models  \,   { \ottsym{(}  \sigma  \ottsym{(}  \Phi_{{\mathrm{2}}}  \ottsym{)}  \ottsym{)} }^\nu   \ottsym{(}  \pi_{{\mathrm{2}}}  \ottsym{)} } $ and
     $ { \models  \,   { \ottsym{(}   \sigma  \ottsym{(}  \ottnt{S_{{\mathrm{2}}}}  \ottsym{)}    [  \ottnt{v_{{\mathrm{1}}}}  /  \mathit{x_{{\mathrm{1}}}}  ]    \ottsym{)} }^\nu (  \pi_{{\mathrm{2}}}  )  } $.
     We have the conclusion by \reflem{ts-valid-intro-intersect}
     with the following.
     \begin{itemize}
      \item We show that $ { \models  \,   { \ottsym{(}  \sigma  \ottsym{(}  \Phi_{{\mathrm{1}}} \,  \tceappendop  \, \Phi_{{\mathrm{2}}}  \ottsym{)}  \ottsym{)} }^\nu   \ottsym{(}  \pi  \ottsym{)} } $, which is
            implied by Lemmas~\ref{lem:ts-eff-compose-consist-trace-compose} and \ref{lem:lr-rel-subst-ty-compose}
            with $ \pi    =    \varpi_{{\mathrm{1}}} \,  \tceappendop  \, \pi_{{\mathrm{2}}} $ and the following.
            \begin{itemize}
             \item We show that $ { \models  \,   { \sigma  \ottsym{(}  \Phi_{{\mathrm{1}}}  \ottsym{)} }^\mu   \ottsym{(}  \varpi_{{\mathrm{1}}}  \ottsym{)} } $.
                   Because $ \emptyset   \vdash   (  \varpi_{{\mathrm{1}}}  , \bot )   \ottsym{<:}  \sigma  \ottsym{(}  \Phi_{{\mathrm{1}}}  \ottsym{)}$,
                   we have $\mathit{x} \,  {:}  \,  \Sigma ^\ast   \models    { \ottsym{(}   (  \varpi_{{\mathrm{1}}}  , \bot )   \ottsym{)} }^\mu   \ottsym{(}  \mathit{x}  \ottsym{)}    \Rightarrow     { \ottsym{(}  \sigma  \ottsym{(}  \Phi_{{\mathrm{1}}}  \ottsym{)}  \ottsym{)} }^\mu   \ottsym{(}  \mathit{x}  \ottsym{)} $.
                   By \reflem{ts-term-subst-valid},
                   $ { \models  \,    { \ottsym{(}   (  \varpi_{{\mathrm{1}}}  , \bot )   \ottsym{)} }^\mu   \ottsym{(}  \varpi_{{\mathrm{1}}}  \ottsym{)}    \Rightarrow     { \ottsym{(}  \sigma  \ottsym{(}  \Phi_{{\mathrm{1}}}  \ottsym{)}  \ottsym{)} }^\mu   \ottsym{(}  \varpi_{{\mathrm{1}}}  \ottsym{)}  } $.
                   Then, Lemmas~\ref{lem:ts-refl-const-eq} and \ref{lem:ts-discharge-assumption-valid}
                   imply the conclusion $ { \models  \,   { \ottsym{(}  \sigma  \ottsym{(}  \Phi_{{\mathrm{1}}}  \ottsym{)}  \ottsym{)} }^\mu   \ottsym{(}  \varpi_{{\mathrm{1}}}  \ottsym{)} } $.
             \item We show that $ { \models  \,   { \ottsym{(}  \sigma  \ottsym{(}  \Phi_{{\mathrm{2}}}  \ottsym{)}  \ottsym{)} }^\nu   \ottsym{(}  \pi_{{\mathrm{2}}}  \ottsym{)} } $, which has been obtained.
            \end{itemize}

      \item We show that $ { \models  \,   { \ottsym{(}  \sigma  \ottsym{(}  \ottnt{S_{{\mathrm{1}}}}  \gg\!\!=  \ottsym{(}   \lambda\!  \, \mathit{x_{{\mathrm{1}}}}  \ottsym{.}  \ottnt{S_{{\mathrm{2}}}}  \ottsym{)}  \ottsym{)}  \ottsym{)} }^\nu (  \pi  )  } $.
            By case analysis on $\ottnt{S_{{\mathrm{1}}}}  \gg\!\!=  \ottsym{(}   \lambda\!  \, \mathit{x_{{\mathrm{1}}}}  \ottsym{.}  \ottnt{S_{{\mathrm{2}}}}  \ottsym{)}$.
            \begin{caseanalysis}
             \case $\ottnt{S_{{\mathrm{1}}}}  \gg\!\!=  \ottsym{(}   \lambda\!  \, \mathit{x_{{\mathrm{1}}}}  \ottsym{.}  \ottnt{S_{{\mathrm{2}}}}  \ottsym{)} \,  =  \,  \square $:
              We must prove $ { \models  \,   \top  } $, which is implied by \reflem{ts-valid-top}.

             \case $  \exists  \, { \mathit{y}  \ottsym{,}  \ottnt{C_{{\mathrm{1}}}}  \ottsym{,}  \ottnt{C_{{\mathrm{2}}}} } . \  \ottnt{S_{{\mathrm{1}}}}  \gg\!\!=  \ottsym{(}   \lambda\!  \, \mathit{x_{{\mathrm{1}}}}  \ottsym{.}  \ottnt{S_{{\mathrm{2}}}}  \ottsym{)} \,  =  \,  ( \forall  \mathit{y}  .  \ottnt{C_{{\mathrm{1}}}}  )  \Rightarrow   \ottnt{C_{{\mathrm{2}}}}  $:
              We must prove $ { \models  \,   { \ottsym{(}  \sigma  \ottsym{(}  \ottnt{C_{{\mathrm{2}}}}  \ottsym{)}  \ottsym{)} }^\nu (  \pi  )  } $.
              By definition,
              \begin{itemize}
               \item $\ottnt{S_{{\mathrm{1}}}} \,  =  \,  ( \forall  \mathit{x_{{\mathrm{1}}}}  .  \ottnt{C_{{\mathrm{3}}}}  )  \Rightarrow   \ottnt{C_{{\mathrm{2}}}} $,
               \item $\ottnt{S_{{\mathrm{2}}}} \,  =  \,  ( \forall  \mathit{y}  .  \ottnt{C_{{\mathrm{1}}}}  )  \Rightarrow   \ottnt{C_{{\mathrm{3}}}} $, and
               \item $\mathit{x_{{\mathrm{1}}}} \,  \not\in  \,  \mathit{fv}  (  \ottnt{C_{{\mathrm{1}}}}  )  \,  \mathbin{\backslash}  \,  \{  \mathit{y}  \} $
              \end{itemize}
              for some $\ottnt{C_{{\mathrm{3}}}}$.
              Because $ \emptyset  \,  |  \, \ottnt{T'_{{\mathrm{1}}}} \,  \,\&\,  \,  (  \varpi_{{\mathrm{1}}}  , \bot )   \vdash   \square   \ottsym{<:}  \sigma  \ottsym{(}  \ottnt{S_{{\mathrm{1}}}}  \ottsym{)}$,
              we have
              \begin{itemize}
               \item $\mathit{x_{{\mathrm{1}}}} \,  {:}  \, \ottnt{T'_{{\mathrm{1}}}}  \vdash   (  \varpi_{{\mathrm{1}}}  , \bot )  \,  \tceappendop  \, \sigma  \ottsym{(}  \ottnt{C_{{\mathrm{3}}}}  \ottsym{)}  \ottsym{<:}  \sigma  \ottsym{(}  \ottnt{C_{{\mathrm{2}}}}  \ottsym{)}$ and
               \item $\mathit{x_{{\mathrm{1}}}} \,  \not\in  \,  \mathit{fv}  (  \sigma  \ottsym{(}  \ottnt{C_{{\mathrm{2}}}}  \ottsym{)}  ) $.
              \end{itemize}
              Because $ \emptyset   \vdash  \ottnt{v_{{\mathrm{1}}}}  \ottsym{:}  \ottnt{T'_{{\mathrm{1}}}}$,
              Lemmas~\ref{lem:ts-val-subst-subtyping} and \ref{lem:ts-val-subst-ty-compose} imply
              $ \emptyset   \vdash   (  \varpi_{{\mathrm{1}}}  , \bot )  \,  \tceappendop  \, \ottsym{(}   \sigma  \ottsym{(}  \ottnt{C_{{\mathrm{3}}}}  \ottsym{)}    [  \ottnt{v_{{\mathrm{1}}}}  /  \mathit{x_{{\mathrm{1}}}}  ]    \ottsym{)}  \ottsym{<:}  \sigma  \ottsym{(}  \ottnt{C_{{\mathrm{2}}}}  \ottsym{)}$.
              Because $ { \models  \,   { \ottsym{(}   \sigma  \ottsym{(}  \ottnt{S_{{\mathrm{2}}}}  \ottsym{)}    [  \ottnt{v_{{\mathrm{1}}}}  /  \mathit{x_{{\mathrm{1}}}}  ]    \ottsym{)} }^\nu (  \pi_{{\mathrm{2}}}  )  } $,
              we have $ { \models  \,   { \ottsym{(}   \sigma  \ottsym{(}  \ottnt{C_{{\mathrm{3}}}}  \ottsym{)}    [  \ottnt{v_{{\mathrm{1}}}}  /  \mathit{x_{{\mathrm{1}}}}  ]    \ottsym{)} }^\nu (  \pi_{{\mathrm{2}}}  )  } $.
              Because $ { \models  \,   { \ottsym{(}   (  \varpi_{{\mathrm{1}}}  , \bot )   \ottsym{)} }^\mu   \ottsym{(}  \varpi_{{\mathrm{1}}}  \ottsym{)} } $ by \reflem{ts-refl-const-eq}, and $ \varpi_{{\mathrm{1}}} \,  \tceappendop  \, \pi_{{\mathrm{2}}}    =    \pi $,
              \reflem{lr-eff-cty-compose-consist-trace-compose}
              implies
              \[
                { \models  \,   { \ottsym{(}   (  \varpi_{{\mathrm{1}}}  , \bot )  \,  \tceappendop  \, \ottsym{(}   \sigma  \ottsym{(}  \ottnt{C_{{\mathrm{3}}}}  \ottsym{)}    [  \ottnt{v_{{\mathrm{1}}}}  /  \mathit{x_{{\mathrm{1}}}}  ]    \ottsym{)}  \ottsym{)} }^\nu (  \pi  )  }  ~.
              \]
              \reflem{ts-extract-subst-val-valid} implies
              $\mathit{y} \,  {:}  \,  \Sigma ^\omega   \models   \ottsym{(}   \pi    =    \mathit{y}   \ottsym{)}    \Rightarrow     { \ottsym{(}   (  \varpi_{{\mathrm{1}}}  , \bot )  \,  \tceappendop  \, \ottsym{(}   \sigma  \ottsym{(}  \ottnt{C_{{\mathrm{3}}}}  \ottsym{)}    [  \ottnt{v_{{\mathrm{1}}}}  /  \mathit{x_{{\mathrm{1}}}}  ]    \ottsym{)}  \ottsym{)} }^\nu (  \mathit{y}  )  $.
              Because $ \emptyset   \vdash   (  \varpi_{{\mathrm{1}}}  , \bot )  \,  \tceappendop  \, \ottsym{(}   \sigma  \ottsym{(}  \ottnt{C_{{\mathrm{3}}}}  \ottsym{)}    [  \ottnt{v_{{\mathrm{1}}}}  /  \mathit{x_{{\mathrm{1}}}}  ]    \ottsym{)}  \ottsym{<:}  \sigma  \ottsym{(}  \ottnt{C_{{\mathrm{2}}}}  \ottsym{)}$,
              \reflem{lr-subtyping-nu} implies
              $\mathit{y} \,  {:}  \,  \Sigma ^\omega   \models   \ottsym{(}   \pi    =    \mathit{y}   \ottsym{)}    \Rightarrow     { \ottsym{(}  \sigma  \ottsym{(}  \ottnt{C_{{\mathrm{2}}}}  \ottsym{)}  \ottsym{)} }^\nu (  \mathit{y}  )  $.
              By \reflem{ts-term-subst-valid},
              $ { \models  \,   \ottsym{(}   \pi    =    \pi   \ottsym{)}    \Rightarrow     { \ottsym{(}  \sigma  \ottsym{(}  \ottnt{C_{{\mathrm{2}}}}  \ottsym{)}  \ottsym{)} }^\nu (  \pi  )   } $.
              Because $ { \models  \,   \pi    =    \pi  } $ by \reflem{ts-refl-const-eq},
              \reflem{ts-discharge-assumption-valid} implies the conclusion
              $ { \models  \,   { \ottsym{(}  \sigma  \ottsym{(}  \ottnt{C_{{\mathrm{2}}}}  \ottsym{)}  \ottsym{)} }^\nu (  \pi  )  } $.
            \end{caseanalysis}
     \end{itemize}
   \end{caseanalysis}
 \end{caseanalysis}
\end{proof}

\begin{lemmap}{Compatibility: Let for Values}{lr-compat-let-val}
 If
 $\ottnt{v_{{\mathrm{1}}}} \,  \in  \,  \LRRel{ \mathcal{O} }{ \Gamma  \, \vdash \,  \ottnt{T_{{\mathrm{1}}}} } $ and
 $\ottnt{e_{{\mathrm{2}}}} \,  \in  \,  \LRRel{ \mathcal{O} }{ \Gamma  \ottsym{,}  \mathit{x_{{\mathrm{1}}}} \,  {:}  \, \ottnt{T_{{\mathrm{1}}}}  \, \vdash \,  \ottnt{C_{{\mathrm{2}}}} } $,
 then
 $\mathsf{let} \, \mathit{x_{{\mathrm{1}}}}  \ottsym{=}  \ottnt{v_{{\mathrm{1}}}} \,  \mathsf{in}  \, \ottnt{e_{{\mathrm{2}}}} \,  \in  \,  \LRRel{ \mathcal{O} }{ \Gamma  \, \vdash \,   \ottnt{C_{{\mathrm{2}}}}    [  \ottnt{v_{{\mathrm{1}}}}  /  \mathit{x_{{\mathrm{1}}}}  ]   } $.
\end{lemmap}
\begin{proof}
 By \T{LetV}, we have
 $\Gamma  \vdash  \mathsf{let} \, \mathit{x_{{\mathrm{1}}}}  \ottsym{=}  \ottnt{v_{{\mathrm{1}}}} \,  \mathsf{in}  \, \ottnt{e_{{\mathrm{2}}}}  \ottsym{:}   \ottnt{C_{{\mathrm{2}}}}    [  \ottnt{v_{{\mathrm{1}}}}  /  \mathit{x_{{\mathrm{1}}}}  ]  $.
 Let $\sigma \,  \in  \,  \LRRel{ \mathcal{G} }{ \Gamma } $.
 Without loss of generality, we can suppose that $\mathit{x_{{\mathrm{1}}}}$ does not occur in $\sigma$.
 We must show that
 \[
  \mathsf{let} \, \mathit{x_{{\mathrm{1}}}}  \ottsym{=}  \sigma  \ottsym{(}  \ottnt{v_{{\mathrm{1}}}}  \ottsym{)} \,  \mathsf{in}  \, \sigma  \ottsym{(}  \ottnt{e_{{\mathrm{2}}}}  \ottsym{)} \,  \in  \,  \LRRel{ \mathcal{E} }{ \sigma  \ottsym{(}   \ottnt{C_{{\mathrm{2}}}}    [  \ottnt{v_{{\mathrm{1}}}}  /  \mathit{x_{{\mathrm{1}}}}  ]    \ottsym{)} }  ~.
 \]
 We have $\mathsf{let} \, \mathit{x_{{\mathrm{1}}}}  \ottsym{=}  \sigma  \ottsym{(}  \ottnt{v_{{\mathrm{1}}}}  \ottsym{)} \,  \mathsf{in}  \, \sigma  \ottsym{(}  \ottnt{e_{{\mathrm{2}}}}  \ottsym{)} \,  \in  \,  \LRRel{ \mathrm{Atom} }{ \sigma  \ottsym{(}   \ottnt{C_{{\mathrm{2}}}}    [  \ottnt{v_{{\mathrm{1}}}}  /  \mathit{x_{{\mathrm{1}}}}  ]    \ottsym{)} } $
 by \reflem{lr-typeability}.
 Because $\mathsf{let} \, \mathit{x_{{\mathrm{1}}}}  \ottsym{=}  \sigma  \ottsym{(}  \ottnt{v_{{\mathrm{1}}}}  \ottsym{)} \,  \mathsf{in}  \, \sigma  \ottsym{(}  \ottnt{e_{{\mathrm{2}}}}  \ottsym{)} \,  \mathrel{  \longrightarrow  }  \,  \sigma  \ottsym{(}  \ottnt{e_{{\mathrm{2}}}}  \ottsym{)}    [  \sigma  \ottsym{(}  \ottnt{v_{{\mathrm{1}}}}  \ottsym{)}  /  \mathit{x_{{\mathrm{1}}}}  ]   \,  \,\&\,  \,  \epsilon $
 by \E{Red}/\R{Let},
 \reflem{lr-anti-red} implies that it suffices to show that
 \[
   \sigma  \ottsym{(}  \ottnt{e_{{\mathrm{2}}}}  \ottsym{)}    [  \sigma  \ottsym{(}  \ottnt{v_{{\mathrm{1}}}}  \ottsym{)}  /  \mathit{x_{{\mathrm{1}}}}  ]   \,  \in  \,  \LRRel{ \mathcal{E} }{ \sigma  \ottsym{(}   \ottnt{C_{{\mathrm{2}}}}    [  \ottnt{v_{{\mathrm{1}}}}  /  \mathit{x_{{\mathrm{1}}}}  ]    \ottsym{)} }  ~,
 \]
 which is implied by
 $\ottnt{e_{{\mathrm{2}}}} \,  \in  \,  \LRRel{ \mathcal{O} }{ \Gamma  \ottsym{,}  \mathit{x_{{\mathrm{1}}}} \,  {:}  \, \ottnt{T_{{\mathrm{1}}}}  \, \vdash \,  \ottnt{C_{{\mathrm{2}}}} } $ and $\ottnt{v_{{\mathrm{1}}}} \,  \in  \,  \LRRel{ \mathcal{O} }{ \Gamma  \, \vdash \,  \ottnt{T_{{\mathrm{1}}}} } $ and $\sigma \,  \in  \,  \LRRel{ \mathcal{G} }{ \Gamma } $.
\end{proof}

\begin{lemmap}{Approximation Decomposition}{lr-approx-decompose}
 \begin{enumerate}
  \item $\ottnt{e} \,  \prec_{ \ottnt{v} }  \, \ottnt{e'} \,  \uparrow \, (  \ottmv{i}  \ottsym{+}  \ottmv{j}  ,  \pi  )  \,  =  \, \ottnt{e} \,  \prec_{ \ottnt{v} }  \, \ottsym{(}  \ottnt{e} \,  \prec_{ \ottnt{v} }  \, \ottnt{e'} \,  \uparrow \, (  \ottmv{i}  ,  \pi_{{\mathrm{0}}}  )   \ottsym{)} \,  \uparrow \, (  \ottmv{j}  ,  \pi  ) $.
  \item $\ottnt{A} \,  \prec_{ \ottnt{v} }  \, \ottnt{A'} \,  \uparrow \, (  \ottmv{i}  \ottsym{+}  \ottmv{j}  ,  \pi  )  \,  =  \, \ottnt{A} \,  \prec_{ \ottnt{v} }  \, \ottsym{(}  \ottnt{A} \,  \prec_{ \ottnt{v} }  \, \ottnt{A'} \,  \uparrow \, (  \ottmv{i}  ,  \pi_{{\mathrm{0}}}  )   \ottsym{)} \,  \uparrow \, (  \ottmv{j}  ,  \pi  ) $.
  \item $\phi \,  \prec_{ \ottnt{v} }  \, \phi' \,  \uparrow \, (  \ottmv{i}  \ottsym{+}  \ottmv{j}  ,  \pi  )  \,  =  \, \phi \,  \prec_{ \ottnt{v} }  \, \ottsym{(}  \phi \,  \prec_{ \ottnt{v} }  \, \phi' \,  \uparrow \, (  \ottmv{i}  ,  \pi_{{\mathrm{0}}}  )   \ottsym{)} \,  \uparrow \, (  \ottmv{j}  ,  \pi  ) $.
 \end{enumerate}
\end{lemmap}
\begin{proof}
 \TS{OK: 2022/01/17}
 Straightforward by simultaneous induction on $\ottnt{e}$, $\ottnt{A}$, and $\phi$.
\end{proof}

\begin{lemmap}{Approximation Consistent with Value Substitution}{lr-approx-val-subst}
 \begin{enumerate}
  \item $ \ottsym{(}  \ottnt{e} \,  \prec_{ \ottnt{v_{{\mathrm{0}}}} }  \, \ottnt{e'} \,  \uparrow \, (  \ottmv{i}  ,  \pi  )   \ottsym{)}    [  \ottnt{v} \,  \prec_{ \ottnt{v_{{\mathrm{0}}}} }  \, \ottnt{v'} \,  \uparrow \, (  \ottmv{i}  ,  \pi  )   /  \mathit{x}  ]   \,  =  \,  \ottnt{e}    [  \ottnt{v}  /  \mathit{x}  ]   \,  \prec_{ \ottnt{v_{{\mathrm{0}}}} }  \,  \ottnt{e'}    [  \ottnt{v'}  /  \mathit{x}  ]   \,  \uparrow \, (  \ottmv{i}  ,  \pi  ) $.
  \item $ \ottsym{(}  \ottnt{A} \,  \prec_{ \ottnt{v_{{\mathrm{0}}}} }  \, \ottnt{A'} \,  \uparrow \, (  \ottmv{i}  ,  \pi  )   \ottsym{)}    [  \ottnt{v} \,  \prec_{ \ottnt{v_{{\mathrm{0}}}} }  \, \ottnt{v'} \,  \uparrow \, (  \ottmv{i}  ,  \pi  )   /  \mathit{x}  ]   \,  =  \,  \ottnt{A}    [  \ottnt{v}  /  \mathit{x}  ]   \,  \prec_{ \ottnt{v_{{\mathrm{0}}}} }  \,  \ottnt{A'}    [  \ottnt{v'}  /  \mathit{x}  ]   \,  \uparrow \, (  \ottmv{i}  ,  \pi  ) $.
  \item $ \ottsym{(}  \phi \,  \prec_{ \ottnt{v_{{\mathrm{0}}}} }  \, \phi' \,  \uparrow \, (  \ottmv{i}  ,  \pi  )   \ottsym{)}    [  \ottnt{v} \,  \prec_{ \ottnt{v_{{\mathrm{0}}}} }  \, \ottnt{v'} \,  \uparrow \, (  \ottmv{i}  ,  \pi  )   /  \mathit{x}  ]   \,  =  \,  \phi    [  \ottnt{v}  /  \mathit{x}  ]   \,  \prec_{ \ottnt{v_{{\mathrm{0}}}} }  \,  \phi'    [  \ottnt{v'}  /  \mathit{x}  ]   \,  \uparrow \, (  \ottmv{i}  ,  \pi  ) $.
 \end{enumerate}
\end{lemmap}
\begin{proof}
 \TS{OK: 2022/01/14}
 By simultaneous induction on $\ottnt{e}$, $\ottnt{A}$, and $\phi$.
 We mention only the interesting cases.
 \begin{caseanalysis}
  \case $ { \ottnt{e} \,  =  \, \ottnt{v_{{\mathrm{0}}}} } \,\mathrel{\wedge}\, {   \exists  \, { \ottmv{j}  \ottsym{,}  \pi_{{\mathrm{0}}} } . \  \ottnt{e'} \,  =  \,  \ottnt{v_{{\mathrm{0}}}} _{  \mu   \ottsym{;}  \ottmv{j}  \ottsym{;}  \pi_{{\mathrm{0}}} }   } $:
   We are given $\ottnt{e} \,  \prec_{ \ottnt{v_{{\mathrm{0}}}} }  \, \ottnt{e'} \,  \uparrow \, (  \ottmv{i}  ,  \pi  )  \,  =  \,  \ottnt{v_{{\mathrm{0}}}} _{  \mu   \ottsym{;}  \ottmv{j}  \ottsym{+}  \ottmv{i}  \ottsym{;}  \pi } $.
   By definition, $\ottnt{v_{{\mathrm{0}}}}$ is closed, which further implies that
   $ \ottnt{v_{{\mathrm{0}}}} _{  \mu   \ottsym{;}  \ottmv{j}  \ottsym{;}  \pi_{{\mathrm{0}}} } $ and $ \ottnt{v_{{\mathrm{0}}}} _{  \mu   \ottsym{;}  \ottmv{j}  \ottsym{+}  \ottmv{i}  \ottsym{;}  \pi } $ are closed.
   Therefore:
   \[\begin{array}{ll} &
     \ottsym{(}  \ottnt{e} \,  \prec_{ \ottnt{v_{{\mathrm{0}}}} }  \, \ottnt{e'} \,  \uparrow \, (  \ottmv{i}  ,  \pi  )   \ottsym{)}    [  \ottnt{v} \,  \prec_{ \ottnt{v_{{\mathrm{0}}}} }  \, \ottnt{v'} \,  \uparrow \, (  \ottmv{i}  ,  \pi  )   /  \mathit{x}  ]   \\ = &
      \ottnt{v_{{\mathrm{0}}}} _{  \mu   \ottsym{;}  \ottmv{j}  \ottsym{+}  \ottmv{i}  \ottsym{;}  \pi }     [  \ottnt{v} \,  \prec_{ \ottnt{v_{{\mathrm{0}}}} }  \, \ottnt{v'} \,  \uparrow \, (  \ottmv{i}  ,  \pi  )   /  \mathit{x}  ]   \\ = &
     \ottnt{v_{{\mathrm{0}}}} _{  \mu   \ottsym{;}  \ottmv{j}  \ottsym{+}  \ottmv{i}  \ottsym{;}  \pi }  \\ = &
    \ottnt{v_{{\mathrm{0}}}} \,  \prec_{ \ottnt{v_{{\mathrm{0}}}} }  \,  \ottnt{v_{{\mathrm{0}}}} _{  \mu   \ottsym{;}  \ottmv{j}  \ottsym{;}  \pi_{{\mathrm{0}}} }  \,  \uparrow \, (  \ottmv{i}  ,  \pi  )  \\ = &
     \ottnt{v_{{\mathrm{0}}}}    [  \ottnt{v}  /  \mathit{x}  ]   \,  \prec_{ \ottnt{v_{{\mathrm{0}}}} }  \,   \ottnt{v_{{\mathrm{0}}}} _{  \mu   \ottsym{;}  \ottmv{j}  \ottsym{;}  \pi_{{\mathrm{0}}} }     [  \ottnt{v'}  /  \mathit{x}  ]   \,  \uparrow \, (  \ottmv{i}  ,  \pi  )  \\ = &
     \ottnt{e}    [  \ottnt{v}  /  \mathit{x}  ]   \,  \prec_{ \ottnt{v_{{\mathrm{0}}}} }  \,  \ottnt{e'}    [  \ottnt{v'}  /  \mathit{x}  ]   \,  \uparrow \, (  \ottmv{i}  ,  \pi  )  ~.
     \end{array}
   \]

  \case $ { \ottnt{A} \,  =  \,  \ottnt{A} _{  \mu  }  } \,\mathrel{\wedge}\, {   \exists  \, { \mathit{X}  \ottsym{,}  \ottmv{j}  \ottsym{,}  \sigma_{\ottmv{j}} } . \  \ottnt{A'} \,  =  \, \sigma_{\ottmv{j}}  \ottsym{(}  \mathit{X}  \ottsym{)}  } $
        (where $ \ottnt{A} _{  \mu  } $, $\mathit{X}$, $\sigma_{\ottmv{j}}$ are determined by $\ottnt{v_{{\mathrm{0}}}}$):
   By definition, $\ottnt{v_{{\mathrm{0}}}}$ is closed, so
   $ \ottnt{A} _{  \mu  } $, $\sigma_{\ottmv{j}}  \ottsym{(}  \mathit{X}  \ottsym{)}$, and $ { \sigma }_{ \ottmv{j}  \ottsym{+}  \ottmv{i} }   \ottsym{(}  \mathit{X}  \ottsym{)}$ (which is also determined by $\ottnt{v_{{\mathrm{0}}}}$) are also closed.
   Therefore:
   \[\begin{array}{ll} &
     \ottsym{(}  \ottnt{A} \,  \prec_{ \ottnt{v_{{\mathrm{0}}}} }  \, \ottnt{A'} \,  \uparrow \, (  \ottmv{i}  ,  \pi  )   \ottsym{)}    [  \ottnt{v} \,  \prec_{ \ottnt{v_{{\mathrm{0}}}} }  \, \ottnt{v'} \,  \uparrow \, (  \ottmv{i}  ,  \pi  )   /  \mathit{x}  ]   \\ = &
      { \sigma }_{ \ottmv{j}  \ottsym{+}  \ottmv{i} }   \ottsym{(}  \mathit{X}  \ottsym{)}    [  \ottnt{v} \,  \prec_{ \ottnt{v_{{\mathrm{0}}}} }  \, \ottnt{v'} \,  \uparrow \, (  \ottmv{i}  ,  \pi  )   /  \mathit{x}  ]   \\ = &
     { \sigma }_{ \ottmv{j}  \ottsym{+}  \ottmv{i} }   \ottsym{(}  \mathit{X}  \ottsym{)} \\ = &
     \ottnt{A} _{  \mu  }  \,  \prec_{ \ottnt{v_{{\mathrm{0}}}} }  \, \sigma_{\ottmv{j}}  \ottsym{(}  \mathit{X}  \ottsym{)} \,  \uparrow \, (  \ottmv{i}  ,  \pi  )  \\ = &
      \ottnt{A} _{  \mu  }     [  \ottnt{v}  /  \mathit{x}  ]   \,  \prec_{ \ottnt{v_{{\mathrm{0}}}} }  \,  \sigma_{\ottmv{j}}  \ottsym{(}  \mathit{X}  \ottsym{)}    [  \ottnt{v'}  /  \mathit{x}  ]   \,  \uparrow \, (  \ottmv{i}  ,  \pi  )  \\ = &
     \ottnt{A}    [  \ottnt{v}  /  \mathit{x}  ]   \,  \prec_{ \ottnt{v_{{\mathrm{0}}}} }  \,  \ottnt{A'}    [  \ottnt{v'}  /  \mathit{x}  ]   \,  \uparrow \, (  \ottmv{i}  ,  \pi  )  ~.
     \end{array}
   \]

  \case $ { \ottnt{A} \,  =  \,  \ottnt{A} _{  \nu  }  } \,\mathrel{\wedge}\, { \ottnt{A'} \,  =  \,  \top^P  } $
        (where $ \ottnt{A} _{  \nu  } $ is determined by $\ottnt{v_{{\mathrm{0}}}}$):
   By definition, $\ottnt{v_{{\mathrm{0}}}}$ is closed, so
   $ \ottnt{A} _{  \nu  } $ is also closed.
   Therefore:
   \[\begin{array}{ll} &
     \ottsym{(}  \ottnt{A} \,  \prec_{ \ottnt{v_{{\mathrm{0}}}} }  \, \ottnt{A'} \,  \uparrow \, (  \ottmv{i}  ,  \pi  )   \ottsym{)}    [  \ottnt{v} \,  \prec_{ \ottnt{v_{{\mathrm{0}}}} }  \, \ottnt{v'} \,  \uparrow \, (  \ottmv{i}  ,  \pi  )   /  \mathit{x}  ]   \\ = &
      \top^P     [  \ottnt{v} \,  \prec_{ \ottnt{v_{{\mathrm{0}}}} }  \, \ottnt{v'} \,  \uparrow \, (  \ottmv{i}  ,  \pi  )   /  \mathit{x}  ]   \\ = &
     \top^P  \\ = &
     \ottnt{A} _{  \nu  }  \,  \prec_{ \ottnt{v_{{\mathrm{0}}}} }  \,  \top^P  \,  \uparrow \, (  \ottmv{i}  ,  \pi  )  \\ = &
      \ottnt{A} _{  \nu  }     [  \ottnt{v}  /  \mathit{x}  ]   \,  \prec_{ \ottnt{v_{{\mathrm{0}}}} }  \,   \top^P     [  \ottnt{v'}  /  \mathit{x}  ]   \,  \uparrow \, (  \ottmv{i}  ,  \pi  )  \\ = &
     \ottnt{A}    [  \ottnt{v}  /  \mathit{x}  ]   \,  \prec_{ \ottnt{v_{{\mathrm{0}}}} }  \,  \ottnt{A'}    [  \ottnt{v'}  /  \mathit{x}  ]   \,  \uparrow \, (  \ottmv{i}  ,  \pi  )  ~.
     \end{array}
   \]

  \case $\ottnt{e} \,  =  \, \mathit{y}$:
   We have $\ottnt{e'} = \ottsym{(}  \ottnt{e} \,  \prec_{ \ottnt{v_{{\mathrm{0}}}} }  \, \ottnt{e'} \,  \uparrow \, (  \ottmv{i}  ,  \pi  )   \ottsym{)} = \mathit{y}$.

   If $ \mathit{x}    =    \mathit{y} $, then:
   \[\begin{array}{ll} &
     \ottsym{(}  \ottnt{e} \,  \prec_{ \ottnt{v_{{\mathrm{0}}}} }  \, \ottnt{e'} \,  \uparrow \, (  \ottmv{i}  ,  \pi  )   \ottsym{)}    [  \ottnt{v} \,  \prec_{ \ottnt{v_{{\mathrm{0}}}} }  \, \ottnt{v'} \,  \uparrow \, (  \ottmv{i}  ,  \pi  )   /  \mathit{x}  ]   \\ = &
     \mathit{x}    [  \ottnt{v} \,  \prec_{ \ottnt{v_{{\mathrm{0}}}} }  \, \ottnt{v'} \,  \uparrow \, (  \ottmv{i}  ,  \pi  )   /  \mathit{x}  ]   \\ = &
    \ottnt{v} \,  \prec_{ \ottnt{v_{{\mathrm{0}}}} }  \, \ottnt{v'} \,  \uparrow \, (  \ottmv{i}  ,  \pi  )  \\ = &
     \ottnt{e}    [  \ottnt{v}  /  \mathit{x}  ]   \,  \prec_{ \ottnt{v_{{\mathrm{0}}}} }  \,  \ottnt{e'}    [  \ottnt{v'}  /  \mathit{x}  ]   \,  \uparrow \, (  \ottmv{i}  ,  \pi  )  ~.
     \end{array}
   \]

   Otherwise, if $ \mathit{x}    \not=    \mathit{y} $, then:
   \[\begin{array}{ll} &
     \ottsym{(}  \ottnt{e} \,  \prec_{ \ottnt{v_{{\mathrm{0}}}} }  \, \ottnt{e'} \,  \uparrow \, (  \ottmv{i}  ,  \pi  )   \ottsym{)}    [  \ottnt{v} \,  \prec_{ \ottnt{v_{{\mathrm{0}}}} }  \, \ottnt{v'} \,  \uparrow \, (  \ottmv{i}  ,  \pi  )   /  \mathit{x}  ]   \\ = &
     \mathit{y}    [  \ottnt{v} \,  \prec_{ \ottnt{v_{{\mathrm{0}}}} }  \, \ottnt{v'} \,  \uparrow \, (  \ottmv{i}  ,  \pi  )   /  \mathit{x}  ]   \\ = &
    \mathit{y} \\ = &
    \mathit{y} \,  \prec_{ \ottnt{v_{{\mathrm{0}}}} }  \, \mathit{y} \,  \uparrow \, (  \ottmv{i}  ,  \pi  )  \\ = &
     \ottnt{e}    [  \ottnt{v}  /  \mathit{x}  ]   \,  \prec_{ \ottnt{v_{{\mathrm{0}}}} }  \,  \ottnt{e'}    [  \ottnt{v'}  /  \mathit{x}  ]   \,  \uparrow \, (  \ottmv{i}  ,  \pi  )  ~.
     \end{array}
   \]

  \case $\ottnt{e} \,  =  \, \ottsym{(}   \lambda\!  \,  \tceseq{ \mathit{y} }   \ottsym{.}  \ottnt{e_{{\mathrm{0}}}}  \ottsym{)}$:
   We have
   $\ottnt{e'} \,  =  \, \ottsym{(}   \lambda\!  \,  \tceseq{ \mathit{y} }   \ottsym{.}  \ottnt{e'_{{\mathrm{0}}}}  \ottsym{)}$ and
   $\ottnt{e} \,  \prec_{ \ottnt{v_{{\mathrm{0}}}} }  \, \ottnt{e'} \,  \uparrow \, (  \ottmv{i}  ,  \pi  )  \,  =  \, \ottsym{(}   \lambda\!  \,  \tceseq{ \mathit{y} }   \ottsym{.}  \ottsym{(}  \ottnt{e_{{\mathrm{0}}}} \,  \prec_{ \ottnt{v_{{\mathrm{0}}}} }  \, \ottnt{e'_{{\mathrm{0}}}} \,  \uparrow \, (  \ottmv{i}  ,  \pi  )   \ottsym{)}  \ottsym{)}$
   for some $\ottnt{e'_{{\mathrm{0}}}}$.
   Without loss of generality, we can suppose that,
   for every $\mathit{y'} \in  \tceseq{ \mathit{y} } $,
   $\mathit{y'} \,  \not\in  \,  \mathit{fv}  (  \ottnt{v}  )  \,  \mathbin{\cup}  \,  \mathit{fv}  (  \ottnt{v'}  )  \,  \mathbin{\cup}  \,  \mathit{fv}  (  \ottnt{v} \,  \prec_{ \ottnt{v_{{\mathrm{0}}}} }  \, \ottnt{v'} \,  \uparrow \, (  \ottmv{i}  ,  \pi  )   )  \,  \mathbin{\cup}  \,  \{  \mathit{x}  \} $.
   Then,
   \[\begin{array}{ll} &
     \ottsym{(}  \ottnt{e} \,  \prec_{ \ottnt{v_{{\mathrm{0}}}} }  \, \ottnt{e'} \,  \uparrow \, (  \ottmv{i}  ,  \pi  )   \ottsym{)}    [  \ottnt{v} \,  \prec_{ \ottnt{v_{{\mathrm{0}}}} }  \, \ottnt{v'} \,  \uparrow \, (  \ottmv{i}  ,  \pi  )   /  \mathit{x}  ]   \\ = &
     \ottsym{(}   \lambda\!  \,  \tceseq{ \mathit{y} }   \ottsym{.}  \ottsym{(}  \ottnt{e_{{\mathrm{0}}}} \,  \prec_{ \ottnt{v_{{\mathrm{0}}}} }  \, \ottnt{e'_{{\mathrm{0}}}} \,  \uparrow \, (  \ottmv{i}  ,  \pi  )   \ottsym{)}  \ottsym{)}    [  \ottnt{v} \,  \prec_{ \ottnt{v_{{\mathrm{0}}}} }  \, \ottnt{v'} \,  \uparrow \, (  \ottmv{i}  ,  \pi  )   /  \mathit{x}  ]   \\ = &
     \lambda\!  \,  \tceseq{ \mathit{y} }   \ottsym{.}  \ottsym{(}   \ottsym{(}  \ottnt{e_{{\mathrm{0}}}} \,  \prec_{ \ottnt{v_{{\mathrm{0}}}} }  \, \ottnt{e'_{{\mathrm{0}}}} \,  \uparrow \, (  \ottmv{i}  ,  \pi  )   \ottsym{)}    [  \ottnt{v} \,  \prec_{ \ottnt{v_{{\mathrm{0}}}} }  \, \ottnt{v'} \,  \uparrow \, (  \ottmv{i}  ,  \pi  )   /  \mathit{x}  ]    \ottsym{)} \\ = &
     \lambda\!  \,  \tceseq{ \mathit{y} }   \ottsym{.}  \ottsym{(}   \ottnt{e_{{\mathrm{0}}}}    [  \ottnt{v}  /  \mathit{x}  ]   \,  \prec_{ \ottnt{v_{{\mathrm{0}}}} }  \,  \ottnt{e'_{{\mathrm{0}}}}    [  \ottnt{v'}  /  \mathit{x}  ]   \,  \uparrow \, (  \ottmv{i}  ,  \pi  )   \ottsym{)} \quad \text{(by the IH)} \\ = &
    \ottsym{(}    \lambda\!  \,  \tceseq{ \mathit{y} }   \ottsym{.}  \ottnt{e_{{\mathrm{0}}}}    [  \ottnt{v}  /  \mathit{x}  ]    \ottsym{)} \,  \prec_{ \ottnt{v_{{\mathrm{0}}}} }  \, \ottsym{(}    \lambda\!  \,  \tceseq{ \mathit{y} }   \ottsym{.}  \ottnt{e'_{{\mathrm{0}}}}    [  \ottnt{v'}  /  \mathit{x}  ]    \ottsym{)} \,  \uparrow \, (  \ottmv{i}  ,  \pi  )  \\ = &
     \ottnt{e}    [  \ottnt{v}  /  \mathit{x}  ]   \,  \prec_{ \ottnt{v_{{\mathrm{0}}}} }  \,  \ottnt{e'}    [  \ottnt{v'}  /  \mathit{x}  ]   \,  \uparrow \, (  \ottmv{i}  ,  \pi  )  ~.
     \end{array}
   \]

  \case $\phi \,  =  \, \ottnt{A_{{\mathrm{0}}}}  \ottsym{(}   \tceseq{ \ottnt{t} }   \ottsym{)}$:
   We have $\phi' \,  =  \, \ottnt{A'_{{\mathrm{0}}}}  \ottsym{(}   \tceseq{ \ottnt{t} }   \ottsym{)}$ and $\phi \,  \prec_{ \ottnt{v_{{\mathrm{0}}}} }  \, \phi' \,  \uparrow \, (  \ottmv{i}  ,  \pi  )  \,  =  \, \ottsym{(}  \ottnt{A_{{\mathrm{0}}}} \,  \prec_{ \ottnt{v_{{\mathrm{0}}}} }  \, \ottnt{A'_{{\mathrm{0}}}} \,  \uparrow \, (  \ottmv{i}  ,  \pi  )   \ottsym{)}  \ottsym{(}   \tceseq{ \ottnt{t} }   \ottsym{)}$ for some $\ottnt{A'_{{\mathrm{0}}}}$.

   We first show that
   \[
     \tceseq{  \ottnt{t}    [  \ottnt{v} \,  \prec_{ \ottnt{v_{{\mathrm{0}}}} }  \, \ottnt{v'} \,  \uparrow \, (  \ottmv{i}  ,  \pi  )   /  \mathit{x}  ]   }  =  \tceseq{  \ottnt{t}    [  \ottnt{v}  /  \mathit{x}  ]   }  =  \tceseq{  \ottnt{t}    [  \ottnt{v'}  /  \mathit{x}  ]   } 
   \]
   by case analysis on whether $\mathit{x}$ occurs in $ \tceseq{ \ottnt{t} } $.
   \begin{caseanalysis}
    \case $\mathit{x}$ occurs in $ \tceseq{ \ottnt{t} } $:
     In this case, (at least) one of $\ottnt{v}$, $\ottnt{v'}$, and $\ottnt{v} \,  \prec_{ \ottnt{v_{{\mathrm{0}}}} }  \, \ottnt{v'} \,  \uparrow \, (  \ottmv{i}  ,  \pi  ) $
     must be a constant; furthermore, it implies that
     $\ottnt{v} = \ottnt{v'} = \ottnt{v} \,  \prec_{ \ottnt{v_{{\mathrm{0}}}} }  \, \ottnt{v'} \,  \uparrow \, (  \ottmv{i}  ,  \pi  )  = \ottnt{c}$ for some $\ottnt{c}$.
     Therefore, $ \tceseq{  \ottnt{t}    [  \ottnt{v} \,  \prec_{ \ottnt{v_{{\mathrm{0}}}} }  \, \ottnt{v'} \,  \uparrow \, (  \ottmv{i}  ,  \pi  )   /  \mathit{x}  ]   }  =  \tceseq{  \ottnt{t}    [  \ottnt{v}  /  \mathit{x}  ]   }  =  \tceseq{  \ottnt{t}    [  \ottnt{v'}  /  \mathit{x}  ]   } $.

    \case $\mathit{x}$ does not occur in $ \tceseq{ \ottnt{t} } $:
     Obviously $ \tceseq{  \ottnt{t}    [  \ottnt{v} \,  \prec_{ \ottnt{v_{{\mathrm{0}}}} }  \, \ottnt{v'} \,  \uparrow \, (  \ottmv{i}  ,  \pi  )   /  \mathit{x}  ]   }  =  \tceseq{  \ottnt{t}    [  \ottnt{v}  /  \mathit{x}  ]   }  =  \tceseq{  \ottnt{t}    [  \ottnt{v'}  /  \mathit{x}  ]   } $.
   \end{caseanalysis}

   Then:
   \[\begin{array}{ll} &
     \ottsym{(}  \phi \,  \prec_{ \ottnt{v_{{\mathrm{0}}}} }  \, \phi' \,  \uparrow \, (  \ottmv{i}  ,  \pi  )   \ottsym{)}    [  \ottnt{v} \,  \prec_{ \ottnt{v_{{\mathrm{0}}}} }  \, \ottnt{v'} \,  \uparrow \, (  \ottmv{i}  ,  \pi  )   /  \mathit{x}  ]   \\ = &
     \ottsym{(}  \ottsym{(}  \ottnt{A_{{\mathrm{0}}}} \,  \prec_{ \ottnt{v_{{\mathrm{0}}}} }  \, \ottnt{A'_{{\mathrm{0}}}} \,  \uparrow \, (  \ottmv{i}  ,  \pi  )   \ottsym{)}  \ottsym{(}   \tceseq{ \ottnt{t} }   \ottsym{)}  \ottsym{)}    [  \ottnt{v} \,  \prec_{ \ottnt{v_{{\mathrm{0}}}} }  \, \ottnt{v'} \,  \uparrow \, (  \ottmv{i}  ,  \pi  )   /  \mathit{x}  ]   \\ = &
     \ottsym{(}  \ottnt{A_{{\mathrm{0}}}} \,  \prec_{ \ottnt{v_{{\mathrm{0}}}} }  \, \ottnt{A'_{{\mathrm{0}}}} \,  \uparrow \, (  \ottmv{i}  ,  \pi  )   \ottsym{)}    [  \ottnt{v} \,  \prec_{ \ottnt{v_{{\mathrm{0}}}} }  \, \ottnt{v'} \,  \uparrow \, (  \ottmv{i}  ,  \pi  )   /  \mathit{x}  ]    \ottsym{(}   \tceseq{  \ottnt{t}    [  \ottnt{v} \,  \prec_{ \ottnt{v_{{\mathrm{0}}}} }  \, \ottnt{v'} \,  \uparrow \, (  \ottmv{i}  ,  \pi  )   /  \mathit{x}  ]   }   \ottsym{)} \\ = &
    \ottsym{(}   \ottnt{A_{{\mathrm{0}}}}    [  \ottnt{v}  /  \mathit{x}  ]   \,  \prec_{ \ottnt{v_{{\mathrm{0}}}} }  \,  \ottnt{A'_{{\mathrm{0}}}}    [  \ottnt{v'}  /  \mathit{x}  ]   \,  \uparrow \, (  \ottmv{i}  ,  \pi  )   \ottsym{)}  \ottsym{(}   \tceseq{  \ottnt{t}    [  \ottnt{v} \,  \prec_{ \ottnt{v_{{\mathrm{0}}}} }  \, \ottnt{v'} \,  \uparrow \, (  \ottmv{i}  ,  \pi  )   /  \mathit{x}  ]   }   \ottsym{)}
     \quad \text{(by the IH)} \\ = &
     \ottnt{A_{{\mathrm{0}}}}    [  \ottnt{v}  /  \mathit{x}  ]    \ottsym{(}   \tceseq{  \ottnt{t}    [  \ottnt{v} \,  \prec_{ \ottnt{v_{{\mathrm{0}}}} }  \, \ottnt{v'} \,  \uparrow \, (  \ottmv{i}  ,  \pi  )   /  \mathit{x}  ]   }   \ottsym{)} \,  \prec_{ \ottnt{v_{{\mathrm{0}}}} }  \,  \ottnt{A'_{{\mathrm{0}}}}    [  \ottnt{v'}  /  \mathit{x}  ]    \ottsym{(}   \tceseq{  \ottnt{t}    [  \ottnt{v} \,  \prec_{ \ottnt{v_{{\mathrm{0}}}} }  \, \ottnt{v'} \,  \uparrow \, (  \ottmv{i}  ,  \pi  )   /  \mathit{x}  ]   }   \ottsym{)} \,  \uparrow \, (  \ottmv{i}  ,  \pi  )  \\ = &
     \ottnt{A_{{\mathrm{0}}}}    [  \ottnt{v}  /  \mathit{x}  ]    \ottsym{(}   \tceseq{  \ottnt{t}    [  \ottnt{v}  /  \mathit{x}  ]   }   \ottsym{)} \,  \prec_{ \ottnt{v_{{\mathrm{0}}}} }  \,  \ottnt{A'_{{\mathrm{0}}}}    [  \ottnt{v'}  /  \mathit{x}  ]    \ottsym{(}   \tceseq{  \ottnt{t}    [  \ottnt{v'}  /  \mathit{x}  ]   }   \ottsym{)} \,  \uparrow \, (  \ottmv{i}  ,  \pi  )  \\ = &
     \phi    [  \ottnt{v}  /  \mathit{x}  ]   \,  \prec_{ \ottnt{v_{{\mathrm{0}}}} }  \,  \phi'    [  \ottnt{v'}  /  \mathit{x}  ]   \,  \uparrow \, (  \ottmv{i}  ,  \pi  )  ~.
     \end{array}
   \]
 \end{caseanalysis}
\end{proof}

\begin{lemmap}{Approximation Consistent with Predicate Substitution}{lr-approx-pred-subst}
 \begin{enumerate}
  \item $ \ottsym{(}  \ottnt{e} \,  \prec_{ \ottnt{v} }  \, \ottnt{e'} \,  \uparrow \, (  \ottmv{i}  ,  \pi  )   \ottsym{)}    [  \ottnt{A} \,  \prec_{ \ottnt{v} }  \, \ottnt{A'} \,  \uparrow \, (  \ottmv{i}  ,  \pi  )   /  \mathit{X}  ]   \,  =  \,  \ottnt{e}    [  \ottnt{A}  /  \mathit{X}  ]   \,  \prec_{ \ottnt{v} }  \,  \ottnt{e'}    [  \ottnt{A'}  /  \mathit{X}  ]   \,  \uparrow \, (  \ottmv{i}  ,  \pi  ) $.
  \item $ \ottsym{(}  \ottnt{A_{{\mathrm{0}}}} \,  \prec_{ \ottnt{v} }  \, \ottnt{A'_{{\mathrm{0}}}} \,  \uparrow \, (  \ottmv{i}  ,  \pi  )   \ottsym{)}    [  \ottnt{A} \,  \prec_{ \ottnt{v} }  \, \ottnt{A'} \,  \uparrow \, (  \ottmv{i}  ,  \pi  )   /  \mathit{X}  ]   \,  =  \,  \ottnt{A_{{\mathrm{0}}}}    [  \ottnt{A}  /  \mathit{X}  ]   \,  \prec_{ \ottnt{v} }  \,  \ottnt{A'_{{\mathrm{0}}}}    [  \ottnt{A'}  /  \mathit{X}  ]   \,  \uparrow \, (  \ottmv{i}  ,  \pi  ) $.
  \item $ \ottsym{(}  \phi \,  \prec_{ \ottnt{v} }  \, \phi' \,  \uparrow \, (  \ottmv{i}  ,  \pi  )   \ottsym{)}    [  \ottnt{A} \,  \prec_{ \ottnt{v} }  \, \ottnt{A'} \,  \uparrow \, (  \ottmv{i}  ,  \pi  )   /  \mathit{X}  ]   \,  =  \,  \phi    [  \ottnt{A}  /  \mathit{X}  ]   \,  \prec_{ \ottnt{v} }  \,  \phi'    [  \ottnt{A'}  /  \mathit{X}  ]   \,  \uparrow \, (  \ottmv{i}  ,  \pi  ) $.
 \end{enumerate}
\end{lemmap}
\begin{proof}
 \TS{OK: 2022/01/14}
 By simultaneous induction on $\ottnt{e}$, $\ottnt{A_{{\mathrm{0}}}}$, and $\phi$.
 We mention only the interesting case.
 \begin{caseanalysis}
  \case $\ottnt{A_{{\mathrm{0}}}} \,  =  \, \mathit{X_{{\mathrm{0}}}}$:
   We have $\ottnt{A'_{{\mathrm{0}}}} \,  =  \, \mathit{X_{{\mathrm{0}}}}$ and $\ottnt{A_{{\mathrm{0}}}} \,  \prec_{ \ottnt{v} }  \, \ottnt{A'_{{\mathrm{0}}}} \,  \uparrow \, (  \ottmv{i}  ,  \pi  )  \,  =  \, \mathit{X_{{\mathrm{0}}}}$.

   If $\mathit{X} \,  =  \, \mathit{X_{{\mathrm{0}}}}$, then:
   \[
   \begin{array}{ll} &
     \ottsym{(}  \ottnt{A_{{\mathrm{0}}}} \,  \prec_{ \ottnt{v} }  \, \ottnt{A'_{{\mathrm{0}}}} \,  \uparrow \, (  \ottmv{i}  ,  \pi  )   \ottsym{)}    [  \ottnt{A} \,  \prec_{ \ottnt{v} }  \, \ottnt{A'} \,  \uparrow \, (  \ottmv{i}  ,  \pi  )   /  \mathit{X}  ]   \\ = &
     \mathit{X}    [  \ottnt{A} \,  \prec_{ \ottnt{v} }  \, \ottnt{A'} \,  \uparrow \, (  \ottmv{i}  ,  \pi  )   /  \mathit{X}  ]   \\ = &
    \ottnt{A} \,  \prec_{ \ottnt{v} }  \, \ottnt{A'} \,  \uparrow \, (  \ottmv{i}  ,  \pi  )  \\ = &
     \ottnt{A_{{\mathrm{0}}}}    [  \ottnt{A}  /  \mathit{X}  ]   \,  \prec_{ \ottnt{v} }  \,  \ottnt{A'_{{\mathrm{0}}}}    [  \ottnt{A'}  /  \mathit{X}  ]   \,  \uparrow \, (  \ottmv{i}  ,  \pi  )  ~.
   \end{array}
   \]

   Otherwise, if $\mathit{X} \,  \not=  \, \mathit{X_{{\mathrm{0}}}}$, then:
   \[
   \begin{array}{ll} &
     \ottsym{(}  \ottnt{A_{{\mathrm{0}}}} \,  \prec_{ \ottnt{v} }  \, \ottnt{A'_{{\mathrm{0}}}} \,  \uparrow \, (  \ottmv{i}  ,  \pi  )   \ottsym{)}    [  \ottnt{A} \,  \prec_{ \ottnt{v} }  \, \ottnt{A'} \,  \uparrow \, (  \ottmv{i}  ,  \pi  )   /  \mathit{X}  ]   \\ = &
    \mathit{X_{{\mathrm{0}}}} \\ = &
    \mathit{X_{{\mathrm{0}}}} \,  \prec_{ \ottnt{v} }  \, \mathit{X_{{\mathrm{0}}}} \,  \uparrow \, (  \ottmv{i}  ,  \pi  )  \\ = &
     \ottnt{A_{{\mathrm{0}}}}    [  \ottnt{A}  /  \mathit{X}  ]   \,  \prec_{ \ottnt{v} }  \,  \ottnt{A'_{{\mathrm{0}}}}    [  \ottnt{A'}  /  \mathit{X}  ]   \,  \uparrow \, (  \ottmv{i}  ,  \pi  )  ~.
   \end{array}
   \]
 \end{caseanalysis}
\end{proof}

\begin{lemmap}{Well-Definedness of Approximation}{lr-approx-wf}
 \begin{enumerate}
  \item If $\ottnt{e} \,  \prec_{ \ottnt{v} }  \, \ottnt{e'} \,  \uparrow \, (  \ottmv{i}  ,  \pi  ) $ is well defined,
        then $\ottnt{e} \,  \prec_{ \ottnt{v} }  \, \ottnt{e'} \,  \uparrow \, (  \ottmv{j}  ,  \pi_{{\mathrm{0}}}  ) $ is also well defined
        for any $\ottmv{j}$ and $\pi_{{\mathrm{0}}}$.
  \item If $\ottnt{A} \,  \prec_{ \ottnt{v} }  \, \ottnt{A'} \,  \uparrow \, (  \ottmv{i}  ,  \pi  ) $ is well defined,
        then $\ottnt{A} \,  \prec_{ \ottnt{v} }  \, \ottnt{A'} \,  \uparrow \, (  \ottmv{j}  ,  \pi_{{\mathrm{0}}}  ) $ is also well defined
        for any $\ottmv{j}$ and $\pi_{{\mathrm{0}}}$.
  \item If $\phi \,  \prec_{ \ottnt{v} }  \, \phi' \,  \uparrow \, (  \ottmv{i}  ,  \pi  ) $ is well defined,
        then $\phi \,  \prec_{ \ottnt{v} }  \, \phi' \,  \uparrow \, (  \ottmv{j}  ,  \pi_{{\mathrm{0}}}  ) $ is also well defined
        for any $\ottmv{j}$ and $\pi_{{\mathrm{0}}}$.
  \item If $\ottnt{E} \,  \prec_{ \ottnt{v} }  \, \ottnt{E'} \,  \uparrow \, (  \ottmv{i}  ,  \pi  ) $ is well defined,
        then $\ottnt{E} \,  \prec_{ \ottnt{v} }  \, \ottnt{E'} \,  \uparrow \, (  \ottmv{j}  ,  \pi_{{\mathrm{0}}}  ) $ is also well defined
        for any $\ottmv{j}$ and $\pi_{{\mathrm{0}}}$.
 \end{enumerate}
\end{lemmap}
\begin{proof}
 \TS{OK: 2022/01/17}
 The first three are proven straightforwardly by mutual induction on the derivations of
 $\ottnt{e} \,  \prec_{ \ottnt{v} }  \, \ottnt{e'} \,  \uparrow \, (  \ottmv{i}  ,  \pi  ) $,
 $\ottnt{A} \,  \prec_{ \ottnt{v} }  \, \ottnt{A'} \,  \uparrow \, (  \ottmv{i}  ,  \pi  ) $, and
 $\phi \,  \prec_{ \ottnt{v} }  \, \phi' \,  \uparrow \, (  \ottmv{i}  ,  \pi  ) $.
 The last is proven by induction on the derivation of
 $\ottnt{E} \,  \prec_{ \ottnt{v} }  \, \ottnt{E'} \,  \uparrow \, (  \ottmv{i}  ,  \pi  ) $ with the first case.
\end{proof}

\begin{lemmap}{Substitution of Approximation Preserves Well-Definedness}{lr-wf-subst-approx}
 If $\ottnt{e_{{\mathrm{0}}}} \,  \prec_{ \ottnt{v} }  \, \ottnt{e'_{{\mathrm{0}}}}$ and $ \tceseq{ \ottnt{v_{{\mathrm{0}}}} \,  \prec_{ \ottnt{v} }  \, \ottnt{v'_{{\mathrm{0}}}} } $ and $ \ottnt{e_{{\mathrm{0}}}}    [ \tceseq{ \ottnt{v_{{\mathrm{0}}}}  /  \mathit{x} } ]  $ is well defined,
 then $ \ottnt{e'_{{\mathrm{0}}}}    [ \tceseq{ \ottnt{v'_{{\mathrm{0}}}}  /  \mathit{x} } ]  $ is well defined.
\end{lemmap}
\begin{proof}
 \TS{OK: 2022/01/14}
 Let $\mathit{x_{{\mathrm{1}}}} \in  \tceseq{ \mathit{x} } $, and $\ottnt{v_{{\mathrm{1}}}} \in  \tceseq{ \ottnt{v_{{\mathrm{0}}}} } $ and $\ottnt{v'_{{\mathrm{1}}}} \in  \tceseq{ \ottnt{v'_{{\mathrm{0}}}} } $ that
 correspond to $\mathit{x_{{\mathrm{1}}}}$.

 Suppose that $\mathit{x_{{\mathrm{1}}}}$ occurs in formulas of $\ottnt{e_{{\mathrm{0}}}}$.
 Because variables occurring in formulas must be replaced with constants,
 the well-formedness of $ \ottnt{e_{{\mathrm{0}}}}    [ \tceseq{ \ottnt{v_{{\mathrm{0}}}}  /  \mathit{x} } ]  $ implies that $\ottnt{v_{{\mathrm{1}}}}$ is a constant.
 Then, because $\ottnt{v_{{\mathrm{1}}}} \,  \prec_{ \ottnt{v} }  \, \ottnt{v'_{{\mathrm{1}}}}$, $\ottnt{v'_{{\mathrm{1}}}}$ is a constant.
 Therefore, the substitution of $\ottnt{v'_{{\mathrm{1}}}}$ for $\mathit{x_{{\mathrm{1}}}}$ in $\ottnt{e'_{{\mathrm{0}}}}$ is well defined.

 Otherwise, if $\mathit{x_{{\mathrm{1}}}}$ does not occur in formulas of $\ottnt{e_{{\mathrm{0}}}}$,
 then $\ottnt{e_{{\mathrm{0}}}} \,  \prec_{ \ottnt{v} }  \, \ottnt{e'_{{\mathrm{0}}}}$ implies that $\mathit{x_{{\mathrm{1}}}}$ does not occur in formulas of $\ottnt{e'_{{\mathrm{0}}}}$.
 Then, the substitution of $\ottnt{v'_{{\mathrm{1}}}}$ for $\mathit{x_{{\mathrm{1}}}}$ in $\ottnt{e'_{{\mathrm{0}}}}$ is well defined
 for any value $\ottnt{v'_{{\mathrm{1}}}}$.
\end{proof}

\begin{lemmap}{Approximation of Evaluation Contexts}{lr-approx-ectx}
 If $ \ottnt{E}  [  \ottnt{e_{{\mathrm{0}}}}  ]  \,  \prec_{ \ottnt{v} }  \, \ottnt{e'}$, then $\ottnt{e'} \,  =  \,  \ottnt{E'}  [  \ottnt{e'_{{\mathrm{0}}}}  ] $ for some $\ottnt{E'}$ and $\ottnt{e'_{{\mathrm{0}}}}$
 such that $\ottnt{E} \,  \prec_{ \ottnt{v} }  \, \ottnt{E'}$ and $\ottnt{e_{{\mathrm{0}}}} \,  \prec_{ \ottnt{v} }  \, \ottnt{e'_{{\mathrm{0}}}}$.
 Furthermore, if $\ottnt{E} \,  =  \, \ottnt{K}$ for some $\ottnt{K}$, then $\ottnt{E'} \,  =  \, \ottnt{K'}$ for some $\ottnt{K'}$.
\end{lemmap}
\begin{proof}
 \TS{OK: 2022/01/17}
 By induction on $\ottnt{E}$.
 \begin{caseanalysis}
  \case $\ottnt{E} \,  =  \, \,[\,]\,$:
   Because $ \ottnt{E}  [  \ottnt{e_{{\mathrm{0}}}}  ]  = \ottnt{e_{{\mathrm{0}}}} \,  \prec_{ \ottnt{v} }  \, \ottnt{e'}$ and $\,[\,]\, \,  \prec_{ \ottnt{v} }  \, \,[\,]\,$,
   we have the conclusion by letting $\ottnt{E'} \,  =  \, \,[\,]\,$ and $\ottnt{e'_{{\mathrm{0}}}} \,  =  \, \ottnt{e'}$.

  \case $  \exists  \, { \mathit{x_{{\mathrm{1}}}}  \ottsym{,}  \ottnt{E_{{\mathrm{1}}}}  \ottsym{,}  \ottnt{e_{{\mathrm{2}}}} } . \  \ottnt{E} \,  =  \, \ottsym{(}  \mathsf{let} \, \mathit{x_{{\mathrm{1}}}}  \ottsym{=}  \ottnt{E_{{\mathrm{1}}}} \,  \mathsf{in}  \, \ottnt{e_{{\mathrm{2}}}}  \ottsym{)} $:
   Because $ \ottnt{E}  [  \ottnt{e_{{\mathrm{0}}}}  ]  \,  =  \, \ottsym{(}  \mathsf{let} \, \mathit{x_{{\mathrm{1}}}}  \ottsym{=}   \ottnt{E_{{\mathrm{1}}}}  [  \ottnt{e_{{\mathrm{0}}}}  ]  \,  \mathsf{in}  \, \ottnt{e_{{\mathrm{2}}}}  \ottsym{)}$ and $ \ottnt{E}  [  \ottnt{e_{{\mathrm{0}}}}  ]  \,  \prec_{ \ottnt{v} }  \, \ottnt{e'}$,
   we have
   \begin{itemize}
    \item $\ottnt{e'} \,  =  \, \mathsf{let} \, \mathit{x_{{\mathrm{1}}}}  \ottsym{=}  \ottnt{e'_{{\mathrm{1}}}} \,  \mathsf{in}  \, \ottnt{e'_{{\mathrm{2}}}}$,
    \item $ \ottnt{E_{{\mathrm{1}}}}  [  \ottnt{e_{{\mathrm{0}}}}  ]  \,  \prec_{ \ottnt{v} }  \, \ottnt{e'_{{\mathrm{1}}}}$, and
    \item $\ottnt{e_{{\mathrm{2}}}} \,  \prec_{ \ottnt{v} }  \, \ottnt{e'_{{\mathrm{2}}}}$
   \end{itemize}
   for some $\ottnt{e'_{{\mathrm{1}}}}$ and $\ottnt{e'_{{\mathrm{2}}}}$.
   By the IH,
   \begin{itemize}
    \item $\ottnt{e'_{{\mathrm{1}}}} \,  =  \,  \ottnt{E'_{{\mathrm{1}}}}  [  \ottnt{e'_{{\mathrm{0}}}}  ] $,
    \item $\ottnt{E_{{\mathrm{1}}}} \,  \prec_{ \ottnt{v} }  \, \ottnt{E'_{{\mathrm{1}}}}$, and
    \item $\ottnt{e_{{\mathrm{0}}}} \,  \prec_{ \ottnt{v} }  \, \ottnt{e'_{{\mathrm{0}}}}$
   \end{itemize}
   for some $\ottnt{E'_{{\mathrm{1}}}}$ and $\ottnt{e'_{{\mathrm{0}}}}$.
   By definition, $\ottnt{E_{{\mathrm{1}}}} \,  \prec_{ \ottnt{v} }  \, \ottnt{E'_{{\mathrm{1}}}}$ and $\ottnt{e_{{\mathrm{2}}}} \,  \prec_{ \ottnt{v} }  \, \ottnt{e'_{{\mathrm{2}}}}$ imply
   $\ottnt{E} = \ottsym{(}  \mathsf{let} \, \mathit{x_{{\mathrm{1}}}}  \ottsym{=}  \ottnt{E_{{\mathrm{1}}}} \,  \mathsf{in}  \, \ottnt{e_{{\mathrm{2}}}}  \ottsym{)} \,  \prec_{ \ottnt{v} }  \, \ottsym{(}  \mathsf{let} \, \mathit{x_{{\mathrm{1}}}}  \ottsym{=}  \ottnt{E'_{{\mathrm{1}}}} \,  \mathsf{in}  \, \ottnt{e'_{{\mathrm{2}}}}  \ottsym{)}$.
   Therefore, we have $\ottnt{e'} \,  =  \,  \ottnt{E'}  [  \ottnt{e'_{{\mathrm{0}}}}  ] $ and $\ottnt{E} \,  \prec_{ \ottnt{v} }  \, \ottnt{E'}$
   by letting $\ottnt{E'} \,  =  \, \mathsf{let} \, \mathit{x_{{\mathrm{1}}}}  \ottsym{=}  \ottnt{E'_{{\mathrm{1}}}} \,  \mathsf{in}  \, \ottnt{e'_{{\mathrm{2}}}}$.

   Finally, suppose that $\ottnt{E} \,  =  \, \ottnt{K}$ for some $\ottnt{K}$.
   Then, $\ottnt{E_{{\mathrm{1}}}} \,  =  \, \ottnt{K_{{\mathrm{1}}}}$ for some $\ottnt{K_{{\mathrm{1}}}}$ such that
   $\ottnt{K} \,  =  \, \ottsym{(}  \mathsf{let} \, \mathit{x_{{\mathrm{1}}}}  \ottsym{=}  \ottnt{K_{{\mathrm{1}}}} \,  \mathsf{in}  \, \ottnt{e_{{\mathrm{2}}}}  \ottsym{)}$.
   By the IH, $\ottnt{E'_{{\mathrm{1}}}} \,  =  \, \ottnt{K'_{{\mathrm{1}}}}$ for some $\ottnt{K'_{{\mathrm{1}}}}$, so
   $\ottnt{E'} \,  =  \, \ottnt{K'}$ for some $\ottnt{K'}$ such that
   $\ottnt{K'} \,  =  \, \ottsym{(}  \mathsf{let} \, \mathit{x_{{\mathrm{1}}}}  \ottsym{=}  \ottnt{K'_{{\mathrm{1}}}} \,  \mathsf{in}  \, \ottnt{e_{{\mathrm{2}}}}  \ottsym{)}$.

  \case $  \exists  \, { \ottnt{E_{{\mathrm{1}}}} } . \  \ottnt{E} \,  =  \,  \resetcnstr{ \ottnt{E_{{\mathrm{1}}}} }  $:
   Because $ \ottnt{E}  [  \ottnt{e_{{\mathrm{0}}}}  ]  \,  =  \,  \resetcnstr{  \ottnt{E_{{\mathrm{1}}}}  [  \ottnt{e_{{\mathrm{0}}}}  ]  } $ and $ \ottnt{E}  [  \ottnt{e_{{\mathrm{0}}}}  ]  \,  \prec_{ \ottnt{v} }  \, \ottnt{e'}$,
   we have
   \begin{itemize}
    \item $\ottnt{e'} \,  =  \,  \resetcnstr{ \ottnt{e'_{{\mathrm{1}}}} } $ and
    \item $ \ottnt{E_{{\mathrm{1}}}}  [  \ottnt{e_{{\mathrm{0}}}}  ]  \,  \prec_{ \ottnt{v} }  \, \ottnt{e'_{{\mathrm{1}}}}$
   \end{itemize}
   for some $\ottnt{e'_{{\mathrm{1}}}}$.
   By the IH,
   \begin{itemize}
    \item $\ottnt{e'_{{\mathrm{1}}}} \,  =  \,  \ottnt{E'_{{\mathrm{1}}}}  [  \ottnt{e'_{{\mathrm{0}}}}  ] $,
    \item $\ottnt{E_{{\mathrm{1}}}} \,  \prec_{ \ottnt{v} }  \, \ottnt{E'_{{\mathrm{1}}}}$, and
    \item $\ottnt{e_{{\mathrm{0}}}} \,  \prec_{ \ottnt{v} }  \, \ottnt{e'_{{\mathrm{0}}}}$
   \end{itemize}
   for some $\ottnt{E'_{{\mathrm{1}}}}$ and $\ottnt{e'_{{\mathrm{0}}}}$.
   By definition, $\ottnt{E_{{\mathrm{1}}}} \,  \prec_{ \ottnt{v} }  \, \ottnt{E'_{{\mathrm{1}}}}$ implies
   $\ottnt{E} =  \resetcnstr{ \ottnt{E_{{\mathrm{1}}}} }  \,  \prec_{ \ottnt{v} }  \,  \resetcnstr{ \ottnt{E'_{{\mathrm{1}}}} } $.
   Therefore, we have the conclusion by letting $\ottnt{E'} \,  =  \,  \resetcnstr{ \ottnt{E'_{{\mathrm{1}}}} } $.
   Note that $\ottnt{E} \,  \not=  \, \ottnt{K}$ for any $\ottnt{K}$.
 \end{caseanalysis}
\end{proof}

\begin{lemmap}{Embedding Expressions into Approximated Continuations}{lr-approx-embed-exp}
 If $\ottnt{E} \,  \prec_{ \ottnt{v} }  \, \ottnt{E'}$, then
 \[
    \forall  \, { \ottmv{i}  \ottsym{,}  \pi } . \   \ottnt{E}  [  \ottnt{e}  ]  \,  \prec_{ \ottnt{v} }  \,  \ottnt{E'}  [  \ottnt{e'}  ]  \,  \uparrow \, (  \ottmv{i}  ,  \pi  )  \,  =  \,  \ottsym{(}  \ottnt{E} \,  \prec_{ \ottnt{v} }  \, \ottnt{E'} \,  \uparrow \, (  \ottmv{i}  ,  \pi  )   \ottsym{)}  [  \ottnt{e} \,  \prec_{ \ottnt{v} }  \, \ottnt{e'} \,  \uparrow \, (  \ottmv{i}  ,  \pi  )   ]   ~.
 \]
\end{lemmap}
\begin{proof}
 \TS{OK: 2022/01/17}
 By induction on $\ottnt{E}$.
 Let
 $\ottmv{i}$ be any natural number and
 $\pi$ be any infinite trace.
 Note that $\ottnt{E} \,  \prec_{ \ottnt{v} }  \, \ottnt{E'} \,  \uparrow \, (  \ottmv{i}  ,  \pi  ) $ is well defined
 by \reflem{lr-approx-wf} with $\ottnt{E} \,  \prec_{ \ottnt{v} }  \, \ottnt{E'}$.
 \begin{caseanalysis}
  \case $\ottnt{E} \,  =  \, \,[\,]\,$:
   Because $\ottnt{E} \,  \prec_{ \ottnt{v} }  \, \ottnt{E'}$ and $\ottnt{E} \,  =  \, \,[\,]\,$,
   we have $\ottnt{E'} \,  =  \, \,[\,]\,$, so $ \ottnt{E}  [  \ottnt{e}  ]  \,  =  \, \ottnt{e}$ and $ \ottnt{E'}  [  \ottnt{e'}  ]  \,  =  \, \ottnt{e'}$ and
   $\ottnt{E} \,  \prec_{ \ottnt{v} }  \, \ottnt{E'} \,  \uparrow \, (  \ottmv{i}  ,  \pi  )  \,  =  \, \,[\,]\,$.
   Therefore, it suffices to show that
   \[
    \ottnt{e} \,  \prec_{ \ottnt{v} }  \, \ottnt{e'} \,  \uparrow \, (  \ottmv{i}  ,  \pi  )  \,  =  \, \ottnt{e} \,  \prec_{ \ottnt{v} }  \, \ottnt{e'} \,  \uparrow \, (  \ottmv{i}  ,  \pi  )  ~,
   \]
   which holds obviously.

  \case $  \exists  \, { \mathit{x_{{\mathrm{1}}}}  \ottsym{,}  \ottnt{E_{{\mathrm{1}}}}  \ottsym{,}  \ottnt{e_{{\mathrm{2}}}} } . \  \ottnt{E} \,  =  \, \ottsym{(}  \mathsf{let} \, \mathit{x_{{\mathrm{1}}}}  \ottsym{=}  \ottnt{E_{{\mathrm{1}}}} \,  \mathsf{in}  \, \ottnt{e_{{\mathrm{2}}}}  \ottsym{)} $:
   $\ottnt{E} \,  \prec_{ \ottnt{v} }  \, \ottnt{E'}$ implies
   \begin{itemize}
    \item $\ottnt{E'} \,  =  \, \mathsf{let} \, \mathit{x_{{\mathrm{1}}}}  \ottsym{=}  \ottnt{E'_{{\mathrm{1}}}} \,  \mathsf{in}  \, \ottnt{e'_{{\mathrm{2}}}}$,
    \item $\ottnt{E_{{\mathrm{1}}}} \,  \prec_{ \ottnt{v} }  \, \ottnt{E'_{{\mathrm{1}}}}$, and
    \item $\ottnt{e_{{\mathrm{2}}}} \,  \prec_{ \ottnt{v} }  \, \ottnt{e'_{{\mathrm{2}}}}$
   \end{itemize}
   for some $\ottnt{E'_{{\mathrm{1}}}}$ and $\ottnt{e'_{{\mathrm{2}}}}$.
   Further, we have
   \[
    \ottnt{E} \,  \prec_{ \ottnt{v} }  \, \ottnt{E'} \,  \uparrow \, (  \ottmv{i}  ,  \pi  )  \,  =  \, \mathsf{let} \, \mathit{x_{{\mathrm{1}}}}  \ottsym{=}  \ottsym{(}  \ottnt{E_{{\mathrm{1}}}} \,  \prec_{ \ottnt{v} }  \, \ottnt{E'_{{\mathrm{1}}}} \,  \uparrow \, (  \ottmv{i}  ,  \pi  )   \ottsym{)} \,  \mathsf{in}  \, \ottsym{(}  \ottnt{e_{{\mathrm{2}}}} \,  \prec_{ \ottnt{v} }  \, \ottnt{e'_{{\mathrm{2}}}} \,  \uparrow \, (  \ottmv{i}  ,  \pi  )   \ottsym{)} ~.
   \]
   Then, we have
   \[\begin{array}{ll} &
     \ottnt{E}  [  \ottnt{e}  ]  \,  \prec_{ \ottnt{v} }  \,  \ottnt{E'}  [  \ottnt{e'}  ]  \,  \uparrow \, (  \ottmv{i}  ,  \pi  )  \\ = &
    \ottsym{(}  \mathsf{let} \, \mathit{x_{{\mathrm{1}}}}  \ottsym{=}   \ottnt{E_{{\mathrm{1}}}}  [  \ottnt{e}  ]  \,  \mathsf{in}  \, \ottnt{e_{{\mathrm{2}}}}  \ottsym{)} \,  \prec_{ \ottnt{v} }  \, \ottsym{(}  \mathsf{let} \, \mathit{x_{{\mathrm{1}}}}  \ottsym{=}   \ottnt{E'_{{\mathrm{1}}}}  [  \ottnt{e'}  ]  \,  \mathsf{in}  \, \ottnt{e'_{{\mathrm{2}}}}  \ottsym{)} \,  \uparrow \, (  \ottmv{i}  ,  \pi  )  \\ = &
    \mathsf{let} \, \mathit{x_{{\mathrm{1}}}}  \ottsym{=}  \ottsym{(}   \ottnt{E_{{\mathrm{1}}}}  [  \ottnt{e}  ]  \,  \prec_{ \ottnt{v} }  \,  \ottnt{E'_{{\mathrm{1}}}}  [  \ottnt{e'}  ]  \,  \uparrow \, (  \ottmv{i}  ,  \pi  )   \ottsym{)} \,  \mathsf{in}  \, \ottsym{(}  \ottnt{e_{{\mathrm{2}}}} \,  \prec_{ \ottnt{v} }  \, \ottnt{e'_{{\mathrm{2}}}} \,  \uparrow \, (  \ottmv{i}  ,  \pi  )   \ottsym{)} \\ = &
    \mathsf{let} \, \mathit{x_{{\mathrm{1}}}}  \ottsym{=}  \ottsym{(}   \ottsym{(}  \ottnt{E_{{\mathrm{1}}}} \,  \prec_{ \ottnt{v} }  \, \ottnt{E'_{{\mathrm{1}}}} \,  \uparrow \, (  \ottmv{i}  ,  \pi  )   \ottsym{)}  [  \ottnt{e} \,  \prec_{ \ottnt{v} }  \, \ottnt{e'} \,  \uparrow \, (  \ottmv{i}  ,  \pi  )   ]   \ottsym{)} \,  \mathsf{in}  \, \ottsym{(}  \ottnt{e_{{\mathrm{2}}}} \,  \prec_{ \ottnt{v} }  \, \ottnt{e'_{{\mathrm{2}}}} \,  \uparrow \, (  \ottmv{i}  ,  \pi  )   \ottsym{)}
     \quad \text{(by the IH)} \\ = &
     \ottsym{(}  \mathsf{let} \, \mathit{x_{{\mathrm{1}}}}  \ottsym{=}  \ottsym{(}  \ottnt{E_{{\mathrm{1}}}} \,  \prec_{ \ottnt{v} }  \, \ottnt{E'_{{\mathrm{1}}}} \,  \uparrow \, (  \ottmv{i}  ,  \pi  )   \ottsym{)} \,  \mathsf{in}  \, \ottsym{(}  \ottnt{e_{{\mathrm{2}}}} \,  \prec_{ \ottnt{v} }  \, \ottnt{e'_{{\mathrm{2}}}} \,  \uparrow \, (  \ottmv{i}  ,  \pi  )   \ottsym{)}  \ottsym{)}  [  \ottnt{e} \,  \prec_{ \ottnt{v} }  \, \ottnt{e'} \,  \uparrow \, (  \ottmv{i}  ,  \pi  )   ]  \\ = &
     \ottsym{(}  \ottsym{(}  \mathsf{let} \, \mathit{x_{{\mathrm{1}}}}  \ottsym{=}  \ottnt{E_{{\mathrm{1}}}} \,  \mathsf{in}  \, \ottnt{e_{{\mathrm{2}}}}  \ottsym{)} \,  \prec_{ \ottnt{v} }  \, \ottsym{(}  \mathsf{let} \, \mathit{x_{{\mathrm{1}}}}  \ottsym{=}  \ottnt{E'_{{\mathrm{1}}}} \,  \mathsf{in}  \, \ottnt{e'_{{\mathrm{2}}}}  \ottsym{)} \,  \uparrow \, (  \ottmv{i}  ,  \pi  )   \ottsym{)}  [  \ottnt{e} \,  \prec_{ \ottnt{v} }  \, \ottnt{e'} \,  \uparrow \, (  \ottmv{i}  ,  \pi  )   ]  \\ = &
     \ottsym{(}  \ottnt{E} \,  \prec_{ \ottnt{v} }  \, \ottnt{E'} \,  \uparrow \, (  \ottmv{i}  ,  \pi  )   \ottsym{)}  [  \ottnt{e} \,  \prec_{ \ottnt{v} }  \, \ottnt{e'} \,  \uparrow \, (  \ottmv{i}  ,  \pi  )   ]  ~.
     \end{array}
   \]

  \case $  \exists  \, { \ottnt{E_{{\mathrm{1}}}} } . \  \ottnt{E} \,  =  \,  \resetcnstr{ \ottnt{E_{{\mathrm{1}}}} }  $:
   $\ottnt{E} \,  \prec_{ \ottnt{v} }  \, \ottnt{E'}$ implies
   \begin{itemize}
    \item $\ottnt{E'} \,  =  \,  \resetcnstr{ \ottnt{E'_{{\mathrm{1}}}} } $ and
    \item $\ottnt{E_{{\mathrm{1}}}} \,  \prec_{ \ottnt{v} }  \, \ottnt{E'_{{\mathrm{1}}}}$
   \end{itemize}
   for some $\ottnt{E'_{{\mathrm{1}}}}$.
   Further, we have
   \[
    \ottnt{E} \,  \prec_{ \ottnt{v} }  \, \ottnt{E'} \,  \uparrow \, (  \ottmv{i}  ,  \pi  )  \,  =  \,  \resetcnstr{ \ottnt{E_{{\mathrm{1}}}} \,  \prec_{ \ottnt{v} }  \, \ottnt{E'_{{\mathrm{1}}}} \,  \uparrow \, (  \ottmv{i}  ,  \pi  )  }  ~.
   \]
   Then, we have
   \[\begin{array}{ll} &
     \ottnt{E}  [  \ottnt{e}  ]  \,  \prec_{ \ottnt{v} }  \,  \ottnt{E'}  [  \ottnt{e'}  ]  \,  \uparrow \, (  \ottmv{i}  ,  \pi  )  \\ = &
     \resetcnstr{  \ottnt{E_{{\mathrm{1}}}}  [  \ottnt{e}  ]  }  \,  \prec_{ \ottnt{v} }  \,  \resetcnstr{  \ottnt{E'_{{\mathrm{1}}}}  [  \ottnt{e'}  ]  }  \,  \uparrow \, (  \ottmv{i}  ,  \pi  )  \\ = &
     \resetcnstr{  \ottnt{E_{{\mathrm{1}}}}  [  \ottnt{e}  ]  \,  \prec_{ \ottnt{v} }  \,  \ottnt{E'_{{\mathrm{1}}}}  [  \ottnt{e'}  ]  \,  \uparrow \, (  \ottmv{i}  ,  \pi  )  }  \\ = &
     \resetcnstr{  \ottsym{(}  \ottnt{E_{{\mathrm{1}}}} \,  \prec_{ \ottnt{v} }  \, \ottnt{E'_{{\mathrm{1}}}} \,  \uparrow \, (  \ottmv{i}  ,  \pi  )   \ottsym{)}  [  \ottnt{e} \,  \prec_{ \ottnt{v} }  \, \ottnt{e'} \,  \uparrow \, (  \ottmv{i}  ,  \pi  )   ]  } 
     \quad \text{(by the IH)} \\ = &
      \resetcnstr{ \ottnt{E_{{\mathrm{1}}}} \,  \prec_{ \ottnt{v} }  \, \ottnt{E'_{{\mathrm{1}}}} \,  \uparrow \, (  \ottmv{i}  ,  \pi  )  }   [  \ottnt{e} \,  \prec_{ \ottnt{v} }  \, \ottnt{e'} \,  \uparrow \, (  \ottmv{i}  ,  \pi  )   ]  \\ = &
     \ottsym{(}  \ottnt{E} \,  \prec_{ \ottnt{v} }  \, \ottnt{E'} \,  \uparrow \, (  \ottmv{i}  ,  \pi  )   \ottsym{)}  [  \ottnt{e} \,  \prec_{ \ottnt{v} }  \, \ottnt{e'} \,  \uparrow \, (  \ottmv{i}  ,  \pi  )   ]  ~.
     \end{array}
   \]
 \end{caseanalysis}
\end{proof}

\begin{lemmap}{Reflexivity of Approximation}{lr-approx-refl}
 \begin{enumerate}
  \item $\ottnt{e} \,  \prec_{ \ottnt{v} }  \, \ottnt{e} \,  \uparrow \, (  \ottmv{i}  ,  \pi  )  \,  =  \, \ottnt{e}$.
  \item $\ottnt{A} \,  \prec_{ \ottnt{v} }  \, \ottnt{A} \,  \uparrow \, (  \ottmv{i}  ,  \pi  )  \,  =  \, \ottnt{A}$.
  \item $\phi \,  \prec_{ \ottnt{v} }  \, \phi \,  \uparrow \, (  \ottmv{i}  ,  \pi  )  \,  =  \, \phi$.
  \item $\ottnt{E} \,  \prec_{ \ottnt{v} }  \, \ottnt{E} \,  \uparrow \, (  \ottmv{i}  ,  \pi  )  \,  =  \, \ottnt{E}$.
 \end{enumerate}
\end{lemmap}
\begin{proof}
 \TS{OK: 2022/01/20}
 The first three cases are proven
 by simultaneous induction on $\ottnt{e}$, $\ottnt{A}$, and $\phi$.
 Note that the following cases do not happen.
 \begin{caseanalysis}
  \case $ { \ottnt{e} \,  =  \, \ottnt{v} } \,\mathrel{\wedge}\, {   \exists  \, { \ottmv{j}  \ottsym{,}  \pi_{{\mathrm{0}}} } . \  \ottnt{e} \,  =  \,  \ottnt{v} _{  \mu   \ottsym{;}  \ottmv{j}  \ottsym{;}  \pi_{{\mathrm{0}}} }   } $:
   We have $\ottnt{v} \,  =  \,  \ottnt{v} _{  \mu   \ottsym{;}  \ottmv{j}  \ottsym{;}  \pi_{{\mathrm{0}}} } $, but it is contradictory
   because $\ottnt{v}$ is a recursive function, while
   $ \ottnt{v} _{  \mu   \ottsym{;}  \ottmv{j}  \ottsym{;}  \pi_{{\mathrm{0}}} } $ is a $\lambda$-abstraction.

  \case $ { \ottnt{A} \,  =  \,  \ottnt{A} _{  \mu  }  } \,\mathrel{\wedge}\, {   \exists  \, { \mathit{X}  \ottsym{,}  \ottmv{j}  \ottsym{,}  \sigma_{\ottmv{j}} } . \  \ottnt{A} \,  =  \, \sigma_{\ottmv{j}}  \ottsym{(}  \mathit{X}  \ottsym{)}  } $
   (where $ \ottnt{A} _{  \mu  } $, $\mathit{X}$, $\sigma_{\ottmv{j}}$ are determined by $\ottnt{v}$):
   Contradictory because $ \ottnt{A} _{  \mu  } $ is implemented by the least fixpoint operator while $\sigma_{\ottmv{j}}  \ottsym{(}  \mathit{X}  \ottsym{)}$
   is by the greatest fixpoint operator;
   this implementation is possible because the bound predicate variable $\mathit{X}$
   does not occur free in the body of $\sigma_{\ottmv{j}}  \ottsym{(}  \mathit{X}  \ottsym{)}$ (see
   \refdef{approx-fixpoint-pred}).

  \case $ { \ottnt{A} \,  =  \,  \ottnt{A} _{  \nu  }  } \,\mathrel{\wedge}\, { \ottnt{A} \,  =  \,  \top^P  } $:
   Contradictory.
 \end{caseanalysis}

 The last case $\ottnt{E} \,  \prec_{ \ottnt{v} }  \, \ottnt{E} \,  \uparrow \, (  \ottmv{i}  ,  \pi  )  \,  =  \, \ottnt{E}$ is proven
 by induction on $\ottnt{E}$ with the first case.
\end{proof}

\begin{lemmap}{Approximation of Values}{lr-approx-val}
 If $\ottnt{v_{{\mathrm{0}}}} \,  \prec_{ \ottnt{v} }  \, \ottnt{e_{{\mathrm{0}}}} \,  \uparrow \, (  \ottmv{i}  ,  \pi  ) $ is well defined,
 then $\ottnt{e_{{\mathrm{0}}}}$ and $\ottnt{v_{{\mathrm{0}}}} \,  \prec_{ \ottnt{v} }  \, \ottnt{e_{{\mathrm{0}}}} \,  \uparrow \, (  \ottmv{i}  ,  \pi  ) $ are values.
\end{lemmap}
\begin{proof}
 \TS{OK: 2022/01/15}
 By case analysis on $\ottnt{v_{{\mathrm{0}}}}$.
 \begin{caseanalysis}
  \case $\ottnt{v_{{\mathrm{0}}}} \,  =  \, \ottnt{c}$, $\mathit{x}$, or $ \Lambda   \ottsym{(}  \mathit{X} \,  {:}  \,  \tceseq{ \ottnt{s} }   \ottsym{)}  \ottsym{.}  \ottnt{e}$ for some
   $\ottnt{c}$, $\mathit{x}$, $\mathit{X}$, $ \tceseq{ \ottnt{s} } $, and $\ottnt{e}$:
   Obvious by the definition of $\ottnt{v_{{\mathrm{0}}}} \,  \prec_{ \ottnt{v} }  \, \ottnt{e_{{\mathrm{0}}}} \,  \uparrow \, (  \ottmv{i}  ,  \pi  ) $.

  \case $ {   \exists  \, {  \tceseq{ \mathit{X} }   \ottsym{,}   \tceseq{ \mathit{Y} }   \ottsym{,}  \mathit{f}  \ottsym{,}   \tceseq{ \mathit{x} }   \ottsym{,}   \tceseq{  \ottnt{A} _{  \mu  }  }   \ottsym{,}   \tceseq{  \ottnt{A} _{  \nu  }  }   \ottsym{,}  \ottnt{e}  \ottsym{,}   \tceseq{ \ottnt{v_{{\mathrm{2}}}} }  } . \  \ottnt{v_{{\mathrm{0}}}} \,  =  \, \ottsym{(}   \mathsf{rec}  \, (  \tceseq{  \mathit{X} ^{  \ottnt{A} _{  \mu  }  },  \mathit{Y} ^{  \ottnt{A} _{  \nu  }  }  }  ,  \mathit{f} ,   \tceseq{ \mathit{x} }  ) . \,  \ottnt{e}   \ottsym{)} \,  \tceseq{ \ottnt{v_{{\mathrm{2}}}} }   } \,\mathrel{\wedge}\, { \ottsym{\mbox{$\mid$}}   \tceseq{ \ottnt{v_{{\mathrm{2}}}} }   \ottsym{\mbox{$\mid$}} \,  <  \, \ottsym{\mbox{$\mid$}}   \tceseq{ \mathit{x} }   \ottsym{\mbox{$\mid$}} } $:
   Because $\ottnt{v_{{\mathrm{0}}}} \,  \prec_{ \ottnt{v} }  \, \ottnt{e_{{\mathrm{0}}}} \,  \uparrow \, (  \ottmv{i}  ,  \pi  ) $ is well defined, we have
   \begin{itemize}
    \item $\ottnt{e_{{\mathrm{0}}}} \,  =  \, \ottnt{v'_{{\mathrm{1}}}} \,  \tceseq{ \ottnt{v'_{{\mathrm{2}}}} } $,
    \item $\ottsym{(}   \mathsf{rec}  \, (  \tceseq{  \mathit{X} ^{  \ottnt{A} _{  \mu  }  },  \mathit{Y} ^{  \ottnt{A} _{  \nu  }  }  }  ,  \mathit{f} ,   \tceseq{ \mathit{x} }  ) . \,  \ottnt{e}   \ottsym{)} \,  \prec_{ \ottnt{v} }  \, \ottnt{v'_{{\mathrm{1}}}} \,  \uparrow \, (  \ottmv{i}  ,  \pi  ) $ is well defined, and
    \item $\ottsym{\mbox{$\mid$}}   \tceseq{ \ottnt{v_{{\mathrm{2}}}} }   \ottsym{\mbox{$\mid$}} \,  =  \, \ottsym{\mbox{$\mid$}}   \tceseq{ \ottnt{v'_{{\mathrm{2}}}} }   \ottsym{\mbox{$\mid$}}$
   \end{itemize}
   for some $\ottnt{v'_{{\mathrm{1}}}}$ and $ \tceseq{ \ottnt{v'_{{\mathrm{2}}}} } $.
   We proceed by case analysis on $\ottnt{v'_{{\mathrm{1}}}}$.
   We have only two cases to be considered because
   $\ottsym{(}   \mathsf{rec}  \, (  \tceseq{  \mathit{X} ^{  \ottnt{A} _{  \mu  }  },  \mathit{Y} ^{  \ottnt{A} _{  \nu  }  }  }  ,  \mathit{f} ,   \tceseq{ \mathit{x} }  ) . \,  \ottnt{e}   \ottsym{)} \,  \prec_{ \ottnt{v} }  \, \ottnt{v'_{{\mathrm{1}}}} \,  \uparrow \, (  \ottmv{i}  ,  \pi  ) $ is well defined.
   \begin{caseanalysis}
    \case $  \exists  \, { \ottmv{j}  \ottsym{,}  \pi_{{\mathrm{0}}} } . \  \ottnt{v'_{{\mathrm{1}}}} \,  =  \,  \ottnt{v} _{  \mu   \ottsym{;}  \ottmv{j}  \ottsym{;}  \pi_{{\mathrm{0}}} }  $:
     In this case, $ \mathsf{rec}  \, (  \tceseq{  \mathit{X} ^{  \ottnt{A} _{  \mu  }  },  \mathit{Y} ^{  \ottnt{A} _{  \nu  }  }  }  ,  \mathit{f} ,   \tceseq{ \mathit{x} }  ) . \,  \ottnt{e}  \,  =  \, \ottnt{v}$, so
     $\ottnt{v'_{{\mathrm{1}}}}$ is a $\lambda$-abstraction that takes
     $\ottsym{\mbox{$\mid$}}   \tceseq{ \mathit{x} }   \ottsym{\mbox{$\mid$}}$ arguments.
     Then, because $\ottsym{\mbox{$\mid$}}   \tceseq{ \ottnt{v'_{{\mathrm{2}}}} }   \ottsym{\mbox{$\mid$}} = \ottsym{\mbox{$\mid$}}   \tceseq{ \ottnt{v_{{\mathrm{2}}}} }   \ottsym{\mbox{$\mid$}} \,  <  \, \ottsym{\mbox{$\mid$}}   \tceseq{ \mathit{x} }   \ottsym{\mbox{$\mid$}}$,
     $\ottnt{e_{{\mathrm{0}}}} \,  =  \, \ottnt{v'_{{\mathrm{1}}}} \,  \tceseq{ \ottnt{v'_{{\mathrm{2}}}} } $ is a value.

     Further, because
     \begin{itemize}
      \item $\ottnt{v_{{\mathrm{0}}}} \,  \prec_{ \ottnt{v} }  \, \ottnt{e_{{\mathrm{0}}}} \,  \uparrow \, (  \ottmv{i}  ,  \pi  )  \,  =  \, \ottsym{(}  \ottsym{(}   \mathsf{rec}  \, (  \tceseq{  \mathit{X} ^{  \ottnt{A} _{  \mu  }  },  \mathit{Y} ^{  \ottnt{A} _{  \nu  }  }  }  ,  \mathit{f} ,   \tceseq{ \mathit{x} }  ) . \,  \ottnt{e}   \ottsym{)} \,  \prec_{ \ottnt{v} }  \, \ottnt{v'_{{\mathrm{1}}}} \,  \uparrow \, (  \ottmv{i}  ,  \pi  )   \ottsym{)} \,  \tceseq{ \ottnt{v_{{\mathrm{2}}}} \,  \prec_{ \ottnt{v} }  \, \ottnt{v'_{{\mathrm{2}}}} \,  \uparrow \, (  \ottmv{i}  ,  \pi  )  } $,
      \item $\ottsym{(}   \mathsf{rec}  \, (  \tceseq{  \mathit{X} ^{  \ottnt{A} _{  \mu  }  },  \mathit{Y} ^{  \ottnt{A} _{  \nu  }  }  }  ,  \mathit{f} ,   \tceseq{ \mathit{x} }  ) . \,  \ottnt{e}   \ottsym{)} \,  \prec_{ \ottnt{v} }  \, \ottnt{v'_{{\mathrm{1}}}} \,  \uparrow \, (  \ottmv{i}  ,  \pi  )  \,  =  \,  \ottnt{v} _{  \mu   \ottsym{;}  \ottmv{j}  \ottsym{+}  \ottmv{i}  \ottsym{;}  \pi } $,
      \item $ \ottnt{v} _{  \mu   \ottsym{;}  \ottmv{j}  \ottsym{+}  \ottsym{1}  \ottsym{;}  \pi } $ is a $\lambda$-abstraction that takes $\ottsym{\mbox{$\mid$}}   \tceseq{ \mathit{x} }   \ottsym{\mbox{$\mid$}}$ arguments, and
      \item $\ottsym{\mbox{$\mid$}}   \tceseq{ \ottnt{v_{{\mathrm{2}}}} \,  \prec_{ \ottnt{v} }  \, \ottnt{v'_{{\mathrm{2}}}} \,  \uparrow \, (  \ottmv{i}  ,  \pi  )  }   \ottsym{\mbox{$\mid$}} \,  <  \, \ottsym{\mbox{$\mid$}}   \tceseq{ \mathit{x} }   \ottsym{\mbox{$\mid$}}$,
     \end{itemize}
     $\ottnt{v_{{\mathrm{0}}}} \,  \prec_{ \ottnt{v} }  \, \ottnt{e_{{\mathrm{0}}}} \,  \uparrow \, (  \ottmv{i}  ,  \pi  ) $ is a value.

    \case $  \exists  \, {  \tceseq{  \ottnt{A'} _{  \mu  }  }   \ottsym{,}   \tceseq{  \ottnt{A'} _{  \nu  }  }   \ottsym{,}  \ottnt{e'} } . \  \ottnt{v'_{{\mathrm{1}}}} \,  =  \,  \mathsf{rec}  \, (  \tceseq{  \mathit{X} ^{  \ottnt{A'} _{  \mu  }  },  \mathit{Y} ^{  \ottnt{A'} _{  \nu  }  }  }  ,  \mathit{f} ,   \tceseq{ \mathit{x} }  ) . \,  \ottnt{e'}  $:
     Because $\ottsym{\mbox{$\mid$}}   \tceseq{ \ottnt{v'_{{\mathrm{2}}}} }   \ottsym{\mbox{$\mid$}} = \ottsym{\mbox{$\mid$}}   \tceseq{ \ottnt{v_{{\mathrm{2}}}} }   \ottsym{\mbox{$\mid$}} \,  <  \, \ottsym{\mbox{$\mid$}}   \tceseq{ \mathit{x} }   \ottsym{\mbox{$\mid$}}$,
     $\ottnt{e_{{\mathrm{0}}}} \,  =  \, \ottnt{v'_{{\mathrm{1}}}} \,  \tceseq{ \ottnt{v'_{{\mathrm{2}}}} } $ is a value.
     Further, because
     \[
      \ottnt{v_{{\mathrm{0}}}} \,  \prec_{ \ottnt{v} }  \, \ottnt{e_{{\mathrm{0}}}} \,  \uparrow \, (  \ottmv{i}  ,  \pi  )  \,  =  \, \ottsym{(}   \mathsf{rec}  \, (  \tceseq{  \mathit{X} ^{  \ottnt{A} _{  \mu  }  \,  \prec_{ \ottnt{v} }  \,  \ottnt{A'} _{  \mu  }  \,  \uparrow \, (  \ottmv{i}  ,  \pi  )  },  \mathit{Y} ^{  \ottnt{A} _{  \nu  }  \,  \prec_{ \ottnt{v} }  \,  \ottnt{A'} _{  \nu  }  \,  \uparrow \, (  \ottmv{i}  ,  \pi  )  }  }  ,  \mathit{f} ,   \tceseq{ \mathit{x} }  ) . \,  \ottsym{(}  \ottnt{e} \,  \prec_{ \ottnt{v} }  \, \ottnt{e'} \,  \uparrow \, (  \ottmv{i}  ,  \pi  )   \ottsym{)}   \ottsym{)} \,  \tceseq{ \ottnt{v_{{\mathrm{2}}}} \,  \prec_{ \ottnt{v} }  \, \ottnt{v'_{{\mathrm{2}}}} \,  \uparrow \, (  \ottmv{i}  ,  \pi  )  }  ~,
     \]
     $\ottnt{v_{{\mathrm{0}}}} \,  \prec_{ \ottnt{v} }  \, \ottnt{e_{{\mathrm{0}}}} \,  \uparrow \, (  \ottmv{i}  ,  \pi  ) $ is a value.
     
   \end{caseanalysis}

  \case $ {   \exists  \, {  \tceseq{ \mathit{x} }   \ottsym{,}  \ottnt{e}  \ottsym{,}   \tceseq{ \ottnt{v_{{\mathrm{2}}}} }  } . \  \ottnt{v_{{\mathrm{0}}}} \,  =  \, \ottsym{(}   \lambda\!  \,  \tceseq{ \mathit{x} }   \ottsym{.}  \ottnt{e}  \ottsym{)} \,  \tceseq{ \ottnt{v_{{\mathrm{2}}}} }   } \,\mathrel{\wedge}\, { \ottsym{\mbox{$\mid$}}   \tceseq{ \ottnt{v_{{\mathrm{2}}}} }   \ottsym{\mbox{$\mid$}} \,  <  \, \ottsym{\mbox{$\mid$}}   \tceseq{ \mathit{x} }   \ottsym{\mbox{$\mid$}} } $:
   Because $\ottnt{v_{{\mathrm{0}}}} \,  \prec_{ \ottnt{v} }  \, \ottnt{e_{{\mathrm{0}}}} \,  \uparrow \, (  \ottmv{i}  ,  \pi  ) $ is well defined, we have
   $\ottnt{e_{{\mathrm{0}}}} \,  =  \, \ottsym{(}   \lambda\!  \,  \tceseq{ \mathit{x} }   \ottsym{.}  \ottnt{e'}  \ottsym{)} \,  \tceseq{ \ottnt{v'_{{\mathrm{2}}}} } $ for some $\ottnt{e'}$ and $ \tceseq{ \ottnt{v'_{{\mathrm{2}}}} } $ such that
   $\ottsym{\mbox{$\mid$}}   \tceseq{ \ottnt{v_{{\mathrm{2}}}} }   \ottsym{\mbox{$\mid$}} \,  =  \, \ottsym{\mbox{$\mid$}}   \tceseq{ \ottnt{v'_{{\mathrm{2}}}} }   \ottsym{\mbox{$\mid$}}$.
   Because $\ottsym{\mbox{$\mid$}}   \tceseq{ \ottnt{v'_{{\mathrm{2}}}} }   \ottsym{\mbox{$\mid$}} = \ottsym{\mbox{$\mid$}}   \tceseq{ \ottnt{v_{{\mathrm{2}}}} }   \ottsym{\mbox{$\mid$}} \,  <  \, \ottsym{\mbox{$\mid$}}   \tceseq{ \mathit{x} }   \ottsym{\mbox{$\mid$}}$, $\ottnt{e_{{\mathrm{0}}}} \,  =  \, \ottsym{(}   \lambda\!  \,  \tceseq{ \mathit{x} }   \ottsym{.}  \ottnt{e'}  \ottsym{)} \,  \tceseq{ \ottnt{v'_{{\mathrm{2}}}} } $ is a value.

   Further, because $\ottnt{v_{{\mathrm{0}}}} \,  \prec_{ \ottnt{v} }  \, \ottnt{e_{{\mathrm{0}}}} \,  \uparrow \, (  \ottmv{i}  ,  \pi  )  \,  =  \, \ottsym{(}   \lambda\!  \,  \tceseq{ \mathit{x} }   \ottsym{.}  \ottsym{(}  \ottnt{e} \,  \prec_{ \ottnt{v} }  \, \ottnt{e'} \,  \uparrow \, (  \ottmv{i}  ,  \pi  )   \ottsym{)}  \ottsym{)} \,  \tceseq{ \ottnt{v_{{\mathrm{2}}}} \,  \prec_{ \ottnt{v} }  \, \ottnt{v'_{{\mathrm{2}}}} \,  \uparrow \, (  \ottmv{i}  ,  \pi  )  } $,
   $\ottnt{v_{{\mathrm{0}}}} \,  \prec_{ \ottnt{v} }  \, \ottnt{e_{{\mathrm{0}}}} \,  \uparrow \, (  \ottmv{i}  ,  \pi  ) $ is a value.
 \end{caseanalysis}
\end{proof}

\begin{lemmap}{Approximation of Evaluation}{lr-approx-eval}
 Suppose that $\ottnt{e_{{\mathrm{1}}}} \,  \prec_{ \ottnt{v} }  \, \ottnt{e'_{{\mathrm{1}}}}$ and $\ottnt{e_{{\mathrm{1}}}} \,  \mathrel{  \longrightarrow  }  \, \ottnt{e_{{\mathrm{2}}}} \,  \,\&\,  \, \varpi$.
 \begin{enumerate}
  \item \label{lem:lr-approx-eval:approx}
        If
        $\ottnt{e_{{\mathrm{1}}}} \,  =  \,  \ottnt{E_{{\mathrm{1}}}}  [  \ottnt{v} \,  \tceseq{ \ottnt{v_{{\mathrm{2}}}} }   ] $ and
        $\ottnt{e'_{{\mathrm{1}}}} \,  =  \,  \ottnt{E'_{{\mathrm{1}}}}  [   \ottnt{v} _{  \mu   \ottsym{;}  \ottsym{0}  \ottsym{;}  \pi_{{\mathrm{0}}} }  \,  \tceseq{ \ottnt{v'_{{\mathrm{2}}}} }   ] $ and
        $ \mathit{arity}  (  \ottnt{v}  )  \,  =  \, \ottsym{\mbox{$\mid$}}   \tceseq{ \ottnt{v_{{\mathrm{2}}}} }   \ottsym{\mbox{$\mid$}}$ ,
        then
        \[
           \exists  \, { \ottnt{e'_{{\mathrm{2}}}} } . \    \forall  \, { \pi } . \    \forall  \, { \ottmv{i} \,  \geq  \, \ottsym{1} } . \  \ottsym{(}  \ottnt{e_{{\mathrm{1}}}} \,  \prec_{ \ottnt{v} }  \, \ottnt{e'_{{\mathrm{1}}}} \,  \uparrow \, (  \ottmv{i}  ,  \pi  )   \ottsym{)} \,  \mathrel{  \longrightarrow  }  \, \ottsym{(}  \ottnt{e_{{\mathrm{2}}}} \,  \prec_{ \ottnt{v} }  \, \ottnt{e'_{{\mathrm{2}}}} \,  \uparrow \, (  \ottmv{i}  \ottsym{-}  \ottsym{1}  ,  \pi  )   \ottsym{)} \,  \,\&\,  \, \varpi    ~.
        \]

  \item \label{lem:lr-approx-eval:non-approx}
        Suppose that,
        for any $\ottnt{E_{{\mathrm{1}}}}$, $\ottnt{E'_{{\mathrm{1}}}}$, $ \tceseq{ \ottnt{v_{{\mathrm{2}}}} } $, $ \tceseq{ \ottnt{v'_{{\mathrm{2}}}} } $, and $\pi_{{\mathrm{0}}}$,
        $\ottnt{e_{{\mathrm{1}}}} \,  \not=  \,  \ottnt{E_{{\mathrm{1}}}}  [  \ottnt{v} \,  \tceseq{ \ottnt{v_{{\mathrm{2}}}} }   ] $ or
        $\ottnt{e'_{{\mathrm{1}}}} \,  \not=  \,  \ottnt{E'_{{\mathrm{1}}}}  [   \ottnt{v} _{  \mu   \ottsym{;}  \ottsym{0}  \ottsym{;}  \pi_{{\mathrm{0}}} }  \,  \tceseq{ \ottnt{v'_{{\mathrm{2}}}} }   ] $ or
        $ \mathit{arity}  (  \ottnt{v}  )  \,  \not=  \, \ottsym{\mbox{$\mid$}}   \tceseq{ \ottnt{v_{{\mathrm{2}}}} }   \ottsym{\mbox{$\mid$}}$.
        Then,
        \[
            \exists  \, { \ottnt{e'_{{\mathrm{2}}}} } . \    \forall  \, { \pi } . \    \forall  \, { \ottmv{i} } . \  \ottsym{(}  \ottnt{e_{{\mathrm{1}}}} \,  \prec_{ \ottnt{v} }  \, \ottnt{e'_{{\mathrm{1}}}} \,  \uparrow \, (  \ottmv{i}  ,  \pi  )   \ottsym{)} \,  \mathrel{  \longrightarrow  }  \, \ottsym{(}  \ottnt{e_{{\mathrm{2}}}} \,  \prec_{ \ottnt{v} }  \, \ottnt{e'_{{\mathrm{2}}}} \,  \uparrow \, (  \ottmv{i}  ,  \pi  )   \ottsym{)} \,  \,\&\,  \, \varpi    ~.
        \]

  \item \label{lem:lr-approx-eval:unified}
        The following holds:
        \[
           \exists  \, { \ottmv{j}  \ottsym{,}  \ottnt{e'_{{\mathrm{2}}}} } . \    \forall  \, { \pi } . \    \forall  \, { \ottmv{i} \,  \geq  \, \ottmv{j} } . \  \ottsym{(}  \ottnt{e_{{\mathrm{1}}}} \,  \prec_{ \ottnt{v} }  \, \ottnt{e'_{{\mathrm{1}}}} \,  \uparrow \, (  \ottmv{i}  ,  \pi  )   \ottsym{)} \,  \mathrel{  \longrightarrow  }  \, \ottsym{(}  \ottnt{e_{{\mathrm{2}}}} \,  \prec_{ \ottnt{v} }  \, \ottnt{e'_{{\mathrm{2}}}} \,  \uparrow \, (  \ottmv{i}  \ottsym{-}  \ottmv{j}  ,  \pi  )   \ottsym{)} \,  \,\&\,  \, \varpi    ~.
        \]
 \end{enumerate}
\end{lemmap}
\begin{proof}
 \TS{OK: 2022/01/17}
 \begin{enumerate}
  \item By induction on the derivation of $\ottnt{e_{{\mathrm{1}}}} \,  \mathrel{  \longrightarrow  }  \, \ottnt{e_{{\mathrm{2}}}} \,  \,\&\,  \, \varpi$.
        Note that $\ottnt{e_{{\mathrm{1}}}} \,  \prec_{ \ottnt{v} }  \, \ottnt{e'_{{\mathrm{1}}}} \,  \uparrow \, (  \ottmv{i}  ,  \pi  ) $ is well defined by \reflem{lr-approx-wf}.
        \begin{caseanalysis}
         \case \E{Red}/\R{RBeta}:
          We are given
          \begin{prooftree}
           \AxiomC{$\ottnt{v} \,  \tceseq{ \ottnt{v_{{\mathrm{2}}}} }  \,  \rightsquigarrow  \,     \ottnt{e_{{\mathrm{0}}}}    [ \tceseq{  \ottnt{A} _{  \mu  }   /  \mathit{X} } ]      [ \tceseq{  \ottnt{A} _{  \nu  }   /  \mathit{Y} } ]      [  \ottnt{v}  /  \mathit{f}  ]      [ \tceseq{ \ottnt{v_{{\mathrm{2}}}}  /  \mathit{x} } ]  $}
           \UnaryInfC{$\ottnt{v} \,  \tceseq{ \ottnt{v_{{\mathrm{2}}}} }  \,  \mathrel{  \longrightarrow  }  \,     \ottnt{e_{{\mathrm{0}}}}    [ \tceseq{  \ottnt{A} _{  \mu  }   /  \mathit{X} } ]      [ \tceseq{  \ottnt{A} _{  \nu  }   /  \mathit{Y} } ]      [  \ottnt{v}  /  \mathit{f}  ]      [ \tceseq{ \ottnt{v_{{\mathrm{2}}}}  /  \mathit{x} } ]   \,  \,\&\,  \,  \epsilon $}
          \end{prooftree}
          for some $ \tceseq{ \ottnt{v_{{\mathrm{2}}}} } $, $ \tceseq{  \ottnt{A} _{  \mu  }  } $, $ \tceseq{  \ottnt{A} _{  \nu  }  } $, $ \tceseq{ \mathit{X} } $, $ \tceseq{ \mathit{Y} } $,
          $\mathit{f}$, $ \tceseq{ \mathit{x} } $, and $\ottnt{e_{{\mathrm{0}}}}$ such that
          \begin{itemize}
           \item $\ottnt{e_{{\mathrm{1}}}} \,  =  \, \ottnt{v} \,  \tceseq{ \ottnt{v_{{\mathrm{2}}}} } $ and
           \item $\ottnt{v} \,  =  \,  \mathsf{rec}  \, (  \tceseq{  \mathit{X} ^{  \ottnt{A} _{  \mu  }  },  \mathit{Y} ^{  \ottnt{A} _{  \nu  }  }  }  ,  \mathit{f} ,   \tceseq{ \mathit{x} }  ) . \,  \ottnt{e_{{\mathrm{0}}}} $ and
           \item $\ottnt{e_{{\mathrm{2}}}} \,  =  \,     \ottnt{e_{{\mathrm{0}}}}    [ \tceseq{  \ottnt{A} _{  \mu  }   /  \mathit{X} } ]      [ \tceseq{  \ottnt{A} _{  \nu  }   /  \mathit{Y} } ]      [  \ottnt{v}  /  \mathit{f}  ]      [ \tceseq{ \ottnt{v_{{\mathrm{2}}}}  /  \mathit{x} } ]  $.
          \end{itemize}
          We also have $\ottnt{E_{{\mathrm{1}}}} \,  =  \, \,[\,]\,$ and $ \varpi    =     \epsilon  $.

          Because
          $\ottnt{e_{{\mathrm{1}}}} \,  \prec_{ \ottnt{v} }  \, \ottnt{e'_{{\mathrm{1}}}}$ and
          $\ottnt{e_{{\mathrm{1}}}} \,  =  \, \ottnt{v} \,  \tceseq{ \ottnt{v_{{\mathrm{2}}}} } $ and
          $\ottnt{e'_{{\mathrm{1}}}} \,  =  \,  \ottnt{E'_{{\mathrm{1}}}}  [   \ottnt{v} _{  \mu   \ottsym{;}  \ottsym{0}  \ottsym{;}  \pi_{{\mathrm{0}}} }  \,  \tceseq{ \ottnt{v'_{{\mathrm{2}}}} }   ] $,
          we have
          \begin{itemize}
           \item $\ottnt{E'_{{\mathrm{1}}}} \,  =  \, \,[\,]\,$,
           \item $\ottnt{e'_{{\mathrm{1}}}} \,  =  \,  \ottnt{v} _{  \mu   \ottsym{;}  \ottsym{0}  \ottsym{;}  \pi_{{\mathrm{0}}} }  \,  \tceseq{ \ottnt{v'_{{\mathrm{2}}}} } $, and
           \item $ \tceseq{ \ottnt{v_{{\mathrm{2}}}} \,  \prec_{ \ottnt{v} }  \, \ottnt{v'_{{\mathrm{2}}}} } $.
          \end{itemize}

          For every $\ottmv{i}$,
          let
          $\sigma_{\ottmv{i}}$ be the substitution as described in \refdef{munu-sim}
          (that is, it satisfying $\ottnt{C}  \triangleright   \emptyset   \ottsym{;}   \tceseq{ \ottsym{(}  \mathit{X}  \ottsym{,}  \mathit{Y}  \ottsym{)} }   \ottsym{;}   \tceseq{ \mathit{z_{{\mathrm{0}}}}  \,  {:}  \,  \iota_{{\mathrm{0}}} }  \,  |  \, \ottnt{C} \,  \vdash _{ \ottsym{0} }  \, \sigma_{\ottmv{i}}$ for some $\ottnt{C}$, $ \tceseq{ \mathit{z_{{\mathrm{0}}}} } $, and $ \tceseq{ \iota_{{\mathrm{0}}} } $).\footnote{The substitutions $\sigma_{\ottmv{i}}$ can be determined uniquely as the proof of \reflem{ts-val-inv-rec}.}.
          \[
           \ottnt{e'_{{\mathrm{2}}}} \,  =  \,     \ottnt{e_{{\mathrm{0}}}}    [ \tceseq{ \sigma_{{\mathrm{0}}}  \ottsym{(}  \mathit{X}  \ottsym{)}  /  \mathit{X} } ]      [ \tceseq{  \top^P   /  \mathit{Y} } ]      [   \ottnt{v} _{  \mu   \ottsym{;}  \ottsym{0}  \ottsym{;}  \pi_{{\mathrm{0}}} }   /  \mathit{f}  ]      [ \tceseq{ \ottnt{v_{{\mathrm{2}}}} \,  \prec_{ \ottnt{v} }  \, \ottnt{v'_{{\mathrm{2}}}} \,  \uparrow \, (  \ottsym{1}  ,  \pi_{{\mathrm{0}}}  )   /  \mathit{x} } ]   ~,
          \]
          and $\ottmv{i} \,  \geq  \, \ottsym{1}$ and $\pi$ be any infinite trace.
          Because
          \begin{itemize}
           \item $\ottnt{e_{{\mathrm{1}}}} \,  \prec_{ \ottnt{v} }  \, \ottnt{e'_{{\mathrm{1}}}} \,  \uparrow \, (  \ottmv{i}  ,  \pi  )  \,  =  \,  \ottnt{v} _{  \mu   \ottsym{;}  \ottmv{i}  \ottsym{;}  \pi }  \,  \tceseq{ \ottnt{v_{{\mathrm{2}}}} \,  \prec_{ \ottnt{v} }  \, \ottnt{v'_{{\mathrm{2}}}} \,  \uparrow \, (  \ottmv{i}  ,  \pi  )  } $,
           \item $\ottmv{i} \,  \geq  \, \ottsym{1}$, and
           \item $ \tceseq{ \ottnt{v_{{\mathrm{2}}}} \,  \prec_{ \ottnt{v} }  \, \ottnt{v'_{{\mathrm{2}}}} \,  \uparrow \, (  \ottmv{i}  ,  \pi  )  } $ are values by \reflem{lr-approx-val},
          \end{itemize}
          we have
          \[
           \ottnt{e_{{\mathrm{1}}}} \,  \prec_{ \ottnt{v} }  \, \ottnt{e'_{{\mathrm{1}}}} \,  \uparrow \, (  \ottmv{i}  ,  \pi  )  \,  \mathrel{  \longrightarrow  }  \,     \ottnt{e_{{\mathrm{0}}}}    [ \tceseq{  { \sigma }_{ \ottmv{i}  \ottsym{-}  \ottsym{1} }   \ottsym{(}  \mathit{X}  \ottsym{)}  /  \mathit{X} } ]      [ \tceseq{  \top^P   /  \mathit{Y} } ]      [   \ottnt{v} _{  \mu   \ottsym{;}  \ottmv{i}  \ottsym{-}  \ottsym{1}  \ottsym{;}  \pi }   /  \mathit{f}  ]      [ \tceseq{ \ottnt{v_{{\mathrm{2}}}} \,  \prec_{ \ottnt{v} }  \, \ottnt{v'_{{\mathrm{2}}}} \,  \uparrow \, (  \ottmv{i}  ,  \pi  )   /  \mathit{x} } ]   \,  \,\&\,  \,  \epsilon 
          \]
          by \E{Red}/\R{Beta}.
          Then, it suffices to show that
          \[
               \ottnt{e_{{\mathrm{0}}}}    [ \tceseq{  { \sigma }_{ \ottmv{i}  \ottsym{-}  \ottsym{1} }   \ottsym{(}  \mathit{X}  \ottsym{)}  /  \mathit{X} } ]      [ \tceseq{  \top^P   /  \mathit{Y} } ]      [   \ottnt{v} _{  \mu   \ottsym{;}  \ottmv{i}  \ottsym{-}  \ottsym{1}  \ottsym{;}  \pi }   /  \mathit{f}  ]      [ \tceseq{ \ottnt{v_{{\mathrm{2}}}} \,  \prec_{ \ottnt{v} }  \, \ottnt{v'_{{\mathrm{2}}}} \,  \uparrow \, (  \ottmv{i}  ,  \pi  )   /  \mathit{x} } ]   \,  =  \, \ottnt{e_{{\mathrm{2}}}} \,  \prec_{ \ottnt{v} }  \, \ottnt{e'_{{\mathrm{2}}}} \,  \uparrow \, (  \ottmv{i}  \ottsym{-}  \ottsym{1}  ,  \pi  )  ~.
          \]
          Because
          \begin{itemize}
           \item $\ottnt{e_{{\mathrm{2}}}} \,  =  \,     \ottnt{e_{{\mathrm{0}}}}    [ \tceseq{  \ottnt{A} _{  \mu  }   /  \mathit{X} } ]      [ \tceseq{  \ottnt{A} _{  \nu  }   /  \mathit{Y} } ]      [  \ottnt{v}  /  \mathit{f}  ]      [ \tceseq{ \ottnt{v_{{\mathrm{2}}}}  /  \mathit{x} } ]  $,
           \item $\ottnt{e'_{{\mathrm{2}}}} \,  =  \,     \ottnt{e_{{\mathrm{0}}}}    [ \tceseq{ \sigma_{{\mathrm{0}}}  \ottsym{(}  \mathit{X}  \ottsym{)}  /  \mathit{X} } ]      [ \tceseq{  \top^P   /  \mathit{Y} } ]      [   \ottnt{v} _{  \mu   \ottsym{;}  \ottsym{0}  \ottsym{;}  \pi_{{\mathrm{0}}} }   /  \mathit{f}  ]      [ \tceseq{ \ottnt{v_{{\mathrm{2}}}} \,  \prec_{ \ottnt{v} }  \, \ottnt{v'_{{\mathrm{2}}}} \,  \uparrow \, (  \ottsym{1}  ,  \pi_{{\mathrm{0}}}  )   /  \mathit{x} } ]  $,
           \item $\ottsym{(}  \ottnt{e_{{\mathrm{0}}}} \,  \prec_{ \ottnt{v} }  \, \ottnt{e_{{\mathrm{0}}}} \,  \uparrow \, (  \ottmv{i}  ,  \pi  )   \ottsym{)} \,  =  \, \ottnt{e_{{\mathrm{0}}}}$ by \reflem{lr-approx-refl},
           \item $ \tceseq{  \ottnt{A} _{  \mu  }  \,  \prec_{ \ottnt{v} }  \, \sigma_{{\mathrm{0}}}  \ottsym{(}  \mathit{X}  \ottsym{)} \,  \uparrow \, (  \ottmv{i}  \ottsym{-}  \ottsym{1}  ,  \pi  )  }  =  \tceseq{  { \sigma }_{ \ottmv{i}  \ottsym{-}  \ottsym{1} }   \ottsym{(}  \mathit{X}  \ottsym{)} } $,
           \item $ \tceseq{  \ottnt{A} _{  \nu  }  \,  \prec_{ \ottnt{v} }  \,  \top^P  \,  \uparrow \, (  \ottmv{i}  \ottsym{-}  \ottsym{1}  ,  \pi  )  }  =  \top^P $,
           \item $\ottnt{v} \,  \prec_{ \ottnt{v} }  \,  \ottnt{v} _{  \mu   \ottsym{;}  \ottsym{0}  \ottsym{;}  \pi_{{\mathrm{0}}} }  \,  \uparrow \, (  \ottmv{i}  \ottsym{-}  \ottsym{1}  ,  \pi  )  =  \ottnt{v} _{  \mu   \ottsym{;}  \ottmv{i}  \ottsym{-}  \ottsym{1}  \ottsym{;}  \pi } $, and
           \item $ \tceseq{ \ottnt{v_{{\mathrm{2}}}} \,  \prec_{ \ottnt{v} }  \, \ottsym{(}  \ottnt{v_{{\mathrm{2}}}} \,  \prec_{ \ottnt{v} }  \, \ottnt{v'_{{\mathrm{2}}}} \,  \uparrow \, (  \ottsym{1}  ,  \pi_{{\mathrm{0}}}  )   \ottsym{)} \,  \uparrow \, (  \ottmv{i}  \ottsym{-}  \ottsym{1}  ,  \pi  )  }  =  \tceseq{ \ottnt{v_{{\mathrm{2}}}} \,  \prec_{ \ottnt{v} }  \, \ottnt{v'_{{\mathrm{2}}}} \,  \uparrow \, (  \ottmv{i}  ,  \pi  )  } $ by \reflem{lr-approx-decompose},
          \end{itemize}
          Lemmas~\ref{lem:lr-approx-val-subst} and \ref{lem:lr-approx-pred-subst}
          imply the conclusion.

         \case \E{Let}:
          We are given
          \begin{prooftree}
           \AxiomC{$\ottnt{e_{{\mathrm{01}}}} \,  \mathrel{  \longrightarrow  }  \, \ottnt{e_{{\mathrm{02}}}} \,  \,\&\,  \, \varpi$}
           \UnaryInfC{$\mathsf{let} \, \mathit{x_{{\mathrm{01}}}}  \ottsym{=}  \ottnt{e_{{\mathrm{01}}}} \,  \mathsf{in}  \, \ottnt{e_{{\mathrm{03}}}} \,  \mathrel{  \longrightarrow  }  \, \mathsf{let} \, \mathit{x_{{\mathrm{01}}}}  \ottsym{=}  \ottnt{e_{{\mathrm{02}}}} \,  \mathsf{in}  \, \ottnt{e_{{\mathrm{03}}}} \,  \,\&\,  \, \varpi$}
          \end{prooftree}
          for some $\mathit{x_{{\mathrm{01}}}}$, $\ottnt{e_{{\mathrm{01}}}}$, $\ottnt{e_{{\mathrm{02}}}}$, and $\ottnt{e_{{\mathrm{03}}}}$ such that
          \begin{itemize}
           \item $\ottnt{e_{{\mathrm{1}}}} \,  =  \, \mathsf{let} \, \mathit{x_{{\mathrm{01}}}}  \ottsym{=}  \ottnt{e_{{\mathrm{01}}}} \,  \mathsf{in}  \, \ottnt{e_{{\mathrm{03}}}}$ and
           \item $\ottnt{e_{{\mathrm{2}}}} \,  =  \, \mathsf{let} \, \mathit{x_{{\mathrm{01}}}}  \ottsym{=}  \ottnt{e_{{\mathrm{02}}}} \,  \mathsf{in}  \, \ottnt{e_{{\mathrm{03}}}}$.
          \end{itemize}
          Because $\ottnt{e_{{\mathrm{1}}}} \,  \prec_{ \ottnt{v} }  \, \ottnt{e'_{{\mathrm{1}}}}$, there exist $\ottnt{e'_{{\mathrm{01}}}}$ and $\ottnt{e'_{{\mathrm{03}}}}$ such that
          \begin{itemize}
           \item $\ottnt{e'_{{\mathrm{1}}}} \,  =  \, \mathsf{let} \, \mathit{x_{{\mathrm{01}}}}  \ottsym{=}  \ottnt{e'_{{\mathrm{01}}}} \,  \mathsf{in}  \, \ottnt{e'_{{\mathrm{03}}}}$ and
           \item $\ottnt{e_{{\mathrm{01}}}} \,  \prec_{ \ottnt{v} }  \, \ottnt{e'_{{\mathrm{01}}}}$ and
           \item $\ottnt{e_{{\mathrm{03}}}} \,  \prec_{ \ottnt{v} }  \, \ottnt{e'_{{\mathrm{03}}}}$.
          \end{itemize}
          By the IH, there exists some $\ottnt{e'_{{\mathrm{02}}}}$ such that
          \[
             \forall  \, { \pi } . \    \forall  \, { \ottmv{i} \,  \geq  \, \ottsym{1} } . \  \ottsym{(}  \ottnt{e_{{\mathrm{01}}}} \,  \prec_{ \ottnt{v} }  \, \ottnt{e'_{{\mathrm{01}}}} \,  \uparrow \, (  \ottmv{i}  ,  \pi  )   \ottsym{)} \,  \mathrel{  \longrightarrow  }  \, \ottsym{(}  \ottnt{e_{{\mathrm{02}}}} \,  \prec_{ \ottnt{v} }  \, \ottnt{e'_{{\mathrm{02}}}} \,  \uparrow \, (  \ottmv{i}  \ottsym{-}  \ottsym{1}  ,  \pi  )   \ottsym{)} \,  \,\&\,  \, \varpi   ~.
          \]

          Let $\pi_{{\mathrm{0}}}$ be an arbitrary infinite trace and
          \[
           \ottnt{e'_{{\mathrm{2}}}} \,  =  \, \mathsf{let} \, \mathit{x_{{\mathrm{01}}}}  \ottsym{=}  \ottnt{e'_{{\mathrm{02}}}} \,  \mathsf{in}  \, \ottsym{(}  \ottnt{e_{{\mathrm{03}}}} \,  \prec_{ \ottnt{v} }  \, \ottnt{e'_{{\mathrm{03}}}} \,  \uparrow \, (  \ottsym{1}  ,  \pi_{{\mathrm{0}}}  )   \ottsym{)} ~.
          \]
          Note that $\ottnt{e_{{\mathrm{03}}}} \,  \prec_{ \ottnt{v} }  \, \ottnt{e'_{{\mathrm{03}}}} \,  \uparrow \, (  \ottsym{1}  ,  \pi_{{\mathrm{0}}}  ) $ is well defined
          by \reflem{lr-approx-wf} with $\ottnt{e_{{\mathrm{03}}}} \,  \prec_{ \ottnt{v} }  \, \ottnt{e'_{{\mathrm{03}}}}$.
          Further, let $\ottmv{i} \,  \geq  \, \ottsym{1}$ and $\pi$ be any infinite trace.
          We have
          \[\begin{array}{ll} &
           \ottnt{e_{{\mathrm{1}}}} \,  \prec_{ \ottnt{v} }  \, \ottnt{e'_{{\mathrm{1}}}} \,  \uparrow \, (  \ottmv{i}  ,  \pi  )  \\ = &
           \mathsf{let} \, \mathit{x_{{\mathrm{01}}}}  \ottsym{=}  \ottsym{(}  \ottnt{e_{{\mathrm{01}}}} \,  \prec_{ \ottnt{v} }  \, \ottnt{e'_{{\mathrm{01}}}} \,  \uparrow \, (  \ottmv{i}  ,  \pi  )   \ottsym{)} \,  \mathsf{in}  \, \ottsym{(}  \ottnt{e_{{\mathrm{03}}}} \,  \prec_{ \ottnt{v} }  \, \ottnt{e'_{{\mathrm{03}}}} \,  \uparrow \, (  \ottmv{i}  ,  \pi  )   \ottsym{)} \\  \longrightarrow  &
           \mathsf{let} \, \mathit{x_{{\mathrm{01}}}}  \ottsym{=}  \ottsym{(}  \ottnt{e_{{\mathrm{02}}}} \,  \prec_{ \ottnt{v} }  \, \ottnt{e'_{{\mathrm{02}}}} \,  \uparrow \, (  \ottmv{i}  \ottsym{-}  \ottsym{1}  ,  \pi  )   \ottsym{)} \,  \mathsf{in}  \, \ottsym{(}  \ottnt{e_{{\mathrm{03}}}} \,  \prec_{ \ottnt{v} }  \, \ottnt{e'_{{\mathrm{03}}}} \,  \uparrow \, (  \ottmv{i}  ,  \pi  )   \ottsym{)} \,  \,\&\,  \, \varpi
            \\ &
            \quad \text{(by the result of the IH and \E{Let})} ~.
            \end{array}
          \]
          Because
          $\ottnt{e_{{\mathrm{03}}}} \,  \prec_{ \ottnt{v} }  \, \ottnt{e'_{{\mathrm{03}}}} \,  \uparrow \, (  \ottmv{i}  ,  \pi  )  \,  =  \, \ottnt{e_{{\mathrm{03}}}} \,  \prec_{ \ottnt{v} }  \, \ottsym{(}  \ottnt{e_{{\mathrm{03}}}} \,  \prec_{ \ottnt{v} }  \, \ottnt{e'_{{\mathrm{03}}}} \,  \uparrow \, (  \ottsym{1}  ,  \pi_{{\mathrm{0}}}  )   \ottsym{)} \,  \uparrow \, (  \ottmv{i}  \ottsym{-}  \ottsym{1}  ,  \pi  ) $ by \reflem{lr-approx-decompose},
          we have
          \[\begin{array}{ll} &
           \ottnt{e_{{\mathrm{1}}}} \,  \prec_{ \ottnt{v} }  \, \ottnt{e'_{{\mathrm{1}}}} \,  \uparrow \, (  \ottmv{i}  ,  \pi  )  \\  \longrightarrow  &
           \mathsf{let} \, \mathit{x_{{\mathrm{01}}}}  \ottsym{=}  \ottsym{(}  \ottnt{e_{{\mathrm{02}}}} \,  \prec_{ \ottnt{v} }  \, \ottnt{e'_{{\mathrm{02}}}} \,  \uparrow \, (  \ottmv{i}  \ottsym{-}  \ottsym{1}  ,  \pi  )   \ottsym{)} \,  \mathsf{in}  \, \ottsym{(}  \ottnt{e_{{\mathrm{03}}}} \,  \prec_{ \ottnt{v} }  \, \ottsym{(}  \ottnt{e_{{\mathrm{03}}}} \,  \prec_{ \ottnt{v} }  \, \ottnt{e'_{{\mathrm{03}}}} \,  \uparrow \, (  \ottsym{1}  ,  \pi_{{\mathrm{0}}}  )   \ottsym{)} \,  \uparrow \, (  \ottmv{i}  \ottsym{-}  \ottsym{1}  ,  \pi  )   \ottsym{)} \,  \,\&\,  \, \varpi \\ = &
           \ottnt{e_{{\mathrm{2}}}} \,  \prec_{ \ottnt{v} }  \, \ottnt{e'_{{\mathrm{2}}}} \,  \uparrow \, (  \ottmv{i}  \ottsym{-}  \ottsym{1}  ,  \pi  )  \,  \,\&\,  \, \varpi ~.
            \end{array}
          \]

         \case \E{Reset}:
          We are given
          \begin{prooftree}
           \AxiomC{$\ottnt{e_{{\mathrm{01}}}} \,  \mathrel{  \longrightarrow  }  \, \ottnt{e_{{\mathrm{02}}}} \,  \,\&\,  \, \varpi$}
           \UnaryInfC{$ \resetcnstr{ \ottnt{e_{{\mathrm{01}}}} }  \,  \mathrel{  \longrightarrow  }  \,  \resetcnstr{ \ottnt{e_{{\mathrm{02}}}} }  \,  \,\&\,  \, \varpi$}
          \end{prooftree}
          for some $\ottnt{e_{{\mathrm{01}}}}$ and $\ottnt{e_{{\mathrm{02}}}}$ such that
          $\ottnt{e_{{\mathrm{1}}}} \,  =  \,  \resetcnstr{ \ottnt{e_{{\mathrm{01}}}} } $ and $\ottnt{e_{{\mathrm{2}}}} \,  =  \,  \resetcnstr{ \ottnt{e_{{\mathrm{02}}}} } $.
          Because $\ottnt{e_{{\mathrm{1}}}} \,  \prec_{ \ottnt{v} }  \, \ottnt{e'_{{\mathrm{1}}}}$ and $\ottnt{e_{{\mathrm{1}}}} \,  =  \,  \resetcnstr{ \ottnt{e_{{\mathrm{01}}}} } $,
          there exists some $\ottnt{e'_{{\mathrm{01}}}}$ such that
          \begin{itemize}
           \item $\ottnt{e'_{{\mathrm{1}}}} \,  =  \,  \resetcnstr{ \ottnt{e'_{{\mathrm{01}}}} } $ and
           \item $\ottnt{e_{{\mathrm{01}}}} \,  \prec_{ \ottnt{v} }  \, \ottnt{e'_{{\mathrm{01}}}}$.
          \end{itemize}
          By the IH, there exists some $\ottnt{e'_{{\mathrm{02}}}}$ such that
          \[
             \forall  \, { \pi } . \    \forall  \, { \ottmv{i} \,  \geq  \, \ottsym{1} } . \  \ottsym{(}  \ottnt{e_{{\mathrm{01}}}} \,  \prec_{ \ottnt{v} }  \, \ottnt{e'_{{\mathrm{01}}}} \,  \uparrow \, (  \ottmv{i}  ,  \pi  )   \ottsym{)} \,  \mathrel{  \longrightarrow  }  \, \ottsym{(}  \ottnt{e_{{\mathrm{02}}}} \,  \prec_{ \ottnt{v} }  \, \ottnt{e'_{{\mathrm{02}}}} \,  \uparrow \, (  \ottmv{i}  \ottsym{-}  \ottsym{1}  ,  \pi  )   \ottsym{)} \,  \,\&\,  \, \varpi   ~.
          \]

          Let $\ottnt{e'_{{\mathrm{2}}}} \,  =  \,  \resetcnstr{ \ottnt{e'_{{\mathrm{02}}}} } $, $\ottmv{i} \,  \geq  \, \ottsym{1}$, and $\pi$ be any infinite trace.
          We have
          \[\begin{array}{ll} &
           \ottnt{e_{{\mathrm{1}}}} \,  \prec_{ \ottnt{v} }  \, \ottnt{e'_{{\mathrm{1}}}} \,  \uparrow \, (  \ottmv{i}  ,  \pi  )  \\ = &
            \resetcnstr{ \ottnt{e_{{\mathrm{01}}}} \,  \prec_{ \ottnt{v} }  \, \ottnt{e'_{{\mathrm{01}}}} \,  \uparrow \, (  \ottmv{i}  ,  \pi  )  }  \\  \longrightarrow  &
            \resetcnstr{ \ottnt{e_{{\mathrm{02}}}} \,  \prec_{ \ottnt{v} }  \, \ottnt{e'_{{\mathrm{02}}}} \,  \uparrow \, (  \ottmv{i}  \ottsym{-}  \ottsym{1}  ,  \pi  )  }  \,  \,\&\,  \, \varpi
            \quad \text{(by the result of the IH and \E{Reset})} \\ = &
           \ottnt{e_{{\mathrm{2}}}} \,  \prec_{ \ottnt{v} }  \, \ottnt{e'_{{\mathrm{2}}}} \,  \uparrow \, (  \ottmv{i}  \ottsym{-}  \ottsym{1}  ,  \pi  )  ~.
            \end{array}
          \]

         \otherwise: Contradictory (in the cases of
           \E{Red}/\R{Const}, \E{Red}/\R{Beta}, \E{Red}/\R{PBeta},
           \E{Red}/\R{IfT}, \E{Red}/\R{IfF}, \E{Red}/\R{Let},
           \E{Red}/\R{Shift}, \E{Red}/\R{Reset}, and \E{Ev}).

        \end{caseanalysis}

  \item By induction on the derivation of $\ottnt{e_{{\mathrm{1}}}} \,  \mathrel{  \longrightarrow  }  \, \ottnt{e_{{\mathrm{2}}}} \,  \,\&\,  \, \varpi$.
        Note that $\ottnt{e_{{\mathrm{1}}}} \,  \prec_{ \ottnt{v} }  \, \ottnt{e'_{{\mathrm{1}}}} \,  \uparrow \, (  \ottmv{i}  ,  \pi  ) $ is well defined by \reflem{lr-approx-wf}.
        \begin{caseanalysis}
         \case \E{Red}/\R{Const}:
          We are given
          \begin{prooftree}
           \AxiomC{$\ottnt{o}  \ottsym{(}   \tceseq{ \ottnt{c} }   \ottsym{)} \,  \rightsquigarrow  \,  \zeta  (  \ottnt{o}  ,   \tceseq{ \ottnt{c} }   ) $}
           \UnaryInfC{$\ottnt{o}  \ottsym{(}   \tceseq{ \ottnt{c} }   \ottsym{)} \,  \mathrel{  \longrightarrow  }  \,  \zeta  (  \ottnt{o}  ,   \tceseq{ \ottnt{c} }   )  \,  \,\&\,  \,  \epsilon $}
          \end{prooftree}
          for some $\ottnt{o}$ and $ \tceseq{ \ottnt{c} } $ such that
          $\ottnt{e_{{\mathrm{1}}}} \,  =  \, \ottnt{o}  \ottsym{(}   \tceseq{ \ottnt{c} }   \ottsym{)}$ and $\ottnt{e_{{\mathrm{2}}}} \,  =  \,  \zeta  (  \ottnt{o}  ,   \tceseq{ \ottnt{c} }   ) $.
          We also have $ \varpi    =     \epsilon  $.
          Because $\ottnt{e_{{\mathrm{1}}}} \,  \prec_{ \ottnt{v} }  \, \ottnt{e'_{{\mathrm{1}}}}$ and $\ottnt{e_{{\mathrm{1}}}} \,  =  \, \ottnt{o}  \ottsym{(}   \tceseq{ \ottnt{c} }   \ottsym{)}$ imply
          $\ottnt{e'_{{\mathrm{1}}}} \,  =  \, \ottnt{o}  \ottsym{(}   \tceseq{ \ottnt{c} }   \ottsym{)}$.
          We have the conclusion by letting $\ottnt{e'_{{\mathrm{2}}}} \,  =  \,  \zeta  (  \ottnt{o}  ,   \tceseq{ \ottnt{c} }   ) $
          because, for any $\ottmv{i}$ and $\pi$,
          \begin{itemize}
           \item $\ottnt{e_{{\mathrm{1}}}} \,  \prec_{ \ottnt{v} }  \, \ottnt{e'_{{\mathrm{1}}}} \,  \uparrow \, (  \ottmv{i}  ,  \pi  )  \,  =  \, \ottnt{o}  \ottsym{(}   \tceseq{ \ottnt{c} }   \ottsym{)}$,
           \item $\ottnt{e_{{\mathrm{2}}}} \,  \prec_{ \ottnt{v} }  \, \ottnt{e'_{{\mathrm{2}}}} \,  \uparrow \, (  \ottmv{i}  ,  \pi  )  \,  =  \,  \zeta  (  \ottnt{o}  ,   \tceseq{ \ottnt{c} }   ) $, and
           \item $\ottnt{o}  \ottsym{(}   \tceseq{ \ottnt{c} }   \ottsym{)} \,  \mathrel{  \longrightarrow  }  \,  \zeta  (  \ottnt{o}  ,   \tceseq{ \ottnt{c} }   )  \,  \,\&\,  \,  \epsilon $.
          \end{itemize}

         \case \E{Red}/\R{Beta}:
          We are given
          \begin{prooftree}
            \AxiomC{$\ottsym{(}   \lambda\!  \,  \tceseq{ \mathit{x} }   \ottsym{.}  \ottnt{e_{{\mathrm{0}}}}  \ottsym{)} \,  \tceseq{ \ottnt{v_{{\mathrm{2}}}} }  \,  \rightsquigarrow  \,  \ottnt{e_{{\mathrm{0}}}}    [ \tceseq{ \ottnt{v_{{\mathrm{2}}}}  /  \mathit{x} } ]  $}
            \UnaryInfC{$\ottsym{(}   \lambda\!  \,  \tceseq{ \mathit{x} }   \ottsym{.}  \ottnt{e_{{\mathrm{0}}}}  \ottsym{)} \,  \tceseq{ \ottnt{v_{{\mathrm{2}}}} }  \,  \mathrel{  \longrightarrow  }  \,  \ottnt{e_{{\mathrm{0}}}}    [ \tceseq{ \ottnt{v_{{\mathrm{2}}}}  /  \mathit{x} } ]   \,  \,\&\,  \,  \epsilon $}
          \end{prooftree}
          for some $ \tceseq{ \mathit{x} } $, $\ottnt{e_{{\mathrm{0}}}}$, and $ \tceseq{ \ottnt{v_{{\mathrm{2}}}} } $ such that
          \begin{itemize}
           \item $\ottnt{e_{{\mathrm{1}}}} \,  =  \, \ottsym{(}   \lambda\!  \,  \tceseq{ \mathit{x} }   \ottsym{.}  \ottnt{e_{{\mathrm{0}}}}  \ottsym{)} \,  \tceseq{ \ottnt{v_{{\mathrm{2}}}} } $,
           \item $\ottnt{e_{{\mathrm{2}}}} \,  =  \,  \ottnt{e_{{\mathrm{0}}}}    [ \tceseq{ \ottnt{v_{{\mathrm{2}}}}  /  \mathit{x} } ]  $, and
           \item $\ottsym{\mbox{$\mid$}}   \tceseq{ \ottnt{v_{{\mathrm{2}}}} }   \ottsym{\mbox{$\mid$}} \,  =  \, \ottsym{\mbox{$\mid$}}   \tceseq{ \mathit{x} }   \ottsym{\mbox{$\mid$}}$.
          \end{itemize}
          We also have $ \varpi    =     \epsilon  $.
          Because $\ottnt{e_{{\mathrm{1}}}} \,  \prec_{ \ottnt{v} }  \, \ottnt{e'_{{\mathrm{1}}}}$ and $\ottnt{e_{{\mathrm{1}}}} \,  =  \, \ottsym{(}   \lambda\!  \,  \tceseq{ \mathit{x} }   \ottsym{.}  \ottnt{e_{{\mathrm{0}}}}  \ottsym{)} \,  \tceseq{ \ottnt{v_{{\mathrm{2}}}} } $,
          we have
          \begin{itemize}
           \item $\ottnt{e'_{{\mathrm{1}}}} \,  =  \, \ottsym{(}   \lambda\!  \,  \tceseq{ \mathit{x} }   \ottsym{.}  \ottnt{e'_{{\mathrm{0}}}}  \ottsym{)} \,  \tceseq{ \ottnt{v'_{{\mathrm{2}}}} } $,
           \item $\ottsym{(}   \lambda\!  \,  \tceseq{ \mathit{x} }   \ottsym{.}  \ottnt{e_{{\mathrm{0}}}}  \ottsym{)} \,  \prec_{ \ottnt{v} }  \, \ottsym{(}   \lambda\!  \,  \tceseq{ \mathit{x} }   \ottsym{.}  \ottnt{e'_{{\mathrm{0}}}}  \ottsym{)}$ (so $\ottnt{e_{{\mathrm{0}}}} \,  \prec_{ \ottnt{v} }  \, \ottnt{e'_{{\mathrm{0}}}}$), and
           \item $ \tceseq{ \ottnt{v_{{\mathrm{2}}}} \,  \prec_{ \ottnt{v} }  \, \ottnt{v'_{{\mathrm{2}}}} } $
          \end{itemize}
          for some $\ottnt{e'_{{\mathrm{0}}}}$ and $ \tceseq{ \ottnt{v'_{{\mathrm{2}}}} } $.

          Let
          $\ottnt{e'_{{\mathrm{2}}}} \,  =  \,  \ottnt{e'_{{\mathrm{0}}}}    [ \tceseq{ \ottnt{v'_{{\mathrm{2}}}}  /  \mathit{x} } ]  $,
          $\ottmv{i}$ be any natural number, and
          $\pi$ be any infinite trace.
          Note that $ \ottnt{e'_{{\mathrm{0}}}}    [ \tceseq{ \ottnt{v'_{{\mathrm{2}}}}  /  \mathit{x} } ]  $ is well defined by \reflem{lr-wf-subst-approx}.
          Because
          \begin{itemize}
           \item $\ottnt{e_{{\mathrm{1}}}} \,  \prec_{ \ottnt{v} }  \, \ottnt{e'_{{\mathrm{1}}}} \,  \uparrow \, (  \ottmv{i}  ,  \pi  )  \,  =  \, \ottsym{(}   \lambda\!  \,  \tceseq{ \mathit{x} }   \ottsym{.}  \ottsym{(}  \ottnt{e_{{\mathrm{0}}}} \,  \prec_{ \ottnt{v} }  \, \ottnt{e'_{{\mathrm{0}}}} \,  \uparrow \, (  \ottmv{i}  ,  \pi  )   \ottsym{)}  \ottsym{)} \,  \tceseq{ \ottnt{v_{{\mathrm{2}}}} \,  \prec_{ \ottnt{v} }  \, \ottnt{v'_{{\mathrm{2}}}} \,  \uparrow \, (  \ottmv{i}  ,  \pi  )  } $ and
           \item $ \tceseq{ \ottnt{v_{{\mathrm{2}}}} \,  \prec_{ \ottnt{v} }  \, \ottnt{v'_{{\mathrm{2}}}} \,  \uparrow \, (  \ottmv{i}  ,  \pi  )  } $ are values by \reflem{lr-approx-val},
          \end{itemize}
          we have
          \[
          \ottnt{e_{{\mathrm{1}}}} \,  \prec_{ \ottnt{v} }  \, \ottnt{e'_{{\mathrm{1}}}} \,  \uparrow \, (  \ottmv{i}  ,  \pi  )  \,  \mathrel{  \longrightarrow  }  \,  \ottsym{(}  \ottnt{e_{{\mathrm{0}}}} \,  \prec_{ \ottnt{v} }  \, \ottnt{e'_{{\mathrm{0}}}} \,  \uparrow \, (  \ottmv{i}  ,  \pi  )   \ottsym{)}    [ \tceseq{ \ottnt{v_{{\mathrm{2}}}} \,  \prec_{ \ottnt{v} }  \, \ottnt{v'_{{\mathrm{2}}}} \,  \uparrow \, (  \ottmv{i}  ,  \pi  )   /  \mathit{x} } ]   \,  \,\&\,  \,  \epsilon 
          \]
          by \E{Red}/\R{Beta}.
          By \reflem{lr-approx-val-subst}, we have the conclusion
          \[
           \ottnt{e_{{\mathrm{1}}}} \,  \prec_{ \ottnt{v} }  \, \ottnt{e'_{{\mathrm{1}}}} \,  \uparrow \, (  \ottmv{i}  ,  \pi  )  \,  \mathrel{  \longrightarrow  }  \, \ottsym{(}   \ottnt{e_{{\mathrm{0}}}}    [ \tceseq{ \ottnt{v_{{\mathrm{2}}}}  /  \mathit{x} } ]   \,  \prec_{ \ottnt{v} }  \,  \ottnt{e'_{{\mathrm{0}}}}    [ \tceseq{ \ottnt{v'_{{\mathrm{2}}}}  /  \mathit{x} } ]   \,  \uparrow \, (  \ottmv{i}  ,  \pi  )   \ottsym{)} \,  \,\&\,  \,  \epsilon  ~.
          \]

         \case \Tr{Red}/\R{RBeta}:
          We are given
          \begin{prooftree}
            \AxiomC{$\ottnt{v_{{\mathrm{1}}}} \,  \tceseq{ \ottnt{v_{{\mathrm{2}}}} }  \,  \rightsquigarrow  \,     \ottnt{e_{{\mathrm{0}}}}    [ \tceseq{  \ottnt{A} _{  \mu  }   /  \mathit{X} } ]      [ \tceseq{  \ottnt{A} _{  \nu  }   /  \mathit{Y} } ]      [  \ottnt{v_{{\mathrm{1}}}}  /  \mathit{f}  ]      [ \tceseq{ \ottnt{v_{{\mathrm{2}}}}  /  \mathit{x} } ]  $}
            \UnaryInfC{$\ottnt{v_{{\mathrm{1}}}} \,  \tceseq{ \ottnt{v_{{\mathrm{2}}}} }  \,  \mathrel{  \longrightarrow  }  \,     \ottnt{e_{{\mathrm{0}}}}    [ \tceseq{  \ottnt{A} _{  \mu  }   /  \mathit{X} } ]      [ \tceseq{  \ottnt{A} _{  \nu  }   /  \mathit{Y} } ]      [  \ottnt{v_{{\mathrm{1}}}}  /  \mathit{f}  ]      [ \tceseq{ \ottnt{v_{{\mathrm{2}}}}  /  \mathit{x} } ]   \,  \,\&\,  \,  \epsilon $}
          \end{prooftree}
          for some $\ottnt{v_{{\mathrm{1}}}}$, $ \tceseq{ \ottnt{v_{{\mathrm{2}}}} } $, $ \tceseq{  \ottnt{A} _{  \mu  }  } $, $ \tceseq{  \ottnt{A} _{  \nu  }  } $, $ \tceseq{ \mathit{X} } $, $ \tceseq{ \mathit{Y} } $,
          $\mathit{f}$, $ \tceseq{ \mathit{x} } $, and $\ottnt{e_{{\mathrm{0}}}}$ such that
          \begin{itemize}
            \item $\ottnt{e_{{\mathrm{1}}}} \,  =  \, \ottnt{v_{{\mathrm{1}}}} \,  \tceseq{ \ottnt{v_{{\mathrm{2}}}} } $,
            \item $\ottnt{v_{{\mathrm{1}}}} \,  =  \,  \mathsf{rec}  \, (  \tceseq{  \mathit{X} ^{  \ottnt{A} _{  \mu  }  },  \mathit{Y} ^{  \ottnt{A} _{  \nu  }  }  }  ,  \mathit{f} ,   \tceseq{ \mathit{x} }  ) . \,  \ottnt{e_{{\mathrm{0}}}} $, and
            \item $\ottnt{e_{{\mathrm{2}}}} \,  =  \,     \ottnt{e_{{\mathrm{0}}}}    [ \tceseq{  \ottnt{A} _{  \mu  }   /  \mathit{X} } ]      [ \tceseq{  \ottnt{A} _{  \nu  }   /  \mathit{Y} } ]      [  \ottnt{v_{{\mathrm{1}}}}  /  \mathit{f}  ]      [ \tceseq{ \ottnt{v_{{\mathrm{2}}}}  /  \mathit{x} } ]  $.
          \end{itemize}
          We also have $ \varpi    =     \epsilon  $.
          Because $\ottnt{e_{{\mathrm{1}}}} \,  \prec_{ \ottnt{v} }  \, \ottnt{e'_{{\mathrm{1}}}}$ and $\ottnt{e_{{\mathrm{1}}}} \,  =  \, \ottnt{v_{{\mathrm{1}}}} \,  \tceseq{ \ottnt{v_{{\mathrm{2}}}} } $,
          there exist some $\ottnt{v'_{{\mathrm{1}}}}$ and $ \tceseq{ \ottnt{v'_{{\mathrm{2}}}} } $ such that
          \begin{itemize}
            \item $\ottnt{e'_{{\mathrm{1}}}} \,  =  \, \ottnt{v'_{{\mathrm{1}}}} \,  \tceseq{ \ottnt{v'_{{\mathrm{2}}}} } $,
            \item $\ottnt{v_{{\mathrm{1}}}} \,  \prec_{ \ottnt{v} }  \, \ottnt{v'_{{\mathrm{1}}}}$, and
            \item $ \tceseq{ \ottnt{v_{{\mathrm{2}}}} \,  \prec_{ \ottnt{v} }  \, \ottnt{v'_{{\mathrm{2}}}} } $.
          \end{itemize}
          We proceed by case analysis on $\ottnt{v'_{{\mathrm{1}}}}$.
          Because $\ottnt{v_{{\mathrm{1}}}} \,  \prec_{ \ottnt{v} }  \, \ottnt{v'_{{\mathrm{1}}}}$ and $\ottnt{v_{{\mathrm{1}}}}$ is a recursive function,
          we have only two cases to be considered.
          \begin{caseanalysis}
           \case $  \exists  \, {  \tceseq{  \ottnt{A'} _{  \mu  }  }   \ottsym{,}   \tceseq{  \ottnt{A'} _{  \nu  }  }   \ottsym{,}  \ottnt{e'_{{\mathrm{0}}}} } . \  \ottnt{v'_{{\mathrm{1}}}} \,  =  \,  \mathsf{rec}  \, (  \tceseq{  \mathit{X} ^{  \ottnt{A'} _{  \mu  }  },  \mathit{Y} ^{  \ottnt{A'} _{  \nu  }  }  }  ,  \mathit{f} ,   \tceseq{ \mathit{x} }  ) . \,  \ottnt{e'_{{\mathrm{0}}}}  $:
            Because $\ottnt{v_{{\mathrm{1}}}} \,  \prec_{ \ottnt{v} }  \, \ottnt{v'_{{\mathrm{1}}}}$, we have
            \begin{itemize}
             \item $ \tceseq{  \ottnt{A} _{  \mu  }  \,  \prec_{ \ottnt{v} }  \,  \ottnt{A'} _{  \mu  }  } $,
             \item $ \tceseq{  \ottnt{A} _{  \nu  }  \,  \prec_{ \ottnt{v} }  \,  \ottnt{A'} _{  \nu  }  } $, and
             \item $\ottnt{e_{{\mathrm{0}}}} \,  \prec_{ \ottnt{v} }  \, \ottnt{e'_{{\mathrm{0}}}}$.
            \end{itemize}
            Let
            \[
             \ottnt{e'_{{\mathrm{2}}}} \,  =  \,     \ottnt{e'_{{\mathrm{0}}}}    [ \tceseq{  \ottnt{A'} _{  \mu  }   /  \mathit{X} } ]      [ \tceseq{  \ottnt{A'} _{  \nu  }   /  \mathit{Y} } ]      [  \ottnt{v'_{{\mathrm{1}}}}  /  \mathit{f}  ]      [ \tceseq{ \ottnt{v'_{{\mathrm{2}}}}  /  \mathit{x} } ]   ~,
            \]
            and $\ottmv{i}$ be any natural number and $\pi$ be any infinite trace.
            Because $ \tceseq{ \ottnt{v_{{\mathrm{2}}}} \,  \prec_{ \ottnt{v} }  \, \ottnt{v'_{{\mathrm{2}}}} \,  \uparrow \, (  \ottmv{i}  ,  \pi  )  } $ are values by \reflem{lr-approx-val},
            we have
            \[\begin{array}{ll} &
             \ottnt{e_{{\mathrm{1}}}} \,  \prec_{ \ottnt{v} }  \, \ottnt{e'_{{\mathrm{1}}}} \,  \uparrow \, (  \ottmv{i}  ,  \pi  )  \\ = &
             \ottnt{v_{{\mathrm{0}}}} \,  \tceseq{ \ottnt{v_{{\mathrm{2}}}} \,  \prec_{ \ottnt{v} }  \, \ottnt{v'_{{\mathrm{2}}}} \,  \uparrow \, (  \ottmv{i}  ,  \pi  )  }  \\  \longrightarrow  &
                 \ottsym{(}  \ottnt{e_{{\mathrm{0}}}} \,  \prec_{ \ottnt{v} }  \, \ottnt{e'_{{\mathrm{0}}}} \,  \uparrow \, (  \ottmv{i}  ,  \pi  )   \ottsym{)}    [ \tceseq{  \ottnt{A} _{  \mu  }  \,  \prec_{ \ottnt{v} }  \,  \ottnt{A'} _{  \mu  }  \,  \uparrow \, (  \ottmv{i}  ,  \pi  )   /  \mathit{X} } ]      [ \tceseq{  \ottnt{A} _{  \nu  }  \,  \prec_{ \ottnt{v} }  \,  \ottnt{A'} _{  \nu  }  \,  \uparrow \, (  \ottmv{i}  ,  \pi  )   /  \mathit{Y} } ]      [  \ottnt{v_{{\mathrm{0}}}}  /  \mathit{f}  ]      [ \tceseq{ \ottnt{v_{{\mathrm{2}}}} \,  \prec_{ \ottnt{v} }  \, \ottnt{v'_{{\mathrm{2}}}} \,  \uparrow \, (  \ottmv{i}  ,  \pi  )   /  \mathit{x} } ]   \,  \,\&\,  \,  \epsilon 
             \\ & \quad \text{(by \E{Red}/\R{RBeta})}
              \end{array}
            \]
            where
            \[
             \ottnt{v_{{\mathrm{0}}}} = \ottnt{v_{{\mathrm{1}}}} \,  \prec_{ \ottnt{v} }  \, \ottnt{v'_{{\mathrm{1}}}} \,  \uparrow \, (  \ottmv{i}  ,  \pi  )  = \ottsym{(}   \mathsf{rec}  \, (  \tceseq{  \mathit{X} ^{  \ottnt{A} _{  \mu  }  \,  \prec_{ \ottnt{v} }  \,  \ottnt{A'} _{  \mu  }  \,  \uparrow \, (  \ottmv{i}  ,  \pi  )  },  \mathit{Y} ^{  \ottnt{A} _{  \nu  }  \,  \prec_{ \ottnt{v} }  \,  \ottnt{A'} _{  \nu  }  \,  \uparrow \, (  \ottmv{i}  ,  \pi  )  }  }  ,  \mathit{f} ,   \tceseq{ \mathit{x} }  ) . \,  \ottsym{(}  \ottnt{e_{{\mathrm{0}}}} \,  \prec_{ \ottnt{v} }  \, \ottnt{e'_{{\mathrm{0}}}} \,  \uparrow \, (  \ottmv{i}  ,  \pi  )   \ottsym{)}   \ottsym{)} ~.
            \]
            Then, it suffices to show that
            \[\begin{array}{ll} &
                 \ottsym{(}  \ottnt{e_{{\mathrm{0}}}} \,  \prec_{ \ottnt{v} }  \, \ottnt{e'_{{\mathrm{0}}}} \,  \uparrow \, (  \ottmv{i}  ,  \pi  )   \ottsym{)}    [ \tceseq{  \ottnt{A} _{  \mu  }  \,  \prec_{ \ottnt{v} }  \,  \ottnt{A'} _{  \mu  }  \,  \uparrow \, (  \ottmv{i}  ,  \pi  )   /  \mathit{X} } ]      [ \tceseq{  \ottnt{A} _{  \nu  }  \,  \prec_{ \ottnt{v} }  \,  \ottnt{A'} _{  \nu  }  \,  \uparrow \, (  \ottmv{i}  ,  \pi  )   /  \mathit{Y} } ]      [  \ottnt{v_{{\mathrm{0}}}}  /  \mathit{f}  ]      [ \tceseq{ \ottnt{v_{{\mathrm{2}}}} \,  \prec_{ \ottnt{v} }  \, \ottnt{v'_{{\mathrm{2}}}} \,  \uparrow \, (  \ottmv{i}  ,  \pi  )   /  \mathit{x} } ]  
             \\ = &
             \ottnt{e_{{\mathrm{2}}}} \,  \prec_{ \ottnt{v} }  \, \ottnt{e'_{{\mathrm{2}}}} \,  \uparrow \, (  \ottmv{i}  ,  \pi  )  ~.
              \end{array}
            \]
            Because
            \begin{itemize}
             \item $\ottnt{e_{{\mathrm{2}}}} \,  =  \,     \ottnt{e_{{\mathrm{0}}}}    [ \tceseq{  \ottnt{A} _{  \mu  }   /  \mathit{X} } ]      [ \tceseq{  \ottnt{A} _{  \nu  }   /  \mathit{Y} } ]      [  \ottnt{v_{{\mathrm{1}}}}  /  \mathit{f}  ]      [ \tceseq{ \ottnt{v_{{\mathrm{2}}}}  /  \mathit{x} } ]  $ and
             \item $\ottnt{e'_{{\mathrm{2}}}} \,  =  \,     \ottnt{e'_{{\mathrm{0}}}}    [ \tceseq{  \ottnt{A'} _{  \mu  }   /  \mathit{X} } ]      [ \tceseq{  \ottnt{A'} _{  \nu  }   /  \mathit{Y} } ]      [  \ottnt{v'_{{\mathrm{1}}}}  /  \mathit{f}  ]      [ \tceseq{ \ottnt{v'_{{\mathrm{2}}}}  /  \mathit{x} } ]  $,
            \end{itemize}
            Lemmas~\ref{lem:lr-approx-val-subst} and \ref{lem:lr-approx-pred-subst}
            imply the conclusion.

           \case $  \exists  \, { \ottmv{j}  \ottsym{,}  \pi_{{\mathrm{0}}} } . \  \ottnt{v'_{{\mathrm{1}}}} \,  =  \,  \ottnt{v} _{  \mu   \ottsym{;}  \ottmv{j}  \ottsym{;}  \pi_{{\mathrm{0}}} }  $:
            In this case, $\ottnt{v_{{\mathrm{1}}}} \,  =  \, \ottnt{v}$.
            Because
            $\ottnt{e_{{\mathrm{1}}}} \,  =  \, \ottnt{v_{{\mathrm{1}}}} \,  \tceseq{ \ottnt{v_{{\mathrm{2}}}} }  = \ottnt{v} \,  \tceseq{ \ottnt{v_{{\mathrm{2}}}} } $ and
            $ \mathit{arity}  (  \ottnt{v}  )  \,  =  \, \ottsym{\mbox{$\mid$}}   \tceseq{ \ottnt{v_{{\mathrm{2}}}} }   \ottsym{\mbox{$\mid$}}$,
            the assumption implies $\ottnt{e'_{{\mathrm{1}}}} \,  \not=  \,  \ottnt{v} _{  \mu   \ottsym{;}  \ottsym{0}  \ottsym{;}  \pi_{{\mathrm{0}}} }  \,  \tceseq{ \ottnt{v'_{{\mathrm{2}}}} } $, that is,
            $ \ottnt{v} _{  \mu   \ottsym{;}  \ottmv{j}  \ottsym{;}  \pi_{{\mathrm{0}}} }  \,  \not=  \,  \ottnt{v} _{  \mu   \ottsym{;}  \ottsym{0}  \ottsym{;}  \pi_{{\mathrm{0}}} } $.
            Therefore, $\ottmv{j} \,  >  \, \ottsym{0}$.

            For every $\ottmv{i}$,
            let
            $\sigma_{\ottmv{i}}$ be the substitution as described in \refdef{munu-sim}, and
            \[
             \ottnt{e'_{{\mathrm{2}}}} \,  =  \,     \ottnt{e_{{\mathrm{0}}}}    [ \tceseq{  { \sigma }_{ \ottmv{j}  \ottsym{-}  \ottsym{1} }   \ottsym{(}  \mathit{X}  \ottsym{)}  /  \mathit{X} } ]      [ \tceseq{  \top^P   /  \mathit{Y} } ]      [   \ottnt{v} _{  \mu   \ottsym{;}  \ottmv{j}  \ottsym{-}  \ottsym{1}  \ottsym{;}  \pi_{{\mathrm{0}}} }   /  \mathit{f}  ]      [ \tceseq{ \ottnt{v'_{{\mathrm{2}}}}  /  \mathit{x} } ]   ~.
            \]
            Furthermore, let $\ottmv{i}$ be any natural number and $\pi$ be any infinite trace.
            Because $ \tceseq{ \ottnt{v_{{\mathrm{2}}}} \,  \prec_{ \ottnt{v} }  \, \ottnt{v'_{{\mathrm{2}}}} \,  \uparrow \, (  \ottmv{i}  ,  \pi  )  } $ are values by \reflem{lr-approx-val},
            we have the conclusion by:
            \[\begin{array}{ll} &
             \ottnt{e_{{\mathrm{1}}}} \,  \prec_{ \ottnt{v} }  \, \ottnt{e'_{{\mathrm{1}}}} \,  \uparrow \, (  \ottmv{i}  ,  \pi  )  \\ = &
              \ottnt{v} _{  \mu   \ottsym{;}  \ottmv{j}  \ottsym{+}  \ottmv{i}  \ottsym{;}  \pi }  \,  \tceseq{ \ottnt{v_{{\mathrm{2}}}} \,  \prec_{ \ottnt{v} }  \, \ottnt{v'_{{\mathrm{2}}}} \,  \uparrow \, (  \ottmv{i}  ,  \pi  )  }  \\  \longrightarrow  &
                 \ottnt{e_{{\mathrm{0}}}}    [ \tceseq{  { \sigma }_{ \ottmv{j}  \ottsym{+}  \ottmv{i}  \ottsym{-}  \ottsym{1} }   \ottsym{(}  \mathit{X}  \ottsym{)}  /  \mathit{X} } ]      [ \tceseq{  \top^P   /  \mathit{Y} } ]      [   \ottnt{v} _{  \mu   \ottsym{;}  \ottmv{j}  \ottsym{+}  \ottmv{i}  \ottsym{-}  \ottsym{1}  \ottsym{;}  \pi }   /  \mathit{f}  ]      [ \tceseq{ \ottnt{v_{{\mathrm{2}}}} \,  \prec_{ \ottnt{v} }  \, \ottnt{v'_{{\mathrm{2}}}} \,  \uparrow \, (  \ottmv{i}  ,  \pi  )   /  \mathit{x} } ]   \,  \,\&\,  \,  \epsilon 
              \\ & \quad \text{(by \E{Red}/\R{Beta})}
              \\ = &
                 \ottnt{e_{{\mathrm{0}}}}    [ \tceseq{  \ottnt{A} _{  \mu  }   /  \mathit{X} } ]      [ \tceseq{  \ottnt{A} _{  \nu  }   /  \mathit{Y} } ]      [  \ottnt{v}  /  \mathit{f}  ]      [ \tceseq{ \ottnt{v_{{\mathrm{2}}}}  /  \mathit{x} } ]   \,  \prec_{ \ottnt{v} }  \,     \ottnt{e_{{\mathrm{0}}}}    [ \tceseq{  { \sigma }_{ \ottmv{j}  \ottsym{-}  \ottsym{1} }   \ottsym{(}  \mathit{X}  \ottsym{)}  /  \mathit{X} } ]      [ \tceseq{  \top^P   /  \mathit{Y} } ]      [   \ottnt{v} _{  \mu   \ottsym{;}  \ottmv{j}  \ottsym{-}  \ottsym{1}  \ottsym{;}  \pi_{{\mathrm{0}}} }   /  \mathit{f}  ]      [ \tceseq{ \ottnt{v'_{{\mathrm{2}}}}  /  \mathit{x} } ]   \,  \uparrow \, (  \ottmv{i}  ,  \pi  )  \,  \,\&\,  \,  \epsilon 
              \\ & \quad \text{(by Lemmas~\ref{lem:lr-approx-refl}, \ref{lem:lr-approx-val-subst}, and \ref{lem:lr-approx-pred-subst})}
              \\ = &
             \ottnt{e_{{\mathrm{2}}}} \,  \prec_{ \ottnt{v} }  \, \ottnt{e'_{{\mathrm{2}}}} \,  \uparrow \, (  \ottmv{i}  ,  \pi  )  \,  \,\&\,  \,  \epsilon ~.
              \end{array}
            \]

          \end{caseanalysis}

         \case \E{Red}/\R{PBeta}:
          We are given
          \begin{prooftree}
           \AxiomC{$\ottsym{(}   \Lambda   \ottsym{(}  \mathit{X} \,  {:}  \,  \tceseq{ \ottnt{s} }   \ottsym{)}  \ottsym{.}  \ottnt{e_{{\mathrm{0}}}}  \ottsym{)} \, \ottnt{A} \,  \rightsquigarrow  \,  \ottnt{e_{{\mathrm{0}}}}    [  \ottnt{A}  /  \mathit{X}  ]  $}
           \UnaryInfC{$\ottsym{(}   \Lambda   \ottsym{(}  \mathit{X} \,  {:}  \,  \tceseq{ \ottnt{s} }   \ottsym{)}  \ottsym{.}  \ottnt{e_{{\mathrm{0}}}}  \ottsym{)} \, \ottnt{A} \,  \mathrel{  \longrightarrow  }  \,  \ottnt{e_{{\mathrm{0}}}}    [  \ottnt{A}  /  \mathit{X}  ]   \,  \,\&\,  \,  \epsilon $}
          \end{prooftree}
          for some $\mathit{X}$, $ \tceseq{ \ottnt{s} } $, $\ottnt{e_{{\mathrm{0}}}}$, and $\ottnt{A}$ such that
          \begin{itemize}
           \item $\ottnt{e_{{\mathrm{1}}}} \,  =  \, \ottsym{(}   \Lambda   \ottsym{(}  \mathit{X} \,  {:}  \,  \tceseq{ \ottnt{s} }   \ottsym{)}  \ottsym{.}  \ottnt{e_{{\mathrm{0}}}}  \ottsym{)} \, \ottnt{A}$ and
           \item $\ottnt{e_{{\mathrm{2}}}} \,  =  \,  \ottnt{e_{{\mathrm{0}}}}    [  \ottnt{A}  /  \mathit{X}  ]  $.
          \end{itemize}
          We also have $ \varpi    =     \epsilon  $.
          Because $\ottnt{e_{{\mathrm{1}}}} \,  \prec_{ \ottnt{v} }  \, \ottnt{e'_{{\mathrm{1}}}}$ and $\ottnt{e_{{\mathrm{1}}}} \,  =  \, \ottsym{(}   \Lambda   \ottsym{(}  \mathit{X} \,  {:}  \,  \tceseq{ \ottnt{s} }   \ottsym{)}  \ottsym{.}  \ottnt{e_{{\mathrm{0}}}}  \ottsym{)} \, \ottnt{A}$,
          we have
          \begin{itemize}
           \item $\ottnt{e'_{{\mathrm{1}}}} \,  =  \, \ottsym{(}   \Lambda   \ottsym{(}  \mathit{X} \,  {:}  \,  \tceseq{ \ottnt{s} }   \ottsym{)}  \ottsym{.}  \ottnt{e'_{{\mathrm{0}}}}  \ottsym{)} \, \ottnt{A'}$,
           \item $\ottnt{e_{{\mathrm{0}}}} \,  \prec_{ \ottnt{v} }  \, \ottnt{e'_{{\mathrm{0}}}}$, and
           \item $\ottnt{A} \,  \prec_{ \ottnt{v} }  \, \ottnt{A'}$
          \end{itemize}
          for some $\ottnt{e'_{{\mathrm{0}}}}$ and $\ottnt{A'}$.

          Let
          $\ottnt{e'_{{\mathrm{2}}}} \,  =  \,  \ottnt{e'_{{\mathrm{0}}}}    [  \ottnt{A'}  /  \mathit{X}  ]  $,
          $\ottmv{i}$ be any natural number, and
          $\pi$ be any infinite trace.
          We have
          \[\begin{array}{ll} &
           \ottnt{e_{{\mathrm{1}}}} \,  \prec_{ \ottnt{v} }  \, \ottnt{e'_{{\mathrm{1}}}} \,  \uparrow \, (  \ottmv{i}  ,  \pi  )  \\ = &
           \ottsym{(}   \Lambda   \ottsym{(}  \mathit{X} \,  {:}  \,  \tceseq{ \ottnt{s} }   \ottsym{)}  \ottsym{.}  \ottsym{(}  \ottnt{e_{{\mathrm{0}}}} \,  \prec_{ \ottnt{v} }  \, \ottnt{e'_{{\mathrm{0}}}} \,  \uparrow \, (  \ottmv{i}  ,  \pi  )   \ottsym{)}  \ottsym{)} \, \ottsym{(}  \ottnt{A} \,  \prec_{ \ottnt{v} }  \, \ottnt{A'} \,  \uparrow \, (  \ottmv{i}  ,  \pi  )   \ottsym{)} \\  \longrightarrow  &
            \ottsym{(}  \ottnt{e_{{\mathrm{0}}}} \,  \prec_{ \ottnt{v} }  \, \ottnt{e'_{{\mathrm{0}}}} \,  \uparrow \, (  \ottmv{i}  ,  \pi  )   \ottsym{)}    [  \ottnt{A} \,  \prec_{ \ottnt{v} }  \, \ottnt{A'} \,  \uparrow \, (  \ottmv{i}  ,  \pi  )   /  \mathit{X}  ]   \,  \,\&\,  \,  \epsilon 
           \quad \text{(by \E{Red}/\R{PBeta})} ~.
            \end{array}
          \]
          Then, it suffices to show that
          \[
            \ottsym{(}  \ottnt{e_{{\mathrm{0}}}} \,  \prec_{ \ottnt{v} }  \, \ottnt{e'_{{\mathrm{0}}}} \,  \uparrow \, (  \ottmv{i}  ,  \pi  )   \ottsym{)}    [  \ottnt{A} \,  \prec_{ \ottnt{v} }  \, \ottnt{A'} \,  \uparrow \, (  \ottmv{i}  ,  \pi  )   /  \mathit{X}  ]   \,  =  \, \ottnt{e_{{\mathrm{2}}}} \,  \prec_{ \ottnt{v} }  \, \ottnt{e'_{{\mathrm{2}}}} \,  \uparrow \, (  \ottmv{i}  ,  \pi  )  ~,
          \]
          which is proven by \reflem{lr-approx-pred-subst}.

         \case \E{Red}/\R{IfT}:
          We are given
          \begin{prooftree}
           \AxiomC{$\mathsf{if} \,  \mathsf{true}  \, \mathsf{then} \, \ottnt{e_{{\mathrm{2}}}} \, \mathsf{else} \, \ottnt{e_{{\mathrm{02}}}} \,  \rightsquigarrow  \, \ottnt{e_{{\mathrm{2}}}}$}
           \UnaryInfC{$\mathsf{if} \,  \mathsf{true}  \, \mathsf{then} \, \ottnt{e_{{\mathrm{2}}}} \, \mathsf{else} \, \ottnt{e_{{\mathrm{02}}}} \,  \mathrel{  \longrightarrow  }  \, \ottnt{e_{{\mathrm{2}}}} \,  \,\&\,  \,  \epsilon $}
          \end{prooftree}
          for some $\ottnt{e_{{\mathrm{02}}}}$ such that
          $\ottnt{e_{{\mathrm{1}}}} \,  =  \, \mathsf{if} \,  \mathsf{true}  \, \mathsf{then} \, \ottnt{e_{{\mathrm{2}}}} \, \mathsf{else} \, \ottnt{e_{{\mathrm{02}}}}$.
          We also have $ \varpi    =     \epsilon  $.
          Because $\ottnt{e_{{\mathrm{1}}}} \,  \prec_{ \ottnt{v} }  \, \ottnt{e'_{{\mathrm{1}}}}$ and $\ottnt{e_{{\mathrm{1}}}} \,  =  \, \mathsf{if} \,  \mathsf{true}  \, \mathsf{then} \, \ottnt{e_{{\mathrm{2}}}} \, \mathsf{else} \, \ottnt{e_{{\mathrm{02}}}}$,
          we have
          \begin{itemize}
           \item $\ottnt{e'_{{\mathrm{1}}}} \,  =  \, \mathsf{if} \,  \mathsf{true}  \, \mathsf{then} \, \ottnt{e'_{{\mathrm{2}}}} \, \mathsf{else} \, \ottnt{e'_{{\mathrm{02}}}}$,
           \item $\ottnt{e_{{\mathrm{2}}}} \,  \prec_{ \ottnt{v} }  \, \ottnt{e'_{{\mathrm{2}}}}$, and
           \item $\ottnt{e_{{\mathrm{02}}}} \,  \prec_{ \ottnt{v} }  \, \ottnt{e'_{{\mathrm{02}}}}$
          \end{itemize}
          for some $\ottnt{e'_{{\mathrm{2}}}}$ and $\ottnt{e'_{{\mathrm{02}}}}$.

          Let
          $\ottmv{i}$ be any natural number and
          $\pi$ be any infinite trace.
          We have
          \[\begin{array}{ll} &
           \ottnt{e_{{\mathrm{1}}}} \,  \prec_{ \ottnt{v} }  \, \ottnt{e'_{{\mathrm{1}}}} \,  \uparrow \, (  \ottmv{i}  ,  \pi  )  \\ = &
           \mathsf{if} \,  \mathsf{true}  \, \mathsf{then} \, \ottsym{(}  \ottnt{e_{{\mathrm{2}}}} \,  \prec_{ \ottnt{v} }  \, \ottnt{e'_{{\mathrm{2}}}} \,  \uparrow \, (  \ottmv{i}  ,  \pi  )   \ottsym{)} \, \mathsf{else} \, \ottsym{(}  \ottnt{e_{{\mathrm{02}}}} \,  \prec_{ \ottnt{v} }  \, \ottnt{e'_{{\mathrm{02}}}} \,  \uparrow \, (  \ottmv{i}  ,  \pi  )   \ottsym{)} \,  \mathrel{  \longrightarrow  }  \, \ottsym{(}  \ottnt{e_{{\mathrm{2}}}} \,  \prec_{ \ottnt{v} }  \, \ottnt{e'_{{\mathrm{2}}}} \,  \uparrow \, (  \ottmv{i}  ,  \pi  )   \ottsym{)} \,  \,\&\,  \,  \epsilon 
            \\ & \quad \text{(by \E{Red}/\R{IfT})} ~.
            \end{array}
          \]
          Therefore, we have the conclusion.

         \case \E{Red}/\R{IfF}:
          We are given
          \begin{prooftree}
           \AxiomC{$\mathsf{if} \,  \mathsf{false}  \, \mathsf{then} \, \ottnt{e_{{\mathrm{01}}}} \, \mathsf{else} \, \ottnt{e_{{\mathrm{2}}}} \,  \rightsquigarrow  \, \ottnt{e_{{\mathrm{2}}}}$}
           \UnaryInfC{$\mathsf{if} \,  \mathsf{false}  \, \mathsf{then} \, \ottnt{e_{{\mathrm{01}}}} \, \mathsf{else} \, \ottnt{e_{{\mathrm{2}}}} \,  \mathrel{  \longrightarrow  }  \, \ottnt{e_{{\mathrm{2}}}} \,  \,\&\,  \,  \epsilon $}
          \end{prooftree}
          for some $\ottnt{e_{{\mathrm{01}}}}$ such that
          $\ottnt{e_{{\mathrm{1}}}} \,  =  \, \mathsf{if} \,  \mathsf{false}  \, \mathsf{then} \, \ottnt{e_{{\mathrm{01}}}} \, \mathsf{else} \, \ottnt{e_{{\mathrm{2}}}}$.
          We also have $ \varpi    =     \epsilon  $.
          Because $\ottnt{e_{{\mathrm{1}}}} \,  \prec_{ \ottnt{v} }  \, \ottnt{e'_{{\mathrm{1}}}}$ and $\ottnt{e_{{\mathrm{1}}}} \,  =  \, \mathsf{if} \,  \mathsf{false}  \, \mathsf{then} \, \ottnt{e_{{\mathrm{01}}}} \, \mathsf{else} \, \ottnt{e_{{\mathrm{2}}}}$,
          we have
          \begin{itemize}
           \item $\ottnt{e'_{{\mathrm{1}}}} \,  =  \, \mathsf{if} \,  \mathsf{false}  \, \mathsf{then} \, \ottnt{e'_{{\mathrm{01}}}} \, \mathsf{else} \, \ottnt{e'_{{\mathrm{2}}}}$,
           \item $\ottnt{e_{{\mathrm{01}}}} \,  \prec_{ \ottnt{v} }  \, \ottnt{e'_{{\mathrm{01}}}}$, and
           \item $\ottnt{e_{{\mathrm{2}}}} \,  \prec_{ \ottnt{v} }  \, \ottnt{e'_{{\mathrm{2}}}}$
          \end{itemize}
          for some $\ottnt{e'_{{\mathrm{01}}}}$ and $\ottnt{e'_{{\mathrm{2}}}}$.

          Let
          $\ottmv{i}$ be any natural number and
          $\pi$ be any infinite trace.
          We have
          \[\begin{array}{ll} &
           \ottnt{e_{{\mathrm{1}}}} \,  \prec_{ \ottnt{v} }  \, \ottnt{e'_{{\mathrm{1}}}} \,  \uparrow \, (  \ottmv{i}  ,  \pi  )  \\ = &
           \mathsf{if} \,  \mathsf{false}  \, \mathsf{then} \, \ottsym{(}  \ottnt{e_{{\mathrm{01}}}} \,  \prec_{ \ottnt{v} }  \, \ottnt{e'_{{\mathrm{01}}}} \,  \uparrow \, (  \ottmv{i}  ,  \pi  )   \ottsym{)} \, \mathsf{else} \, \ottsym{(}  \ottnt{e_{{\mathrm{2}}}} \,  \prec_{ \ottnt{v} }  \, \ottnt{e'_{{\mathrm{2}}}} \,  \uparrow \, (  \ottmv{i}  ,  \pi  )   \ottsym{)} \,  \mathrel{  \longrightarrow  }  \, \ottsym{(}  \ottnt{e_{{\mathrm{2}}}} \,  \prec_{ \ottnt{v} }  \, \ottnt{e'_{{\mathrm{2}}}} \,  \uparrow \, (  \ottmv{i}  ,  \pi  )   \ottsym{)} \,  \,\&\,  \,  \epsilon 
            \\ & \quad \text{(by \E{Red}/\R{IfF})} ~.
            \end{array}
          \]
          Therefore, we have the conclusion.

         \case \E{Red}/\R{Let}:
          We are given
          \begin{prooftree}
           \AxiomC{$\mathsf{let} \, \mathit{x_{{\mathrm{01}}}}  \ottsym{=}  \ottnt{v_{{\mathrm{01}}}} \,  \mathsf{in}  \, \ottnt{e_{{\mathrm{02}}}} \,  \rightsquigarrow  \,  \ottnt{e_{{\mathrm{02}}}}    [  \ottnt{v_{{\mathrm{01}}}}  /  \mathit{x_{{\mathrm{01}}}}  ]  $}
           \UnaryInfC{$\mathsf{let} \, \mathit{x_{{\mathrm{01}}}}  \ottsym{=}  \ottnt{v_{{\mathrm{01}}}} \,  \mathsf{in}  \, \ottnt{e_{{\mathrm{02}}}} \,  \mathrel{  \longrightarrow  }  \,  \ottnt{e_{{\mathrm{02}}}}    [  \ottnt{v_{{\mathrm{01}}}}  /  \mathit{x_{{\mathrm{01}}}}  ]   \,  \,\&\,  \,  \epsilon $}
          \end{prooftree}
          for some $\mathit{x_{{\mathrm{01}}}}$, $\ottnt{v_{{\mathrm{01}}}}$, and $\ottnt{e_{{\mathrm{02}}}}$ such that
          \begin{itemize}
           \item $\ottnt{e_{{\mathrm{1}}}} \,  =  \, \mathsf{let} \, \mathit{x_{{\mathrm{01}}}}  \ottsym{=}  \ottnt{v_{{\mathrm{01}}}} \,  \mathsf{in}  \, \ottnt{e_{{\mathrm{02}}}}$ and
           \item $\ottnt{e_{{\mathrm{2}}}} \,  =  \,  \ottnt{e_{{\mathrm{02}}}}    [  \ottnt{v_{{\mathrm{01}}}}  /  \mathit{x_{{\mathrm{01}}}}  ]  $.
          \end{itemize}
          We also have $ \varpi    =     \epsilon  $.
          Because $\ottnt{e_{{\mathrm{1}}}} \,  \prec_{ \ottnt{v} }  \, \ottnt{e'_{{\mathrm{1}}}}$ and $\ottnt{e_{{\mathrm{1}}}} \,  =  \, \mathsf{let} \, \mathit{x_{{\mathrm{01}}}}  \ottsym{=}  \ottnt{v_{{\mathrm{01}}}} \,  \mathsf{in}  \, \ottnt{e_{{\mathrm{02}}}}$,
          we have
          \begin{itemize}
           \item $\ottnt{e'_{{\mathrm{1}}}} \,  =  \, \mathsf{let} \, \mathit{x_{{\mathrm{01}}}}  \ottsym{=}  \ottnt{v'_{{\mathrm{01}}}} \,  \mathsf{in}  \, \ottnt{e'_{{\mathrm{02}}}}$,
           \item $\ottnt{v_{{\mathrm{01}}}} \,  \prec_{ \ottnt{v} }  \, \ottnt{v'_{{\mathrm{01}}}}$, and
           \item $\ottnt{e_{{\mathrm{02}}}} \,  \prec_{ \ottnt{v} }  \, \ottnt{e'_{{\mathrm{02}}}}$
          \end{itemize}
          for some $\ottnt{v'_{{\mathrm{01}}}}$ and $\ottnt{e'_{{\mathrm{02}}}}$,
          by \reflem{lr-approx-val}.

          Let
          $\ottnt{e'_{{\mathrm{2}}}} \,  =  \,  \ottnt{e'_{{\mathrm{02}}}}    [  \ottnt{v'_{{\mathrm{01}}}}  /  \mathit{x_{{\mathrm{01}}}}  ]  $,
          $\ottmv{i}$ be any natural number, and
          $\pi$ be any infinite trace.
          Note that $ \ottnt{e'_{{\mathrm{02}}}}    [  \ottnt{v'_{{\mathrm{01}}}}  /  \mathit{x_{{\mathrm{01}}}}  ]  $ is well defined by \reflem{lr-wf-subst-approx}.
          Because $\ottnt{v_{{\mathrm{01}}}} \,  \prec_{ \ottnt{v} }  \, \ottnt{v'_{{\mathrm{01}}}} \,  \uparrow \, (  \ottmv{i}  ,  \pi  ) $ is a value by \reflem{lr-approx-val},
          we have
          \[\begin{array}{ll} &
           \ottnt{e_{{\mathrm{1}}}} \,  \prec_{ \ottnt{v} }  \, \ottnt{e'_{{\mathrm{1}}}} \,  \uparrow \, (  \ottmv{i}  ,  \pi  )  \\ = &
           \mathsf{let} \, \mathit{x_{{\mathrm{01}}}}  \ottsym{=}  \ottsym{(}  \ottnt{v_{{\mathrm{01}}}} \,  \prec_{ \ottnt{v} }  \, \ottnt{v'_{{\mathrm{01}}}} \,  \uparrow \, (  \ottmv{i}  ,  \pi  )   \ottsym{)} \,  \mathsf{in}  \, \ottsym{(}  \ottnt{e_{{\mathrm{02}}}} \,  \prec_{ \ottnt{v} }  \, \ottnt{e'_{{\mathrm{02}}}} \,  \uparrow \, (  \ottmv{i}  ,  \pi  )   \ottsym{)} \\
             \longrightarrow  &  \ottsym{(}  \ottnt{e_{{\mathrm{02}}}} \,  \prec_{ \ottnt{v} }  \, \ottnt{e'_{{\mathrm{02}}}} \,  \uparrow \, (  \ottmv{i}  ,  \pi  )   \ottsym{)}    [  \ottnt{v_{{\mathrm{01}}}} \,  \prec_{ \ottnt{v} }  \, \ottnt{v'_{{\mathrm{01}}}} \,  \uparrow \, (  \ottmv{i}  ,  \pi  )   /  \mathit{x_{{\mathrm{01}}}}  ]   \,  \,\&\,  \,  \epsilon 
            \\ & \quad \text{(by \E{Red}/\R{Let})} ~.
            \end{array}
          \]
          Then, it suffices to show that
          \[
            \ottsym{(}  \ottnt{e_{{\mathrm{02}}}} \,  \prec_{ \ottnt{v} }  \, \ottnt{e'_{{\mathrm{02}}}} \,  \uparrow \, (  \ottmv{i}  ,  \pi  )   \ottsym{)}    [  \ottnt{v_{{\mathrm{01}}}} \,  \prec_{ \ottnt{v} }  \, \ottnt{v'_{{\mathrm{01}}}} \,  \uparrow \, (  \ottmv{i}  ,  \pi  )   /  \mathit{x_{{\mathrm{01}}}}  ]   \,  =  \, \ottnt{e_{{\mathrm{2}}}} \,  \prec_{ \ottnt{v} }  \, \ottnt{e'_{{\mathrm{2}}}} \,  \uparrow \, (  \ottmv{i}  ,  \pi  )  ~,
          \]
          which is proven by \reflem{lr-approx-val-subst}.

         \case \E{Red}/\R{Shift}:
          We are given
          \begin{prooftree}
           \AxiomC{$ \resetcnstr{  \ottnt{K_{{\mathrm{0}}}}  [  \mathcal{S}_0 \, \mathit{x_{{\mathrm{0}}}}  \ottsym{.}  \ottnt{e_{{\mathrm{0}}}}  ]  }  \,  \rightsquigarrow  \,  \ottnt{e_{{\mathrm{0}}}}    [   \lambda\!  \, \mathit{x_{{\mathrm{0}}}}  \ottsym{.}   \resetcnstr{  \ottnt{K_{{\mathrm{0}}}}  [  \mathit{x_{{\mathrm{0}}}}  ]  }   /  \mathit{x_{{\mathrm{0}}}}  ]  $}
           \UnaryInfC{$ \resetcnstr{  \ottnt{K_{{\mathrm{0}}}}  [  \mathcal{S}_0 \, \mathit{x_{{\mathrm{0}}}}  \ottsym{.}  \ottnt{e_{{\mathrm{0}}}}  ]  }  \,  \mathrel{  \longrightarrow  }  \,  \ottnt{e_{{\mathrm{0}}}}    [   \lambda\!  \, \mathit{x_{{\mathrm{0}}}}  \ottsym{.}   \resetcnstr{  \ottnt{K_{{\mathrm{0}}}}  [  \mathit{x_{{\mathrm{0}}}}  ]  }   /  \mathit{x_{{\mathrm{0}}}}  ]   \,  \,\&\,  \,  \epsilon $}
          \end{prooftree}
          for some $\ottnt{K_{{\mathrm{0}}}}$, $\mathit{x_{{\mathrm{0}}}}$, and $\ottnt{e_{{\mathrm{0}}}}$ such that
          \begin{itemize}
           \item $\ottnt{e_{{\mathrm{1}}}} \,  =  \,  \resetcnstr{  \ottnt{K_{{\mathrm{0}}}}  [  \mathcal{S}_0 \, \mathit{x_{{\mathrm{0}}}}  \ottsym{.}  \ottnt{e_{{\mathrm{0}}}}  ]  } $ and
           \item $\ottnt{e_{{\mathrm{2}}}} \,  =  \,  \ottnt{e_{{\mathrm{0}}}}    [   \lambda\!  \, \mathit{x_{{\mathrm{0}}}}  \ottsym{.}   \resetcnstr{  \ottnt{K_{{\mathrm{0}}}}  [  \mathit{x_{{\mathrm{0}}}}  ]  }   /  \mathit{x_{{\mathrm{0}}}}  ]  $.
          \end{itemize}
          We also have $ \varpi    =     \epsilon  $.
          Because $\ottnt{e_{{\mathrm{1}}}} \,  \prec_{ \ottnt{v} }  \, \ottnt{e'_{{\mathrm{1}}}}$ and $\ottnt{e_{{\mathrm{1}}}} \,  =  \,  \resetcnstr{  \ottnt{K_{{\mathrm{0}}}}  [  \mathcal{S}_0 \, \mathit{x_{{\mathrm{0}}}}  \ottsym{.}  \ottnt{e_{{\mathrm{0}}}}  ]  } $,
          we have
          \begin{itemize}
           \item $\ottnt{e'_{{\mathrm{1}}}} \,  =  \,  \resetcnstr{  \ottnt{K'_{{\mathrm{0}}}}  [  \mathcal{S}_0 \, \mathit{x_{{\mathrm{0}}}}  \ottsym{.}  \ottnt{e'_{{\mathrm{0}}}}  ]  } $,
           \item $\ottnt{e_{{\mathrm{0}}}} \,  \prec_{ \ottnt{v} }  \, \ottnt{e'_{{\mathrm{0}}}}$, and
           \item $\ottnt{K_{{\mathrm{0}}}} \,  \prec_{ \ottnt{v} }  \, \ottnt{K'_{{\mathrm{0}}}}$
          \end{itemize}
          for some $\ottnt{K'_{{\mathrm{0}}}}$ and $\ottnt{e'_{{\mathrm{0}}}}$,
          by \reflem{lr-approx-ectx}.

          Let
          $\ottnt{e'_{{\mathrm{2}}}} \,  =  \,  \ottnt{e'_{{\mathrm{0}}}}    [   \lambda\!  \, \mathit{x_{{\mathrm{0}}}}  \ottsym{.}   \resetcnstr{  \ottnt{K'_{{\mathrm{0}}}}  [  \mathit{x_{{\mathrm{0}}}}  ]  }   /  \mathit{x_{{\mathrm{0}}}}  ]  $,
          $\ottmv{i}$ be any natural number, and
          $\pi$ be any infinite trace.
          Note that $ \ottnt{e'_{{\mathrm{0}}}}    [   \lambda\!  \, \mathit{x_{{\mathrm{0}}}}  \ottsym{.}   \resetcnstr{  \ottnt{K'_{{\mathrm{0}}}}  [  \mathit{x_{{\mathrm{0}}}}  ]  }   /  \mathit{x_{{\mathrm{0}}}}  ]  $ is well defined by
          \reflem{lr-wf-subst-approx} because
          $\ottsym{(}   \lambda\!  \, \mathit{x_{{\mathrm{0}}}}  \ottsym{.}   \resetcnstr{  \ottnt{K_{{\mathrm{0}}}}  [  \mathit{x_{{\mathrm{0}}}}  ]  }   \ottsym{)} \,  \prec_{ \ottnt{v} }  \, \ottsym{(}   \lambda\!  \, \mathit{x_{{\mathrm{0}}}}  \ottsym{.}   \resetcnstr{  \ottnt{K'_{{\mathrm{0}}}}  [  \mathit{x_{{\mathrm{0}}}}  ]  }   \ottsym{)}$ is well defined
          by \reflem{lr-approx-embed-exp} with $\ottnt{K} \,  \prec_{ \ottnt{v} }  \, \ottnt{K'}$ and $\mathit{x_{{\mathrm{0}}}} \,  \prec_{ \ottnt{v} }  \, \mathit{x_{{\mathrm{0}}}}$.
          We have
          \[\begin{array}{ll} &
           \ottnt{e_{{\mathrm{1}}}} \,  \prec_{ \ottnt{v} }  \, \ottnt{e'_{{\mathrm{1}}}} \,  \uparrow \, (  \ottmv{i}  ,  \pi  ) 
            \\ = &
            \resetcnstr{  \ottnt{K_{{\mathrm{0}}}}  [  \mathcal{S}_0 \, \mathit{x_{{\mathrm{0}}}}  \ottsym{.}  \ottnt{e_{{\mathrm{0}}}}  ]  }  \,  \prec_{ \ottnt{v} }  \,  \resetcnstr{  \ottnt{K'_{{\mathrm{0}}}}  [  \mathcal{S}_0 \, \mathit{x_{{\mathrm{0}}}}  \ottsym{.}  \ottnt{e'_{{\mathrm{0}}}}  ]  }  \,  \uparrow \, (  \ottmv{i}  ,  \pi  ) 
            \\ = &
            \resetcnstr{  \ottsym{(}  \ottnt{K_{{\mathrm{0}}}} \,  \prec_{ \ottnt{v} }  \, \ottnt{K'_{{\mathrm{0}}}} \,  \uparrow \, (  \ottmv{i}  ,  \pi  )   \ottsym{)}  [  \mathcal{S}_0 \, \mathit{x_{{\mathrm{0}}}}  \ottsym{.}  \ottsym{(}  \ottnt{e_{{\mathrm{0}}}} \,  \prec_{ \ottnt{v} }  \, \ottnt{e'_{{\mathrm{0}}}} \,  \uparrow \, (  \ottmv{i}  ,  \pi  )   \ottsym{)}  ]  } 
            \quad \text{(by \reflem{lr-approx-embed-exp})}
            \\  \longrightarrow  &
            \ottsym{(}  \ottnt{e_{{\mathrm{0}}}} \,  \prec_{ \ottnt{v} }  \, \ottnt{e'_{{\mathrm{0}}}} \,  \uparrow \, (  \ottmv{i}  ,  \pi  )   \ottsym{)}    [   \lambda\!  \, \mathit{x_{{\mathrm{0}}}}  \ottsym{.}   \resetcnstr{  \ottsym{(}  \ottnt{K_{{\mathrm{0}}}} \,  \prec_{ \ottnt{v} }  \, \ottnt{K'_{{\mathrm{0}}}} \,  \uparrow \, (  \ottmv{i}  ,  \pi  )   \ottsym{)}  [  \mathit{x_{{\mathrm{0}}}}  ]  }   /  \mathit{x_{{\mathrm{0}}}}  ]   \,  \,\&\,  \,  \epsilon 
            \quad \text{(by \E{Red}/\R{Shift})}
            \\ = &
            \ottsym{(}  \ottnt{e_{{\mathrm{0}}}} \,  \prec_{ \ottnt{v} }  \, \ottnt{e'_{{\mathrm{0}}}} \,  \uparrow \, (  \ottmv{i}  ,  \pi  )   \ottsym{)}    [   \lambda\!  \, \mathit{x_{{\mathrm{0}}}}  \ottsym{.}   \resetcnstr{  \ottsym{(}  \ottnt{K_{{\mathrm{0}}}} \,  \prec_{ \ottnt{v} }  \, \ottnt{K'_{{\mathrm{0}}}} \,  \uparrow \, (  \ottmv{i}  ,  \pi  )   \ottsym{)}  [  \mathit{x_{{\mathrm{0}}}} \,  \prec_{ \ottnt{v} }  \, \mathit{x_{{\mathrm{0}}}} \,  \uparrow \, (  \ottmv{i}  ,  \pi  )   ]  }   /  \mathit{x_{{\mathrm{0}}}}  ]   \,  \,\&\,  \,  \epsilon 
            \\ = &
            \ottsym{(}  \ottnt{e_{{\mathrm{0}}}} \,  \prec_{ \ottnt{v} }  \, \ottnt{e'_{{\mathrm{0}}}} \,  \uparrow \, (  \ottmv{i}  ,  \pi  )   \ottsym{)}    [   \lambda\!  \, \mathit{x_{{\mathrm{0}}}}  \ottsym{.}   \resetcnstr{  \ottnt{K_{{\mathrm{0}}}}  [  \mathit{x_{{\mathrm{0}}}}  ]  \,  \prec_{ \ottnt{v} }  \,  \ottnt{K'_{{\mathrm{0}}}}  [  \mathit{x_{{\mathrm{0}}}}  ]  \,  \uparrow \, (  \ottmv{i}  ,  \pi  )  }   /  \mathit{x_{{\mathrm{0}}}}  ]   \,  \,\&\,  \,  \epsilon 
            \quad \text{(by \reflem{lr-approx-embed-exp})}
            \\ = &
            \ottsym{(}  \ottnt{e_{{\mathrm{0}}}} \,  \prec_{ \ottnt{v} }  \, \ottnt{e'_{{\mathrm{0}}}} \,  \uparrow \, (  \ottmv{i}  ,  \pi  )   \ottsym{)}    [  \ottsym{(}   \lambda\!  \, \mathit{x_{{\mathrm{0}}}}  \ottsym{.}   \resetcnstr{  \ottnt{K_{{\mathrm{0}}}}  [  \mathit{x_{{\mathrm{0}}}}  ]  }   \ottsym{)} \,  \prec_{ \ottnt{v} }  \, \ottsym{(}   \lambda\!  \, \mathit{x_{{\mathrm{0}}}}  \ottsym{.}   \resetcnstr{  \ottnt{K'_{{\mathrm{0}}}}  [  \mathit{x_{{\mathrm{0}}}}  ]  }   \ottsym{)} \,  \uparrow \, (  \ottmv{i}  ,  \pi  )   /  \mathit{x_{{\mathrm{0}}}}  ]   \,  \,\&\,  \,  \epsilon  ~.
            \end{array}
          \]
          Then, it suffices to show that
          \[
            \ottsym{(}  \ottnt{e_{{\mathrm{0}}}} \,  \prec_{ \ottnt{v} }  \, \ottnt{e'_{{\mathrm{0}}}} \,  \uparrow \, (  \ottmv{i}  ,  \pi  )   \ottsym{)}    [  \ottsym{(}   \lambda\!  \, \mathit{x_{{\mathrm{0}}}}  \ottsym{.}   \resetcnstr{  \ottnt{K_{{\mathrm{0}}}}  [  \mathit{x_{{\mathrm{0}}}}  ]  }   \ottsym{)} \,  \prec_{ \ottnt{v} }  \, \ottsym{(}   \lambda\!  \, \mathit{x_{{\mathrm{0}}}}  \ottsym{.}   \resetcnstr{  \ottnt{K'_{{\mathrm{0}}}}  [  \mathit{x_{{\mathrm{0}}}}  ]  }   \ottsym{)} \,  \uparrow \, (  \ottmv{i}  ,  \pi  )   /  \mathit{x_{{\mathrm{0}}}}  ]   \,  =  \, \ottnt{e_{{\mathrm{2}}}} \,  \prec_{ \ottnt{v} }  \, \ottnt{e'_{{\mathrm{2}}}} \,  \uparrow \, (  \ottmv{i}  ,  \pi  )  ~,
          \]
          which is proven by \reflem{lr-approx-val-subst}.

         \case \E{Red}/\R{Reset}:
          We are given
          \begin{prooftree}
           \AxiomC{$ \resetcnstr{ \ottnt{v_{{\mathrm{0}}}} }  \,  \rightsquigarrow  \, \ottnt{v_{{\mathrm{0}}}}$}
           \UnaryInfC{$ \resetcnstr{ \ottnt{v_{{\mathrm{0}}}} }  \,  \mathrel{  \longrightarrow  }  \, \ottnt{v_{{\mathrm{0}}}} \,  \,\&\,  \,  \epsilon $}
          \end{prooftree}
          for some $\ottnt{v_{{\mathrm{0}}}}$ such that
          \begin{itemize}
           \item $\ottnt{e_{{\mathrm{1}}}} \,  =  \,  \resetcnstr{ \ottnt{v_{{\mathrm{0}}}} } $ and
           \item $\ottnt{e_{{\mathrm{2}}}} \,  =  \, \ottnt{v_{{\mathrm{0}}}}$.
          \end{itemize}
          We also have $ \varpi    =     \epsilon  $.
          Because $\ottnt{e_{{\mathrm{1}}}} \,  \prec_{ \ottnt{v} }  \, \ottnt{e'_{{\mathrm{1}}}}$ and $\ottnt{e_{{\mathrm{1}}}} \,  =  \,  \resetcnstr{ \ottnt{v_{{\mathrm{0}}}} } $,
          we have
          \begin{itemize}
           \item $\ottnt{e'_{{\mathrm{1}}}} \,  =  \,  \resetcnstr{ \ottnt{v'_{{\mathrm{0}}}} } $ and
           \item $\ottnt{v_{{\mathrm{0}}}} \,  \prec_{ \ottnt{v} }  \, \ottnt{v'_{{\mathrm{0}}}}$
          \end{itemize}
          for some $\ottnt{v'_{{\mathrm{0}}}}$, by \reflem{lr-approx-val}.

          Let
          $\ottnt{e'_{{\mathrm{2}}}} \,  =  \, \ottnt{v'_{{\mathrm{0}}}}$,
          $\ottmv{i}$ be any natural number, and
          $\pi$ be any infinite trace.
          Because $\ottnt{v_{{\mathrm{0}}}} \,  \prec_{ \ottnt{v} }  \, \ottnt{v'_{{\mathrm{0}}}} \,  \uparrow \, (  \ottmv{i}  ,  \pi  ) $ is a value by \reflem{lr-approx-val},
          we have
          \[\begin{array}{ll} &
           \ottnt{e_{{\mathrm{1}}}} \,  \prec_{ \ottnt{v} }  \, \ottnt{e'_{{\mathrm{1}}}} \,  \uparrow \, (  \ottmv{i}  ,  \pi  )  \\ = &
            \resetcnstr{ \ottnt{v_{{\mathrm{0}}}} \,  \prec_{ \ottnt{v} }  \, \ottnt{v'_{{\mathrm{0}}}} \,  \uparrow \, (  \ottmv{i}  ,  \pi  )  }  \,  \mathrel{  \longrightarrow  }  \, \ottnt{v_{{\mathrm{0}}}} \,  \prec_{ \ottnt{v} }  \, \ottnt{v'_{{\mathrm{0}}}} \,  \uparrow \, (  \ottmv{i}  ,  \pi  )  \,  \,\&\,  \,  \epsilon 
           \quad \text{(by \E{Red}/\R{Reset})} ~.
            \end{array}
          \]
          Therefore, we have the conclusion.

         \case \E{Ev}:
          We are given
          \begin{prooftree}
           \AxiomC{$ \mathsf{ev}  [  \mathbf{a}  ]  \,  \mathrel{  \longrightarrow  }  \,  ()  \,  \,\&\,  \, \mathbf{a}$}
          \end{prooftree}
          for some $\mathbf{a}$ such that
          \begin{itemize}
           \item $\ottnt{e_{{\mathrm{1}}}} \,  =  \,  \mathsf{ev}  [  \mathbf{a}  ] $ and
           \item $ \varpi    =    \mathbf{a} $.
          \end{itemize}
          We also have $\ottnt{e_{{\mathrm{2}}}} \,  =  \,  () $.
          Because $\ottnt{e_{{\mathrm{1}}}} \,  \prec_{ \ottnt{v} }  \, \ottnt{e'_{{\mathrm{1}}}}$ and $\ottnt{e_{{\mathrm{1}}}} \,  =  \,  \mathsf{ev}  [  \mathbf{a}  ] $,
          we have $\ottnt{e'_{{\mathrm{1}}}} \,  =  \,  \mathsf{ev}  [  \mathbf{a}  ] $.

          Let
          $\ottnt{e'_{{\mathrm{2}}}} \,  =  \,  () $,
          $\ottmv{i}$ be any natural number, and
          $\pi$ be any infinite trace.
          We have
          \[
           \ottnt{e_{{\mathrm{1}}}} \,  \prec_{ \ottnt{v} }  \, \ottnt{e'_{{\mathrm{1}}}} \,  \uparrow \, (  \ottmv{i}  ,  \pi  )  =  \mathsf{ev}  [  \mathbf{a}  ]  \,  \mathrel{  \longrightarrow  }  \,  ()  \,  \,\&\,  \, \mathbf{a}
           \quad \text{(by \E{Ev})} ~.
          \]
          Because $ ()  \,  =  \, \ottnt{e_{{\mathrm{2}}}} \,  \prec_{ \ottnt{v} }  \, \ottnt{e'_{{\mathrm{2}}}} \,  \uparrow \, (  \ottmv{i}  ,  \pi  ) $,
          we have the conclusion.

         \case \E{Let}:
          We are given
          \begin{prooftree}
           \AxiomC{$\ottnt{e_{{\mathrm{01}}}} \,  \mathrel{  \longrightarrow  }  \, \ottnt{e_{{\mathrm{02}}}} \,  \,\&\,  \, \varpi$}
           \UnaryInfC{$\mathsf{let} \, \mathit{x_{{\mathrm{01}}}}  \ottsym{=}  \ottnt{e_{{\mathrm{01}}}} \,  \mathsf{in}  \, \ottnt{e_{{\mathrm{03}}}} \,  \mathrel{  \longrightarrow  }  \, \mathsf{let} \, \mathit{x_{{\mathrm{01}}}}  \ottsym{=}  \ottnt{e_{{\mathrm{02}}}} \,  \mathsf{in}  \, \ottnt{e_{{\mathrm{03}}}} \,  \,\&\,  \, \varpi$}
          \end{prooftree}
          for some $\mathit{x_{{\mathrm{01}}}}$, $\ottnt{e_{{\mathrm{01}}}}$, $\ottnt{e_{{\mathrm{02}}}}$, and $\ottnt{e_{{\mathrm{03}}}}$ such that
          \begin{itemize}
           \item $\ottnt{e_{{\mathrm{1}}}} \,  =  \, \mathsf{let} \, \mathit{x_{{\mathrm{01}}}}  \ottsym{=}  \ottnt{e_{{\mathrm{01}}}} \,  \mathsf{in}  \, \ottnt{e_{{\mathrm{03}}}}$ and
           \item $\ottnt{e_{{\mathrm{2}}}} \,  =  \, \mathsf{let} \, \mathit{x_{{\mathrm{01}}}}  \ottsym{=}  \ottnt{e_{{\mathrm{02}}}} \,  \mathsf{in}  \, \ottnt{e_{{\mathrm{03}}}}$.
          \end{itemize}
          Because $\ottnt{e_{{\mathrm{1}}}} \,  \prec_{ \ottnt{v} }  \, \ottnt{e'_{{\mathrm{1}}}}$ and $\ottnt{e_{{\mathrm{1}}}} \,  =  \, \mathsf{let} \, \mathit{x_{{\mathrm{01}}}}  \ottsym{=}  \ottnt{e_{{\mathrm{01}}}} \,  \mathsf{in}  \, \ottnt{e_{{\mathrm{03}}}}$,
          we have
          \begin{itemize}
           \item $\ottnt{e'_{{\mathrm{1}}}} \,  =  \, \mathsf{let} \, \mathit{x_{{\mathrm{01}}}}  \ottsym{=}  \ottnt{e'_{{\mathrm{01}}}} \,  \mathsf{in}  \, \ottnt{e'_{{\mathrm{03}}}}$,
           \item $\ottnt{e_{{\mathrm{01}}}} \,  \prec_{ \ottnt{v} }  \, \ottnt{e'_{{\mathrm{01}}}}$, and
           \item $\ottnt{e_{{\mathrm{03}}}} \,  \prec_{ \ottnt{v} }  \, \ottnt{e'_{{\mathrm{03}}}}$
          \end{itemize}
          for some $\ottnt{e'_{{\mathrm{01}}}}$ and $\ottnt{e'_{{\mathrm{03}}}}$.
          By the IH, there exists some $\ottnt{e'_{{\mathrm{02}}}}$ such that
          \[
             \forall  \, { \ottmv{i}  \ottsym{,}  \pi } . \  \ottsym{(}  \ottnt{e_{{\mathrm{01}}}} \,  \prec_{ \ottnt{v} }  \, \ottnt{e'_{{\mathrm{01}}}} \,  \uparrow \, (  \ottmv{i}  ,  \pi  )   \ottsym{)} \,  \mathrel{  \longrightarrow  }  \, \ottsym{(}  \ottnt{e_{{\mathrm{02}}}} \,  \prec_{ \ottnt{v} }  \, \ottnt{e'_{{\mathrm{02}}}} \,  \uparrow \, (  \ottmv{i}  ,  \pi  )   \ottsym{)} \,  \,\&\,  \, \varpi  ~.
          \]

          Let $\ottnt{e'_{{\mathrm{2}}}} \,  =  \, \mathsf{let} \, \mathit{x_{{\mathrm{01}}}}  \ottsym{=}  \ottnt{e'_{{\mathrm{02}}}} \,  \mathsf{in}  \, \ottnt{e'_{{\mathrm{03}}}}$,
          $\ottmv{i}$ be any natural number, and
          $\pi$ be any infinite trace.
          We have
          \[\begin{array}{ll} &
           \ottnt{e_{{\mathrm{1}}}} \,  \prec_{ \ottnt{v} }  \, \ottnt{e'_{{\mathrm{1}}}} \,  \uparrow \, (  \ottmv{i}  ,  \pi  )  \\ = &
           \mathsf{let} \, \mathit{x_{{\mathrm{01}}}}  \ottsym{=}  \ottsym{(}  \ottnt{e_{{\mathrm{01}}}} \,  \prec_{ \ottnt{v} }  \, \ottnt{e'_{{\mathrm{01}}}} \,  \uparrow \, (  \ottmv{i}  ,  \pi  )   \ottsym{)} \,  \mathsf{in}  \, \ottsym{(}  \ottnt{e_{{\mathrm{03}}}} \,  \prec_{ \ottnt{v} }  \, \ottnt{e'_{{\mathrm{03}}}} \,  \uparrow \, (  \ottmv{i}  ,  \pi  )   \ottsym{)}
            \\  \longrightarrow  &
           \mathsf{let} \, \mathit{x_{{\mathrm{01}}}}  \ottsym{=}  \ottsym{(}  \ottnt{e_{{\mathrm{02}}}} \,  \prec_{ \ottnt{v} }  \, \ottnt{e'_{{\mathrm{02}}}} \,  \uparrow \, (  \ottmv{i}  ,  \pi  )   \ottsym{)} \,  \mathsf{in}  \, \ottsym{(}  \ottnt{e_{{\mathrm{03}}}} \,  \prec_{ \ottnt{v} }  \, \ottnt{e'_{{\mathrm{03}}}} \,  \uparrow \, (  \ottmv{i}  ,  \pi  )   \ottsym{)} \,  \,\&\,  \, \varpi
            \\ &%
            \quad \text{(by the result of the IH and \E{Let})} ~.
            \end{array}
          \]
          Because
          \[
           \ottnt{e_{{\mathrm{2}}}} \,  \prec_{ \ottnt{v} }  \, \ottnt{e'_{{\mathrm{2}}}} \,  \uparrow \, (  \ottmv{i}  ,  \pi  )  \,  =  \, \mathsf{let} \, \mathit{x_{{\mathrm{01}}}}  \ottsym{=}  \ottsym{(}  \ottnt{e_{{\mathrm{02}}}} \,  \prec_{ \ottnt{v} }  \, \ottnt{e'_{{\mathrm{02}}}} \,  \uparrow \, (  \ottmv{i}  ,  \pi  )   \ottsym{)} \,  \mathsf{in}  \, \ottsym{(}  \ottnt{e_{{\mathrm{03}}}} \,  \prec_{ \ottnt{v} }  \, \ottnt{e'_{{\mathrm{03}}}} \,  \uparrow \, (  \ottmv{i}  ,  \pi  )   \ottsym{)} ~,
          \]
          we have the conclusion.

         \case \E{Reset}:
          We are given
          \begin{prooftree}
           \AxiomC{$\ottnt{e_{{\mathrm{01}}}} \,  \mathrel{  \longrightarrow  }  \, \ottnt{e_{{\mathrm{02}}}} \,  \,\&\,  \, \varpi$}
           \UnaryInfC{$ \resetcnstr{ \ottnt{e_{{\mathrm{01}}}} }  \,  \mathrel{  \longrightarrow  }  \,  \resetcnstr{ \ottnt{e_{{\mathrm{02}}}} }  \,  \,\&\,  \, \varpi$}
          \end{prooftree}
          for some $\ottnt{e_{{\mathrm{01}}}}$ and $\ottnt{e_{{\mathrm{02}}}}$ such that
          \begin{itemize}
           \item $\ottnt{e_{{\mathrm{1}}}} \,  =  \,  \resetcnstr{ \ottnt{e_{{\mathrm{01}}}} } $ and
           \item $\ottnt{e_{{\mathrm{2}}}} \,  =  \,  \resetcnstr{ \ottnt{e_{{\mathrm{02}}}} } $.
          \end{itemize}
          Because $\ottnt{e_{{\mathrm{1}}}} \,  \prec_{ \ottnt{v} }  \, \ottnt{e'_{{\mathrm{1}}}}$ and $\ottnt{e_{{\mathrm{1}}}} \,  =  \,  \resetcnstr{ \ottnt{e_{{\mathrm{01}}}} } $,
          we have
          \begin{itemize}
           \item $\ottnt{e'_{{\mathrm{1}}}} \,  =  \,  \resetcnstr{ \ottnt{e'_{{\mathrm{01}}}} } $ and
           \item $\ottnt{e_{{\mathrm{01}}}} \,  \prec_{ \ottnt{v} }  \, \ottnt{e'_{{\mathrm{01}}}}$
          \end{itemize}
          for some $\ottnt{e'_{{\mathrm{01}}}}$.
          By the IH, there exists some $\ottnt{e'_{{\mathrm{02}}}}$ such that
          \[
             \forall  \, { \ottmv{i}  \ottsym{,}  \pi } . \  \ottsym{(}  \ottnt{e_{{\mathrm{01}}}} \,  \prec_{ \ottnt{v} }  \, \ottnt{e'_{{\mathrm{01}}}} \,  \uparrow \, (  \ottmv{i}  ,  \pi  )   \ottsym{)} \,  \mathrel{  \longrightarrow  }  \, \ottsym{(}  \ottnt{e_{{\mathrm{02}}}} \,  \prec_{ \ottnt{v} }  \, \ottnt{e'_{{\mathrm{02}}}} \,  \uparrow \, (  \ottmv{i}  ,  \pi  )   \ottsym{)} \,  \,\&\,  \, \varpi  ~.
          \]

          Let
          $\ottnt{e'_{{\mathrm{2}}}} \,  =  \,  \resetcnstr{ \ottnt{e'_{{\mathrm{02}}}} } $,
          $\ottmv{i}$ be any natural number, and
          $\pi$ be any infinite trace.
          We have
          \[\begin{array}{ll} &
           \ottnt{e_{{\mathrm{1}}}} \,  \prec_{ \ottnt{v} }  \, \ottnt{e'_{{\mathrm{1}}}} \,  \uparrow \, (  \ottmv{i}  ,  \pi  )  \\ = &
            \resetcnstr{ \ottnt{e_{{\mathrm{01}}}} \,  \prec_{ \ottnt{v} }  \, \ottnt{e'_{{\mathrm{01}}}} \,  \uparrow \, (  \ottmv{i}  ,  \pi  )  } 
           \\  \longrightarrow  &
            \resetcnstr{ \ottnt{e_{{\mathrm{02}}}} \,  \prec_{ \ottnt{v} }  \, \ottnt{e'_{{\mathrm{02}}}} \,  \uparrow \, (  \ottmv{i}  ,  \pi  )  }  \,  \,\&\,  \, \varpi
           \quad \text{(by the result of the IH and \E{Reset})} ~.
            \end{array}
          \]
          Because
          \[
           \ottnt{e_{{\mathrm{2}}}} \,  \prec_{ \ottnt{v} }  \, \ottnt{e'_{{\mathrm{2}}}} \,  \uparrow \, (  \ottmv{i}  ,  \pi  )  \,  =  \,  \resetcnstr{ \ottnt{e_{{\mathrm{02}}}} \,  \prec_{ \ottnt{v} }  \, \ottnt{e'_{{\mathrm{02}}}} \,  \uparrow \, (  \ottmv{i}  ,  \pi  )  }  ~,
          \]
          we have the conclusion.
        \end{caseanalysis}

  \item By the cases (\ref{lem:lr-approx-eval:approx}) and (\ref{lem:lr-approx-eval:non-approx}).
 \end{enumerate}
\end{proof}

\begin{lemmap}{Approximation of Evaluation by Multiple Steps}{lr-approx-eval-nsteps}
 If $\ottnt{e_{{\mathrm{1}}}} \,  \prec_{ \ottnt{v} }  \, \ottnt{e'_{{\mathrm{1}}}}$ and $\ottnt{e_{{\mathrm{1}}}} \,  \mathrel{  \longrightarrow  }^{ \ottmv{n} }  \, \ottnt{e_{{\mathrm{2}}}} \,  \,\&\,  \, \varpi$, then
 \[
    \exists  \, { \ottmv{j}  \ottsym{,}  \ottnt{e'_{{\mathrm{2}}}} } . \    \forall  \, { \pi } . \    \forall  \, { \ottmv{i} \,  \geq  \, \ottmv{j} } . \  \ottsym{(}  \ottnt{e_{{\mathrm{1}}}} \,  \prec_{ \ottnt{v} }  \, \ottnt{e'_{{\mathrm{1}}}} \,  \uparrow \, (  \ottmv{i}  ,  \pi  )   \ottsym{)} \,  \mathrel{  \longrightarrow  }^{ \ottmv{n} }  \, \ottsym{(}  \ottnt{e_{{\mathrm{2}}}} \,  \prec_{ \ottnt{v} }  \, \ottnt{e'_{{\mathrm{2}}}} \,  \uparrow \, (  \ottmv{i}  \ottsym{-}  \ottmv{j}  ,  \pi  )   \ottsym{)} \,  \,\&\,  \, \varpi    ~.
 \]
\end{lemmap}
\begin{proof}
 \TS{OK: 2022/01/20}
 By induction on $\ottmv{n}$.
 \begin{caseanalysis}
  \case $\ottmv{n} \,  =  \, \ottsym{0}$:
   We have $\ottnt{e_{{\mathrm{1}}}} \,  =  \, \ottnt{e_{{\mathrm{2}}}}$ and $ \varpi    =     \epsilon  $.
   Let $\ottmv{j} \,  =  \, \ottsym{0}$, $\ottnt{e'_{{\mathrm{2}}}} \,  =  \, \ottnt{e'_{{\mathrm{1}}}}$, $\pi$ be any infinite trace, and $\ottmv{i} \,  \geq  \, \ottsym{0}$.
   It suffices to show that $\ottnt{e_{{\mathrm{1}}}} \,  \prec_{ \ottnt{v} }  \, \ottnt{e'_{{\mathrm{1}}}} \,  \uparrow \, (  \ottmv{i}  ,  \pi  )  \,  =  \, \ottnt{e_{{\mathrm{2}}}} \,  \prec_{ \ottnt{v} }  \, \ottnt{e'_{{\mathrm{2}}}} \,  \uparrow \, (  \ottmv{i}  ,  \pi  ) $,
   which holds because $\ottnt{e_{{\mathrm{1}}}} \,  =  \, \ottnt{e_{{\mathrm{2}}}}$ and $\ottnt{e'_{{\mathrm{1}}}} \,  =  \, \ottnt{e'_{{\mathrm{2}}}}$.

  \case $\ottmv{n} \,  >  \, \ottsym{0}$:
   We are given $\ottnt{e_{{\mathrm{1}}}} \,  \mathrel{  \longrightarrow  }  \, \ottnt{e_{{\mathrm{3}}}} \,  \,\&\,  \, \varpi_{{\mathrm{3}}}$ and $\ottnt{e_{{\mathrm{3}}}} \,  \mathrel{  \longrightarrow  }^{ \ottmv{n}  \ottsym{-}  \ottsym{1} }  \, \ottnt{e_{{\mathrm{2}}}} \,  \,\&\,  \, \varpi_{{\mathrm{2}}}$
   for some $\ottnt{e_{{\mathrm{3}}}}$, $\varpi_{{\mathrm{3}}}$, and $\varpi_{{\mathrm{2}}}$ such that
   $ \varpi    =    \varpi_{{\mathrm{3}}} \,  \tceappendop  \, \varpi_{{\mathrm{2}}} $.
   By \reflem{lr-approx-eval} (\ref{lem:lr-approx-eval:unified}),
   there exist some $ { \ottmv{j} }_{ \ottsym{3} } $ and $\ottnt{e'_{{\mathrm{3}}}}$ such that
   \begin{equation}
      \forall  \, { \pi } . \    \forall  \, { \ottmv{i} \,  \geq  \,  { \ottmv{j} }_{ \ottsym{3} }  } . \  \ottsym{(}  \ottnt{e_{{\mathrm{1}}}} \,  \prec_{ \ottnt{v} }  \, \ottnt{e'_{{\mathrm{1}}}} \,  \uparrow \, (  \ottmv{i}  ,  \pi  )   \ottsym{)} \,  \mathrel{  \longrightarrow  }  \, \ottsym{(}  \ottnt{e_{{\mathrm{3}}}} \,  \prec_{ \ottnt{v} }  \, \ottnt{e'_{{\mathrm{3}}}} \,  \uparrow \, (   { \ottmv{i}  \ottsym{-}  \ottmv{j} }_{ \ottsym{3} }   ,  \pi  )   \ottsym{)} \,  \,\&\,  \, \varpi_{{\mathrm{3}}}   ~.
     \label{lem:lr-approx-eval-nsteps:step:one}
   \end{equation}
   By \reflem{lr-approx-wf}, $\ottnt{e_{{\mathrm{3}}}} \,  \prec_{ \ottnt{v} }  \, \ottnt{e'_{{\mathrm{3}}}}$.
   Because $\ottnt{e_{{\mathrm{3}}}} \,  \mathrel{  \longrightarrow  }^{ \ottmv{n}  \ottsym{-}  \ottsym{1} }  \, \ottnt{e_{{\mathrm{2}}}} \,  \,\&\,  \, \varpi_{{\mathrm{2}}}$, the IH implies that
   there exist some $ { \ottmv{j} }_{ \ottsym{2} } $ and $\ottnt{e'_{{\mathrm{2}}}}$ such that
   \begin{equation}
      \forall  \, { \pi } . \    \forall  \, { \ottmv{i} \,  \geq  \,  { \ottmv{j} }_{ \ottsym{2} }  } . \  \ottsym{(}  \ottnt{e_{{\mathrm{3}}}} \,  \prec_{ \ottnt{v} }  \, \ottnt{e'_{{\mathrm{3}}}} \,  \uparrow \, (  \ottmv{i}  ,  \pi  )   \ottsym{)} \,  \mathrel{  \longrightarrow  }^{ \ottmv{n}  \ottsym{-}  \ottsym{1} }  \, \ottsym{(}  \ottnt{e_{{\mathrm{2}}}} \,  \prec_{ \ottnt{v} }  \, \ottnt{e'_{{\mathrm{2}}}} \,  \uparrow \, (   { \ottmv{i}  \ottsym{-}  \ottmv{j} }_{ \ottsym{2} }   ,  \pi  )   \ottsym{)} \,  \,\&\,  \, \varpi_{{\mathrm{2}}}   ~.
     \label{lem:lr-approx-eval-nsteps:step:IH}
   \end{equation}

   Let $\ottmv{j} \,  =  \,  {  { \ottmv{j} }_{ \ottsym{3} }   \ottsym{+}  \ottmv{j} }_{ \ottsym{2} } $, $\pi$ be any infinite trace, and $\ottmv{i} \,  \geq  \, \ottmv{j}$.
   It suffices to show that
   \[
    \ottsym{(}  \ottnt{e_{{\mathrm{1}}}} \,  \prec_{ \ottnt{v} }  \, \ottnt{e'_{{\mathrm{1}}}} \,  \uparrow \, (  \ottmv{i}  ,  \pi  )   \ottsym{)} \,  \mathrel{  \longrightarrow  }^{ \ottmv{n} }  \, \ottsym{(}  \ottnt{e_{{\mathrm{2}}}} \,  \prec_{ \ottnt{v} }  \, \ottnt{e'_{{\mathrm{2}}}} \,  \uparrow \, (  \ottmv{i}  \ottsym{-}  \ottmv{j}  ,  \pi  )   \ottsym{)} \,  \,\&\,  \, \varpi ~.
   \]
   Because $\ottmv{i} \,  \geq  \, \ottmv{j} =  {  { \ottmv{j} }_{ \ottsym{3} }   \ottsym{+}  \ottmv{j} }_{ \ottsym{2} }  \,  \geq  \,  { \ottmv{j} }_{ \ottsym{3} } $,
   (\ref{lem:lr-approx-eval-nsteps:step:one}) implies
   \[
    \ottsym{(}  \ottnt{e_{{\mathrm{1}}}} \,  \prec_{ \ottnt{v} }  \, \ottnt{e'_{{\mathrm{1}}}} \,  \uparrow \, (  \ottmv{i}  ,  \pi  )   \ottsym{)} \,  \mathrel{  \longrightarrow  }  \, \ottsym{(}  \ottnt{e_{{\mathrm{3}}}} \,  \prec_{ \ottnt{v} }  \, \ottnt{e'_{{\mathrm{3}}}} \,  \uparrow \, (   { \ottmv{i}  \ottsym{-}  \ottmv{j} }_{ \ottsym{3} }   ,  \pi  )   \ottsym{)} \,  \,\&\,  \, \varpi_{{\mathrm{3}}} ~.
   \]
   Because $ { \ottmv{i}  \ottsym{-}  \ottmv{j} }_{ \ottsym{3} }  \,  \geq  \,  { \ottmv{j}  \ottsym{-}  \ottmv{j} }_{ \ottsym{3} }  =  { \ottmv{j} }_{ \ottsym{2} } $,
   (\ref{lem:lr-approx-eval-nsteps:step:IH}) implies
   \[
    \ottsym{(}  \ottnt{e_{{\mathrm{3}}}} \,  \prec_{ \ottnt{v} }  \, \ottnt{e'_{{\mathrm{3}}}} \,  \uparrow \, (   { \ottmv{i}  \ottsym{-}  \ottmv{j} }_{ \ottsym{3} }   ,  \pi  )   \ottsym{)} \,  \mathrel{  \longrightarrow  }^{ \ottmv{n}  \ottsym{-}  \ottsym{1} }  \, \ottsym{(}  \ottnt{e_{{\mathrm{2}}}} \,  \prec_{ \ottnt{v} }  \, \ottnt{e'_{{\mathrm{2}}}} \,  \uparrow \, (   {  { \ottmv{i}  \ottsym{-}  \ottmv{j} }_{ \ottsym{3} }   \ottsym{-}  \ottmv{j} }_{ \ottsym{2} }   ,  \pi  )   \ottsym{)} \,  \,\&\,  \, \varpi_{{\mathrm{2}}} ~.
   \]
   Therefore, we have
   \[
    \ottsym{(}  \ottnt{e_{{\mathrm{1}}}} \,  \prec_{ \ottnt{v} }  \, \ottnt{e'_{{\mathrm{1}}}} \,  \uparrow \, (  \ottmv{i}  ,  \pi  )   \ottsym{)} \,  \mathrel{  \longrightarrow  }^{ \ottmv{n} }  \, \ottsym{(}  \ottnt{e_{{\mathrm{2}}}} \,  \prec_{ \ottnt{v} }  \, \ottnt{e'_{{\mathrm{2}}}} \,  \uparrow \, (   {  { \ottmv{i}  \ottsym{-}  \ottmv{j} }_{ \ottsym{3} }   \ottsym{-}  \ottmv{j} }_{ \ottsym{2} }   ,  \pi  )   \ottsym{)} \,  \,\&\,  \, \varpi_{{\mathrm{3}}} \,  \tceappendop  \, \varpi_{{\mathrm{2}}} ~.
   \]
   Because $\ottmv{j} \,  =  \,  {  { \ottmv{j} }_{ \ottsym{3} }   \ottsym{+}  \ottmv{j} }_{ \ottsym{2} } $ and $ \varpi    =    \varpi_{{\mathrm{3}}} \,  \tceappendop  \, \varpi_{{\mathrm{2}}} $,
   we have the conclusion
   \[
    \ottsym{(}  \ottnt{e_{{\mathrm{1}}}} \,  \prec_{ \ottnt{v} }  \, \ottnt{e'_{{\mathrm{1}}}} \,  \uparrow \, (  \ottmv{i}  ,  \pi  )   \ottsym{)} \,  \mathrel{  \longrightarrow  }^{ \ottmv{n} }  \, \ottsym{(}  \ottnt{e_{{\mathrm{2}}}} \,  \prec_{ \ottnt{v} }  \, \ottnt{e'_{{\mathrm{2}}}} \,  \uparrow \, (  \ottmv{i}  \ottsym{-}  \ottmv{j}  ,  \pi  )   \ottsym{)} \,  \,\&\,  \, \varpi ~.
   \]
 \end{caseanalysis}
\end{proof}

\begin{lemmap}{Anti-Approximation of Values}{lr-approx-anti-val}
 If $\ottnt{e} \,  \prec_{ \ottnt{v} }  \, \ottnt{e'} \,  \uparrow \, (  \ottmv{i}  ,  \pi  ) $ is a value,
 then $\ottnt{e}$ and $\ottnt{e'}$ are values.
\end{lemmap}
\begin{proof}
 \TS{OK: 2022/01/18}
 By case analysis on $\ottnt{e} \,  \prec_{ \ottnt{v} }  \, \ottnt{e'} \,  \uparrow \, (  \ottmv{i}  ,  \pi  ) $.
 If it is a constant, variable, or predicate abstraction,
 then the conclusion is easy to show.
 Otherwise, suppose that $\ottnt{e} \,  \prec_{ \ottnt{v} }  \, \ottnt{e'} \,  \uparrow \, (  \ottmv{i}  ,  \pi  )  \,  =  \, \ottnt{v''_{{\mathrm{1}}}} \,  \tceseq{ \ottnt{v''_{{\mathrm{2}}}} } $ for some $\ottnt{v''_{{\mathrm{1}}}}$
 and $ \tceseq{ \ottnt{v''_{{\mathrm{2}}}} } $ such that
 \begin{itemize}
  \item $\ottnt{v''_{{\mathrm{1}}}}$ is a $\lambda$-abstraction or recursive function, and
  \item $ \mathit{arity}  (  \ottnt{v''_{{\mathrm{1}}}}  )  \,  >  \, \ottsym{\mbox{$\mid$}}   \tceseq{ \ottnt{v''_{{\mathrm{2}}}} }   \ottsym{\mbox{$\mid$}}$.
 \end{itemize}
 Then, $\ottnt{e} \,  =  \, \ottnt{v_{{\mathrm{1}}}} \,  \tceseq{ \ottnt{v_{{\mathrm{2}}}} } $ and $\ottnt{e'} \,  =  \, \ottnt{v'_{{\mathrm{1}}}} \,  \tceseq{ \ottnt{v'_{{\mathrm{2}}}} } $ for some
 $\ottnt{v_{{\mathrm{1}}}}$, $\ottnt{v'_{{\mathrm{1}}}}$, $ \tceseq{ \ottnt{v_{{\mathrm{2}}}} } $, and $ \tceseq{ \ottnt{v'_{{\mathrm{2}}}} } $ such that
 \begin{itemize}
  \item $\ottnt{v_{{\mathrm{1}}}} \,  \prec_{ \ottnt{v} }  \, \ottnt{v'_{{\mathrm{1}}}} \,  \uparrow \, (  \ottmv{i}  ,  \pi  )  \,  =  \, \ottnt{v''_{{\mathrm{1}}}}$ and
  \item $ \tceseq{ \ottnt{v_{{\mathrm{2}}}} \,  \prec_{ \ottnt{v} }  \, \ottnt{v'_{{\mathrm{2}}}} \,  \uparrow \, (  \ottmv{i}  ,  \pi  )  \,  =  \, \ottnt{v''_{{\mathrm{2}}}} } $.
 \end{itemize}
 By definition, $ \mathit{arity}  (  \ottnt{v_{{\mathrm{1}}}}  )  =  \mathit{arity}  (  \ottnt{v'_{{\mathrm{1}}}}  )  =  \mathit{arity}  (  \ottnt{v''_{{\mathrm{1}}}}  ) $, so
 $ \mathit{arity}  (  \ottnt{v_{{\mathrm{1}}}}  )  \,  >  \, \ottsym{\mbox{$\mid$}}   \tceseq{ \ottnt{v_{{\mathrm{2}}}} }   \ottsym{\mbox{$\mid$}}$ and $ \mathit{arity}  (  \ottnt{v'_{{\mathrm{1}}}}  )  \,  >  \, \ottsym{\mbox{$\mid$}}   \tceseq{ \ottnt{v'_{{\mathrm{2}}}} }   \ottsym{\mbox{$\mid$}}$.
 Therefore, $\ottnt{e} \,  =  \, \ottnt{v_{{\mathrm{1}}}} \,  \tceseq{ \ottnt{v_{{\mathrm{2}}}} } $ and $\ottnt{e'} \,  =  \, \ottnt{v'_{{\mathrm{1}}}} \,  \tceseq{ \ottnt{v'_{{\mathrm{2}}}} } $ are values.
\end{proof}

\begin{lemmap}{Anti-Approximation of Evaluation Contexts}{lr-approx-anti-ectx}
 If $\ottnt{e} \,  \prec_{ \ottnt{v} }  \, \ottnt{e'} \,  \uparrow \, (  \ottmv{i}  ,  \pi  )  \,  =  \,  \ottnt{E''}  [  \ottnt{e''_{{\mathrm{0}}}}  ] $,
 then there exist $\ottnt{E}$, $\ottnt{e_{{\mathrm{0}}}}$, $\ottnt{E'}$, and $\ottnt{e'_{{\mathrm{0}}}}$ such that
 \begin{itemize}
  \item $\ottnt{e} \,  =  \,  \ottnt{E}  [  \ottnt{e_{{\mathrm{0}}}}  ] $,
  \item $\ottnt{e'} \,  =  \,  \ottnt{E'}  [  \ottnt{e'_{{\mathrm{0}}}}  ] $,
  \item $\ottnt{E} \,  \prec_{ \ottnt{v} }  \, \ottnt{E'} \,  \uparrow \, (  \ottmv{i}  ,  \pi  )  \,  =  \, \ottnt{E''}$, and
  \item $\ottnt{e_{{\mathrm{0}}}} \,  \prec_{ \ottnt{v} }  \, \ottnt{e'_{{\mathrm{0}}}} \,  \uparrow \, (  \ottmv{i}  ,  \pi  )  \,  =  \, \ottnt{e''_{{\mathrm{0}}}}$.
 \end{itemize}
 Furthermore, if $\ottnt{E''} \,  =  \, \ottnt{K''}$ for some $\ottnt{K''}$, then
 $\ottnt{E} \,  =  \, \ottnt{K}$ and $\ottnt{E'} \,  =  \, \ottnt{K'}$ for some $\ottnt{K}$ and $\ottnt{K'}$.
\end{lemmap}
\begin{proof}
 \TS{OK: 2022/01/18}
 By induction on $\ottnt{E''}$.
 \begin{caseanalysis}
  \case $\ottnt{E''} \,  =  \, \,[\,]\,$:
   We have $\ottnt{e} \,  \prec_{ \ottnt{v} }  \, \ottnt{e'} \,  \uparrow \, (  \ottmv{i}  ,  \pi  )  \,  =  \, \ottnt{e''_{{\mathrm{0}}}}$.
   Then, we have the conclusion
   by letting $\ottnt{E} = \ottnt{E'} =  \,[\,]\, $ and $\ottnt{e_{{\mathrm{0}}}} \,  =  \, \ottnt{e}$ and $\ottnt{e'_{{\mathrm{0}}}} \,  =  \, \ottnt{e'}$.

  \case $  \exists  \, { \mathit{x_{{\mathrm{1}}}}  \ottsym{,}  \ottnt{E''_{{\mathrm{1}}}}  \ottsym{,}  \ottnt{e''_{{\mathrm{2}}}} } . \  \ottnt{E''} \,  =  \, \ottsym{(}  \mathsf{let} \, \mathit{x_{{\mathrm{1}}}}  \ottsym{=}  \ottnt{E''_{{\mathrm{1}}}} \,  \mathsf{in}  \, \ottnt{e''_{{\mathrm{2}}}}  \ottsym{)} $:
   We are given
   $\ottnt{e} \,  \prec_{ \ottnt{v} }  \, \ottnt{e'} \,  \uparrow \, (  \ottmv{i}  ,  \pi  )  \,  =  \, \ottsym{(}  \mathsf{let} \, \mathit{x_{{\mathrm{1}}}}  \ottsym{=}   \ottnt{E''_{{\mathrm{1}}}}  [  \ottnt{e''_{{\mathrm{0}}}}  ]  \,  \mathsf{in}  \, \ottnt{e''_{{\mathrm{2}}}}  \ottsym{)}$.
   By definition, there exist some $\ottnt{e_{{\mathrm{1}}}}$, $\ottnt{e'_{{\mathrm{1}}}}$, $\ottnt{e_{{\mathrm{2}}}}$, and $\ottnt{e'_{{\mathrm{2}}}}$ such that
   \begin{itemize}
    \item $\ottnt{e} \,  =  \, \ottsym{(}  \mathsf{let} \, \mathit{x_{{\mathrm{1}}}}  \ottsym{=}  \ottnt{e_{{\mathrm{1}}}} \,  \mathsf{in}  \, \ottnt{e_{{\mathrm{2}}}}  \ottsym{)}$,
    \item $\ottnt{e'} \,  =  \, \ottsym{(}  \mathsf{let} \, \mathit{x_{{\mathrm{1}}}}  \ottsym{=}  \ottnt{e'_{{\mathrm{1}}}} \,  \mathsf{in}  \, \ottnt{e'_{{\mathrm{2}}}}  \ottsym{)}$,
    \item $\ottnt{e_{{\mathrm{1}}}} \,  \prec_{ \ottnt{v} }  \, \ottnt{e'_{{\mathrm{1}}}} \,  \uparrow \, (  \ottmv{i}  ,  \pi  )  \,  =  \,  \ottnt{E''_{{\mathrm{1}}}}  [  \ottnt{e''_{{\mathrm{0}}}}  ] $, and
    \item $\ottnt{e_{{\mathrm{2}}}} \,  \prec_{ \ottnt{v} }  \, \ottnt{e'_{{\mathrm{2}}}} \,  \uparrow \, (  \ottmv{i}  ,  \pi  )  \,  =  \, \ottnt{e''_{{\mathrm{2}}}}$.
   \end{itemize}
   By the IH, there exist some $\ottnt{E_{{\mathrm{1}}}}$, $\ottnt{e_{{\mathrm{0}}}}$, $\ottnt{E'_{{\mathrm{1}}}}$, and $\ottnt{e'_{{\mathrm{0}}}}$ such that
   \begin{itemize}
    \item $\ottnt{e_{{\mathrm{1}}}} \,  =  \,  \ottnt{E_{{\mathrm{1}}}}  [  \ottnt{e_{{\mathrm{0}}}}  ] $,
    \item $\ottnt{e'_{{\mathrm{1}}}} \,  =  \,  \ottnt{E'_{{\mathrm{1}}}}  [  \ottnt{e'_{{\mathrm{0}}}}  ] $,
    \item $\ottnt{E_{{\mathrm{1}}}} \,  \prec_{ \ottnt{v} }  \, \ottnt{E'_{{\mathrm{1}}}} \,  \uparrow \, (  \ottmv{i}  ,  \pi  )  \,  =  \, \ottnt{E''_{{\mathrm{1}}}}$, and
    \item $\ottnt{e_{{\mathrm{0}}}} \,  \prec_{ \ottnt{v} }  \, \ottnt{e'_{{\mathrm{0}}}} \,  \uparrow \, (  \ottmv{i}  ,  \pi  )  \,  =  \, \ottnt{e''_{{\mathrm{0}}}}$.
   \end{itemize}

   Let $\ottnt{E} \,  =  \, \ottsym{(}  \mathsf{let} \, \mathit{x_{{\mathrm{1}}}}  \ottsym{=}  \ottnt{E_{{\mathrm{1}}}} \,  \mathsf{in}  \, \ottnt{e_{{\mathrm{2}}}}  \ottsym{)}$ and $\ottnt{E'} \,  =  \, \ottsym{(}  \mathsf{let} \, \mathit{x_{{\mathrm{1}}}}  \ottsym{=}  \ottnt{E'_{{\mathrm{1}}}} \,  \mathsf{in}  \, \ottnt{e'_{{\mathrm{2}}}}  \ottsym{)}$.
   We have the conclusion by the following.
   \begin{itemize}
    \item $\ottnt{e} = \ottsym{(}  \mathsf{let} \, \mathit{x_{{\mathrm{1}}}}  \ottsym{=}  \ottnt{e_{{\mathrm{1}}}} \,  \mathsf{in}  \, \ottnt{e_{{\mathrm{2}}}}  \ottsym{)} = \ottsym{(}  \mathsf{let} \, \mathit{x_{{\mathrm{1}}}}  \ottsym{=}   \ottnt{E_{{\mathrm{1}}}}  [  \ottnt{e_{{\mathrm{0}}}}  ]  \,  \mathsf{in}  \, \ottnt{e_{{\mathrm{2}}}}  \ottsym{)} =  \ottnt{E}  [  \ottnt{e_{{\mathrm{0}}}}  ] $.
    \item $\ottnt{e'} = \ottsym{(}  \mathsf{let} \, \mathit{x_{{\mathrm{1}}}}  \ottsym{=}  \ottnt{e'_{{\mathrm{1}}}} \,  \mathsf{in}  \, \ottnt{e'_{{\mathrm{2}}}}  \ottsym{)} = \ottsym{(}  \mathsf{let} \, \mathit{x_{{\mathrm{1}}}}  \ottsym{=}   \ottnt{E'_{{\mathrm{1}}}}  [  \ottnt{e'_{{\mathrm{0}}}}  ]  \,  \mathsf{in}  \, \ottnt{e'_{{\mathrm{2}}}}  \ottsym{)} =  \ottnt{E'}  [  \ottnt{e'_{{\mathrm{0}}}}  ] $.
    \item $\ottnt{E} \,  \prec_{ \ottnt{v} }  \, \ottnt{E'} \,  \uparrow \, (  \ottmv{i}  ,  \pi  )  = \ottsym{(}  \mathsf{let} \, \mathit{x_{{\mathrm{1}}}}  \ottsym{=}  \ottsym{(}  \ottnt{E_{{\mathrm{1}}}} \,  \prec_{ \ottnt{v} }  \, \ottnt{E'_{{\mathrm{1}}}} \,  \uparrow \, (  \ottmv{i}  ,  \pi  )   \ottsym{)} \,  \mathsf{in}  \, \ottsym{(}  \ottnt{e_{{\mathrm{2}}}} \,  \prec_{ \ottnt{v} }  \, \ottnt{e'_{{\mathrm{2}}}} \,  \uparrow \, (  \ottmv{i}  ,  \pi  )   \ottsym{)}  \ottsym{)} = \ottsym{(}  \mathsf{let} \, \mathit{x_{{\mathrm{1}}}}  \ottsym{=}  \ottnt{E''_{{\mathrm{1}}}} \,  \mathsf{in}  \, \ottnt{e''_{{\mathrm{2}}}}  \ottsym{)} = \ottnt{E''}$.
   \end{itemize}

   If $\ottnt{E''} \,  =  \, \ottnt{K''}$ for some $\ottnt{K''}$, then $\ottnt{E''_{{\mathrm{1}}}} \,  =  \, \ottnt{K''_{{\mathrm{1}}}}$ for some $\ottnt{K''_{{\mathrm{1}}}}$
   such that $\ottnt{K''} \,  =  \, \ottsym{(}  \mathsf{let} \, \mathit{x_{{\mathrm{1}}}}  \ottsym{=}  \ottnt{K''_{{\mathrm{1}}}} \,  \mathsf{in}  \, \ottnt{e''_{{\mathrm{2}}}}  \ottsym{)}$.
   Then, the IH implies $\ottnt{E_{{\mathrm{1}}}} \,  =  \, \ottnt{K_{{\mathrm{1}}}}$ and $\ottnt{E'_{{\mathrm{1}}}} \,  =  \, \ottnt{K'_{{\mathrm{1}}}}$ for some $\ottnt{K_{{\mathrm{1}}}}$ and $\ottnt{K'_{{\mathrm{1}}}}$.
   Therefore, $\ottnt{E} \,  =  \, \ottnt{K}$ and $\ottnt{E'} \,  =  \, \ottnt{K'}$ for some $\ottnt{K}$ and $\ottnt{K'}$ such that
   $\ottnt{K} \,  =  \, \ottsym{(}  \mathsf{let} \, \mathit{x_{{\mathrm{1}}}}  \ottsym{=}  \ottnt{K_{{\mathrm{1}}}} \,  \mathsf{in}  \, \ottnt{e_{{\mathrm{2}}}}  \ottsym{)}$ and $\ottnt{K'} \,  =  \, \ottsym{(}  \mathsf{let} \, \mathit{x_{{\mathrm{1}}}}  \ottsym{=}  \ottnt{K'_{{\mathrm{1}}}} \,  \mathsf{in}  \, \ottnt{e_{{\mathrm{2}}}}  \ottsym{)}$.

  \case $  \exists  \, { \ottnt{E''_{{\mathrm{1}}}} } . \  \ottnt{E''} \,  =  \,  \resetcnstr{ \ottnt{E''_{{\mathrm{1}}}} }  $:
   We are given
   $\ottnt{e} \,  \prec_{ \ottnt{v} }  \, \ottnt{e'} \,  \uparrow \, (  \ottmv{i}  ,  \pi  )  \,  =  \,  \resetcnstr{  \ottnt{E''_{{\mathrm{1}}}}  [  \ottnt{e''_{{\mathrm{0}}}}  ]  } $.
   By definition, there exist some $\ottnt{e_{{\mathrm{1}}}}$ and $\ottnt{e'_{{\mathrm{1}}}}$ such that
   \begin{itemize}
    \item $\ottnt{e} \,  =  \,  \resetcnstr{ \ottnt{e_{{\mathrm{1}}}} } $,
    \item $\ottnt{e'} \,  =  \,  \resetcnstr{ \ottnt{e'_{{\mathrm{1}}}} } $, and
    \item $\ottnt{e_{{\mathrm{1}}}} \,  \prec_{ \ottnt{v} }  \, \ottnt{e'_{{\mathrm{1}}}} \,  \uparrow \, (  \ottmv{i}  ,  \pi  )  \,  =  \,  \ottnt{E''_{{\mathrm{1}}}}  [  \ottnt{e''_{{\mathrm{0}}}}  ] $.
   \end{itemize}
   By the IH, there exist some $\ottnt{E_{{\mathrm{1}}}}$, $\ottnt{e_{{\mathrm{0}}}}$, $\ottnt{E'_{{\mathrm{1}}}}$, and $\ottnt{e'_{{\mathrm{0}}}}$ such that
   \begin{itemize}
    \item $\ottnt{e_{{\mathrm{1}}}} \,  =  \,  \ottnt{E_{{\mathrm{1}}}}  [  \ottnt{e_{{\mathrm{0}}}}  ] $,
    \item $\ottnt{e'_{{\mathrm{1}}}} \,  =  \,  \ottnt{E'_{{\mathrm{1}}}}  [  \ottnt{e'_{{\mathrm{0}}}}  ] $,
    \item $\ottnt{E_{{\mathrm{1}}}} \,  \prec_{ \ottnt{v} }  \, \ottnt{E'_{{\mathrm{1}}}} \,  \uparrow \, (  \ottmv{i}  ,  \pi  )  \,  =  \, \ottnt{E''_{{\mathrm{1}}}}$, and
    \item $\ottnt{e_{{\mathrm{0}}}} \,  \prec_{ \ottnt{v} }  \, \ottnt{e'_{{\mathrm{0}}}} \,  \uparrow \, (  \ottmv{i}  ,  \pi  )  \,  =  \, \ottnt{e''_{{\mathrm{0}}}}$.
   \end{itemize}

   Let $\ottnt{E} \,  =  \,  \resetcnstr{ \ottnt{E_{{\mathrm{1}}}} } $ and $\ottnt{E'} \,  =  \,  \resetcnstr{ \ottnt{E'_{{\mathrm{1}}}} } $.
   We have the conclusion by the following.
   \begin{itemize}
    \item $\ottnt{e} =  \resetcnstr{ \ottnt{e_{{\mathrm{1}}}} }  =  \resetcnstr{  \ottnt{E_{{\mathrm{1}}}}  [  \ottnt{e_{{\mathrm{0}}}}  ]  }  =  \ottnt{E}  [  \ottnt{e_{{\mathrm{0}}}}  ] $.
    \item $\ottnt{e'} =  \resetcnstr{ \ottnt{e'_{{\mathrm{1}}}} }  =  \resetcnstr{  \ottnt{E'_{{\mathrm{1}}}}  [  \ottnt{e'_{{\mathrm{0}}}}  ]  }  =  \ottnt{E'}  [  \ottnt{e'_{{\mathrm{0}}}}  ] $.
    \item $\ottnt{E} \,  \prec_{ \ottnt{v} }  \, \ottnt{E'} \,  \uparrow \, (  \ottmv{i}  ,  \pi  )  =  \resetcnstr{ \ottnt{E_{{\mathrm{1}}}} \,  \prec_{ \ottnt{v} }  \, \ottnt{E'_{{\mathrm{1}}}} \,  \uparrow \, (  \ottmv{i}  ,  \pi  )  }  =  \resetcnstr{ \ottnt{E''_{{\mathrm{1}}}} }  = \ottnt{E''}$.
   \end{itemize}
 \end{caseanalysis}
\end{proof}

\begin{lemmap}{$0$-Approximation of Evaluation by Multiple Steps}{lr-approx-eval-nsteps-idx0}
 Suppose that
 \[
     \forall  \, { \ottmv{m} \,  <  \, \ottmv{n} } . \    \forall  \, { \ottnt{E}  \ottsym{,}  \varpi_{{\mathrm{0}}}  \ottsym{,}  \pi_{{\mathrm{0}}}  \ottsym{,}   \tceseq{ \ottnt{v_{{\mathrm{2}}}} }  } . \  \ottsym{(}  \ottnt{e_{{\mathrm{1}}}} \,  \prec_{ \ottnt{v} }  \, \ottnt{e'_{{\mathrm{1}}}} \,  \uparrow \, (  \ottsym{0}  ,  \pi  )   \ottsym{)} \,  \mathrel{  \longrightarrow  }^{ \ottmv{m} }  \,  \ottnt{E}  [   \ottnt{v} _{  \mu   \ottsym{;}  \ottsym{0}  \ottsym{;}  \pi_{{\mathrm{0}}} }  \,  \tceseq{ \ottnt{v_{{\mathrm{2}}}} }   ]  \,  \,\&\,  \, \varpi_{{\mathrm{0}}}    \mathrel{\Longrightarrow}   \mathit{arity}  (  \ottnt{v}  )  \,  \not=  \, \ottsym{\mbox{$\mid$}}   \tceseq{ \ottnt{v_{{\mathrm{2}}}} }   \ottsym{\mbox{$\mid$}}  ~.
 \]
 If $\ottnt{e_{{\mathrm{1}}}} \,  \mathrel{  \longrightarrow  }^{ \ottmv{n} }  \, \ottnt{e_{{\mathrm{2}}}} \,  \,\&\,  \, \varpi$,
 then $\ottsym{(}  \ottnt{e_{{\mathrm{1}}}} \,  \prec_{ \ottnt{v} }  \, \ottnt{e'_{{\mathrm{1}}}} \,  \uparrow \, (  \ottsym{0}  ,  \pi  )   \ottsym{)} \,  \mathrel{  \longrightarrow  }^{ \ottmv{n} }  \, \ottsym{(}  \ottnt{e_{{\mathrm{2}}}} \,  \prec_{ \ottnt{v} }  \, \ottnt{e'_{{\mathrm{2}}}} \,  \uparrow \, (  \ottsym{0}  ,  \pi  )   \ottsym{)} \,  \,\&\,  \, \varpi$
 for some $\ottnt{e'_{{\mathrm{2}}}}$.
\end{lemmap}
\begin{proof}
 \TS{OK: 2022/01/17}
 By induction on $\ottmv{n}$.
 \begin{caseanalysis}
  \case $\ottmv{n} \,  =  \, \ottsym{0}$:
   Obvious by letting $\ottnt{e'_{{\mathrm{2}}}} \,  =  \, \ottnt{e'_{{\mathrm{1}}}}$
   because $\ottnt{e_{{\mathrm{1}}}} \,  =  \, \ottnt{e_{{\mathrm{2}}}}$ and $ \varpi    =     \epsilon  $.

  \case $\ottmv{n} \,  >  \, \ottsym{0}$:
   We have
   \begin{itemize}
    \item $\ottnt{e_{{\mathrm{1}}}} \,  \mathrel{  \longrightarrow  }  \, \ottnt{e_{{\mathrm{3}}}} \,  \,\&\,  \, \varpi_{{\mathrm{1}}}$,
    \item $\ottnt{e_{{\mathrm{3}}}} \,  \mathrel{  \longrightarrow  }^{ \ottmv{n}  \ottsym{-}  \ottsym{1} }  \, \ottnt{e_{{\mathrm{2}}}} \,  \,\&\,  \, \varpi_{{\mathrm{2}}}$, and
    \item $ \varpi_{{\mathrm{1}}} \,  \tceappendop  \, \varpi_{{\mathrm{2}}}    =    \varpi $
   \end{itemize}
   for some $\ottnt{e_{{\mathrm{3}}}}$, $\varpi_{{\mathrm{1}}}$, and $\varpi_{{\mathrm{2}}}$.
   By the assumption,
   \begin{equation}
       \forall  \, { \ottnt{E}  \ottsym{,}  \pi_{{\mathrm{0}}}  \ottsym{,}   \tceseq{ \ottnt{v_{{\mathrm{2}}}} }  } . \  \ottnt{e_{{\mathrm{1}}}} \,  \prec_{ \ottnt{v} }  \, \ottnt{e'_{{\mathrm{1}}}} \,  \uparrow \, (  \ottsym{0}  ,  \pi  )  \,  =  \,  \ottnt{E}  [   \ottnt{v} _{  \mu   \ottsym{;}  \ottsym{0}  \ottsym{;}  \pi_{{\mathrm{0}}} }  \,  \tceseq{ \ottnt{v_{{\mathrm{2}}}} }   ]    \mathrel{\Longrightarrow}   \mathit{arity}  (  \ottnt{v}  )  \,  \not=  \, \ottsym{\mbox{$\mid$}}   \tceseq{ \ottnt{v_{{\mathrm{2}}}} }   \ottsym{\mbox{$\mid$}}  ~.
     \label{lem:lr-approx-eval-nsteps-idx0:step}
   \end{equation}

   We show that
   \begin{equation}
     {  {   \forall  \, { \ottnt{E}  \ottsym{,}  \ottnt{E'}  \ottsym{,}   \tceseq{ \ottnt{v_{{\mathrm{2}}}} }   \ottsym{,}   \tceseq{ \ottnt{v'_{{\mathrm{2}}}} }   \ottsym{,}  \pi_{{\mathrm{0}}} } . \  \ottnt{e_{{\mathrm{1}}}} \,  \not=  \,  \ottnt{E}  [  \ottnt{v} \,  \tceseq{ \ottnt{v_{{\mathrm{2}}}} }   ]   } \,\mathrel{\vee}\, { \ottnt{e'_{{\mathrm{1}}}} \,  \not=  \,  \ottnt{E'}  [   \ottnt{v} _{  \mu   \ottsym{;}  \ottsym{0}  \ottsym{;}  \pi_{{\mathrm{0}}} }  \,  \tceseq{ \ottnt{v'_{{\mathrm{2}}}} }   ]  }  } \,\mathrel{\vee}\, {  \mathit{arity}  (  \ottnt{v}  )  \,  \not=  \, \ottsym{\mbox{$\mid$}}   \tceseq{ \ottnt{v_{{\mathrm{2}}}} }   \ottsym{\mbox{$\mid$}} }  ~.
     \label{lem:lr-approx-eval-nsteps-idx0:assump-approx-eval}
   \end{equation}
   Suppose that $\ottnt{e_{{\mathrm{1}}}} \,  =  \,  \ottnt{E}  [  \ottnt{v} \,  \tceseq{ \ottnt{v_{{\mathrm{2}}}} }   ] $ and $\ottnt{e'_{{\mathrm{1}}}} \,  =  \,  \ottnt{E'}  [   \ottnt{v} _{  \mu   \ottsym{;}  \ottsym{0}  \ottsym{;}  \pi_{{\mathrm{0}}} }  \,  \tceseq{ \ottnt{v'_{{\mathrm{2}}}} }   ] $ and
   $ \mathit{arity}  (  \ottnt{v}  )  \,  =  \, \ottsym{\mbox{$\mid$}}   \tceseq{ \ottnt{v_{{\mathrm{2}}}} }   \ottsym{\mbox{$\mid$}}$
   for some $\ottnt{E}$, $\ottnt{E'}$, $ \tceseq{ \ottnt{v_{{\mathrm{2}}}} } $, $ \tceseq{ \ottnt{v'_{{\mathrm{2}}}} } $, and $\pi_{{\mathrm{0}}}$.
   Because $\ottsym{(}  \ottnt{e_{{\mathrm{1}}}} \,  \prec_{ \ottnt{v} }  \, \ottnt{e'_{{\mathrm{1}}}} \,  \uparrow \, (  \ottsym{0}  ,  \pi  )   \ottsym{)}$ is well defined,
   \reflem{lr-approx-wf} implies $\ottnt{e_{{\mathrm{1}}}} \,  \prec_{ \ottnt{v} }  \, \ottnt{e'_{{\mathrm{1}}}}$, that is,
   $ \ottnt{E}  [  \ottnt{v} \,  \tceseq{ \ottnt{v_{{\mathrm{2}}}} }   ]  \,  \prec_{ \ottnt{v} }  \,  \ottnt{E'}  [   \ottnt{v} _{  \mu   \ottsym{;}  \ottsym{0}  \ottsym{;}  \pi_{{\mathrm{0}}} }  \,  \tceseq{ \ottnt{v'_{{\mathrm{2}}}} }   ] $.
   Then, by \reflem{lr-approx-ectx},
   $\ottnt{E} \,  \prec_{ \ottnt{v} }  \, \ottnt{E'}$ and $\ottnt{v} \,  \tceseq{ \ottnt{v_{{\mathrm{2}}}} }  \,  \prec_{ \ottnt{v} }  \,  \ottnt{v} _{  \mu   \ottsym{;}  \ottsym{0}  \ottsym{;}  \pi_{{\mathrm{0}}} }  \,  \tceseq{ \ottnt{v'_{{\mathrm{2}}}} } $.
   By \reflem{lr-approx-embed-exp},
   \[\begin{array}{ll} &
    \ottnt{e_{{\mathrm{1}}}} \,  \prec_{ \ottnt{v} }  \, \ottnt{e'_{{\mathrm{1}}}} \,  \uparrow \, (  \ottsym{0}  ,  \pi  )  \\ = &
     \ottnt{E}  [  \ottnt{v} \,  \tceseq{ \ottnt{v_{{\mathrm{2}}}} }   ]  \,  \prec_{ \ottnt{v} }  \,  \ottnt{E'}  [   \ottnt{v} _{  \mu   \ottsym{;}  \ottsym{0}  \ottsym{;}  \pi_{{\mathrm{0}}} }  \,  \tceseq{ \ottnt{v'_{{\mathrm{2}}}} }   ]  \,  \uparrow \, (  \ottsym{0}  ,  \pi  )  \\ = &
     \ottsym{(}  \ottnt{E} \,  \prec_{ \ottnt{v} }  \, \ottnt{E'} \,  \uparrow \, (  \ottsym{0}  ,  \pi  )   \ottsym{)}  [   \ottnt{v} _{  \mu   \ottsym{;}  \ottsym{0}  \ottsym{;}  \pi }  \,  \tceseq{ \ottnt{v_{{\mathrm{2}}}} \,  \prec_{ \ottnt{v} }  \, \ottnt{v'_{{\mathrm{2}}}} \,  \uparrow \, (  \ottsym{0}  ,  \pi  )  }   ]  ~.
     \end{array}
   \]
   Then, by the assumption (\ref{lem:lr-approx-eval-nsteps-idx0:step}),
   $ \mathit{arity}  (  \ottnt{v}  )  \,  \not=  \, \ottsym{\mbox{$\mid$}}   \tceseq{ \ottnt{v_{{\mathrm{2}}}} \,  \prec_{ \ottnt{v} }  \, \ottnt{v'_{{\mathrm{2}}}} \,  \uparrow \, (  \ottsym{0}  ,  \pi  )  }   \ottsym{\mbox{$\mid$}} = \ottsym{\mbox{$\mid$}}   \tceseq{ \ottnt{v_{{\mathrm{2}}}} }   \ottsym{\mbox{$\mid$}}$,
   but it is contradictory with $ \mathit{arity}  (  \ottnt{v}  )  \,  =  \, \ottsym{\mbox{$\mid$}}   \tceseq{ \ottnt{v_{{\mathrm{2}}}} }   \ottsym{\mbox{$\mid$}}$.

   Therefore,
   \reflem{lr-approx-eval} (\ref{lem:lr-approx-eval:non-approx}) with
   (\ref{lem:lr-approx-eval-nsteps-idx0:assump-approx-eval}),
   $\ottnt{e_{{\mathrm{1}}}} \,  \prec_{ \ottnt{v} }  \, \ottnt{e'_{{\mathrm{1}}}}$ by \reflem{lr-approx-wf}, and
   $\ottnt{e_{{\mathrm{1}}}} \,  \mathrel{  \longrightarrow  }  \, \ottnt{e_{{\mathrm{3}}}} \,  \,\&\,  \, \varpi_{{\mathrm{1}}}$
   implies
   \[
    \ottsym{(}  \ottnt{e_{{\mathrm{1}}}} \,  \prec_{ \ottnt{v} }  \, \ottnt{e'_{{\mathrm{1}}}} \,  \uparrow \, (  \ottsym{0}  ,  \pi  )   \ottsym{)} \,  \mathrel{  \longrightarrow  }  \, \ottsym{(}  \ottnt{e_{{\mathrm{3}}}} \,  \prec_{ \ottnt{v} }  \, \ottnt{e'_{{\mathrm{3}}}} \,  \uparrow \, (  \ottsym{0}  ,  \pi  )   \ottsym{)} \,  \,\&\,  \, \varpi_{{\mathrm{1}}}
   \]
   for some $\ottnt{e'_{{\mathrm{3}}}}$.
   Further, by the IH with $\ottnt{e_{{\mathrm{3}}}} \,  \mathrel{  \longrightarrow  }^{ \ottmv{n}  \ottsym{-}  \ottsym{1} }  \, \ottnt{e_{{\mathrm{2}}}} \,  \,\&\,  \, \varpi_{{\mathrm{2}}}$,
   \[
    \ottsym{(}  \ottnt{e_{{\mathrm{3}}}} \,  \prec_{ \ottnt{v} }  \, \ottnt{e'_{{\mathrm{3}}}} \,  \uparrow \, (  \ottsym{0}  ,  \pi  )   \ottsym{)} \,  \mathrel{  \longrightarrow  }^{ \ottmv{n}  \ottsym{-}  \ottsym{1} }  \, \ottsym{(}  \ottnt{e_{{\mathrm{2}}}} \,  \prec_{ \ottnt{v} }  \, \ottnt{e'_{{\mathrm{2}}}} \,  \uparrow \, (  \ottsym{0}  ,  \pi  )   \ottsym{)} \,  \,\&\,  \, \varpi_{{\mathrm{2}}}
   \]
   for some $\ottnt{e'_{{\mathrm{2}}}}$.
   Therefore, because $ \varpi_{{\mathrm{1}}} \,  \tceappendop  \, \varpi_{{\mathrm{2}}}    =    \varpi $, we have the conclusion
   \[
    \ottsym{(}  \ottnt{e_{{\mathrm{1}}}} \,  \prec_{ \ottnt{v} }  \, \ottnt{e'_{{\mathrm{1}}}} \,  \uparrow \, (  \ottsym{0}  ,  \pi  )   \ottsym{)} \,  \mathrel{  \longrightarrow  }^{ \ottmv{n} }  \, \ottsym{(}  \ottnt{e_{{\mathrm{2}}}} \,  \prec_{ \ottnt{v} }  \, \ottnt{e'_{{\mathrm{2}}}} \,  \uparrow \, (  \ottsym{0}  ,  \pi  )   \ottsym{)} \,  \,\&\,  \, \varpi ~.
   \]
 \end{caseanalysis}
\end{proof}

\begin{lemmap}{Evaluation of Approximation Modulo Infinite Traces}{lr-approx-eval-modulo-itr}
 If $\ottsym{(}  \ottnt{e} \,  \prec_{ \ottnt{v} }  \, \ottnt{e'} \,  \uparrow \, (  \ottsym{0}  ,  \pi_{{\mathrm{1}}}  )   \ottsym{)} \,  \mathrel{  \longrightarrow  }  \, \ottnt{e_{{\mathrm{1}}}} \,  \,\&\,  \, \varpi$ and $  \forall  \, { \ottnt{E}  \ottsym{,}  \pi } . \  \ottnt{e_{{\mathrm{1}}}} \,  \not=  \,  \ottnt{E}  [   \bullet ^{ \pi }   ]  $,
 then
 \[
   {   \exists  \, { \ottnt{e_{{\mathrm{0}}}}  \ottsym{,}  \ottnt{e'_{{\mathrm{0}}}} } . \  \ottnt{e} \,  \mathrel{  \longrightarrow  }  \, \ottnt{e_{{\mathrm{0}}}} \,  \,\&\,  \, \varpi  } \,\mathrel{\wedge}\, {   \forall  \, { \pi_{{\mathrm{2}}} } . \  \ottsym{(}  \ottnt{e} \,  \prec_{ \ottnt{v} }  \, \ottnt{e'} \,  \uparrow \, (  \ottsym{0}  ,  \pi_{{\mathrm{2}}}  )   \ottsym{)} \,  \mathrel{  \longrightarrow  }  \, \ottsym{(}  \ottnt{e_{{\mathrm{0}}}} \,  \prec_{ \ottnt{v} }  \, \ottnt{e'_{{\mathrm{0}}}} \,  \uparrow \, (  \ottsym{0}  ,  \pi_{{\mathrm{2}}}  )   \ottsym{)} \,  \,\&\,  \, \varpi  }  ~.
 \]
\end{lemmap}
\begin{proof}
 \TS{OK: 2022/01/20}
 By induction on the derivation of $\ottsym{(}  \ottnt{e} \,  \prec_{ \ottnt{v} }  \, \ottnt{e'} \,  \uparrow \, (  \ottsym{0}  ,  \pi_{{\mathrm{1}}}  )   \ottsym{)} \,  \mathrel{  \longrightarrow  }  \, \ottnt{e_{{\mathrm{1}}}} \,  \,\&\,  \, \varpi$.
 Note that $\ottnt{e} \,  \prec_{ \ottnt{v} }  \, \ottnt{e'} \,  \uparrow \, (  \ottsym{0}  ,  \pi_{{\mathrm{2}}}  ) $ is well defined for any $\pi_{{\mathrm{2}}}$
 by \reflem{lr-approx-wf}.
 \begin{caseanalysis}
  \case \E{Red}/\R{Const}:
   We are given
   \begin{prooftree}
    \AxiomC{$\ottnt{o}  \ottsym{(}   \tceseq{ \ottnt{c} }   \ottsym{)} \,  \rightsquigarrow  \,  \zeta  (  \ottnt{o}  ,   \tceseq{ \ottnt{c} }   ) $}
    \UnaryInfC{$\ottnt{o}  \ottsym{(}   \tceseq{ \ottnt{c} }   \ottsym{)} \,  \mathrel{  \longrightarrow  }  \,  \zeta  (  \ottnt{o}  ,   \tceseq{ \ottnt{c} }   )  \,  \,\&\,  \,  \epsilon $}
   \end{prooftree}
   for some $\ottnt{o}$ and $ \tceseq{ \ottnt{c} } $ such that
   $\ottnt{e} \,  \prec_{ \ottnt{v} }  \, \ottnt{e'} \,  \uparrow \, (  \ottsym{0}  ,  \pi_{{\mathrm{1}}}  )  \,  =  \, \ottnt{o}  \ottsym{(}   \tceseq{ \ottnt{c} }   \ottsym{)}$ and $\ottnt{e_{{\mathrm{1}}}} \,  =  \,  \zeta  (  \ottnt{o}  ,   \tceseq{ \ottnt{c} }   ) $.
   We also have $ \varpi    =     \epsilon  $.
   By definition, $\ottnt{e} = \ottnt{e'} = \ottnt{o}  \ottsym{(}   \tceseq{ \ottnt{c} }   \ottsym{)}$.

   Let $\ottnt{e_{{\mathrm{0}}}} = \ottnt{e'_{{\mathrm{0}}}} =  \zeta  (  \ottnt{o}  ,   \tceseq{ \ottnt{c} }   ) $.
   We have the conclusion by the following.
   \begin{itemize}
    \item We have $\ottnt{e} \,  \mathrel{  \longrightarrow  }  \, \ottnt{e_{{\mathrm{0}}}} \,  \,\&\,  \,  \epsilon $ by \E{Red}/\R{Const}.
    \item Let $\pi_{{\mathrm{2}}}$ be any infinite trace.
          Then,
          \[\begin{array}{ll} &
           \ottnt{e} \,  \prec_{ \ottnt{v} }  \, \ottnt{e'} \,  \uparrow \, (  \ottsym{0}  ,  \pi_{{\mathrm{2}}}  )  \\ = &
           \ottnt{o}  \ottsym{(}   \tceseq{ \ottnt{c} }   \ottsym{)} \\  \longrightarrow  &
            \zeta  (  \ottnt{o}  ,   \tceseq{ \ottnt{c} }   )  \,  \,\&\,  \,  \epsilon   \\ = &
           \ottnt{e_{{\mathrm{0}}}} \,  \prec_{ \ottnt{v} }  \, \ottnt{e'_{{\mathrm{0}}}} \,  \uparrow \, (  \ottsym{0}  ,  \pi_{{\mathrm{2}}}  )  \,  \,\&\,  \,  \epsilon  ~.
            \end{array}
          \]
   \end{itemize}

  \case \E{Red}/\R{Beta}:
   We are given
   \begin{prooftree}
    \AxiomC{$\ottsym{(}   \lambda\!  \,  \tceseq{ \mathit{x} }   \ottsym{.}  \ottnt{e''}  \ottsym{)} \,  \tceseq{ \ottnt{v''_{{\mathrm{2}}}} }  \,  \rightsquigarrow  \,  \ottnt{e''}    [ \tceseq{ \ottnt{v''_{{\mathrm{2}}}}  /  \mathit{x} } ]  $}
    \UnaryInfC{$\ottsym{(}   \lambda\!  \,  \tceseq{ \mathit{x} }   \ottsym{.}  \ottnt{e''}  \ottsym{)} \,  \tceseq{ \ottnt{v''_{{\mathrm{2}}}} }  \,  \mathrel{  \longrightarrow  }  \,  \ottnt{e''}    [ \tceseq{ \ottnt{v''_{{\mathrm{2}}}}  /  \mathit{x} } ]   \,  \,\&\,  \,  \epsilon $}
   \end{prooftree}
   for some $ \tceseq{ \mathit{x} } $, $\ottnt{e''}$, and $ \tceseq{ \ottnt{v''_{{\mathrm{2}}}} } $ such that
   \begin{itemize}
    \item $\ottnt{e} \,  \prec_{ \ottnt{v} }  \, \ottnt{e'} \,  \uparrow \, (  \ottsym{0}  ,  \pi_{{\mathrm{1}}}  )  \,  =  \, \ottsym{(}   \lambda\!  \,  \tceseq{ \mathit{x} }   \ottsym{.}  \ottnt{e''}  \ottsym{)} \,  \tceseq{ \ottnt{v''_{{\mathrm{2}}}} } $,
    \item $\ottnt{e_{{\mathrm{1}}}} \,  =  \,  \ottnt{e''}    [ \tceseq{ \ottnt{v''_{{\mathrm{2}}}}  /  \mathit{x} } ]  $, and
    \item $\ottsym{\mbox{$\mid$}}   \tceseq{ \ottnt{v''_{{\mathrm{2}}}} }   \ottsym{\mbox{$\mid$}} \,  =  \, \ottsym{\mbox{$\mid$}}   \tceseq{ \mathit{x} }   \ottsym{\mbox{$\mid$}}$.
   \end{itemize}
   We also have $ \varpi    =     \epsilon  $.

   Because $\ottnt{e} \,  \prec_{ \ottnt{v} }  \, \ottnt{e'} \,  \uparrow \, (  \ottsym{0}  ,  \pi_{{\mathrm{1}}}  )  \,  =  \, \ottsym{(}   \lambda\!  \,  \tceseq{ \mathit{x} }   \ottsym{.}  \ottnt{e''}  \ottsym{)} \,  \tceseq{ \ottnt{v''_{{\mathrm{2}}}} } $,
   there exist some $\ottnt{v_{{\mathrm{1}}}}$, $\ottnt{v'_{{\mathrm{1}}}}$, $ \tceseq{ \ottnt{v_{{\mathrm{2}}}} } $, and $ \tceseq{ \ottnt{v'_{{\mathrm{2}}}} } $ such that
   \begin{itemize}
    \item $\ottnt{e} \,  =  \, \ottnt{v_{{\mathrm{1}}}} \,  \tceseq{ \ottnt{v_{{\mathrm{2}}}} } $,
    \item $\ottnt{e'} \,  =  \, \ottnt{v'_{{\mathrm{1}}}} \,  \tceseq{ \ottnt{v'_{{\mathrm{2}}}} } $,
    \item $\ottnt{v_{{\mathrm{1}}}} \,  \prec_{ \ottnt{v} }  \, \ottnt{v'_{{\mathrm{1}}}} \,  \uparrow \, (  \ottsym{0}  ,  \pi_{{\mathrm{1}}}  )  \,  =  \,  \lambda\!  \,  \tceseq{ \mathit{x} }   \ottsym{.}  \ottnt{e''}$, and
    \item $ \tceseq{ \ottnt{v_{{\mathrm{2}}}} \,  \prec_{ \ottnt{v} }  \, \ottnt{v'_{{\mathrm{2}}}} \,  \uparrow \, (  \ottsym{0}  ,  \pi_{{\mathrm{1}}}  )  \,  =  \, \ottnt{v''_{{\mathrm{2}}}} } $.
   \end{itemize}
   By case analysis on $\ottnt{v_{{\mathrm{1}}}}$.
   We have two cases to be considered because $\ottnt{v_{{\mathrm{1}}}} \,  \prec_{ \ottnt{v} }  \, \ottnt{v'_{{\mathrm{1}}}} \,  \uparrow \, (  \ottsym{0}  ,  \pi_{{\mathrm{1}}}  )  \,  =  \,  \lambda\!  \,  \tceseq{ \mathit{x} }   \ottsym{.}  \ottnt{e''}$.
   \begin{caseanalysis}
    \case $\ottnt{v_{{\mathrm{1}}}} \,  =  \, \ottnt{v}$:
     We can find
     $\ottnt{v'_{{\mathrm{1}}}} \,  =  \,  \ottnt{v} _{  \mu   \ottsym{;}  \ottmv{j}  \ottsym{;}  \pi_{{\mathrm{0}}} } $ and
     $ \lambda\!  \,  \tceseq{ \mathit{x} }   \ottsym{.}  \ottnt{e''} \,  =  \,  \ottnt{v} _{  \mu   \ottsym{;}  \ottmv{j}  \ottsym{;}  \pi_{{\mathrm{1}}} } $
     for some $\ottmv{j}$ and $\pi_{{\mathrm{0}}}$.
     We can suppose that $\ottnt{v} \,  =  \,  \mathsf{rec}  \, (  \tceseq{  \mathit{X} ^{  \ottnt{A} _{  \mu  }  },  \mathit{Y} ^{  \ottnt{A} _{  \nu  }  }  }  ,  \mathit{f} ,   \tceseq{ \mathit{x} }  ) . \,  \ottnt{e} $
     for some $ \tceseq{ \mathit{X} } $, $ \tceseq{ \mathit{Y} } $, $ \tceseq{  \ottnt{A} _{  \mu  }  } $, $ \tceseq{  \ottnt{A} _{  \nu  }  } $, $\mathit{f}$, and $\ottnt{e}$.
     For every $\ottmv{i}$,
     let
     $\sigma_{\ottmv{i}}$ be the substitution as described in \refdef{munu-sim}.
     By case analysis on $\ottmv{j}$.
     \begin{caseanalysis}
      \case $\ottmv{j} \,  =  \, \ottsym{0}$:
       Because $ \lambda\!  \,  \tceseq{ \mathit{x} }   \ottsym{.}  \ottnt{e''} \,  =  \,  \ottnt{v} _{  \mu   \ottsym{;}  \ottsym{0}  \ottsym{;}  \pi_{{\mathrm{1}}} } $,
       we have $\ottnt{e''} \,  =  \,  \bullet ^{ \pi_{{\mathrm{1}}} } $.
       Therefore, $\ottnt{e_{{\mathrm{1}}}} =  \ottnt{e''}    [ \tceseq{ \ottnt{v''_{{\mathrm{2}}}}  /  \mathit{x} } ]   =  \bullet ^{ \pi_{{\mathrm{1}}} } $.
       This is contradictory with the assumption
       $  \forall  \, { \ottnt{E}  \ottsym{,}  \pi } . \  \ottnt{e_{{\mathrm{1}}}} \,  \not=  \,  \ottnt{E}  [   \bullet ^{ \pi }   ]  $.

      \case $\ottmv{j} \,  >  \, \ottsym{0}$:
       Because $ \lambda\!  \,  \tceseq{ \mathit{x} }   \ottsym{.}  \ottnt{e''} \,  =  \,  \ottnt{v} _{  \mu   \ottsym{;}  \ottmv{j}  \ottsym{;}  \pi_{{\mathrm{1}}} } $,
       we have $\ottnt{e''} \,  =  \,    \ottnt{e}    [ \tceseq{  { \sigma }_{ \ottmv{j}  \ottsym{-}  \ottsym{1} }   \ottsym{(}  \mathit{X}  \ottsym{)}  /  \mathit{X} } ]      [ \tceseq{  \top^P   /  \mathit{Y} } ]      [   \ottnt{v} _{  \mu   \ottsym{;}  \ottmv{j}  \ottsym{-}  \ottsym{1}  \ottsym{;}  \pi_{{\mathrm{1}}} }   /  \mathit{f}  ]  $.
       Because $\ottnt{e_{{\mathrm{1}}}} \,  =  \,  \ottnt{e''}    [ \tceseq{ \ottnt{v''_{{\mathrm{2}}}}  /  \mathit{x} } ]  $ and $ \tceseq{ \ottnt{v''_{{\mathrm{2}}}} \,  =  \, \ottnt{v_{{\mathrm{2}}}} \,  \prec_{ \ottnt{v} }  \, \ottnt{v'_{{\mathrm{2}}}} \,  \uparrow \, (  \ottsym{0}  ,  \pi_{{\mathrm{1}}}  )  } $,
       we have
       \[
        \ottnt{e_{{\mathrm{1}}}} \,  =  \,     \ottnt{e}    [ \tceseq{  { \sigma }_{ \ottmv{j}  \ottsym{-}  \ottsym{1} }   \ottsym{(}  \mathit{X}  \ottsym{)}  /  \mathit{X} } ]      [ \tceseq{  \top^P   /  \mathit{Y} } ]      [   \ottnt{v} _{  \mu   \ottsym{;}  \ottmv{j}  \ottsym{-}  \ottsym{1}  \ottsym{;}  \pi_{{\mathrm{1}}} }   /  \mathit{f}  ]      [ \tceseq{ \ottnt{v_{{\mathrm{2}}}} \,  \prec_{ \ottnt{v} }  \, \ottnt{v'_{{\mathrm{2}}}} \,  \uparrow \, (  \ottsym{0}  ,  \pi_{{\mathrm{1}}}  )   /  \mathit{x} } ]   ~.
       \]

       Let
       \begin{itemize}
        \item $\ottnt{e_{{\mathrm{0}}}} \,  =  \,     \ottnt{e}    [ \tceseq{  \ottnt{A} _{  \mu  }   /  \mathit{X} } ]      [ \tceseq{  \ottnt{A} _{  \nu  }   /  \mathit{Y} } ]      [  \ottnt{v}  /  \mathit{f}  ]      [ \tceseq{ \ottnt{v_{{\mathrm{2}}}}  /  \mathit{x} } ]  $ and
        \item $\ottnt{e'_{{\mathrm{0}}}} \,  =  \,     \ottnt{e}    [ \tceseq{  { \sigma }_{ \ottmv{j}  \ottsym{-}  \ottsym{1} }   \ottsym{(}  \mathit{X}  \ottsym{)}  /  \mathit{X} } ]      [ \tceseq{  \top^P   /  \mathit{Y} } ]      [   \ottnt{v} _{  \mu   \ottsym{;}  \ottmv{j}  \ottsym{-}  \ottsym{1}  \ottsym{;}  \pi_{{\mathrm{0}}} }   /  \mathit{f}  ]      [ \tceseq{ \ottnt{v'_{{\mathrm{2}}}}  /  \mathit{x} } ]  $.
       \end{itemize}
       We have the conclusion by the following.
       \begin{itemize}
        \item We have $\ottnt{e} = \ottnt{v} \,  \tceseq{ \ottnt{v_{{\mathrm{2}}}} }  \,  \mathrel{  \longrightarrow  }  \, \ottnt{e_{{\mathrm{0}}}} \,  \,\&\,  \,  \epsilon $ by \E{Red}/\R{RBeta}.
        \item Let $\pi_{{\mathrm{2}}}$ be any infinite trace.
              We show that $\ottsym{(}  \ottnt{e} \,  \prec_{ \ottnt{v} }  \, \ottnt{e'} \,  \uparrow \, (  \ottsym{0}  ,  \pi_{{\mathrm{2}}}  )   \ottsym{)} \,  \mathrel{  \longrightarrow  }  \, \ottsym{(}  \ottnt{e_{{\mathrm{0}}}} \,  \prec_{ \ottnt{v} }  \, \ottnt{e'_{{\mathrm{0}}}} \,  \uparrow \, (  \ottsym{0}  ,  \pi_{{\mathrm{2}}}  )   \ottsym{)} \,  \,\&\,  \,  \epsilon $.
              Because $ \tceseq{ \ottnt{v_{{\mathrm{2}}}} \,  \prec_{ \ottnt{v} }  \, \ottnt{v'_{{\mathrm{2}}}} \,  \uparrow \, (  \ottsym{0}  ,  \pi_{{\mathrm{2}}}  )  } $ are values
              by Lemmas~\ref{lem:lr-approx-wf} and \ref{lem:lr-approx-val},
              we have
              \[\begin{array}{ll} &
               \ottnt{e} \,  \prec_{ \ottnt{v} }  \, \ottnt{e'} \,  \uparrow \, (  \ottsym{0}  ,  \pi_{{\mathrm{2}}}  )  \\ = &
               \ottsym{(}  \ottnt{v} \,  \tceseq{ \ottnt{v_{{\mathrm{2}}}} }   \ottsym{)} \,  \prec_{ \ottnt{v} }  \, \ottsym{(}   \ottnt{v} _{  \mu   \ottsym{;}  \ottmv{j}  \ottsym{;}  \pi_{{\mathrm{0}}} }  \,  \tceseq{ \ottnt{v'_{{\mathrm{2}}}} }   \ottsym{)} \,  \uparrow \, (  \ottsym{0}  ,  \pi_{{\mathrm{2}}}  )  \\ = &
                \ottnt{v} _{  \mu   \ottsym{;}  \ottmv{j}  \ottsym{;}  \pi_{{\mathrm{2}}} }  \,  \tceseq{ \ottnt{v_{{\mathrm{2}}}} \,  \prec_{ \ottnt{v} }  \, \ottnt{v'_{{\mathrm{2}}}} \,  \uparrow \, (  \ottsym{0}  ,  \pi_{{\mathrm{2}}}  )  }  \\  \longrightarrow  &
                   \ottnt{e}    [ \tceseq{  { \sigma }_{ \ottmv{j}  \ottsym{-}  \ottsym{1} }   \ottsym{(}  \mathit{X}  \ottsym{)}  /  \mathit{X} } ]      [ \tceseq{  \top^P   /  \mathit{Y} } ]      [   \ottnt{v} _{  \mu   \ottsym{;}  \ottmv{j}  \ottsym{-}  \ottsym{1}  \ottsym{;}  \pi_{{\mathrm{2}}} }   /  \mathit{f}  ]      [ \tceseq{ \ottnt{v_{{\mathrm{2}}}} \,  \prec_{ \ottnt{v} }  \, \ottnt{v'_{{\mathrm{2}}}} \,  \uparrow \, (  \ottsym{0}  ,  \pi_{{\mathrm{2}}}  )   /  \mathit{x} } ]   \,  \,\&\,  \,  \epsilon 
                \quad \text{(by \E{Red}/\R{Beta})}
                \\ = &
               \ottnt{e_{{\mathrm{0}}}} \,  \prec_{ \ottnt{v} }  \, \ottnt{e'_{{\mathrm{0}}}} \,  \uparrow \, (  \ottsym{0}  ,  \pi_{{\mathrm{2}}}  )  \,  \,\&\,  \,  \epsilon 
                \quad \text{(by Lemmas~\ref{lem:lr-approx-refl}, \ref{lem:lr-approx-val-subst}, and \ref{lem:lr-approx-pred-subst})} ~.
                \end{array}
              \]
       \end{itemize}

      \case $  \exists  \, { \ottnt{e_{{\mathrm{01}}}} } . \  \ottnt{v_{{\mathrm{1}}}} \,  =  \,  \lambda\!  \,  \tceseq{ \mathit{x} }   \ottsym{.}  \ottnt{e_{{\mathrm{01}}}} $:
       We can find
       $\ottnt{v'_{{\mathrm{1}}}} \,  =  \,  \lambda\!  \,  \tceseq{ \mathit{x} }   \ottsym{.}  \ottnt{e'_{{\mathrm{01}}}}$ and
       $\ottnt{e_{{\mathrm{01}}}} \,  \prec_{ \ottnt{v} }  \, \ottnt{e'_{{\mathrm{01}}}} \,  \uparrow \, (  \ottsym{0}  ,  \pi_{{\mathrm{1}}}  )  \,  =  \, \ottnt{e''}$
       for some $\ottnt{e'_{{\mathrm{01}}}}$.

       Let
       $\ottnt{e_{{\mathrm{0}}}} \,  =  \,  \ottnt{e_{{\mathrm{01}}}}    [ \tceseq{ \ottnt{v_{{\mathrm{2}}}}  /  \mathit{x} } ]  $ and
       $\ottnt{e'_{{\mathrm{0}}}} \,  =  \,  \ottnt{e'_{{\mathrm{01}}}}    [ \tceseq{ \ottnt{v'_{{\mathrm{2}}}}  /  \mathit{x} } ]  $.
       We have the conclusion by the following.
       \begin{itemize}
        \item We have $\ottnt{e} = \ottsym{(}   \lambda\!  \,  \tceseq{ \mathit{x} }   \ottsym{.}  \ottnt{e_{{\mathrm{01}}}}  \ottsym{)} \,  \tceseq{ \ottnt{v_{{\mathrm{2}}}} }  \,  \mathrel{  \longrightarrow  }  \, \ottnt{e_{{\mathrm{0}}}} \,  \,\&\,  \,  \epsilon $ by
              \E{Red}/\R{Beta}.

        \item Let $\pi_{{\mathrm{2}}}$ be any infinite trace.
              We show that $\ottsym{(}  \ottnt{e} \,  \prec_{ \ottnt{v} }  \, \ottnt{e'} \,  \uparrow \, (  \ottsym{0}  ,  \pi_{{\mathrm{2}}}  )   \ottsym{)} \,  \mathrel{  \longrightarrow  }  \, \ottsym{(}  \ottnt{e_{{\mathrm{0}}}} \,  \prec_{ \ottnt{v} }  \, \ottnt{e'_{{\mathrm{0}}}} \,  \uparrow \, (  \ottsym{0}  ,  \pi_{{\mathrm{2}}}  )   \ottsym{)} \,  \,\&\,  \,  \epsilon $.
              Because $ \tceseq{ \ottnt{v_{{\mathrm{2}}}} \,  \prec_{ \ottnt{v} }  \, \ottnt{v'_{{\mathrm{2}}}} \,  \uparrow \, (  \ottsym{0}  ,  \pi_{{\mathrm{2}}}  )  } $ are values
              by Lemmas~\ref{lem:lr-approx-wf} and \ref{lem:lr-approx-val},
              we have
              \[\begin{array}{ll} &
               \ottnt{e} \,  \prec_{ \ottnt{v} }  \, \ottnt{e'} \,  \uparrow \, (  \ottsym{0}  ,  \pi_{{\mathrm{2}}}  )  \\ = &
               \ottsym{(}  \ottnt{v_{{\mathrm{1}}}} \,  \prec_{ \ottnt{v} }  \, \ottnt{v'_{{\mathrm{1}}}} \,  \uparrow \, (  \ottsym{0}  ,  \pi_{{\mathrm{2}}}  )   \ottsym{)} \,  \tceseq{ \ottnt{v_{{\mathrm{2}}}} \,  \prec_{ \ottnt{v} }  \, \ottnt{v'_{{\mathrm{2}}}} \,  \uparrow \, (  \ottsym{0}  ,  \pi_{{\mathrm{2}}}  )  }  \\ = &
               \ottsym{(}   \lambda\!  \,  \tceseq{ \mathit{x} }   \ottsym{.}  \ottsym{(}  \ottnt{e_{{\mathrm{01}}}} \,  \prec_{ \ottnt{v} }  \, \ottnt{e'_{{\mathrm{01}}}} \,  \uparrow \, (  \ottsym{0}  ,  \pi_{{\mathrm{2}}}  )   \ottsym{)}  \ottsym{)} \,  \tceseq{ \ottnt{v_{{\mathrm{2}}}} \,  \prec_{ \ottnt{v} }  \, \ottnt{v'_{{\mathrm{2}}}} \,  \uparrow \, (  \ottsym{0}  ,  \pi_{{\mathrm{2}}}  )  }  \\  \longrightarrow  &
                \ottsym{(}  \ottnt{e_{{\mathrm{01}}}} \,  \prec_{ \ottnt{v} }  \, \ottnt{e'_{{\mathrm{01}}}} \,  \uparrow \, (  \ottsym{0}  ,  \pi_{{\mathrm{2}}}  )   \ottsym{)}    [ \tceseq{ \ottnt{v_{{\mathrm{2}}}} \,  \prec_{ \ottnt{v} }  \, \ottnt{v'_{{\mathrm{2}}}} \,  \uparrow \, (  \ottsym{0}  ,  \pi_{{\mathrm{2}}}  )   /  \mathit{x} } ]   \,  \,\&\,  \,  \epsilon 
                \quad \text{(by \E{Red}/\R{Beta})}
                \\  =  &
                \ottnt{e_{{\mathrm{01}}}}    [ \tceseq{ \ottnt{v_{{\mathrm{2}}}}  /  \mathit{x} } ]   \,  \prec_{ \ottnt{v} }  \,  \ottnt{e'_{{\mathrm{01}}}}    [ \tceseq{ \ottnt{v'_{{\mathrm{2}}}}  /  \mathit{x} } ]   \,  \uparrow \, (  \ottsym{0}  ,  \pi_{{\mathrm{2}}}  )  \,  \,\&\,  \,  \epsilon 
                \quad \text{(by \reflem{lr-approx-val-subst})}
                \\  =  &
               \ottnt{e_{{\mathrm{0}}}} \,  \prec_{ \ottnt{v} }  \, \ottnt{e'_{{\mathrm{0}}}} \,  \uparrow \, (  \ottsym{0}  ,  \pi_{{\mathrm{2}}}  )  \,  \,\&\,  \,  \epsilon  ~.
                \end{array}
              \]
       \end{itemize}
     \end{caseanalysis}
   \end{caseanalysis}

  \case \E{Red}/\R{RBeta}:
   We are given
   \begin{prooftree}
    \AxiomC{$\ottnt{v''_{{\mathrm{1}}}} \,  \tceseq{ \ottnt{v''_{{\mathrm{2}}}} }  \,  \rightsquigarrow  \,     \ottnt{e''_{{\mathrm{0}}}}    [ \tceseq{  \ottnt{A''} _{  \mu  }   /  \mathit{X} } ]      [ \tceseq{  \ottnt{A''} _{  \nu  }   /  \mathit{Y} } ]      [  \ottnt{v''_{{\mathrm{1}}}}  /  \mathit{f}  ]      [ \tceseq{ \ottnt{v''_{{\mathrm{2}}}}  /  \mathit{x} } ]  $}
    \UnaryInfC{$\ottnt{v''_{{\mathrm{1}}}} \,  \tceseq{ \ottnt{v''_{{\mathrm{2}}}} }  \,  \mathrel{  \longrightarrow  }  \,     \ottnt{e''_{{\mathrm{0}}}}    [ \tceseq{  \ottnt{A''} _{  \mu  }   /  \mathit{X} } ]      [ \tceseq{  \ottnt{A''} _{  \nu  }   /  \mathit{Y} } ]      [  \ottnt{v''_{{\mathrm{1}}}}  /  \mathit{f}  ]      [ \tceseq{ \ottnt{v''_{{\mathrm{2}}}}  /  \mathit{x} } ]   \,  \,\&\,  \,  \epsilon $}
   \end{prooftree}
   for some $\ottnt{v''_{{\mathrm{1}}}}$, $ \tceseq{ \ottnt{v''_{{\mathrm{2}}}} } $, $ \tceseq{  \ottnt{A''} _{  \mu  }  } $, $ \tceseq{  \ottnt{A''} _{  \nu  }  } $, $ \tceseq{ \mathit{X} } $, $ \tceseq{ \mathit{Y} } $,
   $\mathit{f}$, $ \tceseq{ \mathit{x} } $, and $\ottnt{e''_{{\mathrm{0}}}}$ such that
   \begin{itemize}
    \item $\ottnt{e} \,  \prec_{ \ottnt{v} }  \, \ottnt{e'} \,  \uparrow \, (  \ottsym{0}  ,  \pi_{{\mathrm{1}}}  )  \,  =  \, \ottnt{v''_{{\mathrm{1}}}} \,  \tceseq{ \ottnt{v''_{{\mathrm{2}}}} } $,
    \item $\ottnt{v''_{{\mathrm{1}}}} \,  =  \,  \mathsf{rec}  \, (  \tceseq{  \mathit{X} ^{  \ottnt{A''} _{  \mu  }  },  \mathit{Y} ^{  \ottnt{A''} _{  \nu  }  }  }  ,  \mathit{f} ,   \tceseq{ \mathit{x} }  ) . \,  \ottnt{e''_{{\mathrm{0}}}} $,
    \item $\ottnt{e_{{\mathrm{1}}}} \,  =  \,     \ottnt{e''_{{\mathrm{0}}}}    [ \tceseq{  \ottnt{A''} _{  \mu  }   /  \mathit{X} } ]      [ \tceseq{  \ottnt{A''} _{  \nu  }   /  \mathit{Y} } ]      [  \ottnt{v''_{{\mathrm{1}}}}  /  \mathit{f}  ]      [ \tceseq{ \ottnt{v''_{{\mathrm{2}}}}  /  \mathit{x} } ]  $.
  \end{itemize}
   We also have $ \varpi    =     \epsilon  $.

   Because
   $\ottnt{e} \,  \prec_{ \ottnt{v} }  \, \ottnt{e'} \,  \uparrow \, (  \ottsym{0}  ,  \pi_{{\mathrm{1}}}  )  \,  =  \, \ottnt{v''_{{\mathrm{1}}}} \,  \tceseq{ \ottnt{v''_{{\mathrm{2}}}} } $ and
   $\ottnt{v''_{{\mathrm{1}}}}$ is a recursive function,
   there exist some $\ottnt{v_{{\mathrm{1}}}}$, $ \tceseq{ \ottnt{v_{{\mathrm{2}}}} } $, $\ottnt{v'_{{\mathrm{1}}}}$, $ \tceseq{ \ottnt{v'_{{\mathrm{2}}}} } $, $ \tceseq{  \ottnt{A} _{  \mu  }  } $, $ \tceseq{  \ottnt{A} _{  \nu  }  } $, $\ottnt{e_{{\mathrm{01}}}}$, $ \tceseq{  \ottnt{A'} _{  \mu  }  } $, $ \tceseq{  \ottnt{A'} _{  \nu  }  } $, and $\ottnt{e'_{{\mathrm{01}}}}$ such that
   \begin{itemize}
    \item $\ottnt{e} \,  =  \, \ottnt{v_{{\mathrm{1}}}} \,  \tceseq{ \ottnt{v_{{\mathrm{2}}}} } $,
    \item $\ottnt{e'} \,  =  \, \ottnt{v'_{{\mathrm{1}}}} \,  \tceseq{ \ottnt{v'_{{\mathrm{2}}}} } $,
    \item $\ottnt{v_{{\mathrm{1}}}} \,  =  \,  \mathsf{rec}  \, (  \tceseq{  \mathit{X} ^{  \ottnt{A} _{  \mu  }  },  \mathit{Y} ^{  \ottnt{A} _{  \nu  }  }  }  ,  \mathit{f} ,   \tceseq{ \mathit{x} }  ) . \,  \ottnt{e_{{\mathrm{01}}}} $,
    \item $\ottnt{v'_{{\mathrm{1}}}} \,  =  \,  \mathsf{rec}  \, (  \tceseq{  \mathit{X} ^{  \ottnt{A'} _{  \mu  }  },  \mathit{Y} ^{  \ottnt{A'} _{  \nu  }  }  }  ,  \mathit{f} ,   \tceseq{ \mathit{x} }  ) . \,  \ottnt{e'_{{\mathrm{01}}}} $,
    \item $\ottnt{v_{{\mathrm{1}}}} \,  \prec_{ \ottnt{v} }  \, \ottnt{v'_{{\mathrm{1}}}} \,  \uparrow \, (  \ottsym{0}  ,  \pi_{{\mathrm{1}}}  )  \,  =  \, \ottnt{v''_{{\mathrm{1}}}}$,
    \item $ \tceseq{  \ottnt{A} _{  \mu  }  \,  \prec_{ \ottnt{v} }  \,  \ottnt{A'} _{  \mu  }  \,  \uparrow \, (  \ottsym{0}  ,  \pi_{{\mathrm{1}}}  )  \,  =  \,  \ottnt{A''} _{  \mu  }  } $,
    \item $ \tceseq{  \ottnt{A} _{  \nu  }  \,  \prec_{ \ottnt{v} }  \,  \ottnt{A'} _{  \nu  }  \,  \uparrow \, (  \ottsym{0}  ,  \pi_{{\mathrm{1}}}  )  \,  =  \,  \ottnt{A''} _{  \nu  }  } $,
    \item $\ottnt{e_{{\mathrm{01}}}} \,  \prec_{ \ottnt{v} }  \, \ottnt{e'_{{\mathrm{01}}}} \,  \uparrow \, (  \ottsym{0}  ,  \pi_{{\mathrm{1}}}  )  \,  =  \, \ottnt{e''_{{\mathrm{0}}}}$, and
    \item $ \tceseq{ \ottnt{v_{{\mathrm{2}}}} \,  \prec_{ \ottnt{v} }  \, \ottnt{v'_{{\mathrm{2}}}} \,  \uparrow \, (  \ottsym{0}  ,  \pi_{{\mathrm{1}}}  )  \,  =  \, \ottnt{v''_{{\mathrm{2}}}} } $.
   \end{itemize}

   Let
   $\ottnt{e_{{\mathrm{0}}}} \,  =  \,     \ottnt{e_{{\mathrm{01}}}}    [ \tceseq{  \ottnt{A} _{  \mu  }   /  \mathit{X} } ]      [ \tceseq{  \ottnt{A} _{  \nu  }   /  \mathit{Y} } ]      [  \ottnt{v_{{\mathrm{1}}}}  /  \mathit{f}  ]      [ \tceseq{ \ottnt{v_{{\mathrm{2}}}}  /  \mathit{x} } ]  $ and
   $\ottnt{e'_{{\mathrm{0}}}} \,  =  \,     \ottnt{e'_{{\mathrm{01}}}}    [ \tceseq{  \ottnt{A'} _{  \mu  }   /  \mathit{X} } ]      [ \tceseq{  \ottnt{A'} _{  \nu  }   /  \mathit{Y} } ]      [  \ottnt{v'_{{\mathrm{1}}}}  /  \mathit{f}  ]      [ \tceseq{ \ottnt{v'_{{\mathrm{2}}}}  /  \mathit{x} } ]  $.
   We have the conclusion by the following.
   \begin{itemize}
    \item We have $\ottnt{e} = \ottnt{v_{{\mathrm{1}}}} \,  \tceseq{ \ottnt{v_{{\mathrm{2}}}} }  \,  \mathrel{  \longrightarrow  }  \, \ottnt{e_{{\mathrm{0}}}} \,  \,\&\,  \,  \epsilon $ by \E{Red}/\R{RBeta}.
    \item Let $\pi_{{\mathrm{2}}}$ be any infinite trace.
          We show that $\ottsym{(}  \ottnt{e} \,  \prec_{ \ottnt{v} }  \, \ottnt{e'} \,  \uparrow \, (  \ottsym{0}  ,  \pi_{{\mathrm{2}}}  )   \ottsym{)} \,  \mathrel{  \longrightarrow  }  \, \ottsym{(}  \ottnt{e_{{\mathrm{0}}}} \,  \prec_{ \ottnt{v} }  \, \ottnt{e'_{{\mathrm{0}}}} \,  \uparrow \, (  \ottsym{0}  ,  \pi_{{\mathrm{2}}}  )   \ottsym{)} \,  \,\&\,  \,  \epsilon $.
          Let
          \begin{itemize}
           \item $\ottnt{v'''_{{\mathrm{1}}}} \,  =  \, \ottnt{v_{{\mathrm{1}}}} \,  \prec_{ \ottnt{v} }  \, \ottnt{v'_{{\mathrm{1}}}} \,  \uparrow \, (  \ottsym{0}  ,  \pi_{{\mathrm{2}}}  ) $,
           \item $\ottnt{e'''_{{\mathrm{01}}}} \,  =  \, \ottnt{e_{{\mathrm{01}}}} \,  \prec_{ \ottnt{v} }  \, \ottnt{e'_{{\mathrm{01}}}} \,  \uparrow \, (  \ottsym{0}  ,  \pi_{{\mathrm{2}}}  ) $,
           \item $ \tceseq{  \ottnt{A'''} _{  \mu  }  \,  =  \,  \ottnt{A} _{  \mu  }  \,  \prec_{ \ottnt{v} }  \,  \ottnt{A'} _{  \mu  }  \,  \uparrow \, (  \ottsym{0}  ,  \pi_{{\mathrm{2}}}  )  } $, and
           \item $ \tceseq{  \ottnt{A'''} _{  \nu  }  \,  =  \,  \ottnt{A} _{  \nu  }  \,  \prec_{ \ottnt{v} }  \,  \ottnt{A'} _{  \nu  }  \,  \uparrow \, (  \ottsym{0}  ,  \pi_{{\mathrm{2}}}  )  } $,
          \end{itemize}
          which are well defined
          by Lemmas~\ref{lem:lr-approx-wf} and \ref{lem:lr-approx-val}.
          Because $ \tceseq{ \ottnt{v_{{\mathrm{2}}}} \,  \prec_{ \ottnt{v} }  \, \ottnt{v'_{{\mathrm{2}}}} \,  \uparrow \, (  \ottsym{0}  ,  \pi_{{\mathrm{2}}}  )  } $ are values
          by Lemmas~\ref{lem:lr-approx-wf} and \ref{lem:lr-approx-val},
          we have
          \[\begin{array}{ll} &
           \ottnt{e} \,  \prec_{ \ottnt{v} }  \, \ottnt{e'} \,  \uparrow \, (  \ottsym{0}  ,  \pi_{{\mathrm{2}}}  )  \\ = &
           \ottnt{v'''_{{\mathrm{1}}}} \,  \tceseq{ \ottnt{v_{{\mathrm{2}}}} \,  \prec_{ \ottnt{v} }  \, \ottnt{v'_{{\mathrm{2}}}} \,  \uparrow \, (  \ottsym{0}  ,  \pi_{{\mathrm{2}}}  )  }  \\  \longrightarrow  &
               \ottnt{e'''_{{\mathrm{01}}}}    [ \tceseq{  \ottnt{A'''} _{  \mu  }   /  \mathit{X} } ]      [ \tceseq{  \ottnt{A'''} _{  \nu  }   /  \mathit{Y} } ]      [  \ottnt{v'''_{{\mathrm{1}}}}  /  \mathit{f}  ]      [ \tceseq{ \ottnt{v_{{\mathrm{2}}}} \,  \prec_{ \ottnt{v} }  \, \ottnt{v'_{{\mathrm{2}}}} \,  \uparrow \, (  \ottsym{0}  ,  \pi_{{\mathrm{2}}}  )   /  \mathit{x} } ]   \,  \,\&\,  \,  \epsilon 
            \quad \text{(by \E{Red}/\R{RBeta})} \\  =  &
           \ottnt{e_{{\mathrm{0}}}} \,  \prec_{ \ottnt{v} }  \, \ottnt{e'_{{\mathrm{0}}}} \,  \uparrow \, (  \ottsym{0}  ,  \pi_{{\mathrm{2}}}  )  \,  \,\&\,  \,  \epsilon 
            \quad \text{(by Lemmas~\ref{lem:lr-approx-val-subst} and \ref{lem:lr-approx-pred-subst})} ~.
            \end{array}
          \]
   \end{itemize}

  \case \E{Red}/\R{PBeta}:
   We are given
   \begin{prooftree}
    \AxiomC{$\ottsym{(}   \Lambda   \ottsym{(}  \mathit{X} \,  {:}  \,  \tceseq{ \ottnt{s} }   \ottsym{)}  \ottsym{.}  \ottnt{e''_{{\mathrm{0}}}}  \ottsym{)} \, \ottnt{A''} \,  \rightsquigarrow  \,  \ottnt{e''_{{\mathrm{0}}}}    [  \ottnt{A''}  /  \mathit{X}  ]  $}
    \UnaryInfC{$\ottsym{(}   \Lambda   \ottsym{(}  \mathit{X} \,  {:}  \,  \tceseq{ \ottnt{s} }   \ottsym{)}  \ottsym{.}  \ottnt{e''_{{\mathrm{0}}}}  \ottsym{)} \, \ottnt{A''} \,  \mathrel{  \longrightarrow  }  \,  \ottnt{e''_{{\mathrm{0}}}}    [  \ottnt{A''}  /  \mathit{X}  ]   \,  \,\&\,  \,  \epsilon $}
   \end{prooftree}
   for some $\mathit{X}$, $ \tceseq{ \ottnt{s} } $, $\ottnt{e''_{{\mathrm{0}}}}$, and $\ottnt{A''}$ such that
   \begin{itemize}
    \item $\ottnt{e} \,  \prec_{ \ottnt{v} }  \, \ottnt{e'} \,  \uparrow \, (  \ottsym{0}  ,  \pi_{{\mathrm{1}}}  )  \,  =  \, \ottsym{(}   \Lambda   \ottsym{(}  \mathit{X} \,  {:}  \,  \tceseq{ \ottnt{s} }   \ottsym{)}  \ottsym{.}  \ottnt{e''_{{\mathrm{0}}}}  \ottsym{)} \, \ottnt{A''}$ and
    \item $\ottnt{e_{{\mathrm{1}}}} \,  =  \,  \ottnt{e''_{{\mathrm{0}}}}    [  \ottnt{A''}  /  \mathit{X}  ]  $.
   \end{itemize}
   We also have $ \varpi    =     \epsilon  $.

   Because $\ottnt{e} \,  \prec_{ \ottnt{v} }  \, \ottnt{e'} \,  \uparrow \, (  \ottsym{0}  ,  \pi_{{\mathrm{1}}}  )  \,  =  \, \ottsym{(}   \Lambda   \ottsym{(}  \mathit{X} \,  {:}  \,  \tceseq{ \ottnt{s} }   \ottsym{)}  \ottsym{.}  \ottnt{e''_{{\mathrm{0}}}}  \ottsym{)} \, \ottnt{A''}$,
   there exist some $\ottnt{e_{{\mathrm{01}}}}$, $\ottnt{A}$, $\ottnt{e'_{{\mathrm{01}}}}$, and $\ottnt{A'}$ such that
   \begin{itemize}
    \item $\ottnt{e} \,  =  \, \ottsym{(}   \Lambda   \ottsym{(}  \mathit{X} \,  {:}  \,  \tceseq{ \ottnt{s} }   \ottsym{)}  \ottsym{.}  \ottnt{e_{{\mathrm{01}}}}  \ottsym{)} \, \ottnt{A}$,
    \item $\ottnt{e'} \,  =  \, \ottsym{(}   \Lambda   \ottsym{(}  \mathit{X} \,  {:}  \,  \tceseq{ \ottnt{s} }   \ottsym{)}  \ottsym{.}  \ottnt{e'_{{\mathrm{01}}}}  \ottsym{)} \, \ottnt{A'}$,
    \item $\ottnt{e_{{\mathrm{01}}}} \,  \prec_{ \ottnt{v} }  \, \ottnt{e'_{{\mathrm{01}}}} \,  \uparrow \, (  \ottsym{0}  ,  \pi_{{\mathrm{1}}}  )  \,  =  \, \ottnt{e''_{{\mathrm{0}}}}$, and
    \item $\ottnt{A} \,  \prec_{ \ottnt{v} }  \, \ottnt{A'} \,  \uparrow \, (  \ottsym{0}  ,  \pi_{{\mathrm{1}}}  )  \,  =  \, \ottnt{A''}$.
   \end{itemize}

   Let $\ottnt{e_{{\mathrm{0}}}} \,  =  \,  \ottnt{e_{{\mathrm{01}}}}    [  \ottnt{A}  /  \mathit{X}  ]  $ and $\ottnt{e'_{{\mathrm{0}}}} \,  =  \,  \ottnt{e'_{{\mathrm{01}}}}    [  \ottnt{A'}  /  \mathit{X}  ]  $.
   We have the conclusion by the following.
   \begin{itemize}
    \item We have $\ottnt{e} \,  \mathrel{  \longrightarrow  }  \, \ottnt{e_{{\mathrm{0}}}} \,  \,\&\,  \,  \epsilon $ by \E{Red}/\R{PBeta}.
    \item Let $\pi_{{\mathrm{2}}}$ be any infinite trace.
          Then,
          \[\begin{array}{ll} &
           \ottnt{e} \,  \prec_{ \ottnt{v} }  \, \ottnt{e'} \,  \uparrow \, (  \ottsym{0}  ,  \pi_{{\mathrm{2}}}  )  \\ = &
           \ottsym{(}   \Lambda   \ottsym{(}  \mathit{X} \,  {:}  \,  \tceseq{ \ottnt{s} }   \ottsym{)}  \ottsym{.}  \ottsym{(}  \ottnt{e_{{\mathrm{01}}}} \,  \prec_{ \ottnt{v} }  \, \ottnt{e'_{{\mathrm{01}}}} \,  \uparrow \, (  \ottsym{0}  ,  \pi_{{\mathrm{2}}}  )   \ottsym{)}  \ottsym{)} \, \ottsym{(}  \ottnt{A} \,  \prec_{ \ottnt{v} }  \, \ottnt{A'} \,  \uparrow \, (  \ottsym{0}  ,  \pi_{{\mathrm{2}}}  )   \ottsym{)} \\  \longrightarrow  &
            \ottsym{(}  \ottnt{e_{{\mathrm{01}}}} \,  \prec_{ \ottnt{v} }  \, \ottnt{e'_{{\mathrm{01}}}} \,  \uparrow \, (  \ottsym{0}  ,  \pi_{{\mathrm{2}}}  )   \ottsym{)}    [  \ottnt{A} \,  \prec_{ \ottnt{v} }  \, \ottnt{A'} \,  \uparrow \, (  \ottsym{0}  ,  \pi_{{\mathrm{2}}}  )   /  \mathit{X}  ]   \,  \,\&\,  \,  \epsilon 
            \quad \text{(by \E{Red}/\R{PBeta})}
            \\ = &
           \ottnt{e_{{\mathrm{0}}}} \,  \prec_{ \ottnt{v} }  \, \ottnt{e'_{{\mathrm{0}}}} \,  \uparrow \, (  \ottsym{0}  ,  \pi_{{\mathrm{2}}}  ) 
            \quad \text{(by \reflem{lr-approx-pred-subst})} ~.
            \end{array}
          \]
   \end{itemize}

  \case \E{Red}/\R{IfT}:
   We are given
   \begin{prooftree}
    \AxiomC{$\mathsf{if} \,  \mathsf{true}  \, \mathsf{then} \, \ottnt{e_{{\mathrm{1}}}} \, \mathsf{else} \, \ottnt{e''_{{\mathrm{02}}}} \,  \rightsquigarrow  \, \ottnt{e_{{\mathrm{1}}}}$}
    \UnaryInfC{$\mathsf{if} \,  \mathsf{true}  \, \mathsf{then} \, \ottnt{e_{{\mathrm{1}}}} \, \mathsf{else} \, \ottnt{e''_{{\mathrm{02}}}} \,  \mathrel{  \longrightarrow  }  \, \ottnt{e_{{\mathrm{1}}}} \,  \,\&\,  \,  \epsilon $}
   \end{prooftree}
   for some $\ottnt{e''_{{\mathrm{02}}}}$ such that
   $\ottnt{e} \,  \prec_{ \ottnt{v} }  \, \ottnt{e'} \,  \uparrow \, (  \ottsym{0}  ,  \pi_{{\mathrm{1}}}  )  \,  =  \, \mathsf{if} \,  \mathsf{true}  \, \mathsf{then} \, \ottnt{e_{{\mathrm{1}}}} \, \mathsf{else} \, \ottnt{e''_{{\mathrm{02}}}}$.
   We also have $ \varpi    =     \epsilon  $.

   Because $\ottnt{e} \,  \prec_{ \ottnt{v} }  \, \ottnt{e'} \,  \uparrow \, (  \ottsym{0}  ,  \pi_{{\mathrm{1}}}  )  \,  =  \, \mathsf{if} \,  \mathsf{true}  \, \mathsf{then} \, \ottnt{e_{{\mathrm{1}}}} \, \mathsf{else} \, \ottnt{e''_{{\mathrm{02}}}}$,
   there exist some $\ottnt{e_{{\mathrm{0}}}}$, $\ottnt{e_{{\mathrm{02}}}}$, $\ottnt{e'_{{\mathrm{0}}}}$, and $\ottnt{e'_{{\mathrm{02}}}}$ such that
   \begin{itemize}
    \item $\ottnt{e} \,  =  \, \mathsf{if} \,  \mathsf{true}  \, \mathsf{then} \, \ottnt{e_{{\mathrm{0}}}} \, \mathsf{else} \, \ottnt{e_{{\mathrm{02}}}}$,
    \item $\ottnt{e'} \,  =  \, \mathsf{if} \,  \mathsf{true}  \, \mathsf{then} \, \ottnt{e'_{{\mathrm{0}}}} \, \mathsf{else} \, \ottnt{e'_{{\mathrm{02}}}}$, and
    \item $\ottnt{e_{{\mathrm{0}}}} \,  \prec_{ \ottnt{v} }  \, \ottnt{e'_{{\mathrm{0}}}} \,  \uparrow \, (  \ottsym{0}  ,  \pi_{{\mathrm{1}}}  )  \,  =  \, \ottnt{e_{{\mathrm{1}}}}$.
   \end{itemize}

   We have the conclusion by the following.
   \begin{itemize}
    \item We have $\ottnt{e} \,  \mathrel{  \longrightarrow  }  \, \ottnt{e_{{\mathrm{0}}}} \,  \,\&\,  \,  \epsilon $ by \E{Red}/\R{IfT}.
    \item Let $\pi_{{\mathrm{2}}}$ be any infinite trace.
          Then,
          \[\begin{array}{ll} &
           \ottnt{e} \,  \prec_{ \ottnt{v} }  \, \ottnt{e'} \,  \uparrow \, (  \ottsym{0}  ,  \pi_{{\mathrm{2}}}  )  \\ = &
           \mathsf{if} \,  \mathsf{true}  \, \mathsf{then} \, \ottsym{(}  \ottnt{e_{{\mathrm{0}}}} \,  \prec_{ \ottnt{v} }  \, \ottnt{e'_{{\mathrm{0}}}} \,  \uparrow \, (  \ottsym{0}  ,  \pi_{{\mathrm{2}}}  )   \ottsym{)} \, \mathsf{else} \, \ottsym{(}  \ottnt{e_{{\mathrm{02}}}} \,  \prec_{ \ottnt{v} }  \, \ottnt{e'_{{\mathrm{02}}}} \,  \uparrow \, (  \ottsym{0}  ,  \pi_{{\mathrm{2}}}  )   \ottsym{)} \\  \longrightarrow  &
           \ottnt{e_{{\mathrm{0}}}} \,  \prec_{ \ottnt{v} }  \, \ottnt{e'_{{\mathrm{0}}}} \,  \uparrow \, (  \ottsym{0}  ,  \pi_{{\mathrm{2}}}  )  \,  \,\&\,  \,  \epsilon 
            \quad \text{(by \E{Red}/\R{IfT})} ~.
            \end{array}
          \]
   \end{itemize}

  \case \E{Red}/\R{IfF}:
   We are given
   \begin{prooftree}
    \AxiomC{$\mathsf{if} \,  \mathsf{false}  \, \mathsf{then} \, \ottnt{e''_{{\mathrm{01}}}} \, \mathsf{else} \, \ottnt{e_{{\mathrm{1}}}} \,  \rightsquigarrow  \, \ottnt{e_{{\mathrm{1}}}}$}
    \UnaryInfC{$\mathsf{if} \,  \mathsf{false}  \, \mathsf{then} \, \ottnt{e''_{{\mathrm{01}}}} \, \mathsf{else} \, \ottnt{e_{{\mathrm{1}}}} \,  \mathrel{  \longrightarrow  }  \, \ottnt{e_{{\mathrm{1}}}} \,  \,\&\,  \,  \epsilon $}
   \end{prooftree}
   for some $\ottnt{e''_{{\mathrm{01}}}}$ such that
   $\ottnt{e} \,  \prec_{ \ottnt{v} }  \, \ottnt{e'} \,  \uparrow \, (  \ottsym{0}  ,  \pi_{{\mathrm{1}}}  )  \,  =  \, \mathsf{if} \,  \mathsf{false}  \, \mathsf{then} \, \ottnt{e''_{{\mathrm{01}}}} \, \mathsf{else} \, \ottnt{e_{{\mathrm{1}}}}$.
   We also have $ \varpi    =     \epsilon  $.

   Because $\ottnt{e} \,  \prec_{ \ottnt{v} }  \, \ottnt{e'} \,  \uparrow \, (  \ottsym{0}  ,  \pi_{{\mathrm{1}}}  )  \,  =  \, \mathsf{if} \,  \mathsf{false}  \, \mathsf{then} \, \ottnt{e''_{{\mathrm{01}}}} \, \mathsf{else} \, \ottnt{e_{{\mathrm{1}}}}$,
   there exist some $\ottnt{e_{{\mathrm{01}}}}$, $\ottnt{e_{{\mathrm{0}}}}$, $\ottnt{e'_{{\mathrm{01}}}}$, and $\ottnt{e'_{{\mathrm{0}}}}$ such that
   \begin{itemize}
    \item $\ottnt{e} \,  =  \, \mathsf{if} \,  \mathsf{false}  \, \mathsf{then} \, \ottnt{e_{{\mathrm{01}}}} \, \mathsf{else} \, \ottnt{e_{{\mathrm{0}}}}$,
    \item $\ottnt{e'} \,  =  \, \mathsf{if} \,  \mathsf{false}  \, \mathsf{then} \, \ottnt{e'_{{\mathrm{01}}}} \, \mathsf{else} \, \ottnt{e'_{{\mathrm{0}}}}$, and
    \item $\ottnt{e_{{\mathrm{0}}}} \,  \prec_{ \ottnt{v} }  \, \ottnt{e'_{{\mathrm{0}}}} \,  \uparrow \, (  \ottsym{0}  ,  \pi_{{\mathrm{1}}}  )  \,  =  \, \ottnt{e_{{\mathrm{1}}}}$.
   \end{itemize}

   We have the conclusion by the following.
   \begin{itemize}
    \item We have $\ottnt{e} \,  \mathrel{  \longrightarrow  }  \, \ottnt{e_{{\mathrm{0}}}} \,  \,\&\,  \,  \epsilon $ by \E{Red}/\R{IfF}.
    \item Let $\pi_{{\mathrm{2}}}$ be any infinite trace.
          Then,
          \[\begin{array}{ll} &
           \ottnt{e} \,  \prec_{ \ottnt{v} }  \, \ottnt{e'} \,  \uparrow \, (  \ottsym{0}  ,  \pi_{{\mathrm{2}}}  )  \\ = &
           \mathsf{if} \,  \mathsf{false}  \, \mathsf{then} \, \ottsym{(}  \ottnt{e_{{\mathrm{01}}}} \,  \prec_{ \ottnt{v} }  \, \ottnt{e'_{{\mathrm{01}}}} \,  \uparrow \, (  \ottsym{0}  ,  \pi_{{\mathrm{2}}}  )   \ottsym{)} \, \mathsf{else} \, \ottsym{(}  \ottnt{e_{{\mathrm{0}}}} \,  \prec_{ \ottnt{v} }  \, \ottnt{e'_{{\mathrm{0}}}} \,  \uparrow \, (  \ottsym{0}  ,  \pi_{{\mathrm{2}}}  )   \ottsym{)} \\  \longrightarrow  &
           \ottnt{e_{{\mathrm{0}}}} \,  \prec_{ \ottnt{v} }  \, \ottnt{e'_{{\mathrm{0}}}} \,  \uparrow \, (  \ottsym{0}  ,  \pi_{{\mathrm{2}}}  )  \,  \,\&\,  \,  \epsilon 
            \quad \text{(by \E{Red}/\R{IfF})} ~.
            \end{array}
          \]
   \end{itemize}

  \case \E{Red}/\R{Let}:
   We are given
   \begin{prooftree}
    \AxiomC{$\mathsf{let} \, \mathit{x_{{\mathrm{01}}}}  \ottsym{=}  \ottnt{v''_{{\mathrm{01}}}} \,  \mathsf{in}  \, \ottnt{e''_{{\mathrm{02}}}} \,  \rightsquigarrow  \,  \ottnt{e''_{{\mathrm{02}}}}    [  \ottnt{v''_{{\mathrm{01}}}}  /  \mathit{x_{{\mathrm{01}}}}  ]  $}
    \UnaryInfC{$\mathsf{let} \, \mathit{x_{{\mathrm{01}}}}  \ottsym{=}  \ottnt{v''_{{\mathrm{01}}}} \,  \mathsf{in}  \, \ottnt{e''_{{\mathrm{02}}}} \,  \mathrel{  \longrightarrow  }  \,  \ottnt{e''_{{\mathrm{02}}}}    [  \ottnt{v''_{{\mathrm{01}}}}  /  \mathit{x_{{\mathrm{01}}}}  ]   \,  \,\&\,  \,  \epsilon $}
   \end{prooftree}
   for some $\mathit{x_{{\mathrm{01}}}}$, $\ottnt{v''_{{\mathrm{01}}}}$, and $\ottnt{e''_{{\mathrm{02}}}}$ such that
   \begin{itemize}
    \item $\ottnt{e} \,  \prec_{ \ottnt{v} }  \, \ottnt{e'} \,  \uparrow \, (  \ottsym{0}  ,  \pi_{{\mathrm{1}}}  )  \,  =  \, \ottsym{(}  \mathsf{let} \, \mathit{x_{{\mathrm{01}}}}  \ottsym{=}  \ottnt{v''_{{\mathrm{01}}}} \,  \mathsf{in}  \, \ottnt{e''_{{\mathrm{02}}}}  \ottsym{)}$ and
    \item $\ottnt{e_{{\mathrm{1}}}} \,  =  \,  \ottnt{e''_{{\mathrm{02}}}}    [  \ottnt{v''_{{\mathrm{01}}}}  /  \mathit{x_{{\mathrm{01}}}}  ]  $.
   \end{itemize}
   We also have $ \varpi    =     \epsilon  $.

   Because $\ottnt{e} \,  \prec_{ \ottnt{v} }  \, \ottnt{e'} \,  \uparrow \, (  \ottsym{0}  ,  \pi_{{\mathrm{1}}}  )  \,  =  \, \ottsym{(}  \mathsf{let} \, \mathit{x_{{\mathrm{01}}}}  \ottsym{=}  \ottnt{v''_{{\mathrm{01}}}} \,  \mathsf{in}  \, \ottnt{e''_{{\mathrm{02}}}}  \ottsym{)}$,
   there exist $\ottnt{v_{{\mathrm{01}}}}$, $\ottnt{e_{{\mathrm{02}}}}$, $\ottnt{v'_{{\mathrm{01}}}}$, and $\ottnt{e'_{{\mathrm{02}}}}$ such that
   \begin{itemize}
    \item $\ottnt{e} \,  =  \, \ottsym{(}  \mathsf{let} \, \mathit{x_{{\mathrm{01}}}}  \ottsym{=}  \ottnt{v_{{\mathrm{01}}}} \,  \mathsf{in}  \, \ottnt{e_{{\mathrm{02}}}}  \ottsym{)}$,
    \item $\ottnt{e'} \,  =  \, \ottsym{(}  \mathsf{let} \, \mathit{x_{{\mathrm{01}}}}  \ottsym{=}  \ottnt{v'_{{\mathrm{01}}}} \,  \mathsf{in}  \, \ottnt{e'_{{\mathrm{02}}}}  \ottsym{)}$,
    \item $\ottnt{v_{{\mathrm{01}}}} \,  \prec_{ \ottnt{v} }  \, \ottnt{v'_{{\mathrm{01}}}} \,  \uparrow \, (  \ottsym{0}  ,  \pi_{{\mathrm{1}}}  )  \,  =  \, \ottnt{v''_{{\mathrm{01}}}}$, and
    \item $\ottnt{e_{{\mathrm{02}}}} \,  \prec_{ \ottnt{v} }  \, \ottnt{e'_{{\mathrm{02}}}} \,  \uparrow \, (  \ottsym{0}  ,  \pi_{{\mathrm{1}}}  )  \,  =  \, \ottnt{e''_{{\mathrm{02}}}}$
   \end{itemize}
   with \reflem{lr-approx-anti-val}.

   Let $\ottnt{e_{{\mathrm{0}}}} \,  =  \,  \ottnt{e_{{\mathrm{02}}}}    [  \ottnt{v_{{\mathrm{01}}}}  /  \mathit{x_{{\mathrm{01}}}}  ]  $ and $\ottnt{e'_{{\mathrm{0}}}} \,  =  \,  \ottnt{e'_{{\mathrm{02}}}}    [  \ottnt{v'_{{\mathrm{01}}}}  /  \mathit{x_{{\mathrm{01}}}}  ]  $.
   We have the conclusion by the following.
   \begin{itemize}
    \item We have $\ottnt{e} \,  \mathrel{  \longrightarrow  }  \, \ottnt{e_{{\mathrm{0}}}} \,  \,\&\,  \,  \epsilon $
          by \E{Red}/\R{Let}.
    \item Let $\pi_{{\mathrm{2}}}$ be any infinite trace.
          Because $\ottnt{v_{{\mathrm{01}}}} \,  \prec_{ \ottnt{v} }  \, \ottnt{v'_{{\mathrm{01}}}} \,  \uparrow \, (  \ottsym{0}  ,  \pi_{{\mathrm{2}}}  ) $ is a value
          by Lemmas~\ref{lem:lr-approx-wf} and \ref{lem:lr-approx-val},
          we have
          \[\begin{array}{ll} &
           \ottsym{(}  \ottnt{e} \,  \prec_{ \ottnt{v} }  \, \ottnt{e'} \,  \uparrow \, (  \ottsym{0}  ,  \pi_{{\mathrm{2}}}  )   \ottsym{)} \\ = &
           \mathsf{let} \, \mathit{x_{{\mathrm{01}}}}  \ottsym{=}  \ottsym{(}  \ottnt{v_{{\mathrm{01}}}} \,  \prec_{ \ottnt{v} }  \, \ottnt{v'_{{\mathrm{01}}}} \,  \uparrow \, (  \ottsym{0}  ,  \pi_{{\mathrm{2}}}  )   \ottsym{)} \,  \mathsf{in}  \, \ottsym{(}  \ottnt{e_{{\mathrm{02}}}} \,  \prec_{ \ottnt{v} }  \, \ottnt{e'_{{\mathrm{02}}}} \,  \uparrow \, (  \ottsym{0}  ,  \pi_{{\mathrm{2}}}  )   \ottsym{)} \\  \longrightarrow  &
            \ottsym{(}  \ottnt{e_{{\mathrm{02}}}} \,  \prec_{ \ottnt{v} }  \, \ottnt{e'_{{\mathrm{02}}}} \,  \uparrow \, (  \ottsym{0}  ,  \pi_{{\mathrm{2}}}  )   \ottsym{)}    [  \ottnt{v_{{\mathrm{01}}}} \,  \prec_{ \ottnt{v} }  \, \ottnt{v'_{{\mathrm{01}}}} \,  \uparrow \, (  \ottsym{0}  ,  \pi_{{\mathrm{2}}}  )   /  \mathit{x_{{\mathrm{01}}}}  ]   \,  \,\&\,  \,  \epsilon 
            \quad \text{(by \E{Red}/\R{Let})}
            \\ = &
           \ottnt{e_{{\mathrm{0}}}} \,  \prec_{ \ottnt{v} }  \, \ottnt{e'_{{\mathrm{0}}}} \,  \uparrow \, (  \ottsym{0}  ,  \pi_{{\mathrm{2}}}  )  \,  \,\&\,  \,  \epsilon 
            \quad \text{(by \reflem{lr-approx-val-subst})} ~.
            \end{array}
          \]
   \end{itemize}

  \case \E{Red}/\R{Shift}:
   We are given
   \begin{prooftree}
    \AxiomC{$ \resetcnstr{  \ottnt{K''_{{\mathrm{0}}}}  [  \mathcal{S}_0 \, \mathit{x_{{\mathrm{0}}}}  \ottsym{.}  \ottnt{e''_{{\mathrm{01}}}}  ]  }  \,  \rightsquigarrow  \,  \ottnt{e''_{{\mathrm{01}}}}    [   \lambda\!  \, \mathit{x_{{\mathrm{0}}}}  \ottsym{.}   \resetcnstr{  \ottnt{K''_{{\mathrm{0}}}}  [  \mathit{x_{{\mathrm{0}}}}  ]  }   /  \mathit{x_{{\mathrm{0}}}}  ]  $}
    \UnaryInfC{$ \resetcnstr{  \ottnt{K''_{{\mathrm{0}}}}  [  \mathcal{S}_0 \, \mathit{x_{{\mathrm{0}}}}  \ottsym{.}  \ottnt{e''_{{\mathrm{01}}}}  ]  }  \,  \mathrel{  \longrightarrow  }  \,  \ottnt{e''_{{\mathrm{01}}}}    [   \lambda\!  \, \mathit{x_{{\mathrm{0}}}}  \ottsym{.}   \resetcnstr{  \ottnt{K''_{{\mathrm{0}}}}  [  \mathit{x_{{\mathrm{0}}}}  ]  }   /  \mathit{x_{{\mathrm{0}}}}  ]   \,  \,\&\,  \,  \epsilon $}
   \end{prooftree}
   for some $\ottnt{K''_{{\mathrm{0}}}}$, $\mathit{x_{{\mathrm{0}}}}$, and $\ottnt{e''_{{\mathrm{01}}}}$ such that
   \begin{itemize}
    \item $\ottnt{e} \,  \prec_{ \ottnt{v} }  \, \ottnt{e'} \,  \uparrow \, (  \ottsym{0}  ,  \pi_{{\mathrm{1}}}  )  \,  =  \,  \resetcnstr{  \ottnt{K''_{{\mathrm{0}}}}  [  \mathcal{S}_0 \, \mathit{x_{{\mathrm{0}}}}  \ottsym{.}  \ottnt{e''_{{\mathrm{01}}}}  ]  } $ and
    \item $\ottnt{e_{{\mathrm{1}}}} \,  =  \,  \ottnt{e''_{{\mathrm{01}}}}    [   \lambda\!  \, \mathit{x_{{\mathrm{0}}}}  \ottsym{.}   \resetcnstr{  \ottnt{K''_{{\mathrm{0}}}}  [  \mathit{x_{{\mathrm{0}}}}  ]  }   /  \mathit{x_{{\mathrm{0}}}}  ]  $.
   \end{itemize}

   Because $\ottnt{e} \,  \prec_{ \ottnt{v} }  \, \ottnt{e'} \,  \uparrow \, (  \ottsym{0}  ,  \pi_{{\mathrm{1}}}  )  \,  =  \,  \resetcnstr{  \ottnt{K''_{{\mathrm{0}}}}  [  \mathcal{S}_0 \, \mathit{x_{{\mathrm{0}}}}  \ottsym{.}  \ottnt{e''_{{\mathrm{01}}}}  ]  } $,
   there exist $\ottnt{K_{{\mathrm{0}}}}$, $\ottnt{e_{{\mathrm{01}}}}$, $\ottnt{K'_{{\mathrm{0}}}}$, and $\ottnt{e'_{{\mathrm{01}}}}$ such that
   \begin{itemize}
    \item $\ottnt{e} \,  =  \,  \resetcnstr{  \ottnt{K_{{\mathrm{0}}}}  [  \mathcal{S}_0 \, \mathit{x_{{\mathrm{0}}}}  \ottsym{.}  \ottnt{e_{{\mathrm{01}}}}  ]  } $,
    \item $\ottnt{e'} \,  =  \,  \resetcnstr{  \ottnt{K'_{{\mathrm{0}}}}  [  \mathcal{S}_0 \, \mathit{x_{{\mathrm{0}}}}  \ottsym{.}  \ottnt{e'_{{\mathrm{01}}}}  ]  } $,
    \item $\ottnt{K_{{\mathrm{0}}}} \,  \prec_{ \ottnt{v} }  \, \ottnt{K'_{{\mathrm{0}}}} \,  \uparrow \, (  \ottsym{0}  ,  \pi_{{\mathrm{1}}}  )  \,  =  \, \ottnt{K''_{{\mathrm{0}}}}$, and
    \item $\ottnt{e_{{\mathrm{01}}}} \,  \prec_{ \ottnt{v} }  \, \ottnt{e'_{{\mathrm{01}}}} \,  \uparrow \, (  \ottsym{0}  ,  \pi_{{\mathrm{1}}}  )  \,  =  \, \ottnt{e''_{{\mathrm{01}}}}$
   \end{itemize}
   with \reflem{lr-approx-anti-ectx}.

   Let
   $\ottnt{e_{{\mathrm{0}}}} \,  =  \,  \ottnt{e_{{\mathrm{01}}}}    [   \lambda\!  \, \mathit{x_{{\mathrm{0}}}}  \ottsym{.}   \resetcnstr{  \ottnt{K_{{\mathrm{0}}}}  [  \mathit{x_{{\mathrm{0}}}}  ]  }   /  \mathit{x_{{\mathrm{0}}}}  ]  $ and
   $\ottnt{e'_{{\mathrm{0}}}} \,  =  \,  \ottnt{e'_{{\mathrm{01}}}}    [   \lambda\!  \, \mathit{x_{{\mathrm{0}}}}  \ottsym{.}   \resetcnstr{  \ottnt{K'_{{\mathrm{0}}}}  [  \mathit{x_{{\mathrm{0}}}}  ]  }   /  \mathit{x_{{\mathrm{0}}}}  ]  $.
   We have the conclusion by the following.
   \begin{itemize}
    \item We have $\ottnt{e} \,  \mathrel{  \longrightarrow  }  \, \ottnt{e_{{\mathrm{0}}}} \,  \,\&\,  \,  \epsilon $ by \E{Red}/\R{Shift}.

    \item Let $\pi_{{\mathrm{2}}}$ be any infinite trace.
          Noting that $\ottnt{K_{{\mathrm{0}}}} \,  \prec_{ \ottnt{v} }  \, \ottnt{K'_{{\mathrm{0}}}}$ by \reflem{lr-approx-wf},
          we have
          \[\begin{array}{ll} &
           \ottnt{e} \,  \prec_{ \ottnt{v} }  \, \ottnt{e'} \,  \uparrow \, (  \ottsym{0}  ,  \pi_{{\mathrm{2}}}  )  \\ = &
            \resetcnstr{  \ottnt{K_{{\mathrm{0}}}}  [  \mathcal{S}_0 \, \mathit{x_{{\mathrm{0}}}}  \ottsym{.}  \ottnt{e_{{\mathrm{01}}}}  ]  }  \,  \prec_{ \ottnt{v} }  \,  \resetcnstr{  \ottnt{K'_{{\mathrm{0}}}}  [  \mathcal{S}_0 \, \mathit{x_{{\mathrm{0}}}}  \ottsym{.}  \ottnt{e'_{{\mathrm{01}}}}  ]  }  \,  \uparrow \, (  \ottsym{0}  ,  \pi_{{\mathrm{2}}}  )  \\ = &
            \resetcnstr{  \ottsym{(}  \ottnt{K_{{\mathrm{0}}}} \,  \prec_{ \ottnt{v} }  \, \ottnt{K'_{{\mathrm{0}}}} \,  \uparrow \, (  \ottsym{0}  ,  \pi_{{\mathrm{2}}}  )   \ottsym{)}  [  \mathcal{S}_0 \, \mathit{x_{{\mathrm{0}}}}  \ottsym{.}  \ottsym{(}  \ottnt{e_{{\mathrm{01}}}} \,  \prec_{ \ottnt{v} }  \, \ottnt{e'_{{\mathrm{01}}}} \,  \uparrow \, (  \ottsym{0}  ,  \pi_{{\mathrm{2}}}  )   \ottsym{)}  ]  } 
            \quad \text{(by \reflem{lr-approx-embed-exp})}
            \\  \longrightarrow  &
            \ottsym{(}  \ottnt{e_{{\mathrm{01}}}} \,  \prec_{ \ottnt{v} }  \, \ottnt{e'_{{\mathrm{01}}}} \,  \uparrow \, (  \ottsym{0}  ,  \pi_{{\mathrm{2}}}  )   \ottsym{)}    [   \lambda\!  \, \mathit{x_{{\mathrm{0}}}}  \ottsym{.}   \resetcnstr{  \ottsym{(}  \ottnt{K_{{\mathrm{0}}}} \,  \prec_{ \ottnt{v} }  \, \ottnt{K'_{{\mathrm{0}}}} \,  \uparrow \, (  \ottsym{0}  ,  \pi_{{\mathrm{2}}}  )   \ottsym{)}  [  \mathit{x_{{\mathrm{0}}}}  ]  }   /  \mathit{x_{{\mathrm{0}}}}  ]   \,  \,\&\,  \,  \epsilon 
            \quad \text{(by \E{Red}/\R{Shift})}
            \\ = &
            \ottsym{(}  \ottnt{e_{{\mathrm{01}}}} \,  \prec_{ \ottnt{v} }  \, \ottnt{e'_{{\mathrm{01}}}} \,  \uparrow \, (  \ottsym{0}  ,  \pi_{{\mathrm{2}}}  )   \ottsym{)}    [   \lambda\!  \, \mathit{x_{{\mathrm{0}}}}  \ottsym{.}   \resetcnstr{  \ottnt{K_{{\mathrm{0}}}}  [  \mathit{x_{{\mathrm{0}}}}  ]  \,  \prec_{ \ottnt{v} }  \,  \ottnt{K'_{{\mathrm{0}}}}  [  \mathit{x_{{\mathrm{0}}}}  ]  \,  \uparrow \, (  \ottsym{0}  ,  \pi_{{\mathrm{2}}}  )  }   /  \mathit{x_{{\mathrm{0}}}}  ]   \,  \,\&\,  \,  \epsilon 
            \quad \text{(by \reflem{lr-approx-embed-exp})}
            \\ = &
            \ottsym{(}  \ottnt{e_{{\mathrm{01}}}} \,  \prec_{ \ottnt{v} }  \, \ottnt{e'_{{\mathrm{01}}}} \,  \uparrow \, (  \ottsym{0}  ,  \pi_{{\mathrm{2}}}  )   \ottsym{)}    [  \ottsym{(}   \lambda\!  \, \mathit{x_{{\mathrm{0}}}}  \ottsym{.}   \resetcnstr{  \ottnt{K_{{\mathrm{0}}}}  [  \mathit{x_{{\mathrm{0}}}}  ]  }   \ottsym{)} \,  \prec_{ \ottnt{v} }  \, \ottsym{(}   \lambda\!  \, \mathit{x_{{\mathrm{0}}}}  \ottsym{.}   \resetcnstr{  \ottnt{K'_{{\mathrm{0}}}}  [  \mathit{x_{{\mathrm{0}}}}  ]  }   \ottsym{)} \,  \uparrow \, (  \ottsym{0}  ,  \pi_{{\mathrm{2}}}  )   /  \mathit{x_{{\mathrm{0}}}}  ]   \,  \,\&\,  \,  \epsilon  \\ = &
            \ottnt{e_{{\mathrm{01}}}}    [   \lambda\!  \, \mathit{x_{{\mathrm{0}}}}  \ottsym{.}   \resetcnstr{  \ottnt{K_{{\mathrm{0}}}}  [  \mathit{x_{{\mathrm{0}}}}  ]  }   /  \mathit{x_{{\mathrm{0}}}}  ]   \,  \prec_{ \ottnt{v} }  \,  \ottnt{e'_{{\mathrm{01}}}}    [   \lambda\!  \, \mathit{x_{{\mathrm{0}}}}  \ottsym{.}   \resetcnstr{  \ottnt{K'_{{\mathrm{0}}}}  [  \mathit{x_{{\mathrm{0}}}}  ]  }   /  \mathit{x_{{\mathrm{0}}}}  ]   \,  \uparrow \, (  \ottsym{0}  ,  \pi_{{\mathrm{2}}}  )  \,  \,\&\,  \,  \epsilon 
            \quad \text{(by \reflem{lr-approx-val-subst})}
            \\ = &
           \ottnt{e_{{\mathrm{0}}}} \,  \prec_{ \ottnt{v} }  \, \ottnt{e'_{{\mathrm{0}}}} \,  \uparrow \, (  \ottsym{0}  ,  \pi_{{\mathrm{2}}}  )  \,  \,\&\,  \,  \epsilon  ~.
            \end{array}
          \]
   \end{itemize}

  \case \E{Red}/\R{Reset}:
   We are given
   \begin{prooftree}
    \AxiomC{$ \resetcnstr{ \ottnt{v''_{{\mathrm{0}}}} }  \,  \rightsquigarrow  \, \ottnt{v''_{{\mathrm{0}}}}$}
    \UnaryInfC{$ \resetcnstr{ \ottnt{v''_{{\mathrm{0}}}} }  \,  \mathrel{  \longrightarrow  }  \, \ottnt{v''_{{\mathrm{0}}}} \,  \,\&\,  \,  \epsilon $}
   \end{prooftree}
   for some $\ottnt{v''_{{\mathrm{0}}}}$ such that
   \begin{itemize}
    \item $\ottnt{e} \,  \prec_{ \ottnt{v} }  \, \ottnt{e'} \,  \uparrow \, (  \ottsym{0}  ,  \pi_{{\mathrm{1}}}  )  \,  =  \,  \resetcnstr{ \ottnt{v''_{{\mathrm{0}}}} } $ and
    \item $\ottnt{e_{{\mathrm{1}}}} \,  =  \, \ottnt{v''_{{\mathrm{0}}}}$.
   \end{itemize}
   We also have $ \varpi    =     \epsilon  $.

   Because $\ottnt{e} \,  \prec_{ \ottnt{v} }  \, \ottnt{e'} \,  \uparrow \, (  \ottsym{0}  ,  \pi_{{\mathrm{1}}}  )  \,  =  \,  \resetcnstr{ \ottnt{v''_{{\mathrm{0}}}} } $,
   there exist some $\ottnt{v_{{\mathrm{0}}}}$ and $\ottnt{v'_{{\mathrm{0}}}}$ such that
   \begin{itemize}
    \item $\ottnt{e} \,  =  \,  \resetcnstr{ \ottnt{v_{{\mathrm{0}}}} } $,
    \item $\ottnt{e'} \,  =  \,  \resetcnstr{ \ottnt{v'_{{\mathrm{0}}}} } $, and
    \item $\ottnt{v_{{\mathrm{0}}}} \,  \prec_{ \ottnt{v} }  \, \ottnt{v'_{{\mathrm{0}}}} \,  \uparrow \, (  \ottsym{0}  ,  \pi_{{\mathrm{1}}}  )  \,  =  \, \ottnt{v''_{{\mathrm{0}}}}$
   \end{itemize}
   with \reflem{lr-approx-anti-val}.

   Let $\ottnt{e_{{\mathrm{0}}}} \,  =  \, \ottnt{v_{{\mathrm{0}}}}$ and $\ottnt{e'_{{\mathrm{0}}}} \,  =  \, \ottnt{v'_{{\mathrm{0}}}}$.
   We have the conclusion by the following.
   \begin{itemize}
    \item We have $\ottnt{e} \,  \mathrel{  \longrightarrow  }  \, \ottnt{e_{{\mathrm{0}}}} \,  \,\&\,  \,  \epsilon $ by \E{Red}/\R{Reset}.
    \item Let $\pi_{{\mathrm{2}}}$ be any infinite trace.
          Because $\ottnt{v_{{\mathrm{0}}}} \,  \prec_{ \ottnt{v} }  \, \ottnt{v'_{{\mathrm{0}}}} \,  \uparrow \, (  \ottsym{0}  ,  \pi_{{\mathrm{2}}}  ) $ is a value
          by Lemmas~\ref{lem:lr-approx-wf} and \ref{lem:lr-approx-val},
          we have
          \[\begin{array}{ll} &
           \ottnt{e} \,  \prec_{ \ottnt{v} }  \, \ottnt{e'} \,  \uparrow \, (  \ottsym{0}  ,  \pi_{{\mathrm{2}}}  )  \\ = &
            \resetcnstr{ \ottnt{v_{{\mathrm{0}}}} }  \,  \prec_{ \ottnt{v} }  \,  \resetcnstr{ \ottnt{v'_{{\mathrm{0}}}} }  \,  \uparrow \, (  \ottsym{0}  ,  \pi_{{\mathrm{2}}}  )  \\ = &
            \resetcnstr{ \ottnt{v_{{\mathrm{0}}}} \,  \prec_{ \ottnt{v} }  \, \ottnt{v'_{{\mathrm{0}}}} \,  \uparrow \, (  \ottsym{0}  ,  \pi_{{\mathrm{2}}}  )  }  \\  \longrightarrow  &
           \ottnt{v_{{\mathrm{0}}}} \,  \prec_{ \ottnt{v} }  \, \ottnt{v'_{{\mathrm{0}}}} \,  \uparrow \, (  \ottsym{0}  ,  \pi_{{\mathrm{2}}}  )  \,  \,\&\,  \,  \epsilon 
            \quad \text{(by \E{Red}/\R{Reset})}
            \\ = &
           \ottnt{e_{{\mathrm{0}}}} \,  \prec_{ \ottnt{v} }  \, \ottnt{e'_{{\mathrm{0}}}} \,  \uparrow \, (  \ottsym{0}  ,  \pi_{{\mathrm{2}}}  )  \,  \,\&\,  \,  \epsilon  ~.
            \end{array}
          \]
   \end{itemize}

  \case \E{Ev}:
   We are given
   \begin{prooftree}
    \AxiomC{$ \mathsf{ev}  [  \mathbf{a}  ]  \,  \mathrel{  \longrightarrow  }  \,  ()  \,  \,\&\,  \, \mathbf{a}$}
  \end{prooftree}
   for some $\mathbf{a}$ such that
   \begin{itemize}
    \item $\ottnt{e} \,  \prec_{ \ottnt{v} }  \, \ottnt{e'} \,  \uparrow \, (  \ottsym{0}  ,  \pi_{{\mathrm{1}}}  )  \,  =  \,  \mathsf{ev}  [  \mathbf{a}  ] $ and
    \item $ \varpi    =    \mathbf{a} $.
   \end{itemize}
   We also have $\ottnt{e_{{\mathrm{1}}}} \,  =  \,  () $.

   Because $\ottnt{e} \,  \prec_{ \ottnt{v} }  \, \ottnt{e'} \,  \uparrow \, (  \ottsym{0}  ,  \pi_{{\mathrm{1}}}  )  \,  =  \,  \mathsf{ev}  [  \mathbf{a}  ] $,
   $\ottnt{e} = \ottnt{e'} =  \mathsf{ev}  [  \mathbf{a}  ] $.
   Let $\ottnt{e_{{\mathrm{0}}}} = \ottnt{e'_{{\mathrm{0}}}} =  () $.

   We have the conclusion by the following.
   \begin{itemize}
    \item We have $\ottnt{e} \,  \mathrel{  \longrightarrow  }  \, \ottnt{e_{{\mathrm{0}}}} \,  \,\&\,  \, \mathbf{a}$ by \E{Ev}.
    \item For any $\pi_{{\mathrm{2}}}$, we have
          \[
           \ottnt{e} \,  \prec_{ \ottnt{v} }  \, \ottnt{e'} \,  \uparrow \, (  \ottsym{0}  ,  \pi_{{\mathrm{2}}}  )  =  \mathsf{ev}  [  \mathbf{a}  ]  \,  \mathrel{  \longrightarrow  }  \,  ()  \,  \,\&\,  \, \mathbf{a} = \ottnt{e_{{\mathrm{0}}}} \,  \prec_{ \ottnt{v} }  \, \ottnt{e'_{{\mathrm{0}}}} \,  \uparrow \, (  \ottsym{0}  ,  \pi_{{\mathrm{2}}}  )  \,  \,\&\,  \, \mathbf{a} ~.
          \]
          Therefore, we have the conclusion.
   \end{itemize}

  \case \E{Let}:
   We are given
   \begin{prooftree}
    \AxiomC{$\ottnt{e''_{{\mathrm{01}}}} \,  \mathrel{  \longrightarrow  }  \, \ottnt{e''_{{\mathrm{02}}}} \,  \,\&\,  \, \varpi$}
    \UnaryInfC{$\mathsf{let} \, \mathit{x_{{\mathrm{01}}}}  \ottsym{=}  \ottnt{e''_{{\mathrm{01}}}} \,  \mathsf{in}  \, \ottnt{e''_{{\mathrm{03}}}} \,  \mathrel{  \longrightarrow  }  \, \mathsf{let} \, \mathit{x_{{\mathrm{01}}}}  \ottsym{=}  \ottnt{e''_{{\mathrm{02}}}} \,  \mathsf{in}  \, \ottnt{e''_{{\mathrm{03}}}} \,  \,\&\,  \, \varpi$}
   \end{prooftree}
   for some $\mathit{x_{{\mathrm{01}}}}$, $\ottnt{e''_{{\mathrm{01}}}}$, $\ottnt{e''_{{\mathrm{02}}}}$, and $\ottnt{e''_{{\mathrm{03}}}}$ such that
   \begin{itemize}
    \item $\ottnt{e} \,  \prec_{ \ottnt{v} }  \, \ottnt{e'} \,  \uparrow \, (  \ottsym{0}  ,  \pi_{{\mathrm{1}}}  )  \,  =  \, \ottsym{(}  \mathsf{let} \, \mathit{x_{{\mathrm{01}}}}  \ottsym{=}  \ottnt{e''_{{\mathrm{01}}}} \,  \mathsf{in}  \, \ottnt{e''_{{\mathrm{03}}}}  \ottsym{)}$ and
    \item $\ottnt{e_{{\mathrm{1}}}} \,  =  \, \ottsym{(}  \mathsf{let} \, \mathit{x_{{\mathrm{01}}}}  \ottsym{=}  \ottnt{e''_{{\mathrm{02}}}} \,  \mathsf{in}  \, \ottnt{e''_{{\mathrm{03}}}}  \ottsym{)}$.
   \end{itemize}

   Because $\ottnt{e} \,  \prec_{ \ottnt{v} }  \, \ottnt{e'} \,  \uparrow \, (  \ottsym{0}  ,  \pi_{{\mathrm{1}}}  )  \,  =  \, \ottsym{(}  \mathsf{let} \, \mathit{x_{{\mathrm{01}}}}  \ottsym{=}  \ottnt{e''_{{\mathrm{01}}}} \,  \mathsf{in}  \, \ottnt{e''_{{\mathrm{03}}}}  \ottsym{)}$,
   there exist $\ottnt{e_{{\mathrm{01}}}}$, $\ottnt{e_{{\mathrm{03}}}}$, $\ottnt{e'_{{\mathrm{01}}}}$, and $\ottnt{e'_{{\mathrm{03}}}}$ such that
   \begin{itemize}
    \item $\ottnt{e} \,  =  \, \ottsym{(}  \mathsf{let} \, \mathit{x_{{\mathrm{01}}}}  \ottsym{=}  \ottnt{e_{{\mathrm{01}}}} \,  \mathsf{in}  \, \ottnt{e_{{\mathrm{03}}}}  \ottsym{)}$,
    \item $\ottnt{e'} \,  =  \, \ottsym{(}  \mathsf{let} \, \mathit{x_{{\mathrm{01}}}}  \ottsym{=}  \ottnt{e'_{{\mathrm{01}}}} \,  \mathsf{in}  \, \ottnt{e'_{{\mathrm{03}}}}  \ottsym{)}$,
    \item $\ottnt{e_{{\mathrm{01}}}} \,  \prec_{ \ottnt{v} }  \, \ottnt{e'_{{\mathrm{01}}}} \,  \uparrow \, (  \ottsym{0}  ,  \pi_{{\mathrm{1}}}  )  \,  =  \, \ottnt{e''_{{\mathrm{01}}}}$, and
    \item $\ottnt{e_{{\mathrm{03}}}} \,  \prec_{ \ottnt{v} }  \, \ottnt{e'_{{\mathrm{03}}}} \,  \uparrow \, (  \ottsym{0}  ,  \pi_{{\mathrm{1}}}  )  \,  =  \, \ottnt{e''_{{\mathrm{03}}}}$.
   \end{itemize}
   We have $  \forall  \, { \ottnt{E}  \ottsym{,}  \pi } . \  \ottnt{e''_{{\mathrm{02}}}} \,  \not=  \,  \ottnt{E}  [   \bullet ^{ \pi }   ]  $ because:
   if $\ottnt{e''_{{\mathrm{02}}}} \,  =  \,  \ottnt{E}  [   \bullet ^{ \pi }   ] $ for some $\ottnt{E}$ and $\pi$,
   then $\ottnt{e_{{\mathrm{1}}}} = \ottsym{(}  \mathsf{let} \, \mathit{x_{{\mathrm{01}}}}  \ottsym{=}  \ottnt{e''_{{\mathrm{02}}}} \,  \mathsf{in}  \, \ottnt{e''_{{\mathrm{03}}}}  \ottsym{)} = \ottsym{(}  \mathsf{let} \, \mathit{x_{{\mathrm{01}}}}  \ottsym{=}   \ottnt{E}  [   \bullet ^{ \pi }   ]  \,  \mathsf{in}  \, \ottnt{e''_{{\mathrm{03}}}}  \ottsym{)}$,
   but it is contradictory with the assumption $  \forall  \, { \ottnt{E}  \ottsym{,}  \pi } . \  \ottnt{e_{{\mathrm{1}}}} \,  \not=  \,  \ottnt{E}  [   \bullet ^{ \pi }   ]  $.
   Therefore, by the IH, there exist some $\ottnt{e_{{\mathrm{02}}}}$ and $\ottnt{e'_{{\mathrm{02}}}}$ such that
   \begin{itemize}
    \item $\ottnt{e_{{\mathrm{01}}}} \,  \mathrel{  \longrightarrow  }  \, \ottnt{e_{{\mathrm{02}}}} \,  \,\&\,  \, \varpi$ and
    \item $  \forall  \, { \pi_{{\mathrm{2}}} } . \  \ottsym{(}  \ottnt{e_{{\mathrm{01}}}} \,  \prec_{ \ottnt{v} }  \, \ottnt{e'_{{\mathrm{01}}}} \,  \uparrow \, (  \ottsym{0}  ,  \pi_{{\mathrm{2}}}  )   \ottsym{)} \,  \mathrel{  \longrightarrow  }  \, \ottsym{(}  \ottnt{e_{{\mathrm{02}}}} \,  \prec_{ \ottnt{v} }  \, \ottnt{e'_{{\mathrm{02}}}} \,  \uparrow \, (  \ottsym{0}  ,  \pi_{{\mathrm{2}}}  )   \ottsym{)} \,  \,\&\,  \, \varpi $.
   \end{itemize}

   Let $\ottnt{e_{{\mathrm{0}}}} \,  =  \, \ottsym{(}  \mathsf{let} \, \mathit{x_{{\mathrm{01}}}}  \ottsym{=}  \ottnt{e_{{\mathrm{02}}}} \,  \mathsf{in}  \, \ottnt{e_{{\mathrm{03}}}}  \ottsym{)}$ and $\ottnt{e'_{{\mathrm{0}}}} \,  =  \, \ottsym{(}  \mathsf{let} \, \mathit{x_{{\mathrm{01}}}}  \ottsym{=}  \ottnt{e'_{{\mathrm{02}}}} \,  \mathsf{in}  \, \ottnt{e'_{{\mathrm{03}}}}  \ottsym{)}$.
   We have the conclusion by the following.
   \begin{itemize}
    \item We have $\ottnt{e} \,  \mathrel{  \longrightarrow  }  \, \ottnt{e_{{\mathrm{0}}}} \,  \,\&\,  \, \varpi$ by \E{Let} with $\ottnt{e_{{\mathrm{01}}}} \,  \mathrel{  \longrightarrow  }  \, \ottnt{e_{{\mathrm{02}}}} \,  \,\&\,  \, \varpi$.

    \item Let $\pi_{{\mathrm{2}}}$ be any infinite trace.
          We have
          \[\begin{array}{ll} &
           \ottnt{e} \,  \prec_{ \ottnt{v} }  \, \ottnt{e'} \,  \uparrow \, (  \ottsym{0}  ,  \pi_{{\mathrm{2}}}  )  \\ = &
           \mathsf{let} \, \mathit{x_{{\mathrm{01}}}}  \ottsym{=}  \ottsym{(}  \ottnt{e_{{\mathrm{01}}}} \,  \prec_{ \ottnt{v} }  \, \ottnt{e'_{{\mathrm{01}}}} \,  \uparrow \, (  \ottsym{0}  ,  \pi_{{\mathrm{2}}}  )   \ottsym{)} \,  \mathsf{in}  \, \ottsym{(}  \ottnt{e_{{\mathrm{03}}}} \,  \prec_{ \ottnt{v} }  \, \ottnt{e'_{{\mathrm{03}}}} \,  \uparrow \, (  \ottsym{0}  ,  \pi_{{\mathrm{2}}}  )   \ottsym{)} \\  \longrightarrow  &
           \mathsf{let} \, \mathit{x_{{\mathrm{01}}}}  \ottsym{=}  \ottsym{(}  \ottnt{e_{{\mathrm{02}}}} \,  \prec_{ \ottnt{v} }  \, \ottnt{e'_{{\mathrm{02}}}} \,  \uparrow \, (  \ottsym{0}  ,  \pi_{{\mathrm{2}}}  )   \ottsym{)} \,  \mathsf{in}  \, \ottsym{(}  \ottnt{e_{{\mathrm{03}}}} \,  \prec_{ \ottnt{v} }  \, \ottnt{e'_{{\mathrm{03}}}} \,  \uparrow \, (  \ottsym{0}  ,  \pi_{{\mathrm{2}}}  )   \ottsym{)} \,  \,\&\,  \, \varpi
            \quad \text{(by the result of the IH and \E{Let})}
            \\ = &
           \ottsym{(}  \mathsf{let} \, \mathit{x_{{\mathrm{01}}}}  \ottsym{=}  \ottnt{e_{{\mathrm{02}}}} \,  \mathsf{in}  \, \ottnt{e_{{\mathrm{03}}}}  \ottsym{)} \,  \prec_{ \ottnt{v} }  \, \ottsym{(}  \mathsf{let} \, \mathit{x_{{\mathrm{01}}}}  \ottsym{=}  \ottnt{e'_{{\mathrm{02}}}} \,  \mathsf{in}  \, \ottnt{e'_{{\mathrm{03}}}}  \ottsym{)} \,  \uparrow \, (  \ottsym{0}  ,  \pi_{{\mathrm{2}}}  )  \,  \,\&\,  \, \varpi \\ = &
           \ottnt{e_{{\mathrm{0}}}} \,  \prec_{ \ottnt{v} }  \, \ottnt{e'_{{\mathrm{0}}}} \,  \uparrow \, (  \ottsym{0}  ,  \pi_{{\mathrm{2}}}  )  \,  \,\&\,  \, \varpi ~.
            \end{array}
          \]
   \end{itemize}

  \case \E{Reset}:
   We are given
   \begin{prooftree}
    \AxiomC{$\ottnt{e''_{{\mathrm{01}}}} \,  \mathrel{  \longrightarrow  }  \, \ottnt{e''_{{\mathrm{02}}}} \,  \,\&\,  \, \varpi$}
    \UnaryInfC{$ \resetcnstr{ \ottnt{e''_{{\mathrm{01}}}} }  \,  \mathrel{  \longrightarrow  }  \,  \resetcnstr{ \ottnt{e''_{{\mathrm{02}}}} }  \,  \,\&\,  \, \varpi$}
   \end{prooftree}
   for some $\ottnt{e''_{{\mathrm{01}}}}$ and $\ottnt{e''_{{\mathrm{02}}}}$ such that
   \begin{itemize}
    \item $\ottnt{e} \,  \prec_{ \ottnt{v} }  \, \ottnt{e'} \,  \uparrow \, (  \ottsym{0}  ,  \pi_{{\mathrm{1}}}  )  \,  =  \,  \resetcnstr{ \ottnt{e''_{{\mathrm{01}}}} } $ and
    \item $\ottnt{e_{{\mathrm{1}}}} \,  =  \,  \resetcnstr{ \ottnt{e''_{{\mathrm{02}}}} } $.
   \end{itemize}

   Because $\ottnt{e} \,  \prec_{ \ottnt{v} }  \, \ottnt{e'} \,  \uparrow \, (  \ottsym{0}  ,  \pi_{{\mathrm{1}}}  )  \,  =  \,  \resetcnstr{ \ottnt{e''_{{\mathrm{01}}}} } $,
   there exist $\ottnt{e_{{\mathrm{01}}}}$ and $\ottnt{e'_{{\mathrm{01}}}}$ such that
   \begin{itemize}
    \item $\ottnt{e} \,  =  \,  \resetcnstr{ \ottnt{e_{{\mathrm{01}}}} } $,
    \item $\ottnt{e'} \,  =  \,  \resetcnstr{ \ottnt{e'_{{\mathrm{01}}}} } $, and
    \item $\ottnt{e_{{\mathrm{01}}}} \,  \prec_{ \ottnt{v} }  \, \ottnt{e'_{{\mathrm{01}}}} \,  \uparrow \, (  \ottsym{0}  ,  \pi_{{\mathrm{1}}}  )  \,  =  \, \ottnt{e''_{{\mathrm{01}}}}$.
   \end{itemize}
   We have $  \forall  \, { \ottnt{E}  \ottsym{,}  \pi } . \  \ottnt{e''_{{\mathrm{02}}}} \,  \not=  \,  \ottnt{E}  [   \bullet ^{ \pi }   ]  $ because:
   if $\ottnt{e''_{{\mathrm{02}}}} \,  =  \,  \ottnt{E}  [   \bullet ^{ \pi }   ] $, then $\ottnt{e_{{\mathrm{1}}}} =  \resetcnstr{ \ottnt{e''_{{\mathrm{02}}}} }  =  \resetcnstr{  \ottnt{E}  [   \bullet ^{ \pi }   ]  } $,
   but it is contradictory with the assumption $  \forall  \, { \ottnt{E}  \ottsym{,}  \pi } . \  \ottnt{e_{{\mathrm{1}}}} \,  \not=  \,  \ottnt{E}  [   \bullet ^{ \pi }   ]  $.
   Therefore, by the IH, there exist some $\ottnt{e_{{\mathrm{02}}}}$ and $\ottnt{e'_{{\mathrm{02}}}}$ such that
   \begin{itemize}
    \item $\ottnt{e_{{\mathrm{01}}}} \,  \mathrel{  \longrightarrow  }  \, \ottnt{e_{{\mathrm{02}}}} \,  \,\&\,  \, \varpi$ and
    \item $  \forall  \, { \pi_{{\mathrm{2}}} } . \  \ottsym{(}  \ottnt{e_{{\mathrm{01}}}} \,  \prec_{ \ottnt{v} }  \, \ottnt{e'_{{\mathrm{01}}}} \,  \uparrow \, (  \ottsym{0}  ,  \pi_{{\mathrm{2}}}  )   \ottsym{)} \,  \mathrel{  \longrightarrow  }  \, \ottsym{(}  \ottnt{e_{{\mathrm{02}}}} \,  \prec_{ \ottnt{v} }  \, \ottnt{e'_{{\mathrm{02}}}} \,  \uparrow \, (  \ottsym{0}  ,  \pi_{{\mathrm{2}}}  )   \ottsym{)} \,  \,\&\,  \, \varpi $.
   \end{itemize}

   Let $\ottnt{e_{{\mathrm{0}}}} \,  =  \,  \resetcnstr{ \ottnt{e_{{\mathrm{02}}}} } $ and $\ottnt{e'_{{\mathrm{0}}}} \,  =  \,  \resetcnstr{ \ottnt{e'_{{\mathrm{02}}}} } $.
   We have the conclusion by the following.
   \begin{itemize}
    \item We have $\ottnt{e} \,  \mathrel{  \longrightarrow  }  \, \ottnt{e_{{\mathrm{0}}}} \,  \,\&\,  \, \varpi$ by \E{Reset} with $\ottnt{e_{{\mathrm{01}}}} \,  \mathrel{  \longrightarrow  }  \, \ottnt{e_{{\mathrm{02}}}} \,  \,\&\,  \, \varpi$.

    \item Let $\pi_{{\mathrm{2}}}$ be any infinite trace.
          We have
          \[\begin{array}{ll} &
           \ottnt{e} \,  \prec_{ \ottnt{v} }  \, \ottnt{e'} \,  \uparrow \, (  \ottsym{0}  ,  \pi_{{\mathrm{2}}}  )  \\ = &
            \resetcnstr{ \ottnt{e_{{\mathrm{01}}}} }  \,  \prec_{ \ottnt{v} }  \,  \resetcnstr{ \ottnt{e'_{{\mathrm{01}}}} }  \,  \uparrow \, (  \ottsym{0}  ,  \pi_{{\mathrm{2}}}  )  \\ = &
            \resetcnstr{ \ottnt{e_{{\mathrm{01}}}} \,  \prec_{ \ottnt{v} }  \, \ottnt{e'_{{\mathrm{01}}}} \,  \uparrow \, (  \ottsym{0}  ,  \pi_{{\mathrm{2}}}  )  }  \\  \longrightarrow  &
            \resetcnstr{ \ottnt{e_{{\mathrm{02}}}} \,  \prec_{ \ottnt{v} }  \, \ottnt{e'_{{\mathrm{02}}}} \,  \uparrow \, (  \ottsym{0}  ,  \pi_{{\mathrm{2}}}  )  }  \,  \,\&\,  \, \varpi
            \quad \text{(by the result of the IH and \E{Reset})}
            \\ = &
            \resetcnstr{ \ottnt{e_{{\mathrm{02}}}} }  \,  \prec_{ \ottnt{v} }  \,  \resetcnstr{ \ottnt{e'_{{\mathrm{02}}}} }  \,  \uparrow \, (  \ottsym{0}  ,  \pi_{{\mathrm{2}}}  )  \,  \,\&\,  \, \varpi \\ = &
           \ottnt{e_{{\mathrm{0}}}} \,  \prec_{ \ottnt{v} }  \, \ottnt{e'_{{\mathrm{0}}}} \,  \uparrow \, (  \ottsym{0}  ,  \pi_{{\mathrm{2}}}  )  \,  \,\&\,  \, \varpi ~.
          \end{array}
          \]
   \end{itemize}
 \end{caseanalysis}
\end{proof}

\begin{lemmap}{Evaluation of Approximation Modulo Infinite Traces by Multiple Steps}{lr-approx-eval-modulo-itr-nsteps}
 If $\ottsym{(}  \ottnt{e} \,  \prec_{ \ottnt{v} }  \, \ottnt{e'} \,  \uparrow \, (  \ottsym{0}  ,  \pi_{{\mathrm{1}}}  )   \ottsym{)} \,  \mathrel{  \longrightarrow  }^{ \ottmv{n} }  \, \ottnt{e_{{\mathrm{1}}}} \,  \,\&\,  \, \varpi$ and $  \forall  \, { \ottnt{E}  \ottsym{,}  \pi } . \  \ottnt{e_{{\mathrm{1}}}} \,  \not=  \,  \ottnt{E}  [   \bullet ^{ \pi }   ]  $,
 then
 \[
   {   \exists  \, { \ottnt{e_{{\mathrm{0}}}}  \ottsym{,}  \ottnt{e'_{{\mathrm{0}}}} } . \  \ottnt{e} \,  \mathrel{  \longrightarrow  }^{ \ottmv{n} }  \, \ottnt{e_{{\mathrm{0}}}} \,  \,\&\,  \, \varpi  } \,\mathrel{\wedge}\, {   \forall  \, { \pi_{{\mathrm{2}}} } . \  \ottsym{(}  \ottnt{e} \,  \prec_{ \ottnt{v} }  \, \ottnt{e'} \,  \uparrow \, (  \ottsym{0}  ,  \pi_{{\mathrm{2}}}  )   \ottsym{)} \,  \mathrel{  \longrightarrow  }^{ \ottmv{n} }  \, \ottsym{(}  \ottnt{e_{{\mathrm{0}}}} \,  \prec_{ \ottnt{v} }  \, \ottnt{e'_{{\mathrm{0}}}} \,  \uparrow \, (  \ottsym{0}  ,  \pi_{{\mathrm{2}}}  )   \ottsym{)} \,  \,\&\,  \, \varpi  }  ~.
 \]
\end{lemmap}
\TS{OK: 2022/01/20}
\begin{proof}
 By induction on $\ottmv{n}$.
 \begin{caseanalysis}
  \case $\ottmv{n} \,  =  \, \ottsym{0}$:
   Because $\ottnt{e} \,  \prec_{ \ottnt{v} }  \, \ottnt{e'} \,  \uparrow \, (  \ottsym{0}  ,  \pi_{{\mathrm{1}}}  )  \,  =  \, \ottnt{e_{{\mathrm{1}}}}$ and $ \varpi    =     \epsilon  $,
   we have the conclusion by letting $\ottnt{e_{{\mathrm{0}}}} \,  =  \, \ottnt{e}$ and $\ottnt{e'_{{\mathrm{0}}}} \,  =  \, \ottnt{e'}$.

  \case $\ottmv{n} \,  >  \, \ottsym{0}$:
   We are given
   \begin{itemize}
    \item $\ottsym{(}  \ottnt{e} \,  \prec_{ \ottnt{v} }  \, \ottnt{e'} \,  \uparrow \, (  \ottsym{0}  ,  \pi_{{\mathrm{1}}}  )   \ottsym{)} \,  \mathrel{  \longrightarrow  }  \, \ottnt{e_{{\mathrm{2}}}} \,  \,\&\,  \, \varpi_{{\mathrm{2}}}$,
    \item $\ottnt{e_{{\mathrm{2}}}} \,  \mathrel{  \longrightarrow  }^{ \ottmv{n}  \ottsym{-}  \ottsym{1} }  \, \ottnt{e_{{\mathrm{1}}}} \,  \,\&\,  \, \varpi_{{\mathrm{1}}}$, and
    \item $ \varpi    =    \varpi_{{\mathrm{2}}} \,  \tceappendop  \, \varpi_{{\mathrm{1}}} $
   \end{itemize}
   for some $\ottnt{e_{{\mathrm{2}}}}$, $\varpi_{{\mathrm{2}}}$, and $\varpi_{{\mathrm{1}}}$.

   We first show that $  \forall  \, { \ottnt{E}  \ottsym{,}  \pi } . \  \ottnt{e_{{\mathrm{2}}}} \,  \not=  \,  \ottnt{E}  [   \bullet ^{ \pi }   ]  $.
   Suppose that $\ottnt{e_{{\mathrm{2}}}} \,  =  \,  \ottnt{E}  [   \bullet ^{ \pi }   ] $ for some $\ottnt{E}$ and $\pi$.
   Then, we have no rule to reduce $\ottnt{e_{{\mathrm{2}}}}$, so
   $\ottnt{e_{{\mathrm{2}}}} \,  \mathrel{  \longrightarrow  }^{ \ottmv{n}  \ottsym{-}  \ottsym{1} }  \, \ottnt{e_{{\mathrm{1}}}} \,  \,\&\,  \, \varpi_{{\mathrm{1}}}$ implies $\ottnt{e_{{\mathrm{1}}}} \,  =  \,  \ottnt{E}  [   \bullet ^{ \pi }   ] $, but
   it is contradictory with the assumption $  \forall  \, { \ottnt{E}  \ottsym{,}  \pi } . \  \ottnt{e_{{\mathrm{1}}}} \,  \not=  \,  \ottnt{E}  [   \bullet ^{ \pi }   ]  $.

   Then, by \reflem{lr-approx-eval-modulo-itr},
   there exist some $\ottnt{e_{{\mathrm{02}}}}$ and $\ottnt{e'_{{\mathrm{02}}}}$ such that
   \begin{itemize}
    \item $\ottnt{e} \,  \mathrel{  \longrightarrow  }  \, \ottnt{e_{{\mathrm{02}}}} \,  \,\&\,  \, \varpi_{{\mathrm{2}}}$ and
    \item $  \forall  \, { \pi_{{\mathrm{2}}} } . \  \ottsym{(}  \ottnt{e} \,  \prec_{ \ottnt{v} }  \, \ottnt{e'} \,  \uparrow \, (  \ottsym{0}  ,  \pi_{{\mathrm{2}}}  )   \ottsym{)} \,  \mathrel{  \longrightarrow  }  \, \ottsym{(}  \ottnt{e_{{\mathrm{02}}}} \,  \prec_{ \ottnt{v} }  \, \ottnt{e'_{{\mathrm{02}}}} \,  \uparrow \, (  \ottsym{0}  ,  \pi_{{\mathrm{2}}}  )   \ottsym{)} \,  \,\&\,  \, \varpi_{{\mathrm{2}}} $.
   \end{itemize}
   Because
   $\ottsym{(}  \ottnt{e} \,  \prec_{ \ottnt{v} }  \, \ottnt{e'} \,  \uparrow \, (  \ottsym{0}  ,  \pi_{{\mathrm{1}}}  )   \ottsym{)} \,  \mathrel{  \longrightarrow  }  \, \ottnt{e_{{\mathrm{2}}}} \,  \,\&\,  \, \varpi_{{\mathrm{2}}}$ and
   $\ottsym{(}  \ottnt{e} \,  \prec_{ \ottnt{v} }  \, \ottnt{e'} \,  \uparrow \, (  \ottsym{0}  ,  \pi_{{\mathrm{1}}}  )   \ottsym{)} \,  \mathrel{  \longrightarrow  }  \, \ottsym{(}  \ottnt{e_{{\mathrm{02}}}} \,  \prec_{ \ottnt{v} }  \, \ottnt{e'_{{\mathrm{02}}}} \,  \uparrow \, (  \ottsym{0}  ,  \pi_{{\mathrm{1}}}  )   \ottsym{)} \,  \,\&\,  \, \varpi_{{\mathrm{2}}}$,
   we have $\ottnt{e_{{\mathrm{2}}}} \,  =  \, \ottnt{e_{{\mathrm{02}}}} \,  \prec_{ \ottnt{v} }  \, \ottnt{e'_{{\mathrm{02}}}} \,  \uparrow \, (  \ottsym{0}  ,  \pi_{{\mathrm{1}}}  ) $ by \reflem{lr-eval-det}.
   Therefore, $\ottsym{(}  \ottnt{e_{{\mathrm{02}}}} \,  \prec_{ \ottnt{v} }  \, \ottnt{e'_{{\mathrm{02}}}} \,  \uparrow \, (  \ottsym{0}  ,  \pi_{{\mathrm{1}}}  )   \ottsym{)} \,  \mathrel{  \longrightarrow  }^{ \ottmv{n}  \ottsym{-}  \ottsym{1} }  \, \ottnt{e_{{\mathrm{1}}}} \,  \,\&\,  \, \varpi_{{\mathrm{1}}}$.
   Then, the IH implies that there exist some $\ottnt{e_{{\mathrm{0}}}}$ and $\ottnt{e'_{{\mathrm{0}}}}$ such that
   \begin{itemize}
    \item $\ottnt{e_{{\mathrm{02}}}} \,  \mathrel{  \longrightarrow  }^{ \ottmv{n}  \ottsym{-}  \ottsym{1} }  \, \ottnt{e_{{\mathrm{0}}}} \,  \,\&\,  \, \varpi_{{\mathrm{1}}}$ and
    \item $  \forall  \, { \pi_{{\mathrm{2}}} } . \  \ottsym{(}  \ottnt{e_{{\mathrm{02}}}} \,  \prec_{ \ottnt{v} }  \, \ottnt{e'_{{\mathrm{02}}}} \,  \uparrow \, (  \ottsym{0}  ,  \pi_{{\mathrm{2}}}  )   \ottsym{)} \,  \mathrel{  \longrightarrow  }^{ \ottmv{n}  \ottsym{-}  \ottsym{1} }  \, \ottsym{(}  \ottnt{e_{{\mathrm{0}}}} \,  \prec_{ \ottnt{v} }  \, \ottnt{e'_{{\mathrm{0}}}} \,  \uparrow \, (  \ottsym{0}  ,  \pi_{{\mathrm{2}}}  )   \ottsym{)} \,  \,\&\,  \, \varpi_{{\mathrm{1}}} $.
   \end{itemize}
   Now, we have
   \begin{itemize}
    \item $\ottnt{e} \,  \mathrel{  \longrightarrow  }  \, \ottnt{e_{{\mathrm{0}}}} \,  \,\&\,  \, \varpi_{{\mathrm{2}}} \,  \tceappendop  \, \varpi_{{\mathrm{1}}}$ and
    \item $  \forall  \, { \pi_{{\mathrm{2}}} } . \  \ottsym{(}  \ottnt{e} \,  \prec_{ \ottnt{v} }  \, \ottnt{e'} \,  \uparrow \, (  \ottsym{0}  ,  \pi_{{\mathrm{2}}}  )   \ottsym{)} \,  \mathrel{  \longrightarrow  }^{ \ottmv{n} }  \, \ottsym{(}  \ottnt{e_{{\mathrm{0}}}} \,  \prec_{ \ottnt{v} }  \, \ottnt{e'_{{\mathrm{0}}}} \,  \uparrow \, (  \ottsym{0}  ,  \pi_{{\mathrm{2}}}  )   \ottsym{)} \,  \,\&\,  \, \varpi_{{\mathrm{2}}} \,  \tceappendop  \, \varpi_{{\mathrm{1}}} $.
   \end{itemize}
   Because $ \varpi    =    \varpi_{{\mathrm{2}}} \,  \tceappendop  \, \varpi_{{\mathrm{1}}} $, we have the conclusion.
 \end{caseanalysis}
\end{proof}

\begin{lemmap}{Approximation of Diverging Expressions}{lr-approx-div}
 If $\ottnt{e} \,  \prec_{ \ottnt{v} }  \, \ottnt{e'}$ and $\ottnt{e} \,  \Uparrow  \, \pi$, then
 $  \exists  \, { \pi' } . \  \ottsym{(}  \ottnt{e} \,  \prec_{ \ottnt{v} }  \, \ottnt{e'} \,  \uparrow \, (  \ottsym{0}  ,  \pi'  )   \ottsym{)} \,  \Uparrow  \, \pi $.
\end{lemmap}
\begin{proof}
 \TS{OK: 2022/01/20}
 We consider the two cases on whether there exists some $\pi'_{{\mathrm{0}}}$ such that
 the evaluation of $\ottnt{e} \,  \prec_{ \ottnt{v} }  \, \ottnt{e'} \,  \uparrow \, (  \ottsym{0}  ,  \pi'_{{\mathrm{0}}}  ) $ diverges by calling a
 $0$-approximation function.
 \begin{caseanalysis}
  \case $ {   \exists  \, { \pi'_{{\mathrm{0}}}  \ottsym{,}  \ottmv{n}  \ottsym{,}  \ottnt{E_{{\mathrm{0}}}}  \ottsym{,}  \pi''_{{\mathrm{0}}}  \ottsym{,}   \tceseq{ \ottnt{v_{{\mathrm{02}}}} }   \ottsym{,}  \varpi_{{\mathrm{0}}} } . \  \ottsym{(}  \ottnt{e} \,  \prec_{ \ottnt{v} }  \, \ottnt{e'} \,  \uparrow \, (  \ottsym{0}  ,  \pi'_{{\mathrm{0}}}  )   \ottsym{)} \,  \mathrel{  \longrightarrow  }^{ \ottmv{n} }  \,  \ottnt{E_{{\mathrm{0}}}}  [   \ottnt{v} _{  \mu   \ottsym{;}  \ottsym{0}  \ottsym{;}  \pi''_{{\mathrm{0}}} }  \,  \tceseq{ \ottnt{v_{{\mathrm{02}}}} }   ]  \,  \,\&\,  \, \varpi_{{\mathrm{0}}}  } \,\mathrel{\wedge}\, {  \mathit{arity}  (  \ottnt{v}  )  \,  =  \, \ottsym{\mbox{$\mid$}}   \tceseq{ \ottnt{v_{{\mathrm{02}}}} }   \ottsym{\mbox{$\mid$}} } $:
   We are numbering the assumption to reference:
   \begin{equation}
    \ottsym{(}  \ottnt{e} \,  \prec_{ \ottnt{v} }  \, \ottnt{e'} \,  \uparrow \, (  \ottsym{0}  ,  \pi'_{{\mathrm{0}}}  )   \ottsym{)} \,  \mathrel{  \longrightarrow  }^{ \ottmv{n} }  \,  \ottnt{E_{{\mathrm{0}}}}  [   \ottnt{v} _{  \mu   \ottsym{;}  \ottsym{0}  \ottsym{;}  \pi''_{{\mathrm{0}}} }  \,  \tceseq{ \ottnt{v_{{\mathrm{02}}}} }   ]  \,  \,\&\,  \, \varpi_{{\mathrm{0}}} ~.
     \label{lem:lr-approx-div:idx0}
   \end{equation}
   We can find
   \begin{equation}
       \forall  \, { \ottmv{m} \,  <  \, \ottmv{n} } . \    \forall  \, { \ottnt{E_{{\mathrm{1}}}}  \ottsym{,}  \pi''_{{\mathrm{1}}}  \ottsym{,}   \tceseq{ \ottnt{v_{{\mathrm{12}}}} }   \ottsym{,}  \varpi_{{\mathrm{1}}} } . \  \ottsym{(}  \ottnt{e} \,  \prec_{ \ottnt{v} }  \, \ottnt{e'} \,  \uparrow \, (  \ottsym{0}  ,  \pi'_{{\mathrm{0}}}  )   \ottsym{)} \,  \mathrel{  \longrightarrow  }^{ \ottmv{m} }  \,  \ottnt{E_{{\mathrm{1}}}}  [   \ottnt{v} _{  \mu   \ottsym{;}  \ottsym{0}  \ottsym{;}  \pi''_{{\mathrm{1}}} }  \,  \tceseq{ \ottnt{v_{{\mathrm{12}}}} }   ]  \,  \,\&\,  \, \varpi_{{\mathrm{1}}}    \mathrel{\Longrightarrow}   \mathit{arity}  (  \ottnt{v}  )  \,  \not=  \, \ottsym{\mbox{$\mid$}}   \tceseq{ \ottnt{v_{{\mathrm{12}}}} }   \ottsym{\mbox{$\mid$}} 
     \label{lem:lr-approx-div:TM-0-idx}
   \end{equation}
   because: if $\ottsym{(}  \ottnt{e} \,  \prec_{ \ottnt{v} }  \, \ottnt{e'} \,  \uparrow \, (  \ottsym{0}  ,  \pi'_{{\mathrm{0}}}  )   \ottsym{)} \,  \mathrel{  \longrightarrow  }^{ \ottmv{m} }  \,  \ottnt{E_{{\mathrm{1}}}}  [   \ottnt{v} _{  \mu   \ottsym{;}  \ottsym{0}  \ottsym{;}  \pi''_{{\mathrm{1}}} }  \,  \tceseq{ \ottnt{v_{{\mathrm{12}}}} }   ]  \,  \,\&\,  \, \varpi_{{\mathrm{1}}}$ is derivable and $ \mathit{arity}  (  \ottnt{v}  )  \,  =  \, \ottsym{\mbox{$\mid$}}   \tceseq{ \ottnt{v_{{\mathrm{12}}}} }   \ottsym{\mbox{$\mid$}}$, then we have
   \[
    \ottsym{(}  \ottnt{e} \,  \prec_{ \ottnt{v} }  \, \ottnt{e'} \,  \uparrow \, (  \ottsym{0}  ,  \pi'_{{\mathrm{0}}}  )   \ottsym{)} \,  \mathrel{  \longrightarrow  }^{ \ottmv{m}  \ottsym{+}  \ottsym{1} }  \,  \ottnt{E_{{\mathrm{1}}}}  [   \bullet ^{ \pi''_{{\mathrm{1}}} }   ]  \,  \,\&\,  \, \varpi_{{\mathrm{1}}} ~,
   \]
   but, as $ \ottnt{E_{{\mathrm{1}}}}  [   \bullet ^{ \pi''_{{\mathrm{1}}} }   ] $ cannot be reduced furthermore (\reflem{lr-red-div}),
   \reflem{lr-eval-det} implies that the assumption (\ref{lem:lr-approx-div:idx0}) cannot be
   derived; therefore, there is a contradiction.

   By case analysis on $\ottnt{e} \,  \Uparrow  \, \pi$.
   \begin{caseanalysis}
    \case $ {   \exists  \, { \ottmv{m} \,  \leq  \, \ottmv{n} } . \    \exists  \, { \ottnt{E'}  \ottsym{,}  \pi'  \ottsym{,}  \varpi' } . \  \ottnt{e} \,  \mathrel{  \longrightarrow  }^{ \ottmv{m} }  \,  \ottnt{E'}  [   \bullet ^{ \pi' }   ]  \,  \,\&\,  \, \varpi'   } \,\mathrel{\wedge}\, { \pi \,  =  \, \varpi' \,  \tceappendop  \, \pi' } $:
     By \reflem{lr-approx-eval-nsteps-idx0} with (\ref{lem:lr-approx-div:TM-0-idx}),
     we have
     \[
      \ottsym{(}  \ottnt{e} \,  \prec_{ \ottnt{v} }  \, \ottnt{e'} \,  \uparrow \, (  \ottsym{0}  ,  \pi'_{{\mathrm{0}}}  )   \ottsym{)} \,  \mathrel{  \longrightarrow  }^{ \ottmv{m} }  \, \ottsym{(}   \ottnt{E'}  [   \bullet ^{ \pi' }   ]  \,  \prec_{ \ottnt{v} }  \, \ottnt{e'_{{\mathrm{1}}}} \,  \uparrow \, (  \ottsym{0}  ,  \pi'_{{\mathrm{0}}}  )   \ottsym{)} \,  \,\&\,  \, \varpi'
     \]
     for some $\ottnt{e'_{{\mathrm{1}}}}$.
     Because $ \ottnt{E'}  [   \bullet ^{ \pi' }   ]  \,  \prec_{ \ottnt{v} }  \, \ottnt{e'_{{\mathrm{1}}}}$ by \reflem{lr-approx-wf},
     \reflem{lr-approx-ectx} implies that there exist some $\ottnt{E'_{{\mathrm{1}}}}$ and $\ottnt{e'_{{\mathrm{01}}}}$
     such that
     \begin{itemize}
      \item $\ottnt{e'_{{\mathrm{1}}}} \,  =  \,  \ottnt{E'_{{\mathrm{1}}}}  [  \ottnt{e'_{{\mathrm{01}}}}  ] $,
      \item $\ottnt{E'} \,  \prec_{ \ottnt{v} }  \, \ottnt{E'_{{\mathrm{1}}}}$, and
      \item $ \bullet ^{ \pi' }  \,  \prec_{ \ottnt{v} }  \, \ottnt{e'_{{\mathrm{01}}}}$.
     \end{itemize}
     Then, we have
     \[\begin{array}{ll} &
       \ottnt{E'}  [   \bullet ^{ \pi' }   ]  \,  \prec_{ \ottnt{v} }  \, \ottnt{e'_{{\mathrm{1}}}} \,  \uparrow \, (  \ottsym{0}  ,  \pi'_{{\mathrm{0}}}  )  \\ = &
       \ottnt{E'}  [   \bullet ^{ \pi' }   ]  \,  \prec_{ \ottnt{v} }  \,  \ottnt{E'_{{\mathrm{1}}}}  [  \ottnt{e'_{{\mathrm{01}}}}  ]  \,  \uparrow \, (  \ottsym{0}  ,  \pi'_{{\mathrm{0}}}  )  \\ = &
       \ottsym{(}  \ottnt{E'} \,  \prec_{ \ottnt{v} }  \, \ottnt{E'_{{\mathrm{1}}}} \,  \uparrow \, (  \ottsym{0}  ,  \pi'_{{\mathrm{0}}}  )   \ottsym{)}  [   \bullet ^{ \pi' }  \,  \prec_{ \ottnt{v} }  \, \ottnt{e'_{{\mathrm{01}}}} \,  \uparrow \, (  \ottsym{0}  ,  \pi'_{{\mathrm{0}}}  )   ] 
       \quad \text{(by \reflem{lr-approx-embed-exp})}
       \\ = &
       \ottsym{(}  \ottnt{E'} \,  \prec_{ \ottnt{v} }  \, \ottnt{E'_{{\mathrm{1}}}} \,  \uparrow \, (  \ottsym{0}  ,  \pi'_{{\mathrm{0}}}  )   \ottsym{)}  [   \bullet ^{ \pi' }   ] 
       \end{array}
     \]
     Therefore,
     \[
      \ottnt{e} \,  \prec_{ \ottnt{v} }  \, \ottnt{e'} \,  \uparrow \, (  \ottsym{0}  ,  \pi'_{{\mathrm{0}}}  )  \,  \mathrel{  \longrightarrow  }^{ \ottmv{m} }  \,  \ottsym{(}  \ottnt{E'} \,  \prec_{ \ottnt{v} }  \, \ottnt{E'_{{\mathrm{1}}}} \,  \uparrow \, (  \ottsym{0}  ,  \pi'_{{\mathrm{0}}}  )   \ottsym{)}  [   \bullet ^{ \pi' }   ]  \,  \,\&\,  \, \varpi' ~.
     \]
     Because $ \pi    =    \varpi' \,  \tceappendop  \, \pi' $,
     we have the conclusion $\ottnt{e} \,  \prec_{ \ottnt{v} }  \, \ottnt{e'} \,  \uparrow \, (  \ottsym{0}  ,  \pi'_{{\mathrm{0}}}  )  \,  \Uparrow  \, \pi$.

    \case $ {   \exists  \, { \ottmv{m} \,  >  \, \ottmv{n} } . \    \exists  \, { \ottnt{E'}  \ottsym{,}  \pi'  \ottsym{,}  \varpi' } . \  \ottnt{e} \,  \mathrel{  \longrightarrow  }^{ \ottmv{m} }  \,  \ottnt{E'}  [   \bullet ^{ \pi' }   ]  \,  \,\&\,  \, \varpi'   } \,\mathrel{\wedge}\, { \pi \,  =  \, \varpi' \,  \tceappendop  \, \pi' } $, or $   \forall  \, { \varpi'  \ottsym{,}  \pi' } . \  \pi \,  =  \, \varpi' \,  \tceappendop  \, \pi'   \mathrel{\Longrightarrow}    \exists  \, { \ottnt{e_{{\mathrm{0}}}} } . \  \ottnt{e} \,  \mathrel{  \longrightarrow  ^*}  \, \ottnt{e_{{\mathrm{0}}}} \,  \,\&\,  \, \varpi'  $:
     We first show that there exist some $\ottnt{e_{{\mathrm{1}}}}$, $\varpi'_{{\mathrm{1}}}$, and $\pi'_{{\mathrm{1}}}$ such that
     \begin{itemize}
      \item $\ottnt{e} \,  \mathrel{  \longrightarrow  }^{ \ottmv{n} }  \, \ottnt{e_{{\mathrm{1}}}} \,  \,\&\,  \, \varpi'_{{\mathrm{1}}}$ and
      \item $ \pi    =    \varpi'_{{\mathrm{1}}} \,  \tceappendop  \, \pi'_{{\mathrm{1}}} $.
     \end{itemize}
     \begin{caseanalysis}
      \case $ {   \exists  \, { \ottmv{m} \,  >  \, \ottmv{n} } . \    \exists  \, { \ottnt{E'}  \ottsym{,}  \pi'  \ottsym{,}  \varpi' } . \  \ottnt{e} \,  \mathrel{  \longrightarrow  }^{ \ottmv{m} }  \,  \ottnt{E'}  [   \bullet ^{ \pi' }   ]  \,  \,\&\,  \, \varpi'   } \,\mathrel{\wedge}\, { \pi \,  =  \, \varpi' \,  \tceappendop  \, \pi' } $:
       Because $\ottmv{m} \,  >  \, \ottmv{n}$, we have
       $\ottnt{e} \,  \mathrel{  \longrightarrow  }^{ \ottmv{n} }  \, \ottnt{e_{{\mathrm{1}}}} \,  \,\&\,  \, \varpi'_{{\mathrm{1}}}$ and $\ottnt{e_{{\mathrm{1}}}} \,  \mathrel{  \longrightarrow  }^{ \ottmv{m}  \ottsym{-}  \ottmv{n} }  \,  \ottnt{E'}  [   \bullet ^{ \pi' }   ]  \,  \,\&\,  \, \varpi'_{{\mathrm{2}}}$ such that
       $ \varpi'    =    \varpi'_{{\mathrm{1}}} \,  \tceappendop  \, \varpi'_{{\mathrm{2}}} $.
       Because $\pi = \varpi' \,  \tceappendop  \, \pi' = \varpi'_{{\mathrm{1}}} \,  \tceappendop  \, \varpi'_{{\mathrm{2}}} \,  \tceappendop  \, \pi'$,
       we have the conclusion by letting $ \pi'_{{\mathrm{1}}}    =    \varpi'_{{\mathrm{2}}} \,  \tceappendop  \, \pi' $.

      \case $   \forall  \, { \varpi'  \ottsym{,}  \pi' } . \  \pi \,  =  \, \varpi' \,  \tceappendop  \, \pi'   \mathrel{\Longrightarrow}    \exists  \, { \ottnt{e_{{\mathrm{0}}}} } . \  \ottnt{e} \,  \mathrel{  \longrightarrow  ^*}  \, \ottnt{e_{{\mathrm{0}}}} \,  \,\&\,  \, \varpi'  $:
       Suppose that $\varpi'$ be a prefix of $\pi$ of length $\ottmv{n}$.
       Then, there exist some $\ottmv{m}$ and $\ottnt{e_{{\mathrm{0}}}}$ such that
       $\ottnt{e} \,  \mathrel{  \longrightarrow  }^{ \ottmv{m} }  \, \ottnt{e_{{\mathrm{0}}}} \,  \,\&\,  \, \varpi'$.
       Because emitting $\varpi'$ needs to take at least $\ottmv{n}$ steps,
       $\ottmv{n} \,  \leq  \, \ottmv{m}$.
       Then, there exist some $\ottnt{e_{{\mathrm{1}}}}$, $\varpi'_{{\mathrm{1}}}$, and $\varpi'_{{\mathrm{2}}}$ such that
       $\ottnt{e} \,  \mathrel{  \longrightarrow  }^{ \ottmv{n} }  \, \ottnt{e_{{\mathrm{1}}}} \,  \,\&\,  \, \varpi'_{{\mathrm{1}}}$ and $\ottnt{e_{{\mathrm{1}}}} \,  \mathrel{  \longrightarrow  }^{ \ottmv{m}  \ottsym{-}  \ottmv{n} }  \, \ottnt{e_{{\mathrm{0}}}} \,  \,\&\,  \, \varpi'_{{\mathrm{2}}}$ and $ \varpi'    =    \varpi'_{{\mathrm{1}}} \,  \tceappendop  \, \varpi'_{{\mathrm{2}}} $.
       Because $\varpi'$ is a prefix of $\pi$, $\varpi'_{{\mathrm{1}}}$ is also a prefix of $\pi$.
       Therefore, there exists some $\pi'_{{\mathrm{1}}}$ such that $ \pi    =    \varpi'_{{\mathrm{1}}} \,  \tceappendop  \, \pi'_{{\mathrm{1}}} $.
     \end{caseanalysis}

     Next, we show that $  \forall  \, { \ottnt{E''}  \ottsym{,}  \pi'' } . \   \ottnt{E_{{\mathrm{0}}}}  [   \ottnt{v} _{  \mu   \ottsym{;}  \ottsym{0}  \ottsym{;}  \pi''_{{\mathrm{0}}} }  \,  \tceseq{ \ottnt{v_{{\mathrm{02}}}} }   ]  \,  \not=  \,  \ottnt{E''}  [   \bullet ^{ \pi'' }   ]  $.
     Suppose that $ \ottnt{E_{{\mathrm{0}}}}  [   \ottnt{v} _{  \mu   \ottsym{;}  \ottsym{0}  \ottsym{;}  \pi''_{{\mathrm{0}}} }  \,  \tceseq{ \ottnt{v_{{\mathrm{02}}}} }   ]  \,  =  \,  \ottnt{E''}  [   \bullet ^{ \pi'' }   ] $ for some $\ottnt{E''}$ and $\pi''$.
     Then, because $ \mathit{arity}  (   \ottnt{v} _{  \mu   \ottsym{;}  \ottsym{0}  \ottsym{;}  \pi''_{{\mathrm{0}}} }   )  =  \mathit{arity}  (  \ottnt{v}  )  = \ottsym{\mbox{$\mid$}}   \tceseq{ \ottnt{v_{{\mathrm{02}}}} }   \ottsym{\mbox{$\mid$}}$,
     the expression $ \ottnt{v} _{  \mu   \ottsym{;}  \ottsym{0}  \ottsym{;}  \pi''_{{\mathrm{0}}} }  \,  \tceseq{ \ottnt{v_{{\mathrm{02}}}} } $ is not a value.
     Therefore, $ \ottnt{v} _{  \mu   \ottsym{;}  \ottsym{0}  \ottsym{;}  \pi''_{{\mathrm{0}}} }  \,  \tceseq{ \ottnt{v_{{\mathrm{02}}}} }  \,  =  \,  \bullet ^{ \pi'' } $ by \reflem{lr-decompose-ectx-embed-exp}, but
     it is contradictory.

     Hence, we can apply \reflem{lr-approx-eval-modulo-itr-nsteps} with
     the assumption (\ref{lem:lr-approx-div:idx0});
     then, there exist some $\ottnt{e_{{\mathrm{0}}}}$ and $\ottnt{e'_{{\mathrm{0}}}}$ such that
     \begin{itemize}
      \item $\ottnt{e} \,  \mathrel{  \longrightarrow  }^{ \ottmv{n} }  \, \ottnt{e_{{\mathrm{0}}}} \,  \,\&\,  \, \varpi_{{\mathrm{0}}}$ and
      \item $  \forall  \, { \pi_{{\mathrm{2}}} } . \  \ottsym{(}  \ottnt{e} \,  \prec_{ \ottnt{v} }  \, \ottnt{e'} \,  \uparrow \, (  \ottsym{0}  ,  \pi_{{\mathrm{2}}}  )   \ottsym{)} \,  \mathrel{  \longrightarrow  }^{ \ottmv{n} }  \, \ottsym{(}  \ottnt{e_{{\mathrm{0}}}} \,  \prec_{ \ottnt{v} }  \, \ottnt{e'_{{\mathrm{0}}}} \,  \uparrow \, (  \ottsym{0}  ,  \pi_{{\mathrm{2}}}  )   \ottsym{)} \,  \,\&\,  \, \varpi_{{\mathrm{0}}} $.
     \end{itemize}
     Because $\ottnt{e} \,  \mathrel{  \longrightarrow  }^{ \ottmv{n} }  \, \ottnt{e_{{\mathrm{1}}}} \,  \,\&\,  \, \varpi'_{{\mathrm{1}}}$, we have
     \begin{itemize}
      \item $\ottnt{e_{{\mathrm{0}}}} \,  =  \, \ottnt{e_{{\mathrm{1}}}}$ and
      \item $ \varpi_{{\mathrm{0}}}    =    \varpi'_{{\mathrm{1}}} $
     \end{itemize}
     by \reflem{lr-eval-det}.
     Because of the assumption (\ref{lem:lr-approx-div:idx0}) and \reflem{lr-eval-det},
     we have $\ottsym{(}  \ottnt{e_{{\mathrm{0}}}} \,  \prec_{ \ottnt{v} }  \, \ottnt{e'_{{\mathrm{0}}}} \,  \uparrow \, (  \ottsym{0}  ,  \pi'_{{\mathrm{0}}}  )   \ottsym{)} \,  =  \,  \ottnt{E_{{\mathrm{0}}}}  [   \ottnt{v} _{  \mu   \ottsym{;}  \ottsym{0}  \ottsym{;}  \pi''_{{\mathrm{0}}} }  \,  \tceseq{ \ottnt{v_{{\mathrm{02}}}} }   ] $.
     By \reflem{lr-approx-anti-ectx}, there exist some $\ottnt{E_{{\mathrm{01}}}}$, $\ottnt{e_{{\mathrm{01}}}}$, $\ottnt{E'_{{\mathrm{01}}}}$, and $\ottnt{e'_{{\mathrm{01}}}}$ such that
     \begin{itemize}
      \item $\ottnt{e_{{\mathrm{0}}}} \,  =  \,  \ottnt{E_{{\mathrm{01}}}}  [  \ottnt{e_{{\mathrm{01}}}}  ] $,
      \item $\ottnt{e'_{{\mathrm{0}}}} \,  =  \,  \ottnt{E'_{{\mathrm{01}}}}  [  \ottnt{e'_{{\mathrm{01}}}}  ] $,
      \item $\ottnt{E_{{\mathrm{01}}}} \,  \prec_{ \ottnt{v} }  \, \ottnt{E'_{{\mathrm{01}}}} \,  \uparrow \, (  \ottsym{0}  ,  \pi'_{{\mathrm{0}}}  )  \,  =  \, \ottnt{E_{{\mathrm{0}}}}$, and
      \item $\ottnt{e_{{\mathrm{01}}}} \,  \prec_{ \ottnt{v} }  \, \ottnt{e'_{{\mathrm{01}}}} \,  \uparrow \, (  \ottsym{0}  ,  \pi'_{{\mathrm{0}}}  )  \,  =  \,  \ottnt{v} _{  \mu   \ottsym{;}  \ottsym{0}  \ottsym{;}  \pi''_{{\mathrm{0}}} }  \,  \tceseq{ \ottnt{v_{{\mathrm{02}}}} } $.
     \end{itemize}
     Further, $\ottnt{e_{{\mathrm{01}}}} \,  \prec_{ \ottnt{v} }  \, \ottnt{e'_{{\mathrm{01}}}} \,  \uparrow \, (  \ottsym{0}  ,  \pi'_{{\mathrm{0}}}  )  \,  =  \,  \ottnt{v} _{  \mu   \ottsym{;}  \ottsym{0}  \ottsym{;}  \pi''_{{\mathrm{0}}} }  \,  \tceseq{ \ottnt{v_{{\mathrm{02}}}} } $ implies that
     there exist $\ottnt{v_{{\mathrm{01}}}}$, $ \tceseq{ \ottnt{v_{{\mathrm{002}}}} } $, $\ottnt{v'_{{\mathrm{01}}}}$, and $ \tceseq{ \ottnt{v'_{{\mathrm{002}}}} } $ such that
     \begin{itemize}
      \item $\ottnt{e_{{\mathrm{01}}}} \,  =  \, \ottnt{v_{{\mathrm{01}}}} \,  \tceseq{ \ottnt{v_{{\mathrm{002}}}} } $,
      \item $\ottnt{e'_{{\mathrm{01}}}} \,  =  \, \ottnt{v'_{{\mathrm{01}}}} \,  \tceseq{ \ottnt{v'_{{\mathrm{002}}}} } $,
      \item $\ottnt{v_{{\mathrm{01}}}} \,  \prec_{ \ottnt{v} }  \, \ottnt{v'_{{\mathrm{01}}}} \,  \uparrow \, (  \ottsym{0}  ,  \pi'_{{\mathrm{0}}}  )  \,  =  \,  \ottnt{v} _{  \mu   \ottsym{;}  \ottsym{0}  \ottsym{;}  \pi''_{{\mathrm{0}}} } $, and
      \item $ \tceseq{ \ottnt{v_{{\mathrm{002}}}} \,  \prec_{ \ottnt{v} }  \, \ottnt{v'_{{\mathrm{002}}}} \,  \uparrow \, (  \ottsym{0}  ,  \pi'_{{\mathrm{0}}}  )  \,  =  \, \ottnt{v_{{\mathrm{02}}}} } $.
     \end{itemize}
     By case analysis on $\ottnt{v_{{\mathrm{01}}}}$.
     We have only two cases to be considered because
     $\ottnt{v_{{\mathrm{01}}}} \,  \prec_{ \ottnt{v} }  \, \ottnt{v'_{{\mathrm{01}}}} \,  \uparrow \, (  \ottsym{0}  ,  \pi'_{{\mathrm{0}}}  )  \,  =  \,  \ottnt{v} _{  \mu   \ottsym{;}  \ottsym{0}  \ottsym{;}  \pi''_{{\mathrm{0}}} } $.
     \begin{caseanalysis}
      \case $\ottnt{v_{{\mathrm{01}}}} \,  =  \, \ottnt{v}$:
       We are given $\ottnt{v'_{{\mathrm{01}}}} \,  =  \,  \ottnt{v} _{  \mu   \ottsym{;}  \ottsym{0}  \ottsym{;}  \pi'''_{{\mathrm{0}}} } $ for some $\pi'''_{{\mathrm{0}}}$.
       Then, we have
       \[\begin{array}{ll} &
        \ottnt{e} \,  \prec_{ \ottnt{v} }  \, \ottnt{e'} \,  \uparrow \, (  \ottsym{0}  ,  \pi'_{{\mathrm{1}}}  )  \\  \mathrel{  \longrightarrow  }^{ \ottmv{n} }  &
        \ottnt{e_{{\mathrm{0}}}} \,  \prec_{ \ottnt{v} }  \, \ottnt{e'_{{\mathrm{0}}}} \,  \uparrow \, (  \ottsym{0}  ,  \pi'_{{\mathrm{1}}}  )  \,  \,\&\,  \, \varpi_{{\mathrm{0}}}
         \quad \text{(by the result of applying \reflem{lr-approx-eval-modulo-itr-nsteps})}
         \\ = &
         \ottnt{E_{{\mathrm{01}}}}  [  \ottnt{e_{{\mathrm{01}}}}  ]  \,  \prec_{ \ottnt{v} }  \,  \ottnt{E'_{{\mathrm{01}}}}  [  \ottnt{e'_{{\mathrm{01}}}}  ]  \,  \uparrow \, (  \ottsym{0}  ,  \pi'_{{\mathrm{1}}}  )  \,  \,\&\,  \, \varpi_{{\mathrm{0}}} \\ = &
         \ottsym{(}  \ottnt{E_{{\mathrm{01}}}} \,  \prec_{ \ottnt{v} }  \, \ottnt{E'_{{\mathrm{01}}}} \,  \uparrow \, (  \ottsym{0}  ,  \pi'_{{\mathrm{1}}}  )   \ottsym{)}  [  \ottnt{e_{{\mathrm{01}}}} \,  \prec_{ \ottnt{v} }  \, \ottnt{e'_{{\mathrm{01}}}} \,  \uparrow \, (  \ottsym{0}  ,  \pi'_{{\mathrm{1}}}  )   ]  \,  \,\&\,  \, \varpi_{{\mathrm{0}}}
         \quad \text{(by Lemmas~\ref{lem:lr-approx-wf} and \ref{lem:lr-approx-embed-exp})}
         \\ = &
         \ottsym{(}  \ottnt{E_{{\mathrm{01}}}} \,  \prec_{ \ottnt{v} }  \, \ottnt{E'_{{\mathrm{01}}}} \,  \uparrow \, (  \ottsym{0}  ,  \pi'_{{\mathrm{1}}}  )   \ottsym{)}  [  \ottnt{v_{{\mathrm{01}}}} \,  \tceseq{ \ottnt{v_{{\mathrm{002}}}} }  \,  \prec_{ \ottnt{v} }  \, \ottnt{v'_{{\mathrm{01}}}} \,  \tceseq{ \ottnt{v'_{{\mathrm{002}}}} }  \,  \uparrow \, (  \ottsym{0}  ,  \pi'_{{\mathrm{1}}}  )   ]  \,  \,\&\,  \, \varpi_{{\mathrm{0}}} \\ = &
         \ottsym{(}  \ottnt{E_{{\mathrm{01}}}} \,  \prec_{ \ottnt{v} }  \, \ottnt{E'_{{\mathrm{01}}}} \,  \uparrow \, (  \ottsym{0}  ,  \pi'_{{\mathrm{1}}}  )   \ottsym{)}  [  \ottsym{(}  \ottnt{v_{{\mathrm{01}}}} \,  \prec_{ \ottnt{v} }  \, \ottnt{v'_{{\mathrm{01}}}} \,  \uparrow \, (  \ottsym{0}  ,  \pi'_{{\mathrm{1}}}  )   \ottsym{)} \,  \tceseq{ \ottnt{v_{{\mathrm{002}}}} \,  \prec_{ \ottnt{v} }  \, \ottnt{v'_{{\mathrm{002}}}} \,  \uparrow \, (  \ottsym{0}  ,  \pi'_{{\mathrm{1}}}  )  }   ]  \,  \,\&\,  \, \varpi_{{\mathrm{0}}} \\ = &
         \ottsym{(}  \ottnt{E_{{\mathrm{01}}}} \,  \prec_{ \ottnt{v} }  \, \ottnt{E'_{{\mathrm{01}}}} \,  \uparrow \, (  \ottsym{0}  ,  \pi'_{{\mathrm{1}}}  )   \ottsym{)}  [   \ottnt{v} _{  \mu   \ottsym{;}  \ottsym{0}  \ottsym{;}  \pi'_{{\mathrm{1}}} }  \,  \tceseq{ \ottnt{v_{{\mathrm{002}}}} \,  \prec_{ \ottnt{v} }  \, \ottnt{v'_{{\mathrm{002}}}} \,  \uparrow \, (  \ottsym{0}  ,  \pi'_{{\mathrm{1}}}  )  }   ]  \,  \,\&\,  \, \varpi_{{\mathrm{0}}} ~.
         \end{array}
       \]
       Because
       $ \tceseq{ \ottnt{v_{{\mathrm{002}}}} \,  \prec_{ \ottnt{v} }  \, \ottnt{v'_{{\mathrm{002}}}} \,  \uparrow \, (  \ottsym{0}  ,  \pi'_{{\mathrm{1}}}  )  } $ are values by \reflem{lr-approx-val}, and
       $\ottsym{\mbox{$\mid$}}   \tceseq{ \ottnt{v_{{\mathrm{002}}}} \,  \prec_{ \ottnt{v} }  \, \ottnt{v'_{{\mathrm{002}}}} \,  \uparrow \, (  \ottsym{0}  ,  \pi'_{{\mathrm{1}}}  )  }   \ottsym{\mbox{$\mid$}} = \ottsym{\mbox{$\mid$}}   \tceseq{ \ottnt{v_{{\mathrm{02}}}} }   \ottsym{\mbox{$\mid$}} =  \mathit{arity}  (  \ottnt{v}  )  =  \mathit{arity}  (   \ottnt{v} _{  \mu   \ottsym{;}  \ottsym{0}  \ottsym{;}  \pi'_{{\mathrm{1}}} }   ) $,
       we have
       \[
         \ottsym{(}  \ottnt{E_{{\mathrm{01}}}} \,  \prec_{ \ottnt{v} }  \, \ottnt{E'_{{\mathrm{01}}}} \,  \uparrow \, (  \ottsym{0}  ,  \pi'_{{\mathrm{1}}}  )   \ottsym{)}  [   \ottnt{v} _{  \mu   \ottsym{;}  \ottsym{0}  \ottsym{;}  \pi'_{{\mathrm{1}}} }  \,  \tceseq{ \ottnt{v_{{\mathrm{002}}}} \,  \prec_{ \ottnt{v} }  \, \ottnt{v'_{{\mathrm{002}}}} \,  \uparrow \, (  \ottsym{0}  ,  \pi'_{{\mathrm{1}}}  )  }   ]  \,  \mathrel{  \longrightarrow  }  \,  \ottsym{(}  \ottnt{E_{{\mathrm{01}}}} \,  \prec_{ \ottnt{v} }  \, \ottnt{E'_{{\mathrm{01}}}} \,  \uparrow \, (  \ottsym{0}  ,  \pi'_{{\mathrm{1}}}  )   \ottsym{)}  [   \bullet ^{ \pi'_{{\mathrm{1}}} }   ]  \,  \,\&\,  \,  \epsilon 
       \]
       by \E{Red}/\R{Beta}.
       Therefore, we have
       \[
        \ottsym{(}  \ottnt{e} \,  \prec_{ \ottnt{v} }  \, \ottnt{e'} \,  \uparrow \, (  \ottsym{0}  ,  \pi'_{{\mathrm{1}}}  )   \ottsym{)} \,  \mathrel{  \longrightarrow  }^{ \ottmv{n}  \ottsym{+}  \ottsym{1} }  \,  \ottsym{(}  \ottnt{E_{{\mathrm{01}}}} \,  \prec_{ \ottnt{v} }  \, \ottnt{E'_{{\mathrm{01}}}} \,  \uparrow \, (  \ottsym{0}  ,  \pi'_{{\mathrm{1}}}  )   \ottsym{)}  [   \bullet ^{ \pi'_{{\mathrm{1}}} }   ]  \,  \,\&\,  \, \varpi_{{\mathrm{0}}} ~.
       \]
       Because $ \varpi_{{\mathrm{0}}}    =    \varpi'_{{\mathrm{1}}} $ and $ \varpi'_{{\mathrm{1}}} \,  \tceappendop  \, \pi'_{{\mathrm{1}}}    =    \pi $,
       we have the conclusion $\ottnt{e} \,  \prec_{ \ottnt{v} }  \, \ottnt{e'} \,  \uparrow \, (  \ottsym{0}  ,  \pi'_{{\mathrm{1}}}  )  \,  \Uparrow  \, \pi$.

      \case $  \exists  \, {  \tceseq{ \mathit{x} }   \ottsym{,}  \ottnt{e_{{\mathrm{01}}}} } . \  \ottnt{v_{{\mathrm{01}}}} \,  =  \,  \lambda\!  \,  \tceseq{ \mathit{x} }   \ottsym{.}  \ottnt{e_{{\mathrm{01}}}} $:
       We are given $\ottnt{v'_{{\mathrm{01}}}} \,  =  \,  \lambda\!  \,  \tceseq{ \mathit{x} }   \ottsym{.}  \ottnt{e'_{{\mathrm{01}}}}$ and
       $\ottnt{e_{{\mathrm{01}}}} \,  \prec_{ \ottnt{v} }  \, \ottnt{e'_{{\mathrm{01}}}} \,  \uparrow \, (  \ottsym{0}  ,  \pi'_{{\mathrm{0}}}  )  \,  =  \,  \bullet ^{ \pi''_{{\mathrm{0}}} } $
       for some $\ottnt{e'_{{\mathrm{01}}}}$.
       Because $\ottnt{e_{{\mathrm{01}}}} \,  \prec_{ \ottnt{v} }  \, \ottnt{e'_{{\mathrm{01}}}} \,  \uparrow \, (  \ottsym{0}  ,  \pi'_{{\mathrm{0}}}  )  \,  =  \,  \bullet ^{ \pi''_{{\mathrm{0}}} } $,
       we have $\ottnt{e_{{\mathrm{01}}}} = \ottnt{e'_{{\mathrm{01}}}} =  \bullet ^{ \pi''_{{\mathrm{0}}} } $.
       Then, we have
       \[\begin{array}{ll} &
        \ottnt{e_{{\mathrm{0}}}} \\ = &
         \ottnt{E_{{\mathrm{01}}}}  [  \ottnt{e_{{\mathrm{01}}}}  ]  \\ = &
         \ottnt{E_{{\mathrm{01}}}}  [  \ottnt{v_{{\mathrm{01}}}} \,  \tceseq{ \ottnt{v_{{\mathrm{002}}}} }   ]  \\ = &
         \ottnt{E_{{\mathrm{01}}}}  [  \ottsym{(}   \lambda\!  \,  \tceseq{ \mathit{x} }   \ottsym{.}  \ottnt{e_{{\mathrm{01}}}}  \ottsym{)} \,  \tceseq{ \ottnt{v_{{\mathrm{002}}}} }   ]  \\  \longrightarrow  &
         \ottnt{E_{{\mathrm{01}}}}  [   \ottnt{e_{{\mathrm{01}}}}    [ \tceseq{ \ottnt{v_{{\mathrm{002}}}}  /  \mathit{x} } ]    ]  \,  \,\&\,  \,  \epsilon 
         \quad \text{(by \E{Red}/\R{Beta})}
         \\ = &
         \bullet ^{ \pi''_{{\mathrm{0}}} }  \,  \,\&\,  \,  \epsilon  ~.
         \end{array}
       \]
       Because $\ottnt{e} \,  \mathrel{  \longrightarrow  }^{ \ottmv{n} }  \, \ottnt{e_{{\mathrm{1}}}} \,  \,\&\,  \, \varpi'_{{\mathrm{1}}}$ and $\ottnt{e_{{\mathrm{0}}}} \,  =  \, \ottnt{e_{{\mathrm{1}}}}$,
       we have $\ottnt{e} \,  \mathrel{  \longrightarrow  }^{ \ottmv{n}  \ottsym{+}  \ottsym{1} }  \,  \bullet ^{ \pi''_{{\mathrm{0}}} }  \,  \,\&\,  \, \varpi'_{{\mathrm{1}}}$.
       Because $\ottnt{e} \,  \Uparrow  \, \pi$, we have $ \pi    =    \varpi'_{{\mathrm{1}}} \,  \tceappendop  \, \pi''_{{\mathrm{0}}} $.
       Further, as $ \pi    =    \varpi'_{{\mathrm{1}}} \,  \tceappendop  \, \pi'_{{\mathrm{1}}} $,
       we have $ \pi''_{{\mathrm{0}}}    =    \pi'_{{\mathrm{1}}} $.
       Because
       \begin{itemize}
        \item $\ottsym{(}  \ottnt{e} \,  \prec_{ \ottnt{v} }  \, \ottnt{e'} \,  \uparrow \, (  \ottsym{0}  ,  \pi'_{{\mathrm{0}}}  )   \ottsym{)} \,  \mathrel{  \longrightarrow  }^{ \ottmv{n} }  \,  \ottnt{E_{{\mathrm{0}}}}  [   \ottnt{v} _{  \mu   \ottsym{;}  \ottsym{0}  \ottsym{;}  \pi''_{{\mathrm{0}}} }  \,  \tceseq{ \ottnt{v_{{\mathrm{02}}}} }   ]  \,  \,\&\,  \, \varpi_{{\mathrm{0}}}$,
        \item $ \ottnt{E_{{\mathrm{0}}}}  [   \ottnt{v} _{  \mu   \ottsym{;}  \ottsym{0}  \ottsym{;}  \pi''_{{\mathrm{0}}} }  \,  \tceseq{ \ottnt{v_{{\mathrm{02}}}} }   ]  \,  \mathrel{  \longrightarrow  }  \,  \ottnt{E_{{\mathrm{0}}}}  [   \bullet ^{ \pi''_{{\mathrm{0}}} }   ]  \,  \,\&\,  \,  \epsilon $,
        \item $ \varpi_{{\mathrm{0}}}    =    \varpi'_{{\mathrm{1}}} $, and
        \item $ \pi''_{{\mathrm{0}}}    =    \pi'_{{\mathrm{1}}} $,
       \end{itemize}
       we have
       \[
        \ottsym{(}  \ottnt{e} \,  \prec_{ \ottnt{v} }  \, \ottnt{e'} \,  \uparrow \, (  \ottsym{0}  ,  \pi'_{{\mathrm{0}}}  )   \ottsym{)} \,  \mathrel{  \longrightarrow  }^{ \ottmv{n}  \ottsym{+}  \ottsym{1} }  \,  \ottnt{E_{{\mathrm{0}}}}  [   \bullet ^{ \pi'_{{\mathrm{1}}} }   ]  \,  \,\&\,  \, \varpi'_{{\mathrm{1}}} ~.
       \]
       Because $ \pi    =    \varpi'_{{\mathrm{1}}} \,  \tceappendop  \, \pi'_{{\mathrm{1}}} $,
       we have the conclusion $\ottnt{e} \,  \prec_{ \ottnt{v} }  \, \ottnt{e'} \,  \uparrow \, (  \ottsym{0}  ,  \pi'_{{\mathrm{0}}}  )  \,  \Uparrow  \, \pi$.
     \end{caseanalysis}
   \end{caseanalysis}

  \case $   \forall  \, { \pi'_{{\mathrm{0}}}  \ottsym{,}  \ottnt{E_{{\mathrm{0}}}}  \ottsym{,}  \pi''_{{\mathrm{0}}}  \ottsym{,}   \tceseq{ \ottnt{v_{{\mathrm{02}}}} }   \ottsym{,}  \varpi_{{\mathrm{0}}} } . \  \ottsym{(}  \ottnt{e} \,  \prec_{ \ottnt{v} }  \, \ottnt{e'} \,  \uparrow \, (  \ottsym{0}  ,  \pi'_{{\mathrm{0}}}  )   \ottsym{)} \,  \mathrel{  \longrightarrow  ^*}  \,  \ottnt{E_{{\mathrm{0}}}}  [   \ottnt{v} _{  \mu   \ottsym{;}  \ottsym{0}  \ottsym{;}  \pi''_{{\mathrm{0}}} }  \,  \tceseq{ \ottnt{v_{{\mathrm{02}}}} }   ]  \,  \,\&\,  \, \varpi_{{\mathrm{0}}}   \mathrel{\Longrightarrow}   \mathit{arity}  (  \ottnt{v}  )  \,  \not=  \, \ottsym{\mbox{$\mid$}}   \tceseq{ \ottnt{v_{{\mathrm{02}}}} }   \ottsym{\mbox{$\mid$}} $:
   By case analysis on $\ottnt{e} \,  \Uparrow  \, \pi$.
   \begin{caseanalysis}
    \case $ {   \exists  \, { \varpi'  \ottsym{,}  \pi'  \ottsym{,}  \ottnt{E} } . \  \pi \,  =  \, \varpi' \,  \tceappendop  \, \pi'  } \,\mathrel{\wedge}\, { \ottnt{e} \,  \mathrel{  \longrightarrow  ^*}  \,  \ottnt{E}  [   \bullet ^{ \pi' }   ]  \,  \,\&\,  \, \varpi' } $:
     Let $\pi'_{{\mathrm{0}}}$ be an arbitrary infinite trace.
     By \reflem{lr-approx-eval-nsteps-idx0}, there exists some $\ottnt{e'_{{\mathrm{1}}}}$ such that
     \[
      \ottsym{(}  \ottnt{e} \,  \prec_{ \ottnt{v} }  \, \ottnt{e'} \,  \uparrow \, (  \ottsym{0}  ,  \pi'_{{\mathrm{0}}}  )   \ottsym{)} \,  \mathrel{  \longrightarrow  ^*}  \, \ottsym{(}   \ottnt{E}  [   \bullet ^{ \pi' }   ]  \,  \prec_{ \ottnt{v} }  \, \ottnt{e'_{{\mathrm{1}}}} \,  \uparrow \, (  \ottsym{0}  ,  \pi'_{{\mathrm{0}}}  )   \ottsym{)} \,  \,\&\,  \, \varpi' ~.
     \]
     By Lemmas~\ref{lem:lr-approx-wf}, \ref{lem:lr-approx-ectx} and \ref{lem:lr-approx-embed-exp},
     $ \ottnt{E}  [   \bullet ^{ \pi' }   ]  \,  \prec_{ \ottnt{v} }  \, \ottnt{e'_{{\mathrm{1}}}} \,  \uparrow \, (  \ottsym{0}  ,  \pi'_{{\mathrm{0}}}  )  \,  =  \,  \ottnt{E''}  [   \bullet ^{ \pi' }   ] $
     for some $\ottnt{E''}$.
     Therefore,
     \[
      \ottsym{(}  \ottnt{e} \,  \prec_{ \ottnt{v} }  \, \ottnt{e'} \,  \uparrow \, (  \ottsym{0}  ,  \pi'_{{\mathrm{0}}}  )   \ottsym{)} \,  \mathrel{  \longrightarrow  ^*}  \,  \ottnt{E''}  [   \bullet ^{ \pi' }   ]  \,  \,\&\,  \, \varpi' ~.
     \]
     Because $ \pi    =    \varpi' \,  \tceappendop  \, \pi' $,
     we have the conclusion $\ottnt{e} \,  \prec_{ \ottnt{v} }  \, \ottnt{e'} \,  \uparrow \, (  \ottsym{0}  ,  \pi'_{{\mathrm{0}}}  )  \,  \Uparrow  \, \pi$.

    \case $   \forall  \, { \varpi'  \ottsym{,}  \pi' } . \  \pi \,  =  \, \varpi' \,  \tceappendop  \, \pi'   \mathrel{\Longrightarrow}    \exists  \, { \ottnt{e_{{\mathrm{0}}}} } . \  \ottnt{e} \,  \mathrel{  \longrightarrow  ^*}  \, \ottnt{e_{{\mathrm{0}}}} \,  \,\&\,  \, \varpi'  $:
     Let
     $\pi'_{{\mathrm{0}}}$ be an arbitrary infinite trace, and
     $\varpi'$ be any finite trace and
     $\pi'$ be any infinite trace such that $ \pi    =    \varpi' \,  \tceappendop  \, \pi' $.
     It suffices to show that
     \[
        \exists  \, { \ottnt{e''_{{\mathrm{0}}}} } . \  \ottsym{(}  \ottnt{e} \,  \prec_{ \ottnt{v} }  \, \ottnt{e'} \,  \uparrow \, (  \ottsym{0}  ,  \pi'_{{\mathrm{0}}}  )   \ottsym{)} \,  \mathrel{  \longrightarrow  ^*}  \, \ottnt{e''_{{\mathrm{0}}}} \,  \,\&\,  \, \varpi'  ~.
     \]
     Note that $\ottnt{e} \,  \prec_{ \ottnt{v} }  \, \ottnt{e'} \,  \uparrow \, (  \ottsym{0}  ,  \pi'_{{\mathrm{0}}}  ) $ is well defined by
     \reflem{lr-approx-wf} with $\ottnt{e} \,  \prec_{ \ottnt{v} }  \, \ottnt{e'}$.
     Because $\ottnt{e} \,  \mathrel{  \longrightarrow  ^*}  \, \ottnt{e_{{\mathrm{0}}}} \,  \,\&\,  \, \varpi'$ for some $\ottnt{e_{{\mathrm{0}}}}$,
     \reflem{lr-approx-eval-nsteps-idx0} implies
     $\ottsym{(}  \ottnt{e} \,  \prec_{ \ottnt{v} }  \, \ottnt{e'} \,  \uparrow \, (  \ottsym{0}  ,  \pi'_{{\mathrm{0}}}  )   \ottsym{)} \,  \mathrel{  \longrightarrow  ^*}  \, \ottsym{(}  \ottnt{e_{{\mathrm{0}}}} \,  \prec_{ \ottnt{v} }  \, \ottnt{e'_{{\mathrm{0}}}} \,  \uparrow \, (  \ottsym{0}  ,  \pi'_{{\mathrm{0}}}  )   \ottsym{)} \,  \,\&\,  \, \varpi'$
     for some $\ottnt{e'_{{\mathrm{0}}}}$.
     Therefore, we have the conclusion.
   \end{caseanalysis}
 \end{caseanalysis}
\end{proof}

\begin{lemmap}{Approximation of Logical Relation}{lr-approx-lr}
 \begin{enumerate}
  \item \label{lem:lr-approx-lr:exp}
        If $  \forall  \, { \ottmv{i}  \ottsym{,}  \pi } . \  \ottsym{(}  \ottnt{e_{{\mathrm{0}}}} \,  \prec_{ \ottnt{v} }  \, \ottnt{e'_{{\mathrm{0}}}} \,  \uparrow \, (  \ottmv{i}  ,  \pi  )   \ottsym{)} \,  \in  \,  \LRRel{ \mathcal{E} }{ \ottnt{C} }  $,
        then $\ottnt{e_{{\mathrm{0}}}} \,  \in  \,  \LRRel{ \mathrm{Atom} }{ \ottnt{C} } $ implies $\ottnt{e_{{\mathrm{0}}}} \,  \in  \,  \LRRel{ \mathcal{E} }{ \ottnt{C} } $.

  \item \label{lem:lr-approx-lr:val}
        If $  \forall  \, { \ottmv{i}  \ottsym{,}  \pi } . \  \ottsym{(}  \ottnt{v_{{\mathrm{0}}}} \,  \prec_{ \ottnt{v} }  \, \ottnt{v'_{{\mathrm{0}}}} \,  \uparrow \, (  \ottmv{i}  ,  \pi  )   \ottsym{)} \,  \in  \,  \LRRel{ \mathcal{V} }{ \ottnt{T} }  $,
        then $\ottnt{v_{{\mathrm{0}}}} \,  \in  \,  \LRRel{ \mathrm{Atom} }{ \ottnt{T} } $ implies $\ottnt{v_{{\mathrm{0}}}} \,  \in  \,  \LRRel{ \mathcal{V} }{ \ottnt{T} } $.
 \end{enumerate}
\end{lemmap}
\begin{proof}
 \TS{OK: 2022/01/20}
 By simultaneous induction on $\ottnt{C}$ and $\ottnt{T}$.
 \begin{enumerate}
  \item  We consider each case on the result of evaluating $\ottnt{e_{{\mathrm{0}}}}$.
         \begin{caseanalysis}
          \case $  \exists  \, { \ottnt{v_{{\mathrm{0}}}}  \ottsym{,}  \varpi } . \  \ottnt{e_{{\mathrm{0}}}} \,  \Downarrow  \, \ottnt{v_{{\mathrm{0}}}} \,  \,\&\,  \, \varpi $:
           By \reflem{ts-type-sound-terminating},
           \begin{itemize}
            \item $ \emptyset   \vdash  \ottnt{v_{{\mathrm{0}}}}  \ottsym{:}  \ottnt{T}$,
            \item $ \emptyset   \models   { \Phi }^\mu   \ottsym{(}  \varpi  \ottsym{)}$, and
            \item $ \emptyset   \vdash   \ottnt{T}  \,  \,\&\,  \,  \Phi  \, / \,   \square    \ottsym{<:}  \ottnt{C}$
           \end{itemize}
           for some $\ottnt{T}$ and $\Phi$.

           Now, it suffices to show that
           \[
           \ottnt{v_{{\mathrm{0}}}} \,  \in  \,  \LRRel{ \mathcal{V} }{  \ottnt{C}  . \ottnt{T}  }  ~.
           \]
           Because $ \emptyset   \vdash   \ottnt{T}  \,  \,\&\,  \,  \Phi  \, / \,   \square    \ottsym{<:}  \ottnt{C}$ implies $ \emptyset   \vdash  \ottnt{T}  \ottsym{<:}   \ottnt{C}  . \ottnt{T} $, and
           \reflem{ts-wf-ty-tctx-typing} with $\ottnt{e_{{\mathrm{0}}}} \,  \in  \,  \LRRel{ \mathrm{Atom} }{ \ottnt{C} } $ implies
           $ \emptyset   \vdash   \ottnt{C}  . \ottnt{T} $,
           we have $ \emptyset   \vdash  \ottnt{v_{{\mathrm{0}}}}  \ottsym{:}   \ottnt{C}  . \ottnt{T} $ by \T{Sub}, that is, $\ottnt{v_{{\mathrm{0}}}} \,  \in  \,  \LRRel{ \mathrm{Atom} }{  \ottnt{C}  . \ottnt{T}  } $.
           Therefore, by the IH (\ref{lem:lr-approx-lr:val}), it suffices to show that
           \[
              \exists  \, { \ottnt{v'_{{\mathrm{0}}}} } . \    \forall  \, { \ottmv{i}  \ottsym{,}  \pi } . \  \ottsym{(}  \ottnt{v_{{\mathrm{0}}}} \,  \prec_{ \ottnt{v} }  \, \ottnt{v'_{{\mathrm{0}}}} \,  \uparrow \, (  \ottmv{i}  ,  \pi  )   \ottsym{)} \,  \in  \,  \LRRel{ \mathcal{V} }{  \ottnt{C}  . \ottnt{T}  }    ~.
           \]
           Because $\ottnt{e_{{\mathrm{0}}}} \,  \mathrel{  \longrightarrow  ^*}  \, \ottnt{v_{{\mathrm{0}}}} \,  \,\&\,  \, \varpi$ and $\ottnt{e_{{\mathrm{0}}}} \,  \prec_{ \ottnt{v} }  \, \ottnt{e'_{{\mathrm{0}}}}$,
           Lemmas~\ref{lem:lr-approx-eval-nsteps} and \ref{lem:lr-approx-val} imply that
           there exist $\ottmv{j}$ and $\ottnt{v'_{{\mathrm{0}}}}$ such that
           \[
              \forall  \, { \pi } . \    \forall  \, { \ottmv{i} \,  \geq  \, \ottmv{j} } . \  \ottsym{(}  \ottnt{e_{{\mathrm{0}}}} \,  \prec_{ \ottnt{v} }  \, \ottnt{e'_{{\mathrm{0}}}} \,  \uparrow \, (  \ottmv{i}  ,  \pi  )   \ottsym{)} \,  \mathrel{  \longrightarrow  ^*}  \, \ottsym{(}  \ottnt{v_{{\mathrm{0}}}} \,  \prec_{ \ottnt{v} }  \, \ottnt{v'_{{\mathrm{0}}}} \,  \uparrow \, (  \ottmv{i}  \ottsym{-}  \ottmv{j}  ,  \pi  )   \ottsym{)} \,  \,\&\,  \, \varpi   ~.
           \]

           Let $\pi$ be any infinite trace and $\ottmv{i} \,  \geq  \, \ottmv{j}$.
           Then, we have
           $\ottsym{(}  \ottnt{e_{{\mathrm{0}}}} \,  \prec_{ \ottnt{v} }  \, \ottnt{e'_{{\mathrm{0}}}} \,  \uparrow \, (  \ottmv{i}  ,  \pi  )   \ottsym{)} \,  \mathrel{  \longrightarrow  ^*}  \, \ottsym{(}  \ottnt{v_{{\mathrm{0}}}} \,  \prec_{ \ottnt{v} }  \, \ottnt{v'_{{\mathrm{0}}}} \,  \uparrow \, (  \ottmv{i}  \ottsym{-}  \ottmv{j}  ,  \pi  )   \ottsym{)} \,  \,\&\,  \, \varpi$.
           Because $\ottnt{v_{{\mathrm{0}}}} \,  \prec_{ \ottnt{v} }  \, \ottnt{v'_{{\mathrm{0}}}} \,  \uparrow \, (  \ottmv{i}  \ottsym{-}  \ottmv{j}  ,  \pi  ) $ is a value by \reflem{lr-approx-val},
           we have
           \[
            \ottnt{e_{{\mathrm{0}}}} \,  \prec_{ \ottnt{v} }  \, \ottnt{e'_{{\mathrm{0}}}} \,  \uparrow \, (  \ottmv{i}  ,  \pi  )  \,  \Downarrow  \, \ottnt{v_{{\mathrm{0}}}} \,  \prec_{ \ottnt{v} }  \, \ottnt{v'_{{\mathrm{0}}}} \,  \uparrow \, (  \ottmv{i}  \ottsym{-}  \ottmv{j}  ,  \pi  )  \,  \,\&\,  \, \varpi ~.
           \]
           By the assumption, $\ottsym{(}  \ottnt{e_{{\mathrm{0}}}} \,  \prec_{ \ottnt{v} }  \, \ottnt{e'_{{\mathrm{0}}}} \,  \uparrow \, (  \ottmv{i}  ,  \pi  )   \ottsym{)} \,  \in  \,  \LRRel{ \mathcal{E} }{ \ottnt{C} } $.
           Therefore, $\ottsym{(}  \ottnt{v_{{\mathrm{0}}}} \,  \prec_{ \ottnt{v} }  \, \ottnt{v'_{{\mathrm{0}}}} \,  \uparrow \, (  \ottmv{i}  \ottsym{-}  \ottmv{j}  ,  \pi  )   \ottsym{)} \,  \in  \,  \LRRel{ \mathcal{V} }{  \ottnt{C}  . \ottnt{T}  } $.
           Because $\ottmv{i}  \ottsym{-}  \ottmv{j}$ can be any natural number and $\pi$ can be any infinite trace,
           we have the conclusion
           \[
              \forall  \, { \ottmv{i}  \ottsym{,}  \pi } . \  \ottsym{(}  \ottnt{v_{{\mathrm{0}}}} \,  \prec_{ \ottnt{v} }  \, \ottnt{v'_{{\mathrm{0}}}} \,  \uparrow \, (  \ottmv{i}  ,  \pi  )   \ottsym{)} \,  \in  \,  \LRRel{ \mathcal{V} }{  \ottnt{C}  . \ottnt{T}  }   ~.
           \]

          \case $  \exists  \, { \ottnt{K_{{\mathrm{1}}}}  \ottsym{,}  \mathit{x_{{\mathrm{1}}}}  \ottsym{,}  \ottnt{e_{{\mathrm{1}}}}  \ottsym{,}  \varpi } . \  \ottnt{e_{{\mathrm{0}}}} \,  \Downarrow  \,  \ottnt{K_{{\mathrm{1}}}}  [  \mathcal{S}_0 \, \mathit{x_{{\mathrm{1}}}}  \ottsym{.}  \ottnt{e_{{\mathrm{1}}}}  ]  \,  \,\&\,  \, \varpi $:
           Because $\ottnt{e_{{\mathrm{0}}}} \,  \mathrel{  \longrightarrow  ^*}  \,  \ottnt{K_{{\mathrm{1}}}}  [  \mathcal{S}_0 \, \mathit{x_{{\mathrm{1}}}}  \ottsym{.}  \ottnt{e_{{\mathrm{1}}}}  ]  \,  \,\&\,  \, \varpi$ and $\ottnt{e_{{\mathrm{0}}}} \,  \prec_{ \ottnt{v} }  \, \ottnt{e'_{{\mathrm{0}}}}$,
           \reflem{lr-approx-eval-nsteps} implies that
           there exist $\ottmv{j}$ and $\ottnt{e'_{{\mathrm{01}}}}$ such that
           \[
              \forall  \, { \pi } . \    \forall  \, { \ottmv{i} \,  \geq  \, \ottmv{j} } . \  \ottsym{(}  \ottnt{e_{{\mathrm{0}}}} \,  \prec_{ \ottnt{v} }  \, \ottnt{e'_{{\mathrm{0}}}} \,  \uparrow \, (  \ottmv{i}  ,  \pi  )   \ottsym{)} \,  \mathrel{  \longrightarrow  ^*}  \, \ottsym{(}   \ottnt{K_{{\mathrm{1}}}}  [  \mathcal{S}_0 \, \mathit{x_{{\mathrm{1}}}}  \ottsym{.}  \ottnt{e_{{\mathrm{1}}}}  ]  \,  \prec_{ \ottnt{v} }  \, \ottnt{e'_{{\mathrm{01}}}} \,  \uparrow \, (  \ottmv{i}  \ottsym{-}  \ottmv{j}  ,  \pi  )   \ottsym{)} \,  \,\&\,  \, \varpi   ~.
           \]
           Let $\pi$ be any infinite trace and $\ottmv{i} \,  \geq  \, \ottmv{j}$.
           Then, we have
           \[
            \ottsym{(}  \ottnt{e_{{\mathrm{0}}}} \,  \prec_{ \ottnt{v} }  \, \ottnt{e'_{{\mathrm{0}}}} \,  \uparrow \, (  \ottmv{i}  ,  \pi  )   \ottsym{)} \,  \mathrel{  \longrightarrow  ^*}  \, \ottsym{(}   \ottnt{K_{{\mathrm{1}}}}  [  \mathcal{S}_0 \, \mathit{x_{{\mathrm{1}}}}  \ottsym{.}  \ottnt{e_{{\mathrm{1}}}}  ]  \,  \prec_{ \ottnt{v} }  \, \ottnt{e'_{{\mathrm{01}}}} \,  \uparrow \, (  \ottmv{i}  \ottsym{-}  \ottmv{j}  ,  \pi  )   \ottsym{)} \,  \,\&\,  \, \varpi ~.
           \]
           By Lemmas~\ref{lem:lr-approx-wf} and \ref{lem:lr-approx-ectx},
           there exist some $\ottnt{K'_{{\mathrm{1}}}}$ and $\ottnt{e'_{{\mathrm{1}}}}$ such that
           \begin{itemize}
            \item $\ottnt{e'_{{\mathrm{01}}}} \,  =  \,  \ottnt{K'_{{\mathrm{1}}}}  [  \mathcal{S}_0 \, \mathit{x_{{\mathrm{1}}}}  \ottsym{.}  \ottnt{e'_{{\mathrm{1}}}}  ] $,
            \item $\ottnt{K_{{\mathrm{1}}}} \,  \prec_{ \ottnt{v} }  \, \ottnt{K'_{{\mathrm{1}}}}$, and
            \item $\ottsym{(}  \mathcal{S}_0 \, \mathit{x_{{\mathrm{1}}}}  \ottsym{.}  \ottnt{e_{{\mathrm{1}}}}  \ottsym{)} \,  \prec_{ \ottnt{v} }  \, \ottsym{(}  \mathcal{S}_0 \, \mathit{x_{{\mathrm{1}}}}  \ottsym{.}  \ottnt{e'_{{\mathrm{1}}}}  \ottsym{)}$ (so $\ottnt{e_{{\mathrm{1}}}} \,  \prec_{ \ottnt{v} }  \, \ottnt{e'_{{\mathrm{1}}}}$).
           \end{itemize}
           We have
           \[\begin{array}{ll} &
             \ottnt{K_{{\mathrm{1}}}}  [  \mathcal{S}_0 \, \mathit{x_{{\mathrm{1}}}}  \ottsym{.}  \ottnt{e_{{\mathrm{1}}}}  ]  \,  \prec_{ \ottnt{v} }  \, \ottnt{e'_{{\mathrm{01}}}} \,  \uparrow \, (  \ottmv{i}  \ottsym{-}  \ottmv{j}  ,  \pi  )  \\ = &
             \ottnt{K_{{\mathrm{1}}}}  [  \mathcal{S}_0 \, \mathit{x_{{\mathrm{1}}}}  \ottsym{.}  \ottnt{e_{{\mathrm{1}}}}  ]  \,  \prec_{ \ottnt{v} }  \,  \ottnt{K'_{{\mathrm{1}}}}  [  \mathcal{S}_0 \, \mathit{x_{{\mathrm{1}}}}  \ottsym{.}  \ottnt{e'_{{\mathrm{1}}}}  ]  \,  \uparrow \, (  \ottmv{i}  \ottsym{-}  \ottmv{j}  ,  \pi  )  \\ = &
             \ottsym{(}  \ottnt{K_{{\mathrm{1}}}} \,  \prec_{ \ottnt{v} }  \, \ottnt{K'_{{\mathrm{1}}}} \,  \uparrow \, (  \ottmv{i}  \ottsym{-}  \ottmv{j}  ,  \pi  )   \ottsym{)}  [  \mathcal{S}_0 \, \mathit{x_{{\mathrm{1}}}}  \ottsym{.}  \ottsym{(}  \ottnt{e_{{\mathrm{1}}}} \,  \prec_{ \ottnt{v} }  \, \ottnt{e'_{{\mathrm{1}}}} \,  \uparrow \, (  \ottmv{i}  \ottsym{-}  \ottmv{j}  ,  \pi  )   \ottsym{)}  ] 
             \quad \text{(by \reflem{lr-approx-embed-exp})} ~.
             \end{array}
           \]
           Therefore, we have
           \[
            \ottsym{(}  \ottnt{e_{{\mathrm{0}}}} \,  \prec_{ \ottnt{v} }  \, \ottnt{e'_{{\mathrm{0}}}} \,  \uparrow \, (  \ottmv{i}  ,  \pi  )   \ottsym{)} \,  \Downarrow  \,  \ottsym{(}  \ottnt{K_{{\mathrm{1}}}} \,  \prec_{ \ottnt{v} }  \, \ottnt{K'_{{\mathrm{1}}}} \,  \uparrow \, (  \ottmv{i}  \ottsym{-}  \ottmv{j}  ,  \pi  )   \ottsym{)}  [  \mathcal{S}_0 \, \mathit{x_{{\mathrm{1}}}}  \ottsym{.}  \ottsym{(}  \ottnt{e_{{\mathrm{1}}}} \,  \prec_{ \ottnt{v} }  \, \ottnt{e'_{{\mathrm{1}}}} \,  \uparrow \, (  \ottmv{i}  \ottsym{-}  \ottmv{j}  ,  \pi  )   \ottsym{)}  ]  \,  \,\&\,  \, \varpi ~.
           \]
           By the assumption, $\ottsym{(}  \ottnt{e_{{\mathrm{0}}}} \,  \prec_{ \ottnt{v} }  \, \ottnt{e'_{{\mathrm{0}}}} \,  \uparrow \, (  \ottmv{i}  ,  \pi  )   \ottsym{)} \,  \in  \,  \LRRel{ \mathcal{E} }{ \ottnt{C} } $.
           Therefore, there exist some $\mathit{x}$, $\ottnt{C_{{\mathrm{1}}}}$, and $\ottnt{C_{{\mathrm{2}}}}$ such that
           \begin{itemize}
            \item $ \ottnt{C}  . \ottnt{S}  \,  =  \,  ( \forall  \mathit{x}  .  \ottnt{C_{{\mathrm{1}}}}  )  \Rightarrow   \ottnt{C_{{\mathrm{2}}}} $ and
            \item $  \forall  \, { \ottnt{K} \,  \in  \,  \LRRel{ \mathcal{K} }{   \ottnt{C}  . \ottnt{T}   \, / \, ( \forall  \mathit{x}  .  \ottnt{C_{{\mathrm{1}}}}  )  }  } . \   \resetcnstr{  \ottnt{K}  [  \ottnt{e_{{\mathrm{0}}}} \,  \prec_{ \ottnt{v} }  \, \ottnt{e'_{{\mathrm{0}}}} \,  \uparrow \, (  \ottmv{i}  ,  \pi  )   ]  }  \,  \in  \,  \LRRel{ \mathcal{E} }{ \ottnt{C_{{\mathrm{2}}}} }  $.
           \end{itemize}
           Because $\ottmv{i}$ and $\pi$ are arbitrary, we have
           \begin{equation}
              \forall  \, { \ottmv{i}  \ottsym{,}  \pi }, { \ottnt{K} \,  \in  \,  \LRRel{ \mathcal{K} }{   \ottnt{C}  . \ottnt{T}   \, / \, ( \forall  \mathit{x}  .  \ottnt{C_{{\mathrm{1}}}}  )  }  } . \   \resetcnstr{  \ottnt{K}  [  \ottnt{e_{{\mathrm{0}}}} \,  \prec_{ \ottnt{v} }  \, \ottnt{e'_{{\mathrm{0}}}} \,  \uparrow \, (  \ottmv{i}  ,  \pi  )   ]  }  \,  \in  \,  \LRRel{ \mathcal{E} }{ \ottnt{C_{{\mathrm{2}}}} }   ~.
             \label{lem:lr-approx-lr:exp:shift:approx}
          \end{equation}

          Now, it suffices to show that
          \[
             \forall  \, { \ottnt{K} \,  \in  \,  \LRRel{ \mathcal{K} }{   \ottnt{C}  . \ottnt{T}   \, / \, ( \forall  \mathit{x}  .  \ottnt{C_{{\mathrm{1}}}}  )  }  } . \   \resetcnstr{  \ottnt{K}  [  \ottnt{e_{{\mathrm{0}}}}  ]  }  \,  \in  \,  \LRRel{ \mathcal{E} }{ \ottnt{C_{{\mathrm{2}}}} }   ~.
          \]
          Let $\ottnt{K} \,  \in  \,  \LRRel{ \mathcal{K} }{   \ottnt{C}  . \ottnt{T}   \, / \, ( \forall  \mathit{x}  .  \ottnt{C_{{\mathrm{1}}}}  )  } $.
          Because $\ottnt{e_{{\mathrm{0}}}} \,  \in  \,  \LRRel{ \mathrm{Atom} }{ \ottnt{C} } $, we have
          $ \resetcnstr{  \ottnt{K}  [  \ottnt{e_{{\mathrm{0}}}}  ]  }  \,  \in  \,  \LRRel{ \mathrm{Atom} }{ \ottnt{C_{{\mathrm{2}}}} } $
          by \reflem{lr-typing-cont-exp}.
          Therefore, by the IH (\ref{lem:lr-approx-lr:exp}),
          it suffices to show that
          \[
             \forall  \, { \ottmv{i}  \ottsym{,}  \pi } . \  \ottsym{(}   \resetcnstr{  \ottnt{K}  [  \ottnt{e_{{\mathrm{0}}}}  ]  }  \,  \prec_{ \ottnt{v} }  \,  \resetcnstr{  \ottnt{K}  [  \ottnt{e'_{{\mathrm{0}}}}  ]  }  \,  \uparrow \, (  \ottmv{i}  ,  \pi  )   \ottsym{)} \,  \in  \,  \LRRel{ \mathcal{E} }{ \ottnt{C_{{\mathrm{2}}}} }   ~.
          \]
          By (\ref{lem:lr-approx-lr:exp:shift:approx}),
          \[
             \forall  \, { \ottmv{i}  \ottsym{,}  \pi } . \   \resetcnstr{  \ottnt{K}  [  \ottnt{e_{{\mathrm{0}}}} \,  \prec_{ \ottnt{v} }  \, \ottnt{e'_{{\mathrm{0}}}} \,  \uparrow \, (  \ottmv{i}  ,  \pi  )   ]  }  \,  \in  \,  \LRRel{ \mathcal{E} }{ \ottnt{C_{{\mathrm{2}}}} }   ~.
          \]
          By Lemmas~\ref{lem:lr-approx-refl} and \ref{lem:lr-approx-embed-exp},
          we have the conclusion
          \[
             \forall  \, { \ottmv{i}  \ottsym{,}  \pi } . \   \resetcnstr{  \ottnt{K}  [  \ottnt{e_{{\mathrm{0}}}}  ]  }  \,  \prec_{ \ottnt{v} }  \,  \resetcnstr{  \ottnt{K}  [  \ottnt{e'_{{\mathrm{0}}}}  ]  }  \,  \uparrow \, (  \ottmv{i}  ,  \pi  )  \,  \in  \,  \LRRel{ \mathcal{E} }{ \ottnt{C_{{\mathrm{2}}}} }   ~.
          \]

          \case $  \exists  \, { \pi } . \  \ottnt{e_{{\mathrm{0}}}} \,  \Uparrow  \, \pi $:
           Because $\ottnt{e_{{\mathrm{0}}}} \,  \prec_{ \ottnt{v} }  \, \ottnt{e'_{{\mathrm{0}}}}$,
           \reflem{lr-approx-div} implies that there exists some $\pi'$ such that
           $\ottsym{(}  \ottnt{e_{{\mathrm{0}}}} \,  \prec_{ \ottnt{v} }  \, \ottnt{e'_{{\mathrm{0}}}} \,  \uparrow \, (  \ottsym{0}  ,  \pi'  )   \ottsym{)} \,  \Uparrow  \, \pi$.
           Because the assumption implies $\ottsym{(}  \ottnt{e_{{\mathrm{0}}}} \,  \prec_{ \ottnt{v} }  \, \ottnt{e'_{{\mathrm{0}}}} \,  \uparrow \, (  \ottsym{0}  ,  \pi'  )   \ottsym{)} \,  \in  \,  \LRRel{ \mathcal{E} }{ \ottnt{C} } $,
           we have the conclusion $ { \models  \,   { \ottnt{C} }^\nu (  \pi  )  } $.
         \end{caseanalysis}

  \item By case analysis on $\ottnt{T}$.
        \begin{caseanalysis}
         \case $  \exists  \, { \mathit{x}  \ottsym{,}  \iota  \ottsym{,}  \phi } . \  \ottnt{T} \,  =  \, \ottsym{\{}  \mathit{x} \,  {:}  \, \iota \,  |  \, \phi  \ottsym{\}} $: Obvious.
         \case $  \exists  \, { \mathit{x}  \ottsym{,}  \ottnt{T'}  \ottsym{,}  \ottnt{C'} } . \  \ottnt{T} \,  =  \, \ottsym{(}  \mathit{x} \,  {:}  \, \ottnt{T'}  \ottsym{)}  \rightarrow  \ottnt{C'} $:
          Let $\ottnt{v_{{\mathrm{1}}}} \,  \in  \,  \LRRel{ \mathcal{V} }{ \ottnt{T'} } $.
          Then, we must show that
          \[
           \ottnt{v_{{\mathrm{0}}}} \, \ottnt{v_{{\mathrm{1}}}} \,  \in  \,  \LRRel{ \mathcal{E} }{  \ottnt{C'}    [  \ottnt{v_{{\mathrm{1}}}}  /  \mathit{x}  ]   }  ~.
          \]
          Because $  \forall  \, { \ottmv{i}  \ottsym{,}  \pi } . \  \ottsym{(}  \ottnt{v_{{\mathrm{0}}}} \,  \prec_{ \ottnt{v} }  \, \ottnt{v'_{{\mathrm{0}}}} \,  \uparrow \, (  \ottmv{i}  ,  \pi  )   \ottsym{)} \,  \in  \,  \LRRel{ \mathcal{V} }{ \ottnt{T} }  $
          by the assumption, and $\ottnt{T} \,  =  \, \ottsym{(}  \mathit{x} \,  {:}  \, \ottnt{T'}  \ottsym{)}  \rightarrow  \ottnt{C'}$,
          we have
          \[
             \forall  \, { \ottmv{i}  \ottsym{,}  \pi } . \  \ottsym{(}  \ottnt{v_{{\mathrm{0}}}} \,  \prec_{ \ottnt{v} }  \, \ottnt{v'_{{\mathrm{0}}}} \,  \uparrow \, (  \ottmv{i}  ,  \pi  )   \ottsym{)} \, \ottnt{v_{{\mathrm{1}}}} \,  \in  \,  \LRRel{ \mathcal{E} }{  \ottnt{C'}    [  \ottnt{v_{{\mathrm{1}}}}  /  \mathit{x}  ]   }   ~.
          \]
          With \reflem{lr-approx-refl},
          \[
             \forall  \, { \ottmv{i}  \ottsym{,}  \pi } . \  \ottsym{(}  \ottsym{(}  \ottnt{v_{{\mathrm{0}}}} \, \ottnt{v_{{\mathrm{1}}}}  \ottsym{)} \,  \prec_{ \ottnt{v} }  \, \ottsym{(}  \ottnt{v'_{{\mathrm{0}}}} \, \ottnt{v_{{\mathrm{1}}}}  \ottsym{)} \,  \uparrow \, (  \ottmv{i}  ,  \pi  )   \ottsym{)} \,  \in  \,  \LRRel{ \mathcal{E} }{  \ottnt{C'}    [  \ottnt{v_{{\mathrm{1}}}}  /  \mathit{x}  ]   }   ~.
          \]
          Because $\ottnt{v_{{\mathrm{0}}}} \, \ottnt{v_{{\mathrm{1}}}} \,  \in  \,  \LRRel{ \mathrm{Atom} }{  \ottnt{C'}    [  \ottnt{v_{{\mathrm{1}}}}  /  \mathit{x}  ]   } $ by \T{App},
          the IH (\ref{lem:lr-approx-lr:exp}) implies
          the conclusion $\ottnt{v_{{\mathrm{0}}}} \, \ottnt{v_{{\mathrm{1}}}} \,  \in  \,  \LRRel{ \mathcal{E} }{  \ottnt{C'}    [  \ottnt{v_{{\mathrm{1}}}}  /  \mathit{x}  ]   } $.

         \case $  \exists  \, { \mathit{X}  \ottsym{,}   \tceseq{ \ottnt{s} }   \ottsym{,}  \ottnt{C} } . \  \ottnt{T} \,  =  \,  \forall   \ottsym{(}  \mathit{X} \,  {:}  \,  \tceseq{ \ottnt{s} }   \ottsym{)}  \ottsym{.}  \ottnt{C} $:
          Let $\ottnt{A}$ be a predicate such that $ \emptyset   \vdash  \ottnt{A}  \ottsym{:}   \tceseq{ \ottnt{s} } $.
          Then, we must show that
          \[
           \ottnt{v_{{\mathrm{0}}}} \, \ottnt{A} \,  \in  \,  \LRRel{ \mathcal{E} }{  \ottnt{C}    [  \ottnt{A}  /  \mathit{X}  ]   }  ~.
          \]
          Because $  \forall  \, { \ottmv{i}  \ottsym{,}  \pi } . \  \ottsym{(}  \ottnt{v_{{\mathrm{0}}}} \,  \prec_{ \ottnt{v} }  \, \ottnt{v'_{{\mathrm{0}}}} \,  \uparrow \, (  \ottmv{i}  ,  \pi  )   \ottsym{)} \,  \in  \,  \LRRel{ \mathcal{V} }{ \ottnt{T} }  $
          by the assumption, and $\ottnt{T} \,  =  \,  \forall   \ottsym{(}  \mathit{X} \,  {:}  \,  \tceseq{ \ottnt{s} }   \ottsym{)}  \ottsym{.}  \ottnt{C}$,
          we have
          \[
             \forall  \, { \ottmv{i}  \ottsym{,}  \pi } . \  \ottsym{(}  \ottnt{v_{{\mathrm{0}}}} \,  \prec_{ \ottnt{v} }  \, \ottnt{v'_{{\mathrm{0}}}} \,  \uparrow \, (  \ottmv{i}  ,  \pi  )   \ottsym{)} \, \ottnt{A} \,  \in  \,  \LRRel{ \mathcal{E} }{  \ottnt{C}    [  \ottnt{A}  /  \mathit{X}  ]   }   ~.
          \]
          With \reflem{lr-approx-refl},
          \[
             \forall  \, { \ottmv{i}  \ottsym{,}  \pi } . \  \ottsym{(}  \ottsym{(}  \ottnt{v_{{\mathrm{0}}}} \, \ottnt{A}  \ottsym{)} \,  \prec_{ \ottnt{v} }  \, \ottsym{(}  \ottnt{v'_{{\mathrm{0}}}} \, \ottnt{A}  \ottsym{)} \,  \uparrow \, (  \ottmv{i}  ,  \pi  )   \ottsym{)} \,  \in  \,  \LRRel{ \mathcal{E} }{  \ottnt{C}    [  \ottnt{A}  /  \mathit{X}  ]   }   ~.
          \]
          Because $\ottnt{v_{{\mathrm{0}}}} \, \ottnt{A} \,  \in  \,  \LRRel{ \mathrm{Atom} }{  \ottnt{C}    [  \ottnt{A}  /  \mathit{X}  ]   } $ by \T{PApp},
          the IH (\ref{lem:lr-approx-lr:exp}) implies the conclusion
          $\ottnt{v_{{\mathrm{0}}}} \, \ottnt{A} \,  \in  \,  \LRRel{ \mathcal{E} }{  \ottnt{C}    [  \ottnt{A}  /  \mathit{X}  ]   } $.
        \end{caseanalysis}
 \end{enumerate}
\end{proof}

\begin{lemmap}{Weakening Predicates in Valuation of Formulas}{ts-logic-weak-pred}
 Suppose that $ \mathit{fpv}^\mathbf{p}  (  \phi  ) $ is the set of predicate variables that occur positively
 in a formula $\phi$.
 Assume that $\mathbf{\mathit{f} }_{{\mathrm{1}}} \,  \sqsubseteq_{ \mathcal{T} }  \, \mathbf{\mathit{f} }_{{\mathrm{2}}}$.
 \begin{enumerate}
  \item If $ { [\![  \phi  ]\!]_{  \{  \mathit{Y}  \mapsto  \mathbf{\mathit{f} }_{{\mathrm{1}}}  \}   \mathbin{\uplus}  \theta } }  \,  =  \,  \top $ and $\mathit{Y} \,  \not\in  \,  \mathit{fpv}^\mathbf{n}  (  \phi  ) $, then $ { [\![  \phi  ]\!]_{  \{  \mathit{Y}  \mapsto  \mathbf{\mathit{f} }_{{\mathrm{2}}}  \}   \mathbin{\uplus}  \theta } }  \,  =  \,  \top $ and
  \item If $ { [\![  \phi  ]\!]_{  \{  \mathit{Y}  \mapsto  \mathbf{\mathit{f} }_{{\mathrm{1}}}  \}   \mathbin{\uplus}  \theta } }  \,  =  \,  \bot $ and $\mathit{Y} \,  \not\in  \,  \mathit{fpv}^\mathbf{p}  (  \phi  ) $, then $ { [\![  \phi  ]\!]_{  \{  \mathit{Y}  \mapsto  \mathbf{\mathit{f} }_{{\mathrm{2}}}  \}   \mathbin{\uplus}  \theta } }  \,  =  \,  \bot $.
 \end{enumerate}
\end{lemmap}
\begin{proof}
 \TS{CHECKED (NO CHANGE): 2022/07/05}
 Let $\theta_{{\mathrm{1}}} \,  =  \,  \{  \mathit{Y}  \mapsto  \mathbf{\mathit{f} }_{{\mathrm{1}}}  \}   \mathbin{\uplus}  \theta$ and $\theta_{{\mathrm{2}}} \,  =  \,  \{  \mathit{Y}  \mapsto  \mathbf{\mathit{f} }_{{\mathrm{2}}}  \}   \mathbin{\uplus}  \theta$.
 By induction on $\phi$.
 \begin{caseanalysis}
  \case $\phi \,  =  \,  \top ,  \bot $, or $\ottnt{P}  \ottsym{(}   \tceseq{ \ottnt{t} }   \ottsym{)}$: Obvious.

  \case $\phi \,  =  \,  \neg  \phi_{{\mathrm{0}}} ,  \phi_{{\mathrm{1}}}    \wedge    \phi_{{\mathrm{2}}} ,  \phi_{{\mathrm{1}}}    \vee    \phi_{{\mathrm{2}}} ,  \phi_{{\mathrm{1}}}    \Rightarrow    \phi_{{\mathrm{2}}} ,   \forall  \,  \mathit{x}  \mathrel{  \in  }  \ottnt{s}  . \  \phi_{{\mathrm{0}}} ,   \exists  \,  \mathit{x}  \mathrel{  \in  }  \ottnt{s}  . \  \phi_{{\mathrm{0}}} ,   \forall  \,  \mathit{X} \,  {:}  \,  \tceseq{ \ottnt{s} }   . \  \phi_{{\mathrm{0}}} $, or $  \exists  \,  \mathit{X} \,  {:}  \,  \tceseq{ \ottnt{s} }   . \  \phi_{{\mathrm{0}}} $:
   By the IH(s).

  \case $\phi \,  =  \, \ottsym{(}   \mu\!  \, \mathit{X}  \ottsym{(}   \tceseq{ \mathit{x}  \,  {:}  \,  \ottnt{s} }   \ottsym{)}  \ottsym{.}  \phi_{{\mathrm{0}}}  \ottsym{)}  \ottsym{(}   \tceseq{ \ottnt{t} }   \ottsym{)}$:
   Without loss of generality, we can suppose that $\mathit{X} \,  \not=  \, \mathit{Y}$.
   Let
   \begin{itemize}
    \item $\mathbf{\mathit{F} }_{{\mathrm{1}}} =  \lambda\!  \, \mathit{X}  \ottsym{.}   \lambda\!  \,  \tceseq{ \mathit{x}  \,  {:}  \,  \ottnt{s} }   \ottsym{.}   { [\![  \phi_{{\mathrm{0}}}  ]\!]_{ \theta_{{\mathrm{1}}}  \mathbin{\uplus}   \{  \mathit{X}  \mapsto  \mathit{X}  \}   \mathbin{\uplus}   \{ \tceseq{ \mathit{x}  \mapsto  \mathit{x} } \}  } } $ and
    \item $\mathbf{\mathit{F} }_{{\mathrm{2}}} =  \lambda\!  \, \mathit{X}  \ottsym{.}   \lambda\!  \,  \tceseq{ \mathit{x}  \,  {:}  \,  \ottnt{s} }   \ottsym{.}   { [\![  \phi_{{\mathrm{0}}}  ]\!]_{ \theta_{{\mathrm{2}}}  \mathbin{\uplus}   \{  \mathit{X}  \mapsto  \mathit{X}  \}   \mathbin{\uplus}   \{ \tceseq{ \mathit{x}  \mapsto  \mathit{x} } \}  } } $.
   \end{itemize}
   We prove each case.
   \begin{enumerate}
    \item We first show that
          \begin{equation}
             \forall  \, { \mathbf{\mathit{f} } \,  \in  \,  \mathcal{D}_{  \tceseq{ \ottnt{s} }   \rightarrow   \mathbb{B}  }  } . \  \mathbf{\mathit{F} }_{{\mathrm{1}}}  \ottsym{(}  \mathbf{\mathit{f} }  \ottsym{)} \,  \sqsubseteq_{  \tceseq{ \ottnt{s} }   \rightarrow   \mathbb{B}  }  \, \mathbf{\mathit{F} }_{{\mathrm{2}}}  \ottsym{(}  \mathbf{\mathit{f} }  \ottsym{)}  ~.
            \label{lem:lr-iff-spos-valid:mu:top:order}
          \end{equation}
          Let $\mathbf{\mathit{f} } \,  \in  \,  \mathcal{D}_{  \tceseq{ \ottnt{s} }   \rightarrow   \mathbb{B}  } $ and $ \tceseq{ \ottnt{d} \,  \in  \,  \mathcal{D}_{ \ottnt{s} }  } $.
          It suffices to show that
          \[
            { [\![  \phi_{{\mathrm{0}}}  ]\!]_{ \theta_{{\mathrm{1}}}  \mathbin{\uplus}   \{  \mathit{X}  \mapsto  \mathbf{\mathit{f} }  \}   \mathbin{\uplus}   \{ \tceseq{ \mathit{x}  \mapsto  \ottnt{d} } \}  } }  \,  \sqsubseteq_{  \mathbb{B}  }  \,  { [\![  \phi_{{\mathrm{0}}}  ]\!]_{ \theta_{{\mathrm{2}}}  \mathbin{\uplus}   \{  \mathit{X}  \mapsto  \mathbf{\mathit{f} }  \}   \mathbin{\uplus}   \{ \tceseq{ \mathit{x}  \mapsto  \ottnt{d} } \}  } }  ~.
          \]
          That is, it suffices to show that
          \[
            \ottsym{(}   { [\![  \phi_{{\mathrm{0}}}  ]\!]_{ \theta_{{\mathrm{1}}}  \mathbin{\uplus}   \{  \mathit{X}  \mapsto  \mathbf{\mathit{f} }  \}   \mathbin{\uplus}   \{ \tceseq{ \mathit{x}  \mapsto  \ottnt{d} } \}  } }  \,  =  \,  \top   \ottsym{)}  \mathrel{\Longrightarrow}  \ottsym{(}   { [\![  \phi_{{\mathrm{0}}}  ]\!]_{ \theta_{{\mathrm{2}}}  \mathbin{\uplus}   \{  \mathit{X}  \mapsto  \mathbf{\mathit{f} }  \}   \mathbin{\uplus}   \{ \tceseq{ \mathit{x}  \mapsto  \ottnt{d} } \}  } }  \,  =  \,  \top   \ottsym{)}  ~.
          \]
          This is proven by the IH.

          Next, we show that
          \begin{equation}
            \mathsf{lfp}^{  \tceseq{ \ottnt{s} }   \rightarrow   \mathbb{B}  }   \ottsym{(}  \mathbf{\mathit{F} }_{{\mathrm{1}}}  \ottsym{)} \,  \sqsubseteq_{  \tceseq{ \ottnt{s} }   \rightarrow   \mathbb{B}  }  \,  \mathsf{lfp}^{  \tceseq{ \ottnt{s} }   \rightarrow   \mathbb{B}  }   \ottsym{(}  \mathbf{\mathit{F} }_{{\mathrm{2}}}  \ottsym{)} ~.
            \label{lem:lr-iff-spos-valid:mu:top:lfp}
          \end{equation}
          By definition, it suffices to show that
          $ \mathsf{lfp}^{  \tceseq{ \ottnt{s} }   \rightarrow   \mathbb{B}  }   \ottsym{(}  \mathbf{\mathit{F} }_{{\mathrm{1}}}  \ottsym{)}$ is a lower bound of
          $ \{ \,  \mathbf{\mathit{f} }  \mathrel{  \in  }   \mathcal{D}_{  \tceseq{ \ottnt{s} }   \rightarrow   \mathbb{B}  }   \mid  \mathbf{\mathit{F} }_{{\mathrm{2}}}  \ottsym{(}  \mathbf{\mathit{f} }  \ottsym{)}  \mathrel{  \sqsubseteq_{  \tceseq{ \ottnt{s} }   \rightarrow   \mathbb{B}  }  }  \mathbf{\mathit{f} }  \, \} $.
          Let $\mathbf{\mathit{f} } \,  \in  \,  \mathcal{D}_{  \tceseq{ \ottnt{s} }   \rightarrow   \mathbb{B}  } $ such that $\mathbf{\mathit{F} }_{{\mathrm{2}}}  \ottsym{(}  \mathbf{\mathit{f} }  \ottsym{)} \,  \sqsubseteq_{  \tceseq{ \ottnt{s} }   \rightarrow   \mathbb{B}  }  \, \mathbf{\mathit{f} }$.
          Then, it suffices to show that
          \[
            \mathsf{lfp}^{  \tceseq{ \ottnt{s} }   \rightarrow   \mathbb{B}  }   \ottsym{(}  \mathbf{\mathit{F} }_{{\mathrm{1}}}  \ottsym{)} \,  \sqsubseteq_{  \tceseq{ \ottnt{s} }   \rightarrow   \mathbb{B}  }  \, \mathbf{\mathit{f} } ~.
          \]
          Because $ \mathsf{lfp}^{  \tceseq{ \ottnt{s} }   \rightarrow   \mathbb{B}  }   \ottsym{(}  \mathbf{\mathit{F} }_{{\mathrm{1}}}  \ottsym{)}$ is the greatest lower bound of
          $ \{ \,  \mathbf{\mathit{f} }'  \mathrel{  \in  }   \mathcal{D}_{  \tceseq{ \ottnt{s} }   \rightarrow   \mathbb{B}  }   \mid  \mathbf{\mathit{F} }_{{\mathrm{1}}}  \ottsym{(}  \mathbf{\mathit{f} }'  \ottsym{)}  \mathrel{  \sqsubseteq_{  \tceseq{ \ottnt{s} }   \rightarrow   \mathbb{B}  }  }  \mathbf{\mathit{f} }'  \, \} $,
          it suffices to show that
          \[
           \mathbf{\mathit{F} }_{{\mathrm{1}}}  \ottsym{(}  \mathbf{\mathit{f} }  \ottsym{)} \,  \sqsubseteq_{  \tceseq{ \ottnt{s} }   \rightarrow   \mathbb{B}  }  \, \mathbf{\mathit{f} } ~.
          \]
          Because
          $\mathbf{\mathit{F} }_{{\mathrm{1}}}  \ottsym{(}  \mathbf{\mathit{f} }  \ottsym{)} \,  \sqsubseteq_{  \tceseq{ \ottnt{s} }   \rightarrow   \mathbb{B}  }  \, \mathbf{\mathit{F} }_{{\mathrm{2}}}  \ottsym{(}  \mathbf{\mathit{f} }  \ottsym{)}$ by (\ref{lem:lr-iff-spos-valid:mu:top:order}), and
          $\mathbf{\mathit{F} }_{{\mathrm{2}}}  \ottsym{(}  \mathbf{\mathit{f} }  \ottsym{)} \,  \sqsubseteq_{  \tceseq{ \ottnt{s} }   \rightarrow   \mathbb{B}  }  \, \mathbf{\mathit{f} }$,
          we have (\ref{lem:lr-iff-spos-valid:mu:top:lfp}).

          Because
          $ { [\![  \ottsym{(}   \mu\!  \, \mathit{X}  \ottsym{(}   \tceseq{ \mathit{x}  \,  {:}  \,  \ottnt{s} }   \ottsym{)}  \ottsym{.}  \phi_{{\mathrm{0}}}  \ottsym{)}  \ottsym{(}   \tceseq{ \ottnt{t} }   \ottsym{)}  ]\!]_{ \theta_{{\mathrm{1}}} } }  \,  =  \,  \top $
          by the assumption,
          we have $ \mathsf{lfp}^{  \tceseq{ \ottnt{s} }   \rightarrow   \mathbb{B}  }   \ottsym{(}  \mathbf{\mathit{F} }_{{\mathrm{1}}}  \ottsym{)}  \ottsym{(}   \tceseq{  { [\![  \ottnt{t}  ]\!]_{ \theta } }  }   \ottsym{)} \,  =  \,  \top $.
          Therefore, (\ref{lem:lr-iff-spos-valid:mu:top:lfp}) implies
          $ \mathsf{lfp}^{  \tceseq{ \ottnt{s} }   \rightarrow   \mathbb{B}  }   \ottsym{(}  \mathbf{\mathit{F} }_{{\mathrm{2}}}  \ottsym{)}  \ottsym{(}   \tceseq{  { [\![  \ottnt{t}  ]\!]_{ \theta } }  }   \ottsym{)} \,  =  \,  \top $, that is,
          $ { [\![  \ottsym{(}   \mu\!  \, \mathit{X}  \ottsym{(}   \tceseq{ \mathit{x}  \,  {:}  \,  \ottnt{s} }   \ottsym{)}  \ottsym{.}  \phi_{{\mathrm{0}}}  \ottsym{)}  \ottsym{(}   \tceseq{ \ottnt{t} }   \ottsym{)}  ]\!]_{ \theta_{{\mathrm{2}}} } }  \,  =  \,  \top $, which is
          the conclusion we must show.

    \item We first show that
          \begin{equation}
             \forall  \, { \mathbf{\mathit{f} } \,  \in  \,  \mathcal{D}_{  \tceseq{ \ottnt{s} }   \rightarrow   \mathbb{B}  }  } . \  \mathbf{\mathit{F} }_{{\mathrm{2}}}  \ottsym{(}  \mathbf{\mathit{f} }  \ottsym{)} \,  \sqsubseteq_{  \tceseq{ \ottnt{s} }   \rightarrow   \mathbb{B}  }  \, \mathbf{\mathit{F} }_{{\mathrm{1}}}  \ottsym{(}  \mathbf{\mathit{f} }  \ottsym{)}  ~.
            \label{lem:lr-iff-spos-valid:mu:bot:order}
          \end{equation}
          Let $\mathbf{\mathit{f} } \,  \in  \,  \mathcal{D}_{  \tceseq{ \ottnt{s} }   \rightarrow   \mathbb{B}  } $ and $ \tceseq{ \ottnt{d} \,  \in  \,  \mathcal{D}_{ \ottnt{s} }  } $.
          It suffices to show that
          \[
            { [\![  \phi_{{\mathrm{0}}}  ]\!]_{ \theta_{{\mathrm{2}}}  \mathbin{\uplus}   \{  \mathit{X}  \mapsto  \mathbf{\mathit{f} }  \}   \mathbin{\uplus}   \{ \tceseq{ \mathit{x}  \mapsto  \ottnt{d} } \}  } }  \,  \sqsubseteq_{  \mathbb{B}  }  \,  { [\![  \phi_{{\mathrm{0}}}  ]\!]_{ \theta_{{\mathrm{1}}}  \mathbin{\uplus}   \{  \mathit{X}  \mapsto  \mathbf{\mathit{f} }  \}   \mathbin{\uplus}   \{ \tceseq{ \mathit{x}  \mapsto  \ottnt{d} } \}  } }  ~.
          \]
          By proof-by-contradiction, it suffices to show that
          \[
            \ottsym{(}   { [\![  \phi_{{\mathrm{0}}}  ]\!]_{ \theta_{{\mathrm{1}}}  \mathbin{\uplus}   \{  \mathit{X}  \mapsto  \mathbf{\mathit{f} }  \}   \mathbin{\uplus}   \{ \tceseq{ \mathit{x}  \mapsto  \ottnt{d} } \}  } }  \,  =  \,  \bot   \ottsym{)}  \mathrel{\Longrightarrow}  \ottsym{(}   { [\![  \phi_{{\mathrm{0}}}  ]\!]_{ \theta_{{\mathrm{2}}}  \mathbin{\uplus}   \{  \mathit{X}  \mapsto  \mathbf{\mathit{f} }  \}   \mathbin{\uplus}   \{ \tceseq{ \mathit{x}  \mapsto  \ottnt{d} } \}  } }  \,  =  \,  \bot   \ottsym{)}  ~.
          \]
          This is proven by the IH.

          Next, we show that
          \begin{equation}
            \mathsf{lfp}^{  \tceseq{ \ottnt{s} }   \rightarrow   \mathbb{B}  }   \ottsym{(}  \mathbf{\mathit{F} }_{{\mathrm{2}}}  \ottsym{)} \,  \sqsubseteq_{  \tceseq{ \ottnt{s} }   \rightarrow   \mathbb{B}  }  \,  \mathsf{lfp}^{  \tceseq{ \ottnt{s} }   \rightarrow   \mathbb{B}  }   \ottsym{(}  \mathbf{\mathit{F} }_{{\mathrm{1}}}  \ottsym{)} ~.
            \label{lem:lr-iff-spos-valid:mu:bot:lfp}
          \end{equation}
          By definition, it suffices to show that
          $ \mathsf{lfp}^{  \tceseq{ \ottnt{s} }   \rightarrow   \mathbb{B}  }   \ottsym{(}  \mathbf{\mathit{F} }_{{\mathrm{2}}}  \ottsym{)}$ is a lower bound of
          $ \{ \,  \mathbf{\mathit{f} }  \mathrel{  \in  }   \mathcal{D}_{  \tceseq{ \ottnt{s} }   \rightarrow   \mathbb{B}  }   \mid  \mathbf{\mathit{F} }_{{\mathrm{1}}}  \ottsym{(}  \mathbf{\mathit{f} }  \ottsym{)}  \mathrel{  \sqsubseteq_{  \tceseq{ \ottnt{s} }   \rightarrow   \mathbb{B}  }  }  \mathbf{\mathit{f} }  \, \} $.
          Let $\mathbf{\mathit{f} } \,  \in  \,  \mathcal{D}_{  \tceseq{ \ottnt{s} }   \rightarrow   \mathbb{B}  } $ such that $\mathbf{\mathit{F} }_{{\mathrm{1}}}  \ottsym{(}  \mathbf{\mathit{f} }  \ottsym{)} \,  \sqsubseteq_{  \tceseq{ \ottnt{s} }   \rightarrow   \mathbb{B}  }  \, \mathbf{\mathit{f} }$.
          Then, it suffices to show that
          \[
            \mathsf{lfp}^{  \tceseq{ \ottnt{s} }   \rightarrow   \mathbb{B}  }   \ottsym{(}  \mathbf{\mathit{F} }_{{\mathrm{2}}}  \ottsym{)} \,  \sqsubseteq_{  \tceseq{ \ottnt{s} }   \rightarrow   \mathbb{B}  }  \, \mathbf{\mathit{f} } ~.
          \]
          Because $ \mathsf{lfp}^{  \tceseq{ \ottnt{s} }   \rightarrow   \mathbb{B}  }   \ottsym{(}  \mathbf{\mathit{F} }_{{\mathrm{2}}}  \ottsym{)}$ is the greatest lower bound of
          $ \{ \,  \mathbf{\mathit{f} }'  \mathrel{  \in  }   \mathcal{D}_{  \tceseq{ \ottnt{s} }   \rightarrow   \mathbb{B}  }   \mid  \mathbf{\mathit{F} }_{{\mathrm{2}}}  \ottsym{(}  \mathbf{\mathit{f} }'  \ottsym{)}  \mathrel{  \sqsubseteq_{  \tceseq{ \ottnt{s} }   \rightarrow   \mathbb{B}  }  }  \mathbf{\mathit{f} }'  \, \} $,
          it suffices to show that
          \[
           \mathbf{\mathit{F} }_{{\mathrm{2}}}  \ottsym{(}  \mathbf{\mathit{f} }  \ottsym{)} \,  \sqsubseteq_{  \tceseq{ \ottnt{s} }   \rightarrow   \mathbb{B}  }  \, \mathbf{\mathit{f} } ~.
          \]
          Because
          $\mathbf{\mathit{F} }_{{\mathrm{2}}}  \ottsym{(}  \mathbf{\mathit{f} }  \ottsym{)} \,  \sqsubseteq_{  \tceseq{ \ottnt{s} }   \rightarrow   \mathbb{B}  }  \, \mathbf{\mathit{F} }_{{\mathrm{1}}}  \ottsym{(}  \mathbf{\mathit{f} }  \ottsym{)}$ by (\ref{lem:lr-iff-spos-valid:mu:bot:order}), and
          $\mathbf{\mathit{F} }_{{\mathrm{1}}}  \ottsym{(}  \mathbf{\mathit{f} }  \ottsym{)} \,  \sqsubseteq_{  \tceseq{ \ottnt{s} }   \rightarrow   \mathbb{B}  }  \, \mathbf{\mathit{f} }$,
          we have (\ref{lem:lr-iff-spos-valid:mu:bot:lfp}).

          Because
          $ { [\![  \ottsym{(}   \mu\!  \, \mathit{X}  \ottsym{(}   \tceseq{ \mathit{x}  \,  {:}  \,  \ottnt{s} }   \ottsym{)}  \ottsym{.}  \phi_{{\mathrm{0}}}  \ottsym{)}  \ottsym{(}   \tceseq{ \ottnt{t} }   \ottsym{)}  ]\!]_{ \theta_{{\mathrm{1}}} } }  \,  =  \,  \bot $
          by the assumption,
          we have either of the following:
          \begin{itemize}
           \item $ { [\![  \ottnt{t'}  ]\!]_{ \theta } }  \,  \not\in  \,  \mathcal{D}_{ \ottnt{s'} } $ for some $\ottnt{t'}$ in $ \tceseq{ \ottnt{t} } $
                 and the corresponding $\ottnt{s'}$ in $ \tceseq{ \ottnt{s} } $; or
           \item $ \mathsf{lfp}^{  \tceseq{ \ottnt{s} }   \rightarrow   \mathbb{B}  }   \ottsym{(}  \mathbf{\mathit{F} }_{{\mathrm{1}}}  \ottsym{)}  \ottsym{(}   \tceseq{  { [\![  \ottnt{t}  ]\!]_{ \theta } }  }   \ottsym{)} \,  =  \,  \bot $.
          \end{itemize}
          In the former case,
          $ { [\![  \ottsym{(}   \mu\!  \, \mathit{X}  \ottsym{(}   \tceseq{ \mathit{x}  \,  {:}  \,  \ottnt{s} }   \ottsym{)}  \ottsym{.}  \phi_{{\mathrm{0}}}  \ottsym{)}  \ottsym{(}   \tceseq{ \ottnt{t} }   \ottsym{)}  ]\!]_{ \theta_{{\mathrm{2}}} } }  \,  =  \,  \bot $ clearly.
          In the latter case,
          (\ref{lem:lr-iff-spos-valid:mu:bot:lfp}) implies
          $ \mathsf{lfp}^{  \tceseq{ \ottnt{s} }   \rightarrow   \mathbb{B}  }   \ottsym{(}  \mathbf{\mathit{F} }_{{\mathrm{2}}}  \ottsym{)}  \ottsym{(}   \tceseq{  { [\![  \ottnt{t}  ]\!]_{ \theta } }  }   \ottsym{)} \,  =  \,  \bot $, that is,
          $ { [\![  \ottsym{(}   \mu\!  \, \mathit{X}  \ottsym{(}   \tceseq{ \mathit{x}  \,  {:}  \,  \ottnt{s} }   \ottsym{)}  \ottsym{.}  \phi_{{\mathrm{0}}}  \ottsym{)}  \ottsym{(}   \tceseq{ \ottnt{t} }   \ottsym{)}  ]\!]_{ \theta_{{\mathrm{2}}} } }  \,  =  \,  \bot $.
          Therefore, in either case, we have the conclusion.
   \end{enumerate}

  \case $\phi \,  =  \, \ottsym{(}   \nu\!  \, \mathit{X}  \ottsym{(}   \tceseq{ \mathit{x}  \,  {:}  \,  \ottnt{s} }   \ottsym{)}  \ottsym{.}  \phi_{{\mathrm{0}}}  \ottsym{)}  \ottsym{(}   \tceseq{ \ottnt{t} }   \ottsym{)}$:
   Without loss of generality, we can suppose that $\mathit{X} \,  \not=  \, \mathit{Y}$.
   Let
   \begin{itemize}
    \item $\mathbf{\mathit{F} }_{{\mathrm{1}}} =  \lambda\!  \, \mathit{X}  \ottsym{.}   \lambda\!  \,  \tceseq{ \mathit{x}  \,  {:}  \,  \ottnt{s} }   \ottsym{.}   { [\![  \phi_{{\mathrm{0}}}  ]\!]_{ \theta_{{\mathrm{1}}}  \mathbin{\uplus}   \{  \mathit{X}  \mapsto  \mathit{X}  \}   \mathbin{\uplus}   \{ \tceseq{ \mathit{x}  \mapsto  \mathit{x} } \}  } } $ and
    \item $\mathbf{\mathit{F} }_{{\mathrm{2}}} =  \lambda\!  \, \mathit{X}  \ottsym{.}   \lambda\!  \,  \tceseq{ \mathit{x}  \,  {:}  \,  \ottnt{s} }   \ottsym{.}   { [\![  \phi_{{\mathrm{0}}}  ]\!]_{ \theta_{{\mathrm{2}}}  \mathbin{\uplus}   \{  \mathit{X}  \mapsto  \mathit{X}  \}   \mathbin{\uplus}   \{ \tceseq{ \mathit{x}  \mapsto  \mathit{x} } \}  } } $.
   \end{itemize}
   We prove each case.
   \begin{enumerate}
    \item First, the IH implies (\ref{lem:lr-iff-spos-valid:mu:top:order})
          in this case as well.

          Next, we show that
          \begin{equation}
            \mathsf{gfp}^{  \tceseq{ \ottnt{s} }   \rightarrow   \mathbb{B}  }   \ottsym{(}  \mathbf{\mathit{F} }_{{\mathrm{1}}}  \ottsym{)} \,  \sqsubseteq_{  \tceseq{ \ottnt{s} }   \rightarrow   \mathbb{B}  }  \,  \mathsf{gfp}^{  \tceseq{ \ottnt{s} }   \rightarrow   \mathbb{B}  }   \ottsym{(}  \mathbf{\mathit{F} }_{{\mathrm{2}}}  \ottsym{)} ~.
            \label{lem:lr-iff-spos-valid:nu:top:gfp}
          \end{equation}
          By definition, it suffices to show that
          $ \mathsf{gfp}^{  \tceseq{ \ottnt{s} }   \rightarrow   \mathbb{B}  }   \ottsym{(}  \mathbf{\mathit{F} }_{{\mathrm{2}}}  \ottsym{)}$ is an upper bound of
          $ \{ \,  \mathbf{\mathit{f} }  \mathrel{  \in  }   \mathcal{D}_{  \tceseq{ \ottnt{s} }   \rightarrow   \mathbb{B}  }   \mid  \mathbf{\mathit{f} }  \mathrel{  \sqsubseteq_{  \tceseq{ \ottnt{s} }   \rightarrow   \mathbb{B}  }  }  \mathbf{\mathit{F} }_{{\mathrm{1}}}  \ottsym{(}  \mathbf{\mathit{f} }  \ottsym{)}  \, \} $.
          Let $\mathbf{\mathit{f} } \,  \in  \,  \mathcal{D}_{  \tceseq{ \ottnt{s} }   \rightarrow   \mathbb{B}  } $ such that $\mathbf{\mathit{f} } \,  \sqsubseteq_{  \tceseq{ \ottnt{s} }   \rightarrow   \mathbb{B}  }  \, \mathbf{\mathit{F} }_{{\mathrm{1}}}  \ottsym{(}  \mathbf{\mathit{f} }  \ottsym{)}$.
          Then, it suffices to show that
          \[
           \mathbf{\mathit{f} } \,  \sqsubseteq_{  \tceseq{ \ottnt{s} }   \rightarrow   \mathbb{B}  }  \,  \mathsf{gfp}^{  \tceseq{ \ottnt{s} }   \rightarrow   \mathbb{B}  }   \ottsym{(}  \mathbf{\mathit{F} }_{{\mathrm{2}}}  \ottsym{)} ~.
          \]
          Because $ \mathsf{gfp}^{  \tceseq{ \ottnt{s} }   \rightarrow   \mathbb{B}  }   \ottsym{(}  \mathbf{\mathit{F} }_{{\mathrm{2}}}  \ottsym{)}$ is the least upper bound of
          $ \{ \,  \mathbf{\mathit{f} }'  \mathrel{  \in  }   \mathcal{D}_{  \tceseq{ \ottnt{s} }   \rightarrow   \mathbb{B}  }   \mid  \mathbf{\mathit{f} }'  \mathrel{  \sqsubseteq_{  \tceseq{ \ottnt{s} }   \rightarrow   \mathbb{B}  }  }  \mathbf{\mathit{F} }_{{\mathrm{2}}}  \ottsym{(}  \mathbf{\mathit{f} }'  \ottsym{)}  \, \} $,
          it suffices to show that
          \[
           \mathbf{\mathit{f} } \,  \sqsubseteq_{  \tceseq{ \ottnt{s} }   \rightarrow   \mathbb{B}  }  \, \mathbf{\mathit{F} }_{{\mathrm{2}}}  \ottsym{(}  \mathbf{\mathit{f} }  \ottsym{)} ~.
          \]
          Because
          $\mathbf{\mathit{F} }_{{\mathrm{1}}}  \ottsym{(}  \mathbf{\mathit{f} }  \ottsym{)} \,  \sqsubseteq_{  \tceseq{ \ottnt{s} }   \rightarrow   \mathbb{B}  }  \, \mathbf{\mathit{F} }_{{\mathrm{2}}}  \ottsym{(}  \mathbf{\mathit{f} }  \ottsym{)}$ by (\ref{lem:lr-iff-spos-valid:mu:top:order}), and
          $\mathbf{\mathit{f} } \,  \sqsubseteq_{  \tceseq{ \ottnt{s} }   \rightarrow   \mathbb{B}  }  \, \mathbf{\mathit{F} }_{{\mathrm{1}}}  \ottsym{(}  \mathbf{\mathit{f} }  \ottsym{)}$,
          we have (\ref{lem:lr-iff-spos-valid:nu:top:gfp}).

          Because
          $ { [\![  \ottsym{(}   \nu\!  \, \mathit{X}  \ottsym{(}   \tceseq{ \mathit{x}  \,  {:}  \,  \ottnt{s} }   \ottsym{)}  \ottsym{.}  \phi_{{\mathrm{0}}}  \ottsym{)}  \ottsym{(}   \tceseq{ \ottnt{t} }   \ottsym{)}  ]\!]_{ \theta_{{\mathrm{1}}} } }  \,  =  \,  \top $
          by the assumption,
          we have $ \mathsf{gfp}^{  \tceseq{ \ottnt{s} }   \rightarrow   \mathbb{B}  }   \ottsym{(}  \mathbf{\mathit{F} }_{{\mathrm{1}}}  \ottsym{)}  \ottsym{(}   \tceseq{  { [\![  \ottnt{t}  ]\!]_{ \theta } }  }   \ottsym{)} \,  =  \,  \top $.
          Therefore, (\ref{lem:lr-iff-spos-valid:nu:top:gfp}) implies
          $ \mathsf{gfp}^{  \tceseq{ \ottnt{s} }   \rightarrow   \mathbb{B}  }   \ottsym{(}  \mathbf{\mathit{F} }_{{\mathrm{2}}}  \ottsym{)}  \ottsym{(}   \tceseq{  { [\![  \ottnt{t}  ]\!]_{ \theta } }  }   \ottsym{)} \,  =  \,  \top $, that is,
          $ { [\![  \ottsym{(}   \nu\!  \, \mathit{X}  \ottsym{(}   \tceseq{ \mathit{x}  \,  {:}  \,  \ottnt{s} }   \ottsym{)}  \ottsym{.}  \phi_{{\mathrm{0}}}  \ottsym{)}  \ottsym{(}   \tceseq{ \ottnt{t} }   \ottsym{)}  ]\!]_{ \theta_{{\mathrm{2}}} } }  \,  =  \,  \top $, which is
          the conclusion we must show.

    \item First, the IH implies (\ref{lem:lr-iff-spos-valid:mu:bot:order})
          in this case as well.

          Next, we show that
          \begin{equation}
            \mathsf{gfp}^{  \tceseq{ \ottnt{s} }   \rightarrow   \mathbb{B}  }   \ottsym{(}  \mathbf{\mathit{F} }_{{\mathrm{2}}}  \ottsym{)} \,  \sqsubseteq_{  \tceseq{ \ottnt{s} }   \rightarrow   \mathbb{B}  }  \,  \mathsf{gfp}^{  \tceseq{ \ottnt{s} }   \rightarrow   \mathbb{B}  }   \ottsym{(}  \mathbf{\mathit{F} }_{{\mathrm{1}}}  \ottsym{)} ~.
            \label{lem:lr-iff-spos-valid:nu:bot:gfp}
          \end{equation}
          By definition, it suffices to show that
          $ \mathsf{gfp}^{  \tceseq{ \ottnt{s} }   \rightarrow   \mathbb{B}  }   \ottsym{(}  \mathbf{\mathit{F} }_{{\mathrm{1}}}  \ottsym{)}$ is an upper bound of
          $ \{ \,  \mathbf{\mathit{f} }  \mathrel{  \in  }   \mathcal{D}_{  \tceseq{ \ottnt{s} }   \rightarrow   \mathbb{B}  }   \mid  \mathbf{\mathit{f} }  \mathrel{  \sqsubseteq_{  \tceseq{ \ottnt{s} }   \rightarrow   \mathbb{B}  }  }  \mathbf{\mathit{F} }_{{\mathrm{2}}}  \ottsym{(}  \mathbf{\mathit{f} }  \ottsym{)}  \, \} $.
          Let $\mathbf{\mathit{f} } \,  \in  \,  \mathcal{D}_{  \tceseq{ \ottnt{s} }   \rightarrow   \mathbb{B}  } $ such that $\mathbf{\mathit{f} } \,  \sqsubseteq_{  \tceseq{ \ottnt{s} }   \rightarrow   \mathbb{B}  }  \, \mathbf{\mathit{F} }_{{\mathrm{2}}}  \ottsym{(}  \mathbf{\mathit{f} }  \ottsym{)}$.
          Then, it suffices to show that
          \[
           \mathbf{\mathit{f} } \,  \sqsubseteq_{  \tceseq{ \ottnt{s} }   \rightarrow   \mathbb{B}  }  \,  \mathsf{gfp}^{  \tceseq{ \ottnt{s} }   \rightarrow   \mathbb{B}  }   \ottsym{(}  \mathbf{\mathit{F} }_{{\mathrm{1}}}  \ottsym{)} ~.
          \]
          Because $ \mathsf{gfp}^{  \tceseq{ \ottnt{s} }   \rightarrow   \mathbb{B}  }   \ottsym{(}  \mathbf{\mathit{F} }_{{\mathrm{1}}}  \ottsym{)}$ is the least upper bound of
          $ \{ \,  \mathbf{\mathit{f} }'  \mathrel{  \in  }   \mathcal{D}_{  \tceseq{ \ottnt{s} }   \rightarrow   \mathbb{B}  }   \mid  \mathbf{\mathit{f} }'  \mathrel{  \sqsubseteq_{  \tceseq{ \ottnt{s} }   \rightarrow   \mathbb{B}  }  }  \mathbf{\mathit{F} }_{{\mathrm{1}}}  \ottsym{(}  \mathbf{\mathit{f} }'  \ottsym{)}  \, \} $,
          it suffices to show that
          \[
           \mathbf{\mathit{f} } \,  \sqsubseteq_{  \tceseq{ \ottnt{s} }   \rightarrow   \mathbb{B}  }  \, \mathbf{\mathit{F} }_{{\mathrm{1}}}  \ottsym{(}  \mathbf{\mathit{f} }  \ottsym{)} ~.
          \]
          Because
          $\mathbf{\mathit{F} }_{{\mathrm{2}}}  \ottsym{(}  \mathbf{\mathit{f} }  \ottsym{)} \,  \sqsubseteq_{  \tceseq{ \ottnt{s} }   \rightarrow   \mathbb{B}  }  \, \mathbf{\mathit{F} }_{{\mathrm{1}}}  \ottsym{(}  \mathbf{\mathit{f} }  \ottsym{)}$ by (\ref{lem:lr-iff-spos-valid:mu:bot:order}), and
          $\mathbf{\mathit{f} } \,  \sqsubseteq_{  \tceseq{ \ottnt{s} }   \rightarrow   \mathbb{B}  }  \, \mathbf{\mathit{F} }_{{\mathrm{2}}}  \ottsym{(}  \mathbf{\mathit{f} }  \ottsym{)}$,
          we have (\ref{lem:lr-iff-spos-valid:nu:bot:gfp}).

          Because
          $ { [\![  \ottsym{(}   \nu\!  \, \mathit{X}  \ottsym{(}   \tceseq{ \mathit{x}  \,  {:}  \,  \ottnt{s} }   \ottsym{)}  \ottsym{.}  \phi_{{\mathrm{0}}}  \ottsym{)}  \ottsym{(}   \tceseq{ \ottnt{t} }   \ottsym{)}  ]\!]_{ \theta_{{\mathrm{1}}} } }  \,  =  \,  \bot $
          by the assumption,
          we have either of the following:
          \begin{itemize}
           \item $ { [\![  \ottnt{t'}  ]\!]_{ \theta } }  \,  \not\in  \,  \mathcal{D}_{ \ottnt{s'} } $ for some $\ottnt{t'}$ in $ \tceseq{ \ottnt{t} } $
                 and the corresponding $\ottnt{s'}$ in $ \tceseq{ \ottnt{s} } $; or
           \item $ \mathsf{gfp}^{  \tceseq{ \ottnt{s} }   \rightarrow   \mathbb{B}  }   \ottsym{(}  \mathbf{\mathit{F} }_{{\mathrm{1}}}  \ottsym{)}  \ottsym{(}   \tceseq{  { [\![  \ottnt{t}  ]\!]_{ \theta } }  }   \ottsym{)} \,  =  \,  \bot $.
          \end{itemize}
          In the former case,
          $ { [\![  \ottsym{(}   \nu\!  \, \mathit{X}  \ottsym{(}   \tceseq{ \mathit{x}  \,  {:}  \,  \ottnt{s} }   \ottsym{)}  \ottsym{.}  \phi_{{\mathrm{0}}}  \ottsym{)}  \ottsym{(}   \tceseq{ \ottnt{t} }   \ottsym{)}  ]\!]_{ \theta_{{\mathrm{2}}} } }  \,  =  \,  \bot $ clearly.
          In the latter case,
          (\ref{lem:lr-iff-spos-valid:nu:bot:gfp}) implies
          $ \mathsf{gfp}^{  \tceseq{ \ottnt{s} }   \rightarrow   \mathbb{B}  }   \ottsym{(}  \mathbf{\mathit{F} }_{{\mathrm{2}}}  \ottsym{)}  \ottsym{(}   \tceseq{  { [\![  \ottnt{t}  ]\!]_{ \theta } }  }   \ottsym{)} \,  =  \,  \bot $, that is,
          $ { [\![  \ottsym{(}   \nu\!  \, \mathit{X}  \ottsym{(}   \tceseq{ \mathit{x}  \,  {:}  \,  \ottnt{s} }   \ottsym{)}  \ottsym{.}  \phi_{{\mathrm{0}}}  \ottsym{)}  \ottsym{(}   \tceseq{ \ottnt{t} }   \ottsym{)}  ]\!]_{ \theta_{{\mathrm{2}}} } }  \,  =  \,  \bot $.
          Therefore, in either case, we have the conclusion.
   \end{enumerate}

  \case $\phi \,  =  \, \mathit{Y}  \ottsym{(}   \tceseq{ \ottnt{t} }   \ottsym{)}$:
   We prove each case.
   \begin{enumerate}
    \item Suppose that $ { [\![  \mathit{Y}  \ottsym{(}   \tceseq{ \ottnt{t} }   \ottsym{)}  ]\!]_{ \theta_{{\mathrm{1}}} } }  \,  =  \,  \top $.
          It implies $\mathbf{\mathit{f} }_{{\mathrm{1}}}  \ottsym{(}   \tceseq{  { [\![  \ottnt{t}  ]\!]_{ \theta } }  }   \ottsym{)} \,  =  \,  \top $.
          It suffices to show that $ { [\![  \mathit{Y}  \ottsym{(}   \tceseq{ \ottnt{t} }   \ottsym{)}  ]\!]_{ \theta_{{\mathrm{2}}} } }  \,  =  \,  \top $,
          which is proven by $\mathbf{\mathit{f} }_{{\mathrm{1}}} \,  \sqsubseteq_{ \mathcal{T} }  \, \mathbf{\mathit{f} }_{{\mathrm{2}}}$.
    \item Contradictory.
   \end{enumerate}

  \case $ { \phi \,  =  \, \mathit{X}  \ottsym{(}   \tceseq{ \ottnt{t} }   \ottsym{)} } \,\mathrel{\wedge}\, { \mathit{X} \,  \not=  \, \mathit{Y} } $: Obvious.
 \end{caseanalysis}
\end{proof}

\begin{lemmap}{Weakening Predicates at Positive Positions in Validity}{lr-iff-spos-valid}
 Suppose that $ \ottnt{A_{{\mathrm{1}}}}  \, \Rightarrow \,  \ottnt{A_{{\mathrm{2}}}}  :   \tceseq{ \ottnt{s} }  $.
 If $\Gamma  \models   \phi    [  \ottnt{A_{{\mathrm{1}}}}  /  \mathit{Y}  ]  $ and $\mathit{Y} \,  \not\in  \,  \mathit{fpv}^\mathbf{n}  (  \phi  ) $, then $\Gamma  \models   \phi    [  \ottnt{A_{{\mathrm{2}}}}  /  \mathit{Y}  ]  $.
\end{lemmap}
\begin{proof}
 \TS{CHECKED (NO CHANGE): 2022/07/05}
 First, we show that
 $\ottnt{A_{{\mathrm{1}}}} \, \Leftrightarrow \, \ottnt{A_{{\mathrm{2}}}}  \ottsym{:}   \tceseq{ \ottnt{s} } $ implies
 $ { [\![  \ottnt{A_{{\mathrm{1}}}}  ]\!]_{  \emptyset  } }  \,  \sqsubseteq_{  \tceseq{ \ottnt{s} }   \rightarrow   \mathbb{B}  }  \,  { [\![  \ottnt{A_{{\mathrm{2}}}}  ]\!]_{  \emptyset  } } $.
 Let $ \tceseq{ \ottnt{d} \,  \in  \,  \mathcal{D}_{ \ottnt{s} }  } $.
 It suffices to show that $ { [\![  \ottnt{A_{{\mathrm{1}}}}  ]\!]_{  \emptyset  } }   \ottsym{(}   \tceseq{ \ottnt{d} }   \ottsym{)} \,  \sqsubseteq_{  \mathbb{B}  }  \,  { [\![  \ottnt{A_{{\mathrm{2}}}}  ]\!]_{  \emptyset  } }   \ottsym{(}   \tceseq{ \ottnt{d} }   \ottsym{)}$.
 Because $ \ottnt{A_{{\mathrm{1}}}}  \, \Rightarrow \,  \ottnt{A_{{\mathrm{2}}}}  :   \tceseq{ \ottnt{s} }  $,
 we have $ \tceseq{ \mathit{x}  \,  {:}  \,  \ottnt{s} }   \models   \ottnt{A_{{\mathrm{1}}}}  \ottsym{(}   \tceseq{ \mathit{x} }   \ottsym{)}    \Rightarrow    \ottnt{A_{{\mathrm{2}}}}  \ottsym{(}   \tceseq{ \mathit{x} }   \ottsym{)} $.
 Therefore, by definition,
 $ { [\![  \ottnt{A_{{\mathrm{1}}}}  ]\!]_{  \{ \tceseq{ \mathit{x}  \mapsto  \ottnt{d} } \}  } }   \ottsym{(}   \tceseq{ \ottnt{d} }   \ottsym{)} \,  \sqsubseteq_{  \mathbb{B}  }  \,  { [\![  \ottnt{A_{{\mathrm{2}}}}  ]\!]_{  \{ \tceseq{ \mathit{x}  \mapsto  \ottnt{d} } \}  } }   \ottsym{(}   \tceseq{ \ottnt{d} }   \ottsym{)}$.
 Because $\ottnt{A_{{\mathrm{1}}}}$ and $\ottnt{A_{{\mathrm{2}}}}$ are closed,
 we have the conclusion.

 Let $\theta$ be any substitution such that
 $\Gamma  \vdash  \theta$.
 Then, by Lemmas~\ref{lem:ts-logic-closing-dsubs} and \ref{lem:ts-logic-pred-subst}, it suffices to show that
 $ { [\![  \phi  ]\!]_{  \{  \mathit{Y}  \mapsto   { [\![  \ottnt{A_{{\mathrm{1}}}}  ]\!]_{ \theta } }   \}   \mathbin{\uplus}  \theta } }  \,  =  \,  \top $, then $ { [\![  \phi  ]\!]_{  \{  \mathit{Y}  \mapsto   { [\![  \ottnt{A_{{\mathrm{2}}}}  ]\!]_{ \theta } }   \}   \mathbin{\uplus}  \theta } }  \,  =  \,  \top $.
 By \reflem{ts-logic-weak-pred} with
 $ { [\![  \ottnt{A_{{\mathrm{1}}}}  ]\!]_{  \emptyset  } }  \,  \sqsubseteq_{  \tceseq{ \ottnt{s} }   \rightarrow   \mathbb{B}  }  \,  { [\![  \ottnt{A_{{\mathrm{2}}}}  ]\!]_{  \emptyset  } } $,
 we have the conclusion.
\end{proof}

\begin{lemmap}{Weakening Predicates at Positive Positions in Effects}{lr-iff-spos-eff}
 Suppose that $ \ottnt{A_{{\mathrm{1}}}}  \, \Rightarrow \,  \ottnt{A_{{\mathrm{2}}}}  :   \tceseq{ \ottnt{s} }  $.
 Let $\sigma_{{\mathrm{1}}}$ and $\sigma_{{\mathrm{2}}}$ be closing substitutions.
 If $\mathit{Y} \,  \not\in  \,  \mathit{fpv}^\mathbf{n}  (  \Phi  ) $ and $\Gamma  \vdash   \sigma_{{\mathrm{1}}}  \ottsym{(}  \Phi  \ottsym{)}    [  \ottnt{A_{{\mathrm{1}}}}  /  \mathit{Y}  ]    \ottsym{<:}   \sigma_{{\mathrm{2}}}  \ottsym{(}  \Phi  \ottsym{)}    [  \ottnt{A_{{\mathrm{1}}}}  /  \mathit{Y}  ]  $,
 then $\Gamma  \vdash   \sigma_{{\mathrm{1}}}  \ottsym{(}  \Phi  \ottsym{)}    [  \ottnt{A_{{\mathrm{1}}}}  /  \mathit{Y}  ]    \ottsym{<:}   \sigma_{{\mathrm{2}}}  \ottsym{(}  \Phi  \ottsym{)}    [  \ottnt{A_{{\mathrm{2}}}}  /  \mathit{Y}  ]  $.
\end{lemmap}
\begin{proof}
 \TS{CHECKED (NO CHANGE): 2022/07/05}
 Because $\Gamma  \vdash   \sigma_{{\mathrm{1}}}  \ottsym{(}  \Phi  \ottsym{)}    [  \ottnt{A_{{\mathrm{1}}}}  /  \mathit{Y}  ]    \ottsym{<:}   \sigma_{{\mathrm{2}}}  \ottsym{(}  \Phi  \ottsym{)}    [  \ottnt{A_{{\mathrm{1}}}}  /  \mathit{Y}  ]  $, we have
 \begin{itemize}
  \item $\Gamma  \ottsym{,}  \mathit{x} \,  {:}  \,  \Sigma ^\ast   \models    { \ottsym{(}   \sigma_{{\mathrm{1}}}  \ottsym{(}  \Phi  \ottsym{)}    [  \ottnt{A_{{\mathrm{1}}}}  /  \mathit{Y}  ]    \ottsym{)} }^\mu   \ottsym{(}  \mathit{x}  \ottsym{)}    \Rightarrow     { \ottsym{(}   \sigma_{{\mathrm{2}}}  \ottsym{(}  \Phi  \ottsym{)}    [  \ottnt{A_{{\mathrm{1}}}}  /  \mathit{Y}  ]    \ottsym{)} }^\mu   \ottsym{(}  \mathit{x}  \ottsym{)} $ and
  \item $\Gamma  \ottsym{,}  \mathit{y} \,  {:}  \,  \Sigma ^\omega   \models    { \ottsym{(}   \sigma_{{\mathrm{1}}}  \ottsym{(}  \Phi  \ottsym{)}    [  \ottnt{A_{{\mathrm{1}}}}  /  \mathit{Y}  ]    \ottsym{)} }^\nu   \ottsym{(}  \mathit{y}  \ottsym{)}    \Rightarrow     { \ottsym{(}   \sigma_{{\mathrm{2}}}  \ottsym{(}  \Phi  \ottsym{)}    [  \ottnt{A_{{\mathrm{1}}}}  /  \mathit{Y}  ]    \ottsym{)} }^\nu   \ottsym{(}  \mathit{y}  \ottsym{)} $
 \end{itemize}
 for some $\mathit{x}  \ottsym{,}  \mathit{y} \,  \not\in  \,  \mathit{fv}  (   \sigma_{{\mathrm{1}}}  \ottsym{(}  \Phi  \ottsym{)}    [  \ottnt{A_{{\mathrm{1}}}}  /  \mathit{Y}  ]    )  \,  \mathbin{\cup}  \,  \mathit{fv}  (   \sigma_{{\mathrm{2}}}  \ottsym{(}  \Phi  \ottsym{)}    [  \ottnt{A_{{\mathrm{1}}}}  /  \mathit{Y}  ]    ) $.
 Without loss of generality, we can suppose that $\mathit{x}  \ottsym{,}  \mathit{y} \,  \not\in  \,  \mathit{fv}  (  \ottnt{A_{{\mathrm{2}}}}  ) $.
 Let $\mathit{Y_{{\mathrm{0}}}}$ be a fresh predicate variable.
 Then,
 \begin{itemize}
  \item $ { \ottsym{(}   \sigma_{{\mathrm{2}}}  \ottsym{(}  \Phi  \ottsym{)}    [  \ottnt{A_{{\mathrm{1}}}}  /  \mathit{Y}  ]    \ottsym{)} }^\mu   \ottsym{(}  \mathit{x}  \ottsym{)} \,  =  \,   { \ottsym{(}   \sigma_{{\mathrm{2}}}  \ottsym{(}  \Phi  \ottsym{)}    [  \mathit{Y_{{\mathrm{0}}}}  /  \mathit{Y}  ]    \ottsym{)} }^\mu   \ottsym{(}  \mathit{x}  \ottsym{)}    [  \ottnt{A_{{\mathrm{1}}}}  /  \mathit{Y_{{\mathrm{0}}}}  ]  $ and
  \item $ { \ottsym{(}   \sigma_{{\mathrm{2}}}  \ottsym{(}  \Phi  \ottsym{)}    [  \ottnt{A_{{\mathrm{1}}}}  /  \mathit{Y}  ]    \ottsym{)} }^\nu   \ottsym{(}  \mathit{y}  \ottsym{)} \,  =  \,   { \ottsym{(}   \sigma_{{\mathrm{2}}}  \ottsym{(}  \Phi  \ottsym{)}    [  \mathit{Y_{{\mathrm{0}}}}  /  \mathit{Y}  ]    \ottsym{)} }^\nu   \ottsym{(}  \mathit{y}  \ottsym{)}    [  \ottnt{A_{{\mathrm{1}}}}  /  \mathit{Y_{{\mathrm{0}}}}  ]  $.
 \end{itemize}
 Therefore,
 \begin{itemize}
  \item $\Gamma  \ottsym{,}  \mathit{x} \,  {:}  \,  \Sigma ^\ast   \models   \ottsym{(}    { \ottsym{(}   \sigma_{{\mathrm{1}}}  \ottsym{(}  \Phi  \ottsym{)}    [  \ottnt{A_{{\mathrm{1}}}}  /  \mathit{Y}  ]    \ottsym{)} }^\mu   \ottsym{(}  \mathit{x}  \ottsym{)}    \Rightarrow     { \ottsym{(}   \sigma_{{\mathrm{2}}}  \ottsym{(}  \Phi  \ottsym{)}    [  \mathit{Y_{{\mathrm{0}}}}  /  \mathit{Y}  ]    \ottsym{)} }^\mu   \ottsym{(}  \mathit{x}  \ottsym{)}   \ottsym{)}    [  \ottnt{A_{{\mathrm{1}}}}  /  \mathit{Y_{{\mathrm{0}}}}  ]  $ and
  \item $\Gamma  \ottsym{,}  \mathit{y} \,  {:}  \,  \Sigma ^\omega   \models   \ottsym{(}    { \ottsym{(}   \sigma_{{\mathrm{1}}}  \ottsym{(}  \Phi  \ottsym{)}    [  \ottnt{A_{{\mathrm{1}}}}  /  \mathit{Y}  ]    \ottsym{)} }^\nu   \ottsym{(}  \mathit{y}  \ottsym{)}    \Rightarrow     { \ottsym{(}   \sigma_{{\mathrm{2}}}  \ottsym{(}  \Phi  \ottsym{)}    [  \mathit{Y_{{\mathrm{0}}}}  /  \mathit{Y}  ]    \ottsym{)} }^\nu   \ottsym{(}  \mathit{y}  \ottsym{)}   \ottsym{)}    [  \ottnt{A_{{\mathrm{1}}}}  /  \mathit{Y_{{\mathrm{0}}}}  ]  $.
 \end{itemize}
 Because $\mathit{Y} \,  \not\in  \,  \mathit{fpv}^\mathbf{n}  (  \Phi  ) $, and $\sigma_{{\mathrm{1}}}$ and $\sigma_{{\mathrm{2}}}$ are closing substitutions,
 we have
 \[
  \mathit{Y_{{\mathrm{0}}}} \,  \not\in  \,  \mathit{fpv}^\mathbf{n}  (   { \ottsym{(}   \sigma_{{\mathrm{2}}}  \ottsym{(}  \Phi  \ottsym{)}    [  \mathit{Y_{{\mathrm{0}}}}  /  \mathit{Y}  ]    \ottsym{)} }^\mu   \ottsym{(}  \mathit{x}  \ottsym{)}  )  \,  \mathbin{\cup}  \,  \mathit{fpv}^\mathbf{n}  (   { \ottsym{(}   \sigma_{{\mathrm{2}}}  \ottsym{(}  \Phi  \ottsym{)}    [  \mathit{Y_{{\mathrm{0}}}}  /  \mathit{Y}  ]    \ottsym{)} }^\nu   \ottsym{(}  \mathit{y}  \ottsym{)}  )  ~.
 \]
 Then, \reflem{lr-iff-spos-valid} implies
 \begin{itemize}
  \item $\Gamma  \ottsym{,}  \mathit{x} \,  {:}  \,  \Sigma ^\ast   \models   \ottsym{(}    { \ottsym{(}   \sigma_{{\mathrm{1}}}  \ottsym{(}  \Phi  \ottsym{)}    [  \ottnt{A_{{\mathrm{1}}}}  /  \mathit{Y}  ]    \ottsym{)} }^\mu   \ottsym{(}  \mathit{x}  \ottsym{)}    \Rightarrow     { \ottsym{(}   \sigma_{{\mathrm{2}}}  \ottsym{(}  \Phi  \ottsym{)}    [  \mathit{Y_{{\mathrm{0}}}}  /  \mathit{Y}  ]    \ottsym{)} }^\mu   \ottsym{(}  \mathit{x}  \ottsym{)}   \ottsym{)}    [  \ottnt{A_{{\mathrm{2}}}}  /  \mathit{Y_{{\mathrm{0}}}}  ]  $ and
  \item $\Gamma  \ottsym{,}  \mathit{y} \,  {:}  \,  \Sigma ^\omega   \models   \ottsym{(}    { \ottsym{(}   \sigma_{{\mathrm{1}}}  \ottsym{(}  \Phi  \ottsym{)}    [  \ottnt{A_{{\mathrm{1}}}}  /  \mathit{Y}  ]    \ottsym{)} }^\nu   \ottsym{(}  \mathit{y}  \ottsym{)}    \Rightarrow     { \ottsym{(}   \sigma_{{\mathrm{2}}}  \ottsym{(}  \Phi  \ottsym{)}    [  \mathit{Y_{{\mathrm{0}}}}  /  \mathit{Y}  ]    \ottsym{)} }^\nu   \ottsym{(}  \mathit{y}  \ottsym{)}   \ottsym{)}    [  \ottnt{A_{{\mathrm{2}}}}  /  \mathit{Y_{{\mathrm{0}}}}  ]  $.
 \end{itemize}
 Hence,
 \begin{itemize}
  \item $\Gamma  \ottsym{,}  \mathit{x} \,  {:}  \,  \Sigma ^\ast   \models    { \ottsym{(}   \sigma_{{\mathrm{1}}}  \ottsym{(}  \Phi  \ottsym{)}    [  \ottnt{A_{{\mathrm{1}}}}  /  \mathit{Y}  ]    \ottsym{)} }^\mu   \ottsym{(}  \mathit{x}  \ottsym{)}    \Rightarrow     { \ottsym{(}   \sigma_{{\mathrm{2}}}  \ottsym{(}  \Phi  \ottsym{)}    [  \ottnt{A_{{\mathrm{2}}}}  /  \mathit{Y}  ]    \ottsym{)} }^\mu   \ottsym{(}  \mathit{x}  \ottsym{)} $ and
  \item $\Gamma  \ottsym{,}  \mathit{y} \,  {:}  \,  \Sigma ^\omega   \models    { \ottsym{(}   \sigma_{{\mathrm{1}}}  \ottsym{(}  \Phi  \ottsym{)}    [  \ottnt{A_{{\mathrm{1}}}}  /  \mathit{Y}  ]    \ottsym{)} }^\nu   \ottsym{(}  \mathit{y}  \ottsym{)}    \Rightarrow     { \ottsym{(}   \sigma_{{\mathrm{2}}}  \ottsym{(}  \Phi  \ottsym{)}    [  \ottnt{A_{{\mathrm{2}}}}  /  \mathit{Y}  ]    \ottsym{)} }^\nu   \ottsym{(}  \mathit{y}  \ottsym{)} $.
 \end{itemize}
 Then, \Srule{Eff} implies the conclusion
 \[
  \Gamma  \vdash   \sigma_{{\mathrm{1}}}  \ottsym{(}  \Phi  \ottsym{)}    [  \ottnt{A_{{\mathrm{1}}}}  /  \mathit{Y}  ]    \ottsym{<:}   \sigma_{{\mathrm{2}}}  \ottsym{(}  \Phi  \ottsym{)}    [  \ottnt{A_{{\mathrm{2}}}}  /  \mathit{Y}  ]   ~.
 \]
\end{proof}

\begin{lemmap}{Composing Approximation Substitutions}{lr-approx-compose-subst}
 Assume that
 \begin{itemize}
  \item $ \tceseq{ \ottsym{(}  \mathit{X''}  \ottsym{,}  \mathit{Y''}  \ottsym{)} }   \vdash  \ottnt{C''}  \Rrightarrow  \ottnt{C'}$,
  \item $ \tceseq{ \ottsym{(}  \mathit{X'}  \ottsym{,}  \mathit{Y'}  \ottsym{)} }   \vdash  \ottnt{C'}  \Rrightarrow  \ottnt{C}$,
  \item $\ottnt{C''}  \triangleright   \tceseq{ \ottsym{(}  \mathit{X''}  \ottsym{,}  \mathit{Y''}  \ottsym{)} }   \ottsym{,}   \tceseq{ \ottsym{(}  \mathit{X'}  \ottsym{,}  \mathit{Y'}  \ottsym{)} }   \ottsym{;}  \ottsym{(}  \mathit{X_{{\mathrm{0}}}}  \ottsym{,}  \mathit{Y_{{\mathrm{0}}}}  \ottsym{)}  \ottsym{,}   \tceseq{ \ottsym{(}  \mathit{X}  \ottsym{,}  \mathit{Y}  \ottsym{)} }   \ottsym{;}   \tceseq{ \mathit{z_{{\mathrm{0}}}}  \,  {:}  \,  \iota_{{\mathrm{0}}} }  \,  |  \, \ottnt{C} \,  \vdash _{ \ottmv{i} }  \, \sigma$,
  \item $\ottnt{C''}  \triangleright   \tceseq{ \ottsym{(}  \mathit{X''}  \ottsym{,}  \mathit{Y''}  \ottsym{)} }   \ottsym{;}   \tceseq{ \ottsym{(}  \mathit{X'}  \ottsym{,}  \mathit{Y'}  \ottsym{)} }   \ottsym{;}   \tceseq{ \mathit{z_{{\mathrm{0}}}}  \,  {:}  \,  \iota_{{\mathrm{0}}} }  \,  |  \, \ottnt{C'} \,  \vdash _{ \ottmv{i} }  \, \sigma'$.
 \end{itemize}
 Then, there exist $\sigma_{{\mathrm{0}}}$ and $\sigma'_{{\mathrm{0}}}$ such that
 \begin{itemize}
  \item $\ottnt{C'}  \triangleright   \tceseq{ \ottsym{(}  \mathit{X''}  \ottsym{,}  \mathit{Y''}  \ottsym{)} }   \ottsym{,}   \tceseq{ \ottsym{(}  \mathit{X'}  \ottsym{,}  \mathit{Y'}  \ottsym{)} }   \ottsym{,}  \ottsym{(}  \mathit{X_{{\mathrm{0}}}}  \ottsym{,}  \mathit{Y_{{\mathrm{0}}}}  \ottsym{)}  \ottsym{;}   \tceseq{ \ottsym{(}  \mathit{X}  \ottsym{,}  \mathit{Y}  \ottsym{)} }   \ottsym{;}   \tceseq{ \mathit{z_{{\mathrm{0}}}}  \,  {:}  \,  \iota_{{\mathrm{0}}} }  \,  |  \,  \ottnt{C}  . \ottnt{S}  \,  \vdash _{ \ottmv{i} }  \, \sigma_{{\mathrm{0}}}$,
  \item $\ottnt{C'}  \triangleright   \tceseq{ \ottsym{(}  \mathit{X''}  \ottsym{,}  \mathit{Y''}  \ottsym{)} }   \ottsym{;}   \tceseq{ \ottsym{(}  \mathit{X'}  \ottsym{,}  \mathit{Y'}  \ottsym{)} }   \ottsym{,}  \ottsym{(}  \mathit{X_{{\mathrm{0}}}}  \ottsym{,}  \mathit{Y_{{\mathrm{0}}}}  \ottsym{)}  \ottsym{;}   \tceseq{ \mathit{z_{{\mathrm{0}}}}  \,  {:}  \,  \iota_{{\mathrm{0}}} }  \,  |  \, \ottnt{C'} \,  \vdash _{ \ottmv{i} }  \, \sigma'_{{\mathrm{0}}}$, and
  \item $\sigma  \mathbin{\uplus}  \sigma' \,  =  \, \sigma_{{\mathrm{0}}}  \mathbin{\uplus}  \sigma'_{{\mathrm{0}}}$.
 \end{itemize}
\end{lemmap}
\begin{proof}
 \TS{CHECKED: 2022/07/07}
 Straightforward by induction on the derivation of $ \tceseq{ \ottsym{(}  \mathit{X'}  \ottsym{,}  \mathit{Y'}  \ottsym{)} }   \vdash  \ottnt{C'}  \Rrightarrow  \ottnt{C}$.
\end{proof}

\begin{lemmap}{Folding Types of $ \nu $-Approximation}{lr-approx-fold-nu-XXX}
 Let $\ottmv{i} \,  >  \, \ottsym{0}$ and $\pi$ be an infinite trace.
 Assume that
 \begin{itemize}
  \item $ \tceseq{ \ottsym{(}  \mathit{X'}  \ottsym{,}  \mathit{Y'}  \ottsym{)} }   \vdash  \ottnt{C'}  \Rrightarrow  \ottnt{C}$,
  \item $\ottsym{(}   \tceseq{ \mathit{X'} }   \ottsym{,}   \tceseq{ \mathit{X} }   \ottsym{)} \,  {:}  \,  (   \Sigma ^\ast   ,   \tceseq{ \iota_{{\mathrm{0}}} }   )   \ottsym{,}  \ottsym{(}   \tceseq{ \mathit{Y'} }   \ottsym{,}   \tceseq{ \mathit{Y} }   \ottsym{)} \,  {:}  \,  (   \Sigma ^\omega   ,   \tceseq{ \iota_{{\mathrm{0}}} }   )   \ottsym{,}   \tceseq{ \mathit{z_{{\mathrm{0}}}}  \,  {:}  \,  \iota_{{\mathrm{0}}} }   \vdash  \ottnt{C}$,
  \item $ \tceseq{ \ottsym{(}  \mathit{X}  \ottsym{,}  \mathit{Y}  \ottsym{)} }   \ottsym{;}   \tceseq{ \mathit{z_{{\mathrm{0}}}}  \,  {:}  \,  \iota_{{\mathrm{0}}} }   \vdash  \ottnt{C_{{\mathrm{0}}}}  \succ  \ottnt{C}$,
  \item $ \tceseq{ \ottsym{(}  \mathit{X'}  \ottsym{,}  \mathit{Y'}  \ottsym{)} }   \ottsym{;}   \tceseq{ \ottsym{(}  \mathit{X}  \ottsym{,}  \mathit{Y}  \ottsym{)} }  \,  \vdash ^{\circ}  \, \ottnt{C}$,
  \item $ \tceseq{ \ottsym{(}  \mathit{X'}  \ottsym{,}  \mathit{Y'}  \ottsym{)} }   \ottsym{,}   \tceseq{ \ottsym{(}  \mathit{X}  \ottsym{,}  \mathit{Y}  \ottsym{)} }   \ottsym{;}   \tceseq{ \mathit{z_{{\mathrm{0}}}}  \,  {:}  \,  \iota_{{\mathrm{0}}} }  \,  |  \, \ottnt{C'}  \vdash  \sigma$,
  \item $\ottnt{C'}  \triangleright   \tceseq{ \ottsym{(}  \mathit{X'}  \ottsym{,}  \mathit{Y'}  \ottsym{)} }   \ottsym{;}   \tceseq{ \ottsym{(}  \mathit{X}  \ottsym{,}  \mathit{Y}  \ottsym{)} }   \ottsym{;}   \tceseq{ \mathit{z_{{\mathrm{0}}}}  \,  {:}  \,  \iota_{{\mathrm{0}}} }  \,  |  \, \ottnt{C} \,  \vdash _{ \ottmv{i} }  \, \sigma_{\ottmv{i}}$,
  \item $\ottnt{C'}  \triangleright   \emptyset   \ottsym{;}   \tceseq{ \ottsym{(}  \mathit{X'}  \ottsym{,}  \mathit{Y'}  \ottsym{)} }   \ottsym{;}   \tceseq{ \mathit{z_{{\mathrm{0}}}}  \,  {:}  \,  \iota_{{\mathrm{0}}} }  \,  |  \, \ottnt{C'} \,  \vdash _{ \ottmv{i} }  \, \sigma'_{\ottmv{i}}$,
  \item $\ottnt{C'}  \triangleright   \tceseq{ \ottsym{(}  \mathit{X'}  \ottsym{,}  \mathit{Y'}  \ottsym{)} }   \ottsym{;}   \tceseq{ \ottsym{(}  \mathit{X}  \ottsym{,}  \mathit{Y}  \ottsym{)} }   \ottsym{;}   \tceseq{ \mathit{z_{{\mathrm{0}}}}  \,  {:}  \,  \iota_{{\mathrm{0}}} }  \,  |  \, \ottnt{C} \,  \vdash _{ \ottmv{i}  \ottsym{-}  \ottsym{1} }  \,  { \sigma }_{ \ottmv{i}  \ottsym{-}  \ottsym{1} } $, and
  \item $\ottnt{C'}  \triangleright   \emptyset   \ottsym{;}   \tceseq{ \ottsym{(}  \mathit{X'}  \ottsym{,}  \mathit{Y'}  \ottsym{)} }   \ottsym{;}   \tceseq{ \mathit{z_{{\mathrm{0}}}}  \,  {:}  \,  \iota_{{\mathrm{0}}} }  \,  |  \, \ottnt{C'} \,  \vdash _{ \ottmv{i}  \ottsym{-}  \ottsym{1} }  \,  { \sigma }_{ \ottmv{i}  \ottsym{-}  \ottsym{1} }' $.
 \end{itemize}
 Let $ { \sigma }_{ \mathit{X} }  \,  =  \,  [ \tceseq{ \sigma  \ottsym{(}  \mathit{X'}  \ottsym{)}  /  \mathit{X'} } ]   \mathbin{\uplus}   [ \tceseq{ \sigma  \ottsym{(}  \mathit{X}  \ottsym{)}  /  \mathit{X} } ] $.
 Then,
 \[
   \tceseq{ \mathit{z_{{\mathrm{0}}}}  \,  {:}  \,  \ottnt{T_{{\mathrm{0}}}} }   \vdash   { \sigma }_{ \ottmv{i}  \ottsym{-}  \ottsym{1} }'   \ottsym{(}   { \sigma }_{ \ottmv{i}  \ottsym{-}  \ottsym{1} }   \ottsym{(}   { \sigma }_{ \mathit{X} }   \ottsym{(}  \ottnt{C}  \ottsym{)}  \ottsym{)}  \ottsym{)}  \ottsym{<:}  \sigma'_{\ottmv{i}}  \ottsym{(}  \sigma_{\ottmv{i}}  \ottsym{(}   { \sigma }_{ \mathit{X} }   \ottsym{(}  \ottnt{C_{{\mathrm{0}}}}  \ottsym{)}  \ottsym{)}  \ottsym{)} ~.
 \]
\end{lemmap}
\begin{proof}
 \TS{CHECKED: 2022/07/07}
 By induction on $\ottnt{C}$.
 Note that, because $ \tceseq{ \ottsym{(}  \mathit{X}  \ottsym{,}  \mathit{Y}  \ottsym{)} }   \ottsym{;}   \tceseq{ \mathit{z_{{\mathrm{0}}}}  \,  {:}  \,  \iota_{{\mathrm{0}}} }   \vdash  \ottnt{C_{{\mathrm{0}}}}  \succ  \ottnt{C}$ and $ \tceseq{ \ottsym{(}  \mathit{X'}  \ottsym{,}  \mathit{Y'}  \ottsym{)} }   \ottsym{;}   \tceseq{ \ottsym{(}  \mathit{X}  \ottsym{,}  \mathit{Y}  \ottsym{)} }  \,  \vdash ^{\circ}  \, \ottnt{C}$,
 we have
 \begin{itemize}
  \item $ \tceseq{ \mathit{X} }  \,  =  \, \mathit{X_{{\mathrm{1}}}}  \ottsym{,}   \tceseq{ \mathit{X_{{\mathrm{2}}}} } $,
  \item $ \tceseq{ \mathit{Y} }  \,  =  \, \mathit{Y_{{\mathrm{1}}}}  \ottsym{,}   \tceseq{ \mathit{Y_{{\mathrm{2}}}} } $,
  \item $ \ottnt{C_{{\mathrm{0}}}}  . \Phi  \,  =  \,  (   \lambda_\mu\!   \,  \mathit{x}  .\,  \mathit{X_{{\mathrm{1}}}}  \ottsym{(}  \mathit{x}  \ottsym{,}   \tceseq{ \mathit{z_{{\mathrm{0}}}} }   \ottsym{)}  ,   \lambda_\nu\!   \,  \mathit{y}  .\,  \mathit{Y_{{\mathrm{1}}}}  \ottsym{(}  \mathit{y}  \ottsym{,}   \tceseq{ \mathit{z_{{\mathrm{0}}}} }   \ottsym{)}  ) $,
  \item $  \{   \tceseq{ \mathit{Y'} }   \ottsym{,}  \mathit{Y_{{\mathrm{1}}}}  \}   \, \ottsym{\#} \,   \mathit{fpv}  (   {  \ottnt{C}  . \Phi  }^\mu   )  $, and
  \item $  \{   \tceseq{ \mathit{X_{{\mathrm{2}}}} }   \ottsym{,}   \tceseq{ \mathit{Y_{{\mathrm{2}}}} }   \}   \, \ottsym{\#} \,   \mathit{fpv}  (   \ottnt{C}  . \Phi   )  $
 \end{itemize}
 for some $\mathit{X_{{\mathrm{1}}}}$, $\mathit{Y_{{\mathrm{1}}}}$, $ \tceseq{ \mathit{X_{{\mathrm{2}}}} } $, $ \tceseq{ \mathit{Y_{{\mathrm{2}}}} } $, $\mathit{x}$, and $\mathit{y}$.
 %

 The conclusion is shown by \Srule{Comp} and \Srule{Eff} with the following.
 \begin{itemize}
  \item We show that
        \[
          \tceseq{ \mathit{z_{{\mathrm{0}}}}  \,  {:}  \,  \iota_{{\mathrm{0}}} }   \vdash   { \sigma }_{ \ottmv{i}  \ottsym{-}  \ottsym{1} }'   \ottsym{(}   { \sigma }_{ \ottmv{i}  \ottsym{-}  \ottsym{1} }   \ottsym{(}   { \sigma }_{ \mathit{X} }   \ottsym{(}   \ottnt{C}  . \ottnt{T}   \ottsym{)}  \ottsym{)}  \ottsym{)}  \ottsym{<:}  \sigma'_{\ottmv{i}}  \ottsym{(}  \sigma_{\ottmv{i}}  \ottsym{(}   { \sigma }_{ \mathit{X} }   \ottsym{(}   \ottnt{C_{{\mathrm{0}}}}  . \ottnt{T}   \ottsym{)}  \ottsym{)}  \ottsym{)} ~.
        \]
        We have
        \begin{itemize}
         \item $ \ottnt{C_{{\mathrm{0}}}}  . \ottnt{T}  \,  =  \,  \ottnt{C}  . \ottnt{T} $ implied by the assumption $ \tceseq{ \ottsym{(}  \mathit{X}  \ottsym{,}  \mathit{Y}  \ottsym{)} }   \ottsym{;}   \tceseq{ \mathit{z_{{\mathrm{0}}}}  \,  {:}  \,  \iota_{{\mathrm{0}}} }   \vdash  \ottnt{C_{{\mathrm{0}}}}  \succ  \ottnt{C}$, and
         \item $  \{   \tceseq{ \mathit{X'} }   \ottsym{,}   \tceseq{ \mathit{Y'} }   \ottsym{,}   \tceseq{ \mathit{X} }   \ottsym{,}   \tceseq{ \mathit{Y} }   \}   \, \ottsym{\#} \,   \mathit{fpv}  (   \ottnt{C}  . \ottnt{T}   )  $ implied by the assumption $ \tceseq{ \ottsym{(}  \mathit{X'}  \ottsym{,}  \mathit{Y'}  \ottsym{)} }   \ottsym{;}   \tceseq{ \ottsym{(}  \mathit{X}  \ottsym{,}  \mathit{Y}  \ottsym{)} }  \,  \vdash ^{\circ}  \, \ottnt{C}$.
        \end{itemize}
        Therefore, it suffices to show that
        \[
          \tceseq{ \mathit{z_{{\mathrm{0}}}}  \,  {:}  \,  \ottnt{T_{{\mathrm{0}}}} }   \vdash   \ottnt{C}  . \ottnt{T}   \ottsym{<:}   \ottnt{C}  . \ottnt{T}  ~.
        \]
        By \reflem{ts-elim-unused-binding-wf-ftyping} with $\ottsym{(}   \tceseq{ \mathit{X'} }   \ottsym{,}   \tceseq{ \mathit{X} }   \ottsym{)} \,  {:}  \,  (   \Sigma ^\ast   ,   \tceseq{ \iota_{{\mathrm{0}}} }   )   \ottsym{,}  \ottsym{(}   \tceseq{ \mathit{Y'} }   \ottsym{,}   \tceseq{ \mathit{Y} }   \ottsym{)} \,  {:}  \,  (   \Sigma ^\omega   ,   \tceseq{ \iota_{{\mathrm{0}}} }   )   \ottsym{,}   \tceseq{ \mathit{z_{{\mathrm{0}}}}  \,  {:}  \,  \iota_{{\mathrm{0}}} }   \vdash  \ottnt{C}$,
        we have $ \tceseq{ \mathit{z_{{\mathrm{0}}}}  \,  {:}  \,  \ottnt{T_{{\mathrm{0}}}} }   \vdash   \ottnt{C}  . \ottnt{T} $.
        Therefore, so we have the conclusion by \reflem{ts-refl-subtyping}.

  \item We show that
        \[
          \tceseq{ \mathit{z_{{\mathrm{0}}}}  \,  {:}  \,  \iota_{{\mathrm{0}}} }   \ottsym{,}  \mathit{x} \,  {:}  \,  \Sigma ^\ast   \models    { \ottsym{(}   { \sigma }_{ \ottmv{i}  \ottsym{-}  \ottsym{1} }'   \ottsym{(}   { \sigma }_{ \ottmv{i}  \ottsym{-}  \ottsym{1} }   \ottsym{(}   { \sigma }_{ \mathit{X} }   \ottsym{(}   \ottnt{C}  . \Phi   \ottsym{)}  \ottsym{)}  \ottsym{)}  \ottsym{)} }^\mu   \ottsym{(}  \mathit{x}  \ottsym{)}    \Rightarrow     { \ottsym{(}  \sigma'_{\ottmv{i}}  \ottsym{(}  \sigma_{\ottmv{i}}  \ottsym{(}   { \sigma }_{ \mathit{X} }   \ottsym{(}   \ottnt{C_{{\mathrm{0}}}}  . \Phi   \ottsym{)}  \ottsym{)}  \ottsym{)}  \ottsym{)} }^\mu   \ottsym{(}  \mathit{x}  \ottsym{)}  ~.
        \]
        Because
        \begin{itemize}
         \item $ \ottnt{C_{{\mathrm{0}}}}  . \Phi  \,  =  \,  (   \lambda_\mu\!   \,  \mathit{x}  .\,  \mathit{X_{{\mathrm{1}}}}  \ottsym{(}  \mathit{x}  \ottsym{,}   \tceseq{ \mathit{z_{{\mathrm{0}}}} }   \ottsym{)}  ,   \lambda_\nu\!   \,  \mathit{y}  .\,  \mathit{Y_{{\mathrm{1}}}}  \ottsym{(}  \mathit{y}  \ottsym{,}   \tceseq{ \mathit{z_{{\mathrm{0}}}} }   \ottsym{)}  ) $,
         \item $  \{   \tceseq{ \mathit{Y'} }   \ottsym{,}  \mathit{Y_{{\mathrm{1}}}}  \}   \, \ottsym{\#} \,   \mathit{fpv}  (   {  \ottnt{C}  . \Phi  }^\mu   )  $, and
         \item $  \{   \tceseq{ \mathit{X_{{\mathrm{2}}}} }   \ottsym{,}   \tceseq{ \mathit{Y_{{\mathrm{2}}}} }   \}   \, \ottsym{\#} \,   \mathit{fpv}  (   \ottnt{C}  . \Phi   )  $,
        \end{itemize}
        it suffices to show that
        \[
          \tceseq{ \mathit{z_{{\mathrm{0}}}}  \,  {:}  \,  \ottnt{T} }   \ottsym{,}  \mathit{x} \,  {:}  \,  \Sigma ^\ast   \models   \sigma  \ottsym{(}   { \ottsym{(}   \ottnt{C}  . \Phi   \ottsym{)} }^\mu   \ottsym{)}  \ottsym{(}  \mathit{x}  \ottsym{)}    \Rightarrow    \sigma  \ottsym{(}  \mathit{X_{{\mathrm{1}}}}  \ottsym{)}  \ottsym{(}  \mathit{x}  \ottsym{,}   \tceseq{ \mathit{z_{{\mathrm{0}}}} }   \ottsym{)}  ~.
        \]
        which is implied by \reflem{ts-mu-nu-valid}.
        \TS{A small gap?}

  \item We show that
        \[
          \tceseq{ \mathit{z_{{\mathrm{0}}}}  \,  {:}  \,  \ottnt{T} }   \ottsym{,}  \mathit{y} \,  {:}  \,  \Sigma ^\omega   \models    { \ottsym{(}   { \sigma }_{ \ottmv{i}  \ottsym{-}  \ottsym{1} }'   \ottsym{(}   { \sigma }_{ \ottmv{i}  \ottsym{-}  \ottsym{1} }   \ottsym{(}   { \sigma }_{ \mathit{X} }   \ottsym{(}   \ottnt{C}  . \Phi   \ottsym{)}  \ottsym{)}  \ottsym{)}  \ottsym{)} }^\nu   \ottsym{(}  \mathit{y}  \ottsym{)}    \Rightarrow     { \ottsym{(}  \sigma'_{\ottmv{i}}  \ottsym{(}  \sigma_{\ottmv{i}}  \ottsym{(}   { \sigma }_{ \mathit{X} }   \ottsym{(}   \ottnt{C_{{\mathrm{0}}}}  . \Phi   \ottsym{)}  \ottsym{)}  \ottsym{)}  \ottsym{)} }^\nu   \ottsym{(}  \mathit{y}  \ottsym{)}  ~.
        \]
        Because
        \begin{itemize}
         \item $\ottnt{C'}  \triangleright   \tceseq{ \ottsym{(}  \mathit{X'}  \ottsym{,}  \mathit{Y'}  \ottsym{)} }   \ottsym{;}   \tceseq{ \ottsym{(}  \mathit{X}  \ottsym{,}  \mathit{Y}  \ottsym{)} }   \ottsym{;}   \tceseq{ \mathit{z_{{\mathrm{0}}}}  \,  {:}  \,  \iota_{{\mathrm{0}}} }  \,  |  \, \ottnt{C} \,  \vdash _{ \ottmv{i} }  \, \sigma_{\ottmv{i}}$,
         \item $\mathit{Y_{{\mathrm{1}}}} \,  \not\in  \,  \mathit{fpv}  (   {  \ottnt{C}  . \Phi  }^\mu   ) $, and
         \item $  \{   \tceseq{ \mathit{X_{{\mathrm{2}}}} }   \ottsym{,}   \tceseq{ \mathit{Y_{{\mathrm{2}}}} }   \}   \, \ottsym{\#} \,   \mathit{fpv}  (   \ottnt{C}  . \Phi   )  $,
        \end{itemize}
        we have
        \[
         \sigma_{\ottmv{i}}  \ottsym{(}  \mathit{Y_{{\mathrm{1}}}}  \ottsym{)} \,  =  \, \ottsym{(}   \mu\!  \, \mathit{Y_{{\mathrm{1}}}}  \ottsym{(}  \mathit{y} \,  {:}  \,  \Sigma ^\omega   \ottsym{,}   \tceseq{ \mathit{z_{{\mathrm{0}}}}  \,  {:}  \,  \iota_{{\mathrm{0}}} }   \ottsym{)}  \ottsym{.}   { \sigma }_{ \ottmv{i}  \ottsym{-}  \ottsym{1} }'   \ottsym{(}   { \sigma }_{ \ottmv{i}  \ottsym{-}  \ottsym{1} }   \ottsym{(}   { \ottsym{(}   { \sigma }_{ \mathit{X} }   \ottsym{(}   \ottnt{C}  . \Phi   \ottsym{)}  \ottsym{)} }^\nu   \ottsym{(}  \mathit{y}  \ottsym{)}  \ottsym{)}  \ottsym{)}  \ottsym{)} ~.
        \]
        Then, because $ \ottnt{C_{{\mathrm{0}}}}  . \Phi  \,  =  \,  (   \lambda_\mu\!   \,  \mathit{x}  .\,  \mathit{X_{{\mathrm{1}}}}  \ottsym{(}  \mathit{x}  \ottsym{,}   \tceseq{ \mathit{z_{{\mathrm{0}}}} }   \ottsym{)}  ,   \lambda_\nu\!   \,  \mathit{y}  .\,  \mathit{Y_{{\mathrm{1}}}}  \ottsym{(}  \mathit{y}  \ottsym{,}   \tceseq{ \mathit{z_{{\mathrm{0}}}} }   \ottsym{)}  ) $,
        it suffices to show that
        \[
          \tceseq{ \mathit{z_{{\mathrm{0}}}}  \,  {:}  \,  \ottnt{T} }   \ottsym{,}  \mathit{y} \,  {:}  \,  \Sigma ^\omega   \models    { \sigma }_{ \ottmv{i}  \ottsym{-}  \ottsym{1} }'   \ottsym{(}   { \sigma }_{ \ottmv{i}  \ottsym{-}  \ottsym{1} }   \ottsym{(}   { \ottsym{(}   { \sigma }_{ \mathit{X} }   \ottsym{(}   \ottnt{C}  . \Phi   \ottsym{)}  \ottsym{)} }^\nu   \ottsym{)}  \ottsym{(}  \mathit{y}  \ottsym{)}  \ottsym{)}    \Rightarrow    \ottsym{(}   \mu\!  \, \mathit{Y_{{\mathrm{1}}}}  \ottsym{(}  \mathit{y} \,  {:}  \,  \Sigma ^\omega   \ottsym{,}   \tceseq{ \mathit{z_{{\mathrm{0}}}}  \,  {:}  \,  \iota_{{\mathrm{0}}} }   \ottsym{)}  \ottsym{.}   { \sigma }_{ \ottmv{i}  \ottsym{-}  \ottsym{1} }'   \ottsym{(}   { \sigma }_{ \ottmv{i}  \ottsym{-}  \ottsym{1} }   \ottsym{(}   { \ottsym{(}   { \sigma }_{ \mathit{X} }   \ottsym{(}   \ottnt{C}  . \Phi   \ottsym{)}  \ottsym{)} }^\nu   \ottsym{(}  \mathit{y}  \ottsym{)}  \ottsym{)}  \ottsym{)}  \ottsym{)}  \ottsym{(}  \mathit{y}  \ottsym{,}   \tceseq{ \mathit{z_{{\mathrm{0}}}} }   \ottsym{)}  ~.
        \]
        which is implied by \reflem{ts-mu-nu-valid}.

  \item We show that
        \[
          \tceseq{ \mathit{z_{{\mathrm{0}}}}  \,  {:}  \,  \iota_{{\mathrm{0}}} }  \,  |  \,  \ottnt{C}  . \ottnt{T}  \,  \,\&\,  \,  { \sigma }_{ \ottmv{i}  \ottsym{-}  \ottsym{1} }'   \ottsym{(}   { \sigma }_{ \ottmv{i}  \ottsym{-}  \ottsym{1} }   \ottsym{(}   { \sigma }_{ \mathit{X} }   \ottsym{(}   \ottnt{C}  . \Phi   \ottsym{)}  \ottsym{)}  \ottsym{)}  \vdash   { \sigma }_{ \ottmv{i}  \ottsym{-}  \ottsym{1} }'   \ottsym{(}   { \sigma }_{ \ottmv{i}  \ottsym{-}  \ottsym{1} }   \ottsym{(}   { \sigma }_{ \mathit{X} }   \ottsym{(}   \ottnt{C}  . \ottnt{S}   \ottsym{)}  \ottsym{)}  \ottsym{)}  \ottsym{<:}  \sigma'_{\ottmv{i}}  \ottsym{(}  \sigma_{\ottmv{i}}  \ottsym{(}   { \sigma }_{ \mathit{X} }   \ottsym{(}   \ottnt{C_{{\mathrm{0}}}}  . \ottnt{S}   \ottsym{)}  \ottsym{)}  \ottsym{)}
        \]
        (note that $  \{   \tceseq{ \mathit{X'} }   \ottsym{,}   \tceseq{ \mathit{Y'} }   \ottsym{,}   \tceseq{ \mathit{X} }   \ottsym{,}   \tceseq{ \mathit{Y} }   \}   \, \ottsym{\#} \,   \mathit{fpv}  (   \ottnt{C}  . \ottnt{T}   )  $ as shown above).
        By case analysis on $ \ottnt{C}  . \ottnt{S} $.
        \begin{caseanalysis}
         \case $ \ottnt{C}  . \ottnt{S}  \,  =  \,  \square $:
          The assumption $ \tceseq{ \ottsym{(}  \mathit{X}  \ottsym{,}  \mathit{Y}  \ottsym{)} }   \ottsym{;}   \tceseq{ \mathit{z_{{\mathrm{0}}}}  \,  {:}  \,  \iota_{{\mathrm{0}}} }   \vdash  \ottnt{C_{{\mathrm{0}}}}  \succ  \ottnt{C}$ implies $ \ottnt{C_{{\mathrm{0}}}}  . \ottnt{S}  \,  =  \,  \square $.
          Therefore, we have the conclusion by \Srule{Empty}.

         \case $  \exists  \, { \mathit{x}  \ottsym{,}  \ottnt{C_{{\mathrm{1}}}}  \ottsym{,}  \ottnt{C_{{\mathrm{2}}}} } . \   \ottnt{C}  . \ottnt{S}  \,  =  \,  ( \forall  \mathit{x}  .  \ottnt{C_{{\mathrm{1}}}}  )  \Rightarrow   \ottnt{C_{{\mathrm{2}}}}  $:
          We have the following.
          \begin{itemize}
           \item $ \tceseq{ \ottsym{(}  \mathit{X'}  \ottsym{,}  \mathit{Y'}  \ottsym{)} }   \vdash  \ottnt{C'}  \Rrightarrow  \ottnt{C}$ implies
                 $ \tceseq{ \ottsym{(}  \mathit{X'}  \ottsym{,}  \mathit{Y'}  \ottsym{)} }   \ottsym{,}  \ottsym{(}  \mathit{X_{{\mathrm{1}}}}  \ottsym{,}  \mathit{Y_{{\mathrm{1}}}}  \ottsym{)}  \vdash  \ottnt{C'}  \Rrightarrow  \ottnt{C_{{\mathrm{2}}}}$.
           \item $ \tceseq{ \ottsym{(}  \mathit{X}  \ottsym{,}  \mathit{Y}  \ottsym{)} }   \ottsym{;}   \tceseq{ \mathit{z_{{\mathrm{0}}}}  \,  {:}  \,  \iota_{{\mathrm{0}}} }   \vdash  \ottnt{C_{{\mathrm{0}}}}  \succ  \ottnt{C}$ implies
                 \begin{itemize}
                  \item $ \ottnt{C_{{\mathrm{0}}}}  . \ottnt{S}  \,  =  \,  ( \forall  \mathit{x}  .  \ottnt{C_{{\mathrm{1}}}}  )  \Rightarrow   \ottnt{C_{{\mathrm{02}}}} $ and
                  \item $ \tceseq{ \ottsym{(}  \mathit{X_{{\mathrm{2}}}}  \ottsym{,}  \mathit{Y_{{\mathrm{2}}}}  \ottsym{)} }   \ottsym{;}   \tceseq{ \mathit{z_{{\mathrm{0}}}}  \,  {:}  \,  \iota_{{\mathrm{0}}} }   \vdash  \ottnt{C_{{\mathrm{02}}}}  \succ  \ottnt{C_{{\mathrm{2}}}}$
                 \end{itemize}
                 for some $\ottnt{C_{{\mathrm{02}}}}$.
           \item $\ottsym{(}   \tceseq{ \mathit{X'} }   \ottsym{,}   \tceseq{ \mathit{X} }   \ottsym{)} \,  {:}  \,  (   \Sigma ^\ast   ,   \tceseq{ \iota_{{\mathrm{0}}} }   )   \ottsym{,}  \ottsym{(}   \tceseq{ \mathit{Y'} }   \ottsym{,}   \tceseq{ \mathit{Y} }   \ottsym{)} \,  {:}  \,  (   \Sigma ^\omega   ,   \tceseq{ \iota_{{\mathrm{0}}} }   )   \ottsym{,}   \tceseq{ \mathit{z_{{\mathrm{0}}}}  \,  {:}  \,  \iota_{{\mathrm{0}}} }   \vdash  \ottnt{C}$ implies
                 \begin{itemize}
                  \item $\ottsym{(}   \tceseq{ \mathit{X'} }   \ottsym{,}   \tceseq{ \mathit{X} }   \ottsym{)} \,  {:}  \,  (   \Sigma ^\ast   ,   \tceseq{ \iota_{{\mathrm{0}}} }   )   \ottsym{,}  \ottsym{(}   \tceseq{ \mathit{Y'} }   \ottsym{,}   \tceseq{ \mathit{Y} }   \ottsym{)} \,  {:}  \,  (   \Sigma ^\omega   ,   \tceseq{ \iota_{{\mathrm{0}}} }   )   \ottsym{,}   \tceseq{ \mathit{z_{{\mathrm{0}}}}  \,  {:}  \,  \iota_{{\mathrm{0}}} }   \ottsym{,}  \mathit{x} \,  {:}  \,  \ottnt{C}  . \ottnt{T}   \vdash  \ottnt{C_{{\mathrm{1}}}}$ and
                  \item $\ottsym{(}   \tceseq{ \mathit{X'} }   \ottsym{,}   \tceseq{ \mathit{X} }   \ottsym{)} \,  {:}  \,  (   \Sigma ^\ast   ,   \tceseq{ \iota_{{\mathrm{0}}} }   )   \ottsym{,}  \ottsym{(}   \tceseq{ \mathit{Y'} }   \ottsym{,}   \tceseq{ \mathit{Y} }   \ottsym{)} \,  {:}  \,  (   \Sigma ^\omega   ,   \tceseq{ \iota_{{\mathrm{0}}} }   )   \ottsym{,}   \tceseq{ \mathit{z_{{\mathrm{0}}}}  \,  {:}  \,  \iota_{{\mathrm{0}}} }   \vdash  \ottnt{C_{{\mathrm{2}}}}$.
                 \end{itemize}
           \item $ \tceseq{ \ottsym{(}  \mathit{X'}  \ottsym{,}  \mathit{Y'}  \ottsym{)} }   \ottsym{;}   \tceseq{ \ottsym{(}  \mathit{X}  \ottsym{,}  \mathit{Y}  \ottsym{)} }  \,  \vdash ^{\circ}  \, \ottnt{C}$ implies
                 \begin{itemize}
                  \item $ \tceseq{ \ottsym{(}  \mathit{X'}  \ottsym{,}  \mathit{Y'}  \ottsym{)} }   \ottsym{,}  \ottsym{(}  \mathit{X_{{\mathrm{1}}}}  \ottsym{,}  \mathit{Y_{{\mathrm{1}}}}  \ottsym{)}  \ottsym{;}   \tceseq{ \ottsym{(}  \mathit{X_{{\mathrm{2}}}}  \ottsym{,}  \mathit{Y_{{\mathrm{2}}}}  \ottsym{)} }  \,  \vdash ^{\circ}  \, \ottnt{C_{{\mathrm{2}}}}$ and
                  \item $  \{   \tceseq{ \mathit{X'} }   \ottsym{,}   \tceseq{ \mathit{Y'} }   \ottsym{,}   \tceseq{ \mathit{X} }   \ottsym{,}   \tceseq{ \mathit{Y} }   \}   \, \ottsym{\#} \,   \mathit{fpv}  (  \ottnt{C_{{\mathrm{1}}}}  )  $.
                 \end{itemize}
          \end{itemize}
          We have the conclusion by \Srule{Ans} with the following.
          \begin{itemize}
           \item We show that $ \tceseq{ \mathit{z_{{\mathrm{0}}}}  \,  {:}  \,  \iota_{{\mathrm{0}}} }   \ottsym{,}  \mathit{x} \,  {:}  \,  \ottnt{C}  . \ottnt{T}   \vdash   { \sigma }_{ \ottmv{i}  \ottsym{-}  \ottsym{1} }'   \ottsym{(}   { \sigma }_{ \ottmv{i}  \ottsym{-}  \ottsym{1} }   \ottsym{(}   { \sigma }_{ \mathit{X} }   \ottsym{(}  \ottnt{C_{{\mathrm{1}}}}  \ottsym{)}  \ottsym{)}  \ottsym{)}  \ottsym{<:}  \sigma'_{\ottmv{i}}  \ottsym{(}  \sigma_{\ottmv{i}}  \ottsym{(}   { \sigma }_{ \mathit{X} }   \ottsym{(}  \ottnt{C_{{\mathrm{1}}}}  \ottsym{)}  \ottsym{)}  \ottsym{)}$.
                 By $  \{   \tceseq{ \mathit{X'} }   \ottsym{,}   \tceseq{ \mathit{Y'} }   \ottsym{,}   \tceseq{ \mathit{X} }   \ottsym{,}   \tceseq{ \mathit{Y} }   \}   \, \ottsym{\#} \,   \mathit{fpv}  (  \ottnt{C_{{\mathrm{1}}}}  )  $,
                 it suffices to show that
                 $ \tceseq{ \mathit{z_{{\mathrm{0}}}}  \,  {:}  \,  \iota_{{\mathrm{0}}} }   \ottsym{,}  \mathit{x} \,  {:}  \,  \ottnt{C}  . \ottnt{T}   \vdash  \ottnt{C_{{\mathrm{1}}}}  \ottsym{<:}  \ottnt{C_{{\mathrm{1}}}}$.
                 Because
                 \[
                  \ottsym{(}   \tceseq{ \mathit{X'} }   \ottsym{,}   \tceseq{ \mathit{X} }   \ottsym{)} \,  {:}  \,  (   \Sigma ^\ast   ,   \tceseq{ \iota_{{\mathrm{0}}} }   )   \ottsym{,}  \ottsym{(}   \tceseq{ \mathit{Y'} }   \ottsym{,}   \tceseq{ \mathit{Y} }   \ottsym{)} \,  {:}  \,  (   \Sigma ^\omega   ,   \tceseq{ \iota_{{\mathrm{0}}} }   )   \ottsym{,}   \tceseq{ \mathit{z_{{\mathrm{0}}}}  \,  {:}  \,  \iota_{{\mathrm{0}}} }   \ottsym{,}  \mathit{x} \,  {:}  \,  \ottnt{C}  . \ottnt{T}   \vdash  \ottnt{C_{{\mathrm{1}}}} ~,
                 \]
                 Lemmas~\ref{lem:ts-elim-unused-binding-wf-ftyping} and \ref{lem:ts-refl-subtyping}
                 imply the conclusion.

           \item We show that $ \tceseq{ \mathit{z_{{\mathrm{0}}}}  \,  {:}  \,  \iota_{{\mathrm{0}}} }   \vdash   { \sigma }_{ \ottmv{i}  \ottsym{-}  \ottsym{1} }'   \ottsym{(}   { \sigma }_{ \ottmv{i}  \ottsym{-}  \ottsym{1} }   \ottsym{(}   { \sigma }_{ \mathit{X} }   \ottsym{(}  \ottnt{C_{{\mathrm{2}}}}  \ottsym{)}  \ottsym{)}  \ottsym{)}  \ottsym{<:}  \sigma'_{\ottmv{i}}  \ottsym{(}  \sigma_{\ottmv{i}}  \ottsym{(}   { \sigma }_{ \mathit{X} }   \ottsym{(}  \ottnt{C_{{\mathrm{02}}}}  \ottsym{)}  \ottsym{)}  \ottsym{)}$.

                 By \reflem{lr-approx-compose-subst},
                 \begin{itemize}
                  \item $\ottnt{C'}  \triangleright   \tceseq{ \ottsym{(}  \mathit{X'}  \ottsym{,}  \mathit{Y'}  \ottsym{)} }   \ottsym{,}  \ottsym{(}  \mathit{X_{{\mathrm{1}}}}  \ottsym{,}  \mathit{Y_{{\mathrm{1}}}}  \ottsym{)}  \ottsym{;}   \tceseq{ \ottsym{(}  \mathit{X_{{\mathrm{2}}}}  \ottsym{,}  \mathit{Y_{{\mathrm{2}}}}  \ottsym{)} }   \ottsym{;}   \tceseq{ \mathit{z_{{\mathrm{0}}}}  \,  {:}  \,  \iota_{{\mathrm{0}}} }  \,  |  \, \ottnt{C_{{\mathrm{2}}}} \,  \vdash _{ \ottmv{i} }  \,  { \sigma }_{ \ottsym{0}  \ottsym{;}  \ottmv{i} } $,
                  \item $\ottnt{C'}  \triangleright   \emptyset   \ottsym{;}   \tceseq{ \ottsym{(}  \mathit{X'}  \ottsym{,}  \mathit{Y'}  \ottsym{)} }   \ottsym{,}  \ottsym{(}  \mathit{X_{{\mathrm{1}}}}  \ottsym{,}  \mathit{Y_{{\mathrm{1}}}}  \ottsym{)}  \ottsym{;}   \tceseq{ \mathit{z_{{\mathrm{0}}}}  \,  {:}  \,  \iota_{{\mathrm{0}}} }  \,  |  \, \ottnt{C'} \,  \vdash _{ \ottmv{i} }  \,  { \sigma }_{ \ottsym{0}  \ottsym{;}  \ottmv{i} }' $,
                  \item $\sigma_{\ottmv{i}}  \mathbin{\uplus}  \sigma'_{\ottmv{i}} \,  =  \,  {  { \sigma }_{ \ottsym{0}  \ottsym{;}  \ottmv{i} }   \mathbin{\uplus}  \sigma }_{ \ottsym{0}  \ottsym{;}  \ottmv{i} }' $,
                  \item $\ottnt{C'}  \triangleright   \tceseq{ \ottsym{(}  \mathit{X'}  \ottsym{,}  \mathit{Y'}  \ottsym{)} }   \ottsym{,}  \ottsym{(}  \mathit{X_{{\mathrm{1}}}}  \ottsym{,}  \mathit{Y_{{\mathrm{1}}}}  \ottsym{)}  \ottsym{;}   \tceseq{ \ottsym{(}  \mathit{X_{{\mathrm{2}}}}  \ottsym{,}  \mathit{Y_{{\mathrm{2}}}}  \ottsym{)} }   \ottsym{;}   \tceseq{ \mathit{z_{{\mathrm{0}}}}  \,  {:}  \,  \iota_{{\mathrm{0}}} }  \,  |  \, \ottnt{C_{{\mathrm{2}}}} \,  \vdash _{ \ottmv{i}  \ottsym{-}  \ottsym{1} }  \,  { \sigma }_{ \ottsym{0}  \ottsym{;}  \ottmv{i}  \ottsym{-}  \ottsym{1} } $,
                  \item $\ottnt{C'}  \triangleright   \emptyset   \ottsym{;}   \tceseq{ \ottsym{(}  \mathit{X'}  \ottsym{,}  \mathit{Y'}  \ottsym{)} }   \ottsym{,}  \ottsym{(}  \mathit{X_{{\mathrm{1}}}}  \ottsym{,}  \mathit{Y_{{\mathrm{1}}}}  \ottsym{)}  \ottsym{;}   \tceseq{ \mathit{z_{{\mathrm{0}}}}  \,  {:}  \,  \iota_{{\mathrm{0}}} }  \,  |  \, \ottnt{C'} \,  \vdash _{ \ottmv{i}  \ottsym{-}  \ottsym{1} }  \,  { \sigma }_{ \ottsym{0}  \ottsym{;}  \ottmv{i}  \ottsym{-}  \ottsym{1} }' $, and
                  \item $ {  { \sigma }_{ \ottmv{i}  \ottsym{-}  \ottsym{1} }   \mathbin{\uplus}  \sigma }_{ \ottmv{i}  \ottsym{-}  \ottsym{1} }'  \,  =  \,  {  { \sigma }_{ \ottsym{0}  \ottsym{;}  \ottmv{i}  \ottsym{-}  \ottsym{1} }   \mathbin{\uplus}  \sigma }_{ \ottsym{0}  \ottsym{;}  \ottmv{i}  \ottsym{-}  \ottsym{1} }' $.
                 \end{itemize}
                 Furthermore, we have the following.
                 \begin{itemize}
                  \item $ \tceseq{ \ottsym{(}  \mathit{X'}  \ottsym{,}  \mathit{Y'}  \ottsym{)} }   \ottsym{,}  \ottsym{(}  \mathit{X_{{\mathrm{1}}}}  \ottsym{,}  \mathit{Y_{{\mathrm{1}}}}  \ottsym{)}  \vdash  \ottnt{C'}  \Rrightarrow  \ottnt{C_{{\mathrm{2}}}}$.
                  \item $\ottsym{(}   \tceseq{ \mathit{X'} }   \ottsym{,}   \tceseq{ \mathit{X} }   \ottsym{)} \,  {:}  \,  (   \Sigma ^\ast   ,   \tceseq{ \iota_{{\mathrm{0}}} }   )   \ottsym{,}  \ottsym{(}   \tceseq{ \mathit{Y'} }   \ottsym{,}   \tceseq{ \mathit{Y} }   \ottsym{)} \,  {:}  \,  (   \Sigma ^\omega   ,   \tceseq{ \iota_{{\mathrm{0}}} }   )   \ottsym{,}   \tceseq{ \mathit{z_{{\mathrm{0}}}}  \,  {:}  \,  \iota_{{\mathrm{0}}} }   \vdash  \ottnt{C_{{\mathrm{2}}}}$.
                  \item $ \tceseq{ \ottsym{(}  \mathit{X_{{\mathrm{2}}}}  \ottsym{,}  \mathit{Y_{{\mathrm{2}}}}  \ottsym{)} }   \ottsym{;}   \tceseq{ \mathit{z_{{\mathrm{0}}}}  \,  {:}  \,  \iota_{{\mathrm{0}}} }   \vdash  \ottnt{C_{{\mathrm{02}}}}  \succ  \ottnt{C_{{\mathrm{2}}}}$.
                  \item $ \tceseq{ \ottsym{(}  \mathit{X'}  \ottsym{,}  \mathit{Y'}  \ottsym{)} }   \ottsym{,}  \ottsym{(}  \mathit{X_{{\mathrm{1}}}}  \ottsym{,}  \mathit{Y_{{\mathrm{1}}}}  \ottsym{)}  \ottsym{;}   \tceseq{ \ottsym{(}  \mathit{X_{{\mathrm{2}}}}  \ottsym{,}  \mathit{Y_{{\mathrm{2}}}}  \ottsym{)} }  \,  \vdash ^{\circ}  \, \ottnt{C_{{\mathrm{2}}}}$.
                  \item $ \tceseq{ \ottsym{(}  \mathit{X'}  \ottsym{,}  \mathit{Y'}  \ottsym{)} }   \ottsym{,}   \tceseq{ \ottsym{(}  \mathit{X}  \ottsym{,}  \mathit{Y}  \ottsym{)} }   \ottsym{;}   \tceseq{ \mathit{z_{{\mathrm{0}}}}  \,  {:}  \,  \iota_{{\mathrm{0}}} }  \,  |  \, \ottnt{C'}  \vdash  \sigma$.
                 \end{itemize}
                 Therefore, the IH implies
                 \[
                   \tceseq{ \mathit{z_{{\mathrm{0}}}}  \,  {:}  \,  \ottnt{T_{{\mathrm{0}}}} }   \vdash   { \sigma }_{ \ottsym{0}  \ottsym{;}  \ottmv{i}  \ottsym{-}  \ottsym{1} }'   \ottsym{(}   { \sigma }_{ \ottsym{0}  \ottsym{;}  \ottmv{i}  \ottsym{-}  \ottsym{1} }   \ottsym{(}   { \sigma }_{ \mathit{X} }   \ottsym{(}  \ottnt{C_{{\mathrm{2}}}}  \ottsym{)}  \ottsym{)}  \ottsym{)}  \ottsym{<:}   { \sigma }_{ \ottsym{0}  \ottsym{;}  \ottmv{i} }'   \ottsym{(}   { \sigma }_{ \ottsym{0}  \ottsym{;}  \ottmv{i} }   \ottsym{(}   { \sigma }_{ \mathit{X} }   \ottsym{(}  \ottnt{C_{{\mathrm{02}}}}  \ottsym{)}  \ottsym{)}  \ottsym{)} ~.
                 \]
                 Because
                 $\sigma_{\ottmv{i}}  \mathbin{\uplus}  \sigma'_{\ottmv{i}} \,  =  \,  {  { \sigma }_{ \ottsym{0}  \ottsym{;}  \ottmv{i} }   \mathbin{\uplus}  \sigma }_{ \ottsym{0}  \ottsym{;}  \ottmv{i} }' $ and
                 $ {  { \sigma }_{ \ottmv{i}  \ottsym{-}  \ottsym{1} }   \mathbin{\uplus}  \sigma }_{ \ottmv{i}  \ottsym{-}  \ottsym{1} }'  \,  =  \,  {  { \sigma }_{ \ottsym{0}  \ottsym{;}  \ottmv{i}  \ottsym{-}  \ottsym{1} }   \mathbin{\uplus}  \sigma }_{ \ottsym{0}  \ottsym{;}  \ottmv{i}  \ottsym{-}  \ottsym{1} }' $,
                 we have the conclusion.
          \end{itemize}
        \end{caseanalysis}
 \end{itemize}
\end{proof}

\begin{lemmap}{Well-Formedness of Initial Computation Types}{lr-approx-init-cty-wf}
 If
 \begin{itemize}
  \item $\Gamma  \ottsym{,}  \ottsym{(}   \tceseq{ \mathit{X'} }   \ottsym{,}   \tceseq{ \mathit{X} }   \ottsym{)} \,  {:}  \,  (   \Sigma ^\ast   ,   \tceseq{ \iota_{{\mathrm{0}}} }   )   \ottsym{,}  \ottsym{(}   \tceseq{ \mathit{Y'} }   \ottsym{,}   \tceseq{ \mathit{Y} }   \ottsym{)} \,  {:}  \,  (   \Sigma ^\omega   ,   \tceseq{ \iota_{{\mathrm{0}}} }   )   \ottsym{,}   \tceseq{ \mathit{z_{{\mathrm{0}}}}  \,  {:}  \,  \iota_{{\mathrm{0}}} }   \vdash  \ottnt{C}$ and
  \item $ \tceseq{ \ottsym{(}  \mathit{X}  \ottsym{,}  \mathit{Y}  \ottsym{)} }   \ottsym{;}   \tceseq{ \mathit{z_{{\mathrm{0}}}}  \,  {:}  \,  \iota_{{\mathrm{0}}} }   \vdash  \ottnt{C_{{\mathrm{0}}}}  \succ  \ottnt{C}$,
 \end{itemize}
 then $\Gamma  \ottsym{,}  \ottsym{(}   \tceseq{ \mathit{X'} }   \ottsym{,}   \tceseq{ \mathit{X} }   \ottsym{)} \,  {:}  \,  (   \Sigma ^\ast   ,   \tceseq{ \iota_{{\mathrm{0}}} }   )   \ottsym{,}  \ottsym{(}   \tceseq{ \mathit{Y'} }   \ottsym{,}   \tceseq{ \mathit{Y} }   \ottsym{)} \,  {:}  \,  (   \Sigma ^\omega   ,   \tceseq{ \iota_{{\mathrm{0}}} }   )   \ottsym{,}   \tceseq{ \mathit{z_{{\mathrm{0}}}}  \,  {:}  \,  \ottnt{T_{{\mathrm{0}}}} }   \vdash  \ottnt{C_{{\mathrm{0}}}}$.
\end{lemmap}
\begin{proof}
 Straightforward by induction on the derivation of $ \tceseq{ \ottsym{(}  \mathit{X}  \ottsym{,}  \mathit{Y}  \ottsym{)} }   \ottsym{;}   \tceseq{ \mathit{z_{{\mathrm{0}}}}  \,  {:}  \,  \iota_{{\mathrm{0}}} }   \vdash  \ottnt{C_{{\mathrm{0}}}}  \succ  \ottnt{C}$.
\end{proof}

\begin{lemmap}{Well-Formedness of Effect Approximation Substitution}{lr-approx-eff-subst-wf}
 Assume that
 \begin{itemize}
  \item $\ottsym{(}   \tceseq{ \mathit{X'} }   \ottsym{,}   \tceseq{ \mathit{X} }   \ottsym{,}   \tceseq{ \mathit{X_{{\mathrm{0}}}} }   \ottsym{)} \,  {:}  \,  (   \Sigma ^\ast   ,   \tceseq{ \iota_{{\mathrm{0}}} }   )   \ottsym{,}  \ottsym{(}   \tceseq{ \mathit{Y'} }   \ottsym{,}   \tceseq{ \mathit{Y} }   \ottsym{,}   \tceseq{ \mathit{Y_{{\mathrm{0}}}} }   \ottsym{)} \,  {:}  \,  (   \Sigma ^\omega   ,   \tceseq{ \iota_{{\mathrm{0}}} }   )   \ottsym{,}   \tceseq{ \mathit{z_{{\mathrm{0}}}}  \,  {:}  \,  \iota_{{\mathrm{0}}} }   \vdash  \ottnt{C'}$,
  \item $ \emptyset   \ottsym{;}   \tceseq{ \ottsym{(}  \mathit{X'}  \ottsym{,}  \mathit{Y'}  \ottsym{)} }   \ottsym{,}   \tceseq{ \ottsym{(}  \mathit{X}  \ottsym{,}  \mathit{Y}  \ottsym{)} }   \ottsym{,}   \tceseq{ \ottsym{(}  \mathit{X_{{\mathrm{0}}}}  \ottsym{,}  \mathit{Y_{{\mathrm{0}}}}  \ottsym{)} }  \,  \vdash ^{\circ}  \, \ottnt{C'}$,
  \item $\ottsym{(}   \tceseq{ \mathit{X'} }   \ottsym{,}   \tceseq{ \mathit{X} }   \ottsym{,}   \tceseq{ \mathit{X_{{\mathrm{0}}}} }   \ottsym{)} \,  {:}  \,  (   \Sigma ^\ast   ,   \tceseq{ \iota_{{\mathrm{0}}} }   )   \ottsym{,}  \ottsym{(}   \tceseq{ \mathit{Y'} }   \ottsym{,}   \tceseq{ \mathit{Y} }   \ottsym{,}   \tceseq{ \mathit{Y_{{\mathrm{0}}}} }   \ottsym{)} \,  {:}  \,  (   \Sigma ^\omega   ,   \tceseq{ \iota_{{\mathrm{0}}} }   )   \ottsym{,}   \tceseq{ \mathit{z_{{\mathrm{0}}}}  \,  {:}  \,  \iota_{{\mathrm{0}}} }   \vdash  \ottnt{C}$,
  \item $\ottnt{C'}  \triangleright   \tceseq{ \ottsym{(}  \mathit{X'}  \ottsym{,}  \mathit{Y'}  \ottsym{)} }   \ottsym{;}   \tceseq{ \ottsym{(}  \mathit{X}  \ottsym{,}  \mathit{Y}  \ottsym{)} }   \ottsym{;}   \tceseq{ \mathit{z_{{\mathrm{0}}}}  \,  {:}  \,  \iota_{{\mathrm{0}}} }  \,  |  \, \ottnt{C} \,  \vdash _{ \ottmv{i} }  \, \sigma_{\ottmv{i}}$, and
  \item $ \tceseq{ \ottsym{(}  \mathit{X'}  \ottsym{,}  \mathit{Y'}  \ottsym{)} }   \ottsym{;}   \tceseq{ \ottsym{(}  \mathit{X}  \ottsym{,}  \mathit{Y}  \ottsym{)} }   \ottsym{,}   \tceseq{ \ottsym{(}  \mathit{X_{{\mathrm{0}}}}  \ottsym{,}  \mathit{Y_{{\mathrm{0}}}}  \ottsym{)} }  \,  \vdash ^{\circ}  \, \ottnt{C}$.
 \end{itemize}
 Then,
 \[
   {   \forall  \, { \mathit{Y_{{\mathrm{0}}}} \,  \in  \,  \{   \tceseq{ \mathit{Y} }   \}  } . \   \emptyset   \vdash  \sigma_{\ottmv{i}}  \ottsym{(}  \mathit{Y_{{\mathrm{0}}}}  \ottsym{)}  \ottsym{:}   (   \Sigma ^\omega   ,   \tceseq{ \iota_{{\mathrm{0}}} }   )   } \,\mathrel{\wedge}\, {   \forall  \, { \mathit{X_{{\mathrm{0}}}} \,  \in  \,  \{   \tceseq{ \mathit{X} }   \}  } . \   \emptyset   \vdash  \sigma_{\ottmv{i}}  \ottsym{(}  \mathit{X_{{\mathrm{0}}}}  \ottsym{)}  \ottsym{:}   (   \Sigma ^\ast   ,   \tceseq{ \iota_{{\mathrm{0}}} }   )   }  ~.
 \]
\end{lemmap}
\begin{proof}
 \TS{CHECKED: 2022/07/07}
 Straightforward by lexicographical induction on the pair of $\ottmv{i}$ and the length of $ \tceseq{ \ottsym{(}  \mathit{X}  \ottsym{,}  \mathit{Y}  \ottsym{)} } $
 with Lemmas~\ref{lem:lr-eff-rec-wf-XXX} and \ref{lem:ts-pred-subst-wf-ftyping}.
\end{proof}

\begin{lemmap}{Typing $ \nu $-Approximation}{lr-typing-nu-approx-XXX}
 Assume that
 \begin{itemize}
  \item $\ottnt{v} \,  =  \,  \mathsf{rec}  \, (  \tceseq{  \mathit{X} ^{  \ottnt{A} _{  \mu  }  },  \mathit{Y} ^{  \ottnt{A} _{  \nu  }  }  }  ,  \mathit{f} ,   \tceseq{ \mathit{z} }  ) . \,  \ottnt{e} $,
  \item $ \tceseq{ \mathit{z_{{\mathrm{0}}}}  \,  {:}  \,  \iota_{{\mathrm{0}}} }  \,  =  \,  \gamma  (   \tceseq{ \mathit{z}  \,  {:}  \,  \ottnt{T_{{\mathrm{1}}}} }   ) $,
  \item $ \tceseq{ \ottsym{(}  \mathit{X}  \ottsym{,}  \mathit{Y}  \ottsym{)} }   \ottsym{;}   \tceseq{ \mathit{z_{{\mathrm{0}}}}  \,  {:}  \,  \iota_{{\mathrm{0}}} }   \vdash  \ottnt{C_{{\mathrm{0}}}}  \stackrel{\succ}{\circ}  \ottnt{C}$,
  \item $ \tceseq{ \ottsym{(}  \mathit{X}  \ottsym{,}  \mathit{Y}  \ottsym{)} }   \ottsym{;}   \tceseq{ \mathit{z_{{\mathrm{0}}}}  \,  {:}  \,  \iota_{{\mathrm{0}}} }  \,  |  \, \ottnt{C}  \vdash  \sigma$,
  \item $  \{   \tceseq{ \mathit{X} }   \ottsym{,}   \tceseq{ \mathit{Y} }   \}   \, \ottsym{\#} \,   \mathit{fpv}  (   \tceseq{ \ottnt{T_{{\mathrm{1}}}} }   )  $,
  \item $  \forall  \, { \ottmv{i} } . \  \ottnt{C}  \triangleright   \emptyset   \ottsym{;}   \tceseq{ \ottsym{(}  \mathit{X}  \ottsym{,}  \mathit{Y}  \ottsym{)} }   \ottsym{;}   \tceseq{ \mathit{z_{{\mathrm{0}}}}  \,  {:}  \,  \iota_{{\mathrm{0}}} }  \,  |  \, \ottnt{C} \,  \vdash _{ \ottmv{i} }  \, \sigma_{\ottmv{i}} $,
  \item $ \tceseq{ \mathit{X} }  \,  {:}  \,  (   \Sigma ^\ast   ,   \tceseq{ \iota_{{\mathrm{0}}} }   )   \ottsym{,}   \tceseq{ \mathit{Y} }  \,  {:}  \,  (   \Sigma ^\omega   ,   \tceseq{ \iota_{{\mathrm{0}}} }   )   \ottsym{,}  \mathit{f} \,  {:}  \, \ottsym{(}   \tceseq{ \mathit{z}  \,  {:}  \,  \ottnt{T_{{\mathrm{1}}}} }   \ottsym{)}  \rightarrow  \ottnt{C_{{\mathrm{0}}}}  \ottsym{,}   \tceseq{ \mathit{z}  \,  {:}  \,  \ottnt{T_{{\mathrm{1}}}} }   \vdash  \ottnt{e}  \ottsym{:}  \ottnt{C}$, and
  \item $ { \sigma }_{ \mathit{X} }  \,  =  \,  [ \tceseq{ \sigma  \ottsym{(}  \mathit{X}  \ottsym{)}  /  \mathit{X} } ] $.
 \end{itemize}
 Then,
 \[
    \forall  \, { \ottmv{i}  \ottsym{,}  \pi } . \   \emptyset   \vdash   \ottnt{v} _{  \nu   \ottsym{;}  \ottmv{i}  \ottsym{;}  \pi }   \ottsym{:}  \ottsym{(}   \tceseq{ \mathit{z}  \,  {:}  \,  \ottnt{T_{{\mathrm{1}}}} }   \ottsym{)}  \rightarrow  \sigma_{\ottmv{i}}  \ottsym{(}   { \sigma }_{ \mathit{X} }   \ottsym{(}  \ottnt{C_{{\mathrm{0}}}}  \ottsym{)}  \ottsym{)}  ~.
 \]
\end{lemmap}
\begin{proof}
 \TS{CHECKED: 2022/07/07}
 By induction on $\ottmv{i}$.
 Let $\pi$ be any infinite trace.


 We proceed by case analysis on $\ottmv{i}$.
 \begin{caseanalysis}
  \case $\ottmv{i} \,  =  \, \ottsym{0}$:
   Because $ \ottnt{v} _{  \nu   \ottsym{;}  \ottsym{0}  \ottsym{;}  \pi }  \,  =  \,  \lambda\!  \,  \tceseq{ \mathit{z} }   \ottsym{.}   \bullet ^{ \pi } $,
   \T{Abs} implies that it suffices to show that
   \[
     \tceseq{ \mathit{z}  \,  {:}  \,  \ottnt{T_{{\mathrm{1}}}} }   \vdash   \bullet ^{ \pi }   \ottsym{:}  \sigma_{\ottmv{i}}  \ottsym{(}   { \sigma }_{ \mathit{X} }   \ottsym{(}  \ottnt{C_{{\mathrm{0}}}}  \ottsym{)}  \ottsym{)} ~.
   \]
   By \T{Div}, it suffices to show the following.
   \begin{itemize}
    \item We show that
          \[
            \tceseq{ \mathit{z}  \,  {:}  \,  \ottnt{T_{{\mathrm{1}}}} }   \vdash  \sigma_{\ottmv{i}}  \ottsym{(}   { \sigma }_{ \mathit{X} }   \ottsym{(}  \ottnt{C_{{\mathrm{0}}}}  \ottsym{)}  \ottsym{)} ~.
          \]
          Because $ \tceseq{ \mathit{X} }  \,  {:}  \,  (   \Sigma ^\ast   ,   \tceseq{ \iota_{{\mathrm{0}}} }   )   \ottsym{,}   \tceseq{ \mathit{Y} }  \,  {:}  \,  (   \Sigma ^\omega   ,   \tceseq{ \iota_{{\mathrm{0}}} }   )   \ottsym{,}  \mathit{f} \,  {:}  \, \ottsym{(}   \tceseq{ \mathit{z}  \,  {:}  \,  \ottnt{T_{{\mathrm{1}}}} }   \ottsym{)}  \rightarrow  \ottnt{C_{{\mathrm{0}}}}  \ottsym{,}   \tceseq{ \mathit{z}  \,  {:}  \,  \ottnt{T_{{\mathrm{1}}}} }   \vdash  \ottnt{e}  \ottsym{:}  \ottnt{C}$,
          we have
          $ \tceseq{ \mathit{X} }  \,  {:}  \,  (   \Sigma ^\ast   ,   \tceseq{ \iota_{{\mathrm{0}}} }   )   \ottsym{,}   \tceseq{ \mathit{Y} }  \,  {:}  \,  (   \Sigma ^\omega   ,   \tceseq{ \iota_{{\mathrm{0}}} }   )   \ottsym{,}   \tceseq{ \mathit{z_{{\mathrm{0}}}}  \,  {:}  \,  \iota_{{\mathrm{0}}} }   \vdash  \ottnt{C}$
          by Lemmas~\ref{lem:ts-wf-ty-tctx-typing} and \ref{lem:ts-elim-unused-binding-wf-ftyping}.
          Then, we have the conclusion by Lemmas~\ref{lem:lr-approx-eff-subst-wf}, \ref{lem:lr-eff-rec-wf-XXX}, and \ref{lem:ts-pred-subst-wf-ftyping}.

    \item We show that
          \[
            { \models  \,   { \sigma_{\ottmv{i}}  \ottsym{(}   { \sigma }_{ \mathit{X} }   \ottsym{(}  \ottnt{C_{{\mathrm{0}}}}  \ottsym{)}  \ottsym{)} }^\nu (  \pi  )  }  ~.
          \]
          Because $ \tceseq{ \ottsym{(}  \mathit{X}  \ottsym{,}  \mathit{Y}  \ottsym{)} }   \ottsym{;}   \tceseq{ \mathit{z_{{\mathrm{0}}}}  \,  {:}  \,  \iota_{{\mathrm{0}}} }   \vdash  \ottnt{C_{{\mathrm{0}}}}  \stackrel{\succ}{\circ}  \ottnt{C}$,
          the $ \nu $-parts of the temporal effects occurring strictly positively in $\ottnt{C_{{\mathrm{0}}}}$
          take the form $\mathit{Y_{{\mathrm{0}}}}  \ottsym{(}  \mathit{y}  \ottsym{,}   \tceseq{ \mathit{z_{{\mathrm{0}}}} }   \ottsym{)}$ for some $\mathit{Y_{{\mathrm{0}}}} \,  \in  \,  \{   \tceseq{ \mathit{Y} }   \} $, and
          the assumption $\ottnt{C}  \triangleright   \emptyset   \ottsym{;}   \tceseq{ \ottsym{(}  \mathit{X}  \ottsym{,}  \mathit{Y}  \ottsym{)} }   \ottsym{;}   \tceseq{ \mathit{z_{{\mathrm{0}}}}  \,  {:}  \,  \iota_{{\mathrm{0}}} }  \,  |  \, \ottnt{C} \,  \vdash _{ \ottmv{i} }  \, \sigma_{\ottmv{i}}$ indicates that
          $\mathit{Y_{{\mathrm{0}}}}$ is replaced by $\sigma_{\ottmv{i}}$.
          Because $\ottmv{i} \,  =  \, \ottsym{0}$, $\mathit{Y_{{\mathrm{0}}}}$ is replaced with $ \top $.
          Therefore, we have the conclusion.
   \end{itemize}

  \case $\ottmv{i} \,  >  \, \ottsym{0}$:
   Because $ \ottnt{v} _{  \nu   \ottsym{;}  \ottmv{i}  \ottsym{;}  \pi }  \,  =  \,     \lambda\!  \,  \tceseq{ \mathit{z} }   \ottsym{.}  \ottnt{e}    [ \tceseq{ \sigma  \ottsym{(}  \mathit{X}  \ottsym{)}  /  \mathit{X} } ]      [ \tceseq{  { \sigma }_{ \ottmv{i}  \ottsym{-}  \ottsym{1} }   \ottsym{(}  \mathit{Y}  \ottsym{)}  /  \mathit{Y} } ]      [   \ottnt{v} _{  \nu   \ottsym{;}  \ottmv{i}  \ottsym{-}  \ottsym{1}  \ottsym{;}  \pi }   /  \mathit{f}  ]  $,
   \T{Abs} implies that it suffices to show that
   \[
     \tceseq{ \mathit{z}  \,  {:}  \,  \ottnt{T_{{\mathrm{1}}}} }   \vdash     \ottnt{e}    [ \tceseq{ \sigma  \ottsym{(}  \mathit{X}  \ottsym{)}  /  \mathit{X} } ]      [ \tceseq{  { \sigma }_{ \ottmv{i}  \ottsym{-}  \ottsym{1} }   \ottsym{(}  \mathit{Y}  \ottsym{)}  /  \mathit{Y} } ]      [   \ottnt{v} _{  \nu   \ottsym{;}  \ottmv{i}  \ottsym{-}  \ottsym{1}  \ottsym{;}  \pi }   /  \mathit{f}  ]    \ottsym{:}  \sigma_{\ottmv{i}}  \ottsym{(}   { \sigma }_{ \mathit{X} }   \ottsym{(}  \ottnt{C_{{\mathrm{0}}}}  \ottsym{)}  \ottsym{)} ~.
   \]
   Because
   \begin{itemize}
    \item $ \tceseq{ \mathit{X} }  \,  {:}  \,  (   \Sigma ^\ast   ,   \tceseq{ \iota_{{\mathrm{0}}} }   )   \ottsym{,}   \tceseq{ \mathit{Y} }  \,  {:}  \,  (   \Sigma ^\omega   ,   \tceseq{ \iota_{{\mathrm{0}}} }   )   \ottsym{,}  \mathit{f} \,  {:}  \, \ottsym{(}   \tceseq{ \mathit{z}  \,  {:}  \,  \ottnt{T_{{\mathrm{1}}}} }   \ottsym{)}  \rightarrow  \ottnt{C_{{\mathrm{0}}}}  \ottsym{,}   \tceseq{ \mathit{z}  \,  {:}  \,  \ottnt{T_{{\mathrm{1}}}} }   \vdash  \ottnt{e}  \ottsym{:}  \ottnt{C}$,
    \item $  \forall  \, { \mathit{X_{{\mathrm{0}}}} \,  \in  \,  \{   \tceseq{ \mathit{X} }   \}  } . \   \emptyset   \vdash  \sigma  \ottsym{(}  \mathit{X}  \ottsym{)}  \ottsym{:}   (   \Sigma ^\ast   ,   \tceseq{ \iota_{{\mathrm{0}}} }   )  $ by \reflem{lr-eff-rec-wf-XXX},
    \item $  \forall  \, { \mathit{Y_{{\mathrm{0}}}} \,  \in  \,  \{   \tceseq{ \mathit{Y} }   \}  } . \   \emptyset   \vdash   { \sigma }_{ \ottmv{i}  \ottsym{-}  \ottsym{1} }   \ottsym{(}  \mathit{Y}  \ottsym{)}  \ottsym{:}   (   \Sigma ^\omega   ,   \tceseq{ \iota_{{\mathrm{0}}} }   )  $ by \reflem{lr-approx-eff-subst-wf}, and
    \item $ \emptyset   \vdash   \ottnt{v} _{  \nu   \ottsym{;}  \ottmv{i}  \ottsym{-}  \ottsym{1}  \ottsym{;}  \pi }   \ottsym{:}  \ottsym{(}   \tceseq{ \mathit{z}  \,  {:}  \,  \ottnt{T_{{\mathrm{1}}}} }   \ottsym{)}  \rightarrow   { \sigma }_{ \mathit{X} }   \ottsym{(}   { \sigma }_{ \ottmv{i}  \ottsym{-}  \ottsym{1} }   \ottsym{(}  \ottnt{C_{{\mathrm{0}}}}  \ottsym{)}  \ottsym{)}$ by the IH.
   \end{itemize}
   we have
   \[
     \tceseq{ \mathit{z}  \,  {:}  \,  \ottnt{T_{{\mathrm{1}}}} }   \vdash     \ottnt{e}    [ \tceseq{ \sigma  \ottsym{(}  \mathit{X}  \ottsym{)}  /  \mathit{X} } ]      [ \tceseq{  { \sigma }_{ \ottmv{i}  \ottsym{-}  \ottsym{1} }   \ottsym{(}  \mathit{Y}  \ottsym{)}  /  \mathit{Y} } ]      [   \ottnt{v} _{  \nu   \ottsym{;}  \ottmv{i}  \ottsym{-}  \ottsym{1}  \ottsym{;}  \pi }   /  \mathit{f}  ]    \ottsym{:}   { \sigma }_{ \ottmv{i}  \ottsym{-}  \ottsym{1} }   \ottsym{(}   { \sigma }_{ \mathit{X} }   \ottsym{(}  \ottnt{C}  \ottsym{)}  \ottsym{)}
   \]
   by Lemmas~\ref{lem:ts-pred-subst-typing}, \ref{lem:ts-val-subst-typing}, and
   \ref{lem:ts-unuse-non-refine-vars}.
   Then, we have the conclusion by \T{Sub} with the following.
   \begin{itemize}
    \item We show that
          \[
            \tceseq{ \mathit{z}  \,  {:}  \,  \ottnt{T_{{\mathrm{1}}}} }   \vdash   { \sigma }_{ \ottmv{i}  \ottsym{-}  \ottsym{1} }   \ottsym{(}   { \sigma }_{ \mathit{X} }   \ottsym{(}  \ottnt{C}  \ottsym{)}  \ottsym{)}  \ottsym{<:}   { \sigma }_{ \ottmv{i} }   \ottsym{(}   { \sigma }_{ \mathit{X} }   \ottsym{(}  \ottnt{C_{{\mathrm{0}}}}  \ottsym{)}  \ottsym{)} ~,
          \]
          which is implied by \reflem{lr-approx-fold-nu-XXX}.

    \item We show that
          \[
            \tceseq{ \mathit{z}  \,  {:}  \,  \ottnt{T_{{\mathrm{1}}}} }   \vdash  \sigma_{\ottmv{i}}  \ottsym{(}   { \sigma }_{ \mathit{X} }   \ottsym{(}  \ottnt{C_{{\mathrm{0}}}}  \ottsym{)}  \ottsym{)} ~.
          \]
          By Lemmas~\ref{lem:lr-approx-eff-subst-wf}, \ref{lem:lr-eff-rec-wf-XXX}, and \ref{lem:ts-pred-subst-wf-ftyping},
          it suffices to show that
          \[
            \tceseq{ \mathit{X} }  \,  {:}  \,  (   \Sigma ^\ast   ,   \tceseq{ \iota_{{\mathrm{0}}} }   )   \ottsym{,}   \tceseq{ \mathit{Y} }  \,  {:}  \,  (   \Sigma ^\omega   ,   \tceseq{ \iota_{{\mathrm{0}}} }   )   \ottsym{,}   \tceseq{ \mathit{z}  \,  {:}  \,  \ottnt{T_{{\mathrm{1}}}} }   \vdash  \ottnt{C_{{\mathrm{0}}}} ~.
          \]
          By Lemmas~\ref{lem:ts-wf-ty-tctx-typing} and \ref{lem:ts-elim-unused-binding-wf-ftyping},
          $ \tceseq{ \mathit{X} }  \,  {:}  \,  (   \Sigma ^\ast   ,   \tceseq{ \iota_{{\mathrm{0}}} }   )   \ottsym{,}   \tceseq{ \mathit{Y} }  \,  {:}  \,  (   \Sigma ^\omega   ,   \tceseq{ \iota_{{\mathrm{0}}} }   )   \ottsym{,}   \tceseq{ \mathit{z_{{\mathrm{0}}}}  \,  {:}  \,  \iota_{{\mathrm{0}}} }   \vdash  \ottnt{C}$.
          Then, we have the conclusion by
          Lemmas~\ref{lem:lr-approx-init-cty-wf}, \ref{lem:ts-refine-base-sorts}, and \ref{lem:ts-weak-wf-ftyping}
          with $ \tceseq{ \ottsym{(}  \mathit{X}  \ottsym{,}  \mathit{Y}  \ottsym{)} }   \ottsym{;}   \tceseq{ \mathit{z_{{\mathrm{0}}}}  \,  {:}  \,  \iota_{{\mathrm{0}}} }   \vdash  \ottnt{C_{{\mathrm{0}}}}  \stackrel{\succ}{\circ}  \ottnt{C}$.
   \end{itemize}
 \end{caseanalysis}
\end{proof}

\begin{lemmap}{Folding Types of $ \mu $-Approximation}{lr-approx-fold-mu-XXX}
 Let $\ottmv{i} \,  >  \, \ottsym{0}$ and $\pi$ be an infinite trace.
 Assume that
 \begin{itemize}
  \item $ \tceseq{ \ottsym{(}  \mathit{X'}  \ottsym{,}  \mathit{Y'}  \ottsym{)} }   \vdash  \ottnt{C'}  \Rrightarrow  \ottnt{C}$,
  \item $\ottsym{(}   \tceseq{ \mathit{X'} }   \ottsym{,}   \tceseq{ \mathit{X} }   \ottsym{)} \,  {:}  \,  (   \Sigma ^\ast   ,   \tceseq{ \iota_{{\mathrm{0}}} }   )   \ottsym{,}  \ottsym{(}   \tceseq{ \mathit{Y'} }   \ottsym{,}   \tceseq{ \mathit{Y} }   \ottsym{)} \,  {:}  \,  (   \Sigma ^\omega   ,   \tceseq{ \iota_{{\mathrm{0}}} }   )   \ottsym{,}   \tceseq{ \mathit{z_{{\mathrm{0}}}}  \,  {:}  \,  \iota_{{\mathrm{0}}} }   \vdash  \ottnt{C}$,
  \item $ \tceseq{ \ottsym{(}  \mathit{X}  \ottsym{,}  \mathit{Y}  \ottsym{)} }   \ottsym{;}   \tceseq{ \mathit{z_{{\mathrm{0}}}}  \,  {:}  \,  \iota_{{\mathrm{0}}} }   \vdash  \ottnt{C_{{\mathrm{0}}}}  \succ  \ottnt{C}$,
  \item $ \tceseq{ \ottsym{(}  \mathit{X'}  \ottsym{,}  \mathit{Y'}  \ottsym{)} }   \ottsym{;}   \tceseq{ \ottsym{(}  \mathit{X}  \ottsym{,}  \mathit{Y}  \ottsym{)} }  \,  \vdash ^{\circ}  \, \ottnt{C}$,
  \item $\ottnt{C'}  \triangleright   \tceseq{ \ottsym{(}  \mathit{X'}  \ottsym{,}  \mathit{Y'}  \ottsym{)} }   \ottsym{;}   \tceseq{ \ottsym{(}  \mathit{X}  \ottsym{,}  \mathit{Y}  \ottsym{)} }   \ottsym{;}   \tceseq{ \mathit{z_{{\mathrm{0}}}}  \,  {:}  \,  \iota_{{\mathrm{0}}} }  \,  |  \, \ottnt{C} \,  \vdash _{ \ottmv{i} }  \, \sigma_{\ottmv{i}}$,
  \item $\ottnt{C'}  \triangleright   \emptyset   \ottsym{;}   \tceseq{ \ottsym{(}  \mathit{X'}  \ottsym{,}  \mathit{Y'}  \ottsym{)} }   \ottsym{;}   \tceseq{ \mathit{z_{{\mathrm{0}}}}  \,  {:}  \,  \iota_{{\mathrm{0}}} }  \,  |  \, \ottnt{C'} \,  \vdash _{ \ottmv{i} }  \, \sigma'_{\ottmv{i}}$,
  \item $\ottnt{C'}  \triangleright   \tceseq{ \ottsym{(}  \mathit{X'}  \ottsym{,}  \mathit{Y'}  \ottsym{)} }   \ottsym{;}   \tceseq{ \ottsym{(}  \mathit{X}  \ottsym{,}  \mathit{Y}  \ottsym{)} }   \ottsym{;}   \tceseq{ \mathit{z_{{\mathrm{0}}}}  \,  {:}  \,  \iota_{{\mathrm{0}}} }  \,  |  \, \ottnt{C} \,  \vdash _{ \ottmv{i}  \ottsym{-}  \ottsym{1} }  \,  { \sigma }_{ \ottmv{i}  \ottsym{-}  \ottsym{1} } $, and
  \item $\ottnt{C'}  \triangleright   \emptyset   \ottsym{;}   \tceseq{ \ottsym{(}  \mathit{X'}  \ottsym{,}  \mathit{Y'}  \ottsym{)} }   \ottsym{;}   \tceseq{ \mathit{z_{{\mathrm{0}}}}  \,  {:}  \,  \iota_{{\mathrm{0}}} }  \,  |  \, \ottnt{C'} \,  \vdash _{ \ottmv{i}  \ottsym{-}  \ottsym{1} }  \,  { \sigma }_{ \ottmv{i}  \ottsym{-}  \ottsym{1} }' $.
 \end{itemize}
 Let $ { \sigma }_{ \mathit{Y} } $ be a substitution that maps $\mathit{Y_{{\mathrm{0}}}} \,  \in  \,  \{   \tceseq{ \mathit{Y} }   \ottsym{,}   \tceseq{ \mathit{Y'} }   \} $
 to $\ottsym{(}   \nu\!  \, \mathit{Y_{{\mathrm{0}}}}  \ottsym{(}  \mathit{y} \,  {:}  \,  \Sigma ^\omega   \ottsym{,}   \tceseq{ \mathit{z_{{\mathrm{0}}}}  \,  {:}  \,  \iota_{{\mathrm{0}}} }   \ottsym{)}  \ottsym{.}   \top   \ottsym{)}$.
 Then,
 \[
   \tceseq{ \mathit{z_{{\mathrm{0}}}}  \,  {:}  \,  \ottnt{T_{{\mathrm{0}}}} }   \vdash   { \sigma }_{ \ottmv{i}  \ottsym{-}  \ottsym{1} }'   \ottsym{(}   { \sigma }_{ \ottmv{i}  \ottsym{-}  \ottsym{1} }   \ottsym{(}   { \sigma }_{ \mathit{Y} }   \ottsym{(}  \ottnt{C}  \ottsym{)}  \ottsym{)}  \ottsym{)}  \ottsym{<:}  \sigma'_{\ottmv{i}}  \ottsym{(}  \sigma_{\ottmv{i}}  \ottsym{(}   { \sigma }_{ \mathit{Y} }   \ottsym{(}  \ottnt{C_{{\mathrm{0}}}}  \ottsym{)}  \ottsym{)}  \ottsym{)} ~.
 \]
\end{lemmap}
\begin{proof}
 \TS{CHECKED: 2022/07/07}
 By induction on $\ottnt{C}$.
 Note that, because $ \tceseq{ \ottsym{(}  \mathit{X}  \ottsym{,}  \mathit{Y}  \ottsym{)} }   \ottsym{;}   \tceseq{ \mathit{z_{{\mathrm{0}}}}  \,  {:}  \,  \iota_{{\mathrm{0}}} }   \vdash  \ottnt{C_{{\mathrm{0}}}}  \succ  \ottnt{C}$ and $ \tceseq{ \ottsym{(}  \mathit{X'}  \ottsym{,}  \mathit{Y'}  \ottsym{)} }   \ottsym{;}   \tceseq{ \ottsym{(}  \mathit{X}  \ottsym{,}  \mathit{Y}  \ottsym{)} }  \,  \vdash ^{\circ}  \, \ottnt{C}$,
 we have
 \begin{itemize}
  \item $ \tceseq{ \mathit{X} }  \,  =  \, \mathit{X_{{\mathrm{1}}}}  \ottsym{,}   \tceseq{ \mathit{X_{{\mathrm{2}}}} } $,
  \item $ \tceseq{ \mathit{Y} }  \,  =  \, \mathit{Y_{{\mathrm{1}}}}  \ottsym{,}   \tceseq{ \mathit{Y_{{\mathrm{2}}}} } $,
  \item $ \ottnt{C_{{\mathrm{0}}}}  . \Phi  \,  =  \,  (   \lambda_\mu\!   \,  \mathit{x}  .\,  \mathit{X_{{\mathrm{1}}}}  \ottsym{(}  \mathit{x}  \ottsym{,}   \tceseq{ \mathit{z_{{\mathrm{0}}}} }   \ottsym{)}  ,   \lambda_\nu\!   \,  \mathit{y}  .\,  \mathit{Y_{{\mathrm{1}}}}  \ottsym{(}  \mathit{y}  \ottsym{,}   \tceseq{ \mathit{z_{{\mathrm{0}}}} }   \ottsym{)}  ) $,
  \item $  \{   \tceseq{ \mathit{Y'} }   \ottsym{,}  \mathit{Y_{{\mathrm{1}}}}  \}   \, \ottsym{\#} \,   \mathit{fpv}  (   {  \ottnt{C}  . \Phi  }^\mu   )  $, and
  \item $  \{   \tceseq{ \mathit{X_{{\mathrm{2}}}} }   \ottsym{,}   \tceseq{ \mathit{Y_{{\mathrm{2}}}} }   \}   \, \ottsym{\#} \,   \mathit{fpv}  (   \ottnt{C}  . \Phi   )  $
 \end{itemize}
 for some $\mathit{X_{{\mathrm{1}}}}$, $\mathit{Y_{{\mathrm{1}}}}$, $ \tceseq{ \mathit{X_{{\mathrm{2}}}} } $, $ \tceseq{ \mathit{Y_{{\mathrm{2}}}} } $, $\mathit{x}$, and $\mathit{y}$.

 The conclusion is shown by \Srule{Comp} and \Srule{Eff} with the following.
 \begin{itemize}
  \item We show that
        \[
          \tceseq{ \mathit{z_{{\mathrm{0}}}}  \,  {:}  \,  \iota_{{\mathrm{0}}} }   \vdash   { \sigma }_{ \ottmv{i}  \ottsym{-}  \ottsym{1} }'   \ottsym{(}   { \sigma }_{ \ottmv{i}  \ottsym{-}  \ottsym{1} }   \ottsym{(}   { \sigma }_{ \mathit{Y} }   \ottsym{(}   \ottnt{C}  . \ottnt{T}   \ottsym{)}  \ottsym{)}  \ottsym{)}  \ottsym{<:}  \sigma'_{\ottmv{i}}  \ottsym{(}  \sigma_{\ottmv{i}}  \ottsym{(}   { \sigma }_{ \mathit{Y} }   \ottsym{(}   \ottnt{C_{{\mathrm{0}}}}  . \ottnt{T}   \ottsym{)}  \ottsym{)}  \ottsym{)} ~.
        \]
        We have
        \begin{itemize}
         \item $ \ottnt{C_{{\mathrm{0}}}}  . \ottnt{T}  \,  =  \,  \ottnt{C}  . \ottnt{T} $ implied by the assumption $ \tceseq{ \ottsym{(}  \mathit{X}  \ottsym{,}  \mathit{Y}  \ottsym{)} }   \ottsym{;}   \tceseq{ \mathit{z_{{\mathrm{0}}}}  \,  {:}  \,  \iota_{{\mathrm{0}}} }   \vdash  \ottnt{C_{{\mathrm{0}}}}  \succ  \ottnt{C}$, and
         \item $  \{   \tceseq{ \mathit{X'} }   \ottsym{,}   \tceseq{ \mathit{Y'} }   \ottsym{,}   \tceseq{ \mathit{X} }   \ottsym{,}   \tceseq{ \mathit{Y} }   \}   \, \ottsym{\#} \,   \mathit{fpv}  (   \ottnt{C}  . \ottnt{T}   )  $ implied by the assumption $ \tceseq{ \ottsym{(}  \mathit{X'}  \ottsym{,}  \mathit{Y'}  \ottsym{)} }   \ottsym{;}   \tceseq{ \ottsym{(}  \mathit{X}  \ottsym{,}  \mathit{Y}  \ottsym{)} }  \,  \vdash ^{\circ}  \, \ottnt{C}$.
        \end{itemize}
        Therefore, it suffices to show that
        \[
          \tceseq{ \mathit{z_{{\mathrm{0}}}}  \,  {:}  \,  \ottnt{T_{{\mathrm{0}}}} }   \vdash   \ottnt{C}  . \ottnt{T}   \ottsym{<:}   \ottnt{C}  . \ottnt{T}  ~.
        \]
        By \reflem{ts-elim-unused-binding-wf-ftyping} with $\ottsym{(}   \tceseq{ \mathit{X'} }   \ottsym{,}   \tceseq{ \mathit{X} }   \ottsym{)} \,  {:}  \,  (   \Sigma ^\ast   ,   \tceseq{ \iota_{{\mathrm{0}}} }   )   \ottsym{,}  \ottsym{(}   \tceseq{ \mathit{Y'} }   \ottsym{,}   \tceseq{ \mathit{Y} }   \ottsym{)} \,  {:}  \,  (   \Sigma ^\omega   ,   \tceseq{ \iota_{{\mathrm{0}}} }   )   \ottsym{,}   \tceseq{ \mathit{z_{{\mathrm{0}}}}  \,  {:}  \,  \iota_{{\mathrm{0}}} }   \vdash  \ottnt{C}$,
        we have $ \tceseq{ \mathit{z_{{\mathrm{0}}}}  \,  {:}  \,  \ottnt{T_{{\mathrm{0}}}} }   \vdash   \ottnt{C}  . \ottnt{T} $.
        Therefore, so we have the conclusion by \reflem{ts-refl-subtyping}.

  \item We show that
        \[
          \tceseq{ \mathit{z_{{\mathrm{0}}}}  \,  {:}  \,  \iota_{{\mathrm{0}}} }   \ottsym{,}  \mathit{x} \,  {:}  \,  \Sigma ^\ast   \models    { \ottsym{(}   { \sigma }_{ \ottmv{i}  \ottsym{-}  \ottsym{1} }'   \ottsym{(}   { \sigma }_{ \ottmv{i}  \ottsym{-}  \ottsym{1} }   \ottsym{(}   { \sigma }_{ \mathit{Y} }   \ottsym{(}   \ottnt{C}  . \Phi   \ottsym{)}  \ottsym{)}  \ottsym{)}  \ottsym{)} }^\mu   \ottsym{(}  \mathit{x}  \ottsym{)}    \Rightarrow     { \ottsym{(}  \sigma'_{\ottmv{i}}  \ottsym{(}  \sigma_{\ottmv{i}}  \ottsym{(}   { \sigma }_{ \mathit{Y} }   \ottsym{(}   \ottnt{C_{{\mathrm{0}}}}  . \Phi   \ottsym{)}  \ottsym{)}  \ottsym{)}  \ottsym{)} }^\mu   \ottsym{(}  \mathit{x}  \ottsym{)}  ~.
        \]
        Because
        \begin{itemize}
         \item $ \ottnt{C_{{\mathrm{0}}}}  . \Phi  \,  =  \,  (   \lambda_\mu\!   \,  \mathit{x}  .\,  \mathit{X_{{\mathrm{1}}}}  \ottsym{(}  \mathit{x}  \ottsym{,}   \tceseq{ \mathit{z_{{\mathrm{0}}}} }   \ottsym{)}  ,   \lambda_\nu\!   \,  \mathit{y}  .\,  \mathit{Y_{{\mathrm{1}}}}  \ottsym{(}  \mathit{y}  \ottsym{,}   \tceseq{ \mathit{z_{{\mathrm{0}}}} }   \ottsym{)}  ) $,
         \item $  \{   \tceseq{ \mathit{Y'} }   \ottsym{,}  \mathit{Y_{{\mathrm{1}}}}  \}   \, \ottsym{\#} \,   \mathit{fpv}  (   {  \ottnt{C}  . \Phi  }^\mu   )  $, and
         \item $  \{   \tceseq{ \mathit{X_{{\mathrm{2}}}} }   \ottsym{,}   \tceseq{ \mathit{Y_{{\mathrm{2}}}} }   \}   \, \ottsym{\#} \,   \mathit{fpv}  (   \ottnt{C}  . \Phi   )  $,
        \end{itemize}
        it suffices to show that
        \[
          \tceseq{ \mathit{z_{{\mathrm{0}}}}  \,  {:}  \,  \ottnt{T} }   \ottsym{,}  \mathit{x} \,  {:}  \,  \Sigma ^\ast   \models    { \ottsym{(}   { \sigma }_{ \ottmv{i}  \ottsym{-}  \ottsym{1} }'   \ottsym{(}   { \sigma }_{ \ottmv{i}  \ottsym{-}  \ottsym{1} }   \ottsym{(}   \ottnt{C}  . \Phi   \ottsym{)}  \ottsym{)}  \ottsym{)} }^\mu   \ottsym{(}  \mathit{x}  \ottsym{)}    \Rightarrow    \ottsym{(}   \mu\!  \, \mathit{X_{{\mathrm{1}}}}  \ottsym{(}  \mathit{x} \,  {:}  \,  \Sigma ^\ast   \ottsym{,}   \tceseq{ \mathit{z_{{\mathrm{0}}}}  \,  {:}  \,  \iota_{{\mathrm{0}}} }   \ottsym{)}  \ottsym{.}   {  { \sigma }_{ \ottmv{i}  \ottsym{-}  \ottsym{1} }'   \ottsym{(}   { \sigma }_{ \ottmv{i}  \ottsym{-}  \ottsym{1} }   \ottsym{(}   \ottnt{C}  . \Phi   \ottsym{)}  \ottsym{)} }^\mu   \ottsym{(}  \mathit{x}  \ottsym{)}  \ottsym{)}  \ottsym{(}  \mathit{x}  \ottsym{,}   \tceseq{ \mathit{z_{{\mathrm{0}}}} }   \ottsym{)}  ~.
        \]
        which is implied by \reflem{ts-mu-nu-valid}.

  \item We show that
        \[
          \tceseq{ \mathit{z_{{\mathrm{0}}}}  \,  {:}  \,  \ottnt{T} }   \ottsym{,}  \mathit{y} \,  {:}  \,  \Sigma ^\omega   \models    { \ottsym{(}   { \sigma }_{ \ottmv{i}  \ottsym{-}  \ottsym{1} }'   \ottsym{(}   { \sigma }_{ \ottmv{i}  \ottsym{-}  \ottsym{1} }   \ottsym{(}   { \sigma }_{ \mathit{Y} }   \ottsym{(}   \ottnt{C}  . \Phi   \ottsym{)}  \ottsym{)}  \ottsym{)}  \ottsym{)} }^\nu   \ottsym{(}  \mathit{y}  \ottsym{)}    \Rightarrow     { \ottsym{(}  \sigma'_{\ottmv{i}}  \ottsym{(}  \sigma_{\ottmv{i}}  \ottsym{(}   { \sigma }_{ \mathit{Y} }   \ottsym{(}   \ottnt{C_{{\mathrm{0}}}}  . \Phi   \ottsym{)}  \ottsym{)}  \ottsym{)}  \ottsym{)} }^\nu   \ottsym{(}  \mathit{y}  \ottsym{)}  ~.
        \]
        Because $ \ottnt{C_{{\mathrm{0}}}}  . \Phi  \,  =  \,  (   \lambda_\mu\!   \,  \mathit{x}  .\,  \mathit{X_{{\mathrm{1}}}}  \ottsym{(}  \mathit{x}  \ottsym{,}   \tceseq{ \mathit{z_{{\mathrm{0}}}} }   \ottsym{)}  ,   \lambda_\nu\!   \,  \mathit{y}  .\,  \mathit{Y_{{\mathrm{1}}}}  \ottsym{(}  \mathit{y}  \ottsym{,}   \tceseq{ \mathit{z_{{\mathrm{0}}}} }   \ottsym{)}  ) $,
        it suffices to show that
        \[
          \tceseq{ \mathit{z_{{\mathrm{0}}}}  \,  {:}  \,  \ottnt{T} }   \ottsym{,}  \mathit{y} \,  {:}  \,  \Sigma ^\omega   \models    { \ottsym{(}   { \sigma }_{ \ottmv{i}  \ottsym{-}  \ottsym{1} }'   \ottsym{(}   { \sigma }_{ \ottmv{i}  \ottsym{-}  \ottsym{1} }   \ottsym{(}   { \sigma }_{ \mathit{Y} }   \ottsym{(}   \ottnt{C}  . \Phi   \ottsym{)}  \ottsym{)}  \ottsym{)}  \ottsym{)} }^\nu   \ottsym{(}  \mathit{y}  \ottsym{)}    \Rightarrow    \ottsym{(}   \nu\!  \, \mathit{Y_{{\mathrm{0}}}}  \ottsym{(}  \mathit{y} \,  {:}  \,  \Sigma ^\omega   \ottsym{,}   \tceseq{ \mathit{z_{{\mathrm{0}}}}  \,  {:}  \,  \iota_{{\mathrm{0}}} }   \ottsym{)}  \ottsym{.}   \top   \ottsym{)}  \ottsym{(}  \mathit{y}  \ottsym{,}   \tceseq{ \mathit{z_{{\mathrm{0}}}} }   \ottsym{)}  ~.
        \]
        which is implied by Lemmas~\ref{lem:ts-mu-nu-valid} and \ref{lem:ts-valid-intro-impl-top}.

  \item We show that
        \[
          \tceseq{ \mathit{z_{{\mathrm{0}}}}  \,  {:}  \,  \iota_{{\mathrm{0}}} }  \,  |  \,  \ottnt{C}  . \ottnt{T}  \,  \,\&\,  \,  { \sigma }_{ \ottmv{i}  \ottsym{-}  \ottsym{1} }'   \ottsym{(}   { \sigma }_{ \ottmv{i}  \ottsym{-}  \ottsym{1} }   \ottsym{(}   { \sigma }_{ \mathit{Y} }   \ottsym{(}   \ottnt{C}  . \Phi   \ottsym{)}  \ottsym{)}  \ottsym{)}  \vdash   { \sigma }_{ \ottmv{i}  \ottsym{-}  \ottsym{1} }'   \ottsym{(}   { \sigma }_{ \ottmv{i}  \ottsym{-}  \ottsym{1} }   \ottsym{(}   { \sigma }_{ \mathit{Y} }   \ottsym{(}   \ottnt{C}  . \ottnt{S}   \ottsym{)}  \ottsym{)}  \ottsym{)}  \ottsym{<:}  \sigma'_{\ottmv{i}}  \ottsym{(}  \sigma_{\ottmv{i}}  \ottsym{(}   { \sigma }_{ \mathit{Y} }   \ottsym{(}   \ottnt{C_{{\mathrm{0}}}}  . \ottnt{S}   \ottsym{)}  \ottsym{)}  \ottsym{)}
        \]
        (note that $  \{   \tceseq{ \mathit{X'} }   \ottsym{,}   \tceseq{ \mathit{Y'} }   \ottsym{,}   \tceseq{ \mathit{X} }   \ottsym{,}   \tceseq{ \mathit{Y} }   \}   \, \ottsym{\#} \,   \mathit{fpv}  (   \ottnt{C}  . \ottnt{T}   )  $ as shown above).
        By case analysis on $ \ottnt{C}  . \ottnt{S} $.
        \begin{caseanalysis}
         \case $ \ottnt{C}  . \ottnt{S}  \,  =  \,  \square $:
          The assumption $ \tceseq{ \ottsym{(}  \mathit{X}  \ottsym{,}  \mathit{Y}  \ottsym{)} }   \ottsym{;}   \tceseq{ \mathit{z_{{\mathrm{0}}}}  \,  {:}  \,  \iota_{{\mathrm{0}}} }   \vdash  \ottnt{C_{{\mathrm{0}}}}  \succ  \ottnt{C}$ implies $ \ottnt{C_{{\mathrm{0}}}}  . \ottnt{S}  \,  =  \,  \square $.
          Therefore, we have the conclusion by \Srule{Empty}.

         \case $  \exists  \, { \mathit{x}  \ottsym{,}  \ottnt{C_{{\mathrm{1}}}}  \ottsym{,}  \ottnt{C_{{\mathrm{2}}}} } . \   \ottnt{C}  . \ottnt{S}  \,  =  \,  ( \forall  \mathit{x}  .  \ottnt{C_{{\mathrm{1}}}}  )  \Rightarrow   \ottnt{C_{{\mathrm{2}}}}  $:
          We have the following.
          \begin{itemize}
           \item $ \tceseq{ \ottsym{(}  \mathit{X'}  \ottsym{,}  \mathit{Y'}  \ottsym{)} }   \vdash  \ottnt{C'}  \Rrightarrow  \ottnt{C}$ implies
                 $ \tceseq{ \ottsym{(}  \mathit{X'}  \ottsym{,}  \mathit{Y'}  \ottsym{)} }   \ottsym{,}  \ottsym{(}  \mathit{X_{{\mathrm{1}}}}  \ottsym{,}  \mathit{Y_{{\mathrm{1}}}}  \ottsym{)}  \vdash  \ottnt{C'}  \Rrightarrow  \ottnt{C_{{\mathrm{2}}}}$.
           \item $ \tceseq{ \ottsym{(}  \mathit{X}  \ottsym{,}  \mathit{Y}  \ottsym{)} }   \ottsym{;}   \tceseq{ \mathit{z_{{\mathrm{0}}}}  \,  {:}  \,  \iota_{{\mathrm{0}}} }   \vdash  \ottnt{C_{{\mathrm{0}}}}  \succ  \ottnt{C}$ implies
                 \begin{itemize}
                  \item $ \ottnt{C_{{\mathrm{0}}}}  . \ottnt{S}  \,  =  \,  ( \forall  \mathit{x}  .  \ottnt{C_{{\mathrm{1}}}}  )  \Rightarrow   \ottnt{C_{{\mathrm{02}}}} $ and
                  \item $ \tceseq{ \ottsym{(}  \mathit{X_{{\mathrm{2}}}}  \ottsym{,}  \mathit{Y_{{\mathrm{2}}}}  \ottsym{)} }   \ottsym{;}   \tceseq{ \mathit{z_{{\mathrm{0}}}}  \,  {:}  \,  \iota_{{\mathrm{0}}} }   \vdash  \ottnt{C_{{\mathrm{02}}}}  \succ  \ottnt{C_{{\mathrm{2}}}}$
                 \end{itemize}
                 for some $\ottnt{C_{{\mathrm{02}}}}$.
           \item $\ottsym{(}   \tceseq{ \mathit{X'} }   \ottsym{,}   \tceseq{ \mathit{X} }   \ottsym{)} \,  {:}  \,  (   \Sigma ^\ast   ,   \tceseq{ \iota_{{\mathrm{0}}} }   )   \ottsym{,}  \ottsym{(}   \tceseq{ \mathit{Y'} }   \ottsym{,}   \tceseq{ \mathit{Y} }   \ottsym{)} \,  {:}  \,  (   \Sigma ^\omega   ,   \tceseq{ \iota_{{\mathrm{0}}} }   )   \ottsym{,}   \tceseq{ \mathit{z_{{\mathrm{0}}}}  \,  {:}  \,  \iota_{{\mathrm{0}}} }   \vdash  \ottnt{C}$ implies
                 \begin{itemize}
                  \item $\ottsym{(}   \tceseq{ \mathit{X'} }   \ottsym{,}   \tceseq{ \mathit{X} }   \ottsym{)} \,  {:}  \,  (   \Sigma ^\ast   ,   \tceseq{ \iota_{{\mathrm{0}}} }   )   \ottsym{,}  \ottsym{(}   \tceseq{ \mathit{Y'} }   \ottsym{,}   \tceseq{ \mathit{Y} }   \ottsym{)} \,  {:}  \,  (   \Sigma ^\omega   ,   \tceseq{ \iota_{{\mathrm{0}}} }   )   \ottsym{,}   \tceseq{ \mathit{z_{{\mathrm{0}}}}  \,  {:}  \,  \iota_{{\mathrm{0}}} }   \ottsym{,}  \mathit{x} \,  {:}  \,  \ottnt{C}  . \ottnt{T}   \vdash  \ottnt{C_{{\mathrm{1}}}}$ and
                  \item $\ottsym{(}   \tceseq{ \mathit{X'} }   \ottsym{,}   \tceseq{ \mathit{X} }   \ottsym{)} \,  {:}  \,  (   \Sigma ^\ast   ,   \tceseq{ \iota_{{\mathrm{0}}} }   )   \ottsym{,}  \ottsym{(}   \tceseq{ \mathit{Y'} }   \ottsym{,}   \tceseq{ \mathit{Y} }   \ottsym{)} \,  {:}  \,  (   \Sigma ^\omega   ,   \tceseq{ \iota_{{\mathrm{0}}} }   )   \ottsym{,}   \tceseq{ \mathit{z_{{\mathrm{0}}}}  \,  {:}  \,  \iota_{{\mathrm{0}}} }   \vdash  \ottnt{C_{{\mathrm{2}}}}$.
                 \end{itemize}
           \item $ \tceseq{ \ottsym{(}  \mathit{X'}  \ottsym{,}  \mathit{Y'}  \ottsym{)} }   \ottsym{;}   \tceseq{ \ottsym{(}  \mathit{X}  \ottsym{,}  \mathit{Y}  \ottsym{)} }  \,  \vdash ^{\circ}  \, \ottnt{C}$ implies
                 \begin{itemize}
                  \item $ \tceseq{ \ottsym{(}  \mathit{X'}  \ottsym{,}  \mathit{Y'}  \ottsym{)} }   \ottsym{,}  \ottsym{(}  \mathit{X_{{\mathrm{1}}}}  \ottsym{,}  \mathit{Y_{{\mathrm{1}}}}  \ottsym{)}  \ottsym{;}   \tceseq{ \ottsym{(}  \mathit{X_{{\mathrm{2}}}}  \ottsym{,}  \mathit{Y_{{\mathrm{2}}}}  \ottsym{)} }  \,  \vdash ^{\circ}  \, \ottnt{C_{{\mathrm{2}}}}$ and
                  \item $  \{   \tceseq{ \mathit{X'} }   \ottsym{,}   \tceseq{ \mathit{Y'} }   \ottsym{,}   \tceseq{ \mathit{X} }   \ottsym{,}   \tceseq{ \mathit{Y} }   \}   \, \ottsym{\#} \,   \mathit{fpv}  (  \ottnt{C_{{\mathrm{1}}}}  )  $.
                 \end{itemize}
          \end{itemize}
          We have the conclusion by \Srule{Ans} with the following.
          \begin{itemize}
           \item We show that $ \tceseq{ \mathit{z_{{\mathrm{0}}}}  \,  {:}  \,  \iota_{{\mathrm{0}}} }   \ottsym{,}  \mathit{x} \,  {:}  \,  \ottnt{C}  . \ottnt{T}   \vdash   { \sigma }_{ \ottmv{i}  \ottsym{-}  \ottsym{1} }'   \ottsym{(}   { \sigma }_{ \ottmv{i}  \ottsym{-}  \ottsym{1} }   \ottsym{(}   { \sigma }_{ \mathit{Y} }   \ottsym{(}  \ottnt{C_{{\mathrm{1}}}}  \ottsym{)}  \ottsym{)}  \ottsym{)}  \ottsym{<:}  \sigma'_{\ottmv{i}}  \ottsym{(}  \sigma_{\ottmv{i}}  \ottsym{(}   { \sigma }_{ \mathit{Y} }   \ottsym{(}  \ottnt{C_{{\mathrm{1}}}}  \ottsym{)}  \ottsym{)}  \ottsym{)}$.
                 By $  \{   \tceseq{ \mathit{X'} }   \ottsym{,}   \tceseq{ \mathit{Y'} }   \ottsym{,}   \tceseq{ \mathit{X} }   \ottsym{,}   \tceseq{ \mathit{Y} }   \}   \, \ottsym{\#} \,   \mathit{fpv}  (  \ottnt{C_{{\mathrm{1}}}}  )  $,
                 it suffices to show that
                 $ \tceseq{ \mathit{z_{{\mathrm{0}}}}  \,  {:}  \,  \iota_{{\mathrm{0}}} }   \ottsym{,}  \mathit{x} \,  {:}  \,  \ottnt{C}  . \ottnt{T}   \vdash  \ottnt{C_{{\mathrm{1}}}}  \ottsym{<:}  \ottnt{C_{{\mathrm{1}}}}$.
                 Because
                 \[
                  \ottsym{(}   \tceseq{ \mathit{X'} }   \ottsym{,}   \tceseq{ \mathit{X} }   \ottsym{)} \,  {:}  \,  (   \Sigma ^\ast   ,   \tceseq{ \iota_{{\mathrm{0}}} }   )   \ottsym{,}  \ottsym{(}   \tceseq{ \mathit{Y'} }   \ottsym{,}   \tceseq{ \mathit{Y} }   \ottsym{)} \,  {:}  \,  (   \Sigma ^\omega   ,   \tceseq{ \iota_{{\mathrm{0}}} }   )   \ottsym{,}   \tceseq{ \mathit{z_{{\mathrm{0}}}}  \,  {:}  \,  \iota_{{\mathrm{0}}} }   \ottsym{,}  \mathit{x} \,  {:}  \,  \ottnt{C}  . \ottnt{T}   \vdash  \ottnt{C_{{\mathrm{1}}}} ~,
                 \]
                 Lemmas~\ref{lem:ts-elim-unused-binding-wf-ftyping} and \ref{lem:ts-refl-subtyping}
                 imply the conclusion.

           \item We show that $ \tceseq{ \mathit{z_{{\mathrm{0}}}}  \,  {:}  \,  \iota_{{\mathrm{0}}} }   \vdash   { \sigma }_{ \ottmv{i}  \ottsym{-}  \ottsym{1} }'   \ottsym{(}   { \sigma }_{ \ottmv{i}  \ottsym{-}  \ottsym{1} }   \ottsym{(}   { \sigma }_{ \mathit{Y} }   \ottsym{(}  \ottnt{C_{{\mathrm{2}}}}  \ottsym{)}  \ottsym{)}  \ottsym{)}  \ottsym{<:}  \sigma'_{\ottmv{i}}  \ottsym{(}  \sigma_{\ottmv{i}}  \ottsym{(}   { \sigma }_{ \mathit{Y} }   \ottsym{(}  \ottnt{C_{{\mathrm{02}}}}  \ottsym{)}  \ottsym{)}  \ottsym{)}$.

                 By \reflem{lr-approx-compose-subst},
                 \begin{itemize}
                  \item $\ottnt{C'}  \triangleright   \tceseq{ \ottsym{(}  \mathit{X'}  \ottsym{,}  \mathit{Y'}  \ottsym{)} }   \ottsym{,}  \ottsym{(}  \mathit{X_{{\mathrm{1}}}}  \ottsym{,}  \mathit{Y_{{\mathrm{1}}}}  \ottsym{)}  \ottsym{;}   \tceseq{ \ottsym{(}  \mathit{X_{{\mathrm{2}}}}  \ottsym{,}  \mathit{Y_{{\mathrm{2}}}}  \ottsym{)} }   \ottsym{;}   \tceseq{ \mathit{z_{{\mathrm{0}}}}  \,  {:}  \,  \iota_{{\mathrm{0}}} }  \,  |  \, \ottnt{C_{{\mathrm{2}}}} \,  \vdash _{ \ottmv{i} }  \,  { \sigma }_{ \ottsym{0}  \ottsym{;}  \ottmv{i} } $,
                  \item $\ottnt{C'}  \triangleright   \emptyset   \ottsym{;}   \tceseq{ \ottsym{(}  \mathit{X'}  \ottsym{,}  \mathit{Y'}  \ottsym{)} }   \ottsym{,}  \ottsym{(}  \mathit{X_{{\mathrm{1}}}}  \ottsym{,}  \mathit{Y_{{\mathrm{1}}}}  \ottsym{)}  \ottsym{;}   \tceseq{ \mathit{z_{{\mathrm{0}}}}  \,  {:}  \,  \iota_{{\mathrm{0}}} }  \,  |  \, \ottnt{C'} \,  \vdash _{ \ottmv{i} }  \,  { \sigma }_{ \ottsym{0}  \ottsym{;}  \ottmv{i} }' $,
                  \item $\sigma_{\ottmv{i}}  \mathbin{\uplus}  \sigma'_{\ottmv{i}} \,  =  \,  {  { \sigma }_{ \ottsym{0}  \ottsym{;}  \ottmv{i} }   \mathbin{\uplus}  \sigma }_{ \ottsym{0}  \ottsym{;}  \ottmv{i} }' $,
                  \item $\ottnt{C'}  \triangleright   \tceseq{ \ottsym{(}  \mathit{X'}  \ottsym{,}  \mathit{Y'}  \ottsym{)} }   \ottsym{,}  \ottsym{(}  \mathit{X_{{\mathrm{1}}}}  \ottsym{,}  \mathit{Y_{{\mathrm{1}}}}  \ottsym{)}  \ottsym{;}   \tceseq{ \ottsym{(}  \mathit{X_{{\mathrm{2}}}}  \ottsym{,}  \mathit{Y_{{\mathrm{2}}}}  \ottsym{)} }   \ottsym{;}   \tceseq{ \mathit{z_{{\mathrm{0}}}}  \,  {:}  \,  \iota_{{\mathrm{0}}} }  \,  |  \, \ottnt{C_{{\mathrm{2}}}} \,  \vdash _{ \ottmv{i}  \ottsym{-}  \ottsym{1} }  \,  { \sigma }_{ \ottsym{0}  \ottsym{;}  \ottmv{i}  \ottsym{-}  \ottsym{1} } $,
                  \item $\ottnt{C'}  \triangleright   \emptyset   \ottsym{;}   \tceseq{ \ottsym{(}  \mathit{X'}  \ottsym{,}  \mathit{Y'}  \ottsym{)} }   \ottsym{,}  \ottsym{(}  \mathit{X_{{\mathrm{1}}}}  \ottsym{,}  \mathit{Y_{{\mathrm{1}}}}  \ottsym{)}  \ottsym{;}   \tceseq{ \mathit{z_{{\mathrm{0}}}}  \,  {:}  \,  \iota_{{\mathrm{0}}} }  \,  |  \, \ottnt{C'} \,  \vdash _{ \ottmv{i}  \ottsym{-}  \ottsym{1} }  \,  { \sigma }_{ \ottsym{0}  \ottsym{;}  \ottmv{i}  \ottsym{-}  \ottsym{1} }' $, and
                  \item $ {  { \sigma }_{ \ottmv{i}  \ottsym{-}  \ottsym{1} }   \mathbin{\uplus}  \sigma }_{ \ottmv{i}  \ottsym{-}  \ottsym{1} }'  \,  =  \,  {  { \sigma }_{ \ottsym{0}  \ottsym{;}  \ottmv{i}  \ottsym{-}  \ottsym{1} }   \mathbin{\uplus}  \sigma }_{ \ottsym{0}  \ottsym{;}  \ottmv{i}  \ottsym{-}  \ottsym{1} }' $.
                 \end{itemize}
                 Furthermore, we have the following.
                 \begin{itemize}
                  \item $ \tceseq{ \ottsym{(}  \mathit{X'}  \ottsym{,}  \mathit{Y'}  \ottsym{)} }   \ottsym{,}  \ottsym{(}  \mathit{X_{{\mathrm{1}}}}  \ottsym{,}  \mathit{Y_{{\mathrm{1}}}}  \ottsym{)}  \vdash  \ottnt{C'}  \Rrightarrow  \ottnt{C_{{\mathrm{2}}}}$.
                  \item $\ottsym{(}   \tceseq{ \mathit{X'} }   \ottsym{,}   \tceseq{ \mathit{X} }   \ottsym{)} \,  {:}  \,  (   \Sigma ^\ast   ,   \tceseq{ \iota_{{\mathrm{0}}} }   )   \ottsym{,}  \ottsym{(}   \tceseq{ \mathit{Y'} }   \ottsym{,}   \tceseq{ \mathit{Y} }   \ottsym{)} \,  {:}  \,  (   \Sigma ^\omega   ,   \tceseq{ \iota_{{\mathrm{0}}} }   )   \ottsym{,}   \tceseq{ \mathit{z_{{\mathrm{0}}}}  \,  {:}  \,  \iota_{{\mathrm{0}}} }   \vdash  \ottnt{C_{{\mathrm{2}}}}$.
                  \item $ \tceseq{ \ottsym{(}  \mathit{X_{{\mathrm{2}}}}  \ottsym{,}  \mathit{Y_{{\mathrm{2}}}}  \ottsym{)} }   \ottsym{;}   \tceseq{ \mathit{z_{{\mathrm{0}}}}  \,  {:}  \,  \iota_{{\mathrm{0}}} }   \vdash  \ottnt{C_{{\mathrm{02}}}}  \succ  \ottnt{C_{{\mathrm{2}}}}$.
                  \item $ \tceseq{ \ottsym{(}  \mathit{X'}  \ottsym{,}  \mathit{Y'}  \ottsym{)} }   \ottsym{,}  \ottsym{(}  \mathit{X_{{\mathrm{1}}}}  \ottsym{,}  \mathit{Y_{{\mathrm{1}}}}  \ottsym{)}  \ottsym{;}   \tceseq{ \ottsym{(}  \mathit{X_{{\mathrm{2}}}}  \ottsym{,}  \mathit{Y_{{\mathrm{2}}}}  \ottsym{)} }  \,  \vdash ^{\circ}  \, \ottnt{C_{{\mathrm{2}}}}$.
                 \end{itemize}
                 Therefore, the IH implies
                 \[
                   \tceseq{ \mathit{z_{{\mathrm{0}}}}  \,  {:}  \,  \ottnt{T_{{\mathrm{0}}}} }   \vdash   { \sigma }_{ \ottsym{0}  \ottsym{;}  \ottmv{i}  \ottsym{-}  \ottsym{1} }'   \ottsym{(}   { \sigma }_{ \ottsym{0}  \ottsym{;}  \ottmv{i}  \ottsym{-}  \ottsym{1} }   \ottsym{(}   { \sigma }_{ \mathit{Y} }   \ottsym{(}  \ottnt{C_{{\mathrm{2}}}}  \ottsym{)}  \ottsym{)}  \ottsym{)}  \ottsym{<:}   { \sigma }_{ \ottsym{0}  \ottsym{;}  \ottmv{i} }'   \ottsym{(}   { \sigma }_{ \ottsym{0}  \ottsym{;}  \ottmv{i} }   \ottsym{(}   { \sigma }_{ \mathit{Y} }   \ottsym{(}  \ottnt{C_{{\mathrm{02}}}}  \ottsym{)}  \ottsym{)}  \ottsym{)} ~.
                 \]
                 Because
                 $\sigma_{\ottmv{i}}  \mathbin{\uplus}  \sigma'_{\ottmv{i}} \,  =  \,  {  { \sigma }_{ \ottsym{0}  \ottsym{;}  \ottmv{i} }   \mathbin{\uplus}  \sigma }_{ \ottsym{0}  \ottsym{;}  \ottmv{i} }' $ and
                 $ {  { \sigma }_{ \ottmv{i}  \ottsym{-}  \ottsym{1} }   \mathbin{\uplus}  \sigma }_{ \ottmv{i}  \ottsym{-}  \ottsym{1} }'  \,  =  \,  {  { \sigma }_{ \ottsym{0}  \ottsym{;}  \ottmv{i}  \ottsym{-}  \ottsym{1} }   \mathbin{\uplus}  \sigma }_{ \ottsym{0}  \ottsym{;}  \ottmv{i}  \ottsym{-}  \ottsym{1} }' $,
                 we have the conclusion.
          \end{itemize}
        \end{caseanalysis}
 \end{itemize}
\end{proof}

\begin{lemmap}{Typing $ \mu $-Approximation}{lr-typing-mu-approx-XXX}
 Assume that
 \begin{itemize}
  \item $\ottnt{v} \,  =  \,  \mathsf{rec}  \, (  \tceseq{  \mathit{X} ^{  \ottnt{A} _{  \mu  }  },  \mathit{Y} ^{  \ottnt{A} _{  \nu  }  }  }  ,  \mathit{f} ,   \tceseq{ \mathit{z} }  ) . \,  \ottnt{e} $,
  \item $ \tceseq{ \mathit{z_{{\mathrm{0}}}}  \,  {:}  \,  \iota_{{\mathrm{0}}} }  \,  =  \,  \gamma  (   \tceseq{ \mathit{z}  \,  {:}  \,  \ottnt{T_{{\mathrm{1}}}} }   ) $,
  \item $ \tceseq{ \ottsym{(}  \mathit{X}  \ottsym{,}  \mathit{Y}  \ottsym{)} }   \ottsym{;}   \tceseq{ \mathit{z_{{\mathrm{0}}}}  \,  {:}  \,  \iota_{{\mathrm{0}}} }   \vdash  \ottnt{C_{{\mathrm{0}}}}  \stackrel{\succ}{\circ}  \ottnt{C}$,
  \item $  \{   \tceseq{ \mathit{X} }   \ottsym{,}   \tceseq{ \mathit{Y} }   \}   \, \ottsym{\#} \,   \mathit{fpv}  (   \tceseq{ \ottnt{T_{{\mathrm{1}}}} }   )  $,
  \item $  \forall  \, { \ottmv{i} } . \  \ottnt{C}  \triangleright   \emptyset   \ottsym{;}   \tceseq{ \ottsym{(}  \mathit{X}  \ottsym{,}  \mathit{Y}  \ottsym{)} }   \ottsym{;}   \tceseq{ \mathit{z_{{\mathrm{0}}}}  \,  {:}  \,  \iota_{{\mathrm{0}}} }  \,  |  \, \ottnt{C} \,  \vdash _{ \ottmv{i} }  \, \sigma_{\ottmv{i}} $,
  \item $ \tceseq{ \mathit{X} }  \,  {:}  \,  (   \Sigma ^\ast   ,   \tceseq{ \iota_{{\mathrm{0}}} }   )   \ottsym{,}   \tceseq{ \mathit{Y} }  \,  {:}  \,  (   \Sigma ^\omega   ,   \tceseq{ \iota_{{\mathrm{0}}} }   )   \ottsym{,}  \mathit{f} \,  {:}  \, \ottsym{(}   \tceseq{ \mathit{z}  \,  {:}  \,  \ottnt{T_{{\mathrm{1}}}} }   \ottsym{)}  \rightarrow  \ottnt{C_{{\mathrm{0}}}}  \ottsym{,}   \tceseq{ \mathit{z}  \,  {:}  \,  \ottnt{T_{{\mathrm{1}}}} }   \vdash  \ottnt{e}  \ottsym{:}  \ottnt{C}$, and
  \item $ { \sigma }_{ \mathit{Y} }  \,  =  \,  [ \tceseq{ \ottsym{(}   \nu\!  \, \mathit{Y}  \ottsym{(}  \mathit{y} \,  {:}  \,  \Sigma ^\omega   \ottsym{,}   \tceseq{ \mathit{z_{{\mathrm{0}}}}  \,  {:}  \,  \iota_{{\mathrm{0}}} }   \ottsym{)}  \ottsym{.}   \top   \ottsym{)}  /  \mathit{Y} } ] $.
 \end{itemize}
 Then,
 \[
    \forall  \, { \ottmv{i}  \ottsym{,}  \pi } . \   \emptyset   \vdash   \ottnt{v} _{  \mu   \ottsym{;}  \ottmv{i}  \ottsym{;}  \pi }   \ottsym{:}  \ottsym{(}   \tceseq{ \mathit{z}  \,  {:}  \,  \ottnt{T_{{\mathrm{1}}}} }   \ottsym{)}  \rightarrow  \sigma_{\ottmv{i}}  \ottsym{(}   { \sigma }_{ \mathit{Y} }   \ottsym{(}  \ottnt{C_{{\mathrm{0}}}}  \ottsym{)}  \ottsym{)}  ~.
 \]
\end{lemmap}
\begin{proof}
 \TS{CHECKED: 2022/07/07}
 By induction on $\ottmv{i}$.
 Let $\pi$ be any infinite trace.

 We proceed by case analysis on $\ottmv{i}$.
 \begin{caseanalysis}
  \case $\ottmv{i} \,  =  \, \ottsym{0}$:
   Because $ \ottnt{v} _{  \mu   \ottsym{;}  \ottsym{0}  \ottsym{;}  \pi }  \,  =  \,  \lambda\!  \,  \tceseq{ \mathit{z} }   \ottsym{.}   \bullet ^{ \pi } $,
   \T{Abs} implies that it suffices to show that
   \[
     \tceseq{ \mathit{z}  \,  {:}  \,  \ottnt{T_{{\mathrm{1}}}} }   \vdash   \bullet ^{ \pi }   \ottsym{:}  \sigma_{\ottmv{i}}  \ottsym{(}   { \sigma }_{ \mathit{Y} }   \ottsym{(}  \ottnt{C_{{\mathrm{0}}}}  \ottsym{)}  \ottsym{)} ~.
   \]
   By \T{Div}, it suffices to show the following.
   \begin{itemize}
    \item We show that
          \[
            \tceseq{ \mathit{z}  \,  {:}  \,  \ottnt{T_{{\mathrm{1}}}} }   \vdash  \sigma_{\ottmv{i}}  \ottsym{(}   { \sigma }_{ \mathit{Y} }   \ottsym{(}  \ottnt{C_{{\mathrm{0}}}}  \ottsym{)}  \ottsym{)} ~.
          \]
          Because $ \tceseq{ \mathit{X} }  \,  {:}  \,  (   \Sigma ^\ast   ,   \tceseq{ \iota_{{\mathrm{0}}} }   )   \ottsym{,}   \tceseq{ \mathit{Y} }  \,  {:}  \,  (   \Sigma ^\omega   ,   \tceseq{ \iota_{{\mathrm{0}}} }   )   \ottsym{,}  \mathit{f} \,  {:}  \, \ottsym{(}   \tceseq{ \mathit{z}  \,  {:}  \,  \ottnt{T_{{\mathrm{1}}}} }   \ottsym{)}  \rightarrow  \ottnt{C_{{\mathrm{0}}}}  \ottsym{,}   \tceseq{ \mathit{z}  \,  {:}  \,  \ottnt{T_{{\mathrm{1}}}} }   \vdash  \ottnt{e}  \ottsym{:}  \ottnt{C}$,
          we have
          $ \tceseq{ \mathit{X} }  \,  {:}  \,  (   \Sigma ^\ast   ,   \tceseq{ \iota_{{\mathrm{0}}} }   )   \ottsym{,}   \tceseq{ \mathit{Y} }  \,  {:}  \,  (   \Sigma ^\omega   ,   \tceseq{ \iota_{{\mathrm{0}}} }   )   \ottsym{,}   \tceseq{ \mathit{z_{{\mathrm{0}}}}  \,  {:}  \,  \iota_{{\mathrm{0}}} }   \vdash  \ottnt{C}$
          by Lemmas~\ref{lem:ts-wf-ty-tctx-typing} and \ref{lem:ts-elim-unused-binding-wf-ftyping}.
          Then, we have the conclusion by Lemmas~\ref{lem:lr-approx-eff-subst-wf} and \ref{lem:ts-pred-subst-wf-ftyping}.

    \item We show that
          \[
            { \models  \,   { \sigma_{\ottmv{i}}  \ottsym{(}   { \sigma }_{ \mathit{Y} }   \ottsym{(}  \ottnt{C_{{\mathrm{0}}}}  \ottsym{)}  \ottsym{)} }^\nu (  \pi  )  }  ~.
          \]
          Because $ \tceseq{ \ottsym{(}  \mathit{X}  \ottsym{,}  \mathit{Y}  \ottsym{)} }   \ottsym{;}   \tceseq{ \mathit{z_{{\mathrm{0}}}}  \,  {:}  \,  \iota_{{\mathrm{0}}} }   \vdash  \ottnt{C_{{\mathrm{0}}}}  \stackrel{\succ}{\circ}  \ottnt{C}$,
          the $ \nu $-parts of the temporal effects occurring strictly positively in $\ottnt{C_{{\mathrm{0}}}}$
          take the form $\mathit{Y_{{\mathrm{0}}}}  \ottsym{(}  \mathit{y}  \ottsym{,}   \tceseq{ \mathit{z_{{\mathrm{0}}}} }   \ottsym{)}$ for some $\mathit{Y_{{\mathrm{0}}}} \,  \in  \,  \{   \tceseq{ \mathit{Y} }   \} $, and
          $ { \sigma }_{ \mathit{Y} } $ replaces $\mathit{Y_{{\mathrm{0}}}}$ with $ \top $.
          Therefore, we have the conclusion.
   \end{itemize}

  \case $\ottmv{i} \,  >  \, \ottsym{0}$:
   Because $ \ottnt{v} _{  \mu   \ottsym{;}  \ottmv{i}  \ottsym{;}  \pi }  \,  =  \,     \lambda\!  \,  \tceseq{ \mathit{z} }   \ottsym{.}  \ottnt{e}    [ \tceseq{  { \sigma }_{ \ottmv{i}  \ottsym{-}  \ottsym{1} }   \ottsym{(}  \mathit{X}  \ottsym{)}  /  \mathit{X} } ]      [ \tceseq{  { \sigma }_{ \mathit{Y} }   \ottsym{(}  \mathit{Y}  \ottsym{)}  /  \mathit{Y} } ]      [   \ottnt{v} _{  \mu   \ottsym{;}  \ottmv{i}  \ottsym{-}  \ottsym{1}  \ottsym{;}  \pi }   /  \mathit{f}  ]  $,
   \T{Abs} implies that it suffices to show that
   \[
     \tceseq{ \mathit{z}  \,  {:}  \,  \ottnt{T_{{\mathrm{1}}}} }   \vdash     \ottnt{e}    [ \tceseq{  { \sigma }_{ \ottmv{i}  \ottsym{-}  \ottsym{1} }   \ottsym{(}  \mathit{X}  \ottsym{)}  /  \mathit{X} } ]      [ \tceseq{  { \sigma }_{ \mathit{Y} }   \ottsym{(}  \mathit{Y}  \ottsym{)}  /  \mathit{Y} } ]      [   \ottnt{v} _{  \mu   \ottsym{;}  \ottmv{i}  \ottsym{-}  \ottsym{1}  \ottsym{;}  \pi }   /  \mathit{f}  ]    \ottsym{:}  \sigma_{\ottmv{i}}  \ottsym{(}   { \sigma }_{ \mathit{Y} }   \ottsym{(}  \ottnt{C_{{\mathrm{0}}}}  \ottsym{)}  \ottsym{)} ~.
   \]
   Because
   \begin{itemize}
    \item $ \tceseq{ \mathit{X} }  \,  {:}  \,  (   \Sigma ^\ast   ,   \tceseq{ \iota_{{\mathrm{0}}} }   )   \ottsym{,}   \tceseq{ \mathit{Y} }  \,  {:}  \,  (   \Sigma ^\omega   ,   \tceseq{ \iota_{{\mathrm{0}}} }   )   \ottsym{,}  \mathit{f} \,  {:}  \, \ottsym{(}   \tceseq{ \mathit{z}  \,  {:}  \,  \ottnt{T_{{\mathrm{1}}}} }   \ottsym{)}  \rightarrow  \ottnt{C_{{\mathrm{0}}}}  \ottsym{,}   \tceseq{ \mathit{z}  \,  {:}  \,  \ottnt{T_{{\mathrm{1}}}} }   \vdash  \ottnt{e}  \ottsym{:}  \ottnt{C}$,
    \item $  \forall  \, { \mathit{X_{{\mathrm{0}}}} \,  \in  \,  \{   \tceseq{ \mathit{X} }   \}  } . \   \emptyset   \vdash   { \sigma }_{ \ottmv{i}  \ottsym{-}  \ottsym{1} }   \ottsym{(}  \mathit{X}  \ottsym{)}  \ottsym{:}   (   \Sigma ^\ast   ,   \tceseq{ \iota_{{\mathrm{0}}} }   )  $ by \reflem{lr-approx-eff-subst-wf}, and
    \item $ \emptyset   \vdash   \ottnt{v} _{  \mu   \ottsym{;}  \ottmv{i}  \ottsym{-}  \ottsym{1}  \ottsym{;}  \pi }   \ottsym{:}  \ottsym{(}   \tceseq{ \mathit{z}  \,  {:}  \,  \ottnt{T_{{\mathrm{1}}}} }   \ottsym{)}  \rightarrow   { \sigma }_{ \mathit{Y} }   \ottsym{(}   { \sigma }_{ \ottmv{i}  \ottsym{-}  \ottsym{1} }   \ottsym{(}  \ottnt{C_{{\mathrm{0}}}}  \ottsym{)}  \ottsym{)}$ by the IH.
   \end{itemize}
   we have
   \[
     \tceseq{ \mathit{z}  \,  {:}  \,  \ottnt{T_{{\mathrm{1}}}} }   \vdash     \ottnt{e}    [ \tceseq{  { \sigma }_{ \ottmv{i}  \ottsym{-}  \ottsym{1} }   \ottsym{(}  \mathit{X}  \ottsym{)}  /  \mathit{X} } ]      [ \tceseq{  { \sigma }_{ \mathit{Y} }   \ottsym{(}  \mathit{Y}  \ottsym{)}  /  \mathit{Y} } ]      [   \ottnt{v} _{  \mu   \ottsym{;}  \ottmv{i}  \ottsym{-}  \ottsym{1}  \ottsym{;}  \pi }   /  \mathit{f}  ]    \ottsym{:}   { \sigma }_{ \ottmv{i}  \ottsym{-}  \ottsym{1} }   \ottsym{(}   { \sigma }_{ \mathit{Y} }   \ottsym{(}  \ottnt{C}  \ottsym{)}  \ottsym{)}
   \]
   by Lemmas~\ref{lem:ts-pred-subst-typing}, \ref{lem:ts-val-subst-typing}, and
   \ref{lem:ts-unuse-non-refine-vars}.
   Then, we have the conclusion by \T{Sub} with the following.
   \begin{itemize}
    \item We show that
          \[
            \tceseq{ \mathit{z}  \,  {:}  \,  \ottnt{T_{{\mathrm{1}}}} }   \vdash   { \sigma }_{ \ottmv{i}  \ottsym{-}  \ottsym{1} }   \ottsym{(}   { \sigma }_{ \mathit{Y} }   \ottsym{(}  \ottnt{C}  \ottsym{)}  \ottsym{)}  \ottsym{<:}   { \sigma }_{ \ottmv{i} }   \ottsym{(}   { \sigma }_{ \mathit{Y} }   \ottsym{(}  \ottnt{C_{{\mathrm{0}}}}  \ottsym{)}  \ottsym{)} ~,
          \]
          which is implied by \reflem{lr-approx-fold-mu-XXX}.

    \item We show that
          \[
            \tceseq{ \mathit{z}  \,  {:}  \,  \ottnt{T_{{\mathrm{1}}}} }   \vdash  \sigma_{\ottmv{i}}  \ottsym{(}   { \sigma }_{ \mathit{Y} }   \ottsym{(}  \ottnt{C_{{\mathrm{0}}}}  \ottsym{)}  \ottsym{)} ~.
          \]
          By Lemmas~\ref{lem:lr-approx-eff-subst-wf} and \ref{lem:ts-pred-subst-wf-ftyping},
          it suffices to show that
          \[
            \tceseq{ \mathit{X} }  \,  {:}  \,  (   \Sigma ^\ast   ,   \tceseq{ \iota_{{\mathrm{0}}} }   )   \ottsym{,}   \tceseq{ \mathit{Y} }  \,  {:}  \,  (   \Sigma ^\omega   ,   \tceseq{ \iota_{{\mathrm{0}}} }   )   \ottsym{,}   \tceseq{ \mathit{z}  \,  {:}  \,  \ottnt{T_{{\mathrm{1}}}} }   \vdash  \ottnt{C_{{\mathrm{0}}}} ~.
          \]
          By Lemmas~\ref{lem:ts-wf-ty-tctx-typing} and \ref{lem:ts-elim-unused-binding-wf-ftyping},
          $ \tceseq{ \mathit{X} }  \,  {:}  \,  (   \Sigma ^\ast   ,   \tceseq{ \iota_{{\mathrm{0}}} }   )   \ottsym{,}   \tceseq{ \mathit{Y} }  \,  {:}  \,  (   \Sigma ^\omega   ,   \tceseq{ \iota_{{\mathrm{0}}} }   )   \ottsym{,}   \tceseq{ \mathit{z_{{\mathrm{0}}}}  \,  {:}  \,  \iota_{{\mathrm{0}}} }   \vdash  \ottnt{C}$.
          Then, we have the conclusion by
          Lemmas~\ref{lem:lr-approx-init-cty-wf}, \ref{lem:ts-refine-base-sorts}, and \ref{lem:ts-weak-wf-ftyping}
          with $ \tceseq{ \ottsym{(}  \mathit{X}  \ottsym{,}  \mathit{Y}  \ottsym{)} }   \ottsym{;}   \tceseq{ \mathit{z_{{\mathrm{0}}}}  \,  {:}  \,  \iota_{{\mathrm{0}}} }   \vdash  \ottnt{C_{{\mathrm{0}}}}  \stackrel{\succ}{\circ}  \ottnt{C}$.
   \end{itemize}
 \end{caseanalysis}
\end{proof}

\begin{lemmap}{Substitution of Effect Approximation Substitution}{lr-subst-eff-approx-subst}
 Assume that $ \tceseq{ \mathit{X'} }   \ottsym{,}   \tceseq{ \mathit{X} } $ and $ \tceseq{ \mathit{Y'} }   \ottsym{,}   \tceseq{ \mathit{Y} } $ do not occur in $\sigma$.
 Then, $\ottnt{C'}  \triangleright   \tceseq{ \ottsym{(}  \mathit{X'}  \ottsym{,}  \mathit{Y'}  \ottsym{)} }   \ottsym{;}   \tceseq{ \ottsym{(}  \mathit{X}  \ottsym{,}  \mathit{Y}  \ottsym{)} }   \ottsym{;}   \tceseq{ \mathit{z_{{\mathrm{0}}}}  \,  {:}  \,  \iota_{{\mathrm{0}}} }  \,  |  \, \ottnt{C} \,  \vdash _{ \ottmv{i} }  \, \sigma_{\ottmv{i}}$ implies $\sigma  \ottsym{(}  \ottnt{C'}  \ottsym{)}  \triangleright   \tceseq{ \ottsym{(}  \mathit{X'}  \ottsym{,}  \mathit{Y'}  \ottsym{)} }   \ottsym{;}   \tceseq{ \ottsym{(}  \mathit{X}  \ottsym{,}  \mathit{Y}  \ottsym{)} }   \ottsym{;}   \tceseq{ \mathit{z_{{\mathrm{0}}}}  \,  {:}  \,  \iota_{{\mathrm{0}}} }  \,  |  \, \sigma  \ottsym{(}  \ottnt{C}  \ottsym{)} \,  \vdash _{ \ottmv{i} }  \, \sigma  \ottsym{(}  \sigma_{\ottmv{i}}  \ottsym{)}$.
\end{lemmap}
\begin{proof}
 \TS{CHECKED: 2022/07/07}
 Straightforward by induction on the derivation of $\ottnt{C'}  \triangleright   \tceseq{ \ottsym{(}  \mathit{X'}  \ottsym{,}  \mathit{Y'}  \ottsym{)} }   \ottsym{;}   \tceseq{ \ottsym{(}  \mathit{X}  \ottsym{,}  \mathit{Y}  \ottsym{)} }   \ottsym{;}   \tceseq{ \mathit{z_{{\mathrm{0}}}}  \,  {:}  \,  \iota_{{\mathrm{0}}} }  \,  |  \, \ottnt{C} \,  \vdash _{ \ottmv{i} }  \, \sigma_{\ottmv{i}}$
 with \reflem{lr-subst-eff-subst}.
\end{proof}

\begin{lemmap}{$ \nu $-Simulation}{lr-nu-sim-XXX}
 Assume that
 \begin{itemize}
  \item $\sigma_{{\mathrm{0}}} \,  \in  \,  \LRRel{ \mathcal{G} }{ \Gamma } $,
  \item $\ottnt{v} \,  =  \,  \mathsf{rec}  \, (  \tceseq{  \mathit{X} ^{  \ottnt{A} _{  \mu  }  },  \mathit{Y} ^{  \ottnt{A} _{  \nu  }  }  }  ,  \mathit{f} ,   \tceseq{ \mathit{z} }  ) . \,  \ottnt{e} $,
  \item $ \tceseq{ \mathit{z_{{\mathrm{0}}}}  \,  {:}  \,  \iota_{{\mathrm{0}}} }  \,  =  \,  \gamma  (   \tceseq{ \mathit{z}  \,  {:}  \,  \ottnt{T_{{\mathrm{1}}}} }   ) $,
  \item $ \tceseq{ \ottsym{(}  \mathit{X}  \ottsym{,}  \mathit{Y}  \ottsym{)} }   \ottsym{;}   \tceseq{ \mathit{z_{{\mathrm{0}}}}  \,  {:}  \,  \iota_{{\mathrm{0}}} }   \vdash  \ottnt{C_{{\mathrm{0}}}}  \stackrel{\succ}{\circ}  \ottnt{C}$,
  \item $  \{   \tceseq{ \mathit{X} }   \ottsym{,}   \tceseq{ \mathit{Y} }   \}   \, \ottsym{\#} \,   \mathit{fpv}  (   \tceseq{ \ottnt{T_{{\mathrm{1}}}} }   )  $,
  \item $ \tceseq{ \ottsym{(}  \mathit{X}  \ottsym{,}  \mathit{Y}  \ottsym{)} }   \ottsym{;}   \tceseq{ \mathit{z_{{\mathrm{0}}}}  \,  {:}  \,  \iota_{{\mathrm{0}}} }  \,  |  \, \ottnt{C}  \vdash  \sigma$,
  \item $  \forall  \, { \ottmv{i} } . \  \ottnt{C}  \triangleright   \emptyset   \ottsym{;}   \tceseq{ \ottsym{(}  \mathit{X}  \ottsym{,}  \mathit{Y}  \ottsym{)} }   \ottsym{;}   \tceseq{ \mathit{z_{{\mathrm{0}}}}  \,  {:}  \,  \iota_{{\mathrm{0}}} }  \,  |  \, \ottnt{C} \,  \vdash _{ \ottmv{i} }  \, \sigma_{\ottmv{i}} $,
  \item $\Gamma' \,  =  \,  \tceseq{ \mathit{X} }  \,  {:}  \,  (   \Sigma ^\ast   ,   \tceseq{ \iota_{{\mathrm{0}}} }   )   \ottsym{,}   \tceseq{ \mathit{Y} }  \,  {:}  \,  (   \Sigma ^\omega   ,   \tceseq{ \iota_{{\mathrm{0}}} }   )   \ottsym{,}  \mathit{f} \,  {:}  \, \ottsym{(}   \tceseq{ \mathit{z}  \,  {:}  \,  \ottnt{T_{{\mathrm{1}}}} }   \ottsym{)}  \rightarrow  \ottnt{C_{{\mathrm{0}}}}  \ottsym{,}   \tceseq{ \mathit{z}  \,  {:}  \,  \ottnt{T_{{\mathrm{1}}}} } $,
  \item $\ottnt{e} \,  \in  \,  \LRRel{ \mathcal{O} }{ \Gamma  \ottsym{,}  \Gamma'  \, \vdash \,  \ottnt{C} } $, and
  \item $ { \sigma }_{ \mathit{X} }  \,  =  \,  [ \tceseq{ \sigma  \ottsym{(}  \mathit{X}  \ottsym{)}  /  \mathit{X} } ] $.
 \end{itemize}
 Then,
 \[
    \forall  \, { \ottmv{i}  \ottsym{,}  \pi } . \   \sigma_{{\mathrm{0}}}  \ottsym{(}  \ottnt{v}  \ottsym{)} _{  \nu   \ottsym{;}  \ottmv{i}  \ottsym{;}  \pi }  \,  \in  \,  \LRRel{ \mathcal{V} }{ \sigma_{{\mathrm{0}}}  \ottsym{(}  \ottsym{(}   \tceseq{ \mathit{z}  \,  {:}  \,  \ottnt{T_{{\mathrm{1}}}} }   \ottsym{)}  \rightarrow  \sigma_{\ottmv{i}}  \ottsym{(}   { \sigma }_{ \mathit{X} }   \ottsym{(}  \ottnt{C_{{\mathrm{0}}}}  \ottsym{)}  \ottsym{)}  \ottsym{)} }   ~.
 \]
\end{lemmap}
\begin{proof}
 \TS{CHECKED: 2022/07/07}
 By induction on $\ottmv{i}$.
 Let $\pi$ be any infinite trace.
 Without loss of generality, we can suppose that $ \tceseq{ \mathit{X} } $, $ \tceseq{ \mathit{Y} } $, $\mathit{f}$, and $ \tceseq{ \mathit{z} } $ do not occur in $\sigma_{{\mathrm{0}}}$.
 %

 First, we show that
 \[
   \sigma_{{\mathrm{0}}}  \ottsym{(}  \ottnt{v}  \ottsym{)} _{  \nu   \ottsym{;}  \ottmv{i}  \ottsym{;}  \pi }  \,  \in  \,  \LRRel{ \mathrm{Atom} }{ \sigma_{{\mathrm{0}}}  \ottsym{(}  \ottsym{(}   \tceseq{ \mathit{z}  \,  {:}  \,  \ottnt{T_{{\mathrm{1}}}} }   \ottsym{)}  \rightarrow  \sigma_{\ottmv{i}}  \ottsym{(}   { \sigma }_{ \mathit{X} }   \ottsym{(}  \ottnt{C_{{\mathrm{0}}}}  \ottsym{)}  \ottsym{)}  \ottsym{)} }  ~.
 \]
 By Lemmas~\ref{lem:lr-typeability}, \ref{lem:lr-subst-consist}, \ref{lem:lr-subst-occur}, \ref{lem:lr-subst-eff-subst}, and \ref{lem:lr-subst-eff-approx-subst},
 we have
 \begin{itemize}
  \item $\sigma_{{\mathrm{0}}}  \ottsym{(}  \Gamma'  \ottsym{)}  \vdash  \sigma_{{\mathrm{0}}}  \ottsym{(}  \ottnt{e}  \ottsym{)}  \ottsym{:}  \sigma_{{\mathrm{0}}}  \ottsym{(}  \ottnt{C}  \ottsym{)}$,
  \item $ \tceseq{ \ottsym{(}  \mathit{X}  \ottsym{,}  \mathit{Y}  \ottsym{)} }   \ottsym{;}   \tceseq{ \mathit{z_{{\mathrm{0}}}}  \,  {:}  \,  \iota_{{\mathrm{0}}} }   \vdash  \sigma_{{\mathrm{0}}}  \ottsym{(}  \ottnt{C_{{\mathrm{0}}}}  \ottsym{)}  \stackrel{\succ}{\circ}  \sigma_{{\mathrm{0}}}  \ottsym{(}  \ottnt{C}  \ottsym{)}$,
  \item $ \tceseq{ \ottsym{(}  \mathit{X}  \ottsym{,}  \mathit{Y}  \ottsym{)} }   \ottsym{;}   \tceseq{ \mathit{z_{{\mathrm{0}}}}  \,  {:}  \,  \iota_{{\mathrm{0}}} }  \,  |  \, \sigma_{{\mathrm{0}}}  \ottsym{(}  \ottnt{C}  \ottsym{)}  \vdash  \sigma_{{\mathrm{0}}}  \ottsym{(}  \sigma  \ottsym{)}$, and
  \item $  \forall  \, { \ottmv{i} } . \  \sigma_{{\mathrm{0}}}  \ottsym{(}  \ottnt{C}  \ottsym{)}  \triangleright   \emptyset   \ottsym{;}   \tceseq{ \ottsym{(}  \mathit{X}  \ottsym{,}  \mathit{Y}  \ottsym{)} }   \ottsym{;}   \tceseq{ \mathit{z_{{\mathrm{0}}}}  \,  {:}  \,  \iota_{{\mathrm{0}}} }  \,  |  \, \sigma_{{\mathrm{0}}}  \ottsym{(}  \ottnt{C}  \ottsym{)} \,  \vdash _{ \ottmv{i} }  \, \sigma_{{\mathrm{0}}}  \ottsym{(}  \sigma_{\ottmv{i}}  \ottsym{)} $.
 \end{itemize}
 Therefore, by \reflem{lr-typing-nu-approx-XXX},
 we have the conclusion.

 Next, let
 $\sigma'_{{\mathrm{0}}} \,  =  \,  [ \tceseq{ \ottnt{v'}  /  \mathit{z} } ] $ for some values $ \tceseq{ \ottnt{v'} } $ such that
 $ \tceseq{ \ottnt{v'} \,  \in  \,  \LRRel{ \mathcal{V} }{ \sigma'_{{\mathrm{0}}}  \ottsym{(}  \sigma_{{\mathrm{0}}}  \ottsym{(}  \ottnt{T_{{\mathrm{1}}}}  \ottsym{)}  \ottsym{)} }  } $.
 Then, we must show that
 \[
   \sigma_{{\mathrm{0}}}  \ottsym{(}  \ottnt{v}  \ottsym{)} _{  \nu   \ottsym{;}  \ottmv{i}  \ottsym{;}  \pi }  \,  \tceseq{ \ottnt{v'} }  \,  \in  \,  \LRRel{ \mathcal{E} }{ \sigma'_{{\mathrm{0}}}  \ottsym{(}  \sigma_{{\mathrm{0}}}  \ottsym{(}  \sigma_{\ottmv{i}}  \ottsym{(}   { \sigma }_{ \mathit{X} }   \ottsym{(}  \ottnt{C_{{\mathrm{0}}}}  \ottsym{)}  \ottsym{)}  \ottsym{)}  \ottsym{)} }  ~.
 \]
 We proceed by case analysis on $\ottmv{i}$.
 \begin{caseanalysis}
  \case $\ottmv{i} \,  =  \, \ottsym{0}$:
   By \reflem{lr-anti-red}, it suffices to show that
   \[
     \bullet ^{ \pi }  \,  \in  \,  \LRRel{ \mathcal{E} }{ \sigma'_{{\mathrm{0}}}  \ottsym{(}  \sigma_{{\mathrm{0}}}  \ottsym{(}  \sigma_{\ottmv{i}}  \ottsym{(}   { \sigma }_{ \mathit{X} }   \ottsym{(}  \ottnt{C_{{\mathrm{0}}}}  \ottsym{)}  \ottsym{)}  \ottsym{)}  \ottsym{)} }  ~.
   \]
   By definition, it suffices to show that
   \[
     { \models  \,   { \sigma'_{{\mathrm{0}}}  \ottsym{(}  \sigma_{{\mathrm{0}}}  \ottsym{(}  \sigma_{\ottmv{i}}  \ottsym{(}   { \sigma }_{ \mathit{X} }   \ottsym{(}  \ottnt{C_{{\mathrm{0}}}}  \ottsym{)}  \ottsym{)}  \ottsym{)}  \ottsym{)} }^\nu (  \pi  )  }  ~.
   \]
   Because $ \tceseq{ \ottsym{(}  \mathit{X}  \ottsym{,}  \mathit{Y}  \ottsym{)} }   \ottsym{;}   \tceseq{ \mathit{z_{{\mathrm{0}}}}  \,  {:}  \,  \iota_{{\mathrm{0}}} }   \vdash  \ottnt{C_{{\mathrm{0}}}}  \stackrel{\succ}{\circ}  \ottnt{C}$,
   the $ \nu $-parts of the temporal effects occurring strictly positively in $\ottnt{C_{{\mathrm{0}}}}$
   take the form $\mathit{Y_{{\mathrm{0}}}}  \ottsym{(}  \mathit{y}  \ottsym{,}   \tceseq{ \mathit{z_{{\mathrm{0}}}} }   \ottsym{)}$ for some $\mathit{Y_{{\mathrm{0}}}} \,  \in  \,  \{   \tceseq{ \mathit{Y} }   \} $, and
   the assumption $\ottnt{C}  \triangleright   \emptyset   \ottsym{;}   \tceseq{ \ottsym{(}  \mathit{X}  \ottsym{,}  \mathit{Y}  \ottsym{)} }   \ottsym{;}   \tceseq{ \mathit{z_{{\mathrm{0}}}}  \,  {:}  \,  \iota_{{\mathrm{0}}} }  \,  |  \, \ottnt{C} \,  \vdash _{ \ottmv{i} }  \, \sigma_{\ottmv{i}}$ indicates that
   $\mathit{Y_{{\mathrm{0}}}}$ is replaced by $\sigma_{\ottmv{i}}$.
   Because $\ottmv{i} \,  =  \, \ottsym{0}$, $\mathit{Y_{{\mathrm{0}}}}$ is replaced with $ \top $.
   Therefore, we have the conclusion.

  \case $\ottmv{i} \,  >  \, \ottsym{0}$:
   By \reflem{lr-anti-red} and the definition of $ \sigma  \ottsym{(}  \ottnt{v}  \ottsym{)} _{  \nu   \ottsym{;}  \ottmv{i}  \ottsym{;}  \pi } $,
   it suffices to show that
   \[
    \sigma'_{{\mathrm{0}}}  \ottsym{(}  \sigma_{{\mathrm{0}}}  \ottsym{(}   { \sigma }_{ \ottmv{i}  \ottsym{-}  \ottsym{1} }   \ottsym{(}   { \sigma }_{ \mathit{X} }   \ottsym{(}   \ottnt{e}    [   \ottnt{v} _{  \nu   \ottsym{;}  \ottmv{i}  \ottsym{-}  \ottsym{1}  \ottsym{;}  \pi }   /  \mathit{f}  ]    \ottsym{)}  \ottsym{)}  \ottsym{)}  \ottsym{)} \,  \in  \,  \LRRel{ \mathcal{E} }{ \sigma'_{{\mathrm{0}}}  \ottsym{(}  \sigma_{{\mathrm{0}}}  \ottsym{(}  \sigma_{\ottmv{i}}  \ottsym{(}   { \sigma }_{ \mathit{X} }   \ottsym{(}  \ottnt{C_{{\mathrm{0}}}}  \ottsym{)}  \ottsym{)}  \ottsym{)}  \ottsym{)} }  ~.
   \]
   By the IH, we have
   \[
     \sigma_{{\mathrm{0}}}  \ottsym{(}  \ottnt{v}  \ottsym{)} _{  \nu   \ottsym{;}  \ottmv{i}  \ottsym{-}  \ottsym{1}  \ottsym{;}  \pi }  \,  \in  \,  \LRRel{ \mathcal{V} }{ \sigma_{{\mathrm{0}}}  \ottsym{(}  \ottsym{(}   \tceseq{ \mathit{z}  \,  {:}  \,  \ottnt{T_{{\mathrm{1}}}} }   \ottsym{)}  \rightarrow   { \sigma }_{ \ottmv{i}  \ottsym{-}  \ottsym{1} }   \ottsym{(}   { \sigma }_{ \mathit{X} }   \ottsym{(}  \ottnt{C_{{\mathrm{0}}}}  \ottsym{)}  \ottsym{)}  \ottsym{)} }  ~.
   \]
   Therefore, by Lemmas~\ref{lem:lr-mu-eff-rec-wf-XXX}, \ref{lem:lr-approx-eff-subst-wf}, and \ref{lem:lr-typeability}, we have
   \[
    \sigma_{{\mathrm{0}}}  \mathbin{\uplus}  \sigma_{{\mathrm{0}}}  \ottsym{(}   { \sigma }_{ \mathit{X} }   \ottsym{)}  \mathbin{\uplus}   [ \tceseq{ \sigma_{{\mathrm{0}}}  \ottsym{(}   { \sigma }_{ \ottmv{i}  \ottsym{-}  \ottsym{1} }   \ottsym{(}  \mathit{Y}  \ottsym{)}  \ottsym{)}  /  \mathit{Y} } ]   \mathbin{\uplus}   [   \sigma_{{\mathrm{0}}}  \ottsym{(}  \ottnt{v}  \ottsym{)} _{  \nu   \ottsym{;}  \ottmv{i}  \ottsym{-}  \ottsym{1}  \ottsym{;}  \pi }   /  \mathit{f}  ]   \mathbin{\uplus}  \sigma'_{{\mathrm{0}}} \,  \in  \,  \LRRel{ \mathcal{G} }{ \Gamma  \ottsym{,}  \Gamma' }  ~.
   \]
   Because $\ottnt{e} \,  \in  \,  \LRRel{ \mathcal{O} }{ \Gamma  \ottsym{,}  \Gamma'  \, \vdash \,  \ottnt{C} } $ by the assumption,
   we have
   \[
    \sigma'_{{\mathrm{0}}}  \ottsym{(}  \sigma_{{\mathrm{0}}}  \ottsym{(}   { \sigma }_{ \ottmv{i}  \ottsym{-}  \ottsym{1} }   \ottsym{(}   { \sigma }_{ \mathit{X} }   \ottsym{(}   \ottnt{e}    [   \ottnt{v} _{  \nu   \ottsym{;}  \ottmv{i}  \ottsym{-}  \ottsym{1}  \ottsym{;}  \pi }   /  \mathit{f}  ]    \ottsym{)}  \ottsym{)}  \ottsym{)}  \ottsym{)} \,  \in  \,  \LRRel{ \mathcal{E} }{ \sigma'_{{\mathrm{0}}}  \ottsym{(}  \sigma_{{\mathrm{0}}}  \ottsym{(}   { \sigma }_{ \ottmv{i}  \ottsym{-}  \ottsym{1} }   \ottsym{(}   { \sigma }_{ \mathit{X} }   \ottsym{(}  \ottnt{C}  \ottsym{)}  \ottsym{)}  \ottsym{)}  \ottsym{)} }  ~.
   \]
   Note that $\mathit{f}$ does not occur in $\ottnt{C}$
   by \reflem{ts-unuse-non-refine-vars}.
   Then, by \reflem{lr-subtyping}, it suffices to show that
   \[
     \emptyset   \vdash  \sigma'_{{\mathrm{0}}}  \ottsym{(}  \sigma_{{\mathrm{0}}}  \ottsym{(}   { \sigma }_{ \ottmv{i}  \ottsym{-}  \ottsym{1} }   \ottsym{(}   { \sigma }_{ \mathit{X} }   \ottsym{(}  \ottnt{C}  \ottsym{)}  \ottsym{)}  \ottsym{)}  \ottsym{)}  \ottsym{<:}  \sigma'_{{\mathrm{0}}}  \ottsym{(}  \sigma_{{\mathrm{0}}}  \ottsym{(}  \sigma_{\ottmv{i}}  \ottsym{(}   { \sigma }_{ \mathit{X} }   \ottsym{(}  \ottnt{C}  \ottsym{)}  \ottsym{)}  \ottsym{)}  \ottsym{)} ~.
   \]
   By \reflem{lr-approx-fold-nu-XXX},
   \[
     \tceseq{ \mathit{z}  \,  {:}  \,  \sigma_{{\mathrm{0}}}  \ottsym{(}  \ottnt{T_{{\mathrm{1}}}}  \ottsym{)} }   \vdash  \sigma_{{\mathrm{0}}}  \ottsym{(}   { \sigma }_{ \ottmv{i}  \ottsym{-}  \ottsym{1} }   \ottsym{(}   { \sigma }_{ \mathit{X} }   \ottsym{(}  \ottnt{C}  \ottsym{)}  \ottsym{)}  \ottsym{)}  \ottsym{<:}  \sigma_{{\mathrm{0}}}  \ottsym{(}  \sigma_{\ottmv{i}}  \ottsym{(}   { \sigma }_{ \mathit{X} }   \ottsym{(}  \ottnt{C}  \ottsym{)}  \ottsym{)}  \ottsym{)} ~.
   \]
   Because $\sigma'_{{\mathrm{0}}} \,  \in  \,  \LRRel{ \mathcal{G} }{  \tceseq{ \mathit{z}  \,  {:}  \,  \sigma_{{\mathrm{0}}}  \ottsym{(}  \ottnt{T_{{\mathrm{1}}}}  \ottsym{)} }  } $,
   we have the conclusion by \reflem{lr-typeability}.
 \end{caseanalysis}
\end{proof}

\begin{lemmap}{$ \mu $-Simulation}{lr-mu-sim-XXX}
 Assume that
 \begin{itemize}
  \item $\sigma_{{\mathrm{0}}} \,  \in  \,  \LRRel{ \mathcal{G} }{ \Gamma } $,
  \item $\ottnt{v} \,  =  \,  \mathsf{rec}  \, (  \tceseq{  \mathit{X} ^{  \ottnt{A} _{  \mu  }  },  \mathit{Y} ^{  \ottnt{A} _{  \nu  }  }  }  ,  \mathit{f} ,   \tceseq{ \mathit{z} }  ) . \,  \ottnt{e} $,
  \item $ \tceseq{ \mathit{z_{{\mathrm{0}}}}  \,  {:}  \,  \iota_{{\mathrm{0}}} }  \,  =  \,  \gamma  (   \tceseq{ \mathit{z}  \,  {:}  \,  \ottnt{T_{{\mathrm{1}}}} }   ) $,
  \item $ \tceseq{ \ottsym{(}  \mathit{X}  \ottsym{,}  \mathit{Y}  \ottsym{)} }   \ottsym{;}   \tceseq{ \mathit{z_{{\mathrm{0}}}}  \,  {:}  \,  \iota_{{\mathrm{0}}} }   \vdash  \ottnt{C_{{\mathrm{0}}}}  \stackrel{\succ}{\circ}  \ottnt{C}$,
  \item $  \{   \tceseq{ \mathit{X} }   \ottsym{,}   \tceseq{ \mathit{Y} }   \}   \, \ottsym{\#} \,   \mathit{fpv}  (   \tceseq{ \ottnt{T_{{\mathrm{1}}}} }   )  $,
  \item $  \forall  \, { \ottmv{i} } . \  \ottnt{C}  \triangleright   \emptyset   \ottsym{;}   \tceseq{ \ottsym{(}  \mathit{X}  \ottsym{,}  \mathit{Y}  \ottsym{)} }   \ottsym{;}   \tceseq{ \mathit{z_{{\mathrm{0}}}}  \,  {:}  \,  \iota_{{\mathrm{0}}} }  \,  |  \, \ottnt{C} \,  \vdash _{ \ottmv{i} }  \, \sigma_{\ottmv{i}} $,
  \item $\Gamma' \,  =  \,  \tceseq{ \mathit{X} }  \,  {:}  \,  (   \Sigma ^\ast   ,   \tceseq{ \iota_{{\mathrm{0}}} }   )   \ottsym{,}   \tceseq{ \mathit{Y} }  \,  {:}  \,  (   \Sigma ^\omega   ,   \tceseq{ \iota_{{\mathrm{0}}} }   )   \ottsym{,}  \mathit{f} \,  {:}  \, \ottsym{(}   \tceseq{ \mathit{z}  \,  {:}  \,  \ottnt{T_{{\mathrm{1}}}} }   \ottsym{)}  \rightarrow  \ottnt{C_{{\mathrm{0}}}}  \ottsym{,}   \tceseq{ \mathit{z}  \,  {:}  \,  \ottnt{T_{{\mathrm{1}}}} } $,
  \item $\ottnt{e} \,  \in  \,  \LRRel{ \mathcal{O} }{ \Gamma  \ottsym{,}  \Gamma'  \, \vdash \,  \ottnt{C} } $, and
  \item $ { \sigma }_{ \mathit{Y} }  \,  =  \,  [ \tceseq{ \ottsym{(}   \nu\!  \, \mathit{Y}  \ottsym{(}  \mathit{y} \,  {:}  \,  \Sigma ^\omega   \ottsym{,}   \tceseq{ \mathit{z_{{\mathrm{0}}}}  \,  {:}  \,  \iota_{{\mathrm{0}}} }   \ottsym{)}  \ottsym{.}   \top   \ottsym{)}  /  \mathit{Y} } ] $.
 \end{itemize}
 Then,
 \[
    \forall  \, { \ottmv{i}  \ottsym{,}  \pi } . \   \sigma_{{\mathrm{0}}}  \ottsym{(}  \ottnt{v}  \ottsym{)} _{  \mu   \ottsym{;}  \ottmv{i}  \ottsym{;}  \pi }  \,  \in  \,  \LRRel{ \mathcal{V} }{ \sigma_{{\mathrm{0}}}  \ottsym{(}  \ottsym{(}   \tceseq{ \mathit{z}  \,  {:}  \,  \ottnt{T_{{\mathrm{1}}}} }   \ottsym{)}  \rightarrow  \sigma_{\ottmv{i}}  \ottsym{(}   { \sigma }_{ \mathit{Y} }   \ottsym{(}  \ottnt{C_{{\mathrm{0}}}}  \ottsym{)}  \ottsym{)}  \ottsym{)} }   ~.
 \]
\end{lemmap}
\begin{proof}
 \TS{CHECKED: 2022/07/07}
 By induction on $\ottmv{i}$.
 Let $\pi$ be any infinite trace.
 Without loss of generality, we can suppose that $ \tceseq{ \mathit{X} } $, $ \tceseq{ \mathit{Y} } $, $\mathit{f}$, and $ \tceseq{ \mathit{z} } $ do not occur in $\sigma_{{\mathrm{0}}}$.
 %

 First, we show that
 \[
   \sigma_{{\mathrm{0}}}  \ottsym{(}  \ottnt{v}  \ottsym{)} _{  \mu   \ottsym{;}  \ottmv{i}  \ottsym{;}  \pi }  \,  \in  \,  \LRRel{ \mathrm{Atom} }{ \sigma_{{\mathrm{0}}}  \ottsym{(}  \ottsym{(}   \tceseq{ \mathit{z}  \,  {:}  \,  \ottnt{T_{{\mathrm{1}}}} }   \ottsym{)}  \rightarrow  \sigma_{\ottmv{i}}  \ottsym{(}   { \sigma }_{ \mathit{Y} }   \ottsym{(}  \ottnt{C_{{\mathrm{0}}}}  \ottsym{)}  \ottsym{)}  \ottsym{)} }  ~.
 \]
 By Lemmas~\ref{lem:lr-typeability}, \ref{lem:lr-subst-consist}, \ref{lem:lr-subst-occur}, \ref{lem:lr-subst-eff-subst}, and \ref{lem:lr-subst-eff-approx-subst},
 we have
 \begin{itemize}
  \item $\sigma_{{\mathrm{0}}}  \ottsym{(}  \Gamma'  \ottsym{)}  \vdash  \sigma_{{\mathrm{0}}}  \ottsym{(}  \ottnt{e}  \ottsym{)}  \ottsym{:}  \sigma_{{\mathrm{0}}}  \ottsym{(}  \ottnt{C}  \ottsym{)}$,
  \item $ \tceseq{ \ottsym{(}  \mathit{X}  \ottsym{,}  \mathit{Y}  \ottsym{)} }   \ottsym{;}   \tceseq{ \mathit{z_{{\mathrm{0}}}}  \,  {:}  \,  \iota_{{\mathrm{0}}} }   \vdash  \sigma_{{\mathrm{0}}}  \ottsym{(}  \ottnt{C_{{\mathrm{0}}}}  \ottsym{)}  \stackrel{\succ}{\circ}  \sigma_{{\mathrm{0}}}  \ottsym{(}  \ottnt{C}  \ottsym{)}$,
  \item $ \tceseq{ \ottsym{(}  \mathit{X}  \ottsym{,}  \mathit{Y}  \ottsym{)} }   \ottsym{;}   \tceseq{ \mathit{z_{{\mathrm{0}}}}  \,  {:}  \,  \iota_{{\mathrm{0}}} }  \,  |  \, \sigma_{{\mathrm{0}}}  \ottsym{(}  \ottnt{C}  \ottsym{)}  \vdash  \sigma_{{\mathrm{0}}}  \ottsym{(}  \sigma  \ottsym{)}$, and
  \item $  \forall  \, { \ottmv{i} } . \  \sigma_{{\mathrm{0}}}  \ottsym{(}  \ottnt{C}  \ottsym{)}  \triangleright   \emptyset   \ottsym{;}   \tceseq{ \ottsym{(}  \mathit{X}  \ottsym{,}  \mathit{Y}  \ottsym{)} }   \ottsym{;}   \tceseq{ \mathit{z_{{\mathrm{0}}}}  \,  {:}  \,  \iota_{{\mathrm{0}}} }  \,  |  \, \sigma_{{\mathrm{0}}}  \ottsym{(}  \ottnt{C}  \ottsym{)} \,  \vdash _{ \ottmv{i} }  \, \sigma_{{\mathrm{0}}}  \ottsym{(}  \sigma_{\ottmv{i}}  \ottsym{)} $.
 \end{itemize}
 Therefore, by \reflem{lr-typing-mu-approx-XXX},
 we have the conclusion.

 Next, let
 $\sigma'_{{\mathrm{0}}} \,  =  \,  [ \tceseq{ \ottnt{v'}  /  \mathit{z} } ] $ for some values $ \tceseq{ \ottnt{v'} } $ such that
 $ \tceseq{ \ottnt{v'} \,  \in  \,  \LRRel{ \mathcal{V} }{ \sigma'_{{\mathrm{0}}}  \ottsym{(}  \sigma_{{\mathrm{0}}}  \ottsym{(}  \ottnt{T_{{\mathrm{1}}}}  \ottsym{)}  \ottsym{)} }  } $.
 Then, we must show that
 \[
   \sigma_{{\mathrm{0}}}  \ottsym{(}  \ottnt{v}  \ottsym{)} _{  \mu   \ottsym{;}  \ottmv{i}  \ottsym{;}  \pi }  \,  \tceseq{ \ottnt{v'} }  \,  \in  \,  \LRRel{ \mathcal{E} }{ \sigma'_{{\mathrm{0}}}  \ottsym{(}  \sigma_{{\mathrm{0}}}  \ottsym{(}  \sigma_{\ottmv{i}}  \ottsym{(}   { \sigma }_{ \mathit{Y} }   \ottsym{(}  \ottnt{C_{{\mathrm{0}}}}  \ottsym{)}  \ottsym{)}  \ottsym{)}  \ottsym{)} }  ~.
 \]
 We proceed by case analysis on $\ottmv{i}$.
 \begin{caseanalysis}
  \case $\ottmv{i} \,  =  \, \ottsym{0}$:
   By \reflem{lr-anti-red}, it suffices to show that
   \[
     \bullet ^{ \pi }  \,  \in  \,  \LRRel{ \mathcal{E} }{ \sigma'_{{\mathrm{0}}}  \ottsym{(}  \sigma_{{\mathrm{0}}}  \ottsym{(}  \sigma_{\ottmv{i}}  \ottsym{(}   { \sigma }_{ \mathit{Y} }   \ottsym{(}  \ottnt{C_{{\mathrm{0}}}}  \ottsym{)}  \ottsym{)}  \ottsym{)}  \ottsym{)} }  ~.
   \]
   By definition, it suffices to show that
   \[
     { \models  \,   { \sigma'_{{\mathrm{0}}}  \ottsym{(}  \sigma_{{\mathrm{0}}}  \ottsym{(}  \sigma_{\ottmv{i}}  \ottsym{(}   { \sigma }_{ \mathit{Y} }   \ottsym{(}  \ottnt{C_{{\mathrm{0}}}}  \ottsym{)}  \ottsym{)}  \ottsym{)}  \ottsym{)} }^\nu (  \pi  )  }  ~.
   \]
   Because $ \tceseq{ \ottsym{(}  \mathit{X}  \ottsym{,}  \mathit{Y}  \ottsym{)} }   \ottsym{;}   \tceseq{ \mathit{z_{{\mathrm{0}}}}  \,  {:}  \,  \iota_{{\mathrm{0}}} }   \vdash  \ottnt{C_{{\mathrm{0}}}}  \stackrel{\succ}{\circ}  \ottnt{C}$,
   the $ \nu $-parts of the temporal effects occurring strictly positively in $\ottnt{C_{{\mathrm{0}}}}$
   take the form $\mathit{Y_{{\mathrm{0}}}}  \ottsym{(}  \mathit{y}  \ottsym{,}   \tceseq{ \mathit{z_{{\mathrm{0}}}} }   \ottsym{)}$ for some $\mathit{Y_{{\mathrm{0}}}} \,  \in  \,  \{   \tceseq{ \mathit{Y} }   \} $, and
   $ { \sigma }_{ \mathit{Y} } $ replaces $\mathit{Y_{{\mathrm{0}}}}$ with $ \top $.
   Therefore, we have the conclusion.

  \case $\ottmv{i} \,  >  \, \ottsym{0}$:
   By \reflem{lr-anti-red} and the definition of $ \sigma  \ottsym{(}  \ottnt{v}  \ottsym{)} _{  \mu   \ottsym{;}  \ottmv{i}  \ottsym{;}  \pi } $,
   it suffices to show that
   \[
    \sigma'_{{\mathrm{0}}}  \ottsym{(}  \sigma_{{\mathrm{0}}}  \ottsym{(}   { \sigma }_{ \ottmv{i}  \ottsym{-}  \ottsym{1} }   \ottsym{(}   { \sigma }_{ \mathit{Y} }   \ottsym{(}   \ottnt{e}    [   \ottnt{v} _{  \mu   \ottsym{;}  \ottmv{i}  \ottsym{-}  \ottsym{1}  \ottsym{;}  \pi }   /  \mathit{f}  ]    \ottsym{)}  \ottsym{)}  \ottsym{)}  \ottsym{)} \,  \in  \,  \LRRel{ \mathcal{E} }{ \sigma'_{{\mathrm{0}}}  \ottsym{(}  \sigma_{{\mathrm{0}}}  \ottsym{(}  \sigma_{\ottmv{i}}  \ottsym{(}   { \sigma }_{ \mathit{Y} }   \ottsym{(}  \ottnt{C_{{\mathrm{0}}}}  \ottsym{)}  \ottsym{)}  \ottsym{)}  \ottsym{)} }  ~.
   \]
   By the IH, we have
   \[
     \sigma_{{\mathrm{0}}}  \ottsym{(}  \ottnt{v}  \ottsym{)} _{  \mu   \ottsym{;}  \ottmv{i}  \ottsym{-}  \ottsym{1}  \ottsym{;}  \pi }  \,  \in  \,  \LRRel{ \mathcal{V} }{ \sigma_{{\mathrm{0}}}  \ottsym{(}  \ottsym{(}   \tceseq{ \mathit{z}  \,  {:}  \,  \ottnt{T_{{\mathrm{1}}}} }   \ottsym{)}  \rightarrow   { \sigma }_{ \ottmv{i}  \ottsym{-}  \ottsym{1} }   \ottsym{(}   { \sigma }_{ \mathit{Y} }   \ottsym{(}  \ottnt{C_{{\mathrm{0}}}}  \ottsym{)}  \ottsym{)}  \ottsym{)} }  ~.
   \]
   Therefore, by Lemmas~\ref{lem:lr-approx-eff-subst-wf}, we have
   \[
    \sigma_{{\mathrm{0}}}  \mathbin{\uplus}  \sigma_{{\mathrm{0}}}  \ottsym{(}   { \sigma }_{ \mathit{Y} }   \ottsym{)}  \mathbin{\uplus}   [ \tceseq{ \sigma_{{\mathrm{0}}}  \ottsym{(}   { \sigma }_{ \ottmv{i}  \ottsym{-}  \ottsym{1} }   \ottsym{(}  \mathit{X}  \ottsym{)}  \ottsym{)}  /  \mathit{X} } ]   \mathbin{\uplus}   [   \sigma_{{\mathrm{0}}}  \ottsym{(}  \ottnt{v}  \ottsym{)} _{  \mu   \ottsym{;}  \ottmv{i}  \ottsym{-}  \ottsym{1}  \ottsym{;}  \pi }   /  \mathit{f}  ]   \mathbin{\uplus}  \sigma'_{{\mathrm{0}}} \,  \in  \,  \LRRel{ \mathcal{G} }{ \Gamma  \ottsym{,}  \Gamma' }  ~.
   \]
   Because $\ottnt{e} \,  \in  \,  \LRRel{ \mathcal{O} }{ \Gamma  \ottsym{,}  \Gamma'  \, \vdash \,  \ottnt{C} } $ by the assumption,
   we have
   \[
    \sigma'_{{\mathrm{0}}}  \ottsym{(}  \sigma_{{\mathrm{0}}}  \ottsym{(}   { \sigma }_{ \ottmv{i}  \ottsym{-}  \ottsym{1} }   \ottsym{(}   { \sigma }_{ \mathit{Y} }   \ottsym{(}   \ottnt{e}    [   \ottnt{v} _{  \mu   \ottsym{;}  \ottmv{i}  \ottsym{-}  \ottsym{1}  \ottsym{;}  \pi }   /  \mathit{f}  ]    \ottsym{)}  \ottsym{)}  \ottsym{)}  \ottsym{)} \,  \in  \,  \LRRel{ \mathcal{E} }{ \sigma'_{{\mathrm{0}}}  \ottsym{(}  \sigma_{{\mathrm{0}}}  \ottsym{(}   { \sigma }_{ \ottmv{i}  \ottsym{-}  \ottsym{1} }   \ottsym{(}   { \sigma }_{ \mathit{X} }   \ottsym{(}  \ottnt{C}  \ottsym{)}  \ottsym{)}  \ottsym{)}  \ottsym{)} }  ~.
   \]
   Note that $\mathit{f}$ does not occur in $\ottnt{C}$
   by \reflem{ts-unuse-non-refine-vars}.
   Then, by \reflem{lr-subtyping}, it suffices to show that
   \[
     \emptyset   \vdash  \sigma'_{{\mathrm{0}}}  \ottsym{(}  \sigma_{{\mathrm{0}}}  \ottsym{(}   { \sigma }_{ \ottmv{i}  \ottsym{-}  \ottsym{1} }   \ottsym{(}   { \sigma }_{ \mathit{Y} }   \ottsym{(}  \ottnt{C}  \ottsym{)}  \ottsym{)}  \ottsym{)}  \ottsym{)}  \ottsym{<:}  \sigma'_{{\mathrm{0}}}  \ottsym{(}  \sigma_{{\mathrm{0}}}  \ottsym{(}  \sigma_{\ottmv{i}}  \ottsym{(}   { \sigma }_{ \mathit{Y} }   \ottsym{(}  \ottnt{C}  \ottsym{)}  \ottsym{)}  \ottsym{)}  \ottsym{)} ~.
   \]
   By \reflem{lr-approx-fold-mu-XXX},
   \[
     \tceseq{ \mathit{z}  \,  {:}  \,  \sigma_{{\mathrm{0}}}  \ottsym{(}  \ottnt{T_{{\mathrm{1}}}}  \ottsym{)} }   \vdash  \sigma_{{\mathrm{0}}}  \ottsym{(}   { \sigma }_{ \ottmv{i}  \ottsym{-}  \ottsym{1} }   \ottsym{(}   { \sigma }_{ \mathit{Y} }   \ottsym{(}  \ottnt{C}  \ottsym{)}  \ottsym{)}  \ottsym{)}  \ottsym{<:}  \sigma_{{\mathrm{0}}}  \ottsym{(}  \sigma_{\ottmv{i}}  \ottsym{(}   { \sigma }_{ \mathit{Y} }   \ottsym{(}  \ottnt{C}  \ottsym{)}  \ottsym{)}  \ottsym{)} ~.
   \]
   Because $\sigma'_{{\mathrm{0}}} \,  \in  \,  \LRRel{ \mathcal{G} }{  \tceseq{ \mathit{z}  \,  {:}  \,  \sigma_{{\mathrm{0}}}  \ottsym{(}  \ottnt{T_{{\mathrm{1}}}}  \ottsym{)} }  } $,
   we have the conclusion by \reflem{lr-typeability}.
 \end{caseanalysis}
\end{proof}

\begin{lemmap}{Logical Relation is Closed Filling Evaluation Contexts}{lr-closed-fill-ectx}
 If $\ottsym{(}  \ottnt{E}  \ottsym{,}  \sigma  \ottsym{)}  \ottsym{:}  \ottnt{C}  \Rrightarrow  \ottnt{C'}$ and $\ottnt{e} \,  \in  \,  \LRRel{ \mathcal{E} }{ \sigma  \ottsym{(}  \ottnt{C}  \ottsym{)} } $, then $ \ottnt{E}  [  \ottnt{e}  ]  \,  \in  \,  \LRRel{ \mathcal{E} }{ \sigma  \ottsym{(}  \ottnt{C'}  \ottsym{)} } $.
\end{lemmap}
\begin{proof}
 \TS{CHECKED (NO CHANGE): 2022/07/07}
 By induction on the derivation of $\ottsym{(}  \ottnt{E}  \ottsym{,}  \sigma  \ottsym{)}  \ottsym{:}  \ottnt{C}  \Rrightarrow  \ottnt{C'}$.
 \begin{caseanalysis}
  \case \TC{Empty}: Obvious because $\ottnt{E} \,  =  \, \,[\,]\,$ and $\ottnt{C} \,  =  \, \ottnt{C'}$.
  \case \TC{Cons}: We are given
   \begin{prooftree}
    \AxiomC{$\ottsym{(}  \ottnt{E'}  \ottsym{,}  \sigma  \ottsym{)}  \ottsym{:}  \ottnt{C}  \Rrightarrow   \ottnt{T}  \,  \,\&\,  \,  \Phi  \, / \,   ( \forall  \mathit{x}  .  \ottnt{C_{{\mathrm{1}}}}  )  \Rightarrow   \ottnt{C'}  $}
    \AxiomC{$\ottnt{K} \,  \in  \,  \LRRel{ \mathcal{K} }{ \sigma  \ottsym{(}   \ottnt{T}  \, / \, ( \forall  \mathit{x}  .  \ottnt{C_{{\mathrm{1}}}}  )   \ottsym{)} } $}
    \BinaryInfC{$\ottsym{(}   \resetcnstr{  \ottnt{K}  [  \ottnt{E'}  ]  }   \ottsym{,}  \sigma  \ottsym{)}  \ottsym{:}  \ottnt{C}  \Rrightarrow  \ottnt{C'}$}
   \end{prooftree}
   for some $\ottnt{K}$, $\ottnt{E'}$, $\ottnt{T}$, $\Phi$, $\mathit{x}$, and $\ottnt{C_{{\mathrm{1}}}}$ such that
   $\ottnt{E} \,  =  \,  \resetcnstr{  \ottnt{K}  [  \ottnt{E'}  ]  } $.
   By the IH, $ \ottnt{E'}  [  \ottnt{e}  ]  \,  \in  \,  \LRRel{ \mathcal{E} }{ \sigma  \ottsym{(}   \ottnt{T}  \,  \,\&\,  \,  \Phi  \, / \,   ( \forall  \mathit{x}  .  \ottnt{C_{{\mathrm{1}}}}  )  \Rightarrow   \ottnt{C'}    \ottsym{)} } $.
   By \reflem{lr-closed-reset}, with $\ottnt{K} \,  \in  \,  \LRRel{ \mathcal{K} }{ \sigma  \ottsym{(}   \ottnt{T}  \, / \, ( \forall  \mathit{x}  .  \ottnt{C_{{\mathrm{1}}}}  )   \ottsym{)} } $,
   we have the conclusion $ \resetcnstr{  \ottnt{K}  [   \ottnt{E'}  [  \ottnt{e}  ]   ]  }  \,  \in  \,  \LRRel{ \mathcal{E} }{ \sigma  \ottsym{(}  \ottnt{C'}  \ottsym{)} } $.
 \end{caseanalysis}
\end{proof}

\begin{lemmap}{Approximating $ \nu $-Effects}{lr-approx-nu-eff-XXX}
 Assume that
 \begin{itemize}
  \item $ \emptyset   \ottsym{;}   \tceseq{ \ottsym{(}  \mathit{X}  \ottsym{,}  \mathit{Y}  \ottsym{)} }  \,  \vdash ^{\circ}  \, \ottnt{C}$,
  \item $ \tceseq{ \ottsym{(}  \mathit{X}  \ottsym{,}  \mathit{Y}  \ottsym{)} }   \ottsym{;}   \tceseq{ \mathit{z_{{\mathrm{0}}}}  \,  {:}  \,  \iota_{{\mathrm{0}}} }  \,  |  \, \ottnt{C}  \vdash  \sigma$,
  \item $  \forall  \, { \ottmv{i} } . \  \ottnt{C}  \triangleright   \emptyset   \ottsym{;}   \tceseq{ \ottsym{(}  \mathit{X}  \ottsym{,}  \mathit{Y}  \ottsym{)} }   \ottsym{;}   \tceseq{ \mathit{z_{{\mathrm{0}}}}  \,  {:}  \,  \iota_{{\mathrm{0}}} }  \,  |  \, \ottnt{C} \,  \vdash _{ \ottmv{i} }  \, \sigma_{\ottmv{i}} $,
  \item $  \forall  \, { \ottmv{i} } . \  \sigma'_{\ottmv{i}} \,  =  \,  [ \tceseq{ \sigma_{\ottmv{i}}  \ottsym{(}  \mathit{Y}  \ottsym{)}  /  \mathit{Y} } ]  $, and
  \item $  \forall  \, { \ottmv{i} } . \   { \models  \,   { \ottsym{(}   \sigma'_{\ottmv{i}}  \ottsym{(}  \ottnt{C}  \ottsym{)}    [ \tceseq{ \ottnt{c_{{\mathrm{0}}}}  /  \mathit{z_{{\mathrm{0}}}} } ]    \ottsym{)} }^\nu (  \pi  )  }  $.
 \end{itemize}
 Let $\sigma' \,  =  \,  [ \tceseq{ \sigma  \ottsym{(}  \mathit{Y}  \ottsym{)}  /  \mathit{Y} } ] $.
 Then, $ { \models  \,   { \ottsym{(}   \sigma'  \ottsym{(}  \ottnt{C}  \ottsym{)}    [ \tceseq{ \ottnt{c_{{\mathrm{0}}}}  /  \mathit{z_{{\mathrm{0}}}} } ]    \ottsym{)} }^\nu (  \pi  )  } $.
\end{lemmap}
\begin{proof}
 \TS{CHECKED: 2022/07/07}
 First, we define type contexts, ranged over by $\ottnt{TC}$, are defined as follows:
 \[
  \ottnt{TC} ::=  \,[\,]\,  \mid  \ottnt{T}  \,  \,\&\,  \,  \Phi  \,/\, ( \forall  \mathit{x}  .  \ottnt{C_{{\mathrm{1}}}}  )  \Rightarrow   \ottnt{TC}  ~.
 \]
 We write $ \ottnt{TC}  [  \ottnt{C}  ] $ for the computation type obtained by filling
 the hole in $\ottnt{TC}$ with $\ottnt{C}$.

 Let $ \tceseq{ \Phi } $ be a sequence of temporal effects $\Phi_{{\mathrm{0}}}$ such that
 $\ottnt{C} \,  =  \,  \ottnt{TC}  [   \ottnt{T}  \,  \,\&\,  \,  \Phi_{{\mathrm{0}}}  \, / \,  \ottnt{S}   ] $ for some $\ottnt{TC}$, $\ottnt{T}$, and $\ottnt{S}$.
 It suffices to show that, for any $\Phi_{{\mathrm{0}}}$ in $ \tceseq{ \Phi } $,
 $ { \models  \,   { \ottsym{(}   \sigma'  \ottsym{(}  \Phi_{{\mathrm{0}}}  \ottsym{)}    [ \tceseq{ \ottnt{c_{{\mathrm{0}}}}  /  \mathit{z_{{\mathrm{0}}}} } ]    \ottsym{)} }^\nu   \ottsym{(}  \pi  \ottsym{)} } $.
 We can also assume that, for any $\mathit{Y_{{\mathrm{0}}}}$ in $ \tceseq{ \mathit{Y} } $,
 $\sigma  \ottsym{(}  \mathit{Y_{{\mathrm{0}}}}  \ottsym{)}$ is the greatest fixpoint of some uniquely determined $\Phi_{{\mathrm{0}}}$ in $ \tceseq{ \Phi } $.
 In what follows, $\Phi_{\ottmv{j}}$ and $\mathit{Y_{\ottmv{j}}}$ are the $\ottmv{j}$-the elements of $ \tceseq{ \Phi } $, and
 that of $ \tceseq{ \mathit{Y} } $, and assume that $\sigma  \ottsym{(}  \mathit{Y_{\ottmv{j}}}  \ottsym{)}$ is the greatest fixpoint of $\Phi_{\ottmv{j}}$.

 Assume that $\Phi_{\ottmv{j}} \,  =  \,  (   \lambda_\mu\!   \,  \mathit{x}  .\,   \phi _{  \ottmv{j}   \ottsym{1}  }   ,   \lambda_\nu\!   \,  \mathit{y}  .\,   \phi _{  \ottmv{j}   \ottsym{2}  }   ) $
 for some $\mathit{x}$, $\mathit{y}$, $ \phi _{  \ottmv{j}   \ottsym{1}  } $, and $ \phi _{  \ottmv{j}   \ottsym{2}  } $.
 Then, it suffices to show that
 \[
    \forall  \, { \ottmv{j} } . \   { \models  \,    \sigma'  \ottsym{(}   \phi _{  \ottmv{j}   \ottsym{2}  }   \ottsym{)}    [  \pi  /  \mathit{y}  ]      [ \tceseq{ \ottnt{c_{{\mathrm{0}}}}  /  \mathit{z_{{\mathrm{0}}}} } ]   }   ~.
 \]
 By definition, it suffices to show that
 \[
    \forall  \, { \ottmv{j} } . \   { [\![    \sigma'  \ottsym{(}   \phi _{  \ottmv{j}   \ottsym{2}  }   \ottsym{)}    [  \pi  /  \mathit{y}  ]      [ \tceseq{ \ottnt{c_{{\mathrm{0}}}}  /  \mathit{z_{{\mathrm{0}}}} } ]    ]\!]_{  \emptyset  } }  \,  =  \,  \top   ~.
 \]
 By Lemmas~\ref{lem:ts-logic-val-subst} and \ref{lem:ts-logic-pred-subst},
 it suffices to show that
 \[
    \forall  \, { \ottmv{j} } . \   { [\![   \phi _{  \ottmv{j}   \ottsym{2}  }   ]\!]_{  \{ \tceseq{ \mathit{Y}  \mapsto   { [\![  \sigma'  \ottsym{(}  \mathit{Y}  \ottsym{)}  ]\!]_{  \emptyset  } }  } \}   \mathbin{\uplus}   \{  \mathit{y}  \mapsto   { [\![  \pi  ]\!]_{  \emptyset  } }   \}   \mathbin{\uplus}   \{  \mathit{z_{{\mathrm{0}}}}  \mapsto   { [\![  \ottnt{c_{{\mathrm{0}}}}  ]\!]_{  \emptyset  } }   \}  } }  \,  =  \,  \top   ~.
 \]
 Let $\mathbf{\mathit{F} } \,  =  \,   \lambda\!   \, \tceseq{ \mathit{Y} } .\,   \otimes {  \tceseq{ \mathbf{\mathit{f} } }  }  $
 where $ \otimes {  \tceseq{ \mathbf{\mathit{f} } }  } $ is the tuple with length of $ \tceseq{ \Phi } $ having as its $j$-th element
 the function
 \[
   \lambda\!  \, \mathit{y} \,  {:}  \,  \Sigma ^\omega   \ottsym{.}   \lambda\!  \,  \tceseq{ \mathit{z_{{\mathrm{0}}}}  \,  {:}  \,  \iota_{{\mathrm{0}}} }   \ottsym{.}   { [\![   \phi _{  \ottmv{j}   \ottsym{2}  }   ]\!]_{  \{ \tceseq{ \mathit{Y}  \mapsto  \mathit{Y} } \}   \mathbin{\uplus}   \{  \mathit{y}  \mapsto  \mathit{y}  \}   \mathbin{\uplus}   \{ \tceseq{ \mathit{z_{{\mathrm{0}}}}  \mapsto  \mathit{z_{{\mathrm{0}}}} } \}  } }  ~.
 \]
 By the assumption and \reflem{ts-valid-elim-intersect}, we have
 $  \forall  \, { \ottmv{j}  \ottsym{,}  \ottmv{i} } . \   { \models  \,   { \ottsym{(}   \sigma'_{\ottmv{i}}  \ottsym{(}  \Phi_{\ottmv{j}}  \ottsym{)}    [ \tceseq{ \ottnt{c_{{\mathrm{0}}}}  /  \mathit{z_{{\mathrm{0}}}} } ]    \ottsym{)} }^\nu   \ottsym{(}  \pi  \ottsym{)} }  $, that is,
 \[
    \forall  \, { \ottmv{j}  \ottsym{,}  \ottmv{i} } . \   { \models  \,    \sigma'_{\ottmv{i}}  \ottsym{(}   \phi _{  \ottmv{j}   \ottsym{2}  }   \ottsym{)}    [  \pi  /  \mathit{y}  ]      [ \tceseq{ \ottnt{c_{{\mathrm{0}}}}  /  \mathit{z_{{\mathrm{0}}}} } ]   }   ~,
 \]
 which implies that
 \[
    \forall  \, { \ottmv{i} } . \  \mathbf{\mathit{F} }  \ottsym{(}   \otimes {  \tceseq{  { [\![  \sigma'_{\ottmv{i}}  \ottsym{(}  \mathit{Y}  \ottsym{)}  ]\!]_{  \emptyset  } }  }  }   \ottsym{)}  \ottsym{(}   { [\![  \pi  ]\!]_{  \emptyset  } }   \ottsym{,}   \tceseq{  { [\![  \ottnt{c_{{\mathrm{0}}}}  ]\!]_{  \emptyset  } }  }   \ottsym{)} \,  =  \,  \otimes {  \tceseq{  \top  }  }  
 \]
 with Lemmas~\ref{lem:ts-logic-val-subst} and \ref{lem:ts-logic-pred-subst}.
 By \reflem{ts-logic-domain-lattice},
 \[
    \bigsqcap\nolimits_{  \otimes {  \tceseq{  \Sigma ^\omega   \rightarrow   \tceseq{ \iota_{{\mathrm{0}}} }   \rightarrow   \mathbb{B}  }  }  }     \{ \,  \mathbf{\mathit{F} }  \ottsym{(}   \otimes {  \tceseq{  { [\![  \sigma'_{\ottmv{i}}  \ottsym{(}  \mathit{Y}  \ottsym{)}  ]\!]_{  \emptyset  } }  }  }   \ottsym{)}  \mid  \ottmv{i}  \mathrel{  \in  } \mathbb{N} \, \}   \,  =  \,  \otimes {  \tceseq{ \mathbf{\mathit{f} }' }  } 
 \]
 where $ \otimes {  \tceseq{ \mathbf{\mathit{f} }' }  } $ is the tuple such that its $j$-th element $\mathbf{\mathit{f} }'_{\ottmv{j}}$ takes the form
 \[
   \lambda\!  \, \mathit{y} \,  {:}  \,  \Sigma ^\omega   \ottsym{.}   \lambda\!  \,  \tceseq{ \mathit{z_{{\mathrm{0}}}}  \,  {:}  \,  \iota_{{\mathrm{0}}} }   \ottsym{.}    \bigsqcap\nolimits_{  \mathbb{B}  }     \{ \,   { [\![   \phi _{  \ottmv{j}   \ottsym{2}  }   ]\!]_{  \{ \tceseq{ \mathit{Y}  \mapsto   { [\![  \sigma'_{\ottmv{i}}  \ottsym{(}  \mathit{Y}  \ottsym{)}  ]\!]_{  \emptyset  } }  } \}   \mathbin{\uplus}   \{  \mathit{y}  \mapsto  \mathit{y}  \}   \mathbin{\uplus}   \{ \tceseq{ \mathit{z_{{\mathrm{0}}}}  \mapsto  \mathit{z_{{\mathrm{0}}}} } \}  } }   \mid  \ottmv{i}  \mathrel{  \in  } \mathbb{N} \, \}   ~.
 \]
 Therefore,
 \[
    \forall  \, { \ottmv{j} } . \  \mathbf{\mathit{f} }'_{\ottmv{j}}  \ottsym{(}   { [\![  \pi  ]\!]_{  \emptyset  } }   \ottsym{,}   \tceseq{  { [\![  \ottnt{c_{{\mathrm{0}}}}  ]\!]_{  \emptyset  } }  }   \ottsym{)} \,  =  \,  \top   ~.
 \]
 By \refasm{cocontinuity-inf},
 \[
  \mathbf{\mathit{F} }  \ottsym{(}    \bigsqcap\nolimits_{  \otimes {  \tceseq{  \Sigma ^\omega   \rightarrow   \tceseq{ \iota_{{\mathrm{0}}} }   \rightarrow   \mathbb{B}  }  }  }     \{ \,   \otimes {  \tceseq{  { [\![  \sigma'_{\ottmv{i}}  \ottsym{(}  \mathit{Y}  \ottsym{)}  ]\!]_{  \emptyset  } }  }  }   \mid  \ottmv{i}  \mathrel{  \in  } \mathbb{N} \, \}    \ottsym{)} \,  =  \,  \otimes {  \tceseq{ \mathbf{\mathit{f} }' }  }  ~.
 \]
 By \reflem{ts-logic-domain-lattice},
 \[
    \bigsqcap\nolimits_{  \otimes {  \tceseq{  \Sigma ^\omega   \rightarrow   \tceseq{ \iota_{{\mathrm{0}}} }   \rightarrow   \mathbb{B}  }  }  }     \{ \,   \otimes {  \tceseq{  { [\![  \sigma'_{\ottmv{i}}  \ottsym{(}  \mathit{Y}  \ottsym{)}  ]\!]_{  \emptyset  } }  }  }   \mid  \ottmv{i}  \mathrel{  \in  } \mathbb{N} \, \}   \,  =  \,  \otimes {  \tceseq{   \bigsqcap\nolimits_{  \Sigma ^\omega   \rightarrow   \tceseq{ \iota_{{\mathrm{0}}} }   \rightarrow   \mathbb{B}  }     \{ \,   { [\![  \sigma'_{\ottmv{i}}  \ottsym{(}  \mathit{Y}  \ottsym{)}  ]\!]_{  \emptyset  } }   \mid  \ottmv{i}  \mathrel{  \in  } \mathbb{N} \, \}   }  } 
 \]
 Because $  \forall  \, { \ottmv{j} } . \  \mathbf{\mathit{f} }'_{\ottmv{j}}  \ottsym{(}   { [\![  \pi  ]\!]_{  \emptyset  } }   \ottsym{,}   \tceseq{  { [\![  \ottnt{c_{{\mathrm{0}}}}  ]\!]_{  \emptyset  } }  }   \ottsym{)} \,  =  \,  \top  $,
 the definition of $\mathbf{\mathit{F} }$ implies
 \[
   { [\![   \phi _{  \ottmv{j}   \ottsym{2}  }   ]\!]_{  \{ \tceseq{ \mathit{Y}  \mapsto  \mathbf{\mathit{f} }_{{\mathrm{0}}} } \}   \mathbin{\uplus}   \{  \mathit{y}  \mapsto   { [\![  \pi  ]\!]_{  \emptyset  } }   \}   \mathbin{\uplus}   \{ \tceseq{ \mathit{z_{{\mathrm{0}}}}  \mapsto   { [\![  \ottnt{c_{{\mathrm{0}}}}  ]\!]_{  \emptyset  } }  } \}  } }  \,  =  \,  \top 
 \]
 where $ \tceseq{ \mathbf{\mathit{f} }_{{\mathrm{0}}} \,  =  \,   \bigsqcap\nolimits_{  \Sigma ^\omega   \rightarrow   \tceseq{ \iota_{{\mathrm{0}}} }   \rightarrow   \mathbb{B}  }     \{ \,   { [\![  \sigma'_{\ottmv{i}}  \ottsym{(}  \mathit{Y}  \ottsym{)}  ]\!]_{  \emptyset  } }   \mid  \ottmv{i}  \mathrel{  \in  } \mathbb{N} \, \}   } $.
 Because $ \emptyset   \ottsym{;}   \tceseq{ \ottsym{(}  \mathit{X}  \ottsym{,}  \mathit{Y}  \ottsym{)} }  \,  \vdash ^{\circ}  \, \ottnt{C}$,
 we have $  \{   \tceseq{ \mathit{Y} }   \}   \, \ottsym{\#} \,   \mathit{fpv}^\mathbf{n}  (   \phi _{  \ottmv{j}   \ottsym{2}  }   )  $.
 Therefore, \reflem{ts-logic-weak-pred} implies that
 it suffices to show that
 \[
   \tceseq{ \mathbf{\mathit{f} }_{{\mathrm{0}}} \,  \sqsubseteq_{  \Sigma ^\omega   \rightarrow   \tceseq{ \iota_{{\mathrm{0}}} }   \rightarrow   \mathbb{B}  }  \,  { [\![  \sigma'  \ottsym{(}  \mathit{Y}  \ottsym{)}  ]\!]_{  \emptyset  } }  }  ~.
 \]

 Let
 \[
  \mathbf{\mathit{F} }_{{\mathrm{0}}} \,  =  \,   \lambda\!   \, \tceseq{ \mathit{Y} } .\,   \otimes {  \tceseq{  \lambda\!  \, \mathit{y} \,  {:}  \,  \Sigma ^\omega   \ottsym{.}   \lambda\!  \,  \tceseq{ \mathit{z_{{\mathrm{0}}}}  \,  {:}  \,  \iota_{{\mathrm{0}}} }   \ottsym{.}   { [\![    \phi _{  \ottmv{j}   \ottsym{2}  }     [ \tceseq{ \sigma'  \ottsym{(}  \mathit{X}  \ottsym{)}  /  \mathit{X} } ]    ]\!]_{  \{ \tceseq{ \mathit{Y}  \mapsto  \mathit{Y} } \}   \mathbin{\uplus}   \{  \mathit{y}  \mapsto  \mathit{y}  \}   \mathbin{\uplus}   \{ \tceseq{ \mathit{z_{{\mathrm{0}}}}  \mapsto  \mathit{z_{{\mathrm{0}}}} } \}  } }  }  }   ~.
 \]
 Because
 $ \otimes {  \tceseq{  { [\![  \sigma'  \ottsym{(}  \mathit{Y_{{\mathrm{0}}}}  \ottsym{)}  ]\!]_{  \emptyset  } }  }  }  \,  =  \,  \mathsf{gfp}^{  \otimes {  \tceseq{  \Sigma ^\omega   \rightarrow   \tceseq{ \iota_{{\mathrm{0}}} }   \rightarrow   \mathbb{B}  }  }  }   \ottsym{(}  \mathbf{\mathit{F} }_{{\mathrm{0}}}  \ottsym{)}$
 by definition,
 it suffices to show that
 \[
   \otimes {  \tceseq{ \mathbf{\mathit{f} }_{{\mathrm{0}}} }  }  \,  \sqsubseteq_{  \Sigma ^\omega   \rightarrow   \tceseq{ \iota_{{\mathrm{0}}} }   \rightarrow   \mathbb{B}  }  \, \mathbf{\mathit{F} }_{{\mathrm{0}}}  \ottsym{(}   \otimes {  \tceseq{ \mathbf{\mathit{f} }_{{\mathrm{0}}} }  }   \ottsym{)} = \mathbf{\mathit{F} }_{{\mathrm{0}}}  \ottsym{(}   \otimes {  \tceseq{   \bigsqcap\nolimits_{  \Sigma ^\omega   \rightarrow   \tceseq{ \iota_{{\mathrm{0}}} }   \rightarrow   \mathbb{B}  }     \{ \,   { [\![  \sigma'_{\ottmv{i}}  \ottsym{(}  \mathit{Y}  \ottsym{)}  ]\!]_{  \emptyset  } }   \mid  \ottmv{i}  \mathrel{  \in  } \mathbb{N} \, \}   }  }   \ottsym{)} ~.
 \]
 By \reflem{ts-logic-domain-lattice} and \refasm{cocontinuity-inf}, it suffices to show that
 \[
   \otimes {  \tceseq{ \mathbf{\mathit{f} }_{{\mathrm{0}}} }  }  \,  \sqsubseteq_{  \otimes {  \tceseq{  \Sigma ^\omega   \rightarrow   \tceseq{ \iota_{{\mathrm{0}}} }   \rightarrow   \mathbb{B}  }  }  }  \,   \bigsqcap\nolimits_{  \otimes {  \tceseq{  \Sigma ^\omega   \rightarrow   \tceseq{ \iota_{{\mathrm{0}}} }   \rightarrow   \mathbb{B}  }  }  }     \{ \,  \mathbf{\mathit{F} }_{{\mathrm{0}}}  \ottsym{(}   \otimes {  \tceseq{  { [\![  \sigma'_{\ottmv{i}}  \ottsym{(}  \mathit{Y}  \ottsym{)}  ]\!]_{  \emptyset  } }  }  }   \ottsym{)}  \mid  \ottmv{i}  \mathrel{  \in  } \mathbb{N} \, \}   ~.
 \]
 Because $  \forall  \, { \ottmv{i} } . \  \mathbf{\mathit{F} }_{{\mathrm{0}}}  \ottsym{(}   \otimes {  \tceseq{  { [\![  \sigma'_{\ottmv{i}}  \ottsym{(}  \mathit{Y}  \ottsym{)}  ]\!]_{  \emptyset  } }  }  }   \ottsym{)} \,  =  \,  \otimes {  \tceseq{  { [\![   { \sigma }_{ \ottmv{i}  \ottsym{+}  \ottsym{1} }'   \ottsym{(}  \mathit{Y}  \ottsym{)}  ]\!]_{  \emptyset  } }  }  }  $,
 we have
 \[
   \otimes {  \tceseq{ \mathbf{\mathit{f} }_{{\mathrm{0}}} }  }  \,  =  \,   \otimes {  \tceseq{ \ottsym{(}   \lambda\!  \, \mathit{y} \,  {:}  \,  \Sigma ^\omega   \ottsym{.}   \lambda\!  \,  \tceseq{ \mathit{z_{{\mathrm{0}}}}  \,  {:}  \,  \iota_{{\mathrm{0}}} }   \ottsym{.}   \top   \ottsym{)} }  }   \mathbin{\wedge_{  \otimes {  \tceseq{  \Sigma ^\omega   \rightarrow   \tceseq{ \iota_{{\mathrm{0}}} }   \rightarrow   \mathbb{B}  }  }  } }  \ottsym{(}    \bigsqcap\nolimits_{  \otimes {  \tceseq{  \Sigma ^\omega   \rightarrow   \tceseq{ \iota_{{\mathrm{0}}} }   \rightarrow   \mathbb{B}  }  }  }     \{ \,  \mathbf{\mathit{F} }_{{\mathrm{0}}}  \ottsym{(}   \otimes {  \tceseq{  { [\![  \sigma'_{\ottmv{i}}  \ottsym{(}  \mathit{Y}  \ottsym{)}  ]\!]_{  \emptyset  } }  }  }   \ottsym{)}  \mid  \ottmv{i}  \mathrel{  \in  } \mathbb{N} \, \}    \ottsym{)}  ~.
 \]
 Because $ \lambda\!  \, \mathit{y} \,  {:}  \,  \Sigma ^\omega   \ottsym{.}   \lambda\!  \,  \tceseq{ \mathit{z_{{\mathrm{0}}}}  \,  {:}  \,  \iota_{{\mathrm{0}}} }   \ottsym{.}   \top $ is the greatest element in $ \mathcal{D}_{  \Sigma ^\omega   \rightarrow   \tceseq{ \iota_{{\mathrm{0}}} }   \rightarrow   \mathbb{B}  } $,
 \[
   \otimes {  \tceseq{ \mathbf{\mathit{f} }_{{\mathrm{0}}} }  }  =   \bigsqcap\nolimits_{  \otimes {  \tceseq{  \Sigma ^\omega   \rightarrow   \tceseq{ \iota_{{\mathrm{0}}} }   \rightarrow   \mathbb{B}  }  }  }     \{ \,  \mathbf{\mathit{F} }_{{\mathrm{0}}}  \ottsym{(}   \otimes {  \tceseq{  { [\![  \sigma'_{\ottmv{i}}  \ottsym{(}  \mathit{Y}  \ottsym{)}  ]\!]_{  \emptyset  } }  }  }   \ottsym{)}  \mid  \ottmv{i}  \mathrel{  \in  } \mathbb{N} \, \}   ~.
 \]
 Therefore, we have the conclusion.
\end{proof}

\begin{lemmap}{Typing Evaluation Contexts is Independent of Predicates}{lr-spos-ectx-XXX}
 If $ \tceseq{ \ottsym{(}  \mathit{X'}  \ottsym{,}  \mathit{Y'}  \ottsym{)} }   \ottsym{;}   \tceseq{ \ottsym{(}  \mathit{X}  \ottsym{,}  \mathit{Y}  \ottsym{)} }  \,  \vdash ^{\circ}  \, \ottnt{C}$ and $\ottsym{(}  \ottnt{E}  \ottsym{,}  \sigma  \mathbin{\uplus}  \sigma'  \ottsym{)}  \ottsym{:}  \ottnt{C}  \Rrightarrow  \ottnt{C'}$ and $\mathit{dom} \, \ottsym{(}  \sigma'  \ottsym{)} \,  =  \,  \{   \tceseq{ \mathit{X} }   \ottsym{,}   \tceseq{ \mathit{X'} }   \ottsym{,}   \tceseq{ \mathit{Y} }   \ottsym{,}   \tceseq{ \mathit{Y'} }   \} $,
 then
 \begin{enumerate}
  \item $ \tceseq{ \ottsym{(}  \mathit{X}  \ottsym{,}  \mathit{Y}  \ottsym{)} }  \,  =  \,  \tceseq{ \ottsym{(}  \mathit{X_{{\mathrm{1}}}}  \ottsym{,}  \mathit{Y_{{\mathrm{2}}}}  \ottsym{)} }   \ottsym{,}   \tceseq{ \ottsym{(}  \mathit{X_{{\mathrm{2}}}}  \ottsym{,}  \mathit{Y_{{\mathrm{2}}}}  \ottsym{)} } $,
  \item $ \tceseq{ \ottsym{(}  \mathit{X'}  \ottsym{,}  \mathit{Y'}  \ottsym{)} }   \ottsym{,}   \tceseq{ \ottsym{(}  \mathit{X_{{\mathrm{1}}}}  \ottsym{,}  \mathit{Y_{{\mathrm{2}}}}  \ottsym{)} }   \ottsym{;}   \tceseq{ \ottsym{(}  \mathit{X_{{\mathrm{2}}}}  \ottsym{,}  \mathit{Y_{{\mathrm{2}}}}  \ottsym{)} }  \,  \vdash ^{\circ}  \, \ottnt{C'}$, and
  \item $   \forall  \, { \sigma'' } . \  \mathit{dom} \, \ottsym{(}  \sigma''  \ottsym{)} \,  =  \,  \{   \tceseq{ \mathit{X} }   \ottsym{,}   \tceseq{ \mathit{X'} }   \ottsym{,}   \tceseq{ \mathit{Y} }   \ottsym{,}   \tceseq{ \mathit{Y'} }   \}    \mathrel{\Longrightarrow}  \ottsym{(}  \ottnt{E}  \ottsym{,}  \sigma  \mathbin{\uplus}  \sigma''  \ottsym{)}  \ottsym{:}  \ottnt{C}  \Rrightarrow  \ottnt{C'} $.
 \end{enumerate}
 for some $ \tceseq{ \mathit{X_{{\mathrm{1}}}} } $, $ \tceseq{ \mathit{X_{{\mathrm{2}}}} } $, $ \tceseq{ \mathit{Y_{{\mathrm{1}}}} } $, and $ \tceseq{ \mathit{Y_{{\mathrm{2}}}} } $.
\end{lemmap}
\begin{proof}
 \TS{OK: 2022/01/27}
 Take any $\sigma'$ and $\sigma''$ such that
 $\mathit{dom} \, \ottsym{(}  \sigma'  \ottsym{)} = \mathit{dom} \, \ottsym{(}  \sigma''  \ottsym{)} =  \{   \tceseq{ \mathit{X} }   \ottsym{,}   \tceseq{ \mathit{X'} }   \ottsym{,}   \tceseq{ \mathit{Y} }   \ottsym{,}   \tceseq{ \mathit{Y'} }   \} $.
 By induction on the derivation of $\ottsym{(}  \ottnt{E}  \ottsym{,}  \sigma  \mathbin{\uplus}  \sigma'  \ottsym{)}  \ottsym{:}  \ottnt{C}  \Rrightarrow  \ottnt{C'}$.
 \begin{caseanalysis}
  \case \TC{Empty}: Trivial.
  \case \TC{Cons}: We are given
   \begin{prooftree}
    \AxiomC{$\ottsym{(}  \ottnt{E'}  \ottsym{,}  \sigma  \mathbin{\uplus}  \sigma'  \ottsym{)}  \ottsym{:}  \ottnt{C}  \Rrightarrow   \ottnt{T}  \,  \,\&\,  \,  \Phi  \, / \,   ( \forall  \mathit{x}  .  \ottnt{C_{{\mathrm{1}}}}  )  \Rightarrow   \ottnt{C'}  $}
    \AxiomC{$\ottnt{K} \,  \in  \,  \LRRel{ \mathcal{K} }{ \ottsym{(}  \sigma  \mathbin{\uplus}  \sigma'  \ottsym{)}  \ottsym{(}   \ottnt{T}  \, / \, ( \forall  \mathit{x}  .  \ottnt{C_{{\mathrm{1}}}}  )   \ottsym{)} } $}
    \BinaryInfC{$\ottsym{(}   \resetcnstr{  \ottnt{K}  [  \ottnt{E'}  ]  }   \ottsym{,}  \sigma  \mathbin{\uplus}  \sigma'  \ottsym{)}  \ottsym{:}  \ottnt{C}  \Rrightarrow  \ottnt{C'}$}
   \end{prooftree}
   for some $\ottnt{K}$, $\ottnt{E'}$, $\ottnt{T}$, $\Phi$, $\mathit{x}$, and $\ottnt{C_{{\mathrm{1}}}}$
   such that $\ottnt{E} \,  =  \,  \resetcnstr{  \ottnt{K}  [  \ottnt{E'}  ]  } $.
   By the IH,
   \begin{itemize}
    \item $ \tceseq{ \ottsym{(}  \mathit{X}  \ottsym{,}  \mathit{Y}  \ottsym{)} }  \,  =  \,  \tceseq{ \ottsym{(}  \mathit{X_{{\mathrm{1}}}}  \ottsym{,}  \mathit{Y_{{\mathrm{1}}}}  \ottsym{)} }   \ottsym{,}   \tceseq{ \ottsym{(}  \mathit{X_{{\mathrm{2}}}}  \ottsym{,}  \mathit{Y_{{\mathrm{2}}}}  \ottsym{)} } $,
    \item $ \tceseq{ \ottsym{(}  \mathit{X'}  \ottsym{,}  \mathit{Y'}  \ottsym{)} }   \ottsym{,}   \tceseq{ \ottsym{(}  \mathit{X_{{\mathrm{1}}}}  \ottsym{,}  \mathit{Y_{{\mathrm{1}}}}  \ottsym{)} }   \ottsym{;}   \tceseq{ \ottsym{(}  \mathit{X_{{\mathrm{2}}}}  \ottsym{,}  \mathit{Y_{{\mathrm{2}}}}  \ottsym{)} }  \,  \vdash ^{\circ}  \,  \ottnt{T}  \,  \,\&\,  \,  \Phi  \, / \,   ( \forall  \mathit{x}  .  \ottnt{C_{{\mathrm{1}}}}  )  \Rightarrow   \ottnt{C'}  $, and
    \item $\ottsym{(}  \ottnt{E'}  \ottsym{,}  \sigma  \mathbin{\uplus}  \sigma''  \ottsym{)}  \ottsym{:}  \ottnt{C}  \Rrightarrow   \ottnt{T}  \,  \,\&\,  \,  \Phi  \, / \,   ( \forall  \mathit{x}  .  \ottnt{C_{{\mathrm{1}}}}  )  \Rightarrow   \ottnt{C'}  $
   \end{itemize}
   for some $ \tceseq{ \mathit{X_{{\mathrm{1}}}} } $, $ \tceseq{ \mathit{X_{{\mathrm{2}}}} } $, $ \tceseq{ \mathit{Y_{{\mathrm{1}}}} } $, and $ \tceseq{ \mathit{Y_{{\mathrm{2}}}} } $.
   Because $ \tceseq{ \ottsym{(}  \mathit{X'}  \ottsym{,}  \mathit{Y'}  \ottsym{)} }   \ottsym{,}   \tceseq{ \ottsym{(}  \mathit{X_{{\mathrm{1}}}}  \ottsym{,}  \mathit{Y_{{\mathrm{1}}}}  \ottsym{)} }   \ottsym{;}   \tceseq{ \ottsym{(}  \mathit{X_{{\mathrm{2}}}}  \ottsym{,}  \mathit{Y_{{\mathrm{2}}}}  \ottsym{)} }  \,  \vdash ^{\circ}  \,  \ottnt{T}  \,  \,\&\,  \,  \Phi  \, / \,   ( \forall  \mathit{x}  .  \ottnt{C_{{\mathrm{1}}}}  )  \Rightarrow   \ottnt{C'}  $,
   we have
   \begin{itemize}
    \item $  \{   \tceseq{ \mathit{X'} }   \ottsym{,}   \tceseq{ \mathit{Y'} }   \ottsym{,}   \tceseq{ \mathit{X} }   \ottsym{,}   \tceseq{ \mathit{Y} }   \}   \, \ottsym{\#} \,  \ottsym{(}   \mathit{fpv}  (  \ottnt{T}  )  \,  \mathbin{\cup}  \,  \mathit{fpv}  (  \ottnt{C_{{\mathrm{1}}}}  )   \ottsym{)} $,
    \item $ \tceseq{ \ottsym{(}  \mathit{X_{{\mathrm{2}}}}  \ottsym{,}  \mathit{Y_{{\mathrm{2}}}}  \ottsym{)} }  \,  =  \, \ottsym{(}  \mathit{X_{{\mathrm{21}}}}  \ottsym{,}  \mathit{Y_{{\mathrm{21}}}}  \ottsym{)}  \ottsym{,}   \tceseq{ \ottsym{(}  \mathit{X_{{\mathrm{22}}}}  \ottsym{,}  \mathit{Y_{{\mathrm{22}}}}  \ottsym{)} } $, and
    \item $ \tceseq{ \ottsym{(}  \mathit{X'}  \ottsym{,}  \mathit{Y'}  \ottsym{)} }   \ottsym{,}   \tceseq{ \ottsym{(}  \mathit{X_{{\mathrm{1}}}}  \ottsym{,}  \mathit{Y_{{\mathrm{1}}}}  \ottsym{)} }   \ottsym{,}  \ottsym{(}  \mathit{X_{{\mathrm{21}}}}  \ottsym{,}  \mathit{Y_{{\mathrm{21}}}}  \ottsym{)}  \ottsym{;}   \tceseq{ \ottsym{(}  \mathit{X_{{\mathrm{22}}}}  \ottsym{,}  \mathit{Y_{{\mathrm{22}}}}  \ottsym{)} }  \,  \vdash ^{\circ}  \, \ottnt{C'}$
   \end{itemize}
   for some $\mathit{X_{{\mathrm{21}}}}$, $\mathit{Y_{{\mathrm{21}}}}$, $ \tceseq{ \mathit{X_{{\mathrm{22}}}} } $, and $ \tceseq{ \mathit{Y_{{\mathrm{22}}}} } $.
   Therefore,
   \[
    \ottsym{(}  \sigma  \mathbin{\uplus}  \sigma'  \ottsym{)}  \ottsym{(}   \ottnt{T}  \, / \, ( \forall  \mathit{x}  .  \ottnt{C_{{\mathrm{1}}}}  )   \ottsym{)} = \sigma  \ottsym{(}   \ottnt{T}  \, / \, ( \forall  \mathit{x}  .  \ottnt{C_{{\mathrm{1}}}}  )   \ottsym{)} = \ottsym{(}  \sigma  \mathbin{\uplus}  \sigma''  \ottsym{)}  \ottsym{(}   \ottnt{T}  \, / \, ( \forall  \mathit{x}  .  \ottnt{C_{{\mathrm{1}}}}  )   \ottsym{)} ~.
   \]
   Hence, $\ottnt{K} \,  \in  \,  \LRRel{ \mathcal{K} }{ \ottsym{(}  \sigma  \mathbin{\uplus}  \sigma''  \ottsym{)}  \ottsym{(}   \ottnt{T}  \, / \, ( \forall  \mathit{x}  .  \ottnt{C_{{\mathrm{1}}}}  )   \ottsym{)} } $.
   Therefore, by \TC{Cons}, we have the conclusion
   $\ottsym{(}   \resetcnstr{  \ottnt{K}  [  \ottnt{E'}  ]  }   \ottsym{,}  \sigma  \mathbin{\uplus}  \sigma''  \ottsym{)}  \ottsym{:}  \ottnt{C}  \Rrightarrow  \ottnt{C'}$.
 \end{caseanalysis}
\end{proof}

\begin{lemmap}{Relating Consistency and Occurrence Checking}{lr-rel-consistency-occur}
 If $ \tceseq{ \ottsym{(}  \mathit{X'}  \ottsym{,}  \mathit{Y'}  \ottsym{)} }   \ottsym{;}   \tceseq{ \ottsym{(}  \mathit{X}  \ottsym{,}  \mathit{Y}  \ottsym{)} }  \,  \vdash ^{\circ}  \, \ottnt{C}$ and $ \tceseq{ \ottsym{(}  \mathit{X}  \ottsym{,}  \mathit{Y}  \ottsym{)} }   \ottsym{;}   \tceseq{ \mathit{z_{{\mathrm{0}}}}  \,  {:}  \,  \iota_{{\mathrm{0}}} }   \vdash  \ottnt{C_{{\mathrm{0}}}}  \succ  \ottnt{C}$,
 then $ \tceseq{ \ottsym{(}  \mathit{X'}  \ottsym{,}  \mathit{Y'}  \ottsym{)} }   \ottsym{;}   \tceseq{ \ottsym{(}  \mathit{X}  \ottsym{,}  \mathit{Y}  \ottsym{)} }  \,  \vdash ^{\circ}  \, \ottnt{C_{{\mathrm{0}}}}$.
\end{lemmap}
\begin{proof}
 Straightforward by induction on the derivation of $ \tceseq{ \ottsym{(}  \mathit{X}  \ottsym{,}  \mathit{Y}  \ottsym{)} }   \ottsym{;}   \tceseq{ \mathit{z_{{\mathrm{0}}}}  \,  {:}  \,  \iota_{{\mathrm{0}}} }   \vdash  \ottnt{C_{{\mathrm{0}}}}  \succ  \ottnt{C}$.
\end{proof}

\begin{lemmap}{Relating Application of Recursive Functions}{lr-applied-rec-XXX}
 Assume that
 \begin{itemize}
  \item $\sigma_{{\mathrm{0}}} \,  \in  \,  \LRRel{ \mathcal{G} }{ \Gamma } $,
  \item $\ottnt{v} \,  =  \,  \mathsf{rec}  \, (  \tceseq{  \mathit{X} ^{ \sigma  \ottsym{(}  \mathit{X}  \ottsym{)} },  \mathit{Y} ^{ \sigma  \ottsym{(}  \mathit{Y}  \ottsym{)} }  }  ,  \mathit{f} ,   \tceseq{ \mathit{z} }  ) . \,  \ottnt{e} $,
  \item $ \tceseq{ \mathit{z_{{\mathrm{0}}}}  \,  {:}  \,  \iota_{{\mathrm{0}}} }  \,  =  \,  \gamma  (   \tceseq{ \mathit{z}  \,  {:}  \,  \ottnt{T_{{\mathrm{1}}}} }   ) $,
  \item $\Gamma' \,  =  \,  \tceseq{ \mathit{X} }  \,  {:}  \,  (   \Sigma ^\ast   ,   \tceseq{ \iota_{{\mathrm{0}}} }   )   \ottsym{,}   \tceseq{ \mathit{Y} }  \,  {:}  \,  (   \Sigma ^\omega   ,   \tceseq{ \iota_{{\mathrm{0}}} }   )   \ottsym{,}  \mathit{f} \,  {:}  \, \ottsym{(}   \tceseq{ \mathit{z}  \,  {:}  \,  \ottnt{T_{{\mathrm{1}}}} }   \ottsym{)}  \rightarrow  \ottnt{C_{{\mathrm{0}}}}  \ottsym{,}   \tceseq{ \mathit{z}  \,  {:}  \,  \ottnt{T_{{\mathrm{1}}}} } $, and
  \item $\ottnt{e} \,  \in  \,  \LRRel{ \mathcal{O} }{ \Gamma  \ottsym{,}  \Gamma'  \, \vdash \,  \ottnt{C} } $,
  \item $ \tceseq{ \ottsym{(}  \mathit{X}  \ottsym{,}  \mathit{Y}  \ottsym{)} }   \ottsym{;}   \tceseq{ \mathit{z_{{\mathrm{0}}}}  \,  {:}  \,  \iota_{{\mathrm{0}}} }   \vdash  \ottnt{C_{{\mathrm{0}}}}  \stackrel{\succ}{\circ}  \ottnt{C}$,
  \item $  \{   \tceseq{ \mathit{X} }   \ottsym{,}   \tceseq{ \mathit{Y} }   \}   \, \ottsym{\#} \,   \mathit{fpv}  (   \tceseq{ \ottnt{T_{{\mathrm{1}}}} }   )  $,
  \item $ \tceseq{ \ottsym{(}  \mathit{X}  \ottsym{,}  \mathit{Y}  \ottsym{)} }   \ottsym{;}   \tceseq{ \mathit{z_{{\mathrm{0}}}}  \,  {:}  \,  \iota_{{\mathrm{0}}} }  \,  |  \, \ottnt{C}  \vdash  \sigma$,
  \item $ \tceseq{ \ottnt{v'} } $ be values such that
        $ \tceseq{ \ottnt{v'} \,  \in  \,  \LRRel{ \mathcal{V} }{  \sigma_{{\mathrm{0}}}  \ottsym{(}  \ottnt{T_{{\mathrm{1}}}}  \ottsym{)}    [ \tceseq{ \ottnt{v'}  /  \mathit{z} } ]   }  } $,
  \item $\sigma'_{{\mathrm{0}}} \,  =  \,  [ \tceseq{ \ottnt{v'}  /  \mathit{z} } ] $,
  \item $\sigma' \,  =  \, \sigma_{{\mathrm{0}}}  \mathbin{\uplus}  \sigma_{{\mathrm{0}}}  \ottsym{(}  \sigma  \ottsym{)}  \mathbin{\uplus}   [ \tceseq{ \ottnt{v'}  /  \mathit{z} } ] $.
 \end{itemize}
 Then,
 $\ottsym{(}  \ottnt{E}  \ottsym{,}  \sigma'  \ottsym{)}  \ottsym{:}  \ottnt{C_{{\mathrm{0}}}}  \Rrightarrow   \ottnt{C} _{ \ottnt{E} } $ implies $ \ottnt{E}  [  \sigma_{{\mathrm{0}}}  \ottsym{(}  \ottnt{v}  \ottsym{)} \,  \tceseq{ \ottnt{v'} }   ]  \,  \in  \,  \LRRel{ \mathcal{E} }{ \sigma'  \ottsym{(}   \ottnt{C} _{ \ottnt{E} }   \ottsym{)} } $.
\end{lemmap}
\begin{proof}
 By induction on $ \ottnt{C} _{ \ottnt{E} } $.
 It is easy to show $ \ottnt{E}  [  \sigma_{{\mathrm{0}}}  \ottsym{(}  \ottnt{v}  \ottsym{)} \,  \tceseq{ \ottnt{v'} }   ]  \,  \in  \,  \LRRel{ \mathrm{Atom} }{ \sigma'  \ottsym{(}   \ottnt{C} _{ \ottnt{E} }   \ottsym{)} } $ by induction on
 the derivation of $\ottnt{E}$ with \reflem{lr-typing-cont-exp}.
 We consider each case on the result of evaluating $ \ottnt{E}  [  \sigma_{{\mathrm{0}}}  \ottsym{(}  \ottnt{v}  \ottsym{)} \,  \tceseq{ \ottnt{v'} }   ] $.
 \begin{caseanalysis}
  \case $  \exists  \, { \ottnt{v_{{\mathrm{0}}}}  \ottsym{,}  \varpi_{{\mathrm{0}}} } . \   \ottnt{E}  [  \sigma_{{\mathrm{0}}}  \ottsym{(}  \ottnt{v}  \ottsym{)} \,  \tceseq{ \ottnt{v'} }   ]  \,  \Downarrow  \, \ottnt{v_{{\mathrm{0}}}} \,  \,\&\,  \, \varpi_{{\mathrm{0}}} $:
   Let $\pi_{{\mathrm{0}}}$ be an arbitrary infinite trace.
   Because, by \reflem{lr-approx-refl},
   \begin{itemize}
    \item $  \forall  \, { \ottmv{i}  \ottsym{,}  \pi } . \  \ottnt{E} \,  =  \, \ottnt{E} \,  \prec_{ \sigma_{{\mathrm{0}}}  \ottsym{(}  \ottnt{v}  \ottsym{)} }  \, \ottnt{E} \,  \uparrow \, (  \ottmv{i}  ,  \pi  )  $, and
    \item $  \forall  \, { \ottmv{i}  \ottsym{,}  \pi } . \   \sigma_{{\mathrm{0}}}  \ottsym{(}  \ottnt{v}  \ottsym{)} _{  \mu   \ottsym{;}  \ottmv{i}  \ottsym{;}  \pi }  \,  \tceseq{ \ottnt{v'} }  \,  =  \, \ottsym{(}  \sigma_{{\mathrm{0}}}  \ottsym{(}  \ottnt{v}  \ottsym{)} \,  \tceseq{ \ottnt{v'} }   \ottsym{)} \,  \prec_{ \sigma_{{\mathrm{0}}}  \ottsym{(}  \ottnt{v}  \ottsym{)} }  \, \ottsym{(}   \sigma_{{\mathrm{0}}}  \ottsym{(}  \ottnt{v}  \ottsym{)} _{  \mu   \ottsym{;}  \ottsym{0}  \ottsym{;}  \pi_{{\mathrm{0}}} }  \,  \tceseq{ \ottnt{v'} }   \ottsym{)} \,  \uparrow \, (  \ottmv{i}  ,  \pi  )  $,
   \end{itemize}
   we have
   \[\begin{array}{rll}
    \forall \ottmv{i}, \pi.\,
      \ottnt{E}  [   \sigma_{{\mathrm{0}}}  \ottsym{(}  \ottnt{v}  \ottsym{)} _{  \mu   \ottsym{;}  \ottmv{i}  \ottsym{;}  \pi }  \,  \tceseq{ \ottnt{v'} }   ]  &=&
      \ottsym{(}  \ottnt{E} \,  \prec_{ \sigma_{{\mathrm{0}}}  \ottsym{(}  \ottnt{v}  \ottsym{)} }  \, \ottnt{E} \,  \uparrow \, (  \ottmv{i}  ,  \pi  )   \ottsym{)}  [  \ottsym{(}  \sigma_{{\mathrm{0}}}  \ottsym{(}  \ottnt{v}  \ottsym{)} \,  \tceseq{ \ottnt{v'} }   \ottsym{)} \,  \prec_{ \sigma_{{\mathrm{0}}}  \ottsym{(}  \ottnt{v}  \ottsym{)} }  \, \ottsym{(}   \sigma_{{\mathrm{0}}}  \ottsym{(}  \ottnt{v}  \ottsym{)} _{  \mu   \ottsym{;}  \ottsym{0}  \ottsym{;}  \pi_{{\mathrm{0}}} }  \,  \tceseq{ \ottnt{v'} }   \ottsym{)} \,  \uparrow \, (  \ottmv{i}  ,  \pi  )   ]  \\ &=&
      \ottnt{E}  [  \sigma_{{\mathrm{0}}}  \ottsym{(}  \ottnt{v}  \ottsym{)} \,  \tceseq{ \ottnt{v'} }   ]  \,  \prec_{ \sigma_{{\mathrm{0}}}  \ottsym{(}  \ottnt{v}  \ottsym{)} }  \,  \ottnt{E}  [   \sigma_{{\mathrm{0}}}  \ottsym{(}  \ottnt{v}  \ottsym{)} _{  \mu   \ottsym{;}  \ottsym{0}  \ottsym{;}  \pi_{{\mathrm{0}}} }  \,  \tceseq{ \ottnt{v'} }   ]  \,  \uparrow \, (  \ottmv{i}  ,  \pi  ) 
      \quad \text{(by \reflem{lr-approx-embed-exp})} ~.
     \end{array}
   \]
   Because $ \ottnt{E}  [  \sigma_{{\mathrm{0}}}  \ottsym{(}  \ottnt{v}  \ottsym{)} \,  \tceseq{ \ottnt{v'} }   ]  \,  \Downarrow  \, \ottnt{v_{{\mathrm{0}}}} \,  \,\&\,  \, \varpi_{{\mathrm{0}}}$,
   Lemmas~\ref{lem:lr-approx-eval-nsteps} and \ref{lem:lr-approx-val} imply that
   there exist some $\ottmv{j}$ and $\ottnt{v'_{{\mathrm{0}}}}$ such that
   \begin{equation}
      \forall  \, { \pi } . \    \forall  \, { \ottmv{i} \,  \geq  \, \ottmv{j} } . \   \ottnt{E}  [   \sigma_{{\mathrm{0}}}  \ottsym{(}  \ottnt{v}  \ottsym{)} _{  \mu   \ottsym{;}  \ottmv{i}  \ottsym{;}  \pi }  \,  \tceseq{ \ottnt{v'} }   ]  \,  \Downarrow  \, \ottsym{(}  \ottnt{v_{{\mathrm{0}}}} \,  \prec_{ \sigma_{{\mathrm{0}}}  \ottsym{(}  \ottnt{v}  \ottsym{)} }  \, \ottnt{v'_{{\mathrm{0}}}} \,  \uparrow \, (  \ottmv{i}  \ottsym{-}  \ottmv{j}  ,  \pi  )   \ottsym{)} \,  \,\&\,  \, \varpi_{{\mathrm{0}}}   ~.
     \label{lem:lr-applied-rec:TM:approx-eval}
   \end{equation}

   Take
   $\sigma_{\ottmv{i}}$ and $\sigma'_{\ottmv{i}}$ for every $\ottmv{i}$ such that
   \begin{itemize}
    \item $\ottnt{C}  \triangleright   \emptyset   \ottsym{;}   \tceseq{ \ottsym{(}  \mathit{X}  \ottsym{,}  \mathit{Y}  \ottsym{)} }   \ottsym{;}   \tceseq{ \mathit{z_{{\mathrm{0}}}}  \,  {:}  \,  \iota_{{\mathrm{0}}} }  \,  |  \, \ottnt{C} \,  \vdash _{ \ottmv{i} }  \, \sigma_{\ottmv{i}}$ and
    \item $\sigma'_{\ottmv{i}} \,  =  \, \sigma_{{\mathrm{0}}}  \mathbin{\uplus}   [ \tceseq{ \sigma_{{\mathrm{0}}}  \ottsym{(}  \sigma_{\ottmv{i}}  \ottsym{(}  \mathit{X}  \ottsym{)}  \ottsym{)}  /  \mathit{X} } ]   \mathbin{\uplus}   [ \tceseq{ \ottsym{(}   \nu\!  \, \mathit{Y}  \ottsym{(}  \mathit{y} \,  {:}  \,  \Sigma ^\omega   \ottsym{,}   \tceseq{ \mathit{z_{{\mathrm{0}}}}  \,  {:}  \,  \iota_{{\mathrm{0}}} }   \ottsym{)}  \ottsym{.}   \top   \ottsym{)}  /  \mathit{Y} } ]   \mathbin{\uplus}   [ \tceseq{ \ottnt{v'}  /  \mathit{z} } ] $.
   \end{itemize}
   We show that
   \begin{equation}
      \forall  \, { \ottmv{i}  \ottsym{,}  \pi } . \   \ottnt{E}  [   \sigma_{{\mathrm{0}}}  \ottsym{(}  \ottnt{v}  \ottsym{)} _{  \mu   \ottsym{;}  \ottmv{i}  \ottsym{;}  \pi }  \,  \tceseq{ \ottnt{v'} }   ]  \,  \in  \,  \LRRel{ \mathcal{E} }{ \sigma'_{\ottmv{i}}  \ottsym{(}   \ottnt{C} _{ \ottnt{E} }   \ottsym{)} }   ~.
     \label{lem:lr-applied-rec:TM:approxl-rel}
   \end{equation}
   Using \reflem{lr-mu-sim-XXX}, we have
   \[
      \forall  \, { \ottmv{i}  \ottsym{,}  \pi } . \   \sigma_{{\mathrm{0}}}  \ottsym{(}  \ottnt{v}  \ottsym{)} _{  \mu   \ottsym{;}  \ottmv{i}  \ottsym{;}  \pi }  \,  \tceseq{ \ottnt{v'} }  \,  \in  \,  \LRRel{ \mathcal{E} }{ \sigma'_{\ottmv{i}}  \ottsym{(}  \ottnt{C_{{\mathrm{0}}}}  \ottsym{)} }   ~.
   \]
   \reflem{lr-rel-consistency-occur} with $ \tceseq{ \ottsym{(}  \mathit{X}  \ottsym{,}  \mathit{Y}  \ottsym{)} }   \ottsym{;}   \tceseq{ \mathit{z_{{\mathrm{0}}}}  \,  {:}  \,  \iota_{{\mathrm{0}}} }   \vdash  \ottnt{C_{{\mathrm{0}}}}  \stackrel{\succ}{\circ}  \ottnt{C}$ implies $ \emptyset   \ottsym{;}   \tceseq{ \ottsym{(}  \mathit{X}  \ottsym{,}  \mathit{Y}  \ottsym{)} }  \,  \vdash ^{\circ}  \, \ottnt{C_{{\mathrm{0}}}}$.
   Then, by \reflem{lr-spos-ectx-XXX} with $\ottsym{(}  \ottnt{E}  \ottsym{,}  \sigma'  \ottsym{)}  \ottsym{:}  \ottnt{C_{{\mathrm{0}}}}  \Rrightarrow   \ottnt{C} _{ \ottnt{E} } $,
   we have
   \[
      \forall  \, { \ottmv{i} } . \  \ottsym{(}  \ottnt{E}  \ottsym{,}  \sigma'_{\ottmv{i}}  \ottsym{)}  \ottsym{:}  \ottnt{C_{{\mathrm{0}}}}  \Rrightarrow   \ottnt{C} _{ \ottnt{E} }   ~.
   \]
   Therefore, we have
   (\ref{lem:lr-applied-rec:TM:approxl-rel}) by \reflem{lr-closed-fill-ectx}.

   By (\ref{lem:lr-applied-rec:TM:approx-eval}) and (\ref{lem:lr-applied-rec:TM:approxl-rel}),
   \[
      \forall  \, { \pi } . \    \forall  \, { \ottmv{i} \,  \geq  \, \ottmv{j} } . \  \ottsym{(}  \ottnt{v_{{\mathrm{0}}}} \,  \prec_{ \sigma_{{\mathrm{0}}}  \ottsym{(}  \ottnt{v}  \ottsym{)} }  \, \ottnt{v'_{{\mathrm{0}}}} \,  \uparrow \, (  \ottmv{i}  \ottsym{-}  \ottmv{j}  ,  \pi  )   \ottsym{)} \,  \in  \,  \LRRel{ \mathcal{V} }{ \sigma'_{\ottmv{i}}  \ottsym{(}    \ottnt{C} _{ \ottnt{E} }   . \ottnt{T}   \ottsym{)} }    ~.
   \]
   Hence,
   \[
      \forall  \, { \pi  \ottsym{,}  \ottmv{i} } . \  \ottsym{(}  \ottnt{v_{{\mathrm{0}}}} \,  \prec_{ \sigma_{{\mathrm{0}}}  \ottsym{(}  \ottnt{v}  \ottsym{)} }  \, \ottnt{v'_{{\mathrm{0}}}} \,  \uparrow \, (  \ottmv{i}  ,  \pi  )   \ottsym{)} \,  \in  \,  \LRRel{ \mathcal{V} }{  { \sigma }_{ \ottmv{i}  \ottsym{+}  \ottmv{j} }'   \ottsym{(}    \ottnt{C} _{ \ottnt{E} }   . \ottnt{T}   \ottsym{)} }   ~.
   \]
   \reflem{lr-spos-ectx-XXX} with
   $ \emptyset   \ottsym{;}   \tceseq{ \ottsym{(}  \mathit{X}  \ottsym{,}  \mathit{Y}  \ottsym{)} }  \,  \vdash ^{\circ}  \, \ottnt{C_{{\mathrm{0}}}}$ and
   $\ottsym{(}  \ottnt{E}  \ottsym{,}  \sigma'  \ottsym{)}  \ottsym{:}  \ottnt{C_{{\mathrm{0}}}}  \Rrightarrow   \ottnt{C} _{ \ottnt{E} } $
   implies
   $  \{   \tceseq{ \mathit{X} }   \ottsym{,}   \tceseq{ \mathit{Y} }   \}   \, \ottsym{\#} \,   \mathit{fpv}  (    \ottnt{C} _{ \ottnt{E} }   . \ottnt{T}   )  $.
   Therefore,
   \[
      \forall  \, { \pi  \ottsym{,}  \ottmv{i} } . \  \ottsym{(}  \ottnt{v_{{\mathrm{0}}}} \,  \prec_{ \sigma_{{\mathrm{0}}}  \ottsym{(}  \ottnt{v}  \ottsym{)} }  \, \ottnt{v'_{{\mathrm{0}}}} \,  \uparrow \, (  \ottmv{i}  ,  \pi  )   \ottsym{)} \,  \in  \,  \LRRel{ \mathcal{V} }{  \sigma_{{\mathrm{0}}}  \ottsym{(}    \ottnt{C} _{ \ottnt{E} }   . \ottnt{T}   \ottsym{)}    [ \tceseq{ \ottnt{v'}  /  \mathit{z} } ]   }   ~.
   \]
   Then, because $\ottnt{v_{{\mathrm{0}}}} \,  \in  \,  \LRRel{ \mathrm{Atom} }{  \sigma_{{\mathrm{0}}}  \ottsym{(}    \ottnt{C} _{ \ottnt{E} }   . \ottnt{T}   \ottsym{)}    [ \tceseq{ \ottnt{v'}  /  \mathit{z} } ]   } $ by Lemmas~\ref{lem:ts-type-sound-terminating}, \ref{lem:lr-red-val}, \ref{lem:ts-refl-subtyping-pure-eff}, \ref{lem:ts-wf-ty-tctx-typing}, \ref{lem:ts-wf-TP_val}, and \T{Sub},
   \reflem{lr-approx-lr} implies
   \[
    \ottnt{v_{{\mathrm{0}}}} \,  \in  \,  \LRRel{ \mathcal{V} }{  \sigma_{{\mathrm{0}}}  \ottsym{(}    \ottnt{C} _{ \ottnt{E} }   . \ottnt{T}   \ottsym{)}    [ \tceseq{ \ottnt{v'}  /  \mathit{z} } ]   }  ~,
   \]
   which we must show in this case (note that $  \{   \tceseq{ \mathit{X} }   \ottsym{,}   \tceseq{ \mathit{Y} }   \}   \, \ottsym{\#} \,   \mathit{fpv}  (    \ottnt{C} _{ \ottnt{E} }   . \ottnt{T}   )  $).

  \case $  \exists  \, { \ottnt{K_{{\mathrm{0}}}}  \ottsym{,}  \mathit{x_{{\mathrm{0}}}}  \ottsym{,}  \ottnt{e_{{\mathrm{0}}}}  \ottsym{,}  \varpi_{{\mathrm{0}}} } . \   \ottnt{E}  [  \sigma_{{\mathrm{0}}}  \ottsym{(}  \ottnt{v}  \ottsym{)} \,  \tceseq{ \ottnt{v'} }   ]  \,  \Downarrow  \,  \ottnt{K_{{\mathrm{0}}}}  [  \mathcal{S}_0 \, \mathit{x_{{\mathrm{0}}}}  \ottsym{.}  \ottnt{e_{{\mathrm{0}}}}  ]  \,  \,\&\,  \, \varpi_{{\mathrm{0}}} $:
   By Lemmas~\ref{lem:ts-type-sound-terminating}, \ref{lem:lr-red-shift}, and \ref{lem:lr-decompose-ectx-embed-exp},
   $  \ottnt{C} _{ \ottnt{E} }   . \ottnt{S}  \,  =  \,  ( \forall  \mathit{x}  .  \ottnt{C'_{{\mathrm{1}}}}  )  \Rightarrow   \ottnt{C'_{{\mathrm{2}}}} $ for some $\mathit{x}$, $\ottnt{C'_{{\mathrm{1}}}}$, and $\ottnt{C'_{{\mathrm{2}}}}$.
   Let $\ottnt{K} \,  \in  \,  \LRRel{ \mathcal{K} }{ \sigma'  \ottsym{(}     \ottnt{C} _{ \ottnt{E} }   . \ottnt{T}   \, / \, ( \forall  \mathit{x}  .  \ottnt{C'_{{\mathrm{1}}}}  )   \ottsym{)} } $.
   We must show that
   \[
     \resetcnstr{  \ottnt{K}  [   \ottnt{E}  [  \sigma_{{\mathrm{0}}}  \ottsym{(}  \ottnt{v}  \ottsym{)} \,  \tceseq{ \ottnt{v'} }   ]   ]  }  \,  \in  \,  \LRRel{ \mathcal{E} }{ \sigma'  \ottsym{(}  \ottnt{C'_{{\mathrm{2}}}}  \ottsym{)} }  ~.
   \]
   By the IH, it suffices to show that
   \[
    \ottsym{(}   \resetcnstr{  \ottnt{K}  [  \ottnt{E}  ]  }   \ottsym{,}  \sigma'  \ottsym{)}  \ottsym{:}  \ottnt{C_{{\mathrm{0}}}}  \Rrightarrow  \ottnt{C'_{{\mathrm{2}}}} ~,
   \]
   which is derived by the following derivation.
   \begin{prooftree}
    \AxiomC{$\ottsym{(}  \ottnt{E}  \ottsym{,}  \sigma'  \ottsym{)}  \ottsym{:}  \ottnt{C_{{\mathrm{0}}}}  \Rrightarrow     \ottnt{C} _{ \ottnt{E} }   . \ottnt{T}   \,  \,\&\,  \,    \ottnt{C} _{ \ottnt{E} }   . \Phi   \, / \,   ( \forall  \mathit{x}  .  \ottnt{C'_{{\mathrm{1}}}}  )  \Rightarrow   \ottnt{C'_{{\mathrm{2}}}}  $}
    \AxiomC{$\ottnt{K} \,  \in  \,  \LRRel{ \mathcal{K} }{ \sigma'  \ottsym{(}     \ottnt{C} _{ \ottnt{E} }   . \ottnt{T}   \, / \, ( \forall  \mathit{x}  .  \ottnt{C'_{{\mathrm{1}}}}  )   \ottsym{)} } $}
    \RightLabel{\TC{Cons}}
    \BinaryInfC{$\ottsym{(}   \resetcnstr{  \ottnt{K}  [  \ottnt{E}  ]  }   \ottsym{,}  \sigma'  \ottsym{)}  \ottsym{:}  \ottnt{C_{{\mathrm{0}}}}  \Rrightarrow  \ottnt{C'_{{\mathrm{2}}}}$}
   \end{prooftree}

  \case $  \exists  \, { \pi_{{\mathrm{0}}} } . \   \ottnt{E}  [  \sigma_{{\mathrm{0}}}  \ottsym{(}  \ottnt{v}  \ottsym{)} \,  \tceseq{ \ottnt{v'} }   ]  \,  \Uparrow  \, \pi_{{\mathrm{0}}} $:
   Take $\sigma_{\ottmv{i}}$ and $\sigma'_{\ottmv{i}}$ for every $\ottmv{i}$ such that
   \begin{itemize}
    \item $\ottnt{C}  \triangleright   \emptyset   \ottsym{;}   \tceseq{ \ottsym{(}  \mathit{X}  \ottsym{,}  \mathit{Y}  \ottsym{)} }   \ottsym{;}   \tceseq{ \mathit{z_{{\mathrm{0}}}}  \,  {:}  \,  \iota_{{\mathrm{0}}} }  \,  |  \, \ottnt{C} \,  \vdash _{ \ottmv{i} }  \, \sigma_{\ottmv{i}}$ and
    \item $\sigma'_{\ottmv{i}} \,  =  \, \sigma_{{\mathrm{0}}}  \mathbin{\uplus}   [ \tceseq{ \sigma_{{\mathrm{0}}}  \ottsym{(}  \sigma  \ottsym{(}  \mathit{X}  \ottsym{)}  \ottsym{)}  /  \mathit{X} } ]   \mathbin{\uplus}   [ \tceseq{ \sigma  \ottsym{(}  \sigma_{\ottmv{i}}  \ottsym{(}  \mathit{Y}  \ottsym{)}  \ottsym{)}  /  \mathit{Y} } ]   \mathbin{\uplus}   [ \tceseq{ \ottnt{v'}  /  \mathit{z} } ] $.
   \end{itemize}
   Using \reflem{lr-nu-sim-XXX}, we have
   \[
      \forall  \, { \ottmv{i}  \ottsym{,}  \pi } . \   \sigma_{{\mathrm{0}}}  \ottsym{(}  \ottnt{v}  \ottsym{)} _{  \nu   \ottsym{;}  \ottmv{i}  \ottsym{;}  \pi }  \,  \tceseq{ \ottnt{v'} }  \,  \in  \,  \LRRel{ \mathcal{E} }{ \sigma'_{\ottmv{i}}  \ottsym{(}  \ottnt{C_{{\mathrm{0}}}}  \ottsym{)} }   ~.
   \]
   Because $ \tceseq{ \ottsym{(}  \mathit{X}  \ottsym{,}  \mathit{Y}  \ottsym{)} }   \ottsym{;}   \tceseq{ \mathit{z_{{\mathrm{0}}}}  \,  {:}  \,  \iota_{{\mathrm{0}}} }   \vdash  \ottnt{C_{{\mathrm{0}}}}  \stackrel{\succ}{\circ}  \ottnt{C}$ and $\ottsym{(}  \ottnt{E}  \ottsym{,}  \sigma'  \ottsym{)}  \ottsym{:}  \ottnt{C_{{\mathrm{0}}}}  \Rrightarrow   \ottnt{C} _{ \ottnt{E} } $,
   we have $\ottsym{(}  \ottnt{E}  \ottsym{,}  \sigma'_{\ottmv{i}}  \ottsym{)}  \ottsym{:}  \ottnt{C_{{\mathrm{0}}}}  \Rrightarrow   \ottnt{C} _{ \ottnt{E} } $
   by Lemmas~\ref{lem:lr-rel-consistency-occur} and \ref{lem:lr-spos-ectx-XXX}.
   Therefore, we have
   \begin{equation}
      \forall  \, { \ottmv{i}  \ottsym{,}  \pi } . \   \ottnt{E}  [   \sigma_{{\mathrm{0}}}  \ottsym{(}  \ottnt{v}  \ottsym{)} _{  \nu   \ottsym{;}  \ottmv{i}  \ottsym{;}  \pi }  \,  \tceseq{ \ottnt{v'} }   ]  \,  \in  \,  \LRRel{ \mathcal{E} }{ \sigma'_{\ottmv{i}}  \ottsym{(}   \ottnt{C} _{ \ottnt{E} }   \ottsym{)} }  
     \label{lem:lr-applied-rec:DIV:approx-rel}
   \end{equation}
   by \reflem{lr-closed-fill-ectx}.
   Furthermore, because $ \ottnt{E}  [  \sigma_{{\mathrm{0}}}  \ottsym{(}  \ottnt{v}  \ottsym{)} \,  \tceseq{ \ottnt{v'} }   ]  \,  \Uparrow  \, \pi_{{\mathrm{0}}}$, we can find
   \[
      \forall  \, { \ottmv{i} } . \    \exists  \, { \pi } . \   \ottnt{E}  [   \sigma_{{\mathrm{0}}}  \ottsym{(}  \ottnt{v}  \ottsym{)} _{  \nu   \ottsym{;}  \ottmv{i}  \ottsym{;}  \pi }  \,  \tceseq{ \ottnt{v'} }   ]  \,  \Uparrow  \, \pi_{{\mathrm{0}}}   ~,
   \]
   which depend on lemmas similar to
   Lemmas~\ref{lem:lr-approx-refl}, \ref{lem:lr-approx-embed-exp}, and \ref{lem:lr-approx-div}.
   By (\ref{lem:lr-applied-rec:DIV:approx-rel}),
   \[
      \forall  \, { \ottmv{i} } . \   { \models  \,   { \ottsym{(}  \sigma'_{\ottmv{i}}  \ottsym{(}   \ottnt{C} _{ \ottnt{E} }   \ottsym{)}  \ottsym{)} }^\nu (  \pi_{{\mathrm{0}}}  )  }   ~.
   \]
   By \reflem{lr-spos-ectx-XXX},
   $ \tceseq{ \ottsym{(}  \mathit{X_{{\mathrm{1}}}}  \ottsym{,}  \mathit{Y_{{\mathrm{1}}}}  \ottsym{)} }   \ottsym{;}   \tceseq{ \ottsym{(}  \mathit{X_{{\mathrm{2}}}}  \ottsym{,}  \mathit{Y_{{\mathrm{2}}}}  \ottsym{)} }  \,  \vdash ^{\circ}  \,  \ottnt{C} _{ \ottnt{E} } $ for some $ \tceseq{ \mathit{X_{{\mathrm{1}}}} } $, $ \tceseq{ \mathit{X_{{\mathrm{2}}}} } $, $ \tceseq{ \mathit{Y_{{\mathrm{1}}}} } $, and $ \tceseq{ \mathit{Y_{{\mathrm{2}}}} } $ such that
   $ \tceseq{ \ottsym{(}  \mathit{X}  \ottsym{,}  \mathit{Y}  \ottsym{)} }  \,  =  \,  \tceseq{ \ottsym{(}  \mathit{X_{{\mathrm{1}}}}  \ottsym{,}  \mathit{X_{{\mathrm{2}}}}  \ottsym{)} }   \ottsym{,}   \tceseq{ \ottsym{(}  \mathit{X_{{\mathrm{2}}}}  \ottsym{,}  \mathit{Y_{{\mathrm{2}}}}  \ottsym{)} } $.
   By \reflem{lr-subst-occur}, $ \tceseq{ \ottsym{(}  \mathit{X_{{\mathrm{1}}}}  \ottsym{,}  \mathit{Y_{{\mathrm{1}}}}  \ottsym{)} }   \ottsym{;}   \tceseq{ \ottsym{(}  \mathit{X_{{\mathrm{2}}}}  \ottsym{,}  \mathit{Y_{{\mathrm{2}}}}  \ottsym{)} }  \,  \vdash ^{\circ}  \, \sigma_{{\mathrm{0}}}  \ottsym{(}   \ottnt{C} _{ \ottnt{E} }   \ottsym{)}$.
   Furthermore, because $\sigma_{{\mathrm{0}}}  \ottsym{(}  \sigma  \ottsym{)}$ is closed,
   we can easily show that $ \tceseq{ \ottsym{(}  \mathit{X_{{\mathrm{1}}}}  \ottsym{,}  \mathit{Y_{{\mathrm{1}}}}  \ottsym{)} }   \ottsym{;}   \tceseq{ \ottsym{(}  \mathit{X_{{\mathrm{2}}}}  \ottsym{,}  \mathit{Y_{{\mathrm{2}}}}  \ottsym{)} }  \,  \vdash ^{\circ}  \,  \sigma_{{\mathrm{0}}}  \ottsym{(}   \ottnt{C} _{ \ottnt{E} }   \ottsym{)}    [ \tceseq{ \sigma_{{\mathrm{0}}}  \ottsym{(}  \sigma  \ottsym{(}  \mathit{X}  \ottsym{)}  \ottsym{)}  /  \mathit{X} } ]  $.
   Then, Lemmas~\ref{lem:ts-val-inv-refine-ty}, \ref{lem:ts-unuse-non-refine-vars}, \ref{lem:lr-subst-eff-subst}, \ref{lem:lr-subst-eff-approx-subst}, and \ref{lem:lr-approx-nu-eff-XXX}
   imply the conclusion
   $ { \models  \,   { \ottsym{(}  \sigma'  \ottsym{(}   \ottnt{C} _{ \ottnt{E} }   \ottsym{)}  \ottsym{)} }^\nu (  \pi_{{\mathrm{0}}}  )  } $.
 \end{caseanalysis}
\end{proof}

\begin{lemmap}{Compatibility: Recursive Functions}{lr-compat-rec}
 Assume that
 \begin{itemize}
  \item $\Gamma' \,  =  \,  \tceseq{ \mathit{X} }  \,  {:}  \,  (   \Sigma ^\ast   ,   \tceseq{ \iota_{{\mathrm{0}}} }   )   \ottsym{,}   \tceseq{ \mathit{Y} }  \,  {:}  \,  (   \Sigma ^\omega   ,   \tceseq{ \iota_{{\mathrm{0}}} }   )   \ottsym{,}  \mathit{f} \,  {:}  \, \ottsym{(}   \tceseq{ \mathit{z}  \,  {:}  \,  \ottnt{T_{{\mathrm{1}}}} }   \ottsym{)}  \rightarrow  \ottnt{C_{{\mathrm{0}}}}  \ottsym{,}   \tceseq{ \mathit{z}  \,  {:}  \,  \ottnt{T_{{\mathrm{1}}}} } $,
  \item $\ottnt{e} \,  \in  \,  \LRRel{ \mathcal{O} }{ \Gamma  \ottsym{,}  \Gamma'  \, \vdash \,  \ottnt{C} } $,
  \item $ \tceseq{ \mathit{z_{{\mathrm{0}}}}  \,  {:}  \,  \iota_{{\mathrm{0}}} }  \,  =  \,  \gamma  (   \tceseq{ \mathit{z}  \,  {:}  \,  \ottnt{T_{{\mathrm{1}}}} }   ) $,
  \item $ \tceseq{ \ottsym{(}  \mathit{X}  \ottsym{,}  \mathit{Y}  \ottsym{)} }   \ottsym{;}   \tceseq{ \mathit{z_{{\mathrm{0}}}}  \,  {:}  \,  \iota_{{\mathrm{0}}} }   \vdash  \ottnt{C_{{\mathrm{0}}}}  \stackrel{\succ}{\circ}  \ottnt{C}$,
  \item $  \{   \tceseq{ \mathit{X} }   \ottsym{,}   \tceseq{ \mathit{Y} }   \}   \, \ottsym{\#} \,   \mathit{fpv}  (   \tceseq{ \ottnt{T_{{\mathrm{1}}}} }   )  $, and
  \item $ \tceseq{ \ottsym{(}  \mathit{X}  \ottsym{,}  \mathit{Y}  \ottsym{)} }   \ottsym{;}   \tceseq{ \mathit{z_{{\mathrm{0}}}}  \,  {:}  \,  \iota_{{\mathrm{0}}} }  \,  |  \, \ottnt{C}  \vdash  \sigma$.
 \end{itemize}
 Then,
 \[
   \mathsf{rec}  \, (  \tceseq{  \mathit{X} ^{ \sigma  \ottsym{(}  \mathit{X}  \ottsym{)} },  \mathit{Y} ^{ \sigma  \ottsym{(}  \mathit{Y}  \ottsym{)} }  }  ,  \mathit{f} ,   \tceseq{ \mathit{z} }  ) . \,  \ottnt{e}  \,  \in  \,  \LRRel{ \mathcal{O} }{ \Gamma  \, \vdash \,  \ottsym{(}   \tceseq{ \mathit{z}  \,  {:}  \,  \ottnt{T_{{\mathrm{1}}}} }   \ottsym{)}  \rightarrow  \sigma  \ottsym{(}  \ottnt{C_{{\mathrm{0}}}}  \ottsym{)} }  ~.
 \]
\end{lemmap}
\begin{proof}
 \TS{CHECKED: 2022/07/07}
 Let $\ottnt{v} \,  =  \,  \mathsf{rec}  \, (  \tceseq{  \mathit{X} ^{ \sigma  \ottsym{(}  \mathit{X}  \ottsym{)} },  \mathit{Y} ^{ \sigma  \ottsym{(}  \mathit{Y}  \ottsym{)} }  }  ,  \mathit{f} ,   \tceseq{ \mathit{z} }  ) . \,  \ottnt{e} $.
 By \T{Fun}, we have
 \[
  \Gamma  \vdash  \ottnt{v}  \ottsym{:}  \ottsym{(}   \tceseq{ \mathit{z}  \,  {:}  \,  \ottnt{T_{{\mathrm{1}}}} }   \ottsym{)}  \rightarrow  \sigma  \ottsym{(}  \ottnt{C_{{\mathrm{0}}}}  \ottsym{)} ~.
 \]
 Let $\sigma_{{\mathrm{0}}} \,  \in  \,  \LRRel{ \mathcal{G} }{ \Gamma } $.
 Then, we must show that
 \[
  \sigma_{{\mathrm{0}}}  \ottsym{(}  \ottnt{v}  \ottsym{)} \,  \in  \,  \LRRel{ \mathcal{E} }{  \sigma_{{\mathrm{0}}}  \ottsym{(}  \ottsym{(}   \tceseq{ \mathit{z}  \,  {:}  \,  \ottnt{T_{{\mathrm{1}}}} }   \ottsym{)}  \rightarrow  \sigma  \ottsym{(}  \ottnt{C_{{\mathrm{0}}}}  \ottsym{)}  \ottsym{)}  \,  \,\&\,  \,   \Phi_\mathsf{val}   \, / \,   \square   }  ~.
 \]
 By \reflem{lr-val}, it suffices to show that
 \[
  \sigma_{{\mathrm{0}}}  \ottsym{(}  \ottnt{v}  \ottsym{)} \,  \in  \,  \LRRel{ \mathcal{V} }{ \sigma_{{\mathrm{0}}}  \ottsym{(}  \ottsym{(}   \tceseq{ \mathit{z}  \,  {:}  \,  \ottnt{T_{{\mathrm{1}}}} }   \ottsym{)}  \rightarrow  \sigma  \ottsym{(}  \ottnt{C_{{\mathrm{0}}}}  \ottsym{)}  \ottsym{)} }  ~.
 \]
 Let $ \tceseq{ \ottnt{v'} } $ be values such that $ \tceseq{ \ottnt{v'} \,  \in  \,  \LRRel{ \mathcal{V} }{  \sigma_{{\mathrm{0}}}  \ottsym{(}  \ottnt{T_{{\mathrm{1}}}}  \ottsym{)}    [ \tceseq{ \ottnt{v'}  /  \mathit{z} } ]   }  } $.
 By definition, it suffices to show that
 \[
  \sigma_{{\mathrm{0}}}  \ottsym{(}  \ottnt{v}  \ottsym{)} \,  \tceseq{ \ottnt{v'} }  \,  \in  \,  \LRRel{ \mathcal{E} }{  \sigma_{{\mathrm{0}}}  \ottsym{(}  \sigma  \ottsym{(}  \ottnt{C_{{\mathrm{0}}}}  \ottsym{)}  \ottsym{)}    [ \tceseq{ \ottnt{v'}  /  \mathit{z} } ]   }  ~.
 \]
 By \reflem{lr-applied-rec-XXX}, it suffices to show that
 \[
  \ottsym{(}  \,[\,]\,  \ottsym{,}  \sigma_{{\mathrm{0}}}  \mathbin{\uplus}  \sigma_{{\mathrm{0}}}  \ottsym{(}  \sigma  \ottsym{)}  \mathbin{\uplus}   [ \tceseq{ \ottnt{v'}  /  \mathit{z} } ]   \ottsym{)}  \ottsym{:}  \ottnt{C_{{\mathrm{0}}}}  \Rrightarrow  \ottnt{C_{{\mathrm{0}}}} ~,
 \]
 which is derived by \TC{Empty}.
\end{proof}

\begin{lemmap}{Compatibility: Divergence}{lr-compat-div}
 If $\Gamma  \vdash  \ottnt{C}$ and $\Gamma  \models   { \ottnt{C} }^\nu (  \pi  ) $, then $ \bullet ^{ \pi }  \,  \in  \,  \LRRel{ \mathcal{O} }{ \Gamma  \, \vdash \,  \ottnt{C} } $.
\end{lemmap}
\begin{proof}
 \TS{OK: 2022/01/29}
 By \T{Div}, $\Gamma  \vdash   \bullet ^{ \pi }   \ottsym{:}  \ottnt{C}$.
 Let $\sigma \,  \in  \,  \LRRel{ \mathcal{G} }{ \Gamma } $.
 Then, we must show that
 \[
   \bullet ^{ \pi }  \,  \in  \,  \LRRel{ \mathcal{E} }{ \sigma  \ottsym{(}  \ottnt{C}  \ottsym{)} }  ~.
 \]
 By definition, it suffices to show that
 \[
   { \models  \,   { \sigma  \ottsym{(}  \ottnt{C}  \ottsym{)} }^\nu (  \pi  )  }  ~.
 \]
 Because $\Gamma  \models   { \ottnt{C} }^\nu (  \pi  ) $ and $\sigma \,  \in  \,  \LRRel{ \mathcal{G} }{ \Gamma } $,
 \reflem{lr-typeability} implies that $ { \models  \,  \sigma  \ottsym{(}   { \ottnt{C} }^\nu (  \pi  )   \ottsym{)} } $.
 Because $\sigma  \ottsym{(}   { \ottnt{C} }^\nu (  \pi  )   \ottsym{)} \,  =  \,  { \sigma  \ottsym{(}  \ottnt{C}  \ottsym{)} }^\nu (  \pi  ) $, we have the conclusion.
\end{proof}

\begin{lemmap}{Compatibility: Primitive Operations}{lr-compat-op}
 If $\vdash  \Gamma$ and $ \tceseq{ \ottnt{v} \,  \in  \,  \LRRel{ \mathcal{O} }{ \Gamma  \, \vdash \,  \ottnt{T} }  } $ and $\mathit{ty} \, \ottsym{(}  \ottnt{o}  \ottsym{)} \,  =  \,  \tceseq{(  \mathit{x}  \,  {:}  \,  \ottnt{T}  )}  \rightarrow   \ottnt{T_{{\mathrm{0}}}} $,
 then $\ottnt{o}  \ottsym{(}   \tceseq{ \ottnt{v} }   \ottsym{)} \,  \in  \,  \LRRel{ \mathcal{O} }{ \Gamma  \, \vdash \,   \ottnt{T_{{\mathrm{0}}}}    [ \tceseq{ \ottnt{v}  /  \mathit{x} } ]   } $.
\end{lemmap}
\begin{proof}
 By \T{Op}, $\Gamma  \vdash  \ottnt{o}  \ottsym{(}   \tceseq{ \ottnt{v} }   \ottsym{)}  \ottsym{:}   \ottnt{T_{{\mathrm{0}}}}    [ \tceseq{ \ottnt{v}  /  \mathit{x} } ]  $.
 Let $\sigma \,  \in  \,  \LRRel{ \mathcal{G} }{ \Gamma } $.
 We must show that
 \[
  \ottnt{o}  \ottsym{(}   \tceseq{ \sigma  \ottsym{(}  \ottnt{v}  \ottsym{)} }   \ottsym{)} \,  \in  \,  \LRRel{ \mathcal{E} }{  \sigma  \ottsym{(}   \ottnt{T_{{\mathrm{0}}}}    [ \tceseq{ \ottnt{v}  /  \mathit{x} } ]    \ottsym{)}  \,  \,\&\,  \,   \Phi_\mathsf{val}   \, / \,   \square   }  ~.
 \]
 By \refasm{const}, $ \tceseq{ \ottnt{T} \,  =  \, \ottsym{\{}  \mathit{x} \,  {:}  \, \iota \,  |  \, \phi  \ottsym{\}} } $ and $\ottnt{T_{{\mathrm{0}}}} \,  =  \, \ottsym{\{}  \mathit{x_{{\mathrm{0}}}} \,  {:}  \, \iota_{{\mathrm{0}}} \,  |  \, \phi_{{\mathrm{0}}}  \ottsym{\}}$
 for some $ \tceseq{ \ottsym{\{}  \mathit{x} \,  {:}  \, \iota \,  |  \, \phi  \ottsym{\}} } $, $\mathit{x_{{\mathrm{0}}}}$, $\iota_{{\mathrm{0}}}$, and $\phi_{{\mathrm{0}}}$, and
 $\mathit{ty} \, \ottsym{(}  \ottnt{o}  \ottsym{)}$ is closed.
 By \reflem{lr-typeability}, $ \emptyset   \vdash  \ottnt{o}  \ottsym{(}   \tceseq{ \sigma  \ottsym{(}  \ottnt{v}  \ottsym{)} }   \ottsym{)}  \ottsym{:}  \sigma  \ottsym{(}   \ottsym{\{}  \mathit{x_{{\mathrm{0}}}} \,  {:}  \, \iota_{{\mathrm{0}}} \,  |  \, \phi_{{\mathrm{0}}}  \ottsym{\}}    [ \tceseq{ \ottnt{v}  /  \mathit{x} } ]    \ottsym{)}$.
 By \reflem{ts-progress}, $\ottnt{o}  \ottsym{(}   \tceseq{ \sigma  \ottsym{(}  \ottnt{v}  \ottsym{)} }   \ottsym{)}$ can be evaluated.
 Because only \E{Red}/\R{Const} is applicable,
 $ \tceseq{ \sigma  \ottsym{(}  \ottnt{v}  \ottsym{)} }  \,  =  \,  \tceseq{ \ottnt{c} } $ for some $ \tceseq{ \ottnt{c} } $, and
 $\ottnt{o}  \ottsym{(}   \tceseq{ \sigma  \ottsym{(}  \ottnt{v}  \ottsym{)} }   \ottsym{)} = \ottnt{o}  \ottsym{(}   \tceseq{ \ottnt{c} }   \ottsym{)} \,  \mathrel{  \longrightarrow  }  \,  \zeta  (  \ottnt{o}  ,   \tceseq{ \ottnt{c} }   )  \,  \,\&\,  \,  \epsilon $.
 Therefore, by \reflem{lr-anti-red}, it suffices to show that
 \[
   \zeta  (  \ottnt{o}  ,   \tceseq{ \ottnt{c} }   )  \,  \in  \,  \LRRel{ \mathcal{E} }{  \sigma  \ottsym{(}   \ottsym{\{}  \mathit{x_{{\mathrm{0}}}} \,  {:}  \, \iota_{{\mathrm{0}}} \,  |  \, \phi_{{\mathrm{0}}}  \ottsym{\}}    [ \tceseq{ \ottnt{v}  /  \mathit{x} } ]    \ottsym{)}  \,  \,\&\,  \,   \Phi_\mathsf{val}   \, / \,   \square   }  ~.
 \]
 By Lemmas~\ref{lem:lr-red-val} and \ref{lem:lr-red-inconsist-eval-div},
 it suffices to show that
 $ \zeta  (  \ottnt{o}  ,   \tceseq{ \ottnt{c} }   )  \,  \in  \,  \LRRel{ \mathcal{V} }{ \sigma  \ottsym{(}   \ottsym{\{}  \mathit{x_{{\mathrm{0}}}} \,  {:}  \, \iota_{{\mathrm{0}}} \,  |  \, \phi_{{\mathrm{0}}}  \ottsym{\}}    [ \tceseq{ \ottnt{v}  /  \mathit{x} } ]    \ottsym{)} } $,
 which is derived by \reflem{ts-subject-red-red}.
\end{proof}

\begin{lemmap}{Compatibility: Function Applications}{lr-compat-app}
 If $\ottnt{v_{{\mathrm{1}}}} \,  \in  \,  \LRRel{ \mathcal{O} }{ \Gamma  \, \vdash \,  \ottsym{(}  \mathit{x} \,  {:}  \, \ottnt{T}  \ottsym{)}  \rightarrow  \ottnt{C} } $ and $\ottnt{v_{{\mathrm{2}}}} \,  \in  \,  \LRRel{ \mathcal{O} }{ \Gamma  \, \vdash \,  \ottnt{T} } $,
 then $\ottnt{v_{{\mathrm{1}}}} \, \ottnt{v_{{\mathrm{2}}}} \,  \in  \,  \LRRel{ \mathcal{O} }{ \Gamma  \, \vdash \,   \ottnt{C}    [  \ottnt{v_{{\mathrm{2}}}}  /  \mathit{x}  ]   } $.
\end{lemmap}
\begin{proof}
 By \T{App}, $\Gamma  \vdash  \ottnt{v_{{\mathrm{1}}}} \, \ottnt{v_{{\mathrm{2}}}}  \ottsym{:}   \ottnt{C}    [  \ottnt{v_{{\mathrm{2}}}}  /  \mathit{x}  ]  $.
 Let $\sigma \,  \in  \,  \LRRel{ \mathcal{G} }{ \Gamma } $.
 We must show that
 \[
  \sigma  \ottsym{(}  \ottnt{v_{{\mathrm{1}}}}  \ottsym{)} \, \sigma  \ottsym{(}  \ottnt{v_{{\mathrm{2}}}}  \ottsym{)} \,  \in  \,  \LRRel{ \mathcal{E} }{ \sigma  \ottsym{(}   \ottnt{C}    [  \ottnt{v_{{\mathrm{2}}}}  /  \mathit{x}  ]    \ottsym{)} }  ~.
 \]
 Without loss of generality, we can suppose that $\mathit{x}$ does not occur in $\sigma$.
 By \reflem{lr-typeability}, $\sigma  \ottsym{(}  \ottnt{v_{{\mathrm{1}}}}  \ottsym{)} \, \sigma  \ottsym{(}  \ottnt{v_{{\mathrm{2}}}}  \ottsym{)} \,  \in  \,  \LRRel{ \mathrm{Atom} }{ \sigma  \ottsym{(}   \ottnt{C}    [  \ottnt{v_{{\mathrm{2}}}}  /  \mathit{x}  ]    \ottsym{)} } $.
 Because $\ottnt{v_{{\mathrm{1}}}} \,  \in  \,  \LRRel{ \mathcal{O} }{ \Gamma  \, \vdash \,  \ottsym{(}  \mathit{x} \,  {:}  \, \ottnt{T}  \ottsym{)}  \rightarrow  \ottnt{C} } $ and $\ottnt{v_{{\mathrm{2}}}} \,  \in  \,  \LRRel{ \mathcal{O} }{ \Gamma  \, \vdash \,  \ottnt{T} } $,
 we have $\sigma  \ottsym{(}  \ottnt{v_{{\mathrm{1}}}}  \ottsym{)} \,  \in  \,  \LRRel{ \mathcal{E} }{  \ottsym{(}  \mathit{x} \,  {:}  \, \sigma  \ottsym{(}  \ottnt{T}  \ottsym{)}  \ottsym{)}  \rightarrow  \sigma  \ottsym{(}  \ottnt{C}  \ottsym{)}  \,  \,\&\,  \,   \Phi_\mathsf{val}   \, / \,   \square   } $ and $\sigma  \ottsym{(}  \ottnt{v_{{\mathrm{2}}}}  \ottsym{)} \,  \in  \,  \LRRel{ \mathcal{E} }{  \sigma  \ottsym{(}  \ottnt{T}  \ottsym{)}  \,  \,\&\,  \,   \Phi_\mathsf{val}   \, / \,   \square   } $,
 so $\sigma  \ottsym{(}  \ottnt{v_{{\mathrm{1}}}}  \ottsym{)} \,  \in  \,  \LRRel{ \mathcal{V} }{ \ottsym{(}  \mathit{x} \,  {:}  \, \sigma  \ottsym{(}  \ottnt{T}  \ottsym{)}  \ottsym{)}  \rightarrow  \sigma  \ottsym{(}  \ottnt{C}  \ottsym{)} } $ and $\sigma  \ottsym{(}  \ottnt{v_{{\mathrm{2}}}}  \ottsym{)} \,  \in  \,  \LRRel{ \mathcal{V} }{ \sigma  \ottsym{(}  \ottnt{T}  \ottsym{)} } $ with \reflem{lr-red-val}.
 Therefore, we have the conclusion $\sigma  \ottsym{(}  \ottnt{v_{{\mathrm{1}}}}  \ottsym{)} \, \sigma  \ottsym{(}  \ottnt{v_{{\mathrm{2}}}}  \ottsym{)} \,  \in  \,  \LRRel{ \mathcal{E} }{  \sigma  \ottsym{(}  \ottnt{C}  \ottsym{)}    [  \sigma  \ottsym{(}  \ottnt{v_{{\mathrm{2}}}}  \ottsym{)}  /  \mathit{x}  ]   }  =  \LRRel{ \mathcal{E} }{ \sigma  \ottsym{(}   \ottnt{C}    [  \ottnt{v_{{\mathrm{2}}}}  /  \mathit{x}  ]    \ottsym{)} } $.
\end{proof}

\begin{lemmap}{Compatibility: Predicate Applications}{lr-compat-papp}
 If $\ottnt{v} \,  \in  \,  \LRRel{ \mathcal{O} }{ \Gamma  \, \vdash \,   \forall   \ottsym{(}  \mathit{X} \,  {:}  \,  \tceseq{ \ottnt{s} }   \ottsym{)}  \ottsym{.}  \ottnt{C} } $ and $\Gamma  \vdash  \ottnt{A}  \ottsym{:}   \tceseq{ \ottnt{s} } $,
 then $\ottnt{v} \, \ottnt{A} \,  \in  \,  \LRRel{ \mathcal{O} }{ \Gamma  \, \vdash \,   \ottnt{C}    [  \ottnt{A}  /  \mathit{X}  ]   } $.
\end{lemmap}
\begin{proof}
 By \T{PApp}, $\Gamma  \vdash  \ottnt{v} \, \ottnt{A}  \ottsym{:}   \ottnt{C}    [  \ottnt{A}  /  \mathit{X}  ]  $.
 Let $\sigma \,  \in  \,  \LRRel{ \mathcal{G} }{ \Gamma } $.
 We must show that
 \[
  \sigma  \ottsym{(}  \ottnt{v}  \ottsym{)} \, \sigma  \ottsym{(}  \ottnt{A}  \ottsym{)} \,  \in  \,  \LRRel{ \mathcal{E} }{ \sigma  \ottsym{(}   \ottnt{C}    [  \ottnt{A}  /  \mathit{X}  ]    \ottsym{)} }  ~.
 \]
 Without loss of generality, we can suppose that $\mathit{X}$ does not occur in $\sigma$.
 By \reflem{lr-typeability}, $\sigma  \ottsym{(}  \ottnt{v}  \ottsym{)} \, \sigma  \ottsym{(}  \ottnt{A}  \ottsym{)} \,  \in  \,  \LRRel{ \mathrm{Atom} }{ \sigma  \ottsym{(}   \ottnt{C}    [  \ottnt{A}  /  \mathit{X}  ]    \ottsym{)} } $.
 Because $\ottnt{v} \,  \in  \,  \LRRel{ \mathcal{O} }{ \Gamma  \, \vdash \,   \forall   \ottsym{(}  \mathit{X} \,  {:}  \,  \tceseq{ \ottnt{s} }   \ottsym{)}  \ottsym{.}  \ottnt{C} } $,
 we have $\sigma  \ottsym{(}  \ottnt{v}  \ottsym{)} \,  \in  \,  \LRRel{ \mathcal{E} }{   \forall   \ottsym{(}  \mathit{X} \,  {:}  \,  \tceseq{ \ottnt{s} }   \ottsym{)}  \ottsym{.}  \sigma  \ottsym{(}  \ottnt{C}  \ottsym{)}  \,  \,\&\,  \,   \Phi_\mathsf{val}   \, / \,   \square   } $,
 so $\sigma  \ottsym{(}  \ottnt{v}  \ottsym{)} \,  \in  \,  \LRRel{ \mathcal{V} }{  \forall   \ottsym{(}  \mathit{X} \,  {:}  \,  \tceseq{ \ottnt{s} }   \ottsym{)}  \ottsym{.}  \sigma  \ottsym{(}  \ottnt{C}  \ottsym{)} } $ with \reflem{lr-red-val}.
 Therefore, because $ \emptyset   \vdash  \sigma  \ottsym{(}  \ottnt{A}  \ottsym{)}  \ottsym{:}   \tceseq{ \ottnt{s} } $ by \reflem{lr-typeability},
 we have the conclusion $\sigma  \ottsym{(}  \ottnt{v}  \ottsym{)} \, \sigma  \ottsym{(}  \ottnt{A}  \ottsym{)} \,  \in  \,  \LRRel{ \mathcal{E} }{  \sigma  \ottsym{(}  \ottnt{C}  \ottsym{)}    [  \sigma  \ottsym{(}  \ottnt{A}  \ottsym{)}  /  \mathit{X}  ]   }  =  \LRRel{ \mathcal{E} }{ \sigma  \ottsym{(}   \ottnt{C}    [  \ottnt{A}  /  \mathit{X}  ]    \ottsym{)} } $.
\end{proof}

\begin{lemmap}{Compatibility: If Expressions}{lr-compat-if}
 If $\ottnt{v} \,  \in  \,  \LRRel{ \mathcal{O} }{ \Gamma  \, \vdash \,  \ottsym{\{}  \mathit{x} \,  {:}  \,  \mathsf{bool}  \,  |  \, \phi  \ottsym{\}} } $ and $\ottnt{e_{{\mathrm{1}}}} \,  \in  \,  \LRRel{ \mathcal{O} }{ \Gamma  \ottsym{,}    \ottnt{v}     =     \mathsf{true}    \, \vdash \,  \ottnt{C} } $ and $\ottnt{e_{{\mathrm{2}}}} \,  \in  \,  \LRRel{ \mathcal{O} }{ \Gamma  \ottsym{,}    \ottnt{v}     =     \mathsf{false}    \, \vdash \,  \ottnt{C} } $,
 then $\mathsf{if} \, \ottnt{v} \, \mathsf{then} \, \ottnt{e_{{\mathrm{1}}}} \, \mathsf{else} \, \ottnt{e_{{\mathrm{2}}}} \,  \in  \,  \LRRel{ \mathcal{O} }{ \Gamma  \, \vdash \,  \ottnt{C} } $.
\end{lemmap}
\begin{proof}
 By \T{If}, $\Gamma  \vdash  \mathsf{if} \, \ottnt{v} \, \mathsf{then} \, \ottnt{e_{{\mathrm{1}}}} \, \mathsf{else} \, \ottnt{e_{{\mathrm{2}}}}  \ottsym{:}  \ottnt{C}$.
 Let $\sigma \,  \in  \,  \LRRel{ \mathcal{G} }{ \Gamma } $.
 We must show that
 \[
  \mathsf{if} \, \sigma  \ottsym{(}  \ottnt{v}  \ottsym{)} \, \mathsf{then} \, \sigma  \ottsym{(}  \ottnt{e_{{\mathrm{1}}}}  \ottsym{)} \, \mathsf{else} \, \sigma  \ottsym{(}  \ottnt{e_{{\mathrm{2}}}}  \ottsym{)} \,  \in  \,  \LRRel{ \mathcal{E} }{ \sigma  \ottsym{(}  \ottnt{C}  \ottsym{)} }  ~.
 \]
 By \reflem{lr-typeability}, $\mathsf{if} \, \sigma  \ottsym{(}  \ottnt{v}  \ottsym{)} \, \mathsf{then} \, \sigma  \ottsym{(}  \ottnt{e_{{\mathrm{1}}}}  \ottsym{)} \, \mathsf{else} \, \sigma  \ottsym{(}  \ottnt{e_{{\mathrm{2}}}}  \ottsym{)} \,  \in  \,  \LRRel{ \mathrm{Atom} }{ \sigma  \ottsym{(}  \ottnt{C}  \ottsym{)} } $.
 Because $ \emptyset   \vdash  \sigma  \ottsym{(}  \ottnt{v}  \ottsym{)}  \ottsym{:}  \sigma  \ottsym{(}  \ottsym{\{}  \mathit{x} \,  {:}  \,  \mathsf{bool}  \,  |  \, \phi  \ottsym{\}}  \ottsym{)}$ by \reflem{lr-typeability},
 \reflem{ts-val-inv-refine-ty} implies $\sigma  \ottsym{(}  \ottnt{v}  \ottsym{)} \,  =  \, \ottnt{c}$ for some $\ottnt{c}$ such that $ \mathit{ty}  (  \ottnt{c}  )  \,  =  \,  \mathsf{bool} $.
 By case analysis on $\ottnt{c}$.
 \begin{caseanalysis}
  \case $ \ottnt{c}    =     \mathsf{true}  $:
   By \E{Red}/\R{IfT}, $\mathsf{if} \, \sigma  \ottsym{(}  \ottnt{v}  \ottsym{)} \, \mathsf{then} \, \sigma  \ottsym{(}  \ottnt{e_{{\mathrm{1}}}}  \ottsym{)} \, \mathsf{else} \, \sigma  \ottsym{(}  \ottnt{e_{{\mathrm{2}}}}  \ottsym{)} \,  \mathrel{  \longrightarrow  }  \, \sigma  \ottsym{(}  \ottnt{e_{{\mathrm{1}}}}  \ottsym{)} \,  \,\&\,  \,  \epsilon $.
   By \reflem{lr-anti-red}, it suffices to show that $\sigma  \ottsym{(}  \ottnt{e_{{\mathrm{1}}}}  \ottsym{)} \,  \in  \,  \LRRel{ \mathcal{E} }{ \sigma  \ottsym{(}  \ottnt{C}  \ottsym{)} } $.
   Because $\sigma \,  \in  \,  \LRRel{ \mathcal{G} }{ \Gamma } $ and $\sigma  \ottsym{(}  \ottnt{v}  \ottsym{)} \,  =  \,  \mathsf{true} $,
   we have $\sigma  \mathbin{\uplus}   [   ()   /  \mathit{y}  ]  \,  \in  \,  \LRRel{ \mathcal{G} }{ \Gamma  \ottsym{,}  \mathit{y} \,  {:}  \, \ottsym{\{}  \mathit{y} \,  {:}  \,  \mathsf{unit}  \,  |  \,   \ottnt{v}     =     \mathsf{true}    \ottsym{\}} } $ for fresh variable $\mathit{y}$.
   Then, $\ottnt{e_{{\mathrm{1}}}} \,  \in  \,  \LRRel{ \mathcal{O} }{ \Gamma  \ottsym{,}    \ottnt{v}     =     \mathsf{true}    \, \vdash \,  \ottnt{C} } $ implies the conclusion.

  \case $ \ottnt{c}    =     \mathsf{false}  $:
   By \E{Red}/\R{IfF}, $\mathsf{if} \, \sigma  \ottsym{(}  \ottnt{v}  \ottsym{)} \, \mathsf{then} \, \sigma  \ottsym{(}  \ottnt{e_{{\mathrm{1}}}}  \ottsym{)} \, \mathsf{else} \, \sigma  \ottsym{(}  \ottnt{e_{{\mathrm{2}}}}  \ottsym{)} \,  \mathrel{  \longrightarrow  }  \, \sigma  \ottsym{(}  \ottnt{e_{{\mathrm{2}}}}  \ottsym{)} \,  \,\&\,  \,  \epsilon $.
   By \reflem{lr-anti-red}, it suffices to show that $\sigma  \ottsym{(}  \ottnt{e_{{\mathrm{2}}}}  \ottsym{)} \,  \in  \,  \LRRel{ \mathcal{E} }{ \sigma  \ottsym{(}  \ottnt{C}  \ottsym{)} } $.
   Because $\sigma \,  \in  \,  \LRRel{ \mathcal{G} }{ \Gamma } $ and $\sigma  \ottsym{(}  \ottnt{v}  \ottsym{)} \,  =  \,  \mathsf{false} $,
   we have $\sigma  \mathbin{\uplus}   [   ()   /  \mathit{y}  ]  \,  \in  \,  \LRRel{ \mathcal{G} }{ \Gamma  \ottsym{,}  \mathit{y} \,  {:}  \, \ottsym{\{}  \mathit{y} \,  {:}  \,  \mathsf{unit}  \,  |  \,   \ottnt{v}     =     \mathsf{false}    \ottsym{\}} } $ for fresh variable $\mathit{y}$.
   Then, $\ottnt{e_{{\mathrm{2}}}} \,  \in  \,  \LRRel{ \mathcal{O} }{ \Gamma  \ottsym{,}    \ottnt{v}     =     \mathsf{false}    \, \vdash \,  \ottnt{C} } $ implies the conclusion.
 \end{caseanalysis}
\end{proof}

\begin{lemmap}{Compatibility: Event Expressions}{lr-compat-ev}
 If $\vdash  \Gamma$,
 then $ \mathsf{ev}  [  \mathbf{a}  ]  \,  \in  \,  \LRRel{ \mathcal{O} }{ \Gamma  \, \vdash \,   \ottsym{\{}  \mathit{x} \,  {:}  \,  \mathsf{unit}  \,  |  \,  \top   \ottsym{\}}  \,  \,\&\,  \,   (  \mathbf{a}  , \bot )   \, / \,   \square   } $.
\end{lemmap}
\begin{proof}
 By \T{Event}, $\Gamma  \vdash   \mathsf{ev}  [  \mathbf{a}  ]   \ottsym{:}   \ottsym{\{}  \mathit{x} \,  {:}  \,  \mathsf{unit}  \,  |  \,  \top   \ottsym{\}}  \,  \,\&\,  \,   (  \mathbf{a}  , \bot )   \, / \,   \square  $.
 It suffices to show that
 \[
   \mathsf{ev}  [  \mathbf{a}  ]  \,  \in  \,  \LRRel{ \mathcal{E} }{  \ottsym{\{}  \mathit{x} \,  {:}  \,  \mathsf{unit}  \,  |  \,  \top   \ottsym{\}}  \,  \,\&\,  \,   (  \mathbf{a}  , \bot )   \, / \,   \square   }  ~.
 \]
 We consider each case on the result of evaluating $ \mathsf{ev}  [  \mathbf{a}  ] $.
 \begin{caseanalysis}
  \case $  \exists  \, { \ottnt{v}  \ottsym{,}  \varpi } . \   \mathsf{ev}  [  \mathbf{a}  ]  \,  \Downarrow  \, \ottnt{v} \,  \,\&\,  \, \varpi $:
   Because $ \mathsf{ev}  [  \mathbf{a}  ] $ can be evaluated only by \E{Ev},
   \reflem{lr-red-val} implies that $\ottnt{v} \,  =  \,  () $ and $ \varpi    =    \mathbf{a} $.
   Then, it suffices to show that $ ()  \,  \in  \,  \LRRel{ \mathcal{V} }{ \ottsym{\{}  \mathit{x} \,  {:}  \,  \mathsf{unit}  \,  |  \,  \top   \ottsym{\}} } $,
   which is implied by \T{Const} and \T{Sub}.

  \case $  \exists  \, { \ottnt{K_{{\mathrm{0}}}}  \ottsym{,}  \mathit{x_{{\mathrm{0}}}}  \ottsym{,}  \ottnt{e_{{\mathrm{0}}}}  \ottsym{,}  \varpi } . \   \mathsf{ev}  [  \mathbf{a}  ]  \,  \Downarrow  \,  \ottnt{K_{{\mathrm{0}}}}  [  \mathcal{S}_0 \, \mathit{x_{{\mathrm{0}}}}  \ottsym{.}  \ottnt{e_{{\mathrm{0}}}}  ]  \,  \,\&\,  \, \varpi $:
   Contradictory by \E{Ev} and Lemmas~\ref{lem:lr-eval-det} and \ref{lem:lr-red-val}.

  \case $  \exists  \, { \pi } . \   \mathsf{ev}  [  \mathbf{a}  ]  \,  \Uparrow  \, \pi $:
   Contradictory by \E{Ev} and Lemmas~\ref{lem:lr-red-decomp-div} and \ref{lem:lr-red-val}.
 \end{caseanalysis}
\end{proof}

\begin{lemmap}{Fundamental Property}{lr-fundamental-prop}
 If $\Gamma  \vdash  \ottnt{e}  \ottsym{:}  \ottnt{C}$, then $\ottnt{e} \,  \in  \,  \LRRel{ \mathcal{O} }{ \Gamma  \, \vdash \,  \ottnt{C} } $.
\end{lemmap}
\begin{proof}
 By induction on the typing derivation with
 Lemmas~\ref{lem:lr-compat-cvar}, \ref{lem:lr-compat-var}, \ref{lem:lr-compat-const}, \ref{lem:lr-compat-pabs}, \ref{lem:lr-compat-abs}, \ref{lem:lr-compat-subtyping}, \ref{lem:lr-compat-shift}, \ref{lem:lr-compat-shift-discard-cont}, \ref{lem:lr-compat-reset}, \ref{lem:lr-compat-let}, \ref{lem:lr-compat-let-val}, \ref{lem:lr-compat-rec}, \ref{lem:lr-compat-div}, \ref{lem:lr-compat-op}, \ref{lem:lr-compat-app}, \ref{lem:lr-compat-papp}, \ref{lem:lr-compat-if}, and \ref{lem:lr-compat-ev}.
\end{proof}

\begin{lemmap}{Soundness for Infinite Traces}{lr-sound}
 If $ \emptyset   \vdash  \ottnt{e}  \ottsym{:}  \ottnt{C}$ and $\ottnt{e} \,  \Uparrow  \, \pi$, then $ { \models  \,   { \ottnt{C} }^\nu (  \pi  )  } $.
\end{lemmap}
\begin{proof}
 By \reflem{lr-fundamental-prop}, $\ottnt{e} \,  \in  \,  \LRRel{ \mathcal{E} }{ \ottnt{C} } $.
 We have the conclusion by the definition of $ \LRRel{ \mathcal{E} }{ \ottnt{C} } $.
\end{proof}